\documentclass[12pt]{report}

\usepackage{rotate}
\usepackage[francais]{babel}
\usepackage{psfig}
\usepackage[dvips]{graphics}
\usepackage{fancyheadings}

\def \sefs {sections efficaces }
\def \sef {section efficace }
\def \sm {Standard Model }

\def \sugra {supergravity }
\def \fc {flavor changing }
\def \fcnc {flavor changing neutral current }
\def \ms {Mod\`ele Standard }
\def \susi {supersym\'etrie }
\def \susy {supersymmetry }
\def \susiq {supersym\'etrique }
\def \susiqs {supersym\'etriques }
\def \susyq {supersymmetric }
\def \brs {branching ratios }
\def \br {branching ratio }
\def \L {\Lambda }
\def \l {\lambda }
\def \k {\kappa }
 
\def \t {\theta }

\def \a {\alpha }

\def \d {\delta }
\def \D {\Delta }

\def \g {\gamma }
\def \G {\Gamma }

\def \b {\beta }
\def \S {\Sigma }
\def \s {\sigma }
\def \ep {\epsilon }
\def \e {\epsilon }
 
\def \ud {{1 \over 2} }

\def \bea {\begin{equation} }
\def \eea {\end{equation} }
\def\be{\begin{equation}}
\def\ee{\end{equation}}
\def\ba{\begin{array}}
\def\ea{\end{array}} 

\def \cc {coupling constant }
\def \ccs {coupling constants }
\def \tchi  {{\tilde \chi} }
\def \Eslash {E \kern-.6em\slash }
\def \pslash {p \kern-.5em\slash }
\def \kslash {k \kern-.5em\slash }
\def \qslash {Q \kern-.5em\slash }
\def \ppslash {p' \kern-.5em\slash }
\def \Pslash {P \kern-.5em\slash }
\def \FFL {{ {\l '}_{Jjk}^\star \l ' _{J'jk}\over (4\pi )^2 } }
\def \FFQ {{  {\l '}_{iJk}^\star \l ' _{iJ'k}\over (4\pi )^2 } }
\def \FFP {{ {\l '}_{ijJ'}^\star \l ' _{ijJ}\over (4\pi )^2 } }

\def \FFIL {{  {\l }_{iJk}^\star \l _{iJ'k}\over (4\pi )^2 } }
\def \FF2L {{ {\l }_{ijJ} \l _{ijJ'}^\star \over (4\pi )^2 } }
\def \FLU {{ {\l ' }_{iJ'k} \l  _{iJk}^{'\star } \over (4\pi )^2 } }

\def \pr  {Phys. Rev. }
\def \np {Nucl. Phys. }

\def \prl {Phys. Rev. Lett. }
\def \pl {Phys. Lett. }

\def\chione{\tilde \chi_1^0}

\def\chionepm{\tilde \chi_1^\pm}

\def\chionep{\tilde \chi_1^+}

\def\m0{$m_{0}$}

\def\tchi{{\tilde\chi}}

\def\snm{{\tilde\nu_\mu}}

\def\GeV{{\rm GeV}}
\def\TeV{{\rm TeV}}

\def \RPV {R-parity violating }

\def\beqa{\begin{eqnarray}}
\def\eqar{\end{array}}
\def\beqar{\begin{array}}
\def\eqa{\end{eqnarray}}
\def\bars{\begin{eqnarray*}}
\def\ears{\end{eqnarray*}}

\def \capa {{\cal A }} 
\def \capb {{\cal B }} 
\def \capc {{\cal C }} 
\def \capd {{\cal D }}
\def \cala {{\cal A }}
\def \call {{\cal L }}
\def \cddd {{\cal D }}
\def \calf {{\cal F }}
\def \calr {{\cal R }}

\def \ya  { X_1 }
\def \yb  { X_2 }
\def \yc  { X_3 }
\def \yd  { X_4 }
\def \ye  { X_5 }
\def \yf  { X_6 }
\def \yg  { X_7 }
\def \yh  { X_8 }

\def \etal {et al.}

\def\bentarrow{\:\raisebox{1.3ex}{\rlap{$\vert$}}\!\rightarrow}

\def\dksl#1#2#3#4{
	\begin{equation}
	\begin{array}{r c l }
	#1 & \rightarrow & #2  \\
	 & & \bentarrow #3 \\
         & & \phantom{\bentarrow \;}  \bentarrow #4 
	\end{array}
        \label{eqq}
	\end{equation}
		}

\catcode`@=11 
\def \gsim{\mathrel{\mathpalette\@versim>}}
\def \lsim{\mathrel{\mathpalette\@versim<}}
\def \@versim#1#2{\lower0.4ex\vbox{\baselineskip\z@skip\lineskip\z@skip
     \lineskiplimit\z@\ialign{$\m@th#1\hfil##\hfil$%
     \crcr#2\crcr\sim\crcr}}}
\catcode`@=12 
\newcommand{\LV}{\mbox{$\not \hspace{-0.15cm} L \hspace{0.1cm}$}}
\newcommand{\BV}{\mbox{$\not \hspace{-0.15cm} B \hspace{0.1cm}$}}
\newcommand{\Rp}{\mbox{$\not \hspace{-0.12cm} R_p$ }}
\newcommand{\rpv}{\mbox{$\not \hspace{-0.12cm} R_p$ }}
\newcommand{\rpvi}{\mbox{$\not \hspace{-0.12cm} R_p$}}
\newcommand{\rpi}{\mbox{$\not \hspace{-0.12cm} R_p$}}
\newcommand{\Dslash}{\mbox{$\not \hspace{-0.10cm} D$ }}

\newcommand{\bt}{\begin{tabular}}
\newcommand{\et}{\end{tabular}}
\newcommand{\bd}{\begin{displaymath}}
\newcommand{\ed}{\end{displaymath}\noindent}
\newcommand{\ec}{\end{center}}
\newcommand{\bc}{\begin{center}}

\def\GeV{\hbox{$\;\hbox{\rm GeV}$}}
\def\MeV{\hbox{$\;\hbox{\rm MeV}$}}
\def\TeV{\hbox{$\;\hbox{\rm TeV}$}}

\newcommand{\nanob}{\mbox{{\rm ~nb}~}}
\newcommand{\picob}{\mbox{{\rm ~pb}~}}
\newcommand{\femtob}{\mbox{{\rm ~fb}~}}

\def\ARNS#1#2#3  {{\rm Ann.~Rev.~Nucl.~Sci.}  {\bf#1  \,}  (#2) \, #3}
\def\CPC#1#2#3   {{\rm Comp.~Phys.~Comm.}     {\bf#1  \,}  (#2) \, #3}
\def\EPJC#1#2#3  {{\rm Eur.~Phys.~J}          {\bf #1 \,}  (#2) \, #3}
\def\MPLA#1#2#3  {{\rm Mod.~Phys.~Lett.}      {\bf{A#1} \,}(#2) \, #3}
\def\NIMA#1#2#3  {{\rm Nucl.~Instr.~and~Meth.}{\bf#1  \,}  (#2) \, #3}
\def\NIMB#1#2#3  {{\rm Nucl. Instr. Meth.}    {\bf#1  \,}  (#2) \, #3}
\def\NPB#1#2#3   {{\rm Nucl.~Phys.}           {\bf{B#1} \,}(#2) \ #3}
\def\NPA#1#2#3   {{\rm Nucl.~Phys.}           {\bf{A#1} \,}(#2) \, #3}
\def\NPAPC#1#2#3 {{\rm Nucl.~Phys.}           {\bf{#1A} \,}(#2) \, #3}
\def\NPBPC#1#2#3 {{\rm Nucl.~Phys.}           {\bf{#1B} \,}(#2) \, #3}
\def\PLB#1#2#3   {{\rm Phys.~Lett.}           {\bf{B#1} \,}(#2) \ #3}
\def\PR#1#2#3    {{\rm Phys.~Rep.}            {\bf#1  \,}  (#2) \, #3}
\def\PRC#1#2#3   {{\rm Phys.~Rev.}            {\bf{C#1} \,}(#2) \, #3}
\def\PRD#1#2#3   {{\rm Phys.~Rev.}            {\bf{D#1} \,}(#2) \ #3}
\def\PRL#1#2#3   {{\rm Phys.~Rev.~Lett.}      {\bf#1  \,}  (#2) \ #3}
\def\PTP#1#2#3   {{\rm Prog.~Theor.~Phys.}    {\bf#1  \,}   (#2) \, #3}
\def\RMP#1#2#3   {{\rm Rev.~Mod.~Phys.}       {\bf#1  \,}   (#2) \, #3}
\def\JETPL#1#2#3 {{\rm Sov. Phys. JETP Lett.} {\bf{C#1} \,} (#2) \ #3} 
\def\ZHETF#1#2#3 {{\rm ZhETF Pis. Red.}       {\bf{C#1} \,} (#2) \, #3}
\def\ZPC#1#2#3   {{\rm Z.~Phys.}              {\bf{C#1} \,} (#2) \, #3}
\def\CERNPPE#1   {{\em Preprint CERN,} CERN-PPE/#1}
\def\CERNEP#1    {{\em Preprint CERN,} CERN-EP/#1}

\def \SM {Standard Model }
\def \SUSY {Supersymmetry }

\def \SUGRA {Supergravity }

\def\ch0{{\tilde{\chi}}^0_1}

\def \ud {{1 \over 2} }

\def\met{\mbox{${\hbox{$E$\kern-0.6em\lower-.1ex\hbox{/}}}_T \hspace{.3cm}$}}

\if@twoside\oddsidemargin25mm
\else\oddsidemargin25mm\evensidemargin25mm\marginparwidth25mm\fi
\begin{document}



\begin{titlepage}
\begin{center}
{  }
\end{center}
\vspace{1 mm}
\begin{center}
{\bf \it UNIVERSIT\'E PARIS VII-DENIS DIDEROT \\}
{\bf \it UFR DE PHYSIQUE TH\'EORIQUE}
\end{center}
\vspace{10 mm}
2000  \ \ \ \ \ \ \ \ \ \ \ \ \ \ \ \ \ \ \ \ \ \ \ \ \
\ \ \ \ \ \ \ \ \ \ \ \ \ \ \ \ \ \ \ \ \ \ \ \ \ \ \ \ \
\ \ \ \ \ \ \ \ \ \ \ \ \ \ \ \ \ \ \ \ \ \ \ \ \
\ \ \ \ \ \ \ \ \ \ \ \ \ \ \ \ \ \ \ \ \ \ \ \ \
\ \ \ \ \ \ \
T00/176
\vspace{5 mm}
\begin{center}
\underline{{\bf TH\`ESE}}
\end{center}
\vspace{5 mm}
\begin{center}
pour l'obtention du Dipl\^ome de
\end{center}
\vspace{5 mm}
\begin{center}
{\bf  DOCTEUR DE L'UNIVERSIT\'E PARIS VII}
\end{center}
\vspace{5 mm}
\begin{center}
SP\'ECIALIT\'E: PHYSIQUE TH\'EORIQUE DES PARTICULES
\end{center}
\vspace{5 mm}
\begin{center} 
pr\'esent\'ee et soutenue publiquement 
\end{center}
\begin{center} par \end{center}
\begin{center} Gr\'egory MOREAU \end{center}
\begin{center} le 27 Novembre 2000 \end{center}
\vspace{8 mm}
\begin{center} \underline{{\bf \it Titre:}} \end{center}
\begin{center}
{\bf \Large 
\'Etude ph\'enom\'enologique des interactions violant la
sym\'etrie de R-parit\'e dans les th\'eories supersym\'etriques}
\end{center} 
\vspace{7 mm}
\begin{center}  ------  \end{center}
\begin{center} {\bf \it Directeur de th\`ese:} \end{center}
\begin{center} Marc CHEMTOB \end{center}
\begin{center}  ------  \end{center}
\vspace{3 mm}
\begin{center} JURY \end{center}
\vspace{7 mm}
\begin{center}
\begin{tabular}{lll} 
MM. & Pierre BIN\'ETRUY,  & Rapporteur,            \\     
    & Marc CHEMTOB,       & Directeur de th\`ese,   \\    
    & Herbert DREINER,      &                        \\     
    & Jean-Pierre GAZEAU, &                      \\          
    & Stavros KATSANEVAS, & Rapporteur, \\
    & Carlos SAVOY,       &  \\ 
    & Yves SIROIS,        &  Pr\'esident.  \\ 
\end{tabular}
\end{center}
\end{titlepage}

\clearpage 

\begin{titlepage}

\vspace{10 mm}
\begin{center}
{  }
\end{center}
\vspace{30 mm}

\end{titlepage}

\clearpage

\clearpage

\begin{titlepage}
\begin{flushright}
\begin{figure}[b] 
\centerline{\psfig{figure=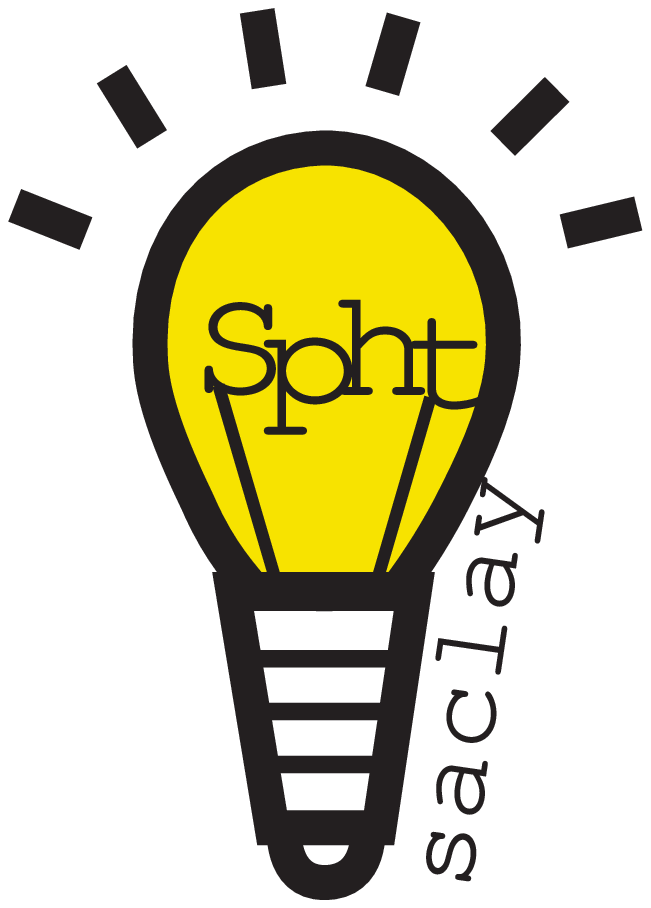,height=1.5in}}
\end{figure}
\end{flushright}
\end{titlepage}

\clearpage

\clearpage 

\begin{titlepage}

\vspace{10 mm}
\begin{center}
{  }
\end{center}
\vspace{30 mm}

\end{titlepage}

\clearpage

\begin{titlepage}

\begin{center}
{  }
\end{center}
\vspace{40 mm}

\begin{flushright} {\bf \huge \it Remerciements} \end{flushright}

\vspace{10 mm}

{\huge \it A}vant tout, 
je tiens \`a remercier P.~Chiappetta, pour ses conseils 
avis\'es dans le choix de ma th\`ese, et mon directeur, M.~Chemtob, 
d'avoir accept\'e de m'encadrer \`a l'occasion de ce doctorat.
M.~Chemtob a toujours su se montrer disponible tout en
respectant ma libert\'e. Ma reconnaissance va aussi 
aux membres du jury de th\`ese, \`a savoir, P.~Bin\'etruy, 
H.~Dreiner, J.-P.~Gazeau, S.~Katsanevas, C.~Savoy et Y.~Sirois. 

\vspace{10 mm}

Mes travaux ont souvent \'et\'e le fruit d'un effort
collectif. Aussi, je tiens \`a mentionner mes collaborateurs 
exp\'erimentateurs et amis: M.~Besan\c{c}on, F.~D\'eliot, 
G.~Polesello, C.~Royon et enfin E.~Perez qui a r\'epondu \`a
mes nombreuses questions avec une constante attention. Je 
souhaite par ailleurs citer 
H.-U.~Martyn et Y.~Sirois qui m'ont fortement encourag\'e
dans mon travail concernant la physique aupr\`es des 
Collisionneurs Lin\'eaires. 

\vspace{10 mm}

Par ailleurs, je tiens \`a exprimer ici
toute ma reconnaissance et ma sympathie 
\`a R.~Balian, F.~Bernard, M.~Boonekamp,  
F.~Boudjema, A.~Djouadi, E.~Dudas, 
S.~Gamblin, R.~Grimm, J.-L.~Kneur, C.~Lachaud, J.~Lamarcq, F.~Ledroit, 
R.~Lopez, P.~Lutz, B.~Machet, P.~Micout, G.~Moultaka, 
S.~Munier, J.-M.~Normand, R.~Peschanski, 
V.~Poireau, G.~Sajot, C.~Savoy, L.~Schoeffel, G.~Servant et X.~Tata, 
qui m'ont accompagn\'e et aid\'e durant ces
trois ann\'ees de th\`ese.

\vspace{10 mm}

Ces trois ann\'ees de th\`ese ont aussi 
\'et\'e pour moi l'occasion d'inoubliables
rencontres. Je pense aux longues, et parfois m\^eme tardives, 
discussions avec N.~Ghodbane, qui ne concernaient pas
seulement SUSYGEN. Je pense d'autre part au magicien \`a
trois corps Y.~Mambrini aupr\`es duquel j'ai toujours \'eprouv\'e
un grand plaisir \`a passer pour ``Monsieur tout le monde''
lors de ses impressionnants tours de magie. Je n'oublie pas
la convivialit\'e et la gentillesse de mes coll\`egues de bureau, 
T.~Duguet et N.~Regnault, qui ont permis des conditions de 
travail plus qu'agr\'eables durant ces trois ann\'ees pass\'ees
\`a l'Orme des Merisiers. N.~Regnault a \'et\'e un assistant scientifique et
informatique de tout premier choix. Quant \`a T.~Duguet, il a toujours
eu \`a l'esprit un sujet original permettant 
d'initier un d\'ebat passionnant et de provoquer ainsi une pause bien 
m\'erit\'ee.

\end{titlepage}

\begin{titlepage}

\begin{center}
{  }
\end{center}
\vspace{4 mm}

D'autre part, je dois en grande partie cette th\`ese \`a ma famille: 
R.~Savoie m'a accompagn\'e, support\'e et soutenu, de par
sa douce pr\'esence, pendant ce petit bout de vie. Mes parents, 
I.~Jaglin, J.-M.~Moreau, F.~Jaglin et C.~Moreau, ont su
\^etre \`a l'\'ecoute de mes aspirations et respecter mes choix, 
rendant ainsi cette th\`ese envisageable.
Mes fr\`eres, Anatole M., Arthur J., Cl\'ement J., Martin J. et 
Th\'eophile M., et ma soeur, Valentine M., ainsi que S.~Jendoubi 
et M.~Villeger m'ont tous communiqu\'e leur joie et leur bonheur.
Enfin, J.~L\'egasse, P.~Bailhache, A.~Aubanel et L.~Savoie ont
\'et\'e d'un grand soutien.

\vspace{10 mm}

Sans oublier mes ami(e)s, M.~Boudet, M.~Descottes, A.~Devaux, 
E.~Duque, F.~Grandjacques, A.~Jeanmet, V.~Knysak, T.~Laval, 
A.~Forgiel, A.~Forgiel, T.~Malbrancq, L.~Meiers, C.~Morel, A.~Mouillaud, 
C.~Vannini, E. de Tayrac, M.~Tourdjman et M.~Werner, 
qui m'ont apport\'e un certain recul par rapport \`a mon travail 
en m'apprenant le sens de la d\'erision, si n\'ecessaire lors 
d'une th\`ese et \`a de nombreux autres moments de la vie.

\end{titlepage}

\clearpage 

\begin{titlepage}

\vspace{10 mm}
\begin{center}
{  }
\end{center}
\vspace{30 mm}

\end{titlepage}

\clearpage

\clearpage 

\begin{titlepage}

\vspace{10 mm}
\begin{center}
{  }
\end{center}
\vspace{30 mm}

\end{titlepage}

\clearpage

\tableofcontents

\clearpage

\vspace{10 mm}
\begin{center}
{  }
\end{center}
\vspace{10 mm}

\clearpage 

\begin{center}
{  }
\end{center}
\vspace{50 mm}

\begin{flushright} {\large \it ``Il n'y a pas de chemin vers le bonheur, \\
le bonheur est le chemin.''} \end{flushright}

\clearpage

\vspace{10 mm}
\begin{center}
{  }
\end{center}
\vspace{10 mm}

\clearpage

\chapter{Supersym\'etrie}
\label{chaSUSY}

\section{Origine des th\'eories supersym\'etriques}

Deux grandes classes de sym\'etries sont \`a distinguer en Physique: Les
sym\'etries d'espace-temps
(r\'eunies dans le groupe de Poincar\'e) et les sym\'etries internes qui se
classent en deux
sous-cat\'egories, \`a savoir les sym\'etries globales (saveur, chirale) 
et les sym\'etries de jauge (couleur,
\'electro\-faible). Les
premi\`eres tentatives d'unification de ces deux genres de sym\'etries
furent non relativistes.
Ces mod\`eles concernaient les quarks et rassemblaient le groupe SU(2) de
spin et SU(3) de
saveur dans un groupe SU(6). En 1967, Coleman et Mandula furent \`a
l'origine d'un th\'eor\`eme \cite{Cole} 
qui est le plus pr\'ecis et le plus puissant d'une s\'erie de ``no-go
theorems''  traitant des
sym\'etries possibles de la matrice S, dans le cadre d'une th\'eorie
quantique des champs:
D'apr\`es ce th\'eor\`eme, si G est un groupe de sym\'etrie connexe de 
la matrice S qui contient le groupe de Poincar\'e, qui met un nombre fini 
de particules dans un supermultiplet et qui a des g\'en\'erateurs pouvant
\^etre repr\'esent\'es comme des op\'erateurs d'int\'egration dans l'espace 
des moments (avec des noyaux qui sont des distributions), si la matrice 
S n'est pas triviale et si les amplitudes de diffusion \'elastique 
sont des fonctions analytiques de $s$ et $t$ dans un voisinage de la 
r\'egion physique, alors G est localement isomorphique au produit
direct du groupe de Poincar\'e et d'un groupe de sym\'etrie interne. 
En 1971, Gol'fand et Likhtman montr\`erent que l'alg\`ebre d'une extension
du groupe de Poincar\'e
devait \^etre gradu\'ee afin de ne pas violer la connection entre le spin et
la statistique \cite{Gol}.
Une alg\`ebre gradu\'ee comprend des g\'en\'erateurs $B_i$, appartenant \`a
une alg\`ebre de Lie, et des
g\'en\'erateurs $F_i$, ob\'eissant \`a des relations d'anti-commutation
entre eux et \`a des relations de
commutation avec les $B_i$:
\begin{eqnarray}
[B_i, B_j] \sim B_k, \
[B_i, F_j]  \sim F_k, \
\{F_i, F_j\} \sim B_k.
\label{ALgrad}
\end{eqnarray}
Si l'on consid\`ere une extension du groupe de Poincar\'e, les $B_i$
repr\'esentent les g\'en\'erateurs
du groupe de Poincar\'e ($SO(1,3) \times Translations$). La seconde relation
permet \`a l'extension
obtenue de ne pas \^etre trivialement un produit direct entre le groupe de
Poincar\'e et le groupe
associ\'e aux g\'en\'erateurs $F_i$. Ces g\'en\'erateurs $F_i$, ne commutant
pas avec les transformations
de Lorentz, changent le spin des particules. Les g\'en\'erateurs $F_i$
poss\`edent donc un spin contrairement
aux $B_i$, d'o\`u la notation ($B_i$ pour bosonique et $F_i$ pour fermionique). 
De plus, le th\'eor\`eme de Coleman-Mandula
supprime la possibilit\'e de prendre les
$F_i$ de spin entier. Ce nouveau groupe \'echange donc les fermions et les
bosons introduisant ainsi une nouvelle
sym\'etrie. Notons que cette sym\'etrie Bose-Fermi a aussi \'et\'e
introduite en 1971 par Neveu, Schwarz et Ramond,
dans des mod\`eles de cordes pour les fermions. Haag, Sohnius et Lopuszanski
prouv\`erent ensuite que l'alg\`ebre
de  supersym\'etrie (ou superalg\`ebre) \'etait la seule alg\`ebre gradu\'ee
g\'en\'eralisant le groupe de Poincar\'e,
compatible avec une th\'eorie quantique des champs \cite{Haag,WeBa}.
L'introduction de repr\'esentations lin\'eaires
de la supersym\'etrie dans le contexte de th\'eories quantiques des champs
fut donn\'ee la premi\`ere fois par Wess
et Zumino en 1974 \cite{wzum}. Peu apr\`es, Ferrara, Salam, Strathdee, Wess
et Zumino invent\`erent le formalisme des superespaces et superchamps 
\cite{Ferra,Ferrb,Ferrc,Brew,Ferr}.
Depuis le d\'ebut des ann\'ees 1980, des efforts importants ont \'et\'e
investis dans le d\'eveloppement de la
supersym\'etrie tant sur le plan exp\'erimental que th\'eorique.
Aucun partenaire \susiq d'une particule du Mod\`ele Standard,
c'est \`a dire aucune particule supersym\'etrique, n'a cependant \'et\'e
d\'ecouvert \`a ce jour.

\section{Alg\`ebre de supersym\'etrie}
\label{AlSUSY}

Adoptons la notation suivante: $C$ est l'op\'erateur conjugaison de charge,
$\eta^{\mu \rho}$ repr\'esente la m\'etrique
de Minkowski  et  enfin,   $\gamma^{\mu \nu} = {1 \over 2}  [\gamma^{\mu},
\gamma^{\nu}]$, o\`u les $\gamma^{\mu}$
sont les matrices de Dirac. L'alg\`ebre de la supersym\'etrie s'\'ecrit
alors:
\begin{eqnarray}
[M^{\mu \nu}, M^{\rho \sigma}] &  = &  -i ( \eta^{\mu \rho} M^{\nu \sigma} 
+ \eta^{\nu \sigma} M^{\mu \rho}
-  \eta^{\mu \sigma} M^{\nu \rho}  -  \eta^{\nu \rho} M^{\mu \sigma} ), \cr
[M^{\mu \nu}, P^{\sigma}]  &  = &  -i ( \eta^{\mu \rho} P^{\nu} - \eta^{\nu \rho}
P^{\mu} ), \cr
[P^{\mu}, P^{\nu}]   &  = &   0, \cr
[M^{\mu \nu}, Q_{\alpha}^i] &  = &  {i \over 2} ( \gamma^{\mu \nu} Q^i )_{\alpha}, \cr
[P^{\mu}, Q_{\alpha}^i] &   = &  0, \cr
\{ Q_{\alpha}^i, Q_{\beta}^j \} &  = &  -2 ( \gamma^{\mu} C )_{\alpha \beta}
P_{\mu} \delta^{ij} + C_{\alpha \beta} (U^{ij}+\gamma_5 V^{ij}).
\label{ALsusy}
\end{eqnarray}
Les $P^{\mu}$ sont les g\'en\'erateurs des translations: $P^{\mu} = - i
\partial^{\mu}$.
L'indice $\alpha$ des $Q_{\alpha}^i$
est un indice spinoriel variant de 1 \`a 4 car les $Q_{\alpha}^i$,
g\'en\'erateurs de la supersym\'etrie, sont des
spineurs de Majorana. L'indice $i$ des $Q_{\alpha}^i$ distingue les
diff\'erents g\'en\'erateurs. Dans le cas
d'une supersym\'etrie \`a N g\'en\'erateurs, l'indice $i$ varie de 1 \`a N.
Les $M^{\mu \nu}$ sont les g\'en\'erateurs
du groupe de Poincar\'e: $M^{\mu \nu} = i ( x^{\mu} \partial^{\nu} - x^{\nu}
\partial^{\mu} )$ pour une repr\'esentation
de champ scalaire, et $ M^{\mu \nu} = {i \over 2} \gamma^{\mu \nu}$ pour une
repr\'esentation de champ spinoriel. Enfin, $U^{ij}$ et $V^{ij}$ 
sont les charges centrales.

Les trois premi\`eres relations de Eq.\ref{ALsusy} ne sont rien d'autre que
l'alg\`ebre de Lie du groupe de Poincar\'e.
Cherchons \`a expliquer la quatri\`eme relation qui d'apr\`es
Eq.\ref{ALgrad} est du type:
\begin{eqnarray}
[M^{\mu \nu}, Q_{\alpha}^i] =  (b^{\mu \nu})_{\alpha}^{\beta} Q_{\beta}^i.
\label{ALgr2}
\end{eqnarray}
En utilisant Eq.(\ref{ALgr2}) et l'identit\'e de Jacobi: $[[B_1, B_2], F_3] +
[[B_2, F_3], B_1] + [[F_3, B_1], B_2] = 0$,
avec $B_1=M^{\mu \nu}$, $B_2 =M^{\rho \sigma}$  et  $F_3 =Q_{\beta}^i$, on
trouve:
\begin{eqnarray}
[ b^{\mu \nu}, b^{\rho \sigma} ] = -i ( \eta^{\mu \rho} M^{\nu \sigma} +
\eta^{\nu \sigma} M^{\mu \rho}
-  \eta^{\nu \rho} M^{\mu \sigma}  -  \eta^{\mu \sigma} M^{\nu \rho} ).
\label{CoefJacob}
\end{eqnarray}
Ce commutateur est celui des g\'en\'erateurs $M^{\mu \nu}$. Ainsi, les
$b^{\mu \nu}$ forment une
repr\'esentation de l'alg\`ebre de Lorentz pour les spineurs. Leur
expression est connue et vaut:
$b^{\mu \nu} = {i \over 2} \gamma^{\mu \nu}$, d'o\`u la quatri\`eme
relation.
Par un raisonnement similaire, on obtient la cinqui\`eme relation, qui
montre l'ind\'ependance
des transformations de supersym\'etrie vis \`a vis des translations.
Nous reviendrons plus tard sur la derni\`ere relation.
\'Etudions une cons\'equence importante de la cinqui\`eme relation:
Soit un champ scalaire $z$ et son partenaire \susiq un champ
spinoriel $\Psi$. La cinqui\`eme relation entraine:
\begin{eqnarray}
[P^{\mu} P_{\mu},  Q_{\alpha}^i]   =  0   \Leftrightarrow   &      
P^2 Q_{\alpha}^i |z> = Q_{\alpha}^i P^2 |z>   
 &   \Leftrightarrow  \cr P^2 |\Psi> =   Q_{\alpha}^i ( m^2 |z> )      
 \Leftrightarrow    &   m' |\Psi> = m |\Psi>   
 &   \Leftrightarrow    m=m'.
\label{consAL}
\end{eqnarray}
Ainsi, les particules membres d'un m\^eme supermultiplet ont la m\^eme
masse.

La transformation de supersym\'etrie infinit\'esimale d'un champ $\phi$
s'\'ecrit:
\begin{eqnarray}
\delta \Phi = i \bar {\cal E}^{\alpha} Q_{\alpha} \Phi,
\label{trSUSYgen}
\end{eqnarray}
o\`u $\delta$ repr\'esente la transformation infinit\'esimale de
supersym\'etrie et ${\cal E}^{\alpha}$ ($\alpha=1,...,4$), qui
est un spineur anti-commutant (variable de Grassmann) satisfaisant \`a la
condition de Majorana,
est le param\`etre de cette transformation. 
Des variables de Grassmann ${\cal E}^\alpha$ ($\alpha=1,...,4$)
sont d\'efinies par
les propri\'et\'es d'anti-commutation suivantes,
\begin{eqnarray}
\{ {\cal E}_1,{\cal E}_2 \}=0, \
\{ {\cal E},Q \}=0,
\label{GRASS}
\end{eqnarray}
o\`u ${\cal E}_1$ et ${\cal E}_2$ sont les param\`etres 
associ\'es \`a deux transformations infinit\'esimales
de supersym\'etrie.  
Notons que le param\`etre de transformation infinit\'esimale de
supersym\'etrie ${\cal E}^\alpha$ est
ind\'ependant de l'espace-temps dans le cas de la supersym\'etrie globale.
Une transformation finie de supersym\'etrie sur un supermultiplet s'obtient
par exponentiation de Eq.(\ref{trSUSYgen}):
\begin{eqnarray}
\Phi' = exp ( i \bar {\cal E}^{\alpha} Q_{\alpha}  ) \Phi.
\label{trSUSYgep}
\end{eqnarray}

\section{Motivations de la supersym\'etrie}

Quelles sont aujourd'hui les motivations pour la supersym\'etrie ? Tout
d'abord, la supersym\'etrie (SUSY)
apporte un cadre particuli\`erement propice aux th\'eories de grande
unification
dites th\'eories GUT (Grand Unification Theory) \cite{gutfirsta,gutfirstb,ano,GGR} ainsi qu'aux
mod\`eles d'unification des forces
incluant la gravitation vers $10^{19} GeV$: Les th\'eories de cordes
(supercordes si elles incluent SUSY).
Dans le mod\`ele GUT supersym\'etrique bas\'e sur le groupe de jauge SU(5),
si l'\'echelle effective de brisure de la supersym\'etrie est de l'ordre du $TeV$,
les trois constantes de couplage du Mod\`ele Standard s'unifient \`a 
l'\'echelle d'unification $M_{GUT}$, ce qui n'est pas le cas dans le mod\`ele 
non supersym\'etrique GUT bas\'e sur le groupe de jauge SU(5). De plus, dans 
le mod\`ele GUT SU(5) supersym\'etrique, l'\'echelle d'unification est repouss\'ee de 
$M_{GUT} \approx 10^{15} GeV$ (cas du mod\`ele GUT SU(5) non supersym\'etrique) \`a 
$M_{GUT} \approx 2 \ 10^{16} GeV$ ce qui augmente le temps de vie du proton. Notons
cependant que dans les th\'eories SUSY, la stabilit\'e du proton est menac\'ee par des 
op\'erateurs non renormalisables. Par ailleurs, SUSY
offre un candidat naturel pour la masse cach\'ee de l'Univers: La LSP
(Lightest
Supersymmetric Particle), qui
est la particule supersym\'etrique la plus l\'eg\`ere. En effet, la LSP est
stable (except\'e dans les sc\'enarios dans lesquels la sym\'etrie de R-parit\'e 
est viol\'ee, comme nous le discuterons au d\'ebut de la Section \ref{4:mot}) 
et interagit faiblement
avec la mati\`ere. Mais la principale motivation pour les th\'eories SUSY
reste la r\'esolution du probl\`eme
des hi\'erarchies d'\'echelles de masse comme nous le discuterons dans la
prochaine Section.

\subsection{Le probl\`eme des hi\'erarchies d'\'echelles}
\label{hier}

Le probl\`eme de hi\'erarchie est la grande diff\'erence entre
l'\'echelle de brisure \'electrofaible qui est de l'ordre du $TeV$ et
l'\'echelle de la physique sous-jacente au Mod\`ele
Standard: l'\'echelle d'unification $M_{GUT}$, l'\'echelle des cordes $M_{string}$,
l'\'echelle de Planck $M_P$,... 
L'avantage des th\'eories supersym\'etriques vis \`a vis de cette
difficult\'e,
comme nous allons le voir maintenant, est la possibilit\'e
de conserver au niveau des corrections radiatives une hi\'erarchie
existant dans le
potentiel effectif obtenu \`a l'ordre des arbres. Cette possibilit\'e
est li\'ee \`a l'existence du th\'eor\`eme de non-renormalisation (voir
Section \ref{TNR})
qui pr\'edit notamment que les termes de masse des champs scalaires ne sont
pas affect\'es par
les corrections radiatives quadratiques dans les th\'eories SUSY.
Cette invariance provient de l'annulation de toutes les divergences
quadratiques dans les th\'eories SUSY qui est due elle \`a la compensation
entre des graphes
\'echangeant des particules du \ms et les graphes associ\'es
qui impliquent leur partenaire supersym\'etrique.
Assurer la stabilit\'e des hi\'erarchies de masse est certes important mais
ne r\'epond
pas \`a l'autre question de nature `dynamique': Pourquoi existe-t-il des
hi\'erarchies
d'\'echelles d'\'energie dans la nature et quelle est la dynamique
responsable de leur apparition ?
En effet, les th\'eories SUSY n'apportent pas d'explication \'evidente
\`a l'origine de
ces deux \'echelles
d'\'energie si diff\'erentes qui apparaissent dans le potentiel \`a l'ordre
des arbres.
Dans certaines th\'eories de supercordes, la valeur pr\'edite de
l'\'echelle de brisure de la sym\'etrie GUT est proche
de l'\'echelle des cordes, qui est elle-m\^eme 
reli\'ee \`a l'\'echelle de Planck. 
Mais dans ce type de th\'eories, l'origine de
l'\'echelle de brisure \'electrofaible reste probl\'ematique. 
Une solution envisageable \`a ce probl\`eme
est que l'\'echelle de brisure \'electrofaible soit d\'etermin\'ee par les
corrections radiatives.
Plus pr\'ecis\'ement, dans des mod\`eles comportant des termes de brisure de la 
supersym\'etrie, la masse au carr\'e du
boson de Higgs du \ms est positive \`a l'\'echelle GUT mais l'\'evolution des
constantes de
couplages par le groupe de renormalisation rend la masse au carr\'e du Higgs
n\'egative \`a
l'\'echelle du $TeV$, provoquant ainsi la brisure \'electrofaible \`a
l'\'echelle d'\'energie souhait\'ee \cite{CASav1,CASav2}.
Le probl\`eme de hi\'erarchie peut \^etre expliqu\'e de deux mani\`eres
diff\'erentes que nous allons d\'ecrire maintenant.

La premi\`ere explication du probl\`eme 
de hi\'erarchie s'appuie sur les valeurs
moyennes dans le vide (VEV) du boson de Higgs et d'un boson
associ\'e \`a la brisure du groupe de jauge d'une th\'eorie de grande
unification.
Dans les th\'eories GUT, une premi\`ere \'echelle d'\'energie d\'etermine
l'\'echelle de brisure $M_{GUT}$ de la sym\'etrie GUT. Cette \'echelle est
donn\'ee par la VEV d'un champ $\Phi$ et vaut typiquement,
\begin{eqnarray}
< 0 \vert \Phi \vert 0> = V = O(10^{15}GeV).
\label{GUTVEV}
\end{eqnarray}
La seconde \'echelle d'\'energie de ces mod\`eles
est l'\'echelle de brisure \'electrofaible qui doit \^etre,
\begin{eqnarray}
< 0 \vert \phi \vert 0> = v \simeq 246GeV.
\label{EWSBVEV}
\end{eqnarray}
Le probl\`eme de hi\'erarchie vient de la grande disparit\'e entre ces deux
\'echelles,
\begin{eqnarray}
{V \over v} = O(10^{13}),
\label{diffVEV}
\end{eqnarray}
ainsi que de la difficult\'e \`a rendre la coexistence de ces 2 \'echelles
naturelle.
Le potentiel effectif au niveau en arbres pour ces champs scalaires $\Phi$
et $\phi$ est,
\begin{eqnarray}
V_0(\Phi,\phi)= - {1 \over 2} A \Phi^2 + {1 \over 4} B \Phi^4 - {1 \over 2}
a \phi^2
+ {1 \over 4} b \phi^4 + {1 \over 2} \l \Phi^2 \phi^2.
\label{potGUT}
\end{eqnarray}
La brisure de la sym\'etrie GUT (associ\'ee \`a l'\'echelle \ref{GUTVEV})
est assur\'ee par la condition,
\begin{eqnarray}
V^2={A \over B},
\label{condGUT}
\end{eqnarray}
qui fixe la masse du champ de Higgs lourd $\Phi$. Le
probl\`eme vient
en fait du terme ${1 \over 2} \l \Phi^2 \phi^2$ qui est pr\'esent
si $A \neq 0$ et $a \neq 0$ et qui
communique l'\'echelle d'\'energie $V$ au secteur du champ de Higgs $\phi$.
Le champ
de Higgs
lourd $\Phi$ d\'ecouple de telle sorte que la condition de brisure
\'electrofaible
devient,
\begin{eqnarray}
v^2={a -\l V^2  \over b}.
\label{condEWSB}
\end{eqnarray}
D'apr\`es Eq.(\ref{condEWSB}), afin d'obtenir l'\'echelle de brisure
\ref{EWSBVEV},
il est n\'ecessaire de faire
un `fine-tuning' du param\`etre $a/b$ qui doit \^etre de l'ordre de
$(V/v)^2 \approx 10^{26} GeV$ (voir Eq.(\ref{diffVEV})).
C'est pr\'ecis\'ement ce r\'eglage fin qui n'est pas consid\'er\'e comme
\'etant naturel.
Le probl\`eme de fine-tuning est accentu\'e par le fait que les
corrections
radiatives produisent des corrections \`a l'ordre des boucles au potentiel
effectif
de telle sorte que le fine-tuning des param\`etres, $a,\l,V,b,$ du potentiel
doit
\^etre refait \`a chaque ordre de la th\'eorie des perturbations. C'est \`a
ce stade
que les th\'eories SUSY sont int\'eressantes, car l'annulation des divergences
quadratiques qui
leur est propre assure que la masse du boson de Higgs $\sqrt a$ (voir
Eq.(\ref{potGUT})) n'est pas modifi\'ee par ces divergences quadratiques 
et donc que le fine-tuning ne doit \^etre effectu\'e qu'une seule fois.
La situation n'est en fait pas aussi claire car la supersym\'etrie doit
\^etre bris\'ee
et les masses des particules du Mod\`ele Standard doivent donc \^etre
diff\'erentes
des masses de leur partenaire supersym\'etrique,
comme nous le discuterons par la suite. Par cons\'equent, il ne doit pas y
avoir d'annulation entre
des graphes \'echangeant des particules du \ms et les graphes associ\'es
qui impliquent leur partenaire supersym\'etrique. Il doit donc exister des
divergences quadratiques
dans les th\'eories SUSY et l'on doit obtenir des corrections radiatives non
nulles \`a la masse
du boson de Higgs de l'ordre de,
\begin{eqnarray}
\delta \sqrt a \sim {\l^2 \over 8 \pi^2} \tilde m^2,
\label{corrrad}
\end{eqnarray}
o\`u $\tilde m$ repr\'esente la diff\'erence de masse typique entre les
particules du Mod\`ele Standard et
et leur partenaire supersym\'etrique,
et, $\l$ est un couplage de Yukawa ou une constante de couplage associ\'ee
\`a un
groupe jauge. Le probl\`eme du fine-tuning \`a chaque ordre de la th\'eorie
des perturbations
peut donc \^etre r\'esolu dans les th\'eories SUSY mais uniquement pour un
\'ecart de masse dans
le multiplet supersym\'etrique de l'ordre de l'\'echelle \'electrofaible
donn\'ee dans
Eq.(\ref{EWSBVEV}), ce qui s'\'ecrit dans nos notations,
\begin{eqnarray}
\tilde m \sim v.
\label{solSUSY}
\end{eqnarray}
En conclusion, les masses des particules \susiqs ainsi que l'\'echelle de
brisure
\'electrofaible doivent \^etre de l'ordre du TeV
si l'on veut que le probl\`eme de hi\'erarchie soit ``r\'esolu'' par la
supersym\'etrie.

Certaines th\'eories de supergravit\'e (voir Chapitre \ref{SUGRA})
permettent d'engendrer des masses pour les particules
\susiq de l'ordre du TeV. Ces mod\`eles assument l'existence d'un
secteur cach\'e
qui n'interagit avec les particules du secteur observable (quarks,
leptons,...) que par les interactions gravitationnelles. 
Dans ces mod\`eles, on peut
obtenir des masses pour les particules SUSY de l'ordre de,
\begin{eqnarray}
\tilde m \sim {M_s^2 \over M_P},
\label{solSUGRA1}
\end{eqnarray}
o\`u $M_s$ est l'\'echelle de brisure de la supersym\'etrie 
et $M_P$ est la masse de Planck. Donc, l'\'echelle du TeV requise pour 
l'\'ecart de masse des supermultiplets est obtenue dans ces mod\`eles si ,
\begin{eqnarray}
M_s = O(10^{10}GeV).
\label{solSUGRA2}
\end{eqnarray}
De plus, dans les mod\`eles de supergravit\'e appel\'es `no-scale'
\cite{nscla,nsclb,nsclc,nscld,BaiLov} cette \'echelle de brisure
SUSY, $M_s$, est obtenue naturellement comme une suppression de l'\'echelle
de Planck, $M_P$.

La seconde explication du probl\`eme de hi\'erarchie
est bas\'ee uniquement sur l'\'etude du secteur scalaire
du Mod\`ele Standard, \`a savoir
le secteur du boson de Higgs. Ce point de vue n'est donc pas restreint au
seul cas des th\'eories GUT
mais est valable pour toute nouvelle physique au-del\`a du Mod\`ele
Standard.
Dans le Mod\`ele Standard, le potentiel du boson de Higgs $H$ est
$V = - \mu^2 HH^\dagger +{1 \over 2} \lambda (H H^\dagger)^2$
ce qui donne une valeur dans le vide \`a $H H^\dagger$ de ${\mu^2 \over
\lambda}$.
Nous connaissons aujourd'hui les masses des bosons de jauge $W^{\pm}$ et
$Z^0$ qui donnent \`a $<H H^\dagger>$
une valeur de l'ordre
de $10^4 GeV^2$. $\lambda$ ne pouvant pas \^etre arbitrairement petit,
$\mu^2$ doit \^etre du m\^eme ordre de grandeur que
$<H H^\dagger>$. Au niveau en arbre, nous pouvons donner \`a $\mu^2$ sa
valeur exacte (inconnue aujourd'hui). Cependant,
les corrections quantiques ne pr\'eserveront pas ce choix. En effet,
l'\'equation de renormalisation \`a une boucle est,
\begin{eqnarray}
\mu_{RE}^2  = \mu^2 + C \alpha^2 M^2,
\label{muloop}
\end{eqnarray}
o\`u  $\mu_{RE}$  est le param\`etre $\mu$ renormalis\'e et $C$ un nombre de
l'ordre de $100 \pm 1$. $M$ est
la coupure ultraviolette, identifi\'ee \`a l'\'echelle de toute nouvelle
physique sous-jacente au Mod\`ele
Standard pouvant exister \`a haute \'energie. Par exemple, $M=M_P=
M_{PLANCK} \approx 10^{19} GeV$, $M = M_{GUT} \approx 2 \ 10^{16} GeV$ 
ou toute autre \'echelle comprise entre
$M_P$ et l'\'echelle de Fermi: $M_W  = 80 GeV$. 
$\alpha$ est un param\`etre reli\'e aux
constantes de couplage des interactions
\'electrofaible et forte. $\alpha$ ne pouvant \^etre arbitrairement petit,
il y a une diff\'erence d'ordre
de grandeur entre $\mu$ et $\mu_{RE}$ d'\`a peu pr\`es $10^{30} GeV^2$. Il
faut donc ajuster $\mu^2$ tr\`es
pr\'ecis\'ement de fa\c{c}on \`a ce que $\mu_{RE}^2$  ait la bonne valeur,
c'est \`a dire de l'ordre de
$10^4 GeV^2$ (fine-tuning). De plus, un nouvel ajustement doit \^etre
effectu\'e \`a chaque ordre de la th\'eorie des perturbations. Il est donc
clair que l'existence de
plusieurs \'echelles de masse n'est pas du tout naturelle. On retrouve le
probl\`eme des hi\'erarchies
d'\'echelles. Dans le cadre d'une th\'eorie de SUSY, l'absence des
divergences quadratiques assure que
$\mu^2$ est au plus logarithmiquement renormalisable et l'expression
\ref{muloop} de
la masse renormalis\'ee du boson de Higgs devient donc,
\begin{eqnarray}
\mu_{RE}^2  = \mu^2  [1 + C \alpha^2 ln ( {M^2 \over \mu^2})].
\label{muSUSY}
\end{eqnarray}
$\mu_{RE}^2$  et $\mu^2$  sont maintenant du m\^eme ordre de grandeur et le
probl\`eme de naturalit\'e
est r\'esolu: Modifier tr\`es peu les param\`etres fondamentaux (ici $M$)
n'affecte plus la physique \`a
basse \'energie.

Remarquons dans ce contexte une analogie
entre le boson de Higgs et les autres particules du Mod\`ele Standard.
La masse des bosons de jauge est `prot\'eg\'ee'
par les sym\'etries de jauge et celle des fermions est `prot\'eg\'ee' par
la nature chirale (par opposition \`a vectorielle) du groupe de jauge
\'electrofaible.
Quant \`a la masse des bosons de Higgs, sans \^etre contrainte \`a prendre
une valeur nulle,
elle est `prot\'eg\'ee' des divergences quadratiques par la supersym\'etrie
(voir \ref{muSUSY}).

\section{Formalisme des th\'eories supersym\'etriques}

\subsection{Exemple du spineur de Weyl}
\label{Wey}

\'Etablissons les transformations supersym\'etriques laissant invariant le
lagrangien d'un spineur de Weyl libre.
Le spineur de Weyl a un spin $1/2$ et une h\'elicit\'e
donn\'ee.  Par cons\'equent, le spineur de Weyl ne poss\`ede que deux
composantes non nulles et n'est pas massif.
Les deux composantes non nulles d'un spineur de Weyl d'h\'elicit\'e gauche
(droite) d\'ecrivent l'h\'elicit\'e gauche
(droite) de la particule et l'h\'elicit\'e droite (gauche) de
l'antiparticule. Le spineur de Weyl est d\'efini par:
\begin{eqnarray}
L \Psi_L = \Psi_L  \   ou  \  R \Psi_R  = \Psi_R, \ avec \
L ={1+\gamma_5 \over 2}  \  et \  R ={1-\gamma_5 \over 2},
\label{Weyl}
\end{eqnarray}
selon qu'il est d'h\'elicit\'e gauche ($\Psi_L$) ou droite ($\Psi_R$).
Consid\'erons le lagrangien d'un spineur de Weyl
d'h\'elicit\'e gauche $\Psi_L$ libre et de son partenaire scalaire
supersym\'etrique $\phi$:
\begin{eqnarray}
{\cal L}=  (\partial_{\mu} \phi)^\star  (\partial^{\mu} \phi) + \bar \Psi_L
( i \gamma_{\mu} \partial^{\mu} )  \Psi_L,
\label{LagWeyl}
\end{eqnarray}
avec selon nos notations,
\begin{eqnarray}
\bar \Psi_L  =  \overline {(\Psi_L)}=\overline {(L \Psi )}=(L \Psi)^\dagger
\gamma_0.
\label{notspin}
\end{eqnarray}
Les partenaires supersym\'etriques des champs fermioniques de spin 1/2
sont des champs scalaires et sont appel\'es
sfermions (s pour scalaire): squarks, sleptons,...,
tandis que les partenaires des champs
bosoniques sont des champs fermioniques
de spin 1/2 et ont un nom prenant un suffixe `ino':
wino, zino, photino, gluino, higgsino,...
Les transformations de la supersym\'etrie N =1 (voir Section \ref{AlSUSY})
les plus g\'en\'erales laissant invariant
le lagrangien \ref{LagWeyl} s'\'ecrivent:
\begin{eqnarray}
\delta \phi &  = &  \sqrt {2} \bar {\cal E}_R  \Psi_L, \cr
\delta \phi^\star &  = &   \sqrt {2}  \bar \Psi_L {\cal E}_R, \cr
\delta \Psi_L &  = &  -i \sqrt {2} \gamma^{\mu} \partial_{\mu} \phi {\cal E}_R,  \cr
\delta \bar \Psi_L &   = &  -i \sqrt {2} {\cal E}_R \gamma^{\mu} \partial_{\mu}
\phi^\star.
\label{tranSUSY1}
\end{eqnarray}
Nous avons adopt\'e la notation:
${\cal E}_R \Psi_L = {\cal E}_R^{\alpha} \Psi_{L \alpha}$ (voir Section \ref{AlSUSY}).
 Les deux premi\`eres relations de Eq.\ref{tranSUSY1}
sont des \'egalit\'es entre scalaires et les deux derni\`eres entre
spineurs.
Le param\`etre ${\cal E}_R$ de cette transformation
doit \^etre un spineur de Weyl d'h\'elicit\'e droite pour \'eviter
$\delta \phi = 0$ et afin d'avoir $L \delta \Psi_L= \delta \Psi_L$.
Le facteur $\sqrt 2$ permet de simplifier les calculs.
Remarquons finalement que le lagrangien \ref{LagWeyl}
n'est invariant par les transformations \ref{tranSUSY1}
qu'\`a une d\'eriv\'ee pr\`es, mais les termes en d\'eriv\'ees
du lagrangien ne modifient pas les \'equations du mouvement.
En effet, l'action s'obtient en int\'egrant le lagrangien,
or ces termes s'annulent apr\`es int\'egration
car ils deviennent alors des valeurs de champs pris \`a l'infini.

\subsection{Exemple du spineur de Majorana}
\label{Majo}

Consid\'erons maintenant un lagrangien d\'ecrivant un spineur de Majorana
libre et
son partenaire supersym\'etrique. Un spineur de Majorana est \'egal \`a son
spineur conjugu\'e de charge,
ce qui s'\'ecrit:
\begin{eqnarray}
\Psi = \Psi^c =C \bar \Psi^T
\label{Majorana}
\end{eqnarray}
o\`u $C$ est l'op\'erateur conjugaison de charge.
D'apr\`es nos notations, nous \'ecrivons,
\begin{eqnarray}
\bar \Psi^c_H  =  \overline {[(\Psi^c)_H]} =  \overline {[P_H (\Psi^c)]}, \
H=L,R.
\label{notspiconj}
\end{eqnarray}
Le spineur de Majorana n'a que 2 composantes ind\'ependantes du fait de la
relation \ref{Majorana}.
Consid\'erons le lagrangien d\'ecrivant un spineur de Majorana $\Psi$ libre
ainsi
que son partenaire supersym\'etrique $\phi$:
\begin{eqnarray}
{\cal L} = {1 \over 2} (\partial_{\mu} B) (\partial^{\mu} B) + {1 \over 2}
(\partial_{\mu} A) (\partial_{\mu} A )
- {1 \over 2} m^2 (A^2 + B^2) + {1 \over 2} \bar \Psi ( i \gamma_{\mu}
{\partial}^{\mu} - m ) \Psi
\label{LagMaj}
\end{eqnarray}
o\`u $A$ et $B$ sont des champs scalaires r\'eels tels que $\phi = (A + i B)
/ \sqrt 2$.
Il est plus transparent de travailler au niveau des champs $A$ et $B$ car
les transformations de
supersym\'etrie doivent respecter l'invariance sous la parit\'e du
lagrangien \ref{Majorana},
c'est \`a dire que l' on doit avoir
$\delta A \to \delta A$ et $\delta B \to - \delta B$ puisque les champs $A$
et $B$ se transforment sous l'action de
l'op\'erateur parit\'e par $A \to A$ et $B \to -B$.
Le spineur de Majorana se transforme sous l'action de l'op\'erateur parit\'e
selon: $\Psi  \to \gamma_0 \Psi$.
Le lagrangien \ref{LagMaj} est invariant sous les transformations de
supersym\'etrie N =1 (voir Section \ref{AlSUSY})
suivantes:
\begin{eqnarray}
\delta A &  = &  \bar {\cal E}  \Psi, \cr
\delta B &  = &  i \bar {\cal E} \gamma_5  \Psi,  \cr
\delta \Psi &  = &  - [ i \gamma^{\mu} \partial_{\mu} (A + i B \gamma_5 ) + m (A +
i B \gamma_5 ) ] {\cal E},
\label{tranSUSY2}
\end{eqnarray}
o\`u le param\`etre ${\cal E}$ de transformation est un spineur de Majorana afin
que sous l'action de l'op\'erateur parit\'e
on ait: $\delta \Psi = \delta \Psi^c =C \overline {(\delta \Psi)}^T$. Nous
remarquons d'apr\`es Eq.(\ref{LagMaj})
que $A$, $B$ et $\Psi$, qui constituent le supermultiplet $(A, B, \Psi)$,
ont la m\^eme masse. On retrouve une des propri\'et\'es de la
supersym\'etrie:
Les membres d'un m\^eme supermultiplet ont des masses identiques.

\subsection{Champs auxiliaires}
\label{aux}

On dit que l'alg\`ebre de \susi ferme lorsque:
\begin{eqnarray}
[ \delta_1, \delta_2 ] = i \Delta_{\mu} P^{\mu}, \
avec \ \Delta_{\mu} = - 2 i (\bar {\cal E}_2 \gamma_\mu {\cal E}_1 ) \ et \
P^\mu = - i \partial^\mu,
\label{SUSYclose}
\end{eqnarray}
$\delta_1$, $\delta_2$ \'etant les variations infinit\'esimales
associ\'ees \`a 2 transformations de supersym\'etrie, et
${\cal E}_1$, ${\cal E}_2$ \'etant les param\`etres de ces 2 transformations.
Remarquons par ailleurs que,
\begin{eqnarray}
[ \delta_1, \delta_2 ] = - [\bar {\cal E}_1 Q, \bar {\cal E}_2 Q] =  \bar {\cal E}_1^\alpha
\bar {\cal E}_2^\beta \{ Q^\alpha, Q^\beta \},
\label{SUSYclop}
\end{eqnarray}
car les ${\cal E}^\alpha$ sont des variables de Grassmann 
qui v\'erifient par d\'efinition les relations de 
Eq.(\ref{GRASS}).
Les relations de Eq.(\ref{SUSYclose}) et Eq.(\ref{SUSYclop}) conduisent \`a
la relation d'anti-commutation
des g\'en\'erateurs $Q$ de la \susi (voir Eq.(\ref{ALsusy})). Donc, la
condition de fermeture de l'alg\`ebre
de SUSY (Eq.(\ref{SUSYclose})) permet de retrouver l'alg\`ebre de SUSY (voir
Eq.(\ref{ALsusy})).

La relation de Eq.(\ref{SUSYclose}) est satisfaite lorsque les particules
sont sur la couche de masse (``on shell''),
c'est \`a dire quand les champs v\'erifient les \'equations de mouvement du
lagrangien correspondant.
La relation de Eq.(\ref{SUSYclose}) est par exemple v\'erifi\'ee dans les
deux cas trait\'es
dans les Sections \ref{Wey} et \ref{Majo} si les particules sont on shell.
Pour le spineur de Weyl du lagrangien \ref{LagWeyl},
l'\'equation du mouvement est: $\gamma_{\mu} \partial^{\mu} \Psi_L = 0$, et
pour le spineur de Majorana du lagrangien
\ref{LagMaj}, l'\'equation du mouvement est $( i \gamma_{\mu} \partial^{\mu}
- m ) \Psi = 0$.\\
Afin que la relation de Eq.(\ref{SUSYclose}) soit aussi v\'erifi\'ee dans le
formalisme hors couche (``off shell''),
on introduit des champs dits ``auxiliaires''. Par exemple, dans le cas du
spineur de Weyl (voir Section \ref{Wey}),
on introduit le champ auxiliaire scalaire complexe $f$ en ajoutant au
lagrangien la partie ${\cal L}_{AUX}= f f^\dagger$, et en adjoignant la
nouvelle loi de transformation pour $f$:
$\delta f = i \sqrt 2 \bar {\cal E}_R  \gamma^{\mu} \partial_{\mu} \Psi_L$. La
relation de transformation du spineur de Weyl
de Eq.(\ref{tranSUSY1}) devient alors $\delta \Psi_L = - \sqrt {2} (i
\gamma^{\mu} \partial_{\mu} \phi +f) {\cal E}_R$.
L'\'equation de mouvement du champ $f$ est $f = 0$. Dans le cas du spineur
de Majorana (voir Section \ref{Majo}),
on introduit les deux champs auxiliaires scalaires r\'eels $F$ et $G$ en
ajoutant au lagrangien le terme:
${\cal L}_{AUX}= (F^2 + G^2)/2 - m(A \ F+B \ G)$. Les lois de transformation
de $F$ et $G$ sont:
$\delta F = i \bar {\cal E} \gamma^{\mu} \partial_{\mu} \Psi$  et  $\delta G = -
\bar {\cal E} \gamma^{\mu} \partial_{\mu} \Psi$.
La relation de transformation du spineur de Majorana de Eq.(\ref{tranSUSY2})
devient alors:
$\delta \Psi = - [ i \gamma^{\mu} \partial_{\mu} (A + i B \gamma_5 ) + (F +
i G \gamma_5 ) ] {\cal E}$.
Les \'equations de mouvement des $F$ et $G$ sont $F = mA$ et $G = mB$.
\begin{table}[t]
\begin{center}
\begin{tabular}{|c||c|c||c|c|}
\hline
          &  BOSONS               &  Degr\'es de libert\'e &  FERMIONS  & 
Degr\'es de libert\'e   \\
\hline
OFF SHELL &  $A$, $B$, $F$ et $G$ &   4 $\times$ 1         &  $\Psi$    &  1
$\times$ 4            \\
\hline
ON SHELL  &  $A$ et $B$           &   2 $\times$ 1         &  $\Psi$    &  1
$\times$ 2            \\
\hline
\end{tabular}
\caption{Degr\'es de libert\'e fermioniques et bosoniques dans
un supermultiplet associ\'e \`a un spineur de Majorana.}
\label{degAUX}
\end{center}
\end{table}
Les champs auxiliaires ont aussi l'int\'er\^et de rendre \'egaux les nombres
de degr\'es de libert\'e
fermioniques et bosoniques dans un supermultiplet chiral. Afin d'illustrer
ce point,
nous pr\'esentons dans la Table \ref{degAUX} les degr\'es de
libert\'e dans un supermultiplet associ\'e \`a un spineur de Majorana:

\subsection{Superespace}
\label{superesp}

Commen\c{c}ons cette section par quelques rappels.
Les transformations infinit\'esimales associ\'ees au groupe de Lorentz et
aux translations s'\'ecrivent,
\begin{eqnarray}
Lorentz: \ \delta \Phi=i{1 \over 2} \ep_{\mu \nu} M^{\mu \nu} \Phi, \\
Translations: \ \delta \Phi=i \Delta_{\mu} P^{\mu} \Phi,
\label{Poinf}
\end{eqnarray}
o\`u les param\`etres de transformation associ\'es au groupe de Lorentz
$\ep_{\mu \nu}$ et aux translations
$\Delta_{\mu}$ sont des nombres r\'eels.
Le facteur $1/2$ apparaissant dans les transformations de Lorentz permet
d'\'eviter les doubles comptages
li\'es au fait que l'on a $M^{\mu \nu}=-M^{\nu \mu}$.
Les transformations totales associ\'ees au groupe de Lorentz et aux
translations sont,
\begin{eqnarray}
Lorentz: \ \Phi'=exp \bigg ( i{1 \over 2} \ep_{\mu \nu} M^{\mu \nu} \bigg )
\Phi, \\
Translations: \ \Phi'=exp \bigg ( i \Delta_{\mu} P^{\mu} \bigg ) \Phi.
\label{Pointot}
\end{eqnarray}
Dans le cas o\`u $\Phi$ est un champ scalaire, les g\'en\'erateurs du groupe
de Lorentz et des translations
sont,
\begin{eqnarray}
M^{\mu \nu} =i(x^{\mu} \partial^{\nu}-x^{\nu} \partial^{\mu}), \ \
P^{\mu}=-i \partial^{\mu}.
\label{Poinsca}
\end{eqnarray}
Si $\Phi$ est un spineur, les g\'en\'erateurs du groupe de Lorentz et des
translations
s'\'ecrivent alors,
\begin{eqnarray}
M^{\mu \nu} ={i \over 4} [ \gamma^{\mu},\gamma^{\nu} ], \ \
P^{\mu}=-i \partial^{\mu}.
\label{Poinspin}
\end{eqnarray}
Enfin, une translation agit sur un champ $\Phi$ comme:
\begin{eqnarray}
\Phi' (  x^{\mu} )=\Phi (  x^{\mu} +\Delta_{\mu}).
\label{translat}
\end{eqnarray}

Le produit anti-commutant des g\'en\'erateurs $Q_{\alpha}$ de l'alg\`ebre de
\susi est proportionnel
au g\'en\'erateur $P^{\mu}$ des translations. 
Il semble donc int\'eressant de tenter d'\'ecrire les
transformations de \susi comme des translations g\'en\'eralis\'ees. Pour
cela l'espace-temps doit
\^etre g\'en\'eralis\'e en un espace comprenant des nouvelles coordonn\'ees
qui soient translat\'ees
par la supersym\'etrie: Le ``superespace''.
Les champs contenus dans le superespace sont appel\'es ``superchamps''. \\
Le param\`etre des transformations \susiqs ${\cal E}^{\alpha}$ (voir
Eq.(\ref{trSUSYgen}))
est un spineur anti-commutant (variable de Grassmann) qui satisfait \`a la
condition de Majorana.
Nous en d\'eduisons d'apr\`es Eq.(\ref{translat}), par analogie entre le
param\`etre de transformation
associ\'e \`a l'alg\`ebre de \susi ${\cal E}^{\alpha}$ (voir Eq.(\ref{trSUSYgen}))
et celui associ\'e aux
translations $\Delta_{\mu}$ (voir Eq.(\ref{Poinf})), que les coordonn\'ees
du superespace sont,
\begin{eqnarray}
( x^{\mu} , \theta, \bar \theta),
\label{newcoord}
\end{eqnarray}
o\`u $\theta$ est aussi un spineur anti-commutant (variable de Grassmann)
qui satisfait \`a la condition
de Majorana. Un spineur de Majorana n'a que 2 composantes ind\'ependantes
(voir Section \ref{Majo}).
Par cons\'equent,
dans les sections suivantes, o\`u nous allons d\'evelopper le formalisme du
superespace qui permet
de construire de fa\c{c}on simple des th\'eories supersym\'etriques, nous
adopterons
la notation \`a 2 composantes des spineurs. Dans la section suivante, nous
rappelons donc le
formalisme de la notation \`a 2 composantes pour les spineurs.

\subsection{Notation des spineurs \`a deux composantes}
\label{2c}

Le champ de spin 1/2 \`a 2 composantes
$\psi_{\alpha}=\psi_L$ ($\alpha=1,2$) appartient \`a la repr\'esentation
(1/2,0) du groupe de Lorentz,
c'est \`a dire qu'il se comporte comme un champ de spin 1/2 sous les
transformations ``gauches'' et comme un champ scalaire sous les
transformations ``droites'',
alors que le spineur \`a 2 composantes $\bar \psi^{\dot \alpha}=\psi_R$
($\dot \alpha=1,2$)
appartient \`a la repr\'esentation (0,1/2) du groupe de Lorentz.
$\psi_{\alpha}=\psi_L$ est donc un spineur de chiralit\'e gauche et $\bar
\psi^{\dot \alpha}=\psi_R$
un spineur de chiralit\'e droite.
Par ailleurs, le tenseur antisym\'etrique \`a 2 indices $\ep^{\alpha \beta}$
permet
d'abaisser ou bien d'\'elever les indices des spineurs \`a 2 composantes:
\begin{eqnarray}
\psi^{\alpha} & = & \ep^{\alpha \beta} \psi_{\beta}, \cr
\psi_{\alpha} & = & \ep_{\alpha \beta} \psi^{\beta}.
\label{epalbe}
\end{eqnarray}
Les champs de spin 1/2 \`a 2 composantes se transforment sous le groupe de
Lorentz $SO(1,3)$
par l'action des matrices $M \in Sl(2,C)$ selon,
\begin{eqnarray}
\psi_{\alpha}^{\prime}=M^\beta_\alpha \psi_{\beta}, \ & 
\psi^{\alpha \ \prime}=(M^{-1})^{\alpha}_\beta \psi^{\beta}, \cr
\bar \psi_{\dot \alpha}^{\prime}=(M^{\star})^{\dot \beta}_{\dot \alpha} \bar
\psi_{\dot \beta}, \ & 
\bar \psi^{\dot \alpha \ \prime}=(M^{\star -1})_{\dot \beta}^{\dot \alpha} \bar
\psi^{\dot \beta},
\label{Lorentz2c}
\end{eqnarray}
l'\'etoile $\star$ signifiant complexe conjugu\'e.
Les produits,
\begin{equation}
\psi_1 \psi_2=\psi_1^{\alpha} \psi_{2 \alpha}, \ \ \
\bar \psi_1 \bar \psi_2=\bar \psi_{1 \dot \alpha} \bar \psi_2^{\dot \alpha}.
\label{inv2c}
\end{equation}
sont donc invariants de Lorentz.
Enfin, le champ $\Psi$ de spin 1/2 \`a 4 composantes et son conjugu\'e de
charge
$\Psi^c$ s'\'ecrivent \`a partir des champs de spin 1/2 \`a 2 composantes
$\chi_{\alpha}$ et $\eta_{\alpha}$, comme suit,
\begin{equation}
\Psi=
\left (
\begin{array}{c}
\chi_{\alpha} \\ \bar \eta^{\dot \alpha}
\end{array}
\right ) , \ \
\bar \Psi=(\eta^{\alpha} , \bar \chi_{\dot \alpha}) , \ \
\Psi^c=
\left (
\begin{array}{c}
\eta_{\alpha} \\ \bar \chi^{\dot \alpha}
\end{array}
\right ) , \ \  \alpha=1,2,  \  \dot \alpha=1,2.
\label{4to2spin}
\end{equation}
Le spineur de Majorana $\Psi$ de spin 1/2 \`a 4 composantes s'\'ecrit quant
\`a lui,
\begin{equation}
\Psi=
\left (
\begin{array}{c}
\chi_{\alpha} \\ \bar \chi^{\dot \alpha}
\end{array}
\right ) , \ \ \ \alpha=1,2,  \  \dot \alpha=1,2.
\label{4to2maj}
\end{equation}
Les produits $\bar \Psi_1 P_{L,R} \Psi_2$ de spineurs \`a 4 composantes
$\Psi_1$ et $\Psi_2$
s'\'ecrivent \`a partir des spineurs \`a 2 composantes $\chi_{\alpha}$ et
$\eta_{\alpha}$,
comme suit,
\begin{eqnarray}
\bar \Psi_1 P_L \Psi_2 =\eta_1 \chi_2,
\label{2to4L}
\end{eqnarray}
\begin{eqnarray}
\bar \Psi_1 P_R \Psi_2 =\bar \eta_2 \bar \chi_1.
\label{2to4R}
\end{eqnarray}
Par cons\'equent, un produit $\bar \Psi_1 \Psi_2$ s'\'ecrit,
\begin{eqnarray}
\bar \Psi_1 \Psi_2 =\eta_1 \chi_2 + \bar \eta_2 \bar \chi_1.
\label{2to4mass}
\end{eqnarray}

\'Ecrivant le g\'en\'erateur de supersym\'etrie $Q$, qui est un spineur de
Majorana, dans la notation
\`a deux composantes:
\begin{equation}
Q=
\left (
\begin{array}{c}
Q_{\alpha} \\ \bar Q^{\dot \alpha}
\end{array}
\right ),
\label{SUSYg2c}
\end{equation}
ainsi que le param\`etre des transformations SUSY ${\cal E}$ qui est aussi un
spineur de Majorana:
\begin{equation}
{\cal E}=
\left (
\begin{array}{c}
{\cal E}_{\alpha} \\ \bar {\cal E}^{\dot \alpha}
\end{array}
\right ),
\label{SUSYg2cp}
\end{equation}
la relation d'anti-commutation des g\'en\'erateurs $Q$ de la \susi (voir
Eq.(\ref{ALsusy})) devient,
\begin{eqnarray}
 \{ Q_{\alpha}, Q_{\beta} \} =  \{ \bar Q_{\dot \alpha}, \bar Q_{\dot \beta}
\} =0, \cr
 \{ Q_{\alpha}, \bar Q_{\dot \beta} \} = -2 P_{\mu} (\sigma^{\mu})_{\alpha
\dot \beta},
\label{ALsu2c}
\end{eqnarray}
et la transformation infinit\'esimale de SUSY (voir Eq.(\ref{trSUSYgen}))
prend la forme,
\begin{eqnarray}
\delta \Phi = i ({\cal E}^{\alpha} Q_{\alpha} + \bar {\cal E}_{\dot \alpha} \bar Q^{\dot
\alpha}) \Phi.
\label{trSUSYg2c}
\end{eqnarray}
En utilisant Eq.(\ref{trSUSYg2c}) ainsi que les relations 
de d\'efinition de variables de Grassmann ${\cal E}^\alpha$
dans la notation des spineurs \`a 2 composantes:
\begin{eqnarray}
 \{ {\cal E}^\alpha, {\cal E}^\beta \} =  
 \{ {\cal E}^\alpha, \bar {\cal E}^{\dot \alpha} \} =  
\{ \bar {\cal E}^{\dot
\alpha}, \bar {\cal E}^{\dot \beta} \} = 0, \cr
 \{ {\cal E}^\alpha, Q^\alpha \} =  \{ {\cal E}^\alpha, \bar Q^{\dot \alpha} \} =  \{
\bar {\cal E}^{\dot
\alpha}, Q^\alpha \}
=  \{ \bar {\cal E}^{\dot \alpha}, \bar Q^{\dot \alpha} \} = 0,
\label{trSUSYg2cp}
\end{eqnarray}
nous obtenons l'\'equivalent de Eq.(\ref{SUSYclose}) dans la notation \`a 2
composantes, \`a savoir,
\begin{eqnarray}
[ \delta_1, \delta_2 ] = - 2 ( {\cal E}_1 \sigma^{\mu} \bar {\cal E}_2 - {\cal E}_2
\sigma^{\mu} \bar {\cal E}_1) P_{\mu}, \
avec \ {\cal E} \sigma^{\mu} \bar {\cal E} = {\cal E}^{\alpha} (\sigma^{\mu})_{\alpha \dot
\alpha} \bar {\cal E}^{\dot \alpha}.
\label{SUSYclo2c}
\end{eqnarray}

\`A partir de maintenant et dans tout ce qui suit,
nous adopterons la notation des spineurs \`a 2 composantes.

\subsection{G\'en\'erateurs de supersym\'etrie dans le superespace}
\label{gediSUSY}

D\'eterminons la forme g\'en\'erale dans le superespace du g\'en\'erateur
$Q_{\alpha}$ de l'alg\`ebre
de la supersym\'etrie $N =1$. Ce g\'en\'erateur $Q_{\alpha}$, qui est
associ\'e \`a des translations
dans le superespace $( x^{\mu} , \theta, \bar \theta)$ (voir Section
\ref{superesp}), doit agir sur
les superchamps par l'interm\'ediaire de d\'eriv\'ees par rapport \`a
$x^{\mu}$, $\theta$ et $\bar \theta$,
de m\^eme que le g\'en\'erateur $P^{\mu}$ des translations dans
l'espace-temps \`a 4 dimensions
agit suivant des d\'eriv\'ees par rapport \`a $x^{\mu}$ (voir Section
\ref{superesp}).
La forme g\'en\'erale du g\'en\'erateur $Q_{\alpha}$ dans le superespace est
donc,
\begin{eqnarray}
Q_{\alpha} & =&   a_\alpha^{\mu} \partial_{\mu} + b {\partial  \over \partial
\theta^\alpha}
+ c_{\alpha \dot \alpha} {\partial  \over \partial \bar \theta_{\dot
\alpha}}, \cr
\bar Q_{\dot \alpha} & =&   \bar a_{\dot \alpha}^{\mu} \partial_{\mu}
+ \bar b {\partial  \over \partial \bar \theta^{\dot \alpha}}
+ \bar c^\alpha_{\dot \alpha} {\partial  \over \partial \theta^\alpha}.
\label{gensup1}
\end{eqnarray}
o\`u $b$ et $\bar b$ sont des nombres complexes ind\'ependants.

Raisonnons par les dimensions.
D'apr\`es Eq.(\ref{ALsu2c}), $[Q_{\alpha}]=[\bar Q_{\dot \alpha}]=1/2$
puisque
$P_{\mu}=-i\partial_{\mu}$ et $[\partial_{\mu}]=1$ (voir Appendice
\ref{dims}).
Par cons\'equent, $[{\cal E}^\alpha]=[\bar {\cal E}_{\dot \alpha}]=-1/2$ selon
Eq.(\ref{trSUSYg2c}).
De plus, d'apr\`es Eq.(\ref{gensup1}),
$[a_\alpha^{\mu}] = [\bar a_{\dot \alpha}^{\mu}]=-1/2$
car $[Q_{\alpha}]=[\bar Q_{\dot \alpha}]=1/2$ et $[\partial_{\mu}]=1$.
Ayant $[\theta^\alpha]=[\bar \theta_{\dot \alpha}]=[{\cal E}^\alpha]=[\bar
{\cal E}_{\dot \alpha}]=-1/2$,
il est naturel de poser:
\begin{eqnarray}
a_\alpha^{\mu} &  = &  a (\sigma^{\mu})_{\alpha \dot \alpha} \bar \theta^{\dot
\alpha}, \cr
\bar a_{\dot \alpha}^{\mu} &  = &  \bar a \theta^\alpha (\sigma^{\mu})_{\alpha \dot
\alpha},
\label{gensup2}
\end{eqnarray}
o\`u $a$ et $\bar a$ sont des nombres complexes ind\'ependants tels que
$[a]=[\bar a]=0$.
De plus, $[c_{\alpha \dot \alpha}]=[\bar c^\alpha_{\dot \alpha}]=0$ car
$[Q_{\alpha}]=[\bar Q_{\dot \alpha}]=1/2$ et
$[\partial / \partial \theta^\alpha]=[\partial / \partial \bar \theta_{\dot
\alpha}]=1/2$.
Une inspection syst\'ematique montre que
$c_{\alpha \dot \alpha}=\bar c^\alpha_{\dot \alpha}=0$.

L'anti-commutateur $ \{ Q_\alpha,\bar Q_{\dot \beta} \}$ vaut,
\begin{eqnarray}
 \{ Q_\alpha,\bar Q_{\dot \beta} \} = (\bar a b + a \bar b)
(\sigma^{\mu})_{\alpha \dot \beta} \partial_{\mu}.
\label{gensup3}
\end{eqnarray}
D'apr\`es Eq.(\ref{ALsu2c}) et sachant que $P_{\mu}=-i\partial_{\mu}$, on
obtient $\bar a b + a \bar b = - 2i$.
L'alg\`ebre de SUSY laisse donc un certain arbitraire dans le choix de $a$,
$\bar a$, $b$ et $\bar b$.
On choisira ces nombres tels que:
\begin{eqnarray}
Q_{\alpha} & =&  -i \bigg ( {\partial  \over \partial \theta^\alpha} -i
(\sigma^{\mu})_{\alpha \dot \alpha} \bar \theta^{\dot \alpha} \partial_{\mu}
\bigg ), \cr
\bar Q_{\dot \alpha} & =&  -i \bigg ( -{\partial  \over \partial \bar
\theta^{\dot \alpha}} +i
\theta^{\alpha} (\sigma^{\mu})_{\alpha \dot \alpha} \partial_{\mu} \bigg ).
\label{gensup4}
\end{eqnarray}

\subsection{Superchamps}
\label{spchp}

Les superchamps $S$ doivent \^etre des fonctions des coordonn\'ees du
superespace
\`a savoir $x^{\mu}$, $\theta_{\alpha}$ et $\bar \theta^{\dot \alpha}$.
Les puissances cubiques des variables de Grassmann s'annulent:
$\theta^{\alpha} \theta^{\beta} \theta^{\gamma}= 0$
($\alpha,\beta,\gamma=1,2$),
car $\theta^1 \theta^1=\theta^2 \theta^2=0$ d'apr\`es Eq.(\ref{GRASS}).
Par cons\'equent, la forme g\'en\'erale d'un
superchamp $S(x^{\mu},\theta_{\alpha},\bar \theta^{\dot \alpha})$ est:
\begin{eqnarray}
S(x,\theta,\bar \theta)&=&z(x)+ \theta \psi (x)+\bar \theta \bar \chi (x)
+ \theta \theta  f(x) + \bar \theta \bar \theta  g(x)
+ \theta \sigma^{\mu} \bar \theta  v_{\mu}(x) \cr && + \theta \theta \bar \theta
\bar \l (x)
+ \bar \theta \bar \theta \theta \eta (x) + \theta \theta \bar \theta \bar
\theta D(x),
\label{supchp1}
\end{eqnarray}
o\`u $v_{\mu}(x)$ est un champ vectoriel complexe, $z(x)$, $f(x)$, $g(x)$ et
$D(x)$
sont des champs scalaires complexes, $\psi (x)$, $\chi (x)$, $\l (x)$, et
$\eta (x)$
sont des champs de spin 1/2 \`a deux composantes,
et o\`u, d'apr\`es nos notations, $\theta  \theta=\theta^{\alpha}
\theta_{\alpha}$,
$\bar \theta  \bar \theta=\bar \theta_{\dot \alpha} \bar \theta^{\dot
\alpha}$,
$\theta \psi = \theta^{\alpha} \psi_{\alpha}$
et $\bar \theta \bar \psi = \bar \theta_{\dot \alpha} \bar \psi^{\dot
\alpha}$ avec
$\alpha = 1,2$ et $\dot \alpha = 1,2$.
L'expression g\'en\'erique \ref{supchp1} d'un superchamp $S(x,\theta,\bar
\theta)$
indique que la dimension d'un superchamp est $[S(x,\theta,\bar
\theta)]=[z(x)]=1$
(voir Appendice \ref{dims}) et que le produit de superchamps est lui-m\^eme
un
superchamp.
Eq.(\ref{supchp1}) montre aussi que le superchamp le plus g\'en\'eral
$S(x,\theta,\bar \theta)$ contient 16 degr\'es de libert\'e fermioniques
ainsi
que 16 degr\'es de libert\'e bosoniques, ce qui est trop pour d\'ecrire par
exemple
le supermultiplet $(A,B,\Psi,F,G)$ contenant le spineur de Majorana (voir
Section \ref{aux}).
Un moyen de r\'eduire le nombre de degr\'es de libert\'e contenus dans un
superchamp
est l'imposition de contraintes.

\subsubsection{Superchamps chiraux}
\label{spchpa}

Une contrainte possible est l'annulation de certaines d\'eriv\'ees
$D_{\alpha},\bar D_{\dot \alpha}$ du
superchamp. Afin que cette contrainte soit consistante, les d\'eriv\'ees
$D_{\alpha},\bar D_{\dot \alpha}$ doivent se transformer de fa\c{c}on
covariante sous la supersym\'etrie:
\begin{eqnarray}
D_{\alpha} (\delta S)=\delta (D_{\alpha} S), \
\bar D_{\dot \alpha} (\delta S)=\delta (\bar D_{\dot \alpha} S)
\label{supchp2}
\end{eqnarray}
D'apr\`es Eq.(\ref{trSUSYg2c}), cette condition s'\'ecrit:
\begin{eqnarray}
 \{ Q_{\alpha},D_{\beta} \}= \{ \bar Q_{\dot \alpha},D_{\beta} \}=
 \{ Q_{\alpha},\bar D_{\dot \beta} \}= \{ \bar Q_{\dot \alpha},\bar D_{\dot
\beta} \}=0.
\label{supchp3}
\end{eqnarray}
D'apr\`es Eq.(\ref{gensup4}), des expressions possibles pour $D_{\alpha}$
et $\bar D_{\dot \alpha}$ sont donc par exemple,
\begin{eqnarray}
D_{\alpha} &=&  {\partial  \over \partial \theta^\alpha} -i
(\sigma^{\mu})_{\alpha \dot \alpha} \bar \theta^{\dot \alpha} \partial_{\mu}, \cr
\bar D_{\dot \alpha} &=& {\partial  \over \partial \bar \theta^{\dot \alpha}}
-i
\theta^{\alpha} (\sigma^{\mu})_{\alpha \dot \alpha} \partial_{\mu}.
\label{supchp4}
\end{eqnarray}
Le superchamp chiral droit $\bar \Phi$ est d\'efini par: $D_{\alpha} \bar
\Phi = 0$
et le superchamp chiral gauche $\Phi$ par $\bar D_{\dot \alpha} \Phi = 0$.
La forme des superchamps chiraux s'obtient facilement en remarquant que:
\begin{eqnarray}
D_{\alpha} \bar \theta = 0, \ et \
D_{\alpha} \bar y^\mu = 0  \ avec  \  \bar y^\mu= x^\mu
+ i \theta \sigma^{\mu} \bar \theta, \cr
\bar D_{\dot \alpha} \theta = 0, \ et \
\bar D_{\dot \alpha} y^\mu = 0  \ avec  \  y^\mu= x^\mu
- i \theta \sigma^{\mu} \bar \theta.
\label{supchp5}
\end{eqnarray}
Par exemple, d'apr\`es Eq.(\ref{supchp5}), un superchamp chiral gauche
est une fonction de $\theta$ et $y^\mu$ exclusivement:
\begin{eqnarray}
\Phi(y,\theta)=z(y)+ \sqrt 2 \theta \psi (y)- \theta \theta f(y).
\label{supchp6}
\end{eqnarray}
D\'eveloppant  $z(y)$, $\psi(y)$ et $f(y)$ autour de $x^\mu$,
l'expression \ref{supchp6} devient,
\begin{eqnarray}
\Phi(x,\theta,\bar \theta)&=&z(x)+ \sqrt 2 \theta \psi (x) - \theta \theta
f(x)
-i (\theta \sigma^{\mu} \bar \theta)  \partial_{\mu} z(x) \cr &&
+{i \over \sqrt 2} \theta \theta (\partial_{\mu} \psi (x) \sigma^{\mu} \bar
\theta)
-{1 \over 4} \theta \theta \bar \theta \bar \theta \partial_{\mu} 
{\partial}^{\mu} z(x).
\label{supchp7}
\end{eqnarray}
De m\^eme, un superchamp chiral droit s'\'ecrit:
\begin{eqnarray}
\bar \Phi(x,\theta,\bar \theta)&=&\bar z(x)
+ \sqrt 2 \bar \theta \bar \psi (x) - \bar \theta \bar \theta \bar f(x)
+i (\theta \sigma^{\mu} \bar \theta)  \partial_{\mu} \bar z(x) \cr &&
-{i \over \sqrt 2} \bar \theta \bar \theta (\theta \sigma^{\mu} 
{\partial}_{\mu} \bar \psi (x))
-{1 \over 4} \theta \theta \bar \theta \bar \theta \partial_{\mu} 
{\partial}^{\mu} \bar z(x).
\label{supchp8}
\end{eqnarray}
Notons qu'un produit de superchamps chiraux est aussi un superchamp chiral
car,
\begin{eqnarray}
\bar D_{\dot \alpha} \Phi^n= n \Phi^{n-1} \bar D_{\dot \alpha} \Phi=0, \cr
D_{\alpha} \bar \Phi^n= n \bar \Phi^{n-1} D_{\alpha} \bar \Phi=0.
\label{supchp8b}
\end{eqnarray}

En calculant
$\delta \Phi = i ({\cal E}^{\alpha} Q_{\alpha} + \bar {\cal E}_{\dot \alpha} \bar
Q^{\dot \alpha}) \Phi$
au moyen des expressions \ref{gensup4} des g\'en\'erateurs de la
supersym\'etrie,
et en comparant le r\'esultat \`a
$\delta \Phi(x,\theta,\bar \theta)=\delta z(x)+ \sqrt 2 \theta \delta \psi
(x)
- \theta \theta \delta f(x)$, on trouve les lois de transformations
supersym\'etriques
suivantes,
\begin{eqnarray}
\delta z(x) &=& \sqrt {2} {\cal E}  \psi (x), \cr
\delta \psi (x)&=& - \sqrt {2} f(x) {\cal E} - i \sqrt {2}
(\sigma^{\mu} \bar {\cal E} ) \partial_{\mu} z(x),  \cr
\delta f(x)&=& - i \sqrt {2} \partial_{\mu} \psi (x) \sigma^{\mu} \bar {\cal E} .
\label{supchp9}
\end{eqnarray}

Les superchamps chiraux gauches $\Phi$ (droits $\bar \Phi$) contiennent donc
les champs
$\psi_{\alpha}=\psi_L$ ($\bar \psi^{\dot \alpha}=\psi_R$) de spin 1/2 et de
chiralit\'e
gauche (droite) ainsi que leur partenaire supersym\'etrique $z$ ($\bar z$)
qui sont des
champs scalaires (voir Section \ref{2c}).
Par cons\'equent, les superchamps chiraux permettent de d\'ecrire les
supermultiplets de mati\`ere, c'est \`a dire les supermultiplets contenant
les leptons,
les sleptons, les quarks, les squarks, les bosons de Higgs et les higgsinos.

\subsubsection{Superchamps vectoriels}

Un autre choix de contrainte covariante consiste \`a imposer la condition de
r\'ealit\'e
\`a un superchamp. Un superchamp r\'eel est dit ``vectoriel''. Il est not\'e
$V$ et
v\'erifie $V=V^\dagger$.
Nous verrons que ce superchamp contient les bosons de jauge si l'on \'ecrit
la transformation de jauge comme,
\begin{eqnarray}
V \to V + \Phi + \Phi^\dagger,
\label{supchp10}
\end{eqnarray}
o\`u $\Phi$ est un superchamp chiral gauche.
Dans la jauge dite de Wess-Zumino, le superchamp vectoriel ne contient que
trois champs:
\begin{eqnarray}
V (x^\mu,\theta,\bar \theta) = \theta \sigma^{\mu} \bar \theta  v_{\mu}(x)
+ i \theta \theta \bar \theta \bar \l - i \bar \theta \bar \theta \theta \l
+{1 \over 2} \theta \theta \bar \theta \bar \theta D(x).
\label{supchp11}
\end{eqnarray}
Notons qu'une combinaison lin\'eaire \`a coefficients r\'eels de superchamps
r\'eels
est \'evidemment aussi un superchamp r\'eel.

Par une m\'ethode identique \`a celle utilis\'ee dans la Section
\ref{spchpa}, on obtient
les lois de transformations supersym\'etriques des champs contenus dans
un superchamp vectoriel:
\begin{eqnarray}
\delta v_{\mu} (x)&=&i {\cal E} \sigma^{\mu} \bar \l(x) -i \l (x) \sigma^{\mu} \bar
{\cal E}, \cr
\delta \l (x)&=&i D(x) {\cal E}  - {1 \over 2} (\sigma^{\mu} \bar \sigma^{\nu} {\cal E} )
(\partial_{\mu} v_{\nu} (x) - \partial_{\nu} v_{\mu} (x)), \cr
\delta D(x) &=& {\cal E} \sigma^{\nu} \partial_{\nu} \bar \l(x) +
\partial_{\nu} \l(x) \sigma^{\nu} \bar {\cal E}.
\label{supchp12}
\end{eqnarray}

Les superchamps vectoriels contiennent donc des champs $v_{\mu} (x)$ de spin
1
et leur partenaire supersym\'etrique: les champs $\l (x)$ de spin 1/2.
Par cons\'equent, les superchamps vectoriels permettent de d\'ecrire les
supermultiplets de jauge, c'est \`a dire les supermultiplets contenant
les bosons de jauges et les jauginos.

\subsection{Lagrangiens supersym\'etriques}
\label{LAGsec}

Remarquons que le champ auxiliaire $f(x)$ du superchamp chiral gauche
$\Phi$, c'est \`a dire la partie en $\theta \theta$
du superchamp chiral gauche $\Phi$ (voir Eq.(\ref{supchp7})), se transforme
en une d\'eriv\'ee sous la
supersym\'etrie d'apr\`es Eq.(\ref{supchp9}). L'int\'egrale sur $d^4 x$ 
de la partie en $\theta \theta$
d'un superchamp chiral gauche est donc
invariante sous les transformations SUSY. Une puissance $\Phi^n$ \'etant
aussi un superchamp chiral (voir Eq.(\ref{supchp8b})),
l'int\'egrale sur $d^4 x$ de sa partie en $\theta \theta$, 
not\'ee $[\Phi^n]_{\theta \theta}$, est aussi
invariante sous les transformations SUSY. Les int\'egrales sur $d^4 x$ 
des parties $[\Phi^{n \dagger}]_{\theta \theta}$ et $[\bar \Phi^n]_{\bar
\theta \bar \theta}$
sont bien s\^ur aussi invariantes sous les transformations SUSY.
De m\^eme, le champ auxiliaire $D(x)$ du superchamp vectoriel $V$, c'est \`a
dire la partie en
$\theta \theta \bar \theta \bar \theta$ du superchamp vectoriel $V$ (voir
Eq.(\ref{supchp11})),
se transforme en une d\'eriv\'ee sous la supersym\'etrie d'apr\`es
Eq.(\ref{supchp12}). L'int\'egrale sur $d^4 x$ 
de la partie en $\theta \theta \bar \theta \bar \theta$ d'un superchamp
vectoriel est donc invariante sous les transformations SUSY.
Le produit $\Phi^\dagger \Phi$ \'etant un superchamp vectoriel, 
l'int\'egrale sur $d^4 x$ de sa partie en
$\theta \theta \bar \theta \bar \theta$,
not\'ee $[\Phi^\dagger \Phi]_{\theta \theta \bar \theta \bar \theta}$, est
aussi invariante sous les transformations SUSY.
Nous pouvons donc \'ecrire un premier lagrangien invariant sous les
transformations supersym\'etriques:
\begin{eqnarray}
{\cal L}= \Sigma_{i,j,k} \bigg ( [\Phi_i^\dagger \Phi_i]_{\theta \theta \bar
\theta \bar \theta} +
[W (\Phi)]_{\theta \theta} + [\bar W (\Phi^{\dagger})]_{\bar \theta \bar
\theta} \bigg ),
\label{LAG1}
\end{eqnarray}
o\`u $W(\Phi)$ est un polyn\^ome des superchamps chiraux gauches appel\'e
superpotentiel et valant
dans une th\'eorie renormalisable,
\begin{eqnarray}
W(\Phi) &=& a_i \Phi_i + {1 \over 2} m_{ij} \Phi_i \Phi_j + {1 \over 3}
\l_{ijk} \Phi_i \Phi_j \Phi_k, \cr
\bar W (\Phi^{\dagger}) &=& a_i^{\dagger} \Phi_i^{\dagger} + {1 \over 2}
m_{ij}^{\dagger} \Phi_i^{\dagger} \Phi_j^{\dagger}
+ {1 \over 3} \l_{ijk}^{\dagger} \Phi_i^{\dagger} \Phi_j^{\dagger}
\Phi_k^{\dagger},
\label{LAG2}
\end{eqnarray}
$a_i$, $m_{ij}$ et $\l_{ijk}$ \'etant des constantes. Des puissances
sup\'erieures de superchamps dans le superpotentiel
ne sont pas interdites mais m\`enent \`a des th\'eories non renormalisables.
Prendre la partie en $\theta \theta$, $\bar \theta \bar \theta$ ou $\theta
\theta \bar \theta \bar \theta$
revient \`a d\'eriver un certain nombre de fois par rapport \`a $\theta$ ou
$\bar \theta$. La d\'erivation \'etant
\'equivalente \`a l'int\'egration pour une variable de Grassmann, le
lagrangien \ref{LAG1} peut aussi s'\'ecrire,
\begin{eqnarray}
{\cal L}= \Sigma_{i,j,k} \bigg ( \int \Phi_i^\dagger \Phi_i d^2 \theta d^2
\bar \theta +
\int W (\Phi) d^2 \theta + \int \bar W (\Phi^{\dagger}) d^2 \bar \theta
\bigg ).
\label{LAG3}
\end{eqnarray}
Le lagrangien obtenu \`a partir du lagrangien \ref{LAG3} en rempla\c{c}ant
$\Phi$ par
$\Phi(x,\theta)=z(x)+ \sqrt 2 \theta \psi (x) - \theta \theta f(x)$ (voir
Eq.(\ref{supchp7})) est:
\begin{eqnarray}
{\cal L} &=& \Sigma_{i,j,k} \bigg ( {i \over 2} (\psi_i \sigma^{\mu} 
{\partial}_{\mu} \bar \psi_i
- \partial_{\mu} \psi_i \sigma^{\mu} \bar \psi_i) + (\partial^{\mu} z_i)
(\partial_{\mu} z_i^\dagger)
+ f_i f_i^\dagger - a_i f_i  - m_{ij} ( z_i f_j + {1 \over 2} \psi_i \psi_j
)
\cr &&- \l_{ijk} ( z_i z_j f_k + z_i \psi_j \psi_k )  +  c.c. \bigg ),
\label{LAG4}
\end{eqnarray}
d'o\`u l'on tire, en d\'erivant par rapport \`a $f_i$, l'\'equation du
mouvement des champs auxiliaires $f_i$:
\begin{eqnarray}
f_i^\dagger =  a_i + m_{ij} z_j + \l_{ijk}  z_j z_k= {\partial W(z) \over
\partial z_i},
\label{LAG5}
\end{eqnarray}
$W(z)$ \'etant d\'efini comme en Eq.(\ref{LAG2}) en rempla\c{c}ant les
superchamps $\Phi$ par
leur composante scalaire $z$. Rempla\c{c}ant $f_i$ par son expression
\ref{LAG5} dans
l'expression \ref{LAG4} du lagrangien, on obtient:
\begin{eqnarray}
{\cal L} &=& \Sigma_{i,j} \bigg ( {i \over 2} (\psi_i \sigma^{\mu} 
{\partial}_{\mu} \bar \psi_i
- \partial_{\mu} \psi_i \sigma^{\mu} \bar \psi_i) + (\partial^{\mu} z_i)
(\partial_{\mu} z_i^\dagger) \cr &&
- \vert {\partial W(z) \over \partial z_i} \vert^2
- {1 \over 2} {\partial^2 W(z) \over \partial z_i \partial z_j} \psi_i
\psi_j 
- {1 \over 2} {\partial^2 \bar W(z^\star) \over \partial z_i^\star \partial
z_j^\star}
\bar \psi_i \bar \psi_j \bigg ).
\label{LAG6}
\end{eqnarray}
En comparant Eq.(\ref{LAG1}) et Eq.(\ref{LAG6}), on voit que
$[\Phi_i^\dagger \Phi_i]_{\theta \theta \bar \theta \bar \theta}$ contient
les termes cin\'etiques des spineurs
\`a 2 composantes $\psi_i$ (quarks et leptons) et de leur partenaire
supersym\'etrique:
les champs scalaires $z_i$ (squarks et sleptons).
On observe aussi que la partie du lagrangien associ\'ee au superpotentiel
s'\'ecrit,
\begin{eqnarray}
{\cal L}_W=  [W (\Phi)]_{\theta \theta}+[\bar W (\Phi^{\dagger})]_{\bar
\theta \bar \theta}
=\int W (\Phi) d^2 \theta + \int \bar W (\Phi^{\dagger}) d^2 \bar \theta
={\cal L}_W^1 + {\cal L}_W^2,
\label{LAG7}
\end{eqnarray}
\begin{eqnarray}
{\cal L}_W^1=
\Sigma_{i} \bigg (
- \vert {\partial W(z) \over \partial z_i} \vert^2
\bigg ),
\label{LAG71}
\end{eqnarray}
\begin{eqnarray}
{\cal L}_W^2=
\Sigma_{i,j} \bigg  (
- {1 \over 2} {\partial^2 W(z)
\over \partial z_i \partial z_j}
\psi_i \psi_j
- {1 \over 2} {\partial^2 \bar W(z^\star)
\over \partial z_i^\star \partial z_j^\star}
\bar \psi_i \bar \psi_j
\bigg ).
\label{LAG72}
\end{eqnarray}
Nous voyons d'apr\`es cette expression que
${\cal L}_W$ contient les couplages du potentiel et permet donc notamment de
d\'ecrire
les interactions de Yukawa qui couplent les bosons de Higgs $z_i$ aux
quarks et leptons $\psi_i$.

\subsection{Th\'eories de jauge supersym\'etriques}

Le but de cette partie est de g\'en\'eraliser le formalisme des th\'eories
invariantes de jauge
au cas supersym\'etrique. Nous nous appuierons pour cela sur le lagrangien
supersym\'etrique g\'en\'eral
obtenu dans Eq.(\ref{LAG1}).

Tout d'abord, nous remarquons qu'il manque au lagrangien de Eq.(\ref{LAG1})
les termes cin\'etiques
des bosons de jauge $v_{a \mu}$ et de leur partenaire supersym\'etrique de
spin 1/2: les jauginos $\l_a$
(\`a 2 composantes). Ces termes cin\'etiques sont contenus dans,
\begin{eqnarray}
{\cal L}&=& \Sigma_{a} \bigg ( {1 \over 4} [W_a^\alpha W_{a \alpha}]_{\theta
\theta }+
{1 \over 4} [\bar W_{a \dot \alpha} \bar W_a^{\dot \alpha}]_{\bar \theta
\bar \theta} \bigg ) \cr &=&
\Sigma_{a} \bigg ( {i \over 2} (\l_a \sigma^{\mu} \partial_{\mu} \bar \l_a
- \bar \l_a \sigma^{\mu} \partial_{\mu} \l_a) + {1 \over 2} D_a^2 + F_{a \mu
\nu} F_a^{\mu \nu} \bigg ),
\label{tjs1}
\end{eqnarray}
o\`u,
\begin{eqnarray}
W_a^\alpha&=& - {1 \over 4} (\bar D \bar D) D_\alpha V_a, \cr
\bar W_{a \dot \alpha}&=& - {1 \over 4} (DD) \bar D_{\dot \alpha} V_a,
\label{tjs2}
\end{eqnarray}
et,
\begin{eqnarray}
F_{a \mu \nu} =\partial_{\mu} v_{a \nu}-\partial_{\nu} v_{a \mu}.
\label{tjs3}
\end{eqnarray}
Le lagrangien \ref{tjs1} est invariant sous les transformations
supersym\'etriques car
$W_a^\alpha$ et $\bar W_{a \dot \alpha}$ sont respectivement des superchamps
chiraux gauche et droit puisque
$\bar D_{\dot \alpha} (\bar D \bar D) =0$ et $D_\alpha (DD) =0$.

D'autre part, la transformation sous un groupe de jauge d'un superchamp
chiral s'\'ecrit,
\begin{eqnarray}
\Phi^{\prime} = exp (i \L) \Phi, \ \Phi^{\prime \dagger} = \Phi^\dagger exp
(-i \L^\dagger) ,
 \ \Phi^{c \prime} = exp (-i \L) \Phi^c,
\ avec \ \L = \Sigma_a \L_a T_a,
\label{tjs4}
\end{eqnarray}
%
%
o\`u les $\L_a$ sont des param\`etres r\'eels (d\'ependants de
l'espace-temps dans le cas d'un groupe de jauge
agissant localement) et les $T_a$ sont les g\'en\'erateurs du groupe de
jauge.
Les $\L_a$ peuvent \^etre choisis comme \'etant des superchamps chiraux
gauches
de telle sorte que Eq.(\ref{tjs4}) soit une relation entre superchamps.
L'\'equation Eq.(\ref{tjs4}) montre que les membres d'un m\^eme superchamp
ont les m\^emes nombres quantiques associ\'es au groupe de jauge.
Afin que le terme $[\Phi_i^\dagger \Phi_i]_{\theta \theta \bar \theta \bar
\theta}$
de Eq.(\ref{LAG1}) soit invariant de jauge, il doit \^etre modifi\'e en
$[\Phi_i^\dagger e^V \Phi_i]_{\theta \theta \bar \theta \bar \theta}$ o\`u
$e^V$ se transforme sous
l'action du groupe de jauge par,
\begin{eqnarray}
exp(V)  \to  exp(i \L^\dagger) exp(V) exp(-i \L), \ avec \ V= \Sigma_a V_a
T_a,
\label{tjs5}
\end{eqnarray}
$V_a$ repr\'esentant les superchamps vectoriels.
Au premier ordre, le produit des exponentielles d'op\'erateurs $A$ et $B$
vaut,
\begin{eqnarray}
exp(A)  exp(B) = exp(A + B +{1 \over 2} [A,B]).
\label{tjs6}
\end{eqnarray}
Cette relation nous permet de v\'erifier qu'au premier ordre le terme
$[\Phi_i^\dagger e^V \Phi_i]_{\theta \theta \bar \theta \bar \theta}$ est
bien invariant sous
l'action d'un groupe de jauge ab\'elien puisque pour un tel groupe $[T_a,
T_b] = 0$.
De plus, selon Eq.(\ref{tjs6}), pour un groupe de jauge ab\'elien,
Eq.(\ref{tjs5}) devient au premier ordre,
\begin{eqnarray}
exp(V)  \to  exp(V + i \L^\dagger -i \L) \ \Leftrightarrow \  V  \to  V  -i
\L + i \L^\dagger .
\label{tjs7}
\end{eqnarray}
On retrouve donc bien la transformation de Eq.(\ref{supchp10}) avec $\Phi =
- i \L$.
Cette introduction de superchamps vectoriels, qui permet de rendre la
th\'eorie invariante de jauge,
est analogue \`a l'introduction de champs de jauge dans les
th\'eories de jauge non supersym\'etriques. Par ailleurs, la transformation
de Eq.(\ref{tjs5})
ne laisse pas invariant les termes $[W_a^\alpha W_{a \alpha}]_{\theta \theta
}$ et
$[\bar W_{a \dot \alpha} \bar W_a^{\dot \alpha}]_{\bar \theta \bar \theta}$
de Eq.(\ref{tjs1}).
En revanche, si l'on red\'efini $W_a^\alpha$ et $\bar W_{a \dot \alpha}$
par,
\begin{eqnarray}
W^\alpha&=& - {1 \over 4} (\bar D \bar D) e^{-V} D_\alpha e^V, \cr
\bar W_{\dot \alpha}&=& - {1 \over 4} (DD) e^{-V} \bar D_{\dot \alpha} e^V,
\label{tjs8}
\end{eqnarray}
ces superchamps chiraux subissent les transformations de jauge,
\begin{eqnarray}
W^\alpha &\to & exp(i \L) W^\alpha exp(-i \L), \cr
\bar W_{\dot \alpha} &\to & exp(i \L) \bar W_{\dot \alpha} exp(-i \L),
\label{tjs8p}
\end{eqnarray}
de telle sorte que les termes $Tr[W_a^\alpha W_{a \alpha}]_{\theta \theta }$
et
$Tr[\bar W_{a \dot \alpha} \bar W_a^{\dot \alpha}]_{\bar \theta \bar
\theta}$,
la trace \'etant prise sur les indices de la repr\'esentation du groupe de
jauge,
sont invariants de jauge.
Finalement, dans une th\'eorie supersym\'etrique invariante de jauge,
le superpotentiel doit avoir chacun de ces termes invariants de jauge.

Il faut aussi introduire les constantes de couplage $g$ associ\'ees aux
groupes de jauge.
Pour cela, on red\'efini les champs des supermultiplets vectoriels par
$\l_a \to 2 g \l_a$,
$v_{a \mu} \to 2 g v_{a \mu}$ et $D_a \to 2 g D_a$, ce qui \'equivaut \`a
red\'efinir
le superchamp vectoriel lui-m\^eme par $V \to 2gV$.

Le lagrangien g\'en\'erique d'une th\'eorie supersym\'etrique invariante
sous un groupe
de jauge a donc la forme suivante,
\begin{eqnarray}
{\cal L}&=& \Sigma_{i,j,k} \bigg ( \bigg [\Phi_i^\dagger (e^{2gV})_{ij} \Phi_j
\bigg ]_{\theta \theta \bar \theta \bar \theta} +
\bigg [ W (\Phi)+{1 \over 16 g^2 n_R} Tr(W^{\alpha} W_{\alpha}) \bigg
]_{\theta \theta} \cr && +
\bigg [\bar W (\Phi^{\dagger}) 
+ {1 \over 16 g^2 n_R} Tr(\bar W_{\dot \alpha} \bar W^{\dot \alpha}) \bigg
]_{\bar \theta \bar \theta} \bigg ),
\label{tjs9}
\end{eqnarray}
o\`u le nombre $n_R$ est d\'efini par $[T_a, T_b] = n_R \delta_{ab}$.
Le facteur $1/g^2$ dans Eq.(\ref{tjs9}) permet
de ne pas g\'en\'erer de puissances de la constante $g$ sup\'erieures ou
\'egales \`a 2.
Apr\`es calcul, le lagrangien \ref{tjs9} devient,
\begin{eqnarray}
{\cal L} &=& \Sigma_{a,i,j,k} \bigg (
{i \over 2} \psi_i \sigma^{\mu} (D_{\mu} \bar \psi)_i
- {i \over 2} (D_{\mu} \psi)_i \sigma^{\mu} \bar \psi_i
+ (D_{\mu} z)_i^{\dagger} (D^{\mu} z)_i
- {1 \over 4} F_{a \mu \nu} F_a^{\mu \nu}
\cr && + {i \over 2} \l_a \sigma^{\mu} (D_{\mu} \bar \l)_a
- {i \over 2} (D_{\mu} \l)_a \sigma^{\mu} \bar \l_a
+ i \sqrt 2  g (\bar \psi_i \bar \l_a) T_{aij} z_j
- i \sqrt 2  g z_i^{\dagger} T_{aij} (\psi_j \l_a) \cr &&
- {1 \over 2} {\partial^2 W(z) \over \partial z_i \partial z_j} \psi_i
\psi_j
- {1 \over 2} {\partial^2 \bar W(z^\dagger) \over \partial z_i^\dagger 
{partial} z_j^\dagger}
\bar \psi_i \bar \psi_j - V(z_i,z_j^\dagger) \bigg ),
\label{tjs10}
\end{eqnarray}
o\`u le superpotentiel $W$ est donn\'e dans Eq.(\ref{LAG2}), les
d\'eriv\'ees covariantes
sont d\'efinies par,
\begin{eqnarray}
(D_{\mu} z)_i&=&\partial_{\mu} z_i +ig v_{a \mu} T_{aij} z_j, \cr
(D_{\mu} \psi)_i&=&\partial_{\mu} \psi_i +ig v_{a \mu} T_{aij} \psi_j, \cr
(D_{\mu} \l)_a&=&\partial_{\mu} \l_a - g f_{a b c} v_{b \mu} \l_c,
\label{tjs11}
\end{eqnarray}
avec,
\begin{eqnarray}
F_a^{\mu \nu}=\partial^{\mu} v_a^{\nu}-\partial^{\nu} v_a^{\mu} - g f_{a b
c} v_b^{\mu} v_c^{\nu},
\label{tjs12}
\end{eqnarray}
\begin{eqnarray}
[T_a, T_b] = i f_{a b c} T^c,
\label{tjs13}
\end{eqnarray}
et le potentiel vaut,
\begin{eqnarray}
V(z_i,z_j^\dagger)= \vert f_i \vert^2 + {1 \over 2} D_a^2,
\label{tjs14}
\end{eqnarray}
avec,
\begin{eqnarray}
f_i = {\partial W(z) \over \partial z_i},
\label{tjs15}
\end{eqnarray}
\begin{eqnarray}
D_a = g z_i^\dagger T_{aij} z_j.
\label{tjs16}
\end{eqnarray}

\subsection{Th\'eor\`eme de non-renormalisation}
\label{TNR}

Introduisons tout d'abord un vocabulaire sp\'ecifique.
On appelle respectivement terme F (F term) et terme D (D term) les termes
$[...]_{\theta \theta}$
(ou $[...]_{\bar \theta \bar \theta}$)
et $[...]_{\theta \theta \bar \theta \bar \theta}$ du lagrangien (voir
Eq.(\ref{tjs9})).
Cette appellation est due au fait que l'on d\'enote usuellement $f(x)$ et
$D(x)$ les champs auxiliaires
qui sont \`a l'origine de tels termes (voir Section \ref{LAGsec}).

Le th\'eor\`eme de non-renormalisation peut \^etre \'enonc\'e comme suit:
Dans les th\'eories supersym\'etriques, les corrections \`a l'ordre des
boucles sont toujours des termes D.

Une cons\'equence du th\'eor\`eme de non-renormalisation est que
les param\`etres du superpotentiel (voir Eq.(\ref{LAG2}) et
Eq.(\ref{tjs9})), et notamment les masses des champs scalaires comme 
les bosons de Higgs, ne re\c{c}oivent pas de corrections quantiques 
quadratiques tant que la \susi est pr\'eserv\'ee. Cette cons\'equence
est \`a l'origine de la r\'esolution du probl\`eme de hi\'erarchie,
comme nous l'avons discut\'e dans la Section \ref{hier}.

Le th\'eor\`eme de non-renormalisation a pour origine l'absence de 
divergences quadratiques dans les th\'eories supersym\'etriques. 
Cette absence de divergences quadratiques
dans les th\'eories \susiqs provient de l'annulation de la somme de graphes
\'echangeant des particules du \ms avec les graphes associ\'es impliquant
leur partenaire supersym\'etrique.

\section{Brisure de la supersym\'etrie}
\label{BRIsec}

Nous avons vu dans Eq.(\ref{consAL}) que dans les th\'eories \susiqs les
particules membres d'un m\^eme
supermultiplet ont une masse identique (cette propri\'et\'e a aussi \'et\'e
illustr\'ee dans la Section \ref{Majo}).
Or aucune particule supersym\'etrique de masse \'egale \`a une masse de
particule du \ms n'a \'et\'e d\'ecouverte
aupr\`es des collisionneurs jusqu'\`a aujourd'hui. Nous en d\'eduisons que
si la nature est effectivement supersym\'etrique, les masses des partenaires
\susiq sont sup\'erieures aux
masses des particules du \ms et la \susi est donc bris\'ee.

Quels sont les crit\`eres de brisure spontan\'ee de la supersym\'etrie ?
La \susi est bris\'ee spontan\'ement si le vide n'est pas invariant sous la
supersym\'etrie,
ce qui peut s'\'ecrire,
\begin{eqnarray}
< \delta \psi > = < 0 \vert i ({\cal E}^{\alpha} Q_{\alpha}
+ \bar {\cal E}_{\dot \alpha} \bar Q^{\dot \alpha}) \psi \vert 0 > \neq 0,
\label{BRISS1}
\end{eqnarray}
o\`u $Q$ est le g\'en\'erateur de la \susi et $\psi$ est un spineur. La
valeur moyenne dans le vide
des champs fermioniques est nulle \`a cause de l'invariance de Lorentz. On
ne peut donc pas exprimer
de condition de brisure spontan\'ee de la \susi sur $< \delta \phi >$,
$\phi$ \'etant un boson,
qui doit toujours \^etre nul.
D'apr\`es les transformations \susiqs des spineurs $\psi (x)$ appartenant aux
superchamps chiraux
(Eq.(\ref{supchp9})) et $\l (x)$ appartenant aux superchamps vectoriels
(Eq.(\ref{supchp12})), la condition
de brisure spontan\'ee de SUSY de Eq.(\ref{BRISS1}) est \'equivalente \`a,
\begin{eqnarray}
< f_i >  \neq 0, \ ou \  < D_a >  \neq 0,
\label{BRISS2}
\end{eqnarray}
$f_i$ et $D_a$ \'etant respectivement les champs auxiliaires des
supermultiplets chiraux et vectoriels
(voir Eq.(\ref{tjs15}) et Eq.(\ref{tjs16})). La condition de brisure
spontan\'ee de SUSY globale peut s'\'ecrire
diff\'eremment: En prenant la valeur moyenne dans le vide de la relation
$ \{ Q_{\alpha}, \bar Q_{\dot \beta} \} = -2 P_{\mu} (\sigma^{\mu})_{\alpha
\dot
\beta}$
(voir Eq.(\ref{ALsu2c})), on obtient,
\begin{eqnarray}
< 0 \vert Q_{\alpha} \bar Q_{\dot \beta} + \bar Q_{\dot \beta} Q_{\alpha}
\vert 0 > =
2 < 0 \vert  H \vert 0 > \delta_{\alpha \dot \beta},
\label{BRISS3}
\end{eqnarray}
soit,
\begin{eqnarray}
\vert Q_{\alpha} \vert 0 > \vert ^2= <V>,
\label{BRISS4}
\end{eqnarray}
o\`u $H$ est l'hamiltonien et $<V>$ la valeur moyenne du potentiel dans le
vide. D'apr\`es Eq.(\ref{BRISS4}),
la condition de brisure spontan\'ee de SUSY globale peut donc aussi s'\'ecrire,
\begin{eqnarray}
<V> \neq 0.
\label{BRISS5}
\end{eqnarray}
Le potentiel \'etant positif ou nul d'apr\`es Eq.(\ref{tjs14}), la condition
\ref{BRISS5}
est \'equivalente \`a,
\begin{eqnarray}
<V> > 0.
\label{BRISS6}
\end{eqnarray}
Nous v\'erifions de plus d'apr\`es Eq.(\ref{tjs14}) que les 2 conditions
\ref{BRISS2} et \ref{BRISS6}
de brisure spontan\'ee de SUSY sont \'equivalentes.

En fait, une brisure spontan\'ee de la \susi dans le secteur observable
engendre une hi\'erarchie entre les masses
des particules qui n'est pas r\'ealiste \cite{Dered}.
Plus pr\'ecis\'ement,
apr\`es une brisure spontan\'ee de la supersym\'etrie, les masses des
particules fermioniques du
\ms sont sup\'erieures \`a certaines des masses de leur partenaire \susiq
scalaire.

La \susi si elle existe doit donc subir une brisure spontan\'ee
dans un secteur diff\'erent du secteur observable, appel\'e secteur ``cach\'e'',
et cette brisure de SUSY doit \^etre m\'edi\'ee au secteur observable. 
Il existe aujourd'hui deux principaux types de mod\`eles dans lesquels ce 
sc\'enario de brisure de SUSY peut \^etre r\'ealis\'e: Les mod\`eles de 
supergravit\'e (voir Chapitre \ref{SUGRA}) et les mod\`eles dits
GMSB (Gauge Mediated Supersymmetry Breaking) \cite{GiuRat}.
Dans les mod\`eles de supergravit\'e, le secteur cach\'e n'interagit 
avec le secteur observable que par les interactions de type gravitationnel,
et par cons\'equent la brisure de SUSY est m\'edi\'ee au secteur observable
par les interactions de gravitationnelles.
Dans les mod\`eles GMSB, le secteur cach\'e n'interagit 
avec le secteur observable que par les interactions de jauge,
et par cons\'equent la brisure de SUSY est m\'edi\'ee au secteur observable
par les interactions de jauge. 

Les termes de brisure de la supersym\'etrie dans le 
secteur observable sont soumis \`a quelques contraintes.
D'une part, les termes de brisure de la supersym\'etrie ne doivent
pas introduire
de divergences quadratiques qui engendreraient des corrections radiatives
\`a la masse des bosons
de Higgs et emp\^echeraient la r\'esolution du probl\`eme de hi\'erarchie
(voir Section \ref{hier}).
Les termes de brisure de la supersym\'etrie ne g\'en\'erant pas de
divergences quadratiques
sont appel\'es termes de brisure douce de la supersym\'etrie ou
termes doux (`soft terms'). Les termes doux sont de mani\`ere
g\'en\'erale,
\begin{eqnarray}
{\cal L}_{doux}= \sum_{ijk}  \bigg (
f(z)+f^\dagger (z^\dagger) + m_0^{ij \ 2} z_i z_j^\dagger
- {1 \over 2} m_{1/2}^{ij} (\l_i \l_j + \bar \l_i \bar \l_j) \bigg ), 
\cr avec \ \ \ 
f(z)=X^i z_i + {1 \over 2} Y^{ij} z_i z_j + {1 \over 3} A^{ijk} z_i z_j z_k,
\label{BRISS7}
\end{eqnarray}
o\`u les $z_i$ sont des champs scalaires et les $\l_i$ des champs de spin
1/2
appartenant \`a un supermultiplet vectoriel. D'autre part, les masses des
particules
\susiqs $m_0$ et $m_{1/2}$ provenant des termes de brisure de la
supersym\'etrie (voir Eq.(\ref{BRISS7}))
repr\'esentent typiquement la diff\'erence not\'ee $\tilde m$ (voir Section
\ref{hier})
entre les masses des particules du \ms et de leur partenaire
supersym\'etrique. Or nous avons vu dans
la Section \ref{hier} que $\tilde m$ doit \^etre typiquement inf\'erieure au
$TeV$ afin que les corrections
radiatives li\'ees aux divergences quadratiques, et provenant de la
non-annulation entre des graphes
\'echangeant des particules du \ms et les graphes associ\'es qui impliquent
leur partenaire supersym\'etrique,
n'engendrent pas des corrections \`a la masse des bosons de Higgs
sup\'erieures au $TeV$
ce qui assure la conservation de la hi\'erarchie de masse existant au
niveau des arbres.

Dans le cadre des mod\`eles de supergravit\'e aussi bien que 
des mod\`eles GMSB, les termes de brisure de la supersym\'etrie dans le 
secteur observable peuvent \^etre des termes de brisure douce de  
SUSY g\'en\'erant des masses de particules \susiqs inf\'erieures 
au $TeV$.

En conclusion, si la nature est r\'eellement supersym\'etrique, la \susi
doit \^etre bris\'ee spontan\'ement dans un secteur cach\'e
et les termes de brisure de SUSY
dans le secteur observable doivent \^etre des termes doux (c'est \`a
dire de la forme de Eq.(\ref{BRISS7}))
g\'en\'erant des masses $\tilde m_z$ et $\tilde m_{\l}$ de particules
\susiqs sup\'erieures aux masses des particules du \ms 
et inf\'erieures au $TeV$. Les mod\`eles de supergravit\'e ainsi que 
les mod\`eles GMSB permettent de r\'ealiser un tel sc\'enario de
brisure de la supersym\'etrie.

\section{Mod\`ele Standard Supersym\'etrique Minimal (MSSM)}

Nous d\'ecrivons dans cette Section le Mod\`ele Standard Supersym\'etrique
Minimal (MSSM).
Le MSSM est minimal en ce sens qu'il contient le nombre minimum de
particules n\'ecessaire
\`a l'extension \susiq du Mod\`ele Standard. Le MSSM est parfaitement
coh\'erent, il est bien d\'efini th\'eoriquement et peut \^etre test\'e
exp\'erimentalement.

\subsection{Le contenu en particules}
\label{MSSMA}

Les superchamps chiraux du MSSM, qui d\'ecrivent les champs de mati\`ere,
sont tous des superchamps chiraux gauches. Ces superchamps sont pr\'esent\'es
dans Eq.(\ref{nombquant})
avec leurs nombres quantiques vis-\`a-vis du groupe de jauge du \ms
$SU(3)_c \times SU(2)_L \times U(1)_Y$.
\begin{equation}
\begin{array}{cccc}
Q:({\bf 3},{\bf 2},1/6), &  U^c:({\bf \bar 3},{\bf 1},-2/3), &  D^c:({\bf \bar
3},{\bf 1},1/3), & 
L:({\bf 1},{\bf 2},-1/2), \\ E^c:({\bf 1},{\bf 1},1), & 
H_1:({\bf 1},{\bf 2},-1/2), &  H_2:({\bf 1},{\bf 2},1/2). & 
\end{array}
\label{nombquant}
\end{equation}
Dans Eq.(\ref{nombquant}), $L$ et $Q$ sont respectivement les superchamps
doublets de $SU(2)_L$
de leptons et de quarks, $E^c$, $U^c$ et $D^c$
sont les superchamps de leptons charg\'es, quarks up et down conjugu\'es de
charge et $H_1$ et $H_2$ sont les 2 superchamps de Higgs doublets de
$SU(2)_L$
s'\'ecrivant,
\begin{equation}
 H_1= \left (\begin{array}{c}  H^0_1 \\ H^-_1  \end{array} \right ), \ \
 H_2= \left (\begin{array}{c}  H^+_2 \\ H^0_2  \end{array} \right ).
\label{wYUKAWA}
\end{equation}
Dans le MSSM, il existe 2 superchamps de Higgs afin de pouvoir assurer
l'annulation des anomalies
du groupe $U(1)_Y$ \cite{ano,GGR,Peskin}. Pour que $U(1)_Y$ n'ait pas
d'anomalies,
il faut que $\sum_{fermions} Y^3 =0$. Or cette relation est vraie dans le
\ms mais n'est \`a priori
plus respect\'ee dans une extension \susiq du Mod\`ele Standard, celle-ci
devant contenir
le partenaire \susiq du boson de Higgs qui est un fermion d'hypercharge
$Y=-1/2$. Afin de r\'etablir
la relation $\sum_{fermions} Y^3 =0$, un second superchamp chiral de Higgs
d'hypercharge $Y=1/2$ peut
\^etre rajout\'e. C'est pr\'ecis\'ement ce qui est fait dans le MSSM (voir
Eq.(\ref{nombquant})).

Les superchamps vectoriels du MSSM contiennent les champs de jauge. Les
superchamps
vectoriels du MSSM associ\'es aux groupes $U(1)_Y$, $SU(2)_L$ et $SU(3)_c$
sont not\'es respectivement
$V_1$, $V_2^a$ et $V_3^b$. D\'efinissons pour la suite les superchamps
vectoriels suivants,
\begin{equation}
 V_2= \sum_{a=1}^3 V_2^a {\sigma^a \over 2}, \ \
 V_3= \sum_{b=1}^8 V_2^b {\l^b \over 2},
\label{mssm1}
\end{equation}
o\`u les $\sigma^a$ sont les matrices de Pauli et les $\l^b$ les matrices de
Gell-Mann.

\subsection{Le lagrangien}
\label{MSSMB}

Le superpotentiel du MSSM, qui est une fonction des superchamps chiraux
gauche,
contient les couplages de Yukawa et le terme de masse du Higgs:
\begin{eqnarray}
W_{MSSM}= h^e_{ij} H_1L_i E^c_j+ h^d_{ij} H_1 Q_i D^c_j+h^u_{ij} H_2 Q_i
U^c_j
+\mu H_1 H_2.
\label{wMSSM}
\end{eqnarray}
Dans Eq.(\ref{wMSSM}), $i$ et $j$ sont des indices de saveur.

Le lagrangien du MSSM est d\'efini par l'expression suivante,
\begin{eqnarray}
{\cal L}_{MSSM}&=& \Sigma_{i,j} \bigg ( \bigg [
Q_i^\dagger e^{{1 \over 6} 2g_1 V_1} e^{2g_2 V_2} e^{2g_3 V_3} Q_i +
U_i^{c \ \dagger} e^{-{2 \over 3} 2g_1 V_1} e^{-2g_3 V_3} U_i^c +
D_i^{c \ \dagger} e^{{1 \over 3} 2g_1 V_1} e^{-2g_3 V_3} D_i^c \cr & + &
L_i^{\dagger} e^{-{1 \over 2} 2g_1 V_1} e^{2g_2 V_2} L_i +
E_i^{c \ \dagger} e^{2g_1 V_1} E_i^c +
H_1^{\dagger} e^{-{1 \over 2} 2g_1 V_1} e^{2g_2 V_2} H_1^c +
H_2^{\dagger} e^{{1 \over 2} 2g_1 V_1} e^{2g_2 V_2} H_2^c
\bigg ]_{\theta \theta \bar \theta \bar \theta} \cr & + &
\bigg [ W_{MSSM} (\Phi) +
{1 \over 16 g_1^2} Tr(W_1^{\alpha} W_{1 \ \alpha}) +
{1 \over  8 g_2^2} Tr(W_2^{\alpha} W_{2 \ \alpha}) +
{1 \over  8 g_3^2} Tr(W_3^{\alpha} W_{3 \ \alpha})
\bigg ]_{\theta \theta} \cr & + &
\bigg [\bar W_{MSSM} (\Phi^{\dagger})  +
{1 \over 16 g_1^2 } Tr(\bar W_{1 \ \dot \alpha} \bar W_1^{\dot \alpha}) +
{1 \over  8 g_2^2 } Tr(\bar W_{2 \ \dot \alpha} \bar W_2^{\dot \alpha}) +
{1 \over  8 g_3^2 } Tr(\bar W_{3 \ \dot \alpha} \bar W_3^{\dot \alpha})
\bigg ]_{\bar \theta \bar \theta} \bigg ). \cr &&
\label{mssm2}
\end{eqnarray}
Pour \^etre totalement complet, le lagrangien ${\cal L}_{MSSM}$ du MSSM doit
aussi contenir les termes
de brisure douce de la \susi qui sont donn\'es dans
Eq.(\ref{BRISS7}).

\subsection{Le spectre supersym\'etrique}
\label{MSSMC}

\subsubsection{Masses des squarks et sleptons}

Nous noterons $\tilde f_{L}$ ($\tilde f_{R}$) le partenaire \susiq de spin
0,
not\'e $z$ ($\bar z$) dans la Section \ref{spchpa},
d'un champ $\psi_{\alpha}=\psi_L$ ($\bar \psi^{\dot \alpha}=\psi_R$) de spin
1/2 et de chiralit\'e
gauche (droite). $\tilde f_{L,R}$ d\'esigne donc les sfermions c'est \`a
dire les squarks
$\tilde q_{L,R}=\tilde u_{L,R},\tilde d_{L,R}$ ainsi que les sleptons
$\tilde l_{L,R}=\tilde e_{L,R},\tilde \nu_L$.

Le lagrangien ${\cal L}_{MSSM}$ du MSSM (voir Eq.(\ref{mssm2}) et
Eq.(\ref{BRISS7})) engendre des termes
de masse de sfermions qui s'\'ecrivent comme suit,
\begin{eqnarray}
-{\cal L}^{scal}_{mass}&=&
(\tilde u_{iL}, \tilde u^{\dagger}_{iR})
\left ( \begin{array}{cc}
(m^{u \ 2}_{LL})_{ij} &  (m^{u \ 2}_{LR})_{ij} \\
(m^{u \ 2 \dagger}_{LR})_{ij} &  (m^{u \ 2}_{RR})_{ij}
\end{array} \right )
\left ( \begin{array}{c}
\tilde u^{\dagger}_{jL} \\
\tilde u_{jR}
\end{array} \right )
\cr  &+&
(\tilde d_{iL}, \tilde d^{\dagger}_{iR})
\left ( \begin{array}{cc}
(m^{d \ 2}_{LL})_{ij} &  (m^{d \ 2}_{LR})_{ij} \\
(m^{d \ 2 \dagger}_{LR})_{ij} &  (m^{d \ 2}_{RR})_{ij}
\end{array} \right )
\left ( \begin{array}{c}
\tilde d^{\dagger}_{jL} \\
\tilde d_{jR}
\end{array} \right )
\cr  &+&
(\tilde e_{iL}, \tilde e^{\dagger}_{iR})
\left ( \begin{array}{cc}
(m^{e \ 2}_{LL})_{ij} &  (m^{e \ 2}_{LR})_{ij} \\
(m^{e \ 2 \dagger}_{LR})_{ij} &  (m^{e \ 2}_{RR})_{ij}
\end{array} \right )
\left ( \begin{array}{c}
\tilde e^{\dagger}_{jL} \\
\tilde e_{jR}
\end{array} \right )
+
(m^{\nu \ 2}_{LL})_{ij} \tilde \nu_{iL} \tilde \nu^{\dagger}_{jL},
\label{MSSMC1}
\end{eqnarray}
o\`u $i$ et $j$ sont des indices de saveur et o\`u,
\begin{eqnarray}
(m^{u \ 2}_{LL})_{ij}&=&m^2_{ij}(\tilde u_L) + (m_u m^\dagger_u)_{ij}
+ \cos 2 \beta m_Z^2 (-{1 \over 2} + {2 \over 3} \sin^2 \theta_W)
\delta_{ij}, \cr
(m^{d \ 2}_{LL})_{ij}&=&m^2_{ij}(\tilde d_L) + (m_d m^\dagger_d)_{ij}
+ \cos 2 \beta m_Z^2 ( {1 \over 2} - {2 \over 3} \sin^2 \theta_W)
\delta_{ij}, \cr
(m^{u \ 2}_{RR})_{ij}&=&m^2_{ij}(\tilde u_R) + (m_u m^\dagger_u)_{ij}
- \cos 2 \beta m_Z^2 {2 \over 3} \sin^2 \theta_W \delta_{ij}, \cr
(m^{d \ 2}_{RR})_{ij}&=&m^2_{ij}(\tilde d_R) + (m_d m^\dagger_d)_{ij}
+ \cos 2 \beta m_Z^2 {1 \over 3} \sin^2 \theta_W \delta_{ij}, \cr
(m^{e \ 2}_{LL})_{ij}&=&m^2_{ij}(\tilde e_L) + (m_e m^\dagger_e)_{ij}
+ \cos 2 \beta m_Z^2 ( {1 \over 2} - \sin^2 \theta_W) \delta_{ij}, \cr
(m^{\nu \ 2}_{LL})_{ij}&=&m^2_{ij}(\tilde \nu_L) + (m_\nu m^\dagger_\nu)_{ij}
- \cos 2 \beta m_Z^2 {1 \over 2} \delta_{ij}, \cr
(m^{e \ 2}_{RR})_{ij}&=&m^2_{ij}(\tilde e_R) + (m_e m^\dagger_e)_{ij}
+ \cos 2 \beta m_Z^2  \sin^2 \theta_W \delta_{ij}, \cr
(m^{u \ 2}_{LR})_{ij}&=&\bigg  ( A_u^{ij} + {\mu \over \tan \beta} \bigg )  (
m_u )_{ij}, \cr
(m^{d \ 2}_{LR})_{ij}&=&\bigg  ( A_d^{ij} + \mu \tan \beta \bigg )  ( m_d
)_{ij}, \cr
(m^{e \ 2}_{LR})_{ij}&=&\bigg  ( A_e^{ij} + \mu \tan \beta \bigg )  ( m_e
)_{ij},
\label{MSSMC2}
\end{eqnarray}
$A$ \'etant d\'efini dans Eq.(\ref{BRISS7}). Dans Eq.(\ref{MSSMC2}), $\sin
\theta_W$ est le sinus
de l'angle \'electrofaible et $\tan \beta=<h^0_2>/<h^0_1>$, $h^0_1$ et
$h^0_2$ \'etant les composantes scalaires
des superchamps $H^0_1$ et $H^0_2$ de Eq.(\ref{wYUKAWA}). Par exemple, les
termes $(m_u m^\dagger_u)_{ij}$
de Eq.(\ref{MSSMC2}) viennent de $\vert {\partial W_{MSSM}(\tilde f) \over
\partial \tilde u_{i(L,R)}} \vert^2$
(voir Eq.(\ref{tjs14}) et Eq.(\ref{tjs15})) o\`u $W_{MSSM}$ est donn\'e dans
Eq.(\ref{wMSSM}).
Le terme ${\mu \over \tan \beta} ( m_u )_{ij}$ de Eq.(\ref{MSSMC2}) vient
lui de $\vert {\partial W_{MSSM}(\tilde f)
\over \partial \tilde h^0_2} \vert^2$ (voir Eq.(\ref{tjs14}) et
Eq.(\ref{tjs15})). Prenons un dernier exemple:
Le terme $\cos 2 \beta m_Z^2 (-{1 \over 2} + {2 \over 3} \sin^2 \theta_W)
\delta_{ij}$ provient des termes de
Eq.(\ref{tjs16}) (voir aussi Eq.(\ref{tjs14})) qui sont des termes D
puisqu'ils sont engendr\'es par des termes
du type $[\Phi^\dagger (e^{2gV}) \Phi ]_{\theta \theta \bar \theta \bar
\theta}$ (voir Eq.(\ref{tjs9})).

Les masses des sfermions not\'ees $m^2_{ij}(\tilde f_{L,R})$ dans
Eq.(\ref{MSSMC2}) sont des masses douces,
c'est \`a dire des masses du type $m_0$ provenant des termes de brisure
douce de la \susi
(voir Eq.(\ref{BRISS7})). Dans les mod\`eles bas\'es sur la supergravit\'e,
un lagrangien effectif \susiq
accompagn\'e de termes de brisure douce de SUSY est engendr\'e \`a l'\'echelle de
Planck
$M_P \sim 10^{19} GeV$ qui est sup\'erieure \`a l'\'echelle
d'unification $M_{GUT} \sim 2 \ 10^{16} GeV$ (voir Chapitre \ref{SUGRA}).
Ces termes de brisure douce peuvent donner une masse universelle $m_0$ \`a
tous
les champs scalaires ainsi qu'une masse universelle $m_{1/2}$ \`a tous les
jauginos. Les masses douces
des champs scalaires et des jauginos \`a des \'energies inf\'erieures \`a
$M_{SUGRA}$ sont obtenues \`a partir des masses $m_0$ et $m_{1/2}$
par les solutions des \'equations du groupe de renormalisation. Si les
masses
douces des diff\'erents champs scalaires sont \'egales entre elles
\`a l'\'echelle de Planck, elles le restent aux
\'energies sup\'erieures \`a l'\'echelle
d'unification $M_{GUT}$ puisque les champs scalaires appartiennent aux
m\^emes repr\'esentations du groupe de jauge de
grande unification et ont donc des couplages identiques. En revanche, \`a
des \'energies inf\'erieures \`a $M_{GUT}$,
les masses douces des diff\'erents champs scalaires deviennent diff\'erentes
car chaque champ scalaire est charg\'e
diff\'eremment vis-\`a-vis du groupe de jauge $SU(3)_c \times SU(2)_L \times
U(1)_Y$ et a donc ses propres couplages.
Il en est de m\^eme pour les masses douces des jauginos. Admettant que les
champs scalaires (jauginos) aient une
masse commune $m_0$ ($m_{1/2}$) \`a l'\'echelle $M_{GUT}$ qui soit identique
pour les 3 saveurs
et n\'egligeant les couplages de Yukawa,
l'int\'egration des \'equations du groupe de renormalisation donne les
expressions suivantes pour les masses douces
des champs scalaires \`a une \'echelle $Q$ inf\'erieure \`a $M_{GUT}$
\cite{1:Iban},
\begin{eqnarray}
m^2(\tilde u_L)&=& m_0^2 + 2 m^2_{1/2}
({1 \over 36} \tilde \alpha_1 f_1 + {3 \over 4} \tilde \alpha_2 f_2 + {4
\over 3} \tilde \alpha_3 f_3), \cr
m^2(\tilde d_L)&=& m_0^2 + 2 m^2_{1/2}
({1 \over 36} \tilde \alpha_1 f_1 + {3 \over 4} \tilde \alpha_2 f_2 + {4
\over 3} \tilde \alpha_3 f_3), \cr
m^2(\tilde u_R)&=& m_0^2 + 2 m^2_{1/2}
({4 \over 9} \tilde \alpha_1 f_1 + {4 \over 3} \tilde \alpha_3 f_3), \cr
m^2(\tilde d_R)&=& m_0^2 + 2 m^2_{1/2}
({1 \over 9} \tilde \alpha_1 f_1 + {4 \over 3} \tilde \alpha_3 f_3), \cr
m^2(\tilde e_L)&=& m_0^2 + 2 m^2_{1/2}
({1 \over 4} \tilde \alpha_1 f_1 + {3 \over 4} \tilde \alpha_2 f_2), \cr
m^2(\tilde \nu_L)&=& m_0^2 + 2 m^2_{1/2}
({1 \over 4} \tilde \alpha_1 f_1 + {3 \over 4} \tilde \alpha_2 f_2), \cr
m^2(\tilde e_R)&=& m_0^2 + 2 m^2_{1/2} \tilde \alpha_1 f_1,
\label{MSSMC3}
\end{eqnarray}
avec,
\begin{eqnarray}
\tilde \alpha_i = { \alpha_i (M_{GUT}) \over 4 \pi}, \
f_i = { ( 2 + b_i \tilde \alpha_i t ) \over ( 1 + b_i \tilde \alpha_i t )^2
} t,
\label{MSSMC4}
\end{eqnarray}
$t$ valant $t=2 \log (M_{GUT}/Q)$, les $\alpha_i$ ($i=1,2,3$) \'etant les 3
constantes de couplage
du groupe de jauge $SU(3)_c \times SU(2)_L \times U(1)_Y$ du Mod\`ele
Standard et les $b_i=(-3,1,11)$
\'etant les coefficients des fonctions $\beta$ \`a une boucle associ\'ees
aux interactions $SU(3)_c$,
$SU(2)_L$ et $U(1)_Y$, respectivement.

\subsubsection{Masses des jauginos et higgsinos}
\label{SpecMSSMjh}

$\bullet$ {\bf Charginos} D\'efinissons avant tout les vecteurs,
\begin{equation}
\psi^+ =
\left ( \begin{array}{c} -i \l^+ \\ \psi_{H^+} \end{array} \right ), \
\psi^- =
\left ( \begin{array}{c} -i \l^- \\ \psi_{H^-} \end{array} \right ),
\label{MSSMC5}
\end{equation}
o\`u $\l^{\pm}$ est le wino c'est \`a dire le partenaire \susiq de spin 1/2
(\`a 2 composantes)
du boson $W^{\pm}$, et o\`u $\psi_{H^+}$ et $\psi_{H^-}$ sont les higgsinos
charg\'es c'est \`a dire les
partenaires supersym\'etriques de spin 1/2 (\`a 2 composantes) des bosons de
Higgs charg\'es.
Plus pr\'ecis\'ement, $\psi_{H^+}$ et $\psi_{H^-}$ sont respectivement les
composantes spinorielles
des superchamps $H^+$ et $H^-$ d\'efinis dans Eq.(\ref{wYUKAWA}).

Le lagrangien ${\cal L}_{MSSM}$ du MSSM (voir Eq.(\ref{mssm2}) et
Eq.(\ref{BRISS7})) engendre des termes
de masse pour le wino et les higgsinos charg\'es qui s'\'ecrivent comme
suit,
\begin{equation}
-{\cal L}^{charg}_{mass}=
{1 \over 2} ((\psi^+)^T,(\psi^-)^T) \left ( \begin{array}{cc} 0 &  X^T \\ X & 
0 \end{array} \right )
\left ( \begin{array}{c} \psi^+ \\ \psi^- \end{array} \right ) + h.c.,
\label{MSSMC6}
\end{equation}
avec,
\begin{equation}
X= \left ( \begin{array}{cc}
M_2 &  M_W \sqrt 2 \sin \beta \\
M_W \sqrt 2 \cos \beta &  \mu
\end{array} \right ).
\label{MSSMC7}
\end{equation}
Dans Eq.(\ref{MSSMC7}), $M_2$ est la masse douce du wino \`a l'\'echelle $Q$
consid\'er\'ee,
c'est \`a dire une masse du type $m_{1/2}$ provenant des termes de
brisure douce de la \susi
(voir Eq.(\ref{BRISS7})). Par exemple, dans un mod\`ele bas\'e sur la
supergravit\'e, $M_2$ peut \^etre obtenue
par le biais des solutions des \'equations du groupe de renormalisation \`a
partir des masses universelles
$m_0$ et $m_{1/2}$ \`a l'\'echelle $M_{GUT}$. Notons aussi que dans
Eq.(\ref{MSSMC7}) $M_W$ est la masse
du boson $W^{\pm}$ qui s'exprime dans le MSSM,
\begin{equation}
M_W^2={1 \over 4} g_2^2 (<h^0_1>^2+<h^0_2>^2).
\label{MSSMC8}
\end{equation}
Le terme de masse en $\mu$ de Eq.(\ref{MSSMC7}) provient du terme
$- {1 \over 2} {\partial^2 W_{MSSM}(\tilde f) \over \partial h^+ \partial
h^-} \psi_{H^+} \psi_{H^-}$
de Eq.(\ref{tjs10}), $h^+$ et $h^-$ \'etant respectivement les composantes
scalaires
des superchamps $H^+$ et $H^-$ d\'efinis dans Eq.(\ref{wYUKAWA}).
Les termes de masse en $M_W \sqrt 2 \sin \beta$ et $M_W \sqrt 2 \cos \beta$
de Eq.(\ref{MSSMC7})
proviennent des termes $[ H_1^{\dagger} e^{-{1 \over 2} 2g_1 V_1} e^{2g_2
V_2} H_1^c +
H_2^{\dagger} e^{{1 \over 2} 2g_1 V_1} e^{2g_2 V_2} H_2^c ]_{\theta \theta
\bar \theta \bar \theta}$
de Eq.(\ref{mssm2}), qui engendrent des termes du type $ i \sqrt 2  g (\bar
\psi_{H_{1,2}} \bar \l_a) T_a h_{1,2}
- i \sqrt 2  g h_{1,2}^{\dagger} T_a (\psi_{H_{1,2}} \l_a)$ (voir
Eq.(\ref{tjs10})), les $T_a$ \'etant
les g\'en\'erateurs du groupe $SU(2)_L$ et les $h_{1,2}$ ($\psi_{H_{1,2}}$)
les composantes scalaires (spinorielles)
des superchamps $H_{1,2}$ de Eq.(\ref{wYUKAWA}) qui sont des doublets de
$SU(2)_L$.

Les termes de masse de Eq.(\ref{MSSMC6}) peuvent s'\'ecrire apr\`es
diagonalisation de la matrice $X$ \cite{PhysRep117},
\begin{equation}
-{\cal L}^{charg}_{mass}=
(\chi^-_1,\chi^-_2) \left ( \begin{array}{cc} \tilde m_1 &  0 \\ 0 &  \tilde
m_2 \end{array} \right )
\left ( \begin{array}{c} \chi^+_1 \\ \chi^+_2 \end{array} \right ) + h.c.,
\label{MSSMC9}
\end{equation}
o\`u,
\begin{equation}
\left ( \begin{array}{c} \chi^+_1 \\ \chi^+_2 \end{array} \right )=V \psi^+,
\
\left ( \begin{array}{c} \chi^-_1 \\ \chi^-_2 \end{array} \right )=U \psi^-,
\
U^\star X V^{-1}= \left ( \begin{array}{cc} \tilde m_1 &  0 \\ 0 &  \tilde m_2
\end{array} \right ),
\label{MSSMC10}
\end{equation}
$U$ et $V$ \'etant des matrices unitaires $2 \times 2$. Les spineurs \`a 2
composantes
$\chi^{\pm}_{1,2}$ sont donc les \'etats propres de masse de la matrice $X$
c'est \`a dire
des m\'elanges entre wino $\l^{\pm}$ et higgsinos $\psi_{H^{\pm}}$.
En utilisant Eq.(\ref{2to4mass}), on peut exprimer le lagrangien
\ref{MSSMC9}
en terme de spineurs \`a 4 composantes comme suit,
\begin{equation}
-{\cal L}^{charg}_{mass}=
(\bar {\tilde {\chi}}_1,\bar {\tilde {\chi}}_2)
\left ( \begin{array}{cc} \tilde m_1 &  0 \\ 0 &  \tilde m_2 \end{array}
\right )
\left ( \begin{array}{c} \tilde {\chi}_1 \\ \tilde {\chi}_2 \end{array}
\right ),
\label{MSSMC11}
\end{equation}
$\tilde {\chi}_1$ et $\tilde {\chi}_2$ \'etant les spineurs \`a 4
composantes appel\'es charginos
et d\'efinis par,
\begin{equation}
\tilde {\chi}_1= \left ( \begin{array}{c} \chi^+_1 \\ \bar \chi^-_1
\end{array} \right ), \
\tilde {\chi}_2= \left ( \begin{array}{c} \chi^+_2 \\ \bar \chi^-_2
\end{array} \right ).
\label{MSSMC12}
\end{equation}

$\bullet$ {\bf Neutralinos} D\'efinissons avant tout le vecteur,
\begin{equation}
(\psi^0)^T = ( -i \l' , -i \l^3 , \psi_{H^0_1} , \psi_{H^0_2} ),
\label{MSSMC13}
\end{equation}
o\`u $\l'$ ($\l^3$) est le bino (wino) c'est \`a dire le partenaire \susiq
de spin 1/2 (\`a 2 composantes)
du boson $B$ ($W^3$), et o\`u $\psi_{H^0_1}$ et $\psi_{H^0_2}$ sont les
higgsinos neutres c'est \`a dire les
partenaires supersym\'etriques de spin 1/2 (\`a 2 composantes) des bosons de
Higgs neutres.
Plus pr\'ecis\'ement, $\psi_{H^0_1}$ et $\psi_{H^0_2}$ sont respectivement
les composantes spinorielles
des superchamps $H^0_1$ et $H^0_2$ d\'efinis dans Eq.(\ref{wYUKAWA}).

Le lagrangien ${\cal L}_{MSSM}$ du MSSM (voir Eq.(\ref{mssm2}) et
Eq.(\ref{BRISS7})) engendre des termes
de masse pour le bino, le wino et les higgsinos neutres qui s'\'ecrivent
comme suit,
\begin{equation}
-{\cal L}^{neut}_{mass} = {1 \over 2} (\psi^0)^T Y \psi^0 + h.c.,
\label{MSSMC14}
\end{equation}
avec,
\begin{equation}
Y = \left ( \begin{array}{cccc}
M_1 &  0 &  - M_Z \cos \beta \sin \theta_W &  M_Z \sin \beta \sin \theta_W  \\
0 &  M_2 &   M_Z \cos \beta \cos \theta_W &  - M_Z \sin \beta \cos \theta_W \\
- M_Z \cos \beta \sin \theta_W &  M_Z \cos \beta \cos \theta_W &  0 &  -\mu \\
M_Z \sin \beta \sin \theta_W &  - M_Z \sin \beta \cos \theta_W &  -\mu &  0
\end{array} \right ).
\label{MSSMC15}
\end{equation}
Dans Eq.(\ref{MSSMC15}), $M_1$ ($M_2$) est la masse douce du bino (wino) \`a
l'\'echelle $Q$ consid\'er\'ee,
c'est \`a dire une masse du type $m_{1/2}$ provenant des termes de
brisure douce de la \susi
(voir Eq.(\ref{BRISS7})).
Les termes de masse en $-\mu$ de Eq.(\ref{MSSMC15}) proviennent du terme
$- {1 \over 2} {\partial^2 W_{MSSM}(\tilde f) \over \partial h^0_1 \partial
h^0_2} \psi_{H^0_1} \psi_{H^0_2}$
de Eq.(\ref{tjs10}), $h^0_1$ et $h^0_2$ \'etant respectivement les
composantes scalaires
des superchamps $H^0_1$ et $H^0_2$ d\'efinis dans Eq.(\ref{wYUKAWA}).
Les termes de masse en $M_Z f(\beta,\theta_W)$ de Eq.(\ref{MSSMC15})
proviennent des termes $[ H_1^{\dagger} e^{-{1 \over 2} 2g_1 V_1} e^{2g_2
V_2} H_1^c +
H_2^{\dagger} e^{{1 \over 2} 2g_1 V_1} e^{2g_2 V_2} H_2^c ]_{\theta \theta
\bar \theta \bar \theta}$
de Eq.(\ref{mssm2}), qui engendrent des termes du type $ i \sqrt 2  g (\bar
\psi_{H_{1,2}} \bar \l_a) T_a h_{1,2}
- i \sqrt 2  g h_{1,2}^{\dagger} T_a (\psi_{H_{1,2}} \l_a)$ (voir
Eq.(\ref{tjs10})), les $T_a$ \'etant
les g\'en\'erateurs du groupe $SU(2)_L$ et les $h_{1,2}$ ($\psi_{H_{1,2}}$)
les composantes scalaires (spinorielles)
des superchamps $H_{1,2}$ de Eq.(\ref{wYUKAWA}) qui sont des doublets de
$SU(2)_L$.

Les termes de masse de Eq.(\ref{MSSMC14}) peuvent s'\'ecrire apr\`es
diagonalisation de la matrice $Y$ \cite{PhysRep117},
\begin{equation}
-{\cal L}^{neut}_{mass}= {1 \over 2}
(\chi^0_1,\chi^0_2,\chi^0_3,\chi^0_4)
\left ( \begin{array}{cccc}
\tilde m^0_1 &  0 &  0 &  0 \\
0 &  \tilde m^0_2 &  0 &  0 \\
0 &  0 &  \tilde m^0_3 &  0 \\
0 &  0 &  0 &  \tilde m^0_4
\end{array} \right )
\left ( \begin{array}{c} \chi^0_1 \\ \chi^0_2 \\ \chi^0_3 \\ \chi^0_4
\end{array} \right ) + h.c.,
\label{MSSMC16}
\end{equation}
o\`u,
\begin{equation}
\left ( \begin{array}{c} \chi^0_1 \\ \chi^0_2 \\ \chi^0_3 \\ \chi^0_4
\end{array} \right )=N \psi^0, \
N^\star Y N^{-1}=
\left ( \begin{array}{cccc}
\tilde m^0_1 &  0 &  0 &  0 \\
0 &  \tilde m^0_2 &  0 &  0 \\
0 &  0 &  \tilde m^0_3 &  0 \\
0 &  0 &  0 &  \tilde m^0_4
\end{array} \right ),
\label{MSSMC17}
\end{equation}
$N$ \'etant une matrice unitaires $4 \times 4$. Les spineurs \`a 2
composantes
$\chi^0_{1,2,3,4}$ sont donc les \'etats propres de masse de la matrice $Y$
c'est \`a dire
des m\'elanges entre bino $\l'$, wino $\l^3$ et higgsinos
$\psi_{H^0_{1,2}}$.
En utilisant Eq.(\ref{2to4mass}), on peut exprimer le lagrangien
\ref{MSSMC16}
en terme de spineurs \`a 4 composantes comme suit,
\begin{equation}
-{\cal L}^{neut}_{mass}= {1 \over 2}
(\bar {\tilde {\chi^0}}_1,\bar {\tilde {\chi^0}}_2,\bar {\tilde
{\chi^0}}_3,\bar {\tilde {\chi^0}}_4)
\left ( \begin{array}{cccc}
\tilde m^0_1 &  0 &  0 &  0 \\
0 &  \tilde m^0_2 &  0 &  0 \\
0 &  0 &  \tilde m^0_3 &  0 \\
0 &  0 &  0 &  \tilde m^0_4
\end{array} \right )
\left ( \begin{array}{c} \tilde \chi^0_1 \\ \tilde \chi^0_2 \\ \tilde
\chi^0_3 \\ \tilde \chi^0_4
\end{array} \right ),
\label{MSSMC18}
\end{equation}
$\tilde \chi^0_1$, $\tilde \chi^0_2$, $\tilde \chi^0_3$ et $\tilde \chi^0_4$
\'etant les spineurs
\`a 4 composantes appel\'es neutralinos et d\'efinis par,
\begin{equation}
\tilde \chi^0_1= \left ( \begin{array}{c} \chi^0_1 \\ \bar \chi^0_1
\end{array} \right ), \
\tilde \chi^0_2= \left ( \begin{array}{c} \chi^0_2 \\ \bar \chi^0_2
\end{array} \right ), \
\tilde \chi^0_3= \left ( \begin{array}{c} \chi^0_3 \\ \bar \chi^0_3
\end{array} \right ), \
\tilde \chi^0_4= \left ( \begin{array}{c} \chi^0_4 \\ \bar \chi^0_4
\end{array} \right ).
\label{MSSMC19}
\end{equation}

\clearpage

\vspace{10 mm}
\begin{center}
{  }
\end{center}
\vspace{10 mm}

\clearpage

\chapter{Supergravit\'e}
\label{SUGRA}

\section{Motivations}
\label{SUGmot}

Les sym\'etries fondamentales en physique des particules
(comme par exemple les sym\'e\-tries de jauge) sont r\'ealis\'ees localement
plut\^ot que globalement. Cela nous sugg\`ere que si la supersym\'etrie
est r\'eellement une sym\'etrie de la nature, elle doit aussi \^etre
effective de fa\c{c}on locale. L'alg\`ebre de supersym\'etrie (voir
Section \ref{AlSUSY}) contient le g\'en\'erateur des translations $P_{\mu}$.
Par cons\'equent, dans une th\'eorie de supersym\'etrie locale, nous devons
consid\'erer les translations qui varient d'un point \`a l'autre
de l'espace-temps. Cela signifie qu'une th\'eorie localement
supersym\'etrique doit \^etre une th\'eorie de transformations g\'en\'erales
des coordonn\'ees de l'espace-temps. En d'autres termes, une th\'eorie
de supersymt\'etrie locale doit contenir la th\'eorie de la gravitation.
Cela ne nous \'etonne que peu puisque
les g\'en\'erateurs de la supersym\'etrie
ne commutent pas avec les g\'en\'erateurs du groupe de Poincar\'e.
C'est donc pour cette raison que les th\'eories
de supersym\'etrie locale sont
aussi appel\'ees th\'eories de supergravit\'e.
En effet, comme nous allons le voir dans ce chapitre, la gravitation
joue naturellement un r\^ole majeur lorsque nous tentons
de rendre la supersym\'etrie locale. Plus pr\'ecis\'ement, le partenaire
supersym\'etrique du graviton, le gravitino, va permettre \`a la
supersym\'etrie d'agir localement, de m\^eme que les bosons de jauge
permettent aux sym\'etries de jauge d'agir localement.
En ce sens, il existe une forte analogie entre la supersym\'etrie locale
(et le gravitino) et les sym\'etries de jauge (et les bosons de jauge).
Le r\^ole naturel de la gravitation dans les th\'eories de
supersym\'etrie locale est une motivation suppl\'ementaire pour
la construction de telles th\'eories.

Outre leurs motivations th\'eoriques, les th\'eories de supergravit\'e
ont de nombreux attraits ph\'enom\'enologiques (que nous avons d\'ej\`a
mentionn\'e dans la Section \ref{BRIsec}).
Effectivement, comme nous allons l'expliquer en d\'etail dans ce chapitre,
dans certaines th\'eories de supergravit\'e la supersym\'etrie locale
est bris\'ee spontan\'ement dans un secteur cach\'e et 
un lagrangien effectif supersym\'etrique accompagn\'e de termes de
brisure douce de SUSY (voir Section \ref{BRIsec}) peut ainsi \^etre engendr\'e 
dans le secteur observable. De plus, les masses
g\'en\'er\'ees pour les partenaires supersym\'etriques par ces termes de
brisure douce peuvent \^etre sup\'erieures aux masses des particules du
Mod\`ele Standard \`a l'\'echelle \'electrofaible. Enfin,
les masses des partenaires supersym\'etriques,
bien que d\'efinies \`a l'\'echelle de Planck (gravitation),
peuvent \^etre inf\'erieures au $TeV$, \`a l'\'echelle \'electrofaible,
ce qui est n\'ecessaire pour r\'esoudre le probl\`eme de hi\'erarchie
(voir Section \ref{hier}).

Nous avons vu dans la Section \ref{BRIsec} que les mod\`eles dits
GMSB (Gauge Mediated Supersymmetry Breaking) \cite{GiuRat} 
peuvent aussi g\'en\'erer, suite \`a une brisure spontan\'ee de la
supersym\'etrie dans un secteur cach\'e, des termes de brisure douce 
de SUSY engendrant des masses pour les
partenaires supersym\'etriques sup\'erieures aux masses des particules du
Mod\`ele Standard et inf\'erieures au $TeV$, \`a l'\'echelle \'electrofaible.
Comparons les int\'er\^ets ph\'enom\'enologiques des mod\`eles
GMSB et des mod\`eles bas\'es sur la supergravit\'e.
\\ D'un point de vue ph\'enom\'enologique,
la principale diff\'erence entre les mod\`eles GMSB et les mod\`eles de
supergravit\'e est li\'ee au probl\`eme de la saveur: Le lagrangien du \ms
sans les couplages
de Yukawa est invariant sous la sym\'etrie globale $U(3)^5$, chaque groupe
$U(3)$ agissant
dans l'espace des saveurs des 5 repr\'esentations irr\'eductibles de
fermions du
groupe de jauge: $(q_L, u_R^c, d_R^c, l_L, e_R^c)_i$ ($i=1,2,3$). Or,
nous ignorons la dynamique
responsable de la brisure de la sym\'etrie de saveur, dont la seule trace
\`a basse
\'energie est visible dans la structure des couplages de Yukawa. Supposons
que
cette dynamique de brisure a lieu au-del\`a d'une \'echelle $\L_F$.
\\ Dans les mod\`eles de supergravit\'e, les termes doux de brisure
sont d\'efinis \`a l'\'echelle de Planck, c'est \`a dire \`a une
\'echelle n\'ecessairement
sup\'erieure \`a $\L_F$. Il n'y a donc pas de raison \'evidente
pour que les termes doux
de brisure soient ind\'ependants de la saveur. 
Cette violation de la saveur dans les termes doux, qui sont
entre autres des termes de masse pour
les squarks et les sleptons, est tr\`es dangereuse car elle engendre
des contributions supersym\'etriques \`a des processus
comme le m\'elange $K^0-\bar K^0$ ou encore
la d\'esint\'egration $\mu  \to e \g$ sur lesquels les bornes
exp\'erimentales sont fortes.
C'est le probl\`eme de changement de saveur en \susi qui est li\'e
au fait que les matrices de masse pour les fermions et leurs partenaires
\susiqs ne sont pas \`a priori
diagonales dans la m\^eme base. Bien s\^ur, cela ne signifie pas que les
sc\'enarios de supergravit\'e ne sont pas r\'ealistes. Il se peut qu'au
niveau de la gravit\'e
quantique les termes doux soient ind\'ependants de la saveur, ou bien que
certaines sym\'etries
de saveur \cite{symsav1,symsav2} ou certains m\'ecanismes dynamiques
\cite{mecdyn}
soient responsables d'un alignement approximatif
entre les matrices de masse des fermions et des sfermions.
\\ En ce qui concerne les mod\`eles GMSB, les termes
doux sont d\'efinis \`a une \'echelle $M$, qui repr\'esente la masse des
messagers.
Cette \'echelle n'est \`a priori pas reli\'ee \`a $\L_F$ et peut donc \^etre
choisie telle que
$M<\L_F$. Pour un tel choix de l'\'echelle $M$, 
les termes doux, qui sont g\'en\'er\'es par des graphes \`a l'ordre des
boucles mettant en jeu les
interactions de jauge, ne ressentiraient pas les effets de violation
de saveur pr\'esents dans ce cas uniquement dans les couplages de Yukawa. 
Le m\'ecanisme de GIM peut alors \^etre g\'en\'eralis\'e au m\'ecanisme
de superGIM incluant les particules et leur partenaire supersym\'etrique. Le
probl\`eme de changement
de saveur est donc naturellement d\'ecoupl\'e dans les mod\`eles GMSB, en
contraste avec les th\'eories de supergravit\'e.
\\ Par ailleurs, les mod\`eles GMSB posent le ``probl\`eme du terme $\mu$''
(qui sera expos\'e dans la Section \ref{pbm})
ce qui n'est pas le cas des th\'eories de supergravit\'e.
\\ Enfin, les versions {\it minimales} des mod\`eles GMSB 
et des mod\`eles de supergravit\'e sont toutes deux
assez pr\'edictives vis \`a vis du spectre de masse
des particules supersym\'etriques, 
ce qui leur permettra d'\^etre test\'ees
aupr\`es des futurs collisionneurs de particules. 
De plus, dans les mod\`eles GMSB,
la LSP est le gravitino ($\equiv$ partenaire \susiq du graviton) 
ce qui engendre des signaux ph\'enom\'enologiques sp\'ecifiques.

\section{Lagrangiens de supergravit\'e}

\subsection{Proc\'edure de Noether}
\label{Noether}

Afin de d\'eriver les lagrangiens localement supersym\'etriques,
nous allons utiliser la proc\'edure de Noether. La proc\'edure de
Noether est une m\'ethode syst\'ematique pour obtenir une action
ayant une sym\'etrie locale \`a partir de l'action ayant une sym\'etrie
globale. Dans cette partie, nous illustrons cette proc\'edure
en prenant l'exemple d'une sym\'etrie de jauge.

Consid\'erons donc l'action d'un champ de Dirac libre et non massif,
\begin{eqnarray}
S_0 = i \int d^4 x \bar \psi \gamma^{\mu} \partial_{\mu} \psi.
\label{ActDirac}
\end{eqnarray}
Cette action est invariante sous la transformation,
\begin{eqnarray}
\psi \to e^{-i \xi} \psi,
\label{transgaug}
\end{eqnarray}
o\`u $\xi$ est une phase constante. $S_0$ a donc une sym\'etrie
globale ab\'elienne. Pour que cette sym\'etrie soit locale,
la phase $\xi$ doit \^etre une fonction des coordonn\'ees
d'espace-temps. La transformation de Eq.(\ref{transgaug}) s'\'ecrit
alors,
\begin{eqnarray}
\psi \to e^{-i \xi(x)} \psi.
\label{transgaugl}
\end{eqnarray}
L'action $S_0$ de Eq.(\ref{ActDirac}) n'est plus invariante sous
cette sym\'etrie de jauge. La variation de l'action $S_0$ par la
transformation locale de Eq.(\ref{transgaug}l) est,
\begin{eqnarray}
\delta S_0 = \int d^4 x \bar \psi \gamma^{\mu} \psi \partial_{\mu} \xi
=\int d^4 x j^{\mu} \partial_{\mu} \xi,
\label{dActDirac}
\end{eqnarray}
o\`u,
\begin{eqnarray}
j^{\mu}= \bar \psi \gamma^{\mu} \psi,
\label{courant}
\end{eqnarray}
est le courant de Noether associ\'e \`a la sym\'etrie (\ref{transgaug})
de $S_0$. Afin de restaurer l'invariance, un champ de jauge $A_{\mu}$
est introduit. Ce dernier doit se transformer sous la sym\'etrie
\ref{transgaugl} par,
\begin{eqnarray}
A_{\mu} \to A_{\mu} + \partial_{\mu} \xi,
\label{transboson}
\end{eqnarray}
et un terme de couplage entre ce champ de jauge et le courant de
Noether doit \^etre ajout\'e \`a l'action $S_0$:
\begin{eqnarray}
S= S_0 - \int d^4 x j^{\mu} A_{\mu}=\int d^4 x i
\bar \psi \gamma^{\mu} (\partial_{\mu} +i A_{\mu}) \psi
=\int d^4 x i
\bar \psi \gamma^{\mu} D_{\mu} \psi,
\label{ActDiracl}
\end{eqnarray}
o\`u $D_{\mu}=\partial_{\mu}+iA_{\mu}$ est une d\'eriv\'ee covariante.
La variation de ce terme sous la sym\'etrie locale \ref{transgaugl} va
exactement compenser la variation de $S_0$ de Eq.(\ref{ActDiracl}).
L'action $S$ est alors invariante sous les transformations de jauge
\ref{transgaugl} et \ref{transboson}.

Plus g\'en\'eralement, pour une sym\'etrie globale initiale non
ab\'elienne, cette m\'ethode n'est applicable qu'au premier ordre
d'un param\`etre donn\'e et doit donc \^etre it\'er\'ee. A chaque
\'etape, un terme suppl\'ementaire doit \^etre ajout\'e \`a l'action
(Eq.(\ref{ActDiracl})) afin de compenser les variations \`a
l'ordre correspondant de cette m\^eme action. En g\'en\'eral,
un terme suppl\'ementaire doit aussi \^etre ajout\'e \`a la loi de
transformation du champ de jauge (Eq.(\ref{transboson})). Apr\`es un
nombre fini d'it\'erations (avec de la chance !) une action est obtenue,
qui est exactement invariante sous la sym\'etrie locale consid\'er\'ee
(Eq.(\ref{transgaugl}) +
forme finale de Eq.(\ref{transboson})).
\\ Afin d'illustrer ces propos, nous allons maintenant appliquer la
proc\'edure de Noether sur une th\'eorie ayant une sym\'etrie
non-ab\'elienne. De plus, nous allons consid\'erer une action
de champs de jauge, ce qui nous sera utile par la suite.
L'action consid\'er\'ee est la suivante,
\begin{eqnarray}
S_0 = - { 1 \over 4} \int d^4 x G_a^{\mu \nu} G_{\mu \nu}^a,
\label{Actgaug}
\end{eqnarray}
o\`u,
\begin{eqnarray}
G^{\mu \nu}_a = \partial^{\mu} A_a^{\nu} - \partial^{\nu} A_a^{\mu},
\label{Gmua}
\end{eqnarray}
et les $A_a^{\mu}, \ a=1,...,r$, sont des champs vectoriels appartenant
\`a la repr\'esentation adjointe de dimension $r$ d'un groupe de Lie
donn\'e. Les g\'en\'erateurs, $T_a$, et les constantes de structure,
$f_{abc}$, de ce groupe de Lie v\'erifient,
\begin{eqnarray}
[T_a,T_b]=i f_{abc} T_c.
\label{structcons}
\end{eqnarray}
L'action de Eq.(\ref{Actgaug}) est invariante sous la transformation {\it
infinit\'esimale} de param\`etre $\xi_a$ \`a l'ordre $g$,
\begin{eqnarray}
A_a^{\mu} \to A_a^{\mu} +g f_{abc} \xi_b A_c^{\mu}.
\label{transgaug1}
\end{eqnarray}
En revanche, si le param\`etre $\xi_a$ de Eq.(\ref{transgaug1}) d\'epend
de l'espace-temps, l'action $S_0$ de Eq.(\ref{Actgaug}) n'est plus
invariante et sa variation vaut,
\begin{eqnarray}
\delta S_0=\int d^4 x j^{\mu}_a \partial_{\mu} \xi_a,
\label{dActgaug}
\end{eqnarray}
o\`u,
\begin{eqnarray}
j^{\mu}_a=g f_{abc} G^{\mu \nu}_b A^c_{\nu},
\label{courant1}
\end{eqnarray}
est le courant de Noether associ\'e \`a la sym\'etrie correspondant
aux g\'en\'erateurs $T_a$. Nous remarquons l'analogie entre les \'equations
Eq.(\ref{dActDirac}) et Eq.(\ref{dActgaug}).
L'invariance est restaur\'ee \`a l'ordre $g$ si
on ajoute \`a l'action de Eq.(\ref{Actgaug}) le terme,
\begin{eqnarray}
S_1=S_0 - { 1 \over 2} \int d^4 x j_a^{\mu} A^a_{\mu},
\label{Actgaugl}
\end{eqnarray}
et si la loi de transformation (\ref{transgaug1}) est modifi\'ee en,
\begin{eqnarray}
A_a^{\mu} \to A_a^{\mu} +g f_{abc} \xi_b A_c^{\mu}+\partial^{\mu} \xi_a.
\label{transgaug1l}
\end{eqnarray}
Remarquons \`a nouveau l'analogie entre les \'equations
Eq.(\ref{ActDiracl}),
Eq.(\ref{transgaugl}) et Eq.(\ref{Actgaugl}), Eq.(\ref{transgaug1l}).
Notons aussi que l'action de Eq.(\ref{Actgaug}) est invariante
sous la transformation locale,
\begin{eqnarray}
A_a^{\mu} \to A_a^{\mu} + \partial^{\mu} \xi_a,
\label{transuniq}
\end{eqnarray}
qui correspond \`a une sym\'etrie ab\'elienne. Lorsque la transformation
du groupe de Lie est consid\'er\'ee au second ordre en
$g$, la variation de l'action $S_1$ sous (\ref{transgaug1l})
(au second ordre) est,
\begin{eqnarray}
\delta S_1 =  \int d^4 x  j_{\mu}^d \partial^{\mu} \xi_d.
\label{dActgaugord}
\end{eqnarray}
avec,
\begin{eqnarray}
j_{\mu}^d = -g^2 f_{abc} f_{bde}
A^a_{\mu} A^c_{\nu} A_e^{\nu}.
\label{courantord}
\end{eqnarray}
Comme pr\'ec\'edemment, l'invariance est restaur\'ee en ajoutant
\`a l'action un terme:
\begin{eqnarray}
S=S_1 - { 1 \over 4} \int d^4 x j_{\mu}^d  A^{\mu}_d.
\label{Actgaugll}
\end{eqnarray}
Il n'est pas n\'ecessaire ici de modifier la loi de transformation
(\ref{transgaug1l}) d\'efinie au second ordre. En fait, un effort
suppl\'ementaire nous montrerait m\^eme que l'action $S$
de Eq.(\ref{Actgaugll}) est invariante \`a tout ordre en $g$.
Cette action $S$ peut aussi s'\'ecrire,
\begin{eqnarray}
S=-{ 1 \over 4} \int d^4 x F_{\mu \nu}^a  F^{\mu \nu}_a,
\label{Actgaugllf}
\end{eqnarray}
avec,
\begin{eqnarray}
F^{\mu \nu}_a=G^{\mu \nu}_a - g f_{abc} A_b^{\mu} A_c^{\nu}
=\partial^{\mu} A_a^{\nu} - \partial^{\nu} A_a^{\mu}
- g f_{abc} A_b^{\mu} A_c^{\nu}.
\label{Fmunu}
\end{eqnarray}
Et nous reconnaissons ici l'action d'une th\'eorie de jauge
non-ab\'elienne pure.

\subsection{Lagrangien localement supersym\'etrique pour
le multiplet de supergravit\'e}

L'action globalement supersym\'etrique pour le multiplet de
supergravit\'e est la suivante,
\begin{eqnarray}
S=-{1 \over 2} \int d^4 x \ep^{\mu \nu \rho \s} \bar \Psi_{\mu}
\gamma_5 \gamma_{\nu} \partial_{\rho} \Psi_{\s}
- {1 \over 2} \int d^4 x
(R^L_{\mu \nu }-{1 \over 2} \eta_{\mu \nu} R^L) h^{\mu \nu},
\label{ActEin}
\end{eqnarray}
o\`u $\ep^{\mu \nu \rho \s}$ d\'esigne le
tenseur antisym\'etrique de Levi-Civita d\'efini tel que
$\ep^{0123}=1$, $R^L_{\mu \nu}$ d\'enote le tenseur de Ricci
donn\'e par,
\begin{eqnarray}
R^L_{\mu \nu}={1 \over 2} \bigg (
-{\partial^2 h_{\mu \nu} \over \partial x^{\lambda} \partial x_{\lambda}}
+{\partial^2 h^{\lambda}_{\nu} \over \partial x^{\mu} \partial x^{\lambda}}
+{\partial^2 h^{\lambda}_{\mu} \over \partial x^{\nu} \partial x^{\lambda}}
-{\partial^2 h^{\lambda}_{\lambda} \over \partial x^{\mu} \partial x^{\nu}}
\bigg ),
\label{Ricci}
\end{eqnarray}
et $R^L$ est la courbure scalaire donnn\'ee par 
$R^L=\eta^{\mu \nu} R^L_{\mu \nu}$. \\
Le premier terme de Eq.(\ref{ActEin}) 
est l'action de Rarita-Schwinger qui fournit un terme
cin\'etique au partenaire supersym\'etrique $\Psi_{\mu}$ du graviton.
Le champ $\Psi_{\mu}$, qui est appel\'e le gravitino, a un spin $3/2$. 
Le partenaire du graviton
aurait pu \^etre choisi comme \'etant un champ de spin $5/2$,
mais les th\'eories contenant des particules de spin plus grand que 2
ont des particularit\'es ind\'esirables. Nous v\'erifierons par la suite
que ce choix est correct puisque le gravitino de spin $3/2$ appara\^{\i}tra
comme le ``champ de jauge" associ\'e \`a la supersym\'etrie locale.
Le second terme de Eq.(\ref{ActEin}) est l'action d'Einstein lin\'earis\'ee,
c'est \`a dire l'action d'Einstein \'ecrite au premier ordre en $\k$
et exprim\'ee en fonction du champ du graviton $h_{\mu \nu}$ d\'efini par,
\begin{eqnarray}
g_{\mu \nu} = \eta_{\mu \nu}+ \k h_{\mu \nu},
\label{hmunu}
\end{eqnarray}
o\`u $g_{\mu \nu}$ est la m\'etrique d'espace-temps,
$\eta_{\mu \nu}$ est la m\'etrique de Minkowski 
et $\k$ est donn\'e par,
\begin{eqnarray}
 8 \pi G_N M^2_P = \k^2 M^2_P =1,
\label{kappa}
\end{eqnarray}
$G_N$ \'etant la constante gravitationnelle de Newton
et $M_P$ la masse de Planck
(il sera parfois utile de choisir l'unit\'e $\k^2=1$).
Le facteur $\k$ est introduit dans Eq.(\ref{hmunu}) afin
que $h_{\mu \nu}$ ait une dimension 1, comme cela est appropri\'e
pour un champ bosonique d\'ecrivant le graviton. 
Nous utilisons ici l'action d'Einstein lin\'earis\'ee (Eq.(\ref{ActEin}))
pour deux raisons. La premi\`ere est que cette forme de 
l'action d'Einstein fait appara\^{\i}tre
des termes cin\'etiques pour le graviton $h_{\mu \nu}$ qui
sont quadratiques (Eq.(\ref{Ricci})). La seconde est que nous 
allons d\'eterminer dans la suite l'action de supergravit\'e pure
(Eq.(\ref{ActEin})) invariante sous la supersym\'etrie locale
au premier ordre en $\k$.

L'action du multiplet de supergravit\'e (Eq.(\ref{ActEin})) est invariante
sous les transformations de supersym\'etrie globale de param\`etre
${\cal E}$ constant,
\begin{eqnarray}
h_{\mu \nu} \to h_{\mu \nu} + \delta_{{\cal E}} h_{\mu \nu}
=  h_{\mu \nu} -{i \over 2} \bar {\cal E} (\gamma_{\mu} \Psi_{\nu}
+\gamma_{\nu} \Psi_{\mu}),
\label{transgg1}
\end{eqnarray}
\begin{eqnarray}
\Psi_{\mu} \to \Psi_{\mu} + \delta_{{\cal E}} \Psi_{\mu}
=  \Psi_{\mu} -i \s^{\rho \tau} \partial_{\rho} h_{\tau \mu} {\cal E}.
\label{transgg2}
\end{eqnarray}
Nous renvoyons le lecteur \`a la r\'ef\'erence \cite{BaiLov}
pour ce qui est de la d\'etermination du lagrangien de
supergravit\'e pure (Eq.(\ref{ActEin})) ainsi que des transformations
de supersym\'etrie globale associ\'ees (Eq.(\ref{transgg1}) et
Eq.(\ref{transgg2})).

Tentons \`a pr\'esent de rendre l'action de supergravit\'e pure
(Eq.(\ref{ActEin})) invariante au premier ordre en $\k$
sous la supersym\'etrie locale,
c'est \`a dire sous l'action de $e^{i {\cal E}(x) Q}$,
o\`u $Q$ est un g\'en\'erateur de SUSY et ${\cal E}(x)$ est un spineur
de Majorana d\'ependant des coordonn\'ees d'espace et de temps.
La variation de l'action de supergravit\'e pure (Eq.(\ref{ActEin}))
sous la supersym\'etrie locale \`a l'ordre $\k$ est,
\begin{eqnarray}
\delta S=\int d^4 x \bar j^{\mu} \partial_{\mu} {\cal E},
\label{dActEin}
\end{eqnarray}
o\`u le spineur de Majorana vectoriel $j^{\mu}$ est le courant
de Noether donn\'e par,
\begin{eqnarray}
\bar j^{\mu} ={i \over 2} \ep^{\mu \nu \rho \s} \bar \Psi_{\rho}
\gamma_5 \gamma_{\nu} \s^{\lambda \tau} \partial_{\lambda} h_{\tau \s}.
\label{courantgrav}
\end{eqnarray}
L'invariance sous la supersym\'etrie locale peut \^etre obtenue
\`a l'ordre $\k$, en modifiant la loi de transformation du gravitino
de Eq.(\ref{transgg2}) en,
\begin{eqnarray}
\Psi_{\mu} \to \Psi_{\mu} + \delta_{{\cal E}} \Psi_{\mu}
=  \Psi_{\mu} -i \s^{\rho \tau} \partial_{\rho} h_{\tau \mu} {\cal E}
+a \k^{-1} \partial_{\mu} {\cal E},
\label{transgg2l}
\end{eqnarray}
o\`u $a$ est une constante, et en ajoutant un terme \`a l'action
(\ref{ActEin}) du type,
\begin{eqnarray}
S_1=S -{\k \over 2a} \int d^4 x \bar j^{\mu} \Psi_{\mu}.
\label{ActEinl}
\end{eqnarray}
Aucune modification de la loi de transformation (\ref{transgg2})
n'est requise \`a ce stade. Nous remarquons l'analogie entre les
\'equations Eq.(\ref{transgaug1l}), Eq.(\ref{Actgaugl}) et
Eq.(\ref{transgg2l}), Eq.(\ref{ActEinl}).

En it\'erant ce processus, nous obtenons comme dans la Section
\ref{Noether},
les transformations finales de supersym\'etrie locale qui sont,
\begin{eqnarray}
e^m_{\mu} \to e_{\mu}^m + \delta_{{\cal E}} e^m_{\mu}
=  e^m_{\mu} -i \k  \bar {\cal E} \gamma^m \Psi_{\mu},
\label{transgg1ll}
\end{eqnarray}
o\`u $e^m_{\mu}$ est le vierbein ($\mu$ \'etant l'index d'univers et $m$
l'index local de Lorentz) satisfaisant \`a,
$h_{\mu \nu} = e^m_{\mu} e^n_{\nu} \eta_{m n}$, et,
\begin{eqnarray}
\Psi_{\mu} \to \Psi_{\mu} + \delta_{{\cal E}} \Psi_{\mu}
=  \Psi_{\mu} + 2 \k^{-1} D_{\mu} {\cal E},
\label{transgg2ll}
\end{eqnarray}
o\`u $D_{\mu}$ est la d\'eriv\'ee covariante,
\begin{eqnarray}
D_{\mu}=\partial_{\mu}-i \tilde w_{\mu m n} {\s^{m n} \over 4}
\label{Dmucov}
\end{eqnarray}
avec,
\begin{eqnarray}
\tilde w_{\mu m n} = w_{\mu m n} + {i \k^2 \over 4}
(\bar \Psi_{\mu} \gamma_m \Psi_n
+\bar \Psi_m \gamma_{\mu} \Psi_n
-\bar \Psi_{\mu} \gamma_n \Psi_m)
\label{wmntil}
\end{eqnarray}
et,
\begin{eqnarray}
w_{\mu m n} ={1 \over 2}
e_m^{\nu} (\partial_{\mu} e_{n \nu}-\partial_{\nu} e_{n \mu})
+{1 \over 2}
e_m^{\rho} e_n^{\s} \partial_{\s} e_{\rho p} e_{\mu}^p - (m \to n),
\label{wmn}
\end{eqnarray}
qui est la connection de spin standard. Et l'action finale
localement supersym\'etrique que l'on obtient est:
\begin{eqnarray}
S=-{1 \over 2 \k^2} \int d^4 x \vert det \ e \vert  R
-{1 \over 2} \int d^4 x \ep^{\mu \nu \rho \s} \bar \Psi_{\mu}
\gamma_5 \gamma_{\nu} D_{\rho} \Psi_{\s}.
\label{ActEinll}
\end{eqnarray}
Le premier terme est l'action d'Einstein sous sa forme g\'en\'erale
(non lin\'earis\'ee) car l'action de Eq.(\ref{ActEinll}) est exactement 
invariante sous la supersym\'etrie locale (non pas seulement
au premier ordre en $\k$, comme l'est l'action de Eq.(\ref{ActEinl})).
Le second terme est l'action de Rarita-Schwinger avec
une d\'eriv\'ee covariante. Cette d\'eriv\'ee
covariante implique le champ du gravitino
(voir Eq.(\ref{Dmucov})) de m\^eme que
les d\'eriv\'ees covariantes des th\'eories de jauge 
impliquent les bosons de jauge (voir Eq.(\ref{ActDiracl})).

En conclusion, la supersym\'etrie locale est r\'ealis\'ee gr\^ace au fait
que le gravitino se comporte comme le ``boson de jauge" de la
supersym\'etrie (voir Eq.(\ref{transgg2l}), Eq.(\ref{ActEinl})
et Eq.(\ref{Dmucov})).
Le gravitino joue donc un r\^ole aussi naturel que majeur
dans les th\'eories de supersym\'etrie locale.

\subsection{Lagrangien localement supersym\'etrique pour
un supermultiplet chiral libre et non massif}

Consid\'erons le lagrangien du supermultiplet chiral libre
et non massif du mod\`ele de Wess Zumino:
\begin{eqnarray}
{\cal L}_0=\partial_{\mu} \phi^\star \partial^{\mu} \phi
+{i \over 2} \bar \Psi \bar \gamma^{\mu} \partial_{\mu} \Psi
\label{WZeg}
\end{eqnarray}
o\`u $\Psi$ est un spineur de Majorana. Le lagrangien de
Eq.(\ref{WZeg}) est invariant sous les transformations de
supersym\'etrie globale suivantes (voir Section \ref{Majo}):
\begin{eqnarray}
\delta A=\bar {\cal E} \Psi,
\label{WZegtr1}
\end{eqnarray}
\begin{eqnarray}
\delta B=i \bar {\cal E} \gamma_5 \Psi,
\label{WZegtr2}
\end{eqnarray}
\begin{eqnarray}
\delta \Psi=-i \gamma^{\mu} \partial_{\mu} (A+i \gamma_5 B) {\cal E},
\label{WZegtr3}
\end{eqnarray}
o\`u $A$ et $B$ sont les champs r\'eels d\'efinis \`a partir
du champ $\phi$ par,
\begin{eqnarray}
\phi =  {1 \over \sqrt 2} (A+i B),
\label{ABphi}
\end{eqnarray}
et ${\cal E}$ est le param\`etre spineur de Majorana donn\'e par,
\begin{equation}
{\cal E}=  \left ( \begin{array}{c} {\cal E}_W \\ \bar {\cal E}_W \end{array} \right ).
\label{SUSYpar}
\end{equation}
Lorsque la supersym\'etrie devient locale, c'est \`a dire
lorsque le param\`etre ${\cal E}$ des transformations
(\ref{WZegtr1})-(\ref{WZegtr3})
d\'epend de l'espace-temps,
la variation du lagrangien (\ref{WZeg}) est
(\`a une d\'eriv\'ee pr\`es),
\begin{eqnarray}
\delta {\cal L}_0=\partial_{\mu} \bar {\cal E} j^{\mu},
\label{dWZeg}
\end{eqnarray}
o\`u $j^{\mu}$ est le spineur vectoriel suivant,
\begin{eqnarray}
j^{\mu}=\Dslash (A-i \gamma_5 B) \gamma^{\mu} \Psi.
\label{Ncour}
\end{eqnarray}
Remarquons l'analogie entre les \'equations
Eq.(\ref{dWZeg}) et Eq.(\ref{dActEin}).
Afin de rendre le lagrangien ${\cal L}_0$ localement supersym\'etrique,
et par analogie avec l'\'equation Eq.(\ref{ActEinl}),
nous ajoutons au lagrangien le terme,
\begin{eqnarray}
{\cal L}={\cal L}_0+ a \bar \Psi_{\mu} j^{\mu},
\label{WZegl}
\end{eqnarray}
o\`u $\Psi_{\mu}$ doit \^etre un spineur vectoriel de m\^eme
que $j^{\mu}$. Nous associons ce spineur vectoriel au champ
du gravitino. Si le gravitino se transforme sous la supersym\'etrie 
locale selon,
\begin{eqnarray}
\Psi_{\mu} \to \Psi_{\mu} +2 \k^{-1} \partial_{\mu} {\cal E},
\label{transWZl}
\end{eqnarray}
alors le lagrangien ${\cal L}$ de Eq.(\ref{WZegl}) 
est invariant sous
la supersym\'etrie locale au premier ordre en $\k$ pour,
\begin{eqnarray}
a=-{\k \over 2}.
\label{acstel}
\end{eqnarray}
Le lagrangien localement supersym\'etrique final, obtenu en
it\'erant cette m\'ethode plusieurs fois, vaut,
\begin{eqnarray}
{\cal L}_f&=&
-{  1 \over 2 \k^2 }  \vert det \ e \vert R
-{ 1 \over 2 }  \ep^{\mu \nu \rho \sigma}  \bar \Psi_{\mu} \gamma_5
\gamma_{\nu}  \tilde D_{\rho}  \Psi_{\sigma}
+ \vert det \ e \vert \partial_{\mu} \phi^\star  \partial^{\mu} \phi
\cr && -{  \k \over 2 }  \vert det \ e \vert \bar \Psi_{\mu}
\phi (A-i \gamma_5 B) \gamma^{\mu} \Psi 
-{ i \k^2 \over 4 }  \ep^{\mu \tau \rho \sigma}  \bar \Psi_{\mu} \gamma_{\tau}
\Psi_{\rho} A
\stackrel{\leftrightarrow}{D}_{\sigma} B \cr &&
-{ \k^2 \over 4 } \vert det \ e \vert \bar \Psi \gamma_5 \gamma^{\mu}
\Psi A
\stackrel{\leftrightarrow}{D}_{\mu} B + \ Termes(4 \ fermions), \cr && 
\label{WZlf}
\end{eqnarray}
o\`u les d\'eriv\'ees $D_{\mu}$ sont des d\'eriv\'ees
covariantes vis-\`a-vis
de la gravitation. Les lois de supersym\'etrie locale
obtenues sont,
\begin{eqnarray}
\delta A=\bar {\cal E} \Psi,
\label{transWZll1}
\end{eqnarray}
\begin{eqnarray}
\delta B=i \bar {\cal E} \gamma_5 \Psi,
\label{transWZll2}
\end{eqnarray}
\begin{eqnarray}
\delta e_{\mu}^m=- i \k \bar {\cal E} \gamma^m \Psi_{\mu}
\label{transWZll3}
\end{eqnarray}
\begin{eqnarray}
\delta \Psi_{\mu} =2 \k^{-1} D_{\mu} {\cal E}
+ i \k {\cal E} A \stackrel{\leftrightarrow}{D}_{\mu} B + \ Termes(2 \ fermions),
\label{transWZll4}
\end{eqnarray}
\begin{eqnarray}
\delta \Psi=-i \gamma^{\mu} D_{\mu} (A+i \gamma_5 B) {\cal E}
+ \ Termes(2 \ fermions).
\label{transWZll5}
\end{eqnarray}
Les deux premiers termes du lagrangien
${\cal L}_f$ de Eq.(\ref{WZlf}) sont
les deux termes du lagrangien de supergravit\'e pure (voir
Eq.(\ref{ActEinll})). Par ailleurs, la transformation
du gravitino \ref{transWZll4}
contient la transformation du gravitino \ref{transgg2ll}
d'une th\'eorie de supergravit\'e pure. Nous en concluons
que le choix du gravitino comme boson de jauge de la
supersym\'etrie locale est coh\'erent.
Une fois encore donc, le gravitino acquiert naturellement un
r\^ole crucial lorsque la supersym\'etrie locale est requise.
\\ Notons finalement qu'une approche similaire
permet de d\'eriver les couplages du
supermultiplet vectoriel \`a la supergravit\'e.

\subsection{Lagrangien localement supersym\'etrique  g\'en\'eral}

Le lagrangien localement supersym\'etrique  g\'en\'eral
peut s'obtenir \`a partir du
lagrangien globalement supersym\'etrique, en utilisant,
comme dans les sections pr\'ec\'edentes, la proc\'edure de Noether.
Cependant, en pratique, ce r\'esultat compliqu\'e a \'et\'e d\'eduit
de la m\'ethode de calcul tensoriel local 
\cite{oritl,west,WessBagger}.
Nous pr\'esentons ici ce lagrangien g\'en\'eral de supergravit\'e.
Le lagrangien  globalement supersym\'etrique le plus g\'en\'eral
s'\'ecrit,
\begin{eqnarray}
{\cal L}_{GLOBAL}=-3 \int d^4 \theta e^{-{1 \over 3}
K(\Phi^{\dagger}  e^{2gV}, \Phi)} + \int d^2 \theta (W(\Phi)+h.c.)
+ \int d^2 \theta (f_{ab}(\Phi) W^{\alpha}_a W_{\alpha b}+h.c.), \cr
\label{SUSYGLOB}
\end{eqnarray}
o\`u $W_a^{\alpha}$ repr\'esente le superchamp de jauge, $\alpha$ \'etant
l'indice spinoriel et $a$ l'indice du groupe de jauge.
$K$ est une fonction g\'en\'erale de $\Phi_i$ et $\Phi^{\dagger}_i e^{2gV}$
appel\'ee potentiel de K\"ahler qui autorise
des termes cin\'etiques non renormalisables et $f_{ab}$ est une fonction
arbitraire de $\Phi_i$ se r\'eduisant \`a $\delta_{ab}$ dans le cas
renormalisable. De m\^eme, le superpotentiel
$W$ peut contenir des puissances arbitraires des superchamps $\Phi_i$.
Il appara\^{\i}t en fait que le lagrangien de supergravit\'e ne d\'epend que
d'une fonction simple des champs scalaires, $\phi_i$ et $\phi_i^\star$,
appel\'ee potentiel de K\"ahler g\'en\'eralis\'e et d\'efinie par,
\begin{eqnarray}
G(\phi,\phi^\star)=K(\phi,\phi^\star) + ln \vert W \vert^2.
\label{Gfonc}
\end{eqnarray}
Les m\^emes lagrangiens de supergravit\'e peuvent \^etre obtenus pour
diff\'erents choix
de $K$ et $W$, car $G$ est invariant sous les transformations suivantes:
\begin{eqnarray}
K & \to &  K + h(\phi) + h^\star(\phi^\star)  \cr
W &  \to &  e^{-h} W
\label{JWtrans}
\end{eqnarray}
pour une fonction $h$ arbitraire.

Le lagrangien de supergravit\'e g\'en\'eral ${\cal L}_{SUGRA}$ peut se
d\'ecomposer en,
\begin{eqnarray}
{\cal L}_{SUGRA}={\cal L}_B+{\cal L}_{FK}+{\cal L}_F+\tilde {\cal L}_B
+\tilde {\cal L}_{FK}+\tilde {\cal L}_F,
\label{LSUGRAGEN}
\end{eqnarray}
o\`u ${\cal L}_B$ contient les champs scalaires, ${\cal L}_{FK}$ contient
les champs
fermioniques ainsi
que des d\'eriv\'ees covariantes (en particulier les termes cin\'etiques
pour les fermions)
et ${\cal L}_F$ contient les champs fermioniques (en particulier les
couplages de
Yukawa)
mais pas de d\'eriv\'ees covariantes. Les lagrangiens $\tilde {\cal L}_B$,
$\tilde {\cal L}_{FK}$ et
$\tilde {\cal L}_F$ contiennent les champs de jauge.
Si $R$ est la courbure scalaire et $\vert det \ e \vert$ le d\'eterminant du
vierbein $e^m_\mu$,
alors la partie bosonique du lagrangien est donn\'ee par (en unit\'es telles
que $\k^2=1$),
\begin{eqnarray}
\vert det \ e \vert^{-1} {\cal L}_B=-{  1 \over 2 } R
+G^i_j D_{\mu} \phi_i  D^{\mu} \phi^{j\star}
+e^G(3-G_i(G^{-1})^i_jG^j),
\label{LrB}
\end{eqnarray}
o\`u les d\'eriv\'ees $D_{\mu}$ sont covariantes vis \`a vis des groupes de
jauge mais aussi de la gravit\'e.
Les d\'eriv\'ees du potentiel de K\"ahler g\'en\'eralis\'e par
rapport aux champs scalaires $\phi_i$ et $\phi^{i\star}$
sont not\'ees,
\begin{eqnarray}
G_i= {\partial G \over \partial \phi^{i\star}},  \ \ \ \
G^i= {\partial G \over \partial \phi_i},    \ \ \ \
G^i_j = {\partial^2 G \over \partial \phi_i \partial \phi^{j\star}}.
\label{Gfoncder}
\end{eqnarray}
De plus, l'inverse $(G^{-1})^i_j$ est d\'efini par,
\begin{eqnarray}
(G^{-1})^i_jG^j_k=\delta^i_k.
\label{Ginv}
\end{eqnarray}
Les termes cin\'etiques fermioniques sont donn\'es par le lagrangien
suivant,
\begin{eqnarray}
\vert det \ e \vert^{-1} {\cal L}_{FK}&=&
-{ 1 \over 2 } \vert det \ e \vert^{-1}
  \ep^{\mu \nu \rho \sigma}  \bar \Psi_{\mu} \gamma_5 \gamma_{\nu}
  D_{\rho} \Psi_{\sigma}
+{  1 \over 4 }  \vert det \ e \vert^{-1}
  \ep^{\mu \nu \rho \sigma}  \bar \Psi_{\mu} \gamma_{\nu} \Psi_{\rho}
   (G^iD_{\sigma} \phi_i- G_iD_{\sigma} \phi^{i\star})
\cr && + \bigg (
        { i \over 2 } G^i_j \bar \Psi_{i L} \gamma^{\mu} D_{\mu} \Psi^j_L
        +{i \over 2} \bar \Psi_{i L}  \Dslash \phi_j \Psi_{k L}
              (-G^{ij}_k+{1 \over 2} G^i_k G^j)
        \cr && +{ 1 \over \sqrt 2 } G^i_j
         \bar \Psi_{\mu L} \Dslash \phi^{i\star} \gamma^{\mu} \Psi_{j R} +
h.c.
\bigg ),
\label{LrFK}
\end{eqnarray}
et les interactions de Yukawa viennent de (voir la litt\'erature originelle
\cite{orilit} et la revue sur la supersym\'etrie \cite{Nilles} pour le lagrangien complet),
\begin{eqnarray}
\vert det \ e \vert^{-1} {\cal L}_{F}&=&{ i \over 2 } e^{G/2}
\bar \Psi_{\mu} \sigma^{\mu \nu} \Psi_{\nu} \cr
&+& \bigg (
{ 1 \over 2 } e^{G/2} (-G^{ij}-G^iG^j+G^{ij}_k(G^{-1})^k_lG^l) \bar \Psi_{i
L} \Psi_{j R}
{ i \over \sqrt 2 } e^{G/2} G^i \bar \Psi_{\mu L} \gamma^{\mu} \Psi_{iL}+
h.c.
\bigg ) \cr &+& \ Termes(4 \ fermions).
\label{LrF}
\end{eqnarray}
$\tilde {\cal L}_B$ est donn\'e par,
\begin{eqnarray}
\vert det \ e \vert^{-1} \tilde {\cal L}_B&=&
-{  1 \over 4 } Re(f_{ab}) (F_a)_{\mu \nu} F_b^{\mu \nu}
-{  i \over 4 } Im(f_{ab}) (F_a)_{\mu \nu} \tilde F_b^{\mu \nu}
\cr && -{  g^2 \over 2 } Re(f^{-1}_{ab}) G^i (T_a)_{ij} \phi_j G^k (T_b)_{kl}
\phi_l, \cr && 
\label{LrBtil}
\end{eqnarray}
o\`u les $(T_a)_{ij}$ sont les g\'en\'erateurs du groupe de jauge,
les champs de jauge sont d\'efinis par,
\begin{eqnarray}
(F_a)_{\mu \nu} = \partial_{\mu} V_{\nu a}-\partial_{\nu}V_{\mu a}-g
f_{abc} V_{\mu b}
V_{\nu c},
\label{Famunu}
\end{eqnarray}
et leur dual par,
\begin{eqnarray}
(\tilde F_a)_{\mu \nu} =  \ep_{\mu \nu \rho \sigma} (F_a)^{\rho \sigma}.
\label{Fdual}
\end{eqnarray}
$\tilde {\cal L}_{FK}$ s'\'ecrit,
\begin{eqnarray}
\vert det \ e \vert^{-1} \tilde {\cal L}_{FK}&=&
{  1 \over 2 } Re(f_{ab}(\phi))
\bigg (
 { 1 \over 2 } \bar \lambda_a \Dslash \lambda_b
+{ 1 \over 2 } \bar \lambda_a \gamma^{\mu} \sigma^{\nu \rho} \Psi_{\mu}
(F_b)_{\nu \rho}
+{ 1 \over 2 } G^i D^{\mu}  \phi_i \bar \lambda_{aL} \gamma_{\mu}
\lambda_{bL}
\bigg ) \cr &&
-{ i \over 8 } Im(f_{ab}(\phi)) D_{\mu} (\vert det \ e \vert
  \bar \lambda_a \gamma_5 \gamma^{\mu} \lambda_b)
- { 1 \over 2 } {\partial f_{ab}(\phi) \over \partial \phi_i}  \bar \Psi_{i
R}
\sigma^{\mu \nu}
(F_a)_{\mu \nu} \lambda_{bL}+h.c., \cr &&
\label{LrFKtil}
\end{eqnarray}
o\`u $\lambda_a$ est le jaugino. Finalement, $\tilde {\cal L}_F$ vaut,
\begin{eqnarray}
\vert det \ e \vert^{-1} \tilde {\cal L}_F&=&
{1 \over 4} e^{G/2} {\partial f_{ab}^\star \over \partial \phi^{j\star}}
(G^{-1})^j_k G^k
\lambda_a \lambda_b 
-{i \over 2} g G^i (T_a)_{ij} \phi_j \bar \Psi_{\mu L} \gamma^{\mu}
\lambda_{a L}
+2 i g G_j^i (T_a)_{ik} \phi_k \bar \lambda_{a R} \Psi_{i L} \cr &&
-{i \over 2} g Re(f_{ab}^{-1}) {\partial f_{bc} \over \partial \phi_k}
G^i(T_a)_{ij}
\phi_j \bar \Psi_{k R}
+\lambda_{c L} + h.c. + \ Termes(4 \ fermions). \cr &&
\label{LrFtil}
\end{eqnarray}
Notons que le lagrangien de Eq.(\ref{LrFtil}) peut donner
des termes de masse pour les jauginos si les champs
scalaires $\phi_i$ d\'eveloppent des valeurs moyennes dans le vide.

Les transformations de supersym\'etrie locale associ\'ees au lagrangien
${\cal L}_{SUGRA}$
(Eq.(\ref{LSUGRAGEN})) sont d\'efinies par les variations infinit\'esimales
suivantes
(voir la litt\'erature originelle \cite{orilit} pour les
expressions compl\`etes),
\begin{eqnarray}
\delta \phi_i= \sqrt 2 \bar {\cal E} \Psi_i = \sqrt 2 \bar {\cal E}_R \Psi_{iL},
\label{trphi}
\end{eqnarray}
\begin{eqnarray}
\delta e_{\mu}^m= -i \k \bar {\cal E} \gamma^m \Psi_{\mu},
\label{tre}
\end{eqnarray}
\begin{eqnarray}
\delta \Psi_{\mu}= 2 \k^{-1} D_{\mu} {\cal E} + \k {\cal E}(G^iD_{\mu} \phi_i-
G_iD_{\mu} \phi^{i\star})
+i e^{G/2} \gamma_{\mu} {\cal E} + \ Termes(2 \ fermions),
\label{trpsimu}
\end{eqnarray}
\begin{eqnarray}
\delta \Psi_i= -i \Dslash(A_i+i \gamma_5 B_i) {\cal E} -\sqrt 2 e^{G/2}
(G^{-1})^j_i G_j {\cal E}
+ Termes(2 \ fermions),
\label{trpsi}
\end{eqnarray}
\begin{eqnarray}
\delta V_a^{\mu}= - \bar {\cal E}_L \gamma^{\mu} \lambda_{a L}+h.c.,
\label{trV}
\end{eqnarray}
\begin{eqnarray}
\delta \lambda_{a L}= \sigma^{\mu \nu} (F_a)_{\mu \nu} {\cal E}_L
+{i \over 2} g Re (f_{ab}^{-1}) G^i (T_b)_{ij} \phi_j {\cal E}_L
+ \ Termes(2 \ fermions),
\label{trlambda}
\end{eqnarray}
o\`u $\phi_i$ se d\'ecompose en deux champs scalaires r\'eels $A_i$ et $B_i$
comme,
\begin{eqnarray}
\phi_i = {1 \over \sqrt 2} (A_i+iB_i).
\label{phidecomp}
\end{eqnarray}

Remarquons que le lagrangien invariant par supergravit\'e
n'incluant pas de champs de jauge (supermultiplets vectoriels),
${\cal L}={\cal L}_B + {\cal L}_{FK} + {\cal L}_F$,
g\'en\`ere le lagrangien invariant sous la supersym\'etrie globale
(accompagn\'e de termes non renormalisables),
pour un superpotentiel de la forme,
\begin{eqnarray}
W(\Phi)=\lambda_{ijk} \Phi_i \Phi_j \Phi_k,
\label{sup1}
\end{eqnarray}
et pour un potentiel de K\"ahler valant,
\begin{eqnarray}
K(\Phi^{\dagger}, \Phi)=-3ln(1-{1 \over 3} \Phi_i^{\dagger} \Phi_i).
\label{Kah1}
\end{eqnarray}
Bien que la constante n'aie pas de signification dans une th\'eorie
globalement supersym\'etrique, celle-ci est importante dans une th\'eorie
de supersym\'etrie locale qui est coupl\'ee \`a la gravit\'e.

Par ailleurs, une th\'eorie incluant la gravit\'e n'a pas de raison 
d'\^etre renormalisable. En particulier, le potentiel de K\"ahler
n'est pas obligatoirement celui qui correspond
uniquement \`a des termes cin\'etiques
renormalisables dans la th\'eorie supersym\'etrique globale.
Il est en revanche utile pour les calculs de d\'efinir la forme
du potentiel de K\"ahler qui donne les termes cin\'etiques minimaux
dans le lagrangien de supergravit\'e. Cette forme
du potentiel de K\"ahler est,
\begin{eqnarray}
K(\Phi^{\dagger}, \Phi)=\Phi_i \Phi^{i\dagger},
\label{Kah2}
\end{eqnarray}
soit,
\begin{eqnarray}
G=\phi_i \phi^{i\star}+ln \vert W \vert ^2,
\label{Gfonc1}
\end{eqnarray}
d'o\`u,
\begin{eqnarray}
G_j^i=\delta_j^i.
\label{Gder1}
\end{eqnarray}
D'apr\`es l'Eq.(\ref{Gder1}), les termes cin\'etiques
sont simplement, $\partial_{\mu} \phi_i \partial^{\mu} \phi^\star_i$, pour
les champs
scalaires (voir ${\cal L}_B$) et, ${i \over 2} \bar \Psi_{iL} \gamma^{\mu}
\partial_{\mu} \Psi_{iL}+ h.c.$, pour les champs fermioniques (voir ${\cal
L}_{FK}$).

\section{Brisure spontan\'ee en supergravit\'e}

Pour avoir une brisure spontan\'ee en supergravit\'e, au moins un
des champs doit avoir une VEV qui n'est pas invariante sous l'action
de la supersym\'etrie locale. Les seules transformations supersym\'etriques
parmi les Eq.(\ref{trphi})-Eq.(\ref{trlambda} pouvant avoir un membre de droite
ayant une VEV non nulle sans briser l'invariance de Lorentz, sont les lois
des Eq.(\ref{trpsi}) et (\ref{trlambda}). Si nous supposons qu'il n'y a pas
de VEV pour les termes impliquant des champs fermioniques, les valeurs
moyennes dans le vide des Eq.(\ref{trpsi}) et (\ref{trlambda}) s'\'ecrivent,
\begin{eqnarray}
<0 \vert \delta \Psi_i \vert 0>=-e^{G/2} (G^{-1})^j_i G_j {\cal E},
\label{VEVP1}
\end{eqnarray}
\begin{eqnarray}
<0 \vert \delta \lambda_a \vert 0>={i \over 2} g Re f^{-1}_{ab} G^i
(T_b)_{ij} \phi_j {\cal E},
\label{VEVl1}
\end{eqnarray}
o\`u les champs scalaires ont \'et\'e utilis\'es afin de
d\'enoter les VEV.

Consid\'erons tout d'abord le cas simple d'un superchamp chiral singlet
de jauge avec les termes cin\'etiques minimaux, c'est \`a dire le cas:
\begin{eqnarray}
G=\phi \phi^\star+ln \vert W \vert ^2.
\label{Gfonc2}
\end{eqnarray}
Les Eq.(\ref{VEVP1}) et (\ref{VEVl1}) deviennent alors,
\begin{eqnarray}
<0 \vert \delta \Psi \vert 0>=-{exp({1 \over 2}(\phi^\star \phi +
ln \vert W \vert ^2)) \over W^\star} \bigg ( {\partial W^\star \over \partial \phi^\star}
+ \phi W^\star \bigg ) {\cal E},
\label{VEVP1s}
\end{eqnarray}
\begin{eqnarray}
<0 \vert \delta \lambda_a \vert 0>=0.
\label{VEVl1s}
\end{eqnarray}
Le crit\`ere de brisure spontan\'ee est donc,
\begin{eqnarray}
{\partial W \over \partial \phi} + \phi^\star W \neq 0.
\label{critere}
\end{eqnarray}
Ceci est la g\'en\'eralisation de la brisure de supersym\'etrie globale
par un terme F.
\\ Qu'en est-il du potentiel ? D'apr\`es l'\'equation Eq.(\ref{Gfonc2})
et le lagrangien ${\cal L}_B$, le potentiel vaut,
\begin{eqnarray}
V=e^{\phi^\star \phi} \bigg ( \bigg \vert {\partial W \over \partial \phi}
+ \phi^\star W \bigg \vert ^2 -3 \vert W \vert ^2 \bigg ),
\label{V1}
\end{eqnarray}
\`a comparer avec, $V= \vert W \vert ^2$,
dans le cas d'une supersym\'etrie globale. D'apr\`es Eq.(\ref{V1}),
l'\'energie du vide a maintenant la possibilit\'e 
d'\^etre n\'egative. Nous remarquons m\^eme d'apr\`es Eq.(\ref{critere})
que l'\'energie du vide supersym\'etrique est n\'egative
puisque le potentiel correspondant vaut,
\begin{eqnarray}
V= -3 e^{\phi^\star \phi } \vert W \vert ^2.
\label{V2}
\end{eqnarray}
D'apr\`es Eq.(\ref{critere}), lorsque la supersym\'etrie est bris\'ee,
une annulation dans le potentiel (Eq.(\ref{V1})) peut donner un vide
d'\'energie nulle. Dans les th\'eories de supergravit\'e, il existe
donc une possibilit\'e d'obtenir un \'etat du vide ayant une constante
cosmologique nulle, si la supersym\'etrie est bris\'ee.

\'Etudions \`a pr\'esent le cas d'un superchamp chiral appartenant \`a une
repr\'esentation non
triviale du groupe de
de jauge avec les termes cin\'etiques minimaux (voir Eq.(\ref{Gfonc1})).
Choisissons aussi les termes cin\'etiques minimums pour les champs de jauge,
c'est \`a dire,
\begin{eqnarray}
f_{ab}(\phi)=\delta_{ab}.
\label{fab}
\end{eqnarray}
Les Eq.(\ref{VEVP1}) et (\ref{VEVl1}) deviennent alors,
\begin{eqnarray}
<0 \vert \delta \Psi_i \vert 0>=-{exp({1 \over 2}(\phi^{j\star} \phi_j +
ln \vert W \vert ^2)) \over W^\star} \bigg ( {\partial W^\star \over 
\partial \phi_i^\star}
+ \phi^i W^\star \bigg ) {\cal E},
\label{VEVP2s}
\end{eqnarray}
\begin{eqnarray}
<0 \vert \delta \lambda_a \vert 0>={i \over 2} g G^i
(T_a)_{ij} \phi_j {\cal E}.
\label{VEVl2s}
\end{eqnarray}
Il y a dans ce cas deux crit\`eres possibles de brisure spontan\'ee:
\begin{eqnarray}
{\partial W \over \partial \phi_i} + \phi_i^\star W \neq 0, \ \ pour \ \ certaines
\ \ valeurs \ \ de \ \ i,
\label{critere2}
\end{eqnarray}
\begin{eqnarray}
G^i(T_a)_{ij} \phi_j \neq 0, \ \ pour \ \ certaines \ \ valeurs \ \ de \ \ a,
\label{critere3}
\end{eqnarray}
qui sont les g\'en\'eralisations de la brisure de supersym\'etrie globale
par un terme F ou par un terme D, respectivement.
\\ D'apr\`es l'\'equation Eq.(\ref{Gfonc1})
et les lagrangiens ${\cal L}_B$ et $\tilde {\cal L}_B$, le potentiel vaut,
\begin{eqnarray}
V=e^{\phi^{j\star} \phi_j} \bigg ( \bigg \vert {\partial W \over \partial
\phi_i}
+ \phi^{i\star} W \bigg \vert ^2 -3 \vert W \vert ^2 \bigg )
+ {g^2 \over 2} G^i(T_a)_{ij} \phi_j G^k(T_a)_{kl} \phi_l,
\label{V3}
\end{eqnarray}
o\`u,
\begin{eqnarray}
G^i=\phi^{i\star}+{1 \over W} {\partial W \over \partial \phi_i}
\label{Gder}
\end{eqnarray}
\`a comparer avec le potentiel semi-d\'efini positif d'une
th\'eorie globalement supersym\'etrique,
\begin{eqnarray}
V= \vert {\partial W \over \partial \phi_i} \vert ^2
+ {g^2 \over 2} \phi^{i\star}(T_a)_{ij} \phi_j \phi^{k\star}(T_a)_{kl}
\phi_l.
\label{Vglob}
\end{eqnarray}

Nous remarquons d'apr\`es Eq.(\ref{Gder}) que
la condition de brisure de Eq.(\ref{critere2}) est \'equiva\-lente \`a
$G^i \neq 0$. Et le champ $G^i \Psi_i$ n'est en fait rien d'autre
que le fermion de Goldstone associ\'e \`a la brisure de supersym\'etrie.
Il y a un m\'elange entre ce fermion
de Goldstone et le gravitino $\Psi_{\mu}$, comme nous le voyons dans
le lagrangien ${\cal L}_F$. Le fermion de Goldstone (de spin ${1 \over 2}$)
est en fait "mang\'e" par le gravitino (de spin ${3 \over 2}$) qui
devient alors massif. C'est ce que l'on nomme le m\'ecanisme de
super-Higgs. Le terme de masse du gravitino est le premier terme
du lagrangien ${\cal L}_F$. La masse du gravitino, $m_{3/2}$, est une
fonction
de la valeur dans le vide de la fonction $G$, not\'ee $G_0$, ainsi
que de la masse de Planck, $M_P$:
\begin{eqnarray}
m_{3/2} \approx e^{G_0/2} M_P,
\label{gravmass}
\end{eqnarray}
puisque nous avions choisi de travailler en unit\'es de $\k^2$
qui est d\'efini dans Eq.(\ref{kappa}).

Remarquons que le gravitino se comporte une fois
encore comme un `boson de jauge' pour la
supersym\'etrie locale dans la mesure o\`u il
acquiert une masse lors de la brisure spontan\'ee de la supersym\'etrie
locale.

\section{Brisure de la supersym\'etrie dans le secteur ``cach\'e''}

Comme nous allons le voir dans la prochaine section,
la brisure de supersym\'etrie locale op\`ere de mani\`ere satisfaisante
lorsqu'elle est transmise au secteur "observable" \`a partir
d'un secteur ``cach\'e''. Par brisure satisfaisante de la supersym\'etrie
locale
nous entendons une brisure engendrant, par le biais de termes
de brisure douce (voir Section \ref{BRIsec}), des
masses pour les particules supersym\'etriques sup\'erieures
aux masses des particules du Mod\`ele Standard et inf\'erieures au TeV
(voir Section \ref{hier}). Par ailleurs,
le secteur cach\'e est un ensemble de particules
interagissant avec les particules du MSSM uniquement via les
forces gravitationnelles.

Consid\'erons donc le mod\`ele le plus simple de brisure de
la supersym\'etrie locale dans un secteur cach\'e: Le mod\`ele de Polonyi.
Dans ce mod\`ele, le secteur cach\'e comprend
un unique superchamp chiral $\Phi$ et le superpotentiel
associ\'e est le suivant,
\begin{eqnarray}
W(\Phi)=m^2 (\Phi+ \beta),
\label{Polonyi}
\end{eqnarray}
$m$ et $\beta$ \'etant des param\`etres r\'eels ayant des dimensions
de masse. Pour des raisons de simplicit\'e nous adoptons aussi
le choix de
sc\'enario dans lequel les termes cin\'etiques sont minimaux
(voir Eq.(\ref{Gfonc2})).
Pour le superpotentiel de Eq.(\ref{Polonyi}), le potentiel
s'\'ecrit donc, d'apr\`es Eq.(\ref{V1}),
\begin{eqnarray}
V= m^4 e^{\phi^\star \phi} ( \vert 1 + \phi^\star (\phi + \beta) \vert ^2
-3 \vert \phi + \beta \vert ^2 \bigg ).
\label{V4}
\end{eqnarray}
Nous pouvons montrer que pour la valeur, $\beta =2-\sqrt 3$, le
potentiel $V$ a un minimum absolu en, $\phi=\phi_0=\sqrt 3-1$,
pour lequel, $V=0$ (constante cosmologique nulle).
De plus, pour ce minimum, la supersym\'etrie est bris\'ee par
l'\'equivalent d'un terme F car on a,
${\partial W \over \partial \phi} + \phi^\star W =\sqrt 3 m^2 \neq 0$
(voir Eq.(\ref{critere})).
D'apr\`es Eq.(\ref{gravmass}), le gravitino acquiert la
masse suivante,
\begin{eqnarray}
m_{3/2}=exp[{1 \over 2}(\sqrt 3 -1)^2] {m^2 \over M_P^2} M_P,
\label{gravmass2}
\end{eqnarray}
apr\`es restauration de la constante $M_P$. Nous voyons ici que
la masse du gravitino peut \^etre tr\`es inf\'erieure \`a la
masse de Planck si, ${m^2 \over M_P^2}<<1$.

D'apr\`es Eq.(\ref{VEVP1}) qui est, rappelons-le, la
g\'en\'eralisation de la brisure de SUSY par un terme F,
l'\'echelle de brisure de la supersym\'etrie est,
\begin{eqnarray}
M_S^2=e^{G/2} (G^{-1})^j_i G_j.
\label{SUSYscale}
\end{eqnarray}
Dans le cas pr\'esent,
\begin{eqnarray}
M_S^2= e^{G/2} \bigg ( \phi^\star +
{1 \over W}{\partial W \over \partial \phi} \bigg )=\sqrt 3 m_{3/2} M_P.
\label{SUSYscale2}
\end{eqnarray}
Ce r\'esultat est g\'en\'eral pour les th\'eories dans
lesquelles le vide brisant la supersym\'etrie est en
$V=0$. L'\'equation Eq.(\ref{SUSYscale2}) donne pour
la masse du gravitino,
\begin{eqnarray}
m_{3/2} = {M_S^2 \over \sqrt 3 M_P}.
\label{gravmass3}
\end{eqnarray}
La masse du gravitino est donc petite par rapport \`a
l'\'echelle de brisure de SUSY, si celle-ci est petite
comparativement \`a la masse de Planck.

D'apr\`es ${\cal L}_B$, les champs scalaires, du seul supermultiplet
chiral du secteur cach\'e, acqui\`erent les masses suivantes,
\begin{eqnarray}
{\cal L}_B^m=-2m^2_{3/2} \phi'^\star \phi'-2(\sqrt 3-1)m^2_{3/2}
(\phi' \phi'+\phi'^\star \phi'^\star),
\label{LBm}
\end{eqnarray}
o\`u nous avons pris, $\phi=\phi_0+\phi'$, $\phi_0$
\'etant la VEV d\'ecrite plus haut. D\'efinissant
les champs $A$ et $B$ par, $\phi'={1 \over \sqrt2 }(A+iB)$,
nous obtenons les masses, $m_A^2=2 \sqrt 3 m^2_{3/2}$ et
$m_B^2=2 (2-\sqrt 3) m^2_{3/2}$.
Nous remarquons donc que les particules
scalaires $A$ et $B$ ont bien des
masses sup\'erieures \`a celles de leur partenaires
supersym\'etriques fermioniques (qui sont ici nulles).
Dans le cas o\`u l'on a un multiplet de
supergravit\'e et $N$ supermultiplets chiraux,
ceci reste vrai comme le montre la supertrace
qui est toujours positive:
\begin{eqnarray}
STr(M^2)=2(N-1)m^2_{3/2}.
\label{STrgrav}
\end{eqnarray}

{\bf Condensation de jauginos}

Dans une th\'eorie de jauge non ab\'elienne de type $SU(N)$,
la brisure spontan\'ee de la supersym\'etrie dans le secteur cach\'e 
peut aussi \^etre assur\'ee par une condensation de jauginos,
autrement dit un produit de deux champs fermioniques de jauginos 
d\'eveloppant une VEV, issue d'un m\'ecanisme non perturbatif.
En effet, la variation infinit\'esimale par supersym\'etrie du champ
fermionique
$\Psi_i$ (Eq.(\ref{trlambda})) contient un terme proportionnel \`a un
produit de jauginos:
\begin{eqnarray}
\delta \lambda_{a L}&=& \sigma^{\mu \nu} (F_a)_{\mu \nu} {\cal E}_L
+{i \over 2} g Re (f_{ab}^{-1}) G^i (T_b)_{ij} \phi_j {\cal E}_L
-{1 \over 8} f_{abj}  (G^{-1})_i^j \lambda_a  \lambda_b \cr
\cr&& + \ other \ \ Termes(2 \ fermions),
\label{trlambdap}
\end{eqnarray}
avec,
\begin{eqnarray}
f_{abj} = { \partial f_{ab} \over \partial \phi^{j\star} },
\label{fabj}
\end{eqnarray}
$f_{ab}$ \'etant d\'efini dans Eq.(\ref{SUSYGLOB}).
Une VEV pour $\lambda_a  \lambda_b$ peut donc briser la supersym\'etrie
en rendant la valeur dans le vide de $\Psi_i$ non invariante sous les
transformations supersym\'etriques. Pour qu'un tel sc\'enario de brisure
soit r\'ealisable,
il est n\'ecessaire que certaines composantes de $f_{abj}$ ne soient pas
nulle,
c'est \`a dire que les termes cin\'etiques pour les champs de jauge ne
soient
pas minimaux. Par ailleurs, il est possible que la condensation de jauginos
ait lieu
dans un secteur cach\'e de la th\'eorie si le groupe de jauge est un produit
direct.
Par exemple, dans les th\'eories de cordes h\'et\'erotiques, le groupe
de jauge peut \^etre $E_8 \times E_8$, le premier groupe exceptionnel
contenant
le groupe de jauge du Mod\`ele Standard et le second contenant le groupe de
jauge
du secteur cach\'e.

Comme dans le cas de la brisure par une VEV de champ scalaire,
le boson de Goldstone, qui est ici $\eta=f^i_{ab}<\lambda_a \lambda_b>
\Psi_i$,
se m\'elange au gravitino. Ce m\'elange est explicite dans le terme de
couplage \`a 4 fermions
du lagrangien \ref{LrFtil}:
\begin{eqnarray}
\vert det \ e \vert^{-1} {\cal L}_{MIX}= {1 \over 2} f^i_{ab} \bar \Psi_{iL}
\sigma^{\mu \nu}
\lambda_{a L} \bar \Psi_{\nu L} \gamma_{\mu} \lambda_{b R} + h.c.
\label{LMIX}
\end{eqnarray}

La masse du gravitino est aussi donn\'ee dans le cas de la condensation de
jauginos
par le premier terme du lagrangien ${\cal L}_F$ et vaut donc toujours
$m_{3/2}
\approx e^{G_0/2} M_P$
(voir Eq.(\ref{gravmass})). Cette masse peut aussi s'exprimer comme
pr\'ec\'edemment en fonction
de l'\'echelle de brisure de \susi, $M_s$, \`a savoir, $m_{3/2} = M_s^2/M_P$
(voir Eq.(\ref{gravmass3})). En revanche, l'\'echelle de brisure de \susi
est diff\'erente.
Dans un sc\'enario de brisure dynamique, le condens\^at de jauginos 
d\'eveloppe la VEV,
\begin{eqnarray}
<\lambda^a \lambda^b> = c \mu^3,
\label{VEVjc}
\end{eqnarray}
o\`u $c$ est de l'ordre de $1$ et o\`u $\mu$ est l'\'echelle dynamique
(analogue \`a $\L_{QCD}$), c'est \`a dire l'\'echelle d'\'energie
\`a laquelle la constante de couplage du groupe 
de jauge associ\'e devient forte. 
Or dans la limite de basse \'energie $M_P \to \infty$, nous
savons que la
supersym\'etrie n'est pas bris\'ee et que l'\'echelle de brisure doit donc
tendre vers z\'ero.
Nous pouvons donc estimer l'\'echelle de brisure de \susi $M_s$ comme
\'etant au maximum,
\begin{eqnarray}
M_s^2 \approx { \mu^3 \over M_P },
\label{VEVjcp}
\end{eqnarray}
avec \'eventuellement une puissance plus grande de $1/M_P$. Notons
finalement que pour
une telle \'echelle de brisure, la masse du gravitino est d'apr\`es
Eq.(\ref{gravmass3}),
\begin{eqnarray}
m_{3/2} \approx {\mu^3 \over M_P^2}.
\label{gravmassJC}
\end{eqnarray}
M\^eme pour une grande valeur de l'\'echelle d'\'energie $\mu$, une petite
masse peut
\^etre obtenue pour le gravitino dans la mesure o\`u $\mu$ est petit devant
l'\'echelle
de Planck.

\section{Effets de la brisure de supersym\'etrie locale dans le
secteur ``observable"}

Maintenant que nous avons discut\'e un m\'ecanisme de brisure
de la supersym\'etrie locale dans le secteur cach\'e, nous devons
\'etudier la mani\`ere dont cette brisure est transmise au
secteur observable.

Nous d\'esignerons les superchamps (champs scalaires) du secteur
cach\'e par, $Z_i$ ($z_i$), et les superchamps (champs scalaires)
du secteur observable par, $Y_r$ ($y_r$).
Le superpotentiel de la th\'eorie contient les superchamps du
secteur cach\'e et du secteur observable de mani\`ere additive
afin qu'il n'y ait pas d'interactions entre eux:
\begin{eqnarray}
W(Z_i, Y_r)=\bar W (Z_i) + \tilde W (Y_r).
\label{Wc+o}
\end{eqnarray}
Supposons une fois de plus que les termes cin\'etiques sont minimaux:
\begin{eqnarray}
G= M_P^{-2} (z^{i\star} z_i+y^{r\star} y_r)
+ ln ( {\vert W \vert ^2 \over M_P^6} ),
\label{Gfoncf}
\end{eqnarray}
o\`u les dimensions sont maintenant explicites. D'apr\`es
Eq.(\ref{V3}), le potentiel sans les termes D s'\'ecrit ici,
\begin{eqnarray}
V=exp({z^{i\star} z_i+y^{r\star} y_r \over M_P^2})
\bigg (
 \bigg \vert {\partial \bar W \over \partial z_i}
+ {z^{i\star} \over M_P^2} ( \bar W + \tilde W) \bigg \vert ^2
 \cr +\bigg \vert {\partial \tilde W \over \partial y_r}
+ {y^{r\star} \over M_P^2} ( \bar W + \tilde W) \bigg \vert ^2
-3 \vert W \vert ^2
\bigg ). \cr
\label{Vf}
\end{eqnarray}
Nous v\'erifions bien d'apr\`es ce potentiel
que si l'on prend la limite, $M_P \to \infty$,
c'est \`a dire si l'on n\'eglige les interactions
gravitationnelles, il n'y a plus d'interactions entre les particules
des secteurs cach\'es et observables. En effet, Dans cette limite,
le potentiel de Eq.(\ref{Vf}) s'\'ecrit,
$V=\vert {\partial \bar W \over \partial z_i} \vert ^2 +
\vert {\partial \tilde W \over \partial y_r} \vert ^2 $.
Autrement dit les seules interactions entre les champs des deux secteurs
sont proportionnelles \`a une puissance n\'egative de la masse de Planck,
c'est \`a dire qu'il s'agit des interactions gravitationnelles.
Afin d'obtenir une forme effective du potentiel (de
Eq.(\ref{Vf})) appropri\'ee
aux basses \'energies, nous allons remplacer le champ scalaire
$z_i$ par sa VEV et prendre la limite de basse \'energie:
$M_P \to \infty$.
En g\'en\'eral, les VEV correspondant \`a une brisure de la
supersym\'etrie dans le secteur cach\'e sont du type,
\begin{eqnarray}
<z_i>=a_i M_P,
\label{zVEV}
\end{eqnarray}
\begin{eqnarray}
<\bar W>=\mu M_P^2,
\label{WVEV}
\end{eqnarray}
\begin{eqnarray}
<{\partial \bar W \over \partial z_i}>=c_i \mu M_P,
\label{dWVEV}
\end{eqnarray}
o\`u $a_i$ et $c_i$ sont des quantit\'es sans dimensions et
$\mu$ est une \'echelle caract\'eristique du superpotentiel
du secteur cach\'e (comme par exemple $m$ dans le superpotentiel
de Polonyi). D'apr\`es Eq.(\ref{gravmass}), la masse du gravitino
est donn\'ee par,
\begin{eqnarray}
m_{3/2} \approx e^{\vert a_i \vert^2 /2} \mu.
\label{gravmassfinale}
\end{eqnarray}
La limite \`a basse \'energie revient \`a faire tendre $M_P$ vers
l'infini tout en gardant la masse du gravitino $m_{3/2}$ fixe.
D'apr\`es Eq.(\ref{gravmassfinale}), cela est \'equivalent
\`a travailler au premier ordre en $\mu / M_P$.
Notons que les \'eventuelles VEV des champs scalaires du secteur observable
sont de l'ordre de l'\'echelle \'electrofaible et sont donc bien
inf\'erieures \`a l'\'echelle de Planck: $<y_r><<M_P$. Le potentiel
de Eq.(\ref{Vf}) s'\'ecrit donc \`a basse \'energie,
\begin{eqnarray}
V=e^{\vert a_i \vert ^2}
\bigg [
\bigg \vert {\partial \tilde W \over \partial y_r} \bigg \vert ^2
+ \mu^2 \vert y_r \vert ^2
+ \mu \bigg ( y_r {\partial \tilde W \over \partial y_r}
+ (A-3) \tilde W + C.C. \bigg )
\bigg ],
\label{VfBE}
\end{eqnarray}
o\`u $A=(c_i^\star+a_i)a_i^\star$. En utilisant Eq.(\ref{gravmassfinale})
et en red\'efinissant le superpotentiel par, $e^{\vert a_i \vert^2/2}
\tilde W \to W'$, le potentiel de Eq.(\ref{VfBE}) peut encore
s'\'ecrire,
\begin{eqnarray}
V=
\bigg \vert {\partial W' \over \partial y_r} \bigg \vert ^2
+ m_{3/2}^2 \vert y_r \vert ^2
+ m_{3/2} \bigg ( y_r {\partial W' \over \partial y_r}
+ (A-3) W' + C.C. \bigg ).
\label{VfBEp}
\end{eqnarray}
Le premier terme est le potentiel d'une th\'eorie
de supersym\'etrie globale avec un superpotentiel $W'$.
Les termes suivants sont des termes
de brisure explicite de la supersym\'etrie globale.
En particulier, le terme de masse pour le champ scalaire $y_r$
vient du terme,
\begin{eqnarray}
e^{({<z^{i\star}><z_i> \over M_P^2})}
\vert { < \bar W > \over M_P^2}  y^{r\star} \vert ^2,
\label{terme}
\end{eqnarray}
du potentiel de Eq.(\ref{Vf}). Ce terme de masse
pour le champ scalaire du secteur observable vient
donc bien d'une interaction gravitationnelle: Nous voyons
clairement ici que la brisure de la supersym\'etrie
est transmise par la gravitation.
\\ Si nous reprenons l'exemple du mod\`ele de Polonyi,
les VEV du champ scalaire appartenant au secteur cach\'e et du
superpotentiel \'etaient, $<z_i>=(\sqrt 3-1) M_P$, et,
$<\bar W>=m^2 M_P$, respectivement. D'apr\`es Eq.(\ref{terme}),
la masse du champ scalaire $y_r$ est par cons\'equent,
$exp((\sqrt 3-1) ^2) { m^4 \over M_P^2}=m_{3/2}^2$
(voir Eq.(\ref{gravmass2})). Nous retrouvons donc bien le fait
que la masse des scalaires $y_r$ est \'egale \`a celle du gravitino
$m_{3/2}$. Par ailleurs, il est facile de montrer que
dans le mod\`ele de Polonyi, la constante $A$ d\'efinie
ult\'erieurement vaut, $A=3-\sqrt 3$.
\\ Il est int\'eressant de noter qu'\`a l'\'echelle de Planck,
il y a universalit\'e des masses des champs scalaires.
C'est \`a dire que les champs scalaires
acqui\`erent tous la m\^eme masse. Effectivement, le potentiel
de Eq.(\ref{VfBEp}) montre que la masse du champ scalaire
est \'egale \`a la masse du gravitino, pour chaque multiplet chiral.
Il est possible d'obtenir des potentiels effectifs dans lesquels
l'universalit\'e est absente, en ne choisissant pas les termes
cin\'etiques minimaux pour les superchamps chiraux.
Cependant l'universalit\'e est int\'eressante puisque c'est une
solution naturelle au probl\`eme ph\'enom\'enologique des courants
neutres changeant la saveur. Notons que l'universalit\'e est
ici pr\'esente \`a l'\'echelle de Planck, ce qui n'implique
pas n\'ecessairement
l'universalit\'e \`a des \'echelles d'\'energie inf\'erieures.
\\ Le cas o\`u le superpotentiel du secteur observable $W'$ est
trilin\'eaire dans les superchamps chiraux est particuli\`erement
int\'eressant car cela \'evite les petits ($\approx 100GeV$)
ajustements de param\`etres de masse. Dans ce cas, le potentiel se
simplifie en,
\begin{eqnarray}
V=
\bigg \vert {\partial W' \over \partial y_r} \bigg \vert ^2
+ m_{3/2}^2 \vert y_r \vert ^2
+ A m_{3/2}  ( W' + W'^\star ).
\label{Vtri}
\end{eqnarray}
Une autre source de brisure de la supersym\'etrie globale dans le secteur
observable appara\^{\i}t dans le premier terme du lagrangien $\tilde {\cal
L}_F$.
Il s'agit d'un terme de masse pour les jauginos:
\begin{eqnarray}
{1 \over 4} e^{G/2} (G^{-1})^j_k G^k
{\partial f^\star_{ab} \over \partial \phi^{j\star}} \lambda_a \lambda_b.
\label{jaugmass}
\end{eqnarray}
Afin que ce terme ne soit pas nul, la fonction $f_{ab}(\phi_j)$ ne doit pas
\^etre une fonction triviale. C'est \`a dire que les termes cin\'etiques
de jauge ne doivent pas \^etre minimaux. La seconde condition est
que l'expression, $e^{G/2} (G^{-1})^j_k G^k$,
ait une VEV non nulle. Ceci est assur\'e par la condition de brisure
de la supersym\'etrie locale par l'\'equivalent d'un terme F
(voir Eq.(\ref{VEVP1})) dans le secteur cach\'e. Notons en effet
qu'ici les indices j et k d\'esignent des champs du secteur cach\'e.
La VEV des champs scalaires du secteur cach\'e
\'etant proportionnelle \`a la masse de Planck,
$<\phi^{j\star}> \approx M_P$, le terme de Eq.(\ref{jaugmass})
est un terme proportionnel \`a une puissance n\'egative de la masse
de Planck. C'est donc un terme d'interaction gravitationnelle.
La transmission de la brisure de SUSY est donc effectu\'ee une fois
encore par les interactions gravitationnelles.
\'Etant dans le cas d'une brisure par l'\'equivalent d'un
terme F, l'\'echelle de brisure de SUSY est donn\'ee par
Eq.(\ref{SUSYscale}).
Par cons\'equent, d'apr\`es Eq.(\ref{jaugmass}) et Eq.(\ref{gravmass3}),
la masse des jauginos vaut,
\begin{eqnarray}
m_{jaug} \approx {M_S^2 \over M_P} \approx m_{3/2}.
\label{jaugmassp}
\end{eqnarray}
La masse des jauginos est donc, comme la masse des champs scalaires, de
l'ordre de la masse du gravitino.
\\ Enfin, de mani\`ere g\'en\'erale, le potentiel effectif de basse
\'energie peut aussi contenir des termes D. D'apr\`es Eq.(\ref{Gfoncf}),
le lagrangien $\tilde {\cal L}_B$ donne un potentiel du type,
\begin{eqnarray}
V&=&
\bigg \vert {\partial W' \over \partial y_r} \bigg \vert ^2
+ m_{3/2}^2 \vert y_r \vert ^2
+ m_{3/2} \bigg ( y_r {\partial W' \over \partial y_r}
+ (A-3) W' + C.C. \bigg )
\cr && + {g^2 \over 2} Re(f^{-1}_{ab}) y^{r\star}(T_a)_{rs}y_s
y^{k\star}(T_b)_{kl}y_l.
\label{V+D}
\end{eqnarray}
Les termes D du potentiel de Eq.(\ref{V+D}) sont des termes D du secteur
observable. Leur
pr\'esence impliquerait
une brisure de la supersym\'etrie locale dans
le secteur observable si $f_{ab}(\phi)=\delta_{ab}$
(voir Eq.(\ref{critere3})).
Notons que nous n'avons pas non plus trait\'e les
termes D du secteur cach\'e puisque nous avons \'etudi\'e
une brisure de la supersym\'etrie locale dans le secteur cach\'e par (une
g\'en\'eralisation)
des
termes F. Et la transmission de cette brisure de SUSY au secteur observable
s'effectue aussi via les termes F du potentiel (voir Eq.(\ref{Vf})).
Rappelons cependant qu'une brisure de SUSY dans le secteur cach\'e
par des termes D est possible ($\neq$ transmission au secteur observable).
La masse du gravitino est aussi
donn\'ee dans ce cas par l'\'equation Eq.(\ref{gravmass}). En revanche,
l'\'echelle de brisure
est alors donn\'ee par Eq.(\ref{VEVl1}).

En conclusion, dans les th\'eories de supergravit\'e la brisure
spontan\'ee de la supersym\'etrie locale dans un secteur cach\'e
permet d'obtenir, \`a l'\'echelle de Planck $M_P$, un secteur observable
r\'egit par un lagrangien globalement supersym\'etrique accompagn\'e
de termes de brisure douce (voir Section \ref{BRIsec}).
Ces termes de brisure peuvent donner une masse universelle $m_0$ aux champs
scalaires et une masse universelle $m_{1/2}$ aux jauginos
du secteur observable \`a l'\'echelle de Planck.
Les masses des champs scalaires $m_0$ et des jauginos $m_{1/2}$
sont de l'ordre de la masse du gravitino $m_{3/2}$ \`a l'\'echelle de
Planck.
Typiquement, la formule de Eq.(\ref{gravmass3}) nous montre que pour une
\'echelle de brisure de la supersym\'etrie de l'ordre de,
$M_S \approx 10^{10}GeV$, la masse du gravitino est de l'ordre de,
$m_{3/2} \approx 100GeV$ \`a l'\'echelle de Planck.
Par cons\'equent, \`a l'\'echelle \'electrofaible,
les masses des particules supersym\'etriques peuvent \^etre inf\'erieures au
TeV
et le probl\`eme de hi\'erarchie (voir Section \ref{hier})
peut ainsi \^etre r\'esolu. De plus, \`a l'\'echelle \'electrofaible,
les masses des particules supersym\'etriques peuvent \^etre
sup\'erieures \`a celles de leur partenaire \susiq du Mod\`ele Standard.

\clearpage

\vspace{10 mm}
\begin{center}
{  }
\end{center}
\vspace{10 mm}

\clearpage

\chapter{Origines et motivations th\'eoriques de la sym\'etrie de 
R-parit\'e}
\label{chaRPV}

\section{G\'en\'eralit\'e du superpotentiel du MSSM}
\label{antisymetrie}

Le superpotentiel de Eq.(\ref{wMSSM}) contient tous les termes n\'ecessaires
\`a
l'extension \susiq minimale du Mod\`ele Standard: le MSSM. Cependant,
ce superpotentiel n'est pas le plus g\'en\'eral dans le sens o\`u il
existe
d'autres couplages invariants de jauge qui n'ont pas \'et\'e pris
en compte. Pour \^etre totalement g\'en\'eral le superpotentiel de
Eq.(\ref{wMSSM})
doit contenir en plus les termes suivants:
\begin{eqnarray}
W_{\rpv}=\sum_{i,j,k} \bigg \{{1 \over 2} \l_{ijk} L_iL_j E^c_k+
\l'_{ijk} L_i Q_j D^c_k+
{1 \over 2} \l''_{ijk} U_i^cD_j^cD_k^c  + \mu_i H_2 L_i \bigg \},
\label{wrpv}
\end{eqnarray}
o\`u les indices $i,j,k$ d\'esignent les 3 g\'en\'erations.
Le symbole $\rpv$ ainsi que les facteurs ${1 \over 2}$ seront
explicit\'es par la suite. Quant aux
$\l_{ijk}$, $\l'_{ijk}$ et $\l''_{ijk}$, ce sont des nouvelles
constantes de couplage s'ajoutant aux param\`etres du MSSM.
Remarquons que les deux premiers termes trilin\'eaires ainsi
que le terme bilin\'eaire de Eq.(\ref{wrpv}) peuvent \^etre obtenus
en rempla\c{c}ant le superchamp de Higgs $H_1$ par le superchamp
leptonique $L_i$ dans le superpotentiel du MSSM (voir Eq.(\ref{wMSSM}))
qui contient notamment les couplages de Yukawa.
En effet, les superchamps $H_1$ et $L_i$ ont les m\^emes nombres
quantiques.

Combien de param\`etres suppl\'ementaires
sont effectivement introduits par le superpotentiel de Eq.(\ref{wrpv}) ?
Le ``tenseur''  $\l_{ijk}$ des constantes de couplage
est antisym\'etrique
par rapport aux indices de famille $i$ et $j$ \`a cause de
du tenseur antisym\'etrique $\ep_{ab}$ de $SU(2)$:
\begin{eqnarray}
\l_{ijk} L_iL_j E^c_k  = &  \l_{ijk} L_i^a L_j^b  E^c_k \ep_{ab}
= &  \l_{jik} L_j^a L_i^b  E^c_k \ep_{ab} = - \l_{jik} L_j^a L_i^b  E^c_k
\ep_{ba} \cr
 = & - \l_{jik} L_i^b L_j^a  E^c_k \ep_{ba} =  & - \l_{jik} L_i L_j  E^c_k. 
\label{wrpvAS}
\end{eqnarray}
De m\^eme, $\l''_{ijk}$ est antisym\'etrique
par rapport aux indices $j$ et $k$ \`a cause de
l'antisym\'etrie du tenseur $\ep_{mnp}$ de $SU(3)$. En effet,
le terme trilin\'eaire en $\l''_{ijk}$ est un couplage
invariant de $SU(3)$ du type ${\bf \bar 3 \bar 3 \bar 3}$
contract\'e par le tenseur antisym\'etrique $\ep_{mnp}$ de $SU(3)$:
$\l''_{ijk} U_i^cD_j^cD_k^c =
\l''_{ijk} U_i^{c \ m} D_j^{c \ n} D_k^{c \ p} \ep_{mnp}$.\\
Par cons\'equent, il n'existe que 9 constantes $\l_{ijk}$
et 9 constantes $\l''_{ijk}$
alors qu'il y a 27 constantes du type $\l'_{ijk}$.
Le superpotentiel de Eq.(\ref{wrpv}) introduit donc
dans le MSSM 45 param\`etres r\'eels suppl\'ementaires.
En effet, le terme bilin\'eaire de Eq.(\ref{wrpv}) peut \^etre
\'elimin\'e par une red\'efinition des superchamps:
Si l'on d\'efini $L_\alpha = (H_1,L_1,L_2,L_3)$ et
$\mu_\alpha = (\mu,\mu_1,\mu_2,\mu_3)$, les termes bilin\'eaires
de Eq.(\ref{wMSSM}) et Eq.(\ref{wrpv})
s'\'ecrivent $\mu_\alpha H_2 L_\alpha$. Or, en effectuant
la transformation $L_\alpha \to U_{\alpha \beta} L^\beta$
o\`u $U_{\alpha \beta}$ est donn\'e en premi\`ere approximation par,
\begin{equation}
U_{\alpha \beta} \approx
\left (
\begin{array}{cc}
1 & -{\mu_m \over \mu} \\ {\mu_m \over \mu} & I_3
\end{array}
\right ) ,
\label{CASava}
\end{equation}
$I_3$ \'etant la matrice identit\'e $3 \times 3$
et $m$ valant $m=1,2,3$, le terme
$\mu_\alpha H_2 L_\alpha$ se r\'eduit au terme bilin\'eaire de
Eq.(\ref{wMSSM}): $\mu H_1 H_2$. Cependant, notons que la red\'efinition 
des champs de Eq.(\ref{CASava}) n'\'elimine pas les termes bilin\'eaires
pr\'esents parmi les termes de brisure douce de SUSY, si ces termes bilin\'eaires
ont une structure g\'en\'erique.

Quelle est la dimension des constantes $\l$, $\l'$ et $\l''$ ?
La dimension des variables de Grassmann est $[\theta^\alpha]=-1/2$
comme nous l'avons mentionn\'e dans la Section \ref{gediSUSY}.
La dimension de la d\'eriv\'ee par rapport \`a une variable de Grassmann
vaut donc $[\partial  / \partial  \theta^{\alpha}]=1/2$.
D'apr\`es Eq.(\ref{LAG7}), le superpotentiel doit avoir une dimension
$[W]=3$
puisque l'int\'egration est \'equivalente \`a la d\'erivation
pour une variable de Grassmann et que la dimension du lagrangien est $[{\cal
L}]=4$
(voir Appendice \ref{dims}).
La dimension d'un superchamp $S$ \'etant \'egale \`a $[S]=1$ (voir Section
\ref{spchp}),
l'expression du superpotentiel de Eq.(\ref{wrpv}) indique
que les constantes $\l_{ijk}$, $\l'_{ijk}$ et $\l''_{ijk}$ sont sans
dimension:
$[\l_{ijk}]=[\l'_{ijk}]=[\l''_{ijk}]=0$.

Nous allons maintenant d\'eterminer le lagrangien associ\'e
aux termes trilin\'eaires du superpotentiel de
Eq.(\ref{wrpv}). La contribution de ces termes au potentiel scalaire,
qui se d\'eduit de Eq.(\ref{LAG71}), donne des op\'erateurs
quadruples dans les champs de sleptons et de squarks. Ces
interactions entre particules \susiqs de spin 0
n'ont pas de cons\'equences ph\'enom\'enologiques
\`a basse \'energie pour des particules \susiqs lourdes.
L'autre contribution des termes trilin\'eaires du superpotentiel de
Eq.(\ref{wrpv}) au lagrangien est donn\'ee par Eq.(\ref{LAG72}).
Nous effectuerons explicitement le calcul de cette contribution uniquement
pour
le terme en $\l_{ijk}$ de Eq.(\ref{wrpv}),
la m\'ethode \'etant tout \`a fait similaire pour
les termes en $\l'_{ijk}$ et $\l''_{ijk}$.
Explicitons tout d'abord le produit sous $SU(2)$
de ces termes en $\l_{ijk}$:
\begin{eqnarray}
{1 \over 2} \l_{ijk} L_i L_j E^c_k
=  {1 \over 2} \l_{ijk} L_i^a L_j^b  E^c_k \ep_{ab}
= {1 \over 2} \l_{ijk} \bigg ( N_i E_j-E_i N_j \bigg ) E^c_k,
\label{wrpvl}
\end{eqnarray}
o\`u $N_i$ et $E_i$ sont les superchamps de chiralit\'e
gauche associ\'es au neutrino
et au lepton charg\'e respectivement.
Dans Eq.(\ref{wrpvl}), la somme sur les indices $i,j,k$ est implicite.
De m\^eme, dans tout ce calcul
nous n'\'ecrirons plus le symbole $\sum_{i,j,k}$ afin
d'all\'eger les \'equations.
Le lagrangien associ\'e aux termes de Eq.(\ref{wrpvl}) et donn\'e
par Eq.(\ref{LAG72}) s'\'ecrit,
\begin{eqnarray}
{\cal L}_{\l}=
& - & {1 \over 2} \sum_{a,b}  { \partial ^2
\bigg [
{1 \over 2} \l_{ijk}
\bigg (\tilde \nu_{iL} \tilde e_{jL}
-\tilde e_{iL} \tilde \nu_{jL} \bigg )
\tilde e^c_{kR}
\bigg ]
\over \partial z_a \ \partial z_b } \ \psi_a \psi_b \cr
& - & {1 \over 2} \sum_{a,b}  { \partial ^2
\bigg [
{1 \over 2} \l^\star_{ijk}
\bigg (\tilde \nu^\star_{iL} \tilde e^\star_{jL}
-\tilde e^\star_{iL} \tilde \nu^\star_{jL} \bigg )
\tilde e^{c\star}_{kR}
\bigg ]
\over \partial z^\star_a \ \partial z^\star_b } \ \bar \psi_a \bar \psi_b,
\label{lagrangl1}
\end{eqnarray}
o\`u l'exposant $\star$ signifie complexe conjugu\'e.
Remarquons que nous avons \'ecrit
le partenaire scalaire du spineur de chiralit\'e
gauche \`a quatre composantes
du positron $e^c_{kL}$:
$scal[e^c_{kL}]=scal[(e_{kR})^c]=\tilde e^c_{kR}=\tilde e^\star_{kR}$.
Le lagrangien de Eq.(\ref{lagrangl1}) s'\'ecrit,
\begin{eqnarray}
{\cal L}_{\l}=
& - &
{1 \over 2} \l_{ijk}
\bigg ( \chi_{\nu_i} \chi_{e_j}  \tilde e^c_{kR}
+ \chi_{\nu_i} \eta_{e_k}  \tilde e_{jL}
+ \chi_{e_j}   \eta_{e_k}  \tilde \nu_{iL}
- (i \leftrightarrow j) \bigg ) \cr
& - &
{1 \over 2} \l^\star_{ijk}
\bigg ( \bar \chi_{\nu_i} \bar \chi_{e_j}  \tilde e^{c\star}_{kR}
+ \bar \chi_{\nu_i} \bar \eta_{e_k}  \tilde e^\star_{jL}
+ \bar \chi_{e_j}   \bar \eta_{e_k}  \tilde \nu^\star_{iL}
- (i \leftrightarrow j) \bigg ),
\label{lagrangl2}
\end{eqnarray}
o\`u les spineurs \`a deux composantes de l'\'electron et du neutrino
sont d\'efinis \`a partir des spineurs \`a quatre composantes
correspondants, not\'es $e$, $\nu$ et $e^c$, $\nu^c$
(conjugu\'es de charge), par,
\begin{equation}
e=
\left (
\begin{array}{c}
\chi_e \\ \bar \eta_e
\end{array}
\right ) ,
\ \ \
e^c=
\left (
\begin{array}{c}
\eta_e \\ \bar \chi_e
\end{array}
\right ) ,
\ \ \
\nu=
\left (
\begin{array}{c}
\chi_{\nu} \\ \bar \eta_{\nu}
\end{array}
\right ) ,
\ \ \
\nu^c=
\left (
\begin{array}{c}
\eta_{\nu} \\ \bar \chi_{\nu}
\end{array}
\right ) .
\label{notspinenu}
\end{equation}
En passant de Eq.(\ref{lagrangl1}) \`a Eq.(\ref{lagrangl2}),
le facteur $1/2$ venant de Eq.(\ref{LAG72}) s'est \'elimin\'e.
L'explication est que la somme sur $a$ et $b$ effectu\'ee dans
Eq.(\ref{lagrangl1}) et provenant aussi de Eq.(\ref{LAG72})
fait appara\^{\i}tre deux fois le m\^eme couplage. En effet,
la d\'eriv\'ee seconde de Eq.(\ref{lagrangl1}) donne par exemple un
couplage identique pour $z_a=\tilde \nu_{iL},z_b= \tilde e_{jL}$
et $z_a=\tilde e_{jL},z_b= \tilde \nu_{iL}$. Notons que la
d\'eriv\'ee seconde $d^2 / dz_a \ dz_b$ prise dans
Eq.(\ref{lagrangl1}) est nulle pour $a=b$.
En revanche, ce facteur $1/2$ venant de Eq.(\ref{LAG72}) appara\^{\i}t
dans le terme de masse du lagrangien pour un spineur de Majorana
qui s'\'ecrit,
${\cal L}={1 \over 2} m \bar \psi \psi$, car la somme sur $a$ et $b$
de Eq.(\ref{LAG72}) n'est alors prise que sur un seul champ puisque
le superpotentiel g\'en\'erant cette masse est
$W=m \Phi^2$, $\Phi$ \'etant le superchamp associ\'e au spineur de
Majorana.\\
En utilisant les formules de Eq.(\ref{2to4L}) et Eq.(\ref{2to4R})
reliant les spineurs \`a deux composantes
$\chi$, $\eta$ aux spineurs \`a quatre composantes $\Psi$, $\Psi^c$,
on peut exprimer le lagrangien de Eq.(\ref{lagrangl2}) en fonction
des spineurs \`a quatre composantes:
\begin{eqnarray}
{\cal L}_{\l}=
& - &
{1 \over 2} \l_{ijk}
\bigg ( \bar \nu^c_i P_L e_j  \tilde e^\star_{kR}
+ \bar e_k     P_L \nu_i       \tilde e_{jL}
+ \bar e_k     P_L e_j         \tilde \nu_{iL}
- (i \leftrightarrow j) \bigg ) \cr
& - &
{1 \over 2} \l^\star_{ijk}
\bigg ( \bar e_j   P_R \nu^c_i   \tilde e_{kR}
+ \bar \nu_i P_R e_k       \tilde e^\star_{jL}
+ \bar e_j   P_R e_k       \tilde \nu^\star_{iL}
- (i \leftrightarrow j) \bigg ).
\label{lagrangl3}
\end{eqnarray}
Afin d'exprimer le premier terme du lagrangien de Eq.(\ref{lagrangl2})
en fonction de spineurs \`a quatre composantes, nous avons consid\'er\'e
$\chi_{\nu}$ comme \'etant
le spineur conjugu\'e de charge du neutrino appartenant
\`a la repr\'esentation $(0,1/2)$ du groupe de Lorentz
(voir Eq.(\ref{notspinenu})).
Les facteurs $1/2$ dans Eq.(\ref{lagrangl3}) s'\'eliminent lorsque
la somme sur $i,j,k$ est effectu\'ee, car des couples de couplages
identiques
appara\^{\i}ssent alors, \`a cause de l'antisym\'etrie de $\l_{ijk}$.\\
En r\'ep\'etant le m\^eme exercice pour les interactions en $\l'_{ijk}$
et $\l''_{ijk}$, nous trouvons l'ensemble
de la contribution au lagrangien d\'ecrite dans Eq.(\ref{LAG72})
des termes trilin\'eaires du superpotentiel de Eq.(\ref{wrpv}):
\begin{eqnarray}
{\cal L}_{\rpv}=\sum_{ijk} \bigg \{  & \l_{ijk} & \ud
\bigg ( \tilde  \nu_{iL}\bar e_{kR}e_{jL} +
\tilde e_{jL}\bar e_{kR}\nu_{iL} + \tilde e^\star _{kR}\bar \nu^c_{iR}
e_{jL}
-(i \leftrightarrow j) \bigg )  \cr
&+\l '_{ijk}& \bigg ( \tilde  \nu_{iL}\bar d_{kR}d_{jL} +
\tilde d_{jL}\bar d_{kR}\nu_{iL} + \tilde d^\star _{kR}\bar \nu^c_{iR}
d_{jL} \cr
&&-\tilde  e_{iL}\bar d_{kR}u_{jL} -
\tilde u_{jL}\bar d_{kR}e_{iL} - \tilde d^\star _{kR}\bar e^c_{iR} u_{jL}
\bigg )  \cr
 & +{\l ''}_{ijk} & \ud \bigg (\tilde  u^\star _{i R}\bar d_{j R}d^c_{k L} +
\tilde  d^\star _{j R}\bar u_{i R}d^c_{k L} \cr &&+
\tilde  d^\star _{k R}\bar u_{i R}d^c_{j L}
-(j \leftrightarrow k) \bigg ) \ \ \ \ \ \ \ \bigg \} + h.c.
\label{eqyuk}
\end{eqnarray}

\section{D\'esint\'egration du proton}
\label{proton}

Les termes du lagrangien \ref{eqyuk} sont des op\'erateurs dangereux
dans le sens o\`u ils peuvent donner lieu \`a une d\'esint\'egration
rapide du proton \cite{Wein,Sakai,Hin,Goity,Carlson}. En effet,
les interactions en $\l_{ijk}$, $\l'_{ijk}$ violent
le nombre leptonique et les interactions en $\l''_{ijk}$ violent
le nombre baryonique. Cette d\'esint\'egration du proton est en fait
li\'ee \`a la pr\'esence simultan\'ee des couplages $\l'_{ijk}$ et
$\l''_{ijk}$.
Les bornes exp\'erimentales sur le temps de vie du proton peuvent
donc \^etre utilis\'ees pour contraindre les produits de
constantes de couplage du type $\l' \l''$.

Par exemple, le proton peut se d\'esint\'egrer en un pion et un positron,
comme $P \to \pi^0 e^+$, via le processus $ud \to \tilde d^\star_k \to e^+
\bar u$
qui implique les couplages $\l''_{11k} \tilde  d^\star _{kR}\bar
u_{R}d^c_{L}$ et
$\l'_{11k} \tilde d^\star _{kR}\bar e^c_{R} u_{L}$.
La largeur de cette d\'esint\'egration peut \^etre estim\'ee  par,
\begin{eqnarray}
\Gamma(P \to \pi^0 e^+) \approx \alpha(\l'_{11k}) \alpha(\l''_{11k})
{M^5_{proton} \over m^4_{\tilde d_k}},
\label{laprot}
\end{eqnarray}
o\`u $\alpha(\l)=\l^2/4\pi$.
L'\'equation Eq.(\ref{laprot}) et la contrainte exp\'erimentale
sur le temps de vie du proton relative \`a la d\'esint\'egration
du proton en pion,
$\tau(P \to \pi^0 e^+)>10^{32} ans$ \cite{PDG},
permettent de borner le produit $\l'_{11k} \l''_{11k}$ par,
\begin{eqnarray}
\l'_{11k} \l''_{11k}
\stackrel{<}{\sim}
2.10^{-27} ({m_{\tilde d_k} \over 100GeV})^2.
\label{boprot}
\end{eqnarray}
Il a m\^eme \'et\'e montr\'e dans \cite{Viss} que l'\'etude de la
d\'esint\'egration du proton pouss\'ee \`a l'ordre d'une boucle
pouvait contraindre chaque
produit $\l' \l''$ de fa\c{c}on conservative par,
\begin{eqnarray}
\l' \l''<10^{-9}.
\label{boloop}
\end{eqnarray}

\subsection{Op\'erateurs de dimension 5}
\label{prot5dim}

Les nombres leptonique L et baryonique B peuvent aussi \^etre
viol\'es par des op\'erateurs non renormalisables. Les
op\'erateurs de dimension 5 violant les
nombres leptonique et baryonique sont en $1/M$ o\`u $M$ est
l'\'echelle de la physique au-del\`a du Mod\`ele Standard
(\susiq minimal) g\'en\'erant une violation de L et B.
L'\'echelle $M$ typique est l'\'echelle
des th\'eories de grande unification
$M_{GUT}$, et l'\'echelle $M$ la plus grande est bien s\^ur celle
de Planck $M_P$. Les
op\'erateurs de dimension 5 du superpotentiel (termes F)
invariants sous le groupe de jauge du MSSM et violant
les nombres leptonique et baryonique sont,
en termes des superchamps,
\begin{equation}
\begin{array}{rlrl}
{\cal O}_1  = &  QQQL & {\cal O}_4   =  &  QU^cE^cH_1 \\
{\cal O}_2  = & U^c U^c D^c E^c & {\cal O}_5  = & LLH_2H_2 \\
{\cal O}_3  = & QQQH_1 & {\cal O}_6 = & LH_1H_2H_2
\label{dim5F}
\end{array}
\end{equation}
De tels op\'erateurs existent aussi parmi les termes en
$\int d^2 \theta d^2 \bar \theta$ (termes D) et s'\'ecrivent
en termes des superchamps,
\begin{equation}
\begin{array}{rlrl}
{\cal O}_7= & H_2H_2E^{c\dagger } & {\cal O}_9= & QU^cL^{\dagger } \cr
{\cal O}_8= & H_2^{c \dagger }H_1E^c & {\cal O}_{10}= & U^cD^{c\dagger }E^c
\label{dim5D}
\end{array}
\end{equation}
D'apr\`es Eq.(\ref{LAG3}), les op\'erateurs de
Eq.(\ref{dim5F}) et Eq.(\ref{dim5D}) sont bien de
dimension 5, puisque $[\partial  / \partial  \theta^{\alpha}]=1/2$
et que la dimension d'un superchamp est \'egale \`a 1.

Les bornes exp\'erimentales sur les temps de vie des nucl\'eons
contraignent la plupart de ces op\'erateurs de dimension 5
de mani\`ere significative,
m\^eme si la violation de L et B appara\^{\i}t \`a une \'echelle
$M=M_P$. Les processus impliquant deux fois un op\'erateur ${\cal O}_i$
sont supprim\'es par un facteur $(m/M_P)^2$,
o\`u $m=m_{SUSY},m_{weak}$, et donnent donc une limite
peu significative sur les coefficients  $\eta_i$ de ces op\'erateurs.
L'op\'erateur ${\cal O}_1$ couple des particules \susiqs et peut donc
contribuer
\`a la d\'esint\'egration du nucl\'eon en un m\'eson et un lepton
par un graphe \`a une boucle impliquant des winos.
L'amplitude du processus correspondant est r\'eduit d'un facteur de boucle
et d'un facteur $m_{SUSY}/M_P$. Les donn\'ees exp\'erimentales
sur les d\'esint\'egrations nucl\'eoniques
impliquent la borne $\eta_1<10^{-7}$ \cite{O1} sur le coefficient de ${\cal
O}_1$
pour $m_{\tilde W} \approx m_W$.
L'op\'erateur ${\cal O}_2$, couplant des particules \susiqs n'interagissant
pas
avec les winos, ne peut contribuer \`a la d\'esint\'egration du nucl\'eon
par un graphe \`a une boucle impliquant des winos.
Le coefficient $\eta_2$ n'est donc pas contraint de fa\c{c}on significative.
L'op\'erateur ${\cal O}_3$ ne peut pas non plus contribuer
\`a la d\'esint\'egration du
nucl\'eon par un graphe \`a une boucle impliquant des winos.
En revanche, les op\'erateurs ${\cal O}_3$, consid\'er\'e
avec la vev $v$ associ\'ee au superchamp $H_1$,
et $LQD^c$ permettent la d\'esint\'egration du
nucl\'eon via un processus du type $q q \to \tilde q \to q l$ qui est
supprim\'e
par un facteur $v/M_P$. La limite qui en d\'ecoule est $\l' \eta_3<10^{-10}$
pour $v \approx 100GeV$.
Les combinaisons entre op\'erateurs ${\cal O}_i$ et op\'erateurs
en $\l$,$\l'$,$\l''$ donnent aussi des limites significatives sur les
coefficients $\eta_i$ des ${\cal O}_i$ ($i=4,...,10$).
Les contraintes relatives aux op\'erateurs ${\cal O}_i$ mentionn\'ees
ci-dessus
sont les seules qui soient significatives.

\section{Sym\'etries du superpotentiel}
\label{sympot}

Les fortes contraintes sur les produits de constantes de couplage
$\l' \l''$ d\'eduites des bornes
exp\'erimentales sur le temps de vie du proton (voir Section \ref{proton})
ainsi que les violations des nombres leptonique et
baryonique engendr\'ees par les interactions en $\l, \l'$
et $\l''$, respectivement,
sugg\`erent que les couplages du lagrangien \ref{eqyuk}
soient interdits par une certaine sym\'etrie.
Cependant, le probl\`eme de la d\'esint\'egration rapide du proton
peut aussi \^etre r\'esolu par l'imposition d'une sym\'etrie
permettant uniquement les couplages en $\l'$ ou en $\l''$
du lagrangien \ref{eqyuk}. Les sym\'etries discr\`etes interdisant les couplages
en $\l$,$\l'$,$\l''$ sont appel\'ees parit\'es de mati\`ere,
en $\l$,$\l'$ (couplages violant le nombre leptonique) parit\'es
leptoniques et en $\l''$ (couplages violant le nombre baryonique)
parit\'es baryoniques. Dans ce chapitre, nous discutons les diverses
sym\'etries du superpotentiel, pouvant prot\'eger le proton de sa
d\'esint\'egration, et leurs caract\'eristiques.

La suppression par des sym\'etries des op\'erateurs dangereux de dimension
5, vus dans la Section \ref{prot5dim}, est moins motiv\'ee.
En effet, la violation de L et B dans une th\'eorie au-del\`a du MSSM
peut \'eventuellement \^etre supprim\'ee par un autre type de m\'ecanisme
propre \`a cette nouvelle th\'eorie.

\subsection{Les R-sym\'etries}

Les R-sym\'etries furent propos\'ees
par A. Salam et J. Strathdee dans \cite{Salam} et par P. Fayet dans
\cite{Fayet1} pour interdire la violation des nombres
leptonique ou baryonique. Les R-sym\'etries apparaissent
naturellement dans les th\'eories les plus simples.

La R-sym\'etrie est une sym\'etrie
$U(1)$ continue agissant sur les superchamps vectoriels
$V$ et chiraux $\Phi$ et $\bar \Phi$ par les transformations,
\begin{eqnarray}
V_k(x,\theta,\bar \theta) & \to &
V_k(x,\theta e^{-i \alpha},\bar \theta e^{i \alpha}),
\label{Rsym1a}
\end{eqnarray}
\begin{eqnarray}
\Phi_l(x,\theta) & \to &
e^{i R_l \alpha} \ \Phi_l(x,\theta e^{-i \alpha}),
\label{Rsym1b}
\end{eqnarray}
\begin{eqnarray}
\bar \Phi_l(x,\bar \theta) & \to &
e^{-i R_l \alpha} \
\bar \Phi_l(x,\bar \theta e^{i \alpha}),
\label{Rsym1c}
\end{eqnarray}
o\`u les charges $R_l$ sont des nouveaux nombres quantiques additifs.
Le choix de ces charges est arbitraire et caract\'erise la
R-sym\'etrie. En rempla\c{c}ant dans Eq.(\ref{Rsym1a}) et
Eq.(\ref{Rsym1b}) les superchamps par leur expression
litt\'erale (voir Eq.(\ref{supchp7}), Eq.(\ref{supchp8})
et Eq.(\ref{supchp11})), nous obtenons,
\begin{eqnarray}
&& \theta \sigma^{\mu} \bar \theta v_{k  \mu}(x)
+i \theta \theta \bar \theta \bar \l_k(x)
- i \bar \theta \bar \theta \theta \l_k(x)
+\ud  \theta \theta \bar \theta \bar \theta D_k(x)
\to  \cr
&& \theta \sigma^{\mu} \bar \theta v_{k  \mu}(x)
+i \theta \theta \bar \theta e^{-i \alpha} \bar \l_k(x)
- i \bar \theta \bar \theta \theta e^{i \alpha} \l_k(x)
+\ud  \theta \theta \bar \theta \bar \theta D_k(x),
\label{Rsym2a}
\end{eqnarray}
\begin{eqnarray}
&& z_l(x)+\sqrt 2 \theta \psi_l(x)- \theta \theta f_l(x)
\to \cr && \ \ \ \ \ \ \ \ \ \ \ \ \ \ \ \ e^{i R_l \alpha} \
\bigg ( z_l(x)+\sqrt 2 \theta e^{-i \alpha} \psi_l(x)
- \theta \theta e^{-2i \alpha} f_l(x) \bigg ),
\label{Rsym2b}
\end{eqnarray}
\begin{eqnarray}
&& \bar z_l(x)+\sqrt 2 \bar \theta \bar \psi_l(x)- \bar \theta \bar \theta
f_l(x)
\to \cr && \ \ \ \ \ \ \ \ \ \ \ \ \ \ \ \ e^{-i R_l \alpha} \
\bigg ( \bar z_l(x)+\sqrt 2 \bar \theta e^{i \alpha} \bar \psi_l(x)
- \bar \theta \bar \theta e^{2i \alpha} \bar f_l(x) \bigg ).
\label{Rsym2c}
\end{eqnarray}
Les formules \ref{Rsym2a}, \ref{Rsym2b} et \ref{Rsym2c}
montrent que les transformations
de R-sym\'etrie \ref{Rsym1a}, \ref{Rsym1b} et \ref{Rsym1c}
d\'efinies sur les superchamps
et les variables de Grassmann
peuvent de mani\`ere \'equivalente \^etre d\'efinies sur
les champs eux-m\^emes par,
\begin{equation}
\begin{array}{rcl}
 v_{k  \mu}(x) & \to & v_{k  \mu}(x), \\
 \bar \l_k(x) & \to & e^{-i \alpha} \ \bar \l_k(x), \\
 \l_k(x) & \to & e^{i \alpha} \ \l_k(x), \\
 D_k(x) & \to & D_k(x),
\end{array}
\begin{array}{rcl}
 z_l(x) & \to & e^{i R_l \alpha} \ z_l(x), \\
 \bar z_l(x) & \to & e^{-i R_l \alpha} \ \bar z_l(x), \\
 \psi_l(x) & \to & e^{i(R_l-1)\alpha} \ \psi_l(x), \cr
 \bar \psi_l(x) & \to & e^{-i(R_l-1)\alpha} \ \bar \psi_l(x), \\
 f_l(x) & \to & e^{i(R_l-2) \alpha} \ f_l(x), \\
 \bar f_l(x) & \to & e^{-i(R_l-2) \alpha} \ \bar f_l(x).
\end{array}
\label{Rsym3}
\end{equation}
Travaillons dans la repr\'esentation chirale des matrices $\gamma$.
La matrice $\gamma_5$ et les projecteurs
de chiralit\'e gauche et droite s'\'ecrivent alors,
\begin{equation}
\gamma_5=
\left (
\begin{array}{cc}
I_2 & 0 \\ 0 & -I_2
\end{array}
\right ) ,
P_L={I_4+\gamma_5 \over 2}=
\left (
\begin{array}{cc}
I_2 & 0 \\ 0 & 0
\end{array}
\right ) ,
P_R={I_4-\gamma_5 \over 2}=
\left (
\begin{array}{cc}
0 & 0 \\ 0 & I_2
\end{array}
\right ) ,
\label{chirep}
\end{equation}
o\`u $I_2$ et $I_4$ sont les matrices identit\'e $2 \times 2$
et $4 \times 4$, respectivement. Les spineurs \`a quatre composantes
de chiralit\'e gauche et droite sont donc donn\'es par,
\begin{equation}
\Psi_L=P_L \Psi= P_L
\left (
\begin{array}{c}
\chi \\ \bar \eta
\end{array}
\right )
=
\left (
\begin{array}{c}
\chi \\ 0
\end{array}
\right ) ,
\\\
\Psi_R=P_R \Psi=P_R
\left (
\begin{array}{c}
\chi \\ \bar \eta
\end{array}
\right )
=
\left (
\begin{array}{c}
0 \\ \bar \eta
\end{array}
\right ) ,
\label{chispi}
\end{equation}
Par cons\'equent, le spineur \`a 2 composantes $\psi$ ($\bar \psi$)
a la m\^eme charge $R$ que le spineur \`a 4 composantes $\Psi_L$
($\Psi_R$). Nous pouvons donc r\'e\'ecrire les transformations
en terme de spineurs \`a 4 composantes comme:
\begin{equation}
\begin{array}{rcl}
v_{k  \mu}(x) & \to & v_{k  \mu}(x), \\
\L_k(x) & \to & e^{i \gamma_5 \alpha} \ \L_k(x), \\
D_k(x) & \to & D_k(x),
\end{array}
\begin{array}{rcl}
 \tilde f_{lL}(x) & \to & e^{i R_l \alpha} \ \tilde f_{lL}(x), \\
 \tilde f_{lR}(x) & \to & e^{-i R_l \alpha} \ \tilde f_{lR}(x), \\
 \Psi_l(x) & \to & e^{i \gamma_5 (R_l-1)\alpha} \ \Psi_l(x), \\
 f_l(x) & \to & e^{i(R_l-2) \alpha} \ f_l(x), \\
 \bar f_l(x) & \to & e^{-i(R_l-2) \alpha} \ \bar f_l(x).
\end{array}
\label{Rsym4}
\end{equation}
Dans Eq.(\ref{Rsym4}), nous avons not\'e les
spineurs \`a 4 composantes $\L(x)$ pour les jauginos et
$\Psi(x)$ pour les quarks et les leptons.
Par ailleurs, les notations pour les champs scalaires
gauche et droit ont \'et\'e chang\'e par rapport
\`a Eq.(\ref{Rsym3}) selon
$z \to \tilde f_L$ et $\bar z \to \tilde f_R$, respectivement.

Les transformations \ref{Rsym4} permettent
de d\'eterminer quels sont les termes du lagrangien
invariants sous la R-sym\'etrie (R-invariants).
Nous constatons d'apr\`es Eq.(\ref{Rsym4})
que les termes cin\'etiques pour les champs de spin 0,
$({\partial}_{\mu} \tilde f_{L,R})^\star ({\partial}^{\mu} \tilde f_{L,R})$,
de spin 1/2, $i \bar \Psi \gamma^{\mu } {\partial}_{\mu} \Psi$,
et de spin 1, $-(1/4)F_{\mu \nu}F^{\mu \nu}$ o\`u
$F_{\mu \nu}={\partial}_{\mu} v_{\nu}-{\partial}_{\nu} v_{\mu}$,
sont tous invariants sous les R-sym\'etries.
L'invariance sous les R-sym\'etries pour les termes du lagrangien
d\'eriv\'es du superpotentiel peut aussi \^etre test\'ee au niveau du
superpotentiel lui-m\^eme. En effet, le lagrangien ${\cal L}$ associ\'e
au superpotentiel $W$ d\'ependant des superchamps chiraux
$\Phi_l(x,\theta)$ est obtenu par (voir Eq.(\ref{LAG7})),
\begin{eqnarray}
{\cal L}= \int d^2 \theta \ W(\Phi_l(x,\theta)).
\label{int2W}
\end{eqnarray}
Ce lagrangien $\cal L$ se transforme sous la R-sym\'etrie
consid\'er\'ee par,
\begin{eqnarray}
\int d^2 \theta \ W(\Phi_l(x,\theta)) \to
\int d^2 \theta \ e^{i R_W \alpha} \ W(\Phi_l(x,\theta e^{-i \alpha})),
\label{RsymW}
\end{eqnarray}
o\`u $R_W$ est la charge associ\'ee au superpotentiel
d\'ependant
des charges $R_l$ des superchamps $\Phi_l(x,\theta)$.
Afin de pouvoir mieux comparer les membres de gauche et de droite de la
transformation \ref{RsymW}, effectuons pour le membre de droite
le changement de variable suivant,
\begin{eqnarray}
\theta \to \theta' e^{i \alpha},
d \theta \to d \theta' e^{-i \alpha},
d^2 \theta \to d^2 \theta' e^{-2i \alpha}.
\label{chvarg}
\end{eqnarray}
La diff\'erence de signe dans les transformations de changement de variable
pour $\theta$ et $d \theta$ vient du fait que pour une variable de Grassmann
l'int\'egration est \'equivalente \`a la d\'erivation.
La transformation de R-sym\'etrie \ref{RsymW} s'\'ecrit
maintenant,
\begin{eqnarray}
\int d^2 \theta \ W(\Phi_l(x,\theta)) \to
\int d^2 \theta' \ e^{-2i \alpha} e^{i R_W \alpha} \ W(\Phi_l(x,\theta')).
\label{RsymWp}
\end{eqnarray}
Finalement, nous voyons clairement
dans la transformation \ref{RsymWp} que
le lagrangien $\cal L$ associ\'e au superpotentiel $W$
(Eq.(\ref{int2W})) est invariant sous
les R-sym\'etries donnant une charge $R_W=2$ \`a $W$.

Les R-sym\'etries permettent d'interdire certains des couplages
en $\l$, $\l'$ et $\l''$ du superpotentiel \ref{wrpv} ou
bien l'ensemble de ces couplages.
\'Etudions un exemple de R-sym\'etrie
interdisant tous les couplages du superpotentiel \ref{wrpv}
(comme le fait une parit\'e de mati\`ere).
Consid\'erons la R-sym\'etrie,
\begin{eqnarray}
V(x,\theta,\bar \theta) & \to &
V(x,\theta e^{-i \alpha},\bar \theta e^{i \alpha}),
\label{Rfay1a}
\end{eqnarray}
\begin{eqnarray}
H_{1,2}(x,\theta) & \to &
H_{1,2}(x,\theta e^{-i \alpha}),
\label{Rfay1b}
\end{eqnarray}
\begin{eqnarray}
\Phi(x,\theta) & \to &
e^{i \alpha} \ \Phi(x,\theta e^{-i \alpha}),
 \ \Phi=Q,U^c,D^c,L,E^c.
\label{Rfay1c}
\end{eqnarray}
Dans Eq.(\ref{Rfay1a}), $V$ d\'esigne
chaque superchamp vectoriel du groupe de jauge
$SU(3) \times SU(2) \times U(1)$ du MSSM.
Les transformations \ref{Rfay1b} et \ref{Rfay1c}
des superchamps chiraux sont associ\'ees aux charges
suivantes: $R=0$ pour $H_1,H_2$ et $R=1$ pour $Q,U^c,D^c,L,E^c$.
Par cons\'equent, les transformations de cette R-sym\'etrie
sur les champs s'\'ecrivent,
\begin{equation}
\begin{array}{rcl}
v_{ \mu}(x) & \to & v_{ \mu}(x), \\
\L(x) & \to & e^{i \gamma_5 \alpha} \ \L(x), \\
D(x) & \to & D(x),
\end{array}
\begin{array}{rcl}
 h_{1,2}(x) & \to & h_{1,2}(x), \\
 \tilde h_{1,2}(x) & \to &  e^{-i \gamma_5 \alpha} \ \tilde h_{1,2}(x), \\
 f_h(x) & \to & e^{-2i \alpha} \ f_h(x), \\
 \bar f_h(x) & \to & e^{2i \alpha} \ \bar f_h(x),
\end{array}
\begin{array}{rcl}
 \tilde f_L(x) & \to & e^{i  \alpha} \ \tilde f_L(x), \\
 \tilde f_R(x) & \to & e^{-i  \alpha} \ \tilde f_R(x), \\
 \Psi(x) & \to &  \Psi(x), \\
 f(x) & \to & e^{-i \alpha} \ f(x), \\
 \bar f(x) & \to & e^{i \alpha} \ \bar f(x),
\end{array}
\label{Rfay2}
\end{equation}
o\`u $v_{ \mu}(x)$ d\'esigne les bosons de jauge,
$\L(x)$ les jauginos,
$h_{1,2}(x)$ les bosons de Higgs, $\tilde h_{1,2}(x)$
les higgsinos, $\tilde f$ les squarks et les sleptons
et $\Psi(x)$ les quarks et les leptons. \\
Les interactions en $\l$, $\l'$ et $\l''$ du lagrangien \ref{eqyuk}
sont interdites par cette R-sym\'etrie, qui agit donc comme
une parit\'e de mati\`ere. En effet, les transformations \ref{Rfay2}
montrent que les couplages du lagrangien \ref{eqyuk}
qui sont du type $\bar \Psi \Psi \tilde f_{L,R}$ sont interdits
par cette R-sym\'etrie.
Nous pouvons aussi raisonner au niveau du superpotentiel:
Les termes trilin\'eaires du superpotentiel \ref{wrpv} sont
interdits car ils
ont une charge $R_W=3$ puisque les superchamps $Q,U^c,D^c,L$ et $E^c$
ont chacun une charge $R=1$. Notons que le terme bilin\'eaire du
superpotentiel \ref{wrpv} a une charge $R_W=1$ et n'est donc
pas non plus invariant.
En revanche, les couplages de Yukawa qui sont du type
$h_{1,2} \bar \Psi \Psi$
sont autoris\'es par cette R-sym\'etrie, d'apr\`es Eq.(\ref{Rfay2}).
Cette R-invariance des couplages de Yukawa
nous est confirm\'ee par le fait que les termes trilin\'eaires correspondant
\`a ces couplages dans le superpotentiel (voir Eq.(\ref{wMSSM}))
ont une charge $R_W=2$, d'apr\`es Eq.(\ref{Rfay1b}) et Eq.(\ref{Rfay1c}).

\subsection{Probl\`emes et int\'er\^ets des R-sym\'etries}
\label{pbm}

Les R-sym\'etries souffrent de deux principaux probl\`emes.
Nous expliquons ici la premi\`ere difficult\'e qui est
essentiellement th\'eorique \cite{Fayet2}. \\
Comme nous l'avons vu dans la Section \ref{SUGmot},
le caract\`ere local des sym\'etries de jauge sugg\`ere
que si la \susi est une sym\'etrie
fondamentale de la nature elle doit \^etre
r\'ealis\'ee de fa\c{c}on locale.
Une th\'eorie dans laquelle la supersym\'etrie est une sym\'etrie
locale, doit contenir la th\'eorie de la gravitation et est par cons\'equent
appel\'ee une th\'eorie de supergravit\'e. Les th\'eories de \susi locale
sont donc aussi motiv\'ees par le r\^ole naturel
que joue la gravitation dans ces th\'eories. Outre ces motivations
th\'eoriques, les th\'eories de supergravit\'e permettent
une brisure spontan\'ee r\'ealiste de la supersym\'etrie
(voir Section \ref{SUGmot}).
Dans ce contexte des th\'eories de supergravit\'e, le gravitino, qui est
la particule de jauge de la \susi locale, acquiert une masse $m_{3/2}$
lors de la brisure spontan\'ee de la supersym\'etrie,
de m\^eme que les bosons de jauge du Mod\`ele Standard acqui\`erent une
masse
lors de la brisure spontan\'ee du groupe de jauge $SU(2)_L \times U(1)_Y$.
Or, les R-sym\'etries n'autorisent
pas les termes de masse pour le gravitino. En effet,
g\'en\'eralisant les transformations \ref{Rsym4}
nous obtenons les transformations sous une R-sym\'etrie
du vierbein $e^m_{\mu}$ et du champ de spin 3/2 du gravitino $\Psi_{\mu}$,
\begin{eqnarray}
e^m_{\mu}(x)  & \to &   e^m_{\mu}(x), \cr
\Psi_{\mu}(x)  & \to &  e^{i \gamma_5 \alpha} \ \Psi_{\mu}(x),
\label{Rgrino}
\end{eqnarray}
qui sont incompatibles avec l'existence du terme de masse pour
le gravitino, $m_{3/2} \bar \Psi_{\mu}
[\gamma^{\mu},\gamma^{\nu}] \Psi_{\nu}$,
qui implique des champs $\Psi_{\mu}$
et $\Psi_{\nu}$ de chiralit\'e oppos\'ees. En conclusion,
la coexistence d'une R-sym\'etrie avec les th\'eories
de supergravit\'e, qui pr\'esentent en outre de nombreux attraits,
para\^{\i}t difficile.

La seconde difficult\'e est davantage ph\'enom\'enologique.
D'apr\`es Eq.(\ref{Rsym4}),
les R-sym\'e\-tries interdisent les termes de brisure douce
de \susi donnant directement une masse aux jauginos:
\begin{eqnarray}
-{i \over 2} M_1 \bar {\tilde B} \tilde B
-{i \over 2} M_2 \bar {\tilde W_3} \tilde W_3
- M_2 \bar {\tilde W} \tilde W
-{i \over 2} M_3 \bar {\tilde g} \tilde g,
\label{massjaug}
\end{eqnarray}
o\`u nous avons d\'enot\'e les jauginos
$\Lambda=\tilde B$ (bino), $\tilde W_3$ (wino neutre),
$\tilde W$ (wino charg\'e), $\tilde g$ (gluino).
$\tilde B$, $\tilde W_3$ et $\tilde g$ sont des spineurs de Majorana.
La raison est qu'un terme de masse pour un spineur \`a 4 composantes
est du type, $\bar \Psi_H \Psi_{H'}$ o\`u $H,H'=L,R$ avec $H \neq H'$,
et que les transformations \ref{Rsym4} distinguent les spineurs
de chiralit\'e gauche et droite lorsqu'il s'agit de jauginos.
Cependant, certaines R-sym\'etries, comme la R-sym\'etrie d\'efinie dans
Eq.(\ref{Rfay2}),
autorisent les termes de m\'elange entre jauginos et higgsinos qui sont
du type $\bar \Lambda \tilde h_{1,2}$ et g\'en\`erent aussi des masses
pour les jauginos.
Dans le MSSM, de tels termes de m\'elange existent pour les winos
et le bino qui peuvent donc acqu\'erir une masse.
Par exemple, les matrices de masse des charginos et neutralinos
dans le MSSM (voir Section \ref{SpecMSSMjh}) munie de la R-sym\'etrie
d\'efinie dans Eq.(\ref{Rfay2}) sont respectivement,
{\small
\begin{equation}
\left ( \begin{array}{cc}
(M_2=0) & M_W \sqrt 2 \sin \beta \\
 M_W \sqrt 2 \cos \beta & (\mu=0) \\
\end{array} \right ),
\label{Rchargma}
\end{equation}
et,
\begin{equation}
\left ( \begin{array}{cccc}
(M_1=0) & 0 & - M_Z \sqrt 2 \cos \beta \sin \theta_W
& M_Z \sqrt 2 \sin \beta \sin \theta_W \\
0 & (M_2=0) &  M_Z \sqrt 2 \cos \beta \cos \theta_W
& - M_Z \sqrt 2 \sin \beta \cos \theta_W \\
- M_Z \sqrt 2 \cos \beta \sin \theta_W
& M_Z \sqrt 2 \cos \beta \cos \theta_W & 0 & (-\mu=0) \\
M_Z \sqrt 2 \sin \beta \sin \theta_W
& - M_Z \sqrt 2 \sin \beta \cos \theta_W & (-\mu=0) & 0 \\
\end{array} \right ).
\vspace*{0.5cm}
\label{Rneutma}
\end{equation}
}
Notons qu'il existe d'autres R-sym\'etries
que celle d\'efinie dans
Eq.(\ref{Rfay2}) pouvant simultan\'ement
interdire les couplages violant les nombres leptonique et baryonique du
lagrangien \ref{eqyuk},
autoriser certains termes de m\'elange entre jauginos et higgsinos
et autoriser le terme $\mu H_1 H_2$,
qui est n\'ecessaire \`a la brisure \'electrofaible
et donne une masse aux higgsinos.
En revanche, le MSSM ne contient pas de termes m\'elangeant higgsinos et
gluinos.
Dans le cadre du MSSM,
les R-sym\'etries sont donc incompatibles avec l'existence d'un terme de
masse
au niveau des arbres
pour les gluinos. Les R-sym\'etries semblent donc \^etre en contradiction
avec les donn\'ees exp\'erimentales
qui donnent une limite inf\'erieure sur la masse
du gluino: $m_{\tilde g}>173GeV$ pour $\mu=-200GeV,\tan \beta=2$
\cite{3:PDG98}.

Quelles solutions s'offrent au probl\`eme de la masse du gluino
dans les th\'eories munies de R-sym\'etries ?
Tout d'abord, la masse du gluino ne peut \^etre g\'en\'er\'ee
spontan\'ement. En effet, les couplages du gluino $\tilde q \bar q \tilde g$
impliquent les champs de squarks qui ne peuvent acqu\'erir une valeur
moyenne
dans le vide (vev), le groupe de jauge $SU(3)_c$ du Mod\`ele Standard
ne devant pas \^etre bris\'e. Cela nous am\`ene aux alternatives
int\'eressantes d'une masse de gluino g\'en\'er\'ee radiativement ou
par un m\'ecanisme dynamique, pour lesquelles la masse du gluino est
pr\'edite \cite{Rand,Gir,Farrar}. Cependant, de tels mod\`eles donnent des
gluinos trop l\'egers et sont exclus \cite{Sark}. Un superchamp chiral
octet de $SU(3)_c$ peut aussi \^etre ajout\'e pour donner une masse au
gluino,
mais ces sc\'enarios ne sont pas naturels et brisent en g\'en\'eral
$SU(3)_c$
\cite{Rand}. Enfin, le gluino peut acqu\'erir une masse lors d'une brisure
spontan\'ee
de R-sym\'etrie \cite{Farrar}. Mais cette masse est petite et la brisure
spontan\'ee
d'une R-sym\'etrie
pose le probl\`eme de l'existence d'un ind\'esirable pseudo boson de
Goldstone
l\'eger appel\'e R-axion \cite{CP}.
En conclusion, le probl\`eme de la masse du gluino n'a pas de solution
\'evidente.

Malgr\'e les deux probl\`emes de l'existence d'une masse
pour le gravitino et le gluino, les R-sym\'etries ont suscit\'e
un grand int\'er\^et dans la litt\'erature. Voici les
int\'er\^ets majeurs des R-sym\'etries, outre la possibilit\'e
de garantir la conservation des nombres leptonique et baryonique.
Tout d'abord, de m\^eme que la sym\'etrie de Peccei-Quinn
\cite{CP},
la R-invariance a \'et\'e propos\'ee comme une solution originale
au probl\`eme de la violation forte de CP \cite{buwy}, ainsi
qu'au probl\`eme du moment dipolaire \'electrique du neutron
\cite{buwy,affleck}.
De plus, il a \'et\'e montr\'e dans \cite{nelsei} que
l'existence d'une R-sym\'etrie est une condition n\'ecessaire \`a
la brisure dynamique de la supersym\'etrie. Les auteurs de \cite{nelsei}
ont aussi montr\'e que la pr\'esence d'une R-sym\'etrie bris\'ee spontan\'ement est
une condition suffisante pour la brisure dynamique de la supersym\'etrie,
dans le cas o\`u le lagrangien effectif 
est un lagrangien g\'en\'erique consistent avec les sym\'etries de la th\'eorie
(pas de fine-tuning) et o\`u la th\'eorie de basse \'energie peut \^etre d\'ecrite 
par un lagrangien effectif supersym\'etrique de Wess-Zumino sans champs de jauge.
Tous les mod\`eles connus aujourd'hui de brisure dynamique de la supersym\'etrie poss\`edent
une telle R-sym\'etrie bris\'ee spontan\'ement et contiennent donc un axion pouvant \^etre
probl\'ematique. Cependant, les auteurs de \cite{nelsei} ont montr\'e que la R-sym\'etrie peut
dans beaucoup de cas \^etre bris\'ee explicitement sans restaurer la \susi de telle sorte que
l'axion puisse acqu\'erir une masse suffisamment grande.      
Enfin, certaines R-sym\'etries interdisent le
terme de m\'elange des superchamps de Higgs dans le superpotentiel:
\begin{eqnarray}
W=\mu H_1 H_2,
\label{muterm}
\end{eqnarray}
comme par exemple la R-sym\'etrie d\'efinie dans
Eq.(\ref{Rfay1a},\ref{Rfay1b},\ref{Rfay1c}) qui assigne une charge $R=0$ \`a
$H_1$ et $H_2$. Or la brisure spontan\'ee de telles R-sym\'etries
permet de contr\^oler naturellement la valeur du param\`etre $\mu$
qui doit \^etre de l'ordre de l'\'echelle \'electrofaible
afin de permettre une brisure \'electrofaible r\'ealiste,
c'est \`a dire donnant une masse permise
par les bornes exp\'erimentales au boson de Higgs l\'eger et donnant
les masses attendues aux particules du Mod\`ele Standard.
(``probl\`eme du terme $\mu$''). Le terme de Eq.(\ref{muterm})
peut effectivement \^etre g\'en\'er\'e par un terme du type,
\begin{eqnarray}
W=\alpha {1 \over M_P^{n-1}} [\prod^n_{i=1} S_i] H_1 H_2,
\label{Sterm}
\end{eqnarray}
o\`u les $S_i$ sont des nouveaux superchamps chiraux
singlets de jauge dont les charges de R-sym\'etrie
v\'erifient $\sum^n_{i=1} R_i=2$.
(dans le cas de la R-sym\'etrie de
Eq.(\ref{Rfay1a},\ref{Rfay1b},\ref{Rfay1c}))
et dont les champs scalaires acqui\`erent une vev.
Le terme $\mu$ de Eq.(\ref{muterm}) sera alors donn\'e par,
\begin{eqnarray}
\mu=\alpha {<S_1S_2...S_n> \over M_P^{n-1}},
\label{Smusol}
\end{eqnarray}
$M_P$ \'etant l'\'echelle de Planck,
et peut \^etre de l'ordre de l'\'echelle \'electrofaible pour $<~S_i~> \ll
M_P$.
Les R-sym\'etries offrent donc un cadre propice \`a la r\'esolution du
probl\`eme du $\mu$ \cite{hara,dine,kim2,kim3}.
Rappelons toutefois que la brisure spontan\'ee d'une
R-sym\'etrie pose le probl\`eme de l'existence
du R-axion.
Notons aussi que cette solution au probl\`eme du $\mu$
via l'imposition d'une R-sym\'etrie est tr\`es analogue
\`a la solution caract\'eris\'ee par l'existence
d'une sym\'etrie de Peccei-Quinn \cite{kim3,kim1}.

\subsection{Les R-sym\'etries discr\`etes et la R-parit\'e}

Les R-sym\'etries discr\`etes permettent d'\'eviter les probl\`emes
de l'existence d'une masse pour le gravitino et le gluino.
Consid\'erons par exemple la R-sym\'etrie discr\`ete obtenue
\`a partir de la R-sym\'etrie d\'efinie dans
Eq.(\ref{Rfay1a},\ref{Rfay1b},\ref{Rfay1c})
et Eq.(\ref{Rfay2}) en prenant $\alpha=\pi$.
Les transformations associ\'ees \`a la R-sym\'etrie discr\`ete ainsi obtenue
s'\'ecrivent sur les superchamps,
\begin{eqnarray}
V(x,\theta,\bar \theta) & \to &
V(x,-\theta ,-\bar \theta ),
\label{Rpa1a}
\end{eqnarray}
\begin{eqnarray}
H_{1,2}(x,\theta) & \to &
H_{1,2}(x,-\theta ),
\label{Rpa1b}
\end{eqnarray}
\begin{eqnarray}
\Phi(x,\theta) & \to &
- \Phi(x,-\theta ),
 \ \Phi=Q,U^c,D^c,L,E^c,
\label{Rpa1c}
\end{eqnarray}
et donc sur les champs,
\begin{equation}
\begin{array}{rcl}
v_{ \mu}(x) & \to & v_{ \mu}(x), \\
\L(x) & \to & - \L(x), \\
D(x) & \to & D(x),
\end{array}
\begin{array}{rcl}
 h_{1,2}(x) & \to & h_{1,2}(x), \\
 \tilde h_{1,2}(x) & \to &  - \tilde h_{1,2}(x), \\
 f_h(x) & \to &  f_h(x), \\
 \bar f_h(x) & \to & \bar f_h(x),
\end{array}
\begin{array}{rcl}
 \tilde f_L(x) & \to & - \tilde f_L(x), \\
 \tilde f_R(x) & \to & - \tilde f_R(x), \\
 \Psi(x) & \to &  \Psi(x), \\
 f(x) & \to & - f(x), \\
 \bar f(x) & \to & - \bar f(x).
\end{array}
\label{Rpa2}
\end{equation}
La R-sym\'etrie discr\`ete de Eq.(\ref{Rpa2}) autorise le terme de masse
du gravitino ainsi que les termes de masse des jauginos exprim\'es dans
Eq.(\ref{massjaug}). En effet, les transformations \ref{Rpa2}
sont identiques pour les champs de jauginos de chiralit\'e gauche et droite,
\`a l'inverse des transformations g\'en\'erales de R-sym\'etrie
donn\'ees dans Eq.(\ref{Rsym4}).

Les R-sym\'etries discr\`etes peuvent \^etre aussi bien
des parit\'es de mati\`ere que des parit\'es leptoniques ou
des parit\'es baryoniques. Une \'etude des diff\'erentes R-sym\'etries
discr\`etes possibles a \'et\'e tra\^{\i}t\'ee dans \cite{Iboss2}.

Nous observons que la R-sym\'etrie discr\`ete de Eq.(\ref{Rpa2})
est une parit\'e de mati\`ere, puisqu'elle interdit les couplages
en $\l$, $\l'$ et $\l''$ de Eq.(\ref{eqyuk}), ce qui est coh\'erent
car la R-sym\'etrie originelle, d\'efinie dans
Eq.(\ref{Rfay2}), agit aussi comme une parit\'e de mati\`ere. En fait,
la R-sym\'etrie discr\`ete de Eq.(\ref{Rpa2}) est la sym\'etrie dite
de {\it R-parit\'e}. 
La sym\'etrie de R-parit\'e a \'et\'e consid\'er\'ee
les premi\`eres fois dans \cite{Wein,FaFa,rp1,rp2,rp3,rp4,rp5}.
Bien que la R-parit\'e ne soit pas la seule parit\'e de mati\`ere,
on associe souvent aujourd'hui, pour des raisons historiques,
la sym\'etrie interdisant les couplages en $\l$ $\l'$ et $\l''$ \`a la
R-parit\'e
plut\^ot qu'\`a la parit\'e de mati\`ere.
D'apr\`es Eq.(\ref{Rpa2}), la R-parit\'e transforme les
champs des particules du Mod\`ele Standard avec un facteur $+1$
et les champs des particules \susiqs avec un facteur $-1$.
La R-parit\'e peut donc \^etre d\'efinie de fa\c{c}on \'equivalente \`a
Eq.(\ref{Rpa2}) par l'action de l'op\'erateur \cite{FaFa},
\begin{equation}
R_p=(-1)^{3B+L+2S},
\label{Rpdef}
\end{equation}
o\`u $B$ est le nombre baryonique, $L$ le nombre leptonique et $S$ le spin
de la particule. Nous remarquons que la R-parit\'e autorise les couplages
de Yukawa, de m\^eme que la R-sym\'etrie originelle d\'efinie
par Eq.(\ref{Rfay2}).

\subsection{Les sym\'etries discr\`etes}
\label{symdis}

De simples sym\'etries discr\`etes peuvent aussi \^etre utilis\'ees pour
supprimer les diff\'erents couplages du lagrangien \ref{eqyuk}
et assurer ainsi un temps de vie suffisamment long au proton \cite{Haki}.
Par exemple, la sym\'etrie,
\begin{equation}
(Q,U^c,D^c,L,E^c) \to - (Q,U^c,D^c,L,E^c), \
(H_1,H_2) \to (H_1,H_2),
\label{pamat}
\end{equation}
est une parit\'e de mati\`ere, la sym\'etrie,
\begin{equation}
(L,E^c) \to - (L,E^c), \
(Q,U^c,D^c,H_1,H_2) \to (Q,U^c,D^c,H_1,H_2),
\label{palept}
\end{equation}
est une parit\'e leptonique et la sym\'etrie,
\begin{equation}
(Q,U^c,D^c) \to - (Q,U^c,D^c), \
(L,E^c,H_1,H_2) \to (L,E^c,H_1,H_2),
\label{pabar}
\end{equation}
est une parit\'e baryonique. Ces trois sym\'etries discr\`etes
sont d\'efinies sur les superchamps chiraux et laissent les superchamps
vectoriels invariants. Notons que les composantes de spin 0, 1/2 et 1
des superchamps se transforment sous des sym\'etries discr\`etes comme
les superchamps eux-m\^emes.
Les couplages de Yukawa sont conserv\'es sous les sym\'etries
\ref{pamat}, \ref{palept} et \ref{pabar}.
Souvent, on appelle ces sym\'etries \ref{pamat}, \ref{palept}
et \ref{pabar} parit\'es de mati\`ere, parit\'es
leptoniques et parit\'es baryoniques, respectivement,
et on nomme les sym\'etries
interdisant les couplages en $\l$,$\l'$,$\l''$ parit\'es de mati\`ere
g\'en\'eralis\'ees, en $\l$,$\l'$ parit\'es leptoniques
g\'en\'eralis\'ees et en $\l''$ parit\'es baryoniques g\'en\'eralis\'ees.

Dans \cite{Iboss2}, la forme g\'en\'erale d'une sym\'etrie $Z_N$
d\'efinie par,
\begin{equation}
\Phi_l \to e^{i q_l {2 \pi \over N}} \ \Phi_l, \
\Phi_l=Q,U^c,D^c,L,E^c,H_1,H_2,
\label{ZNsym}
\end{equation}
et autorisant les couplages de Yukawa a \'et\'e donn\'ee.
Cette forme g\'en\'erale a permis
aux auteurs de \cite{Iboss2} de d\'eterminer la liste compl\`ete des
sym\'etries $Z_2$ et $Z_3$ ind\'ependantes qui autorisent les couplages
de Yukawa. Des sym\'etries $Z_N$ avec de grandes valeurs de $N$ sont peu
naturelles d'un point de vue esth\'etique ainsi que
du point de vue des th\'eories de cordes.
Nous pr\'esentons dans la Table \ref{Z2sym} les charges
associ\'ees \`a toutes les sym\'etries $Z_2$ ind\'ependantes autorisant
les couplages de Yukawa.
\begin{table}
\begin{center}
\begin{tabular}{c|c|c|c|c|c|c|c}
     & $Q$  & $U^c$  & $D^c$  & $L$  & $E^c$  & $H_1$  & $H_2$
 \\
\hline
 PM1 & 0  & -1   & 1    & 0  & 1    & -1   & 1
 \\
\hline
 PM2 & 0  & -1   & 0    & -1 & 1    & 0    & 1
 \\
\hline
 PL1 & 0  & 0    & 0    & -1 & 1    & 0    & 0
 \\
\hline
 PL2 & 0  & 0    & -1   & -2 & 1    & 1    & 0
 \\
\hline
 PB1 & 0  & -1   & 1    & -1 & 2    & -1   & 1
 \\
\hline
 PB2 & 0  & -1   & 0    & -2 & 2    & 0    & 1
 \\
\end{tabular}
\caption{{\small {\it
Charges des superchamps associ\'ees aux sym\'etries
PM1, PM2 (parit\'es de mati\`ere),
PL1, PL2 (parit\'es leptoniques),
PB1 et PB2 (parit\'es baryoniques) qui repr\'esentent les sym\'etries
$Z_2$ ind\'ependantes autorisant les couplages de Yukawa.}}}
\label{Z2sym}
\end{center}
\end{table}
La Table \ref{Z2sym} montre que le nombre de ces sym\'etries $Z_2$
est \'etonnement petit \'etant donn\'e le peu de contraintes impos\'ees.
Le nombre des sym\'etries $Z_2$ ind\'ependantes et compatibles
avec l'existence
des couplages de Yukawa peut m\^eme \^etre r\'eduit davantage si l'on
requiert l'existence un terme $\mu$ du type de Eq.(\ref{muterm})
ainsi que l'absence de toute contribution significative
par des op\'erateurs de dimension 5 (voir Eq.(\ref{dim5F}) et
Eq.(\ref{dim5D}))
\`a la d\'esint\'egration des nucl\'eons.
En effet, les seules sym\'etries satisfaisant \`a ces deux conditions
suppl\'ementaires sont les sym\'etries PL1 et PB1 (voir Table \ref{Z2sym}).
La sym\'etrie PL1 autorise l'op\'erateur ${\cal O}_3$ de Eq.(\ref{dim5F})
mais interdit les couplages en $\l'$ de sorte que ${\cal O}_3$ ne peuvent
contribuer dangereusement \`a la d\'esint\'egration des nucl\'eons
(voir Section \ref{prot5dim}). De m\^eme,
la sym\'etrie PB1 permet l'existence des ${\cal O}_4,...,{\cal O}_{10}$
de Eq.(\ref{dim5F}) et Eq.(\ref{dim5D})
mais interdit les couplages en $\l''$ et donc les ${\cal O}_4,...,{\cal
O}_{10}$
ne peuvent contribuer \`a la d\'esint\'egration des nucl\'eons
de fa\c{c}on significative. En effet, les seules contraintes relevantes
sur les coefficients $\eta_4,...,\eta_{10}$ viennent de combinaisons
entre les op\'erateurs ${\cal O}_4,...,{\cal O}_{10}$ et $U^cD^cD^c$.
Les deux conditions suppl\'ementaires d\'ecrites plus haut ne sont cependant
pas obligatoires.
D'une part, un terme $\mu$ (n\'ecessaire \`a une brisure
\'electrofaible correcte) interdit par une certaine sym\'etrie discr\`ete
peut tout de m\^eme \^etre g\'en\'er\'e par la brisure spontan\'ee de
cette sym\'etrie. Ce m\'ecanisme permet en outre de pr\'edire la valeur
de $\mu$ comme nous l'avons expliqu\'e dans la Section \ref{pbm}
pour le cas de la brisure spontan\'ee d'une R-sym\'etrie.
D'autre part, la suppression de toute contribution significative
par des op\'erateurs de dimension 5 \`a la d\'esint\'egration des
nucl\'eons n'est pas n\'ecessairement assur\'ee par des sym\'etries,
et son origine peut \^etre li\'ee \`a une nouvelle physique, comme
mentionn\'e au d\'ebut de la Section \ref{sympot}.

\section{Motivations th\'eoriques de la parit\'e de mati\`ere ou R-parit\'e}

Si le proton n'est pas prot\'eg\'e de sa d\'esint\'egration
par une parit\'e de mati\`ere, mais seulement par une parit\'e leptonique
ou baryonique ou bien par une sym\'etrie interdisant uniquement
les couplages $\l'$ ou $\l+\l''$, certains couplages
$\l$, $\l'$ ou $\l''$ du lagrangien \ref{eqyuk} peuvent \^etre pr\'esents.
Un tel sc\'enario implique des effets importants sur la ph\'enom\'enologie
de la \susi aupr\`es des collisionneurs de particules.
D'un point de vue ph\'enom\'enologique, l'impact majeur d'une violation
de la parit\'e de mati\`ere, ou violation de la R-parit\'e (\rpv),
est li\'e \`a la stabilit\'e de
la particule \susiq la plus l\'eg\`ere (LSP): En pr\'esence des couplages
\rpv
$\l$, $\l'$ ou $\l''$, la LSP n'est plus stable car elle peut se
d\'esint\'egrer
en particules du Mod\`ele Standard plus l\'eg\`eres qu'elle.
Les signaux caract\'eristiques des r\'eactions \susiq
ne sont alors plus des \'etats finals avec une grande
\'energie manquante mais des signatures multileptoniques ou multi-jets,
suivant le couplage \rpv dominant. En pr\'esence de couplages \rpv,
une particule \susiq peut aussi \^etre produite \`a la r\'esonance
et donner ainsi lieu \`a une d\'ecouverte spectaculaire de la \susi
aupr\`es d'un collisionneur de particules.
Par cons\'equent, \'etant donn\'e les effets importants des couplages
\rpv sur la ph\'enom\'enologie de la supersym\'etrie, il est crucial
de savoir si le proton est prot\'eg\'e par une parit\'e de mati\`ere,
ou en d'autres termes, si la sym\'etrie de R-parit\'e est conserv\'ee.
Dans cette partie, nous nous demanderons si les th\'eories actuelles majeurs
favorisent davantage les mod\`eles avec ou sans parit\'e de mati\`ere.
Nous constaterons qu'aujourd'hui aucun argument th\'eorique ne permet
de trancher clairement entre un sc\'enario $R_p$ et \rpv.

\subsection{Point de vue des sym\'etries discr\`etes de jauge}

Un r\'esultat th\'eorique fort int\'eressant a \'et\'e obtenu dans
\cite{Banks}:
Toute sym\'etrie non jaug\'ee est fortement viol\'ee par
des effets de gravit\'e quantique. Les sym\'etries globales
ainsi viol\'ees ne peuvent pas
garantir la stabilit\'e du proton \cite{Gilbert}.
Les auteurs de \cite{Krau} ont cependant
montr\'e que les sym\'etries discr\`etes
obtenues par brisure spontan\'ee de sym\'etries de jauge
(sym\'etries discr\`etes de jauge)
\'echappent \`a ces effets de gravit\'e quantique et peuvent
donc \^etre conserv\'ees. Cela sugg\`ere que la sym\'etrie
discr\`ete assurant la stabilit\'e du proton soit une
sym\'etrie discr\`ete de jauge. Or les sym\'etries discr\`etes de jauge
sont soumises \`a certaines contraintes provenant des conditions
d'annulation
des anomalies des sym\'etries de jauge originelles.
Ces contraintes sont int\'eressantes dans la mesure o\`u elles
constituent des arguments
th\'eoriques en faveur de l'\'elimination de certaines sym\'etries
discr\`etes
jusqu'alors possibles.

Par exemple, consid\'erons une th\'eorie effective de basse \'energie
comprenant les superchamps non massifs
$\Phi_i$ de charges $q_i \in {\bf Z}$ sous une sym\'etrie $Z_N$ donn\'ee.
Supposons que cette sym\'etrie $Z_N$ descende d'une sym\'etrie de jauge
$U(1)$.
La condition d'annulation de l'anomalie $U(1)-U(1)-U(1)$ de la
la sym\'etrie de jauge $U(1)$ originelle s'\'ecrit,
\begin{equation}
\sum_{\alpha} Q_{\alpha}^3=0,
\label{U1an}
\end{equation}
o\`u les $Q_{\alpha}$ sont les charges associ\'ees \`a la sym\'etrie $U(1)$
de tous les fermions non massifs de la th\'eorie.
Ces charges sont n\'ecessairement du type,
\begin{equation}
Q(\Phi_i) =q_i + m_i N, \ \
Q(H) =m_H N, \ \ m_i,m_H \in {\bf Z},
\label{chargU1}
\end{equation}
pour les superchamps $\Phi_i$ et le superchamp $H$
dont la composante scalaire acquiert une vev
brisant la sym\'etrie de jauge $U(1)$. En effet,
l'invariance sous $U(1)$ du terme
du superpotentiel responsable de la brisure spontan\'ee de $U(1)$,
\begin{equation}
W=\Phi_1  \Phi_2 ... \Phi_n H^h, h \in {\bf Z},
\label{WbriU1}
\end{equation}
implique d'apr\`es Eq.(\ref{chargU1}),
\begin{equation}
\sum_{i=1}^n Q(\Phi_i) + h Q(H)=
\sum_{i=1}^n (q_i + m_i N) + h m_H N=0.
\label{dem0}
\end{equation}
Par cons\'equent, apr\`es que la sym\'etrie $U(1)$ est bris\'ee par
la vev de $H$, il appara\^{\i}t bien que le terme \ref{WbriU1}
qui devient alors,
\begin{equation}
W=\Phi_1  \Phi_2 ... \Phi_n <~H~>^h, h \in {\bf Z},
\label{Wbrnew}
\end{equation}
est invariant sous la sym\'etrie $Z_N$
puisque l'on a d'apr\`es Eq.(\ref{dem0}),
\begin{equation}
\sum_{i=1}^n q_i = - \sum_{i=1}^n m_i N - h m_H N = k N, \ k \in {\bf Z}.
\label{dem1}
\end{equation}
La th\'eorie dans sa totalit\'e contient les superchamps $\Phi_i$ de
la th\'eorie effective mais aussi les superchamps qui acqui\`erent une masse
lors de la brisure du groupe $U(1)$. Or les charges $U(1)$ de ces derniers
doivent
aussi \^etre prises en compte dans Eq.(\ref{U1an}).
Les charges $U(1)$ des superchamps $\Phi_j$
acqu\'erant une masse lors de la brisure de $U(1)$ sont du m\^eme type
que les charges $U(1)$ des superchamps effectifs $\Phi_i$ (voir
Eq.(\ref{chargU1})),
c'est \`a dire du type $Q(\Phi_j) =q_j + m_j N$,
o\`u $q_j$ est la charge de $\Phi_j$ sous la sym\'etrie $Z_N$ et $m_j \in
{\bf Z}$.
Les charges $Z_N$ des superchamps $\Phi_j$
sont contraintes, par l'invariance sous $Z_N$
des termes de masse $W=m \Phi_j^c \Phi_j$, \`a \^etre du type,
\begin{equation}
q_j^c+q_j = l_j N, \ l_j \in {\bf Z}.
\label{dem5}
\end{equation}
Cette \'equation est \'equivalente \`a l'\'equation,
\begin{equation}
q_j^c+m^c_j N+q_j+m_j N = (l_j+m^c_j+m_j) N, \ l_j,m^c_j,m_j \in {\bf Z},
\label{dem6}
\end{equation}
qui donne une relation sur les charges $U(1)$
des superchamps $\Phi_j$:
\begin{equation}
Q(\Phi_j^c)+Q(\Phi_j) = p_j N, \ p_j \in {\bf Z},
\label{dem2}
\end{equation}
qui peut aussi s'\'ecrire,
\begin{equation}
Q^3(\Phi_j^c)+Q^3(\Phi_j) = p_j N \bigg (
3Q^2(\Phi_j)-3p_j N Q(\Phi_j) +p_j^2 N^2 \bigg ), \ p_j \in {\bf Z}.
\label{dem3}
\end{equation}
Les charges $Z_N$ des superchamps de Majorana $S_j$, acqu\'erant eux
une masse par des termes du type
$W=m S_j^2$ lors de la brisure de $U(1)$, sont contraintes par,
\begin{equation}
q^s_j = {k_j N \over 2}, \ k_j \in {\bf Z}.
\label{dem7}
\end{equation}
o\`u $k_j$ est pair si $N$ est impair, soit encore,
\begin{equation}
q^s_j +m^s_j N = {(k_j+2m^s_j) N \over 2}, \ {p'}_j \in {\bf Z},
\label{dem8}
\end{equation}
avec $k_j$ pair si $N$ est impair. L'\'equation Eq.(\ref{dem8})
donne une relation sur les charges $U(1)$
des superchamps $S_j$:
\begin{equation}
Q(S_j) = {p'_j N \over 2}, \ {p'}_j \in {\bf Z}.
\label{dem4}
\end{equation}
o\`u $p'_j$ est pair si $N$ est impair.
Finalement, en rempla\c{c}ant Eq.(\ref{chargU1}),Eq.(\ref{dem3}) et
Eq.(\ref{dem4})
dans Eq.(\ref{U1an}), on obtient une contrainte, sur les charges $q_i$
d'une sym\'etrie $Z_N$ descendant d'une sym\'etrie de jauge $U(1)$,
qui peut s'\'ecrire \cite{Iboss1},
\begin{equation}
\sum_i q_i^3=m N+ \eta n {N^3 \over 8}, \ \ m,n \in {\bf Z},
\label{ZZZan}
\end{equation}
avec $\eta=1,0$ pour $N \equiv pair,impair$ et,
\begin{equation}
\begin{array}{ccl}
m = & -  \sum_i & \bigg ( 3 q_i^2 m_i + 3 q_i m_i^2 N + m_i^3 N^3 \bigg ) \\
    & -  \sum_j & \bigg ( p_j N(3Q^2(\Phi_j)-3p_j N Q(\Phi_j) +p_j^2 N^2)
\bigg ), \\
n=  & -  \sum_j & {p'}_j^3,
\end{array}
\label{ZZZanp}
\end{equation}
la somme sur $i$ \'etant prise sur les superchamps de
la th\'eorie effective et la somme sur $j$ sur les superchamps
acqu\'erant une masse lors de la brisure de $U(1)$.
De m\^eme, les conditions d'annulation des anomalies
$U(1)-Gravitation-Gravitation$,
$U(1)-SU(M)-SU(M)$ donnent respectivement les contraintes,
\begin{equation}
\sum_i q_i=p N+ \eta q {N \over 2}, \ \ p,q \in {\bf Z},
\label{ZGGan}
\end{equation}
\begin{equation}
\sum_i T_i q_i={1 \over 2} r N, \ \ r \in {\bf Z},
\label{ZSUSUan}
\end{equation}
o\`u les $T_i$ sont
les Casimirs quadratiques de SU(M) correspondant \`a chaque repr\'esentation,
avec une normalisation telle que le Casimir est \'egal \`a 1/2 pour
un M-plet. Dans Eq.(\ref{ZGGan}) et Eq.(\ref{ZSUSUan}) les nombres $p,q$ et
$r$
ob\'eissent \`a des \'egalit\'es du type de Eq.(\ref{ZZZanp}) \cite{Iboss1}.
Il est important de r\'ealiser que les \'egalit\'es \ref{ZZZan}, \ref{ZGGan}
et \ref{ZSUSUan} repr\'esentent des conditions n\'ecessaires mais pas
suffisantes
pour que la sym\'etrie $Z_N$ consid\'er\'ee puisse d\'ecouler d'une
sym\'etrie de
jauge $U(1)$. En effet, pour une sym\'etrie $Z_N$ donn\'ee il n'existe
pas n\'ecessairement un mod\`ele acceptable comprenant une sym\'etrie de
jauge
originelle $U(1)$ satisfaisant entre autre \`a Eq.(\ref{ZZZanp}).
Certains mod\`eles r\'ealistes contenant des sym\'etries de jauge
$U(1)$ engendrant des sym\'etries discr\`etes $Z_N$ \`a basse \'energie
ont \'et\'e construits explicitement dans \cite{Iboss1,Iboss2}.

Dans \cite{Iboss2}, il a \'et\'e montr\'e, \`a partir de la forme
g\'en\'erale
d'une sym\'etrie $Z_N$ autorisant les couplages de Yukawa,
que les seules sym\'etries $Z_N$ respectant les contraintes
\ref{ZZZan}, \ref{ZGGan} et \ref{ZSUSUan}, c'est \`a dire les seules
sym\'etries $Z_N$ pouvant d\'ecouler d'une sym\'etrie de jauge $U(1)$,
sont la sym\'etrie PM1 (voir Section \ref{symdis}) et la sym\'etrie $Z_3$
d\'efinie par les charges suivantes,
\begin{equation}
q(Q)=0, \ q(U^c)=-1, \ q(D^c)=1, \ q(L)=-1, \ q(E^c)=2, \ q(H_1)=-1, \
q(H_2)=1.
\label{Z3jau}
\end{equation}
Ces deux sym\'etries autorisent un terme $\mu$ (voir Eq.(\ref{muterm})).
La sym\'etrie de Eq.(\ref{Z3jau}) supprime toute contribution dangereuse
par des op\'erateurs de dimension 5 (voir Eq.(\ref{dim5F}) et
Eq.(\ref{dim5D}))
\`a la d\'esint\'egration des nucl\'eons, pour les m\^emes raisons
que la sym\'etrie PB1 (voir Section \ref{symdis}).
En revanche, la sym\'etrie PM1 autorise l'existence de l'op\'erateur
${\cal O}_1$ qui engendre une d\'esint\'egration rapide du proton
(voir Section \ref{prot5dim}). En conclusion, si la sym\'etrie
prot\'egeant le proton est une sym\'etrie discr\`ete de jauge $Z_N$
provenant d'une sym\'etrie de jauge $U(1)$, elle peut \^etre
aussi bien une parit\'e de mati\`ere (PM1) ou une parit\'e baryonique
(sym\'etrie $Z_3$ d\'efinie dans Eq.(\ref{Z3jau})), avec une
pr\'ef\'erence pour la parit\'e baryonique qui interdit toute contribution
dangereuse par des op\'erateurs de dimension 5 \`a la d\'esint\'egration
des nucl\'eons. Les auteurs de \cite{dinep} ont m\^eme montr\'e que
les r\'esultats de \cite{Iboss2}
d\'ependent du mod\`ele consid\'er\'e, autorisant
ainsi un plus grand nombre de sym\'etries discr\`etes de jauges.

Mentionnons aussi que dans certains mod\`eles \cite{chamdrein1} les
couplages
en $\l$, $\l'$ et $\l''$ sont supprim\'es par un facteur $M_X/M_P$
dans lequel $M_X$ est l'\'echelle de brisure d'une sym\'etrie de jauge
suppl\'ementaire $U(1)$.

Les R-sym\'etries discr\`etes de jauge proviennent de R-sym\'etries
jaug\'ees, or les R-sym\'etries continues posent de s\'erieux
probl\`emes li\'es aux masses du gravitino et des gluinos
comme nous l'avons vu dans la Section \ref{pbm}. De plus,
les R-sym\'etries ne peuvent \^etre jaug\'ees que dans un contexte
de \susi locale \cite{chamdrein2}. En effet, d'apr\`es Eq.(\ref{Rsym1a},
\ref{Rsym1b},\ref{Rsym1c}) les transformations associ\'ees
\`a une R-sym\'etrie de jauge s'\'ecrivent,
\begin{eqnarray}
V_k(x,\theta,\bar \theta) & \to &
V_k(x,\theta e^{-i \alpha(x)},\bar \theta e^{i \alpha(x)}),
\label{Rsymjau1}
\end{eqnarray}
\begin{eqnarray}
\Phi_l(x,\theta) & \to &
e^{i R_l \alpha(x)} \ \Phi_l(x,\theta e^{-i \alpha(x)}),
\label{Rsymjau2}
\end{eqnarray}
\begin{eqnarray}
\bar \Phi_l(x,\bar \theta) & \to &
e^{-i R_l \alpha(x)} \
\bar \Phi_l(x,\bar \theta e^{i \alpha(x)}),
\label{Rsymjau3}
\end{eqnarray}
et sont des formes sp\'eciales de transformations locales
dans le superespace qui peuvent \^etre vues comme des transformations
locales sur les champs. Le fait que les R-sym\'etries de jauge
n'existent que dans les th\'eories de \susi locale peut aussi
\^etre vu de la mani\`ere suivante. Il est clair que les transformations
de R-sym\'etries (voir Eq.(\ref{Rsym1a}), Eq.(\ref{Rsym1b}) et
Eq.(\ref{Rsym1c}))
ne commutent pas avec les transformations de supersym\'etrie.
La relation d'anti-commutation entre les g\'en\'erateurs $R$ de
R-sym\'etries
et les g\'en\'erateurs $Q_\alpha$ de la \susi s'\'ecrit
$[Q_\alpha,R] = i(\gamma_5)_\alpha^\beta Q_\beta$ \cite{west}. Or,
cette relation d'anti-commutation reste vraie quand le g\'en\'erateur $R$
d\'epend des coordonn\'ees d'espace-temps, seulement si
les g\'en\'erateurs $Q_\alpha$ deviennent eux aussi les g\'en\'erateurs
d'une supersym\'etrie locale.
Malgr\'e les difficult\'es engendr\'ees par les R-sym\'etries de jauge
et la n\'ecessit\'e de consid\'erer de telles sym\'etries en \susi
locale, certains mod\`eles bas\'es sur des R-sym\'etries jaug\'ees ont
\'et\'e construits \cite{chamdrein2,Nardi,Cast}.
De tels mod\`eles peuvent pr\'eserver la stabilit\'e du proton soit en
engendrant des R-sym\'etries discr\`etes de jauge soit en g\'en\'erant
des couplages en $\l$, $\l'$, $\l''$ supprim\'es par un facteur $M/M_P$,
$M$ \'etant une \'echelle d'\'energie.
Certains des mod\`eles avec R-sym\'etrie de jauge construits
\cite{chamdrein2} peuvent prot\'eger le proton de sa d\'esint\'egration
tout en autorisant certains couplages \rpv.

\subsection{Point de vue des th\'eories de grande unification (GUT)}
\label{GUT}

Nous consid\'erons dans un premier temps le cas du
mod\`ele de grande unification bas\'e sur le groupe de jauge $SU(5)$
\cite{GUT1} qui est le groupe le plus petit pouvant contenir le groupe
de jauge du Mod\`ele Standard sans rajouter de nouveaux fermions.
Puis nous discuterons les th\'eories de grande
unification en g\'en\'eral.

Rappelons que dans le mod\`ele de grande unification $SU(5)$,
les superchamps $L$ et $D^c$ sont contenus dans une repr\'esentation
$\bf \bar 5$ de $SU(5)$, alors que $Q$, $U^c$ et $E^c$ appartiennent
\`a une repr\'esentation $\bf 10$ de $SU(5)$:
\begin{equation}
{\bf \bar 5} =
\left ( \begin{array}{c}
D^c  \\ i \sigma_2 L
\end{array} \right ), \ \ \
{\bf 10} =
\left ( \begin{array}{cc}
U^c & -Q \\ Q & - E^c i \sigma_2
\end{array} \right ),
 \label{SU5rep}
\end{equation}
o\`u $\sigma_2$ est une des matrices de Pauli.
Quant aux superchamps de Higgs $H_1$ et $H_2$, ils sont dans des
repr\'esentations ${\bf \bar 5}_H$ et ${\bf 5}_H$, respectivement.
Ces deux repr\'esentations contiennent aussi des superchamps de Higgs
additionnels triplets de $SU(3)_c$.
La th\'eorie GUT bas\'ee sur le groupe de jauge $SU(5)$ contient
le superchamp $\Sigma$ appartenant \`a la repr\'esentation
$\bf 24$ (adjointe de $SU(5)$) qui brise spontan\'ement $SU(5)$,
donnant ainsi aux bosons de jauge de $SU(5)$
ainsi qu'aux superchamps de Higgs triplets de $SU(3)_c$ des masses de
l'ordre de
$<~\Sigma~> \sim M_{GUT}$, $M_{GUT}$ \'etant l'\'echelle
de la th\'eorie GUT.
Dans le mod\`ele $SU(5)$, les interactions de Yukawa s'\'ecrivent,
\begin{equation}
W= h_{ij} {\bf 10}^i {\bf 10}^j {\bf 5}_H
+ \bar h_{ij} {\bf 10}^i {\bf \bar 5}^j {\bf \bar 5}_H,
\label{SU5yuk}
\end{equation}
o\`u les $h_{ij}$ et $\bar h_{ij}$ sont les constantes de couplages
de Yukawa. Quant aux interactions \rpv, elles d\'ecoulent
toutes d'un seul et m\^eme op\'erateur:
\begin{equation}
{\bf 10}^i {\bf \bar 5}^j {\bf \bar 5}^k
\   \mapsto \  \l_{ijk}L_jL_k E^c_i, \l'_{ijk}L_k Q_i D^c_j,
\l''_{ijk}U_i^cD_j^cD_k^c.
\label{SU5rpva}
\end{equation}
Par cons\'equent il ne semble pas \`a priori possible
d'interdire uniquement certains termes \rpv, comme par exemple les couplages
en $\l'$, en imposant une sym\'etrie. La seule sym\'etrie prot\'egeant
le proton para\^{\i}t donc \^etre une parit\'e de mati\`ere, dans le
mod\`ele
$SU(5)$. En r\'ealit\'e, cette premi\`ere conclusion est fausse car
certains couplages \rpv sont engendr\'es par des op\'erateurs distincts
non renormalisables de dimension 5 et 6
qui deviennent renormalisables apr\`es brisure de $SU(5)$:
\begin{eqnarray}
\label{SU5rpvb}
({\bf 10}^i \Sigma )_{\bf 10} \ \ ({\bf \bar 5}^j {\bf \bar 5}^k)_{\bf \bar
{10}}
&   \mapsto &  \l_{ijk}L_jL_k E^c_i, \l'_{ijk}L_k Q_i D^c_j,
\l''_{ijk}U_i^cD_j^cD_k^c \\
\label{SU5rpvc}
({\bf 10}^i \Sigma )_{\bf 15} \ \ ({\bf \bar 5}^j {\bf \bar 5}^k)_{\bf \bar
{15}}
&   \mapsto &   \l'_{ijk}L_k Q_i D^c_j \\
\label{SU5rpvd}
({\bf 10}^i \Sigma )_{\bf 15} \ \
( [{\bf \bar 5}^j {\bf \bar 5}^k]_{\bf \bar {10}} \Sigma )_{\bf \bar {15}}
&  \mapsto &   \l'_{ijk}L_k Q_i D^c_j.
\end{eqnarray}
Dans l'op\'erateur de Eq.(\ref{SU5rpvc}) c'est la repr\'esentation
sym\'etrique
($\bf \bar {15}$) du produit ${\bf \bar 5}^j {\bf \bar 5}^k$ qui est
s\'electionn\'ee.
Par cons\'equent cet op\'erateur engendre des couplages $\Lambda_{ijk}$
sym\'etriques dans les indices de saveur $j$ et $k$, puisque l'on a,
\begin{equation}
\Lambda_{ijk} ({\bf \bar 5}_a^j {\bf \bar 5}_b^k)_{\bf \bar {15}}=
\Lambda_{ijk} ({\bf \bar 5}_b^j {\bf \bar 5}_a^k)_{\bf \bar {15}}=
\Lambda_{ikj} ({\bf \bar 5}_b^k {\bf \bar 5}_a^j)_{\bf \bar {15}}=
\Lambda_{ikj} ({\bf \bar 5}_a^j {\bf \bar 5}_b^k)_{\bf \bar {15}},
\label{SU5dem1}
\end{equation}
o\`u $a,b=1,...,5$ sont des indices du groupe $SU(5)$. C'est pour cette
raison que l'op\'erateur de Eq.(\ref{SU5rpvc})
n'engendre pas les couplages $\l_{ijk}$
et $\l''_{ijk}$ qui sont antisym\'etriques dans les indices de saveur $j$
et $k$ (voir Eq.(\ref{wrpvAS}) avec la notation des couplages \rpv de
Eq.(\ref{SU5rpva})). L'op\'erateur de Eq.(\ref{SU5rpvb}) engendre les
couplages
$\l_{ijk}$,$\l'_{ijk}$ et $\l''_{ijk}$ car c'est la repr\'esentation
antisym\'etrique
($\bf \bar {10}$) du produit ${\bf \bar 5}^j {\bf \bar 5}^k$ qui est
s\'electionn\'ee.
L'op\'erateur de Eq.(\ref{SU5rpvd}) ne g\'en\`ere pas $\l_{ijk}$
et $\l''_{ijk}$ car $({\bf 10}^i \Sigma )_{\bf 15}$ ne se proj\`ete que sur
le superchamp $Q$ de la repr\'esenta\-tion $\bf 10$ de Eq.(\ref{SU5rep})
apr\`es brisure de $SU(5)$.
Des op\'erateurs comprenant des puissances $n>2$ de $<~\Sigma~>^n$
se r\'eduisent \`a des combinaisons des op\'erateurs de Eq.(\ref{SU5rpvb}),
Eq.(\ref{SU5rpvc}) et Eq.(\ref{SU5rpvd})
car $<~\Sigma~>^n$ est une combinaison lin\'eaire de $<~\Sigma~>$
et de l'identit\'e pour tout $n$.
Il est donc possible d'imposer une sym\'etrie
supprimant les op\'erateurs de Eq.(\ref{SU5rpva}), Eq.(\ref{SU5rpvb}) et
Eq.(\ref{SU5rpvc})
et autorisant l'op\'erateur de Eq.(\ref{SU5rpvd}), prot\'egeant ainsi le
proton
en permettant uniquement l'existence de couplages $\l'_{ijk}$.
Ce sc\'enario est aussi possible si seul l'op\'erateur de Eq.(\ref{SU5rpvc})
existe. Illustrons cette alternative par un ``toy model''.
Consid\'erons un mod\`ele contenant, en plus des superchamps du mod\`ele
$SU(5)$,
deux superchamps $S$ et $\bar S$ appartenant respectivement \`a des
repr\'esentations
$\bf 15$ et $\bf \bar {15}$ de $SU(5)$ ainsi qu'un superchamp $\Phi$ singlet
de jauge.
La pr\'esence simultan\'ee du superchamp $S$ et de son
conjugu\'e de charge $\bar S$ pr\'eserve l'annulation des anomalies de
$SU(5)$.
Imposons la sym\'etrie discr\`ete $Z_N$ caract\'eris\'ee par les charges,
\begin{equation}
q({\bf 10})=-1, \ q({\bf \bar 5})=3, \ q({\bf 5}_H)=2, \ q({\bf \bar
5}_H)=-2,
\ q(\Sigma)=0, \ q(S)=-6, \ q(\bar S)=1, \ q(\Phi)=5.
\label{SU5ZN}
\end{equation}
Cette sym\'etrie autorise les couplages de Yukawa de Eq.(\ref{SU5yuk}), les
couplages
responsables de la brisure de $SU(5)$ et $SU(3)_c \times SU(2)_L \times
U(1)_Y$
et enfin les termes suivants,
\begin{equation}
W=\bar S ({\bf 10}^i \Sigma)_{\bf 15} +
S ({\bf \bar 5}^j {\bf \bar 5}^k)_{\bf \bar {15}}
+ \bar S S \Phi.
\label{SU5dem2}
\end{equation}
Si $\Phi$ acquiert une vev $<~\Phi~> \sim M_X$
\`a l'\'echelle $M_X>M_{GUT}$, la th\'eorie effective
\`a $M_{GUT}$, obtenue en int\'egrant $S$ et $\bar S$ dans
Eq.(\ref{SU5dem2}),
contient l'op\'erateur de Eq.(\ref{SU5rpvc}). Dans un tel mod\`ele,
les couplages $\l'$ g\'en\'er\'es sont en $\l' \sim M_{GUT}/M_X$.

Il est aussi possible dans le mod\`ele $SU(5)$ de prot\'eger le proton
par une parit\'e leptonique ou baryonique \cite{GUT1}.
En effet, supposons une sym\'etrie
interdisant les op\'erateurs de Eq.(\ref{SU5rpva}) et de
Eq.(\ref{SU5rpvb}), Eq.(\ref{SU5rpvc}) et Eq.(\ref{SU5rpvd}) mais autorisant
les termes
bilin\'eaires de Eq.(\ref{wrpv}) (repr\'esentant des couplages \rpv)
qui s'\'ecrivent dans le mod\`ele $SU(5)$,
\begin{equation}
W=\mu_k {\bf \bar 5}^k {\bf 5}_H.
\label{SU5dem3}
\end{equation}
Ce terme couple le superchamp de Higgs triplet de $SU(3)_c$ contenu dans la
repr\'esentation ${\bf 5}_H$ avec le superchamp $D^c$ de la
repr\'esentation $\bf \bar 5$ de Eq.(\ref{SU5rep}). Or, ce superchamp de
Higgs
triplet de $SU(3)_c$ acqui\`ere une masse de l'ordre de $<~\Sigma~> \sim
M_{GUT}$.
Donc, en int\'egrant ce superchamp dans les superpotentiels \ref{SU5yuk} et
\ref{SU5dem3}, on obtient dans la th\'eorie effective \`a une
\'echelle juste inf\'erieure \`a $M_{GUT}$ uniquement des couplages $\l''$
et
qui sont de l'ordre,
\begin{equation}
\l''_{ijk} \sim \bar h_{ij} { \mu_k \over M_{GUT}}.
\label{SU5dem4}
\end{equation}
De m\^eme, l'int\'egration du superchamp de Higgs du MSSM dans les
superpotentiels
\ref{SU5yuk} et \ref{SU5dem3} permet de ne g\'en\'erer que des couplages
$\l_{ijk}$ et $\l'_{ijk}$, qui seraient alors
de l'ordre de $\bar h_{ij} \mu_k / \mu$,
$\mu$ \'etant le param\`etre de Eq.(\ref{muterm}).

De mani\`ere g\'en\'eral, dans les th\'eories GUT, les superchamps de quarks
et de leptons sont compris dans les m\^emes supermultiplets et
ont par cons\'equent les m\^emes charges associ\'ees aux sym\'etries
discr\`etes.
Il ne semble donc pas possible \`a priori d'imposer dans les th\'eories GUT
une sym\'etrie n'interdisant que certains couplages parmi $\l$,$\l'$ et
$\l''$,
qui distingue typiquement les quarks des leptons, et le seul type de
sym\'etrie
para\^{\i}ssant pouvoir prot\'eger le proton est une parit\'e de mati\`ere.
Cependant, plusieurs mod\`eles GUT prot\'egeant le proton par des
sym\'etries
diff\'erentes de la R-parit\'e ont \'et\'e \'elabor\'es
dans la litt\'erature \cite{Haki,GUT1,GUT13a,GUT13b,GUT15}. En g\'en\'eral,
ces mod\`eles engendrent certains des couplages $\l$,$\l'$, $\l''$ \`a
partir d'op\'erateurs non renormalisables qui
deviennent renormalisables apr\`es brisure
du groupe de jauge de grande unification. De tels
mod\`eles GUT ont par exemple \'et\'e construits pour les groupes de jauge
$SU(5)$ \cite{Haki,GUT1}, $SO(10)$ \cite{GUT1}
et $SU(5) \times U(1)$ \cite{GUT1,GUT13a,GUT13b}.
En conclusion, le choix entre une parit\'e de mati\`ere
et une autre sym\'etrie
pour prot\'eger le proton reste de mani\`ere g\'en\'eral tout \`a fait
arbitraire dans les th\'eories GUT.

\subsection{Point de vue des th\'eories de cordes}

Dans les th\'eories de cordes, l'unification des interactions du \ms
peut \^etre r\'ealis\'ee sans un groupe simple, par contraste avec
les mod\`eles de grande unification mentionn\'es dans la Section \ref{GUT}.
Les superchamps de quarks
et de leptons peuvent donc sans difficult\'e appartenir \`a des multiplets
diff\'erents et avoir ainsi des charges distinctes
sous une sym\'etrie discr\`ete, de telle sorte que cette sym\'etrie puisse
prot\'eger le proton de sa d\'esint\'egration en interdisant seulement
certains couplages \rpv parmi $\l$, $\l'$ et $\l''$.
Des th\'eories de cordes avec conservation ou violation de la R-parit\'e
ont toutes deux \'et\'e construites dans la litt\'erature \cite{Bento,Lazar}.
Il ne semble pas par ailleurs y avoir d'arguments
th\'eoriques en faveur d'une parit\'e de mati\`ere ou d'une autre sym\'etrie
emp\^echant la d\'esint\'egration rapide du proton, du point de vue des
th\'eories de cordes \`a l'heure actuelle.

\subsection{Conclusion}

En conclusion, le cadre th\'eorique actuel ne motive pas
davantage un sc\'enario avec violation ou conservation
de la sym\'etrie de R-parit\'e.
Par cons\'equent, les deux types de mod\`eles $R_p$ et \rpv
doivent \^etre consid\'er\'es au m\^eme titre du point de vue
de la ph\'enom\'enologie de la supersym\'etrie aupr\`es
des prochains collisionneurs.

\clearpage

\vspace{10 mm}
\begin{center}
{  }
\end{center}
\vspace{10 mm}

\clearpage

\chapter{Production simple des particules supersym\'etriques 
par les interactions violant la R-parit\'e}
\label{4:chaprod}

\section{Motivations}
\label{4:mot}

D'apr\`es la conclusion du Chapitre \ref{chaRPV}, d'un point de vue
th\'eorique,
les mod\`eles avec violation de la R-parit\'e sont \`a consid\'erer au
m\^eme titre
que les mod\`eles dans lesquels la R-parit\'e est conserv\'ee. Par
cons\'equent,
les mod\`eles avec violation de la R-parit\'e
doivent \^etre \'etudi\'es dans la ph\'enom\'enologie de la supersym\'etrie
aupr\`es
des collisionneurs de particules actuels et futurs. D'un point de vue
ph\'enom\'enologique,  
il appara\^{\i}t aussi tout \`a fait important d'\'etudier les sc\'enarios de
violation de
la R-parit\'e dans la mesure o\`u les signaux de la \susi dans les
collisionneurs sont
fondamentalement diff\'erents dans les sc\'enarios de violation et de
conservation de
la R-parit\'e. En effet, la production de particules supersym\'etriques dans
les collisionneurs
donnerait lieu typiquement \`a des cascades de d\'esint\'egrations se
terminant par
la production de la particule \susiq la plus l\'eg\`ere appel\'ee LSP
(Lightest Supersymmetric
Particle). Or la LSP ne peut se d\'esint\'egrer ni par des interactions de
jauge ni par des
interactions de Yukawa, qui couplent toutes un nombre pair de particules
supersym\'etriques
(voir Chapitre \ref{chaSUSY}).
En revanche, la LSP a la possibilit\'e de se d\'esint\'egrer uniquement en
particules du \ms
par des interactions violant la R-parit\'e (voir Chapitre \ref{chaRPV}).
Les \'etats finals caract\'eristiques de la
production de particules supersym\'etriques dans les collisionneurs sont
donc des \'etats
contenant une grande \'energie manquante dans les sc\'enarios $R_p$ et des
\'etats multileptoniques
ou multijets dans les sc\'enarios \rpv (si le temps de vie de la LSP est
suffisamment court pour
que celle-ci se d\'esint\`egre dans le d\'etecteur).

Dans l'hypoth\`ese o\`u certains couplages \rpv ne seraient pas nuls, il
serait naturel
de tenter de d\'eterminer exp\'erimentalement l'intensit\'e de ces
couplages.
Le premier probl\`eme ph\'enom\'enologique sur lequel nous nous sommes
pench\'es
a donc \'et\'e le calcul de la sensibilit\'e qui peut \^etre obtenue sur
la valeur
absolue des diff\'erentes constantes de couplage \rpi.

Tout d'abord,
la valeur des constantes de couplage \rpv peut \^etre d\'eduite de la
longueur de ``vol''
de la LSP, c'est \`a dire de la distance entre le vertex de production de la
LSP et
le vertex de d\'esint\'egration de la LSP par des interactions \rpi.
Notons que
cette m\'ethode n'est envisageable que si les positions des vertex de
production et de
d\'esint\'egration la LSP sont identifiables, c'est \`a dire si ces vertex
connectent au
moins 2 particules d\'etectables par les calorim\`etres hadroniques et
leptoniques, \`a savoir
des jets ou des leptons charg\'es mais pas des neutrinos. De plus, cette
m\'ethode est
particuli\`erement d\'elicate dans les collisionneurs hadroniques qui
g\'en\`erent de
nombreuses traces de particules dans les d\'etecteurs.
Cherchons maintenant les valeurs maximums de constantes de couplage \rpv
auxquelles seraient
sensibles une telle \'etude de vertex d\'eplac\'e de la LSP. Ces valeurs
maximums d\'ependant
de la nature de la LSP ainsi que
de la g\'eom\'etrie et des performances du d\'etecteur consid\'er\'e. Nous
consid\'ererons
le d\'etecteur d'un futur collisionneur lin\'eaire \cite{4:Tesla,4:NLC} afin
de travailler
dans un contexte de collisions leptoniques et
dans des conditions de performance optimale. Par ailleurs, nous supposerons
que la LSP
est le plus l\'eger des neutralinos (not\'e $\tilde \chi^0_1$), comme c'est
le cas typiquement
dans les mod\`eles de supergravit\'e.
Partons de la longueur de vol du $\tilde \chi^0_1$ dans le r\'ef\'erentiel
du laboratoire
qui est donn\'ee en m\`etres par \cite{4:Dreinoss},
\begin{eqnarray}
 c \gamma \tau_{LSP} \sim 3 \gamma \ 10^{-3} m ({\tilde M \over 100GeV })^4
       ({1GeV \over M_{LSP}})^5 ({1 \over \L})^2,
\label{chaq1}
\end{eqnarray}
o\`u $\L=\l$, $\l'$ ou $\l''$, $c$ est la vitesse de la lumi\`ere, $\gamma$
le facteur du boost de Lorentz, $\tau_{LSP}$ le temps de vie du $\tilde
\chi^0_1$, $M_{LSP}$
la masse du $\tilde \chi^0_1$ et $\tilde M$ la masse de la particule \susiq
\'echang\'ee dans la
d\'esint\'egration du $\tilde \chi^0_1$ impliquant le couplage $\L$
consid\'er\'e:
La d\'esint\'egration du $\tilde \chi^0_1$ par des interactions \rpv est du
type
$\tilde \chi^0_1 \to \tilde f_i^* f_i$, $\tilde f_i^* \to f_j f_k$ o\`u
$\tilde f_i^*$
est une particule scalaire \susiq virtuelle et $f_i$, $f_j$ et $f_k$ sont
des fermions du Mod\`ele Standard.
Aux collisionneurs lin\'eaires, la pr\'ecision sur les param\`etres
d'impact devrait \^etre nettement am\'elior\'ee par rapport aux performances
de LEP II: Alors que la pr\'ecision sur la position de vertex secondaires
est typiquement de $5 \ 10^{-5}m$ au LEP, 
la pr\'ecision esp\'er\'ee aux collisionneurs
lin\'eaires atteind $5 \ 10^{-6}m$. Par cons\'equent, si l'on prend la convention
de demander quatre \'ecarts standards ($4 \sigma$) et si l'on suppose que
l'erreur sur la distance s\'eparant les vertex primaire et secondaire est
principalement contr\^ol\'ee par l'incertitude sur la position du vertex secondaire, alors
la distance minimum entre deux vertex n\'ecessaire pour distinguer exp\'erimentalement
ces deux vertex, et donc la plus petite longueur de vol mesurable, est de l'ordre de 
$2 \ 10^{-5}m$ aux collisionneurs lin\'eaires. 
D'apr\`es Eq.(\ref{chaq1}), les valeurs de constantes de
couplage \rpv pouvant \^etre test\'ees par la m\'ethode d'analyse de vertex 
d\'eplac\'e sont donc typiquement major\'ees par,
\begin{eqnarray}
 \L < 1.2 \ 10^{-4} \gamma^{1/2} ({\tilde M \over 100GeV })^2
       ({100GeV \over M_{LSP}})^{5/2}.
\label{chaq2}
\end{eqnarray}
Or les limites exp\'erimentales indirectes sur les constantes de couplage
\rpv
obtenues \`a basse \'energie sont de l'ordre de $\L <10^{-1}-10^{-2}$
pour des masses des particules supersym\'etriques de $100GeV$
\cite{4:Drein,4:Bhatt,4:GDR}.
La m\'ethode d'analyse de vertex d\'eplac\'e de la LSP ne permet donc pas
de tester tout l'interval des valeurs de constantes de couplage \rpv
possibles.

La seconde m\'ethode permettant de d\'eterminer la valeur des constantes de
couplage \rpv
est l'\'etude de la production de particules du \ms (MS) ou de particules
supersym\'etriques
par des interactions \rpi. En effet, les sections efficaces de telles
productions
sont proportionnelles \`a des puissances des constantes de couplage $\rpv$
et l'on peut donc
d\'eduire de leur valeur l'intensit\'e des couplages \rpv en fonction
d'autres
param\`etres supersym\'etriques.
L'\'etude de la production de particules du MS ou de particules
supersym\'etriques
par des interactions \rpv permet de tester plus facilement de grandes
valeurs des constantes
de couplage
\rpi, celles-ci favorisant les sections efficaces des productions
\'etudi\'ees par rapport
aux sections efficaces des bruits de fond associ\'es.
Les \'etudes de productions de particules du MS via des interactions
\rpi
dans les collisionneurs hadroniques \cite{4:Dim2}-\cite{4:Rizz}
et leptoniques \cite{4:Dim1}-\cite{4:Deba}
ont montr\'ees que les constantes de couplage \rpv pouvaient \^etre
test\'ees jusqu'\`a des valeurs
inf\'erieures aux limites de basse \'energie \cite{4:Drein,4:Bhatt,4:GDR}.
Les productions de particules du MS via des interactions \rpi, qui sont
en fait
des contributions \rpv \`a des r\'eactions du MS, impliquent 2 vertex \rpv
et ont donc des sections efficaces proportionnelles \`a $\L^4$ ($\L=\l$,
$\l'$ ou $\l''$). Les
productions de particules impliquant un seul vertex \rpv ont des sections
efficaces proportionnelles
\`a $\L^2$ qui sont donc moins r\'eduites \'etant
donn\'e les bornes existant sur les couplages \rpv
\cite{4:Drein,4:Bhatt,4:GDR}.
Les productions de particules impliquant un seul vertex \rpv sont des
productions d'une
seule particule supersym\'etrique appel\'ees productions simples de
particule supersym\'etrique.
L'\'etude des productions simples de particule supersym\'etrique permet,
tout comme
l'\'etude des contributions \rpv \`a des r\'eactions du MS, de tester des
valeurs
des constantes de couplage \rpi.
Les \'etudes de contributions \rpv \`a des r\'eactions du MS permettent
d'\'etablir des limites sur certaines constantes de couplage \rpv qui ne sont pas
testables via les productions simples, comme par exemple la
constante de couplage $\l_{133}$ \cite{4:Chou}. 
De plus, les \'etudes de contributions \rpv \`a des r\'eactions du
MS permettent d'\'etudier des produits de diff\'erents couplages \rpv
ce qui est impossible via les productions simples. En revanche, les
productions simples ayant
des sections efficaces proportionnelles \`a $\L^2$ uniquement, leur \'etude
permet
d'obtenir de fortes sensibilit\'es sur certains couplages \rpv
comparativement
aux limites de basse \'energie mais aussi par rapport aux sensibilit\'es
obtenues
via les contributions \rpv \`a des r\'eactions du MS,
comme nous le montrerons dans les Sections \ref{4:had} et \ref{4:lep}.
De plus, l'\'etude des productions simples permet de tester des valeurs des
constantes de couplage
\rpv pouvant atteindre $\sim 10^{-4}$, comme nous allons aussi le montrer
dans les
Sections \ref{4:had} et \ref{4:lep}, mettant ainsi
en \'evidence une grande compl\'ementarit\'e entre les 2 m\'ethodes de
d\'etermination
des valeurs des constantes de couplage \rpv (analyse de vertex d\'eplac\'e
de la LSP et
\'etude de la production de particules supersym\'etriques par des
interactions \rpi).

Du point de vue de la d\'ecouverte de la supersym\'etrie, la production
simple pr\'esente
aussi certains int\'er\^ets. En effet, dans un contexte de violation de la
R-parit\'e,
l'analyse de la production simple de particule supersym\'etrique permet,
tout comme l'analyse
de la production de paire de particules supersym\'etriques, d'\'etudier les
param\`etres SUSY
et de reconstruire les masses des particules SUSY d'une mani\`ere
ind\'ependante du mod\`ele
th\'eorique. Dans un sc\'enario \rpv o\`u les valeurs des constantes de
couplage \rpv
sont faibles, la production simple de particule SUSY a une section efficace
inf\'erieure
\`a celle de la production de paire de particules SUSY qui n'implique pas de
vertex \rpi.
Dans un tel sc\'enario, la production de paire de particules SUSY est la
r\'eaction favoris\'ee
pour l'\'etude des param\`etres SUSY et de la reconstruction des masses des
particules SUSY
aupr\`es des futurs collisionneurs
leptoniques (collisionneurs lin\'eaires) \cite{4:Godbole} et hadroniques,
\`a savoir le Tevatron
(Run II) \cite{4:Barg,4:tat1,4:tat2,4:runII} et le LHC \cite{4:Atlas}.
En revanche, dans un sc\'enario \rpv o\`u les valeurs des constantes de
couplage \rpv
sont proches de leur borne indirecte actuelle \cite{4:Drein,4:Bhatt,4:GDR},
l'analyse de la production simple de particule SUSY permet d'obtenir une
meilleure
sensibilit\'e sur certains param\`etres SUSY que celle obtenue par la
production de paire
de particules SUSY, comme nous allons le voir dans les Sections \ref{4:had}
et \ref{4:lep}.
La raison est que l'espace de phase de la production simple est moins
r\'eduit que celui de la production de paire. De plus, dans un tel
sc\'enario, la production simple
de particule SUSY permet de reconstruire les masses des particules SUSY avec
un bruit de fond
combinatoire plus faible que celui de la production de paire de particules
SUSY,
comme nous le montrerons aussi dans les Sections \ref{4:had} et \ref{4:lep}.
Ceci est d\^u au fait que la production simple de particule SUSY ne g\'en\`ere
qu'une seule cascade
de d\'esint\'egration de particules SUSY alors que la production de paire en
g\'en\`ere deux
ce qui complique l'identification de l'origine des particules de l'\'etat
final.

Nous nous sommes donc int\'eress\'es \`a l'\'etude de la production simple de
particule
supersym\'etrique. Dans les deux sections suivantes, nous pr\'esentons les
r\'esultats
de cette \'etude dans le cadre des collisionneurs hadroniques (Section
\ref{4:had}) et
leptoniques (Section \ref{4:lep}), en se concentrant sur la sensibilit\'e
qui peut \^etre
obtenue par une telle \'etude sur les constantes de couplage \rpi.

\section{Collisionneurs hadroniques}
\label{4:had}

Nous pr\'esentons dans cette section des \'etudes de production simple de
particule
SUSY aux collisionneurs hadroniques. Pr\'ecisons que dans ces \'etudes, pour
des raisons
de simplification, nous avons toujours suppos\'e
qu'une seule constante de couplage \rpv \'etait dominante par rapport aux
autres.
Cette hypoth\`ese peut \^etre justifi\'ee par l'analogie entre les
structures des interactions
de Yukawa et des interactions \rpv et la forte hi\'erarchie existant parmi
les couplages de Yukawa.

La production simple de particule SUSY aux collisionneurs hadroniques
implique
des interactions $\l'$ ou $\l''$. Dans le cas d'une constante de couplage
$\l''$ dominante,
la particule SUSY cr\'e\'ee dans la production simple donnerait lieu \`a une
cascade de
d\'esint\'egration se terminant par la d\'esint\'egration de la LSP en 3
jets via le couplage $\l''$
dominant. La production simple conduirait donc typiquement \`a des \'etats
finals multijets,
or les \'etats finals multijets ont un grand bruit de fond QCD \cite{4:Dim2,4:Bin}. 
Il est donc plus
prometteur dans un premier temps de se concentrer sur les productions
simples de particule
SUSY aux collisionneurs hadroniques impliquant des interactions $\l'$.

\begin{figure}[t]
\begin{center}
\leavevmode
\centerline{\psfig{figure=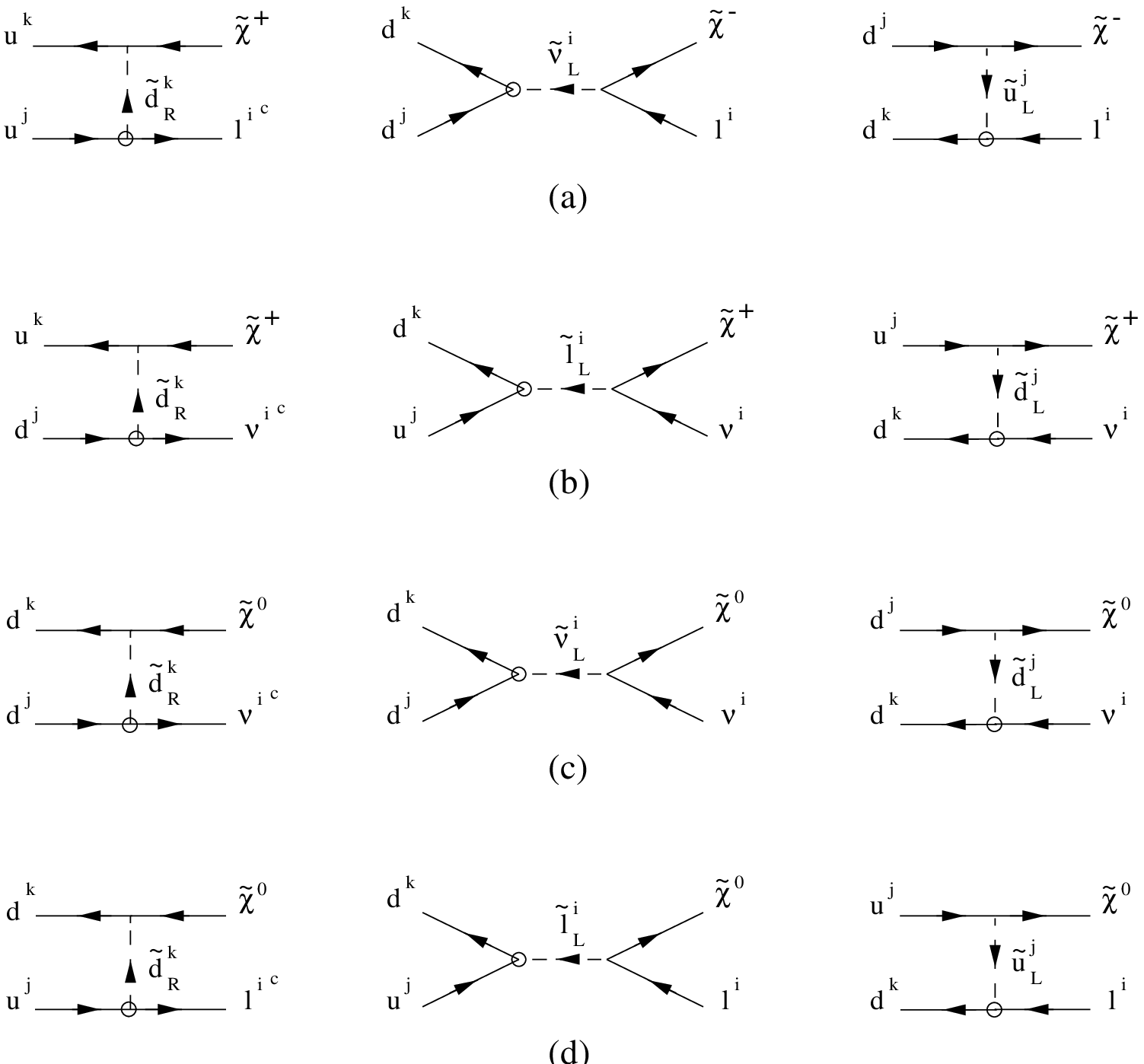,height=5.in}}
\end{center}
\caption{\footnotesize  \it Diagrammes de Feynman des 4 processus de
production simple
de particule SUSY via $\l'_{ijk}$ aux collisionneurs hadroniques qui sont du
type $2 \to 2-corps$
et qui re\c{c}oivent une contribution de la production r\'esonante d'un partenaire
supersym\'etrique. Les couplages $\l'_{ijk}$ sont symbolis\'es par des
cercles
et les fl\`eches repr\'esentent les moments des particules.
\rm \normalsize }
\label{4:graphes}
\end{figure}

Par ailleurs, il est plus int\'eressant de consid\'erer les productions
simples de particule SUSY du type
$2 \to 2-corps$ afin d'optimiser l'espace de phase. Toutes les productions
simples de particule SUSY
aux collisionneurs hadroniques du type $2 \to 2-corps$ et impliquant des
interactions $\l'$
sont pr\'esent\'ees dans la Figure \ref{4:graphes} et dans la liste
suivante,
\begin{itemize}
\item la production de gluino $\bar u_j d_k \to \tilde g l_i$ via
l'\'echange
d'un squark $\tilde u_{jL}$ ($\tilde d_{kR}$) dans la voie $t$ ($u$),
\item la production de squark $\bar d_j g \to \tilde d_{kR}^* \nu_i$ via
l'\'echange
d'un squark $\tilde d_{kR}$ (quark $d_j$) dans la voie $t$ ($s$),
\item la production de $\bar u_j g \to \tilde d_{kR}^* l_i$ via l'\'echange
d'un squark $\tilde d_{kR}$ (quark $u_j$) dans la voie $t$ ($s$),
\item la production de squark $d_k g \to \tilde d_{jL} \nu_i$ via
l'\'echange
d'un squark $\tilde d_{jL}$ (quark $d_k$) dans la voie $t$ ($s$),
\item la production de squark $d_k g \to \tilde u_{jL} l_i$ via l'\'echange
d'un squark $\tilde u_{jL}$ (quark $d_k$) dans la voie $t$ ($s$),
\item la production de sneutrino $\bar d_j d_k \to Z \tilde \nu_{iL}$ via
l'\'echange
d'un quark $d_k$ ou $d_j$ (sneutrino $\tilde \nu_{iL}$) dans la voie $t$
($s$),
\item la production de slepton charg\'e $\bar u_j d_k \to Z \tilde l_{iL}$
via l'\'echange
d'un quark $d_k$ ou $u_j$ (slepton $\tilde l_{iL}$) dans la voie $t$ ($s$),
\item la production de sneutrino $\bar u_j d_k \to W^- \tilde \nu_{iL}$ via
l'\'echange
d'un quark $d_j$ (slepton $\tilde l_{iL}$) dans la voie $t$ ($s$),
\item la production de slepton charg\'e $\bar d_j d_k \to W^+ \tilde l_{iL}$
via l'\'echange
d'un quark $u_j$ (sneutrino $\tilde \nu_{iL}$) dans la voie $t$ ($s$).
\end{itemize}
Les productions simples repr\'esent\'ees dans la Figure 
re\c{c}oivent une contribution 
de la production r\'esonante d'une particule supersym\'etrique, \`a l'inverse
des productions simples list\'ees ci-dessus. Effectivement, les seules productions
simples parmi la liste ci-dessus pouvant \'eventuellement re\c{c}evoir une contribution 
de production r\'esonante de particule SUSY sont les r\'eactions 
$\bar u_j d_k \to \tilde l_{iL} \to W^- \tilde \nu_{iL}$ et 
$\bar d_j d_k \to \tilde \nu_{iL} \to W^+ \tilde l_{iL}$. Or, dans la plupart des 
mod\`eles SUSY, comme par exemple les mod\`eles de supergravit\'e ou les mod\`eles
GMSB, la diff\'erence de masse entre le slepton charg\'e Left et le sneutrino Left 
est d\^ue aux termes D de telle sorte qu'elle est fix\'ee par la relation
$m^2_{\tilde l^{\pm}_L}-m^2_{\tilde \nu_L}=\cos 2 \beta M_W^2$ \cite{1:Iban}
et n'\'exc\`ede donc pas la masse du boson $W^{\pm}$.
Notons cependant que dans les sc\'enarios \`a grand $\tan \beta$, 
les masses Left-Right $(m_{LR})_{ij}$ peuvent atteindre des valeurs non
n\'egligeables pour les particules appartenant \`a la troisi\`eme famille 
(voir Eq.(\ref{MSSMC2})), g\'en\'erant ainsi un m\'elange entre les particules
scalaires Left et Right de la troisi\`eme famille. 
Les particules scalaires de troisi\`eme g\'en\'eration peuvent 
donc avoir un \'etat propre de masse inf\'erieure aux masses des particules scalaires 
Left et Right de premi\`ere et seconde g\'en\'eration et l'autre \'etat propre de masse 
sup\'erieure aux masses des particules scalaires Left et Right de premi\`ere et seconde 
g\'en\'eration. Dans un sc\'enario \`a grand $\tan \beta$, on peut par exemple avoir
la hi\'erarchie suivante parmi les sleptons charg\'es: $m_{\tilde \tau^{\pm}_2} 
> m_{\tilde e^{\pm}_L}, m_{\tilde e^{\pm}_R}, m_{\tilde \mu^{\pm}_L}, 
m_{\tilde \mu^{\pm}_R}> m_{\tilde \tau^{\pm}_1}$, $\tilde \tau^{\pm}_1$ et
$\tilde \tau^{\pm}_2$ \'etant les \'etats propres de masse de stau.
En particulier, dans un sc\'enario \`a grand $\tan \beta$, la diff\'erence de masse
entre le sneutrino $\tilde \nu_{\tau}$ et l'\'etat propre de masse 
$\tilde \tau^{\pm}_1$ peut \^etre sup\'erieure \`a la masse du $W^{\pm}$, 
de telle sorte que la production simple
de stau $\bar d_j d_k \to \tilde \nu_{\tau} \to W^{\pm} \tilde \tau^{\mp}_1$   
re\c{c}oive la contribution de la production r\'esonante de sneutrino.

Discutons maintenant les sections efficaces des productions
simples de particule SUSY aux collisionneurs hadroniques du type 
$2 \to 2-corps$ et impliquant des interactions $\l'$.
Les sections efficaces des productions simples ``non r\'esonantes'' 
list\'ees ci-dessus ne peuvent atteindre de grandes valeurs: 
La section efficace de la production simple de gluino est limit\'ee 
par les bornes exp\'erimentales sur
les masses de squarks et de gluinos qui sont de l'ordre de
$m_{\tilde q},m_{\tilde g} \stackrel{>}{\sim}200GeV$ \cite{3:PDG98}. 
En effet, la production simple de gluino a lieu par l'\'echange de squark
dans les voies $t$ et $u$, comme nous l'avons vu plus haut, et donc
sa section efficace d\'ecroit si les masses de squarks et de gluinos
augmentent. Pour la valeur $m_{\tilde q}=m_{\tilde g}=250GeV$ qui est 
proche de la limite exp\'erimentale, nous trouvons la section efficace 
de production simple de gluino suivante,
$\sigma(p \bar p \to \tilde g \mu) \approx 10^{-2}pb$.
Les sections efficaces donn\'ees dans ce paragraphe ont \'et\'e calcul\'ees
gr\^ace \`a la version 33.18 du programme COMPHEP \cite{4:COMPHEP}
pour une \'energie dans le centre de masse de $\sqrt s=2TeV$ (Run II du Tevatron)
avec la fonction de structure CTEQ4m \cite{4:CTEQ4}
et pour une valeur de la constante de couplage \rpv de $\l'_{211}=0.09$.
De m\^eme, le taux de production simple de squark ne peut \^etre grand.
Pour $m_{\tilde q}=250GeV$, la \sef $\sigma(p \bar p \to \tilde u_L \mu)$
est de l'ordre de $\sim 10^{-3}pb$. De plus,
la production d'un slepton accompagn\'e d'un boson de jauge massif a un petit
espace de phase et n'implique pas d'interactions fortes. Le taux de ce type
de production est donc faible. Pour $m_{\tilde l}=100GeV$, nous obtenons
$\sigma(p \bar p \to Z \tilde \mu_L) \approx 10^{-2}pb$.
Par ailleurs, nous avons calcul\'e toutes les amplitudes
des productions simples ``r\'esonantes''
pr\'esent\'ees dans la Figure \ref{4:graphes}. Dans \cite{PubliB}
(voir Publication III: {\it ``Single superpartner production at 
Tevatron Run II''}),
nous donnons les r\'esultats analytiques du calcul de ces amplitudes et nous 
pr\'esentons une analyse num\'erique de l'\'evolution des valeurs des 
sections efficaces correspondantes dans l'espace des param\`etres 
supersym\'etriques. D'apr\`es cette \'etude,
les sections efficaces des productions simples r\'esonantes atteignent des 
valeurs de l'ordre de la dizaine de picobarns.

Bien que les productions simples non r\'esonantes soient int\'eressantes du fait 
que certaines d'entre elles d\'ependent de peu de param\`etres SUSY, 
\`a savoir une constante de couplage \rpv et 
une masse de particule scalaire supersym\'etrique, 
l'\'etude des productions simples r\'esonantes est plus attrayante car les sections 
efficaces de ces derni\`eres sont plus importantes.

\subsection{Signature trilepton}

La production d'un chargino et d'un lepton charg\'e aux
collisionneurs hadroniques (voir Figure \ref{4:graphes}(a)) 
donne lieu \`a un \'etat final contenant 3 leptons charg\'es si la
cascade de d\'esint\'egration initi\'ee par le chargino produit est 
$\tilde \chi^{\pm} \to \tilde \chi^0_1 l^{\pm} \nu$,
$\tilde \chi^0_1 \to l^{\pm}_i u_j d_k$, les indices $i,j,k$ correspondant 
aux indices de la constante $\l'_{ijk}$.
L'\'etat final \`a 3 leptons charg\'es est particuli\`erement 
int\'eressant 
comme candidat de signal de la \susi car le bruit de fond 
provenant du \ms est faible. Dans \cite{PubliA} 
(voir Publication II: {\it ``Resonant sneutrino production at 
Tevatron Run II''}), \cite{PubliB}, \cite{PubliC} et 
\cite{PubliD} (voir Publication
IV: {\it ``The three-leptons signature from resonant sneutrino 
production at the LHC''}), 
nous avons \'etudi\'e le signal \`a 3 leptons charg\'es 
engendr\'e par la production simple de chargino 
(voir Figure \ref{4:graphes}(a))
et le bruit de fond associ\'e. 
Le processus de production simple de chargino a \'et\'e 
impl\'ement\'e dans une version du g\'en\'erateur 
d'\'ev\`enements SUSYGEN \cite{4:SUSYGEN} 
incluant la simulation des collisions hadroniques. Ceci a permis de
g\'en\'erer le signal avec SUSYGEN et les bruits de fond provenant du \ms
et des r\'eactions \susiqs avec les
g\'en\'erateurs d'\'ev\`enements PYTHIA \cite{4:PYTHIA} (ainsi
que ONETOP \cite{4:onetop} pour certains processus) et
SHERWIG \cite{4:HERWIG}, respectivement. SUSYGEN, PYTHIA et SHERWIG
ont \'et\'e interfac\'es avec le simulateur de d\'etecteur 
SHW \cite{4:SHW} pour l'\'etude dans le cadre de la physique au 
Run II du Tevatron et avec le simulateur de d\'etecteur 
ATLFAST \cite{4:ATLFAST} pour l'\'etude dans le cadre 
de la physique au LHC.
Lors de la g\'en\'eration du signal et du bruit de fond,
des coupures bas\'ees sur des distributions de variables 
cin\'ematiques (angles d\'emission des particules, 
quadri-impulsions,...) ont \'et\'e appliqu\'ees afin
d'augmenter le signal par rapport au bruit de fond.

\subsubsection{Potentiel de d\'ecouverte}

\begin{figure}[t]
\begin{center}
\leavevmode
\centerline{\psfig{figure=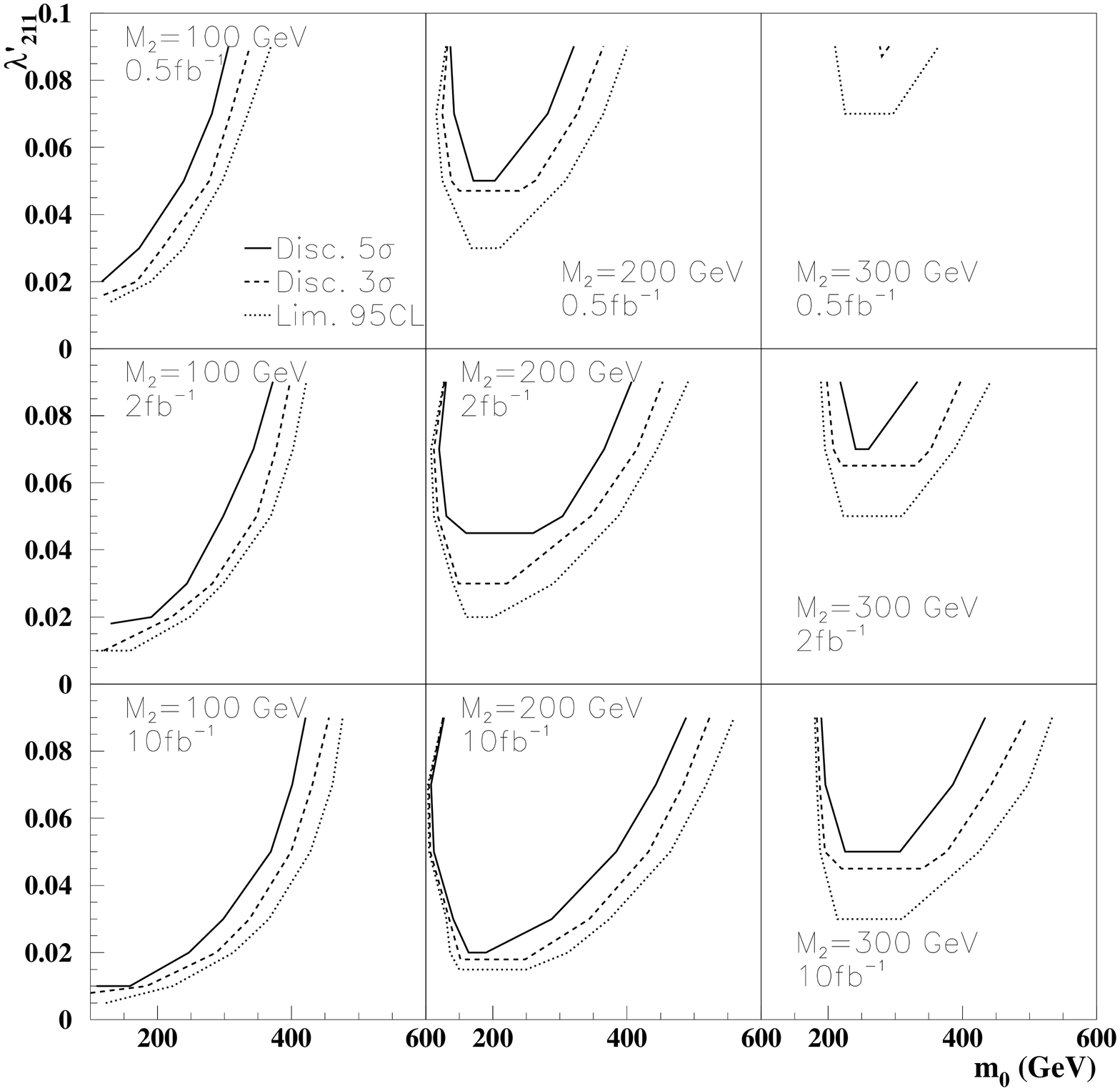,height=5.5in}}
\end{center}
\caption{Contours de d\'ecouverte \`a $5 \sigma$ (ligne),
$3 \sigma$ (tir\'es)
et limites \`a $95 \% \ C.L.$ (pointill\'es) 
associ\'es \`a l'\'etude du canal trilepton au
Run II du Tevatron ($\sqrt s=2 TeV$) et pr\'esent\'es
dans le plan $\l'_{211}$ versus le param\`etre $m_0$
pour $sign(\mu)<0$, $\tan \beta=1.5$
et diff\'erentes valeurs du param\`etre $M_2$ et de
la luminosit\'e.}
\label{fig1t}
\end{figure}

Nous pr\'esentons dans la Figure \ref{fig1t} la sensibilit\'e
dans le plan $\l'_{211}$ versus le param\`etre $m_0$ 
qui pourrait \^etre obtenue \`a partir de 
l'analyse de l'\'etat final \`a 3 leptons charg\'es
(trilepton) au Run II du Tevatron,
dans le cadre d'un mod\`ele de supergravit\'e (SUGRA) 
et pour des valeurs fix\'ees 
des autres param\`etres supersym\'etriques du mod\`ele.
Cette figure a \'et\'e obtenue apr\`es avoir appliqu\'e 
les coupures mentionn\'ees plus haut et en supposant que la 
production simple de chargino avait lieu par l'interm\'ediaire 
d'un couplage du type $\l'_{211}$,
ce qui correspond au cas o\`u le lepton produit avec
le chargino est un muon (voir Figure \ref{4:graphes}(a)). 
Les r\'egions de la Figure \ref{fig1t} se situant au-dessus des
courbes seraient exclues par l'analyse exp\'erimentale
de la signature trilepton. En particulier, le contour
de d\'ecouverte \`a $5 \sigma$ d\'efini la r\'egion 
de l'espace des param\`etres pour lesquels $S/ \sqrt B > 5$
o\`u $S=\sigma_S \times {\cal L} \times {\cal E}_S$ et 
$B=\sigma_B \times {\cal L} \times {\cal E}_B$, 
$\sigma_S$ ($\sigma_B$) \'etant la section efficace du signal
(bruit de fond), ${\cal L}$ la luminosit\'e et 
${\cal E}_S$ (${\cal E}_B$) l'efficacit\'e du signal 
(bruit de fond) apr\`es les coupures.
L'\'evolution de la sensibilit\'e 
avec les param\`etres du mod\`ele SUGRA observ\'ee 
dans la Figure \ref{fig1t} s'explique
par les variations de la \sef du signal
(voir \cite{PubliB}).

D'apr\`es la Figure \ref{fig1t},
la sensibilit\'e obtenue sur $\l'_{211}$ dans certaines
r\'egions de l'espace des param\`etres SUGRA par  
l'\'etude de la signature trilepton au Run II du
Tevatron
permettrait d'am\'eliorer la limite actuelle sur
cette constante de couplage \rpv qui vaut: 
$\l'_{211}<0.09 (m_{\tilde d_R}/100GeV)$ \`a $1 \sigma$ 
($\pi$ decay) \cite{4:Bhatt}. L'\'etude de la signature trilepton
permet aussi d'am\'eliorer les limites indirectes de
nombreuses autres constantes de couplage \rpv si l'on suppose 
que la production simple de chargino 
a lieu de fa\c{c}on dominante par l'interm\'ediaire 
d'une autre constante de couplage 
\mbox{$\not \hspace{-0.10cm} R_p$}. Dans la Table
\ref{tab1}, nous donnons les pr\'evisions des
sensibilit\'es sur toutes les
constantes de couplage \rpv de type $\l'_{ijk}$
obtenues au Run II du Tevatron
par l'\'etude de la signature trilepton  
pour un point de l'espace des param\`etres SUGRA.   
\begin{table}
\begin{center}
\begin{tabular}{|c|c|c|c|c|c|c|c|c|}
\hline
$\l'_{111}$ & 
$\l'_{112}$ & $\l'_{113}$ & $\l'_{121}$ & $\l'_{122}$
& $\l'_{123}$ & $\l'_{131}$ & $\l'_{132}$ & $\l'_{133}$  \\
\hline
0.02 & 
0.04 & 0.07 & 0.05 & 0.12 & 0.21 & 0.10 & 0.36 & 0.63 \\
\hline
$\l'_{211}$ & 
$\l'_{212}$ & $\l'_{213}$ & $\l'_{221}$ & $\l'_{222}$
& $\l'_{223}$ & $\l'_{231}$ & $\l'_{232}$ & $\l'_{233}$  \\
\hline
0.02 & 
0.04 & 0.07 & 0.05 & 0.12 & 0.21 & 0.10 & 0.36 & 0.63 \\
\hline
$\l'_{311}$ & 
$\l'_{312}$ & $\l'_{313}$ & $\l'_{321}$ & $\l'_{322}$
& $\l'_{323}$ & $\l'_{331}$ & $\l'_{332}$ & $\l'_{333}$  \\
\hline
0.06 & 
0.13 & 0.23 & 0.18 & 0.41 & 0.70 & 0.33 & 1.17 & 2.05 \\
\hline
\end{tabular}
\caption{\em
Sensibilit\'es \`a $95 \% CL$ sur toutes les constantes 
de couplage de type $\l'_{ijk}$ dans le cadre de la physique
au Run II du Tevatron
pour une luminosit\'e de 
${\cal L}=2fb^{-1}$ et pour les param\`etres SUGRA suivants,
$m_0=180GeV$, $M_2=200GeV$, $\tan \beta=1.5$ et $\mu=-200GeV$
($m_{\tilde u_L}=601GeV$, $m_{\tilde d_L}=603GeV$,
$m_{\tilde u_R}=582GeV$, $m_{\tilde d_R}=580GeV$,
$m_{\tilde l_L}=253GeV$, $m_{\tilde l_R}=205GeV$
$m_{\tilde \nu_L}=248GeV$,
$m_{\tilde \chi^{\pm}_1}=199GeV$, $m_{\tilde \chi^0_1}=105GeV$).}
\label{tab1}
\end{center}
\end{table}
Les sensibilit\'es sur les constantes de couplage $\l'_{2jk}$ et
$\l'_{3jk}$ pr\'esent\'ees dans la Table \ref{tab1}
repr\'esentent toutes une am\'elioration par rapport aux limites 
de basse \'energie sur ces m\^emes constantes de couplage, 
que nous rappelons ici \cite{4:Bhatt}:
$\l'_{21k}<0.09 (m_{\tilde d_{kR}}/100GeV)$ \`a $1 \sigma$ 
(d\'esint\'egration du $\pi$),
$\l'_{22k}<0.18 (m_{\tilde d_{kR}}/100GeV)$ \`a $1 \sigma$ 
(d\'esint\'egration du $D$),
$\l'_{231}<0.22 (m_{\tilde b_L}/100GeV)$ \`a $2 \sigma$
(diffusion profond\'ement in\'elastique du $\nu_{\mu}$),
$\l'_{232}<0.36 (m_{\tilde q}/100GeV)$ \`a $1 \sigma$ ($R_{\mu}$),
$\l'_{233}<0.36 (m_{\tilde q}/100GeV)$ \`a $1 \sigma$ ($R_{\mu}$),
$\l'_{31k}<0.10 (m_{\tilde d_{kR}}/100GeV)$
\`a $1 \sigma$ ($\tau^- \to \pi^- \nu_{\tau}$),
$\l'_{32k}<0.20$ (pour $m_{\tilde l}=m_{\tilde q}=100GeV$)
\`a $1 \sigma$ (mixing $D^0-\bar D^0$),
$\l'_{33k}<0.48 (m_{\tilde q}/100GeV)$ \`a $1 \sigma$ 
($R_{\tau}$).
Les bornes indirectes sur les constantes de type $\l'_{1jk}$ 
sont typiquement plus fortes que les bornes sur les
constantes $\l'_{2jk}$ et $\l'_{3jk}$ \cite{4:Bhatt}. 
De ce fait, l'\'etude de la signature trilepton permet de tester
un moins grand nombre de constantes $\l'_{1jk}$ que
de constantes $\l'_{2jk}$ ou $\l'_{3jk}$. 
Par exemple, pour le point SUGRA consid\'er\'e dans la Table 
\ref{tab1}, seules les sensibilit\'es sur les constantes de 
couplage $\l'_{112}$, $\l'_{113}$, $\l'_{121}$, $\l'_{131}$ 
et $\l'_{132}$ repr\'esentent une am\'elioration vis \`a 
vis de leur borne indirecte: 
$\l'_{11k}<0.02 (m_{\tilde d_{kR}}/100GeV)$ \`a $2 \sigma$
(universalit\'e du courant charg\'e),
$\l'_{1j1}<0.035 (m_{\tilde q_{jL}}/100GeV)$ \`a $2 \sigma$
(violation de la parit\'e atomique),
$\l'_{132}<0.34$ \`a $1 \sigma$ pour $m_{\tilde q}=100GeV$
($R_e$).

\begin{figure}[t]
\begin{center}
\leavevmode
\centerline{\psfig{figure=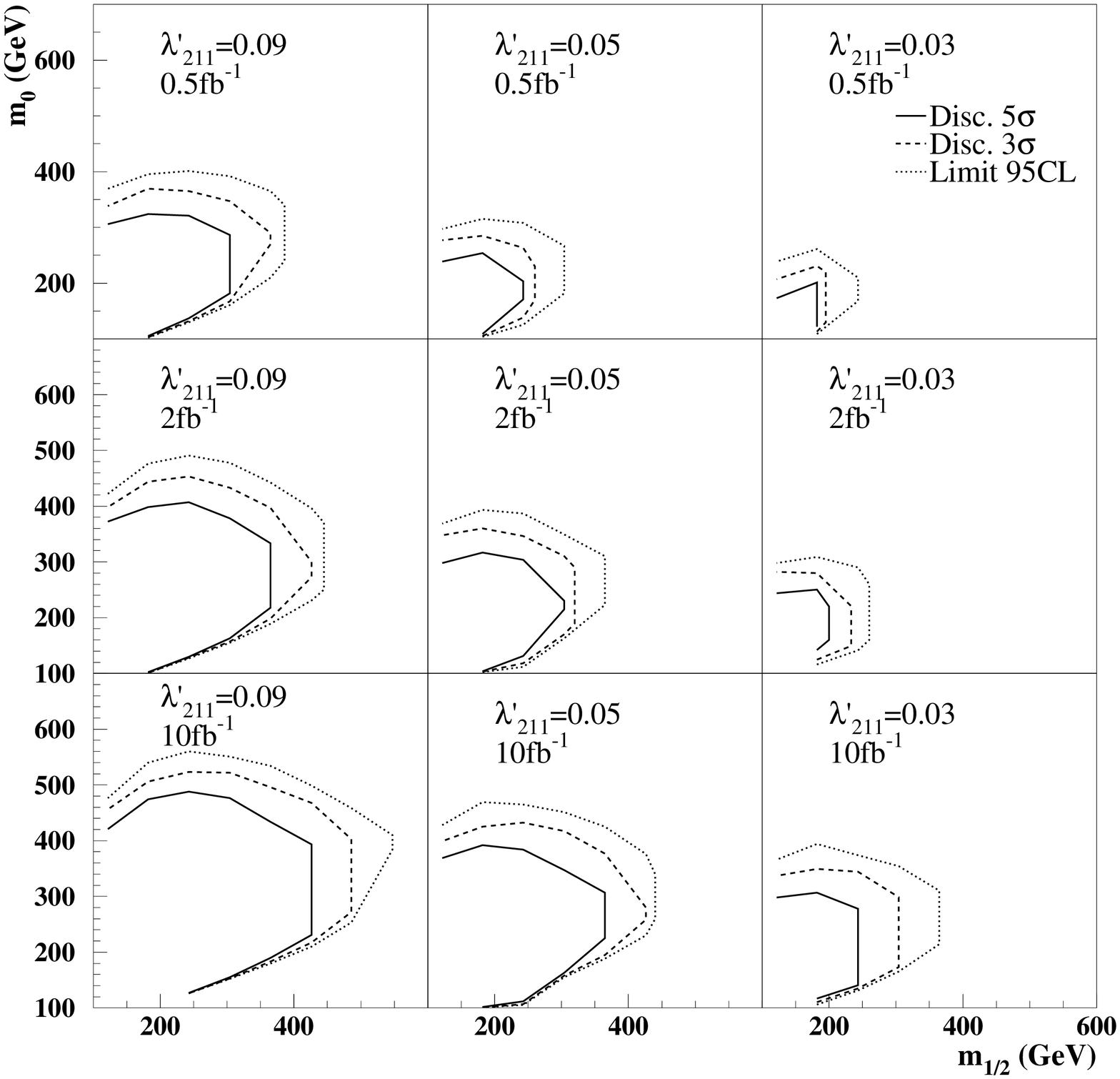,height=5.5in}}
\end{center}
\caption{
Contours de d\'ecouverte \`a $5 \sigma$ (ligne),
$3 \sigma$ (tir\'es)
et limites \`a $95 \% \ C.L.$ (pointill\'es) 
associ\'es \`a l'\'etude du canal trilepton au 
Run II du Tevatron ($\sqrt s=2 TeV$) et pr\'esent\'es
dans le plan $m_0$ versus $m_{1/2}$ 
pour $sign(\mu)<0$, $\tan \beta=1.5$
et diff\'erentes valeurs de $\l'_{211}$ et de
la luminosit\'e.}
\label{fig2t}
\end{figure}

\begin{figure}[t]
\begin{center}
\leavevmode
\centerline{\psfig{figure=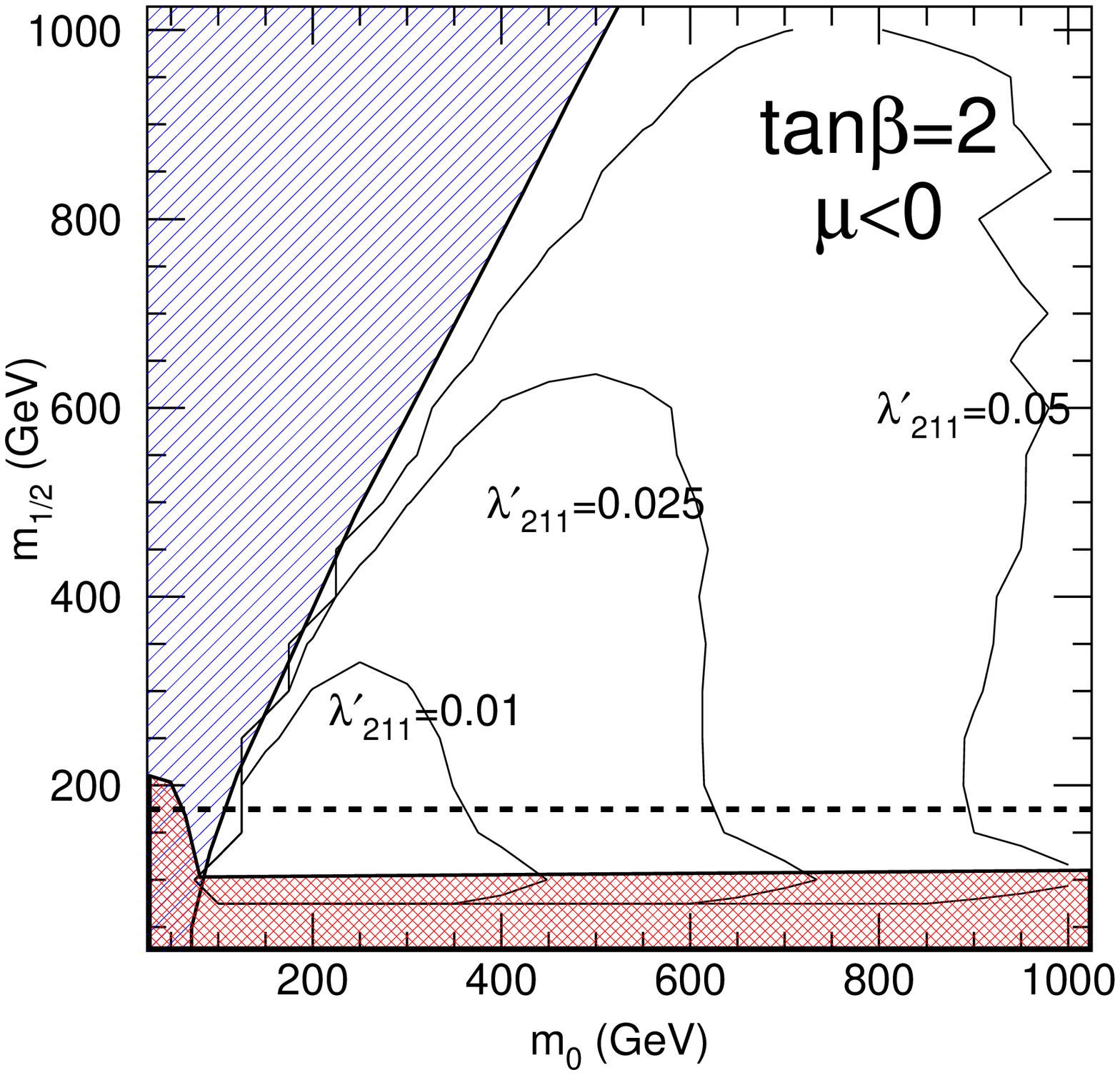,height=5.5in}}
\end{center}
\caption{Limites \`a $95 \% \ C.L.$  
associ\'ees \`a l'\'etude du canal trilepton au 
LHC ($\sqrt s=14 TeV$, ${\cal L}=30 fb^{-1}$) et 
pr\'esent\'ees dans le plan $m_{1/2}$ versus $m_0$ 
pour $sign(\mu)<0$, $\tan \beta=2$
et diff\'erentes valeurs de $\l'_{211}$.
La r\'egion hachur\'ee en haut \`a gauche de la figure
correspond au cas o\`u $m_{\tilde \chi_1^{\pm}}>
m_{\tilde \nu}$. La r\'egion hachur\'ee en bas de la figure
donne la limite cin\'ematique de la production de paires
de $\tilde \chi_1^{\pm}$ et de $\tilde l$ \`a LEP II pour 
une \'energie de $\sqrt s=200GeV$. Dans la r\'egion se situant
en-dessous des pointill\'es, le $\tilde \chi_1^{\pm}$ ne peut
pas se d\'esint\'egrer en un boson $W^{\pm}$ r\'eel.}
\label{fig3t}
\end{figure}

Nous pr\'esentons dans la Figure \ref{fig2t} (\ref{fig3t}) la sensibilit\'e
dans le plan $m_0$ versus $m_{1/2}$ qui pourrait \^etre obtenue \`a partir de 
l'analyse de l'\'etat final trilepton au Run II du Tevatron (LHC)
dans le cadre d'un mod\`ele SUGRA et pour des valeurs fix\'ees 
des autres param\`etres supersym\'etriques du mod\`ele.

En comparant les Figures \ref{fig2t} et \ref{fig3t},
nous observons que le LHC permettra d'obtenir une meilleure 
sensibilit\'e que le Tevatron sur la constante de couplage 
$\l'_{211}$. Cette am\'elioration est d\^ue aux grandes 
\'energies et luminosit\'es qui devraient \^etre atteintes 
par le LHC.

Nous avons montr\'e dans \cite{PubliB} (\cite{PubliD}) 
que l'\'etude du
canal trilepton de la production de paires de particules \susiqs
dans le cadre de la physique au Run II du Tevatron (au LHC)
permettait de tester dans le mod\`ele SUGRA consid\'er\'e 
dans la Figure \ref{fig2t} (\ref{fig3t}) des valeurs du
param\`etre $m_{1/2}$ allant jusqu'\`a $\sim 200 GeV$, et ceci
quelque soit la valeur de la constante de couplage \rpv 
consid\'er\'ee
(pourvu que cette constante soit suffisamment grande pour que
la LSP se d\'esint\`egre dans le d\'etecteur). 
Par cons\'equent, dans la r\'egion $m_{1/2} \stackrel{>}{\sim}
200GeV$ des Figures \ref{fig2t} et \ref{fig3t}, 
la sensibilit\'e obtenue sur les constantes de couplage 
\rpv par la production simple de chargino n'est pas affect\'ee
par la production de paires de superpartenaires. 
En revanche, dans la r\'egion
$m_{1/2} \stackrel{<}{\sim}200GeV$, la production de paires de 
superpartenaires repr\'esente un bruit de fond pour la production
simple de chargino car elle n'implique pas les interactions \rpv
qui sont l'objet de cette \'etude.
Le bruit de fond engendr\'e par la production de paires peut
\^etre r\'eduit par des coupures cin\'ematiques bas\'ees sur
la reconstruction de masse des particules \susiqs (voir Section
\ref{recaa}).

Pla\c{c}ons nous maintenant du point de vue du test des 
param\`etres de brisure douce de la supersym\'etrie (masses
des superpartenaires), et non plus du point de vue du test 
des constantes de couplage \rpvi.
Les limites actuelles sur les masses des particules \susiqs
d\'eduites des donn\'ees exp\'erimentales du collisionneur
LEP II, dans le contexte d'un mod\`ele \rpv contenant des
 couplages
de type $\l'$ non nuls, sont: $m_{\tilde \chi^{0}_1}>26 GeV$ 
(pour $m_0=200 GeV$ et $\tan \beta =\sqrt 2$ \cite{4:neut}), 
$m_{\tilde \chi^{\pm}_1}>100 GeV$, 
$m_{\tilde l}>93 GeV$, $m_{\tilde \nu}> 86 GeV$ \cite{4:Mass}.
Or les valeurs minimums des param\`etres $m_0$ et $m_{1/2}$
de l'espace des param\`etres des Figures \ref{fig1t} et 
\ref{fig2t} correspondent aux valeurs $m_0=100GeV$ et 
$M_2=100GeV$ pour lesquelles le spectre de masse \susiq est
le suivant:
$m_{\tilde \chi^{\pm}_1}= 113GeV$, $m_{\tilde \chi^{0}_1}= 54GeV$,
$m_{\tilde \nu_L}= 127 GeV$, $m_{\tilde l_L}= 137 GeV$,
$m_{\tilde l_R}= 114 GeV$. Puisque ces masses ne sont pas exclues
par les donn\'ees de LEP II \cite{4:neut,4:Mass} et que les 
masses des particules \susiqs augmentent avec $m_0$ et $m_{1/2}$,
l'espace des 
param\`etres des Figures \ref{fig1t} et \ref{fig2t} n'est pas
exclu par les donn\'ees de LEP II \cite{4:neut,4:Mass}.
Par cons\'equent, le potentiel de d\'ecouverte pour le Run II du
Tevatron (Figure \ref{fig2t}) repr\'esente une am\'elioration 
importante par rapport aux limites actuelles sur les 
param\`etres $m_0$ et $m_{1/2}$ du mod\`ele SUGRA consid\'er\'e.

De plus, la Figure \ref{fig2t} montre que la sensibilit\'e
obtenue au Run II du Tevatron, 
dans le plan des param\`etres $m_0$ et $m_{1/2}$ 
du mod\`ele SUGRA consid\'er\'e, par l'\'etude du
canal trilepton de la production simple de chargino 
permettrait d'\'etendre la sensibilit\'e obtenue par l'\'etude du
canal trilepton de la production 
de paires de superpartenaires. En effet, comme nous l'avons
d\'ej\`a mentionn\'e, l'\'etude du canal trilepton de la 
production de paires de superpartenaires au Run II du Tevatron
permettrait de tester dans le mod\`ele SUGRA consid\'er\'e 
dans la Figure \ref{fig2t} des valeurs du
param\`etre $m_{1/2}$ allant jusqu'\`a $\sim 200 GeV$
\cite{PubliB}. \\
De plus, la sensibilit\'e obtenue via la 
production de paires est ind\'ependante du couplage \rpv
consid\'er\'e \cite{PubliB}. En revanche,
d'apr\`es la Figure \ref{fig2t}, la sensibilit\'e dans le plan
$m_0$ versus $m_{1/2}$ obtenue via la production simple
de chargino est d'autant plus forte que la constante de couplage
\rpi, et donc la \sef de cette production, est grande.
Or, de plus grandes valeurs de la 
constante de couplage \rpv auraient pu \^etre
consid\'er\'ees dans la Figure \ref{fig2t}, car la limite
indirecte sur cette constante est $\l'_{211}<0.09 
(m_{\tilde d_R}/100GeV)$ \cite{4:Bhatt} et, 
dans le mod\`ele SUGRA consid\'er\'e 
dans la Figure \ref{fig2t}, la masse du $\tilde d_R$ vaut $m_{\tilde d_R}=304GeV$  
pour $m_0=100.0GeV$ et $m_{1/2}=121.6GeV$.
Par cons\'equent, des sensibilit\'es encore plus fortes que celles
pr\'esent\'ees dans la Figure \ref{fig2t} peuvent \^etre obtenues 
dans le plan $m_0$ versus $m_{1/2}$ via la production simple
de chargino.

En ce qui concerne la physique au LHC, nous voyons sur la 
Figure \ref{fig3t} que les sensibilit\'es pr\'edites, dans le plan
des param\`etres $m_{1/2}$ et $m_0$ du mod\`ele SUGRA 
consid\'er\'e, s'\'etendent aussi bien au-del\`a du domaine
exclu par les limites actuelles de LEP II.
De plus, la Figure \ref{fig3t} montre que la sensibilit\'e
obtenue au LHC, dans le plan des param\`etres $m_{1/2}$ et $m_0$ 
du mod\`ele SUGRA consid\'er\'e, par l'\'etude du
canal trilepton de la production simple de chargino 
permettrait d'\'etendre la sensibilit\'e obtenue par l'\'etude du
canal trilepton de la production 
de paires de superpartenaires. En effet, comme nous l'avons
d\'ej\`a mentionn\'e, l'\'etude du canal trilepton de la 
production de paires de superpartenaires au LHC
permettrait de tester dans le mod\`ele SUGRA consid\'er\'e 
dans la Figure \ref{fig3t} des valeurs du
param\`etre $m_{1/2}$ allant jusqu'\`a $\sim 200 GeV$
\cite{PubliD}. Enfin, de m\^eme que dans le contexte de la 
physique au Tevatron, pour de plus grandes valeurs de la 
constante de couplage \rpv que celles consid\'er\'ees dans la 
Figure \ref{fig3t}, des sensibilit\'es encore plus fortes que 
celles pr\'esent\'ees dans la Figure \ref{fig3t} peuvent \^etre 
obtenues dans le plan $m_{1/2}$ versus $m_0$ via la production 
simple de chargino.

Les constantes de couplage de type $\l'_{ijk}$
peuvent aussi \^etre test\'ees par l'\'etude de la contribution 
des interactions \rpv \`a la production d'un \'etat final dijet
dans un collisionneur hadronique: 
$p \bar p (ou \ p p) \to q q'$ \cite{4:Rizz}. 
Cette contribution implique l'\'echange d'un slepton
dans les voies $s$, $t$ et $u$. Il a \'et\'e montr\'e dans 
\cite{4:Rizz} que les valeurs des
sensibilit\'es obtenues sur les constantes 
de couplage $\l'_{ijk}$ par l'\'etude de cette contribution
dans le cadre de la physique au Run II du
Tevatron sont sup\'erieures \`a $\sim 10^{-1}$ pour une 
luminosit\'e de ${\cal L}= 2 fb^{-1}$.
Ces sensibilit\'es sont donc inf\'erieures \`a la plupart des
sensibilit\'es obtenues sur les m\^emes constantes de couplage 
$\l'_{ijk}$ par l'\'etude de la production simple de chargino
au Run II du Tevatron (voir Table \ref{tab1} et Figure 
\ref{fig1t}). En revanche, si le seul 
superpartenaire pouvant \^etre produit \`a la
r\'esonance via un couplage $\l'_{ijk}$ dans les collisions
hadroniques est la LSP, celui-ci ne peut se d\'esint\'egrer 
une fois produit \`a la r\'esonance que
via le m\^eme couplage $\l'_{ijk}$ en deux jets.
Dans ce cas, la contribution des interactions \rpv \`a la 
production de l'\'etat final dijet aux collisionneurs hadroniques
peut avoir lieu via la
production r\'esonante d'un superpartenaire, \`a l'inverse des
productions simples de la Figure \ref{4:graphes}, 
et elle est donc plus 
sensible aux constantes de couplage $\l'_{ijk}$ que les
productions simples de la Figure \ref{4:graphes}. \\
D'autre part, les constantes de couplage de type $\l'_{ijk}$
pourraient \^etre test\'ees par l'\'etude de la contribution 
des interactions \rpv au processus de Drell-Yan 
$p \bar p (ou \ p p) \to l \nu, l \bar l$ \cite{4:Rizz}.
Cependant,
Cette contribution, qui implique l'\'echange d'un squark
dans les voies $t$ et $u$, 
est en fait peu significative pour des
valeurs des constantes de couplage $\l'_{ijk}$ respectant
leur limite actuelle \cite{4:Rizz}. \\
Finalement, les constantes de couplage de type $\l'_{ijk}$
peuvent \^etre test\'ees par l'\'etude de la contribution 
des interactions \rpv au processus $l^+  l^- \to q \bar q$ 
\cite{4:Chou,4:Deba}. 
Cette contribution implique l'\'echange d'un 
squark dans la voie $t$. Les sensibilit\'es obtenues 
par l'\'etude de cette contribution sur les constantes de 
couplage $\l'_{ijk}$ dans le cadre de la physique aux 
collisionneurs leptoniques sont de l'ordre de $10^{-1}$ 
pour des masses de squarks inf\'erieures au $TeV$ \cite{4:Chou,
4:Deba}.
Ces sensibilit\'es sont donc inf\'erieures \`a la plupart des
sensibilit\'es obtenues sur les m\^emes constantes de couplage 
$\l'_{ijk}$ par l'\'etude de la production simple de chargino
au Run II du Tevatron (voir Table \ref{tab1} et Figure 
\ref{fig1t}).

Remarquons par ailleurs que 
la production simple de chargino et de neutralino
peut avoir lieu au collisionneur HERA par l'interm\'ediaire des 
constantes de couplage de type $\l'_{1jk}$: $e p \to q \tilde 
\chi^{\pm,0}$ \cite{4:ep}. L'\'etude de cette r\'eaction
permet d'obtenir une sensibilit\'e sur les constantes de couplage
$\l'_{1j1}$ de l'ordre de $10^{-2}$ pour une masse de squark de
$m_{\tilde q} \approx 200 GeV$, une \'energie dans le centre de 
masse de $\sqrt s \approx 300 GeV$ et une luminosit\'e de 
${\cal L} = 500 pb^{-1}$ \cite{4:ep}.
Ces sensibilit\'es sont comparables, voir inf\'erieures, 
\`a celles obtenues sur les 
m\^emes constantes de couplage $\l'_{1j1}$ par l'\'etude de la 
production simple de chargino au Run II du Tevatron 
(voir Table \ref{tab1} et Figure \ref{fig1t}).

\subsubsection{Reconstruction de masse}
\label{recaa}

La contribution dominante \`a la production simple de
chargino est la production r\'esonante de sneutrino
(voir Figure \ref{4:graphes}(a)). Or la cascade de 
d\'esint\'egration initi\'ee par la d\'esint\'egration 
du sneutrino et produisant la signature trilepton,
$\tilde \nu_i \to \tilde \chi_1^{\pm} 
l^{\mp}_i$, $\tilde \chi_1^{\pm} \to \tilde \chi^0_1 l^{\pm} \nu$,
$\tilde \chi^0_1 \to l^{\pm}_i u_j d_k$,
peut \^etre int\'egralement reconstruite. Par cons\'equent,
l'\'etude du canal trilepton g\'en\'er\'e par la production
simple de chargino permet de reconstruire les masses des
$\tilde \chi^0_1$, $\tilde \chi_1^{\pm}$ et $\tilde \nu$,
et ce de mani\`ere ind\'ependante du cadre th\'eorique.
Nous avons \'etudi\'e ces reconstructions de masse dans le
cadre de la physique au Run II du Tevatron \cite{PubliA,PubliB} 
et au LHC \cite{PubliC,PubliD}.

Tout d'abord, la masse du $\tilde \chi^0_1$ peut \^etre 
reconstruite en calculant la masse invariante des deux 
jets et du lepton charg\'e produits dans la d\'esint\'egration
$\tilde \chi^0_1 \to l^{\pm}_i u_j d_k$.
Les deux jets sont identifi\'es comme \'etant les deux jets
les plus \'energ\'etiques de l'\'etat final. La raison est que
ces deux jets sont les seuls jets produits dans le processus
supersym\'etrique consid\'er\'e et que les jets provenant des 
radiations QCD des \'etats initial et final ont
typiquement des \'energies inf\'erieures \`a celles des 
jets issus de r\'eactions supersym\'etriques.
Quant \'a l'identification du lepton charg\'e, elle est bas\'ee
sur la saveur ainsi que la charge \'electrique de celui-ci.
En fait, la masse du $\tilde \chi^0_1$ est d\'etermin\'ee
\`a partir du pic apparaissant dans la distribution de la 
masse invariante des deux jets et du lepton charg\'e. 
Ce pic n'est pas un pic de Dirac \`a cause du bruit de fond
combinatoire qui est du dans ce cas au fait que les deux 
jets de l'\'etat final
peuvent \^etre confondus avec les jets irradi\'es par l'\'etat 
initial. La masse du $\tilde \chi^0_1$ peut \^etre reconstruite
au LHC avec une erreur statistique de $\sim 100MeV$ 
\cite{PubliD}.

La masse du $\tilde \chi_1^{\pm}$, qui se d\'esint\`egre via
$\tilde \chi_1^{\pm} \to \tilde \chi^0_1 l^{\pm} \nu$, 
peut alors \^etre d\'etermin\'ee
\`a partir de la reconstruction du $\tilde \chi^0_1$.
Cette d\'etermination est cependant plus pr\'ecise lorsque
la d\'esint\'egration $\tilde \chi_1^{\pm} \to 
\tilde \chi^0_1 l^{\pm} \nu$
a lieu via un boson $W^{\pm}$ r\'eel (voir Figure \ref{fig3t})  
selon la cascade $\tilde \chi_1^{\pm} \to 
\tilde \chi^0_1 W^{\pm}$, $W^{\pm} \to l^{\pm} \nu$.
En effet, dans ce cas la quadri-impulsion du neutrino peut
\^etre calcul\'ee exactement en exprimant la condition 
d'\'egalit\'e entre la masse du boson $W^{\pm}$ et la masse 
invariante des deux leptons provenant de la d\'esint\'egration 
$W^{\pm} \to l^{\pm} \nu$. Le lepton charg\'e produit dans la 
d\'esint\'egration $W^{\pm} \to l^{\pm} \nu$ est une fois de plus
identifi\'e gr\^ace \`a sa saveur et son signe.
Nous avons trouv\'e une valeur de
$\sim 6 GeV$ pour la largeur du pic associ\'e \`a
la masse reconstruite du $\tilde \chi_1^{\pm}$, dans le cadre de
la physique au LHC \cite{PubliD}.

Enfin, la masse du $\tilde \nu$, qui se d\'esint\`egre via
$\tilde \nu_i \to \tilde \chi_1^{\pm} l^{\mp}_i$, 
peut \^etre d\'etermin\'ee
\`a partir de la reconstruction du $\tilde \chi_1^{\pm}$.
Le lepton charg\'e issu de la d\'esint\'egration du sneutrino
est identifi\'e gr\^ace \`a sa saveur et son signe.
Nous avons trouv\'e une valeur de
$\sim 10 GeV$ pour la largeur du pic associ\'e \`a
la masse reconstruite du $\tilde \nu$, dans le cadre de
la physique au LHC \cite{PubliD}.

En conclusion, aux collisionneurs hadroniques, si le 
$\tilde \chi^0_1$ \'etait la LSP (comme c'est le cas dans la 
plupart des mod\`eles supersym\'etriques), sa masse devrait 
\^etre 
facilement reconstruite \cite{4:Atlas,4:NLC}
car il serait alors produit en grand 
nombre dans les productions de paires de gluinos et 
de squarks qui ont d'importantes sections efficaces
\cite{PubliD}. Par cons\'equent, 
pour de faibles valeurs des couplages \rpi,
et donc de la \sef de la production simple de chargino,
la reconstruction du $\tilde \chi^0_1$  
la plus pr\'ecise est \`a priori obtenue 
par l'\'etude des productions de paires de gluinos et 
de squarks. Pour des valeurs des couplages \rpv proches de leur
limite actuelle, les pr\'ecisions sur la reconstruction de la
masse du $\tilde \chi^0_1$ obtenues 
par l'\'etude des productions de paires de gluinos/squarks 
\cite{4:Atlas} et par l'\'etude de la production simple de 
chargino \cite{PubliC,PubliD} sont comparables. 
En revanche,
la production simple de chargino permet d'obtenir
facilement une grande pr\'ecision sur la reconstruction des
masses des $\tilde \chi_1^{\pm}$ et 
$\tilde \nu$ \cite{PubliC,PubliD} ce qui n'est pas le cas 
de la production
de paires de particules supersym\'etriques \cite{4:Atlas,4:NLC}.
Ceci est d\^u au fait que la production simple de particule SUSY 
ne g\'en\`ere qu'une seule cascade de d\'esint\'egration de 
particules SUSY alors que la production de paire en
g\'en\`ere deux ce qui complique l'identification de l'origine 
des particules de l'\'etat final et augmente ainsi 
le bruit de fond combinatoire.
Par ailleurs, m\^eme si le pic associ\'e \`a la 
reconstruction de la masse du $\tilde \chi_1^0$ via la 
production simple de chargino \'etait
rendu invisible par la production de paires de superpartenaires,
la reconstruction des masses des $\tilde \chi_1^{\pm}$ et 
$\tilde \nu$ via la production simple de chargino resterait 
possible connaissant la valeur de la masse du $\tilde \chi_1^0$
d\'etermin\'ee via la production de paires de superpartenaires
\cite{4:Atlas,4:NLC}.

\subsection{Signature dilepton}

L'\'etat final contenant deux leptons charg\'es de m\^eme 
signe et de m\^eme saveur a un faible bruit de fond
provenant du Mod\`ele Standard. 
Cette signature dilepton peut \^etre g\'en\'er\'ee
par les r\'eactions suivantes (voir Figure \ref{4:graphes}):
$p \bar p \to \tilde \chi^0_1 l^{\pm}_i$;
$p \bar p \to \tilde \chi^0_2 l^{\pm}_i$,
$\tilde \chi^0_2 \to \tilde \chi^0_1 + X$ ($X \neq l^{\pm}$);
$p \bar p \to \tilde \chi^{\pm}_1 l^{\mp}_i$,
$\tilde \chi^{\pm}_1 \to \tilde \chi^0_1 q \bar q'$ et
$p \bar p \to \tilde \chi^{\pm}_1 \nu_i$,
$\tilde \chi^{\pm}_1 \to \tilde \chi^0_1 l^{\pm} \nu$,
$i$ correspondant \`a l'indice de saveur de la constante de
couplage $\l'_{ijk}$.
En effet, le $\tilde \chi^0_1$ \'etant une particule de Majorana,
il se d\'esint\`egre via $\l'_{ijk}$ en un lepton selon
$\tilde \chi^0_1 \to l_i u_j \bar d_k$ et en un anti-lepton
selon $\tilde \chi^0_1 \to \bar l_i \bar u_j d_k$ avec la
m\^eme probabilit\'e.
Les r\'eactions $p \bar p \to \tilde \chi^0_{3,4} l^{\pm}_i$,
$p \bar p \to \tilde \chi^{\pm}_2 l^{\mp}_i$ et 
$p \bar p \to \tilde \chi^{\pm}_2 \nu_i$ ne repr\'esentent pas 
des contributions significatives \`a la signature dilepton du
fait de leur relativement faibles sections efficaces 
\cite{PubliB}. Dans \cite{PubliB} nous avons \'etudi\'e dans le
cadre de la physique au Run II du Tevatron le signal dilepton 
engendr\'e par les productions simples de superpartenaires
ainsi que le bruit de fond associ\'e. 
Les quatre processus de production
simple de la Figure \ref{4:graphes} ont \'et\'e 
impl\'ement\'es dans une version du g\'en\'erateur 
d'\'ev\`enements SUSYGEN \cite{4:SUSYGEN} 
incluant la simulation des collisions hadroniques. 
Ceci a permis de g\'en\'erer le signal avec SUSYGEN et 
les bruits de fond provenant du \ms
et des r\'eactions \susiqs avec les
g\'en\'erateurs d'\'ev\`enements PYTHIA \cite{4:PYTHIA} et
SHERWIG \cite{4:HERWIG}, respectivement. SUSYGEN, PYTHIA et 
SHERWIG ont \'et\'e interfac\'es avec le simulateur 
SHW \cite{4:SHW} des d\'etecteurs D0 et CDF du Tevatron (Run II).
Lors de la g\'en\'eration du signal et du bruit de fond,
des coupures bas\'ees sur des distributions de variables 
cin\'ematiques (angles d\'emission des particules, 
quadri-impulsions,...) ont \'et\'e appliqu\'ees afin
d'augmenter le signal par rapport au bruit de fond.

\subsubsection{Potentiel de d\'ecouverte}

\begin{figure}[t]
\begin{center}
\leavevmode
\centerline{
\psfig{figure=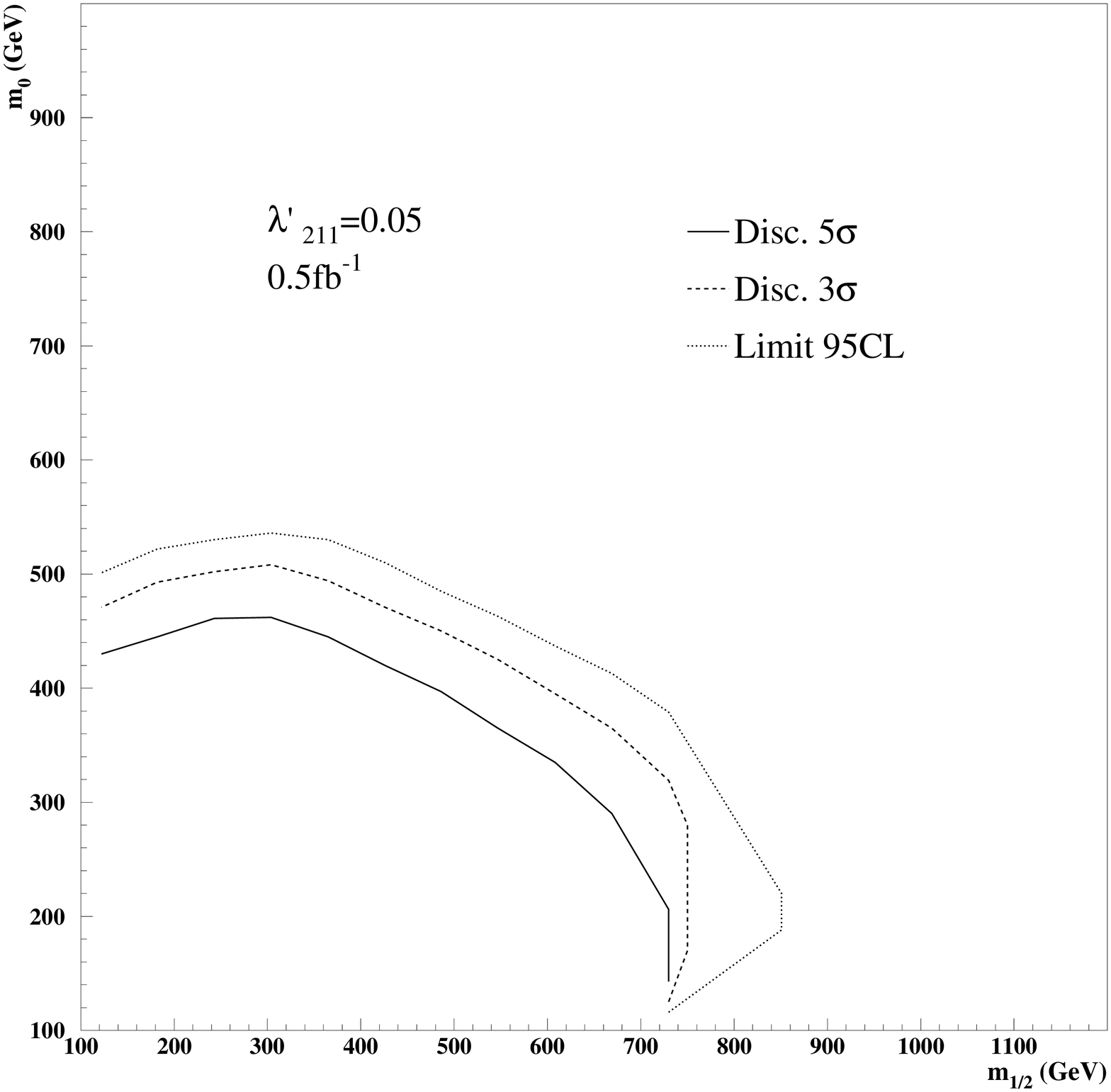,height=3.5in}
\psfig{figure=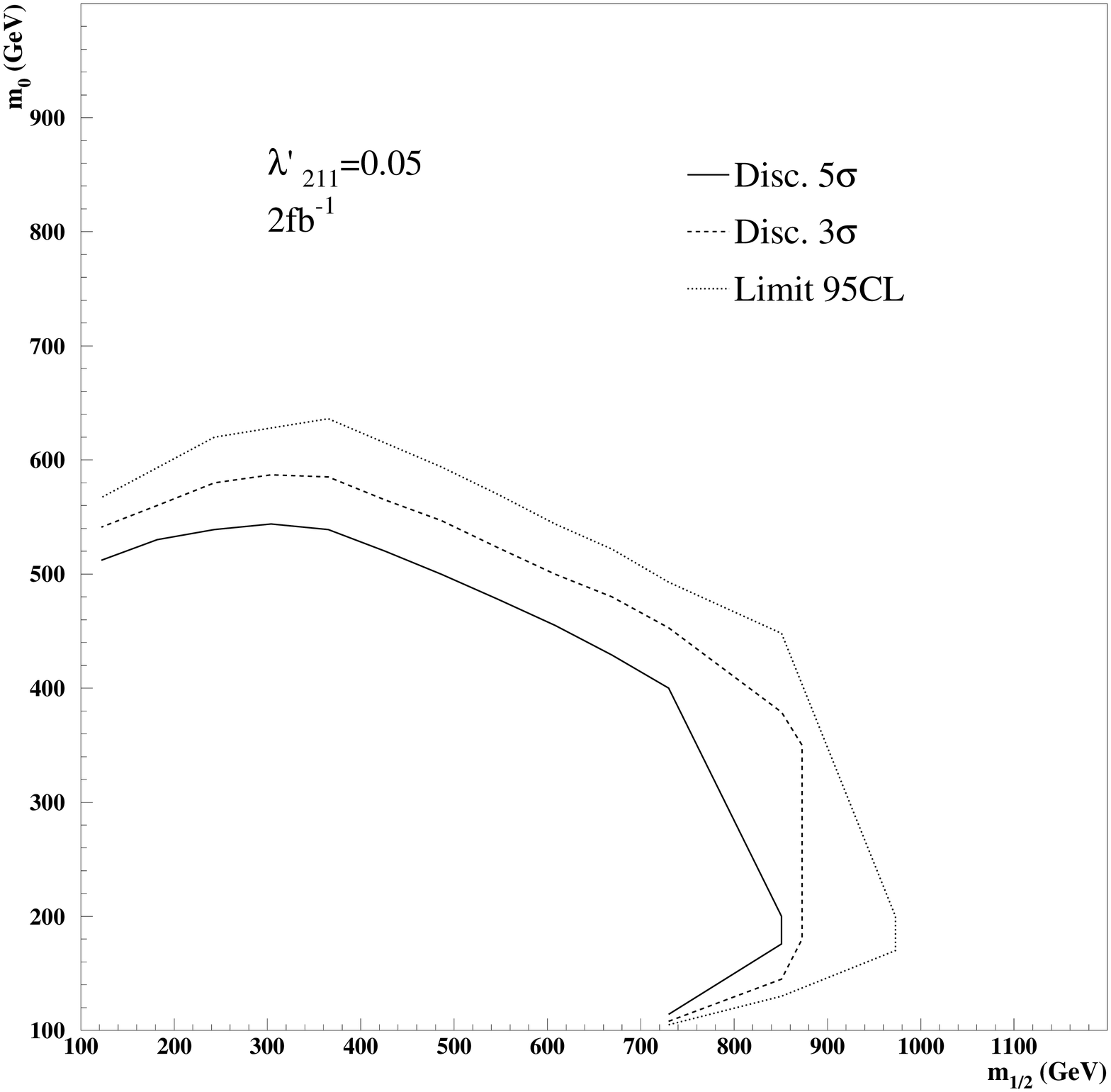,height=3.5in}}
\centerline{\psfig{figure=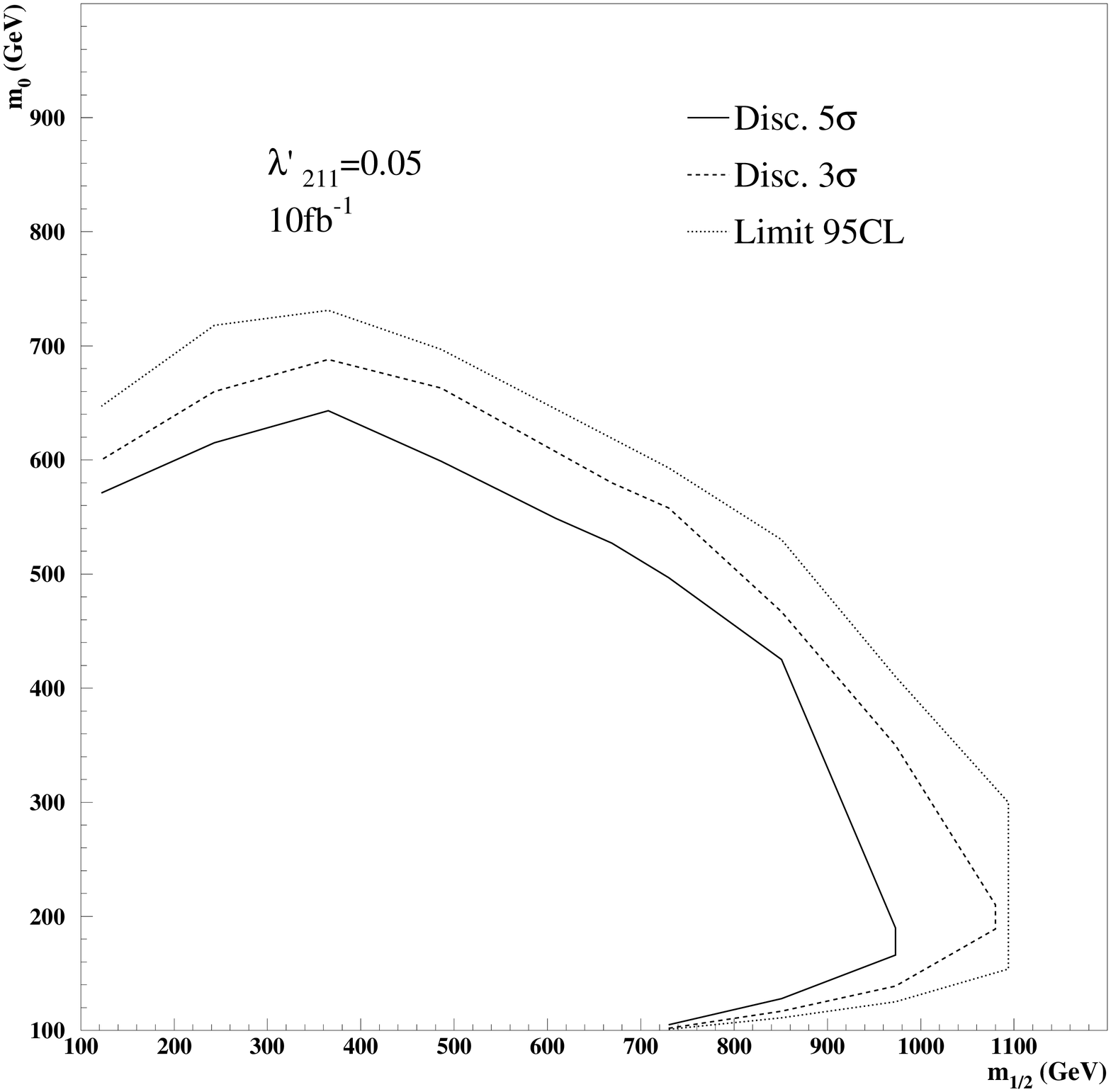,height=3.5in}}
\end{center}
\caption{
Contours de d\'ecouverte \`a $5 \sigma$ (ligne),
$3 \sigma$ (tir\'es)
et limites \`a $95 \% \ C.L.$ (pointill\'es) 
associ\'es \`a l'\'etude du canal dilepton au 
Run II du Tevatron ($\sqrt s=2 TeV$) et pr\'esent\'es
dans le plan $m_0$ versus $m_{1/2}$ 
pour $sign(\mu)<0$, $\tan \beta=1.5$, $\l'_{211}=0.05$
et diff\'erentes valeurs de la luminosit\'e.}
\label{fig4t}
\end{figure}

Nous pr\'esentons dans la Figure \ref{fig4t} la sensibilit\'e
dans le plan des param\`etres $m_0$ et $m_{1/2}$ 
qui pourrait \^etre obtenue \`a partir de 
l'analyse de l'\'etat final dilepton au Run II du Tevatron,
dans le cadre d'un mod\`ele SUGRA et pour des valeurs fix\'ees 
des autres param\`etres supersym\'etriques du mod\`ele.
Cette figure a \'et\'e obtenue apr\`es avoir appliqu\'e 
les coupures mentionn\'ees plus haut et en supposant que la 
production simple avait lieu par l'interm\'ediaire 
d'un couplage du type $\l'_{211}$,
ce qui correspond au cas o\`u le lepton produit avec
le neutralino ou le chargino est un $\mu^{\pm}$ ou un $\nu_\mu$ 
(voir Figure \ref{4:graphes}).

D'apr\`es la Figure \ref{fig4t},
la sensibilit\'e obtenue sur $\l'_{211}$ dans certaines
r\'egions de l'espace des param\`etres SUGRA par  
l'\'etude de la signature dilepton au Run II du
Tevatron permettrait d'am\'eliorer la limite actuelle sur
cette constante de couplage \rpv qui vaut: 
$\l'_{211}<0.09 (m_{\tilde d_R}/100GeV)$ \cite{4:Bhatt}. 
L'\'etude de la signature dilepton
devrait aussi permettre d'am\'eliorer les limites indirectes 
de nombreuses autres constantes de couplage \rpi.

Nous avons montr\'e dans \cite{PubliB} que l'\'etude du
canal dilepton de la production de paires de particules \susiqs
dans le cadre de la physique au Run II du Tevatron
permettait de tester dans le mod\`ele SUGRA consid\'er\'e 
dans la Figure \ref{fig4t} des valeurs du
param\`etre $m_{1/2}$ allant jusqu'\`a $\sim 200 GeV$, et ceci
quelque soit la valeur de la constante de couplage \rpv 
consid\'er\'ee. 
Par cons\'equent, dans la r\'egion $m_{1/2} \stackrel{>}{\sim}
200GeV$ de la Figure \ref{fig4t}, 
la sensibilit\'e obtenue sur les constantes de couplage 
\rpv par la production simple n'est pas affect\'ee
par la production de paires de superpartenaires. 
En revanche, dans la r\'egion
$m_{1/2} \stackrel{<}{\sim}200GeV$, la production de paires de 
superpartenaires repr\'esente un bruit de fond pour la production
simple car elle n'implique pas les interactions \rpv
qui sont l'objet de cette \'etude.
Le bruit de fond engendr\'e par la production de paires peut
\^etre r\'eduit par des coupures cin\'ematiques bas\'ees sur
la reconstruction de masse des particules \susiqs (voir Section
\ref{recbb}).

Pla\c{c}ons nous maintenant du point de vue de la recherche
de la supersym\'etrie aupr\`es des collisionneurs.
De m\^eme que pour les Figures \ref{fig1t} et \ref{fig2t},
l'espace des param\`etres de la Figure \ref{fig4t} n'est pas
exclu par les donn\'ees de LEP II \cite{4:neut,4:Mass}.
Par cons\'equent, le potentiel de d\'ecouverte pour le Run II du
Tevatron (Figure \ref{fig4t}) repr\'esente une am\'elioration 
importante par rapport aux limites actuelles sur les 
param\`etres \susiqs du mod\`ele SUGRA consid\'er\'e.

De plus, la Figure \ref{fig4t} montre que la sensibilit\'e
obtenue au Run II du Tevatron, 
dans le plan des param\`etres $m_0$ et $m_{1/2}$ 
du mod\`ele SUGRA consid\'er\'e, par l'\'etude du
canal dilepton de la production simple
permettrait d'\'etendre la sensibilit\'e obtenue par l'\'etude du
canal dilepton de la production 
de paires de superpartenaires. En effet, comme nous l'avons
d\'ej\`a mentionn\'e, l'\'etude du canal dilepton de la 
production de paires de superpartenaires au Run II du Tevatron
permettrait de tester dans le mod\`ele SUGRA consid\'er\'e 
dans la Figure \ref{fig4t} des valeurs du
param\`etre $m_{1/2}$ allant jusqu'\`a $\sim 200 GeV$
\cite{PubliB}. \\
De plus, la sensibilit\'e obtenue via la 
production de paires est ind\'ependante du couplage \rpv
consid\'er\'e \cite{PubliB}. En revanche,
la sensibilit\'e dans le plan
$m_0$ versus $m_{1/2}$ obtenue via la production simple
est d'autant plus forte que la constante de couplage
\rpi, et donc la \sef de production simple, est grande.
Or, de plus grandes valeurs de la 
constante de couplage \rpv auraient pu \^etre
consid\'er\'ees dans la Figure \ref{fig4t}, car la limite
indirecte sur cette constante est $\l'_{211}<0.09 
(m_{\tilde d_R}/100GeV)$ \cite{4:Bhatt} et, 
dans le mod\`ele SUGRA consid\'er\'e 
dans la Figure \ref{fig4t}, la masse du $\tilde d_R$ vaut $m_{\tilde d_R}=304GeV$  
pour $m_0=100.0GeV$ et $m_{1/2}=121.6GeV$.
Par cons\'equent, des sensibilit\'es encore plus fortes que celles
pr\'esent\'ees dans la Figure \ref{fig4t} peuvent \^etre obtenues 
dans le plan $m_0$ versus $m_{1/2}$ via la production simple.

Remarquons finalement que l'\'etude du canal dilepton (voir 
Figure \ref{fig4t}) permet d'obtenir une plus grande 
sensibilit\'e 
sur la constante de couplage $\l'_{211}$ que l'\'etude du canal 
trilepton (voir Figure \ref{fig2t}) \cite{PubliB}.

\subsubsection{Reconstruction de masse}
\label{recbb}

La contribution dominante du signal \rpv
au canal dilepton est la r\'eaction
$p \bar p \to \tilde l^{\pm}_i \to \tilde \chi^0_1 l^{\pm}_i$,
$\tilde \chi^0_1 \to l^{\pm}_i u_j d_k$ car celle-ci a la plus
grande section efficace \cite{PubliB}.
Or les masses des $\tilde \chi^0_1$ et $\tilde l^{\pm}_i$
peuvent \^etre reconstruites gr\^ace \`a cette r\'eaction,
et ce de mani\`ere ind\'ependante du cadre th\'eorique.
Nous avons \'etudi\'e ces reconstructions de masse dans le
cadre de la physique au Run II du Tevatron \cite{PubliA,PubliB}.

Tout d'abord, la masse du $\tilde \chi^0_1$ peut \^etre 
reconstruite en calculant la masse invariante des deux 
jets et du lepton charg\'e produits dans la d\'esint\'egration
$\tilde \chi^0_1 \to l^{\pm}_i u_j d_k$.
Les deux jets sont simples \'a identifier puisque ce sont
les seuls jets `durs' de l'\'etat final consid\'er\'e.
Quant \'a l'identification du lepton charg\'e, elle ne peut 
\^etre bas\'ee ni sur la saveur ni sur le signe de celui-ci 
car l'\'etat final consid\'er\'e contient deux leptons
charg\'es de m\^eme signe et de m\^eme saveur.
Le lepton produit dans la d\'esint\'egration du neutralino
peut en revanche \^etre identifi\'e par son \'energie 
\cite{PubliB}. Le lepton produit dans la d\'esint\'egration 
du neutralino peut aussi \^etre identifi\'e comme \'etant
le lepton le plus proche dans l'espace ($\eta$,$\phi$) des 
deux jets de l'\'etat final \cite{4:Drr}.
La masse du $\tilde \chi^0_1$ peut \^etre reconstruite
au Tevatron (Run II) avec une pr\'ecision de $\pm 11 GeV$ 
\cite{PubliB}.

La masse du $\tilde l^{\pm}$, qui se d\'esint\`egre via
$\tilde l^{\pm}_i \to \tilde \chi^0_1 l^{\pm}_i$, 
peut alors \^etre d\'etermin\'ee
\`a partir de la reconstruction du $\tilde \chi^0_1$.
Le lepton charg\'e issu de la d\'esint\'egration du slepton
peut \^etre identifi\'e par son \'energie \cite{PubliB}.
Nous avons trouv\'e une pr\'ecision de $\pm 20 GeV$ pour
la masse reconstruite du $\tilde l^{\pm}$, dans le cadre de
la physique au Run II du Tevatron {PubliB}.

En conclusion, aux collisionneurs hadroniques, si le 
$\tilde \chi^0_1$ \'etait la LSP (comme c'est le cas dans la 
plupart des mod\`eles supersym\'etriques), sa masse devrait 
\^etre 
facilement reconstruite car il serait alors produit en grand 
nombre dans les productions de paires de gluinos et 
de squarks qui ont d'importantes sections efficaces
\cite{PubliB}.  
En revanche,
la production simple permet d'obtenir
facilement une grande pr\'ecision sur la reconstruction de
masse du $\tilde l^{\pm}$ \cite{PubliB}  
\`a l'inverse de la production
de paires de particules supersym\'etriques.
Ceci est d\^u au fait que la production simple de particule SUSY 
ne g\'en\`ere qu'une seule cascade de d\'esint\'egration de 
superpartenaires alors que la production de paire en
g\'en\`ere deux ce qui complique l'identification de l'origine 
des particules de l'\'etat final et augmente ainsi 
le bruit de fond combinatoire.
Par ailleurs, m\^eme si le pic associ\'e \`a la 
reconstruction de la masse du $\tilde \chi_1^0$ via la 
production simple \'etait
rendu invisible par la production de paires de superpartenaires,
la reconstruction de masse du $\tilde l^{\pm}$ via la production 
simple resterait possible connaissant la valeur de la masse du 
$\tilde \chi_1^0$ d\'etermin\'ee via la production de paires de 
superpartenaires.

\section{Collisionneurs leptoniques}
\label{4:lep}

Nous pr\'esentons dans cette section des \'etudes de production 
simple de particule
SUSY aux collisionneurs leptoniques. Pr\'ecisons que dans ces 
\'etudes, comme pr\'ec\'edemment, nous avons toujours suppos\'e
qu'une seule constante de couplage \rpv \'etait dominante par 
rapport aux autres.

\begin{figure}[t]
\begin{center}
\leavevmode
\centerline{\psfig{figure=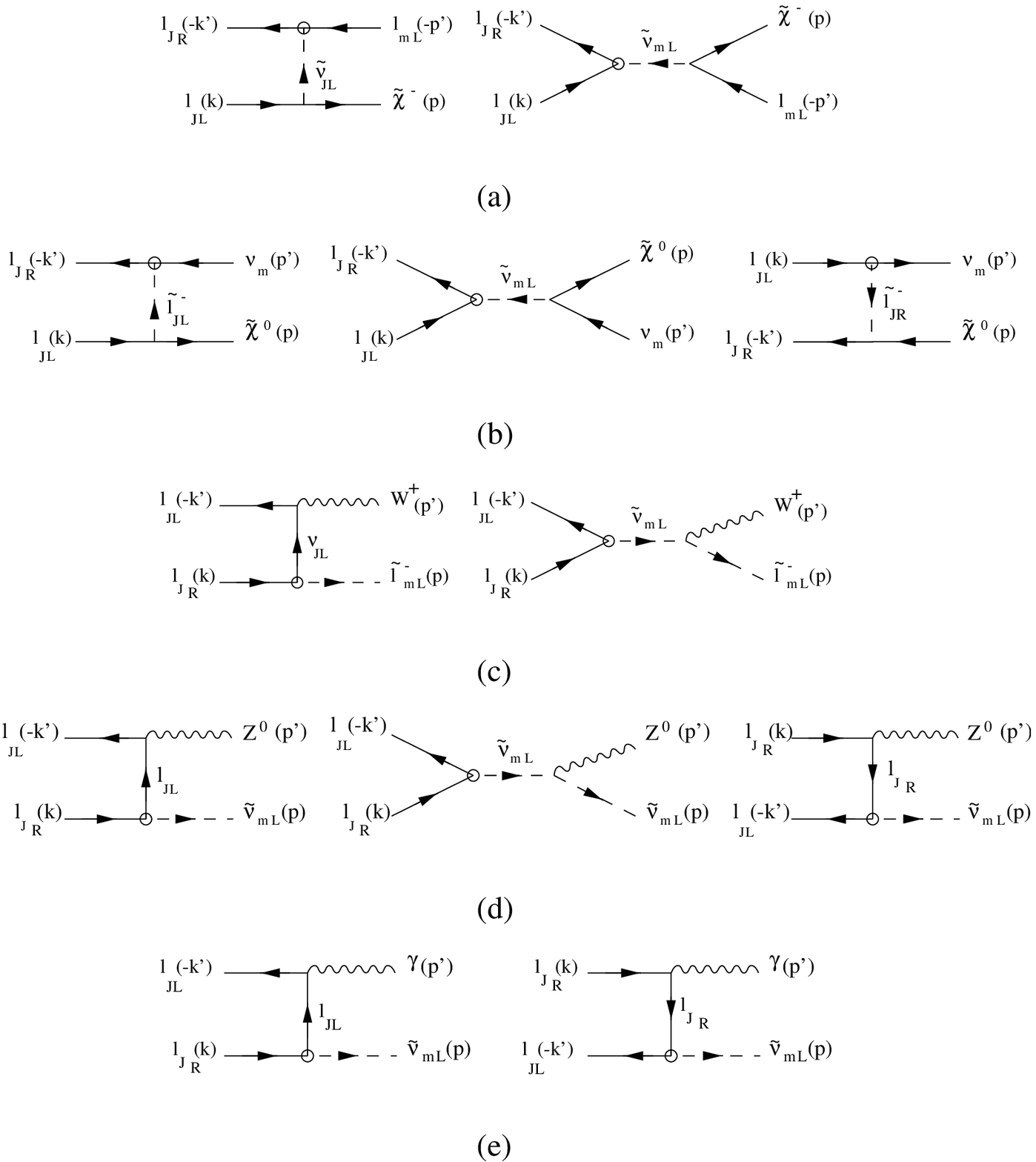,height=7.in}}
\end{center}
\caption{\footnotesize  \it Diagrammes de Feynman des 5 
processus de production simple
de particule SUSY via $\l_{mJJ}$ aux collisionneurs leptoniques 
qui sont du type $2 \to 2-corps$. Les couplages $\l_{mJJ}$ 
impliqu\'es sont symbolis\'es par des cercles
et les fl\`eches repr\'esentent les moments des particules.
\rm \normalsize }
\label{4:graphesp}
\end{figure}

\begin{figure}[t]
\begin{center}
\leavevmode
\centerline{\psfig{figure=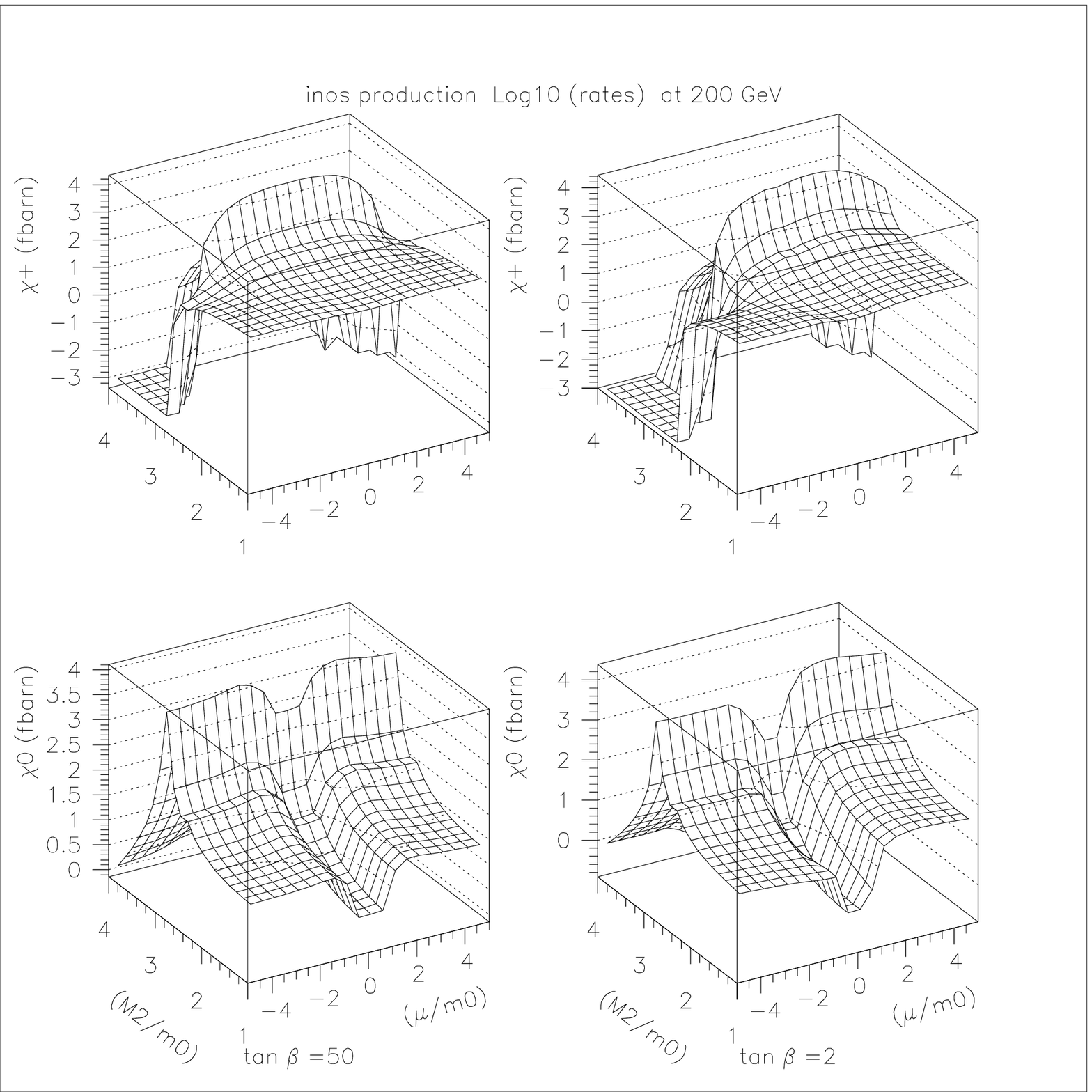,height=6.5in}}
\end{center}
\caption{\footnotesize  \it Sections efficaces (en $fbarns$)
de production simple de $\tilde \chi^{\pm}_1$ et de
$\tilde \chi^0_1$ (voir Figure \ref{4:graphesp}) 
en fonction des param\`etres $M_2/m_0$, 
$\mu/m_0$ (avec $m_0=100GeV$) et $\tan \beta$
du mod\`ele de supergravit\'e d\'ecrit dans \cite{PubliE}, 
pour une \'energie dans le centre de masse de $\sqrt s=200GeV$
et une constante de couplage \rpv \'egale \`a $\l_{121}=0.05$
(limite actuelle pour $m_{\tilde e_R}=100GeV$ \cite{4:Bhatt}). 
\rm \normalsize }
\label{4:singxmc}
\end{figure}

\begin{figure}[t]
\begin{center}
\leavevmode
\centerline{\psfig{figure=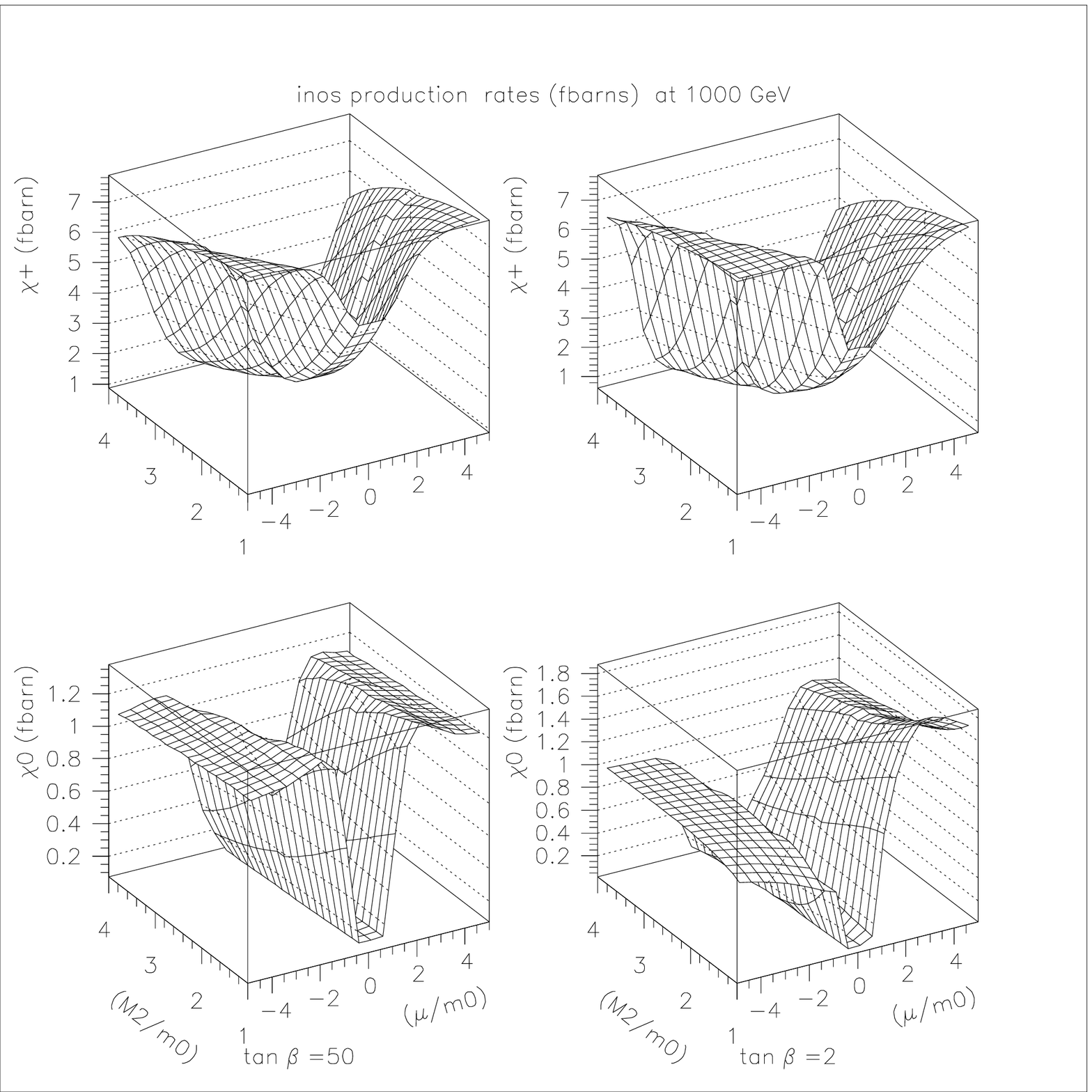,height=6.5in}}
\end{center}
\caption{\footnotesize  \it Sections efficaces (en $fbarns$)
de production simple de $\tilde \chi^{\pm}_1$ et de
$\tilde \chi^0_1$ (voir Figure \ref{4:graphesp}) 
en fonction des param\`etres $M_2/m_0$, 
$\mu/m_0$ (avec $m_0=100GeV$) et $\tan \beta$
du mod\`ele de supergravit\'e d\'ecrit dans \cite{PubliE}, 
pour une \'energie dans le centre de masse de $\sqrt s=1000GeV$
et une constante de couplage \rpv \'egale \`a $\l_{121}=0.05$
(limite actuelle pour $m_{\tilde e_R}=100GeV$ \cite{4:Bhatt}). 
\rm \normalsize }
\label{4:singxmcp}
\end{figure}

La production simple de particule SUSY aux collisionneurs 
leptoniques implique uniquement des interactions $\l$. 
Toutes les productions simples de particule SUSY
aux collisionneurs leptoniques du type $2 \to 2-corps$
sont pr\'esent\'ees dans la Figure \ref{4:graphesp}.
Toutes les amplitudes des productions simples de la Figure
\ref{4:graphesp} ont \'et\'e calcul\'ees analytiquement 
et les r\'esultats sont donn\'es dans \cite{PubliE}
(voir Publication V: {\it ``Systematics of 
single superpartners production at leptonic colliders''}).
Bas\'es sur ces calculs d'amplitudes, nous avons \'etudi\'e
le comportement des \sefs des productions simples dans 
l'espace des param\`etres SUGRA \cite{PubliE} \cite{PubliF}
(voir Publication VI: {\it ``Single chargino production at 
linear colliders''}). Dans les Figures \ref{4:singxmc} et \ref{4:singxmcp}, 
nous pr\'esentons des courbes compl\'ementaires de celles montr\'ees
dans \cite{PubliE,PubliF}: les \sefs 
de production simple de $\tilde \chi^{\pm}_1$ et de
$\tilde \chi^0_1$ (voir Figure \ref{4:graphesp}) en fonction
des param\`etres $M_2/m_0$, $\mu/m_0$ et $\tan \beta$
du mod\`ele de supergravit\'e d\'ecrit dans \cite{PubliE},
pour des \'energies dans le centre de masse de $\sqrt s=200GeV$ 
et $\sqrt s=1000GeV$. 
Par ailleurs, dans \cite{PubliE} nous avons calcul\'e les largeurs
des canaux de d\'esint\'egration des particules \susiqs 
produites et nous avons \'etudi\'e leur \'evolution dans l'espace
des param\`etres SUGRA. Il ressort de cette \'etude que, 
mis \`a part la 
production simple de neutralino (voir Figure \ref{4:graphesp}(b)),
dans la plupart des r\'egions de l'espace des param\`etres SUGRA
les productions simples repr\'esent\'ees 
dans la Figure \ref{4:graphesp}
engendrent des cascades de d\'esint\'egration impliquant des 
interactions de jauge. Ces cascades de d\'esint\'egration
donnent lieu \`a de multiple signatures 
riches en leptons charg\'es et en jets et ayant un faible 
bruit de fond provenant du Mod\`ele Standard.
Il s'agit d'\'etats finals tels que: \newline $4l+\Eslash,\ 6l,
\ 6l+\Eslash,\ 3l+2j+\Eslash,$...

\subsection{Production simple de chargino}

Dans \cite{PubliFv,PubliF}, nous nous sommes concentr\'es sur
la production simple $e^+ e^- \to \tilde \chi^{\pm}_1 l_m^{\mp}$ 
via $\l_{1m1}$ qui a typiquement la plus importante
\sef parmi les productions simples repr\'esent\'ees 
dans la Figure \ref{4:graphesp} \cite{PubliE,PubliF}. 
La production simple de $\tilde \chi^{\pm}_1$ a aussi
l'int\'er\^et de pouvoir engendrer de claires signatures 
multi-jets et multi-leptoniques \cite{PubliE,PubliFv,PubliF}.
En particulier, nous avons consid\'er\'e la r\'eaction 
$e^+ e^- \to \tilde \chi^{\pm}_1 \mu^{\mp}$ qui implique
la constante de couplage $\l_{121}$.

La production simple $e^+ e^- \to \tilde \chi^{\pm}_1 \mu^{\mp}$ 
a une voie $t$ et peut aussi recevoir la contribution
de la production r\'esonante d'un sneutrino (voir Figure 
\ref{4:graphesp}(a)). 
\`A $\sqrt s=200GeV$ la \sef de production simple du
chargino hors du p\^ole du sneutrino est  
de l'ordre de $100 fbarns$ \cite{PubliE} pour une constante de couplage \rpv 
\'egale \`a sa limite actuelle qui est $\l_{121}=0.05$ pour 
$m_{\tilde e_R}=100GeV$ \cite{4:Bhatt}. La \sef hors du p\^ole 
du sneutrino est donc \`a la limite d'observabilit\'e
\`a LEP II pour une luminosit\'e de ${\cal L} \approx 
200 pb^{-1}$. Au p\^ole du sneutrino, la production simple de
chargino atteint de grandes valeurs de la section efficace.
Par exemple pour $\l_{121}=0.05$, la \sef de production simple du
chargino peut atteindre $2 \ 10^{-1} pb$ \`a la r\'esonance
du sneutrino \cite{4:Workshop}. C'est la raison pour laquelle les 
analyses exp\'erimentales de la production simple de chargino au 
LEP \cite{4:ALEPHa,4:DELPHIa,4:DELPHIb,4:Fab1,4:Fab2} 
permettent de tester des valeurs des constantes de 
couplage \rpv inf\'erieures aux limites actuelles seulement 
\`a la r\'esonance $\sqrt s=m_{\tilde \nu_L}$, et, gr\^ace aux 
radiations de l'\'etat initial (ISR), dans un intervalle de
$\sim 50 GeV$ autour du p\^ole du sneutrino. 
Au p\^ole m\^eme du sneutrino, 
les sensibilit\'es sur la constante de couplage $\l_{121}$ 
obtenues \`a LEP II atteignent des valeurs de l'ordre de
$10^{-3}$ \cite{4:ALEPHa,4:DELPHIa,4:DELPHIb,4:Fab1,4:Fab2}.

Les collisionneurs lin\'eaires constitueront un cadre 
propice \`a l'\'etude de la production simple de chargino
de par les grandes \'energies ($\sqrt s \approx 1TeV$) et 
luminosit\'es (${\cal L} \approx 500 fb^{-1}$) atteintes 
\cite{4:Tesla,4:NLC}. En particulier, les luminosit\'es 
attendues aux collisionneurs lin\'eaires permettront d'\^etre
sensible \`a la \sef de production simple du chargino
hors du p\^ole du sneutrino qui est de l'ordre de $10 fbarns$
\cite{PubliE}
pour $1000 GeV> \sqrt s > 500 GeV$ et une constante de couplage 
\rpv \'egale \`a $\l_{121}=0.05$. 
Cependant, aux 
collisionneurs lin\'eaires, la production simple de chargino
risque de souffrir d'un grand bruit de fond provenant de la
production de paires de particules supersym\'etriques.
En effet, aux grandes \'energies esp\'er\'ees dans ces 
collisionneurs, la production de paire de superpartenaires
peut atteindre de grandes sections efficaces.

\subsubsection{Potentiel de d\'ecouverte}
\label{potde}

Dans \cite{PubliF}, nous avons montr\'e que malgr\'e le bruit
de fond provenant de la production de paire de superpartenaires,
l'\'etude de la production simple de chargino aux collisionneurs
lin\'eaires offrira la possibilit\'e d'obtenir une plus forte 
sensibilit\'e sur la constante de couplage $\l_{121}$ que
celle obtenue \`a LEP 
\cite{4:ALEPHa,4:DELPHIa,4:DELPHIb,4:Fab1,4:Fab2}. Pour cela,
nous avons \'etudi\'e l'\'etat final compos\'e de quatre leptons
charg\'es et d'\'energie manquante qui poss\`ede un faible
bruit de fond provenant du \ms \cite{4:Godbole} 
et qui est g\'en\'er\'e par
la production simple de chargino lorsque celui-ci se 
d\'esint\`egre selon $\tilde \chi_1^{\pm} \to \tilde \chi^0_1 
l \nu$, $\tilde \chi_1^0 \to  \bar e e \nu_\mu$,
$\bar e e \bar \nu_\mu$, $\mu \bar e \bar \nu_e$ 
ou $\bar \mu e \nu_e$ via $\l_{121}$.
Nous avons montr\'e que la contribution de la production de 
paire de superpartenaires au signal $4l+\Eslash$ pouvait \^etre
r\'eduite par rapport \`a la contribution de la production simple 
de chargino. Cette r\'eduction est bas\'ee sur deux points.
Tout d'abord, la polarisation des faisceaux d'\'electrons et de 
positrons incidents aux collisionneurs lin\'eaires peut \^etre
utilis\'ee pour r\'eduire la source de bruit de fond que 
repr\'esente la production de paire de superpartenaires.
De plus, la cin\'ematique sp\'ecifique de type $2 \to 2-corps$
de la production simple de chargino permet d'imposer des 
coupures sur les distributions des moments transverses 
des leptons charg\'es de l'\'etat final favorisant la production 
simple de chargino par rapport \`a la production de paire de 
superpartenaires.

\begin{figure}[t]
\begin{center}
\leavevmode
\centerline{\psfig{figure=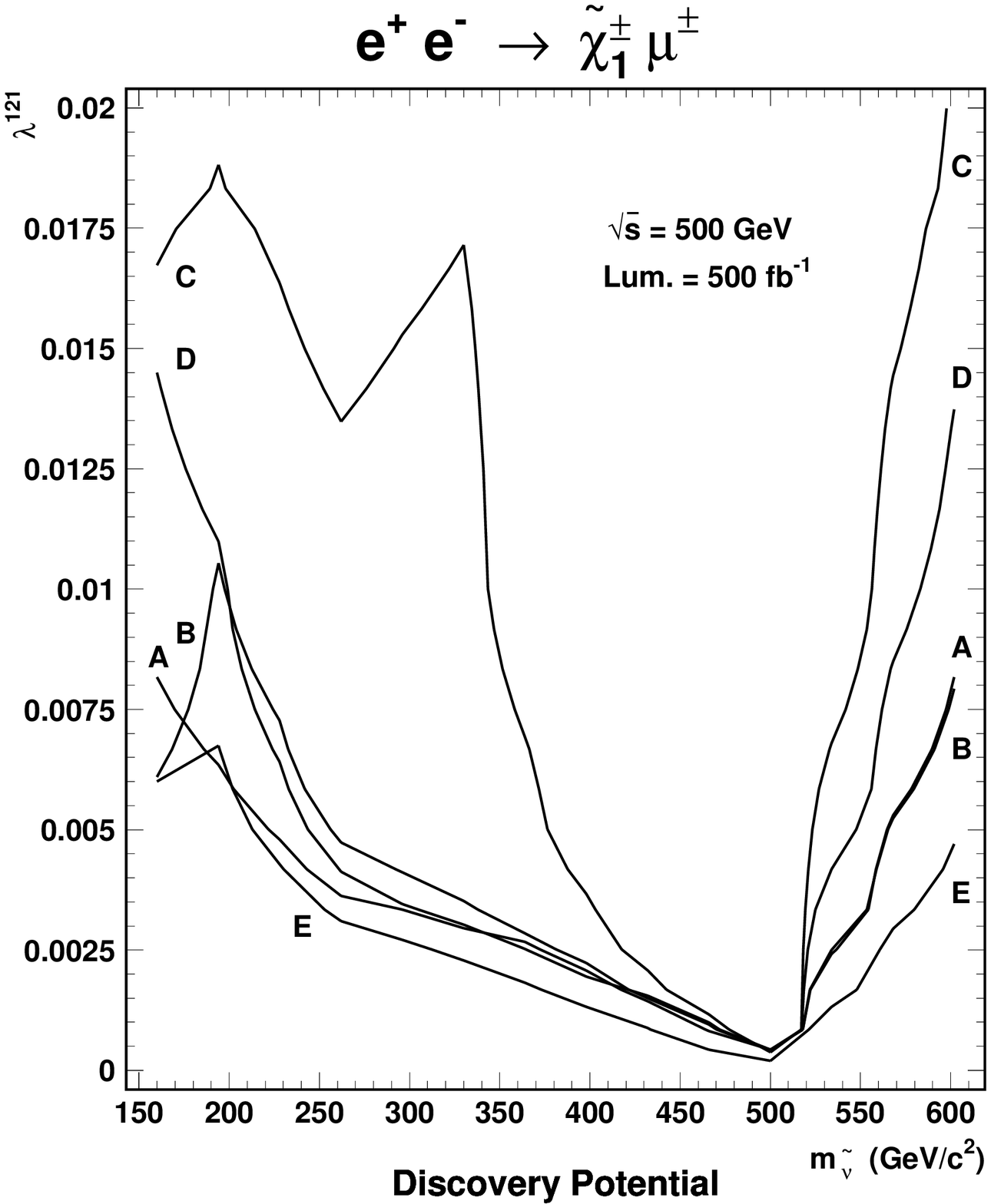,height=5.5in}}
\end{center}
\caption{\footnotesize  \it
Potentiel de d\'ecouverte \`a $5 \sigma$ dans le plan
$\l_{121}$ versus $m_{\tilde \nu}$ (in $GeV/c^2$) 
pour une luminosit\'e de ${\cal L}=500 fb^{-1}$ et une \'energie
dans le centre de masse de $\sqrt s=500 GeV$. Les points A, B, C, 
D et E correspondent aux points de l'espace des param\`etres du 
Mod\`ele Standard Supersym\'etrique Minimal (MSSM) d\'efinis par, 
A: $M_1=200GeV$, $M_2=250GeV$, $\mu=150GeV$, $\tan \beta=3$,
$m_{\tilde l^{\pm}}=300GeV$, $m_{\tilde q}=600GeV$
($m_{\tilde \chi^{\pm}_1}=115.7GeV$, $m_{\tilde
\chi^0_1}=101.9GeV$, $m_{\tilde \chi^0_2}=154.5GeV$);
B: $M_1=100GeV$, $M_2=200GeV$, $\mu=600GeV$, $\tan \beta=3$, 
$m_{\tilde l^{\pm}}=300GeV$, $m_{\tilde q}=600GeV$
($m_{\tilde \chi^{\pm}_1}=189.1GeV$, 
$m_{\tilde \chi^0_1}=97.3GeV$, $m_{\tilde \chi^0_2}=189.5GeV$);
C: $M_1=100GeV$, $M_2=400GeV$, $\mu=400GeV$, $\tan \beta=3$, 
$m_{\tilde l^{\pm}}=300GeV$, $m_{\tilde q}=600GeV$
($m_{\tilde \chi^{\pm}_1}=329.9GeV$, 
$m_{\tilde \chi^0_1}=95.5GeV$, $m_{\tilde \chi^0_2}=332.3GeV$);
D: $M_1=150GeV$, $M_2=300GeV$, $\mu=200GeV$, $\tan \beta=3$, 
$m_{\tilde l^{\pm}}=300GeV$, $m_{\tilde q}=600GeV$
($m_{\tilde \chi^{\pm}_1}=165.1GeV$, 
$m_{\tilde \chi^0_1}=121.6GeV$, $m_{\tilde \chi^0_2}=190.8GeV$);
E: $M_1=100GeV$, $M_2=200GeV$, $\mu=600GeV$, $\tan \beta=3$, 
$m_{\tilde l^{\pm}}=150GeV$, $m_{\tilde q}=600GeV$
($m_{\tilde \chi^{\pm}_1}=189.1GeV$, 
$m_{\tilde \chi^0_1}=97.3GeV$, $m_{\tilde \chi^0_2}=189.5GeV$).
\rm \normalsize }
\label{4:reach1}
\end{figure}

Nous pr\'esentons dans la Figure \ref{4:reach1} la sensibilit\'e 
obtenue dans le plan $\l_{121}$ versus $m_{\tilde \nu}$ gr\^ace 
\`a l'\'etude de la production simple de $\tilde \chi^{\pm}_1$ 
bas\'ee sur la signature $4l+\Eslash$ aux collisionneurs 
lin\'eaires \cite{PubliF}. Afin d'obtenir les r\'esultats 
pr\'esent\'es dans la Figure \ref{4:reach1} la production simple 
de chargino ainsi que le bruit de fond provenant de la production
de paire de superpartenaires ont \'et\'e simul\'es avec la
nouvelle version \cite{4:susynew}
du g\'en\'erateur d'\'ev\`enements SUSYGEN 
\cite{4:SUSYGEN} incluant les effets de polarisation des 
faisceaux incidents. Lors de la g\'en\'eration des \'ev\'enements,
les coupures et polarisations mentionn\'ees ci-dessus ont \'et\'e
appliqu\'ees. Les sensibilit\'es dans le plan 
$\l_{121}$ versus $m_{\tilde \nu}$ pr\'esent\'ees dans la 
Figure \ref{4:reach1} ont \'et\'e obtenues pour des points 
caract\'eristiques de l'espace des param\`etres du 
Mod\`ele Standard Supersym\'etrique Minimal (MSSM). 
Ces points ont \'et\'e choisis tels que la masse du 
$\tilde \chi^0_1$ soit proche de la limite actuelle 
($m_{\tilde \chi^0_1}>52GeV$ pour $\tan \beta =20$ dans le 
cadre d'un
mod\`ele ayant une constante de couplage $\l_{ijk}$ non nulle
\cite{4:DELPHIb}) afin de maximiser la 
\sef de la production de paires de neutralinos qui repr\'esente 
la principale source de bruit de fond issue de la production
de paires de superpartenaires.

Les sensibilit\'es sur la constante de couplage $\l_{121}$ 
pr\'esent\'ees dans la Figure \ref{4:reach1} repr\'e\-sentent une
am\'elioration par rapport \`a la borne indirecte, 
$\l_{121}<0.05 (m_{\tilde e_R}/100GeV)$ \cite{4:Bhatt}, dans un
intervalle de $\sim 500 GeV$ autour de la r\'esonance du 
sneutrino, et atteignent des valeurs de l'ordre de $10^{-4}$ 
au p\^ole du sneutrino. La sensibilit\'e sur la
constante de couplage $\l_{121}$ qui pourra \^etre
obtenue aupr\`es des futurs collisionneurs lin\'eaires 
am\'eliorera donc nettement les r\'esultats de LEP II
\cite{4:ALEPHa,4:DELPHIa,4:DELPHIb,4:Fab1,4:Fab2}
(voir ci-dessus).

Par ailleurs, les sensibilit\'es obtenues via la production 
simple de chargino aux collisionneurs lin\'eaires et 
pr\'esent\'ees dans la Figure \ref{4:reach1} 
repr\'esentent notamment une am\'elioration par rapport \`a la  
limite exp\'erimentale actuelle sur la masse du sneutrino, 
$m_{\tilde \nu}>78 GeV$ (dans le cadre d'un mod\`ele ayant une 
constante de couplage $\l_{ijk}$ non nulle) \cite{4:DELPHIb}.
De plus, dans le domaine $m_{\tilde \nu}>\sqrt s /2$, le 
sneutrino peut \^etre produit comme une r\'esonance 
aux collisionneurs lin\'eaires alors qu'il 
ne peut \^etre produit par paires. Cela signifie que dans ce 
domaine cin\'ematique, la production simple de chargino
aux collisionneurs lin\'eaires, qui 
re\c{c}oit sa principale contribution de la production 
r\'esonante de sneutrino, permet de tester la masse du 
sneutrino, comme l'illustre la Figure \ref{4:reach1} pour 
$\sqrt s = 500 GeV$, \`a l'inverse de la production de paires
de particules supersym\'etriques.

Enfin, notons que la constante de couplage $\l_{121}$ peut
aussi \^etre test\'ee par l'\'etude de la contribution 
des interactions \rpv \`a la diffusion Bhabha $e^+ e^- \to 
e^+ e^-$ \cite{4:Chou}. Cette contribution implique
l'\'echange dans les voies $s$ et $t$ d'un sneutrino.
Les sensibilit\'es obtenues sur la constante de couplage 
$\l_{121}$ par l'\'etude de cette contribution dans le cadre
de la physique au LEP sont de l'ordre
de $10^{-2}$ \`a la r\'esonance du sneutrino 
\cite{4:Chou}. Or ces sensibilit\'es sont inf\'erieures 
\`a celles obtenues sur la m\^eme constante de couplage 
$\l_{121}$ par l'\'etude de la production simple de chargino
au LEP \cite{4:ALEPHa,4:DELPHIa,4:DELPHIb,4:Fab1,4:Fab2}
(voir ci-dessus).

\subsubsection{Reconstruction de masse}

L'\'energie du muon $E(\mu)$ produit dans la r\'eaction de type 
$2 \to 2-corps$, $e^+ e^- \to \tilde \chi^{\pm}_1 \mu^{\mp}$, 
est une fonction de l'\'energie dans le centre de masse 
$\sqrt s$ et de la masse du muon $m_{\mu}$ et du chargino 
$m_{\tilde \chi_1^{\pm}}$: 
\begin{eqnarray}
E(\mu)= {s + m^2_{\mu} - m^2_{\tilde \chi_1^{\pm}} \over 
2 \sqrt s}.
\label{4:enl}
\end{eqnarray}
Par cons\'equent, la masse du 
$\tilde \chi_1^{\pm}$ devrait pouvoir \^etre d\'eduite, 
via Eq.(\ref{4:enl}), de l'\'energie du muon produit
avec le chargino dans les collisionneurs lin\'eaires. 
En fait, \`a cause de l'ISR, un photon est irradi\'e par
l'\'etat initial de telle sorte que la production simple de 
chargino doit \^etre \'etudi\'ee comme la r\'eaction de type 
$2 \to 3 \ corps$, $e^+ e^- \to \tilde \chi^{\pm}_1 \mu^{\mp}
\gamma$. La relation \ref{4:enl} n'est donc pas v\'erifi\'ee. 
Cependant, la masse du $\tilde \chi_1^{\pm}$ peut \^etre 
d\'eduite de l'\'energie du muon dans le centre de masse du 
sneutrino $E^\star(\mu)$ et de la masse 
du sneutrino $m_{\tilde \nu}$ et du muon $m_{\mu}$ par la 
relation \cite{PubliF},  
\begin{eqnarray}
E^\star(\mu)=
{m^2_{\tilde \nu} + m^2_{\mu} - m^2_{\tilde \chi_1^{\pm}}
\over 2 m_{\tilde \nu}}.
\label{4:enmu}
\end{eqnarray}
En effet, la masse du sneutrino peut \^etre d\'eduite de 
l'\'etude du signal $4l+\Eslash$ aux collisionneurs lin\'eaires
en effectuant un scan sur l'\'energie dans le centre de masse 
afin de d\'eterminer la valeur de $\sqrt s$
correspondant au maximum de la \sef de production simple
du chargino associ\'e \`a la r\'esonance du sneutrino 
\cite{PubliF}. De plus, l'\'energie du muon dans le centre de 
masse du sneutrino $E^\star(\mu)$ peut \^etre d\'etermin\'ee
gr\^ace \`a la distribution du moment transverse du muon produit
avec le chargino \cite{PubliF}. La masse du sneutrino pouvant 
\^etre d\'etermin\'ee avec une erreur de $\pm 3.5 GeV$ 
par un scan 
sur l'\'energie dans le centre de masse aux collisionneurs 
lin\'eaires, l'erreur sur la masse du $\tilde \chi_1^{\pm}$ 
reconstruite par la m\'ethode d\'ecrite ci-dessus peut atteindre 
$\pm 5.9 GeV$ \cite{PubliF}. 
La masse du $\tilde \chi_2^{\pm}$ peut aussi \^etre 
reconstruite par la m\'ethode d\'ecrite ci-dessus \cite{PubliF}.

En conclusion, aux collisionneurs lin\'eaires, la production 
simple de chargino permet d'obtenir facilement une grande 
pr\'ecision sur la reconstruction de masse du $\tilde \nu$ et du 
$\tilde \chi_1^{\pm}$ \cite{PubliF} \`a l'inverse de la 
production de paires de particules supersym\'etriques 
\cite{4:Godbole}.
Ceci est d\^u au fait que la production simple de particule SUSY 
ne g\'en\`ere qu'une seule cascade de d\'esint\'egration de 
superpartenaires alors que la production de paire en
g\'en\`ere deux ce qui complique l'identification de l'origine 
des particules de l'\'etat final et augmente ainsi 
le bruit de fond combinatoire. De plus, dans le cadre d'un 
mod\`ele ayant une constante de couplage $\l$ non nulle, si
le $\tilde \chi^0_1$ est la LSP (comme c'est le cas dans la 
plupart des mod\`eles supersym\'etriques) les deux cascades de 
d\'esint\'egration issues de la production de paires de 
superpartenaires se terminent par la d\'esint\'egration 
$\tilde \chi^0_1 \to l \bar l \nu$. Or les neutrinos ainsi 
produits
g\'en\`erent de l'\'energie manquante dans l'\'etat final, ce qui
rend d\'elicat la reconstruction de masse du $\tilde \chi^0_1$ 
et donc des autres particules supersym\'etriques.
Notons par ailleurs 
que les reconstructions des masses du $\tilde \nu$ 
et du $\tilde \chi_1^{\pm}$ bas\'ees sur la production simple de 
chargino restent possibles tant que la distribution du moment 
transverse du muon produit avec le chargino n'est 
pas noy\'ee par le bruit de fond issu de la production de paires 
de particules supersym\'etriques (voir Figure \ref{4:reach1}) 
\cite{PubliF}.

\subsubsection{Extension \`a diff\'erentes constantes
de couplage \rpv}

La production simple de chargino dans les collisions $e^+ e^-$
peut uniquement avoir lieu par les interactions
$\l_{121}$ et $\l_{131}$, \`a cause de l'antisym\'etrie de la
constante de couplage $\l_{ijk}$ (voir Chapitre \ref{antisymetrie}). 
La production simple de chargino via $\l_{131}$ correspond
\`a la production d'un lepton tau: 
$e^+ e^- \to \tilde \chi^{\pm} \tau^{\mp}$ (voir Figure 
\ref{4:graphesp}(a)). Du fait de la nature instable du lepton tau,
la sensibilit\'e sur la constante de couplage $\l_{131}$
obtenue via la production simple de chargino devrait \^etre moins
forte que celle sur $\l_{121}$ (voir Figure \ref{4:reach1})
car les coupures mentionn\'ees dans la Section \ref{potde} sont
moins efficaces dans le cas de l'\'etude de la constante
$\l_{131}$ \cite{PubliF}. D'autre part, les d\'esint\'egrations
du lepton tau rendent les reconstructions de 
masse du $\tilde \nu$ et du $\tilde \chi_1^{\pm}$ difficiles 
\cite{PubliF}.

La production simple de chargino pourrait aussi avoir lieu 
dans les futurs collisionneurs muoniques \cite{4:muon2,4:muon} 
via la constante de couplage
$\l_{2J2}$ ($J=1,3$): $\mu^+ \mu^- \to \tilde \chi^{\pm} 
l_J^{\mp}$ (voir Figure \ref{4:graphesp}(a)).
L'\'etude des constantes de couplage $\l_{212}$ et $\l_{232}$ 
bas\'ee sur la production simple de chargino aux collisionneurs 
muoniques est identique, \`a quelques d\'etails pr\`es
(voir \cite{PubliF}), 
\`a l'\'etude des constantes $\l_{121}$ et $\l_{131}$ 
via la production simple de chargino aux collisionneurs 
lin\'eaires. Par cons\'equent, les sensibilit\'es 
sur les constantes de couplage $\l_{212}$ et 
$\l_{232}$ obtenues aux collisionneurs muoniques devraient \^etre
du m\^eme ordre de grandeur que celles attendues sur les 
constantes $\l_{121}$ et $\l_{131}$ 
aux collisionneurs lin\'eaires. 
De m\^eme, les r\'esultats sur les reconstructions de masse
bas\'ees sur la production simple de chargino devraient \^etre
comparables aux collisionneurs lin\'eaires et aux collisionneurs 
muoniques \cite{PubliF}.

Par ailleurs, notons que les productions simples 
de sleptons $\gamma e^{\pm} \to
e^{\pm} \tilde \nu$ et $\gamma e^{\pm} \to \tilde e^{\pm} \nu$
aux collisionneurs lin\'eaires
permettront de tester les constantes de couplage $\l_{121}$,
$\l_{131}$, $\l_{122}$, $\l_{123}$, $\l_{132}$, $\l_{133}$,
$\l_{231}$, $\l_{232}$ et $\l_{233}$ \cite{4:All,4:Allex}.
Cependant, la production r\'esonante de sneutrino dans les 
collisions $e^+ e^-$ est davantage sensible aux constantes de 
couplage $\l_{121}$ et $\l_{131}$ que ces productions simples de 
sleptons \cite{4:All}.

%
%
%
%
%
%
%
%
%

\clearpage

\vspace{10 mm}
\begin{center}
{  }
\end{center}
\vspace{10 mm}

\clearpage

\chapter{Contributions des interactions \rpv aux taux de 
changement de saveur et aux asym\'etries li\'ees \`a 
la violation de la sym\'etrie CP}
\label{5:bug}

\section{Motivations}
\label{5:mot2}

Comme nous l'avons discut\'e dans le Chapitre \ref{chaRPV} et
la Section \ref{4:mot}, d'un point de vue th\'eorique aussi 
bien que ph\'enom\'enologique, les mod\`eles avec violation et 
conservation de la sym\'etrie de R-parit\'e doivent \^etre 
trait\'es avec la m\^eme attention vis-\`a-vis de la 
ph\'enom\'enologie de la supersym\'etrie.

Dans l'hypoth\`ese o\`u certains couplages \rpv ne seraient pas 
nuls, ces couplages pourraient avoir des phases complexes
et constituer ainsi de nouvelles sources ind\'ependantes de 
violation de la sym\'etrie CP. Notons que m\^eme si les couplages
\rpv existant avaient des phases complexes nulles, ils 
pourraient conduire en combinaison avec d'autres sources de
violation de CP du MSSM \`a de nouveaux tests de violation de CP
\cite{4:29,4:31,4:32,4:38}.

Dans un second travail, 
nous avons d\'evelopp\'e des \'etudes 
\cite{PubliG} (voir Publication VII: 
{\it ``Broken R-parity 
contributions to flavor changing rates and CP asymmetries 
in fermion pair production at leptonic colliders''})
\cite{PubliH} (voir Publication VIII: {\it ``Polarized 
single top production at leptonic colliders from broken 
R-parity interactions incorporating CP violation''})
\cite{PubliI} (voir Publication IX: {\it ``CP violation 
flavor asymmetries in slepton pair production at leptonic 
colliders from broken R-parity''})
\cite{PubliIp} permettant de mettre en \'evidence
la partie imaginaire des constantes de couplage \rpv
par des effets de violation de la sym\'etrie CP
dans le cadre de la physique des collisionneurs
de haute \'energie. L'id\'ee de contributions des interactions 
\rpv \`a des effets de violation de CP a d\'ej\`a motiv\'e de 
nombreuses \'etudes dans le cadre de la physique de basse 
\'energie \cite{4:29,4:31,4:32,4:33}.

Les effets de violation de la sym\'etrie CP offrent un cadre
propice \`a l'\'etude des interactions \rpv et plus 
g\'en\'eralement de toute autre physique au-del\`a du 
Mod\`ele Standard Supersym\'etrique Minimal (MSSM).
La raison est que les contributions des interactions du \ms
aux effets de violation de CP sont faibles car elles
impliquent des diagrammes \`a l'ordre des boucles et 
sont r\'eduites par la quasi d\'eg\'en\'erescence existant entre
les masses des quarks et des leptons.
(voir par exemple l'\'etude des asym\'etries de 
violation de CP li\'ees \`a la d\'esint\'egration des bosons de 
jauge $W^{\pm}$ et $Z^0$ \cite{4:48,4:49}).
De plus, les contributions des interactions du MSSM
aux effets de violation de CP 
(voir par exemple l'\'etude des asym\'etries de 
violation de CP li\'ees \`a la production de paires de sleptons 
\cite{4:ArkaniCP}) sont contraintes par des
donn\'ees exp\'erimentales sur la physique de basse \'energie 
\cite{4:Xconference}.
Dans la plupart des mod\`eles de physique sous-jacente au 
MSSM, des effets importants de violation de CP sont pr\'edits
\cite{4:44,4:52}). Cependant, dans les diff\'erentes \'etudes
effectu\'ees, nous avons suppos\'e que les contributions 
aux effets de violation de CP des interactions \rpv \'etaient 
dominantes.

Nous avons \'etudi\'e les r\'eactions de 
production de paires de fermions \cite{PubliG,PubliH} et de 
sfermions \cite{PubliI} de saveurs diff\'erentes 
dans les collisionneurs leptoniques: 
\begin{eqnarray}
l^+ l^- \to f_J \bar f_{J'} \ \ [J \neq J'], 
\label{chac1}
\end{eqnarray}
\begin{eqnarray}
l^+ l^- \to \tilde f_J \tilde  f^\star_{J'} \ \ [J \neq J'].
\label{chac2}
\end{eqnarray} 
Ces r\'eactions simples et tr\`es \'etudi\'ees ainsi que
l'environnement `propre' des collisionneurs leptoniques offrent
un cadre propice aux tests de violation de CP.

Dans \cite{PubliG,PubliH}, nous avons calcul\'e les asym\'etries 
de violation de CP,
\begin{eqnarray}
{\cal A}_{JJ'}={ \vert M^{JJ'} \vert ^2
- \vert \bar  M^{JJ'} \vert ^2 \over 
\vert M^{JJ'} \vert ^2+ \vert \bar  M^{JJ'} \vert ^2},
\label{chac3}
\end{eqnarray} 
associ\'ees aux contributions des interactions \rpv 
\`a la r\'eaction de Eq.(\ref{chac1}). Dans
Eq.(\ref{chac3}), $M^{JJ'}$ ($\bar M^{JJ'}$) d\'esigne l'amplitude
de la r\'eaction $l^+ l^- \to f_J \bar f_{J'}$   
($l^+ l^- \to f_{J'} \bar f_J$) $[J \neq J']$. $M^{JJ'}$ et
$\bar M^{JJ'}$ sont donc les amplitudes de processus CP
conjugu\'es. Notons que certaines asym\'etries de violation de CP
associ\'ees \`a des r\'eactions impliquant des couplages \rpv 
peuvent \^etre fonction du spin des particules de l'\'etat final
\cite{4:35,4:36}.

Nous avons donc \'et\'e amen\'e \`a \'etudier dans 
\cite{PubliG,PubliH} 
les contributions des interactions \rpv \`a la r\'eaction
changeant la saveur $l^+ l^- \to f_J \bar f_{J'} \ [J \neq J']$ 
\cite{PubliG,PubliH}. Or l'\'etude de ces contributions a aussi 
permis de d\'evelopper un test de l'intensit\'e des couplages 
\rpi. En effet, les contributions des interactions du \ms
aux effets de changement de saveur sont faibles. En particulier,
les contributions des interactions du \ms aux courants neutres 
changeant la saveur sont peu significatives puisqu'elles 
impliquent des diagrammes \`a l'ordre des boucles 
\cite{4:48,4:49}. Les contributions des interactions du \ms
\`a la d\'esint\'egration du boson de jauge $Z^0$ 
en une paire de quarks de diff\'erentes g\'en\'erations sont
particuli\`erement faibles. La raison est que dans le 
lagrangien effectif du Mod\`ele Standard, le couplage 
$Z^0 \bar q_J q_{J'}$ entre le boson de jauge $Z^0$ 
et une paire de quarks de diff\'erentes g\'en\'erations a une 
structure du type $\sum_i V^\star _{iJ} V_{iJ'}  
f(m^2_i/m^2_{Z^0})$, o\`u les $V_{ij}$ sont les \'el\'ements de la
matrice Cabibbo-Kobayashi-Maskawa et $f(m^2_i/m^2_{Z^0})$ est
une fonction issue d'un calcul de boucle et d\'ependant des
masses du boson de jauge $Z^0$ et des quarks propag\'es dans 
la boucle. Le couplage $Z^0 \bar q_J q_{J'}$ est donc fortement 
supprim\'e \`a cause de la 
d\'eg\'en\'erescence des masses des quarks relativement \`a la 
masse du boson de jauge $Z^0$ (valide pour tous les quarks 
\`a l'exception du quark top) et de la propri\'et\'e 
d'unitarit\'e de la matrice Cabibbo-Kobayashi-Maskawa.
De plus, les contributions des interactions du MSSM
aux effets de changement de saveur sont contraintes par des
donn\'ees exp\'erimentales sur la physique de basse \'energie 
\cite{4:Xconference}. Plus pr\'ecis\'ement, ces contributions
sont contraintes par une d\'eg\'en\'erescence des masses des 
particules scalaires supersym\'etriques (qui sont issues des 
termes de brisure douce de SUSY) ou par un alignement entre les 
matrices de transformation de la base de courant \`a la base 
de masse des fermions du \ms et de leur partenaire scalaire.
Les effets de changement de saveur offrent donc un cadre
propice \`a l'\'etude des interactions \rpv mais aussi de toute
autre physique au-del\`a du MSSM. 
La majorit\'e des th\'eories au-del\`a du MSSM pr\'evoit
des effets de changement de saveur significatifs
\cite{4:48,4:49,4:51}. Nous avons
n\'eanmoins travaill\'e dans l'hypoth\`ese selon laquelle les 
principales contributions aux effets de changement de saveur sont 
d\^ues aux couplages \rpvi.

Dans \cite{PubliI}, nous avons calcul\'e les asym\'etries 
de violation de CP,
\begin{eqnarray}
\tilde {{\cal A}}_{JJ'}={ \vert \tilde M^{JJ'} \vert ^2
- \vert \bar {\tilde M}^{JJ'} \vert ^2 \over 
\vert \tilde M^{JJ'} \vert ^2
+ \vert \bar {\tilde M}^{JJ'} \vert ^2},
\label{chac4}
\end{eqnarray} 
associ\'ees aux contributions des interactions \rpv 
\`a la r\'eaction du type de Eq.(\ref{chac2}). Dans
Eq.(\ref{chac4}), $\tilde M^{JJ'}$ ($\bar {\tilde M}^{JJ'}$) 
d\'esigne l'amplitude de la r\'eaction $l^+ l^- \to 
\tilde e_{H J}  \tilde e^\star_{H J'}$  ($l^+ l^- \to 
\tilde e_{H J'}  \tilde e^\star_{H J}$) $[H=L,R; \ J \neq J']$. 
$\tilde M^{JJ'}$ et $\bar {\tilde M}^{JJ'}$ 
sont donc les amplitudes de processus CP conjugu\'es.

Nous avons donc \'et\'e amen\'e \`a \'etudier dans \cite{PubliI} 
les contributions des interactions \rpv \`a la r\'eaction
changeant la saveur $l^+ l^- \to \tilde e_{H J}  
\tilde e^\star_{H J'} \ [H=L,R; \ J \neq J']$ 
\cite{PubliG,PubliH}. Or l'\'etude de ces contributions a aussi 
permis de d\'evelopper un test de l'intensit\'e des couplages 
\rpi. En effet, les contributions des interactions du MSSM
aux effets de changement de saveur sont contraintes par des
donn\'ees exp\'erimentales sur la physique de basse \'energie 
\cite{4:Xconference}. Plus pr\'ecis\'ement, ces contributions
sont contraintes par une d\'eg\'en\'erescence des masses des 
particules scalaires supersym\'etriques (qui sont issues des 
termes de brisure douce de SUSY) ou par un alignement entre les 
matrices de transformation de la base de courant \`a la base 
de masse des fermions du \ms et de leur partenaire scalaire.
Les interactions \rpv pourraient donc apporter une contribution
\`a la r\'eaction changeant la saveur $l^+ l^- \to 
\tilde e_{H J} \tilde e^\star_{H J'} \ [H=L,R; \ J \neq J']$ qui 
soit significative relativement aux contributions des 
interactions du MSSM \cite{4:Arkani}.

\section{Production de paires de fermions}
\label{5:fer}

\subsection{Taux de changement de saveur}
\label{5:ferfcnc}

\begin{figure}[t]
\begin{center}
\leavevmode
\centerline{\psfig{figure=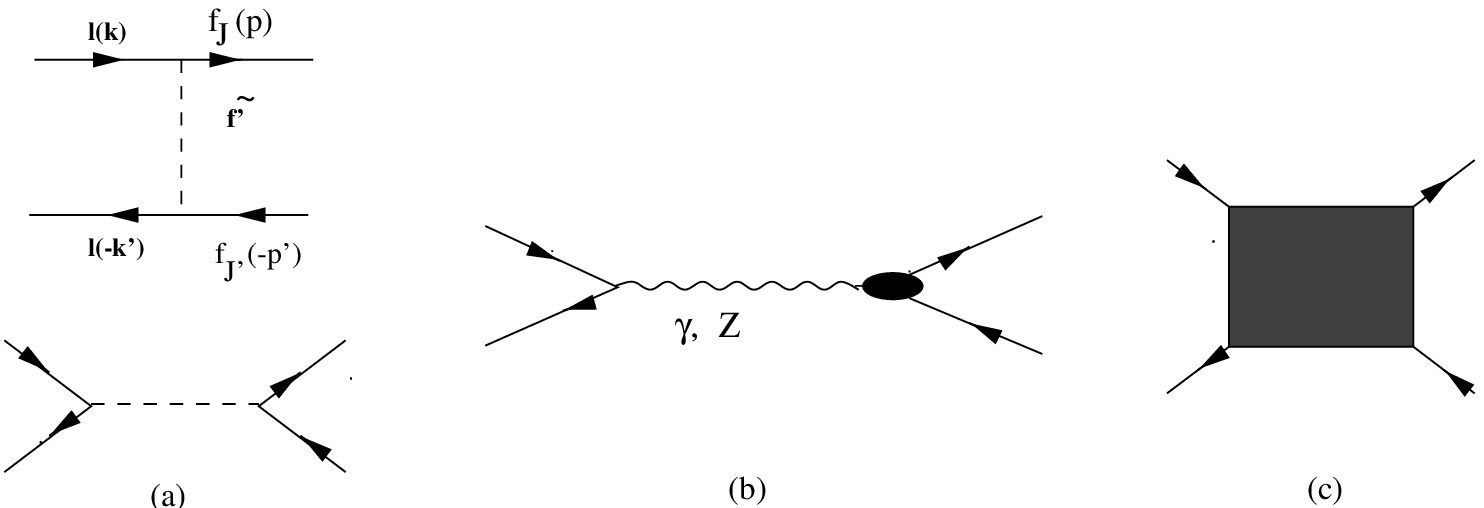,height=2.in}}
\end{center}
\caption{\footnotesize  \it Diagrammes de Feynman
des contributions des interactions \rpv \`a la r\'eaction 
$l^+ l^- \to f_J \bar f_{J'} \ [J \neq J']$.
$f$ d\'enote un fermion et $\tilde f$ un sfermion.
Les diagrammes de Feynman des contributions \`a l'ordre d'une
boucle au vertex $Z^0 f_J \bar f_{J'} \ [J \neq J']$
sont pr\'esent\'es dans la Figure \ref{5:cpcp}.
\rm \normalsize }
\label{5:cpc}
\end{figure}

\begin{figure}[t]
\begin{center}
\leavevmode
\centerline{\psfig{figure=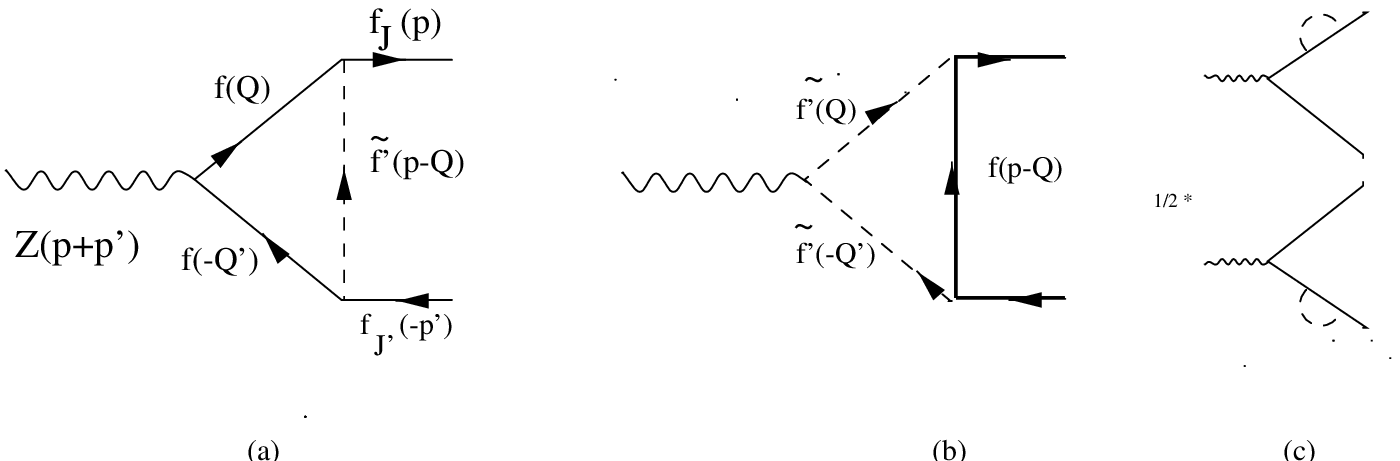,height=2.in}}
\end{center}
\caption{\footnotesize  \it Diagrammes de Feynman
des contributions des interactions \rpv \`a l'ordre d'une
boucle au vertex $Z^0 f_J \bar f_{J'} \ [J \neq J']$.
$f$ d\'enote un fermion et $\tilde f$ un sfermion.
\rm \normalsize }
\label{5:cpcp}
\end{figure}

Les graphes de Feynman des contributions des interactions \rpv
\`a la r\'eaction de Eq.(\ref{chac1}) sont pr\'esent\'es dans les 
Figures \ref{5:cpc} et \ref{5:cpcp}. 
Nous avons consid\'er\'e les contributions
des interactions \rpv \`a l'ordre d'une boucle pour des raisons
qui appara\^{\i}tront claires dans la Section \ref{5:acpf}.

Les contributions des interactions \rpv \`a la production de
paires de leptons charg\'es appartenant \`a 
des familles diff\'erentes
$l^+ l^- \to l^+_J l^-_{J'} \ [J \neq J']$ impliquent les
produits de constantes de couplage 
$\l_{i11} \l^{\star}_{iJJ'}$ (voie $s$),  
$\l_{iJ1} \l^{\star}_{iJ'1}$ (voie $t$) ou
$\l_{i1J} \l^{\star}_{i1J'}$ (voie $t$) au niveau en arbre 
(l'indice $i$ correspond \`a la saveur du sneutrino \'echang\'e) 
et $\l_{iJ'k} \l^{\star}_{iJk}$, $\l_{ijJ} \l^{\star}_{ijJ'}$ 
ou $\l'_{J'jk} \l^{\star \prime}_{Jjk}$ \`a l'ordre d'une boucle
(les indices $i$, $j$ et $k$ correspondent aux saveurs des 
fermions et sfermions \'echang\'es dans la boucle). \\
Les contributions des interactions \rpv \`a la production de
paires de quarks down appartenant \`a 
des familles diff\'erentes
$l^+ l^- \to d_J \bar d_{J'} \ [J \neq J']$ impliquent les
produits de constantes de couplage 
$\l_{i11} \l^{\prime \star}_{iJJ'}$ (voie $s$) ou  
$\l'_{1jJ} \l^{\prime \star}_{1jJ'}$ (voie $t$) au niveau en 
arbre 
(les indices $i$ et $j$ correspondent \`a la saveur du squark
up \'echang\'e) et $\l'_{iJ'k} \l^{\prime \star}_{iJk}$ 
ou $\l'_{ijJ} \l^{\prime \star}_{ijJ'}$ \`a l'ordre d'une boucle
(les indices $i$, $j$ et $k$ correspondent aux saveurs des 
fermions et sfermions \'echang\'es dans la boucle). \\
Les contributions des interactions \rpv \`a la production de
paires de quarks up appartenant \`a 
des familles diff\'erentes
$l^+ l^- \to u_J \bar u_{J'} \ [J \neq J']$ impliquent les
produits de constantes de couplage   
$\l'_{1J'k} \l^{\prime \star}_{1Jk}$ (voie $t$) au niveau en 
arbre 
(l'indice $k$ correspond \`a la saveur du squark
down \'echang\'e) et $\l'_{iJ'k} \l^{\prime \star}_{iJk}$ 
\`a l'ordre d'une boucle
(les indices $i$ et $k$ correspondent aux saveurs des 
fermions et sfermions \'echang\'es dans la boucle).

Nous notons les amplitudes des contributions des interactions \rpv
\`a la r\'eaction de Eq.(\ref{chac1}),
\begin{eqnarray}
M^{JJ'}=a_0^{JJ'}+\sum_\a a_\a ^{JJ'}  F^{JJ'}_\a (s+i\e ), 
\label{chac5bis} \\ 
\bar M^{JJ'}=a^{JJ' \star } _0+\sum_\a a^{JJ'\star } _\a  
F^{J'J} _\a (s+i\e ),
\label{chac5}
\end{eqnarray}
o\`u $a_0^{JJ'}$ repr\'esente l'amplitude des contributions des 
interactions \rpv au niveau en arbre et $\sum_\a a_\a ^{JJ'}  
F^{JJ'}_\a (s+i\e )$ l'amplitude des contributions des 
interactions \rpv \`a l'ordre d'une boucle. $a_0^{JJ'}$ est
proportionnel au produit de constantes de couplage \rpv impliqu\'e
par le processus au niveau en arbre et que l'on note 
$t_i^{JJ'}$, $i$ correspondant \`a la saveur du sfermion 
\'echang\'e au niveau en arbre. L'expression de $a_0^{JJ'}$ fait
intervenir une sommation sur l'indice $i$ dont d\'epend aussi la 
masse du sfermion \'echang\'e. De m\^eme, $a_\a ^{JJ'}$ est
proportionnel au produit de constantes de couplage \rpv impliqu\'e
dans la boucle et que l'on note $l_\a^{JJ'}$, $\a$ correspondant 
aux deux indices de saveur des fermions et sfermions \'echang\'es 
dans la boucle. Enfin, $F^{JJ'}_\a (s+i\e )$ est une fonction
issue d'un calcul de boucle ayant une partie imaginaire (voir
\cite{PubliG}) et d\'ependant notamment des masses des fermions 
et sfermions \'echang\'es dans la boucle et donc de $\a$.

\`A des \'energies dans le centre de masse sup\'erieures \`a la
masse du boson $Z^0$, les contributions dominantes des 
interactions \rpv \`a la r\'eaction de Eq.(\ref{chac1}) 
proviennent de processus \`a l'ordre des arbres (voir Figure 
\ref{5:cpc}). Les sections efficaces $\s_{JJ'}$ de ces 
contributions sont de l'ordre de \cite{PubliG},
\begin{eqnarray}
\s_{JJ'} \approx ({\L \over 0.1})^4 
({100 GeV \over \tilde m})^{2 \ - \ 3} (0.1 \ - \ 10) \  fbarns, 
\label{chac6}
\end{eqnarray}
si l'on suppose que toutes les masses des sfermions (toutes les
valeurs absolues des 
constantes de couplage \rpv de type $\l$, $\l'$ ou $\l''$) sont 
\'egales \`a une m\^eme valeur not\'ee $\tilde m$ ($\L$). \`A
la r\'esonance du sneutrino, les \sefs des contributions des
interactions \rpv \`a la r\'eaction $l^+ l^- \to l^+_J l^-_{J'} 
\ [J \neq J']$ peuvent atteindre 
$(\L / 0.1)^4 10^4 fbarns$.

Au p\^ole du boson de jauge $Z^0$, les canaux de 
d\'esint\'egration $Z \to f_J+\bar f_{J'} \ [J \neq J']$ 
impliquent des diagrammes \`a l'ordre des boucles (voir Figure 
\ref{5:cpcp}). Les rapports de branchement $B_{JJ'}$ de ces 
canaux de d\'esint\'egration s'\'ecrivent, 
\begin{eqnarray}
B_{JJ'}={\G (Z\to f_J+\bar f_{J'} )+ \G (Z\to f_{J'} +\bar f_J )
\over \G (Z \to all)  },
\label{chac7}
\end{eqnarray}
et sont de l'ordre de \cite{PubliG},
\begin{eqnarray}
B_{JJ'} \approx ({\L \over 0.1})^4 
({100 GeV \over \tilde m})^{2.5} \ 10^{-9}, 
\label{chac8}
\end{eqnarray}
si l'on suppose que toutes les masses des sfermions (toutes les 
valeurs absolues des
constantes de couplage \rpv de type $\l$, $\l'$ ou $\l''$) sont 
\'egales \`a une m\^eme valeur not\'ee $\tilde m$ ($\L$).

Les limites actuelles sur les constantes de couplage \rpv
sont typiquement de l'ordre de $10^{-1} (\tilde m / 100GeV)$,
$\tilde m$ \'etant la masse des sfermions \cite{4:Bhatt}. Par 
cons\'equent, les \sefs de Eq.(\ref{chac6}) peuvent \^etre 
sup\'erieures \`a $\sim 10 fbarns$ et les rapports de branchement 
de Eq.(\ref{chac8}) sup\'erieurs \`a $\sim 10^{-9}$.

\'Etant donn\'e les \sefs $\s_{JJ'}$ de Eq.(\ref{chac6}), 
les contributions des interactions \rpv \`a la r\'eaction de 
Eq.(\ref{chac1}) hors du p\^ole du boson de jauge $Z^0$
sont potentiellement observables \`a LEP II ainsi qu'aux 
futurs collisionneurs lin\'eaires pour lesquels les luminosit\'es
attendues sont de l'ordre de $500 fb^{-1}$ \cite{4:Tesla,4:NLC}.

D'apr\`es les rapports de branchement $B_{JJ'}$ de 
Eq.(\ref{chac8}), les limites exp\'erimentales 
actuelles sur les rapports de branchement du boson de jauge 
$Z^0$, $B(Z \to \bar e \mu +
\bar \mu e)<1.7 \ 10^{-6}$,  $B(Z \to \bar e \tau +\bar \tau e)<
9.8 \ 10^{-6} $ et $ B(Z \to \bar \mu \tau +\bar \tau  \mu )<
1.7 \  10^{-5} $ \cite{4:67}, imposent les bornes suivantes sur 
les produits de constantes de couplage \rpv \cite{PubliG}, 
\begin{eqnarray}
\l_{ijJ} \l_{ijJ'}^\star < [0.46,\ 1.1, \ 1.4 ] \ \  pour \ \  
[JJ'= 12,\ 23,\ 13], 
\label{chac9bis} \\
\l'_{J'jk} \l^{\prime \star } _{Jjk} < [0.38,\ 0.91, \ 1.2 ]
\times  10^{-1} \ \  pour \ \  [JJ'= 12,\ 23,\ 13],
\label{chac9}
\end{eqnarray}
dans l'hypoth\`ese d'un produit de constantes de couplage \rpv
dominant et pour une masse des sfermions de $\tilde m=100GeV$.
Les bornes de Eq.(\ref{chac9bis},\ref{chac9}) 
seront am\'elior\'ees dans le cadre
de la physique aux collisionneurs lin\'eaires et pourraient
m\^eme devenir plus fortes que les limites indirectes actuelles
\cite{4:Bhatt}.

Les rapports de branchement de Eq.(\ref{chac8}) sont du m\^eme 
ordre de grandeur que les rapports de branchement des 
d\'esint\'egrations hadroniques du boson de jauge $Z^0$ 
calcul\'ees dans le cadre du \ms qui valent, 
$B(Z \to \bar b s + \bar s b)=10^{-7}$, 
$B(Z \to \bar b d+\bar d b)=10^{-9}$ et 
$B(Z \to \bar s d+\bar d s)=10^{-11}$ 
\cite{4:48,4:49}.


\vspace{5 mm}

\underline{\bf {PRODUCTION SIMPLE DE QUARK TOP:}}

\vspace{5 mm}

Dans le secteur des quarks, les effets de 
changement de saveur li\'es \`a la r\'eaction $l^+ l^- 
\to q_J \bar q_{J'} \ [J \neq J']$ sont difficilement
d\'etectables \`a cause de la mauvaise identification 
exp\'erimentale de la saveur des quarks. Dans ce contexte, la 
production simple de quark top $l^+ l^- 
\to t \bar c \ / \ \bar t c$ para\^{\i}t int\'eressante car 
le quark top donne lieu \`a une signature caract\'eristique. 
En effet, de par sa grande masse, le quark top poss\`ede
un temps de vie $\tau_{top} = [1.56 \ GeV ({m_{top} \over 180 
\ GeV})^3]^{-1}$ qui est plus court que le temps typique 
d'hadronisation $1/ \L _{QCD}$. Par cons\'equent, le quark top,
une fois produit, se d\'esint\'egre, et son principal canal de
d\'esint\'egration est $t \to b W^{\pm}$. Le quark top 
donne donc lieu \`a une signature claire de type $b l \nu$ 
lorsque le boson de jauge $W^{\pm}$ se d\'esint\`egre en leptons.

\begin{figure}[t]
\begin{center}
\leavevmode
\centerline{\psfig{figure=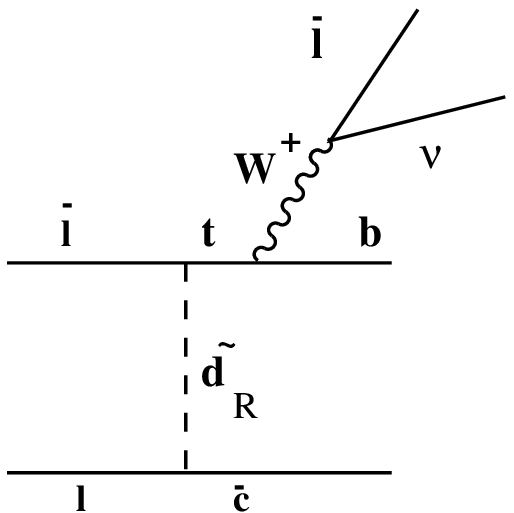,height=3.in}}
\end{center}
\caption{\footnotesize  \it Diagramme de Feynman de
l'amplitude au niveau en arbre
de la contribution des interactions \rpv au processus
$l^+l^- \to \bar c t \to \bar c b \bar l \nu$.
$\tilde d_R$ d\'esigne un squark down droit.
\rm \normalsize }
\label{5:topfey}
\end{figure}

\begin{figure}[t]
\begin{center}
\leavevmode
\centerline{\psfig{figure=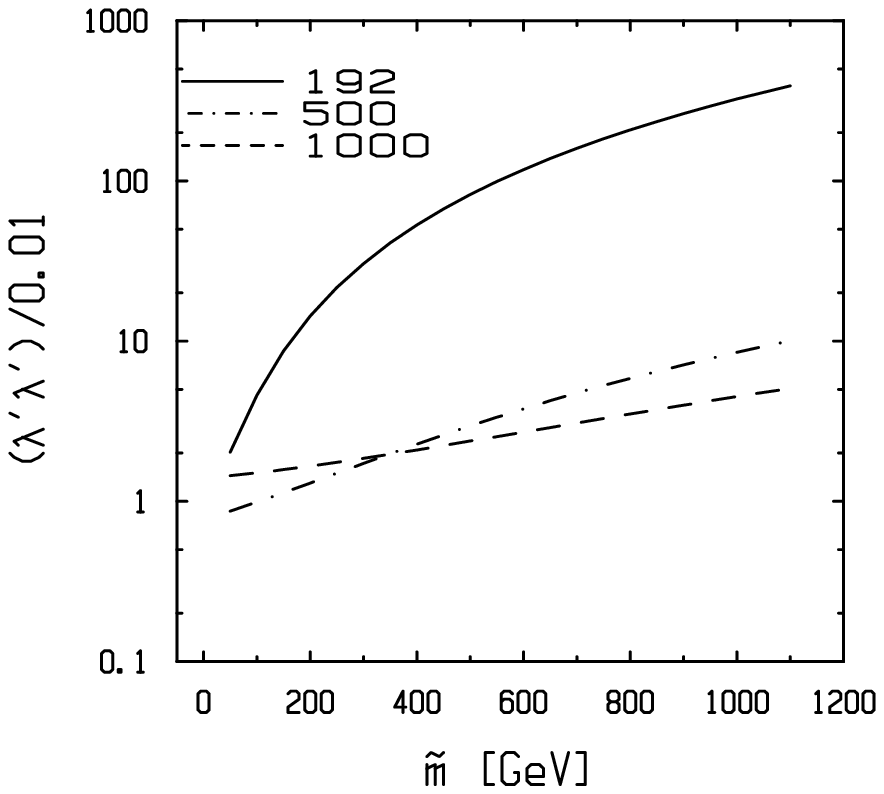,height=4.in}}
\end{center}
\caption{\footnotesize  \it 
Sensibilit\'e sur le produit de constantes de couplage \rpv  
$\l'_{12k} \l^{\prime \star}_{13k} / 0.01$ en fonction de 
la masse du squark down droit $\tilde m$ obtenue par l'\'etude
de la signature $c b l \nu, \ [l=e,\mu]$ \`a des \'energies
dans le centre de masse $s^\ud =[192. \ , 500 , \ 1000 ] \ GeV$ 
et pour les luminosit\'es associ\'ees
${\cal L} = [2., \ 100., \ 100. ]\ fb^{-1}$. Les r\'egions se
situant au-dessus des courbes correspondent \`a une exclusion \`a
$95 \% C.L.$, c'est \`a dire que ces domaines sont tels que 
$S/ \sqrt {S + B}>3$ o\`u $S$
repr\'esente le nombre d'\'ev\`enements du signal \rpv et 
$B$ le nombre d'\'ev\`enements du bruit de fond issu du
Mod\`ele Standard.
\rm \normalsize }
\label{5:topot}
\end{figure}

Dans \cite{PubliH}, nous avons \'etudi\'e l'\'etat final
$c b l \nu, \ [l=e,\mu]$ issu de la contribution des interactions
$\l'_{12k}$ et $\l^{\prime \star}_{13k}$ ($k$ correspondant \`a
l'indice de g\'en\'eration du squark down droit $\tilde d_{k R}$
\'echang\'e dans la voie $t$) \`a la r\'eaction 
$l^+ l^- \to c t \to c b l \nu, \ [l=e,\mu]$ 
(voir Figure \ref{5:topfey}).
Nous avons calcul\'e des distributions de variables 
cin\'ematiques permettant d'appliquer des coupures r\'eduisant 
le bruit de fond issu du \ms et associ\'e \`a la signature 
$c b l \nu, \ [l=e,\mu]$. Ce bruit de fond
provient principalement de la r\'eaction $l^+ l^- \to W^\pm W^\mp
\to c b l \nu, \ [l=e,\mu]$. 
Nous avons obtenu une efficacit\'e pour les
coupures cin\'ematiques mentionn\'ees ci-dessus 
de $0.8$ pour le signal \rpv et de $3 \ 10^{-3}$ pour 
le bruit de fond issu du Mod\`ele Standard. Ces efficacit\'es
d\'ependent faiblement de l'\'energie dans le centre de masse et 
de la masse du $\tilde d_{k R}$. Bas\'es sur ces efficacit\'es, 
nous avons calcul\'e le potentiel de d\'ecouverte associ\'e \`a
l'analyse de la signature $c b l \nu, \ [l=e,\mu]$. 
Dans la Figure 
\ref{5:topot}, nous pr\'esentons ce potentiel de d\'ecouverte 
dans le plan $\l'_{12k} \l^{\prime \star}_{13k}/0.01$ versus la 
masse du $\tilde d_{k R}$. La Figure \ref{5:topot} montre que
l'\'etude de la r\'eaction 
$l^+ l^- \to c t \to c b l \nu, \ [l=e,\mu]$
aupr\`es des futurs collisionneurs lin\'eaires, dont les 
luminosit\'es devraient \^etre de l'ordre de $10^2 fb^{-1}$
\cite{4:Tesla,4:NLC}, permettra d'am\'eliorer les limites
sur (ou bien de d\'etecter) les produits de constantes de couplage 
$\l'_{121} \l^{\prime \star}_{131}$ et $\l'_{122} \l^{\prime \star}_{132}$. 
Effectivement, la borne indirecte actuelle sur le produit 
$\l'_{121} \l^{\prime \star}_{131}$ ($\l'_{122} \l^{\prime \star}_{132}$) 
est $\l'_{121} \l'_{131}<1.225 \ 10^{-3}(m_{\tilde s_L}/100GeV)(m_{\tilde b_L}/100GeV)$
($\l'_{122} \l'_{132}<6.8 \ 10^{-3} (m_{\tilde s} / 100GeV)^{3/2}$)
et vaut donc $\l'_{121} \l'_{131}/0.01<1.225 \ 10^{-1}$ 
($\l'_{122} \l'_{132}/0.01<6.8 \ 10^{-1}$) 
pour $m_{\tilde d_k}=100GeV$ et 
$\l'_{121} \l'_{131}/0.01<12.25$ ($\l'_{122} \l'_{132}/0.01<21.50$) 
pour $m_{\tilde d_k}=1TeV$ \cite{4:Bhatt}. 
En revanche, l'\'etude de la production simple de quark top ne permettra 
pas d'am\'eliorer la limite sur le produit 
$\l'_{123} \l^{\prime \star}_{133}$ dont la borne indirecte actuelle est tr\`es forte: 
$\l'_{123} \l'_{133}<1.4 \ 10^{-4}$ pour $m_{\tilde q}=100GeV$ 
\cite{4:Bhatt}.
L'\'etude des contributions
des interactions \rpv \`a la r\'eaction $l^+ l^- \to c t 
\to c b l \nu, \ [l=e,\mu]$ est la seule \'etude permettant 
de tester 
efficacement le produit de constantes de couplage
$\l'_{12k} \l^{\prime \star}_{13k}$ ($k=1,2$) dans les collisions de haute
\'energie \cite{PubliH}.
Remarquons finalement qu'une \'etude dans le m\^eme contexte de 
la r\'eaction $l^+ l^- \to u t \to u b l \nu, \ [l=e,\mu]$ 
donnerait des r\'esultats semblables sur le produit dominant de 
constantes de couplage \rpv $\l'_{11k} \l^{\prime \star}_{13k}$ 
car l'\'etat final que nous avons effectivement consid\'er\'e 
est $2 \ jets+l+\nu$.


\subsection{Asym\'etries li\'ees \`a la violation de CP}
\label{5:acpf}

En rempla\c{c}ant dans Eq.(\ref{chac3}) les amplitudes par leur
expression explicite (voir Eq.(\ref{chac5bis},\ref{chac5})), 
nous trouvons
apr\`es calcul les expressions suivantes pour les asym\'etries 
li\'ees \`a la violation de CP et associ\'ees aux contributions 
des interactions \rpv \`a la r\'eaction de Eq.(\ref{chac1})
\cite{PubliG},
\begin{eqnarray}
{\cal A}_{JJ'}&=&{2\over \vert a_0^{JJ'} \vert ^2}\bigg [ 
\sum_\a  Im(a_0^{JJ'} a_\a^{JJ' \star} ) 
Im (F_\a^{JJ'} (s+i\e )) \cr 
&-& \sum_{\a < \a ' } Im(a_\a^{JJ'} a_{\a '}^{JJ' \star}) 
Im (F_\a^{JJ'} (s+i\e ) F_{\a '}^{JJ' \star} (s+i\e ) )\bigg ].
\label{chac10}
\end{eqnarray}
Le premier terme de Eq.(\ref{chac10}) correspond \`a une 
interf\'erence entre les amplitudes au niveau en arbre et \`a 
l'ordre d'une boucle. Quant au second terme, il provient 
d'une interf\'erence entre des contributions \`a l'ordre d'une 
boucle impliquant diff\'erentes g\'en\'erations de (s)fermions.
Nous remarquons que les deux termes de Eq.(\ref{chac10}) ne
sont pas nuls 
uniquement si les constantes de couplage \rpv ont une
phase complexe et si l'on consid\`ere les parties imaginaires
issues des calculs de boucle. Hors du p\^ole du boson de jauge 
$Z^0$, la contribution majeur \`a l'asym\'etrie de violation de 
CP d\'efinie dans Eq.(\ref{chac10}) provient du premier terme de 
Eq.(\ref{chac10}).

Au p\^ole du boson de jauge $Z^0$, l'observable li\'ee \`a la
violation de CP correspondant \`a l'asym\'etrie de 
Eq.(\ref{chac3}) est d\'efinie par,
\begin{eqnarray}
{\cal A}_{JJ'}&= & {\G (Z\to f_J+\bar f_{J'} )- \G (Z\to f_{J'} 
+\bar f_J ) \over \G (Z\to f_J+\bar f_{J'} )+ \G (Z\to f_{J'} 
+\bar f_J )}  \cr  & & \cr  & & \cr &=&
- 2 {\sum_{H=L,R}\sum_{\a < \a ' } Im(l_{\a}^{JJ'} 
l_{\a'}^{JJ' \star} )
Im (I_{H\a }  ^{JJ'} (s+i\e ) I_{H \a '}^{JJ'\star }  
(s+i\e ) ) \over \sum_{H=L,R}\vert \sum_\a  l^{JJ'}_{\a} 
I_{H \a}^{JJ'} (s+i\e ) \vert ^2  },
\label{chac11}
\end{eqnarray}
o\`u $I^{JJ'}_{H \a} (s+i\e )$ est une fonction
issue d'un calcul de boucle ayant une partie imaginaire (voir
\cite{PubliG}) et d\'ependant notamment des masses des fermions 
et sfermions \'echang\'es dans la boucle et donc de $\a$ qui,
rappelons-le, correspond aux deux indices de saveur des fermions 
et sfermions \'echang\'es dans la boucle. L'observable d\'efinie
dans Eq.(\ref{chac11}) est bas\'ee sur une interf\'erence entre 
des contributions \`a l'ordre d'une 
boucle impliquant diff\'erentes g\'en\'erations de (s)fermions.
Remarquons qu'au p\^ole du boson de jauge $Z^0$, les asym\'etries
de violation de CP (voir Eq.(\ref{chac11})), tout comme les 
contributions aux taux de changement
de saveur (voir Eq.(\ref{chac7})), impliquent des processus \`a 
l'ordre des boucles.

\`A des \'energies dans le centre de masse sup\'erieures \`a la
masse du boson $Z^0$, les asym\'etries li\'ees \`a la violation 
de CP de Eq.(\ref{chac10}) sont de l'ordre de \cite{PubliG},
\begin{eqnarray}
{\cal A}_{JJ'} \approx (10^{-2} \ - \ 10^{-3}) \ \sin \psi, 
\label{chac12}
\end{eqnarray}
si l'on suppose que toutes les masses des sfermions (toutes les
valeurs absolues des 
constantes de couplage \rpv de type $\l$, $\l'$ ou $\l''$) sont 
\'egales \`a une m\^eme valeur not\'ee $\tilde m$ ($\L$).
Le r\'esultat de Eq.(\ref{chac12}) a \'et\'e obtenu en supposant
que seuls les produits de constantes de couplage 
\rpv impliqu\'ees dans les processus \`a l'ordre d'une boucle ont
une phase complexe et que cette phase complexe $\psi$ est 
identique pour tous ces produits de constantes de couplage 
\rpi, c'est \`a dire dans nos notations $arg(t^{JJ'}_i)=0$ et 
$arg(l^{JJ'}_{\a})=\psi$ pour tout $i$, $\a$, $J$ et $J'$.

Au p\^ole du boson de jauge $Z^0$, les asym\'etries li\'ees \`a 
la violation de CP de Eq.(\ref{chac11}) sont de l'ordre de 
\cite{PubliG},
\begin{eqnarray}
{\cal A}_{JJ'} \approx (10^{-1} \ - \ 10^{-3}) \ \sin \psi, 
\label{chac13}
\end{eqnarray}
si l'on suppose que toutes les masses des sfermions (toutes les
valeurs absolues des 
constantes de couplage \rpv de type $\l$, $\l'$ ou $\l''$) sont 
\'egales \`a une m\^eme valeur not\'ee $\tilde m$ ($\L$).
Le r\'esultat de Eq.(\ref{chac13}) a \'et\'e obtenu en supposant
que seuls les produits de constantes de couplage 
\rpv impliqu\'ees dans les boucles 
dans lesquelles sont \'echang\'es des (s)fermions appartenant \`a 
la troisi\`eme famille ont
une phase complexe et que cette phase complexe $\psi$ est 
identique pour tous ces produits de constantes de couplage 
\rpi, ce qui s'\'ecrit dans nos notations $arg(t^{JJ'}_i)=0$ et 
$arg(l^{JJ'}_{\a})=0 \ (\psi)$ pour tout $i$, $J$, $J'$ et si 
aucun (au moins un) des deux indices de $\a$ n'est (est)
\'egal \`a $3$.

Les asym\'etries d\'efinies dans Eq.(\ref{chac10}) 
et Eq.(\ref{chac11}) ont une 
d\'ependance dans les constantes de couplage \rpv du type,
\begin{eqnarray}
{ \sum_i \sum_\a Im( t^{JJ'}_i l^{JJ' \star}_\a )
\over \vert \sum_i t^{JJ'}_i  \vert^2 }, 
\label{chac14}
\end{eqnarray}
et,
\begin{eqnarray}
{ \sum_{\a < \a'} Im( l^{JJ'}_\a l^{JJ' \star}_{\a'} )
\over \vert \sum_\a l^{JJ'}_\a  \vert^2 }, 
\label{chac15}
\end{eqnarray}
respectivement, qui pourraient conduire \`a 
d'importants facteurs de r\'eduction ou
d'augmentation si les constantes de couplage \rpv 
exhibaient une grande hi\'erarchie dans l'espace 
des saveurs.

Nous avons vu dans \cite{PubliH} que les erreurs statistiques
sur les asym\'etries d\'efinies dans Eq.(\ref{chac10}) sont du 
m\^eme ordre de grandeur que les asym\'etries elles-m\^emes pour
une luminosit\'e de ${\cal L}=100fb^{-1}$ et 
une valeur du produit de constantes de couplage \rpv impliqu\'e 
dans le processus au niveau en arbre de $\L \L=0.1$. Cependant, 
si la structure des constantes de couplage \rpv exhibait une 
forte hi\'erarchie dans l'espace des saveurs, les valeurs des
asym\'etries seraient augment\'ees devenant ainsi sup\'erieures
aux incertitudes statistiques. Par ailleurs, un calcul des 
incertitudes statistiques plus pr\'ecis que celui effectu\'e
dans \cite{PubliH} donnerait des r\'esultats plus optimistes.

Pour des \'energies comprises dans l'intervalle $10^2GeV-10^3GeV$,
la quantit\'e 
$2 \s_{JJ'} \times {\cal A}_{JJ'} \times 500 fb^{-1}$
(${\cal A}_{JJ'}$ \'etant d\'efini dans Eq.(\ref{chac10}))
est de l'ordre de $(1 \ - \ 10)$ \ $(\L / 0.1)^4$ \ 
$(100 GeV / \tilde m)^{2 \ - \ 3}$ \ $\sin \psi$. 
Par cons\'equent, si les constantes de
couplage \rpv ne sont pas effectivement proches de leur limite 
actuelle \cite{4:Bhatt} et si la structure des
constantes de couplage \rpv n'exhibe pas de forte hi\'erarchie 
dans l'espace des saveurs, les asym\'etries de violation de
CP d\^ues aux interactions \rpv seront difficilement observables 
aupr\`es des futurs collisionneurs lin\'eaires 
\cite{4:Tesla,4:NLC}.

Les asym\'etries de Eq.(\ref{chac13}) sont du m\^eme 
ordre de grandeur que les asym\'etries de violation de CP 
associ\'ees aux 
d\'esint\'egrations hadroniques du boson de jauge $Z^0$ 
calcul\'ees dans le cadre du \ms qui valent, 
${\cal A}_{b s}=10^{-5} \ \sin \delta_{CKM}$, 
${\cal A}_{b d}=10^{-3} \ \sin \delta_{CKM}$ et 
${\cal A}_{s d}=10^{-1} \ \sin \delta_{CKM}$ \cite{4:48,4:49}.


\vspace{5 mm}

\underline{\bf {PRODUCTION SIMPLE DE QUARK TOP:}}

\vspace{5 mm}

La r\'eaction $l^+ l^- \to c \bar t \ / \ \bar c t, 
\ \bar t \to \bar b l \bar \nu \ / \ t \to b \bar l \nu \
[l=e,\mu]$
permet de tester les effets de changement de saveur (voir Section
\ref{5:ferfcnc}) et donc les effets de violation de CP, dans le 
secteur des quarks. Pour cette r\'eaction, la quantit\'e 
$2 \s_{23} \times {\cal A}_{23} \times 500 fb^{-1}$ 
(${\cal A}_{JJ'}$ \'etant d\'efini dans Eq.(\ref{chac10})) est 
diminu\'ee d'un facteur $B(W \to l \nu)=21.1 \% \ [l=e,\mu]$ par 
rapport aux r\'eactions du type $l^+ l^- \to f_J \bar f_{J'}, 
\ \bar f_J f_{J'}$, mais elle reste du m\^eme ordre de grandeur.
Les conclusions sur les incertitudes statistiques et sur
l'observabilit\'e des asym\'etries de 
violation de CP d\'efinies dans Eq.(\ref{chac10}) 
sont donc similaires pour les processus  
$l^+ l^- \to c \bar t \ / \ \bar c t, 
\ \bar t \to \bar b l \bar \nu \ / \ t \to b \bar l \nu \
[l=e,\mu]$ et $l^+ l^- \to f_J \bar f_{J'}, \ \bar f_J f_{J'}$.

Dans \cite{PubliH}, nous nous sommes int\'eress\'es \`a une autre
asym\'etrie que celle d\'efinie dans Eq.(\ref{chac10})
en vue d'obtenir une meilleure sensibilit\'e sur les effets de 
violation de CP li\'es 
au processus $l^+ l^- \to c \bar t \ / \ \bar c t, 
\ \bar t \to \bar b l \bar \nu \ / \ t \to b \bar l \nu \
[l=e,\mu]$. L'asym\'etrie de violation de CP que nous avons 
calcul\'e d\'epend du spin du quark top produit 
et est d\'efinie par,
\begin{eqnarray}
{\cal A}^{pol}=  
{{ d \s^+ \over d E_l } - { d \s^- \over d E_l}
\over
{  d \s^+ \over d E_l } + { d \s^- \over d E_l}},
\label{chac15bis}
\end{eqnarray}
o\`u $\s^+$ et $\s^-$ sont respectivement les sections efficaces 
des processus $l^+ l^- \to t \bar c$ et $l^+ l^- \to \bar t c$ et
$E_l$ est l'\'energie du lepton charg\'e produit dans la 
d\'esint\'egration du quark top $t \to b l \nu$. La quantit\'e
$d \s^+ / d E_l$ est fonction de la polarisation du quark top $t$
alors que la quantit\'e $d \s^- / d E_l$ est fonction de la 
polarisation de l'anti-quark top $\bar t$ \cite{PubliH}.

\`A des \'energies dans le centre de masse comprises dans
l'intervalle $10^2GeV-10^3GeV$, les asym\'etries li\'ees 
\`a la violation de CP de Eq.(\ref{chac15bis}) sont de l'ordre de
\cite{PubliH},
\begin{eqnarray}
{\cal A}^{pol} \approx (10^{-2} \ - \ 10^{-3}) \ \sin \psi, 
\label{chac15b2}
\end{eqnarray}
si l'on suppose que toutes les masses des sfermions (toutes les
valeurs absolues 
des constantes de couplage \rpv de type $\l'$) sont 
\'egales \`a une m\^eme valeur $\tilde m=100GeV$ ($\L$).
Le r\'esultat de Eq.(\ref{chac15b2}) a \'et\'e obtenu en supposant
que seuls les produits de constantes de couplage 
\rpv impliqu\'ees dans les processus \`a l'ordre d'une boucle ont
une phase complexe et que cette phase complexe $\psi$ est 
identique pour tous ces produits de constantes de couplage 
\rpi, c'est \`a dire dans nos notations $arg(t^{23}_i)=0$ et 
$arg(l^{23}_{\a})=\psi$ pour tout $i$ et $\a$.

Les asym\'etries d\'efinies dans Eq.(\ref{chac15bis}) ont une 
d\'ependance dans les constantes de couplage \rpv du type,
\begin{eqnarray}
{ \sum_i \sum_\a Im( t^{23}_i l^{23 \star}_\a )
\over \vert \sum_i t^{23}_i  \vert^2 }, 
\label{chac15b3}
\end{eqnarray}
qui pourrait conduire \`a un important facteur de r\'eduction ou
d'augmentation si une grande hi\'erarchie existait parmi les
constantes de couplage \rpv dans l'espace des saveurs.

Nous avons vu dans \cite{PubliH} que les erreurs statistiques
sur les asym\'etries d\'efinies dans Eq.(\ref{chac15bis}) sont du 
m\^eme ordre de grandeur que les asym\'etries elles-m\^emes pour
une luminosit\'e de ${\cal L}=100fb^{-1}$ et 
une valeur du produit de constantes de couplage \rpv impliqu\'e 
dans le processus au niveau en arbre de $\l'_{12k} 
\l^{\prime \star}_{13k}=0.1$. Cependant, 
si la structure des constantes de couplage \rpv exhibait une 
forte hi\'erarchie dans l'espace des saveurs, les valeurs des
asym\'etries seraient augment\'ees devenant ainsi sup\'erieures
aux incertitudes statistiques. Par ailleurs, un calcul des 
incertitudes statistiques plus pr\'ecis que celui effectu\'e
dans \cite{PubliH} donnerait des r\'esultats plus optimistes.

Pour des \'energies comprises dans l'intervalle $10^2GeV-10^3GeV$,
la quantit\'e 
$2 \s_{JJ'} \times {\cal A}^{pol} \times 500 fb^{-1}$
(${\cal A}^{pol}$ \'etant d\'efini dans Eq.(\ref{chac15bis}))
est de l'ordre de $(1 \ - \ 10) (\L / 0.1)^4$ 
($100 GeV / \tilde m)^{2 \ - \ 3}$ \ $\sin \psi$. 
Par cons\'equent, pour le
processus $l^+ l^- \to c \bar t \ / \ \bar c t, 
\ \bar t \to \bar b l \bar \nu \ / \ t \to b \bar l \nu \
[l=e,\mu]$, la 
conclusion concernant les asym\'etries de type ${\cal A}^{pol}$ 
est identique \`a celle concernant les asym\'etries de 
violation de CP d\'efinies dans Eq.(\ref{chac10}): 
si les constantes de
couplage \rpv ne sont pas effectivement proches de leur limite 
actuelle \cite{4:Bhatt} et si la structure des
constantes de couplage \rpv n'exhibe pas de forte hi\'erarchie 
dans l'espace des saveurs, les contributions des interactions 
\rpv aux asym\'etries de violation de CP seront difficilement 
observables aupr\`es des futurs collisionneurs lin\'eaires 
\cite{4:Tesla,4:NLC}.

Remarquons finalement qu'une \'etude dans le m\^eme contexte de 
la r\'eaction $l^+ l^- \to u t \to u b l \nu, \ [l=e,\mu]$ 
donnerait des r\'esultats semblables sur les phases complexes
du produit dominant de 
constantes de couplage $\l'_{11k} \l^{\prime \star}_{13k}$ 
car l'\'etat final que nous avons effectivement consid\'er\'e 
est $2 \ jets+l+\nu$.


\section{Production de paires de sfermions}
\label{5:sfer}

\subsection{Taux de changement de saveur}

\begin{figure}[t]
\begin{center}
\leavevmode
\centerline{\psfig{figure=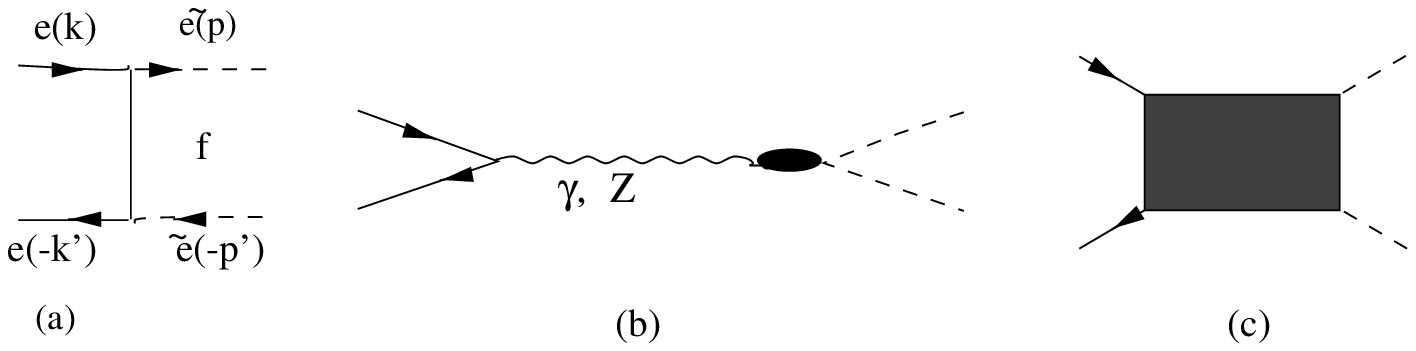,height=1.5in}}
\end{center}
\caption{\footnotesize  \it Diagrammes de Feynman
des contributions des interactions \rpv \`a la r\'eaction
$l^+ l^- \to \tilde e_{H J} \tilde e^\star_{H J'} 
\ [H=L,R; \ J \neq J']$.
$f$ d\'enote un fermion et $\tilde e$ un slepton charg\'e.
Les diagrammes de Feynman des contributions \`a l'ordre d'une
boucle au vertex $Z^0 \tilde e_{H J} \tilde e^\star_{H J'} 
\ [H=L,R; \ J \neq J']$ sont pr\'esent\'es dans la Figure 
\ref{5:scpcp}.
\rm \normalsize }
\label{5:scpc}
\end{figure}

\begin{figure}[t]
\begin{center}
\leavevmode
\centerline{\psfig{figure=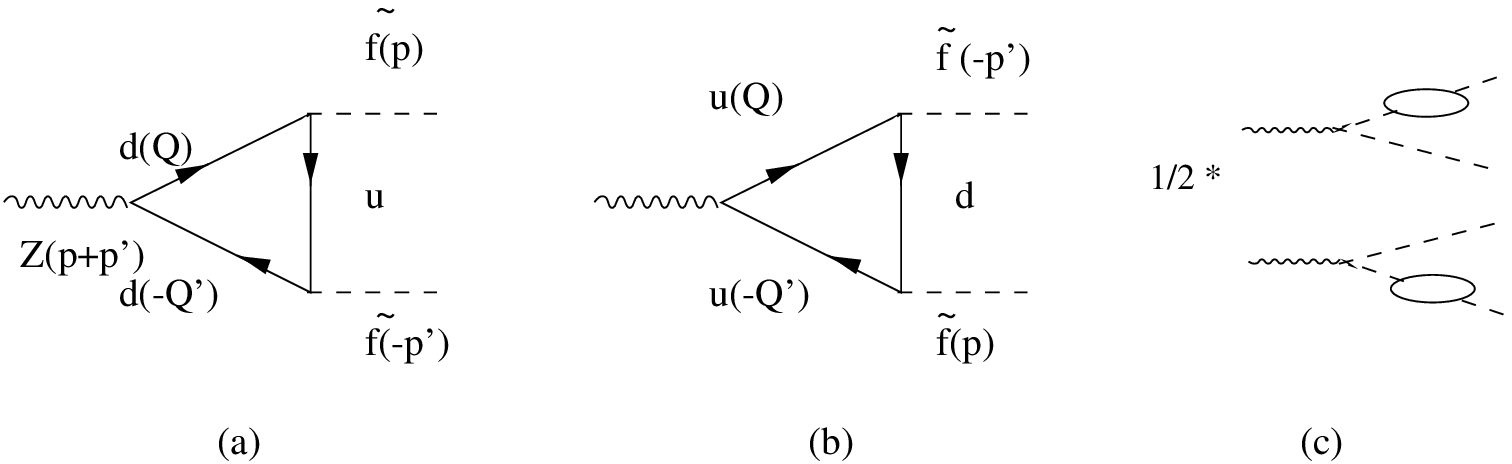,height=1.9in}}
\end{center}
\caption{\footnotesize  \it Diagrammes de Feynman
des contributions des interactions \rpv \`a l'ordre d'une
boucle au vertex $Z^0 \tilde e_{H J} \tilde e^\star_{H J'} 
\ [H=L,R; \ J \neq J']$. $u$ et $d$ d\'enotent des fermions et 
$\tilde f$ un slepton charg\'e.
\rm \normalsize }
\label{5:scpcp}
\end{figure}

Les graphes de Feynman des contributions des interactions \rpv
\`a la r\'eaction du type de Eq.(\ref{chac2}), $l^+ l^- \to 
\tilde e_{H J} \tilde e^\star_{H J'} \ [H=L,R; \ J \neq J']$, 
sont pr\'esent\'es dans les Figures \ref{5:scpc} et 
\ref{5:scpcp}. Nous avons consid\'er\'e les contributions
des interactions \rpv \`a l'ordre d'une boucle pour des raisons
qui appara\^{\i}tront claires dans la Section \ref{5:sapcf}.

Les contributions des interactions \rpv \`a la production de
paires de sleptons charg\'es gauches appartenant \`a 
des familles diff\'erentes $l^+ l^- \to \tilde e_{L J} 
\tilde e^\star_{L J'} \ [J \neq J']$ impliquent les
produits de constantes de couplage 
$\l_{iJ'1} \l^{\star}_{iJ1}$ (voie $t$) au niveau en arbre 
(l'indice $i$ correspond \`a la saveur du neutrino \'echang\'e) 
et $\l_{J'jk} \l^{\star}_{Jjk}$ ou $\l'_{J'jk} 
\l^{\prime \star}_{Jjk}$ \`a l'ordre d'une boucle
(les indices $j$ et $k$ correspondent aux saveurs des 
fermions \'echang\'es dans la boucle). \\
Les contributions des interactions \rpv \`a la production de
paires de sleptons charg\'es droits appartenant \`a 
des familles diff\'erentes $l^+ l^- \to \tilde e_{R J} 
\tilde e^\star_{R J'} \ [J \neq J']$ impliquent les
produits de constantes de couplage 
$\l_{i1J} \l^{\star}_{i1J'}$ (voie $t$) au niveau en arbre 
(l'indice $i$ correspond \`a la saveur du neutrino \'echang\'e) 
et $\l_{ijJ} \l^{\star}_{ijJ'}$ \`a l'ordre d'une boucle
(les indices $i$ et $j$ correspondent aux saveurs des 
leptons \'echang\'es dans la boucle). \\
La production de paires de sleptons charg\'es gauche et droit 
appartenant \`a des familles diff\'erentes 
$l^+ l^- \to \tilde e_{L J} \tilde e^\star_{R J'} \ [J \neq J']$ 
ne re\c{c}oit pas de contributions des interactions \rpv au 
niveau en arbre pour des neutrinos de masses nulles et n'a donc 
pas \'et\'e consid\'er\'ee.

Nous notons les amplitudes des contributions des interactions \rpv
\`a la r\'eaction $l^+ l^- \to \tilde e_{H J} 
\tilde e^\star_{H J'} \ [H=L,R; \ J \neq J']$,
\begin{eqnarray}
\tilde M^{JJ'}=a_0^{JJ'}+\sum_\a a_\a ^{JJ'}  
F^{JJ'}_\a (s+i\e ), \\ 
\bar {\tilde M}^{JJ'}=a^{JJ' \star } _0+\sum_\a a^{JJ'\star } _\a  
F^{J'J} _\a (s+i\e ),
\label{chac16}
\end{eqnarray}
o\`u $a_0^{JJ'}$ repr\'esente l'amplitude des contributions des 
interactions \rpv au niveau en arbre et $\sum_\a a_\a ^{JJ'}  
F^{JJ'}_\a (s+i\e )$ l'amplitude des contributions des 
interactions \rpv \`a l'ordre d'une boucle. $a_0^{JJ'}$ est
proportionnel au produit de constantes de couplage \rpv impliqu\'e
par le processus au niveau en arbre et que l'on note 
$t_i^{JJ'}$, $i$ correspondant \`a la saveur du neutrino 
\'echang\'e au niveau en arbre. L'expression de $a_0^{JJ'}$ fait
intervenir une sommation sur l'indice $i$ dont d\'epend aussi la 
masse du neutrino \'echang\'e. De m\^eme, $a_\a ^{JJ'}$ est
proportionnel au produit de constantes de couplage \rpv impliqu\'e
dans la boucle et que l'on note $l_\a^{JJ'}$, $\a$ correspondant 
aux deux indices de saveur des fermions \'echang\'es 
dans la boucle. Enfin, $F^{JJ'}_\a (s+i\e )$ est une fonction
issue d'un calcul de boucle ayant une partie imaginaire (voir
\cite{PubliI}) et d\'ependant notamment des masses des fermions 
\'echang\'es dans la boucle et donc de $\a$.

Les contributions dominantes des 
interactions \rpv \`a la r\'eaction du type de Eq.(\ref{chac2}),
$l^+ l^- \to \tilde e_{H J} \tilde e^\star_{H J'} 
\ [H=L,R; \ J \neq J']$, 
proviennent de processus \`a l'ordre des arbres (voir Figure 
\ref{5:scpc}). Pour des \'energies dans le centre de masse
comprises dans l'intervalle $10^2 GeV < \sqrt s < 10^3 GeV$, 
les sections efficaces $\tilde \s_{JJ'}$ de ces 
contributions sont de l'ordre de \cite{PubliI},
\begin{eqnarray}
\tilde \s_{JJ'} \approx ({\L \over 0.1})^4 
(2 \ - \ 20) \  fbarns, 
\label{chac17}
\end{eqnarray}
si l'on suppose que toutes les masses des sfermions (toutes les
valeurs absolues des constantes de couplage \rpv de type $\l$, 
$\l'$ ou $\l''$) sont \'egales \`a une m\^eme valeur $\tilde m
< 400 GeV$ ($\L$).

Les limites actuelles sur les constantes de couplage \rpv
sont typiquement de l'ordre de $10^{-1} (\tilde m / 100GeV)$,
$\tilde m$ \'etant la masse des sfermions \cite{4:Bhatt}. Par 
cons\'equent, les \sefs de Eq.(\ref{chac17}) peuvent \^etre 
sup\'erieures \`a $\sim 20 fbarns$.

\'Etant donn\'e les \sefs $\tilde \s_{JJ'}$ de Eq.(\ref{chac17}), 
les contributions des interactions \rpv \`a la r\'eaction 
$l^+ l^- \to \tilde e_{H J} \tilde e^\star_{H J'} 
\ [H=L,R; \ J \neq J']$ seront potentiellement observables aux 
futurs collisionneurs lin\'eaires pour lesquels les luminosit\'es
attendues sont de l'ordre de $500 fb^{-1}$ \cite{4:Tesla,4:NLC}.


Les \sefs de Eq.(\ref{chac17}) sont du m\^eme 
ordre de grandeur que les \sefs des 
contributions des interactions du MSSM \`a la r\'eaction
$l^+ l^- \to \tilde e_{H J} \tilde e^\star_{H J'} 
\ [H=L,R; \ J \neq J']$ qui sont comprises entre $0.1$ ($0.01$) 
et $250$ ($100$) $fbarns$, pour $\sqrt s=190 GeV$ 
($\sqrt s=500 GeV$) \cite{4:Arkani}.


\subsection{Asym\'etries li\'ees \`a la violation de CP}
\label{5:sapcf}

En rempla\c{c}ant dans Eq.(\ref{chac4}) les amplitudes par leur
expression explicite (voir Eq.(\ref{chac16})), nous trouvons
apr\`es calcul les expressions suivantes pour les asym\'etries 
li\'ees \`a la violation de CP et associ\'ees aux contributions 
des interactions \rpv \`a la r\'eaction $l^+ l^- \to 
\tilde e_{H J} \tilde e^\star_{H J'} \ [H=L,R; \ J \neq J']$
\cite{PubliI},
\begin{eqnarray}
\tilde {{\cal A}}_{JJ'} 
\approx {2\over \vert a_0^{JJ'} \vert ^2}\bigg [ 
\sum_\a  Im(a_0^{JJ'} a_\a^{JJ' \star} ) 
Im (F_\a^{JJ'} (s+i\e )) \bigg ].
\label{chac18}
\end{eqnarray}
L'observable d\'efinie dans Eq.(\ref{chac18}) correspond \`a une 
interf\'erence entre les amplitudes au niveau en arbre et \`a 
l'ordre d'une boucle. Nous remarquons que l'asym\'etrie de 
Eq.(\ref{chac18}) n'est pas nulle 
uniquement si les constantes de couplage \rpv ont une
phase complexe et si l'on consid\`ere les parties imaginaires
issues des calculs de boucle.

\`A des \'energies dans le centre de masse comprises dans 
l'intervalle $10^2 GeV < \sqrt s < 10^3 GeV$, les asym\'etries 
li\'ees \`a la violation de CP de Eq.(\ref{chac18}) sont de 
l'ordre de \cite{PubliI},
\begin{eqnarray}
\tilde {{\cal A}}_{JJ'} 
\approx (10^{-2} \ - \ 10^{-3}) \ \sin \psi, 
\label{chac19}
\end{eqnarray}
si l'on suppose que toutes les masses des sfermions (toutes les
valeurs absolues des 
constantes de couplage \rpv de type $\l$, $\l'$ ou $\l''$) sont 
\'egales \`a une m\^eme valeur not\'ee $\tilde m$ ($\L$).
Le r\'esultat de Eq.(\ref{chac19}) a \'et\'e obtenu en supposant
que seuls les produits de constantes de couplage 
\rpv impliqu\'ees dans les processus \`a l'ordre d'une boucle ont
une phase complexe et que cette phase complexe $\psi$ est 
identique pour tous ces produits de constantes de couplage 
\rpi, c'est \`a dire dans nos notations $arg(t^{JJ'}_i)=0$ et 
$arg(l^{JJ'}_{\a})=\psi$ pour tout $i$, $\a$, $J$ et $J'$.

Les asym\'etries d\'efinies dans Eq.(\ref{chac18}) ont une 
d\'ependance dans les constantes de couplage \rpv du type,
\begin{eqnarray}
{ \sum_i \sum_\a Im( t^{JJ'}_i l^{JJ' \star}_\a )
\over \vert \sum_i t^{JJ'}_i  \vert^2 }, 
\label{chac20}
\end{eqnarray}
qui pourrait conduire \`a un important facteur de r\'eduction ou
d'augmentation si une grande hi\'erarchie existait parmi les
constantes de couplage \rpv dans l'espace des saveurs.
\`A titre d'exemple, supposons que les constantes de couplage \rpv
soient \'egales \`a leur limite indirecte actuelle provenant
des contraintes de la physique de basse \'energie \cite{4:Bhatt}.
Dans ce cas, un facteur ${Im(\l_{331} \l^{\star}_{321} 
\l'_{233} \l^{\prime \star}_{333} )
\over \vert \l_{131} \l^{\star}_{121}  \vert^2 }
\approx 90 \sin \psi$, correspondant \`a $J=3$ et $J'=2$, 
appara\^{\i}trait dans les asym\'etries issues des contributions 
des interactions $\l$ (au niveau en arbre) et
$\l'$ (\`a l'ordre d'une boucle) \`a la r\'eaction 
$l^+ l^- \to \tilde e_{L J} \tilde e^\star_{L J'} \ [J \neq J']$.

Pour des \'energies dans le centre de masse sup\'erieures \`a 
$500 GeV$ et une masse universelle des sfermions 
$\tilde m< 400 GeV$, la quantit\'e 
$2 \tilde \s_{JJ'} \times \tilde {{\cal A}}_{JJ'}$ est de l'ordre 
de $(\L / 0.1)^4 \ 10^{-1}$ \ $\sin \psi$ \ $fbarns$. 
Par cons\'equent, Les asym\'etries
li\'ees \`a la violation de CP et associ\'ees aux contributions 
des interactions \rpv \`a la r\'eaction $l^+ l^- \to 
\tilde e_{H J} \tilde e^\star_{H J'} \ [H=L,R; \ J \neq J']$
seront potentiellement d\'etectables aupr\`es des futurs
collisionneurs lin\'eaires dont les luminosit\'es devraient
atteindre $500 fb^{-1}$ pour $\sqrt s> 500 GeV$
\cite{4:Tesla,4:NLC}.

Pour une masse universelle des sfermions $\tilde m< 400 GeV$,
la quantit\'e $2 \tilde \s_{JJ'} \times \tilde {{\cal A}}_{JJ'}$
($\tilde {{\cal A}}_{JJ'}$ 
\'etant d\'efini dans Eq.(\ref{chac4})) 
est de l'ordre de $(\L / 0.1)^4 (10^{-1} \ - \ 10^0)$ 
\ $\sin \psi$ \ $fbarns$. 
Cette valeur typique est inf\'erieure 
\`a la valeur de la quantit\'e 
$\tilde \s_{JJ'} - \tilde \s_{J'J} \approx (3 \ - \ 16) fbarns$ 
issue des contributions \`a la r\'eaction 
$l^+ l^- \to \tilde e_{H J} \tilde e^\star_{H J'} 
\ [H=L,R; \ J \neq J']$ provenant des oscillations des sleptons 
dans le cadre du MSSM \cite{4:ArkaniCP}.
Mentionnons cependant que les contributions provenant de 
l'oscillation des sleptons peuvent d\'ependre davantage du
mod\`ele consid\'er\'e que les contributions calcul\'ees ci-dessus
et que les pr\'edictions de \cite{4:ArkaniCP} ont \'et\'e obtenues
dans le cadre d'hypoth\`eses tendant \`a maximiser les effets de
violation de CP.


\clearpage

\vspace{10 mm}
\begin{center}
{  }
\end{center}

\clearpage

\vspace{10 mm}
\begin{center}
{  }
\end{center}
\vspace{10 mm}

\clearpage 

\addtocontents{toc}{\protect \vspace{1cm} \protect \begin{center} 
\protect {\bf Publications} \protect \end{center} }

\clearpage

\begin{center}
{  }
\end{center}
\vspace{50 mm}
\begin{center}
{\huge \bf \it PUBLICATIONS}
\end{center}

\clearpage

\vspace{10 mm}
\begin{center}
{  }
\end{center}
\vspace{10 mm}

\clearpage 

\input{Publi.sty}

\setcounter{chapter}{0}
\setcounter{section}{0}
\setcounter{subsection}{0}
\setcounter{figure}{0}

\chapter*{Publication I}
\addcontentsline{toc}{chapter}{Production of (s)particles at 
colliders through \rpv interactions}

\newpage

\vspace{10 mm}
\begin{center}
{  }
\end{center}
\vspace{10 mm}

\clearpage 

\begin{center}
{\bf \huge Production of (s)particles at 
colliders through \rpv interactions}
\end{center} 
\vspace{2cm}
\begin{center}
G. Moreau
\end{center} 
\begin{center}
{\em  Service de Physique Th\'eorique \\} 
{ \em  CE-Saclay F-91191 Gif-sur-Yvette, Cedex France \\}
\end{center} 
\vspace{1cm}
\begin{center}
{Part of the review on R-parity violating supersymmetry 
to be submitted
to Phys. Rep. and written by R.~Barbier, C.~B\'erat, 
M.~Besan\c{c}on, M.~Chemtob, A.~Deandrea, E.~Dudas, P.~Fayet, 
S.~Lavignac, F.~Ledroit-Guillon, G.~Moreau and Y.~Sirois}
\end{center}
\vspace{2cm}

\newpage

\section{Singly Produced Sparticles at $e^+e^-$ Colliders} 
\label{sec:singleee}

\setcounter{equation}{0}

%


The measure of the \Rp coupling constants could be 
performed via the detection of the displaced vertex 
associated to the \Rp decay of the LSP. 
The sensitivities on the \Rp couplings obtained 
through this method
depend on the detector geometry and performances.
Let us estimate the largest values of 
the \Rp coupling constants that can be measured 
via the displaced vertex analysis. We suppose that the LSP
is the lightest neutralino ($\tilde \chi^0_1$). 
The flight length of the LSP in the laboratory frame
is then given in meters by \cite{z:dreiner1}, 
\begin{eqnarray}
 c \gamma \tau \sim 3 \gamma \ 10^{-3} m ({\tilde m \over 100GeV })^4
       ({1GeV \over m_{LSP}})^5 ({1 \over \L})^2,
\label{z:Fleng}
\end{eqnarray}
where $\L=\l$, $\l'$ or $\l''$, 
$c$ is the light speed, $\gamma$ the Lorentz boost factor,
$\tau$ the LSP life time, $m_{LSP}$ the LSP mass and  $\tilde m$ the mass 
of the supersymmetric scalar particle 
involved in the three-body \Rp decay of the LSP.
Since the displaced vertex analysis is an experimental challenge at
hadronic colliders, we consider here the futur linear colliders.
Assuming that the minimum distance between two vertex necessary to distinguish 
them experimentally is of order $2 \ 10^{-5}m$ at linear colliders, we see from
Eq.(\ref{z:Fleng}) that the \Rp couplings could be measured up to the values,   
\begin{eqnarray}
 \L < 1.2 \ 10^{-4} \gamma^{1/2} ({\tilde m \over 100GeV })^2
       ({100GeV \over m_{LSP}})^{5/2}.
\label{z:LSP}
\end{eqnarray}
There is a gap between these values and the low-energy 
experimental constraints on the \Rp
couplings which range typically in the interval $\L <10^{-1}-10^{-2}$ 
for superpartners masses of $100GeV$.
However, the domain lying between these low-energy bounds and the
values of Eq.(\ref{z:LSP}) can be tested through another way: The study of the single
production of supersymmetric particles.
Indeed, the cross sections of such reactions are directly proportional
to a power of the relevant \Rp coupling constant(s), which allows
to determine the values of the \Rp couplings.
Therefore, there exists a complementarity between the displaced vertex analysis
and the study of singly produced sparticles, since these
two methods allow to
investigate different ranges of values of the \Rp coupling constants.

Another interest of the single superpartner production 
is the possibility to produce
supersymmetric particles at lower center of mass energies 
than through the superpartner pair production, 
which is the favored reaction in R-parity conserved models.


\subsection{Resonant Production of Sneutrinos}

At leptonic colliders, the 
sneutrinos $\tilde \nu_{\mu}$ and $\tilde \nu_{\tau}$ can be produced at
the resonance
through the couplings $\l_{211}$ and $\l_{311}$, respectively. The
sneutrino may then
decay either via an \Rp interaction, for example
through $\l_{ijk}$ as,
$\tilde \nu^i \to \bar l_j l_k$, or via 
gauge interaction as, $\tilde \nu^i_L \to \tilde \chi^+_a
l^i$,
or, $\tilde \nu^i_L \to \tilde \chi^0_a \nu^i_L$ \cite{z:dimopoulos88}. 
The sneutrino partial widths 
associated to the leptonic and gauge decay channel are given 
in the following equations \cite{z:barger89},


 
\begin{eqnarray}
\Gamma(\tilde \nu^i_L \to \bar l_j l_k)= {\l_{ijk}^2 \over 16 \pi}
m_{\tilde \nu^i_L},
\label{widthform1}
\end{eqnarray}
\begin{eqnarray}
\Gamma(\tilde \nu^i_L \to \tilde \chi^+_a l^i, \ \tilde \chi^0_a \nu^i_L)= 
{ C g^2 \over 16 \pi} 
m_{\tilde \nu^i_L} (1-{m_{\tilde \chi^+_a}^2 \over m_{\tilde \nu^i_L}^2})^2,
\label{widthform2}
\end{eqnarray}  
where $C=\vert V_{a1}\vert ^2$ for the decay into chargino and $C=\vert
N_{a2}\vert ^2$, for the neutralino case, with $V_{a1}$ and $N_{a2}$ 
the mixing matrix elements written
in the notations of \cite{z:haber85}. For reasonable values of $\l_{ijk}$
($\leq 0.1$) 
and most of the region of the \susyq\ parameter space, the decay modes of
Eq.(\ref{widthform2}) are dominant, if kinematically accessible 
\cite{z:lola,z:barger89}. 
In a SUGRA parameter space, if $m_{\tilde \nu^i_L} > 80 \GeV$, with 
$M_2 = 80 \GeV$, $\mu = 150 \GeV$ and $\tan \beta = 2$,
the total sneutrino width is higher than $100 \MeV$ which is comparable to
or greater than 
the typical expected experimental resolutions. 
The cross section formula,
for the sneutrino production in the $s$-channel, is the following
\cite{z:barger89},
\begin{eqnarray}
\s(e^+ e^- \to \tilde \nu^i_L \to X)= { 4 \pi s \over m_{\tilde
\nu^i_L}^2} 
{\Gamma(\tilde \nu^i_L \to e^+ e^-)  \Gamma(\tilde \nu^i_L \to X) \over
(s-m_{\tilde \nu^i_L}^2)^2  + m_{\tilde \nu^i_L}^2 \Gamma_{\tilde
\nu^i_L}^2},
\label{Xsectf1}
\end{eqnarray}
where $\Gamma(X)$ generally denotes the partial width for the sneutrino
decay into the final state $X$. At sneutrino resonance, Eq.(\ref{Xsectf1}) 
takes the form,
\begin{eqnarray}
\s(e^+ e^- \to \tilde \nu^i_L \to X)= { 4 \pi \over m_{\tilde
\nu^i_L}^2} 
B(\tilde \nu^i_L \to e^+ e^-)  B(\tilde \nu^i_L \to X),
\label{Xsectf2}
\end{eqnarray}
where $B(\tilde \nu^i_L \to X)$ 
generally denotes the partial width for sneutrino decay
into a final state $X$. 
\begin{figure}
\begin{center}
\leavevmode
\centerline{\psfig{figure=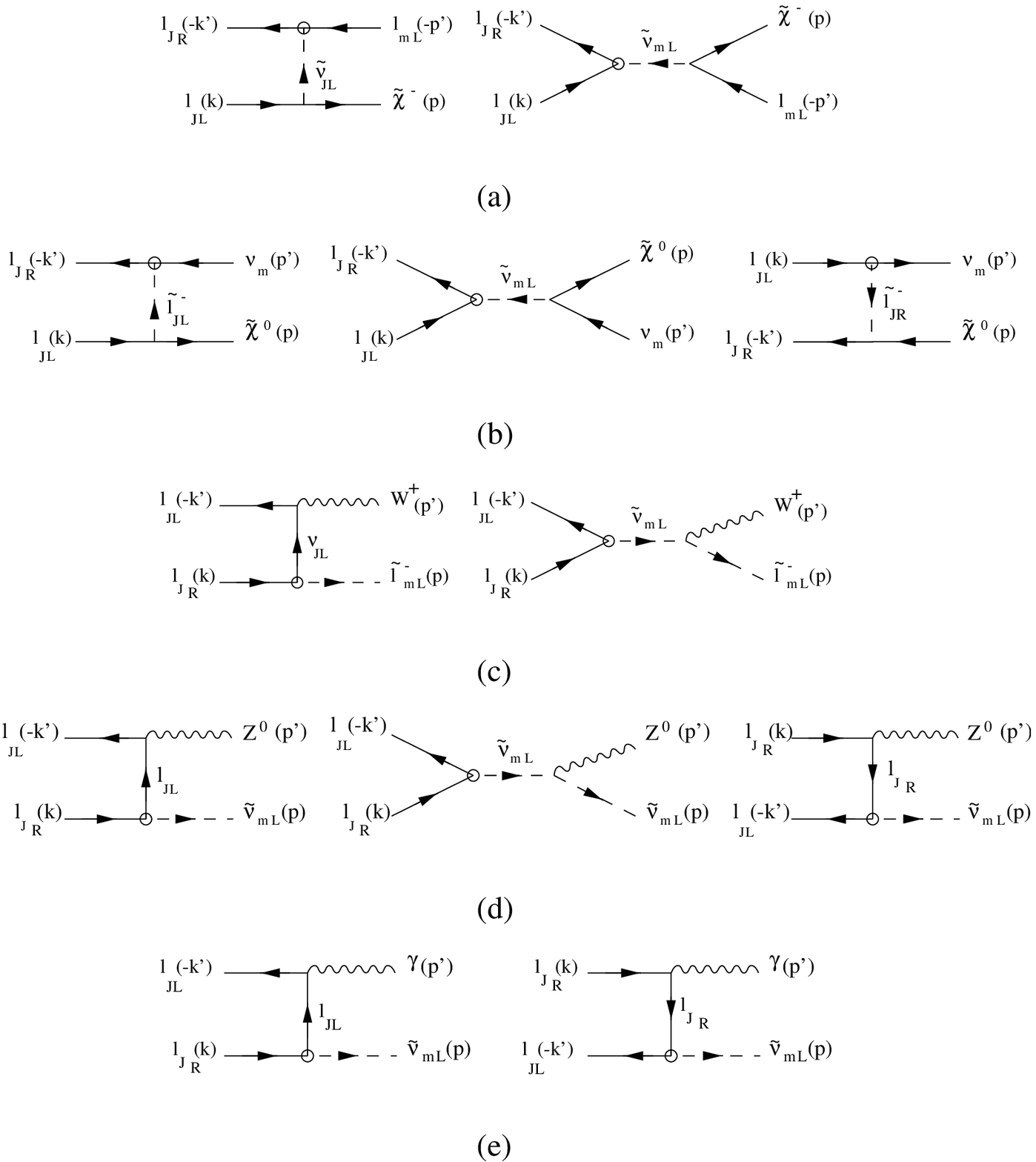}}
\end{center}
\caption{\footnotesize  \it
Feynman diagrams for the single 
production processes at leptonic colliders, namely,
$l_J^+l_J^- \to \tchi^- l_m^+$ (a), 
$l_J^+l_J^- \to \tchi^0 \bar \nu_m$ (b), 
$l_J^+l_J^- \to \tilde l^-_{mL} W^+$ (c),
$l_J^+l_J^- \to \tilde \nu_{mL} Z^0 $ (d) and 
$l_J^+l_J^- \to \tilde \nu_{mL} \ \g $ (e). 
The circled vertex correspond to the \Rp\ interaction, 
with the coupling constant  $\l_{mJJ}$, and
the arrows denote flow of momentum.
\rm \normalsize }
\label{sinprod}
\end{figure}

\subsection{Single Gaugino Production}


Two single superpartner productions receive the
contribution from the resonant sneutrino production at
$e^+ e^-$ colliders: The single chargino  
and neutralino productions (see Fig.\ref{sinprod}(a)(b)).
The single production of chargino, $e^+ e^- \to \tilde \chi^{\pm}_a
l^{\mp}_j$ (via $\l_{1j1}$), receives a contribution from the $s$-channel 
exchange of a $\tilde \nu_{jL}$ sneutrino and another one from the exchange 
of a $\tilde \nu_{eL}$ sneutrino in the $t$-channel (see Fig.\ref{sinprod}(a)). 
The single neutralino production, $e^+ e^- \to \tilde \chi^0_a \nu_j$ 
(via $\l_{1j1}$), occurs through the $s$-channel $\tilde \nu_{jL}$ sneutrino 
exchange and also via the exchange of a $\tilde e_L$ slepton in the $t$-channel 
or a $\tilde e_R $ slepton in the $u$-channel (see Fig.\ref{sinprod}(b)).

\begin{figure}[t]
\begin{center}
\centerline{\psfig{figure=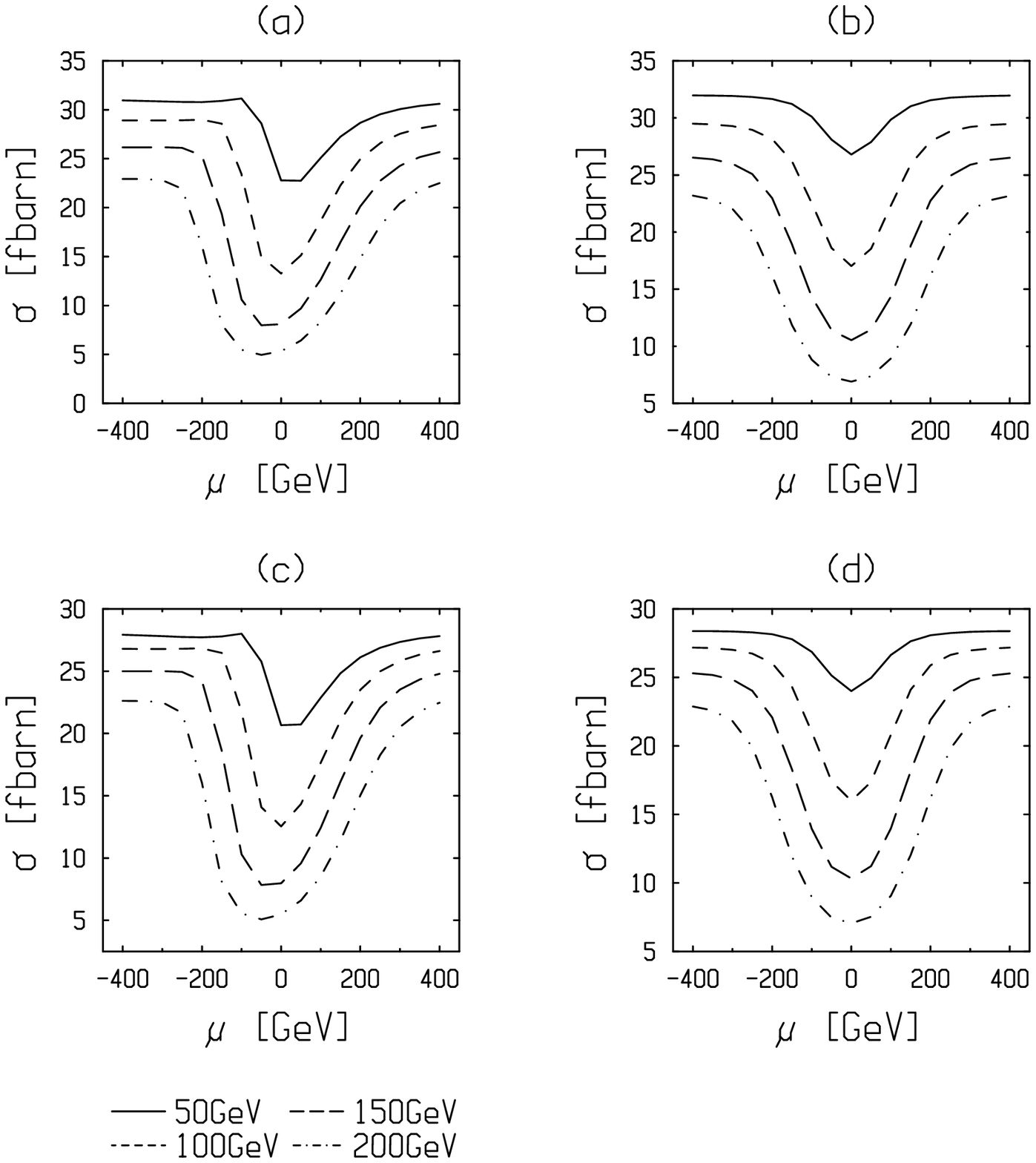,width=10cm}}  
\caption{The integrated cross-sections \cite{z:Chemsin} for the process 
$e^{+} e^{-} \rightarrow {\tilde {\chi}}^{-}_{1} l^{+}_{j}$,
at a center of mass energy of 500 GeV, are shown as
a function of $\mu$ for discrete choices of the remaining parameters:
(a) $\tan \beta = 2$, $m_0 = 50$ GeV, 
(b) $\tan \beta = 50$, $m_0 = 50$ GeV,
(c) $\tan \beta = 2$, $m_0 = 150$ GeV,
and (d) $\tan \beta = 50$, $m_0 = 50$ GeV, with $\lambda_{1j1} = $ 0.05.
The windows conventions are such that $\tan \beta = 2,50$ 
horizontally and  $m_0 = 50,150$ GeV vertically. The different curves
refer to the value of $M_2$ of 50 GeV (continuous line),
100 GeV (dot-dashed line), 150 GeV (dotted line), as indicated at the bottom
of the figure.}
\label{fig:singlechargino}
\end{center}
\end{figure}

\begin{figure}[t]
\begin{center}
\centerline{\psfig{figure=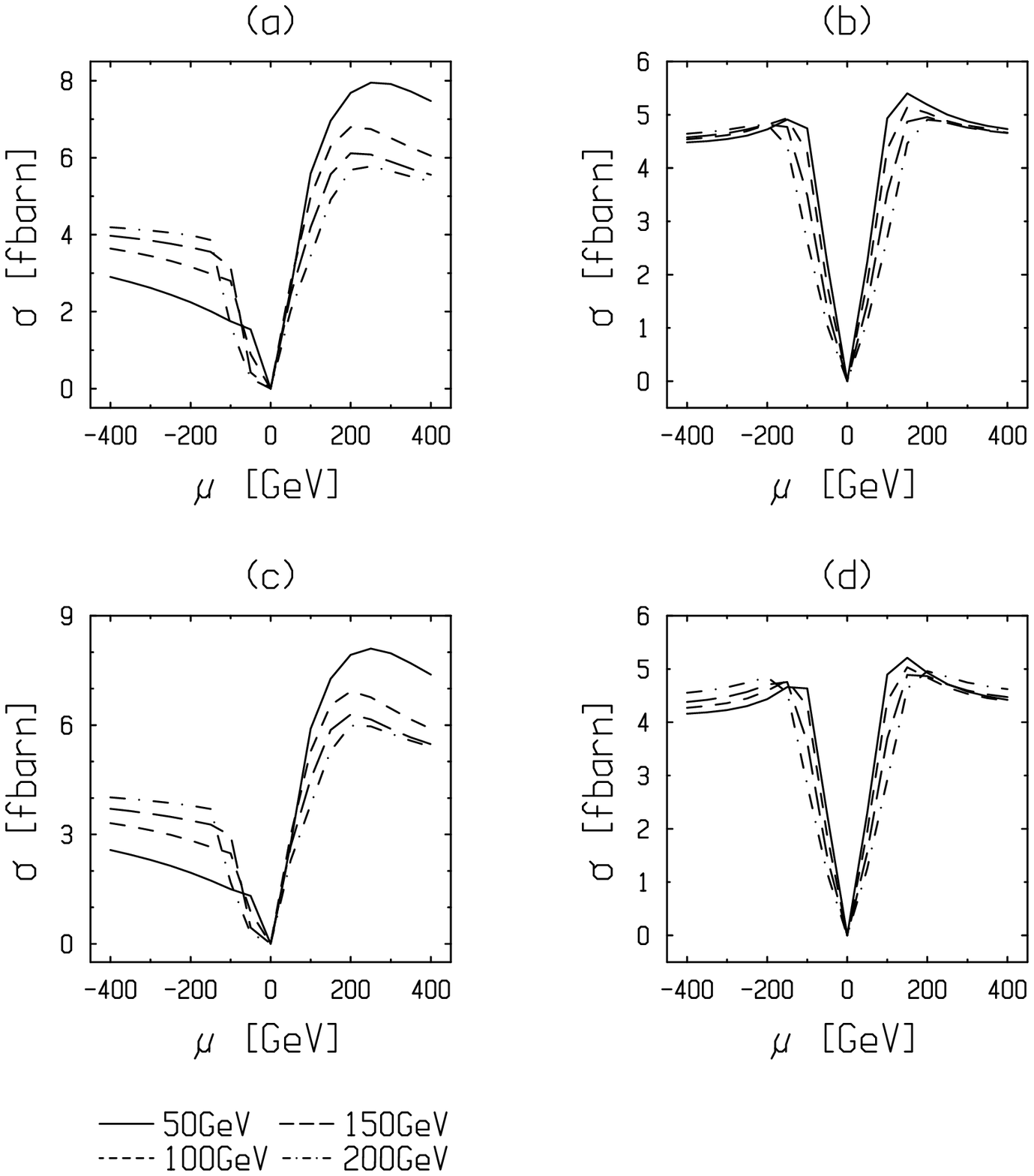,width=10cm}}  
\caption{The integrated cross-sections \cite{z:Chemsin} for the process 
$e^{+} e^{-} \rightarrow {\tilde {\chi}}^{0}_{1} {\bar {\nu}}_{j}$,
at a center of mass energy of 500 GeV, are shown as
a function pf $\mu$ for discrete choices of the remaining parameters:
(a) $\tan \beta = 2$, $m_0 = 50$ GeV, 
(b) $\tan \beta = 50$, $m_0 = 50$ GeV,
(c) $\tan \beta = 2$, $m_0 = 150$ GeV,
and (d) $\tan \beta = 50$, $m_0 = 50$ GeV, with $\lambda_{1j1} = $ 0.05.
The windows conventions are such that $\tan \beta = 2,50$ 
horizontally and  $m_0 = 50,150$ GeV vertically. The different curves
refer to the value of $M_2$ of 50 GeV (continuous line),
100 GeV (dot-dashed line), 150 GeV (dotted line), as indicated at the bottom
of the figure.}
\label{fig:singleneutralino}
\end{center}
\end{figure}

For $\l_{1j1}=0.05$, $50 \GeV < m_0 < 150 \GeV$ 
and $50 \GeV < M_2 < 200 \GeV$ in a SUGRA parameter
space, the off pole values of the cross sections are 
typically of order $100 \femtob$ ($10 \femtob$) 
for the single chargino production and $10 \femtob$ ($1\femtob$) 
for the single neutralino production at $\sqrt s=200 \GeV$ 
($500 \GeV$) \cite{z:Chemsin} (see Fig.\ref{fig:singlechargino} 
and Fig.\ref{fig:singleneutralino}).
\begin{figure}[t]
\begin{center}
\centerline{\psfig{figure=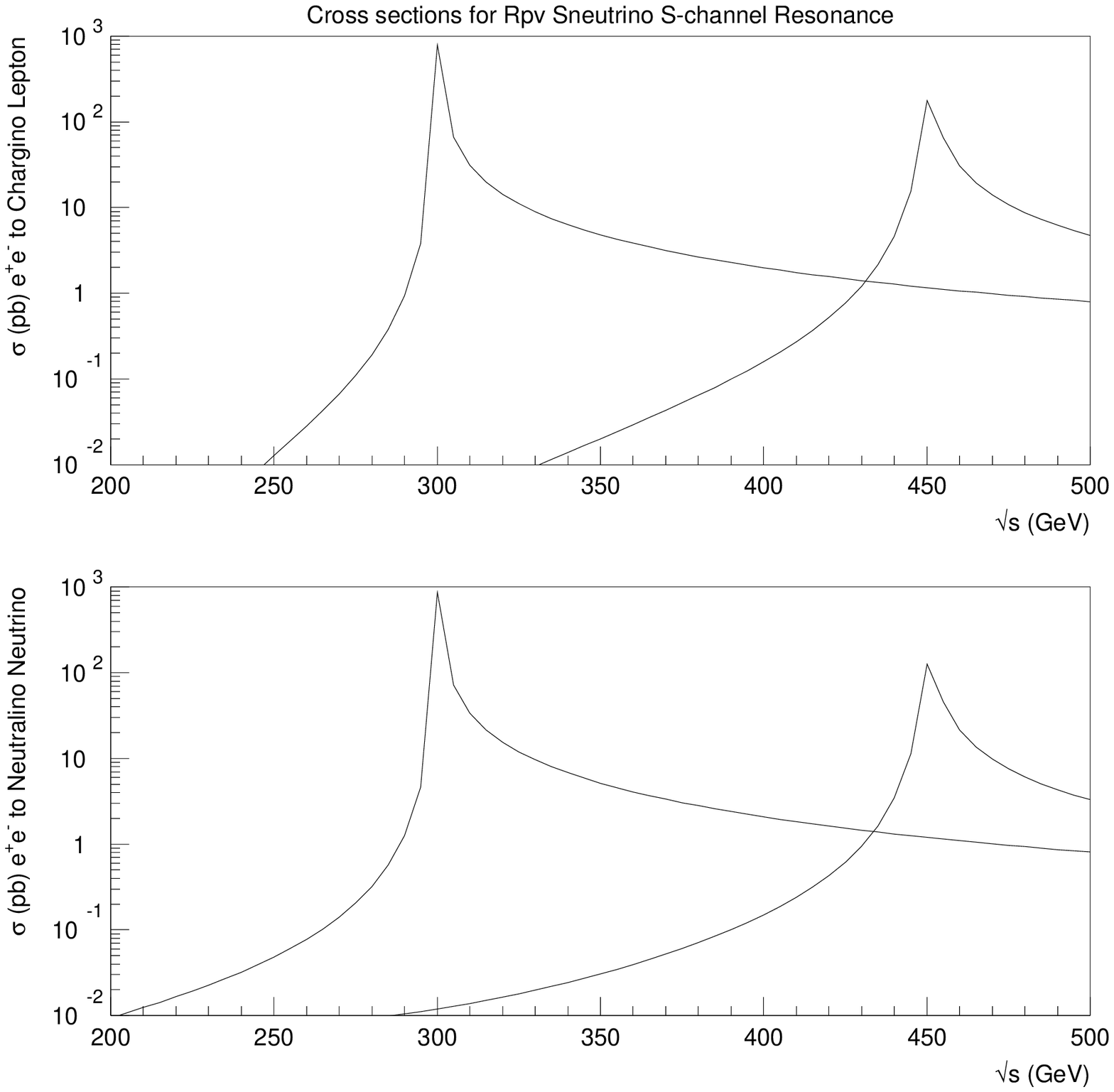,width=10cm}}  
\caption{Cross sections of the single charginos 
and neutralinos productions as a function 
of the center of mass energy for 
the two values $m_{\tilde \nu}=300 GeV, 450 GeV$ with
$m_{\tilde e}=1 TeV$,
$M_2=250 GeV$, $\mu=-200 GeV$, $\tan \beta=2$ and 
$\l_{1j1}=0.1$.
The rates values are calculated by including
the ISR effect and by summing over 
the productions of the different
$\tilde \chi^{\pm}_i$ and $\tilde \chi^0_j$ eigenstates 
which can all be produced for this MSSM point.} 
\label{fig:ISRe}
\end{center}
\end{figure}
At the sneutrino resonance, the cross sections of the
single gaugino productions reach high values:
using Eq.(\ref{Xsectf2}), the rate for the 
neutralino production in association with a neutrino is of order $3 \ 10^3$ in 
units of the QED point cross section, $R=\s_{pt}=4 \pi \alpha^2 / 3 s$, 
for $M_2 = 200 \GeV$, $\mu = 80 \GeV$, $\tan \beta=2$ and $\l_{1j1}=0.1$ at 
$\sqrt s=m_{\tilde \nu^j_L} = 120 \GeV$ \cite{z:barger89}.
The cross section for the single chargino production reaches 
$2 \ 10^{-1} \picob$ at $\sqrt s=m_{\tilde \nu^j_L}=500 \GeV$,
for $\l_{1j1}=0.01$ and $m_{\tilde \chi^{\pm}} = 490 \GeV$  \cite{z:Workshop}.
The Initial State Radiation (ISR) lowers the 
single gaugino production cross section 
at the $\tilde \nu$ pole but increases
greatly the single gaugino production rate in the domain
$m_{gaugino}<m_{\tilde \nu}<\sqrt s$. 
This ISR effect can be observed 
in Fig.\ref{fig:ISRe} which shows the single charginos 
and neutralinos productions cross sections as a function 
of the center of mass energy for a given MSSM point
\cite{z:lola}.\\
The single $\tilde \chi^{\pm}_1$ ($\tilde \chi^0_1$) 
production rate is reduced in 
the higgsino dominated region
$\vert \mu \vert \ll M_1,M_2$ where the $\tilde \chi^{\pm}_1$ 
($\tilde \chi^0_1$) is
dominated by its higgsino component, compared to the wino
dominated domain
$\vert \mu \vert \gg M_1,M_2$ in which the $\tilde \chi^{\pm}_1$ 
($\tilde \chi^0_1$) is
mainly composed by the higgsino \cite{z:Chemsin}. Besides, 
the single $\tilde \chi^{\pm}_1$ ($\tilde \chi^0_1$)
production cross section 
depends weakly on the sign of the $\mu$ parameter 
at large values of $\tan \beta$. 
However, as $\tan \beta$ decreases
the rates increase (decrease) for $sign(\mu)>0$ ($<0$).
This evolution of the rates 
with the $\tan \beta$ and $sign(\mu)$ parameters is 
explained by the evolution of the $\tilde \chi^{\pm}_1$
and $\tilde \chi^0_1$ masses 
in the supersymmetric parameter space \cite{z:Chemsin}.

The experimental searches of the single chargino and 
neutralino productions have been performed at the LEP 
collider at various center of mass energies
\cite{z:Arnoud,z:Fab1,z:ALEPHa,z:LEPSNU,z:DELPHIb}. 
The single $\tilde \chi^{\pm}_1$ production has mainly
been studied 
through the $4l^{\pm}+\Eslash$ final state and the single 
$\tilde \chi^0_1$ production via the $2l^{\pm}+\Eslash$ 
signature. The motivations were 
that if the lightest neutralino is the LSP it 
decays as $\tilde \chi^0_1 \to l \bar l \nu$ via the $\l$ 
coupling and the lightest chargino can decay as 
$\tilde \chi^{\pm}_1 \to \tilde \chi^0_1 l^{\pm} \nu$.
The off pole effects of the single gaugino productions rates 
(see above) are at the limit of observability 
at the LEP collider 
even with the integrated luminosity of LEP II: 
${\cal L } \approx 200 pb^{-1}$. Therefore,
the experimental analyses of the single gaugino 
productions have excluded values of the
$\l_{1j1}$ couplings smaller than the low-energy bounds only 
at the sneutrino resonance point $\sqrt s = m_{\tilde \nu}$ 
and, due to the ISR effect, in a range of typically 
$\Delta m_{\tilde \nu} \approx \sim 50GeV$
around the $\tilde \nu$ pole. Nevertheless,
recall that these analyses have been performed  
at several center of mass energies $\sqrt s$, which has allowed
to cover a wide range of the $\tilde \nu$ mass.
We finally note that at the various sneutrino resonances,
the sensitivities on the $\l_{1j1}$ couplings which have been 
derived from the LEP data reach values of order $10^{-3}$.

The experimental analyses of the single chargino 
and neutralino productions (via both the sneutrino 
resonance study and the off pole effects)
at linear colliders should be interesting due to
the high luminosities and energies expected at these 
futur colliders \cite{z:Tesla,z:NLC}. 
However, the single gaugino productions might suffer
a large supersymmetric background 
at linear colliders. Indeed, due to the
high energies reached at these colliders, the pair productions 
of supersymmetric particles may have large cross sections. 
In \cite{z:Tesgm}, it was shown that the SUSY background of
the $4l^{\pm}+\Eslash$ signature generated by the 
$\tilde \chi^{\pm}_1 \mu^{\mp}$ production via $\l_{121}$ could be
greatly reduced with respect to the signal: First,
this SUSY background can be suppressed by making use of the beam
polarization capability of the linear colliders.
Secondly, the specific kinematics of the single
chargino production reaction allows to put some efficient
cuts on the transverse momentum of the lepton produced 
together with the chargino. 
By consequence of this SUSY background reduction,
the sensitivity on the $\l_{121}$ coupling obtained 
from the $\tilde \chi^{\pm}_1 \mu^{\mp}$
production study at linear colliders 
for $\sqrt s=500GeV$ and ${\cal L}=500 fb^{-1}$
\cite{z:Tesla} would be of order $10^{-4}$
at the sneutrino resonance 
and would improve the low-energy constraint
over a range of $\Delta m_{\tilde \nu} \approx \sim 500GeV$
around the $\tilde \nu$ pole,
assuming the largest SUSY background allowed 
by the experimental limits on the SUSY masses \cite{z:Tesgm}.
We mention that due 
to the high luminosities reached at linear colliders,
the off resonance contributions to the cross section play an 
important role in the single $\tilde \chi^{\pm}_1$ 
production analysis.

Besides, the two-body kinematics of the reactions
$e^+ e^- \to \tilde \chi^{\pm}_{1,2} l^{\mp}$ should
allow to determine the $\tilde \chi^{\pm}_1$ and 
$\tilde \chi^{\pm}_2$ masses \cite{z:Tesgm}. As a matter of fact, 
the energy of the lepton produced together with the chargino 
$E(l^{\mp})$ is completely fixed by the center of mass energy 
$\sqrt s$, the lepton mass $m_{l^{\mp}}$ and the chargino
mass $m_{\tilde \chi^{\pm}_{1,2}}$ via the relation,
\begin{eqnarray}
E(l^{\mp})= {s + m^2_{l^{\mp}} - m^2_{\tilde \chi^{\pm}_{1,2}} 
\over 2 \sqrt s}.
\label{z:enl}
\end{eqnarray}
The lepton momentum $P(l^{\mp})$, which is related to the 
lepton energy by $P(l^{\mp})=(E(l^{\mp})^2
-m^2_{l^{\mp}}c^4)^{1/2}/c$, is thus also fixed. 
Therefore, the experimental momentum value of the
produced lepton should allow to determine the
$\tilde \chi^{\pm}_1$ and
$\tilde \chi^{\pm}_2$ masses through Eq.(\ref{z:enl}).
In fact, a photon is radiated  
from the initial state due to the ISR
so that the single chargino production must be treated 
as the three-body reaction $e^+ e^- \to \tilde \chi^{\pm}_{1,2} 
l^{\mp} \gamma$. However, it was shown in \cite{z:Tesgm}
that the $\tilde \chi^{\pm}_{1,2}$ masses determinations 
would remain 
possible in the case of a large ISR effect and that the 
accuracy on $m_{\tilde \chi^{\pm}_1}$ could reach 
$\sim 6 GeV$ at linear colliders in such a case.


\subsection{Non Resonant Single Production}

The slepton and the sneutrino can also be singly produced via the coupling 
$\l_{1j1}$ in the (non-resonant) reactions 
$e^+ e^- \to \tilde l^{\mp}_{jL} W^{\pm}$,
$e^+ e^- \to \tilde \nu^j_L Z^0$ and $e^+ e^- \to \tilde \nu^j_L \gamma$.
Those reactions receive contributions from the exchange of a charged or neutral
lepton of the first generation in the $t$- or $u$-channel  (see Fig.\ref{sinprod}).
%
%
The single productions of a sneutrino accompanied by a $Z^0$ or a $W^{\pm}$
boson also occur through the exchange in the 
$s$-channel of a $\tilde \nu^j_L$ sneutrino 
which can not be produced on-shell (see Section \ref{Asso}).
When kinematically allowed, 
these processes have some rates of order $100 \femtob$ at $\sqrt s = 200 \GeV$ and 
$10 \femtob$ at $\sqrt s = 500 \GeV$, for $\l_{1j1}=0.05$ and various masses of the
scalar \susyq\ particles~\cite{z:Chemsin}.

\subsection{Fermion Pair Production Via $\Rp$}


The single production involves only one \Rp\ coupling so that the 
corresponding rate is proportional to the Yukawa coupling squared.
This is in contrast to \Rp\ contributions (via additional sparticle
exchange) to \SM\ processes which are suppressed in proportion to
the square of the Yukawa coupling squared.

$\bullet$ {\bf One dominant \Rp\ \cc}
One may first consider the usual case of one dominant \Rp coupling
constant. 
This hypothesis, which is often made for simplifications reasons, has
two main
justifications. First, the low energy indirect bounds are more stringent
on the
products of the \Rp couplings $\l$, $\l'$ of $\l''$ than on the \ccs
separately.
For instance, the bounds imposed on 
the products $\l' \l''$ by the experimental limits on proton
decay are severe, and moreover, apply on all the
different flavor combinations of this
product. Second, from a theoretical point of view, by analogy with the 
Higgs Yukawa couplings structure, a strong
hierarchy
can be assumed on the flavor indices of the \Rp coupling constants.

{\bf Dilepton production}
The resonant sneutrino $\tilde \nu_{\mu}$ or $\tilde \nu_{\tau}$ production 
via $\l_{121}$ or $\l_{131}$ respectively followed by a decay through the 
same coupling constant (i.e. $\tilde \nu^i \to \bar l_j l_k$ via $\l_{ijk}$)
would lead to a spectacular signature, namely, an excess of events in 
the Bhabha scattering \cite{z:dimopoulos88}. 
For $\Gamma_{\tilde \nu_{(\mu,\tau)}}=1 \GeV$ 
and $\l_{1(2,3)1}=0.1$, the cross section of the Bhabha scattering,
including the $\tilde \nu_{(\mu,\tau)} $ sneutrino $s$-channel exchange and 
the interference terms, reaches $3 \picob$ at 
$\sqrt s=m_{\tilde \nu_{(\mu,\tau)}}=200 \GeV$
\cite{z:Kalept,z:Kalseul,z:KalZ}.
Due to these great cross sections, 
the values $\l_{(2,3)11}>0.05$ can be ruled out at $95 \%$ CL for
$\sqrt s=m_{\tilde \nu_{(\mu,\tau)}}=192 \GeV$ with ${\cal L}=500 \picob^{-1}$
\cite{z:Chou}.


In a scenario where none of the \susyq\ particles can be produced with
a significant cross section, due to very heavy \susyq\ particles or too 
weak couplings for the lighter ones with the \SM\ particles, 
the \susyq\ effects could only be virtual. 
In particular, different \Rp\ interactions could manifest themselves as 
contact terms in many contributions to \SM\ processes. 
By calculating the deviations of the cross sections values with respect 
to the Standard Model, bounds can be set on the relevant \Rp\ \ccs values. 

In~\cite{z:Kalept} the analysis of different contributions to \SM\
processes was based on the R-parity violation interpretation of 
the high $x$, high $Q^2$ anomalous HERA events. 
Interpreting the HERA events as top squark production 
\mbox{$\l'_{131}>0.05$}, 
see Eq.(\ref{heranom2})), then the constraints on the \Rp\ \ccs products, 
\mbox{$\l_{121} \l'_{131}<1.8 \ 10^{-4}$}, 
\mbox{$\l_{131} \l'_{131}<2.0$} \ $10^{-3}$ 
and \mbox{$\l_{123} \l'_{131}<2.4 \ 10^{-3}$}, as imposed by the rare $B$ decays, 
put strong bounds on some of the $\l$ couplings which are relevant for 
leptonic processes: \mbox{$\l_{121}<0.0036$}, 
\mbox{$\l_{131}<0.04$} and \mbox{$\l_{123}<0.048$}. 
However, if the HERA data are due to the charm squark production 
($\l'_{121}>0.05$, see Eq.(\ref{heranom2})), the rare $K$ decays do not
constrain the $\l_{131}$ and $\l_{123}$ couplings.
In the situation where the \Rp\ coupling constant $\l_{131}$ dominates, 
the Bhabha scattering receives a contribution from the exchange of 
a $\tilde \nu_{\tau}$ sneutrino in the $s$- and $t$-channels, while the 
$\tau^+ \tau^-$ production can occur via the exchange of a $\tilde \nu_e$ 
sneutrino in the $t$-channel only. For this reason, the impact of the 
\Rp\ diagram on the Bhabha scattering is higher than on the $\tau$ pair
production:
At $\sqrt s = 192 \GeV$, for $m_{\tilde \nu} = 300 \GeV$ and $\l_{131}=0.1$,
the effects on the cross sections are, ${\s(SM+RPV) \over \s(SM) }-1=4
\ 10^{-3}$ and $2 \ 10^{-2}$ for the $\tau^+ \tau^-$ and $e^+ e^-$ 
productions, respectively~\cite{z:Kalept,z:Kalseul,z:KalZ}. 
From the study of the \Rp\ contribution to the Bhabha scattering, the 
coupling $\l_{131}$ could be probed at $95 \%$ CL
at $\sqrt s = 192 \GeV$ with ${\cal L}=500 \picob^{-1}$ down to $0.28,0.57,0.84$ for 
$m_{\tilde \nu_{\tau}} = 400 \GeV, 700 \GeV, 1 \TeV$, respectively \cite{z:Chou}. 
The limit on $\l_{131}$ from the fits to the cross section and 
forward-backward asymmetry, using data collected by the DELPHI detector
during the run of LEP in 1995 (1996) at $\sqrt s =130$ and $136 \GeV$
($\sqrt s =161$ and $172 \GeV$) corresponding to a luminosity of order 
$6 \picob^{-1}$ ($20 \picob^{-1}$), is $\l_{131}<0.74$ for
$m_{\tilde \nu_e}=200 \GeV$ at $95 \%$ CL \cite{z:DELPHI1}
(similar analyses on the \Rp contributions to the
$e^+ e^- \to l^+ l^-$ reaction have been performed
by the L3 \cite{z:L3sm} and OPAL \cite{z:OPALsm} Collaborations). 
In case of a dominant coupling $\l_{123}$, the $t$-channel $\tilde
\nu_{\tau}$ exchange would contribute to $\tau^+ \tau^-$ pair
production. Through this contribution,
the coupling $\l_{123}$ could be probed with a $95 \%$ CL
at $\sqrt s=192 \GeV$ with ${\cal L}=500 \picob^{-1}$ down to $0.3,0.5,0.7$ for 
$m_{\tilde \nu_{\tau}}=400 \GeV, 700 \GeV, 1 \TeV$, respectively \cite{z:Chou}. 

The other \Rp\ \ccs $\l$ relevant for virtual effects in the \SM\ processes,
$e^+ e^- \to l \bar l$, namely, $\l_{121},\l_{231},\l_{122},\l_{132}$ 
and $\l_{133}$, were studied in~\cite{z:Chou}. 
Since the \Rp\ interactions have a spin structure different from the \SM
ones, the angular distribution for the lepton pair production is a sensitive
probe for the existence of the former. Hence, it was proposed to divide the
experimental angular width into bins, and to compare the observed number of 
events in each bin with the \SM\ prediction. 
The more optimistic results hold for the couplings,
$\l_{122},\l_{132}$ and $\l_{133}$, which can be probed with a $95 \%$ CL
at $\sqrt s=192 \GeV$ with ${\cal L}=500 \picob^{-1}$ down to $0.3,0.5,0.7$ for 
$m_{\tilde \nu}=400 \GeV,700 \GeV, 1 \TeV$, respectively.
Besides, the limit on $\l_{121}$ from the fits to the cross section and 
forward-backward asymmetry for the reaction, $e^+ e^- \to \mu^+ \mu^-$,
using data collected by the DELPHI detector,
is $\l_{121}<0.55$ ($\l_{121}<0.68$) for $m_{\tilde \nu_e}=100 \GeV$
($200 \GeV$) at a $95 \%$ confidence level \cite{z:DELPHI1}.
%
\begin{figure}[t]
\begin{center}
\vspace{-0.5cm}
\centerline{\psfig{figure=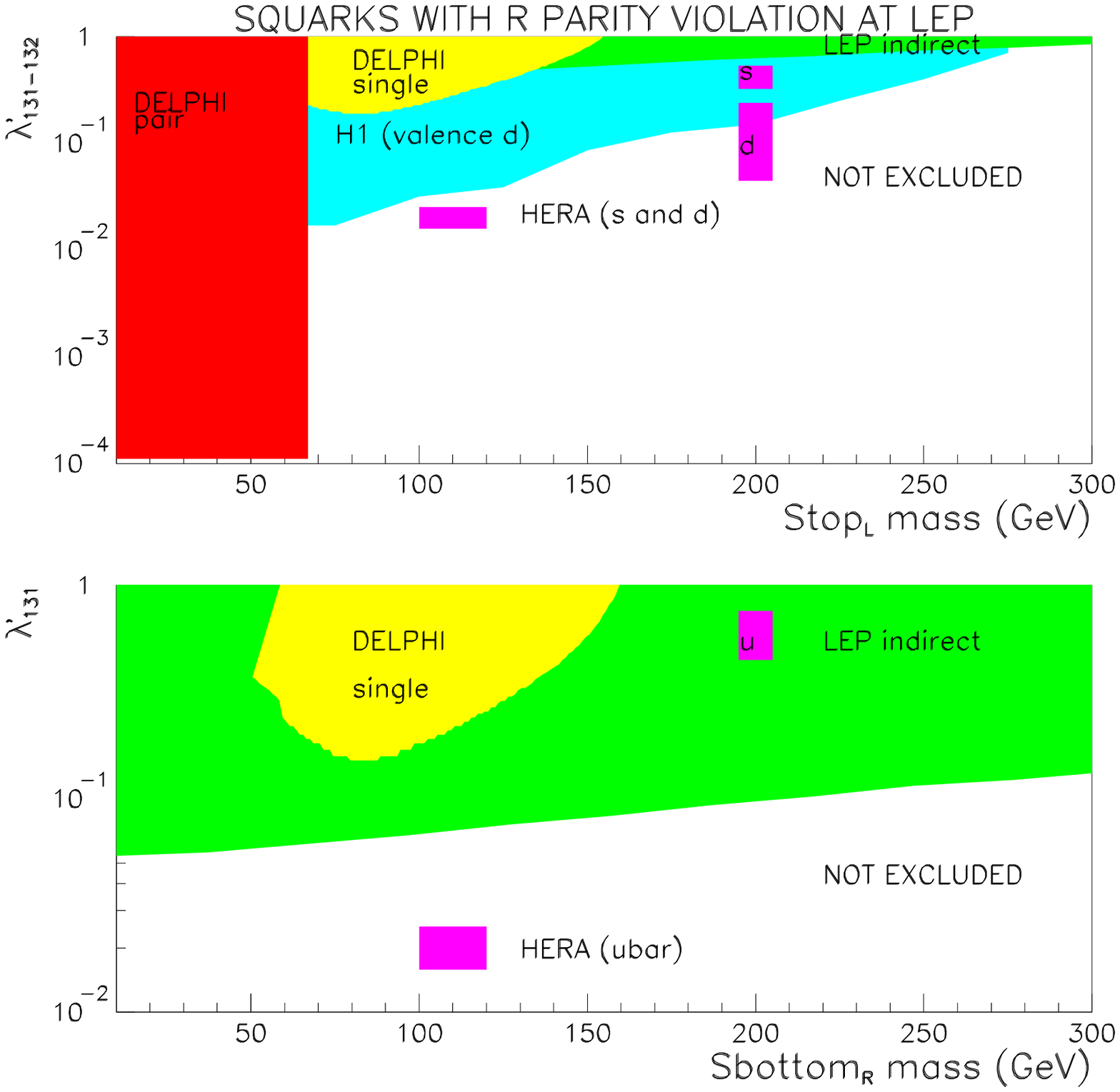,width=17cm}}
 \caption[]{ \label{fig:exclusq}
    { \small Exclusion domain in the $\l'$ versus $m_{\tilde q}$ plane.}}
\end{center}
\end{figure}           

{\bf b) Dijet production}
One may also think of a single dominant $\l'_{ijk}$ coupling. 
This hypothesis allows \Rp\ contributions to quark pair production through the 
exchange of a squark in the $t$-channel. 
For experimental reasons of quark tagging, the \Rp\ contributions to,
$e^+ e^- \to b \bar b, c \bar c$, via, $\l'_{1k3},\l'_{12k}$, respectively, 
are the most easier to analyse at LEP energies. 
Among these couplings $\l'_{1k3}$ and $\l'_{12k}$, the most stringent
constraint, predicted from the study of the quark pair production \cite{z:Chou}, 
arises for the constant $\l'_{123}$ which 
is subject to bounds of order, $0.4,0.67,0.92$, 
for $m_{\tilde q}=400 \GeV, 700 \GeV, 1 \TeV$ at $\sqrt s = 192 \GeV$ with 
a $95 \%$ confidence level and a luminosity, ${\cal L}=500 \picob^{-1}$. 
The effects of R-parity violation in quark pair production have also
been investigated experimentally \cite{z:Arnoud}, using the 1997 LEP data 
and the results of \cite{z:Alt}. 
Assuming that the \Rp\ contribution through the exchange
of a given squark to a given flavour channel is the only source of
deviation from the \SM\ processes,
the limits of Fig.\ref{fig:exclusq} (indicated by `` LEP indirect '') have
been derived. 
%
%
The LEP indirect limit represented in the upper plot of Fig.\ref{fig:exclusq}
derives from the analysis of the process $e^+ e^- \to d \bar d$ 
($e^+ e^- \to s \bar s$) occuring through the $\lambda'_{131}$ 
($\lambda'_{132}$) coupling constant via the exchange of a top-squark in the 
$t$-channel.
The LEP indirect limit in the lower plot of Fig.\ref{fig:exclusq}
is deduced from the analysis of the process $e^+ e^- \to u \bar u$ 
involving $\lambda'_{113}$ and occuring through the exchange of a bottom-squark 
in the $t$-channel.
%
%
The negative interference
term between the squark exchange and the \SM\ amplitudes is maximal for a
down 
type squark \cite{z:Richa}. The sensitivity of the measurement is
therefore reduced 
for the up squark exchange, as can be observed in Fig.\ref{fig:exclusq}.
Note that using bottom, charm or light quarks (u,d,s) tagging, 
the effects on separate quark flavours could also be studied \cite{z:Richa}. 
Besides, if with more data, a rate effect is seen at 
LEP II in $s \bar s$ or $b \bar b$ production, 
the charge asymmetry will help in 
confirming the squark exchange hypothesis.
On Fig.\ref{fig:exclusq}, the relevant exclusion domain from the H1
Collaboration 
is also shown (indicated by `` H1 (valence d) ''), as well as the bands
indicated by 
u,d,s which would have been relevant for the so called HERA anomaly
found in 1997.
These u,d,s bands are respectively associated  
with the three interpretations of the HERA anomalous events,
\begin{eqnarray}
e^+ \bar u \to \tilde s_R,\tilde b_R (\l'_{112},\l'_{113}),
\label{heranom1}
\end{eqnarray}
\begin{eqnarray}
e^+ d \to \tilde c_L,\tilde t_L (\l'_{121},\l'_{131}),
\label{heranom2}
\end{eqnarray}
\begin{eqnarray}
e^+ s \to \tilde c_L,\tilde t_L (\l'_{122},\l'_{132}).
\label{heranom3}
\end{eqnarray}
In particular, we observe on Fig.\ref{fig:exclusq} that the experimental
results
on the search for the indirect effects of R-parity violation in quark
pair production
at LEP completely exclude the HERA interpretation of Eq.(\ref{heranom1})
via $\l'_{113}$.  


$\bullet$ {\bf One dominant product of \Rp\ \ccs} \\ {\bf a) Dilepton
production}
Another interesting hypothesis consist in considering that 
two lepton number violating $\l_{ijk}$ Yukawa couplings 
are much larger than all the others, where both \Rp\ couplings violate
one and 
the same lepton flavour. In this scenario, low energy experiments are not 
restrictive and typically allow for couplings, $\l< 0.1(\tilde m /
200 \GeV)$ 
\cite{z:Kalept}. In case where the couplings $\l_{131}$ and $\l_{232}$ are
simultaneously different of
zero, the process $e^+ e^- \to \mu^+ \mu^-$ receives an additional
contribution from
$s$-channel $\tilde \nu_{\tau}$ exchange. At $\sqrt s=192 \GeV$, for
$m_{\tilde \nu}=300 \GeV$
and $\l_{131}=\l_{232}=0.1$, the effects on the cross sections and on
the forward-backward
asymmetries are, ${\s(SM+RPV) \over \s(SM) }-1=1.5 \
10^{-3}$, and,
$A_{FB}(SM+RPV)-A_{FB}(SM)=9 \ 10^{-4}$
\cite{z:Kalept,z:Kalseul,z:KalZ}. 
The $95 \%$ CL limit on $\l=\l_{131}=\l_{232}$ 
resulting from the fits on cross section 
and forward-backward asymmetry based on 
the 1995 ($\sqrt s =130$ and $136 \GeV$)
and 1996 ($\sqrt s =161$ and $172 \GeV$) LEP data, has been obtained
in \cite{z:DELPHI1}. The best limits on $\l$ are obtained 
for, $m_{\tilde \nu_{\tau}} \approx \sqrt s$, but the radiative 
return process gives a significant sensitivity between those points.
For, $m_{\tilde \nu_{\tau}} =130,136,161,172 \GeV$, one get limits, 
respectively, of order, $\l <0.035,0.03,0.03,0.02$,
assuming a tau-sneutrino width of $1 \GeV$.
Now, if the couplings $\l_{121}$ and $\l_{323}$ do not vanish, 
the $s$-channel exchange of a $\tilde \nu_{\mu}$ sneutrino contributes
to the $\tau^+ \tau^-$ pair production.
%
%
%
%
The $95 \%$ CL limit on $\l=\l_{121}=\l_{323}$ \cite{z:DELPHI1}
resulting from the fits of the $\sigma(e^+ e^- \to \tilde \nu
\to \tau^+ \tau^-)$ cross section based on the 1995 and 1996 LEP data
are of the same order of the limits on $\l=\l_{131}=\l_{232}$. 


{\bf b) Dijet production}
Finally, we can concentrate on the case where one of the dominant
\Rp
\ccs is a $\l_{ijk}$ while the other is a $\l'_{ijk}$. In such a
scenario, 
a $\tilde \nu_i$ sneutrino which is produced at the resonance, 
could decay via $\l'_{ijk}$ into two down squarks, $\tilde q_j \ \tilde
q^*_k$.
Since the $\l'_{3jk}$ are the less constrained constants, the resonant $\tilde
\nu_{\tau}$
production (via $\l_{131}$) is the most promising. The $\tilde
\nu_{\tau}$ 
sneutrino can decay via $\l'_{333}$ into b quarks, which can be tagged
experimentally with a rather good efficiency. 
The effective luminosity required to discover
an excess of $b \bar b$ events at $5 \s$ confidence level at LEPII,
assuming a tagging efficiency of $40 \%$, is 
$1.1 \picob^{-1}/\GeV,0.43 \picob^{-1}/\GeV$ 
for $m_{\tilde \nu_{\tau}}=110 \GeV, 145 \GeV$, $\l_{131}=0.01,0.005$ and 
$\l'_{333}=1.0,0.1$, respectively \cite{z:erler}. 
Assuming that the $\tilde \nu_{\tau}$ sneutrino decays into $b \bar b$,
the limit on $\l_{131}$ at $95 \%$ CL can be derived from
the experimental measurement of the b quark pair production,
using the LEP data at $\sqrt s = 161 \GeV$ and $\sqrt s = 172\GeV$~\cite{z:Arnoud}.
%
%
%
A most interesting window for
the sneutrino resonance exists near the $Z$ boson pole. The sneutrino
resonance
could still be observable since then it would increase the branching
ratio 
$R_b(Z^0 \to b \bar b)$ and reduce the $b$ quark forward-backward
asymmetry 
$A_{FB}(b)$. The resonant $\tilde \nu_{\tau}$ sneutrino could also 
decay into a pair of down quarks,
$d \bar d$, through the coupling $\l'_{311}$.
Since the angular distribution of the $d$ and $\bar d$
jets is nearly isotropic on the sneutrino resonance, the strong
forward-backward asymmetry in the \SM\ continuum, $A_{FB}(b) \approx
0.65$ at $\sqrt s= 200 \GeV$,
is reduced to $\approx 0.03$ on top of the sneutrino resonance
\cite{z:Kal}. 
If the total cross section $\s(e^+ e^- \to hadrons)$ can be measured
with an 
accuracy of about $1 \%$ at $\sqrt s=184(192) \GeV$, the Yukawa couplings
can 
be bounded to $(\l_{131}\l'_{311})^{1/2}<0.072(0.045)$ for a 
$200 \GeV$ tau-sneutrino mass \cite{z:Kal}.
This limit is estimated for energies much below 
the resonance, $ \vert \sqrt s-m_{\tilde \nu_{\tau}} \vert << 
\Gamma_{\tilde \nu_{\tau}}$, and is thus independent of the sneutrino
width.

As a conclusion, the analysis at leptonic colliders of the \Rp\ contributions 
to \SM\ processes lead to bounds on the lepton number violating \ccs which are 
not as stringent as the low energy limits. 

\subsection{Single Production in $e-\gamma$ Collisions}

The charged and neutral sleptons can also be singly produced at
leptonic colliders in the lepton-photon collisions,
$e^{\pm} \gamma \to l^{\pm} \tilde \nu, \tilde l^{\pm} \nu$,
where the photon is an on shell photon radiated by one of the colliding
leptons. Those productions, which were considered in \cite{z:All}, 
allow to study the \Rp\ couplings, 
$\l_{122},\l_{123},\l_{132},\l_{133},\l_{231}$, which are not involved 
in the single productions from $e^+ e^-$ reactions. 
\begin{figure}[t]
\begin{center}
\centerline{\psfig{figure=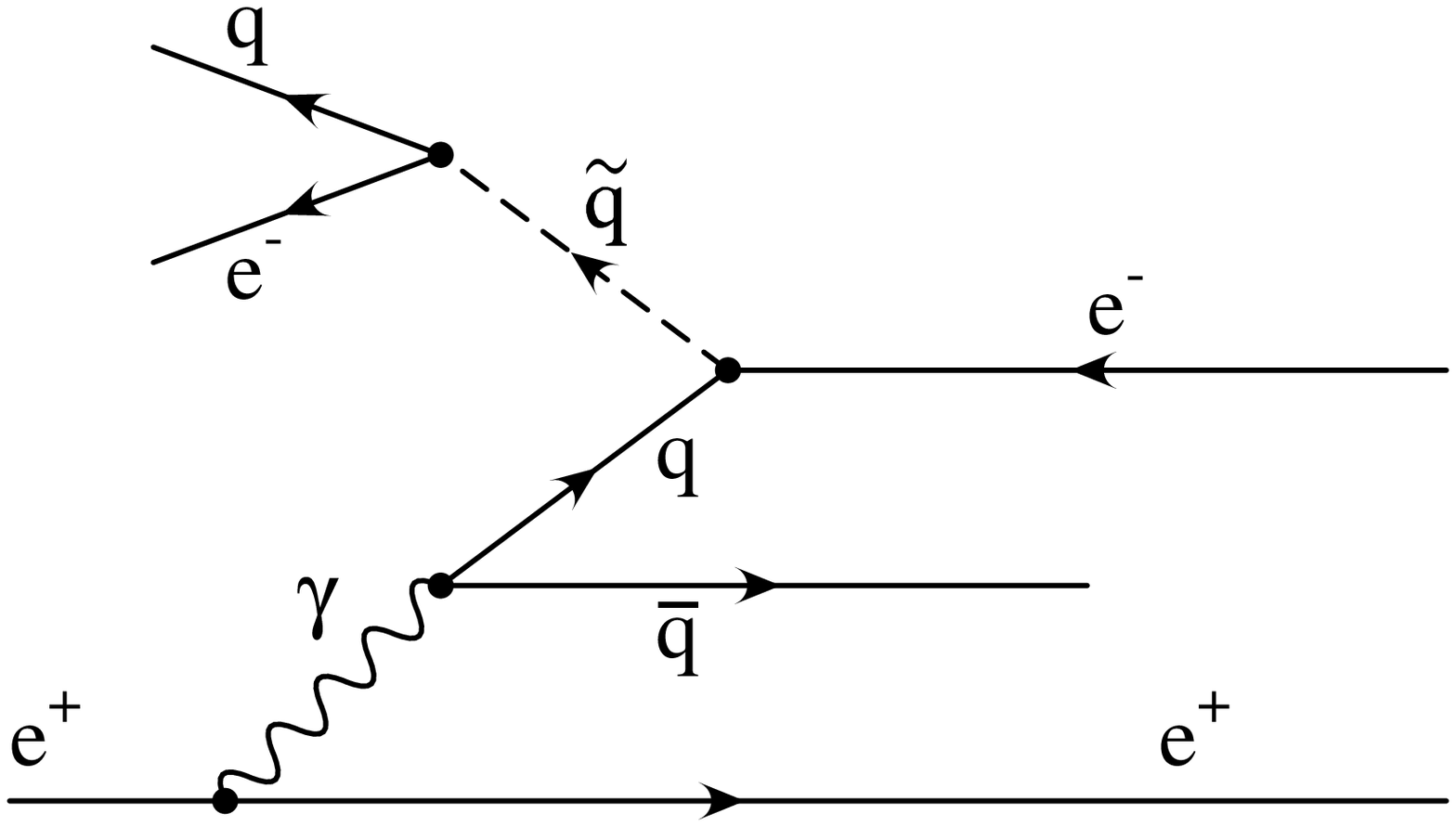,width=10cm}}
\end{center}
\caption{\footnotesize  \it
Single squark production in electron-photon collisions.
\rm \normalsize }
\label{sinsquark}
\end{figure}
The slepton or sneutrino production occurs via the exchange of a charged 
lepton in the $s$-channel, and a charged slepton or lepton, respectively, 
in the $t$-channel. 
Therefore, since the $t$-channel is dominant and $m_{\tilde l}>>m_l$, the 
slepton production is about two order of magnitude less than the 
sneutrino production, which is, 
$\s(e^+ e^- \to \tilde \nu_j e \tau)=30\femtob \ (300\femtob)$ at 
$\sqrt s= 192 \GeV \ (500 \GeV)$, for $m_{\tilde \nu_j}=150 \GeV$ and 
$\l_{j13}=0.05$. 
The produced sneutrino can either decay directly via $\l$ into 2 leptons, 
or indirectly via lighter neutralinos and charginos 
( e.g. $\tilde \nu \to \nu  \tilde \chi^0$ ) leading respectively 
to either $4 l$ or $4 l + \Eslash $ final states. Those various
signatures were simulated together with the associated \SM\ backgrounds 
in order to apply some  kinematic cuts. 
The indirect decay of the sneutrino via $\l_{122}$ leads to the 
signal with the higher sensitivity, allowing to probe the \cc $\l_{122}$
down to the values $0.025,0.04,0.065$ at a $5 \s$ discovery 
level with an energy $\sqrt s =192 \GeV$ and a luminosity 
${\cal L} =100 \picob^{-1}$ for $m_{\tilde \nu} = 100 \GeV, 125 \GeV, 150 \GeV$,
respectively.

The single production of a squark is also possible in the $e \gamma$
interactions, as shown in Fig.\ref{sinsquark}, through $\l'$ couplings. 
Assuming a squark LSP, the produced squark will have a direct decay via
$\l'$ into a lepton plus a quark, so that the final state topologies 
will be energetic mono-jet with one well isolated energetic electron and 
eventually a low energy jet in the forward region of the detector, 
in case where the initial electron which scatters the quasi real photon 
escapes the detection. 
These events have been searched experimentally \cite{z:Arnoud}, 
using the 1997 LEP data at $\sqrt s = 161 \GeV$ and $\sqrt s = 172 \GeV$, 
and since no evidence have been found for the single squark
production, an exclusion domain has been deduced in the plane $\l'$ vs. 
$m_{\tilde q}$ as shown in Fig.\ref{fig:exclusq} (indicated by `` DELPHI
single '').
The difference between the two exclusion plots is explained by the fact 
that the only possible \Rp\ stop decay via $\l'$ is, $\tilde t \to e d$,
while for the sbottom, the charged lepton channel branching ratio is,
$B (\tilde b \to e u) \approx 50 \%$. 
   
\subsection{$\Rp$ Contributions to Flavour Changing Neutral Currents}
\label{rpvfcnc}

In the Standard Model, the \fc effects are exceedingly 
small~\cite{z:Axel,z:Clem,z:Gana}. In particular, it is useful 
to recall that the \fcnc effects 
arise through loop diagrams only.
For instance, the typical structure of the one loop diagram for the
$Z$ boson and quarks pair
vertex $Z q_{J} \bar q_{J'}$, is, $\sum_i V^*_{iJ}  V_{iJ'} f(m_i^2/m_Z^2)$, 
where $V_{ij}$ are the elements of the CKM matrix and $m_i$ are the
masses of the quarks involved in the loop. 
This schematic formula shows explicitly how the CKM matrix unitarity, along 
with the quarks masses degeneracies relative to the $Z^0$ boson mass scale 
(valid for all quarks with the exception of the top quark) strongly suppress 
the \fcnc effects. 
Although \fc effects are expected to attain observable levels in several 
extensions of the Standard Model, the contributions from the MSSM are bounded
by postulating either a degeneracy of the soft \SUSY breaking scalars masses 
or an alignment of the fermion and scalar superpartners mass matrices. 
Those constraints are imposed by some low-energy experimental bounds.
Early calculations of the $Z^0$ boson \fc decay rates, 
$Z \to q_{J} \bar q_{J'}$, through triangle diagrams involving squarks and 
gluinos, found small results compared to that of the Standard 
Model~\cite{z:Dun,z:Mukho}. 
Therefore, the flavor changing effects offer the opportunity
to probe physics beyond the MSSM.
The \Rp\ interaction, because of its non trivial flavour structure, opens up 
the possibility of observable \fc effects at the tree level. 

$\bullet$ {\bf Fermion pair production}
The fermion pair production is a relatively simple and well known reaction 
and hence offers the opportunity of testing new physics.
Besides, the leptonic colliders provide a clean environment 
to study flavor changing physics. These considerations have
motivated the study of
the \Rp\ contributions to the \fcnc reactions, $l^+ l^- \to f_J \bar f_{J'}$, 
with $J \neq J'$ \cite{z:Chemfer}. 
These \Rp\ processes occur at the tree level through the exchange of a 
\susyq\ scalar particle in the $s$- or $t$-channel, and at one loop level at 
the $Z^0$ boson pole. 
Both fermion and scalar particles are running in the triangle loop diagram. 
Considering simple assumptions on the scalars masses and for the flavour 
structures of the \Rp\ coupling constants, the lepton number violating 
interactions contributions to the $Z$-boson flavour off-diagonal decays 
branching ratios, $B_{JJ'}= B(Z \to l^-_J l^+_{J'})$, scale approximately 
as $ B_{JJ'}\approx 
      ({\l \over 0.1})^4 ({100 \GeV \over \tilde m})^{2.5} \ 10^{-9} $,
where $\tilde m$ represents a \susyq\ particle mass. 
At energies well above the $Z$-boson pole, the \fc rates,
$\s_{JJ'}= \s ( l^+ l^-  \to l^-_J l^+_{J'})$, are of order,
$\s_{JJ'}\approx ({\l \over 0.1})^4 ({100 \GeV \over \tilde m})^{2-3}
(1-10) \femtob$, and slowly decrease with the c.m. energy. 
Note that the corresponding results for quarks pair production have an 
extra color factor, $N_c=3$. Besides, the rates are evidently greatly 
enhanced at the resonance of the sneutrino, which is exchanged in the 
$s$-channel.
As a conclusion, due to the strong sensitivity on the \Rp\ couplings 
and on the \susyq\ particles masses, the \fc rates are comparable to the 
\fc rates calculated in the framework of the Standard Model.
Indeed, for the down-quark-antiquark case, for instance, the $Z^0$ decays 
branching fractions were estimated at the values, 
$10^{-7}$ for $(\bar b s  + \bar s b)$, $10^{-9}$ for $(\bar b d + \bar d b)$
and $10^{-11}$ for $(\bar d s  + \bar s d)$. Nevertheless, at energies 
well above the $Z^0$ mass, the \fc rates are high enough to be observable 
with the luminosities expected at the Next Linear Colliders \cite{z:NLC}. 
Finally, the current experimental limits on the flavour non-diagonal 
leptonic branching ratios \cite{z:67}, $B_{JJ'}<[1.7,9.8,17.] \ 10^{-6}$, 
for the family couples $JJ'=[12,23,13]$, constrain the \Rp\ \ccs products 
to be $\l_{ijJ} \l^*_{ijJ'}<[0.46,1.1,1.4]$ and 
${\l'}^*_{Jjk} {\l'}_{J'jk}<[0.38,0.91,1.2] \ 10^{-1}$, for the same flavour 
configurations $JJ'=[12,23,13]$, under the hypothesis of a pair of dominant 
coupling constants and if $\tilde m =100 \GeV$. 
These bounds, which are not competitive with the low energy limits, should
be improved in the context of the physics at linear colliders.

$\bullet$ {\bf Single top quark production}
Of special interest is the case of single top quark production,
$l^+ l^- \to t \bar c, \ \bar t c$~\cite{z:Mahanta,z:JYi,z:chela,z:Chemfer,z:JYigg}. 
The contribution from the \SM\ to the single top quark 
production~\cite{z:Axel,z:Clem,z:Gana} proceeds at one loop level through the 
exchange in the $s$-channel of either an off shell $Z$-boson or photon. 
It is particularly reduced since, unlike $b \bar s$ production for example,
it does not get a large contribution from heavy fermion in the loop. 
Since the MSSM contribution has been shown to be small compared to the
\SM\ one \cite{z:Dun,z:Mukho}, an excess of events in the single top quark 
production would probe the existence of new physics beyond the minimal 
supersymmetric standard model. 
Furthermore, the single top production offers the opportunity to learn about 
\fcnc effects in the up-type quarks sector, since as we will see in
Section\ref{smhad}, the top quark offers some extremely clean
signatures, with a rather energetic lepton, some missing energy and a b jet 
which can be tagged with a good efficiency.

\begin{figure}[t]
\begin{center}
\centerline{\psfig{figure=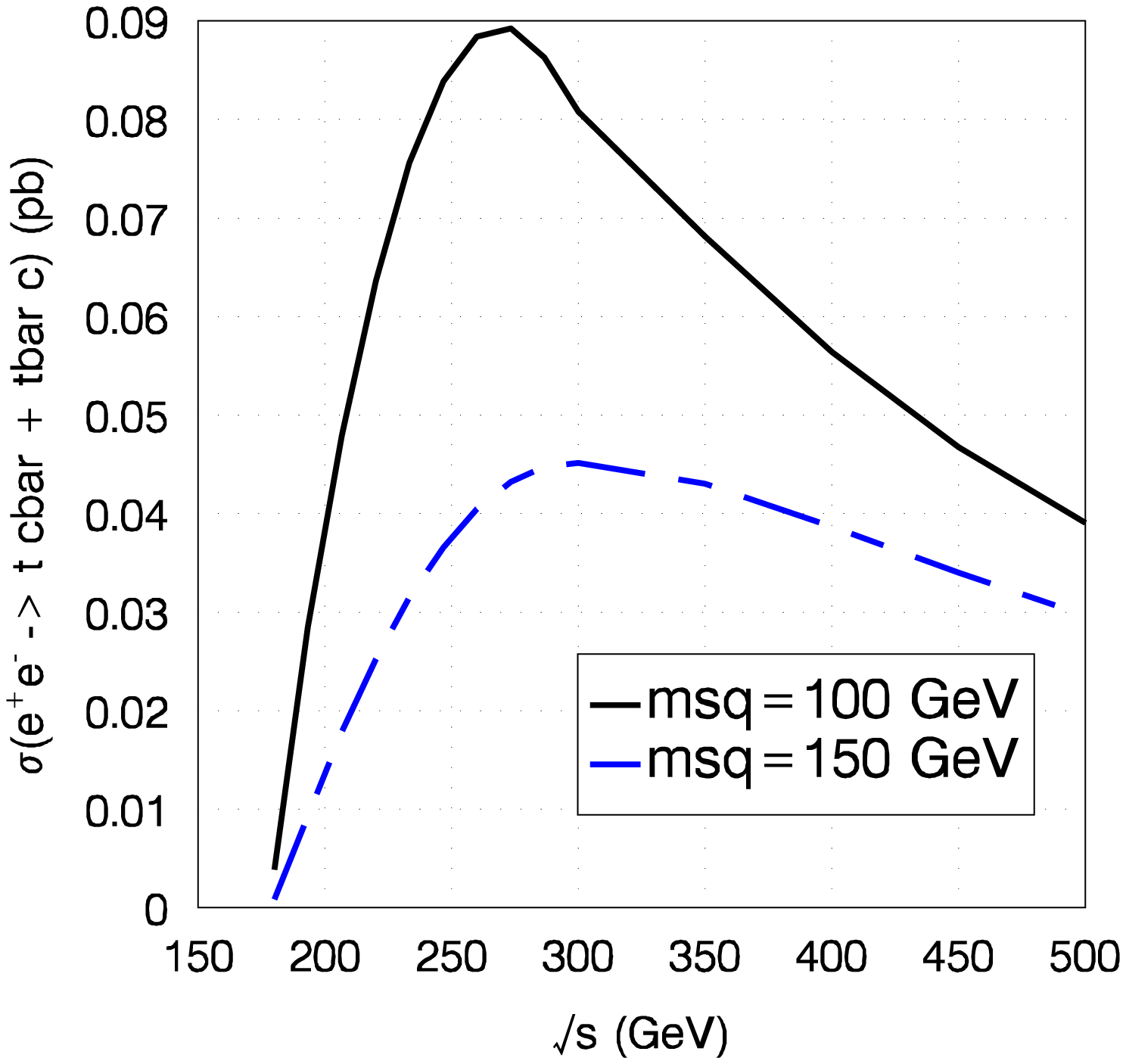,width=11.8cm}}  
\caption{Cross section of the reaction $e^+ e^- \to 
t \bar c + \bar t c$ as a function of the center of mass energy  
for $\l'_{12k} \l'_{13k}=0.01$. The solid line corresponds to
$m_{\tilde d_{kR}}=100GeV$ and the dashed line to 
$m_{\tilde d_{kR}}=150GeV$.} 
\label{fig:eetc}
\end{center}
\end{figure}

The reaction $e^+ e^- \to t \bar c + \bar t c$ occurs via the
exchange of a $\tilde d_{kR}$ squark in the $t$-channel through
the \Rp couplings $\l'_{12k}$ and $\l'_{13k}$. In Fig.\ref{fig:eetc}
\cite{z:JYi} the cross section of this process is shown as a 
function of the center of mass energy for a value of 
$\l'_{12k} \l'_{13k}$ equal to $0.01$ which is the order of 
magnitude of the low-energy 
constraint on this product of \Rp couplings for 
$m_{\tilde f}=100GeV$.\\
Since the top quark mainly decays as $t \to b W$, the reaction
$e^+ e^- \to t c$ leads to the interesting   
final state $b c l \nu$ if the $W$-boson decays leptonically. 
This signature has the Standard Model background $e^+ e^- \to 
W^+ W^- \to b c l \nu$. Nevertheless, this background can
be suppressed by some kinematical cuts \cite{z:chela}. For instance,
an effective cut can be based on the fact that the $c$ quark
produced in the reaction $e^+ e^- \to t c$ has a fixed energy
which is given by $E(c)=(s+m_t^2-m_c^2) / 2 \sqrt s$. Hence,
the study of the final state $b c l \nu$ would allow to probe 
values of the product $\l'_{12k} \l'_{13k}$ down to $\sim 0.1$ 
for $m_{\tilde d_{kR}}=1TeV$ at linear colliders with a center 
of mass energy of $\sqrt s =500 GeV$ and a luminosity of 
${\cal L}=100 fb^{-1}$ \cite{z:chela}.

\begin{figure}[t]
\begin{center}
\centerline{\psfig{figure=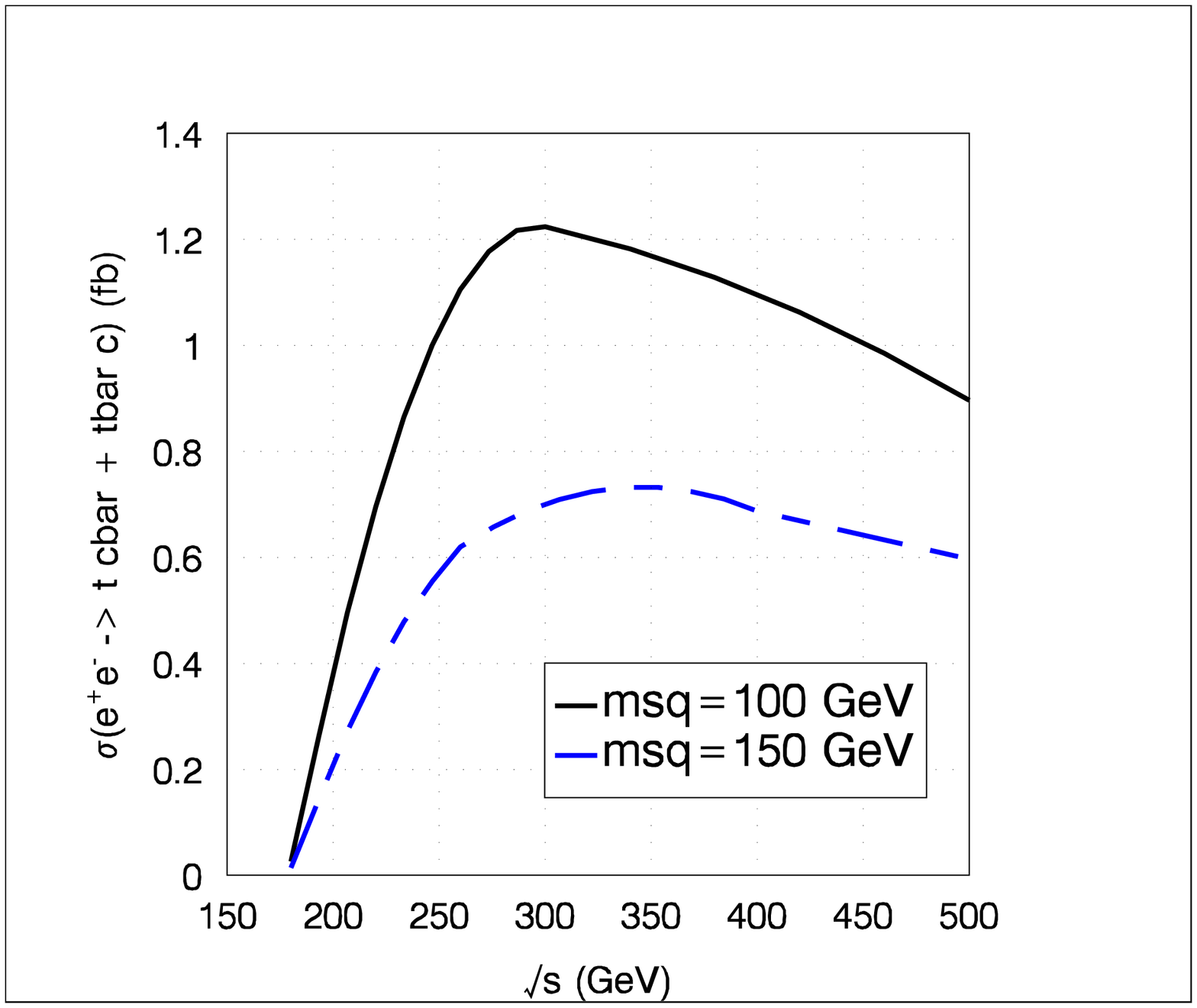,width=11.8cm}}  
\caption{Cross section of the reaction $e^+ e^- \to 
t \bar c + \bar t c$ as a function of the center of mass energy  
for $\l''_{223} \l''_{323}=0.625$. The solid line corresponds to
$m_{\tilde d_R}=100GeV$ and the dashed line to 
$m_{\tilde d_R}=150GeV$.} 
\label{fig:eetcs}
\end{center}
\end{figure}

The reaction $e^+ e^- \to t \bar c + \bar t c$ receives also
contributions at one loop level from the $\l''$ interactions 
\cite{z:Chemfer,z:JYi}.
These contributions exchange a $\tilde d_R$ squark in the loop
and involve the $\l''_{2jk}$ and $\l''_{3jk}$ coupling constants.
In Fig.\ref{fig:eetcs}
\cite{z:JYi} the rate of the sum of these contributions is 
shown as a function of the center of mass energy for a value of 
the involved product of \Rp couplings
$\l''_{223} \l''_{323}$ equal to its low-energy limit for 
$m_{\tilde f}=100GeV$ namely $0.625$. The motivation for 
considering the product $\l''_{223} \l''_{323}$ 
is that it has the less stringent low-energy constraint among the
$\l''_{2jk} \l''_{3jk}$ products.

\begin{figure}[t]
\begin{center}
\centerline{ \rotatebox{-90}{
\psfig{figure=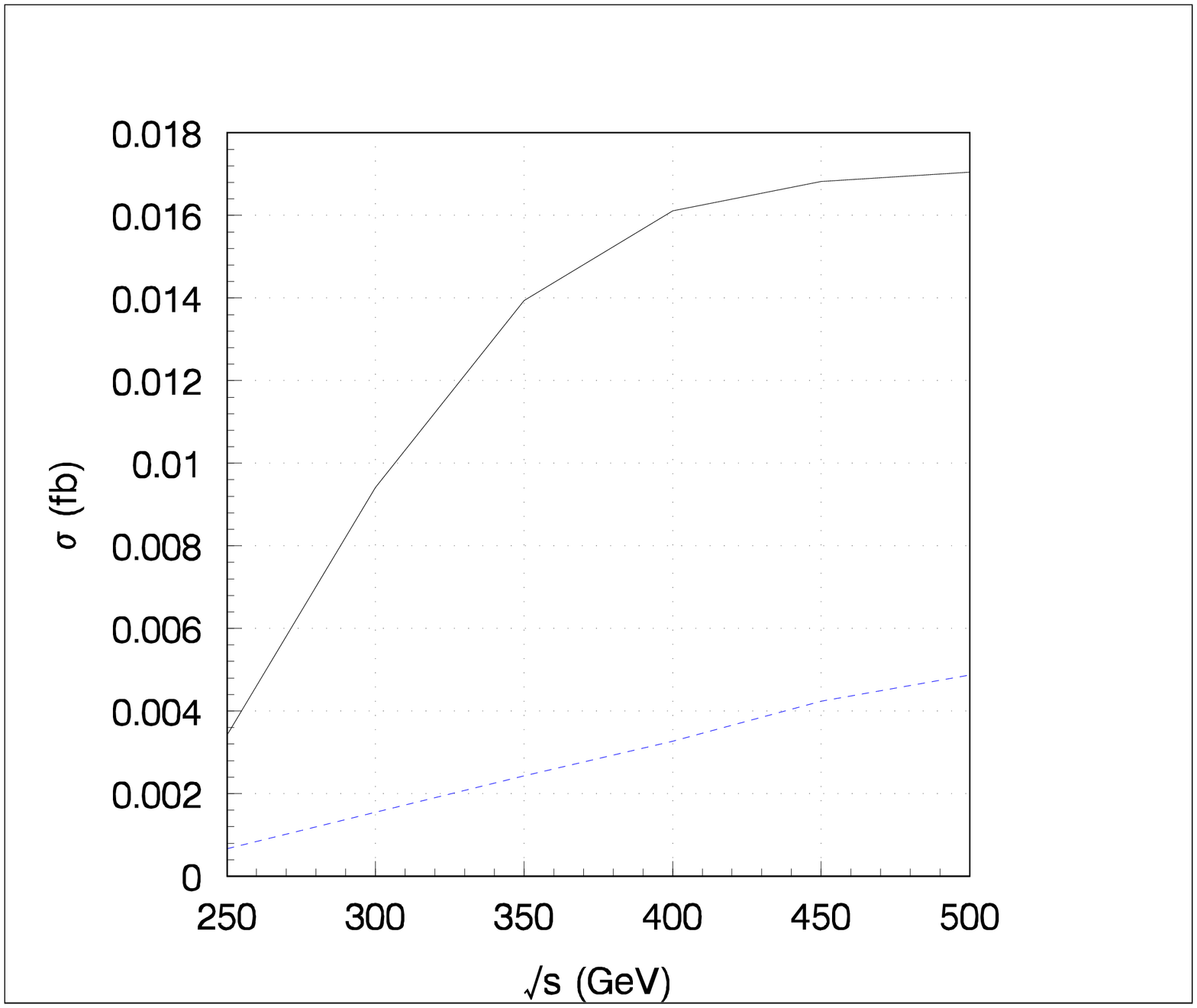,width=9.8cm}}}  
\caption{Cross section of the reaction $e^+ e^- \to 
\gamma \gamma \to
t \bar c + \bar t c$ as a function of the center of mass energy  
for $\l'_{323} \l'_{333}=0.096$. The solid line corresponds to
$m_{\tilde l_{iL}}=m_{\tilde d_{kR}}=100GeV$ and the dashed line 
to $m_{\tilde l_{iL}}=m_{\tilde d_{kR}}=150GeV$.} 
\label{fig:ggtc}
\end{center}
\end{figure}
\begin{figure}[t]
\begin{center}
\centerline{ \rotatebox{-90}{
\psfig{figure=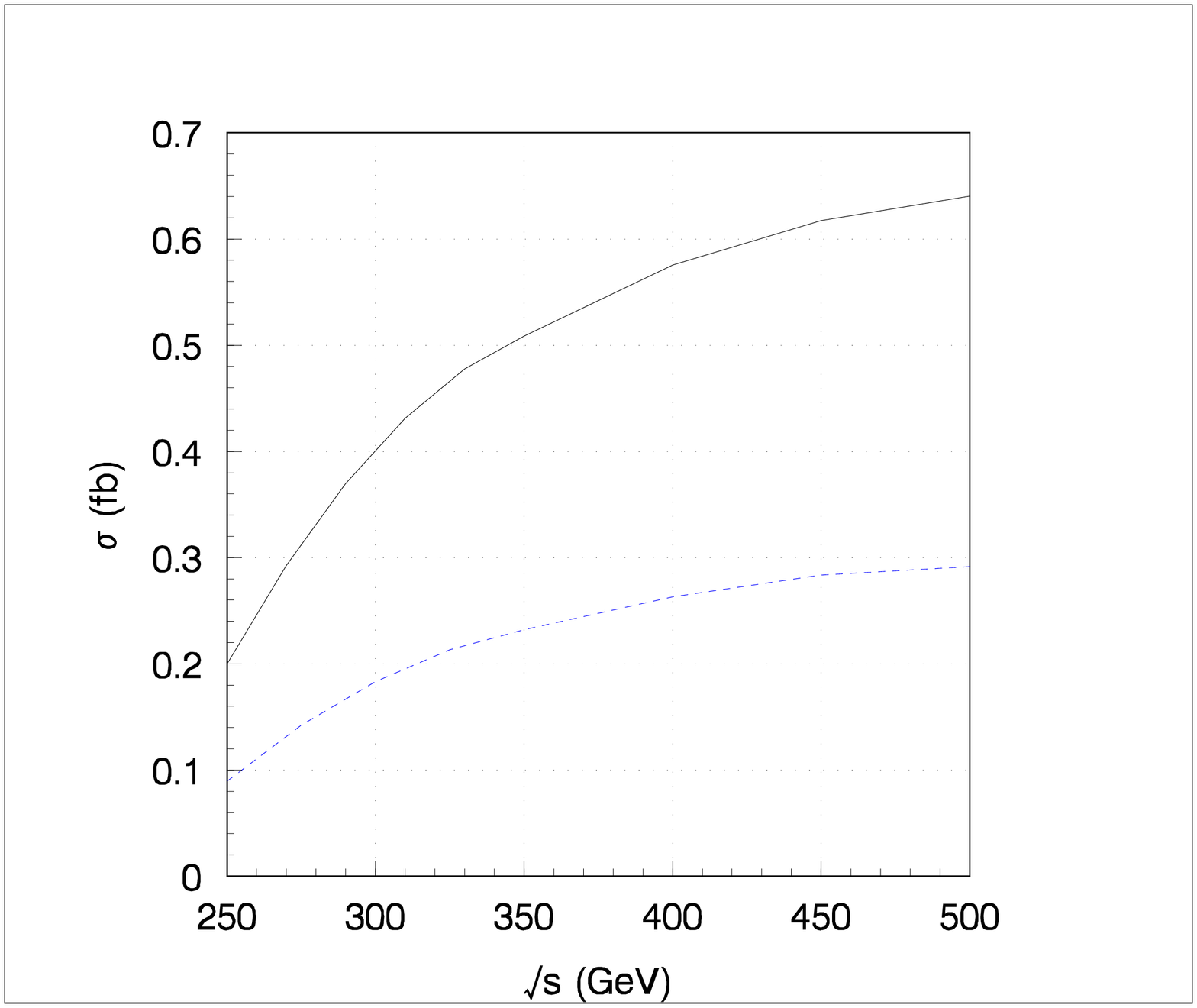,width=9.8cm}}}  
\caption{Cross section of the reaction $e^+ e^- \to 
\gamma \gamma \to
t \bar c + \bar t c$ as a function of the center of mass energy  
for $\l''_{223} \l''_{323}=0.625$. The solid line corresponds to
$m_{\tilde d_R}=100GeV$ and the dashed line to 
$m_{\tilde d_R}=150GeV$.} 
\label{fig:ggtcs}
\end{center}
\end{figure}

The $t \bar c / \bar t c$ production at leptonic colliders 
can also occur at one loop level via photon-photon reactions 
as $e^+ e^- \to 
\gamma \gamma \to t \bar c + \bar t c$. These reactions
involve the products of \Rp couplings 
$\l'_{i2k} \l'_{i3k}$ when $\tilde l_{iL}$ 
sleptons or $\tilde d_{kR}$ squarks are exchanged in the loop
and the products $\l''_{2jk} \l''_{3jk}$
when $\tilde d_R$ squarks run in the loop.
In Fig.\ref{fig:ggtc} (Fig.\ref{fig:ggtcs})
\cite{z:JYigg} the rate of the reaction $e^+ e^- \to 
\gamma \gamma \to t \bar c + \bar t c$ is 
shown as a function of the center of mass energy for a value of 
the involved product of \Rp couplings $\l'_{323} \l'_{333}$ 
($\l''_{223} \l''_{323}$) equal to its low-energy bound for 
$m_{\tilde f}=100GeV$ namely $0.096$ ($0.625$). 
The motivation for choosing the product $\l'_{323} \l'_{333}$ 
($\l''_{223} \l''_{323}$) is that it has the less stringent 
low-energy constraint among the $\l'_{i2k} \l'_{i3k}$ 
($\l''_{2jk} \l''_{3jk}$) products.

By comparing Fig.\ref{fig:eetc} and Fig.\ref{fig:ggtc}, we observe 
that the cross section of the reaction $e^+ e^- \to 
t \bar c + \bar t c$ via $\l'$ interactions is typically
one order of magnitude larger than
the rate of the process $e^+ e^- \to \gamma \gamma 
\to t \bar c + \bar t c$ via the same interactions. 
This is due to the fact that in the case of the $\l'$ 
interactions, the reaction $e^+ e^- \to 
t \bar c + \bar t c$ occurs at tree level while the
process $e^+ e^- \to \gamma \gamma \to t \bar c + \bar t c$
occurs only at one loop level. In contrast, 
it turns out by comparing Fig.\ref{fig:eetcs} and Fig.\ref{fig:ggtcs}
that the $\l''$ couplings give similar effects in the reactions 
$e^+ e^- \to t \bar c + \bar t c$ and 
$e^+ e^- \to \gamma \gamma \to t \bar c + \bar t c$.
Therefore, a combination of the results from the $e^+ e^-$ and 
$\gamma \gamma$ collisions would allow to distinguish between
the $\l'$ and $\l''$ effects on the $t \bar c / \bar t c$ 
production.

\begin{figure}[t]
\begin{center}
\centerline{\psfig{figure=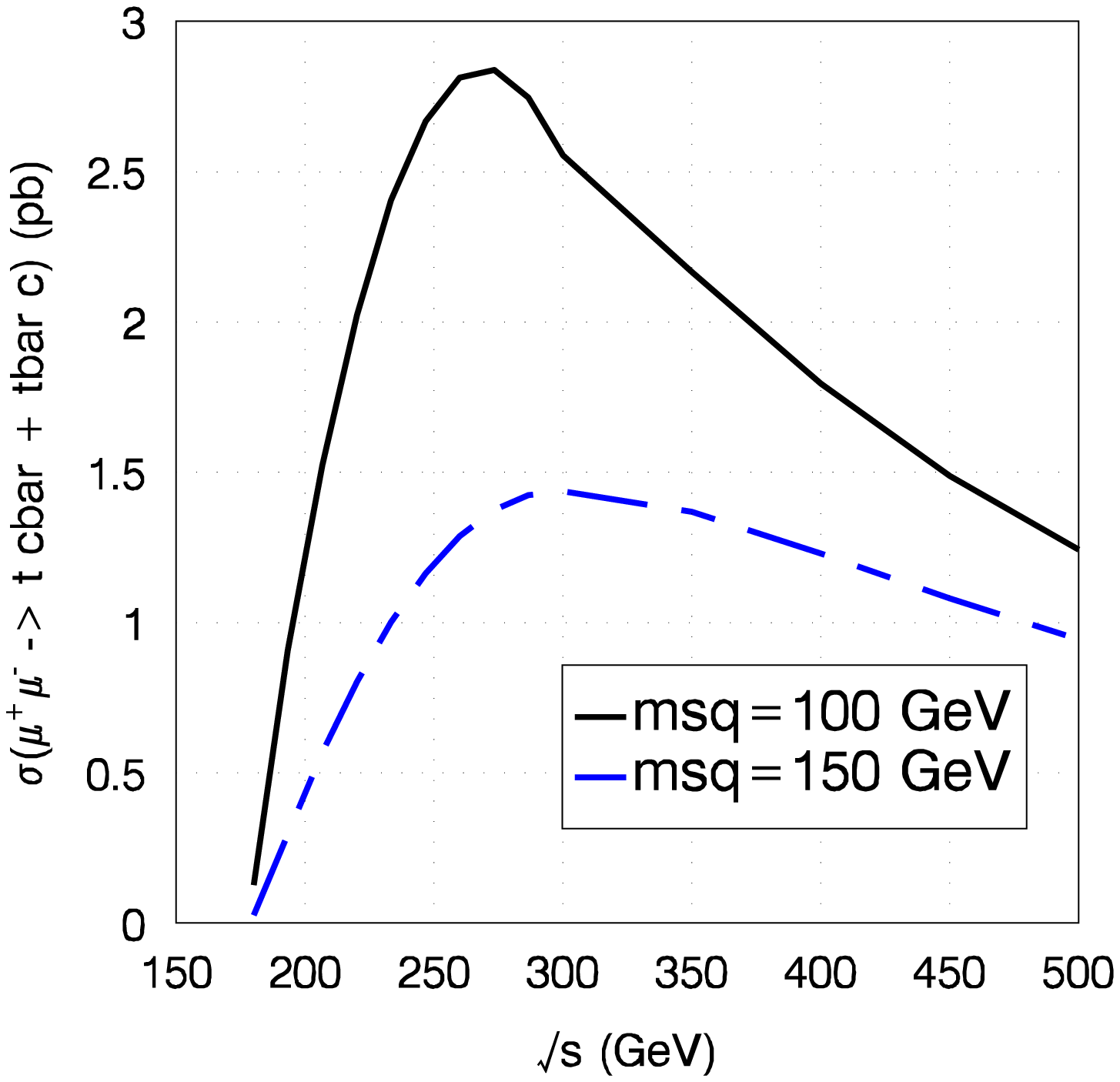,width=11.8cm}}  
\caption{Cross section of the reaction $\mu^+ \mu^- \to 
t \bar c + \bar t c$ as a function of the center of mass energy  
for $\l'_{223} \l'_{233}=0.065$. The solid line corresponds to
$m_{\tilde d_{kR}}=100GeV$ and the dashed line to 
$m_{\tilde d_{kR}}=150GeV$.} 
\label{fig:mmtc}
\end{center}
\end{figure}

Finally, the reaction $\mu^+ \mu^- \to t \bar c / \bar t c$ 
occurs via the
exchange of a $\tilde d_{kR}$ squark in the $t$-channel through
the \Rp couplings $\l'_{22k}$ and $\l'_{23k}$. In Fig.\ref{fig:mmtc}
\cite{z:JYi} the rate of this process is shown as a 
function of the center of mass energy for a value of 
the involved product of \Rp couplings
$\l'_{223} \l'_{233}$ equal to its low-energy limit for 
$m_{\tilde f}=100GeV$ namely $0.065$. The motivation for 
considering the product $\l'_{223} \l'_{233}$  
is that it has the less stringent low-energy constraint among the
$\l'_{22k} \l'_{23k}$ products.


$\bullet$ {\bf Sfermion pair production}
In a version of the MSSM without degeneracies in the sleptons mass
spectra, \fc effects can be induced in the \susyq\ particle pair production,
which should be investigated with high precision measurement in the Next 
Linear Colliders \cite{z:NLC}. 
The \Rp\ interactions could also generate such effects, through the exchange 
of a neutrino in the $t$-channel, in one of the much studied reaction: 
the slepton pair production, 
$e^+ e^-\to \tilde l_J \tilde l^*_{J'}$ ($J \neq J'$). 
The flavour non-diagonal rates vary in the range, 
$\s_{JJ'} \approx ({\L \over 0.1})^4 (2-20) \ \femtob$~\cite{z:Chemsca}, 
with $\L=\l, \l'$, for sleptons masses, $m_{\tilde l} < 400 \GeV$,
as one spans the interval of c.m. energies from the $Z$ boson pole up 
to the $\TeV$. 
Due to the strong dependence on the \Rp\ couplings, the flavour non-diagonal 
rates reach smaller values than the rates obtained in the flavours oscillation
approach~\cite{z:Arkani}, which range between $250 (100)$ and $0.1 (0.01) 
\femtob$ for $\sqrt s=190 (500) \GeV$.

\subsection{$\Rp$ Contributions to CP Violation}
\label{indlept}

The CP violation effects provide also some effective tests for new physics, 
since the contributions to the CP asymmetries in the \SM\ are small.
It is the case, for example, of the vector bosons ($Z$ and/or $W$ boson) 
decay rates CP-odd asymmetries \cite{z:Bern,z:Hou,z:Eil}. 
On the other hand, in most proposals of physics beyond the Standard Model, 
the prospects for observing \fc effects in CP rate asymmetries are on the 
optimistic side. 
The MSSM contributions to CP violation are constrained from experimental
bounds on low energy physics. 

The R-parity odd \ccs could have a complex phase and hence be by
themselves an independent source of CP violation. 
This idea has motivated many studies on low energy \Rp\ physics. 
Furthermore, even if one assumes that the R-parity odd interactions are 
CP conserving, these could still lead, in combination with the other 
possible source of complex phase in the MSSM to new tests of CP 
Violation. For instance, the \Rp\ couplings could bring a dependence on
the CKM matrix elements due to the fermion mass matrix transformation 
from current basis to mass basis. 

The effects of \Rp\ interactions on the CP asymmetries in the processes, 
$l^+ l^- \to f_J \bar f_{J'}$, with $J \neq J'$, were calculated in 
\cite{z:Chemfer}. 
The \Rp\ contributions to these CP asymmetries, are controlled by 
interference terms between tree and loop level amplitudes.
The consideration of loop amplitudes was restricted to the photon and 
$Z$-boson vertex corrections. The flavour off-diagonal CP asymmetries
defined, at the $Z$-boson pole, as
\begin{eqnarray}
A_{JJ'}={ B_{JJ'}-B_{J'J} \over B_{JJ'}+B_{J'J} }, 
\label{CPasymp}
\end{eqnarray}
lie approximately at, $A_{JJ'} \approx (10^{-1}-10^{-3}) sin \psi $,
where $\psi $ is the CP odd phase. The off Z-boson pole asymmetries
given by,
\begin{eqnarray}
A_{JJ'}={ \sigma_{JJ'}-\sigma_{J'J} \over \sigma_{JJ'}+\sigma_{J'J} }, 
\label{CPasym}
\end{eqnarray}
lie at, $A_{JJ'} \approx (10^{-2}-10^{-3}) sin \psi $, for leptons and
quarks, irrespective of whether one deals with light or heavy flavours.
The CP asymmetries depend on a ratio of 
different \Rp\ coupling constants, and are
therefore less sensitive to these couplings than the flavour changing 
rates, which involve higher power of the \Rp\ constants. 
This is the reason why the results are optimistic with respect to the 
CP asymmetries for the $Z$ decays in the Standard Model,
which are, $A_{JJ'} = [10^{-5},10^{-3},10^{-1}] sin \delta_{CKM}$,
$\delta_{CKM}$ being the CP odd phase from the CKM matrix, 
for the production of $(\bar b s  + \bar s b)$,
$(\bar b d  + \bar d b)$ and $(\bar d s  + \bar s d)$, respectively
\cite{z:Bern,z:Hou}. 
The particular rational dependence of the CP asymmetries on the
couplings is of the form, $Im(\l \l^* \l \l^*)/\l^4$, and may thus 
lead to strong enhancement or suppression factors, depending on the
largely unknown flavour hierarchical structure 
of the involved Yukawa couplings.


The study of the reaction $l^+ l^- \to t \bar c$ allows to 
learn about CP violation in the quark sector, 
due to the clear signature of the top quark:
$t \to b W \to b l \nu$. In this reaction, 
the CP violation can be probed through the quantity
defined in Eq.(\ref{CPasym}) or via the following flavour
off-diagonal CP asymmetry \cite{z:chela},
\begin{eqnarray}
   A   ={ {d \sigma^+ \over d E_l}-{d \sigma^- \over d E_l} 
 \over    {d \sigma^+ \over d E_l}+{d \sigma^- \over d E_l} }, 
\label{CPtop}
\end{eqnarray}
where 
$\sigma^+=\sigma(l^+ l^- \to t \bar c \to b \bar c \bar l \nu)$, 
$\sigma^-=\sigma(l^+ l^- \to \bar t c \to \bar b c l \bar \nu)$
and $E_l$ is the energy of the produced charged lepton.
The values of the CP asymmetries defined in Eq.(\ref{CPtop})
range typically in the interval 
$A \approx (10^{-2}-10^{-3}) \sin \psi$ for
$0GeV<E_l<300GeV$ \cite{z:chela}. These CP asymmetries, as
those defined in Eq.(\ref{CPasym}), could be enhanced up to 
$\sim 10^{-1} \sin \psi$ should the \Rp 
coupling constants exhibit
large hierarchies with respect to the generations.


If both non-degeneracies and mixing angles between all slepton
flavours, as well as the CP odd phase, do not vanish, CP violation 
asymmetries could also be observable in \susyq\ particles pair production. 
The R-parity odd interactions could provide an alternative mechanism for
explaining CP violation asymmetries in such productions, through
possible $\psi$ CP odd phase incorporated in the relevant dimensionless 
coupling constant. 
As for the fermion pair production, the \Rp\ contributions to the CP
asymmetries in scalar particles pair production are controlled by 
interference terms between tree and loop level amplitudes. 
The flavour non-diagonal CP asymmetries for the slepton pair production, 
$e^+ e^-\to \tilde l_J \tilde l^*_{J'}$ ($J \neq J'$),
which are defined as in Eq.(\ref{CPasym}), are predicted to be 
of order, $A_{JJ'}\approx (10^{-2}-10^{-3}) sin \psi$~\cite{z:Chemsca}.
 
Finally, the \Rp\ interactions could give rise to CP violation 
effects at tree level in the reaction $e^+ e ^- \to \tau^+ \tau^-$
via the observation of the double spin correlations 
for the produced tau-leptons pair.
This possibility, which was studied in \cite{z:shalom}, stands out as an 
extremely interesting issue by itself, since previous studies of CP 
violating effects in, $e^+ e ^- \to \tau^+ \tau^-$, that can emanate 
from multi-Higgs doublet model, leptoquark, Majorana $\tilde \nu$ and 
supersymmetry, all occur at one loop level. 
Here, the CP asymmetries are generated from the exchange of a
resonant $\tilde \nu_{\mu}$ sneutrino in the $s$-channel, via the real 
coupling $\l_{121}$ and a complex constant $\l_{323}$, if there is a 
$\tilde \nu_{\mu} - \tilde {\bar \nu}_{\mu}$ mixing. 
This sneutrino mixing could generate both CP-even and CP-odd spin 
asymmetries which are forbidden in the Standard Model and that could be 
measured for $\tau$ leptons at leptonic colliders.
The observation of such asymmetries would provide explicit information
about three different aspects of new physics: 
$\tilde \nu_{\mu} - \tilde {\bar \nu}_{\mu}$ mixing, CP violation and 
R-parity violation. The sneutrino-antisneutrino mixing phenomena, 
which have been gaining some interest
recently~\cite{z:hirschosc,z:haber1,z:Mur}, is interesting since it is closely 
related to the generation of neutrino masses \cite{z:hirschosc,z:haber1}. 
At a centre of mass energy, $\sqrt s=192 \GeV$, it is remarkable that 
the maximum value of these spin asymmetries can reach $75 \%$
for $\Delta m_{\tilde \nu^{\mu}}=\Gamma_{\tilde \nu^{\mu}}$ and $10 \%$
for $\Delta m_{\tilde \nu^{\mu}}=\Gamma_{\tilde \nu^{\mu}}/10$, 
where $\Delta m_{\tilde \nu^{\mu}}$ is the mass splitting between 
the CP even $\tilde \nu^{\mu}_+$ and CP odd $\tilde \nu^{\mu}_-$ 
muon-sneutrino mass eigenstates and $\Gamma_{\tilde \nu^{\mu}}$ is the 
sneutrino width.
Note that the condition $\Delta m_{\tilde \nu^{\mu}} \leq \Gamma_{\tilde
\nu^{\mu}}$ is necessary since then, if $\Gamma_{\tilde \nu^{\mu}}=
10^{-2}m_{\tilde \nu^{\mu}_{\pm}}$, which is a viable estimation that
has been assumed in the present study, the constraint, 
${\Delta m_{\tilde \nu^{\mu}}} / m_{\tilde \nu^{\mu}_{\pm}}<<1$,
imposed by bounds on neutrino masses \cite{z:haber1}, is well respected.
Furthermore, in the case, $\Delta m_{\tilde \nu^{\mu}} 
\leq \Gamma_{\tilde \nu^{\mu}}$, the two $\tilde \nu^{\mu}_+$ and 
$\tilde \nu^{\mu}_-$ resonances will overlap and distinguishing between 
them becomes a non trivial experimental task. Thus, the polarization 
asymmetries provide a feasible alternative for establishing the mass
splitting, $m_{\tilde \nu^{\mu}_+} \neq  m_{\tilde \nu^{\mu}_-}$. 
Assuming ${\cal L}=0.5 \femtob^{-1}$ as the total integrated luminosity 
for LEPII at $\sqrt s=192 \GeV$ and setting, $\l_{121}=0.05$, 
$\vert \l_{323} \vert =0.06$, the CP even and CP odd asymmetries may be
detectable, under the best circumstances, around the resonance region
$189.5 \GeV < m_{\tilde \nu^{\mu}} < 194 \GeV$ with a sensitivity of  $3
\s$ for $\Delta m_{\tilde \nu^{\mu}}=\Gamma_{\tilde \nu^{\mu}}/4$.
The polarization asymmetries depend on the relative values of the real 
part, $a$, and the imaginary part, $b$, of the complex \cc $\l_{323}$. 
With the simultaneous measurement of the CP conserving and CP violating
asymmetries, the whole range, $0 \leq {b \over a+b} \leq 1$, can be
covered to at least $3 \s$ for 
$\Delta m_{\tilde \nu^{\mu}}=\Gamma_{\tilde \nu^{\mu}}/4$ and 
$m_{\tilde \nu^{\mu}}= \sqrt s=192 \GeV$. 
At the Next Linear Colliders, with an energy of $\sqrt s=500 \GeV$,
the CP asymmetries could be probed to at least $3 \s$ (best effects are
at the $20 \s$ level) for the range 
$490 \GeV < m_{\tilde \nu^{\mu}} < 510 \GeV$ and for 
$\Delta m_{\tilde \nu^{\mu}}=\Gamma_{\tilde \nu^{\mu}}/5$.
Also, with $\Delta m_{\tilde \nu^{\mu}}=1 \GeV$, the NLC will have 
a sensitivity above $3 \s$ over almost the entire range,
$0 \leq {b \over a+b} \leq 1$.

\section{Singly Produced Sparticles at Hadronic Colliders}
\label{sec:singlepp}

\setcounter{equation}{0}

%
%
%
%
%

\subsection{Resonant Production of Sparticles}


The SUSY particles can be produced as resonances at hadronic colliders
through the \Rp interactions. This is particularly attractive as
hadronic colliders allow to probe for resonances over a
wide mass range given the continuous energy distribution of the colliding 
partons.
If a single $\Rp$-violating coupling is dominant, the resonant
SUSY particle may decay through the same coupling involved in its production,
giving a two quarks final state at the partonic level. However, it is 
also possible that the decay of the resonant SUSY particle is mainly due 
to gauge interactions, giving rise to a cascade decay.


%
%
%


$\bullet$ {\bf Resonant production via $\l'$}
First, a resonant sneutrino can be produced in $d \bar d$ annihilations 
through the constant $\l'_{ijk}$. The associated formula can be written
as follows \cite{z:Qui}:
%
%
\begin{eqnarray}
\s (d_k \bar d_j \to  \tilde \nu^i \to X_1 X_2)= \frac{4}{9} {\frac 
{\pi \Gamma_{d_k \bar d_j} \Gamma_f}{(\hat{s}-m_{\tilde \nu^i}^2)^2 + 
m_{\tilde \nu^i}^2 \Gamma_{\tilde \nu^i}^2}}\;,
\label{snwi1}
\end{eqnarray} 
where $\Gamma_{d_k \bar d_j}$, and $\Gamma_f$ are the partial width of
the channels, $\tilde \nu^i \to d_k \bar d_j$,
and, $\tilde \nu^i \to X_1 X_2$, respectively, $\Gamma_{\tilde \nu^i}$ is
the total width of the sneutrino, 
$m_{\tilde \nu^i}$ is the sneutrino mass and $\hat{s}$ is the square of
the parton center of mass energy. The factor
${1/3}$ in front is from matching the initial colors, and
$\Gamma_{d_k \bar d_j}$ is given by,
\begin{eqnarray}
\Gamma_{d_k \bar d_j} = \frac{3}{4} \alpha_{\l'_{ijk}} m_{\tilde \nu^i},
\label{snwi2}
\end{eqnarray}
where $ \alpha_{{\l'}_{ijk}}= {\l'}^2_{ijk} / 4 \pi $. To compute the rate
at a $p \bar p$ collider,
the usual formalism of the parton model of hadrons can be used
\cite{z:Eich}: 
\begin{eqnarray}
\s (p \bar p \to \tilde \nu^i \to X_1 X_2) = \sum_{j,k} \int_{\tau_0}^1
{\frac {d \tau}{\tau}} ({\frac {1}{s}}{\frac {d L_{jk}}{d \tau}}) \ 
\hat{s} \ \s (d_k \bar d_j \to  \tilde \nu^i \to X_1 X_2),
\label{hsec}
\end{eqnarray}
where $s$ is the center of mass energy squared, $\tau_0$ is given by
$\tau_0=(M_{X_1}+M_{X_2})^2/s$ and $\tau$ is defined by
$\tau=\hat{s} / s=x_1 x_2$, $x_1,x_2$ denoting the longitudinal momentum
fractions of the initial partons $j$ and $k$,
respectively. The quantity ${dL_{jk}/ d \tau}$ is the parton
luminosity defined by,
\begin{eqnarray}
{\frac {dL_{jk}}{d \tau}}= \int_{\tau}^1 {\frac {d x_1}{x_1}} [f_j^{\bar
p}(x_1) f_k^p(\tau / x_1)+
f_j^p(x_1)f_k^{\bar p}(\tau / x_1)],
\label{plum}
\end{eqnarray}
where the parton distribution $f_j^h(x_1)$ denotes the probability of
finding a parton $j$ with momentum fraction $x_1$ inside a hadron h,
and generally depends on the Bjorken variable, $Q^2$, the square of the
characteristic energy scale of the process under consideration. The parton
distributions are measured experimentally at a given scale, and 
are evolved to the very large momentum scales of interest, via the
Altarelli-Parisi equations. 
In order to see the effects of the parton distributions on the resonant sneutrino
production, some values of the rates are given in the following \cite{z:DIMOP90}:
For instance, with an initial state, $d \bar d$, for the hard process,
the cross section value is, $\s(p \bar p \to \tilde \nu^i)=8.5$
nanobarns, for a sneutrino mass of $100 \GeV$ 
and a coupling, $\l'_{i11}=1$, at $\sqrt s =2 \TeV$. For identical values
of the parameters and of the c.m. energy,
the cross section is $\s(p \bar p \to \tilde \nu^i)=4$ nanobarns with
an initial state, $d \bar s$, and 
$\s(p \bar p \to \tilde \nu^i)=0.8$ nanobarns with an initial state,
$d \bar b$.

The charged slepton can also be produced as a
resonance at hadronic colliders from an initial state
$u_j \bar d_k$ and via the constant $\l'_{ijk}$.
%
%
%
The cross section value is, $\s(p \bar p \to \tilde l^i_L)=2$
nanobarns, for $m_{\tilde l^i_L}=100 \GeV$, $\sqrt s=2 \TeV$ and
$\l'_{i11}=1$ (\cite{z:DIMOP90,z:Kal}).


$\bullet$ {\bf Resonant production via $\l''$}
The baryon number violating couplings $\lambda''_{ijk}$ allows 
for resonant production of squarks at hadronic colliders.
Either a squark $\tilde u_i$ or $\tilde d_k$ can be produced at the 
resonance from an initial state, $\bar d_j \bar d_k$ or 
$\bar u_i \bar d_j$, respectively. For $m_{\tilde d^k_R}=100 \GeV$, 
$\sqrt s =2 \TeV$ and $\l''_{11k}=1$, the rate of the down squark 
production at the Tevatron is 
$\s(p \bar p \to \tilde d^k_R)=25$ nanobarns \cite{z:DIMOP90}.
%
%
%
%
For $m_{\tilde t_1}=600 \GeV$, $\sqrt s =2 \TeV$ and $\l''_{323}=0.1$, the
rate of the resonant stop production is 
$\s(p \bar p \to \tilde t_1)=10^{-3}$ picobarns \cite{z:Berg}. Note
that this rate is higher than the stop pair
production rate at the same c.m. energy and for the same stop mass, which 
is of order $\s(p \bar p \to \tilde t_1 \tilde t_1)=10^{-6}$ picobarns.

\subsection{Single Gaugino Production}

$\bullet$ {\bf Single production via $\l''$}
The single superpartner production could also occur as a $2 \to 2$-body process,
through an \Rp\ coupling $\l''$ and  an ordinary gauge interaction vertex. 
In baryon number violating models, any gaugino (including gluino) can be 
produced in association with a quark, in quark-quark scattering, by the
exchange of a squark in the $s$-, $t$- or $u$-channel. 

For example, let us consider the photino and gluino production
\cite{z:DIMOP90}: The rate values in the 
$t$- and $u$-channel are, 
$\s (p \bar p \to \tilde{\gamma} q)=2 \ 10^{-2} \nanob$, and,
$\s (p \bar p \to \tilde g q)=3 \ 10^{-1} \nanob$, for, $m_{\tilde
q}=m_{\tilde g}=m_{\tilde{\gamma} }=100 \GeV$,
$\sqrt s =2 \TeV$ and $\l''_{111}=1$. The photino or gluino which is
produced will then decay into three 
jets via the $\l''$ coupling, resulting in a four jets final state. The
corresponding QCD background is strong:
It is estimated to be about $10 \nanob$ for $\sqrt s =2 \TeV$ \cite{z:Arg}. Of
course, the ratio signal over background can
be enhanced considerably by looking at the mass distribution of the
jets: the QCD 4 jets are produced relatively 
uncorrelated, while the trijet mass distribution of the signal should
peak around the gaugino mass. However, 
frequently one of the three jets will be too soft to be measured and at
other times jet coalescence would occur,
especially for small values of the gaugino mass. The study of this
example bring us to the conclusion that, due to high QCD background, 
the analysis of the single production via $\l''$ remains a challenge.

Nevertheless, there is one case of interest, were the final state could be 
particularly clear \cite{z:Berg,z:DREINER99}: 
A $\tilde \chi^+_1$ chargino can be produced through the resonant production
of a top squark, $\bar d_j \bar d_k \to \tilde t_1 \to b \tilde \chi^+_1$ 
(via $\l''_{3jk}$), and then decay into the lightest neutralino plus leptons
as $\tilde \chi^+_1 \to \bar l_i \nu_i \tilde \chi^0_1$.
Due to the stop resonance, this reaction could reach high rate values.
The cascade decay demands the mass hierarchy, $m_{\tilde t_1} > 
m_{\chi^+_1} > m_{\chi^0_1}$, to be respected,
and by consequence is not allowed in all regions of the \SUGRA parameter
space. If we made the hypothesis that
$\tilde \chi^0_1$ is the LSP and undergo an \Rp\ decay outside of the
detector, the neutralino  should then be
treated as a stable particle. Then, the signal for our process would be
very clear since it would consist of a tagged b quark jet, a lepton
and missing transverse energy. The \SM\ background for such a signature comes
from the the single top
quark production, via $W \ g$ fusion, and the production of a $W$ gauge
boson in association with $b \bar b$,
$c \bar c$ or a jet that is mistaken for a b quark. Applying some cuts
in order to simulate the acceptance of 
the detector and to reduce the background, it has been found that values
of $\l'' > 0.03-0.2$ and $\l''>0.01-0.03$
can be excluded at the $95 \%$ confidence level for, $180 \GeV<m_{\tilde
t_1}<285 \GeV$, at the Run I of the Tevatron 
($\sqrt s=1.8 \TeV$ and $\int {\cal L} dt = 110 \picob^{-1}$) and for, 
$180 <m_{\tilde t_1}<325 \GeV$, at the Run II of the
Tevatron ($\sqrt s=2 \TeV$ and $\int {\cal L} dt = 2 \femtob^{-1}$), respectively.
This result is based on the leading-order
CTEQ-4L parton distribution functions \cite{z:CTEQ4L} and holds for the
normalization,
$\l''=\l''_{312}=\l''_{313}=\l''_{323}$, and for the point of a minimal
\SUGRA model, $m_{1/2}=150 \GeV, \ 
A_0=-300 \GeV$ and $\tan \beta =4$. The constraints obtained on $\l''$
are stronger than the present low energy
bounds.


%


$\bullet$ {\bf Single production via $\l'$}
The single production of SUSY particles via $\l'$ occuring through $2 \to 2$-body
processes, offers the opportunity to study the 
parameter space of the \Rp\ models with a quite high sensitivity at
hadronic colliders.

\begin{figure}[t]
\begin{center}
\centerline{\psfig{figure=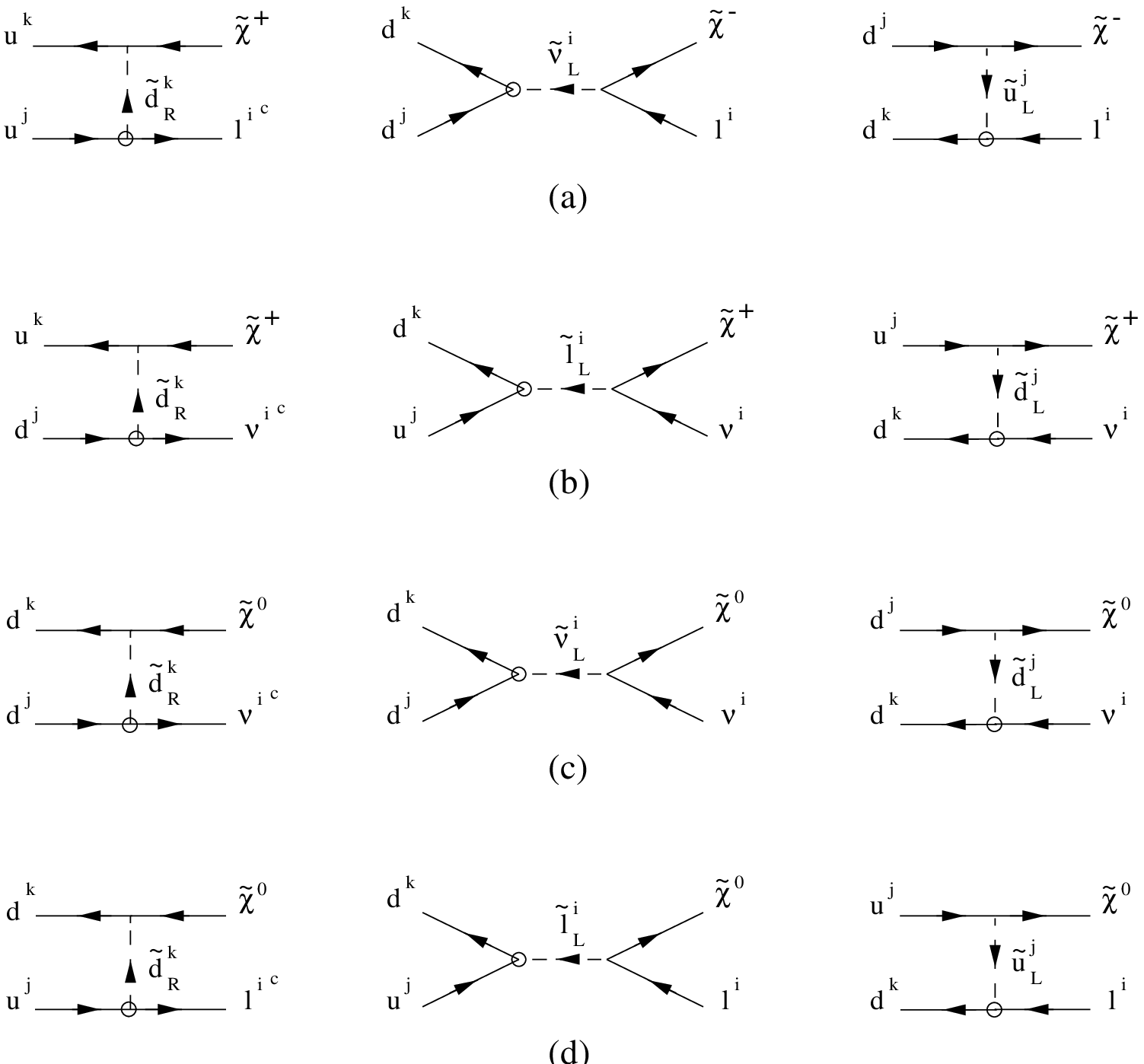,width=15cm}}  
\caption{ Feynman diagrams for the four single 
superpartner production reactions
involving  $\l'_{ijk}$  at hadronic colliders which
receive a contribution from a
resonant \susyq particle production.
The $\l'_{ijk}$ coupling constant is symbolised by a small circle
and the arrows denote the flow of the particle momentum.}
\label{fig:graphes}
\end{center}
\end{figure}

In Fig.\ref{fig:graphes}, we present all the single 
superpartner productions which 
occur via $\l'_{ijk}$ through $2 \to 2$-body processes 
at hadronic colliders and
receive a contribution from a
resonant SUSY particle production \cite{z:More}. 
The spin summed amplitudes of those reactions
including the higgsino contributions
have been calculated in \cite{z:More}. In a SUGRA model, 
the rates of the reactions presented in Fig.\ref{fig:graphes} 
depend mainly on the $m_0$ and $M_2$ parameters.

\begin{figure}[t]
\begin{center}
\centerline{\psfig{figure=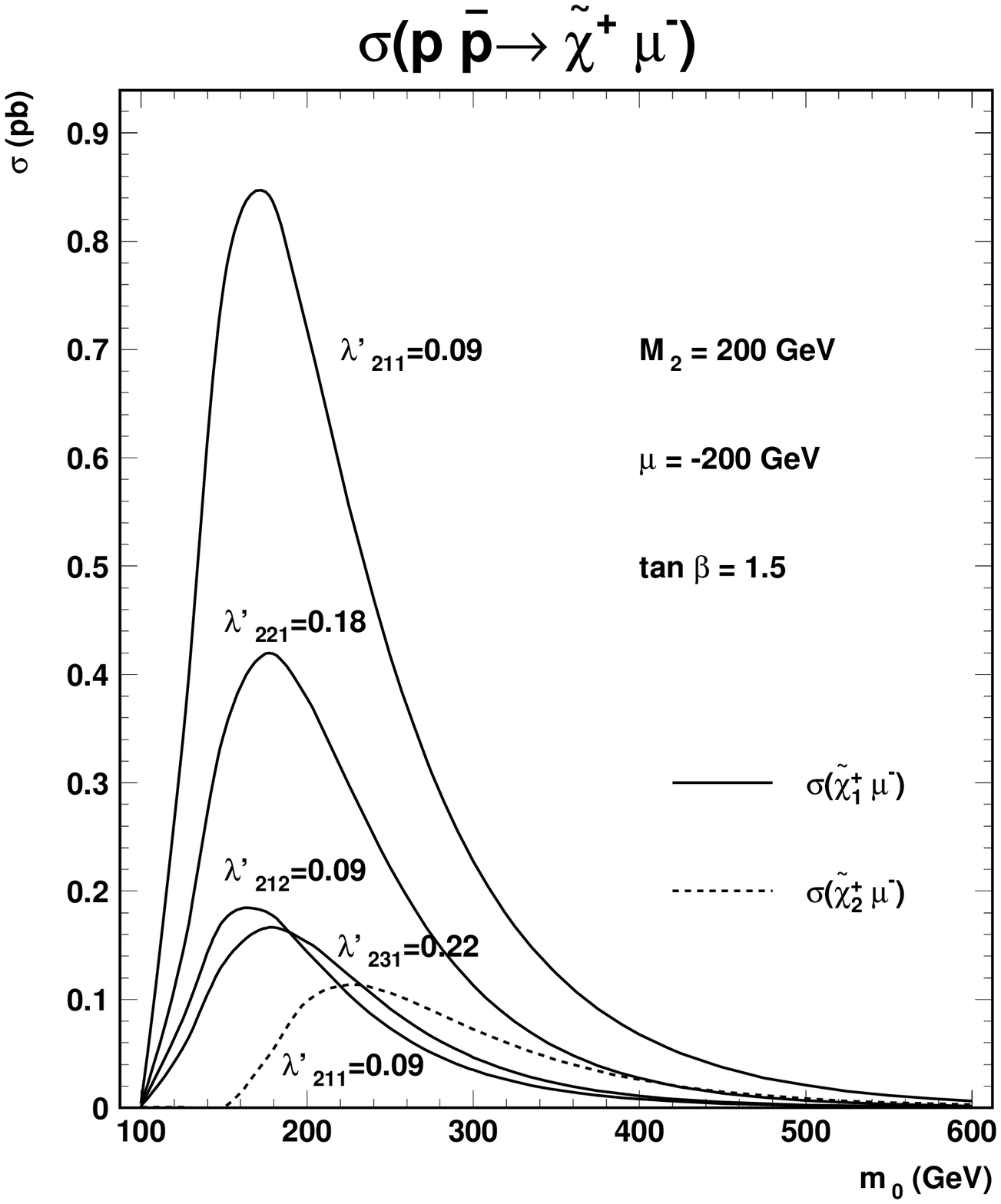,width=13cm}}   
\caption{ Cross sections (in $pb$) of the single chargino productions  
$p \bar p \to \tilde \chi^+_{1,2} \mu^-$   
as a function of the $m_0$ parameter (in $GeV$).
The center of mass energy is taken at $\sqrt s=2 TeV$ 
and the considered set of parameters is: $\l'_{211}=0.09$,
$M_2=200GeV$, $\tan \beta=1.5$ and $\mu=-200GeV$.
The rates for the $\tilde \chi^+_1$ production via the
\Rp couplings $\l'_{212}=0.09$, $\l'_{221}=0.18$ and $\l'_{231}=0.22$ 
are also given. The chosen values of the \Rp couplings correspond 
to the low-energy limits for a squark mass of $100GeV$ \cite{z:reviews}.}
\label{fig:XScl}
\end{center}
\end{figure}

In Fig.\ref{fig:XScl}, we show the variations of the
$\sigma(p \bar p \to \tilde \chi^+_{1,2} \mu^-)$ cross sections with $m_0$
for fixed values of $M_2$, $\mu$ and $\tan \beta$
and various \Rp couplings of the type $\l'_{2jk}$
at Tevatron Run II in a SUGRA model \cite{z:More}. 
The \Rp couplings giving the highest cross sections have been considered.
The $\sigma(p \bar p \to \tilde \chi^+_{1,2} \mu^-)$
rates decrease when $m_0$ increases 
since then the sneutrino becomes heavier 
and more energetic initial partons are
required in order to produce the resonant sneutrino. 
A decrease of the cross sections also occurs at small 
values of $m_0$, the reason being that
when $m_0$ approaches $M_2$ the $\tilde \nu$ mass 
is getting closer
to the $\tilde \chi^{\pm}$ masses so that the phase space
factors associated to the decays
$\tilde \nu_{\mu} \to \chi^{\pm}_{1,2} \mu^{\mp}$ decrease.
The differences between the $\tilde \chi^+_1 \mu^-$
production rates occuring via the various $\l'_{2jk}$ couplings
are explained by the different partonic luminosities. 
Indeed, as shown in Fig.\ref{fig:graphes} 
the hard process associated to the $\tilde \chi^+_1 \mu^-$ 
production occuring through the
$\l'_{2jk}$ coupling constant
has a partonic initial state $\bar q_j q_k$. 
The $\tilde \chi^+_1 \mu^-$ production via the $\l'_{211}$ coupling
has first generation quarks in the initial state 
which provide the maximum partonic luminosity.

\begin{figure}[t]
\begin{center}
\centerline{\psfig{figure=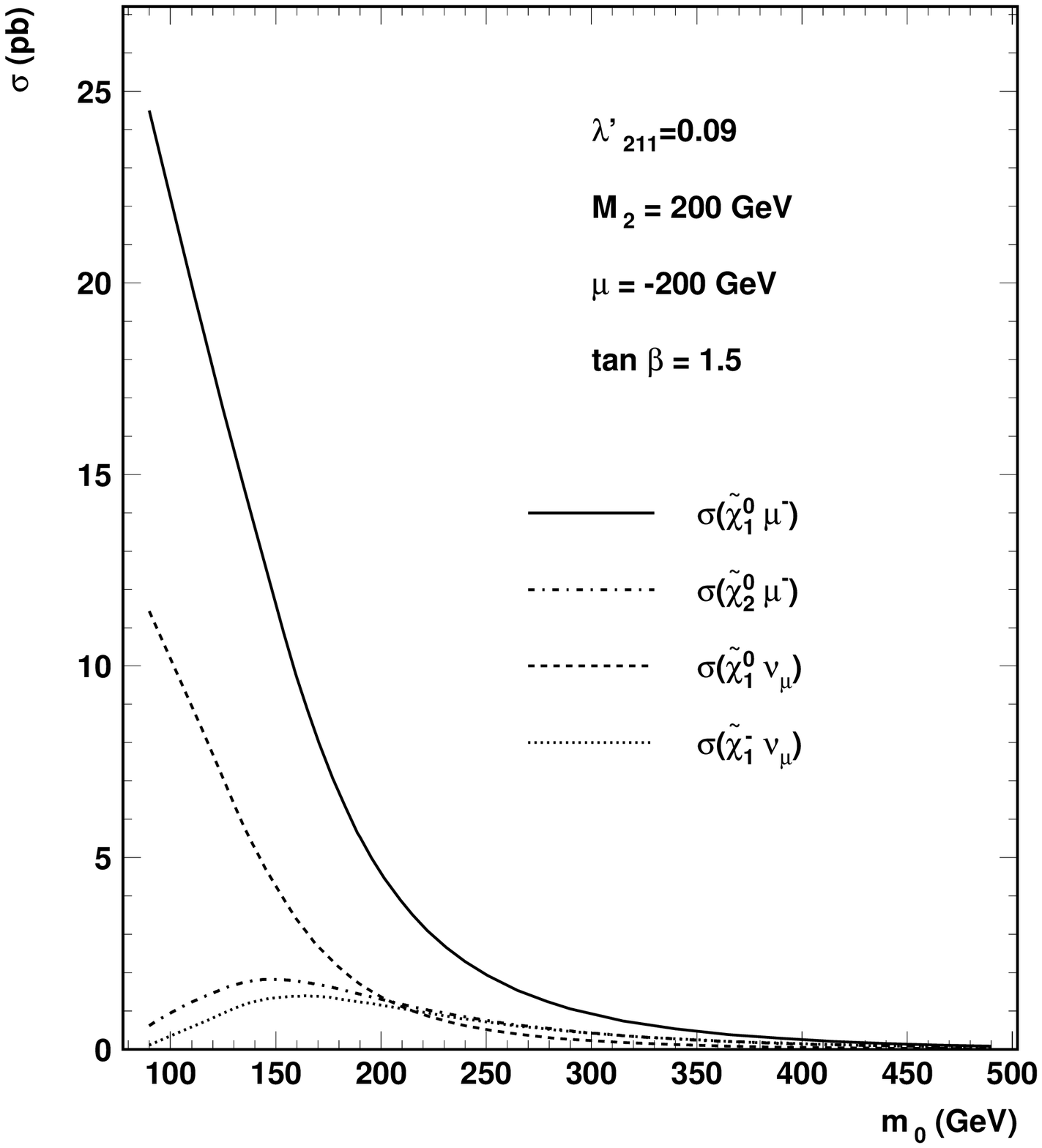,width=13cm}}  
\caption{ Cross sections (in $pb$) of the 
$\tilde \chi^-_1 \nu$, $\tilde \chi^0_{1,2} \mu^-$ 
and $\tilde \chi^0_1 \nu$  productions  
at Tevatron Run II   
as a function of the $m_0$ parameter (in $GeV$).
The center of mass energy is taken at $\sqrt s=2 TeV$ 
and the considered set of parameters is: $\l'_{211}=0.09$,
$M_2=200GeV$, $\tan \beta=1.5$ and $\mu=-200GeV$.}
\label{fig:allXS}
\end{center}
\end{figure}

In Fig.\ref{fig:allXS}, we show the variations
of the rates of the reactions 
$p \bar p \to \tilde \chi^-_1 \nu$, 
$p \bar p \to \tilde \chi^0_{1,2} \mu^-$ 
and $p \bar p \to \tilde \chi^0_1 \nu$
with the $m_0$ parameter in a SUGRA model \cite{z:More}.
We see in this figure that 
the single neutralino productions do not decrease at 
small $m_0$ values in contrast with the 
single chargino productions (see also Fig.\ref{fig:XScl}). 
This is due to the fact that in SUGRA scenarios
the $\tilde \chi^0_1$ and $\tilde l_L$ 
($\tilde l_L=\tilde l^{\pm}_L,\tilde \nu_L$) 
masses are never close enough 
to induce a significant decrease of the  
phase space factor associated to the decay
$\tilde l_L \to \tilde \chi^0_1 l$ ($l=l^{\pm},\nu$). 
By analysing Fig.\ref{fig:XScl} and Fig.\ref{fig:allXS}, 
we also remark that the $\tilde \chi^- \nu$ 
($\tilde \chi^0 \mu^-$) production rate is larger than the
$\tilde \chi^+ \mu^-$ ($\tilde \chi^0 \nu$) production rate. 
The explanation is that in $p \bar p$ collisions
the initial states of the resonant charged slepton production
$u_j \bar d_k, \bar u_j d_k$ have higher partonic luminosities than the 
initial states of the resonant sneutrino production
$d_j \bar d_k, \bar d_j d_k$.


%

The neutralino production in association with a charged lepton via
$\lambda'$ (see Fig.\ref{fig:graphes}(d)) 
is an interesting case at Tevatron \cite{z:DIMOP90}.
The topology of the events consists of an isolated lepton
in one hemisphere balanced by a lepton
plus two jets in the other hemisphere, coming from the neutralino decay
via $\l'$. The \SM\ background arising
from the production of two jets plus a $Z^0$, decaying into two leptons,
has a cross section of order $10^{-3} \nanob$
\cite{z:Eich}, and can be greatly reduced by excluding lepton pairs with
an invariant mass equal to the $Z^0$ mass. 
The other source of \SM\ background, which is the Drell-Yan mechanism
into $2$ leptons accompanied by 2 jets, is
suppressed by a factor, $10^{-6}/ \alpha_{\l}$. Moreover, the signal can
be enhanced by looking at the invariant mass 
of the 2 jets and the lepton in the same hemisphere, which should peak
around the neutralino mass.

The single production via $\lambda'$ of the neutralino together with a 
charged lepton can also generate clean signatures free from large \sm background
at the Tevatron Run II, containing two like-sign charged leptons 
\cite{z:dreiner1,z:Rich1,z:DREINER99,z:Rich2,z:houches,z:More}. As a matter of fact, the neutralino has a
decay channel into a lepton and two jets through 
the coupling $\l'_{ijk}$ and due to its Majorana nature, the neutralino
decays to the charge conjugate
final states with equal probability: $\Gamma (\tilde \chi^0 \to l_i u_j
\bar d_k)=\Gamma(\tilde \chi^0 \to \bar l_i \bar u_j d_k)$. Therefore,
the lepton coming from the production can 
have the same sign than the one coming from the neutralino decay. 
Since $\l'_{111}$ has a strong indirect bound, it is
interesting to consider the \cc $\l'_{211}$,
which corresponds to the dimuons production with an initial state $u
\bar d$ or $\bar u d$ (see Fig.\ref{fig:graphes}(d)) composed of first 
generation quarks.
The analysis of the like sign ditaus signature 
generated by the 
$\tilde \chi^0 \tau^{\pm}$ production 
through the $\l'_{311}$ coupling
(see Fig.\ref{fig:graphes}(d)) suffers a reduction of the 
cuts efficiency due to the tau-lepton decay.
Besides, the study of the $\tilde \chi^0_1 \mu^{\pm}$ 
production via $\l'_{211}$ 
in a scenario where the $\tilde \chi^0_1$ 
is the LSP is particularly attractive since then the
$\tilde \chi^0_1$ can only undergo \Rp decays.
It was found that in a SUGRA model,
such a study could probe
values of the $\l'_{211}$ coupling at the $5 \sigma$
discovery level down to $2 \ 10^{-3}$ ($10^{-2}$) for 
a muon-slepton mass of
$m_{\tilde \mu_L}=100GeV$ ($m_{\tilde \mu_L}=300GeV$) 
with $M_2=100GeV$, $2 <\tan \beta <10$ 
and $\vert \mu \vert < 10^3 GeV$
at Tevatron Run II 
assuming a luminosity of ${\cal L}=2 fb^{-1}$ \cite{z:Rich1,z:DREINER99},
and down to $2 \ 10^{-3}$ ($10^{-2}$) for 
$m_{\tilde \mu_L}=223GeV$ ($m_{\tilde \mu_L}=540GeV$) 
with $m_{1/2}=300GeV$, $A=300GeV$, $\tan \beta=2$
and $sign(\mu)>0$ at the LHC
assuming a luminosity of
${\cal L}=10 fb^{-1}$ \cite{z:Rich2,z:houches}.
It was also shown in \cite{z:More},  
by using a detector response simulation, that
the study of the single LSP production at Tevatron Run II
$p \bar p \to \tilde \chi^0_1 \mu^{\pm}$
would allow to probe $m_{1/2}$ values up to $\sim 850 GeV$
and $m_0$ values up to $\sim 550 GeV$ 
at the $5 \sigma$ discovery level, in a SUGRA scenario
where $sign(\mu)<0$, $A=0$, $\tan \beta=1.5$ and 
$\l'_{211}=0.05$ and assuming 
a luminosity of ${\cal L}=2 fb^{-1}$.

Besides, the like sign dilepton signature analysis 
based on the $\tilde \chi^0_1 \mu^{\pm}$ production 
(see Fig.\ref{fig:graphes}(d)) allows
the $\tilde \chi^0_1$ and $\tilde \mu^{\pm}_L$ mass 
reconstructions since the decay chain 
$\tilde \mu^{\pm}_L \to \tilde \chi^0_1 \mu^{\pm}$, 
$\tilde \chi^0_1 \to \mu^{\pm} u d$ can be fully 
reconstructed \cite{z:More}. 
Based on the like sign dilepton signature analysis,
the $\tilde \chi^0_1$ ($\tilde \mu^{\pm}_L$) mass could be measured with a
statistical error of $\sim  11 GeV$ ($\sim  20 GeV$) at the Tevatron Run II 
\cite{z:More}.

The single $\tilde \chi^{\pm}_1$ production 
in association with a charged lepton 
(see Fig.\ref{fig:graphes}(a)) is another interesting 
reaction at hadronic colliders. 
In a scenario where $\tilde \chi^0_1$ 
is the LSP and $m_{\tilde \nu},
m_{\tilde l},m_{\tilde q}>m_{\tilde \chi^{\pm}_1}$,  
this single production receives a contribution from the 
resonant sneutrino production and the singly produced
chargino decays into quarks and leptons with branching 
ratios respectively of 
$B(\tilde \chi^{\pm}_1 \to \tilde \chi^0_1 d_p u_{p'}) \approx 70\%$ 
($p=1,2,3;p'=1,2$) and 
$B(\tilde \chi^{\pm}_1 \to \tilde \chi^0_1 l^{\pm}_p \nu_p) 
\approx 30\%$ due to the color factor.
The neutralino decays via $\l'_{ijk}$ either into a lepton as, 
$\tilde \chi_1^0 \to l_i u_j \bar d_k,\bar l_i \bar u_j d_k$,
or into a neutrino as, $\tilde \chi_1^0 \to \nu_i d_j \bar d_k,
\bar \nu_i \bar d_j d_k$. Hence, if both the $\tilde \chi^{\pm}_1$ and 
$\tilde \chi^0_1$ decay into charged leptons,
the $\tilde \chi^{\pm}_1 l_i^{\mp}$ 
production can lead to the three charged leptons
signature which has a small Standard Model background
at hadronic colliders \cite{z:Roy,z:gia1,z:houches,z:pol00,z:More}.
The study of the three leptons signature generated by the 
$\tilde \chi^{\pm}_1 \mu^{\mp}$ production via the
$\l'_{211}$ coupling constant is particularly 
interesting for the same reasons as above.
The sensitivity on the $\l'_{211}$ coupling obtained
from this study at Tevatron Run II would reach a 
maximum value of $\sim 0.04$ for $m_0 \approx 200 GeV$ 
in a SUGRA model with $M_2=200 GeV$, $sign(\mu)<0$, $A=0$ 
and $\tan \beta=1.5$, assuming
a luminosity of ${\cal L}=2 fb^{-1}$ \cite{z:More}.
\begin{table}
\begin{center}
\begin{tabular}{|c|c|c|c|c|c|c|c|c|}
\hline
$\l'_{211}$ & $\l'_{212}$ & $\l'_{213}$ & $\l'_{221}$ & $\l'_{222}$ 
& $\l'_{223}$ & $\l'_{231}$ & $\l'_{232}$ & $\l'_{233}$  \\
\hline
0.01 & 0.02 & 0.02 & 0.02 & 0.03 & 0.05 & 0.03 & 0.06 & 0.09 \\

\hline
\end{tabular}
\caption{
Sensitivities on the $\l'_{2jk}$ coupling constants 
for $\tan\beta$=1.5, $M_1=100$~GeV, $M_2=200$~GeV, $\mu=-500$~GeV,
$m_{\tilde q}=m_{\tilde l}=300$~GeV and
$m_{\tilde \nu}=400$~GeV,
assuming an integrated luminosity of ${\cal L}=30 fb^{-1}$.}
\label{tabrev}
\end{center}
\end{table}
In Table \ref{tabrev} we show the sensitivities on
the $\l'_{2jk}$ couplings that could be obtained from the 
trilepton analysis based on the
$\tilde \chi^{\pm}_1 \mu^{\mp}$ production at
the LHC for a given set of MSSM parameters \cite{z:pol00}.
For each of the $\l'_{2jk}$ couplings the sensitivity has been
obtained assuming that the considered coupling
was the single dominant one.
The difference between the various results 
presented in this table is due to the fact that each 
$\l'_{2jk}$ coupling involves a specific initial state 
(see Fig.\ref{fig:graphes}(a)) with its own parton 
density. Besides,
all the sensitivities shown in Table \ref{tabrev} improve 
greatly the present low-energy constraints.
The trilepton analysis based on the
$\tilde \chi^{\pm}_1 e^{\mp}$ ($\tilde \chi^{\pm}_1 \tau^{\mp}$) 
production would allow to test the $\l'_{1jk}$ ($\l'_{3jk}$) 
couplings constants. While the sensitivities obtained on the 
$\l'_{1jk}$ couplings are expected to be of the same order of 
those presented in Table \ref{tabrev}, the sensitivities on the
$\l'_{3jk}$ couplings should be weaker due to
the tau-lepton decay.
The results presented in Table \ref{tabrev} illustrate
the fact that even if some studies on the single superpartner 
production via $\l'$ at hadronic colliders 
(see Fig.\ref{fig:graphes}) only concern the 
$\l'_{211}$ coupling constant, the analysis of a given single 
superpartner production at Tevatron or LHC allows to probe 
many $\l'_{ijk}$ coupling constants down to values
smaller than the associated low-energy limits.

Besides, the three leptons final state study based on the
$\tilde \chi^{\pm}_1 \mu^{\mp}$ production 
(see Fig.\ref{fig:graphes}(a)) allows to reconstruct
the $\tilde \chi^0_1$, $\tilde \chi^{\pm}_1$ and 
$\tilde \nu$ masses \cite{z:Roy,z:gia1,z:houches,z:pol00,z:More}. 
Indeed, the decay chain $\tilde \nu_i \to \tilde \chi^{\pm}_1 l^{\mp}_i$,
$\tilde \chi^{\pm}_1 \to \tilde \chi^0_1 l^{\pm}_p \nu_p$,
$\tilde \chi^0_1 \to l^{\pm}_i u_j d_k$ can be fully 
reconstructed since the produced charged
leptons can be identified thanks to their flavours and signs.
Based on the trilepton signature analysis,
the $\tilde \chi^0_1$ mass could be measured with a
statistical error of $\sim 9 GeV$ at the Tevatron Run II 
\cite{z:Roy,z:More} 
and of $\sim 100 MeV$ at the LHC \cite{z:gia1,z:houches,z:pol00}.
Furthermore,
the width of the gaussian shape of the invariant mass
distribution associated to the $\tilde \chi^{\pm}_1$  
($\tilde \nu$) mass is of $\sim 6 GeV$ ($\sim 10 GeV$) 
at the LHC for the MSSM point defined by $M_1=75 GeV$, 
$M_2=150 GeV$, $\mu=-200 GeV$, $m_{\tilde f}=300 GeV$ 
and $A=0$ \cite{z:gia1,z:houches,z:pol00}.\\
Let us make a general remark concerning the
superpartner mass reconstructions based on
the single superpartner production studies at hadronic colliders:
The combinatorial background associated to these mass
reconstructions is smaller than in the mass reconstructions
analyses based on the supersymmetric particle pair production 
since in the single superpartner production studies only 
one cascade decay must be reconstructed.


\subsection{Non Resonant Single Production}
\label{Asso}

%


At hadronic colliders, some supersymmetric particles can also be singly
produced through $2 \to 2$-body processes 
which generally do not receive contribution from  
resonant superpartner production \cite{z:More}: 
Some single productions of squark
(slepton) in association with a gauge boson 
can occur through the exchange of a quark in the $t$-channel or a squark
(slepton) in the $s$-channel via $\l''$ 
($\l'$). From an initial state $g \ q$, a squark (slepton) can also be
singly produced together with a quark
(lepton) with a \cc $\l''$ ($\l'$) via the exchange of a quark or a
squark in the $t$-channel, and of a quark in
the $s$-channel. Finally, a gluino can be produced in association with
a lepton (quark) through a \cc $\l'$ ($\l''$) via the exchange of a
squark in the $t$-channel (and in the $s$-channel).

Let us enumerate the single scalar particle and gluino 
productions occuring via 
the $2 \to 2$-body processes which involve the 
$\l'_{ijk}$ coupling constants \cite{z:More}
(one must also add the charge conjugate processes):
\begin{itemize}
\item The gluino production $\bar u_j d_k \to \tilde g l_i$ via the exchange
of a $\tilde u_{jL}$ ($\tilde d_{kR}$) squark in the $t$ ($u$) channel.
\item The squark production $\bar d_j g \to \tilde d_{kR}^* \nu_i$ via the
exchange
of a $\tilde d_{kR}$ squark ($d_j$ quark) in the $t$ ($s$) channel.
\item The squark production $\bar u_j g \to \tilde d_{kR}^* l_i$ via the
exchange
of a $\tilde d_{kR}$ squark ($u_j$ quark) in the $t$ ($s$) channel.
\item The squark production $d_k g \to \tilde d_{jL} \nu_i$ via the exchange
of a $\tilde d_{jL}$ squark ($d_k$ quark) in the $t$ ($s$) channel.
\item The squark production $d_k g \to \tilde u_{jL} l_i$ via the exchange
of a $\tilde u_{jL}$ squark ($d_k$ quark) in the $t$ ($s$) channel.
\item The sneutrino production $\bar d_j d_k \to Z \tilde \nu_{iL}$ via the
exchange
of a $d_k$ or $d_j$ quark ($\tilde \nu_{iL}$ sneutrino) in the $t$ ($s$)
channel.
\item The charged slepton production $\bar u_j d_k \to Z \tilde l_{iL}$ via
the exchange
of a $d_k$ or $u_j$ quark ($\tilde l_{iL}$ slepton) in the $t$ ($s$)
channel.
\item The sneutrino production $\bar u_j d_k \to W^- \tilde \nu_{iL}$
via the exchange of a $d_j$ quark ($\tilde l_{iL}$ sneutrino) in the $t$
($s$) channel.
\item The charged slepton production $\bar d_j d_k \to W^+ \tilde l_{iL}$
via the exchange of a $u_j$ quark ($\tilde \nu_{iL}$ sneutrino) in the $t$
($s$) channel.
\end{itemize}
Among these single productions only the $\bar u_j d_k \to W^- \tilde \nu_{iL}$ and  
$\bar d_j d_k \to W^+ \tilde l_{iL}$ reactions could receive
a contribution from a resonant sparticle production.
However, in most of the SUSY models,
as for example the supergravity or the gauge
mediated models, the mass difference
between the Left charged slepton and the Left sneutrino
is due to the D-terms so that it is fixed by the relation
$m^2_{\tilde l^{\pm}_L}-m^2_{\tilde \nu_L}=\cos 2 \beta M_W^2$ \cite{z:Iban}
and thus it does not exceed the $W^{\pm}$-boson mass.
We note that in the scenarios of large $\tan \beta$ values, a
scalar particle of the third generation produced as a resonance 
can generally decay into the $W^{\pm}$-boson due to the large mixing in the
third family sfermions sector. For instance, in the SUGRA model with
a large $\tan \beta$ a tau-sneutrino produced as a resonance 
in $d_k \bar d_j \to \tilde \nu_{\tau}$ through $\lambda'_{3jk}$
can decay as $\tilde \nu_{\tau} \to W^{\pm} \tilde \tau^{\mp}_1$,
$\tilde \tau^{\mp}_1$ being the lightest stau.

Similarly, the single scalar particle and gluino productions occuring via 
the $2 \to 2$-body processes which involve the $\l''_{ijk}$ coupling constants 
cannot receive a contribution from a resonant scalar particle production
for low $\tan \beta$.
Indeed, the only reactions among these $2 \to 2-body$ processes 
which could receive such a contribution are of the type $q q \to \tilde q \to \tilde q W$.
In this type of reaction, the squark produced in the s channel, is produced via 
$\l''_{ijk}$ and is thus either a Right squark $\tilde q_R$, which does not couple to the
$W^{\pm}$-boson, or the squarks $\tilde t_{1,2}$, $\tilde b_{1,2}$.

Therefore, the single scalar particle and gluino productions occuring via 
the $2 \to 2$-body processes are generally non resonant 
single superpartner productions, as we have already mentioned at the begining 
of this section.
These non resonant single superpartner productions have typically smaller
cross sections than the reactions receiving a 
contribution from a resonant superpartner production.
For instance, with $m_{\tilde q}=250GeV$,
the cross section $\sigma(p \bar p \to \tilde u_L \mu)$
is of order $\sim 10^{-3}pb$ at a center of mass
energy of $\sqrt s=2TeV$, assuming an \Rp coupling of
$\l'_{211}=0.09$ \cite{z:More}.
However, the non resonant single productions
could lead to interesting signatures. For instance, the production, 
$q \bar q \to \tilde f W$ leads to the final
state $2l+2j+W$ for a non vanishing \Rp\ \cc $\l'$ and to 
the signature $4 j +W$ for a $\l''$ \cite{z:dreiner1}. 
Furthermore, the non resonant single productions 
are interesting as their cross section
involves few SUSY parameters, namely only one or two scalar
superpartner(s) mass(es) and one \Rp coupling constant. 


\subsection{Fermion Pair Production Via $\Rp$}
\label{smhad}

$\bullet$ {\bf Dijet production}
R-parity violating reactions at hadronic colliders can induce some
contributions to Standard 
Model processes. First, the jets pair production receives contributions
from reactions involving
either $\l'$ or $\l''$ coupling constants. As a matter of fact, 
a pair of quarks can be produced through the $\l''$ couplings
with an initial state $u d$ or $\bar u \bar d$ ($d d$ or $\bar d \bar
d$) by the exchange of a $\tilde d$ ($\tilde u$) squark
 in the $s$-channel, and also with an initial state
$u \bar u$ or $d \bar d$ 
($u \bar d$ or $\bar u d$) by the exchange of a $\tilde u$ or
$\tilde d$ ($\tilde d$) squark in the t
channel. In case where the $s$-channel exchanged particle is produced on
shell, the resonant diagram is 
of course dominant with respect to the $t$-channel diagram. The dijet
channel can also be generated
via the $\l'$ couplings from an initial state $u \bar d, \bar u d$ or $d
\bar d$ through the exchange
of a $\tilde l$ or $\tilde \nu$ slepton, respectively, in the $s$-channel. 
\\ If the dominant mechanism for either the slepton or the squark decay
was into two jets, the resonant production 
of such a scalar particle would result in a bump in the two-jet
invariant mass
plot \cite{z:dimopoulos88,z:DIMOP90}, which would be a 
very clean signature.
\\  However the dijet production through
\Rp\ \ccs will be hard to study at LHC where QCD backgrounds are
expected to be severe for searches 
for narrow resonances which are not strongly produced
\cite{z:dreiner1,z:Bin}.
Similarly, at the Tevatron, 
for both the D0 and CDF detectors,  
the resonant production of the stau $\tilde \tau$, for example, would
lead to observable peak in the
dijet invariant mass distribution only for relatively high values of the
\Rp\ couplings, namely, $0,001 < \l'_{311} \l'_{3jk}
B_{2j} < 0,01$ and 
$300 \GeV < m_{\tilde \tau} < 1200 \GeV$, with an integrated luminosity of
$30 \femtob^{-1}$ during the Run II \cite{z:Rizz}. In these notations,
$B_{2j}$ is the dijet branching ratio and $m_{\tilde \tau}$ the stau mass. 
The study of $\l'_{311}$ is particularly interesting here since this
coupling involves an initial state, for the hard subprocess, composed by
first generation quarks which are valence quarks. Besides, the low energy  
constraint is less stringent for $\l'_{311}$ than for $\l'_{111}$ and
$\l'_{211}$.
In conclusion, due to strong QCD background, the dijet production is
not the best framework to
test the $\l'$ and $\l''$ coupling constants at hadronic colliders.  

$\bullet$ {\bf Dilepton production}
Similarly, some reactions involving both $\l'$ and $\l$ \ccs can mimic
the Drell-Yan signatures. 
In the \SM\ the charged lepton pair production occurs through the 
neutral current channel. In \Rp\ models, it can occur 
trough the exchange of $\tilde d$ squark in
the $t$-channel with an initial state
$u \bar u$, and also via the exchange of $\tilde u$ squark in the t
channel or a $\tilde \nu$ sneutrino 
in the $s$-channel with an initial state $d \bar d$. The charged current 
channel receives also an \Rp\ contribution: there are two Feynman graphs
with a lepton plus a neutrino in the final state and an initial state $u
\bar d$ or $\bar u d$: One exchanges
a $\tilde d$ squark in the $t$-channel while the other exchanges a 
$\tilde l$ charged slepton in the $s$-channel.
The influence on cross sections of the $t$-channels on the resonant ones
is quite small if the indirect bounds are satisfied.
\\The resonant production of a sneutrino could lead to a spectacular
signature: A bump in the dilepton  
invariant mass can be observed for an important branching ratio of the
decay, $\tilde \nu \to l \bar l$
\cite{z:DIMOP90}.
\\ 
Motivated by the weak low-energy constraints 
on the \Rp\ \ccs containing flavor index
from the third generation, the search reach in the resonant
$\tilde \nu_{\tau}$ tau-sneutrino (neutral current)
and $\tilde \tau$ stau (charged current) production channels have 
been obtained in \cite{z:Rizz}, for both the Tevatron and LHC colliders,
in the plane $M$ versus $X$,
where $M$ is the scalar mass and $X=\l'_{311} \l_{3jk} B_{l}$, $B_{l}$
being the leptonic branching ratio. Not only is it important to
notice the very large slepton mass reach (of order $800 \GeV$ at the Run
II Tevatron and $4 \TeV$ at the LHC) 
for sizeable values of $X$ ($\approx 10^{-3}$), but also the small $X$
($X = 10^{-(5-8)}$) reach for relatively small slepton
masses (of order the hundred of $\GeV$). The search reach is greater in the
charged current channel due to
the higher parton luminosities. This analysis is easily extended to the
first and second generation of slepton as well. 
\\ 
Since the charged lepton pair production provides a very clear signature, 
its \Rp\ contributions, from the resonant production of a sneutrino, 
have been treated in more details in the literature.
One may first consider the contribution to $e^+ e^-$ production.
The couplings product $\l'_{311} \l_{311}$ is of particular interest here.
This choice of couplings is motivated by an initial state
with high parton luminosity, $d \bar d$, and by the fact that among the
couplings $\l'_{i11}$ and $\l_{i11}$,
$\l'_{311}$ and $\l_{311}$ have the lowest low-energy bounds.
In this framework, existing Tevatron data \cite{z:data2} 
from the CDF detector on the $e^+ e^-$ production 
have been exploited in \cite{z:Kal} to derive bounds on the product
$\l'_{311} \l_{311}$. In this study, the cross sections for the \Rp
diagrams, in the $s$- and $t$-channel, contributing to $e^+ e^-$ 
production are computed, but only the resonant sneutrino production 
is taken into account. The CTEQ-3L parameterization \cite{z:CTEQ} is used
together with the multiplicative 
K factor calculated for the higher QCD corrections to Drell-Yan pair
production \cite{z:data2}. Since the corresponding K factor 
for slepton production has not been determined yet, the couplings $\l'
\l$ are theoretically uncertain 
at a level of about $10\%$, which is tolerable at a first stage of the
analysis. Assuming the sneutrino 
contribution to be smaller than the experimental error of the data
points in the $e^+ e^-$ invariant mass
distribution, the Yukawa couplings can be estimated to be,
\begin{eqnarray}
(\l'_{311} \l_{311})^{1/2}<0.08 \ \Gamma_{\tilde \nu_{\tau}}^{1/4},
\label{bound1}
\end{eqnarray}
for sneutrino masses in the range $120-250 \GeV$,
where $\Gamma_{\tilde \nu_{\tau}}$ denotes the sneutrino width in units
of $\GeV$.
\\ 
For the \Rp\ contributions via resonant 
sneutrino production to the two 
last families lepton pair production, a different approach, based 
on the total cross section study, has been
adopted in \cite{z:shalom2}.
The motivation, which holds mainly for the $\tau^+ \tau^-$ production, 
is the following: It is experimentally easier to determine the production 
rate above some values of the dilepton invariant mass than to reconstruct 
the dilepton invariant mass itself. In \cite{z:shalom2}, the $\mu^+ \mu^-$
production case was studied in details. By combining the 
low energy limits on individuals \Rp\ couplings with the limits which
exist on products $\l \l'$, we are left with only two relevant \Rp\ \ccs
products, namely, $\l_{232}\l'_{311}$ and $\l_{232}\l'_{322}$, if we neglect
the contributions from the annihilation of the sea quarks, $b \bar b$. 
The $1 \s$-limit contours plot in the plane, $\l_{232}\l'_{311}$ versus
$\l_{232}\l'_{322}$, has been obtained for the reaction, $p \bar p \to
\mu^+ \mu^- +X$. For example, the attainable $1 \s$-limits 
are $-0.003<\l_{232}\l'_{311}<0.003$ and $-0.011<\l_{232}\l'_{322}<0.011$,
for a $\tau$ sneutrino mass of $200 \GeV$, at energy $\sqrt s=2 \TeV$ with a
total integrated luminosity ${\cal L}=2\femtob^{-1}$ appropriate for the Tevatron
Run II. We observe that due to larger valence d quark probability functions, a
significant improvement over the present limits may be obtained for the \Rp\
product $\l_{232}\l'_{311}$.
Note also that the cross sections were calculated
applying an upper cut on the $\mu^+ \mu^-$ system invariant mass of 
$M^+_{\mu^+ \mu^-}=500 \GeV$ and a lower cut of $M^-_{\mu^+
\mu^-}=150 \GeV$. This lower cut practically removes the $s$-channel $Z$
resonance contribution. 
\\ 
In conclusion, dilepton production clearly offers a
unique way to explore the R-parity violating parameter 
space: $\l \l'$ versus the slepton mass.  
\\ 
Finally, if such a resonance were observed, how could we
distinguish between a scalar 
or a new gauge boson resonance ? This point is treated in \cite{z:Rizz}:
In the case of neutral current, one immediate difference would 
be the observation of the very unusual $e \nu$ final states, which would
be a truly remarkable 
signature for R-parity violation. A universality violation, namely, a
substantially different rate 
for the $e^+ e^-$ and the $\mu^+ \mu^-$ final states, would also
eliminate the possibility of a 
resonant $Z'$ boson. If such differences would have not been observed, the
measurement of the 
forward-backward asymmetry ($A_{FB}$) could discriminate between a
scalar resonance, which
produces a null asymmetry, and a boson resonance, which lead to $A_{FB}
\ne 0$ due to parity 
violating fermionic couplings. A more complex situation arises when $Z'$
naturally has 
$A_{FB} = 0$. In such a case, the complete angular distribution analysis
would be conclusive
as to the identity of the resonance. Of course, the required statistics
for these measurements
results in a significant loss in the mass reach: For example, for a
fixed value of $X$
(same notation as above), if the Tevatron were able to discover a
sneutrino with a mass of
$700 \GeV$, the value of $A_{FB}$ could only be extracted for $m_{\tilde
\nu}=500 \GeV$, 
and the angular distribution for $m_{\tilde \nu}=400 \GeV$. In the case
of charge current,
one interesting possibility is to examine the leptonic charge asymmetry
for the electrons and
muons in the final state, which is defined as: 
\begin{eqnarray}
A(\eta)={\frac { dN_+/d \eta - dN_-/d \eta}{dN_+/d \eta + dN_-/d \eta}},
\label{chas}
\end{eqnarray}  
where $N_{\pm}$ are the number of positively/negatively charged leptons
of a given rapidity, $\eta$. The presence of the slepton tends to drive the
asymmetry to smaller absolute values as might be expected in the Standard
Model, while the deviation due to either type of $W'$ substantially increases
the magnitude of the asymmetry.
The minimum value of the product $\l \l'$ for which the asymmetry differs
significantly
from the \SM\ expectation at the Tevatron ($\sqrt{s}=2 \TeV$) is $0.1$, for
a luminosity of $2 \femtob^{-1}$,
assuming $m_{\tilde l} = 750 \GeV$ and $\Gamma_{\tilde l} /
m_{\tilde l} = 0,004$.

\begin{figure}[t]
\begin{center}
\leavevmode
\centerline{\psfig{figure=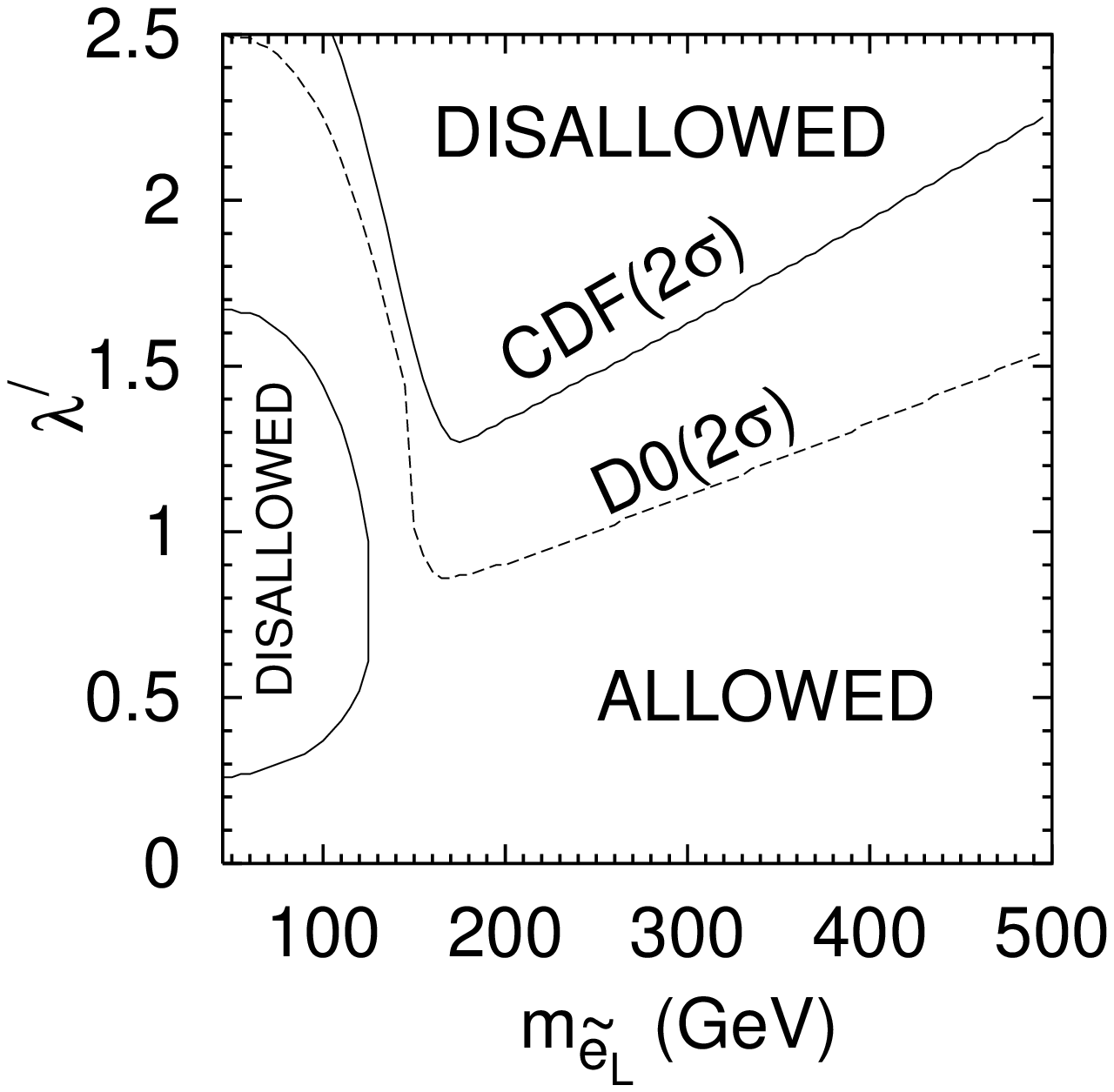,width=5.in}}
\caption{\footnotesize  \it
Allowed regions in the plane of $\lambda'_{i3k}$ and the mass of the
left slepton in a lepton number-violating scenario. Solid 
(dashed) lines correspond to the 2-$\sigma$ bounds from the 
CDF (D0) collaborations. 
\rm \normalsize }
\label{fig1}
\end{center}
\end{figure}

\begin{figure}[t]
\begin{center}
\leavevmode
\centerline{\psfig{figure=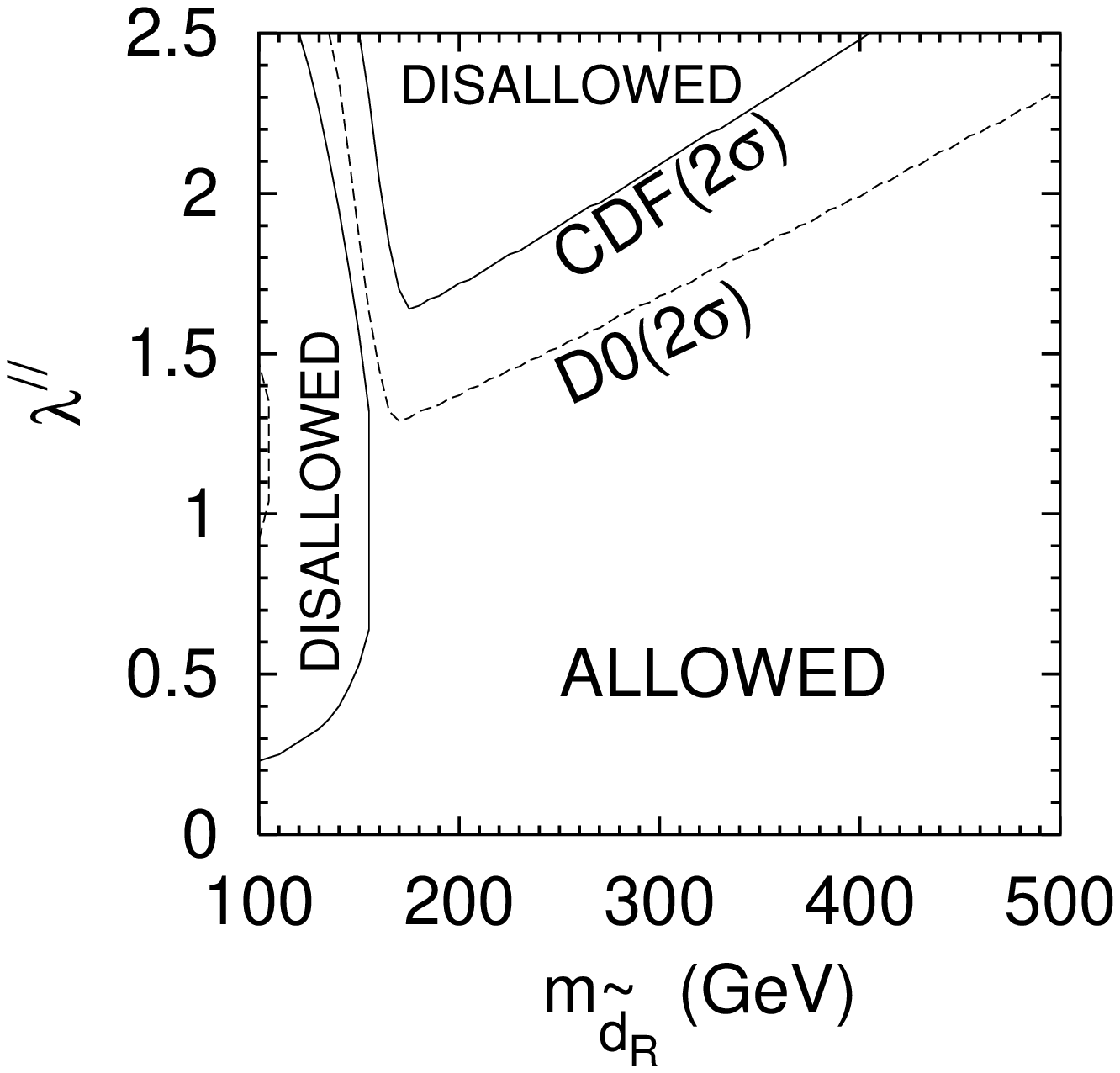,width=5.in}}
\caption{\footnotesize  \it
Allowed regions in the plane of $\lambda''_{3ki}$ and the mass
of the right $d$-squark in a baryon number-violating scenario. Solid 
(dashed) lines correspond to the 2-$\sigma$ bounds from the 
CDF (D0) collaborations. 
\rm \normalsize }
\label{fig2}
\end{center}
\end{figure}

$\bullet$ {\bf Top quark physics}
The large mass of the top quark, $m_{top}$, entails a top lifetime ,
$\tau_{top}=[1.56 \GeV {m_{top}/180 \GeV})^3]^{-1}$, sufficiently
shorter than the typical QCD hadronization time so that decay occurs before
fragmentation. The top decays mainly into a bottom quark in association with
a $W$ gauge boson, which can have a leptonic decay channel, $W \to l \nu$,
with a branching ratio of $2/9$.
Therefore, the top quark production offers some very clean signatures,
with a rather energetic lepton, some missing energy and a b jet which can be
tagged with a good efficiency. This particular behavior of the top
quark, together with the fact that the low energy constraints on $\l'$ and
$\l''$ couplings involving third-generation fields are not very stringent, has
motivated numerous studies on the top physics.

\begin{figure}[t]
\begin{center}
\leavevmode
\centerline{\psfig{figure=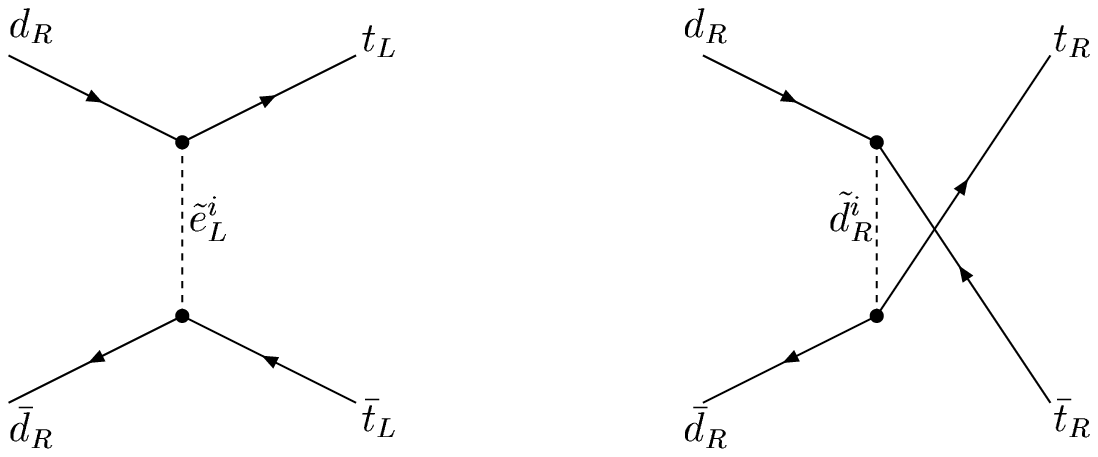,height=3.in}}
\end{center}
\protect\caption{\em Feynman diagrams for the $L$-violating 
process $d{\bar d} \to t_L {\bar t}_L$ via the $\l'$ coupling 
and for the $B$-violating process $d{\bar d} \to t_R {\bar t}_R$ due to
the coupling $\l''$.}
\label{fig:feytt}
\end{figure}

{\bf a) Top quark pair production}
The top quark pair production could receive a contribution from
diagrams with an initial state, $d_k \bar d_k$, and exchanging either a
$\tilde l^i_L$ slepton (via $\l'_{i3k}$) or a $\tilde d^i_R$ squark 
(via $\l''_{3ki}$) in the $t$-channel 
(see Fig.\ref{fig:feytt}). 
Those amplitudes have been
calculated in \cite{z:Sri}, with the CTEQ-3M parton distributions \cite{z:CTEQ}
and folding by the K factor which is extracted from the resummed QCD cross 
sections \cite{z:Ktt} in the $q \bar q$ annihilation channel. The region
of the \susyq\ parameter space allowed at a $ 95 \% $ confidence level by
the D0 and CDF data \cite{z:data1} on $ t \bar t $ production cross section have 
also been obtained in \cite{z:Sri}, and are shown in Fig.\ref{fig1} in the
plane, $ \l'_{i31}/m_{\tilde l^i_L} $, and in Fig.\ref{fig2} in the plane,
$  \l''_{31i}/m_{\tilde d^i_R} $. For squark masses smaller than the top
mass, the bounds on the $\l''_{31i}$ are weakened considerably because the
decay channel, $t \to d \tilde d_R$, opens up, and this tends to dilute 
the signal from the Standard Model decay mode. Simultaneously,
the opening up of the decay channel, $t \to d \tilde l_L$, does not affect
the bounds on the $\l'_{i31}$ quite as much, because the relevant branching
ratio is suppressed by a smaller colour factor. The colour factors that appear
in the  \Rp\ $t \bar t$ productions also explain why the $\l'_{i31}$ are more
constrained than the $\l''_{31i}$. Nevertheless, the bounds obtained on
$\l''_{31i}$ are comparable to the indirect limits, for $m_{\tilde q}=100
\GeV$. Note that since these interactions are chiral, they induce 
polarization of the final state. Polarization can be a useful observable for
probing the \Rp\ couplings \cite{z:hikasa99}. 
At the LHC energy, while the gluon initiated contribution is much
larger, the usual annihilation process in QCD is further suppressed because
it is an $s$-channel process. The $t$-channel \Rp\ subprocess does not suffer
this  suppression and those effects, while smaller than the corresponding
effects at the TeVatron, are still sizeable.


{\bf b) Single top quark production}
Single top quark production is of special importance in the context of
fermion pair production. In contrast to the QCD process of $t \bar t$
production, the single top quark production, $u \bar d \to t \bar b$,
involves only the electroweak interactions and can therefore be used to probe
the electroweak theory and to study models of new physics. The single top
quark production can also come from the $W$-gluon fusion. Since the structure
functions are better known for the quarks than for the gluons, the cross
section accuracy is higher for the single top quark production through the 
exchange of a $W$ gauge boson than for the $W$-gluon fusion. Hence the
process, $u \bar d \to t \bar b$, is a better test for new physics than the
$W$-gluon fusion. The feasibility of single top quark production via squark 
and slepton exchanges to probe several combinations of $R$ parity violating 
couplings at hadron colliders has been studied in \cite{z:Ste,z:Dat,z:Oak,z:Chiap}. 
According to those studies, the LHC is better at probing the $B$ violating
couplings $\lambda^{\prime \prime}$ whereas the Tevatron and the LHC have a
similar sensitivity to $\lambda'$ couplings.
%
%
%
The number of signal events depends on the mass and width of the 
exchanged sparticle, and on the value of the Yukawa couplings. 
%
%
%
The width of the exchanged sparticle is a sum of the widths 
due to $R$-parity conserving and $R$-parity violating decays: 
\begin{equation} 
\Gamma_{tot} = \Gamma_{R_p} + \Gamma_{\Rp} 
\end{equation} 
where $\Gamma_{\Rp}$ is given by  
\begin{equation}
\Gamma_{\Rp}(\tilde d^{*k}_R\longrightarrow u^i d^j)=
\frac{(\lambda''_{ijk})^2}{2\pi} 
\frac{(M^2_{\tilde d^k_R}-M^2_{u^i}-M^2_{d^j})^2}{M^3_{\tilde d^k_R}} 
\end{equation} 
for the squarks, and by 
\begin{equation} 
\Gamma_{\Rp}(\tilde l^i_L\longrightarrow \bar u^j d^k)=
\frac{3(\lambda'_{ijk})^2}{16\pi} 
\frac{(M^2_{\tilde l^i_L}-M^2_{u^j}-M^2_{d^k})^2}{M^3_{\tilde l^i_L}}
\end{equation}
for the sleptons.
%
%
\\
We consider at first a quark-antiquark initial parton state which is
of relevance at the Tevatron as in this case both particles can be
valence quarks in the initial state and may have therefore a considerably
higher cross section in $p \bar p$ collisions. Quark-quark initial parton
states are examined later on in connection with LHC studies. It has been shown
\cite{z:Ste} that the signal for $u \bar d \to t \bar b$ is potentially
observable at the Tevatron with $2-3 \femtob^{-1}$ integrated luminosity,
although at the LHC it will be relatively suppressed by the sea structure
functions and overwhelmed by the large background from $t \bar t$ production
plus $W$-gluon fusion. 
\\ 
The single top quark production, $u_j \bar d_k
\to t \bar b$, can occur through the exchange of a $\tilde l_L^i$ slepton in
the $s$-channel via the couplings product, $\l'_{ijk}\l'_{i33}$. Using a
CTEQ-3L \cite{z:CTEQ} parton distribution, summing over the flavor index $i$
(the sleptons $\tilde l_L^i$ sleptons are supposed degenerate in mass) and
taking for the values of the \Rp\ coupling  constants the values of the
indirect bounds, the ratio of the resonant slepton contribution cross section
over the \SM\ cross section was found \cite{z:Dat} to exceed $ 20 \% $ at
the upgraded Tevatron when slepton mass lies in the narrow range,
$ 200 \GeV - 270 \GeV $, and $ M_2 = - \mu > 200 \GeV$.
This large slepton mass suppression of the ratio can be understood as
follows: When the slepton mass is large, the parton cross section contribution
coming from the slepton resonant production requires
large momenta from the initial partons, which is suppressed by their
structure functions. An additional suppression 
is caused by the increase of the slepton width when the slepton mass
increases. Note also that the ratio increases with the increase of the  
\susyq\ parameters $M_2$ and $\mu$, since then the neutralino and
chargino masses increase and thus the width of the slepton decreases.
The dominant process, $ u \bar d \to \tilde l_L^i \to t \bar b$,
which involves the sum of couplings,
$\l'_{111}\l'_{133}+\l'_{211}\l'_{233}+\l'_{311}\l'_{333}$,  was
considered in \cite{z:Oak}. An interesting signature for this process is an
energetic charged lepton, missing $E_T$ and
double b-quark jets. Using a series of kinematic cuts in order to
reduce the \SM\ background and to consider
the detector acceptance, the values of the slepton mass versus the
couplings to be observable at $95 \%$ confidence level were calculated. The
result shows that for values of the couplings below the low energy bounds,
the slepton mass can only be probed in the range, $ 200 \GeV <m_{\tilde
l_L^i}< 340 \GeV $, for the upgraded TeVatron and in the interval, $200 \GeV
<m_{\tilde l_L^i}< 400 \GeV$,  for the LHC. Although larger parton momenta are
favored at the LHC, the result is not really improved at LHC because of the relative
suppression of the $\bar d$ quark structure function compared to the $d$ quark
one. The difference between the results based on the two sets of structure
functions, MRSA' \cite{z:MRSA} and  CTEQ-3M \cite{z:CTEQ}, has been found to be
small. 
\\ 
The single top quark production, $u_i \bar d_j \to t \bar b$,
receives also a contribution  from the exchange of a $\tilde d_R^k$ squark in
the $t$-channel, through the product of couplings $\l''_{i3k} \l''_{3jk}$.
Since the non-observation of proton decay imposes very stringent
conditions on the simultaneous presence of $\l'$ and $\l''$, only their
separate effects in single top quark production are interesting to study.
Choosing the initial state of the reaction, $u_i \bar d_j \to t \bar b$,
fix the flavor indices of the \ccs product $\l''_{i3k} \l''_{3jk}$ because
of the antisymmetry of the constants $\l''$. Due to either too strong low
energy constraints on the couplings or too low parton luminosities, the only
product of interest is $\l''_{132} \l''_{312}$. If we assume an
observable level of, $\Delta \s / \s_0 > 20\% $, where $\Delta \s$ is the
\Rp\ cross section and $\s_0$ is the \SM\ cross section, the coupling
$\l''_{132}  \l''_{312}$ can be probed at the upgraded Tevatron down to
$0.01,0.02,0.03,0.04,0.06,0.08,0.1$ and $0.13$ for $M_{\tilde s_R}=100, 200,
300, 400, 500, 600, 700$ and $800 \GeV$, respectively \cite{z:Dat}. 
\\ 
%
%
%
Another interesting single top quark production is the reaction, $u_i
d_j \to t b$, which can occur 
through the exchange of a $\tilde d^k_R$ squark in the s channel via the
couplings product $\l''_{ijk} \l''_{33k}$.
Although the corresponding events can have the unique signal of an
energetic charged lepton, missing transverse
eneregy and two same sign $b$ quarks, since the tagging can not
distinguish a b quark from a $\bar b$ quark,
the background is quite as important as that of the reaction, $u_i \bar
d_j \to t \bar b$. 
Quark-quark initial parton states may be valence contributions from the
structure functions at a $pp$ collider as the LHC. Note however that the
strong indirect bound on $\l''_{112}$ makes very small the valence-valence
$ud$ contribution to the cross section.
An example of cross sections that are obtained from different initial parton
states at the LHC is given in table \ref{tavlhc1}. 
\begin{table}[htb] 
\begin{center} 
\begin{tabular}{|c|c|c|c|c|c|}
\hline
Initial partons   & $cd$    &  $cs$    &  $ub$   &  \multicolumn{2}{ c|}{$cb$} \\ 
\hline 
Exchanged particle   
&$\tilde s$ & $\tilde d$ &$ \tilde s$ & $ \tilde d$ & $ \tilde s$ 
\\ \hline 
Couplings    
& $\lambda''_{212}\lambda''_{332}$   & $\lambda''_{212}\lambda''_{331}$   
& $\lambda''_{132}\lambda''_{332}$   & $\lambda''_{231}\lambda''_{331}$   
& $\lambda''_{232}\lambda''_{332}$   \\ \hline 
Cross section in pb    
& 3.98  & 1.45   & 5.01  &  \multicolumn{2}{ c|} {0.659}\\
\hline
\end{tabular} 
\caption{Cross section in pb of
the reaction $u_i d_j \to \tilde d^k_R \to t b$ at LHC 
for a squark of mass of $600$~GeV assuming $\Gamma_{R_p}$ = 0.5 GeV
and $\l''_{ijk}=0.1$.} 
\label{tavlhc1} 
\end{center}
\end{table} 
Using some effective kinematical cuts, the sensibility plot at $95 \%$
confidence level has been obtained in 
the plane $\l''_{212} \l''_{332}$ versus $m_{\tilde s_R}$ in \cite{z:Oak}:
The coupling product $\l''_{212} \l''_{332}$
can be probed up to $0.5,0.1$ and $1.5$ for $m_{\tilde s_R}=700, 800$
and $1000 GeV$ at the upgraded Tevatron, 
and for $m_{\tilde s_R}=2100, 2700$ and $3100 GeV$ at the LHC,
respectively. The sensitivity on the squark mass
is quite higher at the LHC than at the Tevatron. Note that the process,
$c s \to \tilde d_R \to t b$, cannot be
probed as efficiently as, $c d \to \tilde s_R \to t b$, because of the
relative suppression of the strange quark 
structure function compared to the valence down quark.
\begin{figure}[t]
\begin{center}
\centerline{\psfig{figure=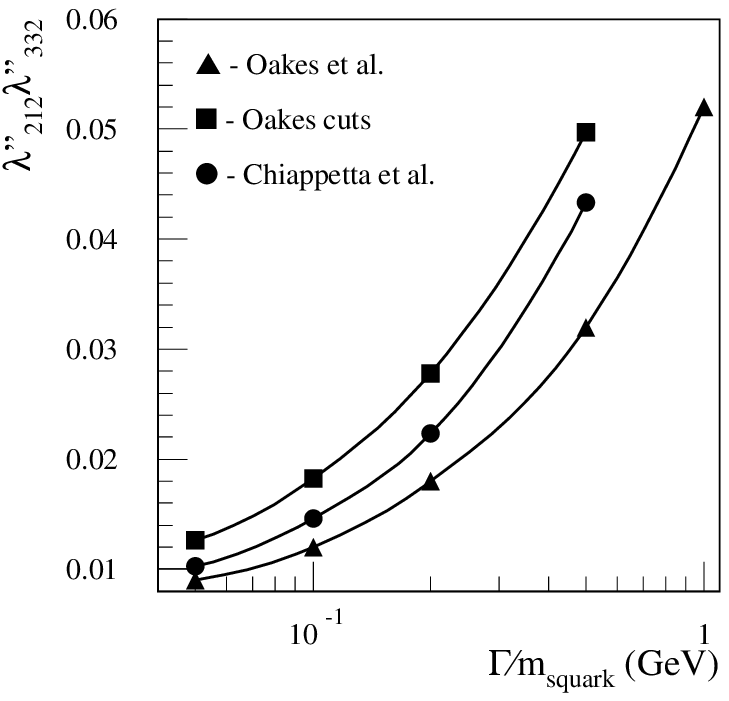,width=10cm}}  
\caption{Sensitivity limits on the $\lambda''_{212}\lambda''_{332}$ 
Yukawa couplings obtained from the analysis of the reaction 
$cd \to \tilde s^* \to t b$ at the LHC 
after 1 year with low luminosity for $m_{\tilde s}$= 
300 GeV, found in \cite{z:Chiap} (circles) and in \cite{z:Oak} (triangles). 
The squares indicate the results obtained in \cite{z:Chiap} by applying 
the simplified cuts used in \cite{z:Oak}.}
\label{fig:deandrea}
\end{center}
\end{figure}
As illustrated in Fig.\ref{fig:deandrea},
the study of \cite{z:Chiap} on the reaction $u_i d_j \to t b$ at LHC, which
was performed at an higher
degree of precison in the simulation process than in \cite{z:Oak}, has lead to
weaker
sensitivities on the $\l''_{ijk} \l''_{33k}$ product of coupling constants
than the analysis
of \cite{z:Oak}.
\\
The reaction $u_j d_k \to t b$ receives also a contribution from the
exchange
of a $\tilde l^{\pm}_{iL}$ slepton in the u channel,
via the $\l'_{ij3}$ and $\l'_{i3k}$ couplings \cite{z:Chiap}.
\\
The single top quark production at hadronic colliders via \Rp couplings
offers the
opportunity to reconstruct supersymmetric particle masses.
For instance, the invariant mass distribution associated to the
the two $b$ quarks, the charged lepton and the reconstructed neutrino
produced in the reaction
$u b \to \tilde s \to t b \to W b b \to l \nu b b$, occuring via the
couplings $\l''_{132}$ and
$\l''_{332}$, allows to reconstruct the $\tilde s$ mass.
The error on this mass reconstruction performed from the events observed in
the ATLAS detector
at LHC would be dominated by the $1 \%$
systematic uncertainty on the jet energy scale in ATLAS, for $m_{\tilde
s}=600GeV$,
$\l''_{132} \l''_{332}=10^{-2}$ and assuming a luminosity of ${\cal
L}=30fb^{-1}$ \cite{z:Chiap}.

To conclude, the top quark physics is more favorable to the tests on
the $\l''$ than on the $\l'$ couplings. Furthermore, it is the only framework
in which the constraints on $\l''$ from physics at colliders are 
comparable or better than the low energy bounds on the $\l''$ coupling
constants.   

\subsection{$\Rp$ Contributions to Flavour Changing Neutral Currents}

As for the leptonic colliders physics, the \Rp\ interactions could
induce \fcnc effects at hadronic colliders in reactions as simple 
as the lepton pair production, $q \bar q \to \bar l_{J} l_{J'}$ 
($J \neq J'$). Nevertheless, the environment is not as clean as for
the leptonic colliders and these \fcnc effects are less easily
observable. Those \fc lepton productions occur from an initial state
$d_j \bar d_k$ ($d_k \bar d_k$) through the exchange
of a $\tilde \nu^i_L$ sneutrino ( $\tilde u^j_L$ squark) in the 
$s$-channel ($t$-channel) via the couplings product $\l'_{ijk}\l_{iJJ'}$
($\l'_{Jjk}\l'_{J'jk}$), or, from
an initial state $u_j \bar u_j$ through the exchange
of a $\tilde d^k_R$ squark  in the 
$t$-channel via the couplings product $\l'_{Jjk}\l'_{J'jk}$.

\subsection{$\Rp$ Contributions to CP violation}

The resonant production of a sneutrino gives rise to the possibility 
of having CP violation effects at tree level, which are therefore quite
important. This was treated in \cite{z:shalom2} for hadronic colliders in
analogy with the case of leptonic colliders \cite{z:shalom}, 
which was exposed above (Section \ref{rpvfcnc}).
If the $\tau$ spins could be measured, the future Tevatron runs will be
capable of detecting CP violation effects in the polarization
asymmetries of the hard process, 
$d_j \bar d_k \to \tilde \nu^{\mu} \to \tau^+ \tau^-$.
The \Rp\ \cc $\l'_{2jk}$ which enters this subprocess is chosen real,
while $\l_{233}$ is taken complex in order to generate CP asymmetries. 
The spin asymmetries study at hadronic colliders must be treated with 
special care.
In fact, the spin asymmetries change sign around 
$\sqrt s \approx m_{\tilde \nu^{\mu}_{\pm}}$.
Since in hadronic collisions one has to integrate over $\sqrt s$, the
spin asymmetries are reduced. 
To compensate this effect, one has to integrate the absolute values of the 
polarization asymmetries. Of course this demands a study, in a previous 
stage, of the $\tau^+ \tau^-$ invariant mass distribution, which is needed 
to determine the $m_{\tilde \nu^{\mu}_{\pm}}$ sneutrino mass.
The maximum of the spin asymmetries reach $20-30 \%$ with a mass splitting 
$\Delta m_{\tilde \nu^{\mu}}=\Gamma_{\tilde \nu^{\mu}}$ and 
$10-13 \%$ for $\Delta m_{\tilde \nu^{\mu}}= \Gamma_{\tilde \nu^{\mu}}/4$,
throughout the mass range $150 \GeV < m_{\tilde \nu^{\mu}_{\pm}} < 450 \GeV$. 
At the Tevatron Run II (III) with ${\cal L}=2 \femtob^{-1}$ ($30 \femtob^{-1}$) 
the CP conserving and CP violating asymmetries may be detected with a 
sensitivity above $3 \s$ over the mass range 
$155 \GeV <m_{\tilde \nu^{\mu}_{\pm}} < 400 \GeV$ 
($155 \GeV<m_{\tilde \nu^{\mu}_{\pm}}<300 \GeV$) if 
$\Delta m_{\tilde \nu^{\mu}}=\Gamma_{\tilde \nu^{\mu}}$
($\Delta m_{\tilde \nu^{\mu}}=\Gamma_{\tilde \nu^{\mu}}/10$).
Finally, the entire range of $r$ (see CP violation in Section~\ref{indlept})
can be practically covered for $m_{\tilde \nu^{\mu}_-}=200 \GeV$ at the 
Tevatron Run II (III) with at least $3 \s$ standard deviations, 
for $\Delta m_{\tilde \nu^{\mu}}=\Gamma_{\tilde \nu^{\mu}}$
($\Delta m_{\tilde \nu^{\mu}}=\Gamma_{\tilde \nu^{\mu}}/4$).
Those computations have been performed taking for $\l'_{2jk}$ and 
$\vert \l_{233} \vert $ the values of the low energy bounds, 
applying an upper cut on the $\tau^+ \tau^-$ system invariant mass of 
$M^+_{\tau^+ \tau^-}=500 \GeV$ and a lower cut of $M^-_{\tau^+
\tau^-}=150 \GeV$ and including all ${j,k}$ combinations 
in $d_j \bar d_k$ fusion.
The results show that, contrary to the leptonic colliders framework
\cite{z:shalom}, the CP odd and CP even spin asymmetries could be observed over 
a wide $\tilde \nu^{\mu}$ sneutrino mass range, of about $300 \GeV$.

\clearpage

\vspace{10 mm}
\begin{center}
{  }
\end{center}
\vspace{10 mm}

\clearpage

\clearpage

\vspace{10 mm}
\begin{center}
{  }
\end{center}
\vspace{10 mm}

\clearpage

\setcounter{chapter}{0}
\setcounter{section}{0}
\setcounter{subsection}{0}
\setcounter{figure}{0}

\chapter*{Publication II}
\addcontentsline{toc}{chapter}{Resonant sneutrino production at Tevatron Run II}

\newpage

\vspace{10 mm}
\begin{center}
{  }
\end{center}
\vspace{10 mm}

\clearpage

\begin{center}
{\bf \huge Resonant sneutrino production at Tevatron Run II}
\end{center} 
\vspace{2cm}
\begin{center}
G. Moreau, M. Chemtob
\end{center} 
\begin{center}
{\em  Service de Physique Th\'eorique \\} 
{ \em  CE-Saclay F-91191 Gif-sur-Yvette, Cedex France\\}
\end{center} 
\begin{center}
F. D\'eliot, C. Royon, E. Perez
\end{center} 
\begin{center}
{\em  Service de Physique des Particules, DAPNIA \\} 
{ \em CE-Saclay F-91191 Gif-sur-Yvette, Cedex France \\}
\end{center}
\vspace{1cm}
\begin{center}
{Phys. Lett. {\bf B475} (2000) 184, hep-ph/9910341}
\end{center}
\vspace{2cm}
\begin{center}
Abstract  
\end{center}
\vspace{1cm}
{\it
We consider the single chargino production at Tevatron 
$p \bar p \to \tilde \nu_i \to \tilde \chi^{\pm}_1 l_i^{\mp}$
as induced by the resonant sneutrino production via a dominant 
\RPV coupling of type $\l'_{ijk} L_i Q_j D_k^c$.
Within a supergravity model, we study the three leptons final state.
The comparison with the expected background 
demonstrate that this signature allows to
extend the sensitivity on the \susyq mass spectrum 
beyond the present LEP limits and to probe 
the relevant \RPV coupling down to values one order of magnitude smaller 
than the most stringent low energy indirect bounds. The trilepton signal 
offers also the opportunity to reconstruct the neutralino mass in a 
model independent way with good accuracy.}

\newpage

In the minimal supersymmetric standard model (MSSM), 
the \susyq (SUSY) particles must be produced in pairs.
In contrast, the single superpartner 
production which benefits from a
larger phase space is allowed in the \RPV (\rpv)
extension of the MSSM.
In particular the SUSY particle resonant production 
can reach high cross-sections
either at leptonic \cite{a:Han} or hadronic colliders \cite{a:Dim2}, 
even taking into account the strongest 
low-energy bounds on \rpv coupling constants \cite{a:Drein}.
Hadronic colliders provide an additional advantage in that they
allow to probe a wide mass range of the new resonant particle,
due to the continuous energy distribution of the colliding partons.
Furthermore, since the resonant production has a cross-section which is 
proportional to the relevant coupling squared,
this could allow an easier
determination of the \rpv coupling than pair production reaction.
Indeed in the latter case,
the sensitivity on the \rpv coupling is
mainly provided by the displaced vertex analysis for the 
Lightest Supersymmetric Particle (LSP) decay,
which is difficult experimentally especially at hadronic colliders.

The SUSY particle produced at the resonance 
mainly decays through 
R-parity conserving interactions into the LSP,
via cascade decays. 
In the case of a dominant 
$\l''_{ijk} U_i^c D_j^c D_k^c$ coupling, the decay of the LSP 
leads to multi-jets final states, wich have an high QCD background
at hadronic colliders. Besides, at hadronic colliders, the 
$\l_{ijk} L_i L_j E_k^c$ 
couplings do not contribute to resonant production.
In this letter, we thus assume a dominant 
$\l'_{ijk} L_i Q_j D_k^c$ coupling which initiates 
the resonant sneutrino production $\bar d_j d_k \to \tilde \nu_i$
and hence the single chargino production at Tevatron through
$ p \bar p \to \tilde \nu_i \to \tilde \chi^{\pm}_1 l_i^{\mp}$.
We focus on the three leptons signature 
associated with the cascade decay 
$\tilde \chi^{\pm}_1 \to \tilde \chi^0_1 l^{\pm}_p \nu_p$, 
$\tilde \chi^0_1 \to l^+_i \bar u_j d_k \ +c.c.$,
assuming the $\tilde \chi^0_1$ to be the LSP. 
The main motivation rests on the possibility to reduce the background.
This is similar in spirit to a recent study \cite{a:Rich} 
of the like sign dilepton signature from the single neutralino 
production at Tevatron via the resonant charged slepton production.


\begin{figure}[b]
\begin{center}
\leavevmode
\centerline{\psfig{figure=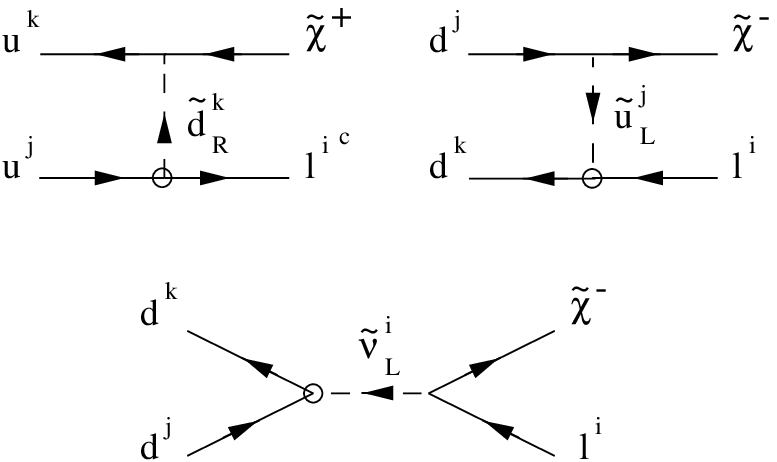,height=2.5in}}
\end{center}
\caption{Feynman diagrams for the single chargino 
production at Tevatron via the $\l'_{ijk}$ coupling 
(symbolised by a circle in the figure). 
The arrows denote flow of the particle momentum.}
\label{fig0}
\end{figure}

We concentrate on the $\l'_{211}$ coupling.
The associated hard scattering processes,
$ d \bar d \to \tilde \nu_{\mu} \to \tilde \chi^{\pm}_1 \mu^{\mp}$,
$d \bar d \to \tilde \chi^{\pm}_1 \mu^{\mp}$ and
$u \bar u \to \tilde \chi^{\pm}_1 \mu^{\mp}$ (see Fig.\ref{fig0}),
involve first generation quarks for the initial partons. 
The indirect constraint on 
this coupling is $\l'_{211}<0.09 (\tilde m /100GeV)$  \cite{a:Drein}.
While $\l'_{111}$ is disfavored due to severe constraints 
\cite{a:Drein}, 
the case of a dominant $\l'_{311}$ could also be of interest.

Our framework is the so-called minimal \sugra model (mSUGRA),
in which the absolute value of the 
Higgsino mixing parameter $\vert \mu \vert$ is determined by the 
radiative electroweak symmetry breaking condition.
We restrict to the infrared fixed point region for the top quark 
Yukawa coupling, in which $\tan \beta$ is fixed \cite{a:Pok}.
We shall present results for the low solution $\tan \beta \simeq 1.5$  
and for $sign(\mu)=-1$, $A=0$. 
In fact the cross-section for the single chargino production
depends smoothly on the $\mu$, $A$ and $\tan \beta$ parameters.
The cross-section can reach values of order a few picobarns. 
For instance, choosing the mSUGRA point, $M_2(m_Z)=200GeV$, 
$m_0=200GeV$, 
and taking $\l'_{211}=0.09$ we find using CTEQ4 \cite{a:SF} 
parametrization for the parton densities a cross-section of 
$\sigma(p \bar p \to \tilde \chi_1^{\pm} \mu^{\mp})=1.45pb$
at a center of mass energy $\sqrt s =2 TeV$.
Choosing other parametrizations does not change significantly   
the results since mainly intermediate Bjorken $x$ partons 
are involved in the studied process.
The cross-section depends mainly on the 
$m_{1/2}$ (or equivalently $M_2$) and $m_0$ 
soft SUSY breaking parameters.
As $M_2$ increases, the chargino mass increases
reducing the single chargino production rate. 
At high values of $m_0$, the sneutrino mass is enhanced so that 
the resonant sneutrino production is reduced. This leads to 
a decrease of the single chargino production rate 
since the $t$ and $u$ channels contributions are small 
compared to the resonant sneutrino contribution.
Finally, for values of $m_{\tilde \nu_{\mu}}$ 
(which is related to $m_0$) approaching
$m_{\tilde \chi^{\pm}_1}$ (which is related to $M_2$), a reduction of the
chargino production is caused by the decrease of the phase space 
factor associated to the decay 
$\tilde \nu_{\mu} \to \chi^{\pm}_1 \mu^{\mp}$.

The single chargino production cross-section must be
multiplied by the leptonic decays branching fractions which
are $B(\tilde \chi^{\pm}_1 \to \tilde \chi^0_1 l^{\pm}_p \nu_p)=33 \%$
(summed over the three leptons species)
and $B(\tilde \chi^0_1 \to \mu u d)=55 \%$, 
for the point chosen above of the mSUGRA parameter space.
The leptonic decay of the chargino is typically of order $30\%$ 
for $m_{\tilde l},m_{\tilde q},m_{\tilde \chi^0_2}>m_{\tilde \chi^{\pm}_1}$, 
and is smaller than the hadronic decay
$\tilde \chi^{\pm}_1 \to \tilde \chi^0_1 \bar q_p q'_p$ 
because of the color factor.   
When $\tilde \chi^0_1$ is the LSP, it decays via $\l'_{211}$ either as
$\tilde \chi^0_1 \to \mu u d$ or as $\tilde \chi^0_1 \to \nu_{\mu} d d$, 
with branching ratios $B(\tilde \chi^0_1 \to \mu u d)$ ranging between 
$\sim 40\%$ and $\sim 70\%$.


The backgrounds for the three leptons signature at Tevatron are:
(1) The top quark pair production followed by the top decays $t \to b W$
where one of the charged leptons is generated in $b$-quark decay.
(2) The $W^{\pm} Z^0$ and $Z^0 Z^0$ productions 
followed by leptonic decays of the gauge bosons.
It has been pointed out recently \cite{a:Match,a:Pai} that non negligible 
contributions can occur through virtual gauge boson, 
as for example the $W^* Z^*$ or $W \g^*$ productions. 
However, these contributions lead at most to one hard jet
in the final state in contrast with the signal and
have not been simulated.
(3) Standard Model productions as for instance the $W t \bar t$ production. 
These backgrounds have been estimated in \cite{a:Barb} 
to be negligible at $\sqrt s=2 TeV$. 
We have checked that the $Zb$ production gives a
negligible contribution to the 3 leptons signature.  
(4) The fake backgrounds as, 
$p \bar p \to \ Z \ + \ X, \ Drell-Yan \ +  \ X, \ b \bar b b$,
where $X$ and $b$-quarks fake a charged lepton.
Monte Carlo simulations using simplified detector
simulation cannot give a reliable estimate of this background.
(5) The \susyq background generated by the superpartner pair production.
This background is characterised by two cascade decays 
ending each with
the decay of the LSP as $\tilde \chi^0_1 \to \mu u d$ 
via the $\l'_{211}$ coupling,
and thus is suppressed compared to the signal due to
the additional branching fraction factors.
Moreover the SUSY background incurs a larger phase space suppression.
In particular its main contribution, namely
the squark and gluino pair productions, is
largely suppressed for large $\tilde{q}$ and
$\tilde{g}$ masses~\cite{a:tat1}.
Although a detailed estimation has not been performed we
expect that this background can be further 
reduced by analysis cuts, since at least four jets
are expected in the final state and leptons should 
appear less isolated than in the signal.


We have simulated the single chargino production 
$p \bar p \to \tilde \chi^{\pm}_1 \mu^{\mp}$ 
with a modified version of the SUSYGEN event generator \cite{a:SUSYGEN3}
and the \sm background
($W^{\pm} \ Z^0$, $Z^0 \ Z^0$ and $t \bar t$ productions)
with the PYTHIA event generator \cite{a:PYTH}. Both 
SUSYGEN and PYTHIA have been interfaced with
the SHW detector simulation package \cite{a:SHW},
which mimics an average of the CDF and D0 Run II detector performance.

The following cuts aimed at enhancing 
the signal-to-background ratio have been applied. First,
we have selected the events with at least three charged leptons
($e^{\pm}$ or $\mu^{\pm}$)
with energies greater than $10GeV$ for the softer of them and 
$20GeV$ for the two others, namely, 
$N_l \geq 3 \ [l=e,\mu]$ with
$E_{min}(l)>10GeV$, $E_{med}(l)>20GeV$ and $E_{max}(l)>20GeV$. 
In addition, since our final state is
$3l+2jets+ \Eslash $ we have required  
that the minimum number of jets should be equal to two, 
where the jets have an energy higher than $10 GeV$, namely, 
$N_j \geq 2$ with $E_j > 10 GeV$. 
This selection criteria suppresses the background from 
the gauge bosons production
which generates at most one hard jet. 
Note that these events requiring high energy charged leptons and jets
are easily triggered at Tevatron. 
In order to eliminate poorly isolated leptons 
originating from the decays of
hadrons (as in the $t \bar t$ production), we have
imposed the isolation cut
$\Delta R=\sqrt{\delta \phi^2+\delta \theta^2}>0.4$, where
$\phi$ is the azimuthal angle and $\theta$ the polar angle, between 
the 3 most energetic charged leptons and the 2 hardest jets. 
We have also demanded that
$\delta \phi>70^{\circ}$ between the leading 
charged lepton and the 2 hardest jets.
With the cuts described above and for an integrated luminosity 
of ${\cal L}=1 fb^{-1}$ at $\sqrt s=2 TeV$ for Tevatron Run II, 
the $Z^0 Z^0$, $W^{\pm} Z^0$, $t \bar t$ productions lead to 
$0.22,0.28,1.1$ events respectively.



\begin{figure}[t]
\begin{center}
\leavevmode
\centerline{\psfig{figure=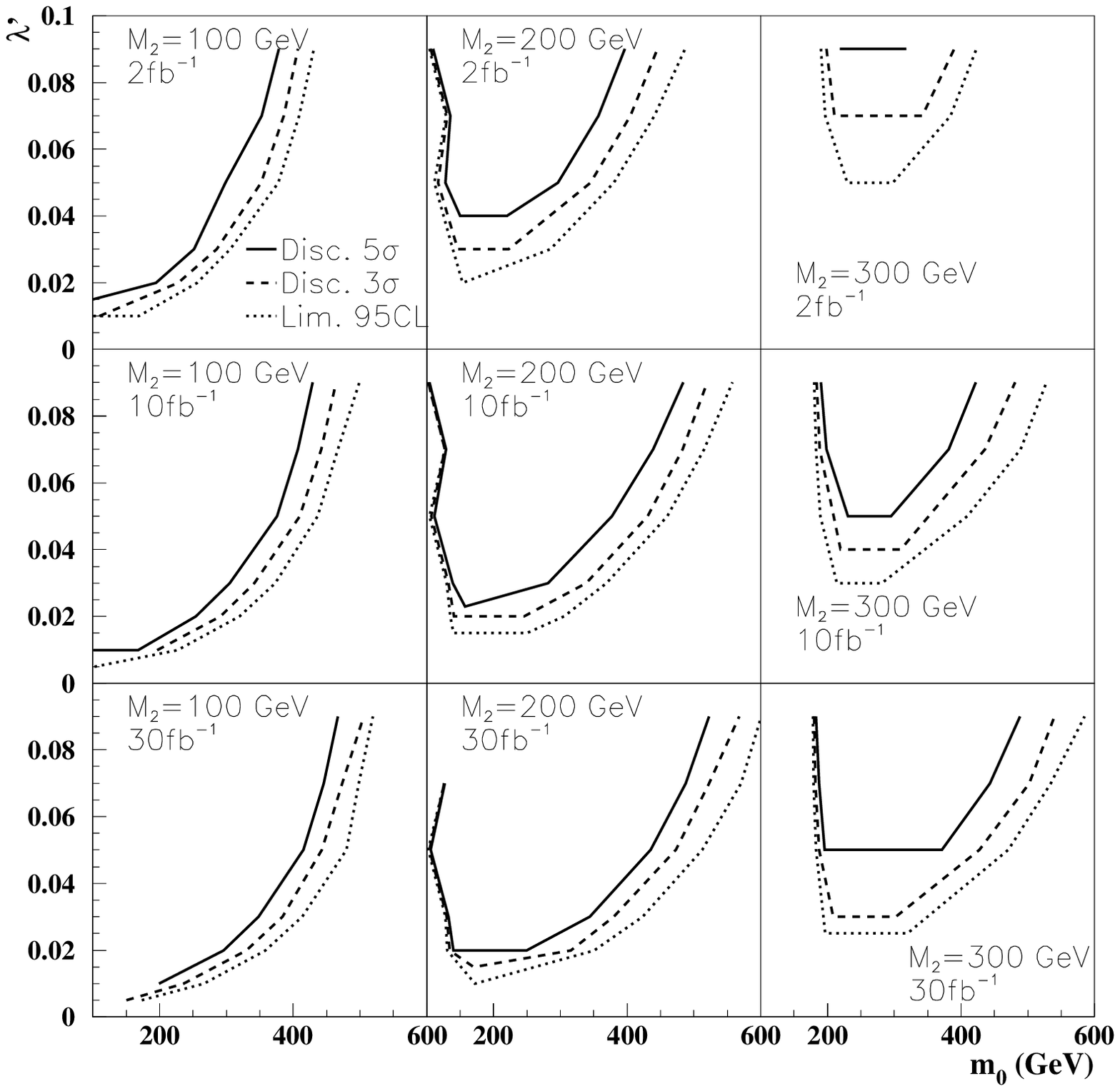,height=5.5in}}
\end{center}
\caption{Discovery contours at $5 \sigma$ (full line), $ 3 \sigma$ 
(dashed line) and limit at $95 \% \ C.L.$ (dotted line) presented
in the plane $\l'_{211}$ versus the $m_0$ parameter,
for different values of $M_2$ and of luminosity.}
\label{a:fig1}
\end{figure}

In Fig.\ref{a:fig1}, we present the $3 \sigma$ and $ 5 \sigma$ discovery  
contours and the limits at $95 \%$
confidence level in the $\l'_{211}$-$m_0$ plane,
using a set of values for $M_2$ and the luminosity. 
For a given value of $M_2$, we note that the sensitivity on the $\l'_{211}$ 
coupling decreases at high and low values of $m_0$. At high values of 
$m_0$, the sneutrino mass is enhanced 
inducing a decrease of the sneutrino production 
cross-section.
At low values of $m_0$, the sneutrino mass decreases leading to
a reduction of the phase space factor for the decay 
$\tilde \nu_{\mu} \to \tilde \chi^{\pm}_1 \mu^{\mp}$ which follows the 
resonant sneutrino production.
Similarly, we note the decrease of the sensitivity on the $\l'_{211}$ 
coupling when $M_2$ increases for a fixed value of $m_0$.
This is due to the increase of the chargino mass which  
results also in a smaller phase space factor for the sneutrino decay.

\begin{figure}[t]
\begin{center}
\leavevmode
\centerline{\psfig{figure=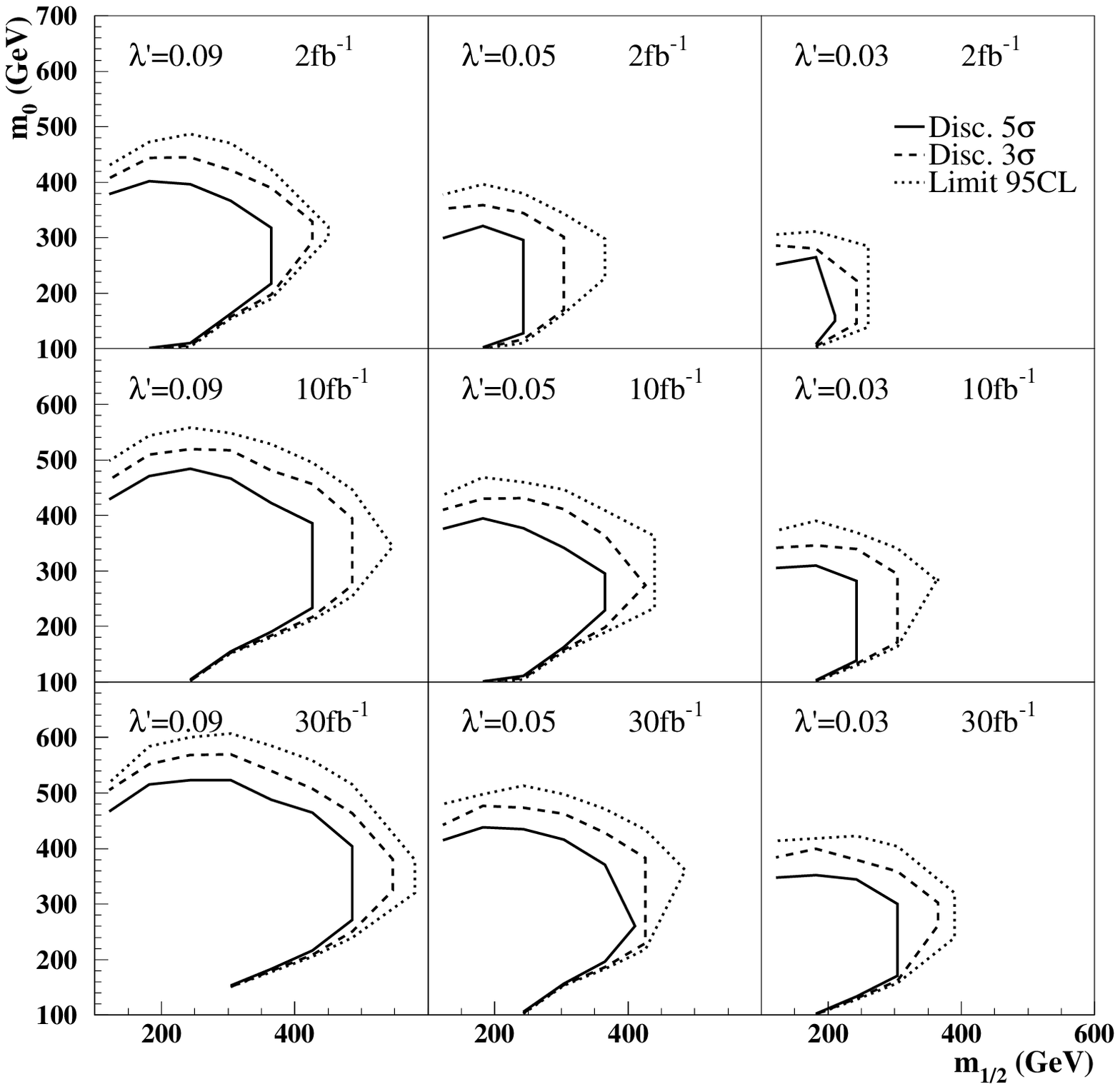,height=5.5in}}
\end{center}
\caption{Discovery contours at $5 \sigma$ (full line), 
$ 3 \sigma$ (dashed line)  
and limit at $95 \% \ C.L.$ (dotted line) presented
in the plane $m_0$ versus $m_{1/2}$,
for different values of $\l'_{211}$ and of luminosity.}
\label{a:fig2}
\end{figure}

In Fig.\ref{a:fig2}, the discovery potential is shown
in the plane $m_0$ versus $m_{1/2}$,
for different values of $\l'_{211}$ and of luminosity.
For the same reasons as above, 
we observe a reduction of the sensitivity on $\l'_{211}$ 
when $m_0$ (respectively, $m_{1/2}$ or equivalently $M_2$) 
increases for a fixed value of $m_{1/2}$ (respectively $m_0$).

\begin{figure}[t]
\begin{center}
\leavevmode
\centerline{\psfig{figure=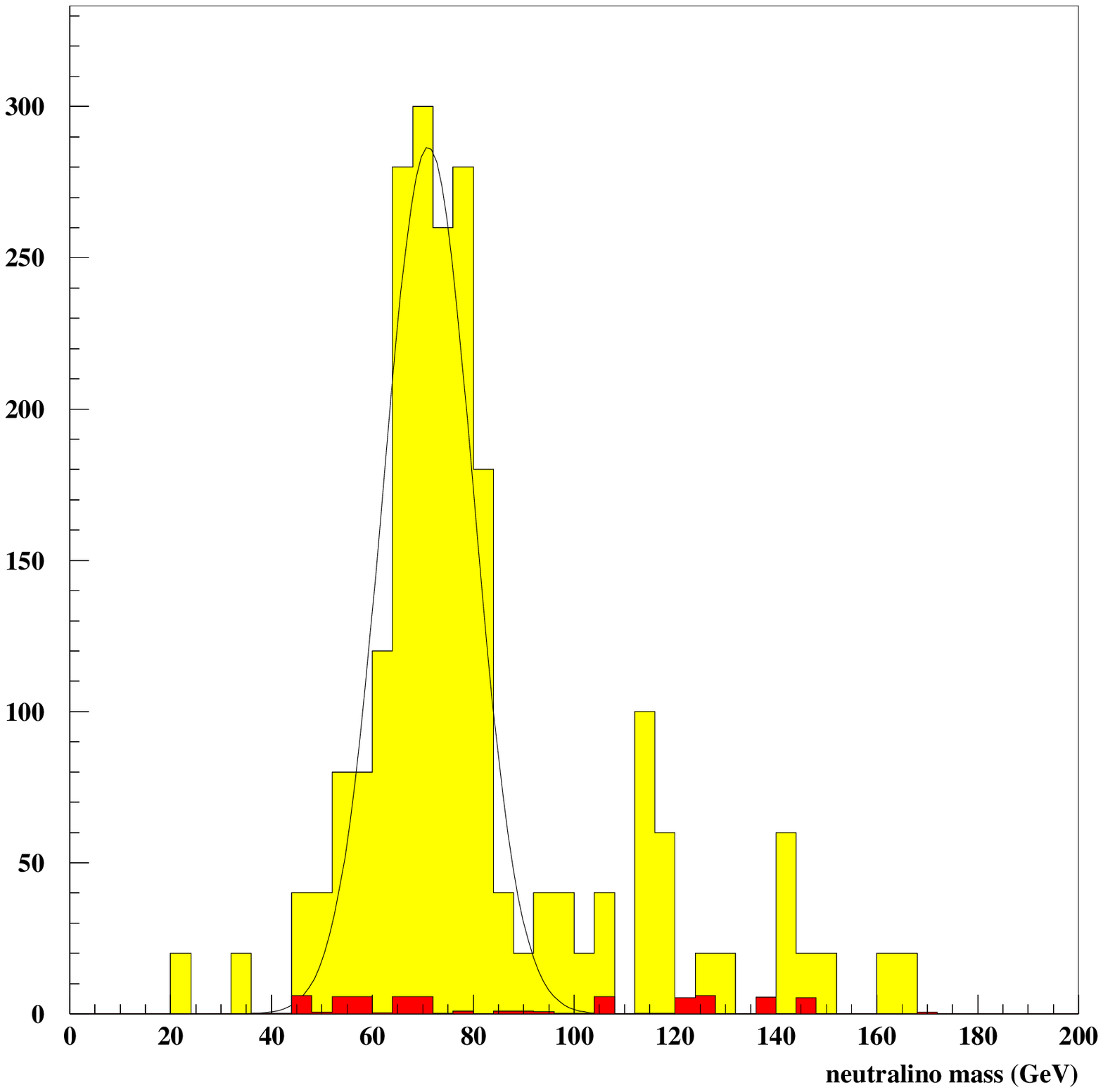,height=5.5in}}
\end{center}
\caption{Distribution for the invariant mass of
the 2 jets and the lower energy muon in the $e \mu \mu$ events, 
for a luminosity of ${\cal L}=30fb^{-1}$. The sum of the 
$WZ,ZZ$ and $t \bar t$ backgrounds is in black and the  
signal is grey. The mSUGRA point taken for this figure is, 
$m_0=200GeV$, $M_2=150GeV$ ($m_{\tilde \chi^0_1}=77GeV$), 
and the \rpv coupling is $\l'_{211}=0.09$.}
\label{fig3}
\end{figure}

An important improvement with respect to
the limits derived recently from LEP data \cite{a:Moriond} can 
already be obtained within the first year of Run II at Tevatron 
(${\cal L}=1 fb^{-1}$). Even
Run I data could probably lead to new limits on the \susyq parameters.
The strongest bounds on the \susyq masses obtained
at LEP in an \rpv model with non-vanishing $\l'$ Yukawa coupling are
$m_{\tilde \chi^{\pm}_1}>94 GeV$, $m_{\tilde \chi^{0}_1}>30 GeV$,
$m_{\tilde l}>81 GeV$ \cite{a:Moriond}. 
Note that for the minimum values of $m_0$ and $m_{1/2}$
spanned by the parameter space described in Fig.\ref{a:fig1} 
and Fig.\ref{a:fig2}, namely 
$m_0=100GeV$ and $M_2=100GeV$, the spectrum is
$m_{\tilde \chi^{\pm}_1}= 113GeV$, $m_{\tilde \chi^{0}_1}= 54GeV$,
$m_{\tilde \nu_L}= 127 GeV$, $m_{\tilde l_L}= 137 GeV$,
$m_{\tilde l_R}= 114 GeV$, so that we are well above these limits.
Since both the scalar and gaugino 
masses increase with $m_0$ and $m_{1/2}$,
the parameter space described in these figures lies outside the present 
forbidden range, in the considered framework.
\\ With the luminosity of ${\cal L}=30 fb^{-1}$ expected at the 
end of the Run II, $m_{1/2}$ values up to $550GeV$ ($350GeV$) 
corresponding to a chargino mass of about 
$m_{\tilde \chi^{\pm}_1} \approx 500 GeV \ (300 GeV)$
can be probed if the $\l'_{211}$ coupling is $0.09$ ($0.03$). 
The sensitivity on $m_0$ reaches
$600GeV$ ($400GeV$), which corresponds to a sneutrino mass of about 
$m_{\tilde \nu_{\mu}} \approx 600GeV \ (450GeV)$,
for a value of the $\l'_{211}$ coupling equal to $0.09$ ($0.03$).   
Couplings down to a value of $0.005$ can also be tested at Tevatron Run II,
in the promising scenario where $m_0 = 200 GeV$ and $M_2=100GeV$, namely,
$m_{\tilde \chi^{\pm}_1} \approx 100GeV$ 
and $m_{\tilde \nu_{\mu}} \approx 200GeV$.

Let us make a few remarks on the model dependence of our results. 
First, as we have discussed above, 
the sensitivity reaches depend on the 
SUSY parameters mainly through the \susyq mass spectrum.
Secondly, in the major part of the mSUGRA parameter space, 
the LSP is the $\tilde \chi^0_1$. Besides, in the mSUGRA model,
the mass difference between
$\tilde \chi^{\pm}_1$ and $\tilde \chi^0_1$ 
is large enough not to induce a dominant \rpv decay for the chargino.  
Notice also that we have chosen the scenario of low $\tan \beta$. 
For high $\tan \beta$, due to the slepton mixing in the third 
generation, the $\tilde \tau$ slepton mass can be reduced
down to $\sim m_{\tilde \chi^{\pm}_1}$ so that the
branching ratio of the $\chi^{\pm}_1$ decay into tau-leptons 
$\tilde \chi^{\pm}_1 \to \tilde \chi^0_1 \tau^{\pm}_p \nu_{\tau}$
increases and exceeds that into $e$ and $\mu$ leptons,
leading to a decrease of the efficiency after cuts.
For example, the efficiency at the mSUGRA point
$m_0=200GeV$, $M_2=150GeV$, $sign(\mu)=-1$, $A=0$,
is $4.93\%$ for $\tan \beta=1.5$ and $1.21\%$ for $\tan \beta=50$.
However, for still decreasing $\tilde{\tau}$ mass,
$\tilde \chi^{\pm}_1 \to \tilde \chi^0_1 \tau^{\pm}_p \nu_{\tau}$
starts to dominate over the hadronic mode
so that the efficiency loss is compensated by the leptonic
decays of the $\tau$, and the branching of the $\chi^{\pm}_1$ into
$e$ and $\mu$ leptons can even increase up to $34 \%$.
For instance, the efficiency for
$m_0=300GeV$, $M_2=300GeV$, $sign(\mu)=-1$, $A=0$,
is $5\%$ for the 2 values $\tan \beta=1.5$ and $\tan \beta=50$.


Another particularly interesting aspect of our signal is
the possibility of a $\tilde \chi^0_1$ neutralino mass reconstruction
in a model independent way. As a matter of fact,
the invariant mass distribution of 
the charged lepton and the 2 jets coming 
from the neutralino decay $\tilde \chi^0_1 \to \mu u d$ 
allows to perform a clear neutralino 
mass reconstruction. The 2 jets found in these
events are generated in the $\tilde \chi^0_1$ decay.   
In order to select the requisite charged lepton, we concentrate on the 
$e \mu \mu$ events. In those events, we know that for a relatively 
important value of the mass difference, 
$m_{\tilde \nu_{\mu}}-m_{\tilde \chi^{\pm}_1}$,
the leading muon comes from the decay, 
$\tilde \nu_{\mu} \to \tilde \chi_1^{\pm} \mu^{\mp}$,
and the other one from the neutralino decay (the electron is generated
in the decay $\tilde \chi^{\pm}_1 \to \tilde \chi^0_1 e^{\pm} \nu_e$).
In Fig.\ref{fig3}, we present the invariant mass distribution of the 
lepton and 2 jets selected through this method. 
The average reconstructed $\tilde \chi^0_1$ mass is about 
$71 \pm 9GeV$ to be compared with 
the generated mass of $\tilde \chi^0_1=77GeV$. 
In a more detailed analysis of this signal \cite{a:Gia,a:next},
the neutralino mass can be reconstructed with higher precision  
using for e.g. constrained fit algorithms.
This mass reconstruction is performed easily
in contrast with the pair production analysis in \rpv scenarios
\cite{a:Atlas} which suffers an higher combinatorial background. 
Moreover, a reconstruction of the chargino and sneutrino
masses is also possible.
This invariant mass distribution would also  
allow to discriminate between the signal and the SUSY background.


As a conclusion, 
we have presented a new possibility of studying resonant sneutrino 
productions in \rpv models at Tevatron. 
Results (see also \cite{a:next}) lead to a sensitivity on the $\l'_{211}$
coupling, on the sneutrino and chargino masses well beyond the 
present limits. Besides, a model-independent reconstruction of the  
neutralino mass can be performed easily with great accuracy.
Our work leads to the interesting conclusion that the three leptons signature
considered as a `gold plated' channel for the discovery of \susy
at hadronic colliders \cite{a:Match,a:Pai,a:Barb}, is also particularly 
attractive in an R-parity violation context.

We acknowledge C. Guyot, R. Peschanski, C. Savoy and X. Tata for 
useful discussions and reading the manuscript.

\clearpage

\vspace{10 mm}
\begin{center}
{  }
\end{center}
\vspace{10 mm}

\clearpage

\setcounter{chapter}{0}
\setcounter{section}{0}
\setcounter{subsection}{0}
\setcounter{figure}{0}

\chapter*{Publication III}
\addcontentsline{toc}{chapter}{Single superpartner production 
at Tevatron Run II}

\newpage

\vspace{10 mm}
\begin{center}
{  }
\end{center}
\vspace{10 mm}

\clearpage

\begin{center}
{\bf \huge Single superpartner production at Tevatron Run II}
\end{center} 
\vspace{2cm}
\begin{center}
G. Moreau
\end{center} 
\begin{center}
{\em  Service de Physique Th\'eorique \\} 
{ \em  CE-Saclay F-91191 Gif-sur-Yvette, Cedex France \\}
\end{center} 
\begin{center}
F. D\'eliot$^1$, C. Royon$^{1,2,3}$   
\end{center} 
\begin{center}
{\em 1: Service de Physique des Particules, DAPNIA  \\} 
{ \em  CE-Saclay F-91191 Gif-sur-Yvette, Cedex France \\}
{\em 2: Brookhaven National Laboratory, Upton, New York, 11973 \\} 
{\em 3: University of Texas, Arlington, Texas, 76019 \\} 
\end{center}
\vspace{1cm}
\begin{center}
{To appear in Eur. Phys. Jour. {\bf C}, hep-ph/0007288}
\end{center}
\vspace{2cm} 
\begin{center}
Abstract  
\end{center}
\vspace{1cm}
{\it
We study the single productions of supersymmetric particles at Tevatron Run
II
which occur in the $2 \to 2-body$ processes involving R-parity violating
couplings
of type $\l'_{ijk} L_i Q_j D_k^c$.
We focus on the single gaugino productions which receive contributions from
the resonant slepton productions.
We first calculate the amplitudes of the single gaugino productions.
Then we perform analyses of the single gaugino productions based on the
three
charged leptons and like sign dilepton signatures. These analyses allow to
probe
supersymmetric particles masses beyond the present experimental limits, and
many of the $\l'_{ijk}$ coupling constants down to values smaller than the
low-energy bounds. Finally, we show that the studies of
the single gaugino productions offer the opportunity to reconstruct
the $\tilde \chi^0_1$, $\tilde \chi^{\pm}_1$, $\tilde \nu_L$ and
$\tilde l^{\pm}_L$ masses with a good accuracy in a model independent way.
}

\newpage

\section{Introduction}
\label{b:intro}

\setcounter{equation}{0}

In the Minimal Supersymmetric Standard Model (MSSM),
the \susyq (SUSY) particles must
be produced in pairs. The phase space 
is largely suppressed in pair production of SUSY particles
due to the large masses of the superpartners.
The R-parity violating (\rpv) extension of the MSSM contains the following
additional terms in the superpotential,
which are trilinear in the quarks and leptons superfields,
\begin{eqnarray}
W_{\rpv}=\sum_{i,j,k} \bigg (\ud \l _{ijk} L_iL_j E^c_k+
\l ' _{ijk} L_i Q_j D^c_k+ \ud \l '' _{ijk} U_i^cD_j^cD_k^c   \bigg ),
\label{b:super}
\end{eqnarray}
where $i,j,k$ are flavour indices.
These \rpv couplings offer the opportunity to produce the scalar
\susyq particles as resonances \cite{b:Dim1,b:Dreinoss}. Although the \rpv
coupling constants are severely constrained by the
low-energy experimental bounds 
\cite{b:Drein,b:Bhatt,b:GDR,b:referbound}, the resonant
superpartner
production reaches high cross sections both at leptonic
\cite{b:Han} and hadronic \cite{b:Dim2} colliders.

The resonant production of SUSY particle has another interest:
since its cross section is proportional
to a power $2$ of the relevant \rpv coupling,
this reaction would allow an easier determination of the
\rpv couplings than the pair production provided the \rpv coupling 
is large enough.
As a matter of fact in the pair production study,
the sensitivity on the \rpv couplings is
mainly provided by the displaced vertex analysis of the
Lightest Supersymmetric Particle (LSP) decay
which is difficult experimentally, especially at hadronic colliders.
Besides, the displaced vertex analysis allows to test a limited
range of couplings which is such that the LSP has a large enough life time 
to have a measurable decay length while still decaying inside the 
detector.

Neither the Grand Unified Theories (GUT), the string theories nor
the study of the discrete gauge symmetries give a strong theoretical
argument in favor of the R-parity violating or
R-parity conserving scenarios \cite{b:Drein}. Hence,
the resonant production of SUSY particle through \rpv couplings
is an attractive possibility which must be considered in the
phenomenology of supersymmetry.

The hadronic colliders have an advantage in detecting
new particles resonance. 
Indeed, due to the wide energy
distribution of the
colliding partons, the resonance can be probed in a wide range of the
new particle mass at hadronic colliders. This is in contrast with the 
leptonic colliders where only large resonances can be probed through 
radiative returns.

At hadronic colliders, either a slepton or a squark can be produced at
the resonance respectively through a $\l'$ or a $\l''$ coupling constant.
In the hypothesis of a single dominant \rpv
coupling constant, the resonant scalar particle
can decay through the same \rpv coupling as in the production,
leading to a two quark final state for the hard process
\cite{b:Dim2,b:Bin,b:Dat,b:Oak,b:Rizz,b:RunII,b:Chiap,b:referparton}.
In the case where both $\l'$ and $\l$ couplings
are non-vanishing, the slepton produced via $\l'$ can
decay through $\l$
giving rise to the same final state as in Drell-Yan process,
namely two leptons
\cite{b:Dim2,b:Rizz,b:RunII,b:Kal,b:SonPart}. However, for reasonable values of the
\rpv coupling constants, the decays
of the resonant scalar particle via gauge interactions are
typically dominant if kinematically allowed \cite{b:Han,b:lola}.\\
The main decay of the resonant scalar particle
through gauge interactions is the decay
into its \sm partner plus a gaugino. Indeed,
in the case where the resonant scalar particle is a squark,
it is produced
through $\l''$ interactions so that it
must be a Right squark $\tilde q_R$ and thus it
cannot decay into the $W^{\pm}$-boson,
which is the only other possible
decay channel via gauge interactions. Besides,
in the case where the resonant scalar particle is a slepton,
it is a Left slepton produced via a $\l'$ coupling
but it cannot generally decay as
$\tilde l^{\pm}_L \to W^{\pm} \tilde \nu_L$ or as
$\tilde \nu_L \to W^{\pm} \tilde l^{\mp}_L$.
The reason is that in most of the SUSY models,
as for example the supergravity or the gauge
mediated models, the mass difference
between the Left charged slepton and the Left sneutrino
is due to the D-terms so that it is fixed by the relation
$m^2_{\tilde l^{\pm}_L}-m^2_{\tilde \nu_L}=\cos 2 \beta M_W^2$
\cite{b:Iban}
and thus it does not exceed the $W^{\pm}$-boson mass.
Nevertheless, we note that in the large $\tan \beta$ scenario, a
resonant
scalar particle of the third generation
can generally decay into the $W^{\pm}$-boson due to the large mixing in the
third family sfermion sector. For instance, in the SUGRA model with
a large $\tan \beta$ a tau-sneutrino produced at the resonance
can decay as $\tilde \nu_{\tau} \to W^{\pm} \tilde \tau^{\mp}_1$,
$\tilde \tau^{\mp}_1$ being the lightest stau.\\
The resonant scalar particle production
at hadronic colliders leads thus mainly
to the single gaugino production,
in case where the decay of the relevant scalar particle
into gaugino is kinematically allowed.
In this paper, we study the single gaugino productions
at Tevatron Run II.
The single gaugino productions at hadronic colliders
were first studied in \cite{b:Dreinoss,b:Dim2}.
Later, studies on the single neutralino \cite{b:Rich}
and single chargino \cite{b:greg}
productions at Tevatron have been performed.
The single neutralino \cite{b:Rich2} 
\footnote{After having submitted our 
paper, we noticed that resonant slepton production is also
studied in \cite{b:referecent}.} and single chargino
\cite{b:Gia1} productions have also been considered
in the context of physics at LHC.
In the present article, we also study the single superpartner
productions at Tevatron Run II which occur via $2 \to 2-body$
processes and do not receive
contributions from resonant SUSY particle productions.
The single slepton production in $2 \to 3-body$
processes has been considered in \cite{b:JLKneur}
in the context of physics at Tevatron and LHC.

The singly produced superpartner
initiates a cascade decay ended typically by
the \rpv decay of the LSP. In case of a single dominant
$\l''$ coupling constant, the LSP decays into quarks so that
this cascade decay leads to multijet final states having a
large QCD background \cite{b:Dim2,b:Bin}. Nevertheless, if some leptonic decays,
as for instance $\tilde \chi^{\pm} \to l^{\pm} \nu \tilde \chi^0$,
$\tilde \chi^{\pm}$ being the chargino and $\tilde \chi^0$ the neutralino,
enter the chain reaction, clearer leptonic signatures can be investigated
\cite{b:Berg}.
In contrast, in the hypothesis of a single dominant $\l'$ coupling constant,
the LSP decay into charged leptons naturally favors leptonic signatures
\cite{b:Dreinoss}. We will thus study
the single superpartner production reaction
at Tevatron Run II within the scenario of a single dominant $\l'_{ijk}$
coupling constant.

In section \ref{theoretical}, we define our theoretical framework.
In section \ref{discussion}, we present the values of the cross sections for
the various single superpartner productions
via $\l'_{ijk}$ at Tevatron Run II and we discuss the interesting
multileptonic signatures that these processes can generate.
In section \ref{analysis1}, we analyse the three lepton signature
induced by the single chargino production.
In section \ref{analysis2}, we study the like sign dilepton
final state generated by the single neutralino and chargino productions.

\section{Theoretical framework}
\label{theoretical}

\setcounter{equation}{0}

Our framework throughout this paper will be
the so-called minimal \sugra model (mSUGRA)
which assumes the existence of a grand unified gauge theory and
family universal boundary conditions on the supersymmetry breaking
parameters.
We choose the 5 following parameters:
$m_0$ the universal scalars mass at the unification scale $M_X$,
$m_{1/2}$ the universal gauginos mass at $M_X$,
$A=A_{t}=A_{b}=A_{\tau}$ the trilinear Yukawa coupling at $M_X$,
$sign(\mu)$ the sign of the $\mu(t)$ parameter
($t=\log (M_X^2/Q^2)$, $Q$ denoting the running scale)
and $\tan \beta=<H_u>/<H_d>$ where $<H_u>$ and
$<H_d>$ denote the vacuum expectation values of the Higgs fields.
In this model, the
higgsino mixing parameter $\vert \mu \vert$ is determined by the
radiative electroweak symmetry breaking condition.
Note also that the parameters $m_{1/2}$ and $M_2(t)$ ($\tilde W$ wino mass)
are related by the solution of the one loop renormalization group
equations $m_{1/2}=(1-\beta_a t)M_a(t)$ with $\beta_a=g_X^2 b_a/(4 \pi)^2$,
where $\beta_a$ are the beta functions,
$g_X$ is the \cc at $M_X$ and
$b_a=[3,-1,-11]$, $a=[3,2,1]$ corresponding to
the gauge group factors $SU(3)_c,SU(2)_L,SU(1)_Y$.
We shall set the unification scale at $M_X=2 \ 10^{16} GeV$
and the running scale at the $Z^0$-boson mass: $Q=m_{Z^0}$.

We also assume the infrared fixed point hypothesis for the top quark
Yukawa coupling \cite{b:Pok} that provides a natural explanation of
a large top quark mass $m_{top}$.
In the infrared fixed point approach,
$\tan \beta$ is fixed up to the ambiguity
associated with large or low $\tan \beta$ solutions.
The low solution of $\tan \beta$ is fixed by
the equation $m_{top}=C \sin \beta$,
where $C \approx 190-210 \ GeV$ for $\alpha_s(m_{Z^0})=0.11-0.13$.
For instance, with a top quark mass of
$m_{top}=174.2GeV$ \cite{b:top}, the low solution is given by $\tan \beta \approx 1.5$.
The second important effect of the infrared fixed point hypothesis is
that the dependence of the electroweak symmetry breaking
constraint on the $A$ parameter becomes
weak so that $\vert \mu \vert$ is a known function of the
$m_0$, $m_{1/2}$ and $\tan \beta$ parameters \cite{b:Pok}.

Finally, we consider the \rpv extension of the mSUGRA model
characterised by a single dominant \rpv coupling constant of type
$\l'_{ijk}$.

\section{Single superpartner productions via $\l'_{ijk}$ at Tevatron Run II}
\label{discussion}

\setcounter{equation}{0}

\subsection{Resonant superpartner production}
\label{resonant}

\begin{figure}[t]
\begin{center}
\leavevmode
\centerline{\psfig{figure=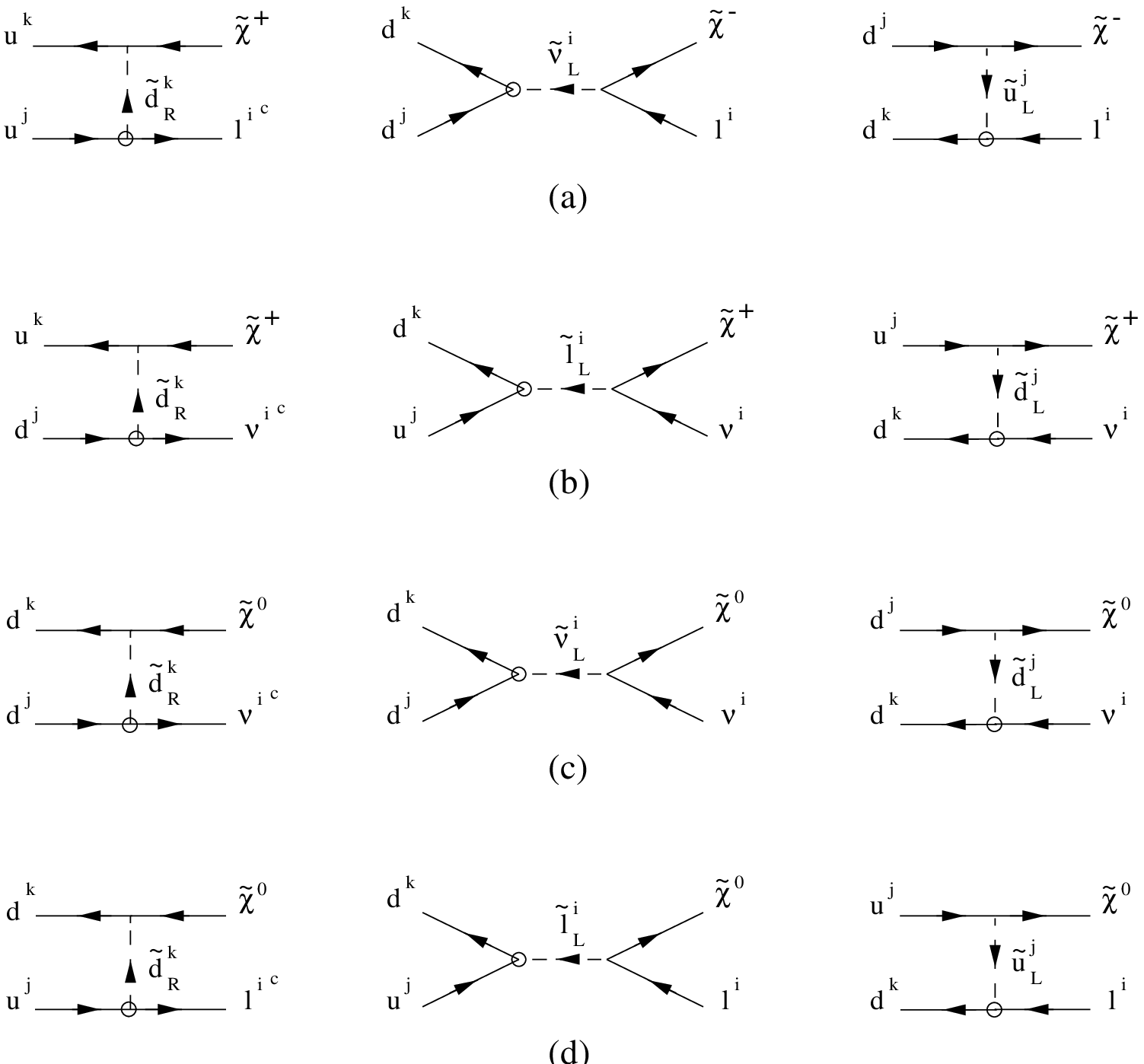,height=5.5in}}
\end{center}
\caption{\footnotesize  \it
Feynman diagrams for the 4 single production reactions
involving  $\l'_{ijk}$  at hadronic colliders which
receive a contribution from a resonant \susyq particle production.
The $\l'_{ijk}$ coupling constant is symbolised by a small circle
and the arrows denote the flow of the particle momentum.
\rm \normalsize }
\label{graphes}
\end{figure}

At hadronic colliders, either a sneutrino ($\tilde \nu$) or a charged
slepton ($\tilde l$) can be produced
at the resonance via the $\l'_{ijk}$ coupling.
As explained in Section \ref{b:intro}, for most of the SUSY models,
the slepton produced at the resonance has two possible gauge
decays, namely a decay into either a chargino or a neutralino.
Therefore, in the scenario of a single dominant
$\l'_{ijk}$ coupling and for most of the SUSY models,
either a chargino or a neutralino
is singly produced together
with either a charged lepton or a neutrino,
through the resonant superpartner production at hadronic colliders.
There are thus four main possible types
of single superpartner production reaction
involving $\l'_{ijk}$ at hadronic colliders
which receive a contribution from resonant SUSY particle production.
The diagrams associated
to these four reactions are drawn in Fig.\ref{graphes}.
As can be seen in this figure, these single superpartner productions receive
also some contributions from both the $t$ and $u$ channels.
Note that all the single superpartner production processes drawn in
Fig.\ref{graphes} have charge conjugated processes.
We have calculated the amplitudes of the processes shown in
Fig.\ref{graphes}
and the results are given in Appendix \ref{formulas}.

\subsubsection{Cross sections}
\label{cross1}

\begin{figure}[t]
\begin{center}
\leavevmode
\centerline{\psfig{figure=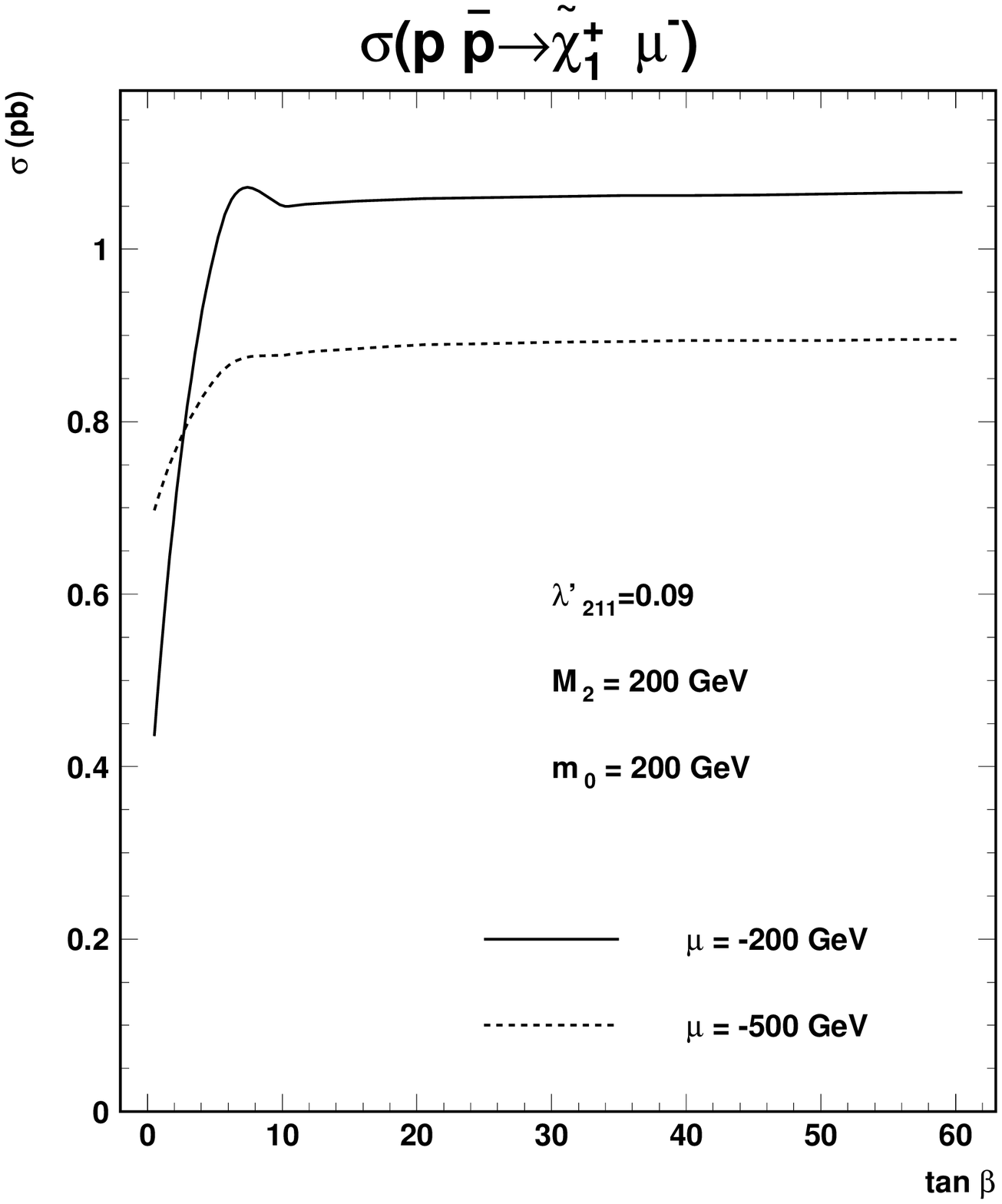,height=5.5in}}
\end{center}
\caption{\footnotesize  \it
Cross sections (in $pb$) of the single chargino production
$p \bar p \to \tilde \chi^+_1 \mu^-$ at a center of mass energy
of $2 TeV$ as a function of the $\tan \beta$ parameter
for $\l'_{211}=0.09$, $M_2=200GeV$,
$m_0=200GeV$ and two values of the $\mu$ parameter:
$\mu=-200GeV,-500GeV$.
\rm \normalsize }
\label{XStan}
\end{figure}

In this section, we discuss the dependence of the
single gaugino production cross sections on the various
\susyq parameters. We will not assume here the
radiative electroweak symmetry breaking condition in order to
study the variations of the cross sections with the
higgsino mixing parameter $\mu$.

First, we study the cross section of the single chargino production
$p \bar p \to \tilde \chi^+ l_i^-$ which
occurs through the $\l'_{ijk}$ coupling
(see Fig.\ref{graphes}(a)).
The differences between the $\tilde \chi^+ e^-$, $\tilde \chi^+ \mu^-$
and $\tilde \chi^+ \tau^-$ production
(occuring respectively through the $\l'_{1jk}$, $\l'_{2jk}$ and
$\l'_{3jk}$ couplings with identical $j$ and $k$ indices)
cross sections involve $m_{l_i}$
lepton mass terms (see Appendix \ref{formulas}) and are thus negligible.
The $p \bar p \to \tilde \chi^+ l_i^-$ reaction receives contributions
from the $s$ channel sneutrino exchange and
the $t$ and $u$ channels squark exchanges as shown in Fig.\ref{graphes}.
However,
the $t$ and $u$ channels represent small contributions
to the whole single chargino production cross section 
when the sneutrino exchanged in the
$s$ channel is real, namely for $m_{\tilde \nu_{iL}}>m_{\tilde
\chi^{\pm}}$.
The $t$ and $u$ channels cross sections will be relevant only
when the produced sneutrino is virtual since the $s$ channel
contribution is small.
In this situation the single chargino production
rate is greatly reduced compared
to the case where the exchanged sneutrino is produced as a resonance.
Hence, The $t$ and $u$ channels do not represent important contributions
to the $\tilde \chi^+ l_i^-$ production rate.

\begin{figure}[t]
\begin{center}
\leavevmode
\centerline{\psfig{figure=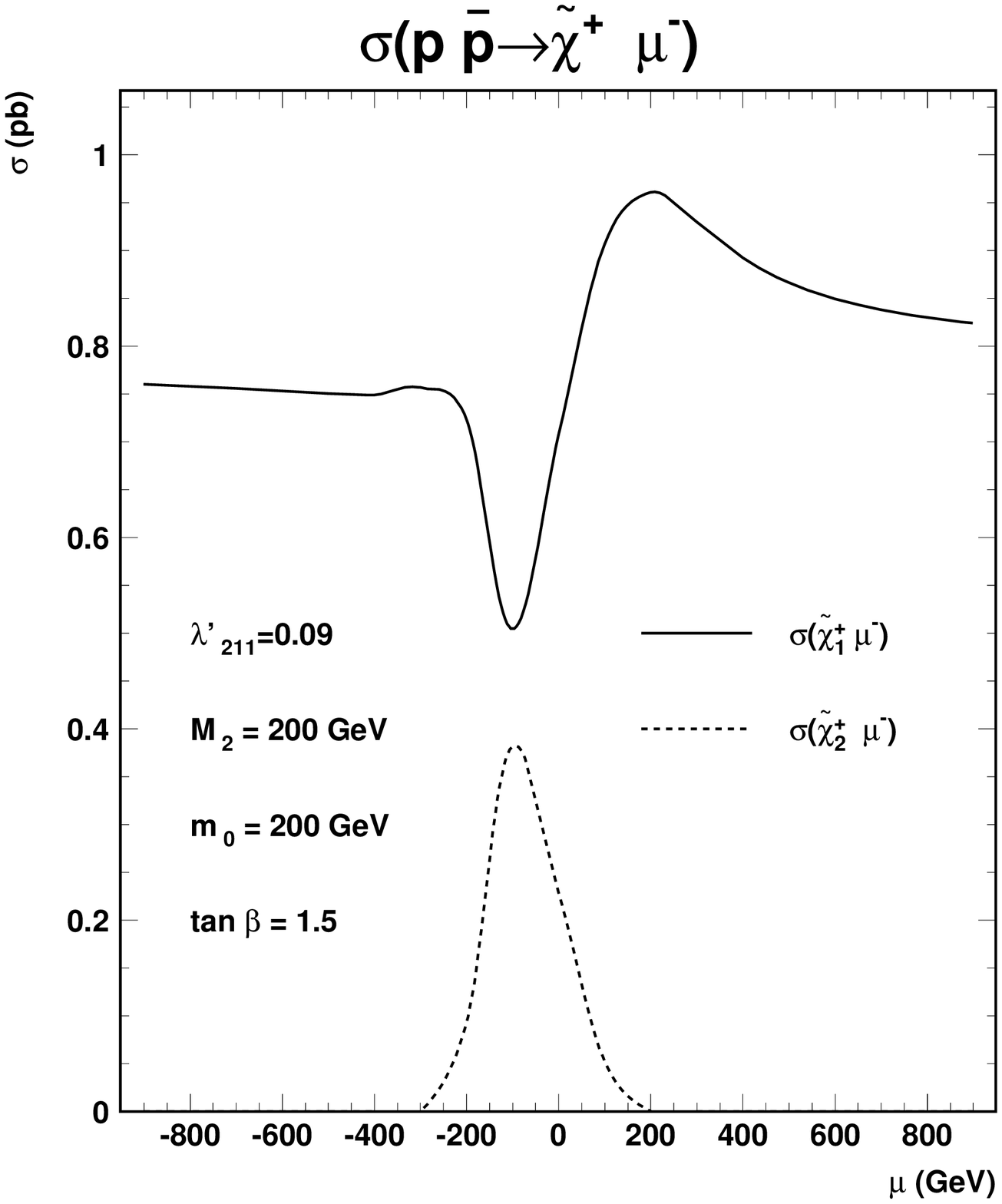,height=5.5in}}
\end{center}
\caption{\footnotesize  \it
Cross sections (in $pb$) of the single chargino productions
$p \bar p \to \tilde \chi^+_{1,2} \mu^-$
as a function of the $\mu$ parameter (in $GeV$)
for $\l'_{211}=0.09$, $M_2=200GeV$, $\tan \beta=1.5$
and $m_0=200GeV$ at a center of mass energy
of $2 TeV$.
\rm \normalsize }
\label{XSmu}
\end{figure}

The dependence of the $\tilde \chi^+ l_i^-$ production rate
on the $A$ coupling is weak. Indeed, the rate depends on the $A$
parameter only through the masses of the third generation squarks
eventually exchanged in the $t$ and $u$ channels
(see Fig.\ref{graphes}). Similarly,
the dependences on the $A$ coupling
of the rates of the other single gaugino productions
shown in Fig.\ref{graphes} are weak.
Therefore, in this article we present the results for
$A=0$. Later, we will discuss the effects
of large $A$ couplings on the cascade decays which
are similar to the effects of large $\tan \beta$ values.

{\bf $\tan \beta$ dependence:}
The dependence of the $\tilde \chi^+ l_i^-$ production rate
on $\tan \beta$ is also weak, except for $\tan \beta<10$.
This can be seen in Fig.\ref{XStan} where the cross section
of the $p \bar p \to \tilde \chi^+_1 \mu^-$ reaction occuring
through the $\l'_{211}$ coupling is shown
as a function of the $\tan \beta$ parameter.
The choice of the $\l'_{211}$ coupling is motivated by the fact
that the analysis in Sections \ref{analysis1} and \ref{analysis2}
are explicitly made for this \rpv coupling. In Fig.\ref{XStan},
we have taken the $\l'_{211}$ value equal to
its low-energy experimental bound
for $m_{\tilde d_R}=100GeV$ which is $\l'_{211}<0.09$ \cite{b:Bhatt}. \\
At this stage, some remarks on the values of the
cross sections presented in this section must be done. First,
the single gaugino production rates must be multiplied
by a factor 2 in order to take into account the charge conjugated
process, which is for example
in the present case $p \bar p \to \tilde \chi^- \mu^+$. Furthermore,
the values of the cross sections for all the single gaugino productions
are obtained using the CTEQ4L structure function \cite{b:CTEQ4}.
Choosing other parametrizations does not change significantly
the results since proton structure functions in our kinematical domain
in Bjorken $x$ are known and have been already measured.
For instance, with the set of parameters $\l'_{211}=0.09$,
$M_2=100GeV$, $\tan \beta=1.5$, $m_0=300GeV$ and $\mu=-500GeV$,
the $\tilde \chi^+_1 \mu^-$ production cross section is
$0.503pb$ for the CTEQ4L structure function \cite{b:CTEQ4},
$0.503pb$ for the BEP structure function \cite{b:BEP},
$0.480pb$ for the MRS (R2) structure function \cite{b:MRS} and
$0.485pb$ for the GRV LO structure function \cite{b:GRV}.

\begin{figure}[t]
\begin{center}
\leavevmode
\centerline{\psfig{figure=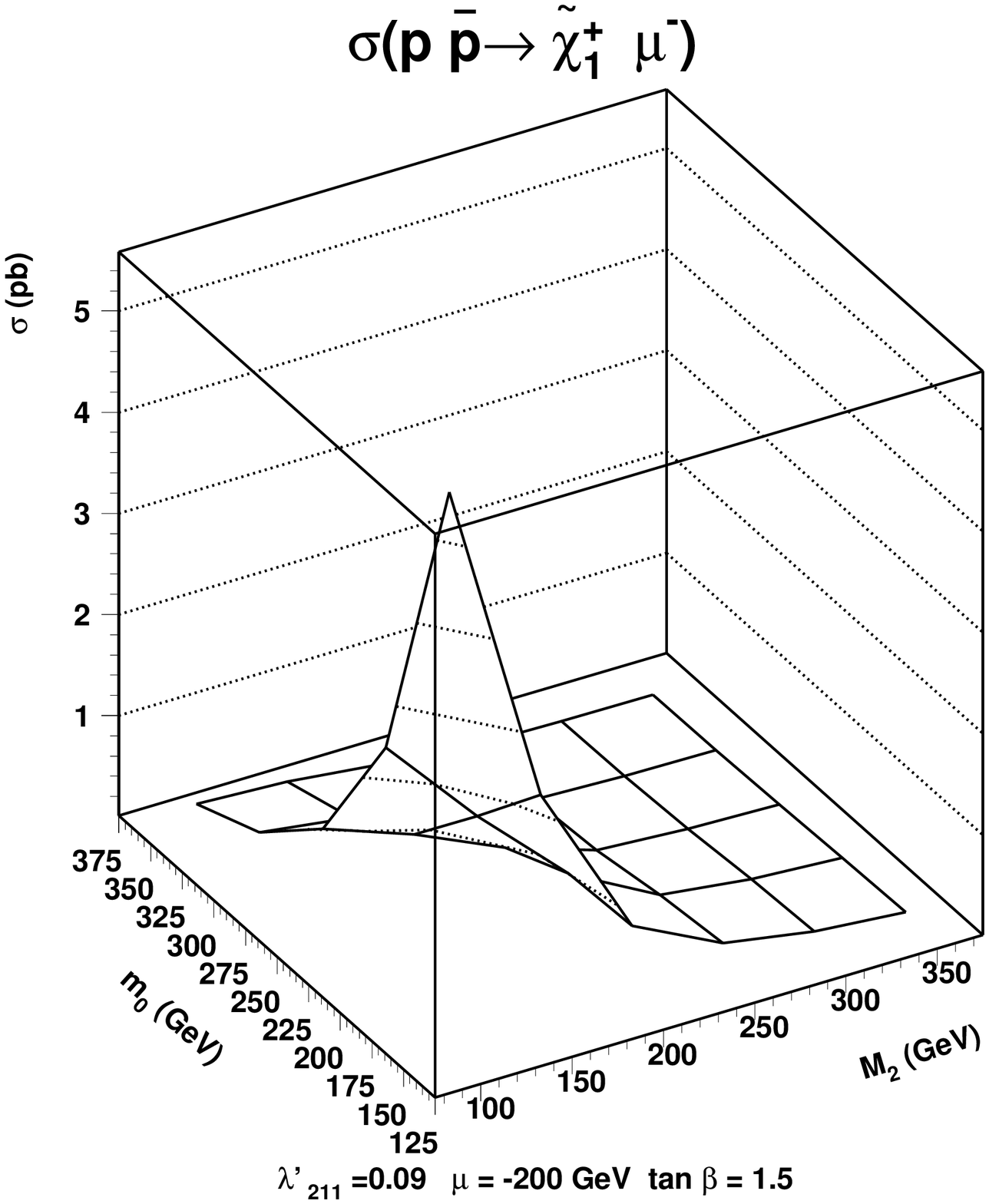,height=5.5in}}
\end{center}
\caption{\footnotesize  \it
Cross section (in $pb$) of the single chargino production
$p \bar p \to \tilde \chi^+_1 \mu^-$
as a function of the $m_0$ (in $GeV$)
and $M_2$ (in $GeV$) parameters.
The center of mass energy is $\sqrt s=2 TeV$
and the other parameters are: $\l'_{211}=0.09$,
$\tan \beta=1.5$ and $\mu=-200GeV$.
\rm \normalsize }
\label{XS02}
\end{figure}

{\bf $\mu$ dependence:}
In Fig.\ref{XSmu},
we present the cross sections of the $\tilde \chi_1^+ \mu^-$ and
$\tilde \chi_2^+ \mu^-$ productions as a function of the $\mu$ parameter.
We observe in this figure the weak dependence of the
cross section $\sigma(p \bar p \to \tilde \chi_1^+ \mu^-)$ on $\mu$
for $| \mu | > M_2$.
The reason is the smooth dependence of the $\tilde \chi_1^{\pm}$
mass
on $\mu$ in this domain.
However, the rate strongly decreases in the region $| \mu | < M_2$
in which the
$\tilde \chi_1^{\pm}$ chargino is mainly composed by the higgsino.
Nevertheless, the small $ \vert \mu \vert$ domain  ($\vert \mu \vert$ smaller than
$\sim 100 GeV$ for $\tan \beta=1.41$, $M_2>100 GeV$, $m_0=500 GeV$ and $\l' \neq 0$)
is excluded by the present experimental limits
derived from the LEP data \cite{b:Mass}. \\
In contrast, the cross section $\sigma(p \bar p \to \tilde \chi_2^+ \mu^-)$
increases in the domain $\vert \mu \vert < M_2$ due to the
fact that the $\tilde \chi_2^{\pm}$ mass is enhanced as $\vert \mu \vert$ increases
and the $\tilde \chi_2^{\pm}$ is primarily wino in the region
$\vert \mu \vert < M_2$.
The region in which $\sigma(p \bar p \to \tilde \chi_2^+ \mu^-)$
becomes important is at small values of $\vert \mu \vert$,
near the LEP limits of \cite{b:Mass}.
We also remark in Fig.\ref{XSmu} that the single $\tilde \chi_1^+$
production rate values remain
above the single $\tilde \chi_2^+$ production rate values in all
the considered range of $\mu$. 
In this figure, we also notice that the cross section is smaller
when $\mu$ is negative. To be conservative, we will take $\mu <0$
in the following.

{\bf $m_0$ and $M_2$ dependences:}
In fact, the cross section
$\sigma(p \bar p \to \tilde \chi^+ l_i^-)$ depends mainly
on the $m_0$ and $M_2$ parameters.
We present in Fig.\ref{XS02} the rate
of the $\tilde \chi^+_1 \mu^-$ production
as a function of the $m_0$ and $M_2$ parameters.
The rate decreases at high values of $m_0$
since the sneutrino becomes heavier as $m_0$ increases 
and more energetic initial partons are
required in order to produce the resonant sneutrino.
The decrease of the rate at large values of $M_2$ is due to
the increase of the chargino mass and thus
the reduction of the phase space factor.

\begin{figure}[t]
\begin{center}
\leavevmode
\centerline{\psfig{figure=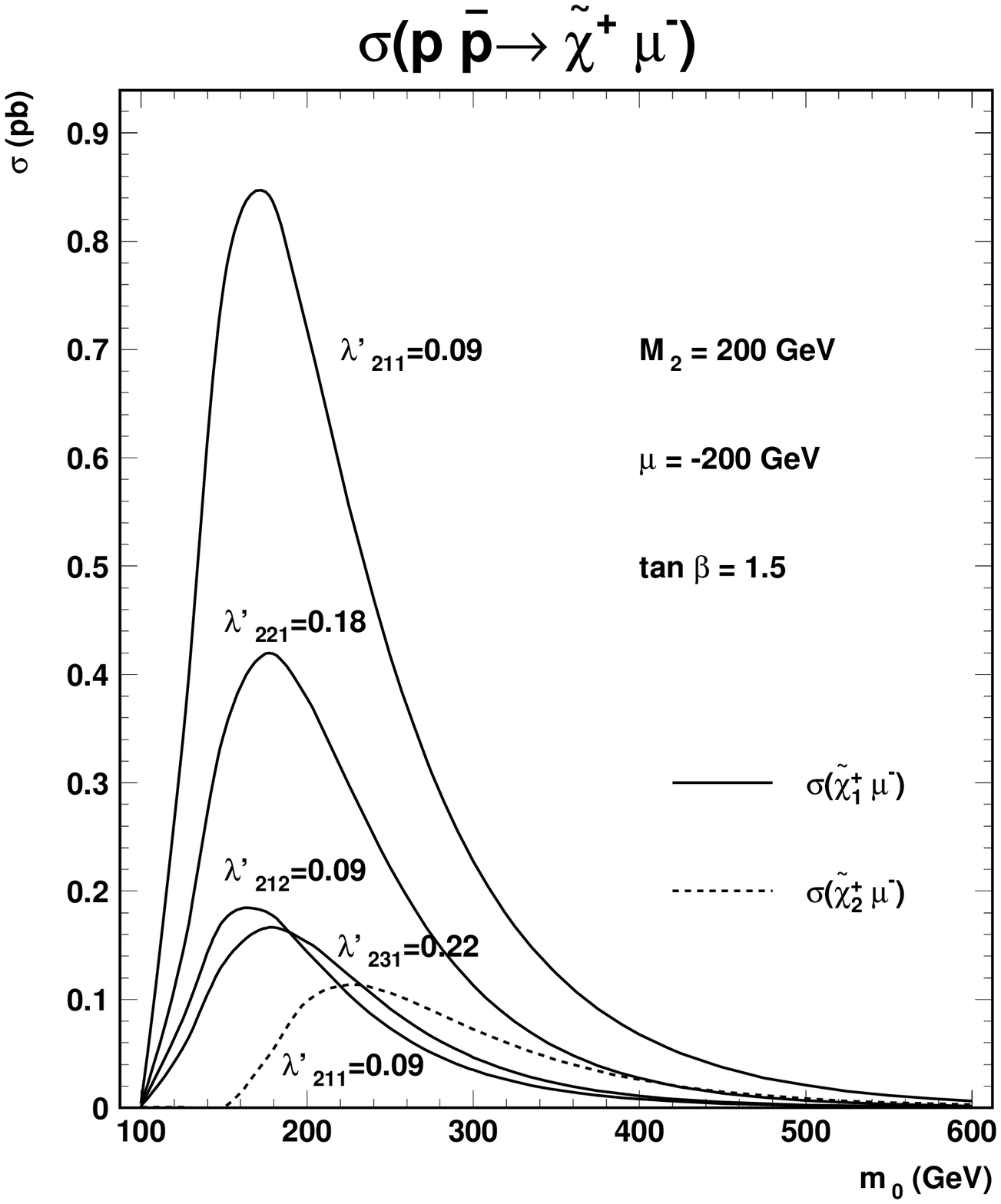,height=5.5in}}
\end{center}
\caption{\footnotesize  \it
Cross sections (in $pb$) of the single chargino productions
$p \bar p \to \tilde \chi^+_{1,2} \mu^-$
as a function of the $m_0$ parameter (in $GeV$).
The center of mass energy is taken at $\sqrt s=2 TeV$
and $\l'_{211}=0.09$,
$M_2=200GeV$, $\tan \beta=1.5$ and $\mu=-200GeV$.
The rates of the single $\tilde \chi^+_1$ production via the
\rpv couplings $\l'_{212}=0.09$, $\l'_{221}=0.18$ and $\l'_{231}=0.22$
are also shown. The chosen values of the \rpv couplings correspond
to the low-energy limits \cite{b:Bhatt} for a squark mass of $100GeV$.
\rm \normalsize }
\label{XScl}
\end{figure}

In Fig.\ref{XScl}, we show the variations of the
$\sigma(p \bar p \to \tilde \chi^+_1 \mu^-)$ cross sections with $m_0$
for fixed values of $M_2$, $\mu$ and $\tan \beta$.
The cross sections corresponding to the $\tilde \chi^+_1 \mu^-$
production through various \rpv couplings of type $\l'_{2jk}$ are
presented. In this figure, we only consider the \rpv couplings giving 
the highest cross sections.
The values of the considered $\l'_{2jk}$ couplings have been taken
at their low-energy limit \cite{b:Bhatt} for a squark
mass of $100GeV$. The rate of the $\tilde \chi^+_2 \mu^-$ production
through $\l'_{211}$ is also shown in this figure.
We already notice that the cross section is significant for many \rpv
couplings and we will come back on this important statement in the following.\\
The $\sigma(p \bar p \to \tilde \chi^+ \mu^-)$
rates decrease as $m_0$ increases for the same reason as in
Fig.\ref{XS02}.
A decrease of the rates also occurs at small values of $m_0$.
The reason is the following.
When $m_0$ decreases, the $\tilde \nu$ mass is getting closer
to the $\tilde \chi^{\pm}$ masses so that the phase space
factor associated to the decay
$\tilde \nu_{\mu} \to \tilde \chi^{\pm} \mu^{\mp}$ decreases.\\
We also observe that the single $\tilde \chi_2^+$
production rate is much smaller than the single
$\tilde \chi_1^+$ production rate, as in Fig.\ref{XSmu}.\\
Since the single chargino production rate scales as $\l'^2$ (see
Appendix \ref{formulas}), we easily see by doing a rescaling of 
the rates that the various $\tilde \chi^+_1 \mu^-$ 
production rates presented 
in Fig.\ref{XScl} would still have different values
for identical values of the involved \rpv coupling constants.
These differences between the $\tilde \chi^+_1 \mu^-$
production rates occuring via the various $\l'_{2jk}$ couplings
are explained by the different parton densities.
Indeed, as shown in Fig.\ref{graphes}
the hard processes associated to the $\tilde \chi^+_1 \mu^-$
production occuring through the
$\l'_{2jk}$ coupling constant
have a partonic initial state $ \bar q_j q_k$.
The influence of the parton density on the single chargino production 
rate can be observed on Fig.\ref{XScl} by comparing the 
$\tilde \chi^+_1 \mu^-$ production rates occuring through the 
$\l'_{211}=0.09$ and $\l'_{212}=0.09$ coupling constants.
For same values of the $\l'_{2jk}$ coupling constants,
the $\tilde \chi^+_1 \mu^-$ production involving the $\l'_{211}$ 
coupling constant has the highest cross section since the associated
hard processes have first generation quarks in the initial 
state which provide the maximum parton density.

\begin{figure}[t]
\begin{center}
\leavevmode
\centerline{\psfig{figure=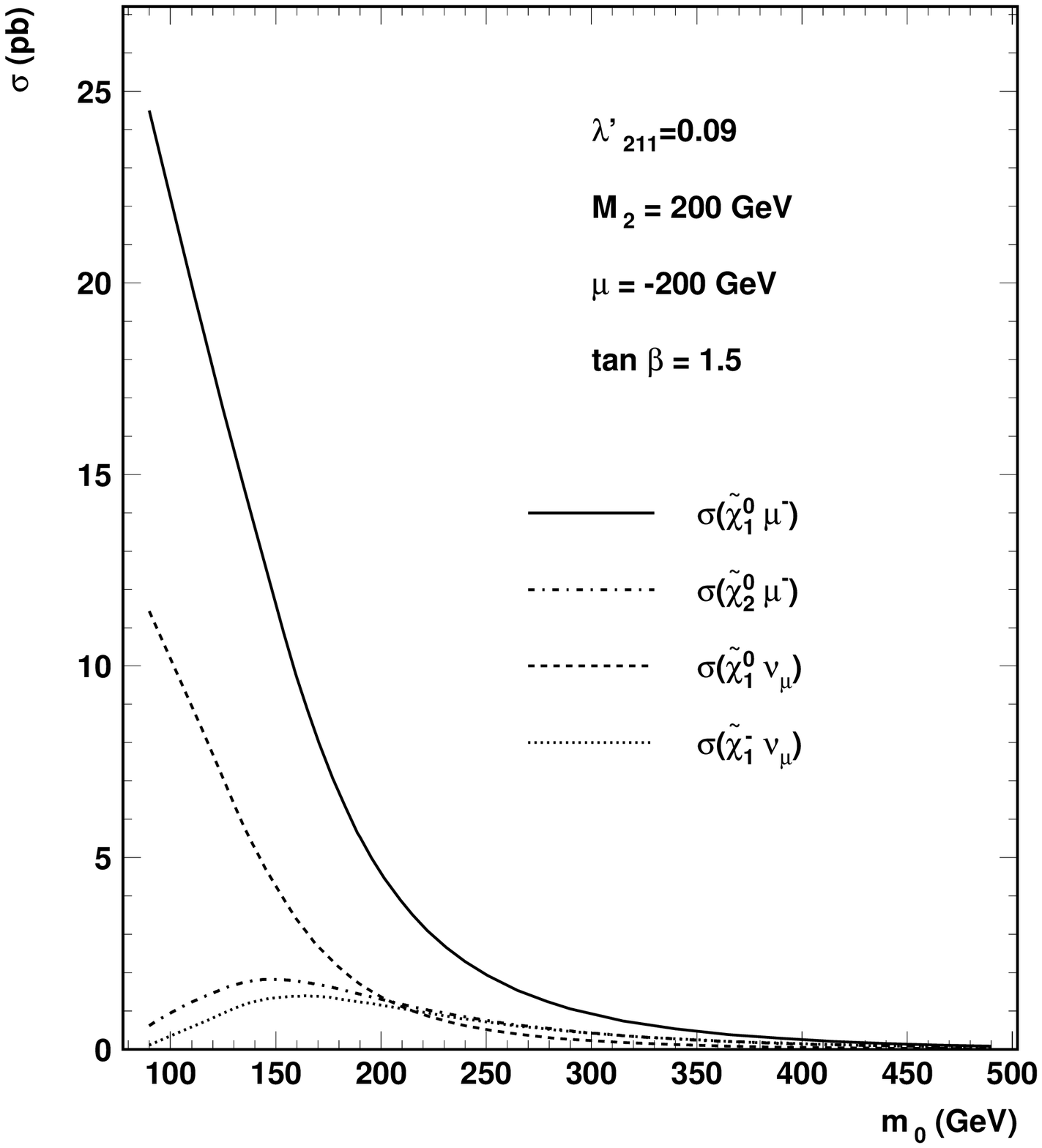,height=5.5in}}
\end{center}
\caption{\footnotesize  \it
Cross sections (in $pb$) of the reactions
$p \bar p \to \tilde \chi^-_1 \nu$,
$p \bar p \to \tilde \chi^0_{1,2} \mu^-$
and $p \bar p \to \tilde \chi^0_1 \nu$
as a function of the $m_0$ parameter (in $GeV$).
The center of mass energy is taken at $\sqrt s=2 TeV$
and the considered set of parameters is: $\l'_{211}=0.09$,
$M_2=200GeV$, $\tan \beta=1.5$ and $\mu=-200GeV$.
\rm \normalsize }
\label{allXS}
\end{figure}

We now discuss the rate behaviours for the reactions
$p \bar p \to \tilde \chi^- \nu_{\mu}$, $p \bar p \to \tilde \chi^0 \mu^-$
and $p \bar p \to \tilde \chi^0 \nu_{\mu}$ which
occur via $\l'_{211}$, in the SUSY parameter space.
The dependences of these rates on the $A$, $\tan \beta$, $\mu$ and $M_2$
parameters are typically the same as for the
$\tilde \chi^+ \mu^-$ production rate.
The variations of the
$\tilde \chi^-_1 \nu_{\mu}$, $\tilde \chi^0_{1,2} \mu^-$
and $\tilde \chi^0_1 \nu_{\mu}$ productions cross sections
with the $m_0$ parameter are shown in Fig.\ref{allXS}.
The $\tilde \chi^-_2 \nu_{\mu}$, $\tilde \chi^0_{3,4} \mu^-$ and
$\tilde \chi^0_{3,4} \nu_{\mu}$ production rates are comparatively
negligible
and thus have not been represented.
We observe in this figure that the cross sections decrease
at large $m_0$ values like the $\tilde \chi^+ \mu^-$ production rate.
However, while
the single $\tilde \chi^{\pm}_1$ productions rates decrease at
small $m_0$ values
(see Fig.\ref{XScl} and Fig.\ref{allXS}),
this is not true for the single $\tilde \chi^0_1$ productions
(see Fig.\ref{allXS}). The reason is that in mSUGRA
the $\tilde \chi^0_1$ and $\tilde l_{iL}$
($l_i=l_i^{\pm},\nu_i$) masses are never close enough
to induce a significant decrease of the
cross section associated to the reaction
$p \bar p \to \tilde l_{iL} \to \tilde \chi^0_1 l_i$,
where $l_i=l_i^{\pm},\nu_i$ (see Fig.\ref{graphes}(c)(d)),
caused by a phase space factor reduction. Therefore,
the resonant slepton contribution to the single $\tilde \chi^0_1$ production
is not reduced at small $m_0$ values like the
resonant slepton contribution to the single $\tilde \chi^{\pm}_1$
production.
For the same reason,
the single $\tilde \chi^0_1$ productions have much higher cross sections
than the single $\tilde \chi^{\pm}_1$ productions in most
of the mSUGRA parameter space, as illustrate Fig.\ref{XScl} and
Fig.\ref{allXS}.
We note that in the particular case of a single dominant $\l'_{3jk}$
coupling
constant and of large $\tan \beta$ values, the rate of the reaction
$p \bar p \to \tilde \tau^{\pm}_1 \to \tilde \chi^0_1 \tau^{\pm}$
(see Fig.\ref{graphes}(d)), where $\tilde \tau^{\pm}_1$ is the lightest
tau-slepton,
can be reduced at low $m_0$ values since
then $m_{\tilde \tau^{\pm}_1}$ can be closed to $m_{\tilde \chi^0_1}$ due to
the large
mixing occuring in the staus sector.
By analysing Fig.\ref{XScl} and Fig.\ref{allXS},
we also remark that the $\tilde \chi^- \nu_{\mu}$
($\tilde \chi^0 \mu^-$) production rate is larger than the
$\tilde \chi^+ \mu^-$ ($\tilde \chi^0 \nu_{\mu}$) one.
The explanation is that in $p \bar p$ collisions
the initial states of the resonant charged slepton production
$u_j \bar d_k, \bar u_j d_k$ have higher partonic densities than the
initial states of the resonant sneutrino production
$d_j \bar d_k, \bar d_j d_k$. 
This phenomenon also increases the difference between the rates of the
$\tilde \chi^0_1 \mu^-$ and $\tilde \chi^+_1 \mu^-$ productions
at Tevatron.
\\ Although the single $\tilde \chi^{\pm}_1$ production cross sections
are smaller than the $\tilde \chi^0_1$ ones, it is interesting to
study both of them since they have quite high values.

\subsection{Non-resonant superpartner production}
\label{non-resonant}

At hadronic colliders, the single productions of SUSY particle
via $\l'_{ijk}$ can occur through some $2 \to 2-body$
processes which do not receive contributions
from any resonant superpartner production.
These non-resonant superpartner productions are
(one must also add the charge conjugated processes):

\begin{itemize}
\item The gluino production $\bar u_j d_k \to \tilde g l_i$ via the exchange
of a $\tilde u_{jL}$ ($\tilde d_{kR}$) squark in the $t$ ($u$) channel.
\item The squark production $\bar d_j g \to \tilde d_{kR}^* \nu_i$ via the
exchange
of a $\tilde d_{kR}$ squark ($d_j$ quark) in the $t$ ($s$) channel.
\item The squark production $\bar u_j g \to \tilde d_{kR}^* l_i$ via the
exchange
of a $\tilde d_{kR}$ squark ($u_j$ quark) in the $t$ ($s$) channel.
\item The squark production $d_k g \to \tilde d_{jL} \nu_i$ via the exchange
of a $\tilde d_{jL}$ squark ($d_k$ quark) in the $t$ ($s$) channel.
\item The squark production $d_k g \to \tilde u_{jL} l_i$ via the exchange
of a $\tilde u_{jL}$ squark ($d_k$ quark) in the $t$ ($s$) channel.
\item The sneutrino production $\bar d_j d_k \to Z \tilde \nu_{iL}$ via the
exchange
of a $d_k$ or $d_j$ quark ($\tilde \nu_{iL}$ sneutrino) in the $t$ ($s$)
channel.
\item The charged slepton production $\bar u_j d_k \to Z \tilde l_{iL}$ via
the exchange
of a $d_k$ or $u_j$ quark ($\tilde l_{iL}$ slepton) in the $t$ ($s$)
channel.
\item The sneutrino production $\bar u_j d_k \to W^- \tilde \nu_{iL}$
via the exchange of a $d_j$ quark ($\tilde l_{iL}$ sneutrino) in the $t$
($s$) channel.
\item The charged slepton production $\bar d_j d_k \to W^+ \tilde l_{iL}$
via the exchange of a $u_j$ quark ($\tilde \nu_{iL}$ sneutrino) in the $t$
($s$) channel.
\end{itemize}

The single gluino production cannot reach high cross sections
due to the strong experimental limits on the squarks and gluinos
masses which are typically about $m_{\tilde q},m_{\tilde g}
\stackrel{>}{\sim}200GeV$ \cite{b:PDG98}. Indeed, the single gluino
production occurs through the exchange of squarks in the $t$
and $u$ channels, as described above, so that the cross section
of this production decreases as the squarks and gluinos
masses increase. For the value
$m_{\tilde q}=m_{\tilde g}=250GeV$ which is close to the
experimental limits, we find the
single gluino production rate
$\sigma(p \bar p \to \tilde g \mu) \approx 10^{-2}pb$
which is consistent with the results of \cite{b:Dim2}.
The cross sections given in this section are computed at a center of mass
energy of
$\sqrt s=2TeV$ using the version 33.18 of the COMPHEP routine \cite{b:COMPHEP}
with the CTEQ4m structure function and an \rpv coupling
$\l'_{211}=0.09$.
Similarly, the single squark production cross section
cannot be large: for $m_{\tilde q}=250GeV$,
the rate $\sigma(p \bar p \to \tilde u_L \mu)$
is of order $\sim 10^{-3}pb$.
The production of a slepton together
with a massive gauge boson has a small phase space factor
and does not involve strong interaction couplings.
The cross section of this type of reaction is thus small.
For instance, with a slepton mass of $m_{\tilde
l}=100GeV$ we find the cross section
$\sigma(p \bar p \to Z \tilde \mu_L)$ to be of order $10^{-2}pb$.

As a conclusion, the non-resonant single superpartner productions
have small rates and will not be considered here.
Nevertheless, some of 
these reactions are interesting as their cross section
involves few SUSY parameters, namely only one scalar
superpartner mass and one \rpv coupling constant.

\section{Three lepton signature analysis}
\label{analysis1}

\setcounter{equation}{0}

\subsection{Signal}
\label{signal1}

In this section, we study the three lepton signature
at Tevatron Run II
generated by the single chargino production through
$\l'_{ijk}$,
$p \bar p \to \tilde \chi^{\pm} l^{\mp}_i$, followed by
the cascade decay,
$\tilde \chi^{\pm} \to \tilde \chi^0_1 l^{\pm} \nu$,
$\tilde \chi^0_1 \to l_i u_j \bar d_k, \ \bar l_i \bar u_j d_k$
(the indices $i,j,k$ correspond to the indices of $\l'_{ijk}$).
In fact, the whole final state is
3 charged leptons + 2 hard jets + missing energy ($\Eslash$).
The two jets and the missing energy come respectively from
the quarks and the neutrino produced in the cascade decay.
In the mSUGRA model, which predicts the $\tilde \chi^0_1$ as the LSP
in most of the parameter space,
the $p \bar p \to \tilde \chi^{\pm} l_i^{\mp}$ reaction
is the only single gaugino production allowing the three lepton signature
to be generated in a significant way.
Since the $\tilde \chi^{\pm}_1 l^{\mp}_i$
production rate is dominant
compared to the $\tilde \chi^{\pm}_2 l^{\mp}_i$ production rate,
as discussed in Section \ref{cross1}, we only consider
the contribution to the three lepton signature
from the single lightest chargino production.

For $m_{\tilde \nu},m_{\tilde l},m_{\tilde q},
m_{\tilde \chi^0_2}>m_{\tilde \chi^{\pm}_1}$,
the branching ratio $B(\tilde \chi^{\pm}_1 \to \tilde \chi^0_1 l^{\pm} \nu)$
is typically of order $30\%$
and is smaller than for the other possible decay
$\tilde \chi^{\pm}_1 \to \tilde \chi^0_1 \bar q_p q'_p$
because of the color factor.\\
Since in our framework the $\tilde \chi^0_1$ is the LSP,
it can only decay via $\l'_{ijk}$, either as
$\tilde \chi^0_1 \to l_i u_j d_k$ or as $\tilde \chi^0_1 \to \nu_i d_j d_k$,
with a branching ratio $B(\tilde \chi^0_1 \to l_i u_j d_k)$ ranging between
$\sim 40\%$ and $\sim 70\%$.

The three lepton signature
is particularly attractive at hadronic colliders because of
the possibility to reduce the associated \sm background.
In Section \ref{back1} we describe this \sm background
and in Section \ref{cut1} we show how it can be reduced.

\subsection{Standard Model
background of the 3 lepton signature at Tevatron}
\label{back1}

The first source of \sm background for the three leptons final state is
the top quark pair production
$q \bar q \to t \bar t$ or $g g \to t \bar t$. Since the top quark life time
is smaller than its hadronisation time, the top decays and its main channel
is the decay into a $W$ gauge boson and a bottom quark as $t \to b W$.
The $t \bar t$ production can thus give rise to a $3l$
final state if the $W$ bosons and one of the b-quarks undergo leptonic
decays simultaneously. The cross section,
calculated at leading order with PYTHIA \cite{b:PYTHIA}
using the CTEQ2L structure function, times the branching fraction is
$\sigma (p \bar p \to t \bar t)
\times B^2(W \to l_p \nu_p) \approx  863fb$ ($704fb$) with $p=1,2,3$
at $\sqrt s= 2 TeV$ for a top quark mass of $m_{top}=170 GeV$ ($175 GeV$).

The other major source of \sm background is the
$W^{\pm} Z^0$ production followed
by the leptonic decays of the gauge bosons, namely $W \to l \nu$ and $Z \to
l \bar l$. The value for
the cross section times the branching ratios is $\sigma (p \bar p \to W Z)
\times B(W \to l_p \nu_p) \times B(Z \to l_p \bar l_p) \approx  82fb$
($p=1,2,3$) at leading
order with a center of mass energy of $\sqrt s= 2 TeV$.
\\ The $W^{\pm} Z^0$ production gives also a small contribution to the 3
leptons
background through the decays: $W \to b u_p$ and $Z \to b \bar b$,
$W \to l \nu$ and $Z \to b \bar b$ or
$W \to b u_p$ and $Z \to l \bar l$, if a lepton is produced in each of the b
jets.

Similarly, the $Z^0 Z^0$ production followed by the decays $Z \to l \bar
l$ ($l=e,\mu$), $Z \to \tau \bar \tau$, where one of the $\tau$ decays into
lepton while
the other decays into jet, leads to three leptons in the final state.
Within the same framework as above,
the cross section is of order $\sigma (p \bar p \to Z Z \to 3 l)
\approx  2fb$.
\\ The $Z^0 Z^0$ production can also contribute weakly to the 3 leptons
background via the decays: $Z \to l \bar l$ and $Z \to b \bar b$
or $Z \to b \bar b$ and $Z \to b \bar b$,
since a lepton can be produced in a b jet.

It has been pointed out recently that the $W Z^*$ (throughout this
paper a star indicates a virtual particle) and the $W \gamma^*$ productions
could represent important contributions to the trilepton background
\cite{b:Lyk,b:Match}.
The complete list of contributions to the 3 leptons final state
from the $WZ$,$W \gamma^*$ and $ZZ$ productions, including cases where
either one or both of the gauge bosons can be virtual, has been calculated
in
\cite{b:Pai}. The authors of \cite{b:Pai} have found that the $WZ$, $W \gamma^*$
and $ZZ$
backgrounds (including virtual boson(s)) at the upgraded Tevatron
have together a cross section
of order $0.5 fb$ after the following cuts have been implemented:
$P_t(l_1)>20GeV$, $P_t(l_2)>15GeV$, $P_t(l_3)>10GeV$;
$\vert \eta (l_1,l_{2,3}) \vert <1.0,2.0$; $ISO_{\delta R=0.4}<2GeV$;
$\Eslash_T >25GeV$;
$81GeV<M_{inv}(l \bar  l)<101GeV$; $12GeV<M_{inv}(l \bar l)$;
$60GeV<m_T(l,\Eslash_T)<85GeV$.
\\ We note that there is at most one hard jet
in the 3 leptons backgrounds generated by the
$WZ$, $W \gamma*$ and $ZZ$ productions (including virtual boson(s)).
Since the number of hard jets is equal to 2 in our signal
(see Section \ref{signal1}), a jet veto can thus
reduce this \sm background with respect to the signal.

Other small sources of \sm background have been estimated in \cite{b:Barb}:
The productions like $Zb$, $Wt$ or $W t \bar t$.
After applying cuts
on the geometrical acceptance, the transverse momentum and the
isolation,
these backgrounds are expected to be at most of order $10^{-4} pb$
in $p \bar p$ collisions with a center of mass energy of $\sqrt s=2 TeV$.
We have checked that the $Zb$ production gives a
negligible contribution to the 3 lepton signature.

There are finally some non-physics sources of background.
First, the 4 leptons signal, which can be generated by the $Z^0 Z^0$
and $t \bar t$ productions, appears as a 3 leptons
signature if one of the leptons is missed.
Besides, the
processes $p \bar p \to \ Z \ + \ X, \ Drell-Yan \ +  \ X$
would mimic a trilepton signal
if $X$ fakes a lepton. Monte Carlo simulations using simplified detector
simulation, like for example SHW \cite{b:SHW}
as in the present study (see Section \ref{cut1}),
cannot give a reliable estimate of this
background.
A knowledge of the details of the detector response as well as the jet
fragmentation
is necessary in order to determinate the probability to fake a lepton.
In \cite{b:Kam}, using standard cuts the background coming from
$p \bar p \to \ Z \ + \ X, \ Drell-Yan \ +  \ X$ has been estimated to be of
order
$2 fb$ at Tevatron with $\sqrt s=2 TeV$. The authors of \cite{b:Kam}
have also estimated the background from the three-jet events faking
trilepton signals
to be around $10^{-3} fb$.

Hence for the study of the \sm background associated to
the 3 lepton signature at Tevatron Run II,
we consider the $W^{\pm} Z^0$ production and both the physics and non-physics
contributions generated by the $Z^0 Z^0$ and $t \bar t$ productions.

\subsection{Supersymmetric background
of the 3 lepton signature at Tevatron}
\label{susyback1}

If an excess of events is
observed in the three lepton channel at Tevatron,
one would wonder what is the origin of those anomalous events.
One would thus have to consider all of the \susyq productions leading to
the three lepton signature.
In the present context of R-parity violation, multileptonic
final states can be generated by the single chargino production
involving \rpv couplings, but
also by the \susyq particle pair production which involves
only gauge couplings \cite{b:Atlas,b:RunII}.
In \rpv models, the superpartner
pair production can even
lead to the trilepton signature \cite{b:Barg,b:tat1,b:tat2}.
As a matter of fact, both of the produced \susyq particles
decay, either directly or through cascade decays, into the LSP
which is the neutralino in our framework. In the hypothesis of a
dominant
$\l'$ coupling constant, each of the 2 produced neutralinos can decay into
a charged lepton and two quarks: at least 
two charged leptons and four jets in the final state are produced.
The third charged lepton can be generated in the cascade decays
as for example at the level of
the chargino decay $\tilde \chi^{\pm} \to \tilde \chi^0 l^{\pm} \nu$.

\begin{table}[t]
\begin{center}
\begin{tabular}{|c|c|c|c|c|c|}
\hline
$m_{1/2} \ \backslash \ m_0$
&                 $100GeV$ & $200GeV$   & $300GeV$  & $400GeV$   & $500GeV$
\\
\hline
$100GeV$ & $6.359$         &   $3.846$        &     $3.369$     &    $3.567$      &   $3.849$         \\
\hline
$200GeV$ & $0.179$         &   $0.149$        &     $0.151$     &    $0.160$      &   $0.170$         \\
\hline
$300GeV$ & $2.2 \ 10^{-2}$ &  $1.6 \ 10^{-2}$ & $1.5 \ 10^{-2}$ & $1.5 \ 10^{-2}$  &  $1.6 \ 10^{-2}$  \\
\hline
\end{tabular}
\caption{Cross section (in $pb$) of the sum of all the
superpartners pair productions at Tevatron Run II
as a function of the $m_0$ and $m_{1/2}$ parameters
for $\tan \beta=1.5$, $sign(\mu)<0$ and $\l'_{211}=0.05$
at a center of mass energy of $\sqrt s=2 TeV$.
These rates have been calculated with HERWIG \cite{b:HERWIG} using
the CTEQ4m structure function.}
\label{xosuper}
\end{center}
\end{table}

In Table \ref{xosuper}, we show
for different mSUGRA points the cross section of
the sum of all superpartner pair productions,
namely the $R_p$ conserving SUSY background
of the 3 lepton signature generated by the
single chargino production. As can be seen in this table, the
summed superpartner pair production rate
decreases as $m_0$ and $m_{1/2}$ increase.
This is due to the increase of the superpartner masses as
the $m_0$ or $m_{1/2}$ parameter increases.
The SUSY background will be important only for low values 
of $m_0$ and $m_{1/2}$ as we will see in the following.

\subsection{Cuts}
\label{cut1}

In order to simulate the single chargino production
$p \bar p \to \tilde \chi^{\pm}_1 l^{\mp}$ at Tevatron,
the matrix elements (see Appendix \ref{formulas}) of this process
have been implemented
in a version of the SUSYGEN event generator \cite{b:SUSYGEN3}
allowing the generation of $p \bar p$ reactions \cite{b:priv}.
The \sm background
($W^{\pm} Z^0$, $Z^0 Z^0$ and $t \bar t$ productions)
has been simulated
using the PYTHIA event generator \cite{b:PYTHIA}
and the SUSY background (all SUSY particles pair productions) using the
HERWIG event generator \cite{b:HERWIG}.
SUSYGEN, PYTHIA and HERWIG have been interfaced with
the SHW detector simulation package \cite{b:SHW},
which mimics an average of the CDF and D0 Run II detector performance.

We have developped a series of cuts
in order to enhance the signal-to-background ratio.\\
First, we have selected the events with at least three leptons where
the leptons are either an electron, a muon or a
tau reconstructed from a jet, namely $N_l \geq 3 \ [l=e,\mu,\tau]$.
We have also considered the case where the selected leptons
are only electrons and muons, namely $N_l \geq 3 \ [l=e,\mu]$.

The selection criteria on the jets
was to have a number of jets greater or equal to two, where the jets have a
transverse momentum
higher than $10 GeV$, namely $N_j \geq 2$ with $P_t(j) > 10 GeV$. This jet
veto
reduces the 3 lepton backgrounds coming from the $W^{\pm} Z^0$ and $Z^0
Z^0$
productions. Indeed, the
$W^{\pm} Z^0$ production generates no hard jets and the $Z^0 Z^0$
production generates at most one hard jet. Moreover,
the hard jet produced in the $Z^0 Z^0$ background is generated by
a tau decay (see Section \ref{back1}) and can thus be identified as a tau.

\begin{figure}[t]
\begin{center}
\leavevmode
\centerline{\psfig{figure=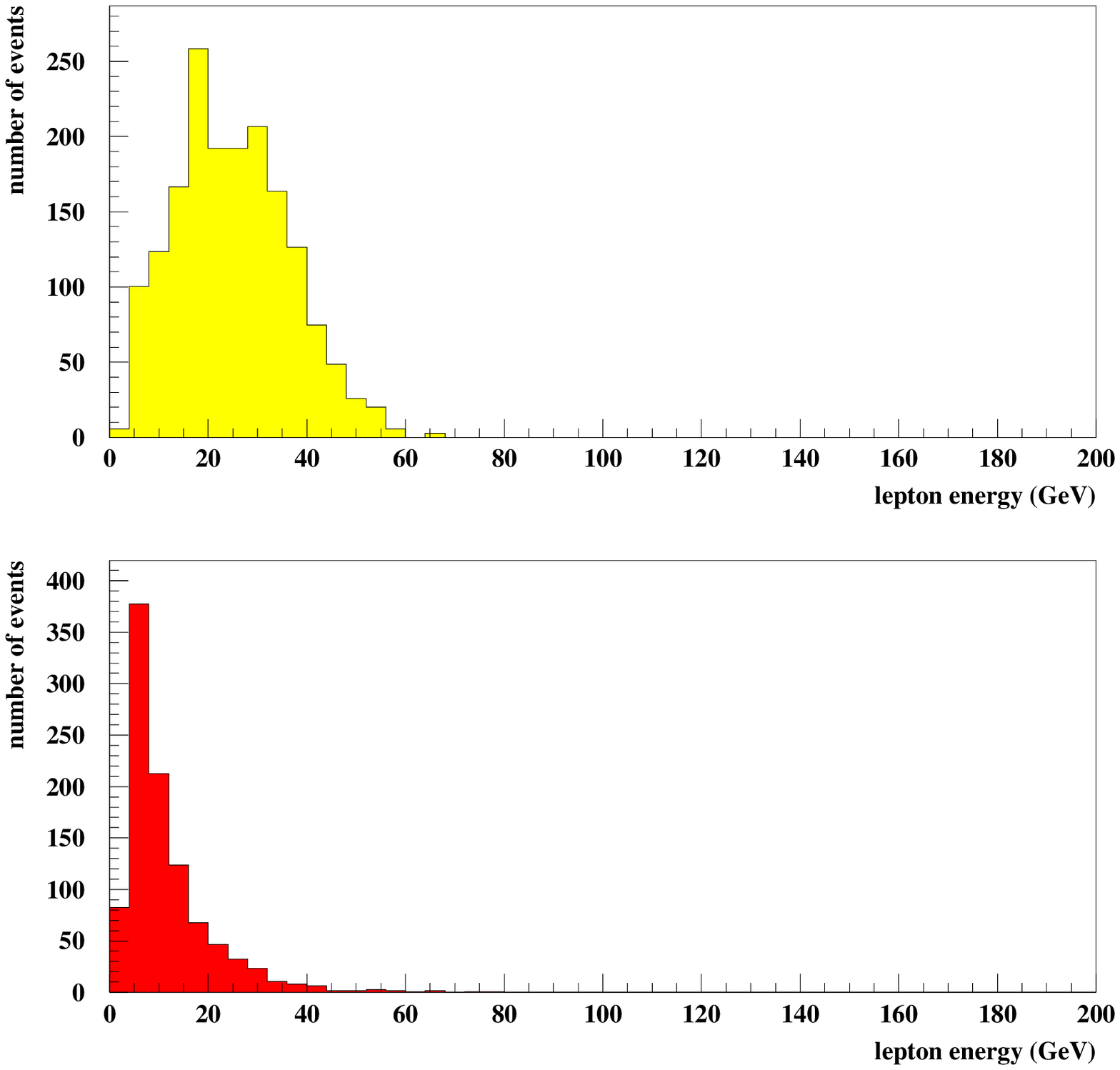,height=5.5in}}
\end{center}
\caption{\footnotesize  \it
Distributions of the lowest lepton energy (in $GeV$) among the energies
of the 3 leading leptons (electrons and muons)
in the events containing at least 3 charged leptons and 2 jets
generated by the \sm background (lower curve),
namely the
$W^{\pm} Z^0$, $Z^0 Z^0$ and $t \bar t$ productions,
and the SUSY signal (upper curve), for $\l'_{211}=0.09$,
$M_2=150GeV$, $m_0=200GeV$, $\tan \beta=1.5$
and $sign(\mu)<0$.
The numbers of events correspond to an integrated luminosity of
${\cal L}=10fb^{-1}$.
\rm \normalsize }
\label{distener}
\end{figure}

Besides, some effective cuts concerning the energies of the produced leptons
have been applied. In Fig.\ref{distener},
we show the distributions of the third leading lepton energy
in the 3 lepton events produced by the \sm background
($W^{\pm} Z^0$, $Z^0 Z^0$ and $t \bar t$) and the SUSY signal.
Based on those kinds of distributions, we have chosen the following cut
on the third leading lepton energy: $E(l_3)>10GeV$.
Similarly, we have required that the energies of the 2 leading leptons
verify $E(l_2)>20GeV$ and $E(l_1)>20GeV$.

We will refer to all the selection criteria described above, namely
$N_l \geq 3 \ [l=e,\mu,\tau]$ with $E(l_1)>20GeV$, $E(l_2)>20GeV$,
$E(l_3)>10GeV$,
and $N_j \geq 2$ with $P_t(j) > 10 GeV$,
as cut $1$.

\begin{figure}[t]
\begin{center}
\leavevmode
\centerline{\psfig{figure=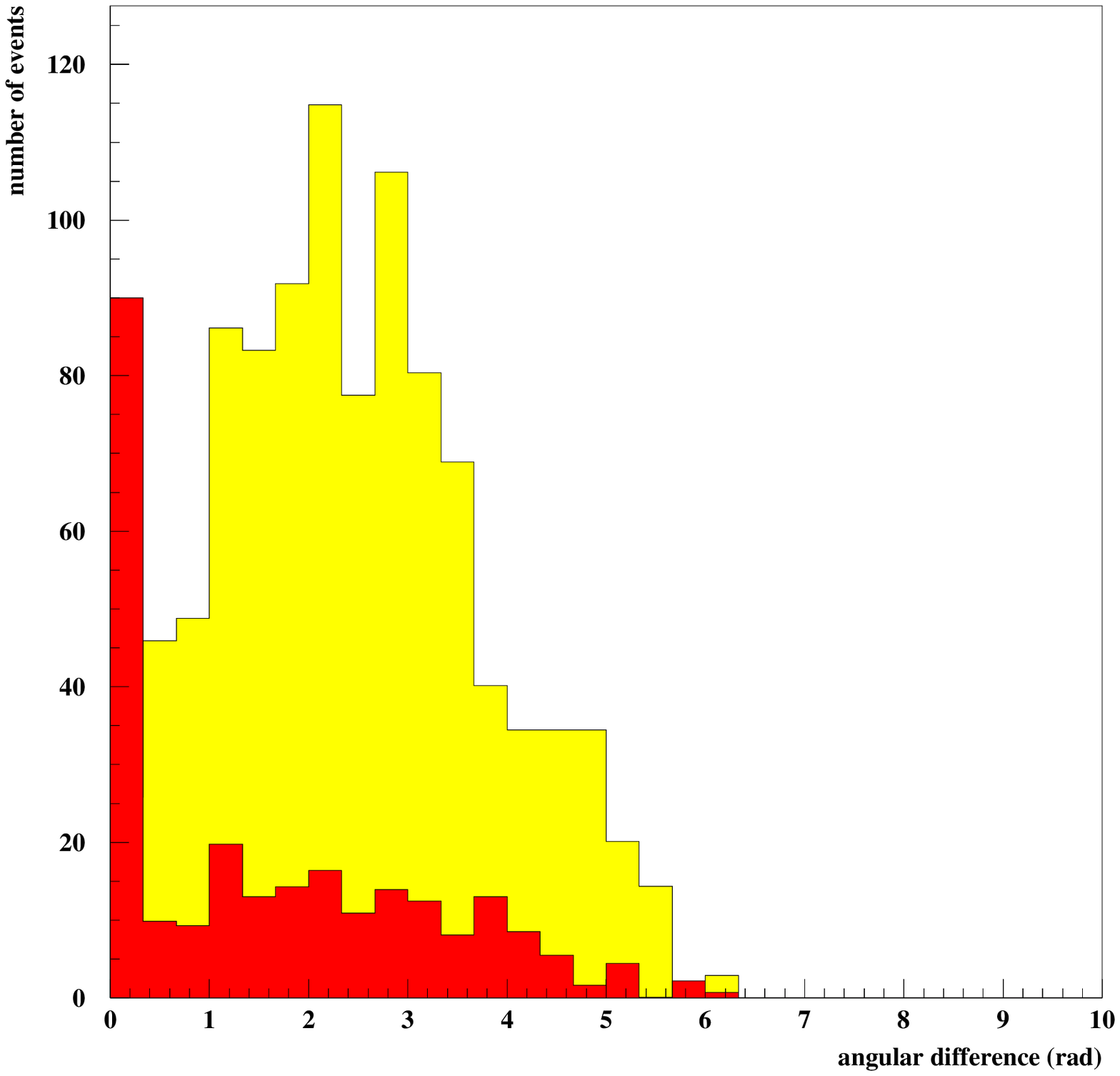,height=5.5in}}
\end{center}
\caption{\footnotesize  \it
Distributions of the $\Delta R$ angular difference (in $rad$)
between the third leading lepton (electron or muon) and the second leading
jet in the 3 leptons events selected by applying cut $1$ and
produced by the \sm background (curve in black),
namely the
$W^{\pm} Z^0$, $Z^0 Z^0$ and $t \bar t$ productions, and the SUSY signal
(curve in grey), for $\l'_{211}=0.09$,
$M_2=150GeV$, $m_0=200GeV$, $\tan \beta=1.5$ and $sign(\mu)<0$.
The numbers of events correspond to an integrated luminosity of
${\cal L}=10fb^{-1}$.
\rm \normalsize }
\label{distang}
\end{figure}

Finally, since the leptons originating from the hadron decays
(as in the $t \bar t$ production) are
not well isolated, we have
applied some cuts on the lepton isolation.
We have imposed the isolation cut
$\Delta R=\sqrt{\delta \phi^2+\delta \theta^2}>0.4$ where
$\phi$ is the azimuthal angle and $\theta$ the polar angle between
the 3 most energetic charged leptons and the 2 hardest jets.
Such a cut is for instance motivated
by the distributions shown in Fig.\ref{distang}
of the $\Delta R$ angular difference
between the third leading lepton and the second leading
jet, in the 3 lepton events generated by the SUSY signal
and \sm background. We call cut $\Delta R>0.4$ together with 
cut $1$, cut $2$.

In order to eliminate poorly isolated leptons,
we have also required that $E<2GeV$, where $E$ represents the
summed energies of the jets being close to a muon or an electron,
namely the jets contained in the cone centered on a muon or an electron
and defined by $\Delta R<0.25$.
This cut is not applied for taus candidates as they have hadronic decays.
It is quite efficient (see Fig.\ref{dianmu} for the 2 lepton case) since
we sum over all jet energies in the cone. The \sm background shows more jets 
and less separation between jets and leptons in $(\theta, \phi)$ in 
final state than the single productions \footnote{This cut will have to be fine tuned
with real events since it will depend on the energy flow inside the detector,
the overlap and minimum biased events.}.
We denote cut $E<2GeV$ plus cut $2$ as cut $3$
\footnote{Although it has not been applied, we mention 
another kind of isolation cut which allows
to further reduce the \sm background:
$\delta \phi>70^{\circ}$ between the leading
charged lepton and the 2 hardest jets.}.

\begin{figure}[t]
\begin{center}
\leavevmode
\centerline{\psfig{figure=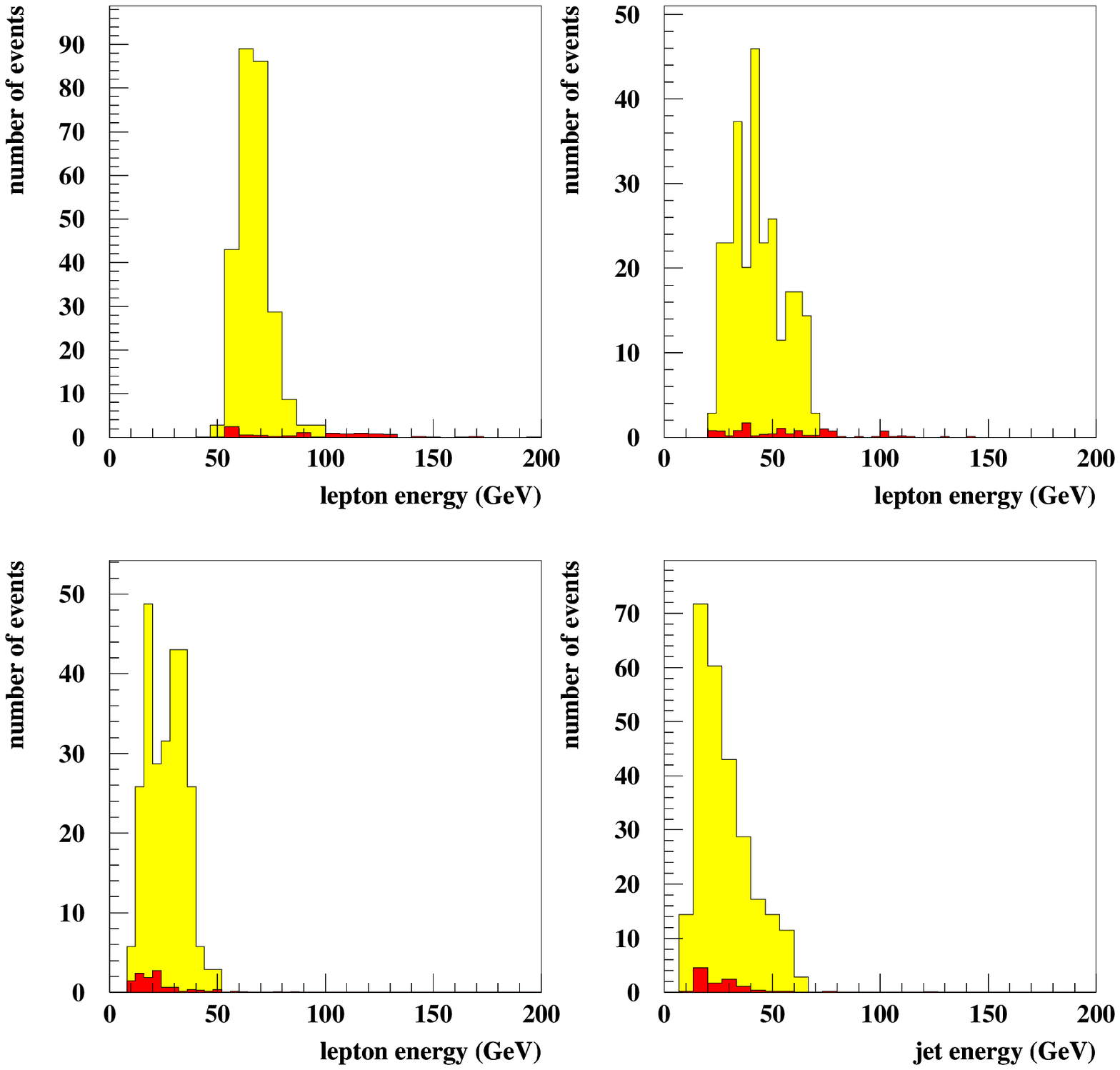,height=5.5in}}
\end{center}
\caption{\footnotesize  \it
Energy distributions (in $GeV$) of the 3 leading charged leptons
and the second leading jet in the events containing at least 3 charged
leptons selected by applying cut $3$ and
produced by the \sm background (curve in black), namely the
$W^{\pm} Z^0$, $Z^0 Z^0$ and $t \bar t$ productions,
and the SUSY signal (curve in grey), for $\l'_{211}=0.09$,
$M_2=150GeV$, $m_0=200GeV$, $\tan \beta=1.5$ and $sign(\mu)<0$.
The upper left plot represents the leading lepton distribution,
the upper right plot the second leading lepton distribution and
the lower left plot the third leading lepton distribution.
The numbers of events correspond to an integrated luminosity of
${\cal L}=10fb^{-1}$.
\rm \normalsize }
\label{apres}
\end{figure}

The selected events require high energy charged leptons and jets
and can thus easily be triggered at Tevatron. This is illustrated in
Fig.\ref{apres} where we show the energy distributions of the 3 leptons
and the second leading jet in the 3 leptons events selected by
applying cut $3$ and generated by the SUSY signal and \sm background.

\begin{table}[t]
\begin{center}
\begin{tabular}{|c|c|c|c|c|}
\hline
& $W^{\pm} Z^0$   & $Z^0 Z^0$       & $t \bar t$
& Total   \\
\hline
cut $1$
& $1.39 \pm 0.11$ & $1.37 \pm 0.11$ & $39.80 \pm 1.00$
& $42.56 \pm 1.01$ \\
\hline
cut $2$
& $0.26 \pm 0.05$ & $0.21 \pm 0.04$ & $4.23  \pm 0.39$
& $4.70 \pm 0.40$ \\
\hline
cut $3$
& $0.24 \pm 0.04$ & $0.17 \pm 0.04$ & $1.14 \pm 0.17$
& $1.55 \pm 0.18$  \\
\hline
cut $1^{\star}$
& $0.51 \pm 0.06$ & $0.73 \pm 0.08$ & $27.80 \pm 0.80$
& $29.04 \pm 0.80$ \\
\hline
cut $2^{\star}$
& $0.26 \pm 0.05$ & $0.21 \pm 0.04$ & $2.92 \pm 0.27$
& $3.39 \pm 0.28$ \\
\hline
cut $3^{\star}$
& $0.23 \pm 0.04$ & $0.17 \pm 0.04$ & $0.64 \pm 0.13$
& $1.04 \pm 0.14$  \\
\hline
\end{tabular}
\caption{Numbers of three lepton events generated by the
\sm background ($W^{\pm} Z^0$, $Z^0 Z^0$ and $t \bar t$ productions)
at Tevatron Run II for the cuts described in the text, assuming an
integrated
luminosity of ${\cal L}=1 fb^{-1}$ and a center of mass energy of $\sqrt s=2
TeV$.
The cuts marked by a $\star$ do not include the reconstruction
of the tau-jets.
These results have been obtained by generating and simulating
$3 \ 10^5$ events for the $W^{\pm} Z^0$ production,
$    10^4$ events for the $Z^0 Z^0$ and
$3 \ 10^5$ events for the $t \bar t$.}
\label{cuteff}
\end{center}
\end{table}

In Table \ref{cuteff}, we give the numbers of three lepton events expected
from the
\sm background at Tevatron Run II with the various cuts described above.
We see in Table \ref{cuteff} that the main source
of \sm background to the three lepton signature at Tevatron
is the $t \bar t$ production.
This is due to the important cross section of the
$t \bar t$ production compared to the other \sm
backgrounds (see Section \ref{back1}).
Table \ref{cuteff} also shows that the $t \bar t$ background
is relatively more suppressed than the other sources of
\sm background by the lepton isolation cuts. The reason is that
in the $t \bar t$ background, one of the 3 charged leptons
of the final state is generated in a $b$-jet and is thus
not well isolated.

\begin{table}[t]
\begin{center}
\begin{tabular}{|c|c|c|c|c|c|}
\hline
$m_{1/2} \ \backslash \ m_0$
&  $100GeV$ & $200GeV$   & $300GeV$  & $400GeV$
&  $500GeV$   \\
\hline
$100GeV$
& $93.94$       & $125.59$       & $80.53$       & $66.62$
& $63.90$ \\
\hline
$200GeV$
& $5.11$        & $4.14$         & $3.86$        & $4.02$
&  $4.26$ \\
\hline
$300GeV$
& $2.26$        & $0.66$         & $0.52$        & $0.55$
&  $0.55$ \\
\hline
\end{tabular}
\caption{Number of 3 lepton events generated by the
SUSY background (all superpartners pair productions)
at Tevatron Run II
as a function of the $m_0$ and $m_{1/2}$ parameters
for $\tan \beta=1.5$, $sign(\mu)<0$ and $\l'_{211}=0.05$.
Cut 3 (see text) has been applied.
These results have been obtained by generating 7500 events
and correspond to an integrated luminosity
of ${\cal L}=1 fb^{-1}$
and a center of mass energy of $\sqrt s=2 TeV$.}
\label{cutSUSY}
\end{center}
\end{table}

In Table \ref{cutSUSY}, we give the number of three lepton
events generated by the
SUSY background (all superpartners pair productions)
at Tevatron Run II as a function of
the $m_0$ and $m_{1/2}$ parameters for the cut 3.
This number of events decreases as $m_0$ and $m_{1/2}$
increase due to the behaviour of the summed superpartners
pair productions cross section in the SUSY parameter space
(see Section \ref{susyback1}).

\subsection{Results}
\label{res}

\subsubsection{Discovery potential for the
$\l'_{2jk}$ coupling constant}
\label{lp211}

We first present the reach in the
mSUGRA parameter space obtained from the analysis of
the trilepton signature at Tevatron Run II generated
by the single chargino production
through the $\l'_{211}$ coupling, namely
$p \bar p \to \tilde \chi^{\pm}_1 \mu^{\mp}$.
The sensitivity that can be obtained on the $\l'_{2jk}$
($j$ and $k$ being not equal to $1$ simultaneously) couplings
based on the $\tilde \chi^{\pm}_1 \mu^{\mp}$ production analysis
will be discussed at the end of this section for a given mSUGRA point.
We give more detailed results for the case of a single dominant
$\l'_{211}$ coupling since this \rpv coupling gives the highest partonic
luminosity to the $\tilde \chi^{\pm}_1 \mu^{\mp}$ production
(see Section \ref{cross1}) and leads thus to
the highest sensitivities.

\begin{figure}[t]
\begin{center}
\leavevmode
\centerline{\psfig{figure=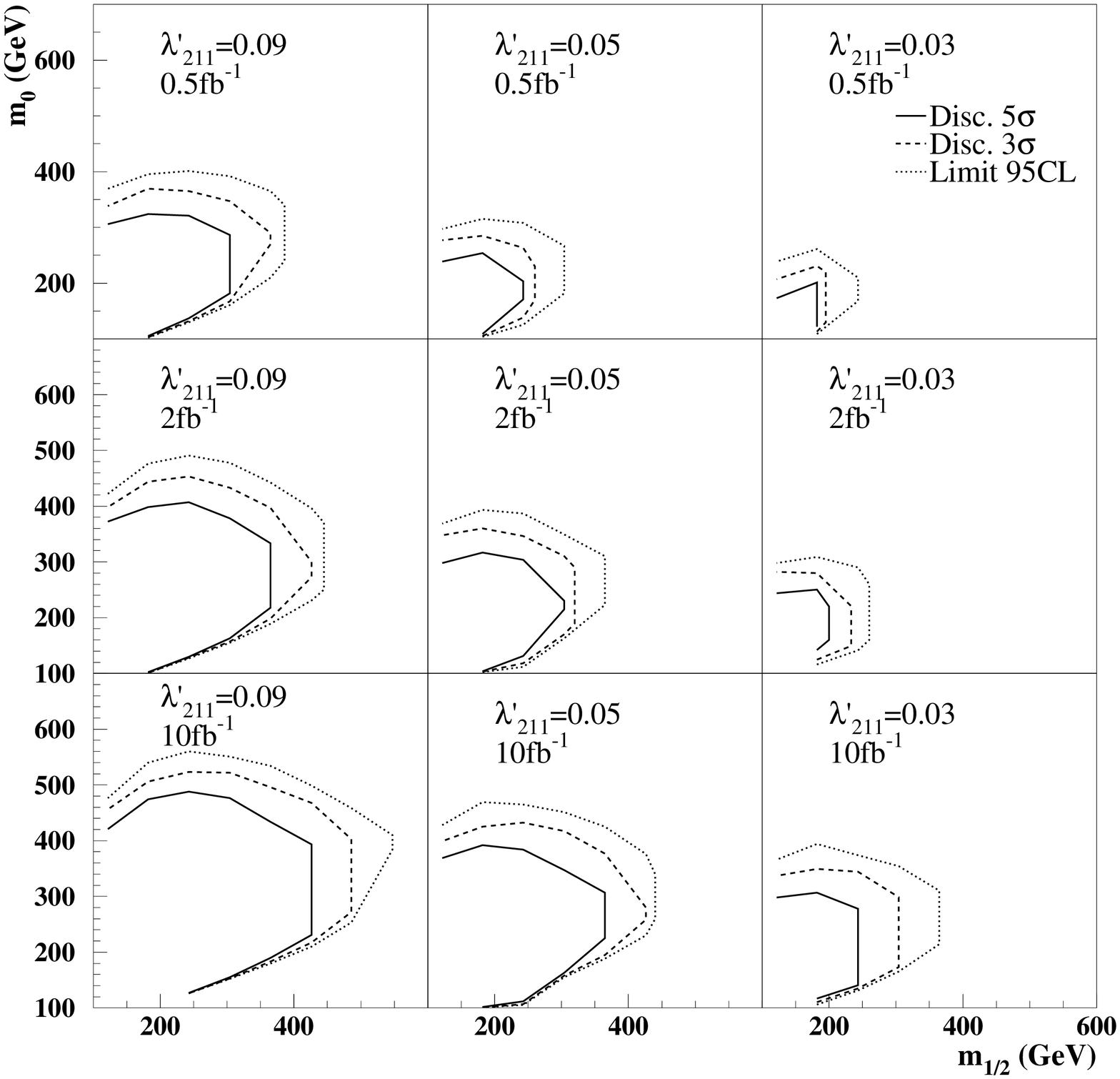,height=5.5in}}
\end{center}
\caption{Discovery contours at $5 \sigma$ (full line),
$ 3 \sigma$ (dashed line)
and limit at $95 \% \ C.L.$ (dotted line)
obtained from the trilepton signature analysis
at Tevatron Run II assuming
a center of mass energy of $\sqrt s=2 TeV$.
This discovery reach is
presented in the plane $m_0$ versus $m_{1/2}$,
for $sign(\mu)<0$, $\tan \beta=1.5$
and different values of $\l'_{211}$ and of luminosity.}
\label{b:fig2}
\end{figure}

\begin{figure}[t]
\begin{center}
\leavevmode
\centerline{\psfig{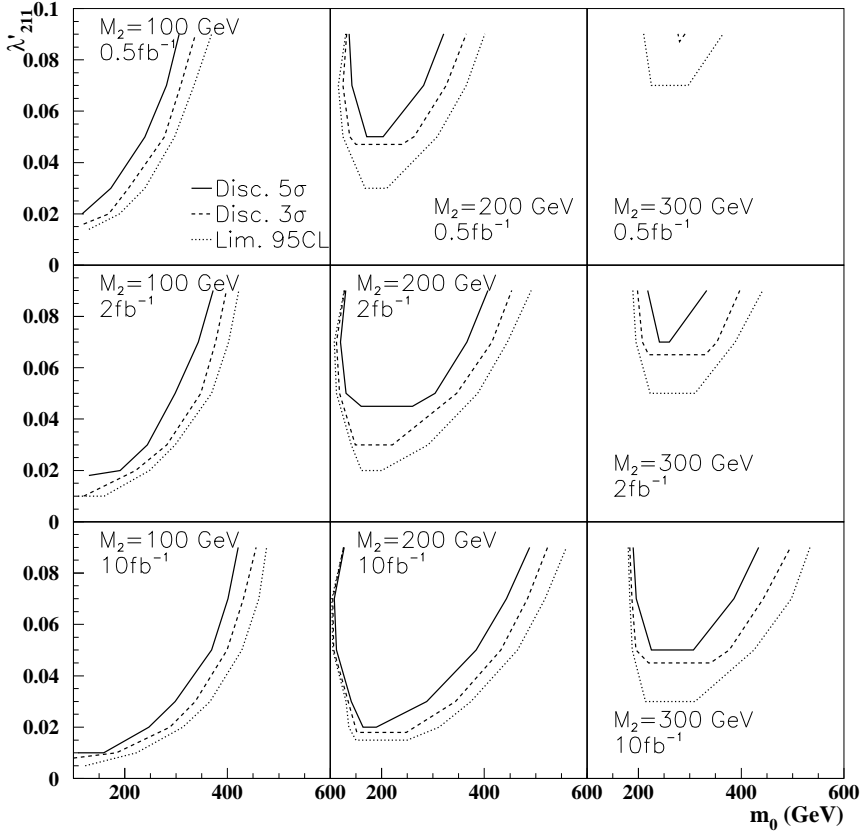}}
\end{center}
\caption{Discovery contours at $5 \sigma$ (full line), $ 3 \sigma$
(dashed line) and limit at $95 \% \ C.L.$ (dotted line) presented
in the plane $\l'_{211}$ versus the $m_0$ parameter,
for $sign(\mu)<0$, $\tan \beta=1.5$
and different values of $M_2$ and of luminosity.}
\label{b:fig1}
\end{figure}

In Fig.\ref{b:fig2},
we present the $3 \sigma$ and $ 5 \sigma$ discovery
contours and the limits at $95 \%$
confidence level in the plane $m_0$ versus $m_{1/2}$,
for $sign(\mu)<0$, $\tan \beta=1.5$ and
using a set of values for $\l'_{211}$ and the luminosity.
This discovery potential was obtained by considering
the $\tilde \chi^{\pm}_1 \mu^{\mp}$ production and
the background originating from the Standard Model.
The signal and background were selected by
using cut $3$ described in Section \ref{cut1}.
The results presented for a luminosity of
${\cal L}=0.5 fb^{-1}$
in Fig.\ref{b:fig2} and Fig.\ref{b:fig1}
were obtained with cut 2 only in order to optimize the
sensitivity on the SUSY parameters.
The reduction of the sensitivity on $\l'_{211}$
observed in Fig.\ref{b:fig2}
when either $m_0$ or $m_{1/2}$ increases is due to the
decrease of the $\tilde \chi^{\pm}_1 \mu^{\mp}$ production cross section
with $m_0$ or $m_{1/2}$ (or equivalently $M_2$), 
which can be observed in Fig.\ref{XS02}.
In Fig.\ref{b:fig2}, we also see that
for all the considered values of $\l'_{211}$ and the luminosity,
the sensitivity on $m_{1/2}$ is reduced to low masses
in the domain $m_0 \stackrel{<}{\sim} 200GeV$. This important reduction
of the sensitivity as $m_0$ decreases is due to
the decrease of the phase space
factor associated to the decay
$\tilde \nu_{\mu} \to \tilde \chi^{\pm} \mu^{\mp}$
(see Section \ref{cross1}).
%
%
Finally, we note from Fig.\ref{XSmu} that for $sign(\mu)>0$
the $\tilde \chi^{\pm}_1 \mu^{\mp}$ production cross section,
and thus the sensitivities presented in Fig.\ref{b:fig2},
would incur a little increase compared to the case $sign(\mu)<0$.

In Fig.\ref{b:fig1}, the discovery potential is shown
in the $\l'_{211}$-$m_0$ plane for different values of $M_2$
and the luminosity.
For a given value of $M_2$, we note that the sensitivity on the $\l'_{211}$
coupling decreases at high and low values of $m_0$.
The main explanation is the decrease of the
$p \bar p \to \tilde \chi^{\pm}_1 \mu^{\mp}$
rate at high and low values of $m_0$
which appears clearly in Fig.\ref{XScl}.
We also observe, as in Fig.\ref{b:fig2},
a decrease of the sensitivity on the $\l'_{211}$
coupling when $M_2$ (or equivalently $m_{1/2}$) increases
for a fixed value of $m_0$.

The strongest bounds on the \susyq masses obtained at LEP 
in an \rpv model with a non-vanishing $\l'$ Yukawa coupling are
$m_{\tilde \chi^{0}_1}>26 GeV$ 
(for $m_0=200 GeV$ and $\tan \beta =\sqrt 2$ \cite{b:neut}), 
$m_{\tilde \chi^{\pm}_1}>100 GeV$, 
$m_{\tilde l}>93 GeV$, $m_{\tilde \nu}> 86 GeV$ \cite{b:Mass}.
For the minimum values of $m_0$ and $m_{1/2}$
spanned by the parameter space described in Figures \ref{b:fig2}
and \ref{b:fig1}, namely
$m_0=100GeV$ and $M_2=100GeV$, the mass spectrum is
$m_{\tilde \chi^{\pm}_1}= 113GeV$, $m_{\tilde \chi^{0}_1}= 54GeV$,
$m_{\tilde \nu_L}= 127 GeV$, $m_{\tilde l_L}= 137 GeV$,
$m_{\tilde l_R}= 114 GeV$, so that we are well above these limits.
Since both the scalar and gaugino
masses increase with $m_0$ and $m_{1/2}$,
the parameter space described in Figures \ref{b:fig2}
and \ref{b:fig1} lies outside the SUSY parameters ranges
excluded by LEP data \cite{b:Mass,b:neut}.
Therefore, the discovery potential of Figures \ref{b:fig2}
and \ref{b:fig1} represents an important improvement with respect to
the \susyq masses limits derived from LEP data \cite{b:Mass,b:neut}.
Figures \ref{b:fig2} and \ref{b:fig1} show also that the low-energy bound
on the considered \rpv coupling,
$\l'_{211}<0.09 (m_{\tilde d_R}/100GeV)$ at $1 \sigma$ (from $\pi$ decay)
\cite{b:Bhatt}, can be greatly improved.

\begin{figure}[t]
\begin{center}
\leavevmode
\centerline{\psfig{figure=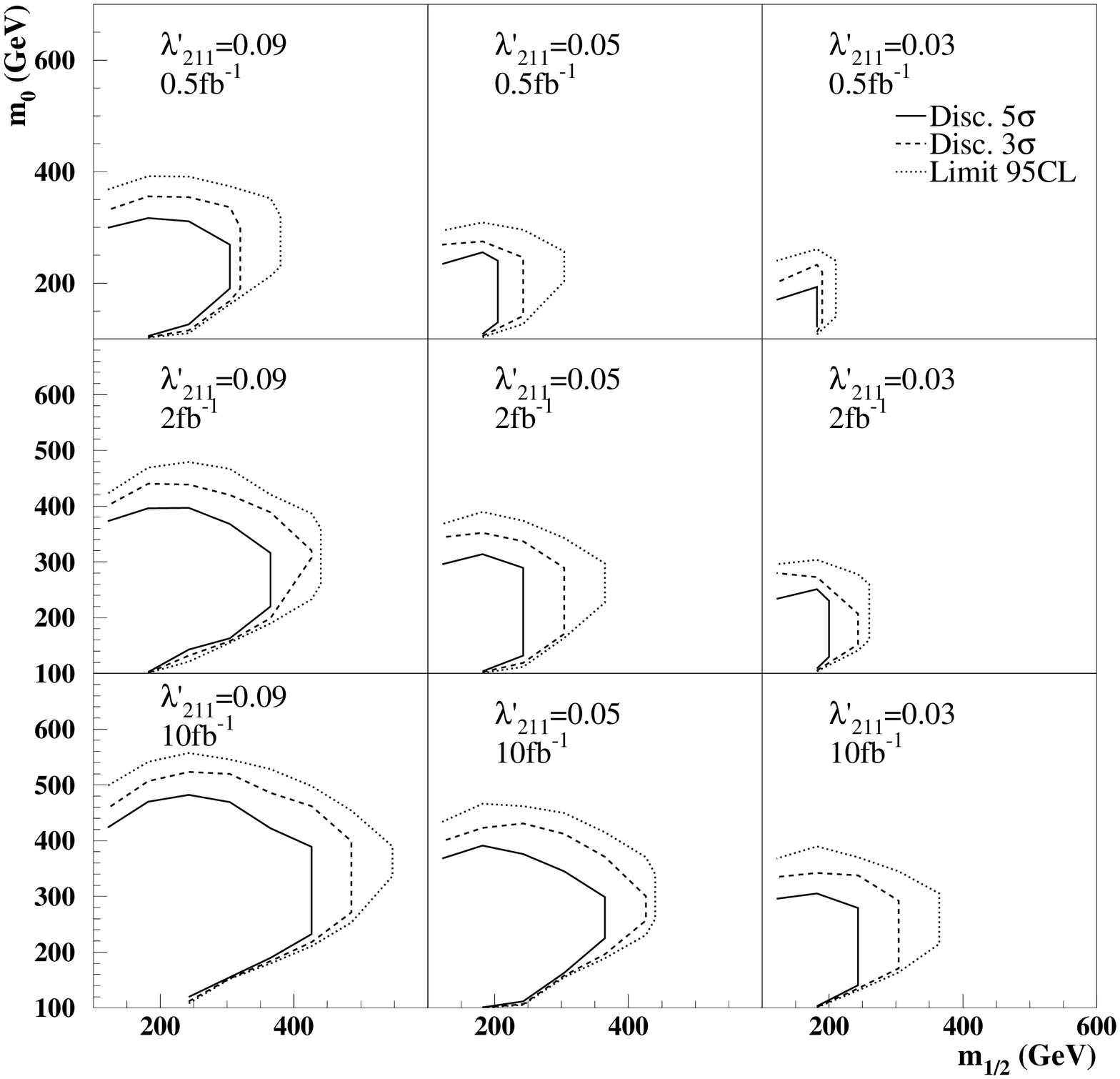,height=5.5in}}
\end{center}
\caption{Discovery contours at $5 \sigma$ (full line),
$ 3 \sigma$ (dashed line)
and limit at $95 \% \ C.L.$ (dotted line) presented
in the plane $m_0$ versus $m_{1/2}$
and obtained without reconstruction of the tau leptons decaying into jets
for $sign(\mu)<0$, $\tan \beta=1.5$
and different values of $\l'_{211}$ and of luminosity.}
\label{b:fig2p}
\end{figure}

\begin{figure}[t]
\begin{center}
\leavevmode
\centerline{\psfig{figure=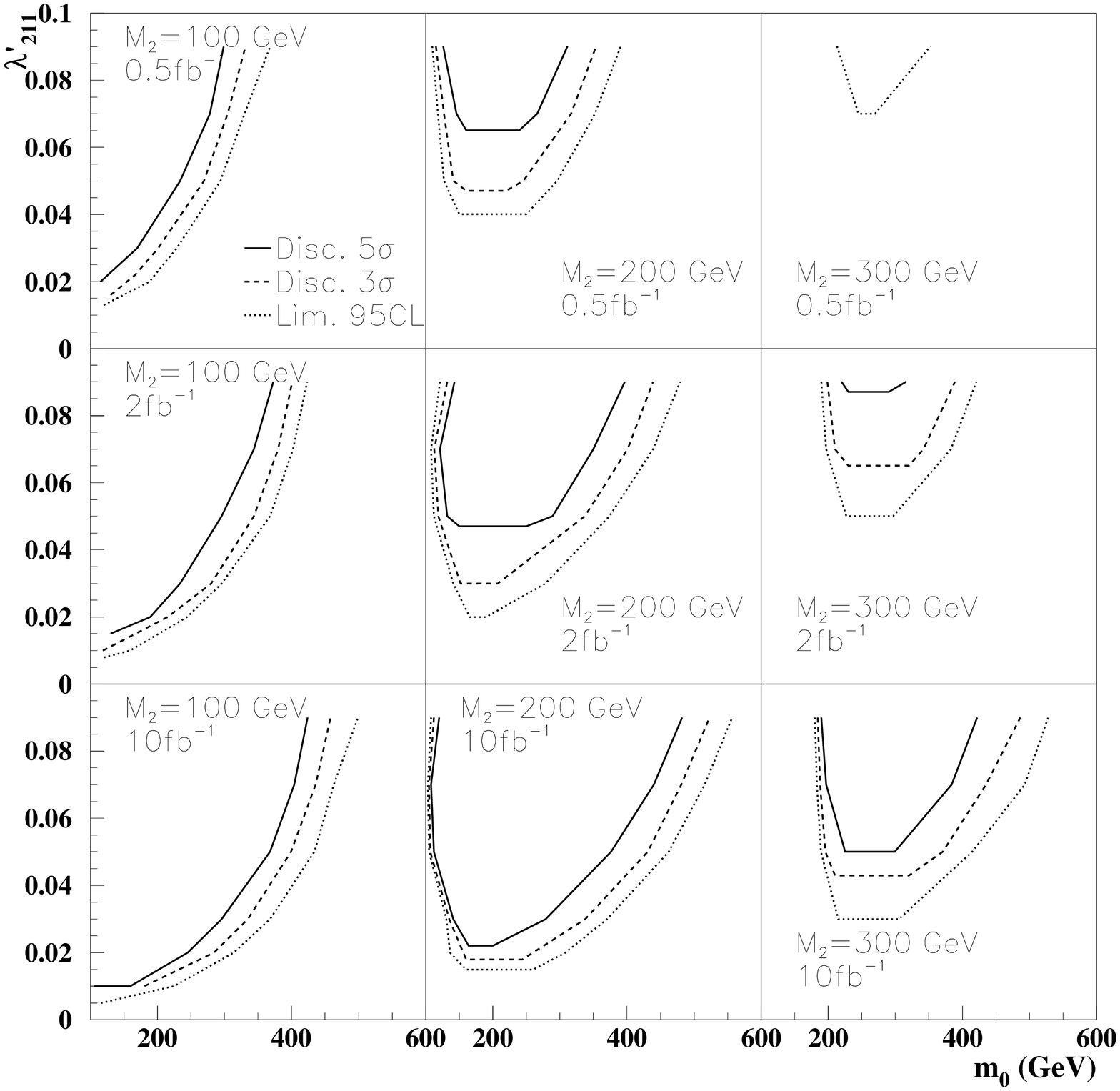,height=5.5in}}
\end{center}
\caption{Discovery contours at $5 \sigma$ (full line), $ 3 \sigma$
(dashed line) and limit at $95 \% \ C.L.$ (dotted line) presented
in the plane $\l'_{211}$ versus the $m_0$ parameter
and obtained without reconstruction of the tau leptons decaying into jets
for $sign(\mu)<0$, $\tan \beta=1.5$
and different values of $M_2$ and of luminosity.}
\label{b:fig1p}
\end{figure}

Interesting sensitivities on the SUSY parameters can
already be obtained within the first year of Run II at Tevatron
with a low luminosity (${\cal L}=0.5 fb^{-1}$)
and no reconstruction of the tau-jets. To illustrate this point,
we present in Fig.\ref{b:fig2p}
and Fig.\ref{b:fig1p} the same discovery potentials as
in Fig.\ref{b:fig2} and Fig.\ref{b:fig1}, respectively,
obtained without reconstruction of the tau leptons decaying into jets.
By comparing Fig.\ref{b:fig2}, Fig.\ref{b:fig1} and
Fig.\ref{b:fig2p}, Fig.\ref{b:fig1p},
we observe that the sensitivity on the SUSY parameters is
weakly affected by the reconstruction of the tau-jets
\footnote{This is actually an artefact of the method:
cut 3 is our most efficient cut to reduce the \sm background with 
electrons and muons but is not applied with taus. Thus, the
relative ratio signal over background is not so good with taus.
Finding another efficient cut could improve our discovery potential
and limits using taus.}.

Using the ratios of the cross sections for the
$\tilde \chi^+_1 \mu^-$ productions via different $\l'_{2jk}$ couplings,
one can deduce
from the sensitivity obtained on $\l'_{211}$ via the 3 lepton
final state analysis an estimation of
the sensitivity on any $\l'_{2jk}$ coupling.
For instance, let us consider the SUSY point
$m_0=180GeV$, $M_2=200GeV$, $\tan \beta=1.5$ and $\mu=-200GeV$
($m_{\tilde u_L}=601GeV$, $m_{\tilde d_L}=603GeV$,
$m_{\tilde u_R}=582GeV$, $m_{\tilde d_R}=580GeV$,
$m_{\tilde l_L}=253GeV$, $m_{\tilde l_R}=205GeV$
$m_{\tilde \nu_L}=248GeV$,
$m_{\tilde \chi^{\pm}_1}=199GeV$, $m_{\tilde \chi^0_1}=105GeV$)
which corresponds, as can be seen in Fig.\ref{b:fig1},
to the point where the sensitivity
on $\l'_{211}$ is maximized for $M_2=200GeV$.
We can see on Fig.\ref{XScl} that for this SUSY point,
the ratio between the rates of the $\tilde \chi^+_1 \mu^-$
productions via $\l'_{211}$ and $\l'_{221}$ is
$\sigma(\l'_{211}) / \sigma(\l'_{221}) \approx 7.9$
for same values of the \rpv couplings. Therefore,
since the single chargino production rate scales as $\l'^2$ (see
Appendix \ref{formulas}),
the sensitivity on $\l'_{221}$ at this SUSY point
is equal to the sensitivity obtained on $\l'_{211}$
($\sim 0.02$ at $95 \% CL$ with ${\cal L}=2fb^{-1}$
as shows Fig.\ref{b:fig1})
multiplied by the factor $\sqrt {7.9}$, namely
$\sim 0.05$. This result represents a significant
improvement with respect to the low-energy
indirect limit $\l'_{221}<0.18 (m_{\tilde d_R}/100GeV)$ \cite{b:Bhatt}.
Using the same method, we find at the same SUSY point the sensitivities
on the $\l'_{2jk}$ coupling constants given in Table \ref{coup}.
\begin{table}[t]
\begin{center}
\begin{tabular}{|c|c|c|c|c|c|c|c|}
\hline
$\l'_{212}$ & $\l'_{213}$ & $\l'_{221}$ & $\l'_{222}$
& $\l'_{223}$ & $\l'_{231}$ & $\l'_{232}$ & $\l'_{233}$  \\
\hline
0.04 & 0.07 & 0.05 & 0.12 & 0.21 & 0.10 & 0.36 & 0.63 \\
\hline
\end{tabular}
\caption{\em
Sensitivities at $95 \% CL$ on the $\l'_{2jk}$
coupling constants derived from the sensitivity
on $\l'_{211}$ for a luminosity of ${\cal L}=2fb^{-1}$
and the following set of SUSY parameters,
$\tan\beta=1.5$,  $M_2=200GeV$, $\mu=-200GeV$ and
$m_0=180GeV$.}
\label{coup}
\end{center}
\end{table}
All the sensitivities on the $\l'_{2jk}$ coupling constants
given in Table \ref{coup} are stronger than
the low-energy bounds of \cite{b:Bhatt} which we rewrite here:
$\l'_{21k}<0.09 (m_{\tilde d_{kR}}/100GeV)$ at $1 \sigma$ ($\pi$ decay),
$\l'_{22k}<0.18 (m_{\tilde d_{kR}}/100GeV)$ at $1 \sigma$ ($D$ decay),
$\l'_{231}<0.22 (m_{\tilde b_L}/100GeV)$ at $2 \sigma$
($\nu_{\mu}$ deep inelastic scattering),
$\l'_{232}<0.36 (m_{\tilde q}/100GeV)$ at $1 \sigma$ ($R_{\mu}$),
$\l'_{233}<0.36 (m_{\tilde q}/100GeV)$ at $1 \sigma$ ($R_{\mu}$).
\\ In the case of a single dominant $\l'_{2j3}$
coupling,
the neutralino decays as $\tilde \chi^0_1 \to \mu u_j b$
and the semileptonic decay of the b-quark could affect
the analysis efficiency. Therefore in this case, the precise sensitivity
cannot be simply calculated by scaling the value obtained for
$\lambda^\prime_{211}$.
Nevertheless, the order of magnitude of the sensitivity
which can be inferred from our analysis should be correct.
\\ The range of SUSY parameters in which the constraint on a given
$\l'_{2jk}$ coupling constant obtained via the three leptons analysis
is stronger than the relevant low-energy bound
depends on the low-energy bound itself as well as
on the values of the cross section for the
single chargino production via the considered $\l'_{2jk}$ coupling.
\\ Finally, we remark that
while the low-energy constraints on the $\l'_{2jk}$ couplings
become weaker as the squark masses increase, the sensitivities
on those couplings obtained from the three leptons analysis
are essentially independent of the squark masses as long as
$m_{\tilde q}>m_{\tilde \chi^{\pm}_1}$ (recall that the
branching ratio of the decay
$\tilde \chi^{\pm}_1 \to q \bar q \tilde \chi^0_1$ is
greatly enhanced when $m_{\tilde q}<m_{\tilde \chi^{\pm}_1}$).

We end this section by some comments on the effect of
the \susyq $R_p$ conserving background to the 3 lepton
signature. In order to illustrate this discussion, we consider
the results on the $\l'_{211}$ coupling constant.\\
We see from Table \ref{cutSUSY} that
the SUSY background to the 3 lepton final
state can affect the sensitivity on the
$\l'_{211}$ coupling constant
obtained by considering only the \sm background,
which is shown in Fig.\ref{b:fig2}, only in the
region of small superpartner masses, namely in the domain
$m_{1/2} \stackrel{<}{\sim} 300GeV$ for $\tan \beta=1.5$,
$sign(\mu)<0$ and assuming a luminosity of ${\cal L}=1fb^{-1}$.\\
In contrast with the SUSY signal amplitude which is increased
if $\l'_{211}$ is enhanced,
the SUSY background amplitude is typically
independent on the value of the $\l'_{211}$ coupling constant
since the superpartner pair production
does not involve \rpv couplings.
Therefore, even if we consider the SUSY background
in addition to the \sm one, it is still true that
large values of the $\l'_{211}$ coupling can
be probed over a wider domain of the SUSY parameter space than
low values, as can be observed in Fig.\ref{b:fig2} for
$m_{1/2} \stackrel{>}{\sim} 300GeV$. Note that in Fig.\ref{b:fig2}
larger values of $\l'_{211}$ could have been considered as the
low-energy bound on this \rpv coupling, namely $\l'_{211}<0.09
(m_{\tilde d_R}/100GeV)$ \cite{b:Bhatt}, is proportional to
the squark mass.\\
Finally, we mention that further cuts,
as for instance some cuts based on the
superpartner mass reconstructions (see Section \ref{recons}),
could allow to reduce the SUSY background to the 3 lepton
signature.

\subsubsection{High $\tan \beta$ scenario}
\label{htanb}

\begin{figure}[t]
\begin{center}
\leavevmode
\centerline{\psfig{figure=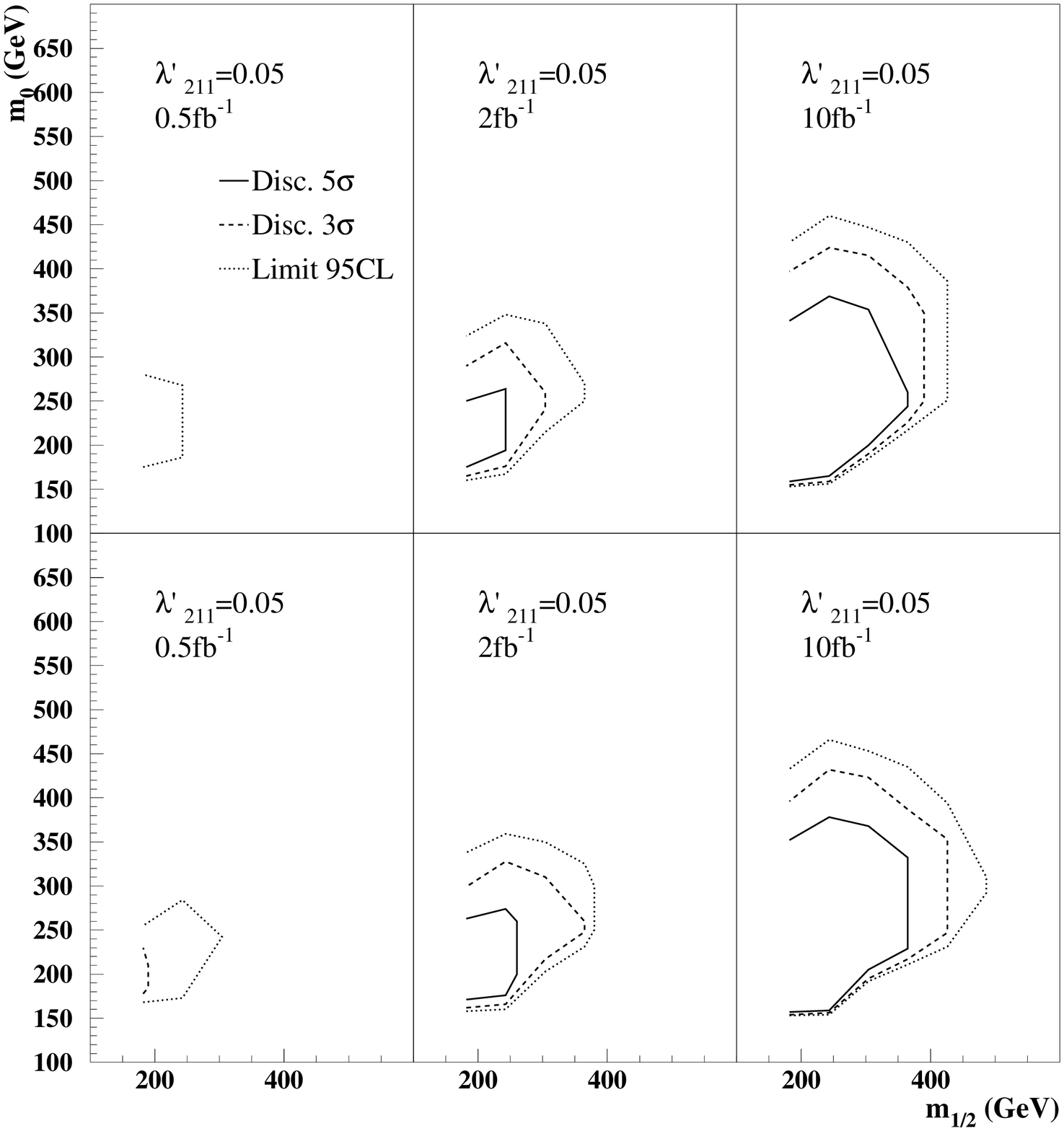,height=5.5in}}
\end{center}
\caption{Discovery contours at $5 \sigma$ (full line),
$ 3 \sigma$ (dashed line)
and limit at $95 \% \ C.L.$ (dotted line) presented
in the plane $m_0$ versus $m_{1/2}$,
for $sign(\mu)<0$, $\tan \beta=50$, $\l'_{211}=0.09$
and different values of luminosity. The upper (lower) curves
are obtained without (with) the reconstruction of the tau-jets.}
\label{b:fig2tanb}
\end{figure}

\begin{figure}[t]
\begin{center}
\leavevmode
\centerline{\psfig{figure=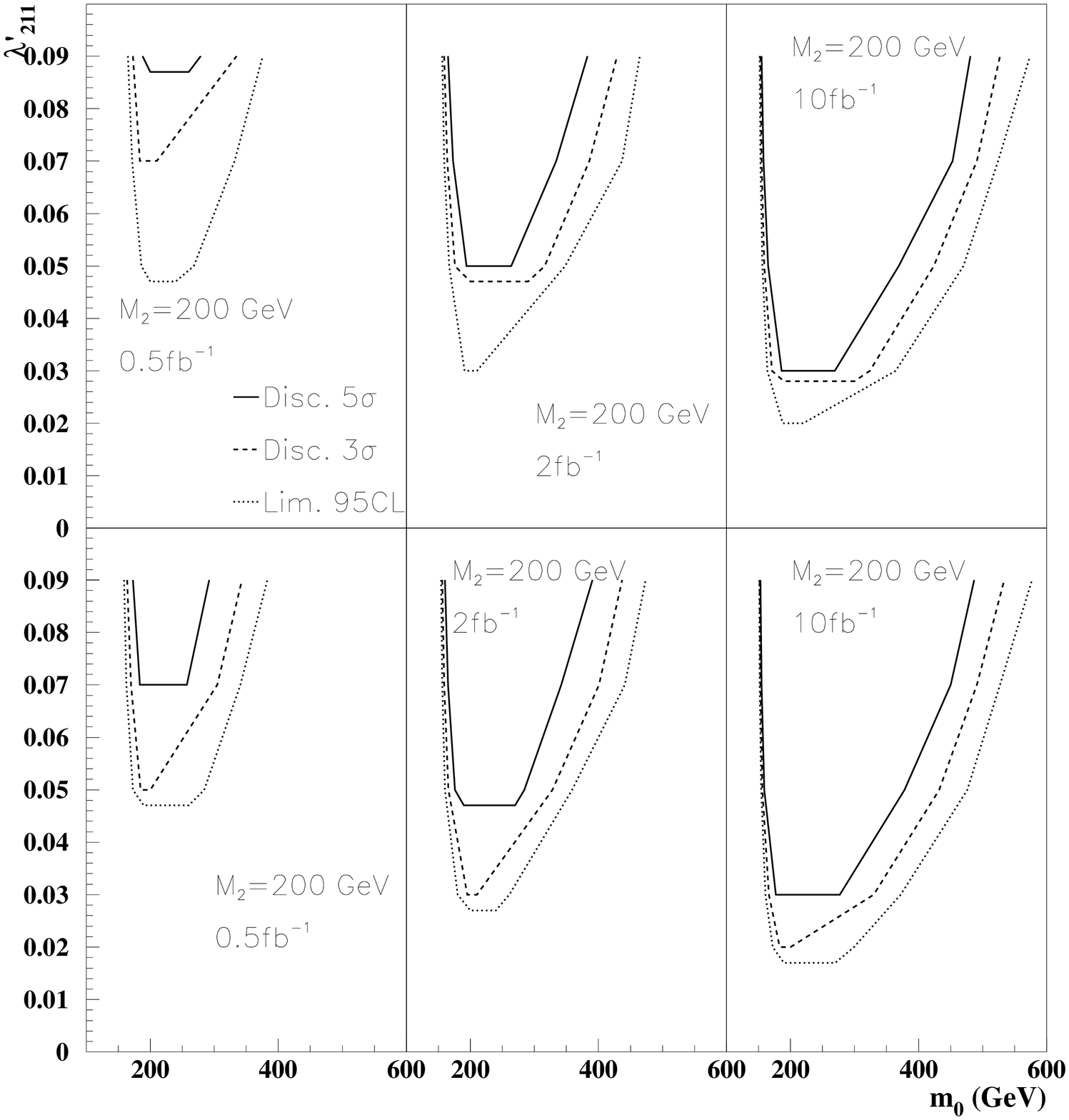,height=5.5in}}
\end{center}
\caption{Discovery contours at $5 \sigma$ (full line), $ 3 \sigma$
(dashed line) and limit at $95 \% \ C.L.$ (dotted line) presented
in the plane $\l'_{211}$ versus the $m_0$ parameter,
for $sign(\mu)<0$, $\tan \beta=50$, $M_2=200GeV$
and different values of luminosity. The upper (lower) curves
are obtained without (with) the reconstruction of the tau-jets.}
\label{b:fig1tanb}
\end{figure}

In mSUGRA, for large values of $\tan \beta$ and small values
of $m_0$ compared to $m_{1/2}$,
due to the large mixing in the third generation sfermions,
the mass of the lighter $\tilde \tau_1$ slepton can become
smaller than $m_{\tilde \chi^{\pm}_1}$,
with the sneutrino remaining heavier than the $\tilde \chi^{\pm}_1$
so that the $\tilde \chi^{\pm}_1 l^{\mp}$ production rate can still
be significant. In this situation, the efficiency for the
3 lepton signature arising mainly through,
$\tilde \chi^{\pm}_1 \to \tilde \tau_1^{\pm} \nu_{\tau},
\ \tilde \tau_1^{\pm} \to \tilde \chi^0_1 \tau^{\pm},
\ \tilde \chi^0_1 \to l^{\pm}_i u_j d_k$, can be enhanced
compared to the case where the 3 lepton signal comes from,
$\tilde \chi^{\pm}_1 \to \tilde \chi^0_1 l^{\pm} \nu,
\ \tilde \chi^0_1 \to l^{\pm}_i u_j d_k$.
Indeed, the branching ratio
$B(\tilde \chi^{\pm}_1 \to \tilde \tau_1^{\pm} \nu_{\tau})$
can reach $\sim 100 \%$,
$B(\tilde \tau_1^{\pm} \to \tilde \chi^0_1 \tau^{\pm}) \approx 100 \%$,
since the $\tilde \chi^0_1$ is the LSP,
$B(\tau \to l \nu_l \nu_{\tau})=35 \%$ ($l=e,\mu$) and
the $\tau$-jets can be reconstructed
at Tevatron Run II. However, in such a scenario
the increased number of tau leptons in the final state
leads to a softer charged lepton spectrum
which tends to reduce the efficiency after cuts.
Therefore, for relatively small values
of $m_0$ compared to $m_{1/2}$, the sensitivity obtained
in the high $\tan \beta$ scenario
is essentially unaffected with respect to the low $\tan \beta$
situation, unless $m_0$ is small enough to render
$m_{\tilde \tau_1}$ and $m_{\tilde \chi^0_1}$
almost degenerate. As a matter of fact,
in such a situation, the energy of the tau produced in the decay
$\tilde \tau_1^{\pm} \to \tilde \chi^0_1 \tau^{\pm}$ often
falls below the analysis cuts. Therefore,
this degeneracy results in a loss of signal efficiency after cuts,
at small values of $m_0$ compared to $m_{1/2}$,
and thus in a loss of sensitivity, with respect to the
low $\tan \beta$ situation.
This can be seen by comparing Fig.\ref{b:fig2}, Fig.\ref{b:fig1}
and Fig.\ref{b:fig2tanb}, Fig.\ref{b:fig1tanb}.
Indeed, the decrease of the sensitivity on
$m_{1/2}$ at low $m_0$ is stronger for high $\tan \beta$ (see
Fig.\ref{b:fig2tanb})
than for low $\tan \beta$ (see Fig.\ref{b:fig2}).
Similarly, the decrease of the sensitivity on
$\l'_{211}$ at low $m_0$ is stronger for high $\tan \beta$ (see
Fig.\ref{b:fig1tanb})
than for low $\tan \beta$ (see Fig.\ref{b:fig1}).

The effect on the discovery potential
of the single chargino production
rate increase at large $\tan \beta$ values
shown in Fig.\ref{XStan} is hidden by the large $\tan \beta$
scenario influences on the cascade decays described above.


In contrast with the low $\tan \beta$ scenario (see Section \ref{lp211}),
the sensitivity on the SUSY parameters depends in a significant way
on the reconstruction of the tau-jets in case where
$\tan \beta$ is large,
as can be seen in Fig.\ref{b:fig2tanb} and Fig.\ref{b:fig1tanb}.
The reason is the increased number of tau leptons among the final state
particles in a large $\tan \beta$ model. This is due to the
decrease of the lighter stau mass which tends to increase the
$B(\tilde \chi^{\pm}_1 \to \tilde \chi^0_1 \tau^{\pm} \nu_{\tau})$
branching ratio.

\subsubsection{Discovery potential for the
$\l'_{1jk}$ and $\l'_{3jk}$ coupling constants}
\label{lp311}

\begin{figure}[t]
\begin{center}
\leavevmode
\centerline{\psfig{figure=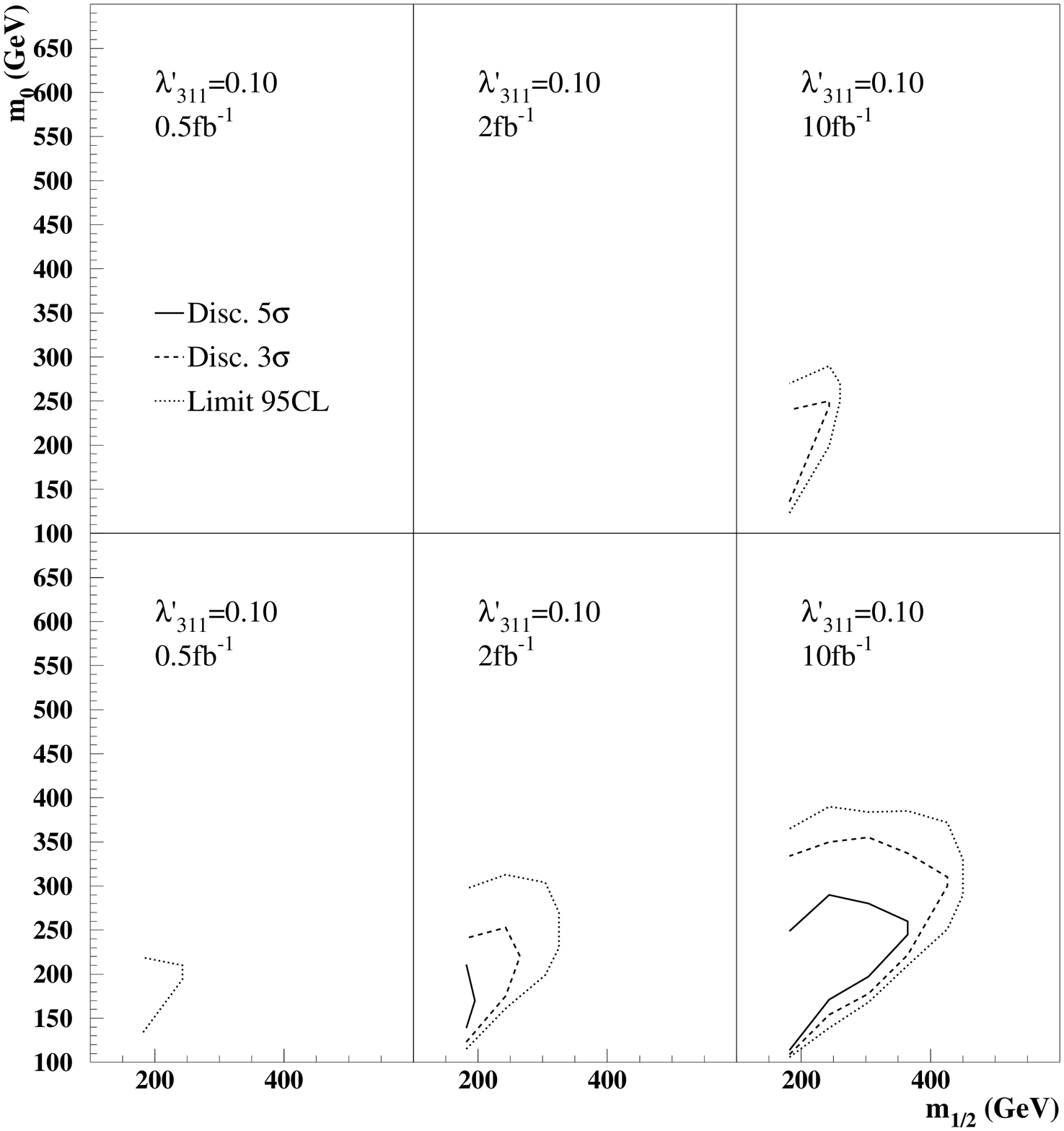,height=5.5in}}
\end{center}
\caption{Discovery contours at $5 \sigma$ (full line),
$ 3 \sigma$ (dashed line)
and limit at $95 \% \ C.L.$ (dotted line) presented
in the plane $m_0$ versus $m_{1/2}$,
for $sign(\mu)<0$, $\tan \beta=1.5$, $\l'_{311}=0.10$
and different values of luminosity. The upper (lower) curves
are obtained without (with) the reconstruction of the tau-jets.}
\label{b:fig2tau}
\end{figure}

\begin{figure}[t]
\begin{center}
\leavevmode
 \centerline{\psfig{figure=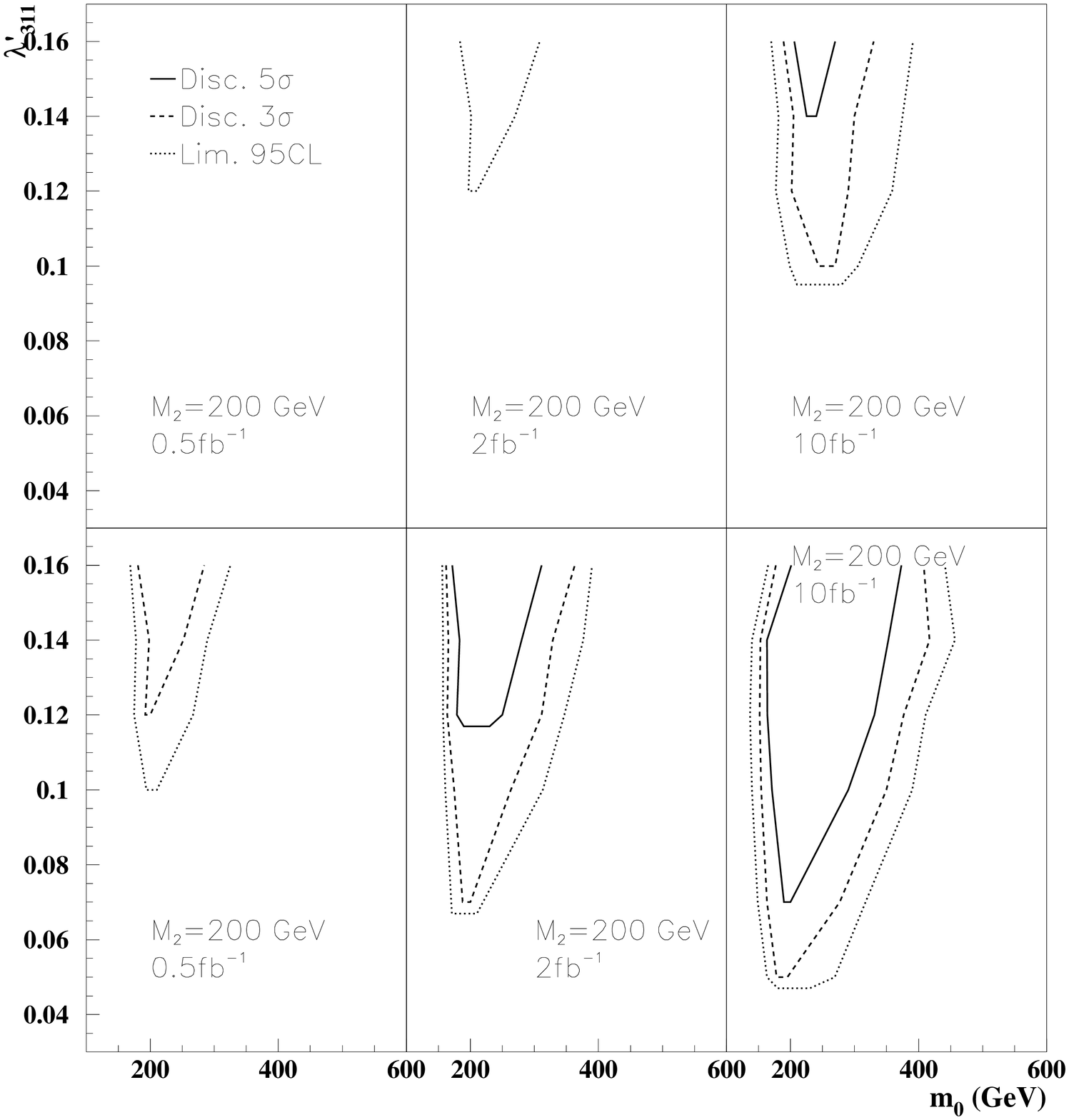,height=5.5in}}
\end{center}
\caption{Discovery contours at $5 \sigma$ (full line), $ 3 \sigma$
(dashed line) and limit at $95 \% \ C.L.$ (dotted line) presented
in the plane $\l'_{311}$ versus the $m_0$ parameter,
for $sign(\mu)<0$, $\tan \beta=1.5$, $M_2=200GeV$
and different values of luminosity. The upper (lower) curves
are obtained without (with) the reconstruction of the tau-jets.}
\label{b:fig1tau}
\end{figure}

In Fig.\ref{b:fig2tau},
we present the $3 \sigma$ and $ 5 \sigma$ discovery
contours and the limits at $95 \%$
confidence level in the plane $m_0$ versus $m_{1/2}$,
for $sign(\mu)<0$, $\tan \beta=1.5$, $\l'_{311}=0.10$
and various values of the luminosity.
In Fig.\ref{b:fig1tau}, the discovery potential is shown
in the $\l'_{311}$-$m_0$ plane for $M_2=200GeV$.
Comparing Fig.\ref{b:fig2tau}, Fig.\ref{b:fig1tau} and
Fig.\ref{b:fig2}, Fig.\ref{b:fig1}, we see that
the sensitivity on the SUSY parameters is weaker
in the case of a single dominant $\l'_{311}$ coupling
than in the case of a single
dominant $\l'_{211}$ coupling. The reason is that in the case of
a single dominant $\l'_{3jk}$ coupling constant,
tau leptons are systematically produced
at the chargino production level
$p \bar p \to \tilde \chi^{\pm}_1 \tau^{\mp}$
(see Fig.\ref{graphes}(a)) as well as in the LSP decay
$\tilde \chi^0_1 \to \tau u_j d_k$ (see Section \ref{signal1}),
so that the number of tau leptons among the 3 charged leptons
of the final state is increased
compared to the dominant $\l'_{2jk}$ case.
The decrease in sensitivity is due to the fact that
a lepton (electron or muon) generated in a tau decay
has an higher probability not to pass the analysis requirements
concerning the particle energy
and that the reconstruction efficiency for a tau decaying into
a jet is limited.\\
Nevertheless, the discovery potentials of Fig.\ref{b:fig2tau}
and Fig.\ref{b:fig1tau}
represent also an important improvement with respect to
the experimental mass limits from LEP measurements \cite{b:Mass,b:neut}
and to the low-energy indirect constraint
$\l'_{311}<0.10 (m_{\tilde d_R}/100GeV)$ at $1 \sigma$
(from $\tau^- \to \pi^- \nu_{\tau}$) \cite{b:Bhatt}.\\
We also observe in Fig.\ref{b:fig2tau} and Fig.\ref{b:fig1tau} that
the results obtained from the $\tilde \chi^{\pm}_1 \tau^{\mp}$
production analysis in the case of a single dominant $\l'_{3jk}$
coupling depend strongly on the reconstruction of
the tau-jets. This is due to the large number of tau leptons
among the 3 charged leptons of the considered final state.

Using the same method and same SUSY point
as in Section \ref{lp211},
we have estimated the sensitivity on all the $\l'_{3jk}$ coupling
constants from the sensitivity obtained on $\l'_{311}$ at
$95 \% CL$ for a luminosity of ${\cal L}=2fb^{-1}$. The
results are given in Table \ref{couptau}.
\begin{table}[t]
\begin{center}
\begin{tabular}{|c|c|c|c|c|c|c|c|}
\hline
$\l'_{312}$ & $\l'_{313}$ & $\l'_{321}$ & $\l'_{322}$
& $\l'_{323}$ & $\l'_{331}$ & $\l'_{332}$ & $\l'_{333}$  \\
\hline
0.13 & 0.23 & 0.18 & 0.41 & 0.70 & 0.33 & 1.17 & 2.05 \\
\hline
\end{tabular}
\caption{\em
Sensitivities at $95 \% CL$ on the $\l'_{3jk}$
coupling constants derived from the sensitivity
on $\l'_{311}$ for a luminosity of ${\cal L}=2fb^{-1}$
and the following set of SUSY parameters,
$\tan\beta=1.5$,  $M_2=200GeV$, $\mu=-200GeV$ and
$m_0=180GeV$.}
\label{couptau}
\end{center}
\end{table}
All the sensitivities on the \rpv couplings
presented in Table \ref{couptau}, except those on $\l'_{32k}$,
are stronger than the present
indirect limits on the same \rpv couplings:
$\l'_{31k}<0.10 (m_{\tilde d_{kR}}/100GeV)$
at $1 \sigma$ ($\tau^- \to \pi^- \nu_{\tau}$),
$\l'_{32k}<0.20$ (for $m_{\tilde l}=m_{\tilde q}=100GeV$)
at $1 \sigma$ ($D^0-\bar D^0$ mix),
$\l'_{33k}<0.48 (m_{\tilde q}/100GeV)$ at $1 \sigma$ ($R_{\tau}$)
\cite{b:Bhatt}.\\
We mention that in the case of a single dominant $\l'_{3j3}$
coupling,
the neutralino decays as $\tilde \chi^0_1 \to \tau u_j b$
so that the b semileptonic decay could affect
a little the analysis efficiency.

We discuss now the sensitivities that could be obtained on a
single dominant
$\l'_{1jk}$ coupling constant via the analysis of the reaction
$p \bar p \to \tilde \chi^{\pm}_1 e^{\mp}$ (see
Fig.\ref{graphes}(a)). Since the cross section of the
$\tilde \chi^{\pm}_1 e^{\mp}$ production through $\l'_{1jk}$
is equal to the rate of
the $\tilde \chi^{\pm}_1 \mu^{\mp}$ production via $\l'_{2jk}$,
for same $j$ and $k$ indices (see Section \ref{cross1}),
the sensitivity obtained on a $\l'_{1jk}$ coupling constant
is expected to be identical to the sensitivity on $\l'_{2jk}$.
If we assume that the sensitivities obtained on the $\l'_{1jk}$
couplings are equal to those presented in Table \ref{coup},
we remark that for the SUSY point chosen in this table
only the sensitivities expected for the $\l'_{112}$,
$\l'_{113}$, $\l'_{121}$, $\l'_{131}$ and $\l'_{132}$ couplings
are stronger than the corresponding low-energy bounds:
$\l'_{11k}<0.02 (m_{\tilde d_{kR}}/100GeV)$ at $2 \sigma$
(Charged current universality),
$\l'_{1j1}<0.035 (m_{\tilde q_{jL}}/100GeV)$ at $2 \sigma$
(Atomic parity violation),
$\l'_{132}<0.34$ at $1 \sigma$ for $m_{\tilde q}=100GeV$
($R_e$) \cite{b:Bhatt}.
The reason is that the low-energy constraints on the $\l'_{1jk}$
couplings are typically more stringent than the limits on the
$\l'_{2jk}$ couplings \cite{b:Bhatt}.

\subsubsection{Mass reconstructions}
\label{recons}

\begin{figure}[t]
\begin{center}
\leavevmode
\centerline{\psfig{figure=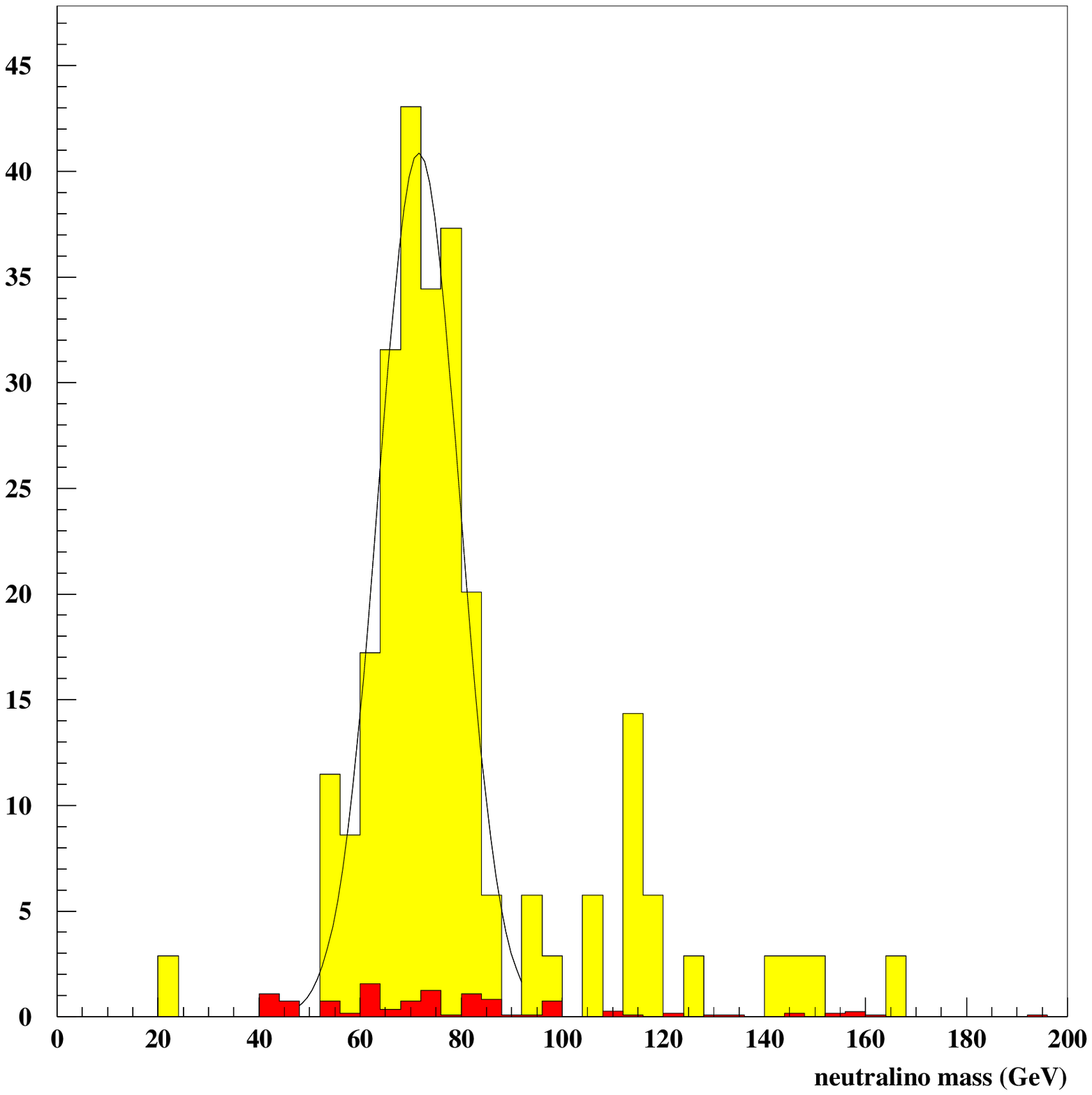,height=5.5in}}
\end{center}
\caption{Distribution of the $softer \ \mu + 2j$
invariant mass in the $e + \mu + \mu + 2j + \nu$ events,
for a luminosity of ${\cal L}=10fb^{-1}$. The sum of the
$WZ$, $ZZ$ and $t \bar t$ backgrounds is in black and the
SUSY signal is in grey. The mSUGRA point taken for this figure is
$m_0=200GeV$, $M_2=150GeV$, $\tan \beta=1.5$ and $sign(\mu)<0$
($m_{\tilde \chi^0_1}=77.7GeV$)
and the considered \rpv coupling is $\l'_{211}=0.09$. The average reconstructed
$\tilde \chi^0_1$ mass is 71 $\pm$ 9 GeV.}
\label{rec3n}
\end{figure}

The $\tilde \chi^0_1$ neutralino decays in our
framework as $\tilde \chi^0_1 \to l_i u_j d_k$ through the
$\l'_{ijk}$ coupling constant.
The invariant mass distribution of the lepton and the 2 jets
coming from this decay channel is peaked at the
$\tilde \chi^0_1$ mass. The experimental analysis of this
invariant mass distribution would thus be particularly interesting
since it would allow a model independent determination of the
lightest neutralino mass.

We have performed the $\tilde \chi^0_1$ mass reconstruction
based on the 3 lepton signature analysis.
The difficulty of this mass reconstruction
lies in the selection of the lepton and the 2 jets coming from the
$\tilde \chi^0_1$ decay.
In the signal we are considering, the only jets come from the
$\tilde \chi^0_1$ decay, and of course
from the initial and final QCD
radiations. Therefore, if there are more than 2 jets
in the final state we have selected the 2 hardest ones.
It is more subtle for the selection of the lepton since
our signal contains 3 leptons.
We have considered the case of a single dominant
coupling of type $\l'_{2jk}$ and focused on the
$e \mu \mu$ final state.
In these events, one of the $\mu^{\pm}$ is generated
in the decay of the produced sneutrino as
$\tilde \nu_{\mu} \to \tilde \chi^{\pm}_1 \mu^{\mp}$
and the other one in the decay of the
$\tilde \chi^0_1$ as $\tilde \chi^0_1 \to \mu^{\pm} u_j d_k$,
while the electron comes from the chargino decay
$\tilde \chi^{\pm}_1 \to \tilde  \chi^0_1 e^{\pm} \nu_e$.
Indeed, the dominant contribution to the single
chargino production is the resonant sneutrino production
(see Fig.\ref{graphes}).
In order to select the muon from the $\tilde \chi^0_1$ decay
we have chosen the
softer muon, since for relatively important values of the
$m_{\tilde \nu_{\mu}}-m_{\tilde \chi^{\pm}_1}$ mass
difference the muon
generated in the sneutrino decay is the most energetic.
Notice that when the $\tilde \nu_{\mu}$ and
$\tilde \chi^{\pm}_1$ masses are close to one another, the sensitivity
on the SUSY parameters
suffers a strong decrease as shown in Section \ref{lp211}.

We present in Fig.\ref{rec3n}
the invariant mass distribution of the muon and the 2 jets
produced in the $\tilde \chi^0_1$ decay.
This distribution has been obtained by using the selection
criteria described above and by considering the mSUGRA point:
$m_0=200GeV$, $M_2=150GeV$, $\tan \beta=1.5$, $sign(\mu)<0$
and $\l'_{211}=0.09$ ($m_{\tilde \chi^0_1} = 77.7 GeV$,
$m_{\tilde \chi^{\pm}_1} = 158.3 GeV$,
$m_{\tilde \nu_L} = 236 GeV$).
We also show on the plot of Fig.\ref{rec3n}
the fit of the invariant mass distribution.
As can be seen from this fit, the distribution is well
peaked around the $\tilde \chi^0_1$ generated mass.
The average reconstructed $\tilde \chi^0_1$ mass is of
$71 \pm 9GeV$.

\begin{figure}[t]
\begin{center}
\leavevmode
\centerline{\psfig{figure=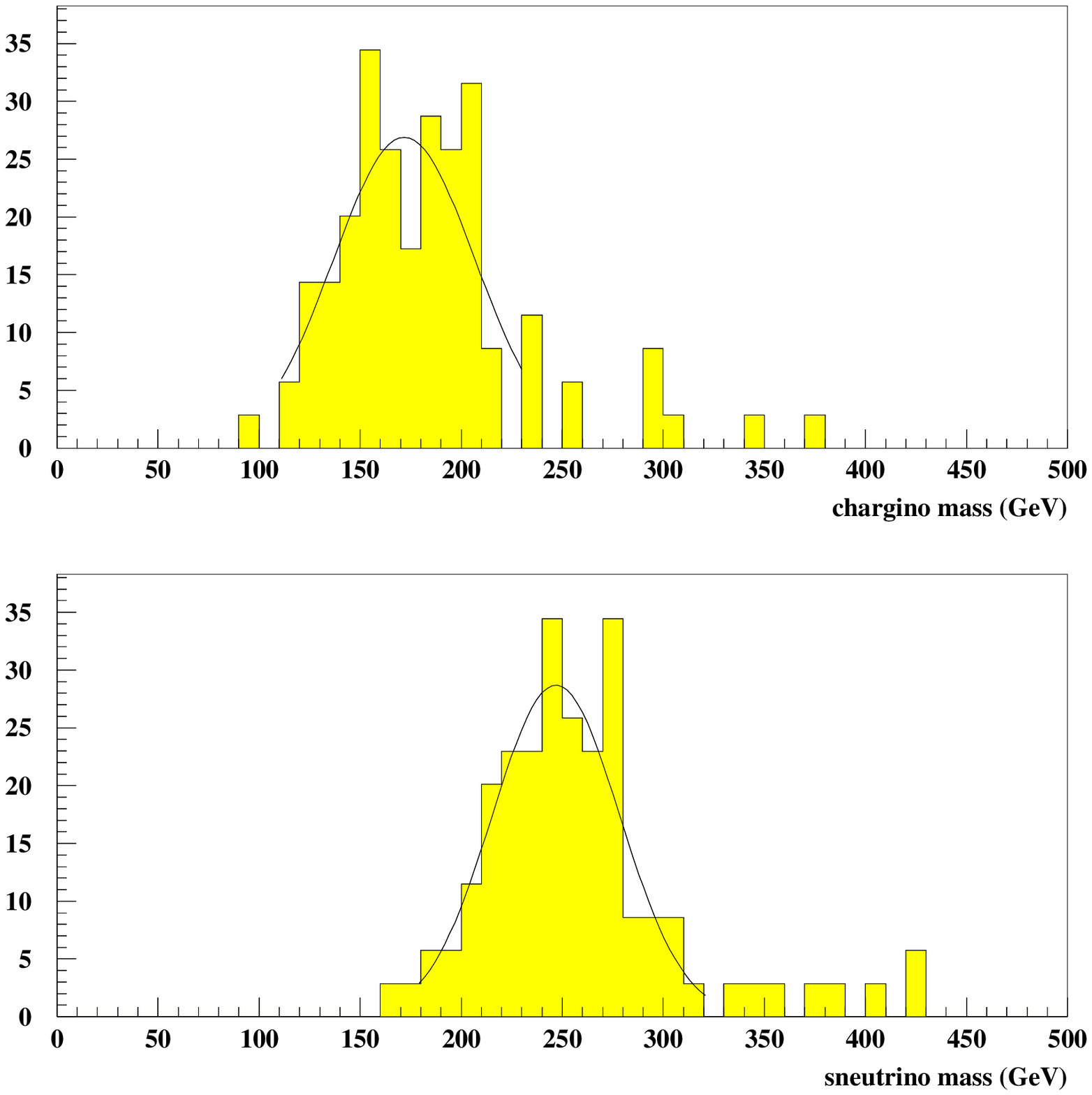,height=5.5in}}
\end{center}
\caption{Distributions of the $e + softer \ \mu + 2j +\nu$
(upper plot) and $e + \mu + \mu + 2j + \nu$ (lower plot)
invariant masses in the $e + \mu + \mu + 2j + \nu$ events,
for a luminosity of ${\cal L}=10fb^{-1}$.
The mSUGRA point taken for these figures is
$m_0=200GeV$, $M_2=150GeV$, $\tan \beta=1.5$ and $sign(\mu)<0$
($m_{\tilde \chi^{\pm}_1}= 158.3 GeV$, $m_{\tilde \nu_{\mu L}}= 236 GeV$)
and the considered \rpv coupling is $\l'_{211}=0.09$. 
The average reconstructed masses are
$m_{\tilde \chi^{\pm}_1} = 171 \pm 35 GeV$ and
$m_{\tilde \nu_{\mu L}} = 246 \pm 32 GeV$.}
\label{rec3c}
\end{figure}

We have also performed the $\tilde \chi^{\pm}_1$ and
$\tilde \nu_{\mu}$ mass reconstructions
based on the 3 lepton signature analysis in the scenario
of a single dominant coupling of type $\l'_{2jk}$.
The $\tilde \chi^{\pm}_1$ and $\tilde \nu_{\mu}$ masses
reconstructions are based on the 4-momentum of the
neutrino present in the $3l+2j+\nu$ final state (see Section
\ref{signal1}). The transverse component
of this momentum can be deduced from the momentum of the
charged leptons and jets present in the final state. However, the
longitudinal component of the neutrino momentum remains unknown
due to the poor detection at small polar angle values.
Therefore, in this study we have assumed a vanishing
longitudinal component of the neutrino momentum.
Besides, we have focused on the $e \mu \mu$ events as in the
$\tilde \chi^0_1$ mass reconstruction study. In this context,
the cascade decay initiated by the produced
lightest chargino is
$\tilde \chi^{\pm}_1 \to \tilde \chi^0_1 e^{\pm} \nu_e$,
$\tilde \chi^0_1 \to \mu^{\pm} u_j d_k$. Therefore, the
$\tilde \chi^{\pm}_1$ has been reconstructed
from the softer muon, the 2 jets, the electron and
the neutrino present in the final state,
since the softer muon is mainly generated
in the $\tilde \chi^0_1$ decay as explained above.
The $\tilde \nu_{\mu}$ has then been reconstructed
from the $\tilde \chi^{\pm}_1$
and the leading muon of the final state.
This was motivated by the fact
that the dominant contribution to the single
chargino production is the reaction
$p \bar p \to \tilde \nu_{\mu} \to \tilde \chi^{\pm}_1 \mu^{\mp}$
(see Fig.\ref{graphes}).

In Fig.\ref{rec3c}, we present the $\tilde \chi^{\pm}_1$ and
$\tilde \nu_{\mu}$ mass reconstructions performed through the
method presented above. We also show on the plots of Fig.\ref{rec3c}
the fits of the invariant mass distributions.
As can be seen from those fits, the distributions are well
peaked around the $\tilde \chi^{\pm}_1$ and $\tilde \nu_{\mu L}$
generated masses. The average reconstructed masses are
$m_{\tilde \chi^{\pm}_1} = 171 \pm 35 GeV$ and
$m_{\tilde \nu_{\mu L}} = 246 \pm 32 GeV$.
This study on the $\tilde \chi^{\pm}_1$ and $\tilde \nu_{\mu L}$
masses shows that based on a simplified mass
reconstruction analysis promising results are obtained from
the 3 lepton signature generated by the single
$\tilde \chi^{\pm}_1$ production. The $\tilde \chi^{\pm}_1$ and
$\tilde \nu_{\mu L}$ mass reconstructions can be
improved using constrained fits.

In the hypothesis of a single dominant coupling constant
of type $\l'_{1jk}$, exactly the same kind of
$\tilde \chi^0_1$, $\tilde \chi^{\pm}_1$
and $\tilde \nu_{\mu}$ mass reconstructions can
be performed by selecting the $e + e + \mu + 2j + \nu$ events.
In contrast, the case of a single dominant
$\l'_{3jk}$ coupling requires more sophisticated methods.

As a conclusion, in the case of a single dominant
coupling constant of type $\l'_{1jk}$ or $\l'_{2jk}$,
the $\tilde \chi^0_1$, $\tilde \chi^{\pm}_1$
and $\tilde \nu_{\mu}$ mass reconstructions based on
the 3 lepton signature generated by the single
$\tilde \chi^{\pm}_1$ production at Tevatron can easily give
precise results, in contrast with the mass reconstructions
performed in the superpartner pair production analysis
at hadronic colliders which suffer a high
combinatorial background \cite{b:Atlas}.

\subsubsection{Model dependence of the results}

In this Section, we discuss qualitatively
the impact on our results of the choice of our theoretical model, namely
mSUGRA with the infrared fixed point hypothesis for the top quark
Yukawa coupling.
We focus on the discovery potentials obtained in Sections \ref{lp211},
\ref{htanb} and \ref{lp311}, since the choice of the theoretical framework
does not influence the study of the neutralino mass reconstruction made
in Section \ref{recons} which is model independent.

The main effect of the infrared fixed point approach is to fix
the value of the $\tan \beta$ parameter, up to the
ambiguity on the low or high solution.
Therefore, the infrared fixed point
hypothesis has no important effects on the results
since the dependences of the single gaugino productions rates
on $\tan \beta$ are smooth,
in the high $\tan \beta$ scenario (see Section \ref{cross1}).

As we have mentioned in Section \ref{theoretical},
in the mSUGRA scenario,
the $\vert \mu \vert$ parameter is fixed. This point does
not influence much our results since the single gaugino
production cross sections vary weakly with $\vert \mu \vert$
as shown in Section \ref{cross1}.

Another particularity of the mSUGRA model is that
the LSP is the $\tilde \chi^0_1$ in most of the parameter space.
For instance, in a model where the LSP would be
the lightest chargino or a squark,
the contribution to the three lepton signature from the
$\tilde \chi^{\pm}_1 l^{\mp}$ production would vanish.

Finally in mSUGRA,
the squark masses are typically larger than the lightest chargino mass
so that the decay $\tilde \chi^{\pm}_1 \to \tilde \chi^0_1 l^{\pm} \nu$
has a branching ratio of at least
$\sim 30\%$ (see Section \ref{signal1}).
In a scenario where $m_{\tilde \chi^{\pm}_1}>m_{\tilde q}$,
the two-body decay $\tilde \chi^{\pm}_1 \to \tilde q q$
would be dominant so that the contribution to the
three lepton signature from the
$\tilde \chi^{\pm}_1 l^{\mp}$ production would be small.
Besides, in mSUGRA, the
$\tilde \chi^{\pm}_1-\tilde \chi^0_1$ mass difference
is typically large enough to avoid
significant branching ratio for the
\rpv decay of the lightest chargino which would result in a decrease
of the sensitivities on the SUSY parameters
presented in Sections \ref{lp211}, \ref{htanb} and \ref{lp311}.

In a model where the contribution to the three lepton signature from the
$\tilde \chi^{\pm} l^{\mp}$ production would be suppressed,
the three lepton final state could be generated in a significant way
by other single gaugino productions, namely the $\tilde \chi^{\pm} \nu$,
$\tilde \chi^0 l^{\mp}$ or $\tilde \chi^0 \nu$ productions.

\section{Like sign dilepton signature analysis}
\label{analysis2}

\setcounter{equation}{0}

\subsection{Signal}
\label{signal2}

Within the context of the mSUGRA model,
three of the single gaugino
productions via $\l'_{ijk}$
presented in Section \ref{resonant} can generate
a final state containing a pair of same sign leptons.
As a matter of fact,
the like sign dilepton signature can be produced
through the reactions
$p \bar p \to \tilde \chi^0_1 l^{\pm}_i$;
$p \bar p \to \tilde \chi^0_2 l^{\pm}_i$,
$\tilde \chi^0_2 \to \tilde \chi^0_1 + X$ ($X \neq l^{\pm}$);
$p \bar p \to \tilde \chi^{\pm}_1 l^{\mp}_i$,
$\tilde \chi^{\pm}_1 \to \tilde \chi^0_1 q \bar q$ and
$p \bar p \to \tilde \chi^{\pm}_1 \nu_i$,
$\tilde \chi^{\pm}_1 \to \tilde \chi^0_1 l^{\pm} \nu$,
$i$ corresponding to the flavour index of the
$\l'_{ijk}$ coupling.
Indeed, since the $\tilde \chi^0_1$ is a Majorana particle,
it decays via
$\l'_{ijk}$ into a lepton, as $\tilde \chi^0_1 \to l_i u_j \bar d_k$,
and into an anti-lepton, as $\tilde \chi^0_1 \to \bar l_i \bar u_j d_k$,
with the same probability.
The $\tilde \chi^0_{3,4} l^{\pm}_i$, $\tilde \chi^{\pm}_2 l^{\mp}_i$ and
$\tilde \chi^{\pm}_2 \nu_i$ productions
do not bring significant contributions to the
like sign dilepton signature due to their relatively small cross sections
(see Section \ref{cross1}).

In mSUGRA, the most important contribution to the
like sign dilepton signature originates from the $\tilde \chi^0_1 l^{\pm}_i$
production since this reaction has a dominant cross section
in most of the mSUGRA parameter space, as shown in
Section \ref{cross1}.
The other reason is that if $\tilde \chi^0_1$ is the LSP,
the $\tilde \chi^0_1 l^{\pm}_i$ production rate
is not affected by branching ratios of any cascade decay since the
$\tilde \chi^0_1$ only decays through \rpv coupling.

\subsection{Standard Model
background of the like sign dilepton signature at Tevatron}
\label{back2}

The $b \bar b$ production can lead to the like sign dilepton signature
if both of the b quarks decay semi-leptonically. The leading order
cross section
of the $\bar b b$ production at Tevatron for an energy of $\sqrt s=2TeV$ is
$\sigma (p \bar p \to b \bar b) \approx 4.654 \ 10^{10}fb$.
This rate has been calculated with PYTHIA \cite{b:PYTHIA}
using the CTEQ2L structure function.

The $t \bar t$ production, followed by the decays
$t \to W^+ b \to l^+ \nu b$,
$\bar t \to W^- \bar b \to \bar q q \bar b \to \bar q q l^+ \nu \bar c$,
or $t \to W^+ b \to \bar q q b \to \bar q q l^- \bar \nu c$,
$\bar t \to W^- \bar b \to l^- \bar \nu \bar b$,
also generates a final state with two same sign leptons.
The leading order cross section of the $t \bar t$ production at
$\sqrt s=2TeV$, including the relevant branching ratios,
is $\sigma (p \bar p \to t \bar t) \times 2
\times B(W \to l_p \nu_p) \times B(W \to q_p \bar q_{p'})
\approx 3181fb$ ($2800fb$) for $m_{top}=170GeV$ ($175GeV$) with
$p,p'=1,2,3$.

The third important source of \sm background is the
$t \bar b / \bar t b$ production
since the (anti-)$b$ quark can undergo a semi-leptonic decay as
$b \to l^- \bar \nu c$ ($\bar b \to l^+ \nu \bar c$)
and the (anti-) top quark can decay simultaneously as $t \to b W^+ \to b l^+
\nu$
($\bar t \to \bar b W^- \to \bar b l^- \bar \nu$).
The leading order cross section at
$\sqrt s=2TeV$ including the branching fraction is
$\sigma(p \bar p \to t q, \bar t q) \times B(W \to l_p \nu_p) \approx 802fb$
($687fb$) for $m_{top}=170GeV$ ($175GeV$) with $p=1,2,3$.

Other small sources of \sm background are
the $W^{\pm} W^{\mp}$ production,
followed by the decays: $W \to l \nu$ and $W \to b u_p$ ($p=1,2$)
or $W \to b u_p$ and $W \to b u_p$ ($p=1,2$),
the $W^{\pm} Z^0$ production,
followed by the decays: $W \to l \nu$ and $Z \to b \bar b$
or $W \to q_p \bar q_{p'}$ and $Z \to b \bar b$,
and the $Z^0 Z^0$ production,
followed by the decays: $Z \to l \bar l$ and $Z \to b \bar b$
or $Z \to q_p \bar q_p$ and $Z \to b \bar b$.

Finally, the 3 lepton final states generated by the $Z^0 Z^0$
and $W^{\pm} Z^0$ productions
(see Section \ref{back1}) can be mistaken
for like sign dilepton events
in case where one of the leptons is lost in the detection.
Non-physics sources of background can also be caused by some fake leptons
or by the misidentification of the charge of a lepton.

Therefore for the study of the \sm background associated to
the like sign dilepton signal at Tevatron Run II,
we consider the $b \bar b$, the $t \bar t$, the
$W^{\pm} W^{\mp}$ and the single top
production and both the physics and non-physics
contributions generated by the $W^{\pm} Z^0$
and $Z^0 Z^0$ productions.

\subsection{Supersymmetric background
of the like sign dilepton signature at Tevatron}
\label{susyback2}

All the pair productions of superpartners are a source of SUSY background
for the like sign dilepton signature originating from the
single gaugino productions. Indeed, both of the produced superpartners
initiate a cascade decay ended by the \rpv decay of the LSP
through $\l'_{ijk}$, and if the two LSP's
undergo the same decay $\tilde \chi^0_1 \to l_i u_j \bar d_k$ or
$\tilde \chi^0_1 \to \bar l_i \bar u_j d_k$,
two same sign charged leptons are generated. Another possible way
for the SUSY pair production to generate the like sign dilepton signature
is that only one of the LSP's decays into a charged lepton of a given sign,
the other decaying as $\tilde \chi^0_1 \to \nu_i d_j d_k$, and
a second charged lepton of the same sign is produced in the cascade decays.

The cross sections of the superpartners pair productions
have been studied in Section \ref{susyback1}.

\subsection{Cuts}
\label{cut2}

In order to simulate the single chargino productions
$p \bar p \to \tilde \chi^{\pm}_1 l^{\mp}$,
$p \bar p \to \tilde \chi^{\pm}_1 \nu$
and the single neutralino production
$p \bar p \to \tilde \chi^0_1 l^{\mp}$ at Tevatron,
the matrix elements (see Appendix \ref{formulas}) of these
processes have been implemented
in a version of the SUSYGEN event generator \cite{b:SUSYGEN3}
allowing the generation of $p \bar p$ reactions \cite{b:priv}.
The \sm background
($W^{\pm} W^{\mp}$, $W^{\pm} Z^0$, $Z^0 Z^0$,
$t \bar b / \bar t b$, $t \bar t$ and $b \bar b$ productions)
has been simulated
using the PYTHIA event generator \cite{b:PYTHIA}
and the SUSY background (all SUSY particles pair productions) using the
HERWIG event generator \cite{b:HERWIG}.
SUSYGEN, PYTHIA and HERWIG have been interfaced with
the SHW detector simulation package \cite{b:SHW}
(see Section \ref{cut1}).

Several selection criteria have been applied
in order to reduce the background.\\
First, we have selected the events containing two
same sign muons. The reason is that in the like sign dilepton
signature analysis we have focused on the
case of a single dominant \rpv coupling constant
of the type $\l'_{2jk}$. In such a scenario, the two same charge
leptons generated in the $\tilde \chi^0_1 l^{\mp}$ production,
which represents the main contribution to the like sign
dilepton final state (see Section \ref{signal2}), are muons
(see Fig.\ref{graphes} and Section \ref{signal2}).
This requirement that the 2 like sign leptons
have the same flavour
allows to reduce the \sm background with respect to the
signal.

We require a number of jets greater or equal to two with a
transverse momentum
higher than $10 GeV$, namely $N_j \geq 2$ with $P_t(j) > 10 GeV$.
This jet veto reduces the non-physics backgrounds
generated by the $W^{\pm} Z^0$ and $Z^0 Z^0$
productions (see Section \ref{back2}) which produce
at most one hard jet (see Section \ref{cut1}).

\begin{figure}[t]
\begin{center}
\leavevmode
\centerline{\psfig{figure=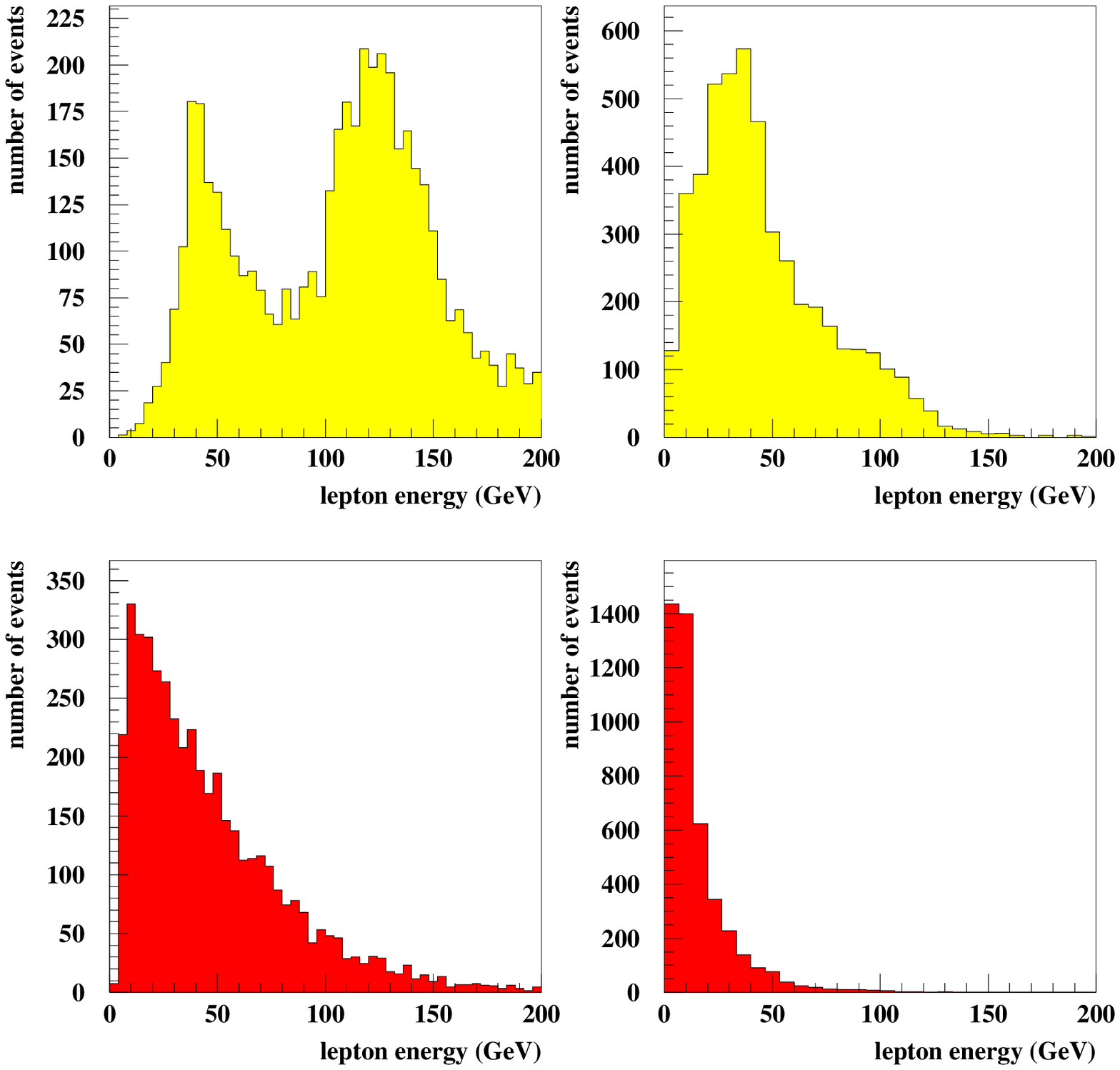,height=5.5in}}
\end{center}
\caption{\footnotesize  \it
Distributions of the 2 muon energies (in $GeV$)
in the events containing 2 same sign muons and at least 2 jets
generated by the \sm background (lower curve),
namely the $W^{\pm} W^{\mp}$,
$W^{\pm} Z^0$, $Z^0 Z^0$, $t \bar t$, $t \bar b / \bar t b$
and $b \bar b$ productions,
and the SUSY signal (upper curve), for $\l'_{211}=0.05$,
$M_2=250GeV$, $m_0=200GeV$, $\tan \beta=1.5$
and $sign(\mu)<0$.
The left plots represent the leading muon distributions and
the right plots the second leading muon distributions.
The numbers of events correspond to an integrated luminosity of
${\cal L}=10fb^{-1}$.
\rm \normalsize }
\label{dienmu}
\end{figure}

Besides, some effective cuts concerning the energies of the 2 selected muons
have been applied. In Fig.\ref{dienmu},
we present the distributions of the 2 muon energies
in the like sign dimuon events generated by the \sm background
($W^{\pm} W^{\mp}$, $W^{\pm} Z^0$, $Z^0 Z^0$, $t \bar t$, $t \bar b / \bar t
b$
and $b \bar b$) and the SUSY signal.
Based on these distributions, we have chosen the following cuts
on the muon energies: $E(\mu_2)>20GeV$ and $E(\mu_1)>20GeV$.

We will refer to all the selection criteria described above, namely
2 same sign muons with $E(\mu_2)>20GeV$ and $E(\mu_1)>20GeV$,
and $N_j \geq 2$ with $P_t(j) > 10 GeV$, as cut $1$.

Let us explain the origin of the two peaks in the upper left
plot of Fig.\ref{dienmu}. This will be helpful for the mass
reconstruction study of Section \ref{reconsp}.\\
The main contribution to the like sign dimuon signature
from the SUSY signal is
the $\tilde \chi^0_1 \mu^{\pm}$ production (see Section
\ref{signal2}) in the case of a single dominant $\l'_{2jk}$
coupling. Furthermore, the dominant contribution to this
production is the reaction $p \bar p \to \tilde \mu^{\pm}_L \to
\tilde \chi^0_1 \mu^{\pm}$. In this reaction, the $\mu^{\pm}$
produced together with the $\tilde \chi^0_1$ has an
energy around $E(\mu^{\pm}) \approx (m_{\tilde \mu^{\pm}_L}^2
+m_{\mu^{\pm}}^2-m_{\tilde \chi^0_1}^2)/2 m_{\tilde \mu^{\pm}_L}
=121.9GeV$ for the SUSY point considered in Fig.\ref{dienmu},
namely $M_2=250GeV$, $m_0=200GeV$, $\tan \beta=1.5$
and $sign(\mu)<0$, which gives rise to the mass spectrum:
$m_{\tilde \chi^0_1}=127.1GeV$, $m_{\tilde \chi^0_2}=255.3GeV$,
$m_{\tilde \chi^{\pm}_1}=255.3GeV$, $m_{\tilde l^{\pm}_L}=298GeV$
and $m_{\tilde \nu^{\pm}_L}=294GeV$.
This energy value corresponds approximatively
to the mean value of the right peak of
the leading muon energy distribution presented in
the upper left plot of Fig.\ref{dienmu}. This is due to the
fact that the leading muon in the dimuon events generated by
the reaction $p \bar p \to \tilde \chi^0_1 \mu^{\pm}$ is the
$\mu^{\pm}$ produced together with the $\tilde \chi^0_1$
for relatively important values of the $m_{\tilde \mu^{\pm}_L}-
m_{\tilde \chi^0_1}$ mass difference. The right peak
in the upper left plot of Fig.\ref{dienmu} is thus associated
to the $\tilde \chi^0_1 \mu^{\pm}$ production. \\
Similarly,
the left peak in the upper left plot of Fig.\ref{dienmu}
corresponds to the reactions $p \bar p \to \tilde \mu^{\pm}_L
\to \tilde \chi^0_2 \mu^{\pm}$ and $p \bar p
\to \tilde \nu_{\mu L} \to \tilde \chi^{\pm}_1 \mu^{\mp}$
which produce $\mu^{\pm}$ of energies around $E(\mu^{\pm})
\approx (m_{\tilde \mu^{\pm}_L}^2+m_{\mu^{\pm}}^2
-m_{\tilde \chi^0_2}^2)/2 m_{\tilde \mu^{\pm}_L}=39.6GeV$ and
$E(\mu^{\pm}) \approx (m_{\tilde \nu_{\mu L}}^2
+m_{\mu^{\pm}}^2-m_{\tilde \chi^{\pm}_1}^2)/2
m_{\tilde \nu_{\mu L}}=36.2GeV$, respectively.
The $\tilde \chi^{\pm}_1 \nu_{\mu}$ production represents a
less important contribution to the like sign dimuon events
compared to the 3 above single gaugino productions
since the 2 same sign leptons generated
in this production are not systematically muons and the
involved branching ratios have smaller values
(see Section \ref{signal2}).

\begin{figure}[t]
\begin{center}
\leavevmode
\centerline{\psfig{figure=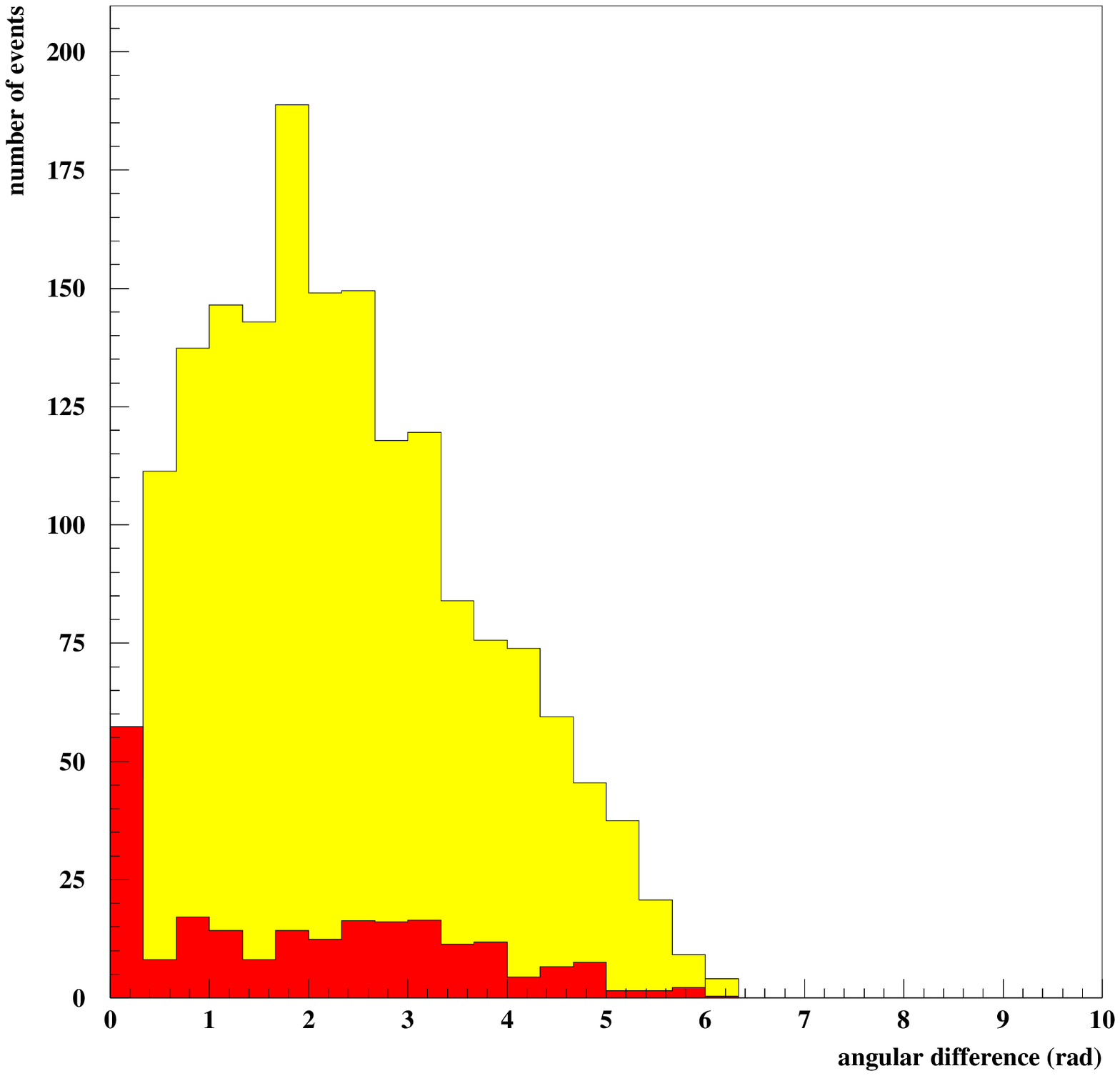,height=5.5in}}
\end{center}
\caption{\footnotesize  \it
Distributions of the $\Delta R$ angular difference (in $rad$)
between the second leading muon and the second leading
jet in the like sign dimuons events selected by applying cut $1$ and
generated by the \sm background (curve in black),
namely the
$W^{\pm} W^{\mp}$, $W^{\pm} Z^0$, $Z^0 Z^0$, $t \bar t$, $t \bar b / \bar t
b$
and $b \bar b$ productions, and the SUSY signal
(curve in grey), for $\l'_{211}=0.05$,
$M_2=250GeV$, $m_0=200GeV$, $\tan \beta=1.5$ and $sign(\mu)<0$.
The numbers of events correspond to an integrated luminosity of
${\cal L}=10fb^{-1}$.
\rm \normalsize }
\label{dianmu}
\end{figure}

Finally, since the leptons produced in the quark $b$
decays are not well isolated (as in the $W^{\pm} W^{\mp}$,
$W^{\pm} Z^0$, $Z^0 Z^0$, $t \bar t$, $t \bar b / \bar t b$
and $b \bar b$ productions),
we have applied some cuts on the lepton isolation.
We have imposed the isolation cut
$\Delta R=\sqrt{\delta \phi^2+\delta \theta^2}>0.4$ where
$\phi$ is the azimuthal angle and $\theta$ the polar angle
between the 2 same sign muons and the 2 hardest jets.
This cut is for example motivated
by the distributions shown in Fig.\ref{dianmu}
of the $\Delta R$ angular difference
between the second leading muon and the second leading
jet, in the like sign dimuons events generated by the SUSY signal
and \sm background.
We call cut $\Delta R>0.4$ together with 
cut $1$, cut $2$.

\begin{figure}[t]
\begin{center}
\leavevmode
\centerline{\psfig{figure=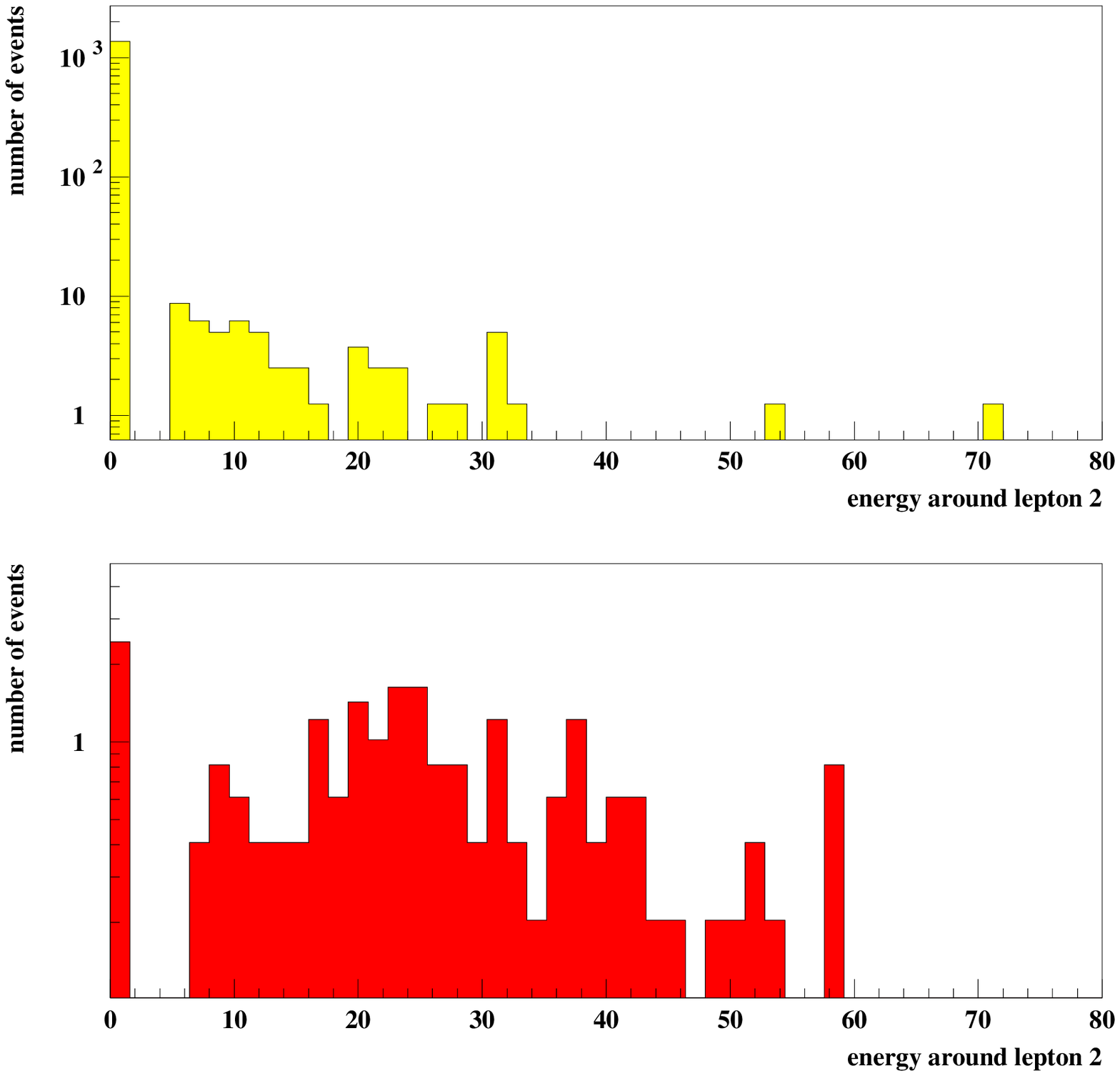,height=5.5in}}
\end{center}
\caption{\footnotesize  \it
Distributions of the summed energies ($E$, in $GeV$) of the jets being
close to the second leading muon, namely the jets contained
in the cone centered on the second leading muon
and defined by $\Delta R<0.25$, in the like sign dimuons events
selected by applying cut $2$ and
generated by the \sm background (lower curve), namely the
$W^{\pm} W^{\mp}$, $W^{\pm} Z^0$, $Z^0 Z^0$, $t \bar t$, $t \bar b / \bar t
b$
and $b \bar b$ productions,
and the SUSY signal (upper curve), for $\l'_{211}=0.05$,
$M_2=250GeV$, $m_0=200GeV$, $\tan \beta=1.5$ and $sign(\mu)<0$.
These distributions were obtained after cut $E<2GeV$, where $E$
represents the
summed energies of the jets being close to the leading muon, has been
applied
in these like sign dimuons events.
The numbers of events correspond to an integrated luminosity of
${\cal L}=10fb^{-1}$.
\rm \normalsize }
\label{ismu}
\end{figure}

In order to eliminate poorly isolated muons,
we have also imposed that $E<2GeV$, where $E$ represents the
summed energies of the jets being close to a muon,
namely the jets contained in the cone centered on a muon
and defined by $\Delta R<0.25$. This cut is for instance motivated by
the distributions shown in Fig.\ref{ismu} which represent
the summed energies $E$ of the jets being
close to the second leading muon in the like sign dimuons events
generated by the SUSY signal and \sm background.
We denote cut $E<2GeV$ plus cut $2$ as cut $3$.


\begin{figure}[t]
\begin{center}
\leavevmode
\centerline{\psfig{figure=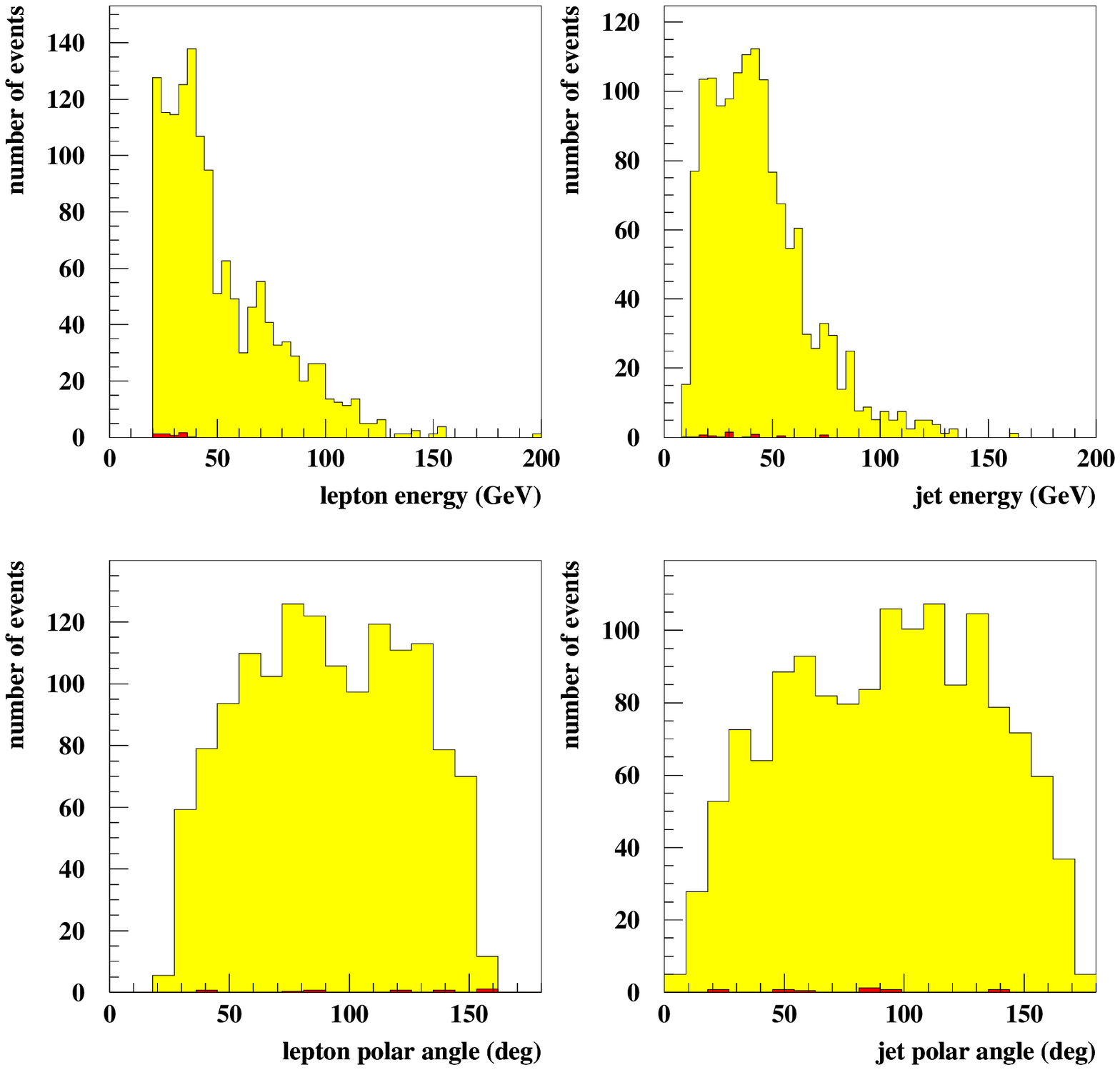,height=5.5in}}
\end{center}
\caption{\footnotesize  \it
Energy (in $GeV$) and polar angle ($\theta$, in $deg$) distributions
of the leading muon and the leading jet in the like sign dimuon events
selected by applying cut $3$ and
generated by the \sm background (curve in black), namely the
$W^{\pm} W^{\mp}$, $W^{\pm} Z^0$, $Z^0 Z^0$, $t \bar t$, $t \bar b / \bar t
b$
and $b \bar b$ productions,
and the SUSY signal (curve in grey), for $\l'_{211}=0.05$,
$M_2=250GeV$, $m_0=200GeV$, $\tan \beta=1.5$ and $sign(\mu)<0$.
The numbers of events correspond to an integrated luminosity of
${\cal L}=10fb^{-1}$.
\rm \normalsize }
\label{aprmu}
\end{figure}

The selected events require high energy charged leptons and jets
and can thus be easily triggered at Tevatron. Moreover, the
considered charged leptons and jets are typically emitted at
intermediate polar angles and would thus be often detected at Tevatron.
These points are illustrated in
Fig.\ref{aprmu} where are shown the energy and polar angle distributions
of the leading muon and the leading jet in the like sign dimuons events
selected
by applying cut $3$ and generated by the SUSY signal and \sm background.

\begin{table}[t]
\begin{center}
\begin{tabular}{|c|c|c|c|c|c|}
\hline
        & $W^{\pm} Z^0$  & $Z^0 Z^0$      & $t \bar t$       & $t \bar b / \bar t b$   & Total       \\
\hline
cut $1$ & $0.21\pm 0.06$ & $0.11\pm 0.04$ & $21.80 \pm 0.70$ & $0.69 \pm 0.13 $        & $22.81 \pm 0.71$  \\
\hline
cut $2$ & $0.05\pm 0.03$ & $0.03\pm 0.03$ & $8.80 \pm 0.50 $ & $0.28 \pm 0.08 $        & $9.16  \pm 0.51$  \\
\hline
cut $3$ & $0.03\pm 0.03$ & $0.01\pm 0.02$ & $0.64 \pm 0.13 $ & $0.10 \pm 0.05 $        & $0.78  \pm 0.14$  \\
\hline
\end{tabular}
\caption{Numbers of like sign dilepton events generated by the
\sm background
($W^{\pm} W^{\mp}$,
$W^{\pm} Z^0$, $Z^0 Z^0$, $t \bar t$, $t \bar b / \bar t b$
and $b \bar b$ productions)
at Tevatron Run II for the cuts described in the text, assuming an
integrated
luminosity of ${\cal L}=1 fb^{-1}$ and a center of mass energy of $\sqrt s=2
TeV$. The numbers of events coming from the
$W^{\pm} W^{\mp}$ and $b \bar b$ backgrounds have been found 
to be negligible after cut 3 is applied.
These results have been obtained by generating
$2 \ 10^4$ events for the $W^{\pm} Z^0$ production,
$    10^4$ events for the $W^{\pm} Z^0$ (non-physics contribution),
$3 \ 10^4$ events for the $Z^0 Z^0$,
$    10^4$ events for the $Z^0 Z^0$ (non-physics contribution),
$3 \ 10^5$ events for the $t \bar t$ and
$    10^5$ events for the $t \bar b / \bar t b$.}
\label{cutefp}
\end{center}
\end{table}

In Table \ref{cutefp}, we give the numbers of like sign dilepton events
expected from the
\sm background at Tevatron Run II with the various cuts described above.
We see in Table \ref{cutefp} that the main source
of \sm background to the like sign dilepton signature at Tevatron
is the $t \bar t$ production. This is due to its important
cross section compared to the other \sm backgrounds
(see Section \ref{back2}) and to the fact that in the $t \bar t$
background, in contrast with the $b \bar b$ background,
only one charged lepton of the final state
is produced in a $b$-jet and is thus not isolated.

\begin{table}[t]
\begin{center}
\begin{tabular}{|c|c|c|c|c|c|}
\hline
$m_{1/2} \ \backslash \ m_0$
&  $100GeV$ & $200GeV$   & $300GeV$  & $400GeV$
&  $500GeV$   \\
\hline
$100GeV$
& $101.64$      & $54.92$        & $44.82$       & $39.26$
& $38.77$ \\
\hline
$200GeV$
& $3.74$        & $4.08$         & $4.33$        & $4.56$
& $4.99$ \\
\hline
$300GeV$
& $1.04$        & $0.63$         & $0.61$        & $0.70$
& $0.66$ \\
\hline
\end{tabular}
\caption{Number of like sign dilepton events generated by the
SUSY background (all superpartner pair productions)
at Tevatron Run II
as a function of the $m_0$ and $m_{1/2}$ parameters
for $\tan \beta=1.5$, $sign(\mu)<0$ and $\l'_{211}=0.05$.
Cut 3 (see text) has been applied.
These results have been obtained by generating 7500 events
and correspond to an integrated luminosity
of ${\cal L}=1 fb^{-1}$
and a center of mass energy of $\sqrt s=2 TeV$.}
\label{cutSUSp}
\end{center}
\end{table}

In Table \ref{cutSUSp}, we give the number of like sign dilepton
events generated by the
SUSY background (all superpartners pair productions)
at Tevatron Run II as a function of
the $m_0$ and $m_{1/2}$ parameters for cut 3.
This number of events decreases as $m_0$ and $m_{1/2}$
increase due to the behaviour of the summed superpartners
pair production cross section in the SUSY parameter space
(see Section \ref{susyback1}).

\subsection{Results}
\label{resp}

\subsubsection{Discovery potential}
\label{lp211p}

We first present the reach in the
mSUGRA parameter space obtained from the analysis of
the like sign dilepton final state at Tevatron Run II produced
by the single neutralino and chargino productions
via $\l'_{211}$:
$p \bar p \to \tilde \chi^0_{1,2} \mu^{\pm}$,
$p \bar p \to \tilde \chi^{\pm}_1 \mu^{\mp}$ and
$p \bar p \to \tilde \chi^{\pm}_1 \nu_{\mu}$.
The sensitivities that can be obtained on the $\l'_{2jk}$
($j$ and $k$ being not equal to $1$ simultaneously), $\l'_{1jk}$
and $\l'_{3jk}$ coupling constants
will be discussed at the end of this section.

\begin{figure}[t]
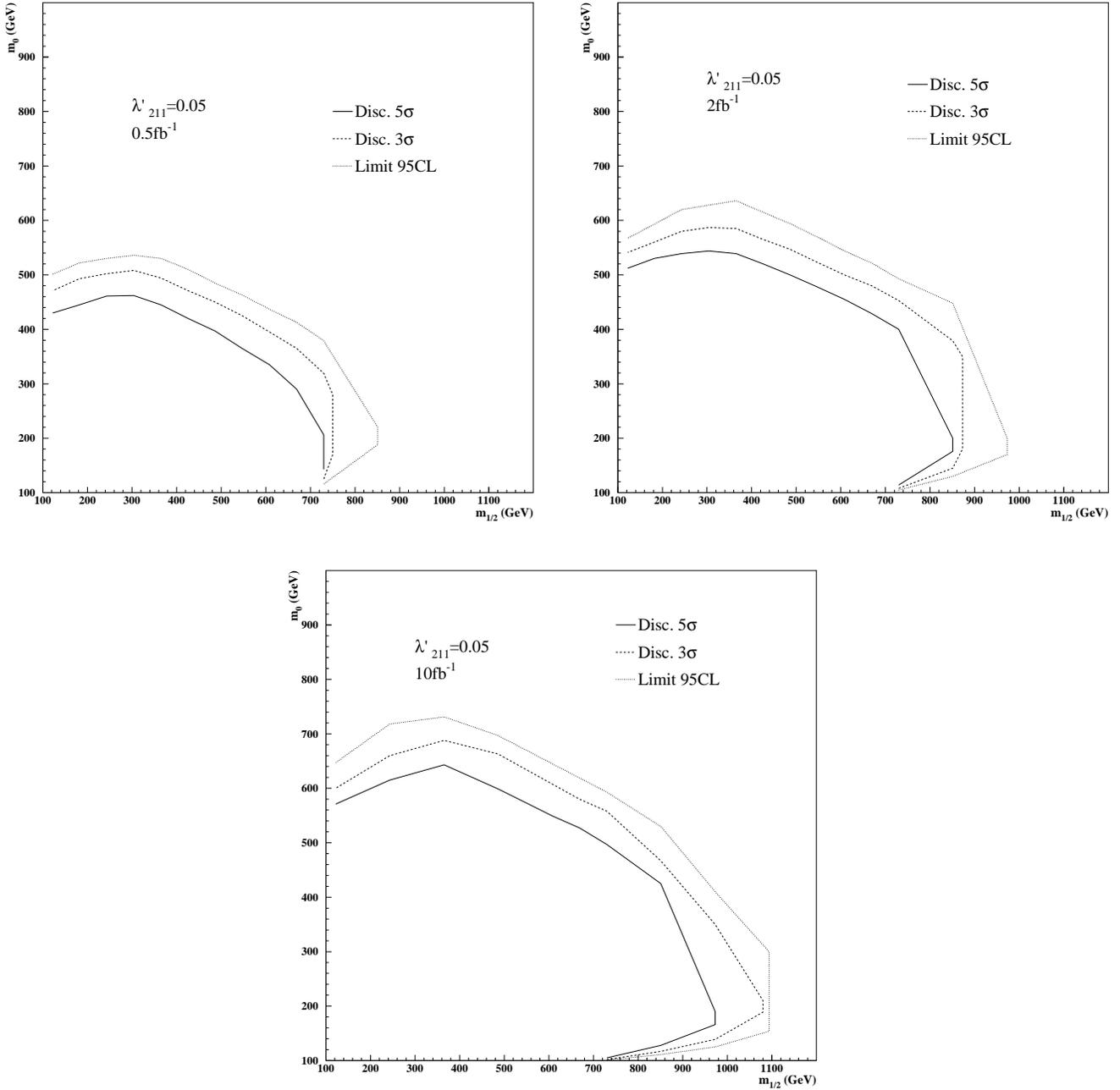

\begin{center}
\leavevmode
\centerline{
\psfig{figure=2lept_m0_m05_05fb.eps,height=3.5in}
\psfig{figure=2lept_m0_m05_2fb.eps,height=3.5in}}
\centerline{\psfig{figure=2lept_m0_m05_10fb.eps,height=3.5in}}
\end{center}
\caption{Discovery contours at $5 \sigma$ (full line),
$ 3 \sigma$ (dashed line)
and limit at $95 \% \ C.L.$ (dotted line)
obtained from the like sign dilepton signature analysis
at Tevatron Run II assuming
a center of mass energy of $\sqrt s=2 TeV$.
These discovery potentials are
presented in the plane $m_0$ versus $m_{1/2}$,
for $sign(\mu)<0$, $\tan \beta=1.5$, $\l'_{211}=0.05$
and different values of luminosity.}
\label{b:fig2d}
\end{figure}

In Fig.\ref{b:fig2d},
we present the $3 \sigma$ and $ 5 \sigma$ discovery
contours and the limits at $95 \%$
confidence level in the plane $m_0$ versus $m_{1/2}$,
for $sign(\mu)<0$, $\tan \beta=1.5$, $\l'_{211}=0.05$ and
using a set of values for the luminosity.
Those discovery potentials were obtained by considering
the $\tilde \chi^0_{1,2} \mu^{\pm}$,
$\tilde \chi^{\pm}_1 \mu^{\mp}$ and
$\tilde \chi^{\pm}_1 \nu_{\mu}$ productions and
the background originating from the Standard Model.
The signal and background were selected by
using cut $3$ described in Section \ref{cut2}.
The reduction of the sensitivity on $m_{1/2}$
observed in Fig.\ref{b:fig2d} as $m_0$ increases is due to the
decrease of the $\tilde \chi^0_{1,2} \mu^{\pm}$,
$\tilde \chi^{\pm}_1 \mu^{\mp}$ and
$\tilde \chi^{\pm}_1 \nu_{\mu}$ productions cross sections
with the $m_0$ increase observed in Fig.\ref{XScl}
and Fig.\ref{allXS}. In Fig.\ref{b:fig2d}, we also see that
the sensitivity on $m_{1/2}$ is reduced
in the domain $m_0 \stackrel{<}{\sim} 200GeV$.
This reduction of the sensitivity is due to the fact that
in mSUGRA at low $\tan \beta$ and
for large values of $m_{1/2}$ and small values
of $m_0$, the LSP is the Right slepton $\tilde l^{\pm}_{iR}$
($i=1,2,3$). Therefore, in this mSUGRA region
the dominant decay channel
of the lightest neutralino is $\tilde \chi^0_1
\to \tilde l^{\pm}_{iR} l^{\mp}_i$ ($i=1,2,3$) so that the
$\tilde \chi^0_1 \mu^{\pm}$ production, which is the main
contribution to the like sign dilepton signature, leads to
the $2 \mu^{\pm}+2 \ jets$ final state only in a few cases.
There are two reasons. First, in this mSUGRA scenario
the charged lepton produced in the main
$\tilde \chi^0_1$ decay is not systematically a muon.
Secondly, if the LSP is the Right slepton $\tilde l^{\pm}_{iR}$
it cannot decay in the case of a single dominant
$\l'_{ijk}$ coupling constant and it is thus a stable particle.

The sensitivities presented in the discovery reach of
Fig.\ref{b:fig2d} which are obtained from the like sign dilepton
signature analysis are higher than the sensitivities shown
in Fig.\ref{b:fig2} which correspond to the trilepton final state
analysis. This is due to the 3 following points.
First, the rate of the $\tilde \chi^0_1 \mu^{\pm}$ production
(recall that it represents the main contribution to the like
sign dilepton final state)
is larger than the $\sigma(p \bar p \to \tilde \chi^{\pm}_1
\mu^{\mp})$ cross section in most of the mSUGRA parameter
space (see Section \ref{cross1}).
Secondly, the $\tilde \chi^0_1$ decay leading to the like sign dilepton
final state
in the case of the $\tilde \chi^0_1 \mu^{\pm}$ production
has a larger branching ratio than the cascade decay initiated by the
$\tilde \chi^{\pm}_1$ which generates the
trilepton final state (see Sections \ref{signal1} and \ref{signal2}).
Finally, at Tevatron Run II
the \sm background of the like sign dilepton
signature is weaker than the trilepton \sm background
(see Tables \ref{cuteff} and \ref{cutefp}).

It is clear from Fig.\ref{b:fig2d} that at low values of the $m_0$ and
$m_{1/2}$ parameters,
high sensitivities can be obtained on the $\l'_{211}$ coupling constant. We
have found that for
instance at the mSUGRA point defined as $m_0=200GeV$, $m_{1/2}=200GeV$,
$sign(\mu)<0$ and
$\tan \beta=1.5$, $\l'_{211}$ values of $\sim 0.03$ can be probed through the
like sign dilepton analysis at
Tevatron Run II assuming a luminosity of ${\cal L}= 1 fb^{-1}$. This result
was obtained by applying
cut $3$ described in Section \ref{cut2} on the SUSY signal ($\tilde
\chi^0_{1,2} \mu^{\pm}$,
$\tilde \chi^{\pm}_1 \mu^{\mp}$ and $\tilde \chi^{\pm}_1 \nu_{\mu}$
productions) and the Standard
Model background.

We expect that, as in the three lepton signature analysis,
interesting sensitivities could be obtained on other
$\l'_{2jk}$ coupling constants.\\
The sensitivities obtained on the $\l'_{3jk}$ couplings
from the like sign dilepton signature analysis should be
weaker than the sensitivities on the $\l'_{2jk}$ couplings
deduced from the same study. Indeed,
in the case of a single dominant $\l'_{3jk}$ coupling
the same sign leptons generated by the
$\tilde \chi^0_1 \tau^{\pm}$ production
would be 2 tau leptons (see Fig.\ref{graphes}(d)
and Section \ref{signal2}). Therefore,
the like sign dileptons ($e^{\pm} e^{\pm}$ or $\mu^{\pm}
\mu^{\pm}$) produced by the \rpv signal
would be mainly generated in tau decays and
would thus have higher probabilities to not pass the analysis cuts
on the particle energy. Moreover, the requirement of
$e^{\pm} e^{\pm}$ or $\mu^{\pm}
\mu^{\pm}$ events would decrease the efficiency after
cuts of the \rpv signal
due to the hadronic decay of the tau.
Finally, the selection of two same flavour like sign dileptons
($e^{\pm} e^{\pm}$ or $\mu^{\pm}
\mu^{\pm}$) would reduce the \rpv signal,
since each of the 2 produced taus could
decay either into an electron or a muon, and hence would not
be an effective cut anymore.\\
The sensitivities obtained on the $\l'_{1jk}$ couplings
from the like sign dilepton signature study
are expected to be identical to the sensitivities on the
$\l'_{2jk}$ couplings obtained from the same study.
Indeed, in the case of a single dominant $\l'_{1jk}$
coupling constant, the only difference in the
like sign dilepton signature analysis
would be that $e^{\pm} e^{\pm}$ events should be
selected instead of $\mu^{\pm} \mu^{\pm}$ events
(see Fig.\ref{graphes}(d) and Section \ref{signal2}).
Nevertheless, a smaller number of $\l'_{1jk}$ couplings
is expected to be probed since the low-energy constraints
on the $\l'_{1jk}$ couplings are generally
stronger than the limits on the
$\l'_{2jk}$ couplings \cite{b:Bhatt}.


In the high $\tan \beta$ case, the
lightest stau $\tilde \tau_1$ can become the LSP
instead of the lightest neutralino,
due to a large mixing in the third generation of charged
sleptons. In such a situation,
the dominant decay channel of the lightest neutralino
is $\tilde \chi^0_1 \to \tilde \tau^{\pm}_1 \tau^{\mp}$.
Two scenarios must then be discussed:
if the single dominant \rpv coupling is not of the type
$\l'_{3jk}$, the $\tilde \tau^{\pm}_1$-LSP is a stable particle
so that the reaction $p \bar p \to \tilde \chi^0_1 l^{\pm}_i$,
representing the main contribution to the like sign dilepton
final state, does not often lead to the
$2 \mu^{\pm}+2 \ jets$ signature.
If the single dominant \rpv coupling is of the type
$\l'_{3jk}$, the $\tilde \chi^0_1 \tau^{\pm}$ production
can receive a contribution from the resonant
$\tilde \tau^{\pm}_2$ production (see Fig.\ref{graphes}(d))
and the $\tilde \tau^{\pm}_1$-LSP decays via $\l'_{3jk}$ as
$\tilde \tau^{\pm}_1 \to u_j d_k$ so that the
$2 \mu^{\pm}+2 \ jets$ signature can still be generated
in a significant way by the $p \bar p \to \tilde \chi^0_1
\tau^{\pm}$ reaction.

We end this Section by some comments on the effect of
the \susyq $R_p$ conserving background to the like sign dilepton
signature. In order to illustrate this discussion, we consider
the results on the $\l'_{211}$ coupling constant.\\
We see from Table \ref{cutSUSp} that
the SUSY background to the like sign dilepton final
state can affect the sensitivity on the
$\l'_{211}$ coupling constant
obtained by considering only the \sm background,
which is shown in Fig.\ref{b:fig2d}, only in the
region of small superpartners masses, namely in the domain
$m_{1/2} \stackrel{<}{\sim} 300GeV$ for $\tan \beta=1.5$,
$sign(\mu)<0$ and assuming a luminosity of ${\cal L}=1fb^{-1}$.\\
In contrast with the SUSY signal amplitude which is increased
if $\l'_{211}$ is enhanced,
the SUSY background amplitude is typically
independent on the value of the $\l'_{211}$ coupling constant
since the superpartner pair production
does not involve \rpv couplings.
Therefore, even if we consider the SUSY background
in addition to the \sm one, it is still true that
large values of the $\l'_{211}$ coupling can
be probed over a wider domain of the SUSY parameter space than
low values, as can be observed in Fig.\ref{b:fig2d} for
$m_{1/2} \stackrel{>}{\sim} 300GeV$. Note that in Fig.\ref{b:fig2d}
larger values of $\l'_{211}$ still respecting the indirect limit
could have been considered.\\
Finally, we mention that further cuts,
as for instance some cuts based on the
superpartners mass reconstructions (see Section \ref{reconsp}),
could allow to reduce the SUSY background to the like sign
dilepton signature.

\subsubsection{Mass reconstructions}
\label{reconsp}

\begin{figure}[t]
\begin{center}
\leavevmode
\centerline{\psfig{figure=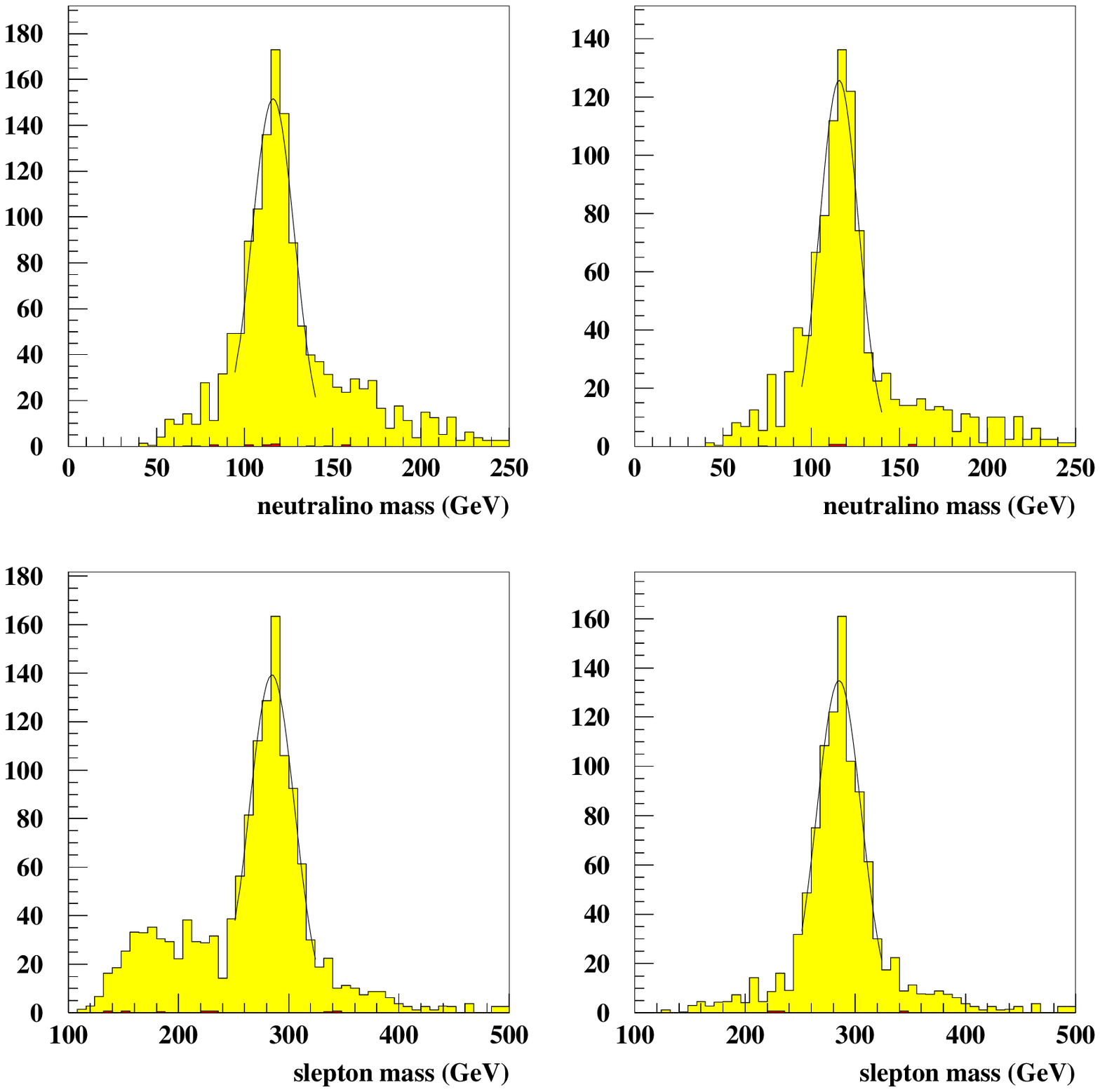,height=5.5in}}
\end{center}
\caption{Distributions of the $softer \ \mu^{\pm}+2 \
leading \ jets$ (upper plots) and $\mu^{\pm}+\mu^{\pm}+2 \
leading \ jets$ (lower plots) invariant masses in the
$\mu^{\pm}+\mu^{\pm}+jets+\Eslash$ events generated
by the SUSY signal ($\tilde \chi^0_{1,2} \mu^{\pm}$,
$\tilde \chi^{\pm}_1 \mu^{\mp}$ and $\tilde \chi^{\pm}_1
\nu_{\mu}$ productions),
for a luminosity of ${\cal L}=10fb^{-1}$.
The 2 right plots are obtained by applying a cut in the
upper left plot of Fig.\ref{dienmu} selecting only
the peak associated to the $\tilde \chi^0_1 \mu^{\pm}$
production.
The mSUGRA point taken for this figure is,
$m_0=200GeV$, $M_2=250GeV$, $\tan \beta=1.5$ and $sign(\mu)<0$
($m_{\tilde \chi^0_1}=127.1GeV$, $m_{\tilde \mu^{\pm}_L}=298.0GeV$)
and the considered \rpv coupling is $\l'_{211}=0.05$.
The average reconstructed masses are
$m_{\tilde \chi^0_1} = 116 \pm 11 GeV$ and
$m_{\tilde \mu^{\pm}_L} = 285 \pm 20 GeV$.}
\label{rec2s}
\end{figure}

The $\tilde \chi^0_1$ and $\tilde l^{\pm}_L$
mass reconstructions can be performed in a model independent
way via the like sign dilepton analysis.
We have simulated these mass reconstructions based on the
like sign dimuon events generated in the scenario
of a single dominant $\l'_{2jk}$ coupling constant. In this
scenario, the main SUSY contribution to the
like sign dilepton signature, namely the $\tilde \chi^0_1
\mu^{\pm}$ production, has the final state
$\mu^{\pm}+\mu^{\pm}+2jets$ (see Section \ref{signal2}).
Indeed, the produced $\tilde \chi^0_1$ decays into
$\mu^{\pm} u_j d_k$ through $\l'_{2jk}$. The muon generated
together with the $\tilde \chi^0_1$ can be identified
as the leading muon for relatively large $m_{\tilde \mu^{\pm}_L}-
m_{\tilde \chi^0_1}$ mass differences (see Section \ref{cut2}).
Note that for nearly degenerate values of $m_{\tilde \mu^{\pm}_L}$
and $m_{\tilde \chi^0_1}$ the $\tilde \chi^0_1
\mu^{\pm}$ production rate and thus the sensitivity on
the SUSY parameters would be reduced (see Section \ref{cross1}).
The muon created in the $\tilde \chi^0_1$
decay can thus be identified as the softer muon so that
the $\tilde \chi^0_1$ can be reconstructed from the
the softer muon and the 2 jets present in the $\tilde \chi^0_1
\mu^{\pm}$ production final state.
The other contributions to the like sign dimuons events
can lead to some missing energy and
at most 4 jets in the final state
(see Section \ref{signal2}). Hence,
we have chosen to reconstruct the
$\tilde \chi^0_1$ from the 2 leading jets when the
final state contains more than 2 jets.
Once the $\tilde \chi^0_1$ has been reconstructed, the
$\tilde \mu^{\pm}_L$ has been reconstructed from the
$\tilde \chi^0_1$ and the leading muon since the dominant
contribution to the $\tilde \chi^0_1 \mu^{\pm}$ production is the
reaction $p \bar p \to \tilde \mu^{\pm}_L \to \tilde \chi^0_1
\mu^{\pm}$. These mass
reconstructions are represented in Fig.\ref{rec2s}.
In this figure, we also represent the same mass
reconstructions obtained by applying a cut in the
upper left plot of Fig.\ref{dienmu} excluding
the peak associated to the
$\tilde \chi^0_2 \mu^{\pm}$ and $\tilde \chi^{\pm}_1
\mu^{\mp}$ productions (see Section \ref{cut2}).
The interest of this cut, as can be seen in Fig.\ref{rec2s},
is to select the $\tilde \chi^0_1
\mu^{\pm}$ production and thus to improve the accuracy on the
$\tilde \chi^0_1$ and $\tilde \mu^{\pm}_L$ reconstructions
which are based on this production.
We observe in Fig.\ref{rec2s} that the $\tilde \chi^0_1$
reconstruction has less combinatorial background than the
$\tilde \mu^{\pm}_L$ reconstruction.
This comes from the fact that the selection of
the softer muon and the 2 leading jets
allows to reconstruct the $\tilde \chi^0_1$ even in the dimuon
events generated by the
$\tilde \chi^0_2 \mu^{\pm}$ and $\tilde \chi^{\pm}_1
\mu^{\mp}$ productions, while the selection of
the 2 muons and the 2 leading jets does not
allow to reconstruct the $\tilde \mu^{\pm}_L$
in the dimuon events generated by the
$\tilde \chi^0_2 \mu^{\pm}$ and $\tilde \chi^{\pm}_1
\mu^{\mp}$ productions (see Section \ref{signal2}).
We have represented on the plots of Fig.\ref{rec2s}
the fits of the invariant mass distributions.
We see from these fits that the distributions are well
peaked around the $\tilde \chi^0_1$ and $\tilde \mu^{\pm}_L$
generated masses. The average reconstructed masses are
$m_{\tilde \chi^0_1} = 116 \pm 11 GeV$ and
$m_{\tilde \mu^{\pm}_L} = 285 \pm 20 GeV$.

We note that the accuracy on the $\tilde \chi^0_1$
(and thus on the $\tilde \mu^{\pm}_L$)
mass reconstruction could be improved if the distributions
in the upper plots of
Fig.\ref{rec2s} were recalculated by selecting the muon giving
the $\tilde \chi^0_1$ mass the closer to the mean value of the
peak obtained in the relevant upper plot of Fig.\ref{rec2s}.

In the hypothesis of a single dominant coupling constant
of type $\l'_{1jk}$ or $\l'_{3jk}$, exactly the same kind of
$\tilde \chi^0_1$
and $\tilde \mu^{\pm}_L$ mass reconstructions can
be performed by selecting the $e^{\pm}+e^{\pm}+jets+\Eslash$
or $l_i^{\pm}+l_j^{\pm}+jets+\Eslash$ events, respectively.

As a conclusion, the $\tilde \chi^0_1$
and $\tilde \mu^{\pm}_L$ mass reconstructions based on
the like sign dilepton signature generated by the
$\tilde \chi^0_{1,2} \mu^{\pm}$,
$\tilde \chi^{\pm}_1 \mu^{\mp}$ and $\tilde \chi^{\pm}_1
\nu_{\mu}$ productions at Tevatron can easily give
precise results, in contrast with the mass reconstructions
performed in the superpartner pair production analysis
at hadronic colliders which suffer an high
combinatorial background \cite{b:Atlas}.

\subsubsection{Model dependence of the results}

In our theoretical framework (see Section \ref{theoretical}),
the values of the $\vert \mu \vert$ and $\tan \beta$
(up to the ambiguity of low/high solution) parameters
are predicted. This has no important effects on the results
presented in Sections \ref{lp211p}
as the single gaugino production cross sections vary weakly
with these parameters (see Section \ref{cross1}).

However, since we have worked within the mSUGRA model,
the $\tilde l_L^{\pm}$ mass was typically larger than the
$\tilde \chi^0_1$ mass. In a situation where
$m_{\tilde l_L^{\pm}}$ would approach $m_{\tilde \chi^0_1}$,
the rate of the $\tilde \chi^0_1 l^{\pm}_i$ production, representing
in mSUGRA
the main contribution to the like sign dilepton signature
(see Section \ref{signal2}),
would decrease. Therefore,
within a model allowing degenerate $\tilde l_L^{\pm}$ and $\tilde \chi^0_1$
masses or even a $\tilde l_L^{\pm}$ lighter than the $\tilde \chi^0_1$,
other single gaugino productions than
the $p \bar p \to \tilde \chi^0_1 l^{\pm}_i$ reaction
could represent the major contribution to the
like sign dilepton signature in some parts of the SUSY
parameter space.

Besides, in a situation
where the LSP would not be the $\tilde \chi^0_1$,
the branching ratios of the $\tilde \chi^0_1$ decays
violating $R_p$ would be reduced with respect to the
case where the LSP is the $\tilde \chi^0_1$,
as often occurs in mSUGRA. However, in such a situation,
the like sign dilepton signature
could receive a significant contribution from
a decay of the $\tilde \chi^0_1$ different from the
\rpv channel. In those kinds of scenarios
where the LSP is not the $\tilde \chi^0_1$,
the $\tilde \chi^0_1 l^{\pm}_i$ production
would not represent systematically the main contribution
to the like sign dilepton signature.

In the several scenarios described above
where the $\tilde \chi^0_1 l^{\pm}_i$ production
is not the major contribution
to the like sign dilepton signature, this signature
could receive quite important contribution from the other
single gaugino productions described in Section \ref{resonant}.

\section{Conclusion}

\setcounter{equation}{0}

The single gaugino productions at Tevatron reach important cross sections
thanks to the contributions of the resonant slepton productions. Hence,
the analysis of the 3 charged leptons and like sign dilepton signatures
generated by the single gaugino productions at Tevatron Run II would allow
to obtain
high sensitivities on many \rpv coupling constants, compared to the
low-energy limits,
in wide domains of the SUSY parameter space. This is also due to the fact
that
the Standard Model backgrounds associated to the 3 charged leptons and like
sign dilepton
final states at Tevatron can be greatly suppressed.

From the supersymmetry discovery point of view, superpartner masses well
beyond the
present experimental limits could be tested through the analysis of the
the 3 charged leptons and like sign dilepton signatures
generated by the single gaugino productions at Tevatron Run II.
If some of the \rpv coupling constants values were close to their low-energy
bounds,
the single gaugino productions study based on the 3 charged leptons and like
sign dilepton
signatures would even allow to
extend the region in the $m_0$-$m_{1/2}$ plane probed by the superpartner
pair
production analyses in the 3 charged leptons and like sign dilepton channels
at Tevatron Run II.
The reason is that the single superpartner production has a larger phase
space factor than
the superpartner pair production.\\
Besides, the 3 charged leptons and like sign dilepton signatures
generated by the single gaugino productions at Tevatron Run II would allow
to reconstruct
in a model independent way the $\tilde \chi^0_1$, $\tilde \chi^{\pm}_1$,
$\tilde \nu_L$ and
$\tilde l^{\pm}_L$ masses with a smaller combinatorial background than in
the superpartner pair
production analysis.

We end this summary by a comparison between the results obtained from
the studies of the 3 charged lepton and like sign dilepton signatures
generated by the single gaugino productions at Tevatron Run II.
In the mSUGRA model, the like sign dilepton signature analysis would give
rise to higher sensitivities on
the SUSY parameters than the study of the 3 charged lepton final state.
This comes notably from the
fact that in mSUGRA, the $\tilde \chi^0_1$ is lighter than the $\tilde
\chi^{\pm}_1$ so that
the cross section of the $\tilde \chi^0_1 l^{\pm}$ production, which is the
main contribution to the like sign
dilepton signature, reaches larger values than the cross section of the
$\tilde \chi^{\pm}_1 l^{\mp}$
production, representing the main contribution to the 3 charged lepton
final state.

Other interesting prospective studies concerning hadronic colliders are 
the analyses of the single gaugino productions
occuring through resonant squark productions via $\lambda''$ 
coupling constants which we will perform in the next future.

\section{Acknowledgments}
We would like to thank Emmanuelle Perez, Robi Peschanski and Auguste Besson for 
fruitful discussions and reading the manuscript.

\newpage

\appendix

\setcounter{subsection}{0}
\setcounter{equation}{0}

\section{Formulas for spin summed amplitudes}
\label{formulas}

In this Appendix, we give the amplitudes for all the single productions
of \susyq particle at hadronic colliders, which can receive a contribution
from a slepton resonant production. These single productions occur
via the \rpv coupling $\l'_{ijk}$
and correspond to the four reactions,
$q \bar q \to \tilde \chi_a^+ \bar \nu_i$, $q \bar q \to \tilde \chi_a^0
\bar \nu_i$,
$ q \bar q \to \tilde \chi_a^0 \bar l_i$, $ q \bar q \to \tilde \chi_a^-
\bar l_i$.
Each of those four processes receives contributions from both the t and u
channel
(see Fig.\ref{graphes}) and have charge conjugated diagrams. Note also that
the contributions coming from the exchange of a right squark in the
u channel involve the higgsino components of the gauginos. These
contributions,
in the case of the single chargino production, do not interfere with the s
channel
slepton exchange since the initial or final states are different
(see Fig.\ref{graphes}).
In the following, we give the formulas for the probability amplitudes,
squared and summed over the polarizations. Our notations closely follow
the notations of \cite{b:Gunion}. In particular, the matrix elements $N'_{ij}$
are defined in the basis of the photino and the zino, as in \cite{b:Gunion}.

{
\hoffset-1in
\voffset-1in
\if@twoside\oddsidemargin25mm
\evensidemargin25mm\marginparwidth25mm
\else\oddsidemargin25mm\evensidemargin25mm\marginparwidth25mm\fi
\footskip30pt
\textwidth 16cm
\baselineskip15pt
\textheight 45\baselineskip

\normalsize

\begin{eqnarray}
& & \vert M_s (u^j \bar d^k \to \tilde \chi_a^+ \bar \nu_i) \vert ^2   =  \cr & & 
{ {\l'_{ijk}}^2  g^2 \vert U_{a1} \vert ^2
\over  12  (s-m_{\tilde l^i_L}^2)^2 }
 (m_{u^j}^2+m_{d^k}^2-s)(m_{\tilde \chi_a^+}^2-s)   \cr
&&  \cr &&  \cr
&& \vert M_t (u^j \bar d^k \to \tilde \chi_a^+ \bar \nu_i) \vert ^2  = \cr & &  {
{\l'_{ijk}}^2  g^2
 \over 12  (t-m_{\tilde d^j_L}^2)^2 }
(m_{d^k}^2-t)
\bigg [ (\vert U_{a1} \vert ^2 +{m_{u^j}^2 \vert V_{a2} \vert ^2 \over
2  m_W^2 \sin^2 \beta} ) (m_{u^j}^2+m_{\tilde \chi_a^+}^2-t) \cr
& - & {4  m_{u^j}^2 m_{\tilde \chi_a^+} Re(U_{a1}V_{a2})
\over \sqrt{2} m_W \sin \beta} \bigg ] \cr
&&  \cr &&  \cr 
&& \vert M_u (u^k \bar d^j \to \tilde \chi_a^+ \nu_i) \vert ^2  = \cr & & 
{ {\l'_{ijk}}^2  g^2 m_{d^k}^2 \vert U_{a2} \vert^2
 \over 24 m_W^2 \cos \beta ^2  (u-m_{\tilde d^k_R}^2)^2 }
  (m_{\tilde \chi_a^+}^2+m_{u^k}^2-u) (m_{d^j}^2-u) \cr
&&  \cr &&  \cr
&& 2 Re [M_s M_t^* (\tilde \chi_a^+ \bar \nu_i)]  = \cr & &  { {\l'_{ijk}}^2  g^2
 \over  6  (s-m_{\tilde l^i_L}^2) (t-m_{\tilde d^j_L}^2) }
\bigg [ { \vert U_{a1} \vert ^2 \over 2}  [(m_{u^j}^2+m_{\tilde
\chi_a^+}^2-t)
 (m_{d^k}^2-t) \cr
 &  +  & (m_{u^j}^2+m_{d^k}^2-s) (m_{\tilde \chi_a^+}^2-s)
-(m_{u^j}^2-u) (m_{\tilde \chi_a^+}^2+m_{d^k}^2-u)] \cr
 &  - &  (m_{d^k}^2-t)
{Re(U_{a1}V_{a2}) m_{\tilde \chi_a^+} m_{u^j}^2  \over \sqrt{2} m_W \sin
\beta} \bigg ],
\label{fchanu}
\end{eqnarray}

where, $s=(p(u^j)-p(\bar d_k))^2$, $t=(p(u^j)-p(\tilde \chi_a^+))^2$
and $u=(p(\bar d^j)-p(\nu_i))^2$.

\begin{eqnarray}
&&  \vert M_s (d_j \bar d_k \to  \tilde \chi_a^0 \bar \nu_i) \vert ^2  = \cr & & 
{ {\l'_{ijk}}^2  g^2 \vert N'_{a2} \vert ^2
  \over 24  \cos^2 \theta_W (s-m_{\tilde \nu^i_L}^2)^2 }
   (s-m_{d^k}^2-m_{d^j}^2) (s-m_{\tilde \chi_a^0}^2)  \cr
&&  \cr &&  \cr
&&  \vert M_t (d_j \bar d_k \to \tilde \chi_a^0 \bar \nu_i) \vert ^2  = \cr & &  {
{\l'_{ijk}}^2  g^2
  \over 6  (t-m_{\tilde d^j_L}^2)^2 } (m^2_{d^k}-t)
 \bigg [ (m_{d^j}^2+m_{\tilde \chi_a^0}^2-t)
\bigg ( { g^2 m_{d^j}^2 \vert N'_{a3} \vert ^2 \over 4  m_W^2 \cos^2 \beta }
 +{e^2 \over 9} \vert N'_{a1} \vert ^2 \cr
& + & {g^2 \vert N'_{a2} \vert ^2 (\sin^2 \theta_W/3 -1/2)^2 \over \cos^2
\theta_W }
 - { 2  e g Re(N'_{a1}N'_{a2}) (\sin^2 \theta_W/3 -1/2) \over 3  \cos
\theta_W } \bigg ) \cr
 & + & {2 m_{\tilde \chi_a^0} m_{d^j}^2 g \over m_W \cos \beta}
\bigg ( -{e Re(N'_{a1}N'_{a3})\over 3} +
 {g Re(N'_{a2}N'_{a3}) \over \cos \theta_W}
({\sin^2 \theta_W \over 3} -{1 \over 2}) \bigg ) \bigg ] \cr
&&  \cr &&  \cr
&&  \vert M_u (d_j \bar d_k \to \tilde \chi_a^0 \bar \nu_i) \vert ^2  = \cr & &  {
{\l'_{ijk}}^2
  \over 6  (u-m_{\tilde d^k_R}^2)^2 }
   (m_{d^j}^2-u)
 \bigg [ (m_{\tilde \chi_a^0}^2+m_{d^k}^2-u) \bigg
( {g^2 m_{d^k}^2 \vert N'_{a3} \vert ^2 \over 4  m_W^2 \cos^2 \beta}
 +{e^2 \vert N'_{a1} \vert ^2 \over 9} \cr
& + & {g^2 \sin^4 \theta_W \vert N'_{a2} \vert ^2 \over 9  \cos^2 \theta_W}
 - { 2  e g Re(N'_{a1}N'_{a2}) \sin^2 \theta_W \over 9  \cos \theta_W} \bigg
) \cr
& - & { 2  m_{\tilde \chi_a^0} m_{d^k}^2 g \over m_W \cos \beta}
\bigg ( -{e Re(N'_{a1}N'_{a3}) \over 3} +{g \sin^2 \theta_W
Re(N'_{a2}N'_{a3})
\over3  \cos \theta_W} \bigg ) \bigg ] \cr
&&  \cr &&  \cr
&&  2 Re [M_s M_t^* (\tilde \chi_a^0 \bar \nu_i)]  =  \cr & &  - { {\l'_{ijk}}^2 g
 \over 12  \cos \theta_W (s-m_{\tilde \nu^i_L}^2) (t-m_{\tilde d^j_L}^2) }
 \bigg [ (m_{d^k}^2-t) {m_{\tilde \chi_a^0} m_{d^j}^2 g Re(N'_{a2}N'_{a3})
\over m_W \cos \beta} \cr
& + & \bigg (-{e Re(N'_{a1}N^*_{a2}) \over 3} +{g \vert N'_{a2}
\vert ^2 \over \cos \theta_W} ({\sin^2 \theta_W \over 3} -{1 \over 2}) \bigg
)
 [ (m_{d^j}^2+m_{\tilde \chi_a^0}^2-t) (m_{d^k}^2-t) \cr
& + & (m_{d^j}^2+m_{d^k}^2-s) (m_{\tilde \chi_a^0}^2-s)
 -(m_{\tilde \chi_a^0}^2+m_{d^k}^2-u) (m_{d^j}^2-u) ] \bigg ] \cr
&&  \cr &&  \cr
&&  2 Re [M_t M_u^* (\tilde \chi_a^0 \bar \nu_i)]  =  \cr & & { {\l'_{ijk}}^2
 \over 6  (u-m_{\tilde d^k_R}^2) (t-m_{\tilde d^j_L}^2) }
 \bigg [ (m_{d^k}^2-t) {  g m_{\tilde \chi_a^0} m_{d^j}^2
\over m_W \cos \beta} \bigg ( {g \sin^2 \theta_W Re(N'_{a2}N'_{a3})
\over 3  \cos \theta_W}-{e Re(N'_{a1}N'_{a3}) \over 3 } \bigg ) \cr
& + & [(m_{d^j}^2-u) (m_{\tilde \chi_a^0}^2+m_{d^k}^2-u)
  +  (m_{d^k}^2-t) (m_{d^j}^2+m_{\tilde \chi_a^0}^2-t)
 -(m_{\tilde \chi_a^0}^2-s) (m_{d^j}^2+m_{d^k}^2-s)] \cr
& & \bigg (  -{egRe(N'_{a1}N'_{a2}) \over 3 \cos \theta_W}
( {2 \sin^2 \theta_W \over 3} -{1 \over 2} )
+{e^2 \vert N'_{a1} \vert ^2 \over 9 }+{g^2 \sin^2 \theta_W \vert N'_{a2}
\vert ^2
\over 3 \cos^2 \theta_W} ({\sin^2 \theta_W \over 3}-{1 \over 2}) \bigg ) \cr
& - & {m_{\tilde \chi_a^0} m_{d^k}^2 g \over m_W \cos \beta}
 \bigg ( -{e Re(N'_{a1}N'_{a3}) \over 3} +{g Re(N'_{a2}N'_{a3})
 \over \cos \theta_W} ({\sin^2 \theta_W \over 3}-{1 \over 2}) \bigg )
 (m_{d^j}^2-u)  \cr
& + & {m_{d^j}^2 m_{d^k}^2  g^2 \vert N'_{a3} \vert ^2
 \over 2  m_W^2 \cos^2 \beta}(m_{\tilde \chi_a^0}^2-s) \bigg ] \cr
&&  \cr &&  \cr
&& 2 Re [M_s M_u^* (\tilde \chi_a^0 \bar \nu_i)]  = \cr & &  { {\l'_{ijk}}^2 g
\over 12  \cos \theta_W (s-m_{\tilde \nu^i_L}^2) (u-m_{\tilde d^k_R}^2) }
\bigg [-  {m_{\tilde \chi_a^0} m_{d^k}^2 g Re(N'_{a2}N'_{a3})
\over m_W \cos \beta} (m_{d^j}^2-u) \cr
&  + & \bigg ( -{e Re(N^*_{a1}N'_{a2}) \over 3} +{\vert N'_{a2} \vert ^2
g \sin^2 \theta_W \over 3  \cos \theta_W} \bigg )
[(m_{d^j}^2+m_{d^k}^2-s) (m_{\tilde \chi_a^0}^2-s) \cr
& + & (m_{\tilde \chi_a^0}^2+m_{d^k}^2-u) (m_{d^j}^2-u)
-(m_{d^j}^2+m_{\tilde \chi_a^0}^2-t) (m_{d^k}^2-t)]  \bigg ],
\label{fnenu}
\end{eqnarray}

where, $s=(p(d^j)-p(\bar d_k))^2$, $t=(p(d^j)-p(\tilde \chi_a^0))^2$
and $u=(p(d^j)-p(\bar \nu_i))^2$.

\begin{eqnarray}
&& \vert M_s (u_j \bar d_k \to  \tilde \chi_a^0 \bar l_i) \vert ^2  =  \cr & & {
{\l'_{ijk}}^2
 \over 6  (s-m_{\tilde l^i_L}^2)^2 }
 (s-m_{u^j}^2-m_{d^k}^2)
 \bigg [ \bigg (  {g^2 m_{l^i}^2 \vert N'_{a3} \vert ^2 \over 4  m_W^2
\cos^2 \beta}
+e^2 \vert N'_{a1} \vert ^2
 + {g^2 \vert N'_{a2} \vert ^2  \over \cos^2 \theta_W}
(\sin^2 \theta_W- {1 \over 2})^2 \cr
& - &{2  e g Re(N'_{a1}N'_{a2}) \over \cos \theta_W}(\sin^2 \theta_W-{1
\over 2}) \bigg )
 (s-m_{l^i}^2-m_{\tilde \chi_a^0}^2)
 - {2  g m_{\tilde \chi_a^0} m_{l^i}^2 \over m_W \cos \beta}
\bigg ( -e Re(N'_{a1}N'_{a3}) \cr
 & + & {g Re(N'_{a2}N'_{a3}) \over \cos \theta_W}
(\sin^2 \theta_W-{1 \over 2}) \bigg ) \bigg ] \cr
&&  \cr &&  \cr
&&  \vert M_t (u_j \bar d_k \to \tilde \chi_a^0 \bar l_i) \vert ^2  = \cr & & {
{\l'_{ijk}}^2
 \over 6  (t-m_{\tilde u^j_L}^2)^2 }
 (-t+m_{l^i}^2+m_{d^k}^2)
 \bigg [ \bigg (  {g^2 m_{u^j}^2 \vert N'_{a4} \vert ^2 \over 4  m_W^2
\sin^2 \beta}
 +{4  e^2 \vert N'_{a1} \vert ^2 \over 9} \cr
 & + & {g^2 \vert N'_{a2} \vert ^2  \over \cos^2 \theta_W}
({1 \over 2}-{2  \sin^2 \theta_W \over 3} )^2
 +{4  e g Re(N'_{a1}N'_{a2})  \over 3  \cos \theta_W}
 ({1 \over 2}-{2  \sin^2 \theta_W \over 3}) \bigg )
 (-t+m_{u^j}^2+m_{\tilde \chi_a^0}^2) \cr
 &  +  & {2  g m_{u^j}^2 m_{\tilde \chi_a^0} \over m_W \sin \beta}
\bigg ( {2  e Re(N'_{a1}N'_{a4}) \over 3} +{g Re(N'_{a2}N'_{a4})
  \over \cos \theta_W} ({1 \over 2}-{2  \sin^2 \theta_W \over 3} ) \bigg )
\bigg ] \cr
&&  \cr &&  \cr
&&  \vert M_u (u_j \bar d_k \to  \tilde \chi_a^0 \bar l_i) \vert ^2  = \cr & & {
{\l'_{ijk}}^2
  \over 6  (u-m_{\tilde d^k_R}^2)^2 }
  (m_{u^j}^2+m_{l^i}^2-u)
 \bigg [ \bigg ( {e^2 \vert N'_{a1} \vert ^2 \over 9}
 + {g^2 \sin^4 \theta_W \vert N'_{a2} \vert ^2 \over 9  \cos^2 \theta_W}
  -{2  e g Re(N'_{a1}N'_{a2}) \sin^2 \theta_W \over 9  \cos \theta_W} \cr
 & + & {g^2 m_{d^k}^2 \vert N'_{a3} \vert ^2 \over 4  m_W^2 \cos^2 \beta}
\bigg )
 (m_{\tilde \chi_a^0}^2+m_{d^k}^2-u)
 - {2  g m_{d^k}^2 m_{\tilde \chi_a^0}  \over m_W \cos \beta}
\bigg ( -{e Re(N'_{a1}N'_{a3}) \over 3} \cr
& + & {g \sin^2 \theta_W Re(N'_{a2}N'_{a3}) \over 3  \cos \theta_W} \bigg )
\bigg ] \cr
&&  \cr &&  \cr
&& 2 Re [M_s M_t^* (\tilde \chi_a^0 \bar l_i)]  = \cr & & - {   {\l'_{ijk}}^2
  \over 6  (s-m_{\tilde l^i_L}^2) (t-m_{\tilde u^j_L}^2) }
  \bigg [  -{m_{l^i}^2 m_{u^j}^2   g^2 Re(N'_{a3}N^*_{a4})
  \over 2  m_W^2 \sin \beta \cos \beta} (m_{\tilde \chi_a^0}^2+m_{d^k}^2-u)
\cr
 & + & \bigg ( {-2  e^2 \vert N'_{a1} \vert ^2 \over 3} +{  e g
Re(N^*_{a1}N'_{a2})
  \over 3  \cos \theta_W}  (4 \sin^2 \theta_W-{5 \over 2})  \cr
 & + &  {g^2 \vert N'_{a2} \vert ^2
 \over \cos^2 \theta_W} ({1 \over 2}-{2  \sin^2 \theta_W \over 3} )
 (\sin^2 \theta_W-{1 \over 2}) \bigg ) \cr
 & & [ (m_{u^j}^2+m_{d^k}^2-s) (m_{\tilde \chi_a^0}^2+m_{l^i}^2-s)
 +(m_{u^j}^2+m_{\tilde \chi_a^0}^2-t) (m_{l^i}^2+m_{d^k}^2-t) \cr
 & - & (m_{u^j}^2+m_{l^i}^2-u) (m_{\tilde \chi_a^0}^2+m_{d^k}^2-u) ]
 + {g m_{u^j}^2 m_{\tilde \chi_a^0} \over m_W \sin \beta}
 \bigg ( -e Re(N'_{a1}N'_{a4})+{g Re(N'_{a2}N'_{a4})  \over \cos \theta_W}
\cr
& & ( \sin^2 \theta_W-{1 \over 2}) \bigg )
 (m_{l^i}^2+m_{d^k}^2-t)
 -(s-m_{u^j}^2-m_{d^k}^2) {g m_{l^i}^2 m_{\tilde \chi_a^0} \over m_W \cos
\beta}
\bigg ( {2  e Re(N'_{a1}N'_{a3}) \over 3} \cr
& + & {g Re(N'_{a2}N'_{a3}) \over \cos \theta_W}
({1 \over 2}-{2 \sin^2 \theta_W \over 3}) \bigg ) \bigg ] \cr
&&  \cr &&  \cr
&& 2 Re [M_t M_u^* (\tilde \chi_a^0 \bar l_i)]  = \cr & &  {   {\l'_{ijk}}^2
 \over 6  (u-m_{\tilde d^k_R}^2) (t-m_{\tilde u^j_L}^2) }
\bigg  [ {g m_{u^j}^2 m_{\tilde \chi_a^0}
     \over m_W \sin \beta} ( m_{l^i}^2+m_{d^k}^2-t )
   \bigg ( -{e Re(N'_{a1}N'_{a4}) \over 3} \cr
    & + & {g \sin^2 \theta_W Re(N'_{a2}N'_{a4}) \over 3  \cos \theta_W}
\bigg )
 - {m_{\tilde \chi_a^0} g m_{d^k}^2  \over m_W \cos \beta}
  \bigg ( {2  e Re(N'_{a1}N'_{a3}) \over 3} +{g Re(N'_{a2}N'_{a3})
  \over \cos \theta_W} ({1 \over 2}-{2  \sin^2 \theta_W \over 3}) \bigg )
\cr
      & &  ( m_{l^i}^2+m_{u^j}^2-u )
 - {g^2 Re(N'_{a3}N^*_{a4}) m_{u^j}^2 m_{d^k}^2
   \over 2  m_W^2 \cos \beta \sin \beta} (s-m_{l^i}^2-m_{\tilde \chi_a^0}^2)
+ \bigg ( -{2  e^2 \vert N'_{a1} \vert ^2 \over 9} \cr
 & + & {e g Re(N^*_{a1}N'_{a2}) \over 3  \cos \theta_W}
 (-{1 \over 2}+{ 4 \sin^2 \theta_W \over 3})
  + {g^2 \sin^2 \theta_W \vert N'_{a2} \vert ^2 \over 3  \cos^2 \theta_W}
\cr
& &  ( {1 \over 2}-{2  \sin^2 \theta_W \over 3} )  \bigg )
 [  (m_{l^i}^2+m_{u^j}^2-u) (m_{\tilde \chi_a^0}^2+m_{d^k}^2-u)
 +(m_{l^i}^2+m_{d^k}^2-t) (m_{\tilde \chi_a^0}^2+m_{u^j}^2-t) \cr
 & - & (m_{l^i}^2+m_{\tilde \chi_a^0}^2-s) (m_{d^k}^2+m_{u^j}^2-s) ]
\bigg ] \cr
&&  \cr &&  \cr
&&  2 Re [M_s M_u^* (\tilde \chi_a^0 \bar l_i)]  = \cr & &  {   {\l'_{ijk}}^2
  \over 6  (s-m_{\tilde l^i_L}^2) (u-m_{\tilde d^k_R}^2) }
 \bigg [ -{g m_{l^i}^2 m_{\tilde \chi_a^0} \over m_W \cos \beta}
 \bigg ( -{e Re(N'_{a1}N'_{a3}) \over 3} +{g \sin^2 \theta_W
Re(N'_{a2}N'_{a3})
\over 3  \cos \theta_W} \bigg ) \cr
  & & ( s-m_{d^k}^2-m_{u^j}^2 )
 - {g m_{d^k}^2 m_{\tilde \chi_a^0}  \over m_W \cos \beta}
 \bigg ( -e Re(N'_{a1}N'_{a3})+{g Re(N'_{a2}N'_{a3}) \over \cos \theta_W}
(\sin
\theta_W^2-{1 \over 2}) \bigg ) \cr
  & & ( m_{l^i}^2+m_{u^j}^2-u )
 + {g^2 m_{l^i}^2 m_{d^k}^2 \vert N'_{a3} \vert ^2
  \over 2  m_W^2 \cos^2 \beta} (m_{\tilde \chi_a^0}^2+m_{u^j}^2-t)
 + \bigg ( {e^2 \vert N'_{a1} \vert ^2 \over 3 } \cr
 & - & {e g Re(N^*_{a1}N'_{a2}) \over 3  \cos \theta_W}
( 2 \sin \theta_W^2-{1 \over 2})
 + {g^2 \vert N'_{a2} \vert ^2 \sin^2 \theta_W  \over 3 \cos^2 \theta_W}
  (\sin^2 \theta_W-{1 \over 2}) \bigg ) \cr
& &  [  (m_{l^i}^2+m_{u^j}^2-u) (m_{\tilde \chi_a^0}^2+m_{d^k}^2-u)
 -(m_{l^i}^2+m_{d^k}^2-t) (m_{\tilde \chi_a^0}^2+m_{u^j}^2-t) \cr
& + & (m_{l^i}^2+m_{\tilde \chi_a^0}^2-s) (m_{d^k}^2+m_{u^j}^2-s) ]
\bigg ],
\label{fnel}
\end{eqnarray}

where, $s=(p(u^j)-p(\bar d_k))^2$, $t=(p(u^j)-p(\tilde \chi_a^0))^2$
and $u=(p(u^j)-p(\bar l_i))^2$.

\begin{eqnarray}
&&       \vert M_s (d_j \bar d_k \to \tilde \chi_a^- \bar l_i) \vert ^2  = \cr & &
{ g^2 {\l'_{ijk}}^2
      \over 6  (s-m_{\tilde \nu^i_L}^2)^2 }
     (s-m_{d^j}^2-m_{d^k}^2)
      \bigg [ ( {m_{l^i}^2 \vert U_{a2} \vert^2 \over
     4  m_W^2 \cos^2 \beta}+ { \vert V_{a1} \vert ^2 \over 2} )
      ( s-m_{\tilde \chi_a^+}^2-m_{l^i}^2 ) \cr
       & + & {\sqrt{2} Re(V_{a1}U_{a2}) m_{l^i}^2 m_{\tilde \chi_a^+}
     \over m_W \cos \beta} \bigg ] \cr
&&  \cr &&  \cr
&&       \vert M_t (d_j \bar d_k \to \tilde \chi_a^- \bar l_i) \vert ^2  = \cr & &
    {     g^2 {\l'_{ijk}}^2
      \over 3  (t-m_{\tilde u^j_L}^2)^2 }
      ( t-m_{d^k}^2-m_{l^i}^2 ) \bigg [ ( t-m_{\tilde \chi_a^+}^2-m_{d^j}^2
)
       ( {\vert V_{a1} \vert ^2 \over 4} + {m_{d^j}^2 \vert U_{a2} \vert^2
     \over 8  M^2_W \cos^2 \beta} ) \cr
     & + & {Re(V_{a1}U_{a2}) m_{\tilde \chi_a^+} m^2_{d^j}
     \over \sqrt{2} m_W \cos \beta}  \bigg ] \cr
&&  \cr &&  \cr
&&       \vert M_u (\bar u_k u_j \to \tilde \chi_a^- \bar l_i) \vert ^2  = \cr & & {
g^2 {\l'_{ijk}}^2
      \over 24  (u-m_{\tilde d^k_R}^2)^2  }
      (m_{\tilde \chi_a^+}^2+m_{u^k}^2-u)
      (m_{l^i}^2+m_{u^j}^2-u)
     { \vert U_{a2} \vert^2  m_{d^k}^2 \over m_W^2 \cos^2 \beta} \cr
&&  \cr &&  \cr
&&        2 Re [M_s M_t^* (\tilde \chi_a^- \bar l_i)]  = \cr & & {  g^2
{\l'_{ijk}}^2
      \over 12  (s-m_{\tilde \nu^i_L}^2) (t-m_{\tilde u^j_L}^2) }
     \bigg [  \vert V_{a1} \vert ^2
      [ -(m_{l^i}^2+m_{d^j}^2-u) (m_{\tilde \chi_a^+}^2+m_{d^k}^2-u) \cr
         & + & (m_{l^i}^2+m_{d^k}^2-t) (m_{\tilde \chi_a^+}^2+m_{d^j}^2-t)
         +(m_{l^i}^2+m_{\tilde \chi_a^+}^2-s) (m_{d^k}^2+m_{d^j}^2-s) ] \cr
 & + & {Re(V_{a1}U_{a2}) m_{\tilde \chi_a^+} \sqrt{2} \over m_W \cos \beta }
 [m^2_{l^i} (s-m_{d^j}^2 - m_{d^k}^2)
- m^2_{d^j} (m_{l^i}^2+m_{d^k}^2-t)] \cr
 & - &  { \vert U_{a2} \vert^2  m_{l^i}^2  m^2_{d^j} \over m_W^2 \cos^2
\beta}
(m_{\tilde \chi_a^+}^2+m_{d^k}^2-u) \bigg ],
\label{fchal}
\end{eqnarray}

where, $s=(p(d^j)-p(\bar d_k))^2$, $t=(p(d^j)-p(\tilde \chi_a^-))^2$
and $u=(p(u^j)-p(\bar l_i))^2$.}

\clearpage

\vspace{10 mm}
\begin{center}
{  }
\end{center}
\vspace{10 mm}

\clearpage 

\setcounter{chapter}{0}
\setcounter{section}{0}
\setcounter{subsection}{0}
\setcounter{figure}{0}

\chapter*{Publication IV}
\addcontentsline{toc}{chapter}{Resonant sneutrino production in
Supersymmetry with R-parity violation at the LHC}

\newpage

\vspace{10 mm}
\begin{center}
{  }
\end{center}
\vspace{10 mm}

\clearpage

\begin{center}
{\bf \huge Resonant sneutrino production in
Supersymmetry with R-parity violation at the LHC}
\end{center} 
\vspace{2cm}
\begin{center}
G. Moreau
\end{center} 
\begin{center}
{\em  Service de Physique Th\'eorique \\} 
{ \em  CE-Saclay F-91191 Gif-sur-Yvette, Cedex France \\}
\end{center} 
\begin{center}
E. Perez
\end{center} 
\begin{center}
{\em  Service de Physique des Particules, DAPNIA \\} 
{ \em  CE-Saclay, F-91191, Gif-sur-Yvette, Cedex France \\}
\end{center} 
\begin{center}
G. Polesello
\end{center} 
\begin{center}
{\em  INFN, Sezione di Pavia \\} 
{ \em  Via Bassi 6, Pavia, Italy \\}
\end{center}
\vspace{1cm}
\begin{center}
{To appear in Nucl. Phys. {\bf B}, hep-ph/0003012}
\end{center}
\vspace{2cm}
\begin{center}
Abstract  
\end{center}
\vspace{1cm}
{\it
The resonant production of sneutrinos at the LHC via the 
R-parity violating couplings $\l ' _{ijk} L_i Q_j D^c_k$ 
is studied through its three-leptons signature.
A detailed particle level study of signal and background
is performed using a fast simulation of the ATLAS detector. 
Through the full reconstruction of
the cascade decay, a model-independent and precise measurement 
of the masses of the involved sparticles can be performed.
Besides, this  signature can be detected for a broad class of supersymmetric
models, and  for a wide range of values of several $\l ' _{ijk}$
coupling constants. 
Within the MSSM, the production of
a 900~GeV sneutrino for $\lambda^{\prime}_{211}>0.05$, and of a 350~GeV sneutrino for
$\lambda^{\prime}_{211}>0.01$ can be observed within
the first three years of LHC running.
}

\newpage

\section{Introduction}

\setcounter{equation}{0}

The most general superpotential respecting the gauge 
symmetries of the Standard Model (SM) contains bilinear and trilinear terms
which are not taken into account in the 
Minimal Supersymmetric Standard Model (MSSM).
Restricting to the trilinear part, these additional terms read as~:
\begin{eqnarray}
W \supset \sum_{i,j,k} \bigg (\ud \l _{ijk} L_iL_j E^c_k+
\l ' _{ijk} L_i Q_j D^c_k+ \ud \l '' _{ijk} U_i^cD_j^cD_k^c   \bigg ), 
\label{d:super}
\end{eqnarray}
where $i,j,k$ are generation indices,
$L$ ($Q$) denote the left-handed leptons (quarks) superfields,
and $E^c$, $D^c$ and $U^c$ are right-handed superfields
for charged leptons, down and up-type quarks, respectively.

The first two terms in Eq.(\ref{d:super}) lead to violation of
the lepton number (\LV), while the last one implies violation
of the baryon number (\BV).
Since the simultaneous presence of \LV and \BV couplings
could lead to a too fast proton decay, a discrete multiplicative symmetry
which forbids the above terms in the superpotential
has been imposed by hand in the MSSM. 
This symmetry, called R-parity ($R_p$),
is defined as $R_p = (-1)^{3B + L + 2S}$, where
$B$, $L$ and $S$ respectively denote the baryon number,
the fermion number and the spin, such that $R_p=-1$ ($R_p = 1$)
for all supersymmetric (SM) particles.
However other solutions can ensure the proton stability,
e.g. if $L$ only is violated, or if only $U_i^cD_j^cD_k^c$ interactions
are allowed and the proton is lighter than
the Lightest Supersymmetric Particle (LSP).
Moreover, on the theoretical point of view, there is no clear preference,
e.g. in models inspired by Grand Unified or string theories,
between \rpv and $R_p$ conservation~\cite{d:Drein}.
It is thus mandatory to search for SUSY in both scenarios.

On the experimental side, the main consequence of \rpv
lies in the possibility for the LSP to decay into ordinary
matter.
This is in contrast to scenarios where $R_p$ is conserved,
in which the LSP is stable and escapes detection, leading to the
characteristic search for missing energy signals in direct 
collider searches.
Moreover, while in $R_p$ conserved models, 
the \susyq (SUSY) particles must be produced in pairs,
\rpv allows the single production of 
superpartners, thus enlarging the mass domain where
SUSY could be discovered.
In particular,
\rpv couplings offer the opportunity to resonantly
produce \susyq particles~\cite{d:Dim1,d:Dreinoss}.
Although the \rpv
coupling constants are severely constrained by the low-energy 
experimental bounds \cite{d:Drein,d:Han,d:Bhatt,d:rapport,d:GDR,d:Alla}, the superpartner
resonant production can have significant cross-sections both at 
leptonic 
\cite{d:Han} and hadronic \cite{d:Dim2} colliders.
This is this possibility which is exploited throughout this paper. \\
The resonant production of \susyq particles is attractive for another reason:
Since its rate is proportional to a power $2$ of the relevant \rpv coupling, 
this reaction would allow an easier determination of the  
\rpv couplings than the pair production. 
In fact in the latter case,
the sensitivity on the \rpv coupling is
mainly provided by the displaced vertex analysis for the LSP decay,
which is difficult experimentally especially at hadronic colliders.

In this paper, we focus on the resonant SUSY particle production at 
the Large Hadron Collider (LHC) 
operating at a center of mass energy of $14$~TeV
with special reference to  the ATLAS detector.
At the LHC due to the continuous 
distribution of the centre of mass  energy of the
colliding partons, a parton-parton resonance can be probed over a wide  mass domain.
This is a distinct advantage over the situation at
lepton colliders, where the search for narrow resonances requires 
lengthy scans over the centre of mass energy of the machine.

At hadronic colliders, either a slepton or a squark can be produced at
the resonance through a $\l'$ or a $\l''$ coupling constant, respectively. 
In the hypothesis of a single dominant \rpv 
coupling constant, the resonant SUSY particle 
could decay through the same \rpv coupling as in the production, 
leading then to a two quarks final state for the hard process \cite{d:Bin,d:Dat,d:Oak,d:Rizz,d:Chiap}. 
In the case where both $\l'$ and $\l$ couplings 
are non-vanishing, the slepton produced via $\l'$ can 
decay through $\l$
giving rise to the same final state as in Drell-Yan process, 
namely two leptons  
\cite{d:Rizz,d:Kal,d:Kalp,d:SonPart}. However, for most of the values of the 
\rpv coupling constants allowed by present indirect searches, the decays  
of the resonant SUSY particle via gauge interactions are 
dominant if kinematically accessible \cite{d:Han}. 
In this favoured situation, typically, the
produced superpartner initiates a cascade decay ended by
the \rpv decay of the LSP. In case of a dominant
$\l''$ coupling constant, 
due to the \rpv decay of the LSP into quarks,
this cascade decay leads to multijet final states which have a
large QCD background \cite{d:Dim2,d:Bin}. Only  if  leptonic decays 
such as for instance $\tilde \chi^+_1 \to \bar l_i \nu_i \tilde \chi^0_1$ 
enter the cascade clearer signatures can be investigated \cite{d:Berg}. 
The situation is more favourable in the hypothesis of 
a single $\l'$ coupling constant,
where the LSP can decay into a charged lepton,
allowing then multileptonic final states to be easily obtained.\\ 
We will thus assume a dominant $\l'_{ijk}$ coupling constant. 
At hadronic colliders, either a $\tilde \nu_i$ 
sneutrino or a $\tilde l_i$ charged slepton can be produced  
at the resonance via $\l'_{ijk}$ and the initial states are $d_j \bar d_k$
and $u_j \bar d_k$, respectively. 
The slepton produced at the resonance has two possible gauge 
decays, either into a chargino or a neutralino. 
In both cases particularly clean signatures can be observed. 
For example, the production of a neutralino together with a charged lepton 
resulting from the resonant charged slepton production
can lead 
to the interesting like-sign dilepton topology \cite{d:Rich,d:Rich2}
since, due to its Majorana nature, the neutralino decays via
$\l'_{ijk}$ into a lepton as $\tilde \chi^0 \to l_i u_j \bar d_k$ 
and into an anti-lepton as $\tilde \chi^0 \to \bar l_i \bar u_j d_k$ 
with the same probability.

\begin{figure}[t]
\begin{center}
\leavevmode
\centerline{\psfig{figure=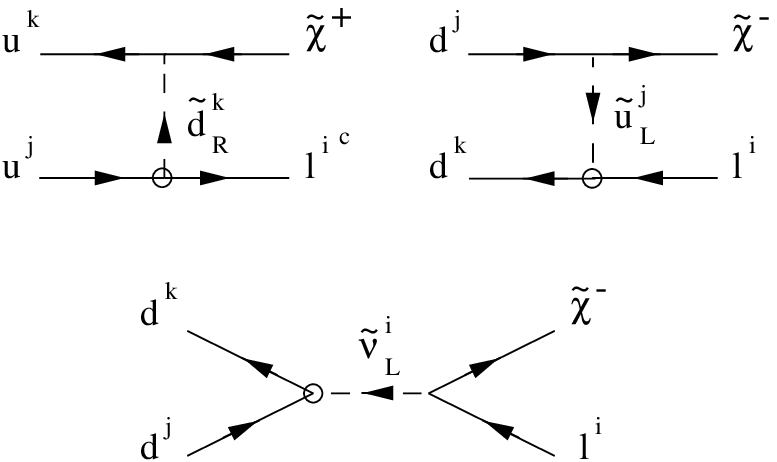,height=2.in}}
\end{center}
\protect\caption{\em Feynman diagrams for the single chargino 
production at hadronic colliders via the $\l'_{ijk}$ coupling 
(symbolised by a circle in the figure). 
The arrows denote the flow of the particle momentum.}
\label{d:fig1}
\end{figure}

In this article, we consider the single lightest chargino
production at LHC as induced by the resonant sneutrino production 
$ p p \to \tilde \nu_i \to \tilde \chi^+_1 l_i$. The single 
$\tilde \chi^{\pm}_1$ production also  receives contributions from the
$t$ and $u$ channel squark exchange diagrams shown in Figure~\ref{d:fig1}.
In many models, the $\tilde \chi^0_1$ neutralino is the LSP 
for most of the SUSY parameter space. 
In the hypothesis of a $\tilde \chi^0_1$ LSP, 
the produced $\tilde \chi^{\pm}_1$ chargino mainly decays into 
the neutralino as $\tilde \chi^{\pm}_1 \to \tilde \chi^0_1 q_p \bar q'_p$
or as $\tilde \chi^{\pm}_1 \to \tilde \chi^0_1 l^{\pm}_p \nu_p$. 
The neutralino then decays via $\l'_{ijk}$ as 
$\tilde \chi^0_1 \to l_i u_j \bar d_k, \ \bar l_i \bar u_j d_k$ or as 
$\tilde \chi^0_1 \to \nu_i d_j \bar d_k, \ \bar \nu_i \bar d_j d_k$. 
We concentrate on the decays of both the chargino and the neutralino 
into charged leptons, which lead to a three leptons final state.
This signature has a low Standard Model background, and 
allows the reconstruction of the whole decay chain, thus providing
a measurement of some parameters of the SUSY model.

\section{The signal}
\label{secmssm}

\setcounter{equation}{0}


\subsection{Theoretical framework}

Our theoretical framework in sections \ref{secmssm} and \ref{analysis}
will be the \rpv extension of the Minimal Supersymmetric Standard Model.
In Section \ref{d:reach} we will also give results in the Minimal 
Supergravity (mSUGRA) model.
The MSSM parameters are the following.
$M_1$, $M_2$ and $M_3$ are the soft-SUSY breaking mass terms
for the bino,  the wino and the gluino, respectively.
$\mu$ is the Higgs mass parameter.
$\tan \beta=<H_u>/<H_d>$ is the ratio of the vacuum expectation values (vev)
for the two-Higgs doublet fields. $A_t$, $A_b$ and $A_{\tau}$ are the
third generation soft-SUSY breaking trilinear couplings. In fact, since
these
trilinear couplings are proportional to the fermion masses one
can neglect the first two generations couplings without any phenomenological
consequence in this context. Finally, $m_{\tilde q}$, $m_{\tilde l}$
and $m_{\tilde \nu}$ are the squark, slepton
and sneutrino mass, respectively.
The value of the squark mass enters our study mainly
in the determination
of the relative branching ratios of the $\tilde \chi^0$ into lepton or
neutrino and of the $\tilde \chi^{\pm}$
into $\tilde \chi^0$ + quarks or $\tilde \chi^0$ + leptons. The remaining
three
parameters $m_{H_u}^2$, $m_{H_d}^2$ and the
soft-SUSY breaking bilinear coupling $B$ are
determined through the electroweak symmetry breaking conditions
which are two necessary minimisation conditions of the Higgs potential.

We choose to study the case of a single dominant
$\lambda_{2jk}^{\prime}$ allowing the reactions
$p p \rightarrow {\tilde{\chi}}^{\pm} \mu^{\mp}$.
In section~\ref{analysis} 
the analysis will be performed explicitly for the
$\lambda_{211}^{\prime}$ coupling, since it corresponds to
the hard subprocess $d \bar d \to \tilde \chi^{\pm}_1 \mu^{\mp}$
which offers the highest partonic luminosity.
We will 
take $\lambda_{211}^{\prime}$=0.09, the upper value
allowed by indirect bound: 
$\lambda_{211}^{\prime}<0.09(m_{\tilde d_R}/100GeV)$  \cite{d:Drein}
for a squark mass of 100~GeV.
A 
quantitative discussion will be given below for the general case
of a single dominant $\lambda_{2jk}^{\prime}$ coupling constant.
We will not treat explicitly the $\lambda^\prime_{1jk}$ couplings
which are associated to the $\tilde \chi^{\pm}$-$e^{\mp}$
production, 
since the low-energy bounds on these couplings 
are rather more stringent than the constraints
on $\lambda^\prime_{2jk}$ and $\lambda^\prime_{3jk}$ \cite{d:Drein}.
However, the three-leptons analysis from sneutrino production
should give similar sensitivities on the $\lambda^\prime_{1jk}$ and
$\lambda^\prime_{2jk}$
couplings since isolation cuts will be included in the selection criteria
for the leptons.
We will not perform the
analysis of the $\lambda^\prime_{3jk}$ couplings which
correspond to the $\tilde \chi^{\pm}$-$\tau^{\mp}$ production.
A technique for mass reconstruction in the ATLAS detector
using the hadronic decays of the $\tau$
has been demonstrated in \cite{d:ianfrank}. The detailed
experimental analysis needed to extract a signal is
beyond the scope of this work.
Besides, in this case the sneutrino and chargino mass reconstruction
studied in Section \ref{secana} is spoiled by the neutrinos produced
in the $\tau$ decay.

\subsection{Single chargino production cross-section}
\label{xsec}

In order to establish the set of models in which
the analysis presented below can be performed, 
we need to study the variations of the single chargino production
rate $\sigma(p p \to \tilde \chi^{\pm} \mu^{\mp})$ with the
MSSM parameters.

\begin{figure}[t]
\begin{center}
\centerline{\psfig{figure=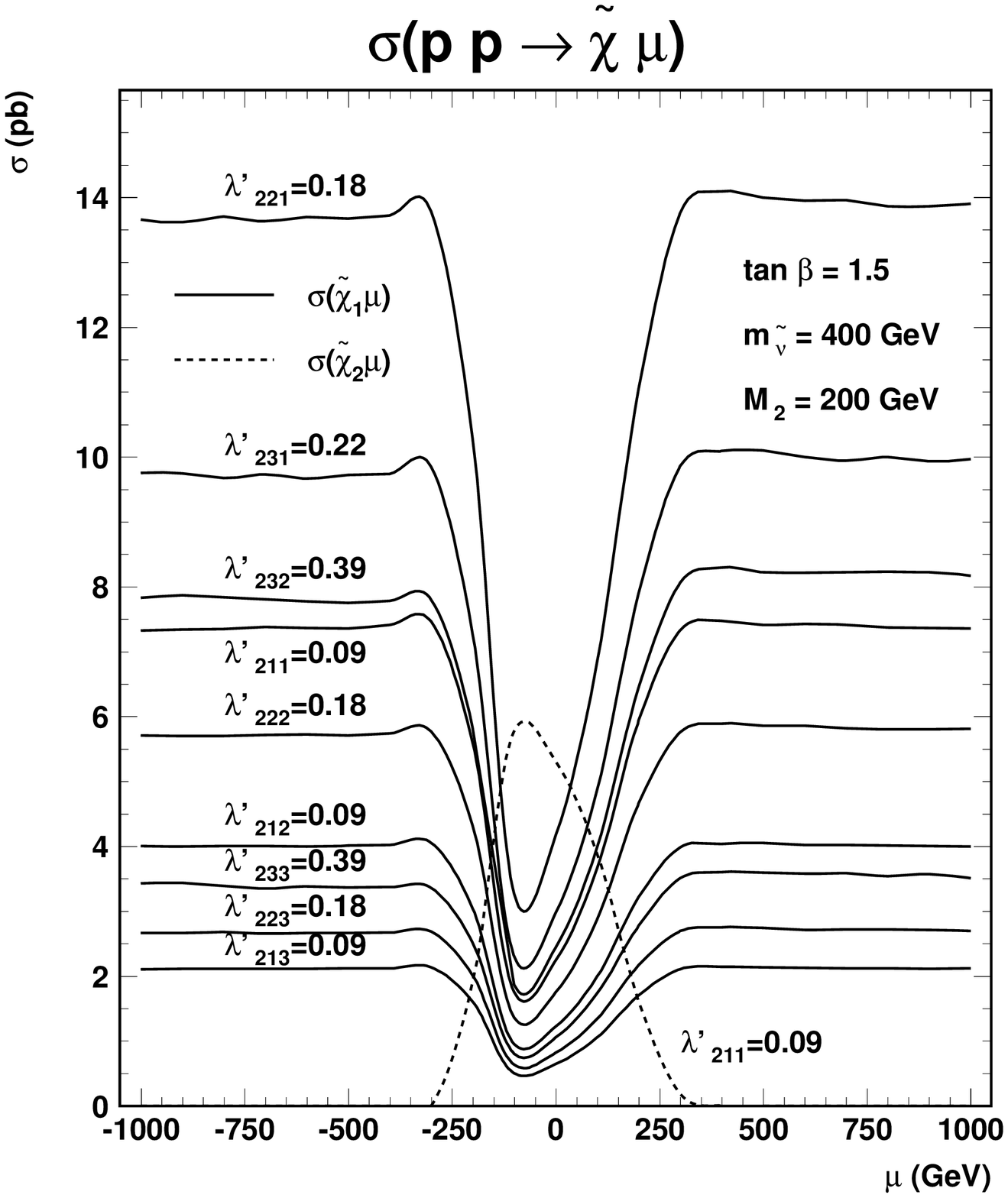,height=5.5in}}
\protect\caption{\em
Cross-sections for the $\tilde \chi_1^{\pm}$-$\mu^{\mp}$
production as a function of the $\mu$ parameter
through various $\lambda'_{2jk}$ couplings,
for $\tan\beta$=1.5, $M_2=200$~GeV and
$m_{\tilde \nu}=400$~GeV. In the case of
$\lambda'_{211}$ the cross-section for $\tilde \chi_2^{\pm}$-$\mu^{\mp}$
is also shown as the dashed line. The values of the
\rpv couplings have been chosen equal to:
$\lambda^\prime_{211}=0.09$,
$\lambda^\prime_{212}=0.09$,
$\lambda^\prime_{213}=0.09$, $\lambda^\prime_{221}=0.18$,
$\lambda^\prime_{222}=0.18$, $\lambda^\prime_{223}=0.18$,
$\lambda^\prime_{231}=0.22$, $\lambda^\prime_{232}=0.39$ 
and $\lambda^\prime_{233}=0.39$, which
correspond to the low-energy limits
for a sfermion mass of $100$~GeV \cite{d:Drein}.}
\label{mssmu}
\end{center}
\end{figure}

In Figure~\ref{mssmu}, we present the cross-sections
for the $\tilde \chi_1^{\pm}$-$\mu^{\mp}$ production
through several $\lambda^\prime_{2jk}$ couplings
as a function of the $\mu$ parameter
for the fixed values: $\tan\beta$=1.5, $M_2=200$~GeV, and
$m_{\tilde \nu}=400$~GeV. For this choice of parameters
and independently of $\mu$,
the chargino ${\tilde{\chi}}^{\pm}_{1}$ is 
lighter than the $\tilde{\nu}$. In this case the contributions of
squark exchange in the $t$ and $u$ channels
are negligible compared to the resonant process
so that the $\tilde \chi_1^{\pm}$-$\mu^{\mp}$ production cross-section
does not depend on the squark mass.
The values for the considered \rpv coupling constants have been 
conservatively taken equal
to the low-energy limits for a sfermion mass of $100$~GeV \cite{d:Drein}.
The cross-sections scale as $\lambda_{2jk}^{\prime 2}$. \\
We see on this Figure that the dependence of the rates on $\mu$ is smooth
for $| \mu | > M_2$.
This is due to the weak dependence of the $\tilde \chi_1^{\pm}$ 
mass
on $\mu$ in this domain.
In contrast, we observe a strong decrease of the rate in the region 
$| \mu | < M_2$
where the
$\tilde \chi_1^{\pm}$ chargino is mainly composed by the higgsino.
Most of the small $ |\mu |$ domain  ($\vert \mu \vert$ smaller than
$\sim 100$~GeV for $\tan \beta=1.41$ and $m_0=500$~GeV)
is however excluded by the present LEP limits~\cite{d:aleph2}. \\
We also show as a dashed line on the plot
the rate for the $\tilde \chi_2^{\pm}$-$\mu^{\mp}$ production through the
$\lambda^\prime_{211}$ coupling.
The decrease of the $\tilde \chi_2^{\pm}$ production rate with increasing
$\vert \mu \vert$
is due to an increase of the $\tilde \chi_2^{\pm}$ mass.
We will not consider the contribution to the three-leptons final state from
the
$\tilde \chi_2^{\pm}$ production since the rate 
becomes important only for a very limited 
range of small $\vert \mu \vert$ values not yet excluded by LEP data. \par
Figure~\ref{mssmu} also allows to compare the sensitivities that
can be reached on various $\lambda'_{2jk}$ couplings
using the single chargino production.
If we  compare for instance the cross-sections of the
$\tilde \chi_1^{\pm}$ production
via $\lambda^\prime_{211}$ and $\lambda^\prime_{221}$
at $\mu=-500$~GeV,
we can see that for equal values of the \rpv
couplings the ratios between the cross-sections
associated to $\lambda^\prime_{211}$ and
$\lambda^\prime_{221}$
is $\sim 2.17$. Therefore, the sensitivity that can be obtained on
$\lambda^\prime_{221}$ is only $\sim \sqrt {2.17}$ times
weaker than the sensitivity on $\lambda^\prime_{211}$,
for a $400$~GeV sneutrino.
Note that the cross-section ratio, and hence the
scaling to be applied, in order to infer
from the reach on $\lambda'_{211}$ the sensitivity on
another coupling $\lambda'_{2jk}$, depends on the 
sneutrino mass.
The reason is that the evolution of the parton densities
with the $x$-Bjorken variable is different for sea quark and valence quark
and for different quark flavours.

\begin{figure}[t]
\begin{center}
\centerline{\psfig{figure=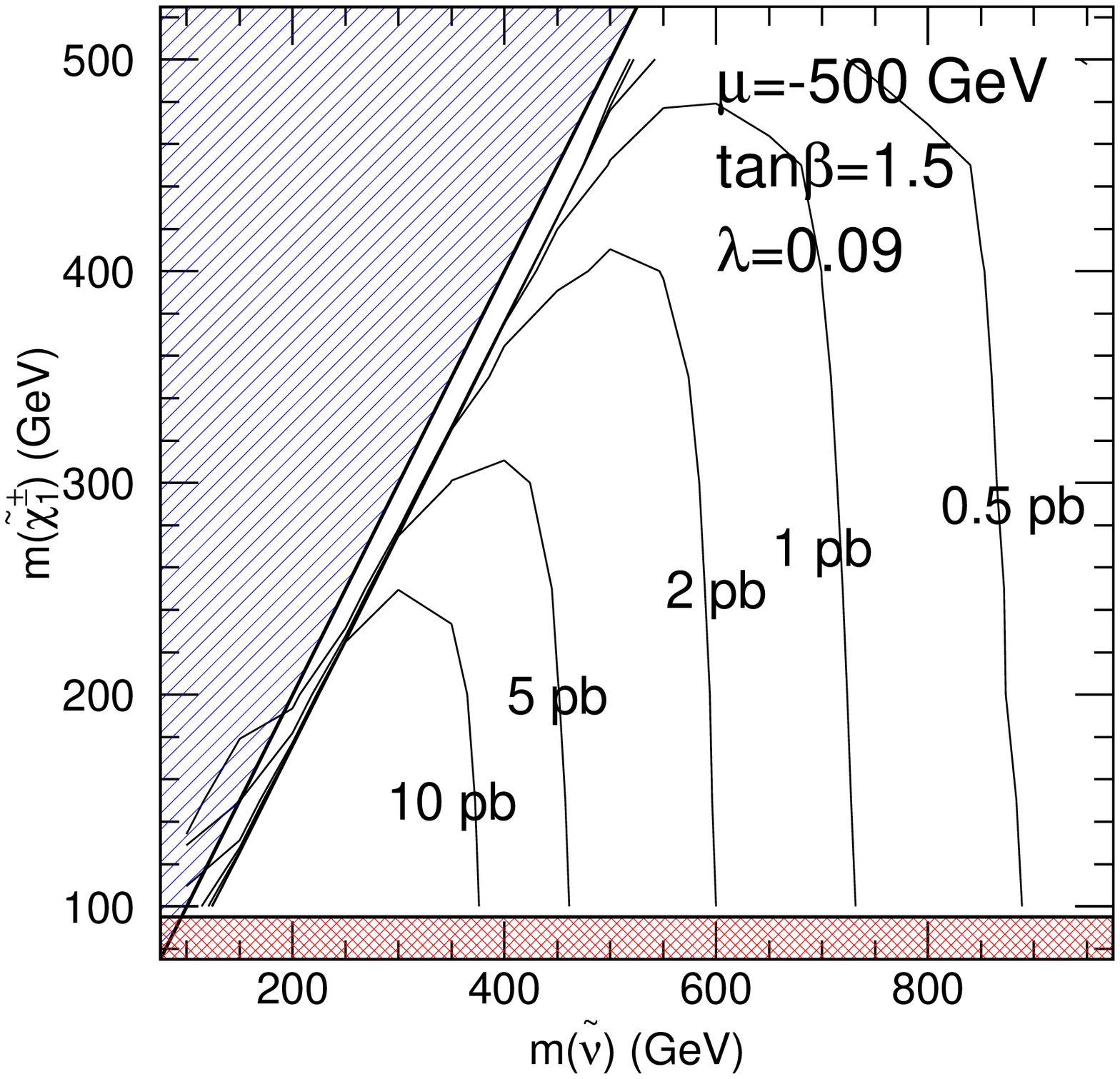,height=5.5in}}
\protect\caption{\em Cross-section for $\chionepm$-$\mu^{\mp}$
production as a function of $m_{\tilde\nu}$ and $m_{\chionepm}$
in the MSSM for the choice of values
$\mu=-500$~GeV, $\tan\beta=1.5$ and
$\lambda^\prime_{211}=0.09$.
The hatched region at the upper
left corresponds to $m_{\tilde \nu}<m_{\tilde\chi^{\pm}_1}$.
The cross-hatched region at low $m_{\chionepm}$ is excluded by the
preliminary LEP results at $\sqrt{s}=196$~GeV \cite{d:aleph2}.}
\label{mssmxse}
\end{center}
\end{figure}

In order to study the dependence of the cross-section on
the masses of the involved sparticles,
the parameters $m_{\tilde \nu}$ and $M_2$ were varied,
and the other model parameters
affecting the cross-section were fixed at the values:
$\lambda^\prime_{211}=0.09$,
$\mu=-500$~GeV and $\tan \beta=1.5$. 
The cross-section for $\chionepm$-$\mu^{\mp}$ production
as a function of
$m_{\tilde\nu}$ and $m_{\chionepm}$ is shown in Figure~\ref{mssmxse}.
Since the $\chionepm$ mass is approximately
equal to $M_2$ as long as $M_2< \vert \mu \vert$, and becomes equal 
to $\vert \mu \vert $ for $M_2 > \vert \mu \vert$,
we studied $\tilde\nu$ masses between 100 and 950 GeV, and values
of
$M_2$ between 100 and 500~GeV. 
For increasing $m_{\tilde\nu}$ the cross-section decreases
due to a reduction of the partonic luminosity.
A decrease of the cross-section is also observed for 
$m_{\chionepm}$  approaching $m_{\tilde\nu}$,
since the phase space factor of the decay $\tilde\nu\to\chionepm\mu^{\mp}$
following the resonant sneutrino production
is then suppressed. 
In the region $m_{\chionepm}>m_{\tilde\nu}$,
the chargino production still receives contributions from 
the $s$ channel exchange of a virtual sneutrino, as well as
from the $t$ and $u$ channels squark exchange 
which in that case also contribute significantly.
However, in this phase space domain where the resonant sneutrino 
production is not accessible, the cross-section is considerably 
reduced.

Finally, the single chargino production rate depends weakly on the
$A$ trilinear couplings. 
Indeed, only the $t$ and $u$ channels squark exchange,
varying with the squark mass which can be influenced by $A$,
depends on these couplings.
The dependence of the rate on the $\tan \beta$ parameter
is also weak.

\subsection{Three leptons branching ratio}

We calculate the total three leptons rate by multiplying
the single chargino cross-section by the 
chargino branching ratio, since we neglect the width of the chargino.
The three-leptons  final state is generated by the cascade decay
$\tilde \chi^{\pm}_1 \to \tilde \chi^0_1 l^{\pm}_p \nu_p$,
$\tilde \chi^0_1 \to \mu u d$.
For $m_{\tilde \nu},m_{\tilde l},m_{\tilde q},
m_{\tilde \chi^0_2}>m_{\tilde \chi^{\pm}_1}$,
the chargino decays mainly 
into a real or virtual $W$ and a $\chione$ and hence 
its branching fraction for the decay into leptons (lepton=$e,\mu$)
is $\sim 22 \%$.

In particular kinematic configurations, the \rpv
modes can compete with the gauge couplings, affecting the
$\chionepm$ branching fractions. However, this does not happen
as long as the chargino is sufficiently heavier than the neutralino,
as is the case for example 
in supergravity inspired models.
When $\tilde \chi^0_1$ is the LSP, 
the branching ratio $B(\tilde \chi^0_1 \to \mu u d)$ ranges between
$\sim 40\%$ and $\sim 70\%$.
For values of $\vert \mu \vert$ much smaller than $M_2$
the other allowed decay $\tilde \chi^0_1 \to \nu_{\mu} d d$  becomes dominant,
spoiling the three-leptons signature.

\section{Experimental analysis}
\label{analysis}

\setcounter{equation}{0}

%
\subsection{Mass reconstruction}
\label{secana}
%
The analysis strategy is based on the exploitation of the decay chain:
\dksl{\snm}{\chionep \;\mu^-}{\chione\;\;W^+\rightarrow e^+(\mu^+)\nu}{\mu^{\pm} \; q\; \bar q^{\prime}}
which presents a sequence of three decays which can be fully reconstructed. 
The strong kinematic constraint provided by the masses of the three sparticles in 
the cascade is sufficient to reduce the contribution of the different
background sources well below the signal rate.

The signal events were generated with a version of the
SUSYGEN MonteCarlo~\cite{d:susygen} modified to allow the generation
of $pp$ processes. The hard-subprocess
$q \bar{q'} \rightarrow \tilde{\chi}^{\pm} \mu^{\mp}$ is first
generated according to the full lowest order matrix elements
corresponding to the diagrams depicted in Figure~\ref{d:fig1}.
Cascade decays of the $\tilde{\chi}$'s are performed according
to the relevant matrix elements. The parton showers
approach~\cite{d:JETSET74} relying on the DGLAP~\cite{d:DGLAP}
evolution equations is used to simulate QCD radiations
in the initial and final states,
and 
the non-perturbative part of the hadronization
is modeled using string fragmentation~\cite{d:JETSET74}.
The 
events were then processed through the program ATLFAST \cite{d:ATLFAST}, 
a parameterized simulation of the ATLAS detector response.

In this section, the analysis will be performed for the
\rpv coupling $\l'_{211}=0.09$ and for the following MSSM point:\\
$M_1=75$ GeV, $M_2=150$ GeV, $\mu=-200$ GeV, $\tan \beta=1.5$, 
$A_t=A_b=A_{\tau}=0$, $m_{\tilde f}=300$ GeV. \\
For this set of MSSM parameters, the masses of the relevant gauginos are:
$$
m_{\chione}=79.9~{\mathrm GeV} \;\; m_{\tilde \chi_1^{\pm}}=162.3~{\mathrm GeV}
$$
and the $\tilde \chi_1^{\pm}$ decay into an on shell $W$ has
a branching ratio of order $100 \%$.
The total cross-section for the resonant sneutrino 
production $pp \to \tilde \nu$ is 37~pb. If we include 
the branching fractions into the three leptons,  the cross-section
is 3.3~pb, corresponding to $\sim 100000$ events for the standard integrated
luminosity of 30~fb$^{-1}$ for the first three years of LHC data taking.

The signal is  characterised by the presence of three isolated leptons and two jets.
For the initial sample selection we require that:
\begin{itemize}
\item
Exactly three isolated leptons are found in the event, with 
$p_T^1>20$~GeV, $p_T^{2,3}>10$~GeV, where $p_T$ is the momentum
component in the plane perpendicular to the beam direction,
and pseudorapidity  $|\eta|<2.5$.
\item
At least two of the three leptons must be muons.
\item
At least two jets with $p_T>15$~GeV are found. 
\item
The invariant mass of any $\mu^+\mu^-$ pair is 
outside $\pm6.5$~GeV of the $Z$ mass. 
\end{itemize}
The isolation prescription on the leptons is necessary to reduce the 
background from the semileptonic decays of heavy quarks, 
and consists in requiring an energy deposition of less than 10~GeV not
associated with the lepton in a pseudorapidity-azimuth ($\eta-\phi$) cone 
of opening $\Delta R=0.2$ around the lepton direction.\par
The efficiency for  these cuts, after the branching fractions have been taken 
into account,  is $\sim 25\%$, where half of the loss comes from requiring
three isolated leptons, 
and the other half is the loss of jets from 
$\chione$ decay either because they are not reconstructed, or
because the two jets from the decay are reconstructed 
as a single jet . The $Z$ mass cut gives a 10$\%$ loss in statistics. 
In order to avoid the combinatorial background from additional 
QCD events we further require that no third jet with $p_T>15$~GeV
is reconstructed in the event. The efficiency after this cut is $\sim15\%$.\par
The reconstruction of the sparticle masses could be performed either
starting from the $\chione$ reconstruction and going up the decay chain,
or trying to exploit the three mass constraints at the same time.
We choose the first approach which is not optimal, but allows a clearer
insight into the kinematics of the events.
%
\begin{figure}[t]
\begin{center}
\centerline{
\psfig{figure=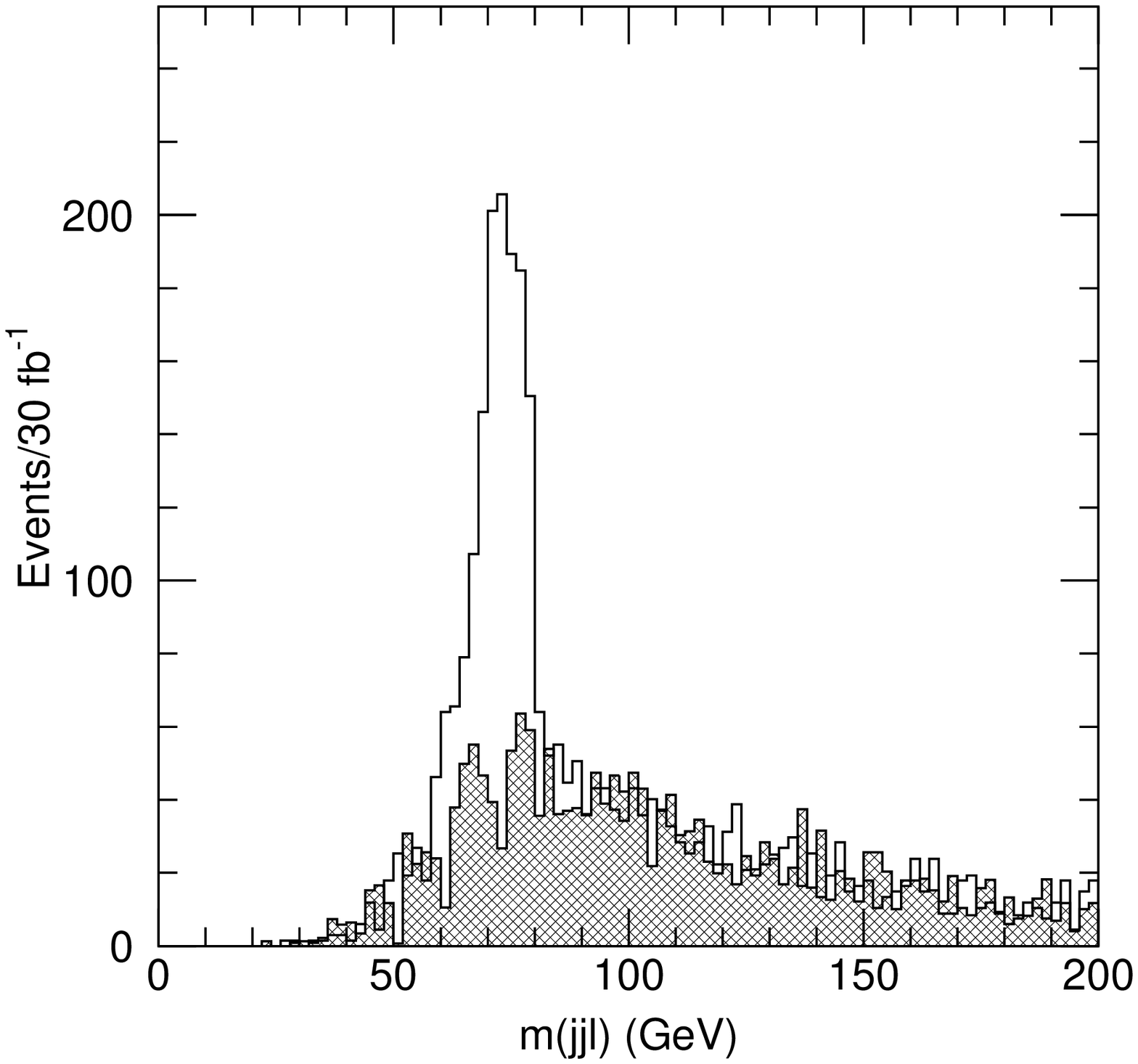,height=3.5in}
\psfig{figure=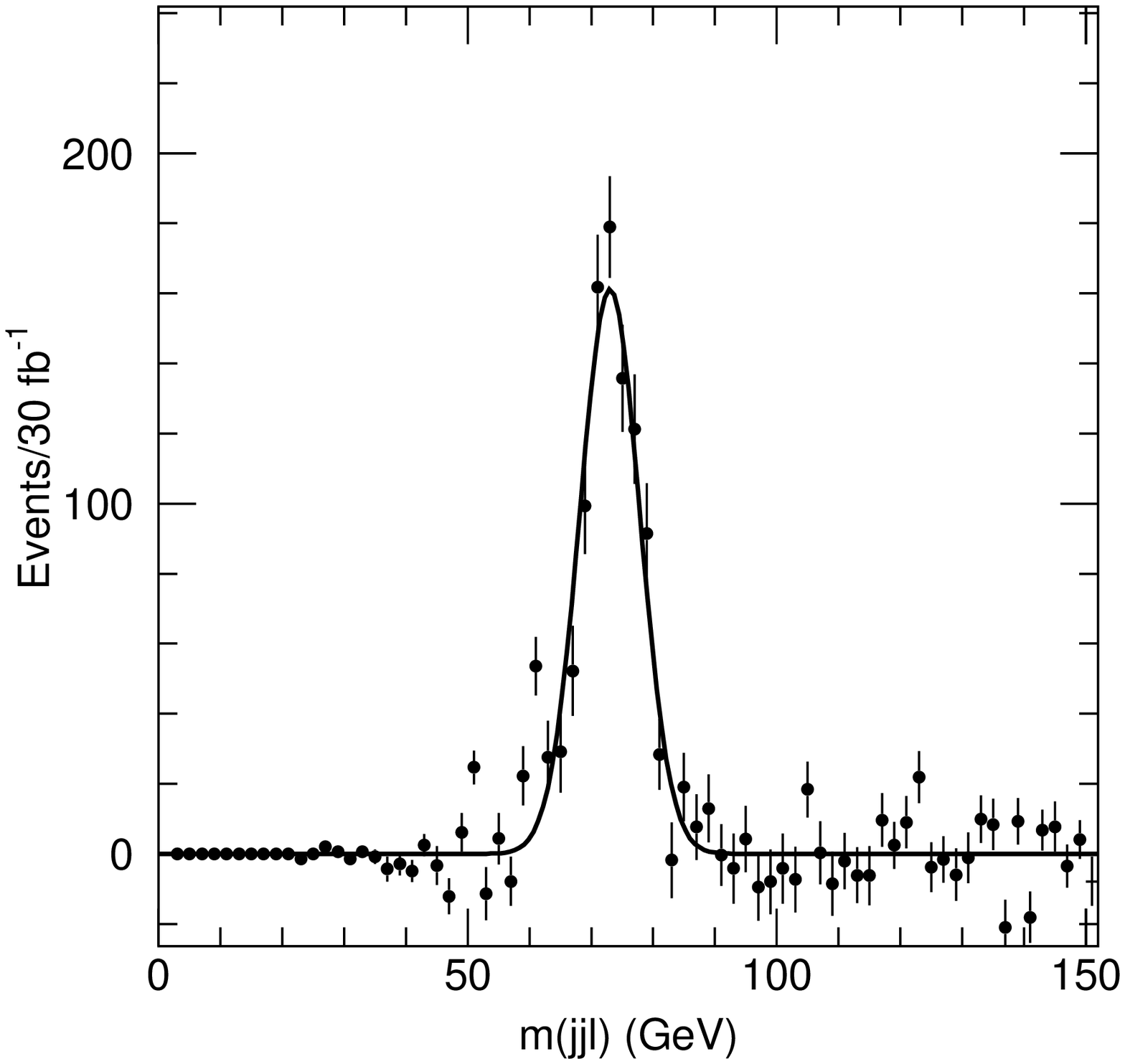,height=3.5in}}
\protect\caption{\em $\mu$-jet-jet invariant mass for events in configuration 1. (see text)
Left: exclusive two jet events with superimposed (hatched) 
the combinatorial background 
Right: $\chione$ peak after background
subtraction.}
\label{figchi01}
\end{center}
\end{figure}

The first step in reconstruction of the $\chione \rightarrow \mu$~jet~jet 
is the choice of the correct muon to attempt the reconstruction. 
The three leptons come in the following  flavour-sign configurations (+ charge conjugates):
\begin{enumerate}
\item
$\mu^- e^+\mu^+$
\item
$\mu^- e^+\mu^-$
\item
$\mu^-\mu^+\mu^+$
\item
$\mu^-\mu^+\mu^-$
\end{enumerate}
where the first lepton comes from the $\snm$, the second one from 
the $W$, and the third one from the $\chione$ decay, corresponding to
three final state signatures : 1)  two opposite sign muons and an electron
\footnote{Here and in the following, ``electron'' stands for both $e^+$ and $e^-$.}, 
2) two same-sign muons and an electron, 3-4) three muons. The configuration 
with three same-sign muons does not correspond to the required signature
and is rejected in the analysis.
For signature 1) 
the muon produced in the $\chione$ decay is defined as the one which has 
the same sign as the electron. For configuration 2) both muons must be tested
to reconstruct the $\chione$. For configuration 3-4), the $\chione$ muon must be
one of the two same-sign ones.\par
%
\begin{figure}[t]
\begin{center}
\centerline{\psfig{figure=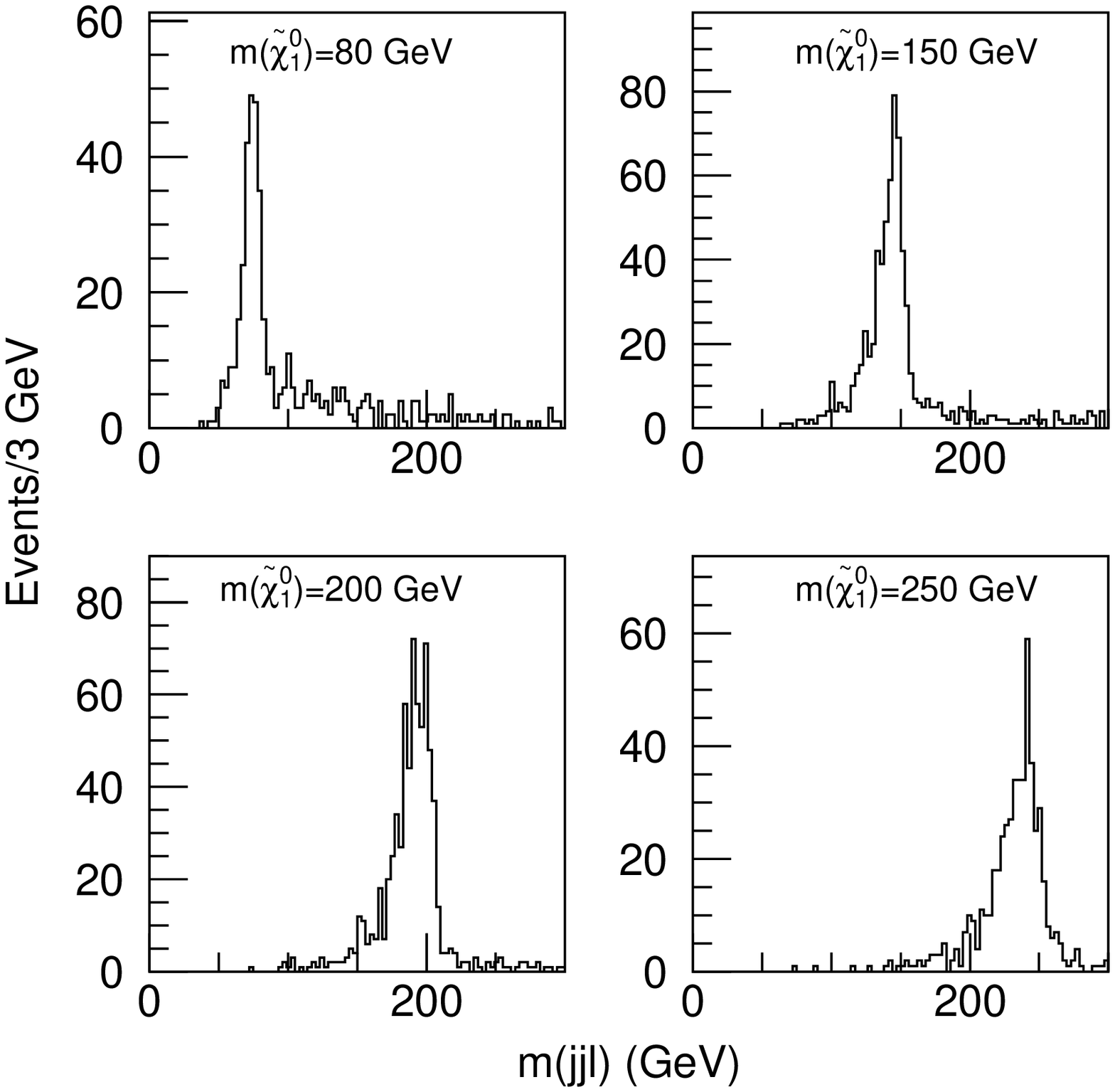,height=5.5in}}
\protect\caption{\em Lepton-jet-jet invariant mass for exclusive two-jet 
events where the $\chione$ lepton is
uniquely defined, for four different values of the $\chione$ mass:
$m_{\chione}$=80, 150, 200 and 250~GeV.
In all cases the sneutrino mass is set at 500~GeV, \mbox{$-\mu=M_2=2M_1$}, 
and \mbox{$\tan\beta=1.5$},
yielding a $\chionepm$ mass twice the $\chione$ mass. 
All the sfermion masses are set to 500~GeV.\
The normalisation is arbitrary.}
\label{figchi}
\end{center}
\end{figure}

In order to minimise the combinatorial background 
we start the reconstruction from signature
1) where each lepton is unambiguously attributed to a step in the decay.
The distribution of the $\mu$-jet-jet invariant mass 
is shown in the left plot of Figure~\ref{figchi01}.
A clear peak is visible corresponding to the $\chione$
mass superimposed to a combinatorial background of events where one of the
two jets from the $\chione$ was lost and a jet from initial state radiation
was picked up. The combinatorial background can be evaluated
using three-jet events, where at least one jet is guaranteed to come
from initial state radiation. The shape of the combinatorial background estimated 
with this method is shown as the
shaded histogram superimposed to the signal peak. 
After background subtraction, an approximately gaussian  peak 
with a width of $\sim 4.5$~GeV, and a statistics of about 1050 events
is reconstructed, shown in the right of Figure~\ref{figchi01}. 
If we consider a window of $\pm12$~GeV around the peak,
corresponding  to $\sim2.5\sigma$  of the gaussian, $\sim1500$ events are observed
in the sample, and the combinatorial contamination is approximately $\sim30\%$.
A tail towards low mass values is observed, corresponding to events
where a fraction of the parton energy is lost in the jet reconstruction.
From this distribution the $\chione$ mass can be measured 
with a statistical error of $\sim 100$~MeV.
The measurement error will in this case be dominated by the systematic
error on the jet energy scale which in ATLAS is estimated to be at the level
of $1\%$ \cite{d:TDR}.\\ 
The $30\%$ combinatorial background is due to 
the 'soft' kinematics of the chosen example point, with a $\chione$ which is both
light and produced with a small boost. In order to show the effect of the
mass hierarchy of the involved sparticles,
the shape of the $\chione$ mass peak is shown in Figure~\ref{figchi} for a sneutrino 
mass of 500~GeV and different choices for the $\chione$ mass. In all cases 
the $\chionepm$ mass is twice the $\chione$ mass, corresponding to 
the gauge unification condition and to $|\mu|$ values of 
the same order as $M_2$. The combinatorial
background is in general smaller than for a 300~GeV sneutrino, due to the higher boost
imparted to the $\chione$, and it decreases  with increasing 
$\chione$ masses, due to the 
higher efficiency for reconstructing both jets from the $\chione$ decay. 
For this analysis no attempt has been done for the recalibration of the
jet energy. This results in the skewing of the distributions towards 
low masses, and in the peak value being slightly displaced 
with respect to the nominal mass value.\\
Once the position of the $\chione$ mass peak is known, 
the reconstructed $\chione$ statistics 
can be increased  by also considering signatures 
2) and 3-4).
For events coming from signatures 2 to 4, the $\chione$ candidate is 
defined as the muon-jet-jet combination which gives a mass 
nearest to the mass peak determined from signature 1) events. In all cases 
the reconstructed mass is required to be within $\pm12$~GeV of the peak position
to define a $\chione$ candidate. 
In $83\%$ of the events containing at least a combination satisfying 
this requirement,  only one $\chione$ candidate is found,
and this sample can be used
to improve the statistical precision on the $\chione$ mass measurement.\par
%
\begin{figure}[t]
\begin{center}
\centerline{\psfig{figure=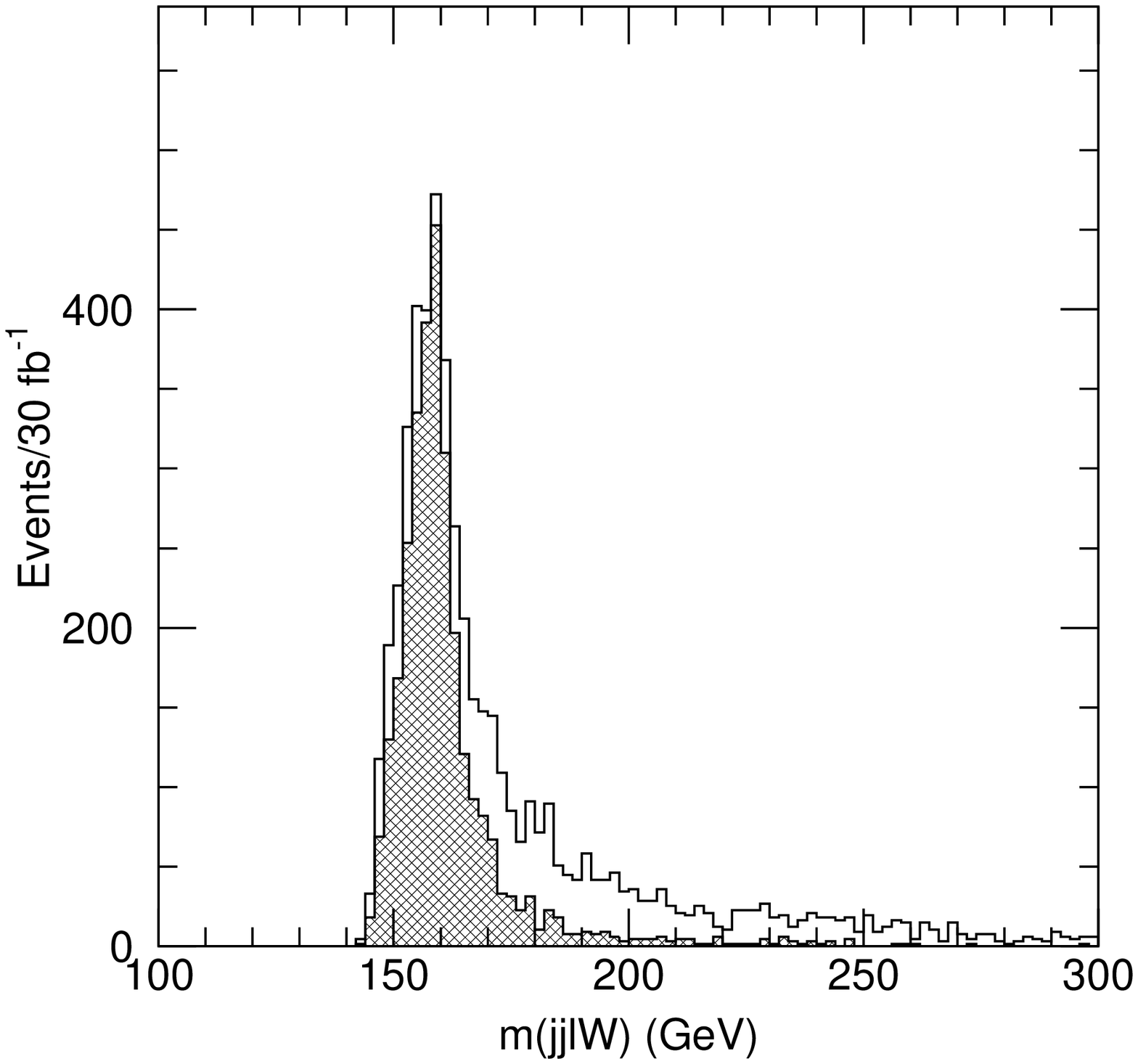,height=5.5in}}
\protect\caption{\em Invariant mass of the $\chione$ with the W candidate. The full
line histogram includes both solutions for the neutrino longitudinal
momentum, the grey one only includes the solution which gives the mass
nearest
to the measured peak.}
\label{figchip}
\end{center}
\end{figure}
%
%
Using the above definition of the $\chione$, we can go further in 
the mass reconstruction of the involved sparticles.
Only configurations 1) and 2) are used,
i.e. the events containing two muons and an electron 
in order to avoid ambiguities in the choice of the lepton from the $W$ decay.
The preliminary step for the reconstruction of the $\chionepm$ is 
the reconstruction of the $W$ boson from its leptonic decay.
The longitudinal momentum of the neutrino from the $W$ decay is calculated from the
missing transverse momentum of the event (considered as $p_T^{\nu}$) and the
requirement that the electron-neutrino invariant mass gives the $W$ mass.
The resulting neutrino longitudinal momentum, 
has a twofold ambiguity. We therefore build the invariant $W-\chione$ mass 
candidate using both solutions for the $W$ boson momentum.
The resulting spectrum is shown in Figure~\ref{figchip},
as the full line histogram.
A clear peak is seen, superimposed on a combinatorial background. If only
the solution yielding the  $\chionepm$ 
mass nearest to the measured mass peak is retained, 
the mass spectrum corresponding to the shaded histogram is obtained.
The peak in the unbiased histogram can be fitted with a gaussian shape, 
with a width of \mbox{$\sim6$~GeV}. \\
The combination with the mass nearest to the measured peak is
taken as $\chionepm$ candidate,  provided that the reconstructed
mass is within 15~GeV of the peak. For $80\%$ of the $e \mu \mu$ events where a 
$\chione$ candidate is found, a $W-\chione$ combination satisfying 
this requirement is reconstructed.\par
%
\begin{figure}[t]
\begin{center}
\centerline{\psfig{figure=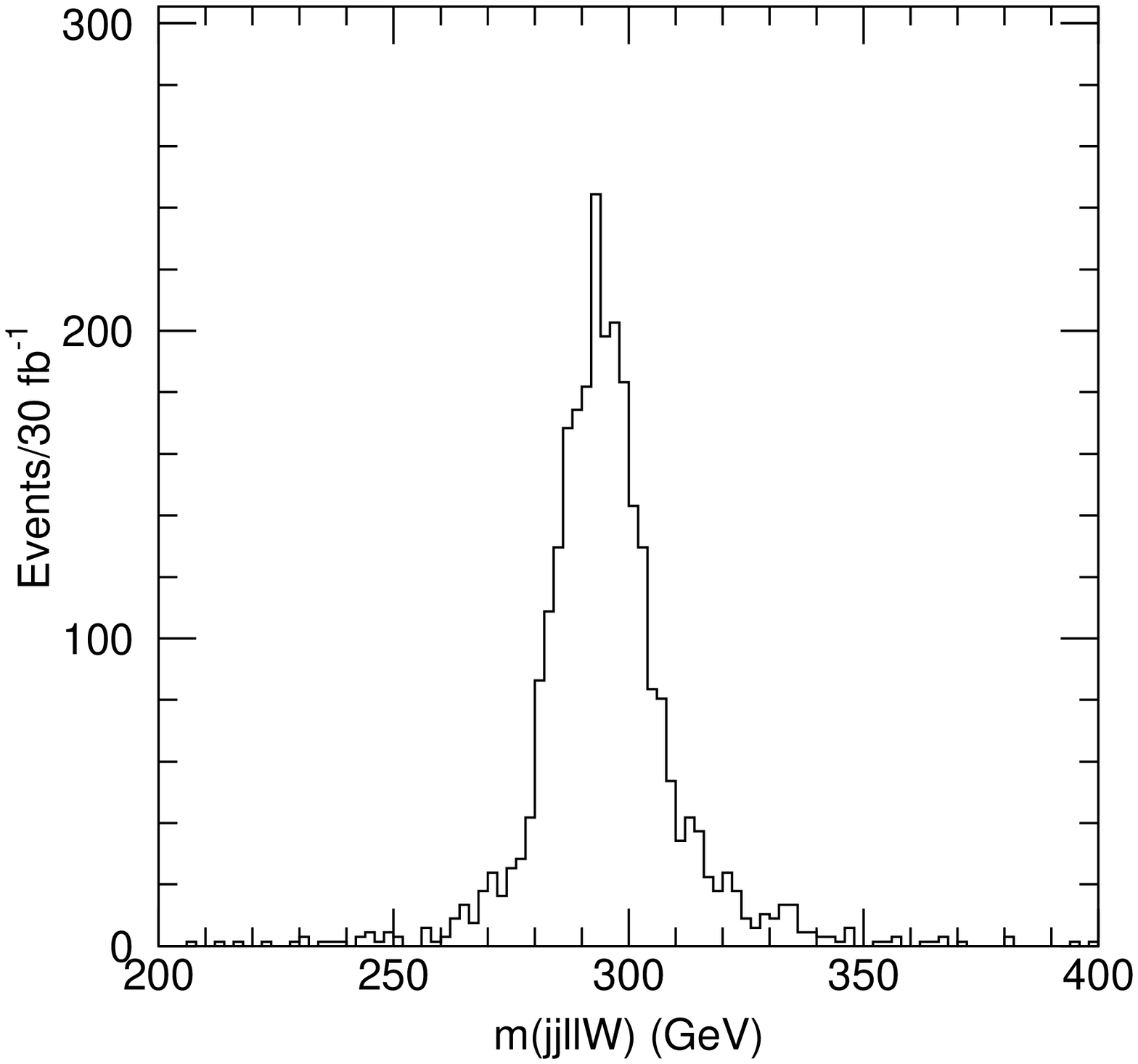,height=5.5in}}
\protect\caption{\em Invariant mass of the third lepton in the event with the
$\chionepm$ candidate.}
\label{figsl}
\end{center}
\end{figure}
%
Finally the $\chionepm$  candidates  are combined with the leftover muon, 
yielding the mass spectrum shown in Figure~\ref{figsl}. The $\tilde\nu$ 
mass peak 
at this point presents very limited tails, and has a width
of \mbox{$\sim10$~GeV}. 
We define fully reconstructed events as those
for which this mass lies within 25 GeV of the measured
$\tilde{\nu}$ peak.  From the estimate of the combinatorial under
the $\chione$ peak, we expect approximately 
$2 \times 1050 = 2100$ events where all the jets
and leptons are correctly assigned over a total of 2450 events observed 
in the peak. The difference between the two numbers are events for which 
one of the two jets used for the $\chione$ reconstruction comes 
from initial state radiation. These jets are typically soft, and therefore
the reconstructed $\chione$ candidate very often has a momentum which both in magnitude
and direction is close to the momentum of the original $\chione$.
Therefore for such events the  reconstructed $\chione$ behaves in the
further steps in the reconstruction as the real one, only inducing 
some widening in the $\chionepm$ and $\tilde\nu$ peaks.
\\
The statistics available at the different steps in the analysis for an integrated
luminosity of 30~fb$^{-1}$ is given in the first column of Table~\ref{tbg}. 
%
\begin{table}
\begin{center}
\begin{tabular}{|l|r|r|r|r|r|r|}
\hline
Process & Signal & $\bar tt$ & $WZ$ & $Wbb$ & $Wt$ & $Zb$ \\
\hline
$\sigma$ (pb) & 3.3 & 590 & 26 & 300 & 60 & 7000 \\
N$_{ev}$(30 fb$^{-1}$) & $1\times 10^5$ & $1.7\times 10^7$ & $8\times 10^5$ & $9\times 10^6$ &
$1.8 \times 10^6$ & $2.1 \times 10^8$ \\
Loose cuts    & 23600 & 2900 & 53   & 2.4 & 3.5 & 56 \\
Jet veto      & 14200 & 1450 & 38   &  -  &  -  & 30  \\
$\chione$     &  6750 & 158  &  4   &  -  &  -  & -   \\
$\chionepm$   &  2700 &  8   &  0.4 &  -  &  -  & -   \\
$\snm$        &  2450 &  0   & 0.25 &  -  &  -  & -  \\
\hline
\end{tabular}
\protect\caption{\em Cross-sections and expected numbers of events after cuts
for the signal and the different Standard Model
background contributions considered in the analysis.
The ``Loose cuts" are described at the beginning of section
\ref{secana}, and the ``Jet veto" consists in adding the requirement that no third
jet with $p_T>15$~GeV is reconstructed in the event.
The line labelled ``${\tilde{\chi}}^0_1$" gives the number of events
from signatures 1 to 4 ($e \mu \mu$ and $\mu \mu \mu$)
for which a $\tilde{\chi}^0_1$ candidate is found.
The line labelled  ``${\tilde{\chi}}^{\pm}_1$" shows the
number of events from signatures 1 and 2 ($e \mu \mu$)
where a $\tilde{\chi}^{\pm}_1$ candidate is found in
addition, and the last line indicates the number of
fully reconstructed events.
In the case of the signal, we give the cross-section for the resonant
sneutrino production multiplied by the branching ratios into three leptons.
}
\label{tbg}
\end{center}
\end{table}
%
For the assumed value of the coupling, $\lambda^{\prime}_{211}=0.09$, 
the uncertainty on the measurement of all the three masses involved will be 
dominated by the 1$\%$ uncertainty on the jet energy scale.\\
The efficiency for the reconstruction of the full decay chain 
with the analysis described above is \mbox{$\sim2.5\%$}. A more sophisticated analysis 
using also the three-muons events should approximately double this efficiency.\\
From the observed number of events and the 
$\tilde\nu$ mass a measurement of the quantity
$\lambda^{\prime 2}_{211}\times~BR$, where $BR$ is the product of the branching ratios
of the decays shown in equation \ref{eqq}, is possible. 
The measurement of additional SUSY processes 
will be needed to disentangle the two terms of this product. 
%
%
\subsection{Standard Model Background}
%
\label{secback}
The requirement of three isolated leptons in the events 
strongly reduces the possible background sources.  
The following processes were considered as a background:
\begin{itemize}
\item
$\bar tt$ production, followed by $t\rightarrow Wb$, where the two $W$ and one 
of the $b$ quarks decay leptonically. 
\item
$WZ$ production, where both bosons decay leptonically. 
\item
$Wt$ production
\item
$Wbb$ production
\item 
$Zb$ production
\end{itemize}
These backgrounds were generated with the PYTHIA MonteCarlo~\cite{d:PYTHIA}, except 
$Wt$ and $Wbb$
for which the ONETOP parton level generator~\cite{d:ONETOP} was used, 
interfaced to PYTHIA for 
hadronisation and fragmentation.
The cross-sections for the various processes, and the number of
total expected events 
for an integrated luminosity of 30~fb$^{-1}$ 
are given in Table~\ref{tbg}, according to the cross-section numbers used in 
the ATLAS physics performance TDR \cite{d:TDR}. 
In particular, even when the cross-section 
is known at NLO, as in the case of the top, the 
Born cross-section is taken for internal
consistency of the study.\\
For each of the background processes a sample  of events between one seventh and
a few times the expected statistics was generated and passed through the
simplified simulation of the ATLAS detector.\par       
%
\begin{figure}[t]
\begin{center}
\centerline{\psfig{figure=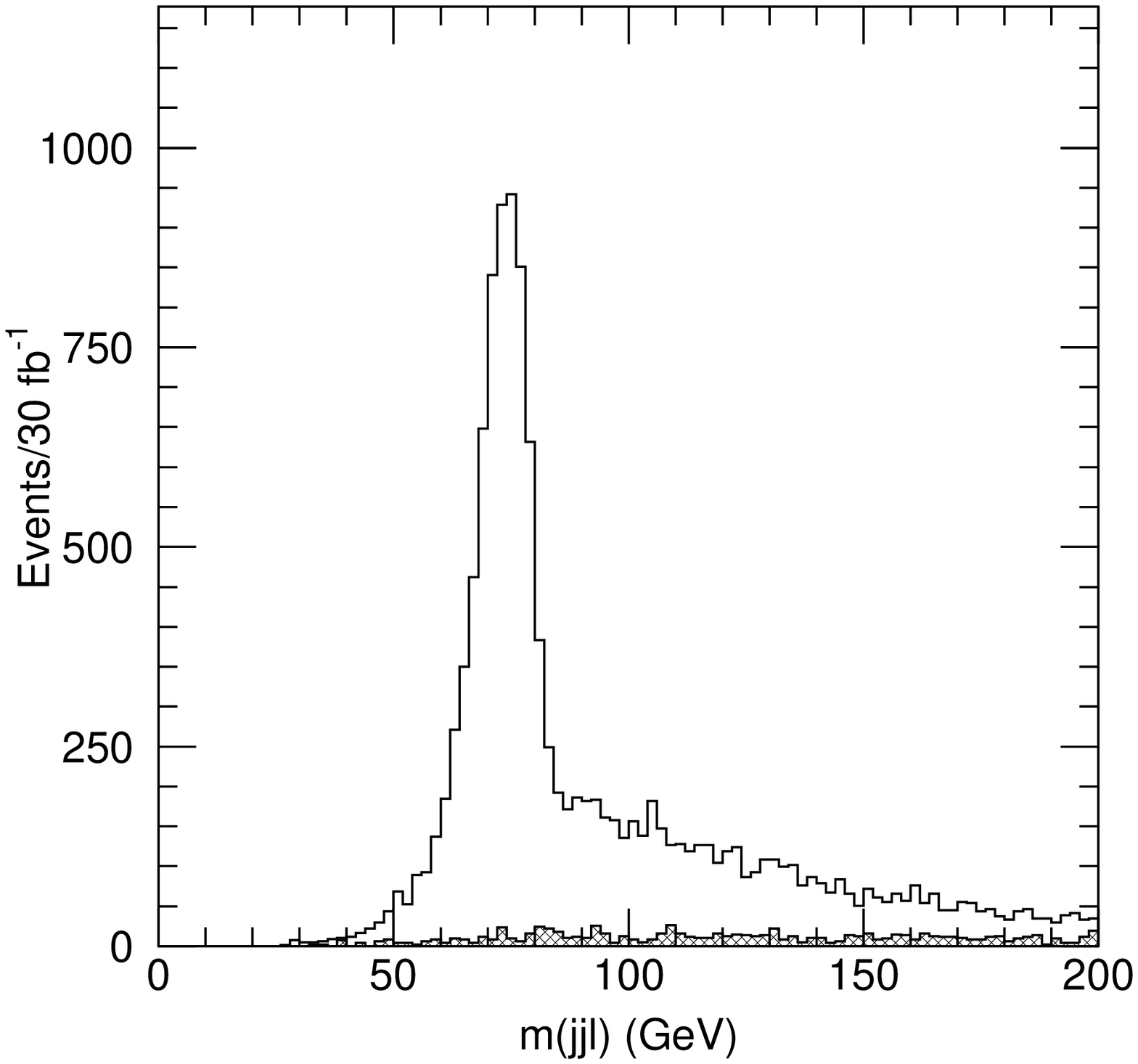,height=5.5in}}
\protect\caption{\em Invariant  mass of the $\chione$ candidates entering the 
kinematic analysis superimposed to the 
Standard Model background (hatched).}
\label{figbg}
\end{center}
\end{figure}
%
After the loose selection cuts described in Section~\ref{secana}, 
the background is dominated
by top production, as can be seen from the numbers shown in Table~\ref{tbg}. 
The distribution of the $\mu$-jet-jet invariant mass for 
background events, obtained as in Section \ref{secana} 
and corresponding to the $\chione$ candidates selection, 
is shown as the hatched histogram in Figure~\ref{figbg}.
In this figure we have superimposed the same distribution for the signal. 
Already at this level, the signal stands out very clearly from the background, and in the
following steps of the reconstruction the background becomes almost
negligible. The numbers of 
background and signal events expected at the various steps of the reconstruction
can be compared in Table~\ref{tbg}. 
The full analysis was performed only for the $\bar tt$ and
$WZ$ background because for the other channels the background is essentially 
negligible compared to top production, and  in most cases the MonteCarlo statistics
after the initial selection was too low to allow a detailed study. 
For the SUSY model considered and the chosen value of the $\lambda^{\prime}$
coupling constant, even the loose selection applied allows to efficiently
separate the signal from the background. 
%
\subsection{Sensitivity on $\lambda^\prime$}
\label{seclam}
%
From these results, it is possible to evaluate the minimum value of the 
$\lambda_{211}^{\prime}$ coupling for which it will be possible to discover the signal.
The starting point in the analysis is the observation of a peak in the 
muon-jet-jet invariant mass
over an essentially flat background. All of the further analysis steps
of the cascade reconstruction rely on the possibility of selecting the
events with a mass around the  $\chione$ peak.\\
For the observation of the peak, the best signal/background ratio is obtained using the 
three-muons sample (configurations 3 and 4 above). 
In the Standard Model, which incorporates lepton universality, 
about one eight of the
three-leptons  events present a three-muons configuration, 
whereas about half of the signal 
events come in this configuration, thereby granting an improvement of a factor 4 
in signal over background, with respect to the full sample.
The three muons come either in the  '$-++$' or in the '$-+-$' sign configuration,
because the two muons from the decay chain
$\tilde\nu (\overline{\tilde\nu} )\to\chionepm\mu^{\mp}\to\chione\mu^{\pm}\mu^{\mp}$ must
have opposite sign, whereas the $\chione$ can decay to muons of either sign.
Therefore the muon for the $\chione$ reconstruction must be chosen 
between the two same-sign ones.
%
\begin{figure}[t]
\begin{center}
\centerline{\psfig{figure=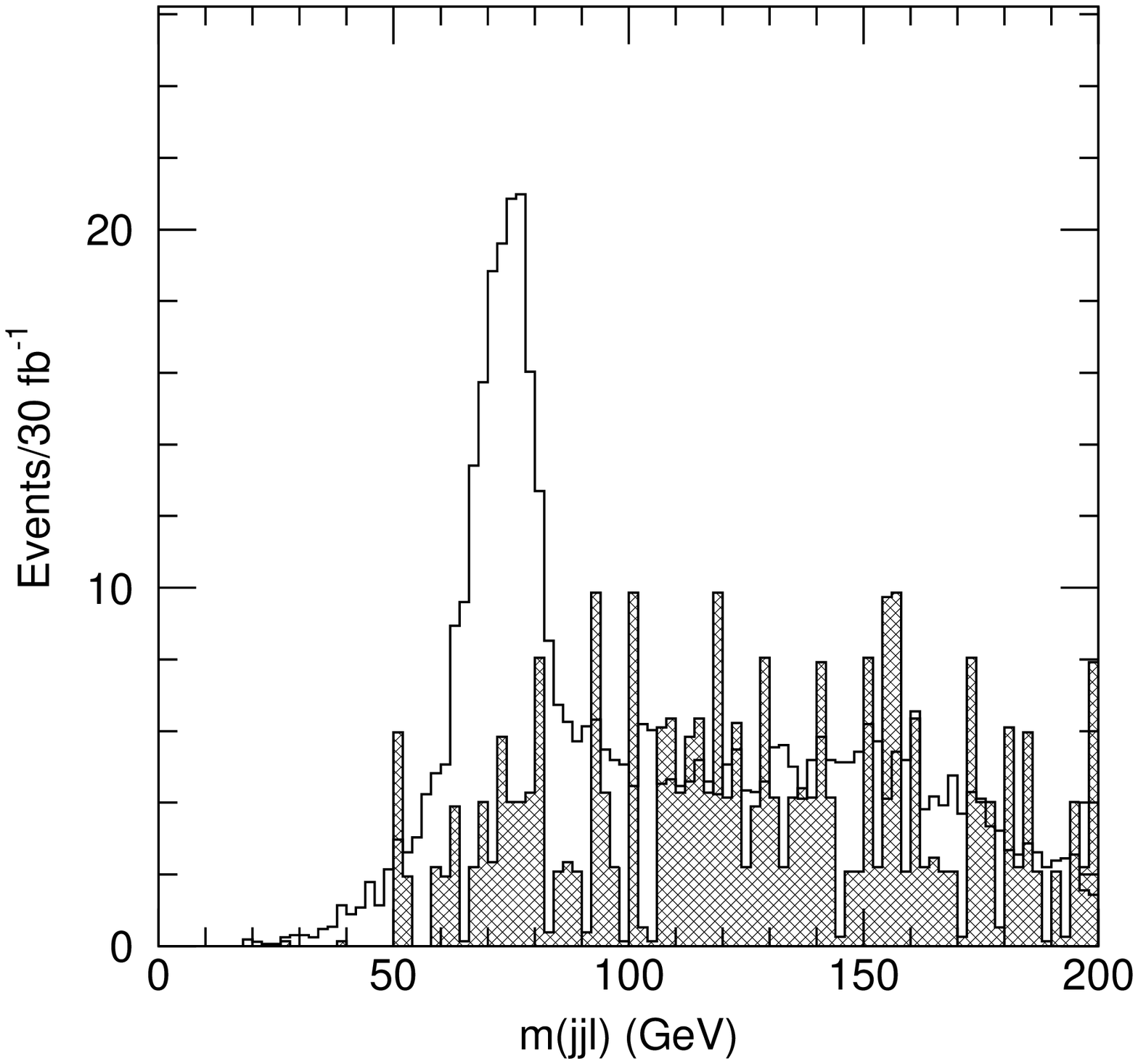,height=5.5in}}
\protect\caption{\em Invariant  mass of the $\chione$ candidates from three-muon
events scaled down by a factor 25,
corresponding to a $\lambda^{\prime}$ value of 0.018,
superimposed to the
Standard Model background (hatched).}
\label{figbg1}
\end{center}
\end{figure}
%
The distribution for the $\mu$-jet-jet invariant mass, for events containing two
jets and three muons is shown in Figure~\ref{figbg1}, scaled down 
by a factor 25,  corresponding to a $\lambda^{\prime}$ value of 0.018, 
superimposed to the expected  top background. In the distribution
each event enters twice, for each of the two same-sign muons which can be 
used  to reconstruct the $\chione$.
We expect, however,  that the
combination with the ``wrong" muon gives in most cases a reconstructed
mass outside of the $\chione$ peak.\\
A statistical prescription is needed to 
define the fact that a peak structure is seen in the signal+background distribution.
Given the exploratory nature of the work, we adopt the naive approach
of calculating the $\lambda^\prime$ value for which $S/\sqrt{B}=5$,
where $S$ and $B$ are respectively the number of signal and background candidates 
counted in an interval of $\pm 15$~GeV around the measured $\chione$ peak. 
The window for the definition of a $\chione$ candidate is enlarged with
respect to the analysis described in Section~\ref{secana}, 
in order to recover the non-gaussian tail of the signal peak, thus increasing
the analysis efficiency.
In this interval, for the chosen point, for an integrated luminosity of 30~fb$^{-1}$, 
$S=580000\times (\lambda^\prime)^2$ and $B=46$ events. In the hypothesis that
the $\bar tt$ background can be precisely measured from data, the 
lower limit on $\lambda^\prime_{211}$ is:
$$\lambda^\prime_{211}>0.0075$$
\\
The pair production of SUSY particles through standard $R_p$-conserving
processes is another possible source of background, due to the possibility 
to obtain final states with high lepton multiplicity, and the high production cross-sections.
This background 
can only be evaluated inside  models  providing predictions 
for the whole SUSY spectrum. As a preliminary study, a sample of 
events were generated with the HERWIG 6.0 MonteCarlo~\cite{d:herwig} by setting the 
slepton masses at 300~GeV, the masses of squarks and gluinos at 1000~GeV
and the chargino-neutralino spectrum as for the example model.
The total $R_p$-conserving cross-section is in this case \mbox{$\sim6$~pb}.
A total of 60 SUSY background events which satisfy the requirements 
used above to define $S$ and $B$ are observed.
All the events surviving the 
cuts are from direct chargino and neutralino
production, with a small contribution
from Drell-Yan slepton production. Since the contributions 
from squark and gluino decays are strongly suppressed by the jet veto requirements,
this result can be considered as a correct order of 
magnitude estimate, independently 
from the assumed values for the squark and gluino masses.
Moreover, the reconstructions of the chargino and sneutrino masses
can also be used in order to reduce the SUSY background.  
A more thorough discussion of the SUSY  
background 
will be given below in the framework of the mSUGRA model.
%
%
%
\section{Analysis reach in various models}
\label{d:reach}

\setcounter{equation}{0}

%
For the example case studied in Section~\ref{secana} it was shown  
that the sneutrino production 
signal can be easily separated from the background, and allows to 
perform precision measurements of the masses of the sparticles 
involved in the decay chain. \\
The analysis can be generalised to investigate the range of SUSY 
parameters in which this kind of analysis is possible, and 
to define the minimum value of the $\lambda^\prime$ constant
which gives a detectable signal in a given SUSY scenario.
The different model parameters enter the definition of the detectability 
at different levels:
\begin{itemize}
\item
The sneutrino production cross-section is a function only of the
sneutrino mass and of the square of the R-parity violating coupling constant.
\item
The branching fraction of the sneutrino decay into three 
leptons is a function of all the SUSY parameters. 
\item
The analysis efficiency is a function of the 
masses of the three supersymmetric particles involved in the decay.
\end{itemize} 
%
\begin{figure}[t]
\begin{center}
\centerline{\psfig{figure=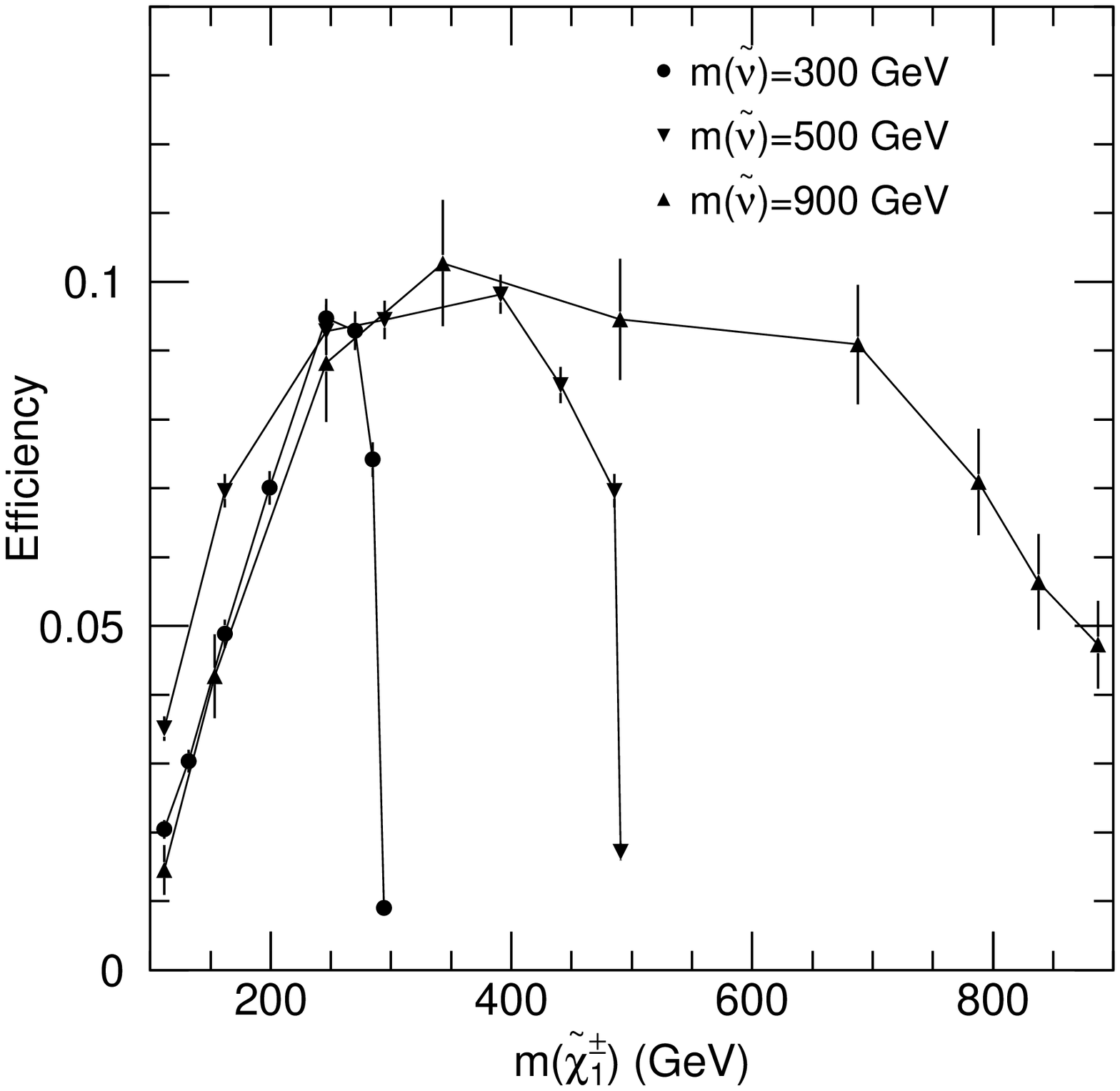,height=5.5in}}
\protect\caption{\em Efficiency for reconstructing  a $\mu$-jet-jet
invariant mass within 15 GeV of the $\chione$ mass in three-muons events,
as a function of the $\chionepm$ mass. The shown points were generated 
with the following parameters: \mbox{$-\mu=M_2=2M_1$}, \mbox{$\tan\beta=1.5$}. 
All the sfermion masses are set equal to the sneutrino mass.
Points  for $m_{\tilde\nu}= 300, 500$ and $900$~GeV are shown.}
\label{pleff}
\end{center}
\end{figure}
%
The dependences of the cross-section and branching ratios 
on the SUSY parameters were discussed in Section \ref{secmssm} for the MSSM, 
and are summarised in Figures~\ref{mssmu} and \ref{mssmxse}.
We only need at this point to parameterize the analysis efficiency
as a function of the sparticle masses. 
The number of signal events for each considered model will then be obtained 
by multiplying the expected number of three-lepton events by the
parameterized efficiency.
\subsection{Efficiency of the three-muon analysis}
According to the discussion presented 
in Section~\ref{seclam}, we need to calculate 
the efficiency for the signal process to satisfy the following requirements:
\begin{itemize}
\item
to pass the initial selection cuts described in Section~\ref{secana}
(Loose cuts), including
the veto on the third jet (Jet veto);
\item
to contain three reconstructed muons, with one of the $\mu$-jet-jet invariant masses 
within 15 GeV of the $\chione$ mass.
\end{itemize}
Three sneutrino  masses, $m_{\tilde\nu}=~300, 500$ and $900$~GeV
were considered, and for each of these the evolution of the efficiency 
with the $\chionepm$ mass was studied.  The mass of the $\chione$ was 
assumed to be half of the mass of the $\chionepm$, relation 
which is in general valid in SUGRA inspired models and correspond
to a choice of values for $|\mu|$ of the same order as $M_2$. \\
The analysis efficiency 
is shown in Figure~\ref{pleff} as a function of the $\tilde\nu_{\mu}$ mass and
of the $\chionepm$ mass. The efficiency values are 
calculated with respect to the number of events 
which at generation level did contain the three leptons, therefore 
they only depend on the event kinematics and not on the
branching ratios.
The loss of efficiency at the lower end of the  $\chionepm$
mass spectrum is due to the inefficiency for detecting two jets
from the $\chione$ decay, either because the two jets are 
reconstructed as a single jet, or because one of the two jets 
is below the detection threshold of 15 GeV. 
The efficiency then becomes approximately independent of the masses of
the sneutrino and of the $\chionepm$, up to the point
where the $\tilde\nu$ and $\chionepm$ masses become close enough to affect 
the efficiency for the detection of the muon from the 
\mbox{$\tilde\nu\to\chionepm\mu$} decay; for $m_{\tilde\nu}-m_{\chionepm}<10$~GeV
the analysis efficiency rapidly drops to zero. The moderate decrease 
in efficiency at high $\chionepm$
masses for $m_{\tilde\nu}=900$~GeV can be ascribed to the fact that one of two
energetic jets from the $\chione$ decay radiates a hard gluon, three jets 
are reconstructed, and the event is rejected by the jet veto.
\\
At this point all the ingredients are available  to study 
the reach in the parameter space for the analysis presented in Section~\ref{analysis}
within different  SUSY models. \\
%
\subsection{Analysis reach in the MSSM}
\label{mssm}
\begin{figure}[t]
\begin{center}
\centerline{\psfig{figure=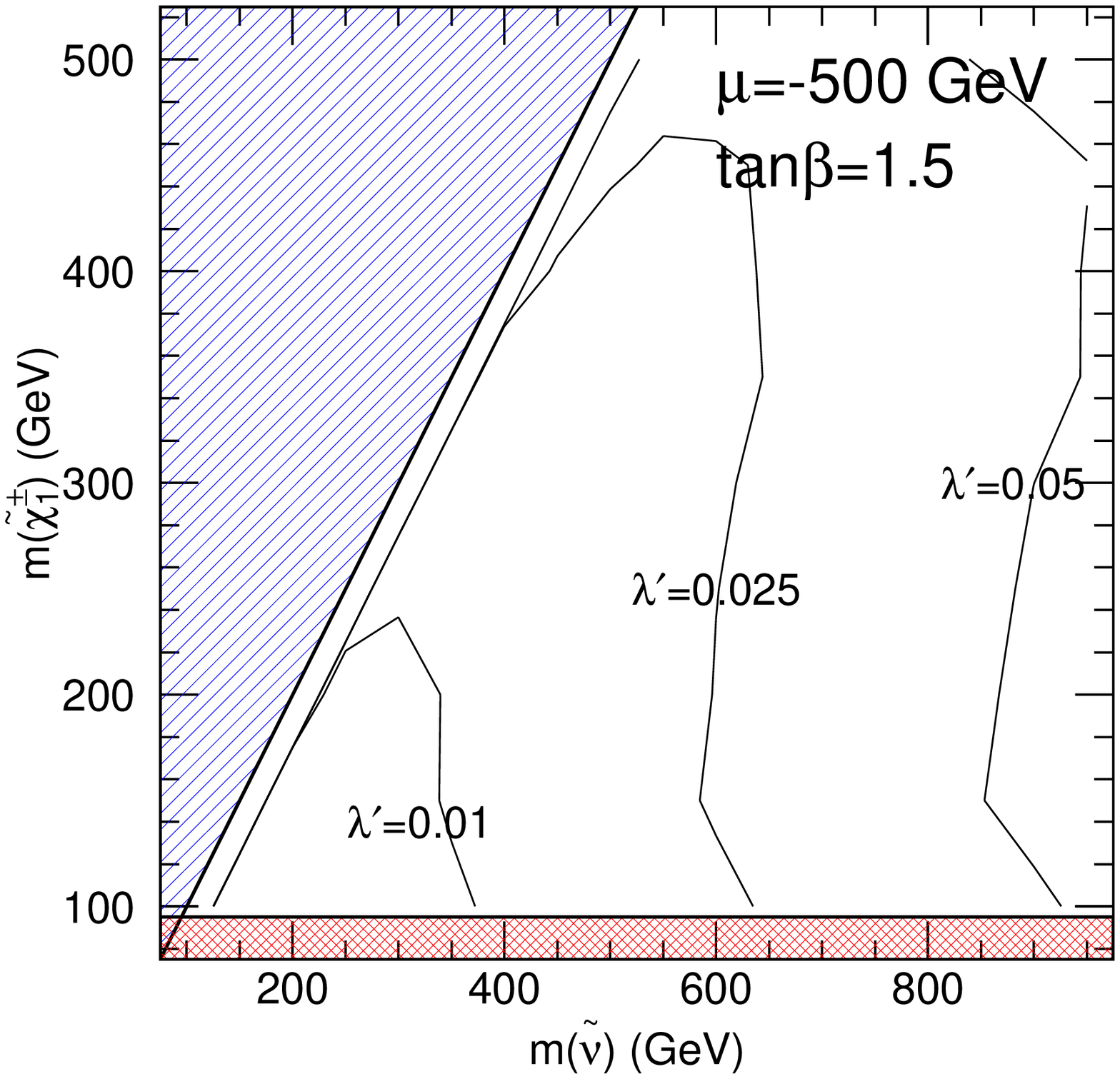,height=5.5in}}
\protect\caption{\em
$5\sigma$ reach in the $m_{\tilde\nu}$-$m_{\chionepm}$
plane for three
different choices of the $\lambda^\prime_{211}$ coupling for 
an integrated luminosity of 30~fb$^{-1}$ at the LHC.
The chosen model parameters were:
\mbox{$\mu=-500$~GeV}, \mbox{$\tan\beta$=1.5}, 
\mbox{$m_{\tilde q}=m_{\tilde l}=300$~GeV},
\mbox{$A_t=A_b=A_{\tau}=0$}, \mbox{$M_2=2M_1$}.
The significance is defined
only considering the Standard Model background, and a signal
of at least ten events is required. 
The hatched region at the upper
left corresponds to \mbox{$m_{\tilde \nu}<m_{\tilde\chi^{\pm}_1}$}.
The cross-hatched region at low $m_{\chionepm}$ is excluded by the
preliminary LEP results at $\sqrt{s}=196$~GeV \cite{d:aleph2}.}
\label{mssmreach}
\end{center}
\end{figure}

%
The region in the $m_{\tilde\nu}$-$m_{\chionepm}$ plane for which the
signal significance is  greater than 5$\sigma$, as defined  in Section~\ref{seclam},
and at least 10 signal events
are observed for an integrated luminosity of
30~fb$^{-1}$ is shown in Figure~\ref{mssmreach}
for different choices of the $\lambda^{\prime}_{211}$ constant.
%
%
The behaviours of the sensitivity curves in the $m_{\tilde\nu}$-$m_{\chionepm}$ plane
are well explained by the variations of the single chargino production 
cross-section shown in Figure \ref{mssmxse} in the same plane of parameters.   
The SUSY background is not considered in the plot, as it depends on 
all the model parameters. It was however verified in a few example cases that 
for our analysis cuts this background is dominated 
by direct chargino and neutralino production, and  it
becomes negligible in the limit of high 
$\chione$ and $\tilde \chi^{\pm}$ masses. The main effect
of taking into account this background will be to reduce the significance 
of the signal for $\chionepm$ masses lower  than 200~GeV.\par
From the curves in Figure~\ref{mssmreach} we can conclude that  
within the MSSM, the production of
a 900~GeV sneutrino for $\lambda^{\prime}_{211}>0.05$, and of a 350~GeV sneutrino for
$\lambda^{\prime}_{211}>0.01$ can be observed 
within the first three years of LHC running,
provided that the sneutrino is heavier than 
the lightest chargino.\par
The sensitivity on an \rpv coupling of type $\l'_{2jk}$ can be derived from 
the sensitivity obtained for $\l'_{211}$, as explained in Section~\ref{xsec}.
For example, we have seen that the sensitivity on $\l'_{221}$ was
$\sim 1.5$ times weaker than the sensitivity on $\l'_{211}$,
for $\tan\beta$=1.5, $M_2=200$~GeV, $\mu=-500$~GeV and
$m_{\tilde \nu}=400$~GeV.
This set of parameters leads to a sensitivity on 
$\l'_{211}$ of about $0.015$ as can be seen in Figure~\ref{mssmreach},
and hence to a sensitivity on $\l'_{221}$ of $\sim 0.022$.  
In Table \ref{couplg}, we present the sensitivity on any $\l'_{2jk}$ coupling 
estimated using the same method and for the same MSSM parameters. 
Those sensitivities represent an important
improvement with respect to the low-energy limits of \cite{d:Drein}. \\
In the case of a single dominant $\l'_{2j3}$
coupling 
the neutralino decays as $\tilde \chi^0_1 \to \mu u_j b$
and the semileptonic decay of the b-quark could affect
the analysis efficiency. Hence in this case, the precise sensitivity
cannot be simply calculated by scaling the value obtained for
$\lambda^\prime_{211}$. The order of magnitude of the sensitivity
which can be inferred from our analysis should however be correct.  
%
\begin{table}
\begin{center}
\begin{tabular}{|c|c|c|c|c|c|c|c|c|}
\hline
$\l'_{211}$ & $\l'_{212}$ & $\l'_{213}$ & $\l'_{221}$ & $\l'_{222}$ 
& $\l'_{223}$ & $\l'_{231}$ & $\l'_{232}$ & $\l'_{233}$  \\
\hline
0.01 & 0.02 & 0.02 & 0.02 & 0.03 & 0.05 & 0.03 & 0.06 & 0.09 \\

\hline
\end{tabular}
\protect\caption{\em
Sensitivities on the $\l'_{2jk}$ coupling constants deduced from the sensitivity
on $\l'_{211}$ for $\tan\beta$=1.5, $M_1=100$~GeV, $M_2=200$~GeV, $\mu=-500$~GeV,
$m_{\tilde q}=m_{\tilde l}=300$~GeV and
$m_{\tilde \nu}=400$~GeV.}
\label{couplg}
\end{center}
\end{table}
%
\subsection{Analysis reach in mSUGRA}
\label{msugra}
%
\begin{figure}[t]
\begin{center}
\centerline{
\psfig{figure=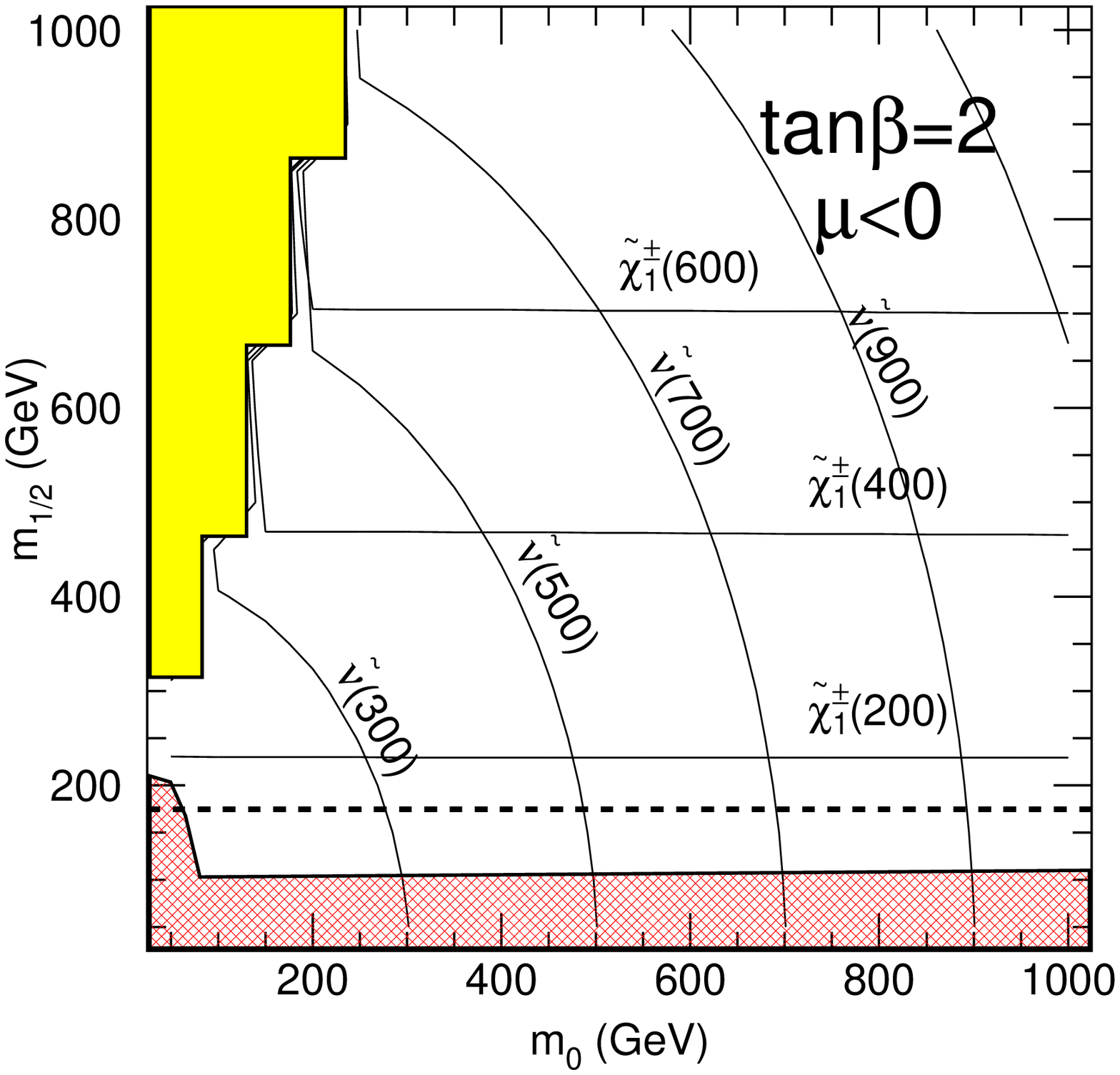,height=3.5in}
\psfig{figure=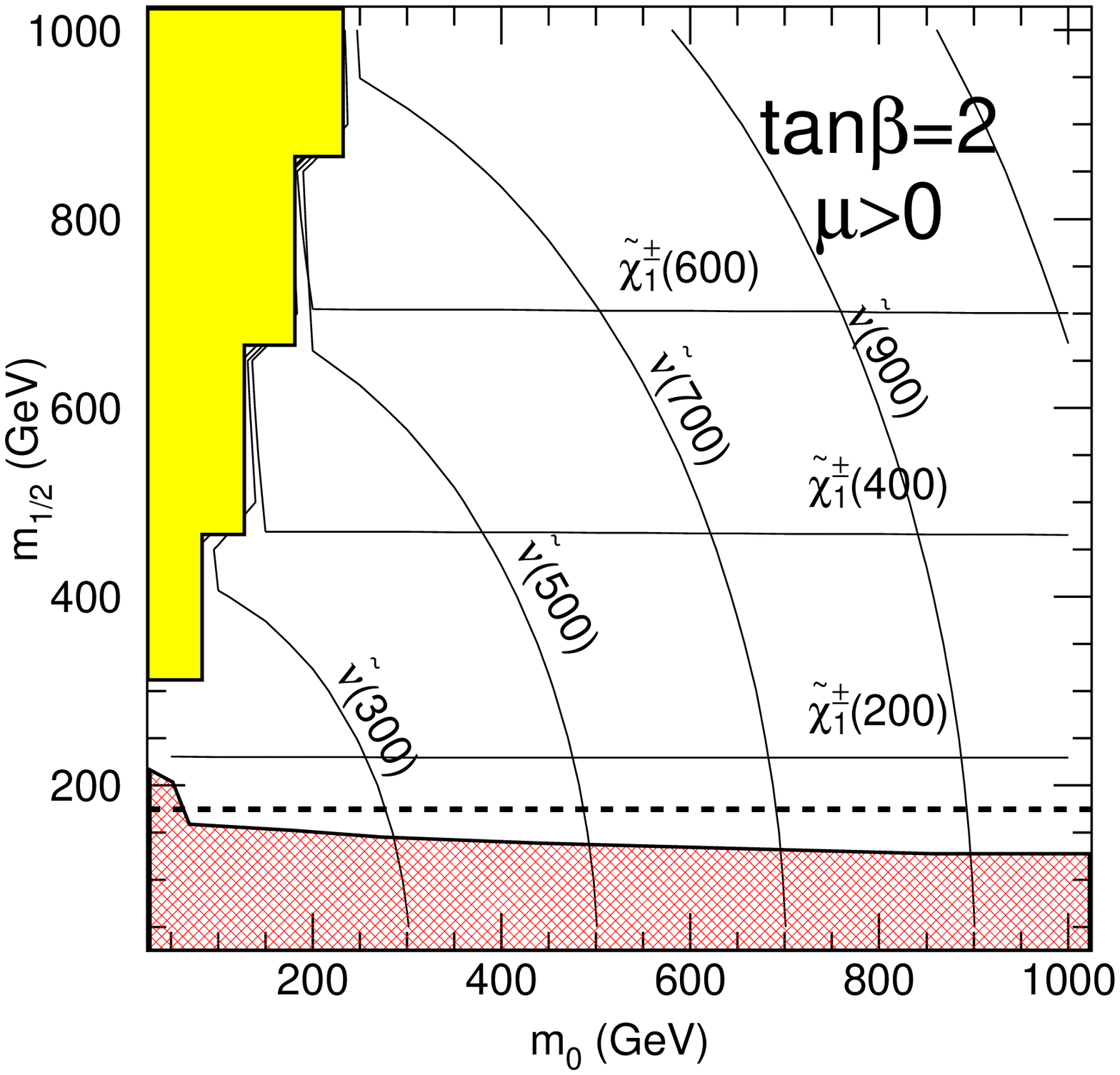,height=3.5in}}
\protect\caption{\em Curves of equal mass for $\tilde\nu$ and $\chionepm$
in the $m_0-m_{1/2}$ plane for $\tan\beta=2$.
The grey region at the upper left indicates the domain where the $\chione$ is
not the LSP.
The cross-hatched region for low $m_{1/2}$ gives the kinematic
limit for the discovery of $\chionepm$  or $\tilde l$ by LEP
running at  $\sqrt{s}=200$~GeV.
The dotted line shows the
region below which the $\chionepm$ decays to a virtual W.
}
\label{plmass}
\end{center}
\end{figure}
%
Our framework throughout this Section will be 
the so-called minimal supergravity model.
In this model the parameters obey a set of boundary conditions at the
Grand Unification Theory (GUT) scale $M_x$. These conditions appear
to be natural in supergravity scenario since the supersymmetry breaking 
occurs in an hidden sector which communicates with the visible sector
only through gravitational interactions. 
First, mSUGRA contains the gauge coupling unification at $M_x$, 
such an unification being suggested by the
experimental results obtained at LEP I.
One can view the gauge coupling unification assumption 
as a fixing of the GUT scale $M_x$. 
Second, the gaugino (bino, wino and gluino) masses at $M_x$ 
are given by the universal mass $m_{1/2}$.
the parameters $m_{1/2}$ and $M_i \ [i=1,2,3]$ 
are thus related by the solutions of the renormalization group 
equations (RGE). Besides, since the gaugino masses and 
the gauge couplings are governed by the
same RGE, one has the well-known relation: 
$M_1={5 \over 3} \tan^2 \theta_W M_2$.
Similarly, at $M_x$, the universal scalars mass is $m_0$ and the 
trilinear couplings are all equal to $A_0$. Finally, in mSUGRA the
absolute value of the higgsino mixing parameter 
$\vert \mu \vert$ as well as the bilinear coupling $B$ are determined by the 
radiative electroweak symmetry breaking conditions.
Therefore, mSUGRA contains only the five following parameters:
$sign(\mu)$, $\tan\beta$, $A_0$, $m_0$ and $m_{1/2}$.\par 
Due to the small  dependence of the single chargino production rate
on the $\mu$ parameter for $M_2\le|\mu|$ (see Section \ref{secmssm}), 
the study of the mSUGRA model in which $\vert \mu \vert$ is fixed 
by the electroweak symmetry breaking condition provides information on 
a broader class of models. 
The single chargino production rate depends mainly on the values 
of $m_0$ and $m_{1/2}$ (see Section \ref{secmssm}).
We will set $A_0=0$,  and study the detectability
of the signal in the $m_0-m_{1/2}$  plane. 
We show in Figure~\ref{plmass} the curves of equal mass for
$\tilde\nu$ and $\chionepm$ for $\tan\beta=2$ calculated with
the ISASUSY \cite{d:isajet} package which uses one-loop  RGE
to get the SUSY spectrum from the mSUGRA parameters.
\par
%
The signal reach can be easily evaluated from the sparticle mass
spectrum and branching fractions by using the parameterization of the
analysis efficiency shown in Figure~\ref{pleff}.
%
\subsubsection{Supersymmetric background}
%
In the case of a well constrained model as mSUGRA, the SUSY background
can in principle be evaluated in each considered point in the parameter space.
For this evaluation the full SUSY sample must be 
generated for each point, requiring 
the generation of a large number of events. \par
The sparticle masses for the model studied in detail in Section~\ref{analysis}
uniquely define a model in the mSUGRA  framework. Therefore, as a first approach
to the problem, a full analysis was performed for this model,  
corresponding to the parameter values: 
$$
m_{0}=275~{\mathrm{GeV}},~~~~m_{1/2}=185~{\mathrm{GeV}},~~~~\tan\beta=1.5,~~~~ \mu<0,~~~ A_0=0.  
$$
%
\begin{figure}[t]
\begin{center}
\centerline{\psfig{figure=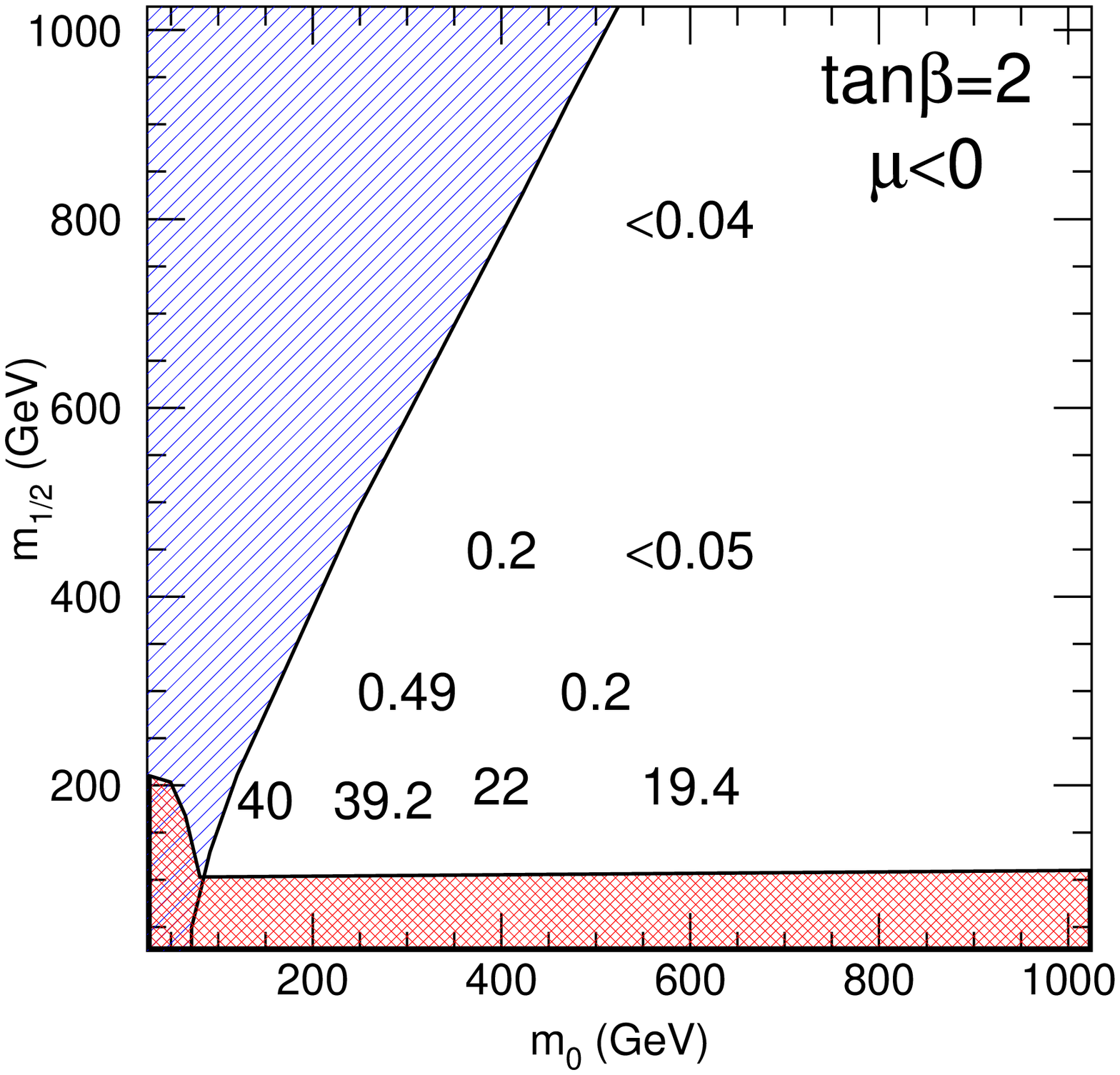,height=5.5in}}
\protect\caption{\em Number of SUSY background events for an integrated
luminosity of $30$~fb$^{-1}$ in the $m_0-m_{1/2}$ plane with $\tan\beta$=2
for a few selected test models. The hatched region at the upper
left corresponds to $m_{\tilde\nu}<m_{\tilde\chi^{\pm}_1}$.
The cross-hatched region for low $m_{1/2}$ gives the kinematic 
limit for the discovery of $\chionepm$  or $\tilde l$ by LEP 
running at  $\sqrt{s}=200$~GeV.}
\label{plback}
\end{center}
\end{figure}
%
For this mSUGRA point, the mass scale for squarks/gluinos is
in the proximity of 500~GeV, and the total cross-section 
for all the SUSY particles pair productions is
approximately 130~pb, yielding a signal of \mbox{$\sim4 \ 10^6$} events for
the first three years of data-taking at the LHC.
A total of 400k events were generated and analysed.
The number of surviving events after cuts in the three-muons sample was $47\pm21$
for an integrated luminosity of 30~fb$^{-1}$. All the background events 
come from direct chargino and neutralino production, as it was the case for the 
MSSM point studied in Section~\ref{seclam}. As a cross-check, we generated
for the same mSUGRA point only the processes of the type $pp\to\tilde\chi+X$,
where $\tilde\chi$ denotes either $\tilde\chi^0$ or $\tilde\chi^{\pm}$, 
and $X$ any other SUSY particle.
The cross-section is in this case 
\mbox{$\sim6$~pb}, and the number of background events is $39\pm7$ events,
in good agreement with the number evaluated generating all the SUSY processes.\par
Based on this result, we have performed
a scan in the $m_0-m_{1/2}$ plane the fixed values $\tan\beta=2$, $\mu<0$.
On a grid of points  we generated  event samples 
for the $pp\to\tilde\chi+X$ processes with the 
HERWIG 6.0 MonteCarlo~\cite{d:herwig}. 
The number of SUSY events with a $\mu$-jet-jet combination with an invariant
mass within 15~GeV of the $\chione$ mass is shown in 
Figure~\ref{plback} in the $m_0-m_{1/2}$ plane for an integrated
luminosity of 30~fb$^{-1}$. The background is 
significant for a $\chionepm$ mass of 175~GeV ($m_{1/2}=$200~GeV), and becomes 
essentially negligible for $\chionepm$ mass of 260~GeV ($m_{1/2}=$300~GeV).
This behaviour is due to the combination of two effects: the $\tilde \chi^{\pm} \tilde \chi^{\pm}$
production cross-section decreases with increasing $\tilde \chi^{\pm}$ mass, and the probability
of losing two of the four jets from the decay of the two $\chione$ in the event
becomes very small for a $\chionepm$ mass of $\sim220$~GeV. Indeed,
the suppression of the SUSY background is mainly due to the Jet veto cut. \par
Given the high SUSY cross-section, and the high lepton multiplicity
from $\chione$ decays, a prominent signal should manifest itself 
through R-conserving sparticle pair production 
in this scenario. Single resonant sneutrino production will then be used as 
a way of extracting information on the value of the $R_p$-violating
coupling constant, and of precisely measuring the masses of $\tilde\nu_\mu$,
$\chionepm$, $\chione$. Moreover,
thanks to the very high number of produced $\chione$
expected from $\tilde q / \tilde g$ pair production, the
$\chione$ mass will be directly reconstructed from $\tilde q$ and
$\tilde g$ decays, as shown in \cite{d:TDR}. 
So, for the present analysis it can be assumed 
that the $\chione$ mass is approximately known, and an attempt to reconstruct
the $\chionepm$ peak can be performed even if the $\chione$ reconstruction
does not yield a significant peak over the SUSY+SM background.  
In order to perform the full reconstruction, one just needs 
to observe a statistically significant 
excess of events over what is expected from the Standard Model 
background in the mass region corresponding to the known $\chione$ mass.
The full kinematic reconstruction described in Section~\ref{secana} above
will then easily allow to separate the process of interest from the SUSY background.
%
\subsubsection{Results}

\begin{figure}[t]
\begin{center}
\centerline{
\psfig{figure=twoneg.eps,height=3.5in}
\psfig{figure=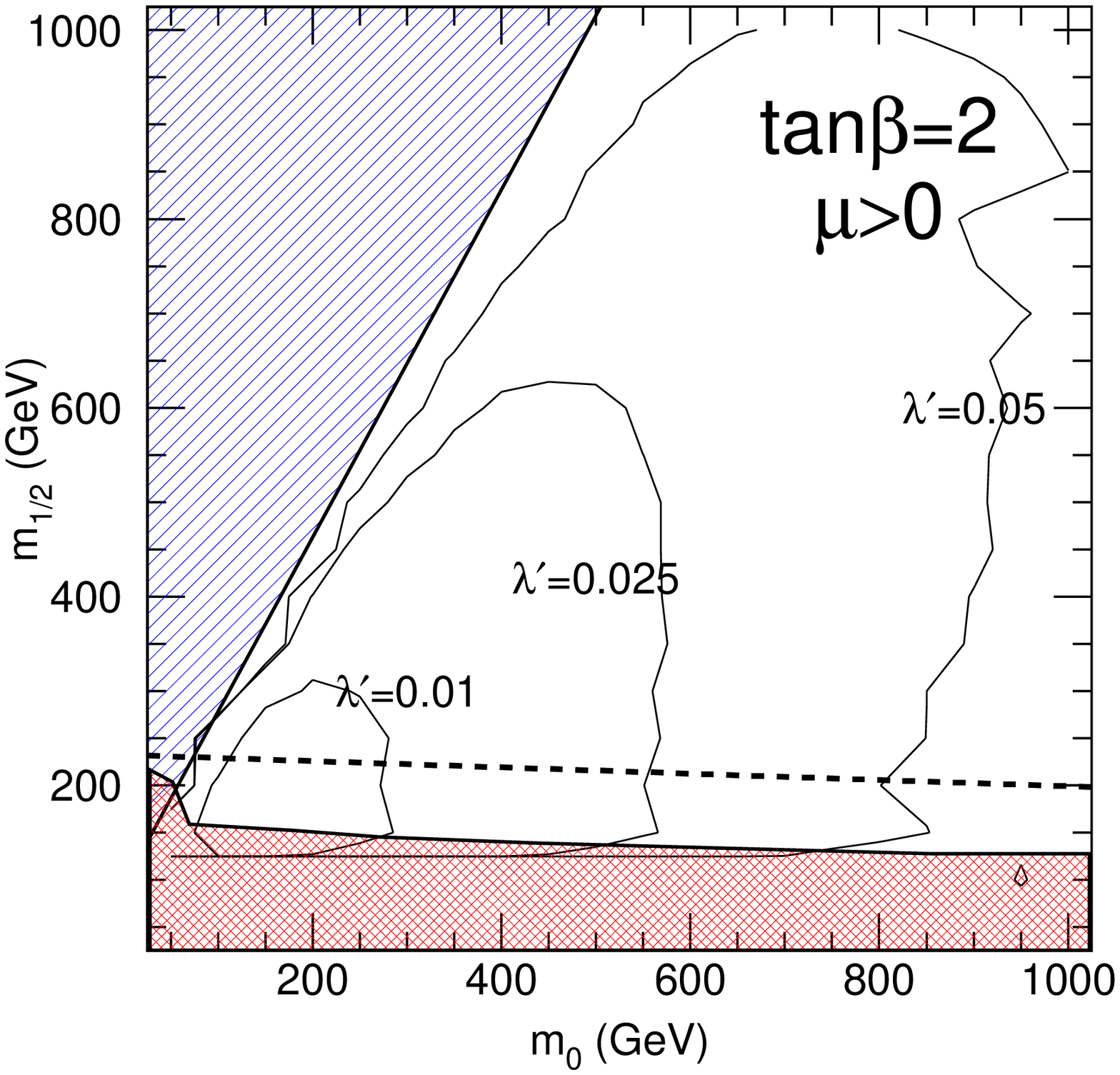,height=3.5in}}
\protect\caption{\em $5\sigma$ reach in the $m_0-m_{1/2}$ plane for $\tan\beta$=2 and three 
different choices of the $\lambda^\prime_{211}$ coupling for an integrated
luminosity of 30~fb$^{-1}$  at the LHC. The significance is defined 
only considering the Standard Model background, and a signal
of at least ten events is required. The dotted line shows the
region below which the $\chionepm$ decays to a virtual W.
}
\label{plreach}
\end{center}
\end{figure}

Based on the discussion in the previous section we calculate the signal significance as
$S/\sqrt{B}$, where for the
signal $S$ we only consider the resonant sneutrino production, 
and for the background $B$ we only consider the SM background.
We show in Figure~\ref{plreach} for $\tan\beta=2$ and for the two 
signs of $\mu$ the regions 
in the $m_0-m_{1/2}$ plane for which the signal significance exceeds 5$\sigma$
and the number of signal events is larger than 10, for an integrated 
luminosity of 30~fb$^{-1}$. The reach is shown for three different
choices of the $\lambda^{\prime}_{211}$ parameter: 
$\lambda^{\prime}_{211}=0.01,0.025,0.05$. Even for the lowest 
considered coupling the signal can be detected in a significant 
fraction of the parameter space. The dotted line shows the
region below which the $\chionepm$ decays to the $\chione$ and a virtual $W$, 
thus making the kinematic reconstruction of the decay chain described
in Section~\ref{secana} impossible.
The reconstruction of the $\chione$ is however still possible,
but the reconstruction efficiency  drops rapidly due to the difficulty 
to separate the two soft jets from the $\chione$ decay. A detailed study 
involving careful consideration of jet identification algorithms 
is needed to assert the LHC reach in that region. \par

\begin{figure}[t]
\begin{center}
\centerline{
\psfig{figure=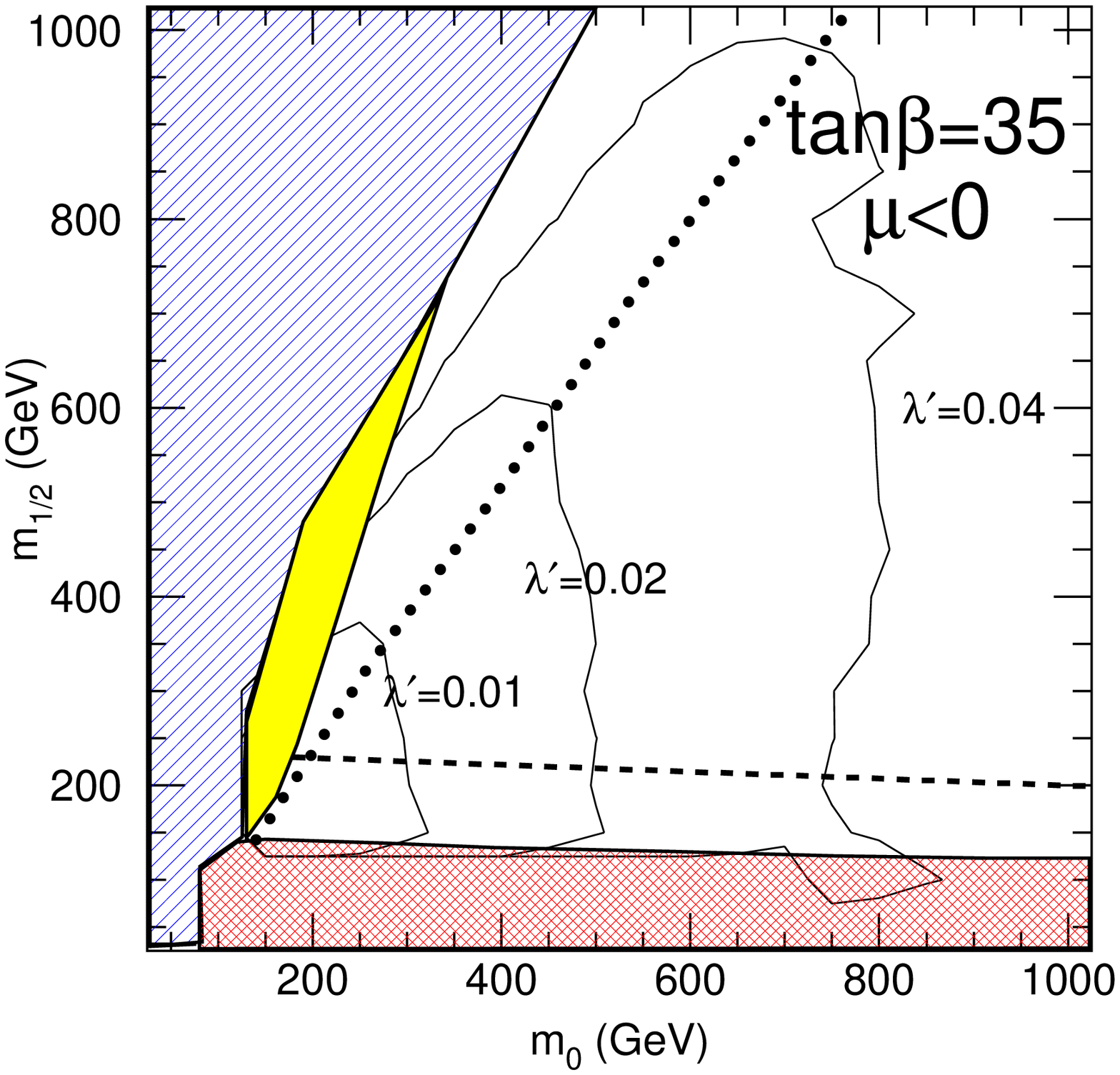,height=3.5in}
\psfig{figure=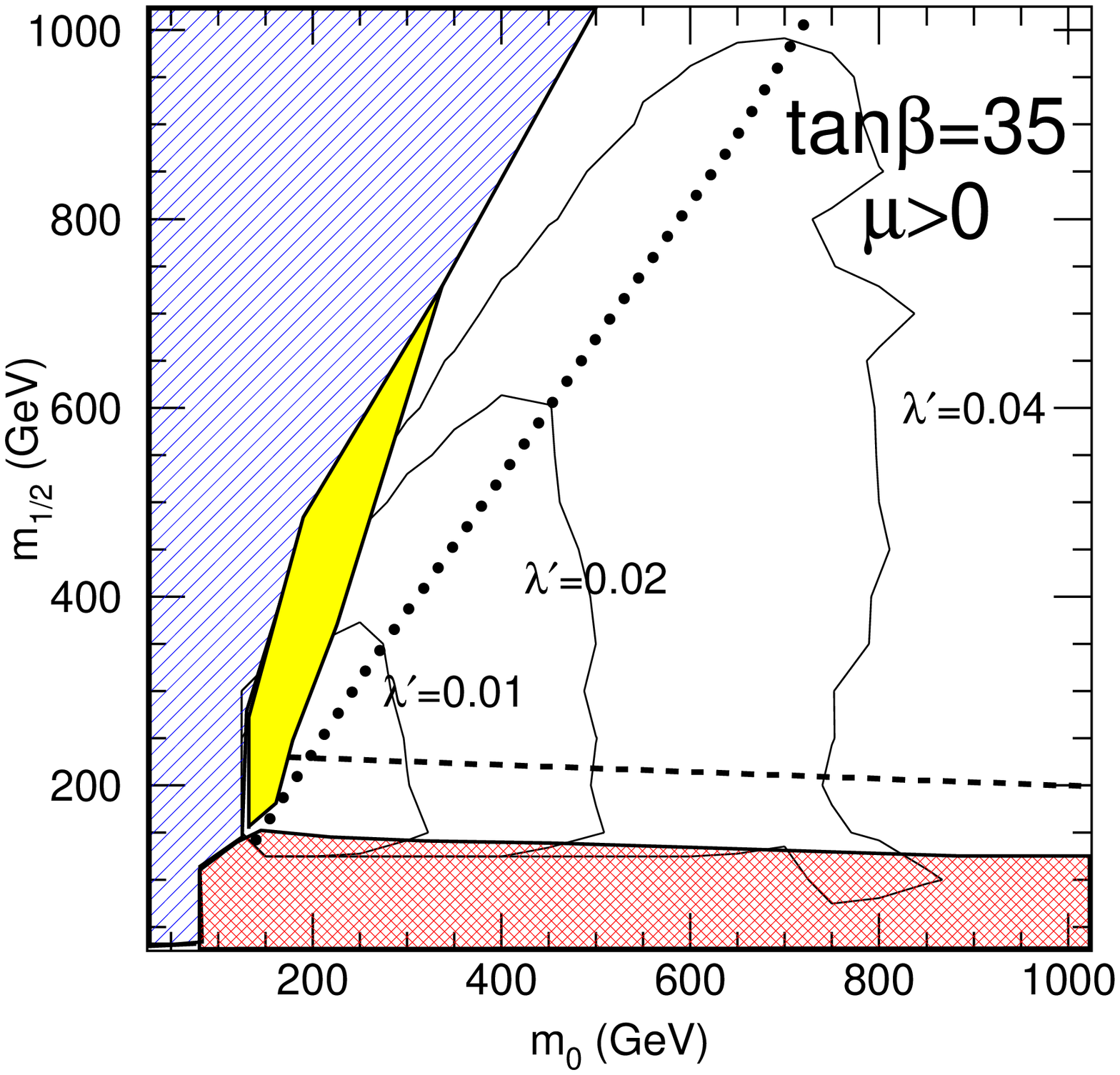,height=3.5in}}
\protect\caption{\em $5\sigma$ reach in the $m_0-m_{1/2}$ plane for $\tan\beta$=35 and three
different choices of the $\lambda^\prime_{211}$ coupling for an integrated
luminosity of 30~fb$^{-1}$  at the LHC. The significance is defined
only considering the Standard Model background, and a signal
of at least ten events is required. The dashed line shows the
region below which the $\chionepm$ decays to a virtual W.
The region to the left of the dotted lines has a branching ratio
for $\chionepm\to\tilde\tau_1\nu_{\tau}$ larger than 50$\%$,
and the grey area indicates the region for which a low signal efficiency is expected.
}
\label{plreach35}
\end{center}
\end{figure}

As observed in \cite{d:greg}, the efficiencies  quoted for this analysis 
rely  on a branching ratio of $\sim100\%$ for the decay
$\chionepm\to W\chione$. This is in general true in SUGRA models
as long as the $\tilde\tau_1$ is heavier  than the $\chionepm$, corresponding
to moderate values for $\tan\beta$.  
For high $\tan\beta$, the decay  $\chionepm\to\tilde\tau_1\nu_{\tau}$
can become kinematically possible,
and its branching ratio can dominate the standard
$\chionepm\to W\chione$. 
The stau in turns typically decays as $\tilde \tau_1 \to \tau \tilde \chi^0_1$.
The three-leptons  signature is in this case
even enhanced, due to the higher branching fraction 
into electrons and muons for the $\tau$ compared to the $W$, 
at the price of a softer lepton spectrum.
The $\chione$ reconstruction is still possible
but the presence of three neutrinos 
(two additional neutrinos come from the leptonic $\tau$ decay) 
renders the reconstruction of the particles earlier in the decay 
chain difficult. The analysis efficiency is essentially unaffected
with respect to the low $\tan\beta$ case as long as the mass difference between
the $\tilde\tau_1$ and the $\chione$ is larger than $\sim 50$~GeV.
For $\tilde\tau_1$ and $\chione$ masses too much degenerate,
the transverse momentum of the charged lepton coming from the $\tau$ 
decay would often fall below the analysis requirements,
leading thus to a reduction of the signal efficiency. 
The reach in the $m_0-m_{1/2}$ plane is shown in Figure~\ref{plreach35}
for $\tan\beta=35$ and three 
different choices of the $\lambda^\prime_{211}$ coupling.
The branching fraction for the decay $\chionepm\to\tilde\tau_1\nu_{\tau}$ is
higher than 50$\%$ to the left of the dotted line, and the region 
for which a reduced signal efficiency is expected is displayed as a grey area.
The reach for $\chione$ detection is similar 
to the low $\tan\beta$ case, 
but the region in which the full reconstruction of the 
sneutrino decay chain is possible is severely restricted.

\section{Conclusions}

\setcounter{equation}{0}

%
We have analysed the resonant sneutrino production at LHC in supersymmetric
models with R-parity violation. We have focused on the three-leptons   
signature which has a small Standard Model background, and 
allows a model-independent mass reconstruction
of the full sneutrino decay chain.\par  
A detailed study for an example MSSM point has shown that
the mass reconstruction analysis has an efficiency of
a few percent, and that a precise measurement of the masses of 
$\tilde\nu$, $\chionepm$, $\chione$
can be performed.  
Both the Standard Model background and the backgrounds
from other SUSY pair productions were studied in detail, and shown to be well 
below the expected signal for a value of the considered \rpv coupling 
$\lambda^{\prime}_{211}$ taken at the present low-energy limit. \par
The trilepton signal from sneutrino production 
was then studied as a function of the model parameters under different
model assumptions, and sensitivity over a significant part of the
parameter space was found. 
Within the MSSM, the production of
a 900~GeV sneutrino for $\lambda^{\prime}_{211}>0.05$, and of a 350~GeV sneutrino for
$\lambda^{\prime}_{211}>0.01$ can be observed 
in the first three years of LHC running. 
In the framework of the mSUGRA model, the region in the $m_0-m_{1/2}$ 
space accessible to the analysis was mapped as a function of the 
value of the $R_p$-violating coupling for representative values 
of $\tan\beta$. A significant part of the $m_0-m_{1/2}$ plane 
will be accessible for $\lambda^{\prime}_{211}>0.01$.\par
Although the detailed study was focused on the case of a single dominant 
$\lambda^{\prime}_{211}$ coupling,
we have found that the resonant sneutrino production analysis can bring
interesting sensitivities on all the \rpv couplings
of the type $\l'_{2jk}$,
compared to the low-energy constraints. 
The resonant sneutrino production 
should also allow to test most
of the $\l'_{1jk}$ coupling constants. \par 
In conclusion we have demonstrated that if minimal supersymmetry 
with R-parity violation is realised in nature, the three-leptons   signature
from sneutrino decay will be a privileged channel for the precision
measurement of sparticle masses and for studying the SUSY parameter space, 
over a broad spectrum of models.
Analyses based on the study of events including three leptons,  
were often advocated in the literature \cite{d:Nath}-\cite{d:Pai}  as a particularly 
sensitive way of attacking the search for SUSY at the LHC in the
standard R-conserving scenario. The higher lepton multiplicity and
the possibility to perform precise measurements of the model parameters
make this kind of analyses an even more attractive possibility 
in the case of R-parity violation with dominant $\lambda^{\prime}$
couplings.
\section{Acknowledgements}
This work was initiated during a workshop held in Les Houches. We warmly thank
Patrick Aurenche and all of the organising team for the stimulating
program, the excellent atmosphere and the outstanding computing
facilities.
We are also deeply indebted to the HERWIG team for allowing us to use a
prerelease of the HERWIG 6.0  MonteCarlo.

\clearpage

\vspace{10 mm}
\begin{center}
{  }
\end{center}
\vspace{10 mm}

\clearpage

\setcounter{chapter}{0}
\setcounter{section}{0}
\setcounter{subsection}{0}
\setcounter{figure}{0}

\chapter*{Publication V}
\addcontentsline{toc}{chapter}{Systematics of single superpartners
production at leptonic colliders}

\newpage

\vspace{10 mm}
\begin{center}
{  }
\end{center}
\vspace{10 mm}

\clearpage

\begin{center}
{\bf \huge Systematics of single superpartners
production at leptonic colliders}
\end{center} 
\vspace{2cm}
\begin{center}
M. Chemtob, G. Moreau
\end{center} 
\begin{center}
{\em  Service de Physique Th\'eorique \\} 
{ \em  CE-Saclay F-91191 Gif-sur-Yvette, Cedex France \\}
\end{center}
\vspace{1cm}
\begin{center}
{Phys. Rev. {\bf D59} (1999) 055003, hep-ph/9807509}
\end{center}
\vspace{2cm} 
\begin{center}
Abstract  
\end{center}
\vspace{1cm}
{\it
We examine the effects of the lepton number violating R parity odd superpotential,
 $W= \ \l_{ijk} L_iL_jE_k^c$, on single production of fermion (charginos and neutralinos)
and scalar (sleptons and sneutrinos) superpartners at leptonic colliders for center of
 mass energies up to $500GeV \ -\  1TeV$. The probability 
amplitudes for all the five $2 \to 2$ body processes:
$l_J^+l_J^- \to \tchi_1^{\pm} l_m^{\mp}, \ \tchi_1^0 \nu_m  \ (\tchi_1^0 \bar \nu_m),
\  \tilde l_m^{\mp} W^{\pm}, \ \tilde \nu_m Z^0   (\tilde {\bar \nu}_m  Z^0) , 
\ \tilde \nu_m \g (\tilde {\bar \nu}_m  \g ) $, 
and the decays branching ratios for the produced superpartners are calculated at tree level.
The rates for all five reactions are proportionnal 
to $\lambda_{mJJ} ^2 $ where $J=1,2$ for 
$e^-e^+ $ and $\mu^-\mu^+$ colliders, respectively.
A semi-quantitative discussion is presented within a supergravity model assuming grand unification 
of gauge interactions and universal (flavor independent) soft supersymmetry 
breaking parameters $m_0$ (scalars), $m_{1/2}$ (gauginos) at the unification scale.
 The predictions obtained for 
the total and partial rates show that the single production reactions have a good potential
 of observability at the future $e^-e^+ $ and $\mu^-\mu^+$ supercolliders. 
For values of the
 R parity violating coupling constant of order $0.05$, the  $\tchi^{\pm,0}$ productions 
could probe all the relevant intervals for $\tan \beta $ and $m_0$ and
broad regions of the parameter space for the $\mu$ (Higgs mixing) and $m_{1/2}$ 
parameters ($\vert \mu \vert <400GeV, \ m_{1/2}<240GeV$),
while  the $\tilde \nu$ and $\tilde l$ productions could probe sneutrinos and sleptons masses up to
the kinematical limits ($m_{\tilde \nu}<500GeV, \ m_{\tilde l}<400GeV$). Using the hypothesis
 of a single dominant R parity violating coupling constant, a Monte Carlo events simulation
for the reactions, $l_J^+l_J^- \to \tchi_1^{\pm}l^{\mp}_m,\tchi_{1,2}^0 \nu_m,
\tchi_{1,2}^0 \bar \nu_m $,   
is employed to deduce some characteristic dynamical distributions of the final states.
}

\newpage

\section{Introduction}

\setcounter{equation}{0}

Should R parity turn out to be an approximate symmetry of
the minimal  supersymmetric
standard model, the truly quantitative tests of such a possibility would have to   
be sought in high energy colliders physics, as was first emphasized in
 \cite{e:Dimo,e:Barg,e:Ross}. The great majority of the existing theoretical
studies for the LEP or the Tevatron accelerators physics have focused on signals 
associated with the LSP (lightest supersymmetric particle)  decays and certain rare 
decays of the standard model particles (gauge \cite{e:Barg,e:Valle,e:Brahm,e:Lola} or Higgs \cite{e:Barg}
 bosons or top-quark \cite{e:Phill}). A few experimental searches have been attempted 
for $Z^0$ boson decays \cite{e:EX1,e:EX2}, for inos decays \cite{e:EX3,e:EX4}   and also in
more general settings \cite{e:EX5,e:EX6}.
Proceeding one step further, interesting proposals were made recently to explain the
so-called ALEPH anomalous four-jets events \cite{e:Aleph} on the basis of R parity violating 
decays of  neutralinos or charginos \cite{e:fourjet1,e:ghosh3,e:chank3}, squarks 
\cite{e:fourjet2,e:farrar1}, sleptons \cite{e:fourjet3}  or sneutrinos \cite{e:barger3} produced 
in pairs through the  two-body processes, $e^+ e^- \to
 \tchi^{0,+}  \tchi^{0,-} $ or $e^+e^- \to \tilde f \tilde {\bar f}$. 
(See \cite{e:expupdate} for recent updates and lists of references.)

Apart from precursor studies devoted to the HERA collider \cite{e:Butt,e:perez,e:aid},   little consideration was given  
in the past to single production of supersymmetric particles in spite
of the potential interest of a discovery of 
supersymmetry that might be accessible at 
 lower incident energies.
 The reason, of course, is the lack of information about the size of
 the R parity odd coupling constants other than the large number of 
indirects bounds deduced from low and intermediate energy phenomenology \cite{e:Bhatt}.
Therefore, for obvious reasons, the existing single production studies have rather
 focused on resonant production of sneutrinos, charged sleptons 
\cite{e:Dimo,e:Barg,e:EX5,e:Esmail,e:Zhang1,e:Zhang2,e:Feng,e:Kal1,e:Kal2}
or squarks \cite{e:Ross,e:Butt,e:Esmail,e:Zhang1,e:Zhang2}. The interpretation of the 
anomalous high $Q^2$ events recently observed at HERA by the ZEUS
 \cite{e:Zeus} and H1 \cite{e:H1} Collaborations, in terms of squark resonant production,
 has also stimulated a renewed interest in
 R parity violation phenomenology \cite{e:anomhera}.

The collider physics tests of supersymmetric models without R parity entail an important change
 in focus with respect to the conventional tests: degraded missing energy, diluted
 signals, additional background from the minimal supersymmetric standard model interactions and uncertainties 
from the R parity violation coupling constants compounded with those from the superpartners
 mass spectra. Our purpose in this work is to discuss semi 
quantitatively the potential for a discovery and the tests of supersymmetry with $2 \to 2$ body 
single superpartner production. Although several order of magnitudes in rates
 are lost with respect to the resonant single production, one can dispose here
  of a rich variety of phenomena with multilepton final states non diagonal in flavor.
 Besides, one may also test larger ranges of the sneutrino mass since this need not be 
restricted by the center of mass energy value.
Encouraged by the recent developments on R parity violation and by the prospects
of high precision measurements  at supercolliders \cite{e:Pesk},
 we propose to study single production at leptonic (electron and muon) 
colliders for the set of five $2 \to 2$ body reactions,
$l_J^+l_J^- \to \tchi^{\pm}l_m^{\mp}$,
$l_J^+l_J^- \to \tchi^0 \nu_m \ (\tchi^0 \bar \nu_m)$,
$l_J^+l_J^- \to \tilde  l_m^{\mp} W^{\pm} $,
$l_J^+l_J^- \to \tilde \nu_{mL} Z^0 \ (\tilde  {\bar \nu}_{mL} Z^0)$,
$l_J^+l_J^- \to \tilde \nu_{mL} \gamma 
\ (\tilde  {\bar \nu}_{mL} \gamma) \ [J=1,2],$
 in a more systematic way 
than has been attempted so far.
We limit ourselves to the lowest inos eigenstates.  
Let us note here that precursor indicative studies 
of the inos single production reactions were already presented in 
\cite{e:DESY1,e:DESY2} and that recent discussions concerning the reactions, 
$e^{\pm} \gamma  \to e^{\pm} \tilde \nu$ and $e^\pm \gamma 
\to \tilde l^{\pm} \nu$,
where the photon flux is radiated by one of the two beams, were presented in  
\cite{e:All}. We shall restrict our study to the lepton
 number violating interactions $L_iL_jE_k^c$ in association with 
the familiar gauge and Yukawa couplings of the minimal supersymmetric
standard model. The final states consist then of multileptons with or without 
hadronic jets. 
 
This paper contains four sections.
In section \ref{sectionF}, we present the main formalism for superpartners production cross
sections and decay rates. In section \ref{sectionR}, based on the supergravity approach
to supersymmetry soft breaking parameters, we present numerical results for the total
rates and the various branching ratios in wide regions of the parameter space. In section \ref{sectionCD},
 we show results for final states distributions of the processes,
 $l_J^+l_J^- \to \tchi^{\pm} l^{\mp},\tchi^0 \nu,\tchi^0 \bar \nu$, obtained by means 
of a Monte Carlo events simulation,
using the SUSYGEN routine \cite{e:Kats}. In section \ref{sectionC}, we state our conclusions.   

\section{General Formalism}
\label{sectionF}

\setcounter{equation}{0}

Five $2 \to 2$ body single production reactions may be observed
at leptonic colliders. We shall use the following short hand notation to denote 
the associated probability amplitudes:
\begin{eqnarray}
M(\tchi_a^-+l_m^+)&=& M(l^-_J+l^+_J\to \tchi_a^- +l^+_m ), \cr 
M(\tchi_a^0+\bar \nu _m)&=& 
M(l^-_J+l^+_J \to \tchi_a^0 +\bar \nu_m ),  \cr 
M(\tilde l_{mL}^- +W^+)&=&
M(l^-_J+l^+_J\to  \tilde l_{mL}^- +W^+ ),\cr
M(\tilde \nu_m +Z)&=&
M(l^-_J+l^+_J\to  \tilde \nu_m + Z), \cr
M(\tilde \nu_m +\g )&=&
M(l^-_J+l^+_J\to  \tilde \nu_m +\g ),
\label{eqrc1}
\end{eqnarray}
where $J= 1,2$ is a flavor index for the initial state leptons (electrons 
and muons, respectively), 
the index  $a$  labels the charginos or neutralinos 
eigenvalues and  the index $ m $ the sleptons or sneutrinos families.
Our theoretical framework is the minimal supersymmetric standard model supplemented by the lepton number violating
R parity odd superpotential, $W= \ud \sum_{ijk} \l_{ijk} L_iL_jE_k^c$. This 
yields the sfermion-fermion Yukawa interactions,
\begin{eqnarray}
L&=&  \ud \sum_{[i\ne j, k]=1}^3 \l_{ijk}[\tilde  \nu_{iL}\bar e_{kR}e_{jL} +
\tilde e_{jL}\bar e_{kR}\nu_{iL} + \tilde e^\star _{kR}\bar \nu^c_{iR} e_{jL}
-(i\to j) ] +h.c.
\label{eqrc2}
\end{eqnarray}
where the sums labelled by indices, $i,j,k,$ run over the three 
leptons and neutrinos families with the condition $i\ne j$ following from the 
antisymmetry property, $\l_{ijk}=-\l_{jik}$..  

\subsection{Production Cross Sections}

\begin{figure}
\begin{center}
\leavevmode
\psfig{figure=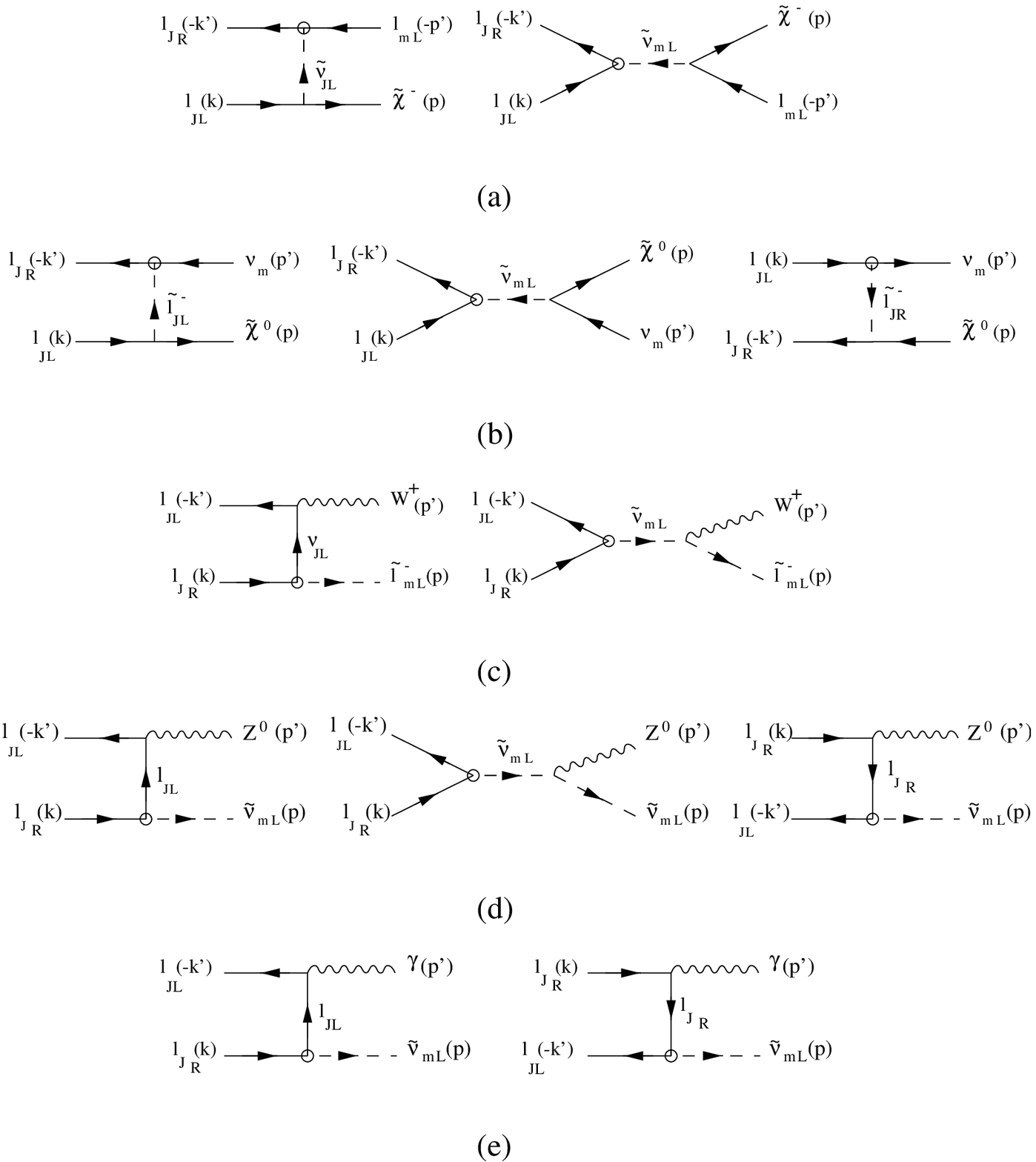}
\end{center}
\caption{\footnotesize  \it
Feynman diagrams for the processes, $l_J^+l_J^- \to \tchi^- l_m^+$ (a), 
 $l_J^+l_J^- \to \tchi^0 \bar \nu_m$ (b),  $l_J^+l_J^- \to \tilde l^-_{mL} W^+$ (c),
 $ l_J^+l_J^- \to \tilde \nu_{mL} Z^0 $ (d) and  $ l_J^+l_J^- 
 \to \tilde \nu_{mL} \ \g $ (e). 
The circled vertex correspond to the RPV interaction, with the coupling constant  $\l_{mJJ}$, and
the arrows denote flow of momentum.
\rm \normalsize }
\label{e:fig1}
\end{figure}

Each of the processes in eq.(\ref{eqrc1}) has a charge conjugate partner such that the
transformation between pairs of conjugate amplitudes can be formally described by
applying a CP transformation to the S-matrix. The relationship is most 
easily described at the level of
the amplitudes squared obtained after summation over the spins. Because of the simple 
action of CP on the initial state, $l_J^+(k)l_J^-(k')$, it can be seen that 
the amplitudes for the pairs of charge conjugated processes are related by the 
substitutions, $k \leftrightarrow k'$ 
and  $\l_{ijk} \leftrightarrow \l_{ijk} ^\star $.
The  tree level probability amplitudes   
are easily calculated by inspection of the Feynman diagrams given in
Fig.\ref{e:fig1}.    
The formulas for the amplitudes are consigned in
Appendix \ref{seca}.  
A few observations are in order at this point. First, the same configurations of 
lepton flavor indices, namely, $\l_{mJJ}$ with $J=1, \ m=2,3$  for $e^-e^+$ 
colliders and $J=2, \ m=1,3$  for $\mu^-\mu ^+$ colliders,
occur in all cases. Second, the amplitude for   
right chirality slepton $\tilde l_{mR}$ production has not been included in the 
above list of formulas for the reason that this is proportional to the  coupling constants
$\l_{11m}$ which vanishes by the antisymmetry property, $\l_{ijk}=-\l_{jik}$. 
Thirdly, all five processus can appear only in a single helicity configuration 
for the initial fermions (assumed massless), corresponding to identical helicities,
 namely, either $l^+_L l^-_L$ or $l^+_R l^-_R$ (recall that physical helicity 
for antiparticle is opposite to chirality).
Lastly, we observe that the relative signs between the $s,t$ and $u$ channels 
contributions are dictated by  both the structure of the interaction
Lagrangian and the signs of the Wick contractions for fermions. The results
for the spin summed squared amplitudes are given by somewhat
 complicated formulas which we have
 assembled in Appendix \ref{seca}. We have checked that our formulas for $\tchi^0$ and 
 $\tchi^{\pm}$ productions agree with the results provided
 in \cite{e:DESY1,e:DESY2} and \cite{e:Kats1}.

\subsection{Decays}

\label{sectionD}

\begin{table}[t]
\begin{center}
\begin{tabular}{|c|c|c|c|}
\hline
 Mass Intervals & Decays & Final State &  $\l_{m11}$\\ 
\hline 
& & & \\
$m_{\tilde l^-} > m_{\tchi^-}    $ (1) &  $\bullet  \tchi^-\to 
\bar \nu_i \bar \nu_j l_k$ & (A)$l_m^+ l_k^-\Eslash $ & $l^+_m e^-$ \\
$m_{\tilde l^-} < m_{\tchi^-}   $ (2)&   $\bullet  
\tchi^-\to \bar \nu_i \tilde l^-_i \to \bar \nu_i l_k \bar \nu_j  $ \ & & \\ 
\hline
& & & \\ 
$ m_{\tilde \nu }> m_{\tchi^-}   $ (3)&  $\bullet  
\tchi^-\to  l_j \bar l_k l_i $ &(B)$l_m^+ l_k^+  l_i^- l_j^- $ & $l^+_ml^-_me^+e^-$ \\   
$ m_{\tilde \nu }< m_{\tchi^-}   $  (4)&  $\bullet
\tchi^-\to l_i \tilde {\bar \nu}_i   \to  l_i l_j \bar l_k $& & \\ 
\hline
& & & \\ 
$m_{\tilde l},m_{\tilde \nu} > 
m_{\tchi^-} > m_{\tchi^0}   $  (5)&  $\bullet
\tchi^-\to \tchi^0 l_p \bar \nu_p  \to l_p \bar \nu_p \nu_i l_j\bar l_k $ & (C)$l_m^+ l^-_p  
l_k^{\pm}  l^{\mp}_i\Eslash $& $l^+_ml^-_pe^+e^-,$ \\
$m_{\tchi^-}  > m_{\tilde l}>m_{\tchi^0}  $  (6)& $\bullet \tchi^-\to \bar \nu_p
\tilde l_p^- \to   \bar \nu_p l_p \tchi^0  $& & $ l^+_ml^-_pe^{\pm}l_m^{\mp}$\\
 & $  \to  l_p \bar \nu_p \nu_i l_j \bar l_k $ & & \\
$m_{\tchi^-} > m_{\tilde \nu }>m_{\tchi^0}   $  (7) & $\bullet  \tchi^-\to
 l_p  \tilde{\bar \nu}_p   \to  l_p \bar \nu_p  \tchi^0
$ & & \\
& $  \to l_p \bar \nu_p \nu_i   l_j\bar l_k $ & & \\ 
\hline
& & & \\ 
$m_{\tilde q} > 
m_{\tchi^-} > m_{\tchi^0}   $ (8)&  $\bullet
\tchi^-\to \tchi^0 q_p \bar q_p, \  
 \to  q_p \bar q_p \nu_i l_j\bar l_k  $ & (D)$l_m^+ l_k^{\pm} l^{\mp}_i
 \Eslash +2jet$ & $l^+_me^+e^-,$ \\
$m_{\tchi^-}  > m_{\tilde q}>m_{\tchi^0}  $ (9) & $\bullet  \tchi^-\to 
\bar q_p \tilde q_p, \   \to q_p \bar q_p  \tchi^0  $& &$ l^+_me^{\pm}l_m^{\mp}$ \\
& $  \to  q_p \bar q_p \nu_i l_j \bar l_k $ & & \\ 
\hline 
& & & \\
$m_{\tchi^-} > m_{\tchi^0} +m_W  $ (10) & $\bullet
\tchi^-\to \tchi^0 W^-  \to  W^- \nu_i l_j\bar l_k $ & (E)$ l_m^+
l_k^{\pm} l^{\mp}_i W^-\Eslash $ & $l^+_me^+e^-,$ \\
&&&$ \ l^+_me^{\pm}l_m^{\mp}$\\ 
\hline 
\end{tabular}
\caption{\footnotesize  \it
The allowed chargino decays for different relative orderings of the superpartners 
masses. The column fields give the mass intervals, the decay schemes, the 
final states corresponding to the process, $l_J^+l_J^- \to \tchi^- l^+_m$,
with a single dominant \cc $\l_{ijk}$
 and the leptonic components of the final states  in the case of a single 
dominant coupling constant $\l_{m11} \ [m=2,3]$. The notation $\Eslash $ stands for missing energy associated with neutrinos. \rm \normalsize}
\label{tabloC}
\end{center}
\end{table}
 
\begin{table}[t]
\begin{center}
\begin{tabular}{|c|c|c|c|}
\hline
Mass Intervals & Decays & Final State &  $\l_{m11}$\\
\hline
& & & \\
$m_{\tilde \nu} < m_{\tchi^+}   $ (1) &  $ \bullet
\tilde \nu_m \to l_k \bar l_j $& (A)$l^-_k l^+_jZ^0$ & $e^+e^-$ \\
\hline
& & & \\
$m_{\tilde \nu} > m_{\tchi^+}   $ (2) &  $ \bullet
\tilde \nu_{m} \to l_{m} \tchi^+ \to l_m \bar l_i \bar l_j l_k$&
(B)$l^+_i l^+_j l^-_k l_{m}^- Z^0$ & $e^+e^-l^+_ml^-_m$ \\
\hline
& & & \\
$m_{\tilde \nu},\ m_{\tilde l} > m_{\tchi^+}   $ (3) &  $ \bullet
\tilde \nu_{m} \to l_{m} \tchi^+ \to l_m \nu_i \nu_j \bar l_k$&(C)
$l_k^+ l^-_{m} \Eslash  Z^0$ & $e^+l_m^-$ \\
$m_{\tilde \nu} > m_{\tchi^+} > m_{\tilde l} $ (4) &  $ \bullet
\tilde \nu_{m} \to l_{m} \tchi^+ \to l_m\tilde l^+_j \nu_j$
& & \\
& $ \to l_m \nu_j \nu_i \bar l_k$ & & \\
\hline 
& & & \\
$m_{\tilde \nu} > m_{\tchi^0}   $ (5)&  $ \bullet
\tilde \nu_m \to \nu_m \tchi^0 \to \nu_m \nu_i l_j \bar 
l_k $&(D)$ l_k^{\pm} l^{\mp}_i\Eslash Z^0$ & $e^+e^-,$ \\
&&& $ \ e^{\pm}l_m^{\mp}$ \\
\hline
& & & \\
$m_{\tilde \nu} > m_{\tchi^+} > m_{\tchi^0}   $  (6) & $ \bullet 
\tilde \nu_{m} \to \tchi^+ l_{m} \to l_m \tchi^0 \bar l_p \nu_p$ &
(E)$l_p^+ l_{m}^- l_k^{\pm} l^{\mp}_i\Eslash Z^0$ & $l^+_p l^-_m e^+e^-,$ \\
& $   \to l_m \bar l_p \nu_p \nu_i l_j \bar l_k$ & &  $  l^+_p l^-_m e^{\pm} l_m^{\mp}$ \\
$m_{\tilde \nu} > m_{\tchi^+}  > m_{\tilde l}>m_{\tchi^0}  $  (7)&  $ \bullet
\tilde \nu_{m} \to \tchi^+ l_{m} \to l_m \nu_p \tilde l_p^+$ & & \\
&  $  \to l_m \nu_p \bar l_p \tchi^0 \to l_m \nu_p \bar l_p \nu_i l_j \bar l_k $ & & \\ 
\hline
$m_{\tilde \nu} > m_{\tchi^+} > m_{\tchi^0}  $ (8) &$  \bullet  
\tilde \nu_{m} \to \tchi^+ l_{m} \to l_m \tchi^0 q_p \bar q_p$ &
(F) $ l^-_{m} l^{\pm}_k l^{\mp}_i Z^0+ 2jet $ & $l^-_me^+e^-,$ \\
& $  \to l_m q_p \bar q_p \nu_i l_j \bar l_k $ & & $  l^-_me^{\pm}l_m^{\mp}$\\
$m_{\tilde \nu} > m_{\tchi^+}  > m_{\tilde q}>m_{\tchi^0}  $ (9) & $ \bullet  
\tilde \nu_{m} \to \tchi^+ l_{m} \to l_m \bar q_p \tilde q_p  $ & & \\
& $ \to l_m \bar q_p q_p \tchi^0 \to l_m q_p \bar q_p \nu_i l_j \bar l_k $ & & \\
\hline
$m_{\tilde \nu} > m_{\tchi^+} > m_{\tchi^0}+m_W $ (10) & $ \bullet   
\tilde \nu_{m} \to \tchi^+ l_{m} \to l_m \tchi^0 W^+$ & (G)
$l^-_{m} l^{\pm}_k l^{\mp}_i \Eslash   W^+ Z^0$ & $l^-_me^+e^-,$ \\
& $  \to l_m W^+ \nu_i l_j \bar l_k $ & & $ l^-_me^{\pm}l_m^{\mp}$\\
\hline
\end{tabular}
\caption{\footnotesize  \it
The allowed sneutrino decays for different relative orderings of the superpartners 
masses. The column fields give the mass intervals, the decay schemes, the 
final states corresponding to the process, $l_J^+l_J^- \to Z^0 \tilde \nu_m$,
with a single dominant \cc $\l_{ijk}$
 and the leptonic components of the final states  in the case of a single 
dominant coupling constant $\l_{m11} \ [m=2,3]$. The notation $\Eslash $ stands for missing energy associated with neutrinos.
\rm \normalsize}
\label{tabloA}
\end{center}
\end{table}

\begin{table}[t]
\begin{center}
\begin{tabular}{|c|c|c|c|}
\hline
Mass Intervals & Decays & Final State &  $\l_{m11}$ \\
\hline
$m_{\tilde l^-} < m_{\tchi^-}   $ (1) &  $ \bullet
\tilde l^-_m\to l_k \bar \nu_i $& (A)$l^-_k \Eslash W^+$ & $e^-$ \\
\hline
$m_{\tilde l^-} > m_{\tchi^-}   $ (2) &  $ \bullet
\tilde l_m^-\to \nu_m \tchi^- \to \nu_m l_k \bar \nu_j \bar \nu_i$&
(B)$l^-_k \Eslash W^+$ & $e^-$ \\
\hline
$m_{\tilde l^-},\ m_{\tilde \nu} > m_{\tchi^-}   $ (3) &  $ \bullet
\tilde l_m^-\to \nu_m \tchi^- \to \nu_m l_j \bar l_k l_i$&(C)$l_k^+ l^-_i
 l_j^- \Eslash W^+$ & $e^+e^-l_m^-$ \\
$m_{\tilde l^-} > m_{\tchi^-} > m_{\tilde \nu}  $ (4) &  $ \bullet
\tilde l_m^-\to \nu_m \tchi^- \to \nu_m \tilde {\bar \nu}_i l_i $& & \\
& $ \to \nu_m l_i l_j \bar l_k $ &&\\
$m_{\tilde l^-} > m_{\tchi^0}   $ (5)&  $ \bullet
\tilde l_m^-\to l_{m} \tchi^0 \to l_m \nu_i l_j \bar 
l_k $&(D)$l_{m}^- l_k^{\pm} l^{\mp}_i\Eslash W^+$ & $l^-_me^+e^-,$ \\
&&& $ \ l^-_me^{\pm}l_m^{\mp}$ \\
\hline
$m_{\tilde l^-} > m_{\tchi^-} > m_{\tchi^0}   $  (6) & $ \bullet
\tilde l_m^- \to \tchi^- \nu_m \to \nu_m \tchi^0 l_p \bar \nu_p$ &
(E)$l_p^- l_k^{\pm} l^{\mp}_i\Eslash W^+$ & $l^-_pe^+e^-,$ \\
& $ \to \nu_m l_p \bar \nu_p \nu_i l_j \bar l_k $ &&$ l^-_pe^{\pm}l_m^{\mp}$  \\
$m_{\tilde l^-} > m_{\tchi^-}  > m_{\tilde \nu}>m_{\tchi^0} $  (7)&  $ \bullet 
\tilde l_m^- \to \tchi^- \nu_m \to \nu_m l_p \tilde {\bar \nu}_p$ 
& & \\
& $ \to \nu_m l_p \bar \nu_p \tchi^0 \to \nu_m l_p \bar \nu_p \nu_i l_j \bar l_k$ 
& & \\ 
\hline
$m_{\tilde l^-} > m_{\tchi^-} > m_{\tchi^0}  $ (8) & $\bullet  
\tilde l_m^- \to \tchi^- \nu_m \to \nu_m \tchi^0 q_p \bar q_p$ & (F)
$ l^{\pm}_k l^{\mp}_i \Eslash W^++ 2jet$ & $e^+e^-,$ \\
 & $ \to \nu_m q_p \bar q_p \nu_i l_j \bar l_k $ & & $ e^{\pm}l_m^{\mp}$  \\
$m_{\tilde l^-} > m_{\tchi^-}  > m_{\tilde q}>m_{\tchi^0}  $ (9) & $\bullet 
\tilde l_m^- \to \tchi^- \nu_m \to \nu_m \bar q_p \tilde q_p $ & & \\
& $\to \nu_m \bar q_p q_p \tchi^0 \to \nu_m \bar q_p q_p  \nu_i l_j \bar l_k $ & & \\
\hline
$m_{\tilde l^-} > m_{\tchi^-} > m_{\tchi^0}+m_W  $ (10) & $\bullet  
\tilde l_m^- \to \tchi^- \nu_m \to \nu_m \tchi^0 W^-$ & (G)
$l^{\pm}_k l^{\mp}_i W^-  \Eslash   W^+$ & $e^+e^-,$ \\
& $ \to \nu_m W^- \nu_i l_j \bar l_k $ & & $ e^{\pm}l_m^{\mp}$ \\
\hline 
\end{tabular}
\caption{\footnotesize  \it
The allowed slepton decays for different relative orderings of the superpartners 
masses. The column fields give the mass intervals, the decay schemes, the 
final states corresponding to the process, $l_J^+l_J^- \to W^+ \tilde l_m^-$,
with a single dominant \cc $\l_{ijk}$
 and the leptonic components of the final states  in the case of a single 
dominant coupling constant $\l_{m11} \ [m=2,3]$. The notation $\Eslash $ stands for missing energy associated with neutrinos.
\rm \normalsize}
\label{tabloB}
\end{center}
\end{table}

In order to exhibit the possible physical final states, we need to consider
 the decays of 
the produced  supersymmetric particles, taking into account both the  
minimal supersymmetric standard model interactions (denoted RPC or R parity  conserving) and the 
R parity odd interactions (denoted RPV or R parity  violating). \\
A number of hypotheses and approximations, which we list below, will be
 employed in the evaluation of partial rates.

\begin{itemize}
\item[1)] Supersymmetric particles  decays are assumed to have
narrow widths  (compared to their masses) and 
are produced on-shell with negligible spin correlations 
between the production and decay stages.
This allows us to apply the familiar  phase space
factorisation formula for the production cross sections.
\item[2)] Spin correlations are neglected at all stages of the cascade decays
 such that the branching ratios in single or double cascades can be obtained by applying recursively 
the standard factorisation formula.
\item[3)] Sleptons belonging to all  three families  and squarks  
belonging to the first two families  are degenerate in mass.
 Therefore, for a given decay process as, for instance, $\tchi^- \to \tilde
 {\bar \nu}_p l_p$, either all three generations will be energetically
 allowed or forbidden. Furthermore, flavor off-diagonal channels 
such as, $\tilde l_1 \to \tilde l_2+Z^0$,... are closed.
 \item[4)] The lowest eigenstates of neutralinos $\tchi_a^0$ and charginos
 $\tchi_a^{\pm}$ ($a=1$) are excited in the cascade chains. 
 \item[5)] All superpartners decay
 inside the detector volume. 
In the presence of broken R parity, the condition  
for  electric charge neutral LSPs  to decay inside the detector
yields comfortable  lower bounds  of order 
$\l > 10^{-7}$ \cite{e:Ross,e:Daw}.
\item[6)]  Either a single RPV coupling constant is dominant in both the production 
 and decay stages, or a pair of RPV coupling constants are dominant, one in the production
 stage ($\l_{mJJ}$) and the other in the decay stage ($\l_{ijk}$).
The latter case with two  dominant RPV \ccs may be of interest since strong bounds 
on quadratic products exist only for a few family configurations. The strongest
 bounds arise from the $\mu \to 3e$ decay \cite{e:quadbound}:
 $\l_{p11} \l_{p12}<6.5 \ 10^{-7},
 \  \l_{p21} \l_{p11}<6.5 \ 10^{-7} \ [p=2,3]$, while other quadratic product 
bounds are of 
order $10^{-3},10^{-4}$. Besides, as long as the coupling constant $\l$, which
controls the RPV decays, is small in comparison with the gauge coupling constants but not very much smaller (so that the LSP decays inside the detector), then the \br will
depend weakly on $\l$ since the last stage of the decay chain (LSP decay) is
 independent of $\l$.
\item[7)] The widths for the decays with four and higher body final states are neglected,
such as those which occur in slepton (sneutrino) decays for, $m_{\tchi^-},m_{\tchi^0}>
m_{\tilde l}$ ($m_{\tchi^-},m_{\tchi^0}>m_{\tilde \nu}$), mediated by virtual charginos
or neutralinos, namely, $\tilde l^-_m \to \nu_m l_k \bar \nu_j \bar \nu_i$ and
$\tilde {l^-}_m \to l_m \bar {l_k} l_j \nu_i $ ($\tilde {\nu}_m \to l_m l_k \bar {l_j} 
\bar {l_i} $ and $ \tilde {\nu}_m \to \nu_m \bar {l_k} l_j \nu_i)$.
\item[8)] A supergravity model for the soft \susy breaking parameters is used where, 
generically, $\tchi^0_1$ is the LSP.
\end{itemize}

The consideration of the various order relations in the superpartners mass spectrum
 leads to a list of decay schemes for the initially produced superparticles. These are
displayed in Tables \ref{tabloC},  \ref{tabloA} and \ref{tabloB}, for $\tchi^{\pm}$,
$\tilde \nu_L$ and $\tilde l_L$, respectively. The signals for the $\tchi^0_1$ decays
are very few in number and will be discussed separately in 
section \ref{neutralinosection}. 
Some comments on these tables are in order.
Except for hadronic dijet pairs from the decay processes, 
$\tchi^{\pm} \to \tchi^0 \bar q q'$, all other final particles will consists of
 multileptons and missing energy associated with neutrinos.
 In the hypothesis of a single dominant RPV coupling constant in the decays stage, 
one can deduce the various final states flavor configurations by an inspection
of the tables (cf. tables captions).
The produced $\tchi^0, \ \tchi^{\pm}, \ \tilde \nu, \ \tilde l^{\pm}$ will
 decay according to cascade schemes
 dictated by the superpartners mass spectrum. It is important to distinguish the 
direct RPV induced decays: $ \tchi^-\to \bar \nu_i \bar \nu_j l_k$,
 $ \tchi^-\to l_i  l_j \bar l_k$, 
$\tchi^0 \to \nu_i l_j  \bar l_k$, $\tchi^0 \to \bar \nu_i \bar l_j  l_k$,
 $\tilde \nu_i \to l_k \bar l_j$, $\tilde l_{kR}^-\to l_j \nu_i $,
 and  $\tilde l_{jL}^-\to l_k \bar \nu_i $, from
the indirect RPC induced decays: $\tilde l_L^- \to \tchi^- \bar \nu$, $\tilde \nu_L
 \to \tchi^+l$, $\tilde l^-_{L,R} \to \tchi^0  l$, $\tilde \nu_L \to \tchi^0 \nu$,
$\tchi^0 \to \tilde l^- {\bar l}$, $\tchi^0 \to l \tilde {l^+}$,
$\tchi^- \to \tilde l^- \bar \nu$, $\tchi^- \to l \tilde {\bar \nu}$ and $\tchi^{\pm} \to 
\tchi^0 W^{\pm}$ for two-body final states and, $\tchi^- \to \tchi^0 f \bar f$
 (f=leptons or quarks), for three-body final states.
All the formulas needed to evaluate the partial decay widths are quoted in the Appendix \ref{secappa}.
As can be seen from Tables \ref{tabloC},\ref{tabloA},\ref{tabloB}, 
a given final state can arise from different
processes, depending on the relative orderings of the masses. A reaction chain 
occuring through an intermediate particle which is produced on-shell leads obviously
 to the same final state when the production of this particle is
 kinematically forbidden and it must then occur through a virtual intermediate state.
In the approximation of family  degenerate sleptons, sneutrinos and squarks, 
the index $p$ in the tables runs over the three generations. A single exception
 is the hadronic decay, $\tchi^{\pm} \to \tchi^0 u_p \bar d_p (\tchi^0 d_p \bar u_p)$,
 which is restricted to the first two families because of the large top-quark mass.

Another subtle point concerns the multiplicity of a given signal, namely, the number of 
different configurations which can lead to the same final state. Due to the 
antisymmetry property of $\l_{ijk}$, the final states from chargino RPV 
decays (cf. A and B in Table \ref{tabloC}) have a multiplicity of two. The reason is that 
these decays proceed through the exchanges of the sleptons (or sneutrinos) in
 families $i$ and $j$, for a given $\l_{ijk}$.
This fact is already accounted for in the virtual $\tchi^{\pm}$ three-body decays 
(A(1),B(1),Table \ref{tabloC}), but must be put by hand in the $\tchi^{\pm}$ cascade decays 
proceeding through the on-shell production of sleptons or sneutrinos which decay
 subsequentially (A(2),B(2),Table \ref{tabloC}).

To get a better understanding of the interplay between RPC and RPV decays,
it is helpful to note that the \brs  can be written formally as,
 $B_D={\l^2 \over c \ g^2 +\l^2}$, for direct decays 
and, $B_I={g^2 B\over c_0 \ g^2 +\l^2}$, for two-stages indirect decays, where 
 $B={\l^2 \over c \ g^2 +\l^2}$ is the LSP branching ratio, $\l$ and $g$ are symbolic notations for the RPV 
and RPC coupling constants and $c,c_0$ are calculable constants.
Of course, $B=1$, if the last decaying particle is the LSP, 
which is the generical case. For values of the 
RPV coupling constants small with respect to the gauge coupling constants 
($\l \leq 0.05$), namely, $g^2 \gg \l^2$,
the dependence on $\l$ of the indirect decays \brs is weak and we have, $B_I \gg B_D$. For large enough RPV \cc (for example, $\l=0.1$) or for suppressed
indirect decays (due for example to kinematical reasons), the direct decays may become
competitive and both the direct and indirect \brs depend strongly on $\l$.
 
\section{The model and its parameter space}
\label{sectionmodel}

\setcounter{equation}{0}

\begin{figure}
\begin{center}
\leavevmode
\psfig{figure=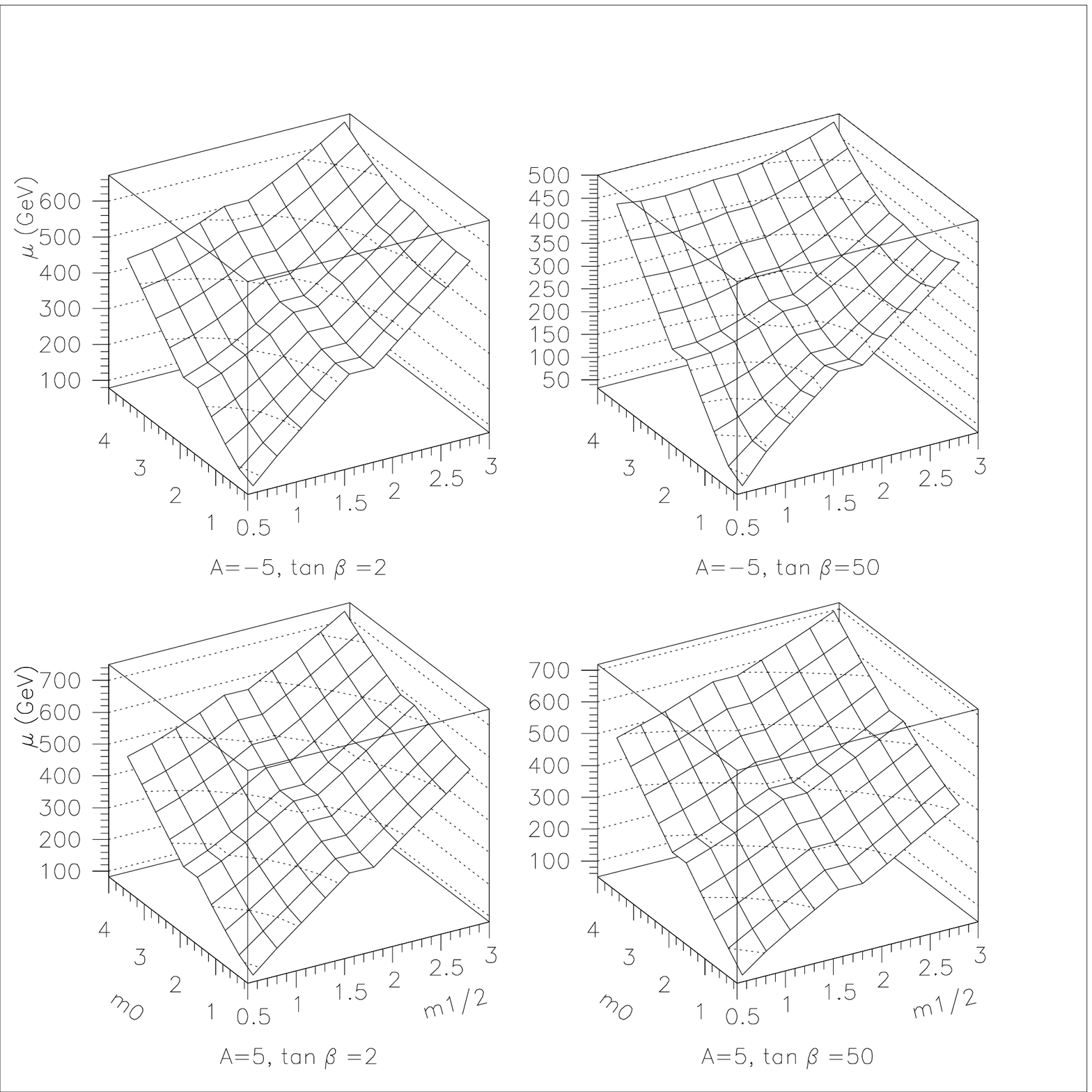,width=7.in}
\end{center}
\caption{\footnotesize  \it
The solution $\mu(t_Z)$, at scale $m_Z$, for the electroweak symmetry radiative breaking 
equations, at running top-quark mass, $m_{t}(m_{t})= 171  $ GeV 
$(m_t^{pole}= 180 $ GeV), is plotted as a function 
of ${m_0 \over 100GeV}$ and ${m_{1/2} \over 100GeV}$ for four 
values of the pair of  parameters, $A$ and $\tan \beta$.  
\rm \normalsize }
\label{efbr}
\end{figure}

\begin{figure}[t]
\begin{center}
\leavevmode
\psfig{figure=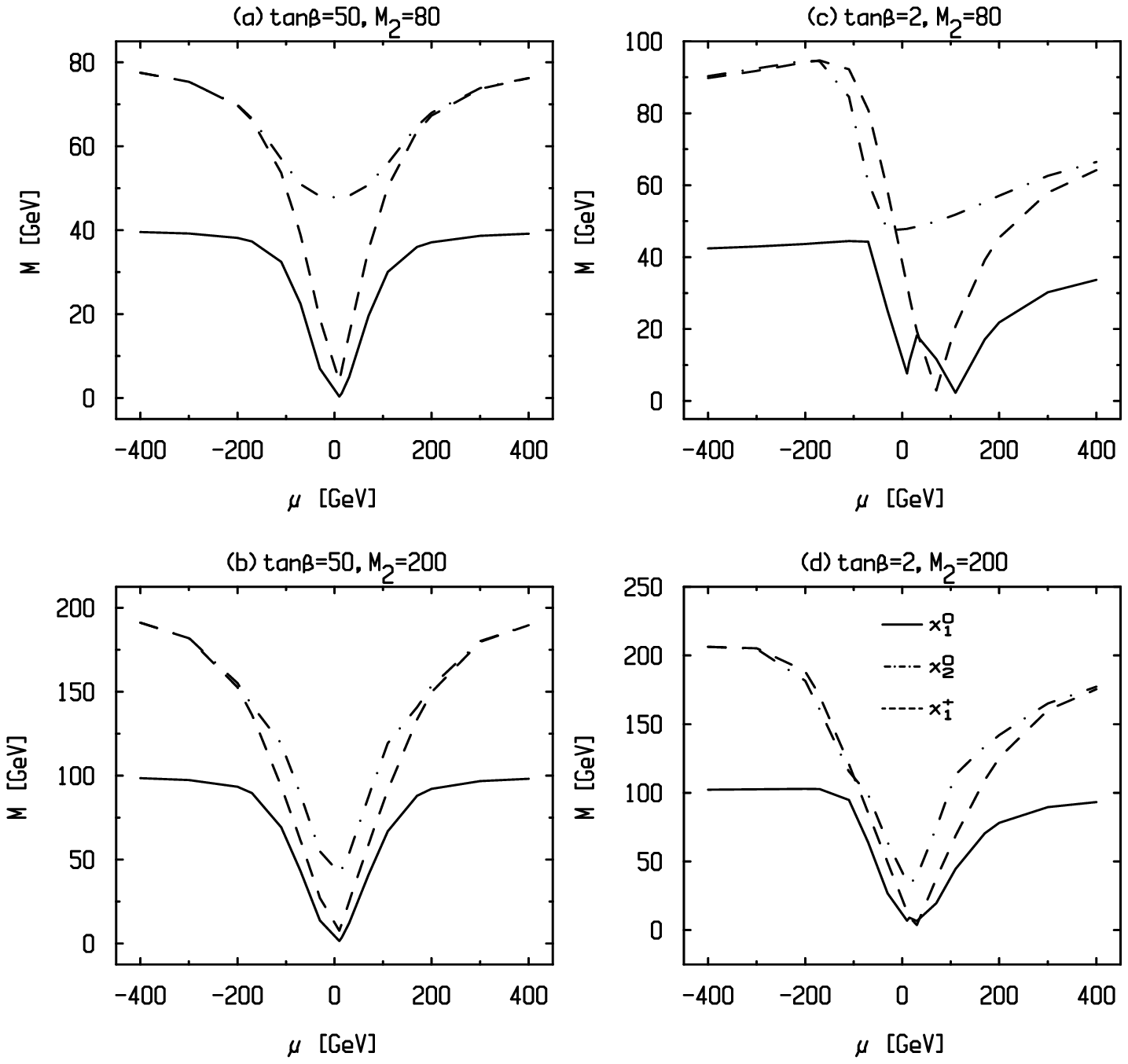}
\end{center}
\caption{\footnotesize  \it
Mass spectrum for the chargino $\tchi^{\pm}_1$ and the first 
two lowest mass neutralinos, $\tchi^0_1$ and $\tchi^0_2$, as a function of $\mu$.
Four choices of the parameters, $\tan \beta$ and $M_2$ (in $GeV$), are used, 
as indicated on top of each window. 
\rm \normalsize }
\label{spec}
\end{figure}

We shall develop the study of single superpartner production within a non minimal supergravity
framework, assuming the existence of a grand unified gauge theory and of family universal 
boundary conditions on 
the supersymmetry breaking parameters. The renormalization group improved classical spectrum
 of the scalar superpartners is determined in principle by the full set of soft supersymmetry breaking
 parameters at the unification scale, $M_X$, namely, $m_0$ (common scalars mass), $m_{1/2}$ 
(common gauginos mass), $A$ (trilinear Yukawa interactions), $B_{\mu}$ (bilinear Higgs
interaction); by the parameters $\tan \beta={v_u \over v_d }
={<H_u> \over <H_d>}$ and $ \mu(t)$, where $t$ denotes the running scale; and by the gauge 
coupling constants, $g_a(t)$, along with fermions masses, $m_f^2(t)$. 
If one neglects the Yukawa interactions of quarks and leptons with the Higgs bosons, then
the running masses of all sfermions remain family degenerate down to the electroweak breaking 
scale where they are described by the familiar additive formula,

\begin{eqnarray}
m_{\tilde f}^2(t)=m_{f}^2(t)+m_0^2+c_f(t)m_{1/2}^2 \pm m_{Z^0}^2 \cos (2 \beta) 
(T^f_3-Q^f x_W),
\label{specsugra}
\end{eqnarray}
  
where $c_f(t)$ are calculable coefficients depending on the gauge interactions parameters and the last term represents the D-term contribution,
the upper and lower sign being for the left and right chirality sfermions, respectively.
 The most relevant Yukawa coupling constants, namely, those of the third family of up-quarks or, for large $\tan \beta$, of d-quarks and leptons,
are expected to induce downwards shifts for the third family squarks (up and down) and sleptons,
which depend non trivially on the parameters $A$ and $\mu$. In this work, we shall restrict
consideration to the simple case of family independent running masses
and employ the 
approximate representation in eq.(\ref{specsugra}) with the numerical values 
quoted in \cite{e:Drees}.
Note that the total rates do not depend on the squarks masses and, as already remarked
in section \ref{sectionD}, the third families of squarks are not considered in the cascade decays.
The charginos and neutralinos classical mass spectra are determined by the subset
 of parameters: $M_1(t), \ M_2(t), \ \mu(t)$ and $\tan \beta$. For fixed $m_{1/2}$, 
the solution of the one loop renormalization group equations is given explicitly by, 
$m_{1/2} =  (1- \beta_a t) M_a(t),$
where  $t=\log({M_X^2 \over Q^2})$, $Q$ denoting the running scale,
$\beta_a = {g_X^2 b_a \over (4 \pi)^2 }$,
$b_a=(3,-1,-11)$ with $a=(3,2,1)$, corresponding to the
beta functions parameters for the  gauge group factors, 
$SU(3),SU(2)_L,U(1)_Y$, and $g_X$ is the coupling constant at unification
scale.
Note that the wino and bino masses are related as, $ M_1(t) = {5 \over 3}M_2(t) 
\tan^2 \theta _W $.
It is useful here to comment on the relation of our framework with the so-called minimal
supergravity framework in which one assumes a constrained parameter space compatible with 
electroweak symmetry breaking. Let us follow
here the so-called ambidextrous minimal supergravity
approach \cite{e:ewsb2}, where one chooses
$[m_0,m_{1/2},A, sign(\mu),\tan \beta]$ as the free parameters set and
 derives $\mu(t_Z),B_{\mu}(t_Z)$, at the electroweak symmetry breaking scale,
 $t_Z= \ln M_X^2/m_Z^2$, 
through the minimisation equations for the Higgs bosons potential. 
For fixed $m_0,m_{1/2}$ 
and $\tan \beta$, varying $A$ will let the parameter $\mu(t_Z)$ span finite intervals of
relatively restricted sizes. In figure \ref{efbr}, we  give results of 
a numerical resolution of the renormalization group equations which show
 the variation of $\vert \mu(t_Z) \vert $ as a function of $m_0$ and $m_{1/2}$, 
and also exhibit its dependence on $A$. Note that the equations admit the 
symmetry, $\mu (t_Z) \to - \mu  (t_Z)$.
Observing that  $\mu(t_Z)$ is typically  a monotonous increasing function of $A$, we see 
from Fig.\ref{efbr} that the  corresponding  incremental increase, 
$\d \mu (t_Z)/\mu (t_Z) $, as  one spans the wide interval, $A\in [-5, +5]$, 
is  small  and of order 20\%. 

In the  infrared fixed point  approach for the top-quark Yukawa coupling, 
$\tan \beta$ is fixed (up to the ambiguity associated with  large 
or low $\tan \beta$ solutions) in terms of the top-quark mass,
 $m_{t}=C \ \sin \beta, $ 
with, $C \simeq 190-210GeV$, for, $\alpha_3(m_{Z^0})=0.11-0.13$ \cite{e:Pok}.
The dependence on $A$ of
 the electroweak constraint also becomes very weak, so that 
  $\mu (t_Z)$ is a known function of $m_0,m_{1/2}$ and $\tan \beta$ \cite{e:Pok}:

\begin{eqnarray}
\mu^2+{m_{Z}^2 \over 2}=m_0^2{1+0.5\tan^2 \beta \over \tan^2\beta -1}+
m_{1/2}^2{0.5+3.5\tan^2\beta \over \tan^2\beta -1}.
\label{ewsb}
\end{eqnarray}

In section \ref{sectionBR}, we will discuss 
results for the branching ratios in this constrained model. The total rates are not affected in any
significant way by which version of the supergravity models is used, 
since, as we will see, their dependence on $\tan \beta$ and $m_0$ turns out to be smooth.

The main uncertain inputs are the superpartners mass spectrum and the coupling constants
$\l_{ijk}$. To survey the characteristic properties of single production over a broad
region of parameter space, we found it convenient to consider
a continuous interval of variation for $\mu(t_Z)$, namely, $\mu(t_Z) \in [-400,+400]GeV$, while 
choosing suitable discrete values for the other parameters: $M_2(t_Z)=50,80,100,150,200 \ GeV$,
$m_0=20,50,150 \ GeV$ and $\tan \beta=2,50$. We shall set the unification scale at 
$M_X=2 \ 10^{16}GeV$ and the running scale at $Q^2=m_Z^2$.
 For definiteness, we choose 
the coupling constant, which controls the size of the 
production cross section, at the reference value: $\l_{mJJ}=0.05$. This is the strongest
bound for a slepton mass of $100GeV$ \cite{e:Bhatt}. The dependence of integrated total
rates on $\l_{mJJ}$ is then given by a simple rescaling $({\l_{mJJ} \over 0.05})^2$ but
that of branching ratios on $\l_{ijk}$ (which may or may not be identified with $\l_{mJJ}$)
is more complicated because of the interplay between the RPC and RPV contributions which
add up in the total decay widths.
The reference value used here, $\l_{ijk}=0.05$, is also an interesting borderline
value since below this value the dependence of branching fractions on $\l_{ijk}$
becomes negligible in generic cases.

It will prove helpful in the following discussion to keep within 
sight the spectrum for the 
low-lying inos. We display in Figure \ref{spec} the results obtained by solving 
numerically the eigenvalues problem for the charginos and neutralinos mass matrices. 
Recall the current experimental bounds \cite{e:PartData}, $m_{\tchi^0_1}>23GeV$,
 $m_{\tchi^{\pm}_1}>45GeV$, $m_{\tilde \nu}>37.1GeV$ and $m_{\tilde l}>45GeV$. 
The following remarks about Figure \ref{spec} will 
prove relevant for the discussion on branching fractions: (i) The symmetry of the spectra
 under, $\mu \leftrightarrow - \mu$, is spoilt  at low $\tan \beta$ as can be seen on the 
explicit expression for the inos masses in \cite{e:haber}; (ii) The mass 
differences $\tchi^+-\tchi^0$ increase with $\vert \mu \vert$ with a steep rise appearing 
at, $\mu=M_2$, the borderline between the Higgsino and gaugino dominant regimes; (iii)
 The spacings $\tchi^0_2-\tchi^0_1$ and $\tchi^+_1-\tchi^0_1$ decrease in magnitudes,
 relatively to the $\tchi^0_1$ mass, with increasing $M_2$. Although we show here the 
results for $\tchi^0_2$ mass, the interesting possibility of exciting 
the second neutralino $\tchi^0_2$ is not considered in the subsequent discussion.

\section{Results and discussion}
\label{sectionR} 

\setcounter{equation}{0}

\subsection{Total production rates}

\begin{figure}
\begin{center}
\leavevmode
\psfig{figure=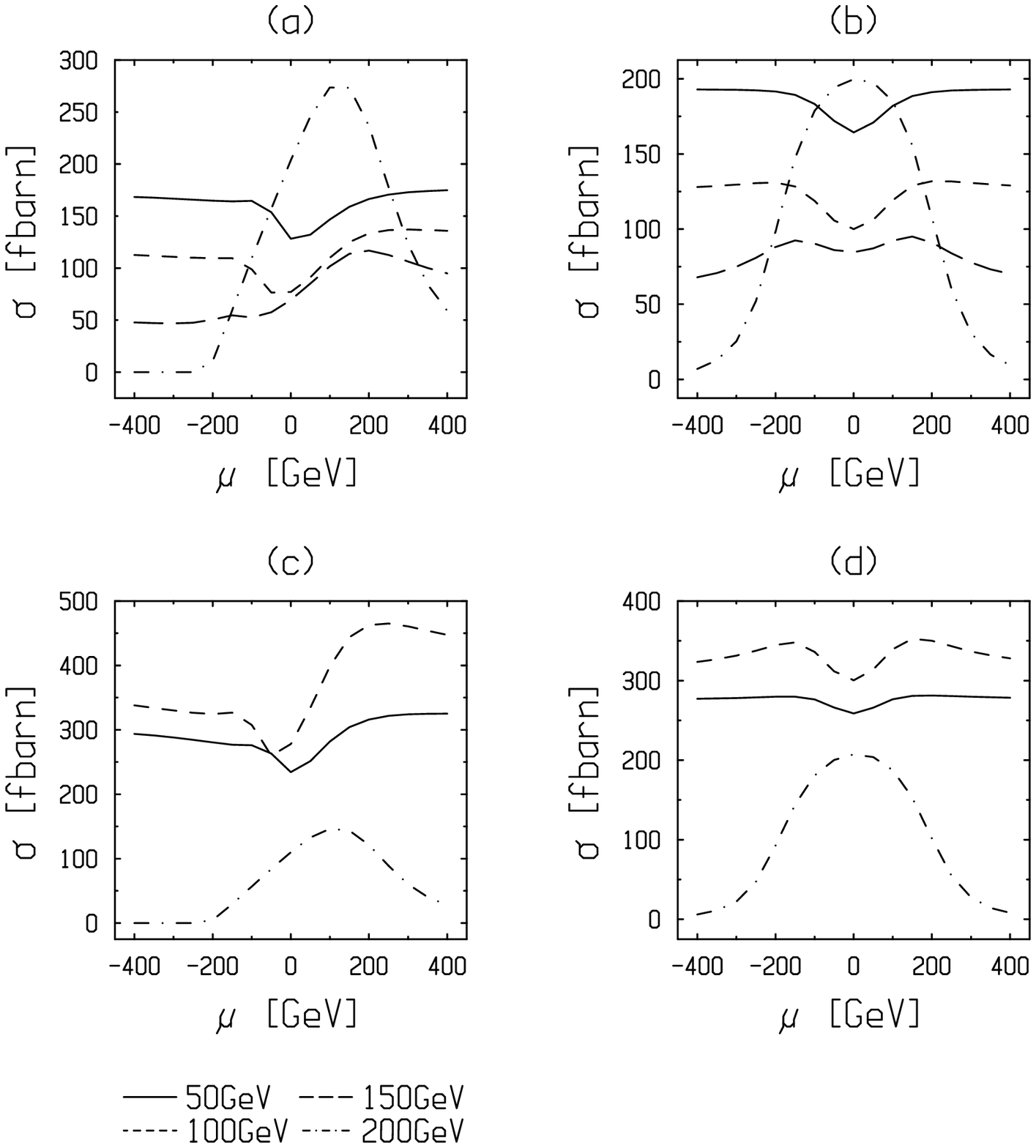,width=5.5in}
\end{center}
\caption{\footnotesize  \it
The integrated cross sections for the process, $l_J^+l_J^- \to \tchi_1^- l_m^+$,
at a center of mass energy of $200GeV$, are shown as a function of $\mu$ for discrete
 choices of the remaining parameters: (a) 
$\tan \beta =2, \ m_0=50GeV$, (b) $\tan \beta =50, \ m_0=50GeV$, 
(c) $\tan \beta =2, \ m_0=150GeV$ and  (d) $\tan \beta =50, \ m_0=150GeV$,
with $\l_{mJJ}=0.05$. The windows conventions are such that $\tan \beta=2,50$
horizontally and $m_0=50,150GeV$ vertically.
The different curves refer to the values of $M_2$ of $50GeV$
(continuous line), $100GeV$ (dot-dashed line), $150GeV$ (dashed line) and $200GeV$
 (dotted line), as indicated  at the bottom of the figure.  
\rm \normalsize }
\label{chsect1}
\end{figure}

\begin{figure}
\begin{center}
\leavevmode
\psfig{figure=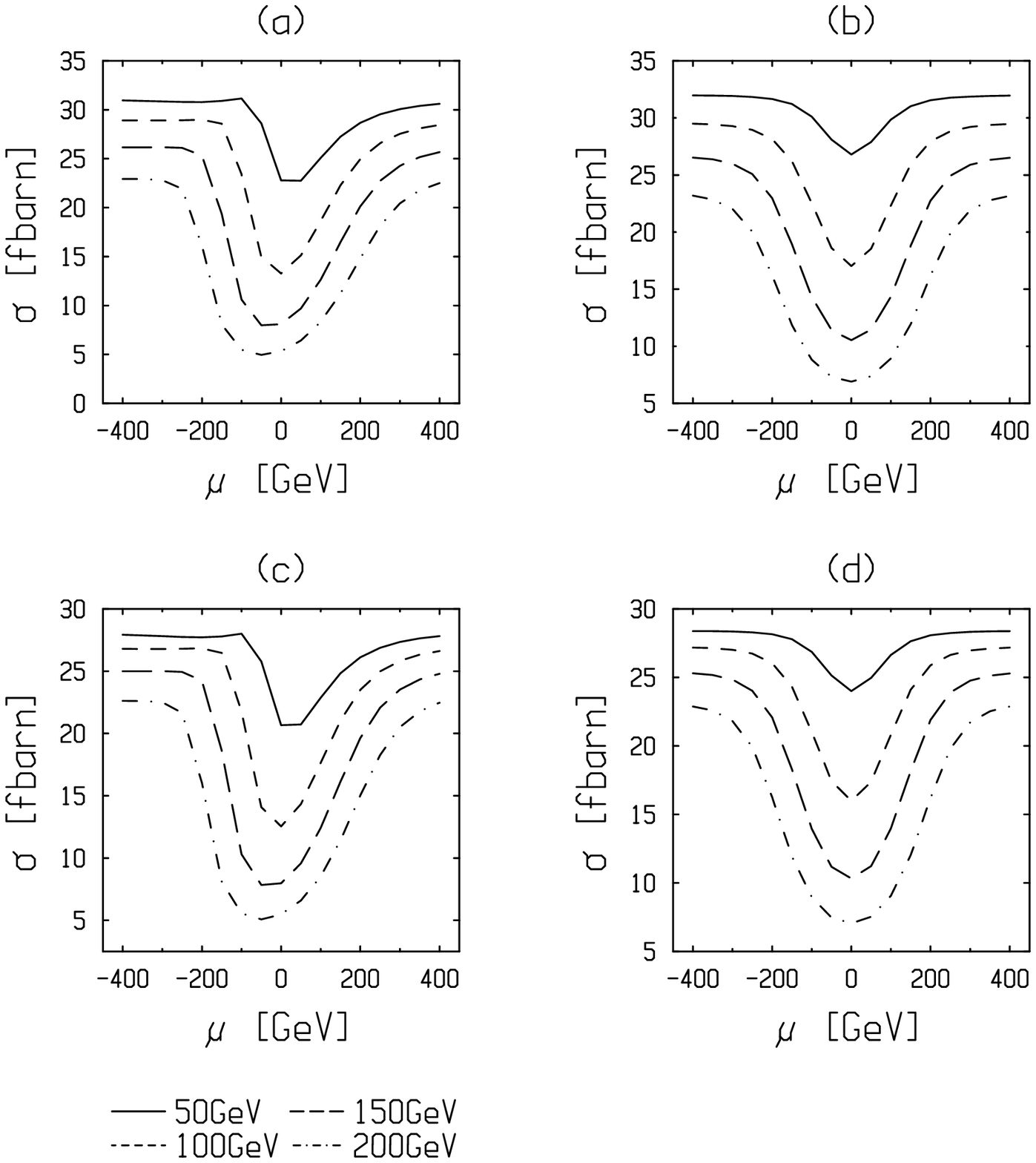,width=5.5in}
\end{center}
\caption{\footnotesize  \it
The integrated cross sections for the process, $l_J^+l_J^- \to \tchi_1^- l_m^+$,
at a center of mass energy of $500GeV$, are shown as a function of $\mu$ for discrete
 choices of the remaining parameters: (a) 
$\tan \beta =2, \ m_0=50GeV$, (b) $\tan \beta =50, \ m_0=50GeV$, 
(c) $\tan \beta =2, \ m_0=150GeV$ and  (d) $\tan \beta =50, \ m_0=150GeV$,
with $\l_{mJJ}=0.05$. The windows conventions are such that $\tan \beta=2,50$
horizontally and $m_0=50,150GeV$ vertically.
The different curves refer to the values of $M_2$ of $50GeV$
(continuous line), $100GeV$ (dot-dashed line), $150GeV$ (dashed line) and $200GeV$
 (dotted line), as indicated  at the bottom of the figure.  
\rm \normalsize }
\label{chsect2}
\end{figure}

\begin{figure}
\begin{center}
\leavevmode
\psfig{figure=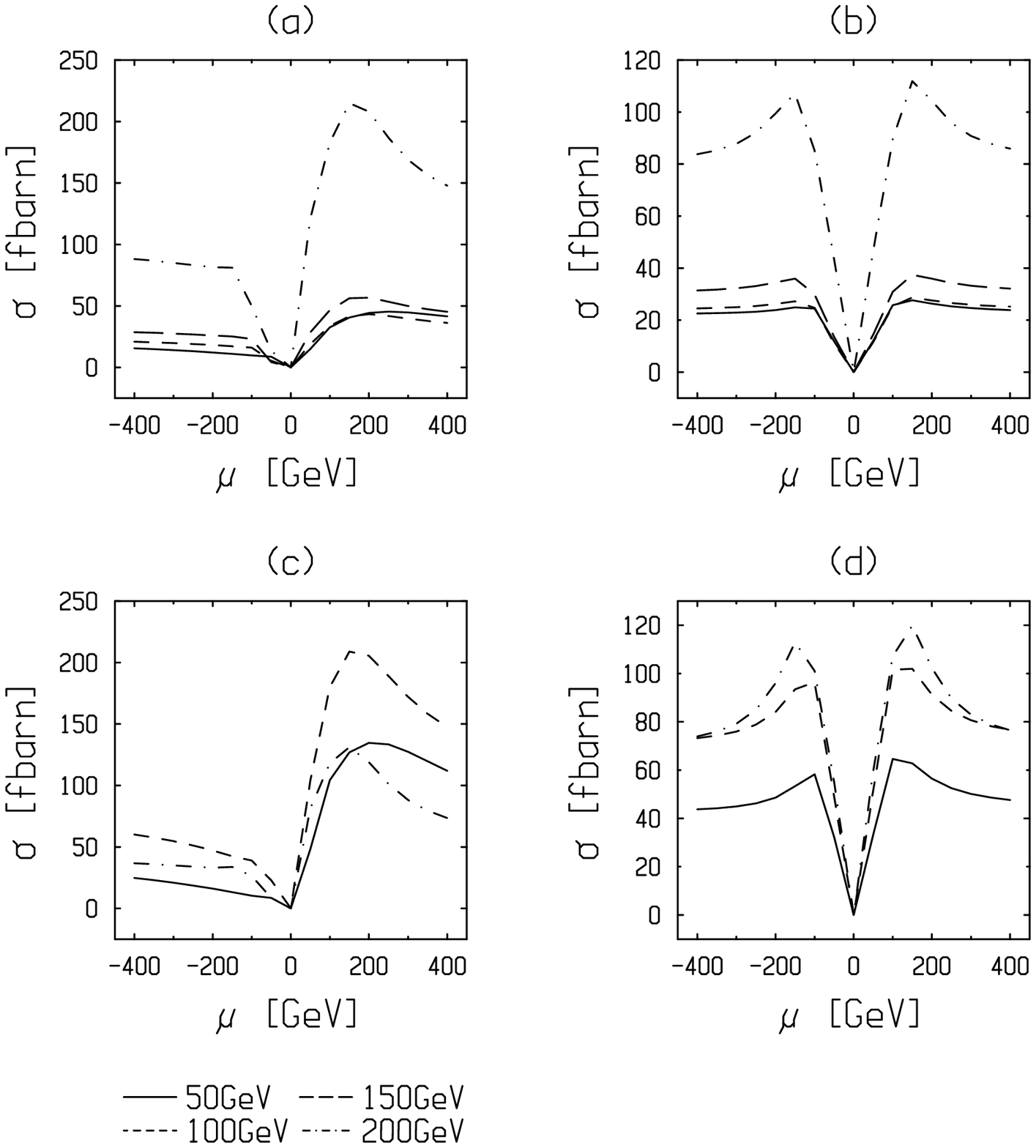,width=5.5in}
\end{center}
\caption{\footnotesize  \it 
The integrated cross sections for the process, $l_J^+l_J^- \to \tchi_1^0 \bar \nu_m$,
at a center of mass energy of $200GeV$, are shown as a function of $\mu$ for discrete
 choices of the remaining parameters: (a) 
$\tan \beta =2, \ m_0=50GeV$, (b) $\tan \beta =50, \ m_0=50GeV$, 
(c) $\tan \beta =2, \ m_0=150GeV$ and  (d) $\tan \beta =50, \ m_0=150GeV$,
with $\l_{mJJ}=0.05$. The windows conventions are such that $\tan \beta=2,50$
horizontally and $m_0=50,150GeV$ vertically.
The different curves refer to the values of $M_2$ of $50GeV$
(continuous line), $100GeV$ (dot-dashed line), $150GeV$ (dashed line) and $200GeV$
 (dotted line), as indicated at the bottom of the figure.
\rm \normalsize}
\label{neusect1}
\end{figure}

\begin{figure}
\begin{center}
\leavevmode
\psfig{figure=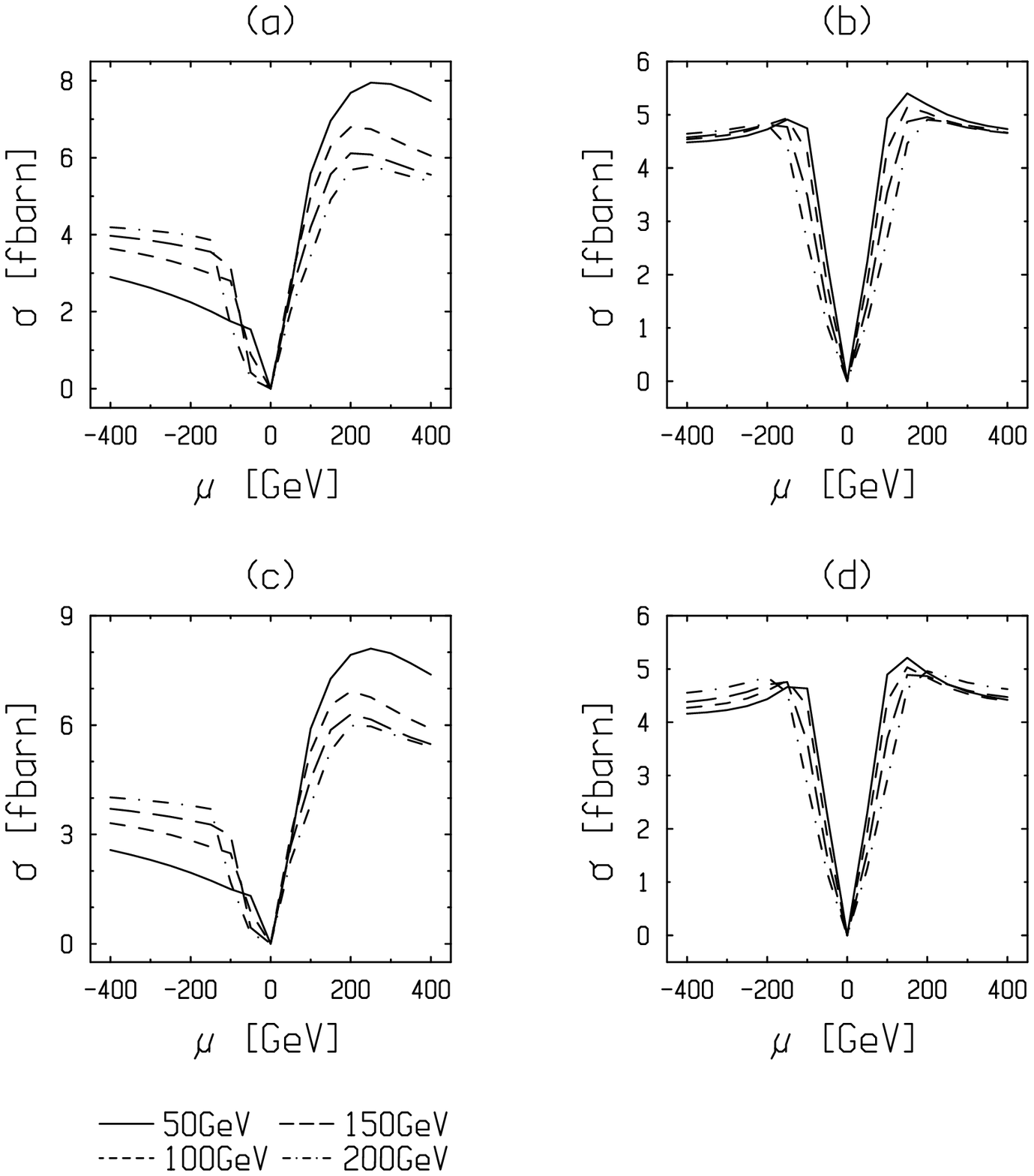,width=5.5in}
\end{center}
\caption{\footnotesize  \it 
The integrated cross sections for the process, $l_J^+l_J^- \to \tchi_1^0 \bar \nu_m$,
at a center of mass energy of $500GeV$, are shown as a function of $\mu$ for discrete
 choices of the remaining parameters: (a) 
$\tan \beta =2, \ m_0=50GeV$, (b) $\tan \beta =50, \ m_0=50GeV$, 
(c) $\tan \beta =2, \ m_0=150GeV$ and  (d) $\tan \beta =50, \ m_0=150GeV$,
with $\l_{mJJ}=0.05$. The windows conventions are such that $\tan \beta=2,50$
horizontally and $m_0=50,150GeV$ vertically.
The different curves refer to the values of $M_2$ of $50GeV$
(continuous line), $100GeV$ (dot-dashed line), $150GeV$ (dashed line) and $200GeV$
 (dotted line), as indicated at the bottom of the figure.
\rm \normalsize}
\label{neusect2}
\end{figure}

\begin{figure}
\begin{center}
\leavevmode
\psfig{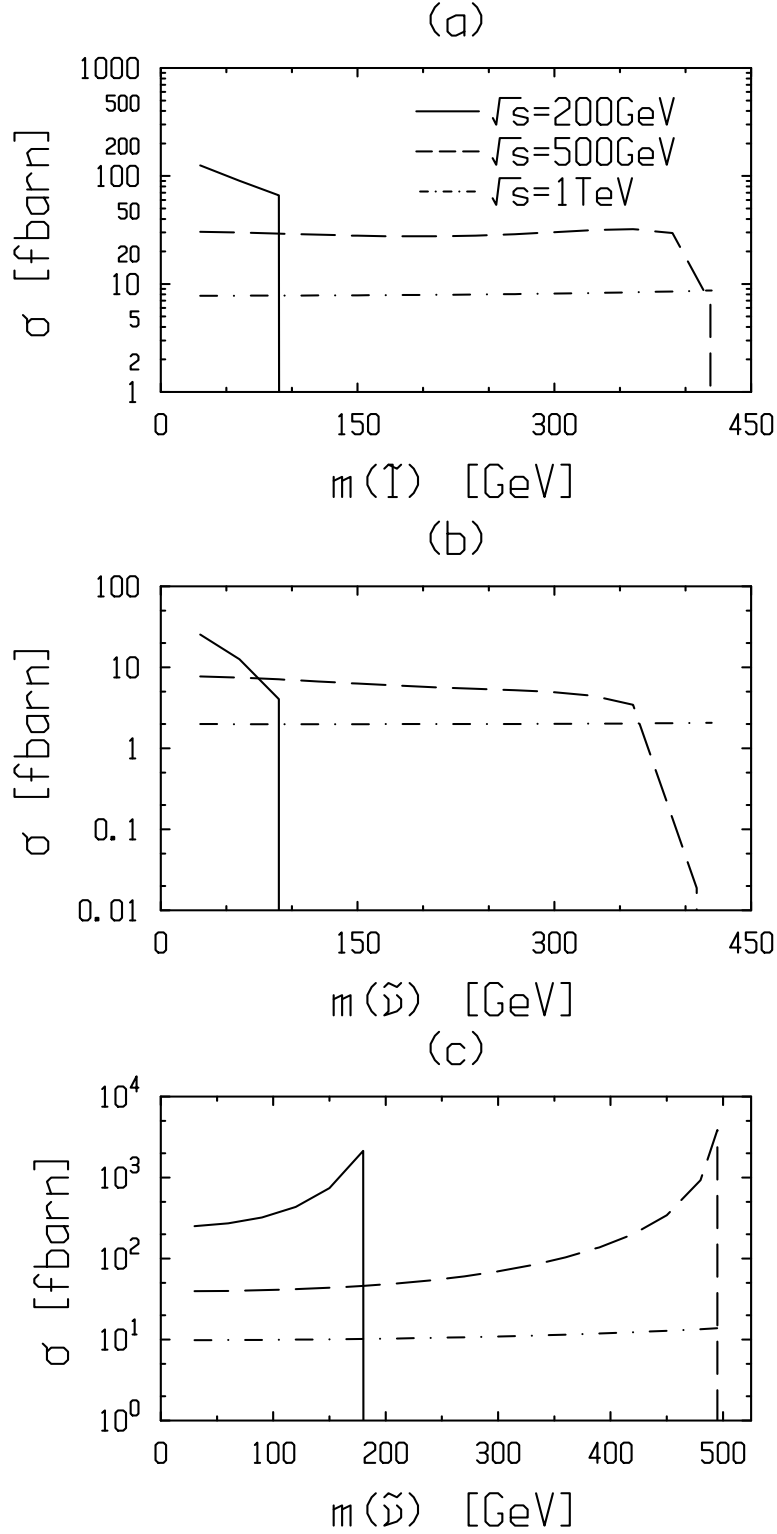}
\end{center}
\caption{\footnotesize  \it
The cross sections for the processes, $l_J^+l_J^- \to \tilde l_m^- W^+$ (a),
$l_J^+l_J^- \to \tilde \nu_m Z^0$ (b) and $l_J^+l_J^-
\to \tilde {\nu}_m  \ \gamma$ (c), 
are shown as a function of the slepton mass and the sneutrino mass,
for $\l_{mJJ}=0.05$. The three values of the center of mass energies 
considered are $200, \ 500$ and $1000GeV$, as quoted in the top window.
\rm \normalsize}
\label{slsect}
\end{figure}

The total production rates are evaluated by taking the angular integral,
$\s=\int_{-x_m}^{+x_m} {d \s \over dx} dx, \ [x=\cos \theta]$,
over the differential cross sections which are given explicitly in
 eqs.(\ref{eqrc4a})-(\ref{eqrc4e}) in Appendix \ref{seca}.
 To follow the usual practice we shall set an angular cut-off to account for 
 the poor detection condition along the beam pipe: $170^o > \theta_m > 10^o  $,
 corresponding to $x_m =\cos \theta_m = 0.9848$.

\subsubsection{Inos production}

The results for the integrated rates of the production of the lowest mass eigenstates
 $\tchi_1^-$ and $\tchi_1^0$, 
 at LEPII energies, are displayed in Figures \ref{chsect1} and \ref{neusect1}, respectively. The inos production rates 
depend smoothly on $\tan \beta$, and on the mass parameters, $\mu, \ m_0, \ M_2$, in a way 
which closely reflects on the mass spectrum. Thus, the symmetry under $\mu \leftrightarrow -\mu$
 is upset only for low $\tan \beta$ and the rates decrease with increasing $M_2$. The only
cases where fast variations of rates arise are 
 for values of $m_0$ and $M_2$ at which the center of mass energy hits on 
the sneutrino s-channel pole, $\sqrt s=m_{\tilde \nu}$. As $m_0$ increases, the resonance occurs
 at smaller values of $M_2$ since the sneutrino mass depends on $ M_2, \ m_0 $ and $ \tan \beta$
 (see eq.(\ref{specsugra})). The pole cross sections themselves, as parametrized by the 
conventional formula, 

\begin{eqnarray}
\s (l^+l^- \to X)&=&{8 \pi s \over m_{\tilde \nu}^2}{\G(\tilde \nu \to l^+l^-) 
\G(\tilde \nu \to X) \over (s-m_{\tilde \nu}^2)^2+\G_{\tilde \nu}^2} \cr
& \approx & 4 \ 10^8 ({100GeV\over m_{\tilde \nu}})^2 B(\tilde \nu \to l^+l^-)
B(\tilde \nu \to X) \ fbarns,
\label{sigmapole}
\end{eqnarray}

can grow to values several order of magnitudes higher. For clarity, 
we have refrained from 
drawing the cross sections close to the resonant energy in the same plot. This is the reason
why  the curves corresponding to $M_2=150GeV$ do not appear in Figures \ref{chsect1}(c)(d) and
 \ref{neusect1}(c)(d). The effect of the pole can be seen for $M_2=200GeV$ in Figures
\ref{chsect1}(a)(b) and \ref{neusect1}(a)(b). We note also that for 
$\mu=0$, $\tchi^0_1$ is a pure higgsino and the $\tchi^0_1$ production cross section vanishes.
The results for inos production rates at NLC or $\mu^+ \mu^- $ colliders center of mass energies are displayed in Figures \ref{chsect2} and \ref{neusect2}. 
The drop with respect to the LEPII energies is nearly by one order of magnitude.
The second neutralino production rates, $\sigma(\tchi_2^0)=
\sigma(l^+_J l^-_J \to \tchi_2^0 \nu_m)$, when this is kinematically allowed, turns out
 to be of the same order of magnitude as $\sigma(\tchi_1^0)$. For $\sqrt s=500GeV$,
$\sigma(\tchi_1^0)$ and $\sigma(\tchi_2^0)$ are numerically close throughout the parameter
 space of our model. However, for $\sqrt s=200GeV$, there are regions (large $\tan \beta$,
 $\mu<0$) where one has $\sigma(\tchi_2^0) \approx 2 \sigma(\tchi_1^0)$ and other regions
 (low $\tan \beta$, $\mu>0$) where one rather has $\sigma(\tchi_2^0) \approx {1 \over 2}
 \sigma(\tchi_1^0)$. As for the production rate of the second chargino,
 $\sigma(\tchi_2^-)$, this is always nearly an order of magnitude below
 $\sigma(\tchi_1^-)$.

\subsubsection{Sleptons production}

The slepton and sneutrino production rates depend solely on the sleptons masses and $\l_{mJJ}$.
 The results, obtained by setting $m_{\tilde l}=m_{\tilde \nu}$, are displayed 
in Figure \ref{slsect} for three values of the center of mass energies. 
An account of the mass difference between
 $m_{\tilde l}$ and $m_{\tilde \nu }$ would not change the numerical results in 
any significant way.

 The differential cross section
for the reaction $l_J^+l_J^- \to \tilde \nu \gamma$ 
must be treated with special care because
 of its extreme sensitivity at the end points, $x= \pm 1$, in the limit of vanishing
 electron mass, $m_e \to 0$. As appears clearly on the expression of the squared momentum
 transfer variable, $t=(k'-p')^2=m_{\g}^2-{1 \over 2} (s-m_{\tilde \nu}^2+m_{\g}^2)
(1-{k \over E_k}{p \over E_p}x)$, for $m_{\g}=0$, the t-channel amplitude 
has a collinear  singularity, 
$t \to 0$ as $x \to 1$. An analogous  collinear  
singularity occurs for the u-channel amplitude,
 $u=(k-p')^2 \to 0$ as $x \to -1$. Imposing the cut-off on the center of mass 
angle $\t$ makes the regularisation of collinear singularities pointless.

 In the limit of vanishing $m_{\g}$, 
independently of $x$ and $m_e$, the sneutrino production  cross section
becomes infinite at the limiting energy point, $\sqrt s = m_{\tilde \nu}$. 
This accounts for the property of the numerical results for the 
integrated cross section to rise with $m_{\tilde \nu}$, as seen in Figure
 \ref{slsect}(c). However, if one were to set $m_{\g}$
 at, say, the $\rho$-meson mass, in line with the vector meson dominance hypothesis, 
one would rather find the opposite behaviour with respect to the dependence on
 $m_{\tilde \nu}$. Observe that the increase of the cross section with $m_{\tilde \nu}$
corresponds to the fact that, for $m_{\tilde \nu} \approx \sqrt s$, the 
process $l_J^+l_J^- \to \tilde \nu \g$ behaves like a sneutrino resonant production,
 accompagnied by the initial state radiation of a soft photon. 

\subsubsection{Discussion}

In summary, the single production rates range from 
several 10 's of $fbarns$ to a few 100's 
of $fbarns$ at LEP energies and several units to a few 10's of $fbarns$ at NLC energies.
 Therefore, the superpartners single production are at the limit of observability
 for LEPII assuming an 
 integrated luminosity per year of $200pb^{-1}$  at $\sqrt s=200GeV$. The prospects 
for single production should be rather good at 
NLC \cite{e:Lumi} and $\mu^+ \mu^- $ colliders \cite{e:Lumi2}
 since the assumed integrated luminosity per
year is expected to be about $50fb^{-1}$ at $\sqrt s=500GeV$.
 Moreover, it is important to note here
 that had we considered for the RPV coupling constants, constant values for the product
 $\l_{mJJ}({m_{\tilde l_R}\over 100GeV})$, rather than for $\l_{mJJ}$, the rates would get 
an important amplification factor $({m_{\tilde l_R}\over 100GeV})^2$ for increasing 
superpartners masses. Note that the slepton involved in the  bound is of right chirality
and thus is of opposite chirality than the slepton involved in the rate. Of course, the masses 
of $\tilde l_L$ and $\tilde l_R$ are related in a given model.
At $\sqrt s=500GeV$ and assuming $\l_{ijk} \geq 0.05$, all the single 
production reactions should be potentially observable over a broad region of parameter space.
The slepton production reactions could then probe slepton masses up to $400GeV$
 (Figure \ref{slsect}(a)) and sneutrino masses up to $500GeV$ (Figure \ref{slsect}(c)).
 The ino production reactions could probe 
a large region of the parameter plane $(\mu,M_2)$,
 since the dependence on the parameters $m_0$ and $\tan \beta$ is smooth.
To strengthen our conclusions, it is necessary to examine the 
signatures associated with the final states, which is the subject of the next section.

\subsection{Branching Ratios}

\label{sectionBR}

\begin{figure}
\begin{center}
\leavevmode
\psfig{figure=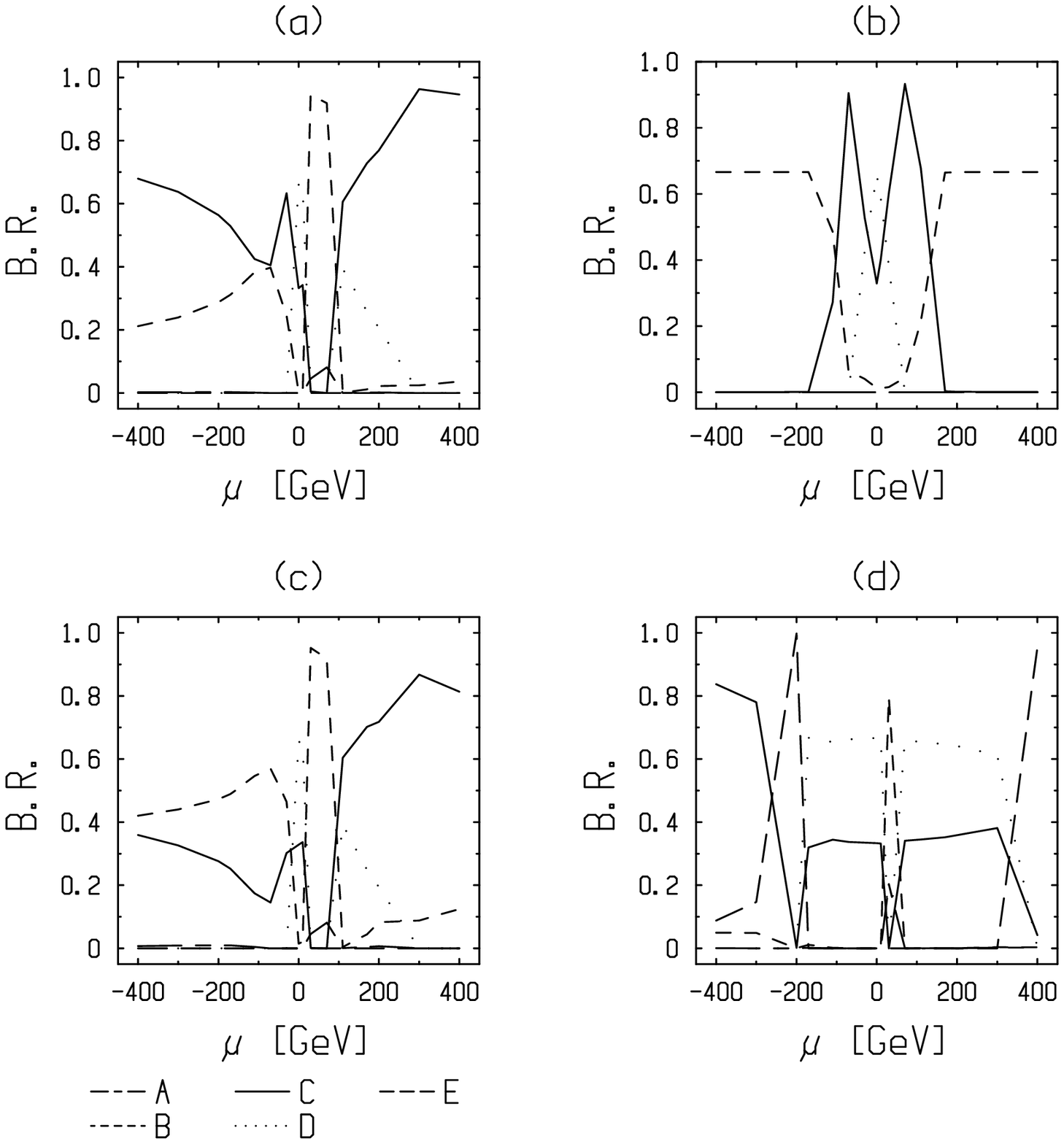,width=6.in}
\end{center}
\caption{\footnotesize  \it
Branching ratios for the chargino $\tchi^-_1$ decays  
as a function of $\mu$. The results in the four windows are obtained 
with the following choices  for the parameters, $[(M_2(GeV),m_0(GeV),\tan 
\beta,\l_{ijk})$, $m_{\tilde \nu_L}(GeV),m_{\tilde l_L}(GeV)]$: 
(a) $[(80,20,2,0.05)$, $53.19,81.66]$,  (b) $[(80,20,50,0.05)$, $34.20,86.97]$, 
(c) $[(80,20,2,0.1)$, $53.19,81.66]$, (d) $[(200,100,2,0.05)$, $195.6,205.2]$.
The final states are labeled by the letters, A,B,C,D,E, which have the same  meaning
as in Table \ref{tabloC}. 
\rm \normalsize }
\label{br1}
\end{figure}

\begin{figure}
\begin{center}
\leavevmode
\psfig{figure=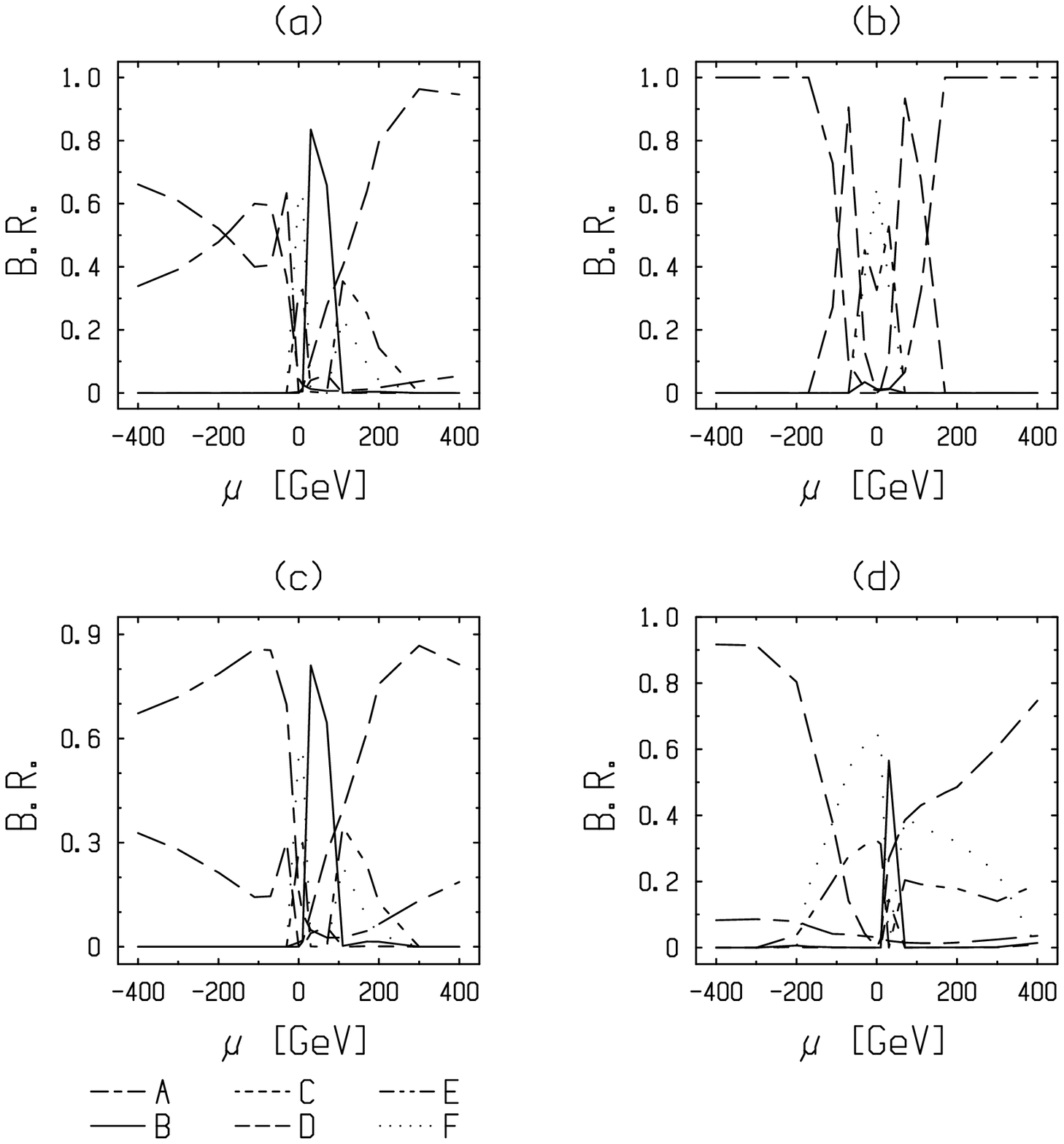,width=6.in}
\end{center}
\caption{\footnotesize  \it
Branching ratios of the sneutrino decays  
as a function of $\mu$. The results in the four windows are obtained 
with the following choices  for the parameters, $[(M_2(GeV),m_0(GeV),\tan 
\beta,\l_{ijk})$, $m_{\tilde \nu_L}(GeV),m_{\tilde l_L}(GeV)]$: 
(a) $[(80,20,2,0.05)$, $53.19,81.66]$,  (b) $[(80,20,50,0.05)$, $34.20,86.97]$, 
(c) $[(80,20,2,0.1)$, $53.19,81.66]$,  (d) $[(200,100,2,0.05)$, $195.6,205.2]$.
The final states are labeled by the letters, A,B,C,D,E,F, which have the same  meaning
as in Table \ref{tabloA}.
\rm \normalsize}
\label{br2}
\end{figure}

\begin{figure}
\begin{center}
\leavevmode
\psfig{figure=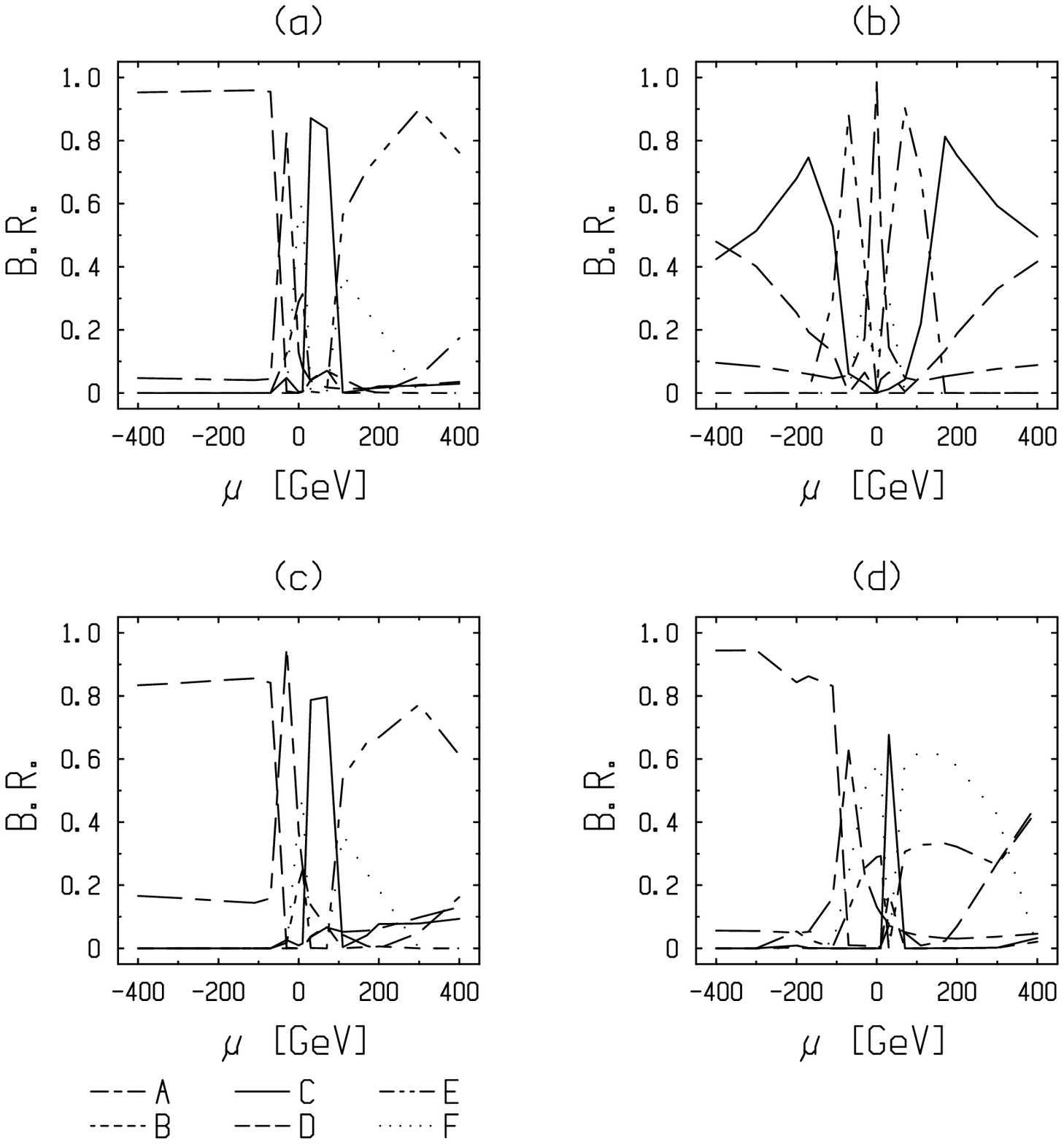,width=6.in}
\end{center}
\caption{\footnotesize  \it
Branching ratios of the slepton decays 
as a function of $\mu$. The results in the four windows are obtained 
with the following choices  for the parameters, $[(M_2(GeV),m_0(GeV),\tan 
\beta,\l_{ijk})$, $m_{\tilde \nu_L}(GeV),m_{\tilde l_L}(GeV)]$:  
(a) $[(80,20,2,0.05)$,  $53.19,81.66]$,  (b) $[(80,20,50,0.05)$,  $34.20,86.97]$, 
(c) $[(80,20,2,0.1)$,  $53.19,81.66]$, (d) $[(200,100,2,0.05)$,  $195.6,205.2]$.
The final states are labeled by the letters, A,B,C,D,E,F, which have the same meaning
as in Table \ref{tabloB}.
\rm \normalsize}
\label{br3}
\end{figure}

\begin{figure}
\begin{center}
\leavevmode
\psfig{figure=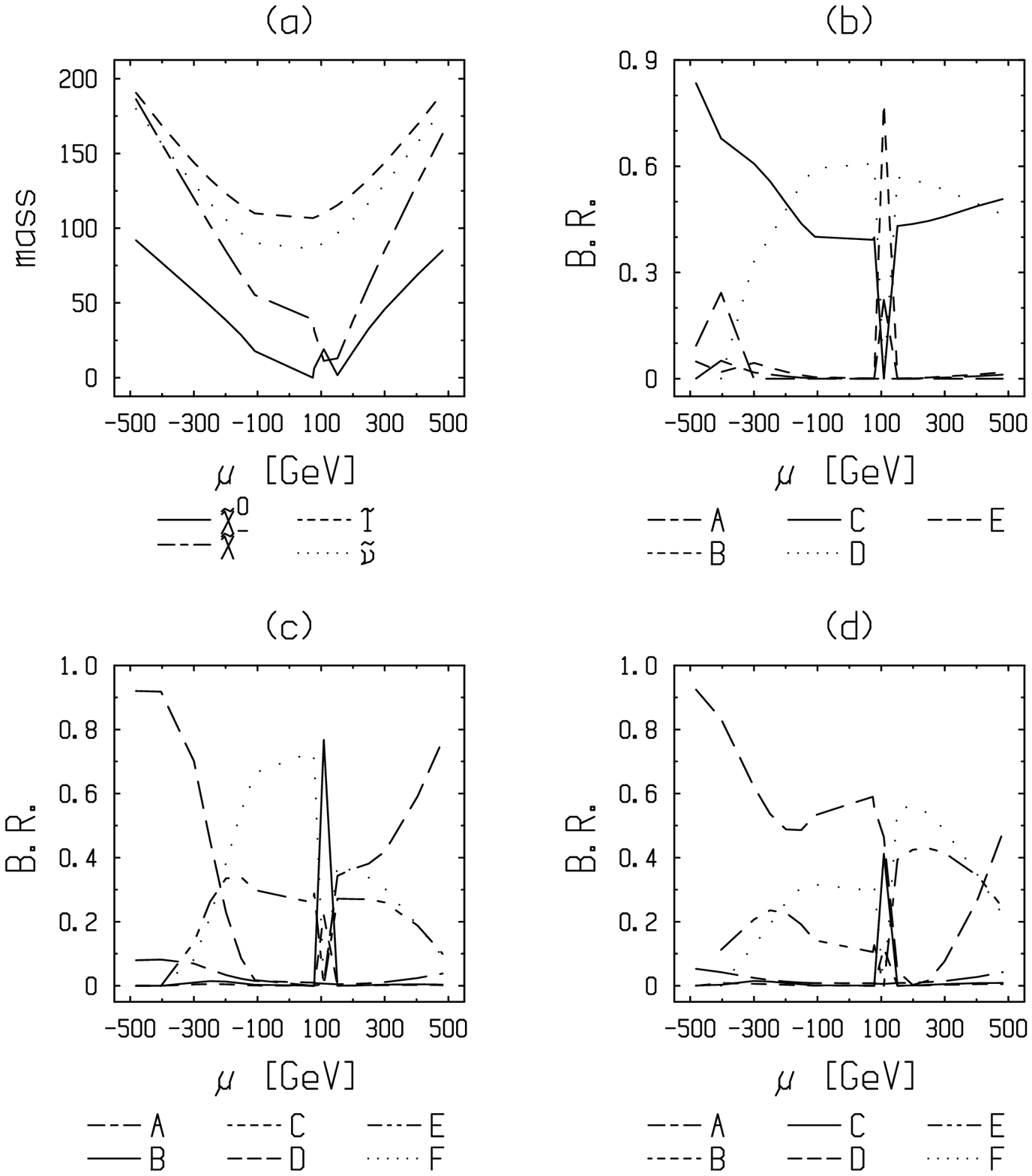,width=6.in}
\end{center}
\caption{\footnotesize  \it
Mass spectrum of the supersymmetric particles (a), in GeV, and branching ratios for the decays
of the chargino (b), sneutrino (c) and slepton (d), as a function of $\mu$.
The results are obtained for $m_0=100GeV$, using equation (\ref{ewsb}).
 The final states in figures (b),(c),(d) are labeled by the letters, A,B,... 
which have the same meaning as in tables \ref{tabloC},\ref{tabloA},\ref{tabloB}, respectively.
\rm \normalsize}
\label{bratewsb}
\end{figure}

In the narrow resonance approximation, the partial transition rates are readily 
obtained by multiplying the total rates for each reaction with the decay branching
fractions. The various final states for each of the $2 \to 2$ single production reactions
have been listed in Tables \ref{tabloC}, \ref{tabloA} and \ref{tabloB}. The leptons
family configurations in the final states will depend on the hypothesis for the
RPV coupling constant (single or pair dominance).

With the purpose of testing characteristic points of the parameter space,
we have evaluated the branching ratios for the decays of the superpartners, namely,
$\tchi_1^{\pm},\tilde l^{\pm}$ and $\tilde \nu$,
for variable $\mu$ at discrete choices of $M_2$, $m_0$ and $\tan \beta$, such that 
the main typical cases in the ordering of the masses $m_{\tchi_1^0}$, $m_{\tchi_1^{\pm}}$,
$m_{\tilde l}$, $m_{\tilde \nu}$, can be explored. The results are shown in Figures
\ref{br1}, \ref{br2} and \ref{br3}, for the chargino, the
sneutrino and the slepton decays, respectively. The curves for the various branching ratios
are distinguished by the same letters (numbers) as those used in Tables \ref{tabloC}, 
\ref{tabloA} and \ref{tabloB} to label the various final states (decay processes).
We shall now discuss in turn the various superpartner decay schemes corresponding
 to the five single production reactions.

\subsubsection{Lowest mass Neutralino}
\label{neutralinosection}

The branching ratios for the $\tchi_1^0$ desintegrations are best analysed separately.
For convenience, we do not treat the cases, $m_{\tchi_1^0}>m_{\tilde q}$ 
and  $m_{\tchi_1^0}>m_{\tchi_1^{\pm}}$, since these arise marginally 
in most of the currently favored models (supergravity or gauge mediated soft supersymmetry breaking).
The cascade decays which occur if $m_{\tchi_1^{\pm}}<m_{\tchi^0_1}$ are also not considered
since the corresponding region of the parameter space (Figure \ref{spec}(c)) is 
forbidden by the experimental constraints on the inos masses. Thus,
the process $l_J^+l_J^- \to \tchi_1^0 \nu_m$ will only generate events with 2 leptons +
 $\Eslash $. At this point, it is necessary to specialize our discussion to a single 
dominant coupling constant hypothesis, assuming $\l_{ijk} \neq 0$ not necessarily
 identical to $\l_{mJJ}$. One may distinguish the following four distinct cases.
For an LSP $\tchi_1^0$, namely, $m_{\tchi_1^0}<m_{\tilde l}, 
m_{\tilde \nu}$ (Case 1), only the direct RPV three-body decays, $\tchi_1^0 \to
\bar \nu_i \bar l_j  l_k$, $\tchi_1^0 \to \nu_i l_j  \bar l_k$, are allowed.
The branching ratios are then determined on the basis of simple combinatoric arguments.
For a dominant coupling constant, say, $\l_{m11}$, there are four final states:
 $\nu_1 l^-_m e^+$, $\bar \nu_1 l^+_m e^-$, $\nu_m e^- e^+$ and $\bar \nu_m e^+ e^-$. 
Accordingly, the branching ratios of $\tchi_1^0$ into two charged leptons  will depend on the type
(flavor,charge) of the final state: The branching ratios equal ${1 \over 2}$ for the flavor diagonal
$e^+e^-$ or flavor non diagonal $l^{\pm}e^{\mp}$ channels, ${1 \over 4}$ for the
fixed charges and flavors $l^+e^-$ or $l^-e^+$ channels and $1$ for the lepton-antilepton
 pairs of unspecified flavors.
For a dominant coupling constant $\l_{ijk} \neq \l_{m11}$, an analogous result is obtained.
For $m_{\tchi_1^0}>m_{\tilde l},m_{\tilde \nu}$ (Case 2), the branching ratio for $\tchi_1^0$ decay is,
\begin{eqnarray}
B(\tchi_1^0 \to \bar \nu_i \bar l_j  l_k)={\G (\tchi_1^0 \to \tilde l_j \bar l_j)
B(\tilde l_j \to \bar \nu_i l_k) + \G(\tchi_1^0 \to \tilde \nu_i \bar \nu_i) 
B(\tilde \nu_i \to \bar l_j l_k) 
\over
3\G (\tchi_1^0 \to \tilde l_j l_j) +3\G (\tchi_1^0 \to \tilde \nu_i \nu_i )}=
{1 \over 3},
\label{eqap3}
\end{eqnarray}
where we have used the fact that in the present case, assuming a dominant coupling constant $\l_{ijk}$,
 $B(\tilde l_j \to \bar \nu_i l_k)=B(\tilde \nu_i \to \bar l_j l_k)=1$. The factors $3$ 
in the denominator account for the number of families. For the 
intermediate case, $m_{\tilde \nu}>m_{\tchi_1^0}>m_{\tilde l}$ (Case 3), there occur 
contributions from 2-body RPC decays and 3-body RPV decays, such that:
\begin{eqnarray} 
B(\tchi_1^0 \to \bar \nu_i \bar l_j  l_k)={\G' (\tchi_1^0 \to \bar \nu_i \bar l_j l_k )
+ \G (\tchi_1^0 \to \tilde l_j \bar l_j)
B(\tilde l_j \to \bar \nu_i l_k) 
\over
\G' (\tchi_1^0 \to \bar \nu_i \bar l_j l_k ) + 3\G (\tchi_1^0 \to \tilde l_j l_j) } \simeq
{1 \over 3},
\label{eqap4}
\end{eqnarray} 
where the prime on $\G'$ is a reminder to indicate that the decay width includes only the 
contribution from a virtual sneutrino exchange. The approximate equality in eq.(\ref{eqap4})
 derives from the fact that $\G' (\tchi_1^0 \to \nu_i  \bar l_j l_k)<<\G 
(\tchi_1^0 \to \tilde l_j \bar l_j)$, based on the expectation that an RPV 3-body 
decay should be much smaller than an RPC 2-body decay. An analogous argument to 
that of case 3 holds 
for the other intermediate case, $m_{\tilde l}>m_{\tchi_1^0}>m_{\tilde \nu}$ (Case 4).
For the cases 2, 3 and 4, the multiplicity factors are the same as for the case 1.
The $\tchi_1^0$ process may occur at the end stage in the decays of $\tchi_1^-$, $\tilde l$ 
and $\tilde \nu,$ to be discussed below. The associated
$\tchi_1^0$ decay multiplicity factors
 for the two leptons final states will then take the same values 
 as quoted above for the various 
selection criteria. In quoting numerical results below, we shall, for convenience, 
assume the case of unspecified lepton flavor and charge and thus will set the
 multiplicity factors to unity.

\subsubsection{Lowest mass Chargino}
The results  in Fig. \ref{br1} for the  high $\tan \beta $ case 
show a high degree of symmetry with respect to  $\mu \leftrightarrow -\mu $,
which arises from the symmetry in
 the inos mass spectrum (Figure \ref{spec}(a)(b)).  
As can be seen from Figure \ref{br1}(a), a dominant mode for the chargino 
at high values of $\vert \mu \vert$ is the cascade decay, 
$\tchi^- \to \tchi^0 l^- \bar \nu$, since this  occurs via the two-body decay,
$\tchi^- \to l^- \tilde {\bar \nu}$ (C(7)). Indeed, for these high values of $\mu$,
one has $m_{\tilde {\bar \nu }}<m_{\tchi^-}$. This two-body decay competes with 
the other two-body decay E, $\tchi^- \to \tchi^0 W^-$, when the latter 
is kinematically allowed,
as is the case for $\mu < -200GeV$ in Figure \ref{br1}(d). The difference between the
values of the branching ratios C(7) and E is explained by the relative phase space 
factors of the associated rates. The RPV direct decays (A and B) are three-body decays
with small coupling constant and are thus suppressed. In the case, $m_{\tilde 
{\bar \nu}}<m_{\tchi^0}<m_{\tchi^{\pm}}$, the only open channel for the sneutrino is,
$\tilde {\bar \nu} \to l \bar l$, so that the dominant mode for the chargino decay
is the RPV decay B(4) (high values of $\vert \mu \vert$ in Figure \ref{br1}(b)).
Even for $m_{\tilde {\bar \nu}} \approx m_{\tchi^0}$, the channel B(4) is competitive
due to a small phase space ($\mu \simeq -100GeV$ in Figure \ref{br1}(a)). In this case, 
for $\l_{ijk}=0.1$, the direct RPV decay B(4) can become dominant (moderate negative 
values of $\mu$ in Figure \ref{br1}(c)). For small values 
of $\vert \mu \vert$, the difference between the two dominant leptonic (C) 
and hadronic (D) cascade decays is due to the flavor and 
color factors. We note also that in a small interval of $\mu$ near $\mu=0$,
$m_{\tchi^0_1}>m_{\tchi^{\pm}_1}$ (see Figure \ref{spec}), and consequently the 
only open channels are the direct RPV decays (Figure \ref{br1}(a)(c)(d)). 
In this region, the direct RPV decay A(1) is
negligible  because the branching ratio depends on $U_{11}$ which is small \cite{e:Pol}.
 In conclusion,
the highest branching ratios are associated with the cascade decays, C,D and E, except
 for the case
in which the sneutrino is the LSP, where they are associated with the RPV decays B. 
The range of $\mu$ for which the chargino $\tchi_1^-$ is the LSP is excluded by the 
experimental constraints on the inos masses  (see Figure \ref{spec}).

\subsubsection{Sneutrino}
We turn now our attention to the sneutrino decays. 
For high values of $\mu$, the cascade decay D has the highest probability
(Figure \ref{br2}(a)(d)) since the decay into chargino is either kinematically forbidden
or suppressed by a small phase space.
As for the chargino study, the RPV direct decay A is of course small except when 
the competitive channel is reduced by a small phase space factor ($\mu \approx -100GeV$ 
in Figure \ref{br2}(a)). In such a case, the RPV direct decay A may be important for values 
of $\l_{ijk}$ near $0.1$ (negative $\mu$ in Figure \ref{br2}(c)). 
 When the sneutrino is the LSP,
the RPV direct decay has a branching ratio equal to unity (Figure \ref{br2}(b)).
 For small $\vert \mu \vert$,
the decays E(6) and F(8) through charginos dominate the decay D through neutralinos.
 The reason is that for $\mu=0$, $\tchi^0_1$ is a pure higgsino, whose couplings are weak. 
In the so called higgsino limit, $\mu \to 0$ \cite{e:Pol}, the decays B and C(3)
 are small since they occur through the
$\tchi^-$ RPV direct decays. However, they have the highest probability if, $m_{\tchi^0_1}>
m_{\tchi^{\pm}_1}$ (Figure \ref{br2}(a)(c)(d)). The relation between the leptonic (E) and 
hadronic (F) cascade decays
can be explained in the light of the study on the chargino. We conclude that the 
cascade decays, B, D, E and F, are always the dominant modes, except when the 
sneutrino is the LSP.

\subsubsection{Slepton}

Finally, we concentrate on slepton decays. For high values of $\mu$, the cascade decays 
via charginos are reduced because of a small phase space ($\vert \mu \vert \approx 400GeV$ 
in Figure \ref{br3}(b) or for $\mu<0$ in Figures \ref{br3}(c)(d)) or even closed 
(for $\mu<0$ in Figure \ref{br3}(a)). In these cases,
the decay D via neutralinos dominates. Elsewhere, the decays via charginos have higher 
branching ratios (for $\mu>0$ in Figures \ref{br3}(a)(d))
since larger coupling constants are  involved. 
In the higgsino limit, the slepton cascade decay D via $\tchi^0_1$ is suppressed for the same 
reason as in the sneutrino study. 
The decay via $\tchi^-_1$ is then dominating. The interpretation of the difference
between the decays, B, C, E and F, via charginos is based on the specific behaviours of the chargino
 branching ratios which have already been described above. We see in Figure \ref{br3}(c), that for 
$\l_{ijk}=0.1$, the RPV direct decay A is still very reduced.
 This is due to the important phase space
for the slepton decay into neutralino. Lastly, a new phenomenon appears for the slepton case.
In the higgsino limit ($\mu \to 0$) at large $\tan \beta$, the matrix element $U_{11} \to 0$
which forces the vertex $\tilde l \ \tchi^{\pm} \ \nu $ (see eq.\ref{eqpr6} in Appendix
 \ref{secappa}) and the branching ratios for the cascade decays through the chargino to vanish \cite{e:Pol}.
 This is the explanation of the fact that for $\mu \simeq 0$, one observes a peak of the 
direct RPV decay branching fraction (Figure \ref{br3}(b)). Similar peaks are also observed
 at shifted $\mu <0$ for the low $\tan \beta$ cases (Figures \ref{br3}(a)(c)(d)). 
However this behaviour appears for ranges of the parameters which are 
forbidden by the bounds on the inos masses (Figure \ref{spec}).
 The conclusion is that the cascade decays have always the highest probability for the reason 
that the L-chirality slepton cannot be the LSP in generic supergravity models.

\subsubsection{Discussion}
In summary, we have learned that the general behaviour of branching ratios is mainly
determinated by the phase space and thus by the ordering of the supersymmetric particles 
masses. We have explored all the characteristic cases,
$m_{\tilde \nu}>m_{\tchi_1^-}>m_{\tchi_1^0}$,
$m_{\tchi_1^-}>m_{\tilde \nu}>m_{\tchi_1^0}$ and  $m_{\tchi_1^-}>m_{\tchi_1^0}>m_{\tilde \nu}$.
For high values of $m_0 $ lying above $M_2$, the sleptons would have masses greater
than the inos masses. We have not analysed this case since
one has then the same situation 
in the mass ordering as for the case of small values of
$\vert \mu \vert$ (except for large
enough values of $m_0$ where the on-shell $W^{\pm}$ production
can take place in $\tilde l$
and $\tilde \nu$ decays G). In this situation, as we have explained
above, the charginos principally decay into neutralinos,
while the sleptons and sneutrinos decay
into charginos. The main conclusion is that the cascade decays are the dominant modes except 
if the sneutrino is the LSP. In this case, the RPV decay, $\tchi_1^- \to l_i \tilde {\bar \nu}_i 
\to l_i l_j \bar l_k$, is dominant for the chargino decays, and the only open channel for 
the sneutrino is of course the direct RPV decay. Besides, for values of $\l_{ijk}$ higher 
than $0.05$, the RPV direct decay branching ratios can reach significant levels for the case where the 
cascade decays are suppressed due to small phase space factors.

The excitation of the second neutralino $\tchi_2^0$ deserves some attention since this may
 have in certain regions of the parameter space comparable, if not larger, production 
rates than 
the excitation of $\tchi_1^0$. Assuming that the direct RPV widths are small enough so 
that the decay chain is initiated by the RPC contributions, then the desintegration mode,
 $\tchi^0_2 \to (\tchi_1^0+l^+ l^-),(\tchi_1^0+\bar \nu \nu)$, will also yield 
2l+$\Eslash$ and 4l+$\Eslash$ final states, respectively, and the other
 desintegration modes, $\tchi^0_2 \to (\tchi_1^+ +l^- \bar \nu,\tchi_1^- +l^+ \nu),
(\tilde {\bar \nu} \nu,\tilde \nu \bar \nu),(\tilde l^{\pm} l^{\mp})$, will yield 
2l+$\Eslash$ and 4l+$\Eslash$ final states according to decay schemes similar to those
 given in Tables \ref{tabloA},\ref{tabloB},\ref{tabloC}. In our supergravity models,
 the $\tchi^0_2$ decay into $\tchi^{\pm}_1$ should be suppressed by a small phase space
 (Fig.\ref{spec}). To determine which of the decay modes, $\tchi^0_2 \to \tchi^0_1, \
 \tilde l$ or $ \tilde \nu$, leads to the dominant signal would require a detailed
 comparison of \brs at the initial as well as the subsequent stages.

Let us ask in what way would alternate hypotheses 
on the family dependence affect our conclusions.
 Especially regarding the multiplicities of final states, this is relevant for the cases,
$m_{\tchi_1^-}>m_{\tilde l},m_{\tilde \nu}$, where the chargino can  
cascade decay to on-shell
 sleptons or sneutrinos (A(2) and B(4) in Table \ref{tabloC}).
As we have emphasized in  the last paragraph of 
Section \ref{sectionD}, the chargino decays have a multiplicity of 2 
for three degenerate families of sleptons. 
For the case of two degenerate families, labeled by the indices, $m,n$, assuming a dominant RPV
coupling constant $\l_{ijk}$, the multiplicity equals 2 for $(m,n)=(i,j)$, 
since the two sleptons from families $i$ and $j$ 
can be produced on-shell, and equals  1 for $m=k$ or $n=k$.
For the physically interesting case of a single low mass family, labeled by the index, $m$,
one finds that the multiplicity equals 1 for $m \neq k$ and 0 otherwise. 
The conclusion is that the RPV contributions A and B (in Table \ref{tabloC}) to 
the chargino branching ratios increase as the number of slepton families, which are lower in mass 
than the chargino, becomes higher. 
This effect, which is quite small, would affect the branching ratios in parameters regions for which the RPV
contributions A and B are not weak, that is for $\mu<0$ in Figures \ref{br1}(a)(b)(c).

In Figure \ref{bratewsb}, we present results for the branching fractions for fixed $m_0$ in the 
infrared fixed point model with electroweak symmetry breaking. 
In this constrained version, 
where $m_{1/2}$ varies with $\mu$, the 
dependence on $\mu$ is rather similar to that of the non minimal 
model where we worked  instead with fixed
 $m_0$ and $m_{1/2}$.  However, as we see from the mass spectrum, here
the LSP is the neutralino $\tchi_1^0$ for all the physical ranges of the parameters.
Due to the large mass difference between the $\tchi_1^0$ LSP and the NLSP (next to LSP), 
the cascade decays are the only dominant modes and the branching ratios for the RPV direct decays are 
very weak.

Let us add a few qualitative remarks on the predictions of
 gauge mediated \susy breaking models. In order for the production rates in
 the minimal model \cite{e:Dimo97} to have the same order
 of magnitude as those obtained in the
 supergravity model of section \ref{sectionR}, one needs a parameter $\Lambda
 ={F\over M} \simeq 10^{4}GeV$, using familiar notations for the \susy
 breaking scale ($\sqrt F$) and messenger scale ($M$). 
Concerning the signals, by
 comparing the mean free paths for
 $\tchi^0_1$ (favourite candidate for LSP) in both models, one finds that the
 decay channel to the gravitino, $\tchi_1^0 \to \gamma \ \tilde G$,  becomes competitive
 with the RPV decay channel, $\tchi_1^0 \to  \nu l \bar l$,   for,
 $ {\sqrt {<F>} \over 100TeV} \leq { 10^{-2} \over \sqrt \l }$.

Let us also comment briefly on some of the experimental issues.
A given final state can possibly arise
 simultaneously from several of the single production processes.
 The important 4l+$\Eslash $ 
signal 
which occurs for $\tchi^{\pm},\tilde l^{\pm},\tilde \nu$ productions is one 
such example where one 
may be forced to add all three types of cross sections in comparing with some given
experimental data sample.
Similarly, for most signals, one must typically add the contributions 
from the two charge conjugate partner processes.
Concerning the competition with the standard model background, 
one expects that the most important contributions
to the final states, 2l+$\Eslash $ and 4l, will arise  from the reactions,  $l_J^+l_J^- \to
W^+l^-\bar \nu , \  W^-l^+\nu , \   W^+W^-,\   Z^0 l^+l^-,\   Z^0Z^0,\  Z^0 \gamma$. 
In spite of the large standard model rates of order one picobarn 
at $\sqrt s = 500 GeV$ \cite{e:DESY2},  one should be able to distinguish
the single production signals by exploiting their specific non diagonal
 flavor character (final state B in Table \ref{tabloB} and A in Table \ref{tabloA}). 
The other multileptons final states, 
generated by the cascade decays, 4l+$\Eslash $, 4l+$Z^0$,
 3l+$Z^0$+$W^{\pm}$+$\Eslash $,... have a standard model background which is negligible. 
The potentially large two photons background processes, induced by 
$\gamma \gamma$ photons pairs radiated by the initial
 leptons, can be significantly reduced by imposing suitable
 cuts on the leptons transverse momenta. Finally, we note that the selection by the
RPV single production of identical helicities for the initial state, $l^+_Hl^-_H$, 
can be exploited to discriminate against the minimal supersymetric standard model
and also the standard model, for which the identical helicities configuration only
appears with the t-channel Z-boson exchange.

\section{Dynamical distributions}
\label{sectionCD}

\setcounter{equation}{0}

\begin{figure}
\begin{center}
\leavevmode
\psfig{figure=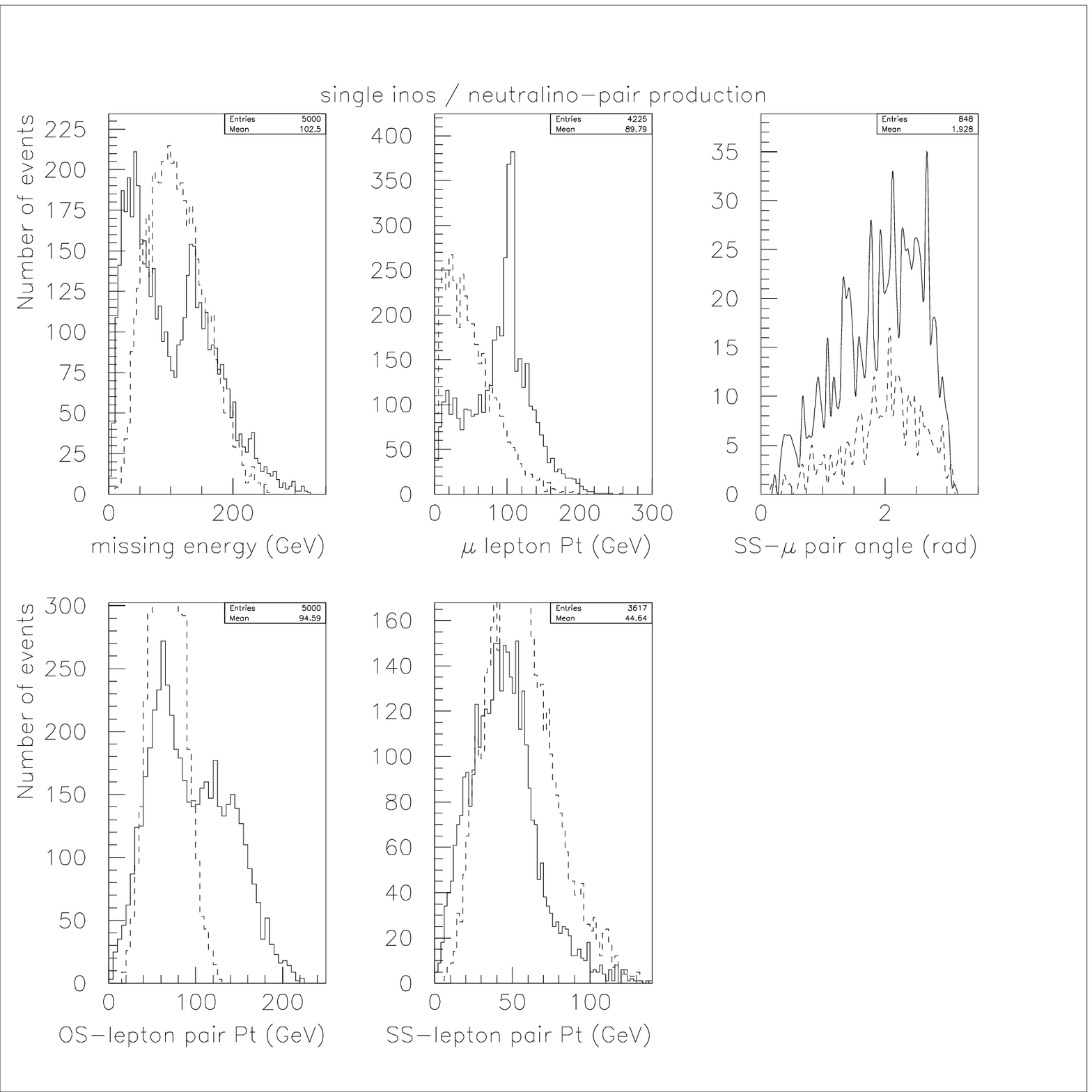,width=6.in}
\end{center}
\caption{\footnotesize  \it
Distributions of missing energy, muon transverse momentum, same sign muon pair angle,
summed transverse momentum for opposite sign (OS) and 
same sign (SS) leptons pairs 
(electrons and muons) for the single production processes, $l_J^+l_J^- \to \tchi_1^{\pm} \mu^{\mp},
\tchi_1^0 \nu_{\mu},\tchi_1^0 {\bar \nu}_{\mu},\tchi_2^0 \nu_{\mu},
\tchi_2^0 {\bar \nu}_{\mu}$ (solid line), and the pair production process,
 $l_J^+l_J^- \to \tchi_1^0 \tchi_1^0$ (dashed line), at a center of mass energy of $350GeV$. 
The parameters values  are, $M_2=250GeV, \ m_0=70GeV, \ \mu=400GeV, \ \tan \beta =2, \ \l_{211}=0.05$.
Events samples, consisting of 5000 events each, are generated for the inos single production 
and neutralino pair production, respectively. 
\rm \normalsize}
\label{MC}
\end{figure}

The distributions of rates with respect to kinematical variables 
associated with the final states offer
helpful means to characterize the underlying production processes. As an indicative study we shall 
present here some characteristic dynamical distributions obtained for the production reactions, 
$l_J^+l_J^- \to \tchi_1^{\pm}l^{\mp},\tchi^0_1 \nu,\tchi^0_1 \bar \nu,
\tchi^0_2 \nu,\tchi^0_2 \bar \nu$, from a Monte Carlo events simulation 
for which we have used the event generator SUSYGEN \cite{e:Kats1}.
We concentrate on the final state signals of 2l+$\Eslash $,  4l and 4l+$\Eslash $.
Note that for high values of $\mu$, the final state 4l+$\Eslash $ is the dominant mode for the
 chargino, slepton and sneutrino decays. This signal also receives contributions from the reactions,
 $l_J^+l_J^- \to \tilde \nu Z^0 (\tilde {\bar \nu} Z^0),\tilde \nu \gamma (\tilde {\bar \nu}
 \gamma),\tilde l^{\pm} W^{\mp}$, which however are not included in the simulation. 
The standard model background is expected to be small for the 4l+$\Eslash $ signal. The main background 
from the minimal supersymmetric standard model interactions arises from the neutralino RPC pair production, 
$l_J^+l_J^- \to \tchi^0_1 \tchi^0_1$.  Following the analysis in \cite{e:FUJII},
we consider an incident energy of $\sqrt s =350GeV$ and use a non minimal supergravity model
 for which we choose the set of parameters, $M_2=250GeV,\mu=400GeV,m_0=70GeV,\tan \beta=2$,
 which yields the spectrum,
 $m_{\tchi_1^0}=118.5GeV,m_{\tchi_2^0}=221.4GeV,m_{\tchi^{\pm}_1}=219.1GeV,
m_{\tilde \nu_L}=225GeV,m_{\tilde l_L}=233GeV,
m_{\tilde l_R}=141GeV$. The integrated rates (ignoring acceptance cuts) are,
 for $\l_{mJJ}=0.05 \ [m=2]$,
 $\sigma (\tchi^+_1 \mu^-)=30.9fb,\sigma (\tchi^0_1 \nu_{\mu})=4.8fb,
\sigma (\tchi^0_2 \nu_{\mu})=12.1fb$ and $\sigma (\tchi^0_1 \tchi^0_1)=238.9fb$.
 We consider the following five dynamical variables for  all types of final states:
Invariant missing energy, 
$E_m= \sum_{i \in \nu} E_{i}$ where the sum is over the neutrinos, 
as appropriate to a broken
 R parity situation; Average per event of the $\mu^{\pm}$
 lepton transverse momentum, $P_t(\mu^{\pm})={\sum_i \vert p_t(\mu^{\pm}_i) \vert \over N_{\mu}}$,
where $N_{\mu}$ is the number of muons; Angle between the
momenta of same electric charge sign (SS) muons pairs; Average per event of the summed transverse 
momenta for leptons pairs  of same sign (SS) or opposite sign (OS),  $P_t^{SS,OS}(ll)=\sum_{(i,j)}
 {p_t(l_i^{\pm,\mp})+p_t(l_j^{\pm}) \over N}$, where $N$ is the number of configurations and
 $l=e,\mu $.

We have generated the inos single  production and the  $\tchi^0_1$
 pair production in separate samples of 5000 events each. Our choice of
 using equal number of events  for both reactions
has been made on the basis of the following three 
somewhat qualitative considerations, none of which is compelling.
 First, the single production reactions occur in company of their charge conjugate partners, which multiplies rates by a factor 2. Second, the other $\tilde \nu$ 
and $\tilde l$ single production reactions, which have not 
been included, would be expected to add contributions of
similar size to the leptonic distributions. Third, assuming 
for the RPV coupling constant the alternative
bound, $\l _{mJJ} { 100 GeV \over m_{\tilde l_R}  } <0.05$, there would  result a 
relative enhancement for single production  over pair 
production by a factor, $ ({ m_{\tilde l_R} \over 100 GeV }) ^2$.
The above three points motivate our rough guess that 
the number of events  chosen for the five single production processes 
(together with their charge conjugate partners) 
should be of comparable size to that of the $\tchi^0_1$ pair production process.

The results are shown in Figure \ref{MC}. 
The single production reactions present 
certain clear  characteristic features: Concentration of 
missing energy $E_m$ at low energies; pronounced peaks in the muon transverse 
momentum $P_t(\mu^{\pm})$ 
and in the angular distribution for the same sign muons pairs;  and 
a double peak in the transverse 
momentum distribution for the opposite sign leptons pairs,
$P^{OS}_t(ll)$. The large transverse 
momentum components  present in the single production distributions in
$P^{OS}_t(ll)$ and in $P_t(\mu^{\pm})$
 are explained by the fact that
 one of the two  leptons (namely, $l^{\pm}_m$)  is created at the production stage. Similarly, the existence of a strong
 angular correlation between same sign muons pairs is interpreted naturally
 by the momentum conservation balance between the 
 lepton $l^{\pm}_m$ produced in the initial stage and 
the other lepton produced at the decay stage. 
Although there are certain distinguishing properties
 between the single and pair production processes, the discrimination between the two may depend
 crucially on the relative sizes of the associated event samples. 
 Of course, the best  possible situation would
 be for an energetically forbidden neutralino pair production.

Finally, we comment on the effect of eventually excluding the $\tchi^0_2$ single
production component. In that case, most of the signals for single production would 
become less diluted in comparison with neutralino pair production, the large missing 
energy signal would be removed while the large OS lepton pair $P^{OS}_t(ll)$ signal
would become amplified.

\section{Conclusions}
\label{sectionC}

\setcounter{equation}{0}

We have analysed the full set of $2 \to 2$ single production processes at leptonic colliders 
induced by the RPV interactions $LLE^c$, within a supergravity model. Although our
 approximate study has obvious limitations (factorisation and narrow resonance approximation,
 neglect of the spin correlations and omission of acceptance cuts), it uncovers the general
 trends of all the 5 single production reactions. Over the whole 
parameter space, for an RPV coupling constant $\l_{mJJ}$ of order $0.05$, 
the integrated rates are of comparable order
 of magnitudes although $\tchi_1^{\pm}$
 and $\tilde \nu$ are typically larger by factors of 2
 compared to $\tilde l$ production and   by 
factors of 5 compared to $\tchi_1^0$ production.
 The detectability for each single production
 separately is modest at LEPII  but comfortable at 
NLC, corresponding to a few events and a few thousands of events
 per year, respectively. 
A wide region of the parameter space
can be probed  at  $\l_{mJJ} > 0.05$. 
 In spite of the rich variety of final states, 
the dominant signals arise from the single
 or double RPC induced cascade decays to the LSP, which are 
also favored by phase space arguments. 
For the minimal supergravity model, assuming electroweak symmetry breaking,
 large mass differences occur between the scalar
 superpartners and the $\tchi_1^0$ LSP, leading to dominant 
cascade decays modes with weak competitivity from
 the RPV direct decays. The signals, 4l+$\Eslash $,6l,6l+$\Eslash $,
 arising from cascade decays are 
free from standard model background which make them quite interesting signatures 
for the discovery of supersymmetry and 
R parity violation. Even the 4l signal arising from direct RPV decays
 could be observable due to a
characteristic non diagonal flavor configuration. For center of mass 
energies well above all the
 thresholds, the 4l+$\Eslash $ signal receives contributions from
 all five single production processes and hence should
 be strongly amplified. We have presented some dynamical distributions for 
the final states which could characterise the single production reactions.

\section{Acknowledgments}

\setcounter{equation}{0}

We are grateful to R. Barbier and S. Katsanevas for 
guidance in using the events generator SUSYGEN,
to P. Lutz and the Direction of the DAPNIA at Saclay  
for providing  us, through the 
DELPHI Collaboration,  an  access to  the ANASTASIE 
computer network at Lyon and  to P. Micout for his technical 
help on  that matter. 
We thank M. Besan\c con, F. Ledroit, R. Lopez and 
G. Sajot for helpful discussions. 

\clearpage

\appendix

\setcounter{subsection}{0}
\setcounter{equation}{0}

\section{Formulas for spin summed amplitudes}
\label{seca}

We  discuss the five $2 \to 2$ body single production processes 
given by eq.(\ref{eqrc1}).  
The  formulas for the probability amplitudes  are:
\begin{eqnarray}
M(\tchi_a^-+ l^+_m)&=& {g\l_{mJJ} V^\star _{a1} \over s-m^2_{\tilde \nu_{mL}} } 
\bar v(k') P_L u(k) \bar u^c (p)P_Lv(p') 
- {g\l_{mJJ}V^\star _{a1}\over t-m^2_{\tilde \nu_{JL} } } 
\bar u^c (p) P_L u(k) \bar v(k') P_L v(p') ,\cr 
M(\tchi_a^0+ \bar \nu_m)&=& + { \sqrt 2 g\l_{mJJ} \over 
s-m^2_{\tilde \nu _{m L} }  } 
{1 \over 2}(N^*_{a2}-tg\t_W N^*_{a1})\bar u(p) P_L v(p') \bar v(k') P_L u(k)\cr
&&+{ \sqrt 2 g\l_{mJJ} \over t-m^2_{\tilde l_{JL} }   } 
{1 \over 2}(N^*_{a2}+tg\t_W N^*_{a1})\bar u(p) P_L u(k) \bar v(k') P_L v(p')\cr
&&+{ \sqrt 2 g\l_{mJJ} \over u-m^2_{\tilde l_{JR} }   } 
(tg\t_W N^\star _{a2})\bar v^c(p') P_L u(k) \bar v(k') P_L v(p),\cr
M(\tilde l_{mL}^-(p) +W^+(p'))&=&  { g \l^*_{mJJ} \over 
\sqrt 2 (s-m^2_{\tilde \nu_{m L} } )  } 2 p\cdot \e (p') 
 \bar v(k') P_R u(k) 
+ {g\l_{mJJ}^\star  \over \sqrt 2 t} \bar v(k') 
\g \cdot \e (p') (\pslash-\kslash ) P_R u(k) , \cr
M(\tilde \nu_{mL}(p) + Z(p'))&=&   {g \l^*_{mJJ} \over 2 \cos \t_W }
[{\bar v(k') \g \cdot \e (p') (\kslash-\pslash )  a_L(e) P_R u(k)
\over t-m^2_{ l_J } } \cr
&&+ {\bar v(k')a_R(e) P_R (\kslash-\pslash ')
 \g \cdot \e (p') u(k)\over u-m^2_{ l_J } } 
+ {\bar v(k')a_L(\tilde \nu) P_R u(k) 2p \cdot \e (p')
\over s-m^2_{ \tilde \nu_{mL} } } ], \cr
M(\tilde \nu_{mL}(p) + \g(p'))&=&- e \l^*_{mJJ} [{\bar v(k')\g \cdot \e (p')
(\kslash-\pslash ) P_R u(k) \over t-m^2_{ l_J } }  
+{\bar v(k') (\kslash - \pslash ' ) \g \cdot \e (p') P_R u(k) 
\over u-m^2_{ l_J } }]. \cr & & 
\end{eqnarray}
In deriving the results for the inos  production  amplitudes,
we have systematically neglected their higgsino components.
The parameters in the $Z^0 f {\bar f}$ and $Z^0 \tilde f \tilde {\bar f}$ vertices  
denoted as, $a_H(f)=a(f_H)$ and $a_H(\tilde f)=a(\tilde f_H)$, 
are defined by, $a(f_H)=a(\tilde f_H)=2T_3^H(f)-2Qx_W$, 
with $H=[L,R]$ and $x_W=\sin ^2 \t_W$. Throughout this work, 
our notations follow closely the Haber-Kane conventions \cite{e:haber}.

The  unpolarized cross sections in the 
center of mass frame are given by  the familiar formula, 
$d\s /d\cos \t =p /(128 \pi k s) \sum_{pol} \vert M\vert^2 $,  where the sums over 
polarizations  for the probability amplitudes squared are given by:

\begin{eqnarray}
\sum_{pol} &&\vert M(\tchi^-_a +l_m^+ )  \vert^2 = 
 \vert \l_{mJJ} g V_{a1}^\star \vert^2  
\bigg [ { s(s-m_{\tchi^-_a}^2-m_{l_m}^2)  \over \vert R_s(\tilde \nu_{mL} )  \vert^2}
+ { (m_{\tchi^-_a}^2-t)(m_{l_m}^2-t) 
\over \vert R_t(\tilde \nu_{JL} ) \vert^2} \cr &-& 
Re \bigg ( { (s(s-m_{\tchi^-_a}^2-m_{l_m}^2) +
(m_{\tchi^-_a}^2-t)(m_{l_m}^2-t) - (m_{\tchi^-_a}^2-u)(m_{l_m}^2-u))  
\over R_s(\tilde \nu_{mL} ) R^\star _t 
(\tilde \nu_{JL}) } \bigg ) \bigg ], \cr & &
\label{eqrc4a}
\end{eqnarray}

\begin{eqnarray}
\sum_{pol} &&\vert M(\tchi^0_a +\bar \nu_m )  
\vert^2 = { g^2 \over 2 } \vert \l_{mJJ}\vert ^2 
 \bigg [ \vert N_{a2}+tg\t_W N_{a1} \vert ^2 {t(t-m_{\tchi^0_a}^2) \over 
\vert R_t(\tilde l_{JL} )\vert^2 }     \cr
&+&4 \vert tg\t_W N_{a2} \vert ^2 { u(u-m_{\tchi^0_a}^2) \over 
\vert R_u(\tilde l_{JR} )\vert^2 }
+ \vert N_{a2}-tg\t_W N_{a1} \vert ^2 { s(s-m_{\tchi^0_a}^2) \over 
\vert R_s(\tilde \nu_{mL})\vert^2  }  \cr
&-& Re \bigg ( (N^*_{a2}-tg\t_W N^*_{a1})(-N_{a2}-tg\t_W N_{a1}) {
(s(s-m_{\tchi^0_a}^2) -t
(m_{\tchi^0_a}^2-t) +u (m_{\tchi^0_a}^2-u))
 \over 
 R_s(\tilde \nu_{mL}) R^*_t(\tilde l_{JL} )} \cr
&+& 2  (N^*_{a2}-tg\t_W N^*_{a1})(-tg\t_W N_{a2}) {
(s(s-m_{\tchi^0_a}^2)-u(m_{\tchi^0_a}^2-u)+t(m_{\tchi^0_a}^2-t))
 \over 
R_s(\tilde \nu_{mL})R^*_u(\tilde l_{JR}) } \cr
&+&2  (-N^*_{a2}-tg\t_W N^*_{a1})(-tg\t_W N_{a2}) {
(-u(m_{\tchi^0_a}^2-u)-t(m_{\tchi^0_a}^2-t)-s(s-m_{\tchi^0_a}^2))
 \over 
R_t(\tilde l_{JL} )R^*_u(\tilde l_{JR}) } \bigg ) \bigg ], \cr & &
\label{eqrc4b}
\end{eqnarray}

\begin{eqnarray}
\sum_{pol} &&\vert M(\tilde l^-_{mL} +W^+)\vert ^2 =
{s g^2 \vert \l_{mJJ}\vert ^2  \over 2 \vert R_s(\tilde \nu_{mL}) \vert ^2 } 
  ( { (s-m^2_{\tilde {l}^-_{mL}}-m_W^2)^2\over m_W^2} -4 m^2_{\tilde {l}^-_{mL}} ) 
- {g^2 \vert \l_{mJJ}\vert ^2  \over  2 \vert t \vert ^2 }\cr & \times & 
[(m^2_{\tilde {l}^-_{mL}}-t)(m_W^2-t) +st  
+{m_W^2-t\over m_W^2} 
\bigg ( (m^2_{\tilde {l}^-_{mL}}-t)(m_W^2+t) + t(m_W^2-u)\bigg )  ] \cr
&-& g^2 Re {\l_{mJJ} \l ^\star _{mJJ}  \over t R^*_s(\tilde \nu_{mL})   } 
[(m^2_{\tilde {l}^-_{mL}}-t)(m^2_{\tilde {l}^-_{mL}}-u) +s (m^2_{\tilde {l}^-_{mL}}-u)+ 
 (m^2_{\tilde {l}^-_{mL}}-t)(m^2_W-t)\cr
&-& {s(s-m^2_W-m^2_{\tilde {l}^-_{mL}})(m^2_W-t)\over m_W^2} ],
\label{eqrc4c}
\end{eqnarray}

\begin{eqnarray}
\sum_{pol} &&\vert M(\tilde \nu_{mL} + Z)\vert ^2 ={g^2 \vert \l_{mJJ}\vert ^2
 \over  \cos^2\t_W } Re \bigg [{s \over \vert R_s(\tilde \nu_{mL})\vert^2 }   
( { (s-m_{\tilde \nu_{mL}}^2-m_Z^2)^2\over 4 m_Z^2} -m^2_{\tilde \nu_{mL}} ) \cr
&-&{( \sin^2\t_W)^2 \over  \vert R_u(l_J)\vert^2}
\bigg ( (m_{\tilde \nu_{mL}}^2-u)(m_Z^2-u) + su 
+{ m_Z^2-u \over m_Z^2} 
 ( (m_{\tilde \nu_{mL}}^2-u)(m_Z^2+t) + u (m_Z^2-t)  ) \bigg ) \cr & - &
{(2 \sin^2\t_W-1)^2 \over 4 \vert R_t(l_J)\vert^2 } \bigg (  (m_{\tilde \nu_{mL}}^2-t)(m_Z^2-t) + st+{m_Z^2-t \over m_Z^2} 
 ( (m_{\tilde \nu_{mL}}^2-t)(m_Z^2+t) + t (m_Z^2-u) )    \bigg ) \cr  
&-&  { ( \sin^2 \t_W) \over   R_u(l_J)  R^*_s(\tilde \nu_{mL}) } 
\bigg ( (m_{\tilde \nu_{mL}}^2-t)(m_{\tilde \nu_{mL}}^2-u)+s (m_{\tilde \nu_{mL}}^2-t)   
+ (m_{\tilde \nu_{mL}}^2-u)(m_Z^2-u) \cr
&-&{s\over m_Z^2}(s-m_{\tilde \nu_{mL}}^2-m_Z^2)(m_Z^2-u) 
\bigg ) 
+ {(2 \sin^2\t_W-1)\over  2 R_t(l_J) R^*_s(\tilde \nu_{mL})}
 \bigg ( (m_{\tilde \nu_{mL}}^2-t)(m_{\tilde \nu_{mL}}^2-u) +s(m_{\tilde \nu_{mL}}^2-u) \cr
&+& 
(m_Z^2-t)(m_{\tilde \nu_{mL}}^2-t)  -{s\over m_Z^2} 
(m_Z^2-t)(s-m_{\tilde \nu_{mL}}^2-m_Z^2) \bigg )  
+ {(2 \sin^2\t_W-1)( \sin^2\t_W)\over R_t(l_J) R^*_u(l_J)} \cr && \times
\bigg ( (m_{\tilde \nu_{mL}}^2-u)(m_{\tilde \nu_{mL}}^2-t) + s \ m_{\tilde \nu_{mL}}^2 
-{1 \over M_Z^2}(-{(s-m_{\tilde \nu_{mL}}^2-m_Z^2)\over 2}((m_{\tilde \nu_{mL}}^2-u)(m_Z^2-u)
 \cr &+& (m_{\tilde \nu_{mL}}^2-t)(m_Z^2-t)
- s (s-m_{\tilde \nu_{mL}}^2-m_Z^2))+(m_Z^2-u)(m_Z^2-t)m_{\tilde \nu_{mL}}^2)
\bigg )  \bigg ],
\label{eqrc4d}
\end{eqnarray}

\begin{eqnarray}
\sum_{pol} &&\vert M(\tilde \nu_{mL} + \g)\vert ^2 =
 2 e^2\vert \l_{mJJ}\vert ^2((m_{\tilde \nu_{mL}}^2-t)(m_{\tilde \nu_{mL}}^2-u)-sm_{\nu_m}^2)
[{1 \over \vert R_t(l_J) \vert ^2}
+{1 \over \vert R_u(l_J) \vert ^2}] \cr
&+& 4 e^2 \vert \l_{mJJ} \vert ^2 Re { (m_{\tilde \nu_{mL}}^2-t)
(m_{\tilde \nu_{mL}}^2-u) \over R_t(l_J) R_u^*(l_J)},
\label{eqrc4e}
\end{eqnarray}

where $ Re $ stands for the real part,  $ R_s(\tilde \nu_i) =s-m^2_{\tilde \nu_{i} }
+i m_{\tilde \nu} \Gamma_{\tilde \nu}, \ R_t(\tilde \nu_i ) =t-m^2_{\tilde \nu_{i} }$  
and  $R_u(\tilde \nu_i ) =u-m^2_{\tilde \nu_{i} },   
\ [s= (k+k')^2, \ t=(k-p)^2, \ u= (k-p')^2] $, with similar 
definitions applying for the propagator factors $R_{s,t,u}(l_i,\tilde l_i)$.

\clearpage

\section{Formulas for partial decay widths}
\label{secappa}

\setcounter{subsection}{0}
\setcounter{equation}{0}

The formulas for the various two-body decay widths are quoted below. 

\begin{eqnarray}
\G (\tilde \nu \to \tchi^+_a +l^-)= {g^2\over 16 \pi }
\vert V_{a1}\vert ^2 m_{\tilde \nu } 
\bigg (1-{m^2_{\tchi_a} \over m^2_{\tilde \nu } }\bigg )^2
\label{eqpr4}
\end{eqnarray}
\begin{eqnarray}
\G (\tilde \nu \to \tchi_a^0 +\nu )= {g^2\over 32 \pi }
\vert N_{a2} -N_{a1} \tan \t_W \vert ^2 m_{\tilde \nu } 
\bigg (1-{m^2_{\tchi_a} \over m^2_{\tilde \nu } }\bigg )^2
\label{eqpr5}
\end{eqnarray}
\begin{eqnarray}
\G (\tilde l^+_L \to \tchi^+_a \bar \nu)= {g^2\over 16 \pi }
\vert U_{a1}\vert ^2 m_{\tilde l_L } 
\bigg (1-{m^2_{\tchi_a} \over m^2_{\tilde l_L } }\bigg )^2
\label{eqpr6}
\end{eqnarray}
\begin{eqnarray}
\G (\tilde l^-_{[L,R]} \to \tchi_a^0 +l^- )= {g^2\over 32 \pi }
[\vert N_{a2} +N_{a1} \tan \t_W \vert ^2, \vert N_{a2} \tan \t_W \vert^2] 
m_{\tilde l_H} 
\bigg (1-{m^2_{\tchi_a} \over m^2_{\tilde l_H } }\bigg )^2
\label{eqpr7}
\end{eqnarray}

\begin{eqnarray}
\G (\tilde \nu_i (M) \to l^-_k (m_1) +l^+_j (m_2)) 
&=&\G (\tilde l^-_{jL} (M) \to \bar \nu_i (m_1) +l^-_k (m_2)) \cr 
&=&\G (\tilde l^-_{kR} (M) \to \nu_i (m_1) +l^-_j (m_2)) \cr 
&=& {\vert \l _{ijk}\vert^2 \over 8 \pi } 
k (1-{m_1^2+m_2^2\over M^2 } ) 
\label{eqpr8}
\end{eqnarray}
\begin{eqnarray}
\G (\tchi^{\pm}_m (M_{\pm}) \to \tchi_l^0  (M_0) +W^{\pm} (m_W) )&=& {g^2 \vert k \vert \over  
16\pi  M_{\pm}^2} \bigg [(\vert O_L\vert^2 + \vert O_R\vert^2 ) 
\bigg ( (M_+^2+M_0^2-m_W^2) \cr && +{1\over m_W^2} (M_{\pm}^2-M_0^2-m_W^2) 
(M_{\pm}^2-M_0^2+m_W^2) \bigg ) \cr && -
12  M_0 M_{\pm} Re\bigg (O_L O_R^\star \bigg )  \bigg ]
\label{eqpr9}
\end{eqnarray}
\begin{eqnarray}
\G (\tchi^0_a\to \tilde f_{[L,R]} \bar f' )&=& {g^2 M_0\over  16\pi }
(1-{m^2_{\tilde f} \over M^2_0})^2 \cr &&
[\vert T^f_3 N_{a2} - \tan \t_W (T^f_3 - Q^f) N_{a1} \vert ^2 , 
\vert \tan \t_W  N_{a2} \vert ^2 ]
\label{eqpr10}
\end{eqnarray}
\begin{eqnarray}
\G (\tchi^{\pm}_a\to \tilde f_{[T^f_3=-1/2, 1/2]} \bar f' )&=& 
{g^2 M_{\pm}\over 32 \pi }
(1-{m^2_{\tilde f} \over M^2_{\pm}})^2 [\vert U_{a1} \vert ^2, 
\vert V_{a1} \vert ^2].
\label{eqpr11}
\end{eqnarray}

We use the notations: $ O^L=O^L_{lm}= N_{l2} V_{m1}^\star -{1\over \sqrt 2} 
N_{l4} V_{m2}^\star, \ O^R=O^R_{lm}= N_{l2} U_{m1} +
{1\over \sqrt 2} N_{l3}^\star  U_{m2}, \ M_{\pm}=m_{\tchi^{\pm}_a}, \
M_0=m_{\tchi^0_a}$ and $ k= \l^\ud (M^2, m_1^2,m_2^2)/2M$ 
with $\l(a,b,c)=a^2+b^2+c^2+ab+bc+ac$. The notations, $T_3^f, \ Q^f$, stand for the third 
component of the $SU(2)_L$ group and the electric charge of the fermion $f$.
We have omitted the higgsino components of the inos.
We shall use the simplified formulas for the RPC three-body decays, 
$ \tchi_m^-\to \tchi_l^0 +l\bar \nu , \ q \bar q  $,  obtained by neglecting the 
three-momenta in the W and $\tilde l$ propagators, as  quoted in \cite{e:fe}. We have set
 in these formulas, the flavor and color parameters to, $N_f=2, \ N_c=3$ for quarks and
 $N_f=3, \ N_c=1$ for leptons.
The formulas for the spin summed amplitudes of the 
RPV decays   $\tchi^-_a\to \bar \nu_i \bar \nu_j l_k^- , \
\tchi^-_a \to  l_k^+ l_j^- l_i^- $, associated to the coupling constants $\l_{ijk}$, were first 
derived in  \cite{e:fourjet1} (see the appendix). The integrated decay rates are given by 
familiar formulas \cite{e:PartData} involving twofold integrals over the
final state three-body phase space. 
If we neglect the final particles masses, an analytic formula can be 
derived for the integral giving the contributions to the charginos partial 
rates associated with the gauginos components only 
(neglecting the higgsino components contribution). For completeness, we display 
the final results:

\begin{eqnarray}
\G (\tchi^-_a)&=&M_{\tchi^-_a} {g^2 X^2_{a1} \vert \l _{ijk}\vert^2 \over 128 \pi^3} \bigg
[{1 \over8 } \bigg (-5+6\mu_i+(2-8 \mu_i+6 \mu^2_i) \log (1-{1 \over \mu_i}) \cr 
&-&5+6\mu_j+(2-8 \mu_j+6 \mu^2_j) \log (1-{1 \over \mu_j}) \bigg ) \cr
&+& {1 \over 2 } \bigg (  \mu_i + \mu_j -{1 \over 2} 
+(\mu^2_i- \mu_i) \log (1-{1 \over \mu_i})+(\mu^2_j- \mu_j) \log (1-{1 \over \mu_j}) \cr
&-& \mu_i \mu_j \log (1-{1 \over \mu_i}) \log ({ \mu_i +\mu_j-1\over \mu_j }) \cr
&-& \mu_i \mu_j \log (1-{1 \over \mu_j}) \log ({ \mu_i +\mu_j-1\over \mu_i }) \cr
&+& \mu_i \mu_j  [ Sp({ \mu_i\over \mu_j})+Sp({ \mu_j\over \mu_i})
- Sp({ 1-\mu_i\over \mu_j}) - Sp({ 1-\mu_j\over \mu_i}) ]\bigg ) \bigg ],
\label{eqpr111}
\end{eqnarray}

where $Sp(x)=Polylog(x)=Li_2(x)$ is the Spence or Polylog function. We use the notations 
$\mu_{\a}= m^2_{\tilde \nu_{\a}}/M^2_{\tchi^-_a}, \ [\a =i,j] $, $X_{a1}=U_{a1}$ for the decay
$\tchi^-_a\to  l_k^+ l_j^- l_i^-$, and  $\mu_{\a}= m^2_{\tilde l_{\a}}/ M^2_{\tchi^-_a}, 
 \ [\a =i,j] $, $X_{a1}=V_{a1}$ for the decay $\tchi^-_a\to  \bar \nu_i \bar \nu_j l_k^- $.

\clearpage

\vspace{10 mm}
\begin{center}
{  }
\end{center}
\vspace{10 mm}

\clearpage

\setcounter{chapter}{0}
\setcounter{section}{0}
\setcounter{subsection}{0}
\setcounter{figure}{0}

\chapter*{Publication VI}
\addcontentsline{toc}{chapter}{Single chargino production at linear colliders}

\newpage

\vspace{10 mm}
\begin{center}
{  }
\end{center}
\vspace{10 mm}

\clearpage

\begin{center}
{\bf \huge Single chargino production at linear colliders}
\end{center} 
\vspace{2cm}
\begin{center}
G. Moreau
\end{center} 
\begin{center}
{\em  Service de Physique Th\'eorique \\} 
{ \em  CE-Saclay F-91191 Gif-sur-Yvette, Cedex France \\}
\end{center} 
\vspace{1cm}
\begin{center}
{Linear Collider note LC-TH-2000-040, hep-ph/0009140}
\end{center}
\vspace{2cm}
\begin{center}
Abstract  
\end{center}
\vspace{1cm}
{\it
We study the single chargino production 
$e^+ e^- \to \tilde \chi^{\pm} \mu^{\mp}$  
at linear colliders which occurs through the
$\l_{121}$ R-parity violating coupling constant. 
We focus on the final state containing 4 leptons and some missing energy. 
The largest background is \susyq and 
can be reduced using the initial beam polarization and
some cuts based on the specific kinematics of the single chargino production.
Assuming the highest allowed supersymmetric background,
a center of mass energy of $\sqrt s=500GeV$ 
and a luminosity of ${\cal L}=500fb^{-1}$, the sensitivities on the
$\l_{121}$ coupling constant obtained from the single chargino production study
improve the low-energy experimental limit over a range of $\Delta m_{\tilde \nu}
\approx 500GeV$ around the sneutrino resonance, and reach values of
$\sim 10^{-4}$ at the $\tilde \nu$ pole.
The single chargino production also allows 
to reconstruct the $\tilde \chi_1^{\pm}$,
$\tilde \chi_2^{\pm}$ and $\tilde \nu$ masses.
The initial state radiation plays a fundamental role in this study.}

\newpage

\section{Introduction}
\label{intro}

\setcounter{equation}{0}

In \susyq theories, there is no clear theoretical argument 
in favor of the conservation of the so-called 
R-parity symmetry, either from the point of view of 
grand unified models, string theories
or scenarios with discrete gauge symmetries \cite{f:Drein}.
The phenomenology of \susy (SUSY) at futur colliders would change
fundamentally if the R-parity symmetry were violated. 
Indeed, in such a scenario the
typical missing energy signature caused by the stable nature of the Lightest
Supersymmetric Particle (LSP) would be replaced by multijet or
multileptonic signals, depending on what are the dominant R-parity violating
(\rpv) couplings. The reason is that
the \rpv terms of the superpotential trilinear in the
quarks and leptons superfields (see Eq.(\ref{super})) allow the LSP,
whatever it is, to decay.
\begin{eqnarray}
W_{\rpv}=\sum_{i,j,k} \bigg (\ud \l _{ijk} L_iL_j E^c_k+
\l ' _{ijk} L_i Q_j D^c_k+ \ud \l '' _{ijk} U_i^cD_j^cD_k^c   \bigg ).
\label{super}
\end{eqnarray}
The effects of the \rpv decays of the LSP on the study of 
SUSY particles pair production have been considered in
the context of linear colliders \cite{f:Godbole}
and futur hadronic colliders, namely the Tevatron (Run II) 
\cite{f:Barg,f:tat1,f:tat2,f:runII} and the LHC \cite{f:Atlas}.

The measure of the \rpv coupling constants of Eq.(\ref{super}) could be 
performed via the detection of the displaced vertex associated to the decay
of the LSP. 
The sensitivities on the \rpv couplings obtained through this method
depend on the detector geometry and performances.
Let us estimate the largest values of 
the \rpv coupling constants that can be measured 
via the displaced vertex analysis. We suppose that the LSP
is the lightest neutralino ($\tilde \chi^0_1$). 
The flight length of the LSP in the laboratory frame
is then given in meters by \cite{f:Dreinoss}, 
\begin{eqnarray}
 c \gamma \tau \sim 3 \gamma \ 10^{-3} m ({\tilde m \over 100GeV })^4
       ({1GeV \over m_{LSP}})^5 ({1 \over \L})^2,
\label{Fleng}
\end{eqnarray}
where $\L=\l$, $\l'$ or $\l''$, 
$c$ is the light speed, $\gamma$ the Lorentz boost factor,
$\tau$ the LSP life time, $m_{LSP}$ the LSP mass and  $\tilde m$ the mass 
of the supersymmetric scalar particle 
involved in the three-body decay of the LSP.
Since the displaced vertex analysis is an experimental challenge at
hadronic colliders, we consider here the linear colliders.
Assuming that the minimum distance between two vertex necessary to distinguish 
them experimentally is of order $2 \ 10^{-5}m$ at linear colliders, we see from
Eq.(\ref{Fleng}) that the \rpv couplings could be measured up to the values,   
\begin{eqnarray}
 \L < 1.2 \ 10^{-4} \gamma^{1/2} ({\tilde m \over 100GeV })^2
       ({100GeV \over m_{LSP}})^{5/2}.
\label{LSP}
\end{eqnarray}
There is a gap between these values and the low-energy 
experimental constraints on the \rpv
couplings which range typically in the interval $\L <10^{-1}-10^{-2}$ 
for superpartners masses of $100GeV$ \cite{f:Drein,f:Bhatt,f:GDR}.
However, the domain lying between these low-energy bounds and the
values of Eq.(\ref{LSP}) can be tested through another way: The study of the
production of either \sm or SUSY particles involving \rpv couplings.
Indeed, the cross sections of such productions are directly proportional
to a power of the relevant \rpv coupling constant(s), which allows
to determine the values of the \rpv couplings.
Therefore, there exists a complementarity between the displaced vertex analysis
and the study of reactions involving \rpv couplings, since these
two methods allow to
investigate different ranges of values of the \rpv coupling constants.

The studies of the \rpv contributions to \sm particle productions
have been performed both at leptonic \cite{f:Dim1}-\cite{f:Deba} and hadronic
\cite{f:Dim2}-\cite{f:Rizz} colliders.
Those contributions generally involve two \rpv vertex and have thus some
rates proportional to $\Lambda^4$. The processes involving 
only one \rpv vertex are less suppressed since 
the \rpv couplings incur stringent experimental limits \cite{f:Drein,f:Bhatt,f:GDR}.
Those reactions correspond to the single production of \susyq particle.
Another interest of the single superpartner production 
is the possibility to produce
SUSY particles at lower energies than through the superpartner pair production, 
which is the favored reaction in R-parity conserved models.

At hadronic colliders, the single superpartner 
production involves either $\l'$ or $\l''$
\ccs \cite{f:Dreinoss,f:Dim2,f:Rich,f:Berg,f:greg,f:Rich2,f:Gia1,f:Gia2}. 
The test of the $\l$ couplings via the single superpartner 
production can only be performed  at leptonic colliders.
At leptonic colliders, either gaugino (not including gluino) 
\cite{f:Han,f:Workshop,f:lola,f:chemsin} or slepton (charged or
neutral) \cite{f:chemsin} can be singly produced in the simple $e^+ e^- \to
2-body$ reactions.
The single production of slepton has a reduced phase space, since the 
slepton is
produced together with a $Z$ or $W$ gauge boson. In contrast,
the $\tilde \chi^0_1$ is
produced together with a neutrino.
Nevertheless, if the $\tilde \chi^0_1$ is the LSP, as in
many \susyq scenarios, it undergoes an \rpv decay which reads as
$\tilde \chi^0_1 \to l \bar l \nu$ if one assumes a dominant $\l$ Yukawa
coupling. Therefore, the single $\tilde \chi^0_1$ production 
leads typically to the $2l + \Eslash$ final state which has a large
\sm background. In this paper, we focus on the single
chargino production
$e^+ e^- \to \tilde \chi^{\pm} l^{\mp}_m$, which occurs through
the \rpv \ccs $\l_{1m1}$ ($m=2$ or $3$).
There are several motivations. First, the single chargino production has an
higher
cross section than the single neutralino production \cite{f:chemsin}. Moreover, it
can lead to
particularly interesting multileptonic signatures due to the cascade decay
initiated by the chargino.

The single gaugino productions at $e^+ e^-$ colliders
have $t$ and $u$ channels
and can also receive a contribution from the resonant sneutrino production.
At $\sqrt s=200 GeV$, the off-resonance rates of the
single chargino and neutralino productions are typically of order $100 fb$
and $10 fb$, respectively \cite{f:chemsin},  for a
value of the relevant \rpv \cc equal to its low-enery bound for
$m_{\tilde e_R}=100GeV$: $\l_{1m1}=0.05$ \cite{f:Bhatt}. 
The off-pole effects of the
single gaugino production are thus at the limit of observability at LEP II
assuming an
integrated luminosity of ${\cal L } \approx 200 pb^{-1}$. At the sneutrino
resonance, the single gaugino production has higher cross section values. 
For instance with $\l_{1m1}=0.01$, the chargino production
rate can reach $2 \ 10^{-1} pb$ at the resonance \cite{f:Workshop}. This
is the reason why the experimental analysis of the single gaugino 
production at the LEP collider 
\cite{f:ALEPHa,f:DELPHIa,f:Fab1,f:Fab2,f:DELPHIb} allows
to test \rpv couplings values smaller than the low-energy bounds only at the
sneutrino resonance $\sqrt s = m_{\tilde \nu}$ and, due to the
Initial State Radiation (ISR) effect, in a range of typically $\sim 50GeV$
around the $\tilde \nu$ pole. The sensitivities on the $\l_{1m1}$
couplings obtained at LEP reach values of order $10^{-3}$ at the
sneutrino resonance.

The experimental analysis of the single gaugino production at futur
linear colliders should be interesting due to 
the high luminosities and energies expected at these futur colliders
\cite{f:Tesla,f:NLC}. However, the single gaugino production might suffer
a large SUSY background at linear colliders. Indeed, due to the
high energies reached at these colliders, the pair productions 
of SUSY particles may have important cross sections.

In this article, we study the single chargino production via the 
$\l_{121}$ coupling at linear colliders and we consider the final 
state containing 4 leptons plus some missing energy ($\Eslash$). 
We show that the SUSY background can be
greatly reduced with respect to the signal.
This discrimination is based on the two following points:
First, the SUSY background can be suppressed by making use of the beam
polarization capability of the linear colliders.
Secondly, the specific kinematics of the single
chargino production reaction allows to put some efficient
cuts on the transverse momentum of the lepton produced together with the
chargino. We find that, by consequence of this background reduction,
the sensitivity on the $\l_{121}$ coupling obtained at the $\tilde \nu$ resonance
at linear colliders 
for $\sqrt s=500GeV$ and ${\cal L}=500 fb^{-1}$
\cite{f:Tesla} would be of order $10^{-4}$, namely one order of 
magnitude better than the results of the LEP analysis 
\cite{f:ALEPHa,f:DELPHIa,f:Fab1,f:Fab2,f:DELPHIb},
assuming the largest supersymmetric background allowed 
by the experimental limits on the SUSY masses.

Besides, in the scenario of a single dominant \rpv coupling of type $\l$
with $\tilde \chi^0_1$ as the LSP,
the experimental superpartner mass reconstruction
from the SUSY particle pair production
suffers an high combinatorial background
both at linear colliders \cite{f:Godbole} and LHC \cite{f:Atlas}.
The reason is that all the masses reconstructions are based on
the $\tilde \chi^0_1$ reconstruction which is degraded by the
imperfect identification of the charged leptons generated in the decays of the 2
neutralinos as $\tilde \chi^0_1 \to l \bar l \nu$, and the
presence of missing energy in the final state.
In particular, the chargino mass reconstruction in the leptonic channel
is difficult since the presence of an additional neutrino, coming from
the decay $\tilde \chi^{\pm} \to \tilde \chi^0 l \nu$,
renders the control on the missing energy more challenging.
In this paper, we show that
through the study of the $4l+\Eslash$ final state, the specific
kinematics of the single chargino production reaction
at linear colliders allows to determine
the $\tilde \chi^{\pm}_{1,2}$ and $\tilde \nu$ masses.

In Section \ref{frame} we define the theoretical framework.
In Sections \ref{signal} and \ref{background} we describe the signal
and the several backgrounds.
In Section \ref{An}, we present the sensitivity on the SUSY parameters
that can be obtained in an $e^+ e^-$ machine
allowing an initial beam polarization.
In this last section, we also show how some information on the SUSY mass spectrum
can be derived from the study of the single chargino production,
and we comment on another kind of signature: the $3l+2jets+\Eslash$ final state.

\section{Theoretical framework}
\label{frame}

\setcounter{equation}{0}

We work within the
Minimal Supersymmetric Standard Model (MSSM) which has the
minimal particle content and the
minimal gauge group $SU(3)_C \times SU(2)_L \times U(1)_Y$,
namely the \sm gauge symmetry.
The \susyq parameters defined at the electroweak scale are
the Higgsino mixing parameter $\mu$,
the ratio of the vacuum expectation values of the two Higgs doublet
fields $\tan \beta=<~H_u~>~/~<~H_d~>$ and the soft SUSY breaking parameters,
namely the bino ($\tilde B$) mass $M_1$,
the wino ($\tilde W$) mass $M_2$, the gluino ($\tilde g$) mass $M_3$,
the sfermion masses $m_{\tilde f}$ and the trilinear Yukawa couplings $A$.
The remaining three
parameters $m_{H_u}^2$, $m_{H_d}^2$ and the
soft SUSY breaking bilinear coupling $B$ are
determined through the electroweak symmetry breaking conditions,
which are two necessary minimization conditions of the Higgs potential.

We assume that all phases in the soft SUSY breaking potential are equal to
zero in order to eliminate all new sources of CP violation which are
constrained by extremely tight experimental limits on the electron and
neutron electric moments. Furthermore, to avoid any problem of
Flavor Changing Neutral Currents, we take the matrices in flavor space
of the sfermion masses and $A$ couplings
close to the unit matrix. In particular, for simplification reason
we consider vanishing $A$ couplings.
This last assumption concerns the splitting
between the Left and Right sfermions masses
and does not affect our analysis which depends mainly on the relative
values of the sleptons, squarks and gauginos masses as we will discuss
in next sections.

Besides, we suppose the R-parity symmetry to be violated so that
the interactions written in the superpotential of Eq.(\ref{super})
are present. 
The existence of a hierarchy among the values of the \rpv couplings
is suggested by the analogy of those couplings
with the Yukawa couplings.
We thus assume a single dominant \rpv coupling of the type $\l_{1m1}$
which allows the single chargino production at $e^+e^-$ colliders.

Finally, we choose the LSP to be the $\tilde\chi^0_1$ neutralino
since it is often real in many models, such as the supergravity
inspired models.

\section{Signal}
\label{signal}

\setcounter{equation}{0}

\begin{figure}[t]
\begin{center}
\leavevmode
\centerline{\psfig{figure=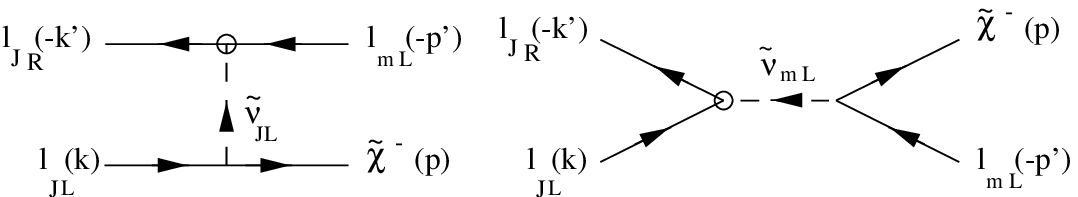,width=6.in}}
\end{center}
\caption{\footnotesize  \it
Feynman diagrams of the single chargino production process at
leptonic colliders.
The \rpv coupling constant $\l_{JmJ}$
($J=1$, $m=2 \ or \ 3$ for $e^+ e^-$ colliders and
$J=2$, $m=1 \ or \ 3$ for $\mu^+ \mu^-$ colliders)
is symbolised by a small circle,
and the arrows denote the flow of momentum of the corresponding particles.
The Left and Right helicities of the particles are denoted by L and R,
respectively, and the momentum of the particles by k,k',p and p'.
The higgsino contribution as well as the charge conjugated process
are not represented.
\rm \normalsize }
\label{graphe}
\end{figure}

The single chargino production $e^+ e^- \to \tilde \chi^{\pm} l^{\mp}_m$
occurs via the $\l_{1m1}$ coupling constant 
either through the exchange of a $\tilde \nu_{mL}$ sneutrino in the $s$
channel or a $\tilde \nu_{eL}$ sneutrino in the $t$ channel (see
Fig.\ref{graphe}).
Due to the antisymmetry of the $\l_{ijk}$ Yukawa couplings in the $i$ and
$j$ indices,
only a muon or a tau-lepton can be produced together with the chargino,
namely $m=2$ or $3$.
The produced chargino can decay into the LSP as
$\tilde \chi^{\pm} \to \tilde \chi^0_1  l \nu$
or as $\tilde \chi^{\pm} \to \tilde \chi^0_1 u d$.
Then the LSP decays as $ \tilde \chi_1^{0} \to  \bar e e \nu_m$,
$\bar e e \bar \nu_m$, $l_m \bar e \bar \nu_e$ 
or $ \bar l_m e \nu_e $ through $\l_{1m1}$. 
Each of these 4 $ \tilde \chi_1^{0}$ decay channels   
has a branching ratio exactly equal to $25\%$.
In this paper,
we study the $4l + \Eslash$ final state generated by the decay
$\tilde \chi^{\pm} \to \tilde \chi^0_1  l \nu$ of the produced chargino.

The single chargino production has a specific feature which
is particularly attractive: Because of 
the simple kinematics of $2 \to 2-body$ type reaction,
the energy of the lepton produced with the chargino, which we write $E(l_m)$,
is completely fixed by the center of mass energy $\sqrt s$,
the lepton mass $m_{l_m}$ and the chargino
mass $m_{\tilde \chi^{\pm}}$:
\begin{eqnarray}
E(l_m)= {s + m^2_{l_m} - m^2_{\tilde \chi^{\pm}} \over 2 \sqrt s}.
\label{enl}
\end{eqnarray}
The lepton momentum $P(l_m)$, which is related to the lepton energy by 
$P(l_m)=(E(l_m)^2-m^2_{l_m}c^4)^{1/2}/c$, is thus also fixed. 
Therefore, the momentum distribution 
of the produced lepton is peaked and offers the opportunity to
put some cuts allowing an effective discrimination between the
signal and the several backgrounds. Besides, the momentum value of the
produced lepton gives the $\tilde \chi^{\pm}$ mass through Eq.(\ref{enl}).
Nevertheless, for a significant ISR effect the radiated photon 
carries a non negligible energy and the single chargino production 
must be treated as a three-body reaction of the type 
$e^+ e^- \to \tilde \chi^{\pm} l^{\mp}_m \gamma$. Therefore,
the energy of the produced lepton is not fixed anymore by the SUSY parameters.
As we will show in Section \ref{kin}, in this situation the transverse
momentum of the produced lepton is more appropriate to apply some cuts
and reconstruct the $\tilde \chi^{\pm}$ mass.
\\ In the case where the produced lepton is a tau, the momentum of the lepton 
($e^{\pm}$ or $\mu^{\pm}$) coming from the $\tau$-decay is of
course not peaked at a given value. Moreover, 
in contrast with the muon momentum, the precise determination
of the tau-lepton momentum is difficult experimentally  
due to the unstable nature of the $\tau$.
From this point of view, the case of a single dominant $\l_{121}$
coupling constant is a more promising scenario
than the situation in which $\l_{131}$ is the dominant coupling.
Hence, from now on we focus on the single dominant \rpv coupling
$\l_{121}$ and consider the single chargino production 
$e^+ e^- \to \tilde \chi^{\pm} \mu^{\mp}$.

At this stage, an important remark can be made concerning the
initial state.
As can be seen from Fig.\ref{graphe}, in the single $\tilde \chi^-$
production
the initial electron and positron have the same helicity, namely, they are
both Left-handed.
Similarly in the $\tilde \chi^+$ production, the electron and the positron
are
both Right-handed. The incoming lepton and anti-lepton 
have thus in any case the same helicity.
This is due to the particular structure of the trilinear \rpv couplings
which flips the chirality of the fermion. 
The incoming electron and positron could also have identical helicities.
However, this contribution involves the higgsino component of the chargino
and is thus suppressed by the coupling of the higgsino to leptons which
is proportional to $m_l / (m_W \ \cos \beta )$.
This same helicity property, which is characteristic
of all type of single superpartner
production at leptonic colliders \cite{f:chemsin},
allows to increase the single chargino production rate by selecting
same helicities initial leptons.

\section{Background}
\label{background}

\setcounter{equation}{0}

\subsection{Non-physic background}
\label{NPback}

\begin{figure}[t]
\begin{center}
\leavevmode
\centerline{\psfig{figure=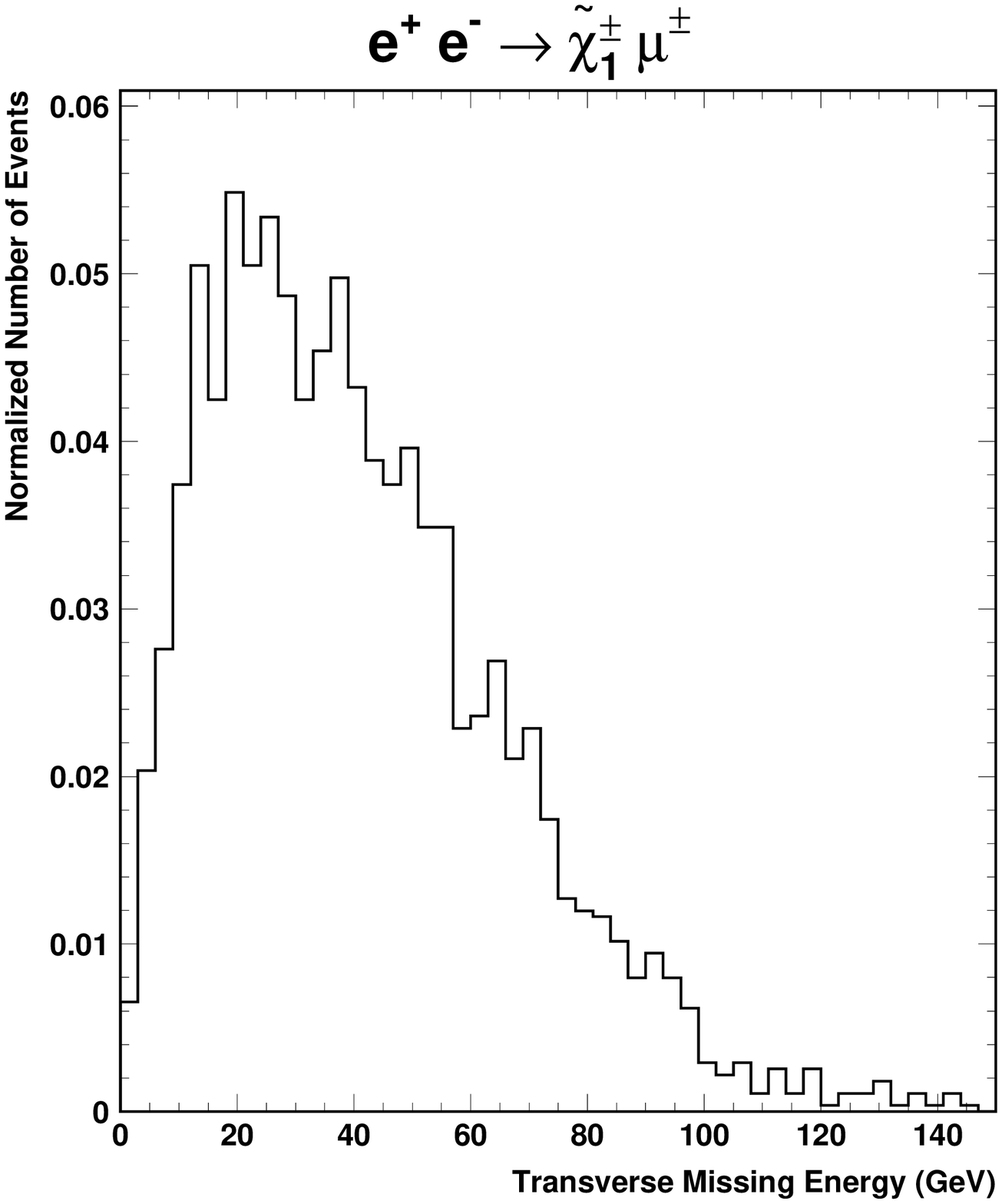,height=5.5in}}
\end{center}
\caption{\footnotesize  \it
Distribution of the transverse missing energy (in $GeV$)
in the $4l+\Eslash$ events generated
by the single $\tilde \chi^{\pm}_1$ production 
at a center of mass energy of $500GeV$ and
for the point A of the SUSY parameter
space with $m_{\tilde \nu}=240GeV$.
The number of events has been normalized to unity.
\rm \normalsize }
\label{distrib}
\end{figure}

First, one has to consider the non-physic background for the $4l +
\Eslash$ signature.
The main source of such a background is the $Z$-boson pair production with
ISR.
Indeed, the ISR photons have an high probability
to be colinear
to the beam axis. They will thus often be missed experimentally becoming
then
a source of missing energy. The four leptons can come from the leptonic
decays of
the 2 $Z$-bosons.
This background can be greatly reduced by excluding the same flavor
opposite-sign dileptons
which have an invariant mass in the range, $10GeV < \vert M_{inv}(l_p \bar
l_p)- M_Z \vert$, $p=1,2,3$ being the generation indice.
Furthermore, the missing energy coming from the ISR is mainly
present at small angles
with respect to the beam axis. This point can be exploited to perform a
better selection of
the signal. As a matter of fact, the missing energy
coming from the signal has a significant transverse component since it
is caused by the presence of neutrinos in the final state.
In Fig.\ref{distrib}, we present the distribution of the transverse
missing energy
in the $4l +\Eslash$ events generated by the single $\tilde \chi^{\pm}_1$
production for a sneutrino mass of $m_{\tilde \nu} = 240 GeV$ and for the 
point A of the SUSY parameter space defined as:
$M_1=200GeV$, $M_2=250GeV$, $\mu=150GeV$, $\tan \beta=3$,
$m_{\tilde l^{\pm}}=300GeV$ and $m_{\tilde q}=600GeV$
($m_{\tilde \chi^{\pm}_1}=115.7GeV$, $m_{\tilde
\chi^0_1}=101.9GeV$, $m_{\tilde \chi^0_2}=154.5GeV$).
The cut on the lepton momentum mentioned in Section \ref{signal} can also
enhance the signal-to-background ratio. Finally, the beam
polarization may be useful in reducing this source of background. 
Indeed, the initial leptons in the
$ZZ$ production process have opposite helicities. The reason is that the
$Z-f-\bar f$
vertex conserves the fermion chirality. Furthermore, recall that in our signal, the
initial electron and positron have similar helicities. Thus if the beam
polarization at
linear collider were chosen in order to favor opposite helicities initial states,
the signal would be enhanced while 
the $ZZ$ background would be suppressed. Since the polarization
expected at linear colliders is of order
$85 \%$ for the electron and $60 \%$ for the positron \cite{f:Tesla}, the
$ZZ$ background would not be entirely eliminated by the beam
polarization effect.

\subsection{\sm background}
\label{SMback}

The \sm backgrounds for the $4l + \Eslash$ signal come from $2 \to 3-body$
processes.
For example, the $e^+ e^- \to Z^0 Z^0 Z^0$ reaction can give rise to the
four leptons plus
missing energy signature. Another source of \sm background is the $e^+ e^-
\to Z^0 W^+ W^-$ reaction.
The rates for those backgrounds have been calculated in \cite{f:Godbole} after
the cuts on the charged lepton rapidity, $\vert \eta (l)\vert<3$,
the charged lepton transverse momentum, $P_t(l)>10GeV$, and
the missing transverse energy, $\Eslash_t>20GeV$, have been applied.
The results are the followings.
The $Z Z Z$ production cross section does not
exceed $1fb$. The $Z W W$ production has a cross section
of $0.4fb$ ($0.1fb$) at $\sqrt s=500GeV$ ($350GeV$) after convoluting
with the branching fractions for the leptonic decays of the gauge bosons.
This rate value is obtained without the ISR effect which should reduce the
cross section of the $Z W W$ production dominated
by the $t$ channel exchange contribution \cite{f:Godbole}.
As before, the $Z W W$ production can be reduced using simultaneously
the cut on the lepton momentum indicated in Section \ref{signal},
the dilepton invariant mass cut and the beam polarization effect. Indeed, 
the initial leptons have opposite helicities in both the $s$ and $t$ channels
of the process $e^+ e^- \to Z^0 W^+ W^-$, as in the $ZZ$ production process.

\subsection{Supersymmetric background}
\label{SUSYB}

The main sources of \susyq background to the $4l+
\Eslash$ signature are the neutralinos and sneutrinos pair productions.

First, the reaction,
\begin{eqnarray}
e^+ e^- \to \tilde \chi^0_1 + \tilde \chi^0_1,
\label{chi1}
\end{eqnarray}
represents a background for the $4l+\Eslash$ signature since
the $\tilde \chi^0_1$, which is the LSP in our
framework, decays as $ \tilde \chi_1^{0} \to  \bar e e \nu_{\mu}$,
$\bar e e \bar \nu_{\mu}$, $\mu \bar e \bar \nu_e$ 
or $ \bar \mu e \nu_e $ 
through the $\l_{121}$ coupling. Therefore, the
$\tilde \chi^0_i \tilde \chi^0_j$  productions
(where $i=1,...,4$ and $j=1,...,4$ are not equal to $1$ simultaneously)
leading through cascade decays
to a pair of $\tilde \chi^0_1$ accompanied by some
neutrinos give also rise to the final state
with 4 leptons plus some missing energy.
These reactions are the following,
\begin{eqnarray}
e^+ e^- \to \tilde \chi^0_i + \tilde \chi^0_1 \to \tilde \chi^0_1 \nu_p \bar
\nu_p + \tilde \chi^0_1,
\label{chi2}
\end{eqnarray}
\begin{eqnarray}
e^+ e^- \to \tilde \chi^0_i + \tilde \chi^0_j \to \tilde \chi^0_1 \nu_p \bar
\nu_p + \tilde \chi^0_1 \nu_{p'} \bar \nu_{p'},
\label{chi3}
\end{eqnarray}
where $i,j=2,3,4$ and $p,p'=1,2,3$.
Besides, the pair productions
of neutralinos followed
by their \rpv decays through $\l_{121}$:
\begin{eqnarray}
e^+ e^- \to \tilde \chi^0_i + \tilde \chi^0_1 \to  l  \bar l  \nu  + \tilde
\chi^0_1,
\label{chrp1}
\end{eqnarray}
\begin{eqnarray}
e^+ e^- \to \tilde \chi^0_i + \tilde \chi^0_j \to  l  \bar l  \nu +  \tilde
\chi^0_1 \nu_p \bar
\nu_p,
\label{chrp2}
\end{eqnarray}
\begin{eqnarray}
e^+ e^- \to \tilde \chi^0_i + \tilde \chi^0_j \to  l  \bar l  \nu  + l  \bar
l  \nu,
\label{chrp3}
\end{eqnarray}
where $i,j=2,3,4$, are also a source of background.

Secondly, the $4l+\Eslash$ signature can arise
in the sneutrino pair production through the reactions,
\begin{eqnarray}
e^+ e^- \to \tilde \nu_p + \tilde \nu_p^* \to \tilde \chi^0_i \nu_p + \tilde
\chi^0_j \bar \nu_p,
\label{sneu}
\end{eqnarray}
where $i,j=1,...,4$, if the produced neutralinos decay as above,
namely either as 
$\tilde \chi^0_i \to \tilde \chi^0_1 \nu_p \bar \nu_p$ (if $i \neq 1$) or as 
$\tilde \chi^0_i \to l  \bar l  \nu$.

The neutralinos and sneutrinos pair productions also lead to the
$4l+\Eslash$ final state via more complex cascade decays such as
$\tilde \chi^0_i \to \tilde \chi^0_j \nu_p \bar \nu_p \to
\tilde \chi^0_1 \nu_{p'} \bar \nu_{p'}  \nu_p \bar \nu_p$ ($i,j=2,3,4$ and
$i>j$) or $\tilde \nu \to l^{\mp} \tilde \chi^{\pm}_1 \to l^{\mp} e^{\pm}
\nu_e \nu_{\mu}$
where the chargino decays through the $\l_{121}$ coupling.

At the high energies available at linear colliders \cite{f:Tesla,f:NLC},
the rates of the neutralinos and sneutrinos pair productions (see Section
\ref{SUSYX})
are typically quite larger than the cross sections of the \sm background
reactions (see Section \ref{SMback}).
We however note that in the regions of large $\tilde \chi^0_i$ and
$\tilde \nu_p$ masses
with respect to the center of mass energy $\sqrt s$,
the $\tilde \chi^0_i \tilde \chi^0_j$ and
$\tilde \nu_p \tilde \nu_p^*$ productions cross sections, respectively,
are
suppressed by the phase space factor.
In the kinematic domain $2m_{\tilde \chi^0_i},2m_{\tilde \nu_p}>\sqrt
s>m_{\tilde
\chi^{\pm}_1}>m_{\tilde \chi^0_1}$ ($i=1,...,4$ and $p=1,2,3$), all
the neutralinos and sneutrinos pair productions are even kinematically
closed while the
single chargino production remains possible.
In such a situation, the only background to the 
$4l+\Eslash$ final state would be the
\sm background so that the sensitivity on the $\lambda_{121}$ coupling would
be greatly improved. Nevertheless, the kinematic domain described above is
restrictive
and not particularly motivated. As a conclusion, except in some particular
kinematic domains,
the main background to the $4l+\Eslash$ final state is supersymmetric.\\
In Section \ref{An}, we will focus on the largest allowed
SUSY background.
Indeed, the various considered domains of the SUSY parameter
space will be chosen such that the $\tilde \chi^0_1$
mass is around the value
$m_{\tilde \chi^0_1} \approx 100GeV$ which is close to the experimental limit
$m_{\tilde \chi^0_1}>52GeV$ (for $\tan \beta =20$ and any $\l_{ijk}$
coupling)
\cite{f:DELPHIb}
in order to maximize the phase space factors of the 
$\tilde \chi^0_i \tilde
\chi^0_j$ productions and the decays $\tilde \nu_p \to 
\tilde \chi^0_i \nu_p$.

The SUSY background could be suppressed by the cut on the lepton momentum 
mentioned in Section \ref{signal}. 
The selection of initial leptons of same helicities 
would also reduce the SUSY background. Indeed,
the neutralino (sneutrino) pair production occurs through the
exchange of either a
$Z$-boson in the $s$ channel or a charged slepton (chargino) in the $t$
channel, and
the helicities of the initial electron and positron are opposite in all
those channels. The $t$ channels of the SUSY background processes 
allow however same helicity leptons in the initial state but this
contribution entails the higgsino component of the relevant gaugino
and is thus suppressed by powers of $m_l /( m_W \ \cos \beta )$.

\section{Analysis}
\label{An}

\setcounter{equation}{0}

In this part, we determine the sensitivity on the $\l_{121}$
\cc expected from the study of the single chargino production
$e^+ e^- \to \tilde \chi_1^{\pm} \mu^{\mp}$ based on
the analysis of the $4l+\Eslash$ final state 
at linear colliders, assuming a center of mass energy of 
$\sqrt s=500GeV$ and a luminosity of ${\cal L}=500fb^{-1}$.
For this purpose, we study the SUSY background and show how it
can be reduced with respect to the signal. At the end of this part, 
we also discuss the determination of the 
$\tilde \chi^{\pm}_1$, $\tilde \chi^{\pm}_2$ and $\tilde \nu$ masses,
the single chargino production at different center of mass energies
and via other \rpv couplings,
the $3l+2j+\Eslash$ final state and the single neutralino production.

We have simulated the signal and the \susyq background with
the new version of the SUSYGEN event generator \cite{f:SUSYGEN3}
including the beam polarization effects.

\subsection{Polarization}
\label{polar}

The signal-to-noise ratio could be enhanced by 
making use of the capability of the linear colliders to polarize the
inital beams. 

First, as we have seen in Section \ref{SUSYB} the SUSY
background would be reduced if initial
leptons of similar helicities were selected.  
Hence, we consider in this study the selection 
of the $e^-_L e^+_L$ initial state, 
namely Left-handed initial electron and positron.
In order to illustrate the effect of this 
polarization on the SUSY background, let us
consider for example the $\tilde \chi^0_1 \tilde \chi^0_2$ production.
At the point A of the SUSY parameter space, the unpolarized cross section
is $\sigma(e^+ e^- \to \tilde \chi^0_1 \tilde \chi^0_2)=153.99fb$
for a center of mass energy of $\sqrt s=500GeV$.
Selecting now the $e^-_L e^+_L$ initial beams 
and assuming a value of $85 \%$  ($60 \%$)
for the electron (positron) polarization, the
production rate becomes
$\sigma_{polar}(e^+ e^- \to \tilde \chi^0_1 \tilde \chi^0_2 )=82.56fb$.

Secondly, the selection of same helicities initial
leptons would increase the signal rate as mentioned in Section \ref{signal}.
We discuss here the effect of the considered
$e^-_L e^+_L$ beam polarization on the signal rate.
As explained in Section \ref{signal}, the single chargino production
occurs through the process
$e^-_L e^+_L \to \tilde \chi^-_1 \mu^+$ (see Fig.\ref{graphe}) or through
its charge conjugated process
$e^-_R e^+_R \to \tilde \chi^+_1 \mu^-$.
These charge conjugated processes have the same cross section:
$\sigma(e^-_L e^+_L \to \tilde \chi^-_1 \mu^+)=\sigma(e^-_R e^+_R \to \tilde
\chi^+_1 \mu^-)$.
The whole cross section is thus
$\sigma=[\sigma(e^-_L e^+_L \to \tilde \chi^-_1 \mu^+)
+\sigma(e^-_R e^+_R \to \tilde \chi^+_1 \mu^-)] / 4
=\sigma(e^-_L e^+_L \to \tilde \chi^-_1 \mu^+) / 2$.
The chosen beam polarization favoring Left-handed initial leptons,
only the process $e^-_L e^+_L \to \tilde \chi^-_1 \mu^+$
would be selected if the polarization were total.
The polarized cross section would then be $\sigma_{polar}^{total}=\sigma(e^-_L e^+_L
\to \tilde \chi^-_1 \mu^+)=2 \ \sigma$. Hence,
the signal would be increased by a factor of $2$
with respect to the whole cross section.
In fact, due to the limited expected efficiency of the beam polarization,
the chargino production can also receive a small contribution from the
process
$e^-_R e^+_R \to \tilde \chi^+_1 \mu^-$.
The polarized cross section is thus replaced by
$\sigma_{polar}=\sigma_{polar}(\tilde \chi^-_1 \mu^+)+\sigma_{polar}(\tilde
\chi^+_1 \mu^-)$
where $\sigma_{polar}(\tilde \chi^-_1 \mu^+)=P(e^-_L)P(e^+_L)\sigma(e^-_L
e^+_L \to \tilde \chi^-_1 \mu^+)$
and $\sigma_{polar}(\tilde \chi^+_1 \mu^-)=P(e^-_R)P(e^+_R)\sigma(e^-_R
e^+_R \to \tilde \chi^+_1 \mu^-)$,
$P(e^+_L)$ being for example the probability to have a Left-handed initial
positron.
Assuming a polarization efficiency of $85 \%$  ($60 \%$) for the electron
(positron)
and selecting Left-handed initial leptons,
the probabilities are $P(e^-_L)=(1+0.85) / 2=0.925$, $P(e^+_L)=(1+0.60)
/ 2=0.8$,
$P(e^-_R)=(1-0.85)/ 2=0.075$ and $P(e^+_R)=(1-0.60)/ 2=0.2$. Hence,
the polarized cross section reads as
$\sigma_{polar}=[P(e^-_L)P(e^+_L)+P(e^-_R)P(e^+_R)]\sigma(e^-_L e^+_L \to
\tilde \chi^-_1 \mu^+)=
0.755 \ \sigma(e^-_L e^+_L \to \tilde \chi^-_1 \mu^+)=1.51 \ \sigma$.
The signal rate is thus enhanced by a factor of $1.51$
with respect to the whole cross section if the 
$e^-_L e^+_L$ beam polarization is applied.

\subsection{Cross sections and branching ratios}

\subsubsection{Signal}
\label{signalX}

The single $\tilde \chi^{\pm}_1$ chargino production has a cross section
typically of order $10fb$ at $\sqrt s=500GeV$ for $\l_{121}=0.05$
\cite{f:chemsin}. 
The single $\tilde \chi^{\pm}_1$ production rate is reduced in 
the higgsino dominated region
$\vert \mu \vert \ll M_1,M_2$ where the $\tilde \chi^{\pm}_1$ is
dominated by its higgsino component, compared to the wino
dominated domain
$\vert \mu \vert \gg M_1,M_2$ in which the $\tilde \chi^{\pm}_1$ is
mainly composed by the charged higgsino \cite{f:chemsin}. Besides, 
the single $\tilde \chi^{\pm}_1$ production cross section 
depends weakly on the sign of the $\mu$ parameter 
at large values of $\tan \beta$. However, as $\tan \beta$ decreases
the rate increases (decreases) for $sign(\mu)>0$ ($<0$).
This evolution of the rate 
with the $\tan \beta$ and $sign(\mu)$ parameters is 
explained by the evolution of the $\tilde \chi^{\pm}_1$ mass 
in the SUSY parameter space \cite{f:chemsin}.
Finally, when the sneutrino mass approaches
the resonance ($m_{\tilde \nu} =\sqrt s$) by lower values,
the single $\tilde \chi^{\pm}_1$ production
cross section is considerably increased due to the ISR effect \cite{f:lola}.
For instance, at the point A of the SUSY parameter space, the rate is
equal to
$\sigma (e^+ e^- \to \tilde \chi_1^{\pm} \mu^{\mp})=353.90fb$ 
for a center of mass energy of
$\sqrt s=500GeV$ and a sneutrino mass of $m_{\tilde \nu} = 240 GeV$.
At the resonance, the single chargino production rate reaches high values.
For example at $m_{\tilde \nu} =\sqrt s=500GeV$, the cross section is  
$\sigma (e^+ e^- \to \tilde \chi_1^{\pm} \mu^{\mp})=30.236pb$ 
for the same MSSM point A.

Since we consider the $4l+ \Eslash$ final state and the chargino
width is neglected,
the single chargino production rate must be multiplied by
the branching ratio of the leptonic chargino decay
$B(\tilde \chi_1^{\pm} \to \tilde \chi^0_1  l_p \nu_p)$ ($p=1,2,3$).
The leptonic decay of the chargino is typically of order $30\%$
for $m_{\tilde \nu},m_{\tilde l^{\pm}},
m_{\tilde q}>m_{\tilde \chi^{\pm}_1}$.
This leptonic decay is suppressed compared to the hadronic decay
$\tilde \chi_1^{\pm} \to \tilde \chi^0_1 d_p u_{p'}$ ($p=1,2,3;p'=1,2$)
because of the color factor.
Indeed for $m_{\tilde \nu},m_{\tilde l^{\pm}},m_{\tilde q}
>m_{\tilde \chi^{\pm}_1}$
the hadronic decay is typically
$B(\tilde \chi_1^{\pm} \to \tilde \chi^0_1 d_p u_{p'}) \approx 70\%$
($p=1,2,3;p'=1,2$).
In the case where $m_{\tilde q}>m_{\tilde \chi^{\pm}_1}>
m_{\tilde \nu},m_{\tilde l^{\pm}}$,
the decay $\tilde \chi_1^{\pm} \to \tilde \chi^0_1  l_p \nu_p$
occurs through the two-body decays
$\tilde \chi_1^{\pm} \to \tilde \nu l^{\pm}$ and
$\tilde \chi_1^{\pm} \to \tilde l^{\pm} \nu$
and is thus the dominant channel. In such a scenario,
the branching ratio of the $4l+ \Eslash$ final state 
for the single $\tilde \chi_1^{\pm}$ production is close to
$100\%$. In contrast, for
$m_{\tilde \nu},m_{\tilde l^{\pm}}>m_{\tilde \chi^{\pm}_1}>
m_{\tilde q}$, the decay
$\tilde \chi_1^{\pm} \to \tilde \chi^0_1 d_p u_{p'}$ dominates,
as it occurs through the two-body decay
$\tilde \chi_1^{\pm} \to \tilde q q'$, and the signal is
negligible. The \rpv decays of the chargino via $\l_{121}$, 
$\tilde \chi_1^{\pm} \to e^{\pm} \mu^{\pm}
e^{\mp}, \ e^{\pm} \nu_e \nu_{\mu}$,
have generally negligible branching fractions due to the
small values of the \rpv coupling constants.
Nevertheless, in the case of nearly degenerate $\tilde \chi^{\pm}_1$
and $\tilde \chi^0_1$
masses, those \rpv decays can become dominant spoiling then
the $4l+ \Eslash$ signature.

\subsubsection{SUSY background}
\label{SUSYX}

In this part, we discuss the variations and the order of magnitude of the
cross sections and branching ratios on which the whole SUSY background
depends.

$\bullet$ {\bf Neutralino pair production}
The neutralinos pair productions represent a SUSY background
for the present study of the R-parity violation.
A detailed description of the neutralinos pair productions cross sections at
linear colliders
for an energy of $\sqrt s=500GeV$ has recently been performed in
\cite{f:Godbole}.
In order to consider in our analysis the main variations of the neutralinos
pair productions rates,
we have generated all the $\tilde \chi^0_i \tilde \chi^0_j$ productions
($i$ and $j$ both varying between $1$ and $4$)
at some points belonging to characteristic regions of the MSSM parameter
space.
The points chosen for the analysis respect the experimental limits
derived from the LEP data on the lightest chargino and neutralino masses,
$m_{\tilde \chi^0_1}>52GeV$ ($\tan \beta=20$)
and $m_{\tilde \chi^{\pm}_1}>94GeV$  \cite{f:DELPHIb},
as well as the excluded regions in the $\mu - M_2$ plane \cite{f:DELPHIb}.
Besides, since the neutralinos pair productions 
rates depend weakly on $\tan \beta$ and $sign(\mu)$ \cite{f:Godbole},
we have fixed those SUSY parameters at $\tan \beta=3$ and $sign(\mu)>0$.

We present now the characteristic domains of the MSSM parameter space
considered in our analysis. For each of those
domains, we will describe the behaviour of all the
$\tilde \chi^0_i \tilde \chi^0_j$
productions cross sections except the $\tilde \chi^0_i \tilde \chi^0_4$
productions ($i=1,...,4$).
This is justified by the fact that
at a center of mass energy of $500GeV$,
the dominant neutralinos productions are the $\tilde \chi^0_1 \tilde
\chi^0_1$,
$\tilde \chi^0_1 \tilde \chi^0_2$ and $\tilde \chi^0_2 \tilde \chi^0_2$
productions
in most parts of the SUSY parameter space.
We will also discuss in each of the considered regions
the values of the branching ratios $B(\tilde
\chi^0_2 \to \tilde \chi^0_1 \nu_p \bar
\nu_p)$ and $B(\tilde \chi^0_2 \to l \bar l \nu)$ 
(for $\l_{121}=0.05$) which determine the contribution of the
$\tilde \chi^0_1 \tilde \chi^0_2$ and $\tilde \chi^0_2 \tilde \chi^0_2$ productions 
to the $4l+ \Eslash$ signature.
The restriction of the discussion to the 
$\tilde \chi^0_2$ decays is justified by the hierarchy
mentioned above between the $\tilde \chi^0_i \tilde \chi^0_j$ productions rates,
and by the fact that the $\tilde \chi^0_3$ and $\tilde \chi^0_4$ cascade decays 
have many possible combinations,
due to the large $\tilde \chi^0_3$ and $\tilde \chi^0_4$ masses
with respect to the $\tilde \chi^0_1$ and $\tilde \chi^0_2$ masses, so that
the contributions of the $\tilde \chi^0_i \tilde \chi^0_j$ productions
(with $i$ or $j$ equal to $3$ or $4$) to the $4l+\Eslash$ signature
are suppressed by small branching ratio factors.
We will also not discuss the complex cascade decays mentioned in
Section \ref{SUSYB} since the associated branching ratios are typically
small.
However, all the $\tilde \chi^0_i \tilde \chi^0_j$ productions as well as
all the cascade decays of the four neutralinos
are taken into account in the analysis.

First, we consider the higgsino dominated region characterised by
$\vert \mu \vert \ll M_1,M_2$ where the $\tilde \chi^0_1$
and $\tilde \chi^0_2$ neutralinos are predominantly composed by the higgsinos.
In this higgsino region, due to the weak couplings of the higgsinos to
charged leptons,
the $\tilde \chi^0_1 \tilde \chi^0_i$
and $\tilde \chi^0_2 \tilde \chi^0_i$ productions ($i=1,2,3$) are reduced.
However,
the $\tilde \chi^0_1 \tilde \chi^0_2$ production rate reaches its larger
values in this domain because of the $Z \tilde \chi^0_i \tilde \chi^0_j$
coupling.
At $\sqrt s=500GeV$, the
$\tilde \chi^0_3 \tilde \chi^0_3$
production is suppressed by a small phase space factor (when
kinematically allowed).
For the MSSM point A which belongs to this higgsino region,
the cross sections including the beam polarization described in Section
\ref{polar} are
$\sigma(\tilde \chi^0_1 \tilde \chi^0_1)=0.29fb$,
$\sigma(\tilde \chi^0_1 \tilde \chi^0_2)=82.56fb$,
$\sigma(\tilde \chi^0_2 \tilde \chi^0_2)=0.11fb$,
$\sigma(\tilde \chi^0_1 \tilde \chi^0_3)=6.62fb$,
$\sigma(\tilde \chi^0_2 \tilde \chi^0_3)=6.31fb$,
$\sigma(\tilde \chi^0_3 \tilde \chi^0_3)=13.57fb$,
$\sigma(\tilde \chi^0_1 \tilde \chi^0_4)=1.11fb$
and $\sigma(\tilde \chi^0_2 \tilde \chi^0_4)=9.59fb$
for the kinematically open neutralinos productions at $\sqrt s=500GeV$.\\
The $\tilde \chi^0_1 \tilde \chi^0_2$ production gives rise to the $4l+
\Eslash$ final state
if the second lightest neutralino decays as $\tilde \chi^0_2 \to \tilde
\chi^0_1 \bar \nu_p \nu_p$.
In the higgsino region, due to the various decay channels of the $\tilde
\chi^0_2$, the branching ratio
$B(\tilde \chi^0_2 \to \tilde \chi^0_1 \bar \nu_p \nu_p)$ reaches values
only about $15 \%$.
At the point A of the MSSM parameter space, this branching is
$B(\tilde \chi^0_2 \to \tilde \chi^0_1 \bar \nu_p \nu_p)=15.8\%$
for $m_{\tilde \nu}=450GeV$.
For a $\tilde \nu$ lighter than the $\tilde \chi^0_2$, the branching
$B( \tilde \chi^0_2\to \tilde \chi^0_1 \bar \nu_p \nu_p)$
is enhanced as this decay occurs via
the two-body decay $\tilde \chi^0_2 \to \tilde \nu \nu$.
At the point A with $m_{\tilde \nu}=125GeV$, 
$B( \tilde \chi^0_2\to \tilde \chi^0_1 \bar \nu_p \nu_p)=62.3\%$.

Secondly, we consider the
wino dominated region $\vert \mu \vert \gg M_1,M_2$ where $\tilde \chi^0_1$ and
$\tilde \chi^0_2$ are primarily bino and wino, respectively.
In this wino region, the $\tilde \chi^0_3 \tilde \chi^0_i$ productions
are reduced since the
$\tilde \chi^0_3$ mass strongly increases with the absolute value of the
parameter
$\vert \mu \vert $. Therefore in the wino region, the main neutralinos
productions are
the $\tilde \chi^0_1 \tilde \chi^0_1$, $\tilde \chi^0_1 \tilde \chi^0_2$ and
$\tilde \chi^0_2 \tilde \chi^0_2$
productions. In this part of the MSSM parameter space
and for $m_{\tilde l^{\pm}}=300GeV$, the $\tilde \chi^0_1
\tilde \chi^0_1$
production can reach $\sim 30fb$, after the beam polarization described in
Section \ref{polar} has been applied. 
Moreover in this region, while the $\tilde \chi^0_1 \tilde \chi^0_2$
production cross section is typically smaller than the $\tilde \chi^0_1
\tilde \chi^0_1$
production rate, the $\tilde \chi^0_2 \tilde \chi^0_2$
production cross section can reach higher values.
At the point B belonging to this wino region and
defined as $M_1=100GeV$, $M_2=200GeV$, $\mu=600GeV$, $\tan
\beta=3$, $m_{\tilde l^{\pm}}=300GeV$
and $m_{\tilde q}=600GeV$
($m_{\tilde \chi^{\pm}_1}=189.1GeV$, $m_{\tilde \chi^0_1}=97.3GeV$,
$m_{\tilde \chi^0_2}=189.5GeV$),
the cross sections (including the beam
polarization described in Section \ref{polar})
for the allowed neutralinos productions at $\sqrt s=500GeV$ are 
$\sigma(\tilde \chi^0_1 \tilde \chi^0_1)=32.21fb$,
$\sigma(\tilde \chi^0_1 \tilde \chi^0_2)=25.60fb$
and $\sigma(\tilde \chi^0_2 \tilde \chi^0_2)=26.40fb$.\\
In the wino region and for $M_1<M_2$, the difference between
the $\tilde \chi^0_2$
and the $\tilde \chi^0_1$ masses increases with $\vert \mu \vert$
\cite{f:chemsin} and can reach high values. Thus, the decay
$\tilde \chi^0_2 \to \tilde \chi^0_1 + Z^0$
is often dominant. At the point B and with $m_{\tilde \nu}=450GeV$,
this channel has a branching fraction of
$B(\tilde \chi^0_2 \to \tilde \chi^0_1 Z^0)=79.9\%$.
The decay of the $Z$-boson into neutrinos has a branching fraction 
of $B(Z^0 \to \nu_p \bar \nu_p)=20.0\%$.  
In the wino region and for $M_1>M_2$, the mass difference
$m_{\tilde \chi^0_2}-m_{\tilde \chi^{\pm}_1}$ is larger than the
$W^{\pm}$ mass as long as $M_1-M_2$ is larger than $\sim 75GeV$.
In this case, the dominant channel is
$\tilde \chi^0_2 \to \tilde \chi^{\pm}_1 W^{\mp}$.
For $M_1-M_2$ smaller than $\sim 75GeV$,
$m_{\tilde \chi^0_2}-m_{\tilde \chi^{\pm}_1}$
remains large enough to allow a dominant decay of type
$\tilde \chi^0_2 \to \tilde \chi^{\pm}_1 f \bar f$, $f$ being a fermion.
As a conclusion, in the wino region the decay
$\tilde \chi^0_2 \to \tilde \chi^0_1 \bar \nu_p \nu_p$ can have a significant 
branching ratio for $M_1<M_2$.

We also consider some intermediate domains. At the point C
defined as $M_1=100GeV$, $M_2=400GeV$,
$\mu=400GeV$, $\tan \beta=3$, $m_{\tilde l^{\pm}}=300GeV$
and $m_{\tilde q}=600GeV$
($m_{\tilde \chi^{\pm}_1}=329.9GeV$, $m_{\tilde \chi^0_1}=95.5GeV$,
$m_{\tilde \chi^0_2}=332.3GeV$) 
with $m_{\tilde \nu}=450GeV$, the cross
sections (including beam polarization)
for the allowed neutralinos productions at $\sqrt s=500GeV$
are
$\sigma(\tilde \chi^0_1 \tilde \chi^0_1)=32.18fb$,
$\sigma(\tilde \chi^0_1 \tilde \chi^0_2)=6.64fb$
and $\sigma(\tilde \chi^0_1 \tilde \chi^0_3)=0.17fb$,
and the branching ratio of the $\tilde \chi^0_2$ decay into neutrinos is
$B(\tilde \chi^0_2 \to \tilde \chi^0_1 \bar \nu_p \nu_p)=0.9\%$.
The branching ratio of the decay 
$\tilde \chi^0_2 \to \tilde \chi^0_1 \bar \nu_p \nu_p$ is
small for this point C since the $\tilde \chi^0_2$ mass is large which favors
other $\tilde \chi^0_2$ decays.

At the point D given by $M_1=150GeV$, $M_2=300GeV$,
$\mu=200GeV$, $\tan \beta=3$, $m_{\tilde l^{\pm}}=300GeV$
and $m_{\tilde q}=600GeV$
($m_{\tilde \chi^{\pm}_1}=165.1GeV$, $m_{\tilde \chi^0_1}=121.6GeV$,
$m_{\tilde \chi^0_2}=190.8GeV$),
the cross sections (including beam polarization)
for the allowed neutralinos productions at $\sqrt s=500GeV$
are
$\sigma(\tilde \chi^0_1 \tilde \chi^0_1)=7.80fb$,
$\sigma(\tilde \chi^0_1 \tilde \chi^0_2)=13.09fb$,
$\sigma(\tilde \chi^0_2 \tilde \chi^0_2)=10.08fb$,
$\sigma(\tilde \chi^0_1 \tilde \chi^0_3)=35.69fb$,
$\sigma(\tilde \chi^0_2 \tilde \chi^0_3)=41.98fb$,
$\sigma(\tilde \chi^0_3 \tilde \chi^0_3)=0.02fb$
and $\sigma(\tilde \chi^0_1 \tilde \chi^0_4)=0.42fb$.
Since this point D lies in a particular region between
the higgsino region and the domain of large $\vert \mu \vert$
(or equivalently large $m_{\tilde \chi^0_3}$), the
$\tilde \chi^0_1 \tilde \chi^0_3$ and $\tilde \chi^0_2 \tilde \chi^0_3$
productions become relatively important. At the MSSM point D
and for $m_{\tilde \nu}=450GeV$,
the branching ratio of the $\tilde \chi^0_2$ decay into neutrinos is
$B(\tilde \chi^0_2 \to \tilde \chi^0_1 \bar \nu_p \nu_p)=10.6\%$.

Finally, in the domain of low charged slepton masses the $\tilde \chi^0_i
\tilde \chi^0_j$
productions are increased due to the $t$ channel $\tilde l^{\pm}$
exchange contribution.
At the point E, defined as the point B with a lower
$\tilde l^{\pm}$ mass, namely $M_1=100GeV$, $M_2=200GeV$,
$\mu=600GeV$, $\tan \beta=3$, $m_{\tilde l^{\pm}}=150GeV$ 
and $m_{\tilde q}=600GeV$, 
the polarized neutralinos pair productions rates which read as
$\sigma(\tilde \chi^0_1 \tilde \chi^0_1)=67.06fb$,
$\sigma(\tilde \chi^0_1 \tilde \chi^0_2)=57.15fb$
and $\sigma(\tilde \chi^0_2 \tilde \chi^0_2)=69.56fb$
are increased compared to the point B.\\
For $m_{\tilde \chi^0_2}>m_{\tilde l^{\pm}}$ 
or $m_{\tilde \chi^0_2}>m_{\tilde q}$,
the dominant $\tilde \chi^0_2$ decays are 
$\tilde \chi^0_2 \to \tilde \chi^0_1 \bar l_p l_p$
or $\tilde \chi^0_2 \to \tilde \chi^0_1 \bar q_p q_p$,
respectively, and the decay $\tilde \chi^0_2 \to \tilde \chi^0_1 \bar \nu_p \nu_p$ 
is typically negligible except for $m_{\tilde \chi^0_2}>m_{\tilde \nu}$. 
At the point E for $m_{\tilde \nu}>m_{\tilde \chi^0_2}$, the $\tilde \chi^0_2$
mainly decays into $\tilde \chi^0_1 \bar l_p l_p$ 
via an on shell $\tilde l_p^{\pm}$ and
the decay into neutrinos has a branching ratio of  
$B(\tilde \chi^0_2 \to \tilde \chi^0_1 \bar \nu_p \nu_p) \sim 0\%$. 
In contrast, at the point E with $m_{\tilde \nu}=160GeV$, 
$B(\tilde \chi^0_2 \to \tilde \chi^0_1 \bar \nu_p \nu_p) =32.0\%$.
Of course for $m_{\tilde l^{\pm}},m_{\tilde q}>m_{\tilde \chi^0_2}>m_{\tilde \nu}$,
the decay $\tilde \chi^0_2 \to \tilde \chi^0_1 \bar \nu_p \nu_p$ is dominant. 

$\bullet$ {\bf Sneutrino pair production}
The other source of SUSY background is the sneutrino pair production.
We discuss here the main variations of this background 
in the SUSY parameter space. First, we note that 
the $\tilde \nu_1$ pair production has the highest cross section 
among the $\tilde \nu_p \tilde \nu^*_p$ 
($p=1,2,3$ being the family index) productions
since it receives a contribution from the $t$ channel exchange of charginos.
Besides, the main contribution to the $4l+\Eslash$ signature
from the sneutrino pair production 
is the reaction $e^+ e^- \to \tilde \nu_p + \tilde \nu_p^* 
\to \tilde \chi^0_1 \nu_p + \tilde \chi^0_1 \bar \nu_p$.
Indeed, this contribution has the simplest cascade decays and furthermore
the decay $\tilde \nu_p \to \tilde \chi^0_1 \nu_p$ is favored by the phase space 
factor. Hence, we restrict the discussion of 
the $\tilde \nu_p \tilde \nu_p^*$ background to this reaction,
although all contributions from the sneutrino pair productions 
to the $4l+\Eslash$ signature are included in our analysis.

The $\tilde \nu_p$ pair production rate is reduced in the higgsino region like the 
$\tilde \chi^0_1 \tilde \chi^0_1$ production. For instance, at the point A with 
$m_{\tilde \nu}=175GeV$ the sneutrino pair production
cross section including the beam polarization effect
described in Section \ref{polar} is 
$\sigma(e^+ e^- \to \tilde \chi^0_1 \nu_p \tilde \chi^0_1 \bar \nu_p)=75.26fb$
($1.96fb$) for $p=1$ ($2,3$),
while it is 
$\sigma(e^+ e^- \to \tilde \chi^0_1 \nu_p \tilde \chi^0_1 \bar \nu_p)=501.55fb$
($10.29fb$) for $p=1$ ($2,3$)
at the point B with the same sneutrino mass.
These values of the cross sections are obtained with SUSYGEN \cite{f:SUSYGEN3} 
and include the spin correlations effect.
Besides, the $\tilde \nu_p$ pair production rates strongly decrease 
as the sneutrino mass increases. 
Considering once more the point B, we find that the rate is reduced to
$\sigma(e^+ e^- \to \tilde \chi^0_1 \nu_p \tilde \chi^0_1 \bar \nu_p)=232.92fb$
($4.59fb$) for $p=1$ ($2,3$) if we take now $m_{\tilde \nu}=200GeV$. \\
The branching ratio $B(\tilde \nu \to \tilde \chi^0_1 \nu)$ 
decreases as the sneutrino mass increases, since the phase space factors
associated to the decays of the sneutrino into other SUSY particles than the
$\tilde \chi^0_1$, like $\tilde \nu \to \tilde \chi^{\pm}_1 l^{\mp}$, 
increase with $m_{\tilde \nu}$. For example, at the point B 
the branching ratio $B(\tilde \nu_e \to \tilde \chi^0_1 \nu_e)$ is equal to $93.4\%$
for $m_{\tilde \nu}=175GeV$ and to $82.6\%$ for $m_{\tilde \nu}=200GeV$.

\subsection{Cuts}
\label{cuts}

\subsubsection{General selection criteria}
\label{gsc}

First, we select the events without jets containing 4 charged leptons
and missing energy.
In order to take into account the observability
of leptons at a $500GeV \ e^+ e^-$ machine, we apply the following cuts
on the transverse momentum and rapidity of all the charged leptons:
$P_t(l)>3 GeV$ and $\vert \eta(l) \vert <3$. This should
simulate the detector acceptance effects in a first approximation.
In order to reduce the \susyq background, we also demand that the number of
muons is at least equal to one. Since we consider the $\tilde \chi_1^{\pm}
\mu^{\mp}$ production, this does not affect the signal.

\subsubsection{Kinematics of the muon produced with the chargino}
\label{kin}

\begin{figure}[t]
\begin{center}
\leavevmode
\centerline{\psfig{figure=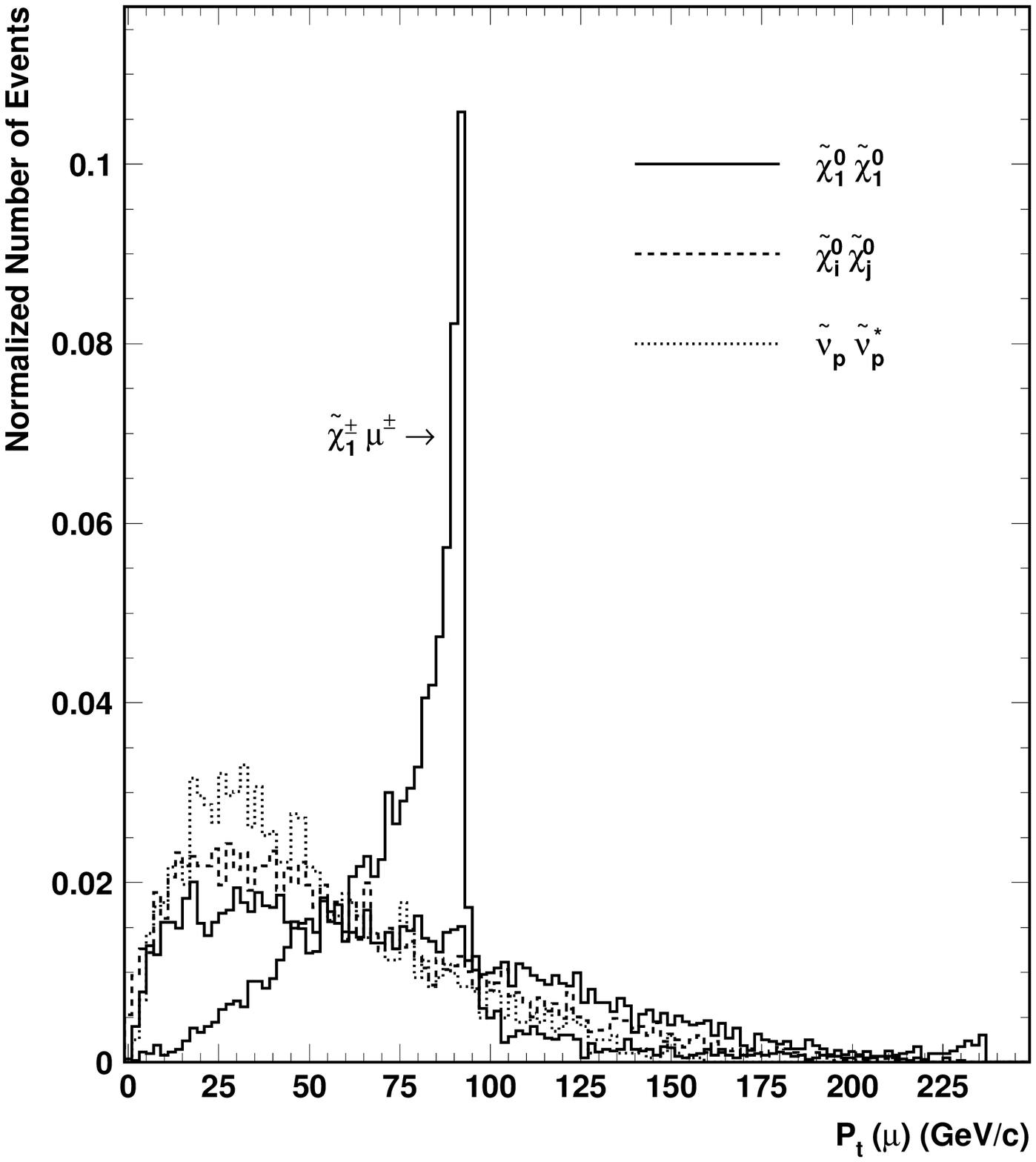,height=5.5in}}
\end{center}
\caption{\footnotesize  \it
Distributions of the highest muon transverse momentum $P_t(\mu)$ (in $GeV/c$)
for the $4l + \Eslash$ events generated by
the single $\tilde \chi^{\pm}_1$ production and the SUSY background
which is
divided into the $\tilde \chi^0_1 \tilde \chi^0_1$, $\tilde \chi^0_i
\tilde \chi^0_j$ and $\tilde \nu_p \tilde \nu_p^*$ productions.
The number of events for each distribution is normalized to the unity,
the center of mass energy is fixed at $500 GeV$ and
the point A of the SUSY parameter space is considered
with $\l_{121}=0.05$ and $m_{\tilde \nu}=240GeV$.
\rm \normalsize }
\label{isr}
\end{figure}

We now discuss in details the most important cut concerning
the momentum of the muon which is produced together with the chargino.
For a negligible ISR effect, 
this muon momentum is completely determined
by the values of the chargino mass and the center of mass energy
through Eq.(\ref{enl}), as explained in Section \ref{signal}. 
Hence, some cuts on the muon momentum should
be efficient to enhance the signal-to-noise ratio
since the muon momentum distribution is perfectly peaked at a given value.
\\ For a significant ISR effect, the energy of the photon generated via the ISR 
must be taken into account. Hence, the muon energy $E(\mu)$ must be calculated
through the three-body kinematics of the reaction
$e^+ e^- \to \tilde \chi_1^{\pm} \mu^{\mp} \gamma$. Since this kinematics
depends on the angle between the muon and the photon, 
the muon energy $E(\mu)$ is not completely fixed by the SUSY parameters anymore.
Therefore, the distribution obtained experimentally of the muon momentum $P(\mu)
= ( E(\mu)^2-m_{\mu^{\pm}}^2 c^4)^{1/2}/c$ would
appear as a peaked curve instead of a Dirac peak. Although the
momentum
of the produced muon remains a good selection criteria, we have
found that in this case the transverse momentum distribution of the produced
muon was
more peaked, and thus more appropriate to apply some cuts.
We explain the reasons why in details below.\\
We have thus chosen to apply cuts on the distribution of the
muon transverse momentum instead of the whole muon momentum, even if
for a negligible ISR effect the transverse momentum distribution is not peaked as
well as the whole momentum distribution. 
The amplitude of the ISR effect depends on the SUSY mass spectrum
as we will discuss below.
\\ The cuts on the muon transverse momentum have been applied on the 
muon with the highest transverse momentum which can be identified
as the muon produced together with the chargino in the case of the signal. 
Indeed, the muon produced together with the chargino 
is typically more energetic than the 3 other leptons generated in the chargino 
cascade decay. We will come back on this point later.

In Fig.\ref{isr}, we show the distribution
of the highest muon transverse momentum $P_t(\mu)$ in the $4l + \Eslash$ events
generated by the single $\tilde \chi^{\pm}_1$ production
and the SUSY background at $\sqrt s=500 GeV$ for the point A of
the SUSY parameter space with $\l_{121}=0.05$ and $m_{\tilde \nu}=240GeV$.
For these SUSY parameters, the cross sections, polarized as in Section 
\ref{polar}, and the branching ratios are
$\sigma(\tilde \chi^{\pm}_1 \mu^{\mp})=534.39fb$, 
$B(\tilde \chi_1^{\pm} \to \tilde \chi^0_1  l_p \nu_p)=34.3\%$,
$\sigma(e^+ e^- \to \tilde \nu_p + \tilde \nu_p^* 
\to \tilde \chi^0_1 \nu_p + \tilde \chi^0_1 \bar \nu_p)=3.81fb$ and
$B(\tilde \chi^0_2 \to \tilde \chi^0_1 \bar \nu_p \nu_p)=16\%$
(see Section \ref{SUSYX} for
the values of $\sigma(\tilde \chi^0_i \tilde \chi^0_j$)).
In Fig.\ref{isr}, the cuts described in Section \ref{gsc}
have not been applied and in order to perform a separate analysis 
for each of the main considered backgrounds,
the SUSY background has been decomposed into
three components: the $\tilde \chi^0_1 \tilde \chi^0_1$, $\tilde \chi^0_i
\tilde \chi^0_j$ ($i,j=1,...,4$ with $i$ and $j$ not
equal to $1$ simultaneously) and $\tilde \nu_p \tilde \nu_p^*$ productions.
The number of $4l + \Eslash$ events 
for each of those 3 SUSY backgrounds and the single chargino
production has been normalized to the unity.
Our motivation for such a normalization is that the relative amplitudes of 
the 3 SUSY backgrounds and the single chargino
production, which depend strongly on the SUSY parameters, 
were discussed in detail in 
Sections \ref{signalX} and \ref{SUSYX}, and in this section we focus on
the shapes of the various distributions.\\ 
We see in the highest muon transverse momentum distribution 
of Fig.\ref{isr} that, as expected,
a characteristic peak arises for the single chargino production. 
Therefore, some cuts
on the highest muon transverse momentum $P_t(\mu)$
would greatly increase the signal with respect to the backgrounds.
For instance, the selection criteria suggested by the Fig.\ref{isr}
are something like 
$60GeV \stackrel{<}{\sim} P_t(\mu) \stackrel{<}{\sim} 100GeV$.
We can also observe in Fig.\ref{isr} 
that the distributions of the $\tilde \chi^0_i
\tilde \chi^0_j$ and $\tilde \nu_p \tilde \nu_p^*$ productions are
concentrated at lower transverse momentum values than 
the $\tilde \chi^0_1 \tilde \chi^0_1$ distribution.
This is due to the energy carried away by the neutrinos in the cascade
decays of the reactions (\ref{chi2}), (\ref{chi3}),
(\ref{chrp2}) and (\ref{sneu}). 
The main variation of the highest muon transverse momentum 
distributions of the SUSY backgrounds with the SUSY parameters
is the following: The SUSY backgrounds distributions
spread to larger values of the muon transverse momentum
as the $\tilde \chi^0_1$ mass increases,
since then the charged leptons coming from the
decay $\tilde \chi^0_1 \to l \bar l \nu$ reach higher energies.
\\ Fig.\ref{isr} shows also clearly that
the peak in the transverse momentum distribution of the muon produced with
the chargino exhibits an upper limit which we note $P^{lim}_t(\mu)$.
This bound $P^{lim}_t(\mu)$ is a kinematic limit
and thus its value depends on the SUSY masses. The consequence on our
analysis is that the cuts on the muon transverse momentum are
modified as the SUSY mass spectrum is changing. Note that the kinematic
limit of the whole muon momentum depends also on the
SUSY masses, so that our choice of working with
the transverse momentum remains judicious.
For a better understanding of the analysis and in view of the study
on the SUSY mass spectrum of Section \ref{marec},
we now determine the value 
of the muon transverse momentum kinematic limit $P^{lim}_t(\mu)$
as a function of the SUSY parameters.
For this purpose we divide the discussion into 2 scenarios.

\begin{figure}[t]
\begin{center}
\leavevmode
\centerline{\psfig{figure=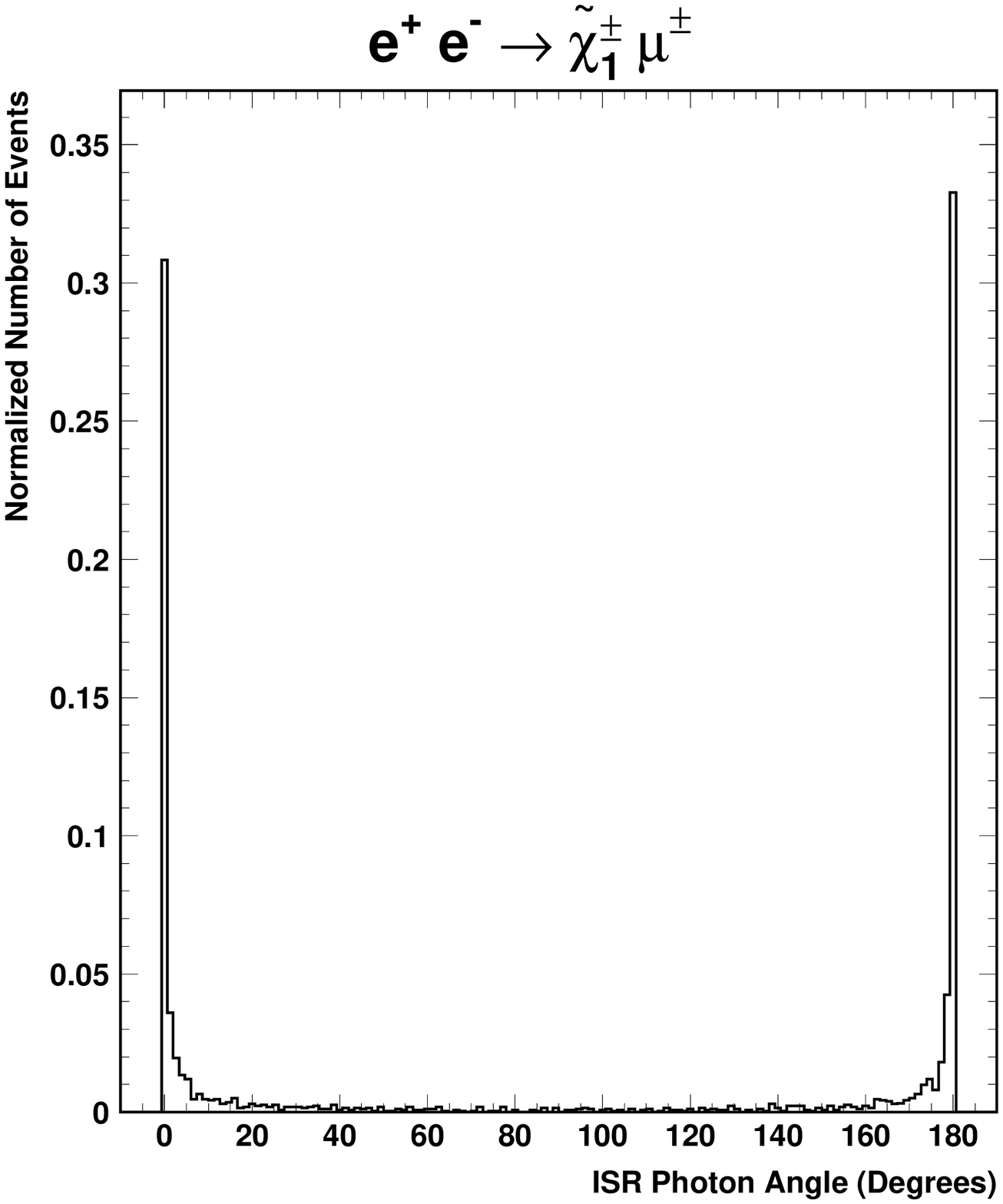,height=5.5in}}
\end{center}
\caption{\footnotesize  \it
Distribution of the angle (in Degrees) between the beam axis and
the photon radiated from the initial state
of the single $\tilde \chi_1^{\pm}$ production process 
at a center of mass energy of $500GeV$ and for
the point A of the SUSY parameter
space with $m_{\tilde \nu}=240GeV$.
The number of events has been normalized to the unity.
\rm \normalsize }
\label{distribII}
\end{figure}

First, we consider the situation where $m_{\tilde \nu_{\mu}}>m_{\tilde
\chi_1^{\pm}}+m_{\mu^{\mp}}$ and $m_{\tilde \nu_{\mu}}<\sqrt s$.
In this case, the dominant
$s$ channel contribution to the three-body 
reaction $e^+ e^- \to \tilde \chi_1^{\pm}
\mu^{\mp} \gamma$, where the photon is generated by the ISR, 
can be decomposed into two levels. First,
a sneutrino is produced together with a photon in the two-body reaction
$e^+ e^- \to \gamma \tilde \nu_{\mu}$
and then the sneutrino decays as $\tilde \nu_{\mu} \to \tilde \chi_1^{\pm}
\mu^{\mp}$.
Thus, the muon energy in the center of mass of the sneutrino $E^\star(\mu)$
(throughout this article a $\star$ indicates that the variable is defined
in the center of mass of the produced sneutrino) is equal to,
\begin{eqnarray}
E^\star(\mu)=
{m^2_{\tilde \nu_{\mu}} + m^2_{\mu} - m^2_{\tilde \chi^{\pm}}
\over 2 m_{\tilde \nu_{\mu}}}.
\label{enmu}
\end{eqnarray}
The transverse momentum limit of the produced muon in the center
of mass of the sneutrino $P^{lim \star}_t(\mu)$ is given by 
$P^{lim \star}_t(\mu) =( E^{\star}(\mu)^2-m_{\mu^{\pm}}^2 c^4)^{1/2}/c$ since
$P_t^\star(\mu) \leq P^\star(\mu)$ and 
$P^\star(\mu) =( E^{\star}(\mu)^2-m_{\mu^{\pm}}^2 c^4)^{1/2}/c$.
The important point is that the sneutrino rest frame
is mainly boosted in the direction of the beam axis.
This is due to the fact that the produced photon
is mainly radiated at small angles with respect to the initial colliding
particles direction (see Fig.\ref{distribII}).
The consequence is that the transverse momentum of the 
produced muon is mainly the same in
the sneutrino rest frame and in the laboratory frame. In conclusion, the
transverse momentum limit of the produced muon in the laboratory frame is
given by $P^{lim}_t(\mu) \approx P^{lim \star}_t(\mu)
=( E^{\star}(\mu)^2-m_{\mu^{\pm}}^2 c^4)^{1/2}/c$,
$E^\star(\mu)$ being calculated through Eq.(\ref{enmu}).
We see explicitly through Eq.(\ref{enmu}) that the value of $P^{lim}_t(\mu)
\approx ( E^{\star}(\mu)^2-m_{\mu^{\pm}}^2 c^4)^{1/2}/c$ 
increases with the sneutrino mass.
\\ In this first situation, where $m_{\tilde \nu_{\mu}}>m_{\tilde
\chi_1^{\pm}}+m_{\mu^{\mp}}$
and $m_{\tilde \nu_{\mu}}<\sqrt s$, the ISR effect is large so that the
single chargino production cross section is enhanced.
The distributions of Fig.\ref{isr} are obtained for the point A
(for which $m_{\tilde \chi_1^{\pm}}=115.7GeV$) 
with $m_{\tilde \nu_{\mu}}=240GeV$
which corresponds to this situation of large ISR.
The distribution of Fig.\ref{isr} gives a value for
the transverse momentum kinematic limit of 
$P^{lim}_t(\mu)=92.3GeV/c$ which is well approximately equal to  
$( E^{\star}(\mu)^2-m_{\mu^{\pm}}^2 c^4)^{1/2}/c=92.1GeV/c$
where $E^\star(\mu)$ is calculated using Eq.(\ref{enmu}). 
\\ Based on this explanation,
we can now understand why for large ISR the transverse momentum
of the produced muon $P_t(\mu)$ is more peaked than its whole momentum
$P(\mu)$:
As we have discussed above, the muon transverse momentum $P_t(\mu)$
is controlled by the two-body kinematics of
the decay $\tilde \nu_{\mu} \to \tilde \chi_1^{\pm} \mu^{\mp}$.
For a set of SUSY parameters, 
the muon transverse momentum is thus fixed if only the
absolute value of the cosinus of the angle that makes the muon
with the initial beam direction in the sneutrino rest frame 
is given. In contrast, the whole muon momentum $P(\mu)$ 
is given by the three-body kinematics of the reaction
$e^+ e^- \to \tilde \chi_1^{\pm} \mu^{\mp} \gamma$ and thus depends on the
cosinus of the angle between the muon and the photon 
(recall that the photon
is mainly emitted along the beam axis) in the laboratory frame. 
Therefore, the muon transverse momentum $P_t(\mu)$ is
less dependent on the muon angle than the whole muon momentum $P(\mu)$,
leading thus to a more peaked distribution.

\begin{figure}[t]
\begin{center}
\leavevmode
\centerline{\psfig{figure=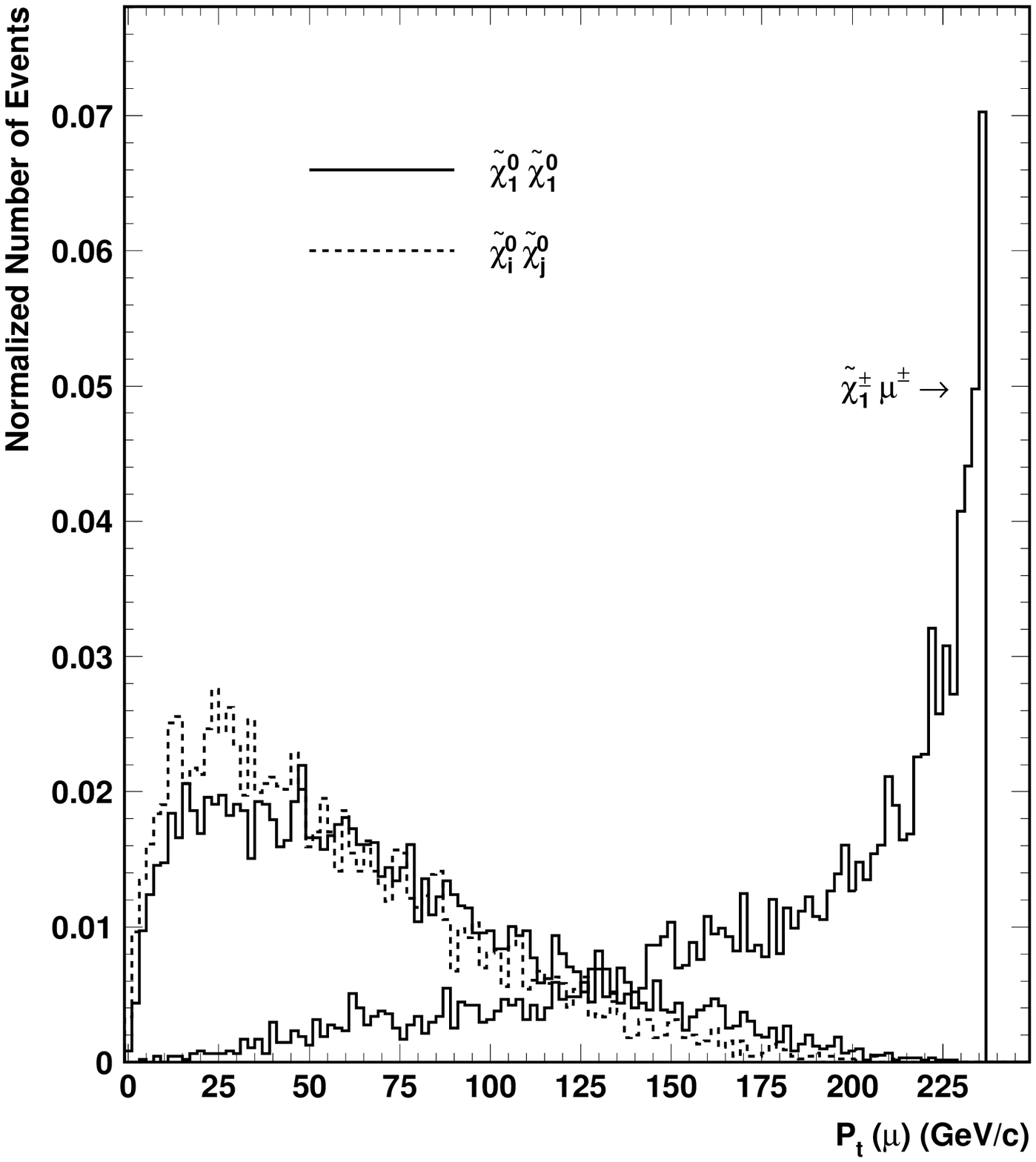,height=5.5in}}
\end{center}
\caption{\footnotesize  \it
Distributions of the highest muon transverse momentum $P_t(\mu)$ (in $GeV/c$)
for the $4l + \Eslash$ events generated by
the single $\tilde \chi^{\pm}_1$ production and the SUSY background
which is
divided into the $\tilde \chi^0_1 \tilde \chi^0_1$ and $\tilde \chi^0_i
\tilde \chi^0_j$ productions.
The number of events for each distribution is normalized to the unity,
the center of mass energy is fixed at $500 GeV$ and
the point A of the SUSY parameter space is considered
with $\l_{121}=0.05$ and $m_{\tilde \nu}=550GeV$.
\rm \normalsize }
\label{noisr}
\end{figure}

We consider now the scenario where $m_{\tilde \nu_{\mu}}<m_{\tilde
\chi_1^{\pm}}+m_{\mu^{\mp}}$ or
$m_{\tilde \nu_{\mu}}>\sqrt s$. In such a situation,
the kinematics is different since the sneutrino
cannot be produced on shell.
The single chargino production $e^+ e^- \to \tilde \chi_1^{\pm}
\mu^{\mp} \gamma$ can thus not occur through the two-body
reaction $e^+ e^- \to \gamma \tilde \nu_{\mu}$. In fact in this situation,
the energy of the radiated photon becomes negligible so that the muon energy
$E(\mu)$ is given in a good approximation by
the two-body kinematics formula of Eq.(\ref{enl}).
The kinematic limit of the muon transverse momentum is related to
this energy through 
$P^{lim}_t(\mu) = P(\mu) =( E(\mu)^2-m_{\mu^{\pm}}^2 c^4)^{1/2}/c$.
\\ In this second scenario where $m_{\tilde \nu_{\mu}}<m_{\tilde
\chi_1^{\pm}}+m_{\mu^{\mp}}$ or $m_{\tilde \nu_{\mu}}>\sqrt s$, 
the ISR effect is negligible
and the single chargino production rate is thus not increased.
In Fig.\ref{noisr}, we show the transverse momentum distribution
(without the cuts of Section \ref{gsc}) 
of the produced muon in such a situation, 
namely at $\sqrt s=500 GeV$ for the point A of
the SUSY parameter space with $\l_{121}=0.05$ and $m_{\tilde
\nu_{\mu}}=550GeV$. For these SUSY parameters, 
the cross sections, polarized as in Section 
\ref{polar}, and the branching ratios are
$\sigma(\tilde \chi^{\pm}_1 \mu^{\mp})=91.05fb$, 
$B(\tilde \chi_1^{\pm} \to \tilde \chi^0_1  l_p \nu_p)=33.9\%$ and
$B(\tilde \chi^0_2 \to \tilde \chi^0_1 \bar \nu_p \nu_p)=15.7\%$
(see Section \ref{SUSYX} for
the values of $\sigma(\tilde \chi^0_i \tilde \chi^0_j$).
We check that the value of the muon transverse
momentum kinematic limit $P^{lim}_t(\mu)=236.7GeV/c$  
obtained from the distribution of Fig.\ref{noisr} 
is well approximately equal to 
$( E(\mu)^2-m_{\mu^{\pm}}^2 c^4)^{1/2}/c=236.6GeV/c$
where $E(\mu)$ is obtained from Eq.(\ref{enl}).

At this stage, we can make a comment on the variation 
of the muon transverse momentum
distribution associated to the 
single $\tilde \chi^{\pm}_1$ production with
the value of the transverse momentum limiy $P^{lim}_t(\mu)$. 
As can be seen by comparing Fig.\ref{isr} and Fig.\ref{noisr},
this distribution is more peaked for small values of
$P^{lim}_t(\mu)$ due
to an higher concentration of the distribution 
at low muon transverse momentum values.
This effect is compensated by the fact that at low muon transverse
momentum values the SUSY background is larger, as can be seen in
Fig.\ref{isr}.

The determination of $P^{lim}_t(\mu)$ performed in this section
allows us to verify
that the muon produced with the chargino is well often
the muon of highest transverse momentum. Indeed,
we have checked that in the two situations of large
(point A with $m_{\tilde
\nu_{\mu}}=240GeV$) and negligible (point A with $m_{\tilde
\nu_{\mu}}=550GeV$) ISR effect,
the calculated values of $P^{lim}_t(\mu)$ for the produced muon
were consistent with the values of $P^{lim}_t(\mu)$ obtained with
the highest muon transverse momentum distributions
(Fig.\ref{isr} and Fig.\ref{noisr}).
Therefore, the identification of the produced muon with the muon of
highest transverse momentum is correct for the two MSSM points considered
above.
The well peaked shapes of the highest muon transverse momentum distributions
for the point A with
$m_{\tilde \nu_{\mu}}=240GeV$ and $m_{\tilde
\nu_{\mu}}=550GeV$ are another confirmation of the validity of this
identification.
Nevertheless, one must wonder what is the domain of validity of this
identification.
In particular, the transverse momentum of the produced muon can be small
since $P^{lim}_t(\mu)$ which is determined via Eq.(\ref{enl}) or
Eq.(\ref{enmu}) can reach low values.
As a matter of fact, $P^{lim}_t(\mu)$ can reach small values
in the scenario where $m_{\tilde \nu_{\mu}}>m_{\tilde \chi_1^{\pm}}+m_{\mu^{\mp}}$
and $m_{\tilde \nu_{\mu}}<\sqrt s$
for $m_{\tilde \nu_{\mu}}$ close to $m_{\tilde \chi^{\pm}_1}$
(see Eq.(\ref{enmu})),
as well as in the scenario where 
$m_{\tilde \nu_{\mu}}<m_{\tilde \chi_1^{\pm}}+m_{\mu^{\mp}}$ or
$m_{\tilde \nu_{\mu}}>\sqrt s$
for $\sqrt s$ close to $m_{\tilde \chi^{\pm}_1}$ (see Eq.(\ref{enl})).
Moreover, the leptons generated in the cascade decay initiated by the
$\tilde \chi_1^{\pm}$
become more energetic, and can thus have larger transverse momentum,
for either larger $\tilde \chi_1^0$ masses or higher mass differences
between the $\tilde \chi_1^{\pm}$ and $\tilde \chi_1^0$.
We have found that for produced muons of small transverse momentum
corresponding to $P^{lim}_t(\mu) \approx 10 GeV$ the identification
remains valid for neutralino masses up to $m_{\tilde \chi_1^0} \approx
750GeV$ and mass differences up to
$m_{\tilde \chi_1^{\pm}}-m_{\tilde \chi_1^0} \approx 750GeV$.

\subsection{Discovery potential}
\label{reach}

\begin{figure}[t]
\begin{center}
\leavevmode
\centerline{\psfig{figure=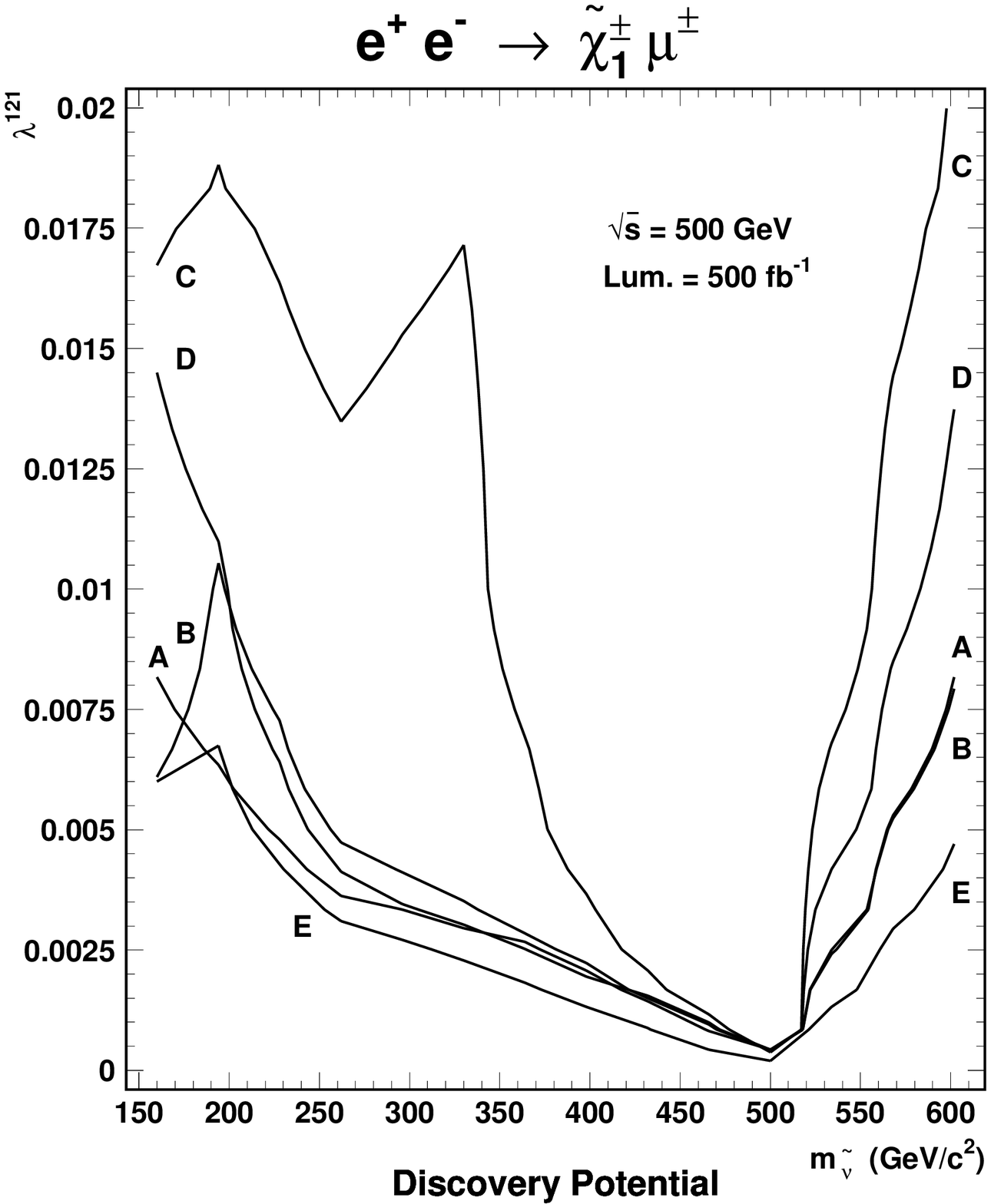,height=5.5in}}
\end{center}
\caption{\footnotesize  \it
Discovery potential at the $5 \sigma$ level
in the plane $\l_{121}$ versus $m_{\tilde \nu}$ (in $GeV/c^2$) 
for the points A, B, C, D, E
of the SUSY parameter space (see text) at a center of mass energy of
$500GeV$ and assuming a luminosity of ${\cal L}=500 fb^{-1}$.
The domains of the SUSY parameter space situated above the curves
correspond to $S / \sqrt B \geq 5$ where $S$ is the $4l+\Eslash$ signal 
generated by the single $\tilde \chi^{\pm}_1$ production
and $B$ is the $R_p$-conserving SUSY background.
\rm \normalsize }
\label{reach1}
\end{figure}

In Fig.\ref{reach1}, we present exclusion plots at the $5 \sigma$ level
in the plane $\l_{121}$ versus $m_{\tilde \nu_{\mu}}$ based on the 
study of the $4l + \Eslash$ final state for several points
of the SUSY parameter space and at a center of mass energy of
$\sqrt s=500GeV$ assuming a luminosity of ${\cal L}=500 fb^{-1}$
\cite{f:Tesla}. The considered signal and backgrounds are respectively the
single $\tilde \chi^{\pm}_1$ production and
the pair productions of all the neutralinos and sneutrinos. 
The curves of Fig.\ref{reach1} correspond to
a number of signal events larger than 10. 
An efficiency of $30\%$ has been assumed for the reconstruction  
of the tau-lepton from its hadronic decay. 
The experimental LEP limit on the sneutrino mass $m_{\tilde \nu}>78GeV$
\cite{f:DELPHIb} has been respected.
We have included the ISR effects as well as the effects of the
polarization described in Section \ref{polar},
assuming an electron (positron) polarization efficiency of $85 \%$  ($60
\%$). The cuts described in Section \ref{cuts} have also been applied. 
In particular, we have
applied some cuts on the transverse momentum of the produced muon. Since
the value of the muon transverse momentum depends on the sneutrino 
and chargino masses (see Section \ref{kin}), 
different cuts have been chosen
which were appropriate to the different values
of $m_{\tilde \nu_{\mu}}$ and $m_{\tilde \chi_1^{\pm}}$ 
considered in Fig.\ref{reach1}.

We can observe in Fig.\ref{reach1} that
the sensitivity on the $\l_{121}$ \cc typically increases as
the sneutrino mass approaches the resonance point 
$m_{\tilde \nu}=\sqrt s =500GeV$. 
This is due to the ISR effect which increases 
the single chargino production rate as discussed in Section \ref{signalX}.
While the ISR effect is significant in the domain 
where $m_{\tilde \nu_{\mu}}>m_{\tilde \chi_1^{\pm}}+m_{\mu^{\mp}}$
and $m_{\tilde \nu_{\mu}}<\sqrt s$, it is small
for $m_{\tilde \nu_{\mu}}<m_{\tilde \chi_1^{\pm}}+m_{\mu^{\mp}}$ or
$m_{\tilde \nu_{\mu}}>\sqrt s$, the reason being that in 
the former case the single chargino production occurs through
the production of an on shell sneutrino as explained in Section \ref{kin}.
This results in a decrease of the sensitivity on the $\l_{121}$ coupling 
at $m_{\tilde \nu_{\mu}} \stackrel{>}{\sim} \sqrt s$ and 
$m_{\tilde \nu_{\mu}} \stackrel{>}{\sim} m_{\tilde \chi_1^{\pm}}$
as illustrate
the various curves of Fig.\ref{reach1}. The
decrease of the sensitivity corresponding to 
$m_{\tilde \nu_{\mu}} \stackrel{>}{\sim} m_{\tilde \chi_1^{\pm}}$
occurs at larger sneutrino masses for the point 
C compared to the other SUSY points, 
since for this set of SUSY parameters  
the chargino mass is larger: $m_{\tilde \chi_1^{\pm}}=329.9GeV$. 
\\ For $m_{\tilde \nu}<m_{\tilde \chi_1^{\pm}}$ the 
decay $\tilde \chi_1^{\pm} \to \tilde \chi_1^0 l_p \nu_p$ 
becomes dominant as explained in Section \ref{signalX}. 
This results in an increase of the sensitivity on $\l_{121}$ which can
be seen for the point C between $m_{\tilde \nu} \approx 330GeV$ 
and $m_{\tilde \nu} \approx 260GeV$ and
for the points B and E between $m_{\tilde \nu} \approx 190 GeV$ 
and $m_{\tilde \nu} \approx 160GeV$. 
\\ Additional comments must be made concerning 
the exclusion curve obtained for the point C:
The decrease of sensitivity in the interval 
$260GeV \stackrel{>}{\sim} m_{\tilde \nu} \stackrel{>}{\sim} 200GeV$  
comes from the increase of the
sneutrino pair production cross section (see Section \ref{SUSYX}),
while the increase of sensitivity in the range
$200GeV \stackrel{>}{\sim} m_{\tilde \nu} \stackrel{>}{\sim} 160GeV$ 
is due to an increase of the
single chargino production rate 
which receives in this domain an important $t$ channel contribution.
The significant sensitivity on the $\l_{121}$ coupling 
obtained for the point C in the interval
$330GeV \stackrel{>}{\sim} m_{\tilde \nu} \stackrel{>}{\sim} 160GeV$ 
emphasizes the importance of the off-resonance contribution
in the single chargino production study at linear colliders.
For $m_{\tilde \nu}>500GeV$, the sensitivity obtained with the point C
is weak with respect to the other SUSY points due to the high chargino
mass which suppresses the signal cross section. 
\\ We also see in Fig.\ref{reach1} that 
there are no important differences between the exclusion curves
obtained for the SUSY points belonging to the higgsino region (point A), 
the wino region (point B) and the intermediate domain (point D).
The reason is the following.
The single chargino production has a cross section which decreases
as going from the wino region to the 
higgsino region (see Section \ref{signalX}). 
However, this is also true for the $\tilde \nu_p \tilde \nu_p^*$ and most of
the $\tilde \chi_i^0 \tilde \chi_j^0$ productions (see Section \ref{SUSYX}). 
Therefore, the sensitivities on the $\l_{121}$ coupling 
obtained in the higgsino and wino region are of the same order. 
\\ Finally, we discuss the sensitivity obtained for the point E
which is defined as the point B but with a smaller charged slepton mass.
Although the neutralino pair production rate is larger for the point E
than for the point B (see Section \ref{SUSYX}),
the sensitivity obtained on $\l_{121}$ 
is higher for the point E in all the considered range of sneutrino mass. 
This is due to the branching ratio 
$B(\tilde \chi_1^{\pm} \to \tilde \chi_1^0 l_p \nu_p)$ 
which becomes important at the point E due to the hierarchy 
$m_{\tilde \chi_1^{\pm}}>m_{\tilde l^{\pm}}$
(see Section \ref{signalX}).

We mention that the sensitivity on the $\l_{121}$ coupling 
tends to decrease as $\tan \beta$ increases 
(for $sign(\mu)>0$) and is weaker
for $sign(\mu)<0$ (with $\tan \beta=3$) due to the evolution of the
single chargino production rate which was described in Section \ref{signalX}.
However, the order of magnitude of the sensitivity on $\l_{121}$ 
found in Fig.\ref{reach1} remains correct for either large $\tan \beta$,
as for instance $\tan \beta=50$, or negative $\mu$.

Let us make some concluding remarks. We see in Fig.\ref{reach1} that
the sensitivity on the $\l_{121}$ coupling reaches values typically of 
order $10^{-4}$ at the sneutrino resonance.
We also observe that for each SUSY point the $5 \sigma$ limit on the 
$\l_{121}$ coupling remains more stringent than
the low-energy limit at $2 \sigma$, 
namely $\l_{121}<0.05 \ (m_{\tilde e_R} / 100GeV)$ \cite{f:Bhatt},
over an interval of the sneutrino mass of $\Delta m_{\tilde \nu} 
\approx 500 GeV$ around the $\tilde \nu$ pole.
Therefore, the sensitivities on the SUSY parameters obtained via 
the single chargino production analysis at linear colliders would 
greatly improve the 
results derived from the LEP analysis (see Section \ref{intro}).
Besides, the range of $\l_{121}$ \cc values investigated at linear colliders
through the single chargino production analysis (see Fig.\ref{reach1}) 
would be complementary to the sensitivities obtained 
via the displaced vertex analysis (see Eq.(\ref{LSP})).

\subsection{SUSY mass spectrum}
\label{marec}

\subsubsection{Lightest chargino and sneutrino}
\label{lightest}

First, the sneutrino mass can be determined 
through the study of the $4l+\Eslash$ final state by 
performing a scan on the
center of mass energy in order to find the value of $\sqrt s$
at which hold the peak of the cross section
associated to the sneutrino resonance.
The accuracy on the measure of the sneutrino mass should be of order
$\delta m_{\tilde \nu} \sim \sigma_{\sqrt s}$, where $\sigma_{\sqrt s}$ is
the root mean square spread in center of mass energy
given in terms of the beam resolution $R$ by \cite{f:muon2},
\begin{eqnarray}
\sigma_{\sqrt s} = (7 MeV) ({R \over 0.01 \%}) ({\sqrt s \over 100GeV}).
\label{beam}
\end{eqnarray}
The values of the beam resolution
at linear colliders are expected to verify $R>1\%$.

Besides, the $\tilde \chi^{\pm}_1$
mass can be deduced from the transverse momentum
distribution of the muon produced with the chargino.
The reason is that, as we have explained in Section \ref{kin}, the
value of the muon transverse momentum limit $P^{lim}_t(\mu)$ is a
function of the $\tilde \chi^{\pm}_1$ and $\tilde \nu$ masses.
In order to discuss the experimental 
determination of the chargino mass, we
consider separately the scenarios of negligible and significant
ISR effect.

In the scenario where the ISR effect is negligible 
($m_{\tilde \nu_{\mu}}<m_{\tilde \chi_1^{\pm}}+m_{\mu^{\mp}}$ or $m_{\tilde
\nu_{\mu}}>\sqrt s$),
the value of the muon transverse momentum
limit $P^{lim}_t(\mu)$ is equal to
$( E(\mu)^2-m_{\mu^{\pm}}^2 c^4)^{1/2}/c$, $E(\mu)$ being calculated
in a good approximation with Eq.(\ref{enl}),
as described in Section \ref{kin}.
Since Eq.(\ref{enl}) gives $E(\mu)$ as a function of the center of mass
energy and the chargino mass, the experimental value of $P^{lim}_t(\mu)$
would allow to determine the $\tilde \chi_1^{\pm}$ mass. \\
We now estimate the accuracy on the chargino mass measured through this
method. In this method,
the first source of error comes from the fact that to calculate the value
of $E(\mu)$ in $P^{lim}_t(\mu)=( E(\mu)^2-m_{\mu^{\pm}}^2 c^4)^{1/2}/c$,
we use the two-body kinematics formula of
Eq.(\ref{enl}) so that the small ISR effect is neglected.
In order to include this error, we rewrite the transverse momentum kinematic
limit as
$P^{lim}_t(\mu) \pm \delta P^{lim}_t(\mu)=( E(\mu)^2-m_{\mu^{\pm}}^2
c^4)^{1/2}/c$,
$E(\mu)$ being calculated using Eq.(\ref{enl}).
By comparing the value of $E(\mu)$ calculated with Eq.(\ref{enl}) and the
value of
$P^{lim}_t(\mu)$ obtained from the transverse momentum distribution, we have
found that
$\delta P^{lim}_t(\mu) \sim 1 GeV$.
One must also take into account the experimental error on
the measure of the muon transverse momentum expected at linear colliders
which is given by:
$\delta P^{exp}_t(\mu) / P_t(\mu)^2 \leq 10^{-4} (GeV / c)^{-1} $
\cite{f:Tesla}.
Hence, we take $\delta P^{lim}_t(\mu) \sim 1 GeV + P^{lim}_t(\mu)^2 
10^{-4}
(GeV / c)^{-1}$. The experimental error on the muon transverse momentum
limit
depends on the value of $P^{lim}_t(\mu)$ itself and thus on the SUSY
parameters
and on the center of mass energy. However,
we have found that in the single chargino production analysis at $\sqrt
s=500GeV$,
the experimental error on the muon transverse momentum
limit never exceeds $\sim 5 GeV$.
Another source of error in the measure of the $\tilde \chi_1^{\pm}$ mass
is the root mean square spread in center of mass energy $\sigma_{\sqrt s}$
which is given by Eq.(\ref{beam}).
\\ For instance, at the point A with $m_{\tilde \nu}=550GeV$
and an energy of $\sqrt s=500GeV$,
the muon transverse momentum distribution, which is shown in Fig.\ref{noisr},
gives a value for the transverse momentum limit of $P^{lim}_t(\mu)=236.7GeV/c$. 
This value leads through
Eq.(\ref{enl}) to a chargino mass of $m_{\tilde \chi_1^{\pm}}=115.3
\pm 29.7GeV$ taking into account the different sources of error and
assuming a beam resolution of $R=1\%$ at linear colliders.
Here, the uncertainty on the chargino mass is important due to the large value 
of $P^{lim}_t(\mu)$ which increases the error $\delta P^{lim}_t(\mu)$.

In the scenario where the ISR effect is important
($m_{\tilde \nu_{\mu}}>m_{\tilde \chi_1^{\pm}}+m_{\mu^{\mp}}$ and $m_{\tilde
\nu_{\mu}}<\sqrt s$), the three-body kinematics of the reaction
$e^+ e^- \to \tilde \chi_1^{\pm} \mu^{\mp} \gamma$ leaves the muon energy
and thus the whole muon momentum $P(\mu) = ( E(\mu)^2-m_{\mu^{\pm}}^2
c^4)^{1/2}/c$ unfixed by the SUSY parameters, 
since the muon energy depends also on the angle between 
the muon and the photon. Hence, $P(\mu)$ cannot bring
any information on the SUSY mass spectrum. In contrast,
the experimental measure of the muon transverse momentum limit
$P^{lim}_t(\mu)$ would give the $\tilde \chi_1^{\pm}$
mass as a function of the $\tilde \nu_{\mu}$ mass
since $P^{lim}_t(\mu) \approx ( E^{\star}(\mu)^2-m_{\mu^{\pm}}^2
c^4)^{1/2}/c$ (see Section \ref{kin})
and $E^\star(\mu)$ is a function of $m_{\tilde \chi_1^{\pm}}$ and $m_{\tilde
\nu_{\mu}}$ (see Eq.(\ref{enmu}).
Assuming that the $\tilde \nu_{\mu}$ mass has been deduced from
the scan on the center of mass energy mentioned above,
one could thus determine the $\tilde \chi_1^{\pm}$ mass. \\
Let us evaluate the degree of precision in the measure of $m_{\tilde
\chi_1^{\pm}}$ through this calculation.
First, we rewrite the transverse momentum limit as
$P^{lim}_t(\mu) \pm  \delta P^{lim}_t(\mu)=
( E^{\star}(\mu)^2-m_{\mu^{\pm}}^2 c^4)^{1/2}/c$ to take into account
the error $\delta P^{lim}_t(\mu)$ coming from the fact that the
emission angle of the radiated photon with
the beam axis is neglected (see Section \ref{kin}). 
By comparing the value of $E^\star(\mu)$ calculated with Eq.(\ref{enmu}) 
and the value of
$P^{lim}_t(\mu)$ obtained from the transverse momentum distribution, we have
found that $\delta P^{lim}_t(\mu) \sim 1 GeV$.  
In order to consider the experimental error on $P^{lim}_t(\mu)$, we take
as before $\delta P^{lim}_t(\mu) \sim 1 GeV + P^{lim}_t(\mu)^2 
10^{-4} (GeV / c)^{-1}$.
One must also take into account the error on the sneutrino mass.
We assume that this mass has been preliminary determined up to
an accuracy of $\delta m_{\tilde \nu} \sim \sigma_{\sqrt s}$.
\\ For example, the transverse momentum distribution of Fig.\ref{isr},
which has been obtained for the point A 
with $m_{\tilde \nu}=240GeV$ and $\sqrt s=500GeV$,
gives a value for the transverse momentum limit of  
$P^{lim}_t(\mu)=92.3GeV$. The chargino mass derived from this value 
of $P^{lim}_t(\mu)$ via
Eq.(\ref{enmu}) is $m_{\tilde \chi_1^{\pm}}=115.3\pm 5.9GeV$, if 
the various uncertainties are considered and assuming that
$\sigma_{\sqrt s} \sim
3.5GeV$ which corresponds to $R=1\%$ and $\sqrt s=500GeV$ 
(see Eq.(\ref{beam})).


Hence, the $\tilde \chi_1^{\pm}$ mass can be determined
from the study of the peak in
the muon transverse momentum distribution
in the two scenarios of negligible and significant ISR effect.
The regions in the SUSY parameter space where
a peak associated to the signal can be observed over the SUSY background
at the $5 \sigma$ level are shown in Fig.\ref{reach1}.
These domains indicate for which SUSY parameters
$P^{lim}_t(\mu)$ and thus 
the chargino mass can be determined, if we assume
that as soon as a peak is observed in
the muon transverse momentum distribution
the upper kinematic limit of this peak
$P^{lim}_t(\mu)$ can be measured.

\subsubsection{Heaviest chargino}

The single $\tilde \chi^{\pm}_2$ production
$e^+ e^- \to \tilde \chi^{\pm}_2 \mu^{\mp}$ is also
of interest at linear colliders. This reaction, when kinematically open,
has a smaller phase space factor than the single $\tilde \chi^{\pm}_1$
production. Hence,
the best sensitivity on the $\l_{121}$ 
coupling is obtained from the study of the
single $\tilde \chi^{\pm}_1$ production.
However, for sufficiently large values of the $\l_{121}$ coupling,
the $e^+ e^- \to \tilde \chi^{\pm}_2 \mu^{\mp}$ reaction would
allow
to determine either the $\tilde \chi^{\pm}_2$ mass or a relation between
the $\tilde \chi^{\pm}_2$ and $\tilde \nu$ masses.
As described in Section \ref{lightest},
those informations could be derived from the upper kinematic limit
$P^{lim}_t(\mu)$ of the peak associated 
to the single $\tilde \chi^{\pm}_2$ production
observed in the muon transverse momentum distribution.
Indeed, the method presented in the study of 
the single $\tilde \chi^{\pm}_1$ production 
remains valid for the single $\tilde \chi^{\pm}_2$ 
production analysis.

\begin{figure}[t]
\begin{center}
\leavevmode
\centerline{\psfig{figure=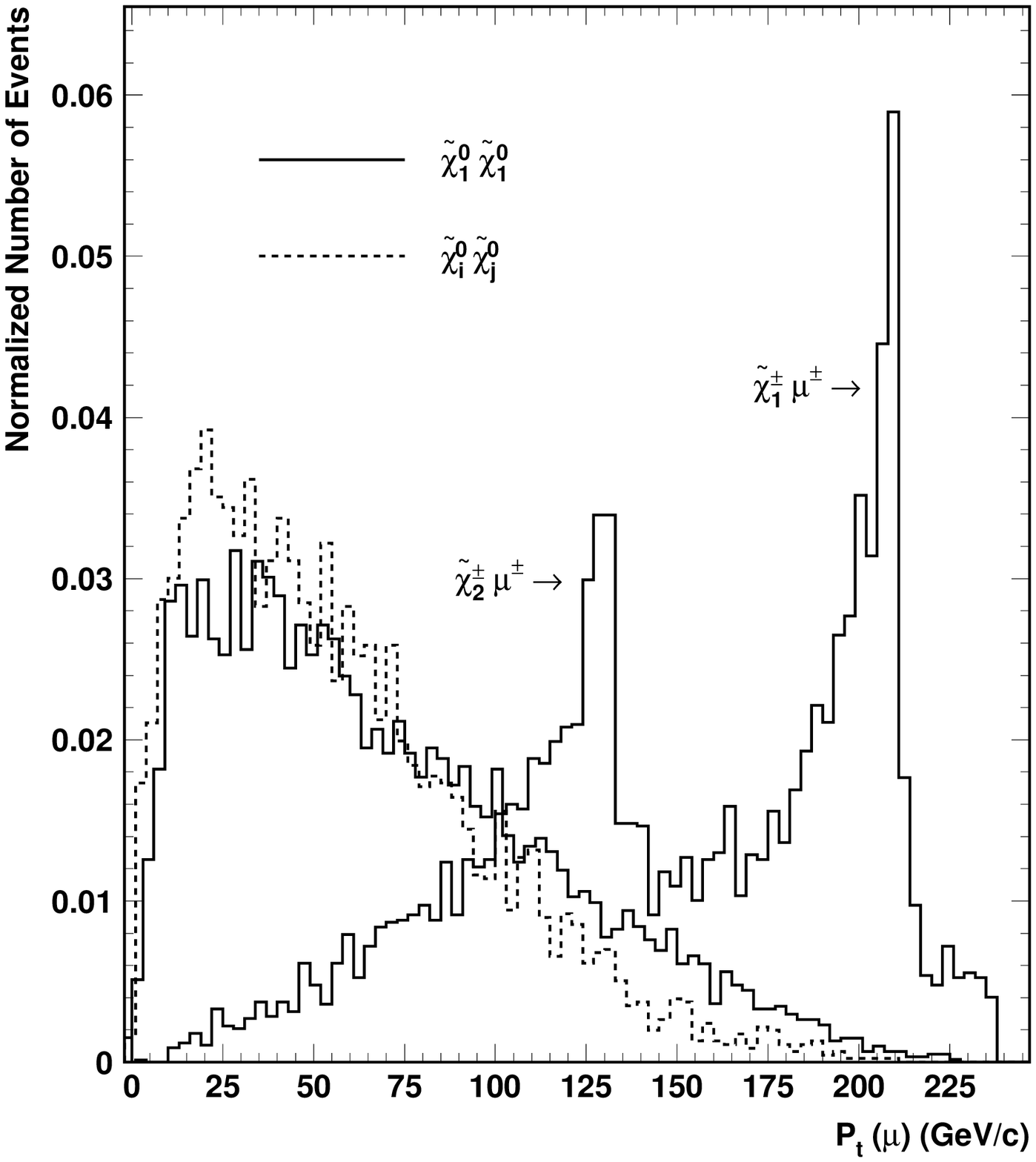,height=5.5in}}
\end{center}
\caption{\footnotesize  \it
Distributions of the highest muon transverse momentum $P_t(\mu)$ (in $GeV/c$)
for the $4l + \Eslash$ events generated by
the single chargino productions
($\tilde \chi^{\pm}_1 \ + \ \tilde \chi^{\pm}_2$) 
and the SUSY background which is
divided into the $\tilde \chi^0_1 \tilde \chi^0_1$ and $\tilde \chi^0_i
\tilde \chi^0_j$ productions.
The number of events for each distribution is normalized to the unity,
the center of mass energy is fixed at $500 GeV$ and
the point A of the SUSY parameter space is considered
with $\l_{121}=0.05$ and $m_{\tilde \nu}=450GeV$.
\rm \normalsize }
\label{isr2}
\end{figure}

The simultaneous determination of the $\tilde \chi^{\pm}_1$ and 
$\tilde \chi^{\pm}_2$ masses is possible since
the peaks in the muon transverse momentum distribution
corresponding to the single $\tilde \chi^{\pm}_1$ and 
$\tilde \chi^{\pm}_2$ productions can be distinguished and identified.
In order to discuss this point,
we present in Fig.\ref{isr2} the muon transverse momentum 
distribution for the $4l + \Eslash$ events generated by
the reactions $e^+ e^- \to \tilde \chi^{\pm}_1 \mu^{\mp}$ and
$e^+ e^- \to \tilde \chi^{\pm}_2 \mu^{\mp}$ 
and by the SUSY background at the MSSM point A 
(for which $m_{\tilde \chi_1^{\pm}}=115.7GeV$ and $m_{\tilde
\chi_2^{\pm}}=290.6GeV$) with
$\l_{121}=0.05$ and $m_{\tilde \nu}=450GeV$.
The cuts described in Section \ref{gsc} have not been applied.
In this Figure, we observe that the peak associated to
the single 
$\tilde \chi^{\pm}_2$ production appears at smaller values
of the transverse momentum
than the peak caused by the single
$\tilde \chi^{\pm}_1$ production.
This is due to the hierarchy of the chargino masses. Indeed,
the chargino masses enter
the formula of Eq.(\ref{enmu}) which gives the values of 
$P^{lim}_t(\mu) \approx
( E^{\star}(\mu)^2-m_{\mu^{\pm}}^2 c^4)^{1/2}/c$
for the point A with $m_{\tilde \nu}=450GeV$. 
The same difference between
the two peaks is observed in the scenario where the values of
$P^{lim}_t(\mu) \approx ( E(\mu)^2-m_{\mu^{\pm}}^2 c^4)^{1/2}/c$ 
are calculated using the formula
of Eq.(\ref{enl}), namely for $m_{\tilde \nu_{\mu}}<m_{\tilde
\chi_1^{\pm}}+m_{\mu^{\mp}}$ and  $m_{\tilde \nu_{\mu}}<m_{\tilde
\chi_2^{\pm}}+m_{\mu^{\mp}}$, or $m_{\tilde \nu_{\mu}}>\sqrt s$.
Fig.\ref{isr2} also illustrates the fact
that the peaks associated to the single
$\tilde \chi^{\pm}_1$ and $\tilde \chi^{\pm}_2$ productions can be
easily identified thanks to their relative heights.
The difference between the heights of the two peaks is due to the relative 
values of the cross sections and branching ratios which read 
for instance with 
the SUSY parameters of Fig.\ref{isr2} as
(including the beam polarization described in Section \ref{polar}):
$\sigma(\tilde \chi^{\pm}_1 \mu^{\mp})=620.09fb$, 
$\sigma(\tilde \chi^{\pm}_2 \mu^{\mp})=605.48fb$, 
$B(\tilde \chi_1^{\pm} \to \tilde \chi^0_1  l^{\pm} \ \Eslash)=33.9\%$ and
$B(\tilde \chi_2^{\pm} \to \tilde \chi^0_1  l^{\pm} \ \Eslash)=10.8\%$.

\subsection{Extension of the analysis to different center of mass energies}

In this section, we comment on a similar study of the
reaction $e^+ e^- \to \tilde \chi^{\pm} \mu^{\mp}$ 
based on the $4l+\Eslash$ events at center of mass energies
different from $\sqrt s=500GeV$.

First, the muon transverse momentum distributions depend mainly on the
relative values of the center of mass energy and of the various SUSY masses,
so that
the discussion on the cuts given in Section \ref{cuts} still hold for
different energies.
The reconstruction of the $\tilde \chi^{\pm}_{1,2}$ and
$\tilde \nu$ masses through the methods exposed in
Section \ref{marec} is possible at any center of mass energy.
We thus discuss in this section the sensitivity on the $\l_{121}$
coupling that would be 
obtained at other center of mass energies than $\sqrt s=500GeV$.

Similarly, the values of the branching ratios are function of the SUSY mass spectrum,
as shown in Sections \ref{signalX} and \ref{SUSYX},
and thus do not change if the center of mass energy is modified.

Besides, the amplitude of the ISR effect on the signal
cross section depends on the relative values of 
$m_{\tilde \nu}$, $m_{\tilde \chi^{\pm}_1}$ and $\sqrt s$
(see Section \ref{kin}). The shapes of the exclusion curves 
obtained at different center of mass energies would thus 
be similar to the shapes of the exclusion plots presented 
in Fig.\ref{reach1} for same relative values of 
$m_{\tilde \nu}$, $m_{\tilde \chi^{\pm}_1}$ and $\sqrt s$.

However, the cross sections of the SUSY backgrounds, namely 
the $\tilde \nu_p \tilde \nu^*_p$ and $\tilde \chi^0_i
\tilde \chi^0_j$ productions, depend strongly on the 
center of mass energy which determines the phase space factors 
of the superpartners pair productions.

Therefore, the sensitivity on the $\l_{121}$ coupling tends to 
decrease (increase) at higher (smaller) center of mass energies
due to the increase (decrease) of the SUSY background rates.

\subsection{Study based on the $3l+2jets+\Eslash$ final state}
\label{hadron}

In this Section, we would like to emphasize the interest of the
$3l+2jets+\Eslash$ final state for the study of the reaction 
$e^+ e^- \to \tilde \chi^{\pm} \mu^{\mp}$ at linear colliders.

First, the $3l+2jets+\Eslash$ final state is generated by the
decay $\tilde \chi^{\pm} \to \tilde \chi^0_1 d_p u_{p'}$ 
which has a larger branching ratio than the decay
$\tilde \chi^{\pm} \to \tilde \chi^0_1  l_p \nu_p$
for the hierarchy $m_{\tilde \nu},m_{\tilde l^{\pm}},
m_{\tilde q}>m_{\tilde \chi^{\pm}_1}$,
as mentioned in Section \ref{signalX}.
\\ Secondly, 
the \sm background of the $3l+2jets+\Eslash$ signature
is the $WWZ$ production. The rate of the
$3l+2jets+\Eslash$ production 
from the $e^+ e^- \to WWZ $ reaction is
$1.5fb$ ($0.5fb$) at $\sqrt s=500GeV$ ($350GeV$)
including the cuts $\vert \eta (l)\vert<3$, $P_t(l)>10GeV$
and neglecting the ISR \cite{f:Godbole}.
This background can be further reduced as explained in Section
\ref{SMback}. Besides, the $3l+2jets+\Eslash$ signature
has no $R_p$-conserving SUSY background
if one assumes that the LSP is the $\tilde \chi^0_1$
and that the single dominant \rpv coupling is $\l_{121}$.
Indeed, the pair productions of SUSY particles typically
lead to final states wich contain at least 4 charged
leptons due to the decay of the two LSP's
through $\l_{121}$ as $\tilde \chi^0_1 \to l \bar l \nu$.
\\ Therefore, the sensitivity on the $\l_{121}$ 
coupling obtained from the study of the
single chargino production based on the 
$3l+2jets+\Eslash$ final state should be greatly
enhanced with respect to the analysis 
of the $4l+\Eslash$ signature.

The $3l+2jets+\Eslash$ final state is also attractive from the mass
reconstruction point of view. Indeed, the full decay chain
$\tilde \chi^{\pm}_1 \to \tilde \chi^0_1 d_p u_{p'}$,
$\tilde \chi^0_1 \to l \bar l \nu$ can be reconstructed.
The reason is that, since the muon produced together with the chargino in
the reaction
$e^+ e^- \to \tilde \chi^{\pm}_1 \mu^{\mp}$ 
can be identified (see Section \ref{kin})
the origin of each particle in the final state can 
be known. First, the $\tilde \chi^0_1$ 
mass can be measured with the distribution of
the invariant mass of the two charged leptons coming from the $\tilde
\chi^0_1$ decay:
The value of the $\tilde \chi^0_1$ mass is readen at the upper endpoint of
this distribution. Similarly, the upper endpoint 
of the invariant mass distribution of the two jets coming
from the chargino decay gives the mass difference $m_{\tilde
\chi^{\pm}_1}-m_{\tilde \chi^0_1}$.
Since $m_{\tilde \chi^0_1}$ has already been determined from the
dilepton invariant mass,
we can deduce from this mass diffference the $\tilde \chi^{\pm}_1$ mass.
\\ Besides, the $\tilde \nu$ and $\tilde \chi^{\pm}_{1,2}$ 
masses can be measured as explained
in Section \ref{marec}.
\\ Hence, in the case of a non-vanishing $\l$ coupling with $\tilde
\chi^0_1$ as the LSP,
the combinatorial background for the $\tilde \chi^0_1$
mass reconstruction from the study of the single
chargino production based on the $3l+2jets+\Eslash$ final state is expected
to be greatly reduced
with respect to the analysis based on the pair production
of SUSY particles \cite{f:Atlas}, due to the better
identification of the charged leptons coming from the $\tilde \chi^0_1$
decay.
Furthermore, while the $\tilde \chi^{\pm}_{1,2}$ and $\tilde \nu$
masses reconstructions are possible
via the single chargino production, these 
reconstructions appear to be more difficult
with the pair production of SUSY particles.
Indeed, the $\tilde \chi^{\pm}$ and $\tilde \nu$
masses reconstructions from the superpartner pair production
are based on the $\tilde \chi^0_1$ reconstruction, and
moreover in the pair prodution 
the $\tilde \chi^{\pm}$ or $\tilde \nu$
decays lead to either an additional uncontroled missing energy
or an higher number of charged particles (or both)
in the final state with respect to the single chargino 
production signature.

\subsection{Study of the single chargino production
through various \rpv couplings}

The single chargino production at muon colliders
$\mu^+ \mu^- \to \tilde \chi^{\pm} l^{\mp}_m$ 
(see Fig.\ref{graphe}) would allow
to study the $\l_{2m2}$ coupling, $m$ being equal to either 
$1$ or $3$ due to the antisymmetry of the $\l_{ijk}$ couplings.

The study of the $\l_{212}$ coupling would be essentially identical
to the study of $\l_{121}$.
Only small modifications would enter the analysis:
Since an electron/positron would be produced together with the chargino,
one should require that the final state contains at least one electron
instead of one muon as in the $\l_{121}$ case (Section \ref{gsc}).
Besides, one should study the distribution of the highest electron
transverse
momentum instead of the highest muon transverse momentum. 
The particularities of the muon colliders 
would cause other differences in the study:
First, the analysis would suffer an additional background
due to the $\mu$ decays in the detector. Secondly,
large polarization would imply sacrifice in luminosity
since at muon colliders this is achieved by keeping only the
larger $p_z$ muons emerging from the target \cite{f:muon}. Therefore,
one expects an interesting sensitivity
on the $\l_{212}$ coupling 
from the single $\tilde \chi^{\pm}_1$ production at muon
colliders
although this sensitivity should not be as high
as in the study of the $\l_{121}$ 
coupling at linear colliders. \\
Besides, the $\mu^+ \mu^- \to \tilde \chi^{\pm} e^{\mp}$ reaction occuring
through
$\l_{212}$ would allow to determine the $\tilde \chi^0_1$, $\tilde \chi^{\pm}_1$,
$\tilde \chi^{\pm}_2$ and $\tilde \nu$ masses
as described in Sections \ref{marec} and \ref{hadron}
for the single chargino production
at linear colliders. The main difference would be the following.
The accuracy in the determination of the sneutrino mass
performed through a scan over the center of mass energy
would be higher at muon colliders than at linear colliders.
The reason is the high beam resolution $R$ expected
at muon colliders (see Section \ref{lightest}).
For instance, at muon colliders the beam resolution should reach $R \sim 0.14\%$
for a luminosity of ${\cal L}=10fb^{-1}$ and
an energy of $\sqrt s=300-500GeV$ \cite{f:muon}.

The production of a single chargino together with a tau-lepton
occurs through the $\l_{131}$ coupling at linear colliders and
via $\l_{232}$ at muon coliders.
Due to the $\tau$-decay, the transverse momentum
distribution of the produced $\tau$-lepton
cannot be obtained with a good accuracy.
The strong cut on the transverse momentum described
in Section \ref{kin} is thus difficult to apply in that case, and one
must think of other discrimination variables such as
the total transverse momentum of the event, the total missing energy,
the rapidity, the polar angle, the isolation angle,
the acoplanarity, the acollinearity
or the event sphericity. Therefore,
the sensitivities on the $\l_{131}$ and $\l_{232}$ couplings obtained
from the $\tilde \chi^{\pm} \tau^{\mp}$ production should be weaker than the
sensitivities expected for $\l_{121}$ and $\l_{212}$
respectively. Furthermore, the 
lack of precision in the $\tau$ transverse momentum distribution
renders the reconstructions of the $\tilde \chi^0_1$, $\tilde \chi^{\pm}_1$
and $\tilde \chi^{\pm}_2$ masses  
from the $\tilde \chi^{\pm} \tau^{\mp}$
production (see Sections \ref{marec} and \ref{hadron}) difficult.

\subsection{Single neutralino production}

We have seen in Sections \ref{kin} and \ref{reach} that when
the sneutrino mass is smaller than the lightest chargino mass
(case of small ISR effect),
the single chargino production cross section is greatly reduced
at linear colliders.
The reason is that in this situation the radiative return to the sneutrino
resonance allowed by the ISR is not possible anymore.
The interesting point is that for 
$m_{\tilde \chi^0_1}<m_{\tilde \nu}<m_{ \tilde \chi^{\pm}_1}$,
the radiative return to the sneutrino
resonance remains possible in the single $\tilde \chi^0_1$ production
through $\l_{1m1}$: $e^+ e^- \to \tilde \chi^0_1 \nu_m$.
Therefore, in the region 
$m_{\tilde \chi^0_1}<m_{\tilde \nu}<m_{ \tilde \chi^{\pm}_1}$ the
single $\tilde \chi^0_1$ production has a larger rate than
the single $\tilde \chi^{\pm}_1$ production and is thus attractive
to test the $\l_{1m1}$ couplings. 
\\ We give now some qualitative comments
on the backgrounds of the single neutralino production.
If we assume that the $\tilde \chi^0_1$ is the LSP,
the $\tilde \chi^0_1 \nu_m$
production leads to the $2l+\Eslash$ final state due to the \rpv decay
$\tilde \chi^0_1 \to l \bar l \nu$. This signature
is free of $R_p$-conserving SUSY background since the pair production
of SUSY particles produces at least 4 charged
leptons due to the presence of the LSP at the end of each of 
the 2 cascade decays.
The \sm background is strong as it comes from the
$WW$ and $ZZ$ productions but it can be reduced by some kinematic cuts.
\\ Therefore, in the region 
$m_{\tilde \chi^0_1}<m_{\tilde \nu}<m_{ \tilde \chi^{\pm}_1 }$,
one expects an interesting sensitivity on $\l_{1m1}$
from the single neutralino production study 
which was for instance considered in \cite{f:feng2}
for muon colliders.

The single $\tilde \chi^0_1$ production allows also to reconstruct
the $\tilde \chi^0_1$ mass. Indeed, 
the neutralino mass is given by the
endpoint of the distribution of the 2 leptons invariant mass
in the $2l+\Eslash$ final state generated by the single 
$\tilde \chi^0_1$ production.
The combinatorial background is extremely weak since
the considered final state contains only 2 charged 
leptons.

If the sneutrino is the LSP, when it is produced at the resonance
through $\l_{1m1}$ as $e^+ e^- \to \tilde \nu_m$,
it can only decay as $\tilde \nu_m \to e^+ e^-$
assuming a single dominant \rpv coupling constant.
Hence, in this scenario, the sneutrino resonance must be studied trhough
its effect on the Bhabha scattering \cite{f:KalZ,f:Kalept,f:Kalp,f:Kalseul}.
This conclusion also holds in the case of nearly degenerate $\tilde \nu$
and $\tilde \chi^0_1$ masses since then the decay $\tilde \nu_m \to  e^+
e^-$
is dominant compared to the decay $\tilde \nu_m \to \tilde \chi^0_1 \nu_m $.

\section{Conclusion}

The study at linear colliders of the single chargino production 
via the single dominant $\l_{121}$ coupling
$e^+ e^- \to \tilde \chi^{\pm} \mu^{\mp}$ 
is promising, due to the high luminosities and energies
expected at these colliders. 
Assuming that the $\tilde \chi^0_1$ is the LSP, the
singly produced chargino has 2 main decay channels: 
$\tilde \chi^{\pm} \to \tilde \chi^0_1 l^{\pm} \nu$ and 
$\tilde \chi^{\pm} \to \tilde \chi^0_1 u d$.

The leptonic decay of the produced chargino 
$\tilde \chi^{\pm} \to \tilde \chi^0_1 l^{\pm} \nu$ 
(through, virtual or real, sleptons and $W$-boson) leads to the clean 
$4l+\Eslash$ final state which is almost free of \sm background. 
This signature suffers a large SUSY background
in some regions of the MSSM parameter space but
this SUSY background can be controled
using the initial beam polarization and some 
cuts based on the specific kinematics of the single chargino production.
Therefore, considering a luminosity of ${\cal L}=500fb^{-1}$
at $\sqrt s=500GeV$ and assuming the largest SUSY background,
values of the $\l_{121}$ coupling smaller
than the present low-energy bound could be probed over a range of the sneutrino 
mass of $\Delta m_{\tilde \nu} \approx 500GeV$ 
around the sneutrino resonance and at the $\tilde \nu$ 
pole the sensitivity on $\l_{121}$ could reach values of order $10^{-4}$.
Besides, the $4l+\Eslash$ channel could allow 
to reconstruct the $\tilde \chi_1^{\pm}$,
$\tilde \chi_2^{\pm}$ and $\tilde \nu$ masses.

The hadronic decay of the produced chargino
$\tilde \chi^{\pm} \to \tilde \chi^0_1 u d$
(through, virtual or real, squarks and $W$-boson)
gives rise to the $3l+2j+\Eslash$ final state.
This signature, free from SUSY background,
has a small \sm background and should thus also
give a good sensitivity
on the $\l_{121}$ coupling constant.
This hadronic channel should allow to reconstruct the $\tilde \chi_1^0$,
$\tilde \chi_1^{\pm}$, $\tilde \chi_2^{\pm}$ and $\tilde \nu$ masses.

The sensitivity on the $\l_{131}$ coupling constant, obtained from 
the $\tilde \chi^{\pm} \tau^{\mp}$ production study, is expected to be 
weaker than the sensitivity found on $\l_{121}$ due to the decays of 
the tau-lepton.

\section{Acknowledgments}

The author is grateful to M. Chemtob, H.-U. Martyn 
and Y. Sirois for helpful discussions and having encouraged this work. 
It is also a pleasure to thank N. Ghodbane and M. Boonekamp for 
instructive conversations.

\clearpage

\vspace{10 mm}
\begin{center}
{  }
\end{center}
\vspace{10 mm}

\clearpage

\clearpage

\vspace{10 mm}
\begin{center}
{  }
\end{center}
\vspace{10 mm}

\clearpage 

\setcounter{chapter}{0}
\setcounter{section}{0}
\setcounter{subsection}{0}
\setcounter{figure}{0}

\chapter*{Publication VII}
\addcontentsline{toc}{chapter}{Broken R parity  contributions to
 flavor changing  rates and CP   asymmetries 
 in fermion pair production at leptonic  colliders}

\newpage

\vspace{10 mm}
\begin{center}
{  }
\end{center}
\vspace{10 mm}

\clearpage

\begin{center}
{\bf \huge Broken R parity  contributions to
 flavor changing  rates and CP   asymmetries 
 in fermion pair production at leptonic  colliders }
\end{center} 
\vspace{2cm}
\begin{center}
M. Chemtob, G. Moreau
\end{center} 
\begin{center}
{\em  Service de Physique Th\'eorique \\} 
{ \em  CE-Saclay F-91191 Gif-sur-Yvette, Cedex France \\}
\end{center} 
\vspace{1cm}
\begin{center}
{Phys. Rev. {\bf D59} (1999) 116012, hep-ph/9806494}
\end{center}
\vspace{2cm}
\begin{center}
Abstract  
\end{center}
\vspace{1cm}
{\it We examine the 
effects of the  R parity odd renormalizable interactions 
on flavor changing rates  and CP asymmetries  in the production of 
fermion-antifermion pairs at leptonic  (electron and muon)  colliders.
In the reactions, $l^-+l^+\to f_J +\bar f_{J'}, \
[l=e , \ \mu ;  \  J\ne J' ]$ 
the  produced fermions may be  leptons, down-quarks   or up-quarks,
and the center of mass energies may range  from
the Z-boson  pole up to $ 1000$ GeV.
Off the  Z-boson pole, the flavor changing rates  are controlled by tree  level 
amplitudes and the  CP asymmetries
by interference  terms between tree and loop level  amplitudes. At the 
Z-boson pole, both observables involve loop amplitudes.
The  lepton number violating interactions,
associated with the coupling constants, $\l_{ijk} , \ \l '_{ijk}$, are only taken into account.
The consideration of  loop amplitudes  is
restricted  to the  photon and Z-boson vertex corrections.
We  briefly review flavor violation physics at colliders.
We present numerical  results  using a single, species and  family independent, 
mass parameter, $\tilde m$,  
for all the scalar superpartners and considering simple assumptions for
the family dependence of the R parity odd coupling constants. 
Finite non diagonal  rates (CP asymmetries) entail non vanishing  
products of two (four) different coupling constants in different family configurations.  
For lepton pair  production,   the Z-boson decays  branching ratios, 
$B_{JJ'}= B(Z\to l^-_J+l^+_{J'})$, scale in order of magnitude as, 
$ B_{JJ'}\approx ({\l \over 0.1})^4 ({100 GeV \over \tilde m} )^{2.5}
 \ 10^{-9} $, with coupling constants $\l = \l_{ijk} $ or 
 $\l '_{ijk}$ in appropriate family 
configurations.  The corresponding results for d- and u-quarks 
are larger, due to an extra color factor, $N_c=3$.   
The  flavor non diagonal rates, at  energies  well above the 
Z-boson pole,  slowly decrease  with  the  center of mass energy and scale  with  the 
mass parameter approximately as,
 $ \s_{JJ'}\approx  ({\l \over 0.1})^4 ({100 GeV \over \tilde m})^{2 \ - \ 3}
(1 \ - \ 10)  fbarns$.  Including the contributions from 
an sneutrino s-channel exchange 
could raise the rates  for leptons or d-quarks by one order of magnitude.  
The CP-odd asymmetries at the Z-boson pole, 
${\cal A}_{JJ'}={B_{JJ'}-B_{J'J}\over B_{JJ'}+B_{J'J} } $, vary inside the range, 
 $(10^{-1}\ -\  10^{-3}) \sin \psi $, where $\psi $ is the CP-odd phase.
At energies  higher   than the Z-boson pole,  CP-odd asymmetries  for 
 leptons,  d-quarks and u-quarks pair production  lie 
approximately  at,  $(10^{-2} \ - \ 10^{-3}) \sin \psi $,
irrespective  of whether  one deals with   light or heavy 
flavors.}

\newpage

\section{Introduction}
\label{secintro}

\setcounter{equation}{0}

An approximate  R parity  symmetry could greatly enhance our insight 
into the  supersymmetric flavor problem. 
As is known,  the  dimension four R parity odd 
superpotential trilinear  in the quarks and leptons superfields, 
\begin{equation}
W_{R-odd}=\sum_{i,j,k} \bigg (\ud \l _{ijk} L_iL_j E^c_k+
\l ' _{ijk} Q_i L_j D^c_k+
\ud \l '' _{ijk} U_i^cD_j^cD_k^c   \bigg )  \  , 
\label{eqi1}
\end{equation}
adds new  dimensionless  couplings
in the family  spaces of the  quarks and leptons and their superpartners.  
Comparing with  the analogous situation for the  Higgs-meson-matter
Yukawa interactions, one naturally expects  the set of 45  dimensionless 
 coupling constants, $\l_{ijk} = -\l_{jik}, \ 
\l '_{ijk} , \ \l ''_{ijk} =- \l ''_{ikj}$, to exhibit  a non-trivial  
hierarchical structure  in  the families spaces.
Our goal in this work will be  to examine a particular  class of tests 
at  high energy colliders by which one could 
access a  direct information on the family structure of  these coupling constants.

The R parity symmetry has inspired a vast literature since the 
pioneering period  of the early 80's
\cite{g:aulak,g:zwirner,g:hallsuz,g:lee,g:ellisvalle,g:rossvalle,g:dawson,g:masiero} 
and the maturation period of the   late 80's  and early 90's
\cite{g:dimohall,g:bargerg,g:dreiner1,g:ibross,g:hinchliffe,g:moha}. This subject 
 is currently witnessing a renewed  interest \cite{g:reviews,g:reviews2}. 
As is well known, the R parity odd interactions can contribute at tree level,
by exchange of the scalar superpartners, 
to processes   which violate the  baryon and lepton  numbers  as well as the  leptons and 
quarks flavors. The major part of the  existing  experimental
constraints on coupling constants is formed from the  indirect  bounds 
gathered  from the low energy phenomenology.
Most often, these have been derived on the basis 
of the  so-called  single coupling  hypothesis, 
where a single  one of the  coupling constants is assumed to 
dominate over all the others, so that
each  of the  coupling constants  contributes once at a time.
Apart from a few  isolated  cases, 
the typical bounds derived under this assumption,  
 assuming  a linear dependence  on the superpartner masses, are of order,   
$[ \l ,\  \l' ,\  \l '' ] <  ( 10^{-1} - 10^{-2}) 
  { \tilde  m\over 100 GeV }$. 

One important variant of the 
single coupling  hypothesis can be defined by assuming that the 
dominance of single operators applies at the level of the gauge (current) 
basis fields rather than the mass eigenstate fields, as was implicit in the 
above original version. 
This  appears as a more natural  assumption in models where 
the presumed hierarchies in coupling constants originate from physics at higher 
 scales (gauge, flavor, or strings).  Flavor changing  
contributions  may then be induced even when 
a single  R parity odd coupling constant is assumed to  dominate \cite{g:agashe}. 
While the redefined mass basis superpotential  may then  depend
on the various  unitary transformation matrices, $V_{L,R}^{u, d} $,
\cite{g:ellis98}, two distinguished predictive 
choices  are those where the  generation mixing  is represented
solely  in terms of the CKM (Cabibbo-Kobayashi-Maskawa)  matrix, with 
flavor changing effects  appearing in  either up-quarks 
or down-quarks flavors
\cite{g:agashe}. A similar situation holds for leptons with respect to 
the couplings, $\l _{ijk}$, and transformations, $V_{L,R}^{l,\nu } .$

A large set of constraints has also been obtained 
by  applying an extended hypothesis
of  dominance  of coupling constants by pairs (or more).
 Several analyses  dealing with  hadron flavor changing effects 
 (mixing parameters for the neutral light and
 heavy  flavored  mesons, rare mesons   decays  such as,
 $K\to \pi +\nu +\bar \nu $, ...)  \cite{g:agashe};
 lepton flavor changing  effects (leptons decays, 
 $l_l^\pm \to l_k^\pm+l_n^-+l^+_p, $ \cite{g:roy}
 $ \quad \mu ^- \to  e^-$  conversion  processes, \cite{g:kim}, 
 neutrinos Majorana mass \cite{g:godbole}, ...); lepton 
number violating effects (neutrinoless double beta decay 
\cite{g:hirsch,g:babu,g:hirsch1}); or  baryon number violating  
effects (proton decay partial branchings \cite{g:smirnov},
 rare non-leptonic decays of heavy mesons \cite{g:carlson},
nuclei desintegration \cite{g:carlson2},...)   have   led to 
strong bounds on a large number of quadratic  products  of the 
coupling constants. All of the above low energy works, however, suffer  
from one  or other form of  model dependence, whether they   rely   
on the consideration of loop diagrams  \cite{g:smirnov},
on additional assumptions concerning the flavor mixing \cite{g:agashe,g:roy,g:kim},
or  on  hadronic  wave functions inputs \cite{g:carlson,g:carlson2}.

Proceeding  further with a linkage of R parity with 
physics beyond the standard model, our main  observation  in this work 
is that the R parity odd coupling constants  could by themselves 
be an independent  source of CP violation. 
Of course, the idea that the RPV interactions could act as a source of 
superweak CP violation is not a new one in the supersymmetry literature.
The principal motivation is that, whether the  RPV interactions operate 
by themselves or in  association with the gauge interactions, by 
exploiting  the  absence   of  strong  constraints  on 
violations with respect to the flavors of  quarks, leptons  and the
scalar superpartners by the RPV interactions, one could greatly enhance  
the potential for observability of CP violation. 
To our knowledge, one of the earliest discussion of this possibility  
is contained in  ref.\cite{g:masiero}, where the r\^ole
 of a relative  complex phase in a pair 
of $\l'_{ijk}$ coupling constants was analyzed
in connection with  the neutral $ K, \ \bar K $ 
mesons  mixing and decays  and also with the neutron 
electric dipole moment.   This subject has attracted
increased interest in the recent literature 
\cite{g:grossman,g:kaplan,g:abel,g:guetta,g:jang,g:chen,g:adhikari,g:shalom,g:shalom2,g:handoko}. 
Thus, 
the r\^ole of complex  $\l'_{ijk} $ coupling constants 
was considered in an analysis of the muon 
polarization in  the decay, $ K^+ \to \mu^ + +\nu +\g $  \cite{g:chen}, 
and also of  the neutral $ B, \ \bar B$ meson CP-odd decays asymmetries 
\cite{g:kaplan,g:guetta,g:jang}; 
that of complex $\l_{ijk} $ interactions was considered 
in a study  of the spin-dependent asymmetries of 
sneutrino-antisneutrino resonant production of $\tau -$lepton pairs,
$l^-l^+\to \tilde \nu , \bar {\tilde \nu } \to \tau ^+ \tau ^-$  
\cite{g:shalom};  and that of  
complex  $\l''_{ijk}$   interactions was considered 
as a possible explanation for 
the cosmological baryon asymmetry \cite{g:adhikari},  
as well as  in the  neutral $B , \ \bar B$  decays asymmetries \cite{g:jang}. 
An interesting alternative proposal \cite{g:abel} 
is to embed the CP-odd phase in the 
scalar superpartner interactions corresponding  to interactions of 
$A'_{ijk} \l'_{ijk}$ type.
Furthermore, even if one  assumes that the 
R parity odd interactions are  CP conserving, these could  still lead, 
in combination with the  other possible  sources  of complex phases 
in the minimal supersymmetric standard model, to new tests of  CP violation.  
Thus, in the hypothesis  of pair of dominant coupling constants 
new contributions involving the   coupling constants $\l '_{ijk}$ 
and the CKM complex phase can arise for  CP-odd  observables associated with  
the  neutral mesons mixing parameters and decays
\cite{g:kaplan,g:guetta,g:jang}.  Also, 
through the interference with the  extra CP-odd phases present 
in the soft supersymmetry parameters,  $A$, 
the interactions $\l_{ijk} $ and $\l '_{ijk}$
may induce  new contributions to  electric dipole moments  \cite{g:hamidian}.

We propose in this work to examine  the effect that 
R parity odd CP violating  interactions could have  on  flavor non-diagonal rates
and CP asymmetries in  the production at high energy colliders 
of fermion-antifermion pairs of different 
families. We consider the two-body  reactions, $l^-(k)+l^+(k')\to f_J(p)
+\bar f_{J'}(p'), \ [J\ne J']$ where $l $ stands for electron or muon, 
the produced fermions are leptons, down-quarks   or up-quarks and
the center of mass energies span the 
relevant range of existing and planned leptonic  (electron  or muon)  colliders, namely,
from the Z-boson  pole up to $ 1000 $ GeV.  High energy colliders tests of the 
RPV contributions to the flavor diagonal reactions were
recently examined in \cite{g:choud96,g:kalin12,g:kalin13,g:kalin14} and for 
flavor non-diagonal reactions in \cite{g:mahanta}.

The physics of CP non conservation at high energy colliders  has motivated 
a  wide variety  of proposals in the past \cite{g:cpcoll}  and is currently
the focus of important activity.
In this work we shall limit ourselves to 
the simplest kind of observable,  namely,  the spin independent observable
involving  differences in rates  between a given flavor
 non-diagonal process and its CP  conjugated process. 
While the R parity odd interactions contribute to 
flavor changing amplitudes already at tree level, their contribution to 
spin  independent CP-odd observables entails the consideration of loop diagrams. Thus, the 
  CP asymmetries in the  Z-boson pole branching fractions,
$B (Z\to f_J+\bar f_{J'}) $, are controlled by a complex phase 
interference between non-diagonal flavor contributions to loop amplitudes,  
whereas the off Z-boson pole asymmetries  are controlled instead  by 
a complex phase 
interference  between tree and loop  amplitudes. Finite contributions 
at tree level order can arise for spin dependent CP-odd observables, as 
discussed in refs.~\cite{g:shalom,g:shalom2}.

It is useful to recall at this point that contributions
in the standard model  to the  flavor
changing rates and/or CP asymmetries   can only appear
through  loop diagrams involving the
quarks-gauge bosons interactions.  Corresponding contributions involving
squarks-gauginos or  sleptons-gauginos interactions also arise
in the minimal supersymmetric standard model. In 
studies performed some time ago
within the  standard model,  the flavor non diagonal  vector bosons 
(Z-boson and/or W-bosons) decay rates  asymmetries   
\cite{g:axelrod,g:clements,g:ganapathi} and  CP-odd 
asymmetries \cite{g:bernabeu,g:hou} were found to be  exceedingly small.
(Similar conclusions were reached  in top-quark phenomenology \cite{g:eilam}.)
On the other hand, in most proposals of physics beyond the standard model, 
the prospects for observing flavor changing effects in 
rates \cite{g:axelrod,g:clements,g:ganapathi,g:bernabeu,g:hou,g:atwood} or 
in CP asymmetries \cite{g:cpcoll,g:atwood1} are on the  optimistic side. Large effects
  were  reported for the supersymmetric  corrections
in flavor changing Z-boson decay rates arising from squarks flavor mixings  
\cite{g:duncan},   but the conclusions from this initial work have been 
challenged in a subsequent work \cite{g:mukho} involving a more complete calculation.

The possibility that the R parity odd interactions could  contribute 
to the CP asymmetries at observable  levels 
depends  in the first place on the accompanying mechanisms 
responsible for the flavor changing rates. Our working assumption in this work will  be
that the R parity odd interactions are the dominant contributors 
to flavor non-diagonal amplitudes.

The contents of this paper  are organized into 
4 sections. In Section \ref{sec1}, 
we develop the basic formalism for describing 
the scattering  amplitudes 
at tree and one-loop levels. We discuss the case of leptons, 
down-quarks and up-quarks 
successively in subsections \ref{subsec1}, \ref{subsec2} and  \ref{subsec3}. 
The evaluation of  the one-loop  loop 
diagrams is based on the standard formalism 
of \cite{g:pasvel}.  Our calculations here  closely parallel
similar ones developed \cite{g:ellis,g:bhatta3} in  connection with corrections  to the 
Z-boson partial widths.  In Section  \ref{sec5}, we first  
briefly review the physics of flavor violation  and next present our numerical results
for the  integrated cross sections (rates)  and   CP asymmetries for fermion pair 
production at and off the Z-boson pole. In Section \ref{seconc}, we state our main conclusions 
and  discuss the impact of 
our results on possible experimental measurements.  

\section{Production of fermion pairs of different flavors}
\vskip 0.5cm
\label{sec1}

\setcounter{equation}{0}

\begin{figure}[t]
\begin{center}
\leavevmode
\psfig{figure=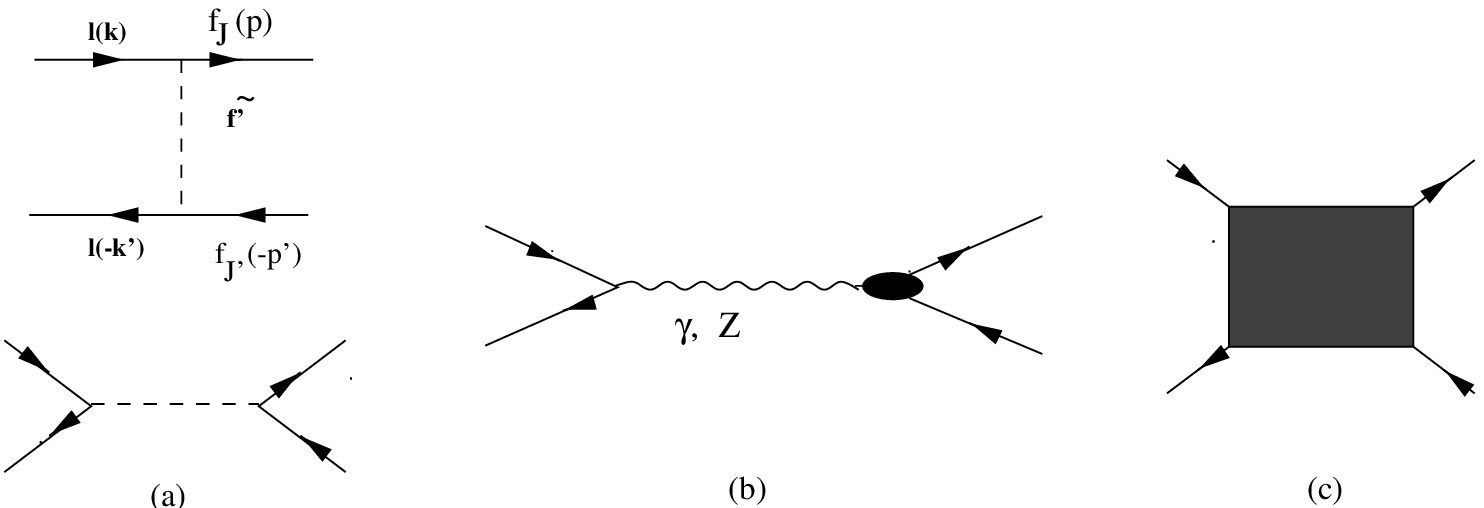}
\end{center}
\caption{Flavor non-diagonal process of $l^-l^+$ production of 
a fermion-antifermion pair, $l^-(k)+l^+(k')\to f_J (p) +\bar f_{J'} 
(p') $. The tree level diagrams in $(a)$ represent t- and s-channel exchange 
amplitudes. 
The  loop   level diagrams represent
$\g $  and $ Z $  gauge boson exchange amplitudes with dressed vertices 
in $(b)$  and box amplitudes in $(c)$.}
\label{g:fig1}
\end{figure}

In this section  we shall examine the contributions induced by 
the RPV (R parity violating)  couplings on the flavor changing processes, 
$l^-(k)+l^+(k')\to f_J(p)
+\bar f_{J'}(p'), \ [l=e,\ \mu ; \ J\ne J']$ where $f$ stands for leptons or quarks and 
$J,  \   J' $ are family  indices.  The relevant 
tree and one-loop level diagrams are shown schematically in Fig. \ref{g:fig1}.
At one-loop order, there arise $\g  -$ and  $Z-$
boson exchange triangle diagrams as well as  box  diagrams.  
In the sequel, for clarity,  we shall 
present the formalism for the one-loop contributions
only for the dressed $Z f\bar f$ vertex in the
Z-boson exchange amplitude.  The dressed  $\g $-exchange amplitude has 
a similar structure and  will be added in together 
with the Z-boson exchange  amplitude at the level of the  numerical results.
Since we shall repeatedly refer in the text to the 
R parity odd effective Lagrangian  
for the fermions-sfermion Yukawa interactions, we quote below its full expression,
\begin{eqnarray}
L&=&\sum_{ijk} \bigg \{ \ud \l_{ijk}[\tilde  \nu_{iL}\bar e_{kR}e_{jL} +
\tilde e_{jL}\bar e_{kR}\nu_{iL} + \tilde e^\star _{kR}\bar \nu^c_{iR} e_{jL}
-(i\to j) ]  \cr
&+&\l '_{ijk}[\tilde  \nu_{iL}\bar d_{kR}d_{jL} +
\tilde d_{jL}\bar d_{kR}\nu_{iL} + \tilde d^\star _{kR}\bar \nu^c_{iR} d_{jL}
-\tilde  e_{iL}\bar d_{kR}u_{jL} -
\tilde u_{jL}\bar d_{kR}e_{iL} - \tilde d^\star _{kR}\bar e^c_{iR} u_{jL}
]  \cr 
&+&\ud {\l ''}_{ijk}\e_{\a \b \g }[\tilde  u^\star _{i\a R}\bar d_{j\b R}d^c_{k\g L} +
\tilde  d^\star _{j\b R}\bar u_{i\a R}d^c_{k\g L} +
\tilde  d^\star _{k\g R}\bar u_{i\a R}d^c_{j\b L}  -(j\to k)] \bigg \} +h.c. \ ,
\cr & &
\label{g:eqyuk}
\end{eqnarray}
noting that the summations run  over the (quarks and leptons) 
families indices, 
$i,j,k  =[(e,\mu , \tau ); \  (d,s,b); \ (u,c,t)]$, subject to the 
antisymmetry properties, $\l_{ijk}=-\l_{jik}, \   \l ''_{ijk}=-\l '' _{ikj}$.
We use precedence conventions for operations on Dirac spinors such that
charge conjugation acts first, chirality projection second and Dirac bar 
third, so that, $\bar \psi ^c_{L,R}= \overline {(\psi ^c)_{L,R} }.$
\subsection{ Charged lepton-antilepton pairs}
\label{subsec1}
\subsubsection{General formalism}
The process  $l^-(k)+l^+(k')\to e^-_J(p) +e^+_{J'}(p')$, for 
$l=e,\  \mu ; \ J\ne J'$, can pick up a finite contribution at tree level 
from  the R parity odd couplings, 
$\l_{ijk}$, only. For clarity, we treat in the following the case 
of electron colliders, noting that the case of muon colliders is 
easily deduced by replacing all occurrences in the RPV coupling constants 
of the index $1$ by the index $2$.
There occur both  t-channel and s-channel  $\tilde \nu_{iL}$ 
exchange contributions, of the type shown by the Feynman diagrams in $(a)$ 
of  Fig. \ref{g:fig1}. 
The scattering amplitude at tree level, $M_t$,  reads: 
\begin{eqnarray}
&M^{JJ'}_t&= -{1\over 2(t-m^2_{\tilde \nu_{iL} } )  } 
\bigg [ \l_{i1J} \l^\star_{i1J'}
\bar u_R(p) \g_\mu v_R(p')\bar v_L(k')\g_\mu u_L(k)  \cr
&+&\l ^\star _{iJ1} \l _{iJ'1}
\bar u_L(p) \g_\mu v_L(p')\bar v_R(k')\g_\mu u_R(k) \bigg ] \cr
&-& {1 \over s-m^2_{\tilde \nu_{iL} } }\bigg [\l _{i11}\l ^\star _{iJJ'}
\bar v_R(k') u_L(k) \bar u_L(p) v_R(p') +
\l _{i11}^\star \l _{iJJ'}
\bar v_L(k') u_R(k) \bar u_R(p) v_L(p') \bigg ] , \cr & &
\label{eq1}
\end{eqnarray}
where to obtain the saturation structure in the 
Dirac spinors indices for the t-channel terms, 
we have applied the  Fierz rearrangement formula, 
\hfill  $\bar u_R(p) u_L(k) \bar v_L(k')v_R(p') =\ud 
\bar u_R(p) \g_\mu v_R(p')\bar v_L(k')\g_\mu u_L(k) .$
The   t-channel (s-channel) exchange terms
on the right hand side of eq.(\ref{eq1}) include two terms each, 
called ${\cal R}$- and ${\cal L}$-type, respectively. These two terms   differ 
by a chirality flip,  $L\leftrightarrow  R $,  and correspond 
to the distinct diagrams  where the  exchanged
sneutrino is emitted or absorbed at the upper (right-handed) vertex.

The Z-boson exchange amplitude (diagram  $(b)$ in Fig. \ref{g:fig1})
 at loop level, $M_l$,  reads:
\begin{eqnarray}
M^{JJ'}_l = \bigg ({g\over 2 \cos \t_W}\bigg )^2 \bar v (k')\g_\mu \bigg  
(a(e_L) P_L+a(e_R)P_R \bigg )u(k) 
{1\over s-m_Z^2+im_Z\G_Z} \G_\mu ^Z (p,p') , 
\label{eq2}
\end{eqnarray}
where the Z-boson current amplitude   vertex function, $\G^Z_\mu  (p,p')$, 
is defined through  the effective Lagrangian density, 
$$ L= -{g\over 2 \cos \t_W } Z^\mu  \G_\mu ^Z (p,p').$$ 
For later convenience, we  record for the  processes, 
$Z(P=p+p') \to f(p)+\bar f'(p')$ and $Z(P)\to \tilde f_H(p)  +
\tilde f_H ^\star  (p')$, 
the familiar definitions of the Z-boson bare vertex functions, 
\begin{eqnarray}
\G _\mu ^Z(p,p') =  \bigg [\bar f(p)\g_\mu \bigg  (
a(f_L)P_L +a(f_R) P_R\bigg )
f'(p')  + (p-p')_\mu  
\tilde f_H^\star (p') 
a(\tilde  f_H)\tilde f_H(p) \bigg ], 
\label{eqm1}
\end{eqnarray}
where  the quantities denoted,   $a(f_H)\equiv a_H(f)$  and 
$a(\tilde f_H) $, taking equal values for both fermions and sfermions, 
are defined by,  $ a(f_H)=a(\tilde f_H)=2T_3^H (f)-2Qx_W, $ where  
$ H=(L,R), \ x_W= \sin^2 \t_W , \ T_3^H $ are $ SU(2)_H $  Cartan subalgebra 
generators, and 
$Q= T_3^L +Y , \ Y  $ are  electric charge and weak hypercharge.  These 
parameters satisfy  the useful  relations:  
$a(\tilde f_H^\star ) =-a(\tilde f_H), \  a_L(f^c)=-a_R(f), 
\  a_R(f^c)=-a_L(f).$ 
Throughout this paper we shall use the conventions in Haber-Kane review 
\cite{g:haber} (metric signature  $(+---)$, 
$P_{L\choose R}=(1\mp \g_5)/2,   $ etc...) and adopt 
the  familiar summation convention on dummy indices. 

The  Lorentz covariant  structure of the dressed Z-boson  current amplitude
in the process,
$Z(P)\to f_J(p) +\bar f_{J'}(p')$,  for a generic value of 
the Z-boson invariant mass $s=P^2$, involves 
three pairs  of vectorial and tensorial vertex functions, 
which are defined in terms of  the general decomposition:
\begin{eqnarray}
 \G _\mu ^Z(p,p')&=&
\bar u(p) \bigg [\g_\mu 
\bigg (\tilde A_L^{JJ'} (f)P_L+
\tilde A_R^{JJ'}(f) P_R\bigg ) \cr 
&+& {1\over m_J+m_{J'}} 
\s_{\mu \nu } \bigg ( (p+p')^\nu  [ia^{JJ'}  + \g_5 d^{JJ'} ] 
+(p-p')^\nu  [ ib^{JJ'} +\g_5 e^{JJ'} ] 
\bigg ) \bigg ]  v(p')  \ , \cr
&&
\label{eq3}
\end{eqnarray}
where,  $ \s_{\mu \nu } = {i\over 2} [\g_\mu , \g_\nu ] $. 
The vector vertex functions  separate additively into
the classical (bare)  and loop contributions, 
$\tilde A_H^{JJ'}(f)=a_H(f)\d_{JJ'}  +A_H^{JJ'}(f), \ [H=L,R]$. 
The  tensor vertex functions, associated with $\s^{\mu \nu } (p+p')_\nu $,  
include the familiar magnetic and 
electric $Z \ f\  \bar f$ couplings, such that  the flavor diagonal vertex functions,
$-{g\over 2 \cos \t_W} {1\over 2m_J} [a^{JJ}, d^{JJ} ]$,   
 identify, in the small momentum transfer limit, 
with  the fermions Z-boson current 
magnetic and  (P and CP-odd) electric dipole moments, respectively. 
In working with  the spinors matrix elements,
 it is helpful to recall the mass shell relations,
\hfill    $ \bar u (p)  \pslash = m_J \bar u(p) , \ \quad  
\pslash '  v(p') =- m_{J'}   v(p')$, and
the Gordon type identities, appropriate to the saturation of 
the Dirac spinor indices, $\bar u(p) \cdots v(p')$,  
$$ \bigg [(p\pm p')_\mu  {\g_5 \choose 1} 
+i\s_{\mu \nu }
(p\mp p')^\nu   {\g_5 \choose 1} \bigg ]= (m_J+m_{J'}) \g_\mu {\g_5 
\choose 1} ,$$
$$ \bigg [(p\mp p')_\mu  {\g_5 \choose 1} 
+i\s_{\mu \nu }
(p\pm p')^\nu   {\g_5 \choose 1} \bigg ]= (m_J-m_{J'}) \g_\mu {\g_5 
\choose 1 } . $$
Based on these identities, one also checks that the additional 
vertex functions,  $[b^{JJ'} , \ e^{JJ'}]$,
associated with the Lorentz covariants,  $\s^{\mu \nu } (p-p')_\nu  [1, \g_5]$, 
can be expressed as linear combinations of the 
vector  or  axial covariants, $\g_\mu \  [1,\g_5] $,  and the total 
momentum covariants, $(p+p')_\mu \  [1,\g_5] $. The latter will yield, upon contraction 
with the initial state $Zl^-l^+$ vertex function,
to  negligible mass terms in the initial leptons.

Let us now perform the summation 
over the  initial and final states  polarizations for the summed tree and 
loop amplitudes, $M^{JJ'}= M_t^{JJ'}+M_l^{JJ'}$,
 where the lower suffices $t , l $ stand 
for tree and loop, respectively. 
(We shall not be interested in this work  in spin observables.)
A straightforward calculation, carried out for the
squared sum of the tree and loop amplitudes, yields the result 
(a useful textbook to consult  here is ref. \cite{g:peskder}): 
\begin{eqnarray}
\sum_{pol}&& \vert M^{JJ'}_t+M^{JJ'}_l\vert^2= \bigg 
\vert -{  \l_{i1J} \l^\star _{i1J'} \over 
2(t-m^2_{\tilde \nu_{iL} } )}  +\bigg ({g\over 2\cos \t_W}\bigg )^2
{  a(e_L) A_R^{JJ'}(e, s+i\e ) \over 
s-m_Z^2+im_Z\G_Z   } \bigg \vert^2 16 (k\cdot p)(k'\cdot p') \cr &+&8m_Jm_{J'} 
(k\cdot k') \varphi_{LL} ({\cal R})   +8m_e^2 (p\cdot p') 
\varphi_{RR}({\cal R}) + \cr 
&+& \bigg \vert -{  \l_{iJ1} \l^\star _{iJ'1} \over 
2(t-m^2_{\tilde \nu_{iL} } )}  +\bigg ({g\over 2\cos \t_W}\bigg )^2
{  a(e_R) A_L^{JJ'}(e, s+i\e ) \over 
s-m_Z^2+im_Z\G_Z   } \bigg \vert^2 16 (k\cdot p)(k'\cdot p')\cr 
&+&8m_Jm_{J'} 
(k\cdot k') \varphi_{RR}({\cal L})   +
8m^2_e 
(p\cdot p') \varphi_{LL}({\cal L})  
+8\bigg  \vert {\l_{i11} \l ^\star _{iJJ'} \over 
s-m^2_{\tilde \nu_{iL} } } \bigg \vert ^2 (k\cdot k') (p\cdot p'),  
\label{eqr4}
\end{eqnarray}
where we  have introduced the following functions,
associated with the ${\cal R} $- and ${\cal L}$-type contributions:
\begin{eqnarray}
\varphi_{HH'} ({\cal R}) &=& -\bigg ({g\over 2\cos \t_W}\bigg )^2
\bigg ({ a(e_H) A_{H'}^{JJ'}(e, s+i\e ) \over 
s-m_Z^2+im_Z\G_Z }\bigg )^\star  \bigg ({ \l_{i1J}\l_{i1J'}^\star \over 
2(t-m^2_{\tilde \nu_{iL} } )  } \bigg ) +c.\ c\  \ , \cr  
\varphi_{HH'} ({\cal L}) &=& -\bigg ({g\over 2\cos \t_W}\bigg )^2
\bigg ({ a(e_H) A_{H'}^{JJ'}(e, s+i\e ) \over 
s-m_Z^2+im_Z\G_Z }\bigg )^\star  \bigg ({ \l_{iJ1}\l_{iJ'1}^\star \over 
2(t-m^2_{\tilde \nu_{iL} } )  } \bigg ) +c.\ c\  \ . 
\label{eq4p}
\end{eqnarray}
The two sets of terms in
 eqs.(\ref{eq4p}) and (\ref{eqr4}), labelled by the letters, $ {\cal R},  {\cal L}$, are
 associated with the two t-channel exchange contributions 
in the tree amplitude, eq.(\ref{eq1}),  which  differ
by the  spinors chirality structure  and the substitutions, 
$\l _{i1J} \l ^\star _{i1J'} \to  
\l^\star _{iJ1} \l _{iJ'1} .$
The terminology,  ${\cal L, \ R } $, is motivated  by the fact that these
contributions are controlled by the Z-boson left and right 
chirality vertex functions, $A_L $ and $A_R$, respectively, in the massless limit.

The  imaginary shift  in the 
argument,  $s+ i \e $ (representing the upper lip of 
the cut  real axis in the complex $s$-plane) of the vertex functions,
$A_H^{JJ'}(f,s+i\e )$, has been appended 
to remind us that the
one-loop  vertex functions are complex functions 
in the complex plane of  the virtual Z-boson mass squared, 
$s=(p+p')^2$, with  branch cuts  starting at the physical thresholds 
where  the production processes, such as, $ Z\to f+\bar f $ or $Z\to \tilde  f
+\tilde f^\star $,  are raised on-shell.
For notational simplicity, we have omitted writing
several terms proportional to the initial leptons masses 
and also some of the small  subleading 
terms  arising from  the loop amplitude squared.  
At the energies of interest,  whose scale is set 
by the initial center of mass energy or by the  Z-boson mass,
 the  terms involving factors of the initial leptons masses $m_e$, 
are entirely negligible, of course.  Thus, 
the  contributions associated with $\varphi_{RR} ({\cal R}),\ 
\varphi_{LL}({\cal L})   $ 
can  safely be dropped.
Also, the contribution from $\varphi_{LL} ({\cal R}) $ and 
$\varphi_{RR} ({\cal L}) $
which are proportional to
the final state leptons masses, $m_J, $ and $  m_{J'}$, 
can to a good approximation be  neglected for leptons production.
Always in the same approximation, we find also that interference terms 
are absent between the s-channel 
exchange and     the 
t-channel amplitudes and between  the s-channel  tree and Z-boson  exchange 
loop    amplitudes.  Similarly,  because of the opposite 
chirality structure of the first two terms in  $M_t^{JJ'}$, 
their cross-product contributions give negligibly small mass terms.
\subsubsection{CP asymmetries}
Our main concern in this work bears on  the comparison of 
the pair of CP conjugate reactions,
$l^-(k)+l^+(k')\to e^-_J(p) +e^+_{J'}(p')$ and 
$l^-(k)+l^+(k')\to e^-_{J'}(p) +e^+_{J}(p')$. 
Denoting the summed tree and one-loop  probability  amplitudes for these reactions as,
$ M^{JJ'}= M_t^{JJ'}+M_l^{JJ'} , \ 
\bar  M^{JJ'}= M_t^{J'J}+M_l^{J'J} = M^{J'J}$, we observe that these amplitudes are simply
related  to one another  by means of a specific  complex conjugation operation. 
The general structure  of this relationship can be expressed schematically as:
\begin{eqnarray}
M^{JJ'}=a_0^{JJ'}+\sum_\a a_\a ^{JJ'}  F^{JJ'}_\a (s+i\e ), \ \ 
\bar M^{JJ'}=a^{JJ' \star } _0+\sum_\a a^{JJ'\star } _\a  
F^{J'J} _\a (s+i\e ),
\label{eqst1}
\end{eqnarray}
where for each of the  equations above,  referring to amplitudes 
for pairs of CP conjugate processes, 
the first and second terms  correspond to  the tree
and loop level contributions, with $a^{JJ'}_0 , \ a_0^{J'J}= a_0^{JJ'\star }$, 
representing   the tree  amplitudes and
$a_\a ^{JJ'},  \ a_\a ^{J'J} = a_\a ^{JJ' \star }$ 
and $F_\a ^{JJ'} , \ F_\a ^{J'J} =F_\a ^{JJ'} $ 
representing the  complex valued  coupling constants products and 
momentum integrals in the loop amplitudes.  The functions 
$F^{JJ'} $ must  be symmetric under the interchange, $J\leftrightarrow J'$.
The summation index $\a  $ labels  the  family
configurations for the intermediate  fermions-sfermions which can
run inside the  loops. 
Defining  the CP asymmetries  by the normalized  differences,
$${\cal A}_{JJ'}={ \vert M^{JJ'} \vert ^2- \vert \bar  M^{JJ'} \vert ^2 \over 
\vert M^{JJ'} \vert ^2+ \vert \bar  M^{JJ'} \vert ^2},$$ 
and inserting the decompositions in eq.(\ref{eqst1}), the result
separates additively into two types of terms:
\begin{eqnarray}
{\cal A}_{JJ'}&=&{2\over \vert a_0\vert ^2}\bigg [ 
\sum_\a  Im(a_0a_\a ^\star ) Im (F_\a (s+i\e )) \cr 
&-& \sum_{\a < \a ' } Im(a_\a a_{\a '}^\star ) Im (F_\a (s+i\e )
F_{\a '}^\star (s+i\e ) )\bigg ],
\label{eq5}
\end{eqnarray}
where, for  notational simplicity, we have suppressed the fixed external  family
indices on  $ a_{0 }^{JJ'} , \ a_{\a } ^{JJ'} $ and $ F_\a ^{JJ'}$,
and   replaced the
full denominator by the tree level amplitude, since this is expected 
to dominate over the loop amplitude.
The first term  in (\ref{eq5}) is associated with an interference between 
tree and loop amplitudes  and the second  with an
interference between  terms arising from different family contributions 
in the loop amplitude.
In the second term of eq.(\ref{eq5}),
the two imaginary parts  factors are antisymmetric  under  the 
interchange of indices,  $\a $ and $\a '$,  so that their  product is 
symmetric  and allows one to  
write, $\sum_{\a < \a ' } = \ud \sum_{\a \ne \a '} .$
To obtain a more explicit formula, let us  specialize 
to the specific case where the Z-boson  vertex functions  decompose as, 
$ A_H^{JJ'}(f,s+i\e ) = \sum_\a  b_{JJ'}^{H\a }
I_{H\a }^{JJ'}  (s+i\e )$.  The  first factors,   $b^{H\a }_{JJ'} = 
\l_{ijJ} \l^\star _{ijJ'}  $ (using $\a =(ij) $ and 
notations for the one-loop contributions to be 
described in the next subsection), include the CP-odd phase from 
the  R parity odd coupling constants. The  second factors,
$I_{H \a } ^{JJ'} $,  include the CP-even phase from the  unitarity cuts 
associated to the physical on-shell  intermediate states.
In the notations  of eq.(\ref{eqst1}), 
$$  a_{\a }^{JJ'}= ({g\over 2\cos \t_W })^2   a(e_{H'})  b^{H\a } _{JJ'} , \  
F_{\a }^{JJ'} =  I_{H\a } ^{JJ'}  (s+i\e )/(s-m_Z^2+im_Z\G_Z),$$
where the right hand sides
incorporate appropriate sums over the chirality indices, $H', \ H$ of the initial and final 
fermions, respectively.

Applying eq.(\ref{eqst1}) to the 
  asymmetry  integrated  with respect to the scattering angle,
one derives for the corresponding integrated  tree-loop interference contribution,
\begin{eqnarray}
{\cal A}_{JJ'}&=&-4\bigg ({ g\over 2\cos \t_W }\bigg )^2
a(e_L) Im(\l_{i1J}\l^\star _{i1J'} a_{JJ'}^{\a \star }(f_R) )
Im\bigg ({  I^R_\a (s+i\e )\over s-m_Z^2+im_Z \G_Z }\bigg  ) \cr 
&\times &\int_{-1}^{1}dx\  {(1- x)^2\over 
(2(t-m_{\tilde \nu_{iL}}^2 ) } 
\bigg [ \sum_i \vert \l_{i1J}\l_{i1J'}^\star \vert^2 
\int_{-1}^1 dx \  { (1- x)^2\over  4(t-m_{\tilde \nu^2_{iL} })^2 } 
\bigg ] ^{-1}\ , 
\label{eq6}
\end{eqnarray}
where,   $ \theta , \ [x=\cos \theta ]$ denotes  the scattering angle variable
in  the center of mass frame
and the  Mandelstam variables in the  case of 
massless final state fermions take the simplified expressions, 
$ s \equiv (k+k')^2 , \ t  \equiv (k-p)^2=  -\ud s (1- x), 
\ u \equiv  (k-p')^2   = -\ud s (1+ x)$.  
Useful kinematical relations in  the  general  case with 
final fermions masses, $m_J, \ m_{J'}$,  are: 
$\sqrt s=2 k= E_p+E_{p'}, \  t=  m_J^2 -s E_p (1-\b x) , \ u= m_{J'}^2 -s E_{p'}(1 +\b ' x), $
where, $E_p =(s+m_J^2-m_{J'}^2)/(2\sqrt s ),\ 
E_{p'}  =(s+m_{J'}^2-m_{J}^2)/(2\sqrt s ), \ \b = p/E_p,\ \b ' =p/E_{p'}$,
with  $k, \ p $  denoting  the 
center of mass  momenta of the two-body initial and final states, respectively.
The unpolarized differential cross section reads then, 
$ d\s /dx = {\vert p \vert \over 
128 \pi s \vert k \vert } \sum_{pol} \vert M \vert^2.$ 

For the Z-boson pole  observables, 
the flavor non-diagonal  branching ratios and
CP asymmetries (where one sets, $s=m_Z^2$)  are
defined   in terms of the  notations specified in the preceeding paragraph by the equations, 
\begin{eqnarray}
B_{JJ'}&\equiv & {\G (Z\to f_J+\bar f_{J'} )+ \G (Z\to f_{J'} +\bar f_J ) \over \G (Z \to all)  }
=2  {  \vert A_L^{JJ'} (f) \vert^2 +  \vert A_R^{JJ'} (f) \vert^2 
\over \sum_f \vert a_L (f) \vert^2 +  \vert a_R (f) \vert^2 } ,\cr
{\cal A}_{JJ'}&\equiv & {\G (Z\to f_J+\bar f_{J'} )- \G (Z\to f_{J'} +\bar f_J ) \over
\G (Z\to f_J+\bar f_{J'} )+ \G (Z\to f_{J'} +\bar f_J )}  \cr   &=&
- 2 {\sum_{H=L,R}\sum_{\a < \a ' } Im(b^{H\a  } _{JJ'} b^{ H\a '\star }_{JJ'} )
Im (I_{H\a }  ^{JJ'} (s+i\e ) I_{H \a '}^{JJ'\star }  (s+i\e ) ) \over 
\sum_{H=L,R}\vert \sum_\a  b_{JJ'}^{H\a } (f)
F_H^\a (s+i\e ) \vert ^2  }  \ .
\label{eq6p}
\end{eqnarray}
For completeness, we  recall the formula for the Z-boson decay width in 
fermion pairs (massless limit), 
$$\G (Z\to f_J+\bar f_{J'}) = {G_F m_Z^3 c_f\over 12 \sqrt {2} \pi }
(\vert A_L^{JJ'}(f) \vert ^2 +\vert A_R^{JJ'}(f) \vert ^2 ),$$
where, $c_f=[1,\ N_c], $ for $[f=l,q] \ (N_c=3  $ 
is the number of colors in the $ SU(3)_c $ color  group) and the 
experimental value  for the total width,
$\G (Z \to  all )_{exp}= 2.497 $ GeV.

The expressions in eqs.(\ref{eq6}) and (\ref{eq6p})  for the CP asymmetries
explicitly incorporate the property of these observables of  depending on 
combinations of the RPV coupling constants, such as, 
$Arg ( \l _{i1J} \l ^\star _{i1J'} 
\l^\star _{i'jJ} \l _{i'jJ'} ), $ or $Arg ( \l _{ijJ} \l ^\star _{ijJ'} 
\l^\star _{i'j'J} \l _{i'j'J'} )$,  which are invariant 
under complex phase redefinitions of the fields.  This freedom
under rephasings of the quarks and leptons superfields actually removes 
$21$ complex phases from the complete general set of $ 45$  complex 
RPV coupling constants.

\subsubsection{One-loop  amplitudes}
\begin{figure}[t]
\begin{center}
\leavevmode
\psfig{figure=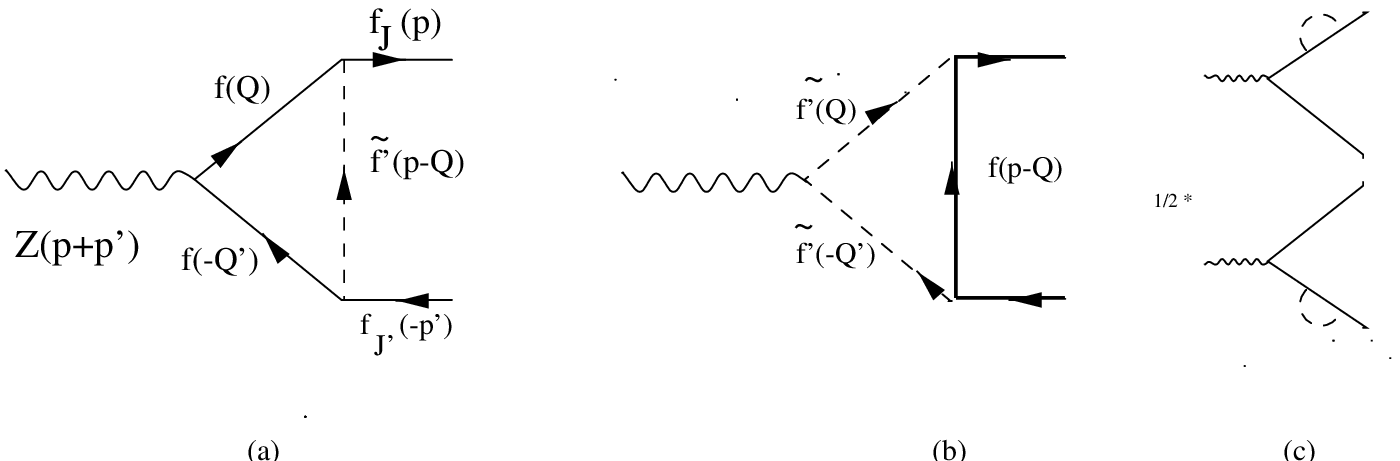}
\end{center}
\caption{One-loop diagrams for the dressed $Z(P) \ f(p) \bar f(p') $ vertex.
The flow of four-momenta for the intermediate  fermions  in $(a)$ 
is  denoted  as, 
$ Z(P)\to f(Q)+\bar f(Q') \to f_J(p) +\bar 
f_{J'} (p') $.  Similar notations are used for the sfermions diagram 
in $(b)$ where,  $ Z(P)\to \tilde f'(Q) +\tilde f ^{'\star } (Q') $, 
and for the self-energy diagrams in $(c)$.  }
\label{g:fig2}
\end{figure}

The relevant triangle Feynman diagrams,
which contribute to the dressed  Z-boson  leptonic vertex, 
$Z(P) l^- (p)  l^+ (p')$, appear in three types,  fermionic, scalar and self-energy, 
as shown in Fig.  \ref{g:fig2}. 
We consider first the contributions induced by the R parity odd
couplings, $\l '_{ijk} $.
The intermediate lines  can assume two distinct
configurations which  contribute both,
in the limit of vanishing external fermions masses, 
to the left-chirality vertex functions only.  We shall refer 
to such contributions by the name  ${\cal L}$- type contributions, 
reserving the name ${\cal R}$-type to contributions to the
right-chirality vector couplings. 
The two  allowed configurations  for the internal fermions and 
sfermions are:
$ f= {d_k \choose u_j^c} ; \tilde f'= 
{\tilde u_{jL}^\star  \choose  \tilde d_{kR}} .$
Our calculations of the triangle diagrams employ the kinematical conventions  
for the flow of electric charge  and momenta 
indicated in Fig.\ref{g:fig2}, where $ P=p+p'=Q+Q'=k+k'$.
The  summed fermion and scalar  Z-boson current contributions are given by: 
\begin{eqnarray}
\G_\mu ^Z ({\cal L}) &=&+i  N_c {\l '}^\star _{Jjk}{\l '}_{J'jk} 
\bigg [ \int_Q 
{  \bar u(p)[P_R(\qslash +m_f )\g_\mu (a(f_L)P_L+a(f_R)P_R) (-\qslash '+m_f) P_L]v(p')
\over (Q^2-m_f^2)((Q-p-p')^2-m_f^2)((Q-p)^2-m_{\tilde f'}^2) }\cr
&+&  \int_Q 
{ a(\tilde f'_L)(Q-Q')_\mu  \bar u(p)[P_R(\pslash -\qslash +m_f )P_L] v(p')
\over (Q^2-m_{\tilde {f}'}^2)((Q-p-p')^2-m_{\tilde {f}' }^2)
((Q-p)^2-m_f^2) } \bigg ].
\label{eq7}
\end{eqnarray}
The integration measure is defined as, $ \int_Q = {1\over (2\pi )^4} \int d^4 Q$.
For a convenient derivation   of the  self-energy diagrams, one may invoke 
the on shell  renormalization condition which relates these to the
fields renormalization constants.
Defining schematically   the self-energy  vertex functions for a Dirac fermion field $\psi $
by the  Lagrangian density, $ L=i \bar \psi 
(\pslash  -m +\S (p) )  \psi , \ \S (p)=
m \s_0 +\pslash  (\s^{L} P_L+\s^{R}P_R )$,
the transition from bare  to renormalized fields and mass terms  may be 
effected by the replacements, 
$$ \psi_H \to {\psi_H\over (1+\s_{H})^\ud } = \psi_HZ_H^\ud , \ 
\  m \to  m {(1+\s^{L})^\ud (1+\s^R)^\ud \over (1-\s_0)   } .\ \  [H=L,R]$$ 
By a straightforward generalization to the case of several fields, labelled by a family index $J$,
the fields renormalization constants become matrices, 
$Z^H_{JJ'}= (1+\s^H)_{JJ'}^{-1} $. The 
self-energy contributions  to 
the dressed Z-boson vertex function is then described as,
\begin{eqnarray}
\G ^Z _{\mu }(p,p')_{SE} &=& \sum_{H=L,R}\bigg ( (Z^H_{JJ'}
Z^{H\star }_{J'J} )^\ud -1 \bigg ) \bar u(p) \g_\mu a(f_H) P_H v(p')
\cr  &=&- \sum_{H=L,R} \ud (\s^H_{JJ'} (p)+
\s^{H\star }_{J'J}(p') )  \bar u(p) \g_\mu a(f_H) P_H v(p'),
\label{eqp8}
\end{eqnarray}
where for the case at hand,
\begin{eqnarray}
\S_{JJ'}(p)&=& -i N_c {\l '}^\star _{Jjk}{\l '}_{J'jk} 
\int_Q { P_R(\qslash +m_f)P_L\over (-Q^2+m_f^2)
(-(Q-p)^2+m_{\tilde {f}'}^2 ) } ,
\label{eq8}
\end{eqnarray}
so that $\s_{JJ'}^R=0$ and $\s_0=0$. Similar
Feynman graphs to those of  Fig. \ref{g:fig2}, 
and similar formulas to those of eqs.(\ref{eq7}) and (\ref{eqp8}),
obtain for the dressed photon  current case, $\g (P) l^-(p)  l^+(p')$.

We organize our one-loop calculations 
in line with the  approach  developed by 
't Hooft and Veltman \cite{g:hooft}  and Passarino and  Veltman\cite{g:pasvel},
keeping in  mind that our spacetime  metric has an 
opposite signature  to theirs, $(-+++)$.
For definiteness, we recall  the conventional  notations for 
the two-point and three-point integrals,
\begin{eqnarray}
{i\pi^2\over {(2\pi )^4}} 
[B_0, -p_\mu B_1 ]=
\int_Q { [1, Q_\mu ] \over  (-Q^2+m_1^2) (-(Q-p)^2
+m_2^2)   } ,
\label{eq9p}
\end{eqnarray}
\begin{eqnarray}
&{i\pi^2\over {(2\pi )^4}}&
[C_0,\  -p_\mu C_{11}-p'_{\mu} C_{12} ,  \ 
p_\mu p_\nu C_{21}+ p'_\nu p'_\mu C_{22} +(p_\mu p'_\nu + 
p_\nu p'_\mu ) C_{23}
\   -g_{\mu \nu } C_{24} ] \cr 
&=& \int_Q  {  [1,Q_\mu ,  Q_\mu Q_\nu ] \over 
(-Q^2+m_1^2) (-(Q-p)^2+m_2^2)  (-(Q-p-p')^2
+m_3^2)  } ,
\label{eq9pp}
\end{eqnarray}
where the arguments for the $B- $ and $C-$ functions are defined as: 
$B_A (-p,m_1,m_2),  \  \ [A=0,1] $ and $   \  
C_B (-p, -p',m_1,m_2,m_3),  \ \  [B=0,11,12, 21,22,23,24] $.
In the algebraic derivation of the  one-loop amplitudes, 
we find it convenient to introduce the  definitions: 
$ p_\mu =\ud P_\mu +\rho_\mu , \ p'_\mu =\ud P_\mu -\rho_\mu , $ where $P=p+p', \ 
\rho = \ud (p-p')$.  The terms 
proportional to the Lorentz covariant $P^\mu = (p+p')^\mu $   will then  reduce, for the  full
Z-boson exchange amplitude in eq. (\ref{eq2}),  to negligible mass terms 
in the initial leptons. 

Dropping mass terms for all external fermions, the tensorial couplings cancel out and 
we need keep track of the vector couplings only, with the result:
\begin{eqnarray}
A_L^{JJ'}({\cal L}) &=& N_c\FFL \bigg [ a (f_L) m_f^2 C_0 
+a(f_R)\bigg (B_0^{(1)}-2C_{24}-m_{\tilde {f}'}^2C_0\bigg  )
+2a(\tilde {f}'_L) \tilde C_{24} + a(f_L) B_1^{(2)} \bigg ], \cr
A_R^{JJ'}({\cal L})&= &0. 
\label{eq9}
\end{eqnarray}
The cancellation of the right chirality vertex function  in this case 
is the reason behind our naming these contributions as ${\cal L}$-type.
The  two-point and three-point integrals functions  without 
a tilde symbol arise through the  fermion current  triangle contribution 
and  the self-energy contribution (represented by the term proportional 
to $ B_1^{(2)}$). These
 involve the argument variables  according to the following conventions, 
\hfill $B_A^{(1)}= B_A(-p-p', m_f, m_f), \ B_A^{(2)}= B_A(-p , m_f, m_{\tilde f'}),
\ B_A^{(3)}= B_A(-p', m_{\tilde f'}, m_f),  $ and 
$C_B = C_A (-p,-p',m_f,m_{\tilde f'} , m_f) .$
The integral functions  with a tilde arise in 
the  sfermion  current  diagram and are described by 
the argument variables, 
$\tilde C_A =  C_A (-p,-p',m_{\tilde f'}, m_f,m_{\tilde f'}).$ 

A very useful check on the above  results
concerns  the cancellation of ultraviolet divergencies. This
is indeed expected on the basis of the  general rule
that those interaction terms which are absent
from the classical  action, as is the case for the  flavor changing currents,
 cannot undergo renormalization. A  detailed discussion of this property
 is developed in \cite{g:soares}.  The  logarithmically divergent 
terms in eq.(\ref{eq9}), proportional to the quantity, 
$\D =-{2\over D-4}+\g -\ln \pi  $, as defined in \cite{g:pasvel}, 
arise from the two- and three-point integrals as, 
$ B_0 \to \D ,\ B_1 \to -\ud \D , \ C_{24}\to {1\over 4} \D  $, 
all other integrals  being finite. 
Performing these substitutions,  we indeed find that $\D $ comes accompanied  
by the overall factors, $ [-a(e_L)+a(\tilde {u_L^\star }) +a(d_R)], $ 
or $ [-a(e_L)+a(\tilde {d_R}) +a_R(u^c)], $ 
which  both  do  vanish  in the relevant configurations for $f, \tilde f'$.

Let us now consider the R parity odd Yukawa interactions 
involving the  $\l_{ijk}$. These contribute through the same triangle
diagrams as in Fig. \ref{g:fig2}. There arise
contributions of   ${\cal L}$-type,  in  the 
single configuration, $ f= e_k, 
\tilde f'=\tilde \nu^\star _{iL}$ and of 
${\cal R}$-type in the two  configurations, 
$f= {e_j\choose \nu_i} , 
\tilde f'={\tilde \nu_{iL} \choose \tilde e_{jL} }$. 
Following the same derivation as above, and neglecting  all of the external mass terms,  
we obtain  the  following results for the one-loop vector coupling vertex functions:  
\begin{eqnarray}
A_L^{JJ'}({\cal L})&=&  \FFIL [ a (f_L) m_f^2 C_0 
+a(f_R)(B_0^{(1)}-2C_{24}-m_{\tilde {f}'}^2 C_0 ) 
+2a(\tilde {f }'_L) \tilde C_{24} + a(f_L) B_1^{(2)} ], \cr
A_R^{JJ'}({\cal R})&= &\FF2L [ a (f_R) m_f^2 C_0 
+a(f_L)(B_0^{(1)}-2C_{24}-m_{\tilde {f}'}^2 C_0 )
+2a(\tilde {f}'_L) \tilde C_{24} + a(f_R) B_1^{(2)} ],  \cr 
&&
\label{eq10}
\end{eqnarray}
with $A_R^{JJ'}({\cal L})= 0, \  A_L^{JJ'}({\cal R})= 0.$
We note that the ${\cal {L} , \ {R} } $ contributions 
are related by a mere  chirality flip transformation and that the color 
factor, $N_c$, is absent in the present case.
\subsection{Down-quark-antiquark pairs}
\label{subsec2}
The processes involving  flavor  non-diagonal 
final down-quark-antiquark pairs, $l^-(k)+l^+(k')\to  d_J(p)+\bar d_{J'}(p')$,
pick up non vanishing contributions  only from  the $\l '_{ijk}$
interactions.  Our discussion here will be brief 
since this case is formally
similar to the leptonic case treated  in subsection \ref{subsec1}. 
In particular, the external fermions  masses, 
 for all three families,  can be neglected 
to a good approximation at the energy scales of interest.
The tree  level amplitude comprises an ${\cal R}$-type 
single t-channel  $\tilde u$-squark 
exchange diagram and  two   s-channel diagrams involving 
 $\tilde \nu  $ and   $\tilde  {\bar \nu  } $ sneutrinos
of the type shown in $(a) $ of  Fig. \ref{g:fig1},
\begin{eqnarray}
M^{JJ'}_t&=& -   
{ \l '_{1jJ} \l ^{'\star}_{1jJ'}\over 2(t-m^2_{\tilde u_{jL} }) } 
\bar u_R(p) \g_\mu v_R(p')\bar v_L(k')\g_\mu u_L(k) \cr
&-& {1 \over s-m^2_{\tilde \nu_{iL} } }\bigg [\l _{i11}\l ^{'\star } _{iJJ'}
\bar v_R(k') u_L(k) \bar u_L(p) v_R(p') +
\l _{i11}^\star \l ' _{iJJ'}
\bar v_L(k') u_R(k) \bar u_R(p) v_L(p')\bigg ] ,\cr 
&&
\label{eqp1}
\end{eqnarray}
where a  Kronecker symbol factor, $\d_{ab} $,  expressing the dependence
on the  final state quarks color indices, $d^a \bar d_b $,  
has been suppressed. This dependence will induce in 
the analog of the formula in  eq.(\ref{eqr4}) expressing  the rates,
an extra color factor,  $N_c$. 

At one-loop level, 
the  dressed   $Z\  d_J\  \bar d_{J'} $ vertex functions in the   
Z-boson s-channel exchange amplitude can be described by the 
same type of triangle diagrams as in Fig. \ref{g:fig2}. 
The fields configurations circulating in the loop
correspond now  to  quarks-sleptons of ${\cal L}$-type,
$ f=d_k;  \tilde l'= \tilde \nu^\star _{iL}, $ and of ${\cal R}$-type,  
$ f={ d_j \choose u_j}; \ \tilde l'= 
{\tilde \nu _{iL} \choose \tilde e_{iL} }$. There also  occurs
corresponding leptons-squarks fields  configurations  of ${\cal L}$-type,  
$ l=\nu_i^c  ; \tilde f'= \tilde d_{kR}$, and ${\cal R}-$type, 
$ l= {\nu_i \choose e_i};  \tilde f'= 
{\tilde d_{jL} \choose \tilde u_{jL} }$. 
The ${\cal L }-$ and ${\cal  R  }-$   type contributions
differ by a  chirality flip, the first contributing  to 
$A_L^{JJ'}$ and the second to $A_R^{JJ'}$.
The calculations are formally similar  to those in subsection \ref{subsec1} and 
the final results  have a nearly identical structure to those 
given in (\ref{eq9}). For clarity, we quote the final 
formulas  for the one-loop vector coupling vertex functions, 
\begin{eqnarray}
A_L^{JJ'}({\cal L})&= &\FFQ [ a (f_L) m_f^2 C_0 
+a(f_R)(B_0^{(1)}-2C_{24}-m_{\tilde {f}'}^2 C_0 ) 
+2a(\tilde {f}'_L) \tilde C_{24} + a(d_L) B_1^{(2)} ], \cr
A_R^{JJ'}({\cal R})&= &\FFP [ a (f_R) m_f^2 C_0 
+a(f_L)(B_0^{(1)}-2C_{24}-m_{\tilde {f}'}^2 C_0 )
+2a(\tilde {f}'_L) \tilde C_{24} + a(d_R) B_1^{(2)} ],  \cr 
&&
\label{eq10p}
\end{eqnarray}
where the intermediate fermion-sfermion fields are labelled  by the 
indices $ f, \ \tilde f' $.  There are implicit sums
 in eq.(\ref{eq10p})   over the 
above quoted leptons-squarks and  quarks-sleptons 
configurations. The attendant ultraviolet divergencies  are accompanied 
 again with vanishing factors, $ a(\tilde d_R) -a(d_L)+a(\nu^c_R) =0,  \  \ 
 a(\tilde d_L) -a(d_R)+a(\nu_L) =0.$ 

\subsection{Up-quark-antiquark pairs}
\label{subsec3}
 The production processes of  
up-quark-antiquark pairs of different families, 
$l^-(k)+l^+(k')\to u_J(p)+\bar u_{J'}(p')$,  
may be  controlled by the $\l'_{ijk}$ interactions only.
The tree amplitude is associated with an u-channel 
$\tilde d$-squark exchange,  of  type  similar to that
 shown by $(a)$ in  Fig. \ref{g:fig1}, and  can be expressed  as,
\begin{eqnarray}
M^{JJ'}_{t}= -{\l ^{'\star }_{1Jk} \l '_{1J'k}
\over 2(u-m^2_{\tilde d_{kR} }) } 
\bar v_L(k') \g^\mu u_L(k)\bar u_L(p)\g_\mu v_L(p'),
\label{eq1pp}
\end{eqnarray}
after using the   Fierz reordering identity, appropriate to commuting 
Dirac (rather than  anticommuting field)
spinors, $\bar u^c(k) P_L v(p')  \bar u(p) P_R v^c(k')  =+\ud  
\bar v_L(k') \g^\mu u_L(k)\bar u_L(p)\g_\mu v_L(p') .$ 
We have  omitted the  Kronecker symbol
$\d_{a b} $ on the $u^a \bar u_b$ color indices, which  will 
result in an extra  color factor $N_c=3$  
for the rates, as shown  explicitly in eq.(\ref{equ4}) below.
The present case is formally similar to the leptonic case treated 
in subsection \ref{subsec1}, except for  a chirality flip in the final fermions.
We are especially interested 
here in final states containing a top-quark, such as $t\bar c$ or $t \bar u$,
for which external particles mass terms cannot obviously  be ignored. 
The  equation, analogous to (\ref{eqr4}), which  expresses the 
summations over the initial and final  polarizations 
in the total  (tree and loop) amplitude, 
takes now  the form, 
\begin{eqnarray}
\sum_{pol} \vert M^{JJ'}_t+M^{JJ'}_l\vert^2&=&  N_c\bigg [ \bigg 
\vert -{  \l'_{1J'k} \l {'^\star } _{1Jk} \over 
2(u-m^2_{\tilde d_{kR} } )}  +\bigg ({g\over 2\cos \t_W}\bigg )^2
{  a(e_L) A_L^{JJ'}(u,s+i\e ) \over 
s-m_Z^2+im_Z\G_Z   } \bigg \vert^2 \cr &\times & 
16 (k\cdot p')(k'\cdot p ) + 8m_Jm_{J'} 
(k\cdot k') \varphi_{LR} ({\cal L})  \bigg ]  \ , 
\label{equ4}
\end{eqnarray}
where  $O(m_e^2)$ terms were ignored and  we  have denoted,
\begin{eqnarray}
\varphi_{LR}({\cal L}) &=& +\bigg ({g\over 2\cos \t_W}\bigg )^2
\bigg ({ a(e_L) A_{R}^{JJ'}(i\e ) \over 
s-m_Z^2+im_Z\G_Z }\bigg )^\star  \bigg ({ \l'_{1J'k}\l _{1Jk}^{'\star }\over 
2(u-m^2_{\tilde d_{kR} } )  } \bigg ) +c.\ c\ . 
\label{equ4p}
\end{eqnarray}
The modified  structure for the kinematical factors  in the above 
up-quarks case, eq.(\ref{equ4}),  in comparison with the leptons  
and d-quarks case, eq.(\ref{eqr4}),  reflects the  difference in 
chiral structure for the RPV  tree  level amplitude. 

In the massless limit  for both  the initial and final fermions 
(where helicity, $h=(-1,+1)$, and chirality, $H=(L,R)$, coincide) 
the RPV interactions contribute to the helicity amplitudes for the  process, 
$l^-+l^+\to f_J +\bar f_{J'}$, in  the mixed type helicity configurations, 
 $ h_{l^-} =-h_{l^+},  \ h_{f_J} =-h_{\bar f_{J'} }$, (same as for the RPC 
gauge interactions)  which are further  restricted  by the conditions, 
$h_{l^-}= -h_{f_J}$, for leptons and d-quarks production, and
 $h_{l^-}= h_{f_J}$, for up-quarks production.   The dependence of the  RPV scattering
amplitudes on scattering angle has a kinematical
factor in the numerator of the form,  $[ 1+h_{l^-} h_{f_J} \cos \theta ] ^2.$ 
[The parts in our formulas in 
eqs. (\ref{equ4}) and (\ref{eqr4}), containing the interference terms between 
RPV and RPC contributions,  partially agree with the published results 
\cite{g:kalin12,g:kalin13}. We disagree
with \cite{g:kalin12,g:kalin13} on the relative signs of RPV and RPC contributions 
and  with \cite{g:kalin13} on the helicity structure for the up-quarks case.
Concerning the latter up-quarks case, our results concur with those 
reported in a recent study \cite{g:mahanta}.]

The states in the internal loops of the triangle diagrams  occur 
in two distinct ${\cal L}$-type  configurations, 
$ f={d_k  \choose e_i^c} ; \ \ \tilde f'= 
{\tilde e_{iL}^\star  \choose   \tilde d_{kR}}$. 
The calculations involved in keeping track of 
the mass terms are rather tedious.  They
were performed by means of the mathematica software package, 
``Tracer" \cite{g:jamin} whose results   were checked against those obtained 
by means of  ``FeynCalc" \cite{g:calc}.
The relevant formulas for the vertex functions  read:
\begin{eqnarray}
 A_L^{JJ'}({\cal L})& =&
 \FLU \bigg [  {a_L(u) }\, {B_1^{(2)}} + 
      {a(f_L)}\, {{ {m_f}}^2}  C_0 + 
      {a(\tilde  f')}\bigg (    
     2 {\tilde C_{24}} + 
      2 \   m_J^2\, ( {\tilde C_{12}}\, - 
      {\tilde C_{21}}\,  + 
      {\tilde C_{23}}\, - 
      {\tilde C_{11}}\, ) \bigg )  \cr &+& 
         {a(f_R)}\, \bigg ( 
       {B_0^{(1)}} - 2\, {C_{24}} - 
         {{ {m_{\tilde  f'}}}^2} {C_0}  + 
         {{ {m_{J}}}^2}   
         \bigg ( C_0 +3 C_{11} -2C_{12} +2C_{21}
	 -2C_{23}\bigg )   -m_{J'}^2 C_{12} \bigg ) \bigg ], \cr 
 A_R^{JJ'}({\cal L})& = & \FLU  m_J m_{J'}  \bigg [ 
      2 {a(\tilde  f')}\,\bigg  ( - {\tilde C_{23}} +  {\tilde C_{22}}\bigg ) + 
  {a(f_R)}\,
        \bigg  (  -C_{11} +C_{12} -2C_{23} +2 {C_{22}} \bigg )
 \, \bigg ] . \cr
 &&
\label{eqm11}
\end{eqnarray}
The above formulas include an implicit sum over the  
two allowed configurations for the  internal fermion-sfermions, namely,
$a(d_ {kH}), \   a(\tilde e^\star _{iL} ) $     and 
$a(e_{iH}^c ), \ a(\tilde d_{kR} ) $.
For completeness, we also display the formula  expressing the tensorial covariants,
\begin{eqnarray}
&&\G_\mu ^Z (p,p')_{tensorial}= \FLU i\s_{\mu \nu }p^\nu \bigg [ m_J P_L \bigg (
a(f_R) (C_{11}-C_{12}+C_{21}-C_{23}) \cr &-&
a(\tilde f ') (\tilde C_{11}+\tilde C_{21}-\tilde C_{12}-\tilde C_{23})
\bigg )  
 + m_{J'} P_R \bigg (
+a(f_R) (C_{22}-C_{23})
+a(\tilde f ') (\tilde C_{23}-\tilde C_{22}) \bigg ) \bigg ]. \cr
&&
\label{eqm11t}
\end{eqnarray}
The complete  $Z f_J\bar f_{J'}$ vertex function, $\G^\mu = \G_{vectorial}^\mu  + 
\G_{tensorial}^\mu $, should (after extracting the external Dirac spinors and the RPV coupling
constant  factors) be symmetrical under the interchange, $J\leftrightarrow J'$, 
or more specifically,  under the interchange,
$m_J\leftrightarrow  m_{J'}$. This property is not explicit on the 
expressions in eqs. (\ref{eqm11}) and (\ref{eqm11t}), but can be established by 
reexpressing the Lorentz covariants by means of the Gordon identity. The
naive use of eq.(\ref{eq6p}) to compute CP-odd asymmetries would seem to 
yield finite contributions  (even in the absence of a CP-odd phase) from the mass terms 
in the vectorial vertex functions, $A_{L}^{JJ'}$,  
owing to their lack of symmetry under, $m_J\leftrightarrow m_{J'}$. 
Clearly, this cannot hold true and is an artefact of restricting to the vectorial couplings.
Including the tensorial couplings is necessary  for a consistent  treatment
of the contributions depending on the external  fermions masses.
Nevertheless, we emphasize that the tensorial vertex  contributions will not included in 
our numerical results.

Finally, we add a general comment  concerning the photon vertex functions,
$ A_{L,R}^{\g JJ'} $,  and the way to incorporate the 
$\g $-exchange contributions in the total
amplitudes, eqs.(\ref{eqr4}) and (\ref{equ4}).
One needs to add terms obtained by  substituting,
${g\over 2 \cos \t _W } \to {e\over 2} = { g \sin \t_W \over 2},
\ a_{L,R}(f)\to 2Q(f), \ (s-m_Z^2  +im_Z \G_Z)^{-1}\to s^{-1},$ along
with the substitution of Z-boson by photon vertex functions,
$ A_{L,R}^{JJ'} (\tilde e, s+i\e ) \to A_{L,R}^{\g JJ'} (\tilde e, s+i\e )$.
The substitution which adds in both  Z-boson and photon exchange 
contributions reads explicitly:
$$ [a_{R,L}(e) A^{JJ'}_{L,R} ]\to
\bigg [ a_{R,L}(e) \sum_f a(f) C_f+ 2Q(e)  \sin^2 \t_W \cos^2 \t_W
[(s-m_Z^2+im_Z \G_Z) / s] \sum_f 2Q(f) C_f   \bigg ]  , $$
where we have used the schematic representation, 
$A^{JJ'}_{L,R}=\sum_f a(f) C_f$.
\section{ Basic assumptions and results}
\label{sec5}

\setcounter{equation}{0}

\subsection{General  context of flavor changing physics}
To place the discussion of the RPV  effects  in perspective, we  
briefly review the current situation of flavor changing physics.
In the standard model, non-diagonal effects with respect 
to  the quarks flavor arise through loop
diagrams. The typical structure of one-loop contributions to, say,
 the $Zf\bar f$ vertex function, $\sum_i
V^\star _{iJ} V_{iJ'}  I(m_i^{f2}/m_Z^2) $, involves   a
summation over quark families of
CKM matrices factors times a 
loop integral. This schematic  formula shows explicitly how 
the CKM matrix unitarity,  along with the near quarks  masses  degeneracies 
relative to the Z-boson mass scale 
(valid for all quarks with the  exception of the top-quark) strongly
suppresses flavor changing  effects. Indeed, for the  down-quark-antiquark case,
the Z-boson decays branching fractions, $B_{JJ'}$, 
 were estimated at the values, 
 $10^{-7}$ for $(\bar b s + \bar s b)$, $10^{-9}$ for $(\bar b d+\bar d b) $, 
 $10^{-11}$ for $(\bar s d+\bar d s)$,    and the  corresponding 
CP  asymmetries, ${\cal A}_{JJ'}$, at  the values,
$ [10^{-5}\ , 10^{-3}, \  10^{-1}]  \sin \d_{CKM} $
\cite{g:bernabeu,g:hou}, respectively. 

By contrast,  flavor changing effects are  expected to  attain
observable levels in several  extensions of the standard model.  
Thus,  one to three order of magnitudes can be  gained on rates $B_{JJ'}$ 
in models accommodating a  fourth  quark family \cite{g:bernabeu,g:hou}.
For the  two Higgs doublets extended standard model,   a recent comprehensive study of
fermion-antifermion  pair production at 
leptonic colliders \cite{g:atwood}  quotes 
for the  flavor changing rates,  $B_{JJ'}\approx 10^{-6} - 10^{-8} $ for 
$Z\to (\bar b +s) +(\bar s +b)$ and    $\s_{JJ'}\approx 10^{-5} - 10^{-6} R$, where,  
$ R =\s (e^++e^- \to \mu^++\mu ^-)=  4\pi \a^2/(3s)= 86.8 /(\sqrt s )^ 2 
fbarns \ (TeV)^{-2}$. Large CP violation signals  are also found in the 
reaction, $p\bar p \to t\bar b X$,  in  the two Higgs doublets and supersymmetric 
models \cite{g:atwood1}.
 
For  the minimal supersymmetric standard model, 
due to the  expected nearness of
superpartners  masses to $m_Z$,
flavor changing  loop corrections 
can become  threateningly large,  unless 
their contributions are bounded  by  postulating either  a degeneracy of the 
soft supersymmetry breaking scalars masses parameters 
 or  an alignment of the  fermion and  scalar 
superpartners current-mass bases transformation matrices.
An early calculation of the  contribution to  Z-boson decay
 flavor changing rates,   $Z\to q_J \bar q_{J'} $, induced by  
 radiative corrections from gluino-squark triangle diagrams of squarks flavor mixing, 
found \cite{g:duncan}:
 $B_{JJ'} \approx 10^{-5}$.
This result is suspect  since a more complete calculation of the effect 
performed subsequently \cite{g:mukho} obtained considerably smaller 
contributions.   Both calculations rely  on  unrealistic inputs, 
including a wrong mass for the top-quark and too low values for 
the  superpartners mass 
parameters. It is hoped that  a complete  updated study 
could be soon performed. 
In fact, during the last few years, the study  of loop corrections in 
extended versions of the standard model has evolved into  a streamlined activity.
For instance, calculations of loop contributions to 
the magnetic moment of the  $\tau $-lepton or of 
the heavy quarks, such as those reported in \cite{g:bernabeu2} (two-Higgs doublets model) or in 
\cite{g:hollik} (minimal supersymmetric standard model) could be usefully transposed to the case  
of fermion pair production observables.

The  information from experimental searches  on flavor changing 
physics at high energy colliders is rather meager \cite{g:branch1}. 
Upper bounds for  the leptonic Z-boson branching ratios, $B_{JJ'}$, 
are reported \cite{g:branch2} at, 
$ 1.7 \ 10^{-6}$ for $( \bar e \mu +\bar \mu e) $,  $9.8 \ 10^{-6} $ for 
$( \bar e \tau +\bar \tau e) $ and $  1.7 \  10^{-5} $ for $(\bar \mu \tau +\bar \tau  \mu )$. 
No results have been quoted  so far   for  $d-$ or $u-$ quark 
pairs production, reflecting the hard experimental problems faced in 
identifying quarks flavors at high energies.   The prospect  for experimental measurements
at the future leptonic colliders is  brightest
 for cases involving one top-quark owing  
to the easier  kinematical identification offered by the large  mass 
disparity  in  the final  state jets.
For leptonic colliders at energies above those of LEP,
the  reactions involving the 
production of  Higgs or heavy $Z'$-gauge  bosons 
which subsequently decay to fermion pairs
could be effective   sources  of flavor non-diagonal
effects, especially when a 
top-quark is produced. At still higher energies, in the TeV regime, the 
production subprocesses involving collisions  of gauge bosons  pairs 
radiated by the incident leptons, as in  $l^-+l^+ \to W^-+W^++\nu  +\bar \nu $, 
could lead to  flavor non-diagonal final states, such as,
 $\nu +\bar \nu + t + \bar c $  with rates of order a few  fbarns \cite{g:hou2}.

\subsection{Choices of parameters and models}
Our main  assumption in this work  is that
no other sources besides  the R parity odd interactions 
contribute significantly to the flavor changing rates and CP asymmetries.  
However, to infer useful information from possible  future  experimental 
results, we must deal with two main types of 
uncertainties. The first concerns the family structure of 
the coupling constants.  On this issue,   one can only 
postulate specific hypotheses or make  model-dependent statements. 
At this point, we may note that the experimental 
indirect  upper bounds  on single coupling constants  are typically, 
 $\l < 0.05  $ or $ \l '  < 0.05$ times ${\tilde m\over 100 GeV}$, 
 except for three special  cases where strong bounds exist: 
 $\l'_{111} < 3.9 \ 10^{-4} ({\tilde m_q\over 100 GeV} )^2
 ({\tilde m_g\over 100 GeV} )^\ud , \quad  (0\nu \b \b -$  decay \cite{g:hirsch}) 
 $\l '_{133}< 2 \ 10^{-3} \ (\nu_e $ mass \cite{g:godbole}) and 
 $\l'_{imk} <  2. \ 10^{-2} ({ m_{\tilde d_{kR}  } \over 100 GeV} ), 
 \ [i, \ k =1,2,3; \ m=1,2]  $, ($K\to \pi \nu \bar \nu $ \cite{g:agashe}).
Strong  bounds  have  been derived
 for products of coupling constants pairs in specific
 family configurations. For instance,  a valuable  source 
 for  the $\l_{ijk}$ coupling constants is provided by the rare decays, $e^-_l\to e^-_m+
e^-_n+e^+_p$ \cite{g:roy}, which  probe the combinations of coupling constants, 
$F_{abcd} = \sum_i
({100 GeV \over m_{\tilde \nu_{iL} } })^2 
\l_{iab} \l ^\star _{icd}$. Except for the  strong bound,
$F_{1112}^2 + F^2_{2111}< 4.3 \ 10^{-13},\   [\mu \to 3e]  $
 the other combinations  of coupling constants involving the third generation are less 
strongly bound, as for instance, 
$ F_{1113}^2 + F^2_{3111}< 3.1 \ 10^{-5} \ [\tau  \to 3e]$
\cite{g:roy}.  
 Another useful source   is provided by the
neutrinoless double beta decay  process \cite{g:babu,g:hirsch,g:hirsch1}.
The strongest bounds occur  for the following configurations of  flavour indices
(using the reference value ${\tilde m = 100 GeV} $): 
$\l'_{113}\l'_{131}< 7.9 \times 10^{-8}, \ 
\l'_{112}\l'_{121}< 2.3 \times 10^{-6}, \ \l ^{'2}_{111}< 4.6 \times 10^{-5} $,
quoting from \cite{g:hirsch1} where
the  initial analysis of \cite{g:babu} was updated. 
Finally,  the strongest bounds deduced from  neutral mesons  $(B\bar B, \ K\bar K)$ 
mixing parameters are:
$ F'_{1311}< 2\ 10^{-5}, F'_{1331}< 3.3\ 10^{-8}, \ 
F'_{1221}< 4.5\ 10^{-9},$ \cite{g:roy}, where 
$F'_{abcd} = \sum_i ({100 GeV \over m_{\tilde \nu_{iL} } })^2 
\l '_{iab} \l ^{'\star }_{icd}$.

The second type  of uncertainties concerns the spectrum of scalar superpartners.
At one extreme, are  the experimental lower bounds, which   reach   for sleptons, 
$40-65 $ GeV,  and for   squarks, $90 - 200$ GeV,  and at the other extreme,  
the theoretical naturalness requirement   which sets
an  upper bound at $1 $ TeV.

In order to estimate  the uncertainties on predictions emanating from 
the above  two sources, it is necessary to delineate the dependence of amplitudes  
on  sfermion masses.
Examining the structure of the  relevant  contributions to flavor changing rates
for, say,  the lepton case,  we note 
that the t-channel exchange   tree amplitudes are given by  a onefold summation 
over sfermions  families, $\sum_i \vert t^i_{JJ'}\vert /\tilde m_i^2$, involving the  
combination  of coupling constants,  $ t^i_{JJ'}=   \l _{i1J} \l_{i1J'}^\star $. 
The typical structure  for the leptonic loop amplitudes is a twofold summation
over fermions and sfermions families,
$\sum_{ij}  l^{ij}_{JJ'} F_{JJ'}^{ij} (s+i\e ) , \ $ where $ l^{ij}_{JJ'}=  \l_{ijJ}
\l^\star _{ijJ'} $, and the   loop integrals, $F^{ij}_{JJ'} $,  have a  non-trivial
dependence on the fermions and sfermions masses, as exhibited 
on the formulas  derived in subsections \ref{subsec1},
\ref{subsec2} and \ref{subsec3} [see, e.g., eq.(\ref{eq10p})].

The effective dependence on the superparticle masses 
involves  ratios, $m_f^2/\tilde m^2 $ or $s/\tilde m^2 $, in such a way that 
the dependence is suppressed for large $\tilde m$. 
(Obviously, $s= m_Z^2$  for Z-boson pole observables.)
In  applications such as ours where,     $s\ge m_Z^2$,
all the  fermions, with the exception of the top-quark,   can be  regarded as 
being massless.
In particular, the  first two light families (for either $  l, \  d,\  u $)
should have  comparable contributions, the third 
family behaving most  distinctly in the  top-quark case. A quick analysis,  
taking the explicit mass factors into account, indicates that  
 loop amplitudes should  scale with sfermions masses as, $(s/\tilde m^2)^n$, with 
a variable  exponent ranging in the interval,  $1< n< 2$. 
Any possible enhancement effect from  the explicit sfermions 
mass factors in  eq.(\ref{eq10p}) is moderated 
in the full result by  the fact that 
the accompanying loop integral factor  has itself
a power decrease with increasing  $\tilde m^2$.
Thus, the Z-boson pole rates should depend on the masses $\tilde m$  roughly 
as $(1/\tilde m^2)^{2n}$, while the off Z-boson  pole rates, being determined 
by the tree amplitudes, should behave  more nearly as $(1/\tilde m^2)^2$.
As for the asymmetries, since these are given by ratios of squared amplitudes, 
one expects them to have a  weak sensitivity on the  sfermion
masses.

To infer   the physical implications on the RPV coupling constants, 
we avoid making too   detailed model-dependent 
assumptions on the scalar superpartners spectrum.
Thus, we shall neglect mass splittings 
between all the sfermions and   set uniformly  all   
 sleptons, sneutrinos and squarks masses
at a unique  family (species)  independent value, $\tilde m$, 
chosen to vary in  the wide  variation interval, 
$100 < \tilde m <  1000$ GeV. This prescription should suffice for the kind of 
semi-realistic  predictions at which we are aiming. This approximation
makes more transparent the dependence on the RPV 
coupling constants, which then involves the 
quadratic products designated by $t^i_{JJ'}$ (tree)  and $l^{ij}_{JJ'}$ (loop),
where the dummy family indices refer to sfermions (tree)
and fermion-sfermions (loop).
For  off Z-boson-pole observables,  flavor non diagonal
rates  are controlled by products of  two  
different couplings, $\vert t_{JJ'}^{i} \vert ^2 ,$ and  asymmetries  by 
normalized products of four different couplings, 
$Im(  t^{i' \star }_{JJ'} l_{JJ'}^{ij} ) / \vert t_{JJ'}^{i''} \vert ^2.$
For Z-boson pole observables,  rates and asymmetries 
are again  controlled by  products of two and
 four different coupling constants, 
$\vert l_{JJ'}^{ij} \vert ^2 $  and  $Im(  l^{i'j' \star }_{JJ'} l_{JJ'}^{ij} ) 
/ \vert l_{JJ'}^{i'' j''} \vert ^2,$  respectively.  Let us note that  if
the off-diagonal rates were dominated by some alternative mechanism, 
the asymmetries  would then  involve products
of four different coupling constants rather than the above ratio. 
 
It is useful here to  set up  a catalog of the 
species and families   configurations for the 
sfermions (tree)  or fermion-sfermions (loop)  involved in the various cases.
 In the tree level amplitudes, these configurations are  for leptons: 
$t^i_{JJ'}=  \l_{iJ1}\l ^\star _{iJ'1},\ \tilde \nu_{iL}  \  ({\cal L}$-type),
$t^i_{JJ'}=  \l_{i1J} \l ^\star _{i1J'},\ \tilde \nu _{iL}  \  ({\cal R}$-type);
for d-quarks,  
$ t_{JJ'}^j = \l ' _{1jJ}\l ^{'\star } _{1jJ'},   \ \tilde u_{jL}\  ({\cal R}$-type);
for u-quarks, 
$t_{JJ'}^k= \l ^{'\star }_{1Jk}\l ' _{1J'k},  \  \tilde d_{kR} \  ({\cal L}$-type).
In the loop level  amplitudes,  the  
coupling constants and internal fermion-sfermion configurations  are for leptons: 

$l^{jk}_{JJ'}= \l ^{'\star }_{Jjk} \l '_{J'jk}, \ 
[{d_k \choose \tilde u^\star _{jL}  } ,
\ {u _j^c \choose \tilde d_{kR}}];  \quad 
l^{ik}_{JJ'}= \l ^{\star }_{iJk} \l _{iJ'k}, \ 
{e_k \choose \tilde \nu^\star _{iL}  } \ 
({\cal L}$-type);    

$l^{ij}_{JJ'}= \l _{ijJ} \l ^{\star }_{ijJ'}, \ 
[{e_j \choose {\tilde  \nu  }_{iL} } ,
\ [{\nu _i \choose \tilde e_{jL}}]  \  ({\cal R}$-type);

for d-quarks:

$l^{ik} _{JJ'}=\l ^{'\star }_{iJk} \l ' _{iJ'k}, \ 
[{d_k \choose \tilde \nu^\star _{iL}} ,
\ {\nu^c_i \choose \tilde d_{kR}}]  \  ({\cal L}$-type);

$l^{ij} _{JJ'}=\l ^{'\star }_{ijJ'} \l ' _{ijJ}, \ 
[{d_j \choose \tilde \nu_{iL}} , \ {u_j \choose \tilde e_{iL}}; 
\quad {\nu_i \choose \tilde d_{jL}} , \ {e_i \choose \tilde u_{jL}}] \   ({\cal R}$-type);

 for u-quarks,

$l_{JJ'}^{ik} = \l ^{'\star }_{iJk} \l'_{iJ'k}, \ 
[{d_k \choose \tilde e^\star _{iL}} , \ {e^c_i \choose \tilde d_{kR}}]  \ 
({\cal L}$-type).

We shall present numerical results for a subset of the above list of cases. 
For leptons and d-quarks, we  shall restrict  consideration 
to the ${\cal R}$-type  terms which contribute to the Z-boson  vertex 
function,  $A_R$.  We also retain the sleptons-quarks internal states for 
d-quark production  (involving $\l^ {'\star }_{ijJ'} \l '_{ijJ}$) 
and the sleptons-leptons for lepton  production  
(involving $\l^ {\star }_{ijJ'} \l _{ijJ}$) 
For the up-quark production, we consider the ${\cal L}$-type terms
(involving $\l^ {'\star }_{iJk} \l '_{iJ'k}$) and, for the  off Z-boson pole
case, omit the term $\varphi_{LR} $ in eq.(\ref{equ4}) in the numerical results.

Since  the running family index in the  parameters relevant to  tree level 
amplitudes refers to sfermions,  
consistently with the approximation of a uniform family independent
mass spectrum, we may as well consider that index as being fixed.
Accordingly, we  shall set  
these parameters at the reference value,  $t^i_{JJ'}=10^{-2}$.  
In contrast to the off Z-boson pole  rates, 
the asymmetries depend non trivially on the  fermion mass spectrum through  
one of the two family indices  in $l^{ij}_{JJ'}$  ($i $ or $j$)
associated to fermions. To discuss our 
predictions, rather than going through the list of four distinct coupling 
constants, we shall make certain general hypotheses regarding the generation 
dependence of the RPV interactions for the fermionic index.
At one extreme is the case where all three 
generations are treated alike, the other extreme being
the case where only one generation 
dominates. We shall  consider three different 
cases which are distinguished  by the interval
over which the fermions family indices are 
allowed  to range in the quantities,  $l_{JJ'}^{ij}$. We define  Case {\bf A } 
by the  prescription  of  equal values for
all three families of fermions ($i=1,2,3$); Case {\bf  B},  for
the second and third  families ($i=2,3$);
and  Case {\bf C},  for the third family only ($i=3$).
For all three cases, we set the
relevant parameters uniformly  at the reference values, 
$l_{JJ'}^{ij}=10^{-2}$. While the   results in Case {\bf C} reflect directly 
on the situation associated with the 
hypothesis of dominant third family configurations, the corresponding 
results in situations where the  first or second family  are assumed dominant, 
can be deduced by taking the differences between the results in 
Cases {\bf A} and {\bf B} and  Cases {\bf B} and {\bf C}, respectively.

In order  to obtain non-vanishing 
CP asymmetries, we still  need to specify   a prescription 
for introducing a relative 
CP-odd complex  phase, denoted  $\psi $, between the 
various RPV coupling constants. 
We shall set  this at the reference value,  $\psi =\pi /2$.
Since the CP asymmetries are proportional
to  the  imaginary part of the phase factor,
the  requisite dependence is simply 
reinstated by inserting the overall factor, $\sin \psi $.
Different prescriptions must be implemented  depending on
whether one considers observables at or off the Z-boson pole.
The Z-boson pole asymmetries are controlled   by a relative complex 
phase  between the  combinations of coupling constants denoted,
$l_{JJ'}^{ij}$ only. For definiteness, 
we choose here to assign a non-vanishing complex phase only 
to the third fermion family, namely, $arg(l^{ij}_{JJ'}) =[0,0,\pi /2]$,
for $[i \ or  \  j =1,2,3]$.  In fact, a relative phase between light families only 
would contribute insignificantly to the 
 Z-boson pole asymmetries,  because of the antisymmetry in $\a \to 
\a '$  in eq.(\ref{eq5}) and the fact that $F^{JJ'} _\a (m_Z^2)$   
are  approximately  equal  when the fermion index in $\a =(i,j)$ belongs 
to the two first  families. 
The off Z-boson pole asymmetries  are controlled by a relative complex 
phase between the tree 
and loop level amplitudes. For definiteness,  we  choose here to 
assign a vanishing argument to the  coupling constants combination, $t^i_{JJ'}$ 
appearing at  tree level and   non-vanishing arguments to the full set of   
loop amplitude combinations, namely,  $arg(l_{JJ'}^{ij}) =\pi/2 , $  
where  the  fermion index ($i \ or \ j $  as the case may be) 
varies  over   the  ranges  relevant  to each of the three  cases 
{\bf A, \ B, \ C}.
 
\subsection{Numerical results and discussion}
\subsubsection{Z-boson decays observables}
\begin{figure}[t]
\begin{center}
\leavevmode
\psfig{figure=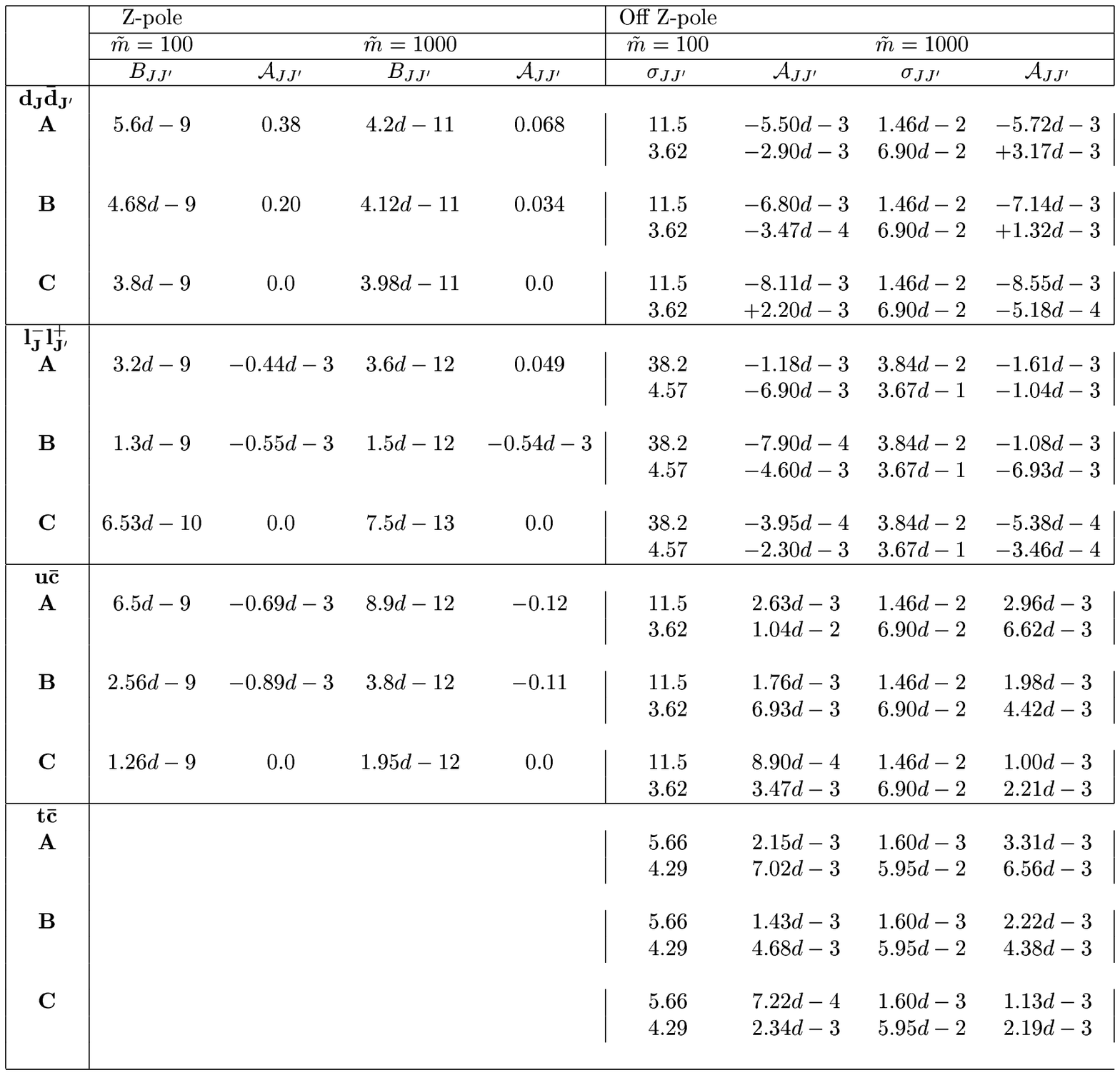}
\end{center}
\caption{Flavor changing
rates and CP asymmetries for  d-quarks, leptons  and u-quarks pair production in 
the three cases,  appearing in line entries as Cases   {\bf A}, {\bf B} 
and {\bf C}, which correspond to internal lines belonging to all three families, 
the second and third families 
and the third family, respectively.     
The results for d-quarks and leptons, unlike those for up-quarks, 
are obtained in the approximation where one
neglects the final fermions masses.
The first four column fields (Z-pole column entry)  show results for
the Z-boson pole  branching fractions $B_{JJ'} $  and 
asymmetries, ${\cal A}_{JJ'}$. The last four column fields  (off Z-pole column entry) 
show results for the flavor non-diagonal cross sections,  $\sigma_{JJ'}$,
in fbarns  and for  the asymmetries, ${\cal A}_{JJ'}$, with photon and Z-boson
exchanges added in. The results in the  two lines for the off Z-boson pole 
are associated to the two values for the center of mass energy,
$s^{1/2} = 200, \ 500$ GeV. The columns subentries indicated by $\tilde m$
correspond to the sfermions mass parameter, $\tilde m=100, 
\ 1000 $  GeV. The notation $ d - n$ stands for $ \ 10^{-n}$.}
\label{table1}
\end{figure}

We  start by presenting  the numerical results for the integrated 
rates associated with Z-boson decays into fermion pairs.  
These   are given 
for the d-quarks,   leptons and u-quarks cases 
in Table \ref{table1}.  We observe a fast decrease  of rates with increasing
values  of the  mass parameter, $\tilde m$.  Our results can be approximately   
fitted by a  power law  dependence 
 which is intermediate between $\tilde m^{-2} $ 
and $\tilde m^{-3}$.  Explicitly, 
 the Z-boson flavor non diagonal   decay rates to d-quarks, leptons and u-quarks, 
are found to scale approximately 
as, $ B_{JJ'} \approx ({\l_{ijJ} \l_{ijJ'} \over 0.01})^2 
({100 GeV \over \tilde m})^{2.5} \times 10^{-9} [5. , \ 1., \ 2. ] $, respectively. 
When a top-quark  intermediate state is allowed  in the loop amplitude,
this dominates over the contributions from the light families. This is 
clearly seen on the  d-quarks results which are somewhat larger than  those 
for up-quarks  and  significantly larger 
 than  those for leptons, the more so for larger $\tilde m$.  This result is
explained partly by the color factor, partly by the
presence of the  top-quark contribution  only 
for the down-quarks case. For contributions involving other intermediate 
states than up-quarks,  whether the internal 
fermion generation index in the RPV coupling constants, $\l_{ijk}$, 
runs over all three generations (Case {\bf A}), 
the second and third generations (Case {\bf B}) 
or the third generation only (Case {\bf  C}),  we find  that   
rates get  reduced by factors  roughly less than 
2 in each  of these stages.  Therefore,  this comparison indicates  
a certain degree of family independence for  the Z-boson branching fractions 
for the cases where  either leptons or d-quarks propagate inside the loops. 
 
Proceeding next to the CP-odd asymmetries, since  these are proportional to
ratios of the  RPV coupling constants,  it follows 
in our prescription of  using uniform values for these, that asymmetries
must be independent of the specific reference value chosen. As for their 
dependence on $\tilde m$,  we  see on Table \ref{table1} 
that this is rather  strong and that the sense of variation with increasing 
$\tilde m$ corresponds (for absolute values of ${\cal A}_{JJ'}$) to a decrease 
for d-quarks and  an increase for u-quarks and leptons.
The comparison of different production cases 
shows that the CP asymmetries are largest, $O(10^{-1})$, for d-quarks at
small $\tilde m \simeq 100 GeV$, and for u-quarks at 
large $\tilde m \simeq 1000 GeV$.  For leptons, the asymmetries are systematically
small, $O(10^{-3} \ - \ 10^{-4})$. The above features are explained by the occurrence for
d-quarks production of an intermediate top-quark contribution  and also 
by the larger  values of the rates  at large $\tilde m$ in  this case.
The comparison of results in Cases {\bf A} and   {\bf B} indicates that the first
two light families give roughly equal contributions in all cases.

For Case {\bf C}, the CP-odd asymmetries are vanishingly small,
as expected from our prescription of assigning the CP-odd phase, since Case {\bf C}
corresponds then to a situation where only single pairs of coupling constants 
dominate. Recall that for  the specific cases considered in the numerical applications, 
namely, ${\cal R}$-type  for d-quarks and leptons and 
${\cal L}$-type  for u-quarks,  the relevant products of RPV coupling constants are, 
$\l^{'\star }_{ijJ'} \l '_{ijJ},  \  \l^{\star }_{ijJ'} \l _{ijJ},  \  
\l^{'\star }_{iJ k} \l '_{iJ'k}$, respectively, where  the
fermions generation index amongst the
dummy indices pairs, $(ij), \  (ik)$, refers to the third family.
Non vanishing contributions to ${\cal A}_{JJ'}$  could arise in Case {\bf C} if 
one assumed that two pairs of the above coupling constants products  
with different sfermions indices dominate, and   further requiring that these sfermions are
not mass degenerate. Another interesting possibility is
by assuming that  the hypothesis of  single pair of RPV coupling constants dominance 
applies  for the  fields current basis.
Applying then to the quark superfields
the transformation matrices relating these to mass basis fields, say, in the 
distinguished choice \cite{g:agashe} where the flavor changing effects bear on 
u-quarks, amounts to perform the substitution, $\l ' _{ijk} \to 
\l ^{'B} _{ink} V^\dagger _{nj}  $, where $ V$ is the CKM matrix.  The CP-odd 
factor, for the d-quark case, say, acquires then the form, 
$Im (l_{JJ'}^{ij \star } l_{JJ'}^{ij'})  \to 
\vert \l ^{'B} _{inJ'} \l ^{'B\star } _{imJ}\vert ^2  Im((V^\dagger )_{nj}
(V^\dagger )^\star _{mj} (V^\dagger )^\star _{nj'} (V^\dagger )_{mj'}),$
where the second factor on the right-hand side is recognized as the familiar
plaquette term, proportional to the products of sines of  all the 
CKM rotation angles times that of the CP-odd phase.

It may be useful to examine the bounds  on the RPV coupling constants 
implied by the current experimental limits on the 
flavor non diagonal leptonic widths \cite{g:branch2}, 
$B_{JJ'}^{exp} < [1.7,\ 9.8, \ 17. ] \ 10^{-6}$ for  the family 
couples, $[JJ'= 12,\ 23,\ 13]$.
The contributions associated with the $\l $ interactions can be  directly
deduced  from the results in Table \ref{table1}. Choosing the value, 
$\tilde m =100 $ GeV,  and  writing our numerical result
as, $ B_{JJ'}\approx ( {\l_{ijJ} \l _{ijJ'}^\star  \over 0.01 })^2  4 \ 10^{-9}$, then
under the hypothesis of a pair of dominant coupling constants, 
one deduces, $\l_{ijJ}\l_{ijJ'}^\star < [0.46,\ 1.1, \ 1.4 ]$, for  all fixed  choices 
of the family couples, $i,  \ j$.  (An extra factor $2$ in $B_{JJ'}$ 
has been included to account for the antisymmetry property of $\l_{ijk}$.)
For the $\l '$ interactions, stronger bounds obtain because of the 
extra color factor and of the internal top-quark  contributions. 
A numerical  calculation (not reported in Table \ref{table1}) 
performed with the choice, $\tilde m =100 GeV $ for Case {\bf C}, gives us:
$B_{JJ'}\approx ({ \l ^{'\star } _{Jjk} \l'_{J'jk} \over 0.01 })^2 1.17 \ 10^{-7}$,
which,  by comparison with the experimental limits, yields the bounds: 
$\l ^{'\star } _{Jjk} \l'_{J'jk} < [0.38,\ 0.91, \ 1.2 ] 10^{-1}$, for the same 
family configurations,  $[J \ J'= 12,\  23,\  13] $, as above. 
These  results agree  in size to within a factor of 2 with results
reported in a recently published  work \cite{g:anwar}.
\subsubsection{Fermion anti-fermion pair production rates}
\begin{figure}[t]
\begin{center}
\leavevmode
\psfig{figure=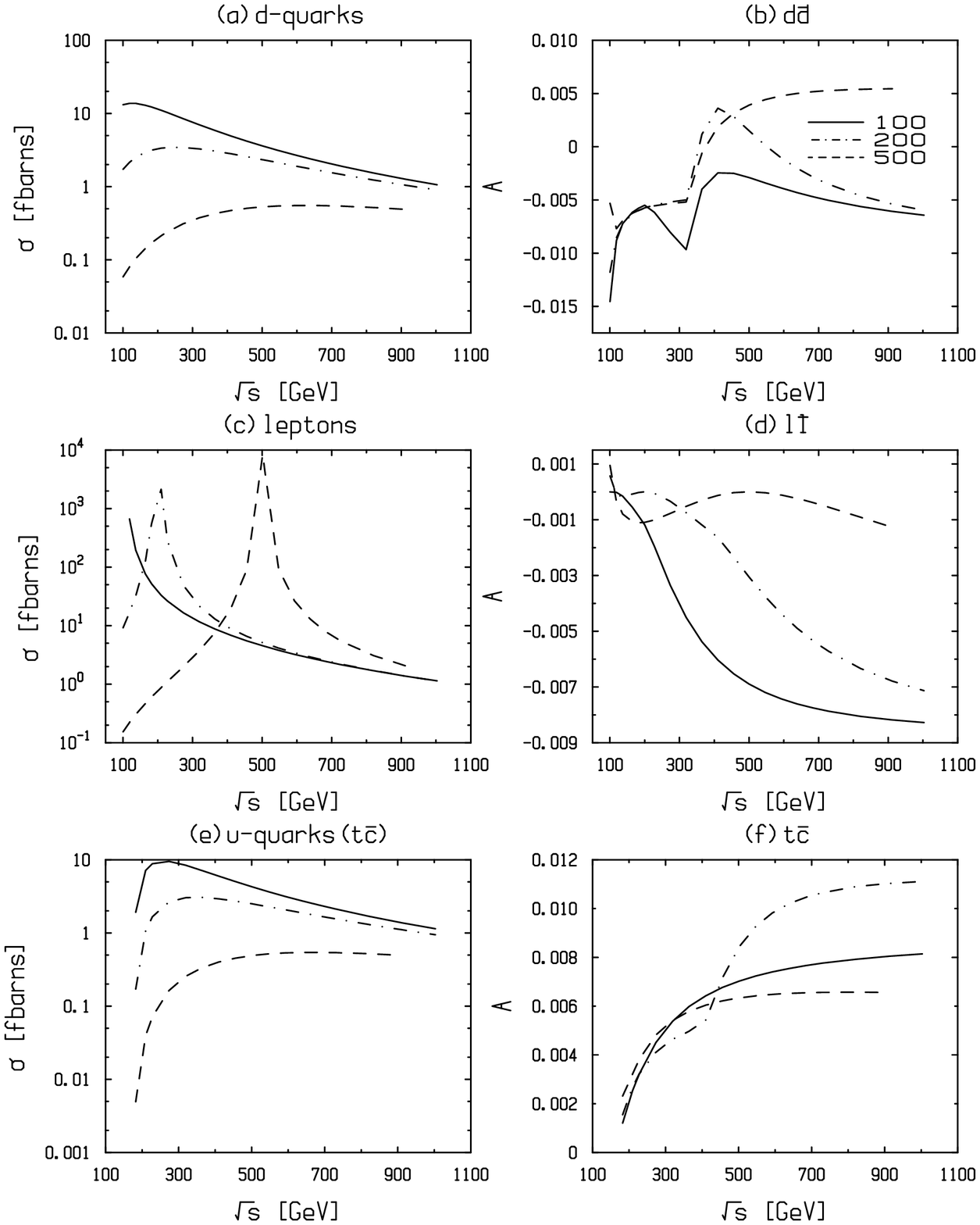,width=5.5in}
\end{center}
\caption{Integrated flavor non-diagonal cross sections  
(left hand side windows) and asymmetries  (right hand side windows) as functions of the 
center of mass  energy in the production of 
down-quark-antiquark pairs (two upper figures  $(a)$ and $(b)$), 
lepton-antilepton pairs (two intermediate figures  $(c) $and $(d)$)
and up-quark-antiquark pairs of type $\bar t c+\bar c t$ 
(two lower figures (e) and (f)). 
The tree level amplitude 
includes only the t-channel contribution for the d-quark case, 
both t- and s-channel exchange contributions for the lepton 
case, and the u-channel exchange for the up-quark case. The one-loop 
amplitudes, with both  Z-boson  and photon exchange contributions, 
include all three internal fermions 
generations, corresponding  to Case {\bf A}. 
Three choices for the 
superpartners uniform mass parameter,  $\tilde m$, 
are considered: $ 100 GeV$ (continuous lines),
$ 200 GeV$ (dashed-dotted lines),  
$ 500 GeV$ (dashed lines).   }
\label{fig4}
\end{figure}

\begin{figure}[t]
\begin{center}
\leavevmode
\psfig{figure=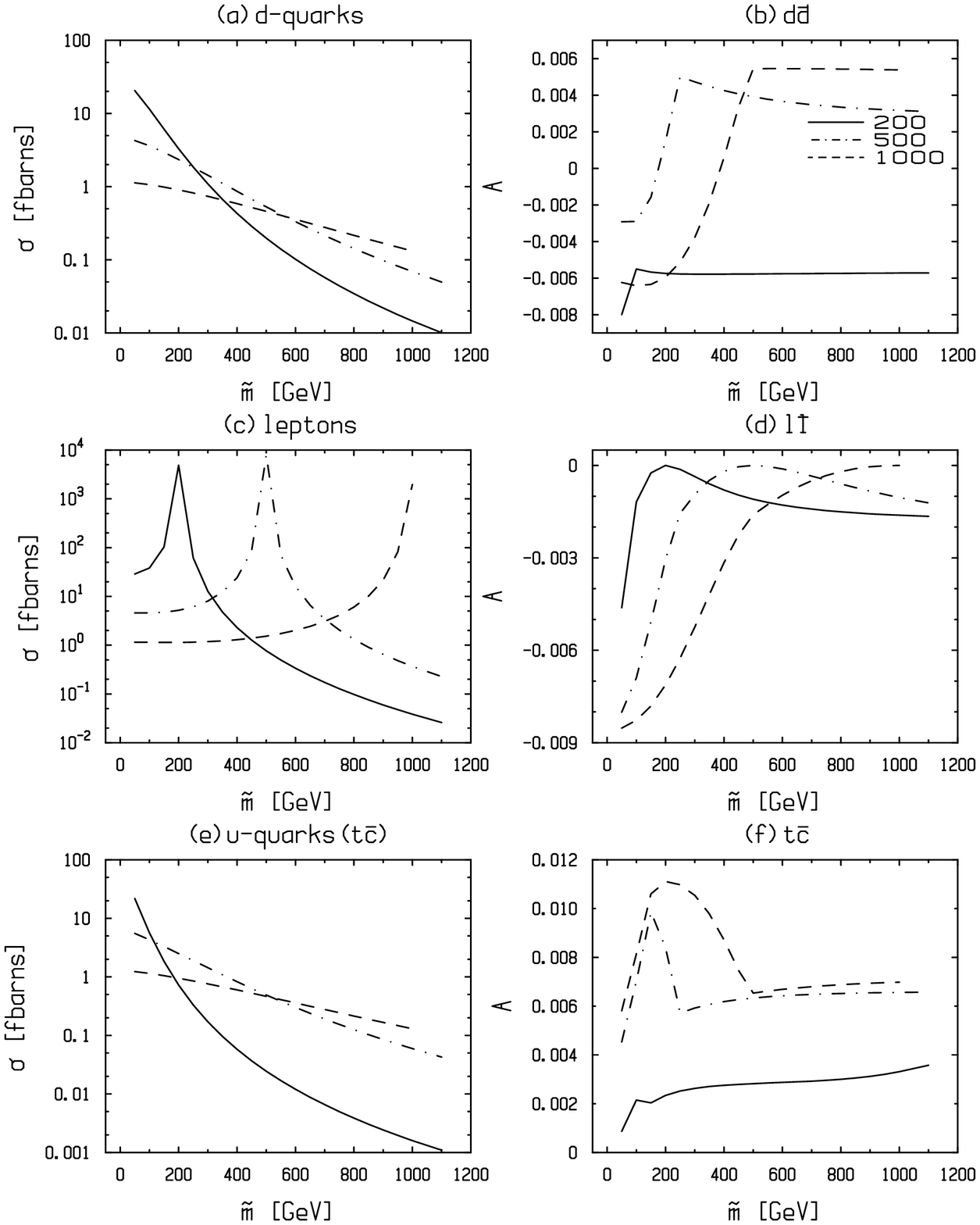,width=5.5in}
\end{center}
\caption{Integrated flavor non-diagonal cross sections  (left hand side windows)  and  CP-odd 
asymmetries (right hand side windows)  as functions of the 
scalar superpartners mass parameter, $\tilde m$, in the production of 
down-quark-antiquark pairs (two upper figures  $(a)$ and $(b)$), 
lepton-antilepton pairs (two intermediate figures  $(c) $and $(d)$)
and up-quark-antiquark pairs of type $\bar t c  $ or $ \bar c t$ 
(two lower figures (e) and (f)). 
The tree level amplitude 
includes only the t-channel contribution for the d-quark case, 
both t- and s-channel exchange contributions for the lepton 
case, and the u-channel exchange for the up-quark case. The one-loop 
amplitudes, with  both  photon and Z-boson exchange contributions,
include all three internal fermions 
generations, corresponding  to Case {\bf A}, with three families running inside loops. 
Three choices for the  center of mass energy, $s^{1/2}$, 
are considered: $ 200 GeV$ (continuous lines),
$ 500 GeV$ (dashed-dotted lines),  
$ 1000 GeV$ (dashed lines).   }
\label{fig5p}
\end{figure}

Let us now proceed to the off Z-boson pole observables.
The numerical results for
the flavor non-diagonal integrated cross sections and CP asymmetries 
are shown in Table \ref{table1} for two selected values
of the center of mass energy, $\sqrt s =  200 $ and $  500 $ 
GeV.  The numerical  results displaying  the 
variation of these observables with the center of  mass of energy  
(fixed $\tilde m$) and with the superpartners mass parameter (fixed $\sqrt s$) 
are given in  Fig.\ref{fig4} and   Fig.\ref{fig5p}, respectively.
All the results presented in this work 
include both photon and Z-boson exchange contributions. 
We observe  here that the predictions for  asymmetries are 
sensitive to the interference effects between photon and Z-boson
exchange contributions.

We discuss first the predictions for flavor  non diagonal rates. We 
observe a strong decrease with increasing values of $\tilde m$ and a 
slow decrease with increasing values of $\sqrt s$. Following a rapid initial 
rise at threshold, the rates settle at values ranging between
$  (10 \ - 10^{-1})$  fbarns for a wide interval of $\tilde m$ values.  
The  dependence on $\tilde m$ can be approximately represented as,
$ \s_{JJ'}/ [  \vert {t^i_{JJ'}\over 0.01} \vert ^2 ({100 \over \tilde m})^{2 \ - \ 3} ] 
\approx  (1 \ - \ 10) $ fbarns  $ \approx R 
({\sqrt s  \over ( 1 \ TeV })^2 (10^{-1} \ - \ 1)  .$ 
The rate of decrease of $\s _{JJ'}$ with $\tilde m$ 
slows down with increasing $s$. It is   interesting to note
that if we had considered here  constant values of
the product $\l  \ (\tilde m/100 GeV) $, 
 rather than constant values of  $\l $, the 
 power dependence of rates  on  $\tilde m$ would  be such as to lead to
interestingly  enhanced rates at large $\tilde m$.

The marked differences exhibited by the results for lepton pair production, 
apparent on windows  $(c)$  and $(d)$ in Figures \ref{fig4} and \ref{fig5p} 
are due to our deliberate choice of adding  
the s-channel $\tilde \nu $ pole term   for the lepton case  while omitting it 
for the d-quark case.   The larger rates found  for leptons as compared 
to d-quarks, in spite of the extra color factor present  for d-quarks 
(recall that the $l^+l^-\to 
f_J\bar f_{J'} $ reactions rates  for down-quarks  and up-quarks 
pick up an extra color factor $ N_c$ with respect to those for  leptons) 
is  thus  explained by the strong enhancement induced by adding in
the sneutrino exchange contribution.
This choice  was made here 
for illustrative purposes, setting for orientation the relevant coupling constant  
at the value, $\l_{1JJ'}=0.1$.
The  $\tilde \nu $ propagator pole  was 
smoothed out by employing  the familiar  shifted propagator 
prescription,  $( s-m^2_{\tilde 
\nu } +im_{\tilde \nu } \G_{\tilde \nu })^{-1}$, while describing approximately
 the   sneutrinos decay width in terms of the RPV contributions alone,
namely,  $\G (\tilde \nu_i \to l^-_k+ l^+_j)= { \l_{ijk}^2 {\tilde  m}_i
\over 16\pi }$  and 
$\G (\tilde \nu_i \to d_k+ \bar d_j)=N_c { {\l  '}_{ijk}^2 {\tilde  m}_i
\over 16\pi }$. 

Proceeding next to the CP-odd asymmetries, we note that since these
 scale as a function  of the RPV coupling constants  as, 
$ Im (l^{ij}_{JJ'} l^{i'j'\star }_{JJ'} )/\vert t^{i''}_{JJ'}
\vert ^2$, our present predictions are independent of 
the uniform reference  value assigned to these coupling constants. 
If the generational dependence of the 
RPV coupling constants  were to exhibit strong hierarchies, this peculiar 
rational dependence could induce strong suppression or enhancement factors.

The cusps in the dependence of ${\cal A}_{JJ'}$ on $\sqrt s$ (Fig. \ref{fig4}) 
occur at values of the center of mass energy where one crosses thresholds for 
fermion-antifermion (for the energies under consideration, $t\bar t$) pair production, 
$\sqrt s =2m_f$, and  scalar superpartners pair production, 
$\sqrt s =2\tilde m$. These are the thresholds for the processes,
$ l^-+l^+\to f\bar f $ or $l^-+l^+\to \tilde f' \tilde f^{'\star }$, 
at which the associated loop  amplitudes acquire finite imaginary parts. 
Correspondingly, in the dependence of  ${\cal A}_{JJ'}$ on $\tilde m$ 
(Fig. \ref{fig5p}) the cusps appear at $\tilde m =\sqrt s /2$. 
We note on the results  that the $t\bar t$  
contributions act to suppress the asymmetries whereas 
the $\tilde f\tilde f^\star $ contributions  rather act to  enhance them. 
Sufficiently beyond these two-particle thresholds, the asymmetries vary
weakly with $\tilde m$.
A  more rapid variation as a function of energy occurs
in the leptons production case
due to the addition there of the sneutrino pole contribution. 

The comparison  of results for asymmetries in  Cases {\bf A, \ 
B, \ C} reflects on the dependence of loop integrals  
with respect to the internal fermions masses.
An examination of Table \ref{table1} reveals that for leptons and
up-quarks, where intermediate states involve leptons or d-quarks,
all three families have nearly equal contributions. 
The results for down-quarks production  are
enhanced because of  the intermediate  top-quark  contribution, which dominates over 
that of lighter families. However, this effect is depleted when the  finite 
imaginary part from $t\bar t$ sets in.
The asymmetries for up-quarks production  
assume values in the range, $10^{-2}- 10^{-3}$,
irrespective of the fact that the final fermions belong to light 
or heavy families.   
\section{Conclusions}
\label{seconc}

\setcounter{equation}{0}

The two-body  production at high energy leptonic colliders
of fermion pairs of different families 
could  provide valuable information on the flavor structure of the
R parity odd Yukawa interactions.  One can only wish that  an  experimental  
identification of    lepton and quark flavors  at high energies 
becomes accessible in the future.
Although the  supersymmetric  loop corrections to these processes 
may not be as strongly  
suppressed as their  standard model counterparts,  
one expects that the  degeneracy or alignment constraints on  the scalar 
superpartners masses and flavor mixing  should 
severely bound their contributions.
Systematic studies  of the supersymmetry corrections  to the 
 flavor changing rates and CP asymmetries in fermion pair production should be 
strongly encouraged.
 
An important characteristic of the R parity odd interactions  is that 
they can  contribute to integrated  rates 
at tree level and to CP asymmetries through 
interference terms between the  tree and loop amplitudes. While we have 
restricted ourselves to the subset of loop contributions 
associated with Z-boson exchange,  a large number of 
contributions,  involving quark-sleptons  or 
lepton-squarks intermediate states in various families configurations,
could still  occur. 
The contributions to rates and asymmetries depend
strongly  on the values of the 
R parity odd  coupling constants. Only the rates   are directly 
sensitive   to  the supersymmetry  breaking scale.
To  circumvent the uncertainties from the sparticles spectrum, we have
resorted to the simplifying   assumption that  the scalar 
superpartners mass differences and mixings 
can be neglected. We have  set the  RPV coupling constants  at
a uniform value while sampling a set of  cases
from which one might reconstruct the family dependence of the RPV coupling constants. 
We have also embedded  a CP complex
phase  in the RPV coupling constants in a specific way, meant to serve
mainly as an illustrative 
example. Although  the  representative  cases that we have considered 
 represent a small fraction of  the host of  possible variations, 
they give a fair idea of the sizes to expect.
Since these processes cover a wide range of family configurations, one 
optimitistic  possibility  could be that  one specific entry for 
the family configurations  would enter with 
a sizeable RPV coupling constant. 

The contributions  to the  flavor changing rates have a 
strong sensitivity on the  RPV coupling constants and the superpartners  mass,
 involving high powers of these  parameters.
We find a generic dependence for the 
flavor changing  Z-boson decay branching ratios  of form, 
$({\l \l  \over 0.01})^2 ({100\over \tilde m})^{2.5} \ 10^{-9}$.
For the typical bounds on the RPV coupling constants,
it  appears that these branchings  are
three order of magnitudes below the current experimental sensitivity.  
At higher energies, the 
flavor changing rates are in  order of magnitude, 
$({\l \l  \over 0.01})^2 ({100\over \tilde m})^{2\ -\  3 }  \ (1\ - 10 )$ fbarns. 
Given  the size for the typical   
integrated luminosity, ${\cal L}= 50 fb^{-1}$/year, 
anticipated  at the future leptonic  machines, one can be moderately optimistic
on the observation of clear signals. 

The Z-boson pole  CP-odd 
asymmetries  are of order,  $ (10^{-1} \ -\ 10^{-3}) \sin \psi $.
For the off Z-boson pole reactions, a CP-odd  phase, $\psi $,  embedded in the 
RPV coupling constants shows up in  asymmetries with  reduced strength, 
$(10^{-2} - 10^{-3})\sin \psi $ for leptons,
d-quarks and u-quarks. 
The largely unknown structure of the RPV coupling constants in flavor space
leaves room for good or bad surprises,  since the 
peculiar rational dependence  on the coupling constants,
$ Im (\l \l ^\star  \l \l  ^\star )/\l ^4 $, 
and similarly with $\l \to \l '$,  may lead to strong enhancement or suppression factors. 


\clearpage

\vspace{10 mm}
\begin{center}
{  }
\end{center}
\vspace{10 mm}

\clearpage

\clearpage

\vspace{10 mm}
\begin{center}
{  }
\end{center}
\vspace{10 mm}

\clearpage

\setcounter{chapter}{0}
\setcounter{section}{0}
\setcounter{subsection}{0}
\setcounter{figure}{0}

\chapter*{Publication VIII}
\addcontentsline{toc}{chapter}{Polarized single top production at leptonic colliders from
broken R parity interactions incorporating CP violation}

\newpage

\vspace{10 mm}
\begin{center}
{  }
\end{center}
\vspace{10 mm}

\clearpage

\begin{center}
{\bf \huge Polarized single top production at leptonic colliders from
broken R parity interactions incorporating CP violation}
\end{center} 
\vspace{2cm}
\begin{center}
M. Chemtob, G. Moreau
\end{center} 
\begin{center}
{\em  Service de Physique Th\'eorique \\} 
{ \em  CE-Saclay F-91191 Gif-sur-Yvette, Cedex France \\}
\end{center} 
\vspace{1cm}
\begin{center}
{Phys. Rev. {\bf D61} (2000) 116004, hep-ph/9910543}
\end{center}
\vspace{2cm}
\begin{center}
Abstract  
\end{center}
\vspace{1cm}
{\it
The contribution from the R parity violating interaction, $\l '_{ijk}
L_i Q_j D_k^c $, in the associated production of a top quark
(antiquark) with a charm antiquark (quark) is examined for high energy
leptonic colliders.  We concentrate on the reaction, $l^-+l^+\to ( t
\bar c) + (c \bar t ) \to (b \bar l \nu \bar c) +(\bar b l \bar \nu c
) $, associated with the semileptonic top decay. A set of
characteristic dynamical distributions for the signal events is
evaluated and the results contrasted against those from the standard
model W-boson pair production background. The sensitivity to
parameters (R parity violating coupling constants and down-squark
mass) is studied at the energies of the CERN LEP-II collider and the
future linear colliders.  Next, we turn to a study of a CP-odd
observable, associated with the top spin, which leads to an asymmetry
in the energy distribution of the emitted charged leptons for the pair
of CP-conjugate final states, $b \bar l \nu \bar c$ and $\bar b l \bar
\nu c $.  A non vanishing asymmetry arises from a CP-odd phase,
embedded in the R parity violating coupling constants, through
interference terms between the R parity violating amplitudes at both
the tree and loop levels.  The one-loop amplitude is restricted to the
contributions from vertex corrections to the photon and Z-boson
exchange diagram. We predict unpolarized and polarized rate
asymmetries of order $ O(10 ^{-3}) - O(10 ^{-2}) $. An order of
magnitude enhancement may be possible, should the R parity violating
coupling constants $\l '_{ijk}$ exhibit a hierarchical structure in
the quarks and leptons generation spaces.
}

\newpage

\section{Introduction}
\label{h:secintro}

\setcounter{equation}{0}

The flavor non diagonal fermion-antifermion pair production,
$l^-l^+\to f_J \bar f_{J'},$ where $ J\ne J' $ are flavor labels,
represents a class of reactions where the high energy colliders could
contribute their own share in probing new physics incorporating flavor
changing and/or CP violation effects.  As is known, the standard model
contributions here are known to be exceedingly small, whereas
promising contributions are generally expected in the standard model
extensions.  (Consult ref.  \cite{h:chemtob} for a survey of the
literature.)  Of special interest is the case where a top quark
(antiquark) is produced in association with a lighter (charm or up)
antiquark (quark). The large top mass entails a top lifetime, $
\tau_{top} = [1.56 \ GeV ({m_t\over 180 \ GeV})^3]^{-1} $,
significantly shorter than the QCD hadronization time, $ 1/\L _{QCD}$,
which simplifies the task of jet reconstruction. \cite{h:toprev} The top
polarization effects also constitute a major attraction.
\cite{h:donoghue,h:topol0,h:kane,h:schmidt,h:chang1} The large top mass entails
a spin depolarization time of the top which is longer than its
lifetime, $\tau _{depol} = [1.7 MeV ({180\over m_t}) ]^{-1} >
\tau_{top} $, thus providing an easy access to top polarization
observables.  Polarization studies for the top-antitop pair production
reaction, in both production and decay, have been actively pursued in
recent years.  \cite{h:toppol,h:bartl,h:rigolin} (An extensive literature
can be consulted from these references.)

It appears worthwhile applying similar ideas to the flavor non
diagonal fermion pair production process involving a single top
production.  This reaction has motivated several theoretical studies
aimed at both leptonic ($ l^- l^+ , \ e \g $ and $\g \g $) and
hadronic ($ p \bar p , \ pp$) colliders.  Exploratory theoretical
studies have been pursued at an implicit level, via the consideration
of higher dimension contact interactions \cite{h:hikasa,h:han,h:abraham},
and at an explicit level, via the consideration of mechanisms
involving leptoquarks, \cite{h:barger} an extended Higgs doublet sector,
\cite{h:atwood,h:atwood1} supersymmetry based on the minimal
supersymmetric standard model with an approximately broken R parity,
\cite{h:chemtob,h:datta,h:oakes1,h:mahanta,h:litte} quark flavor mixing,
\cite{h:koide} standard model loops and four matter generations,
\cite{h:chang,h:huang99} or higher order standard model processes with
multiparticle final states, $ l^-l^+ \to t\bar c \nu \bar \nu
$. \cite{h:hou2} A survey of the current studies is provided in
ref.\cite{h:han}.

In this work, pursuing an effort started in our previous paper,
\cite{h:chemtob} we consider a test of the R parity violating (RPV)
interactions aimed at the top-charm associated production.  Our study
will focus on the contributions to the process, $ l^-l^+\to (t\bar c)
+(\bar t c) $, arising at the tree level from the trilinear RPV
interactions, $\l '_{ijk} L_i Q_j D_k^c $, via a $ \tilde d_{kR} $
squark exchange.  We examine the signal associated with the (electron
and muon) charged semileptonic decay channel of the top, $ t\to b W ^+
\to b l^+ \nu $.  The final states, $ (b l^+ \nu \bar c) + (\bar b l^-
\bar \nu c ), \ [l=e,\ \mu ] $ consist of an isolated energetic
charged lepton, accompanied by a pair of $ b $ and $ c $ quark
hadronic jets and missing energy.  The standard model background may
arise from the W-boson pair production reaction, $l^+ l ^-\to W^+W^-$,
and possibly, in the case of an imperfect $ b $ quark tagging, from
the $ b - \bar b $ quark pair production reaction, $l^+ l ^-\to b \bar
b $, followed by a semileptonic decay of one of the b quarks, $b \to c
l^- \bar \nu $.

The present work consists of two main parts. In the first part, we
discuss the signal associated with the top semileptonic decay channel.
We evaluate a set of characteristic dynamical distributions for the
signal and for the standard model background and obtain predictions
for the effective rates based on a judicious choice of selection cuts
on the final state kinematical variables.  Our discussion will develop
along similar lines as in a recent work of Han and Hewett, \cite{h:han}
which was focused on the contributions initiated by the dimension, $
\cddd =6$, four fields couplings of the Z-boson with fermion pairs and
the neutral Higgs boson.  In the second part of the paper, we examine
a specific CP-odd top polarization observable which corresponds to an
asymmetry in the energy distribution of the final state charged lepton
with respect to the sign of its electric charge.

The contents of the paper are organised into 3 sections.  In Section
\ref{sec2}, we focus on the total and partial semileptonic decay rates
for both the signal and standard model background, allowing for the
case of an imperfect $ b $ quark tagging.  We discuss the constraints
from the indirect bounds on the RPV coupling constants, study the
dependence of rates on the down-squark mass parameter and evaluate a
set relevant dynamical distributions that are of use in devising an
appropriate set of selection cuts.  In Section \ref{sec3}, we discuss
a test of CP violation involving top polarization effects. The CP
violating observable arises through interference terms between the
tree and one-loop contributions to the amplitude and a CP-odd phase
which is embodied in the RPV coupling constants.  Following an
approach similar to one used in earlier proposals,
\cite{h:schmidt,h:chang1} we describe the top production and decay by
means of a factorization approximation and examine the induced charge
asymmetry in the energy distribution of the final state charged
leptons. The production amplitudes are evaluated in the helicity
basis.  Our main conclusions are summarized in Section \ref{sec4}.

\section{Top-charm associated production}
\label{sec2}

\setcounter{equation}{0}

\subsection{ Integrated rates}

\begin{figure}[t]
\begin{center}
\leavevmode \psfig{figure=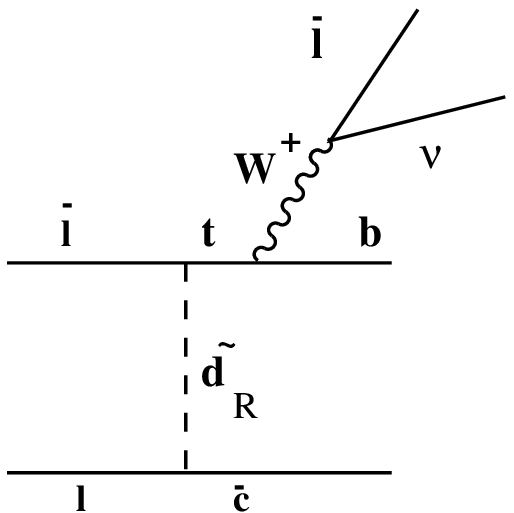,width=3.in}
\end{center}
\caption{Feynman diagram for the tree level amplitude of the process,
$ l^+l^- \to \bar c t \to \bar c b l^+ \nu $.}
\label{h:fig1}
\end{figure}

In a $ l^- l^+$ collision, the tree level transition amplitude for
single top production, as initiated by the RPV interactions, $ \l
'_{ijk} L_iQ_jD_k^c $, proceeds via the $u$-channel exchange of a
right-handed down-squark, $\tilde d_{kR}$, as represented in
Fig.\ref{h:fig1}.  By use of a Fierz ordering identity, the transition
amplitude for the flavor non diagonal production of an up
quark-antiquark pair, $ l^- (k) + l^+ (k') \to u_J (p) +\bar u_{J'}
(p') $, can be written in the form of a Lorentz covariant vectorial
coupling,
\begin{eqnarray}
M^{JJ'}_{t}= -{\l ^{'\star }_{lJk} \l '_{lJ'k} \over 2(u-m^2_{\tilde
d_{kR} }) } \bar v_L(k') \g^\mu u_L(k)\bar u_L(p)\g_\mu v_L(p').
\label{h:eq1pp}
\end{eqnarray}
We shall specialize henceforth to the case of electron-positron
colliders, corresponding to the choice $ l=1$ for the generation
index.  The squared amplitude, summed over the initial and final
fermion spins, reads: \cite{h:chemtob}
\begin{eqnarray}
\sum_{pol} \vert M^{JJ'}_t \vert ^2 &=& N_c \bigg \vert -{ \l'_{1J'k}
\l {'^\star } _{1Jk} \over 2(u-m^2_{\tilde d_{kR} } )} \bigg \vert ^2
16 (k\cdot p')(k'\cdot p ) .
\label{equ1p}
\end{eqnarray}
The production rate for unpolarized initial leptons, integrated over
the scattering angle in the interval, $ 0\le \vert \cos \t \vert \le
x_c$, is given by the analytic formula,
\begin{eqnarray}
\s &= &{ N_c \vert \l'_{1J'k} \l^{'\star } _{1Jk} \vert ^2 \over 64
\pi s^{2} } [(u_--u_+) +(2\tilde m^2-m_J^2 -m_{J'} ^2 )\ln \vert
{u_--\tilde m^2 \over u_+-\tilde m^2} \vert \cr & -& (\tilde
m^2-m_J^2)(\tilde m^2-m_{J'}^2) ({1\over u_--\tilde m^2 } -{1\over
u_+-\tilde m^2 } ) ],
\label{equ1q}
\end{eqnarray}
where, $ u_\pm = m_J^2 -\sqrt s (E_p \pm p x_c )$.  For the top-charm
associated production case, in the limit, $ m_J = m_t >> m_{J'} = m_c
$, one has, $ u_+ \simeq m_t^2 -s, \ u_- \simeq 0$.  For fully
polarized initial beams, since the RPV amplitude selects a single
helicity configuration for the initial state leptons, $l^-_L l^+_R$,
(left handed $l^-$ and right handed $ l^+$) the corresponding
polarized rate would be still described by the same formula as above,
only with an extra enhancement factor of $4$.  The predicted rates for
$ t\bar c $ production are controlled by quadratic products of the RPV
coupling constants, $\l ^{'\star }_{13k} \l '_{12k},\ [ k=1,2,3]$ and
the squark mass, $m_{\tilde d_{kR} }$, denoted for short as, $\tilde
m$. Allowing for the existence in the RPV interactions of an up-quark
flavor mixing, such as would be induced by the transformation from
flavor to mass basis, one may express the amplitude in terms of a
single RPV coupling constant and the CKM matrix, $V$, by rewriting the
coupling constant dependence as, $ \l'_{12k} \l {'^\star } _{13k} \to
\l _{1Mk}^{'\star } \l '_{1M'k} (V^\dagger )_{M'2}(V^\dagger ) ^\star
_{M3} $, and selecting the maximal contribution associated with the
configurations, $M=M'=2 $ or $ 3$.  This yields the order of magnitude
estimate, $\l _{13 k} ^{'\star } \l '_{12 k} \to \vert \l '_{12k}\vert
^2 (V ^\dagger )_{22} (V^\dagger )^\star _{23} \approx 2 \vert \l
'_{12k}\vert ^2 \l^2 $ or $ 2 \vert \l '_{13k}\vert ^2 \l^2 $,
respectively, where, $\l \approx \sin \t_c \approx 0.22 $, denotes the
Cabibbo angle parameter.

We pause briefly to recall the current bounds on the RPV coupling
constants of interest in the present study. \cite{h:reviews} The
relevant single coupling constant bounds are, $ \l '_{12k} < 4.
\times 10^{-2} ,\ \l '_{13k} < 0.37 $ (charged current universality);
$ \l '_{1j1} < 3 \times 10^{-2} $ (atomic physics parity violation); $
\l '_{12k} < 0.3-0.4, \ \l '_{13k} < 0.3 \ - \ 0.6 $ (neutral current
universality); and $\l '_{122} < 7. \times 10^{-2},\ \l '_{133} < 3.5
\times 10^{-3} $ (neutrino Majorana mass).  \cite{h:godbole} The
superpartners scalar particles masses are set at $ 100 \ GeV$. Unless
otherwise stated, all the dummy flavor indices for quarks and leptons
are understood to run over the three generations.  Using the above
results for individual coupling constants bounds, we may deduce for
the following upper bounds on the relevant quadratic products,
\cite{h:grossman} $ \l '_{13k} \l '_{12k} < [O(10^{-3}), \ O(10^{-2}), \
O(10^{-4})] , [k=1,2,3]$.  The indirect quadratic products bounds, $
\l '_{ijk} \l '_{i' 3k} < 1.1 \times 10^{-3} , \ \l '_{ijk} \l '_{i' j
3} < 1.1 \times 10^{-3} , \ [ i' =1,2 ] $ ($B \to X_q \nu \bar \nu $)
are roughly comparable to these single coupling constants bounds.  We
also note that using the CKM flavor mixing along with a single
dominant coupling constant in the current basis, as described at the
end of the previous paragraph, may not be especially beneficial in
avoiding the above stronger pair product bound.  The bound on the
corresponding coupling constant factor, $ 2 \vert \l '_{13k}\vert ^2
\l^2 < O(10^{-2}) $, is competitive for the generation indices, $
k=1,3$.

Numerical results for the integrated rates have already been reported
in previous works \cite{h:mahanta,h:chemtob}.  Setting the relevant RPV
coupling constants at the reference value, $\l '= 0.1 $, one predicts
rates of order $ 1 \ - \ 10 \ fb$, for $ \tilde m = O (100) \ GeV$.
As the center of mass energy varies in the interval, $\sqrt s = 192 \
- \ 1000 \ GeV$, the rates rise sharply from threshold, reaching
smoothly a plateau around $ \sqrt s \simeq 400 \ GeV$. This contrasts
with the predictions from gauge boson mediated higher dimension
interactions \cite{h:han} where the rise of the rates with incident
energy is a more gradual one.  The rates are also found to have a
strong dependence on $\tilde m$, which weakens for increasing center
of mass energies.  One may roughly parametrize the dependence on $ s $
and $ \tilde m$ by the approximate scaling law, $\s \approx ({\l ' \l
' \over 0.01 } )^2 ({100 \ GeV \over \tilde m })^{x (s)},$ where the
power exponent is a fastly decreasing function of energy, taking the
approximate values, $ x (s) \approx [3.65, \ 1.86, \ 0.94] $, at,
$\sqrt s =[0. 192, \ 0.5 , \ 1. ] \ TeV$.

\begin{figure} [t]
\begin{center}
\leavevmode \psfig{figure=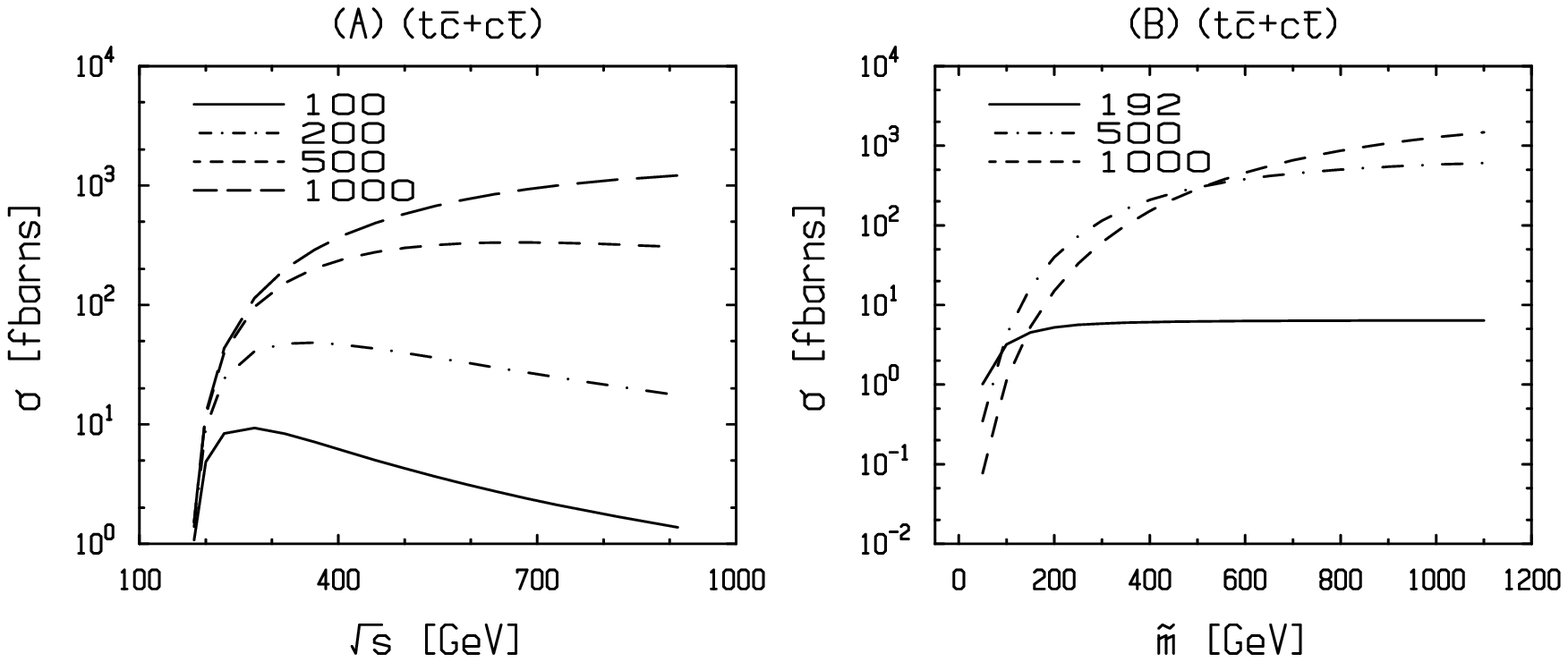,width=17cm}
\end{center}
\caption{The total integrated rate for the RPV induced reaction, $
l^+l^- \to( t \bar c) +(\bar t c) $, setting the values of the
relevant coupling constants as, $ \l '_{12k} = \l '_{ 13k} =0.1 \ (
\tilde m /100 \ GeV ) $, is plotted in window $(A)$ as a function of
center of mass energy, $ s^{1/2} $ for fixed down squark mass, $
\tilde m =[ 100, \ 200, \ 500, \ 1000 ] \ GeV $ and in window $(B)$ as
a function of $\tilde m$ for fixed $s^\ud =[ 192, \ 500, \ 1000 ] \
GeV $. We integrate over an interval of the scattering angle, $ 0\le
\vert \cos \t \vert \le 0.9848 $, corresponding to an opening angle
with respect to the beams axis larger than $10^{o } $. }
\label{h:fig2}
\end{figure}

Although the predicted rates seem to be severely constrained by the
above indirect bounds, one could envisage an optimistic scenario where
the supersymmetry decoupling limit, $\tilde m \to \infty $, is
realized with fixed values for the products, $ \l '_{ijk} ({100 \
GeV\over \tilde m }) \approx 0.1 $, consistently with the current
indirect bounds.  The results obtained with this prescription are
displayed in Figure \ref{h:fig2}.  The integrated rates now depend on $
\tilde m $ as, $ \s \propto ({100 \ GeV \over \tilde m })^{-4+x (s)}$,
which leads at high energies to an enhancement by up to three orders
of magnitudes, compared to the case where the RPV coupling constants
are taken independent of $\tilde m$. The initial energy of LEP-II
falls right in the regime where the cross section is sharply rising
with increasing initial energy.  The decrease with increasing $\tilde
m$ is stronger at LEP-II energies than at the future linear colliders
energies.  Note that at the largest values of the superpartner mass,
$\tilde m \simeq 1.  \ TeV$, the RPV coupling constants in our
prescription enter a strong coupling regime ($\l ' =O( 1) $) and it is
not clear then whether the tree level prediction makes sense.

Next, we consider the process incorporating the top semileptonic
decay, as pictured by the Feynman diagram shown in Fig.\ref{h:fig1}. We
assume that the top decay is dominated by the electroweak semileptonic
decay channel, with branching fraction, $ B (t\to b +W^+) \approx
1$. We also include the pair of CP-conjugate final states, $t\bar c $
and $c \bar t$ production, which multiplies the rate by a factor of
$2$. Note, however, that we restrict ourselves to the $ u_{J'} = c$
charm quark mode only.  The numerical results for rates, including a
branching fraction factor of ${2\over 9} $ (experimental value, $
21.1\% $) to account for the $W \to l \nu , \ [l= e, \ \mu ]$ decay
channels, are displayed in Table \ref{h:table1}.  We also show the
standard model background rate from the $W $-boson pair production,
$l^-l^+\to (W^+ W^-) \to ( \ l^+ \nu \bar u_i d_j ) + ( l^- \bar \nu
\bar d_j u_i) $, with one $ W$-boson decay leptonically and the other
hadronically, where $ i, \ j $ are generation indices.  The
irreducible background from, $ W ^- \to \bar c b $ or $ W ^+ \to \bar
b c$, is strongly suppressed, due to the small branching factor, given
approximately by, $ 0.32 \vert V_{cb} \vert ^2 \approx 5 \ 10 ^{-4}$.
It is safer, however, to allow for the possibility where the light
quark hadronic jets could be misidentified as $ b $ quark hadronic
jets.  Accounting for the leptonic decay for one of the $W$-boson and
the hadronic decay for the other $ W$-boson, introduces for the total
rate, which includes all the subprocesses, the branching fraction
factor, $ 2\times (21.1 \pm 0.64 )\% \times (67.8\pm 1.0)\% = 0.286
\pm 0.024$.  Our numerical results in Table \ref{h:table1} for the
standard model background rates are in qualitative agreement with
those quoted ($ \s = [2252, \ 864] \ fb$ at $ s^\ud = [0.5, \ 1. ] \
TeV$) by Han and Hewett.  \cite{h:han} One should be aware of the
existence of large loop corrections to the $ W^+ W^-$ production rate,
especially at high energies. The predictions including the electroweak
and QCD standard model one-loop contributions read, \cite{h:beenakker}
$\s = [4624, \ 1647, \ 596 ] \ fb$ at $ s^\ud = [0.192, \ 0.5, \ 1. ]
\ TeV$.  We conclude therefore that our use of the tree level
predictions for the $ W^+W^-$ background overestimates the true cross
sections by $ [9\%, 20 \%, 32 \%] $ at the three indicated energies.

\begin{table} 
\begin{center}
\vskip 1 cm
\vskip 0.5cm
\begin{tabular}{|c|c|c|c|}
\hline
$ Energy (TeV) $ & $ 0.192 $& $0.5 $ & $1.0 $ \\ 
\hline
Total rate $ \s (fb) $ & $ 4.099 $& $ 4.291 $ & $1.148$ \\ 
\hline
Signal (fb) & $ 0.68 $ & $ 0.91 $ & $ 0.24 $ \\
\hline  
$W^+W^-$ Background (fb) & $ 5076$ & $ 2080 $ & $ 876 $ \\  
\hline
Cut Signal (fb) & $ 0.54 $ & $ 0.74 $ & $ 0.21$ \\  
\hline
$W^+W^-$ Cut Background (fb) & $ 17. $ & $5.0 $ & $ 2.6 $ \\ 
\hline
\end{tabular}
\end{center}
\caption{ Production rates for the top-charm production signal and the
W-boson pair production background.  The line entries give
successively the total integrated rate for the reaction, $l^+l^- \to (
t \bar c) + (c \bar t )$, using, $\l '= 0.1, \ \tilde m =100 \ GeV$,
the rate for signal events, $(b \bar l \nu \bar c) +(\bar b l \bar \nu
c ) $, associated with the top semileptonic decay, the W-boson pair
production background rate, $l^-l^+\to W^+ W^-\to (l^+ \nu \bar u_i
d_j ) + (l^-\bar \nu \bar d_j u_i ) $, and the corresponding cut
signal and background rates, as obtained by applying the selection
cuts quoted in the text.  The results include the first two
generations of charged leptons, $ \ l=e, \ \mu $. }
\label{h:table1}
\end{table}
%
%

Let us discuss briefly other possible sources of background. The next
important contribution is that arising from the non resonant W-boson
virtual propagation in the amplitude with the intermediate $ W^+ W^-$
bosons branching into four fermions ($l \nu q\bar q'$).  This could be
possibly estimated by subtracting the resonant contribution from the
total background cross sections, weighted by the suitable branching
factors, as independently evaluated by numerical methods in the
literature.  The results for the integrated total cross sections, $
l^-l^+ \to (4 f) + (4 f + \g )$, \cite{h:bardin} including the initial
state radiation and Coulomb corrections, indicate that the off-shell
contributions amount to a small relative correction lower than $
O(10\% ) $.  Alternatively, one may consider, after reconstructing the
neutrino momentum from the missing energy, a procedure to impose
suitable cuts on the $ bW$ invariant mass, aimed at suppressing the
non resonant production background.

 One other potentially important background is that arising from the
$b- \bar b$ quark pair production reaction, $l^+l^- \to \g ^\star / Z
\to b \bar b \to \bar b ( cl^-\bar \nu ) + b (\bar cl^+\nu ) $.
\cite{h:mahanta} The numerically derived predictions for the rates, as
obtained by means of the PYTHIA generator, are: $ \s = [1.631 \ 10^4 ,
\ 2.12 \ 10^ 3 , \ 5.35 \ 10^2 ] \ fb$, at $ s^\ud = [0.192, \ 0.5, \
1. ] TeV$.  It would appear desirable, in view of these large
predicted rates, to eliminate this background by performing a double $
b $ quark tagging analysis on the events sample. This can be performed
at a reasonably low cost, given that the detection efficiency of $b$
quark jets is currently set at $ 50\%$.  If one performs a single $ b
$ quark tagging, the rates for the corresponding events, $l^+l^- \to
\g ^\star / Z \to b \bar b \to ( cl^-\bar \nu ) \bar b $, are reduced
by a branching fraction, $ B(b \to c l \nu ) \approx 10\% $, but this
is compensated by the probability of misidentifying a light quark jet
as a $ b $ quark jet, which lies at the small value of $ 0.4 \% $ with
the current silicon vertex techniques.  If no $ b $ quark tagging is
performed at all, then the above large rates may make it necessary to
resort to an analysis of isolation cuts of the type to be discussed in
the next subsection.

\subsection{Distributions  for the semileptonic top decay events}
\label{sec21} 
\begin{figure} [t]
\centerline{ \psfig{figure=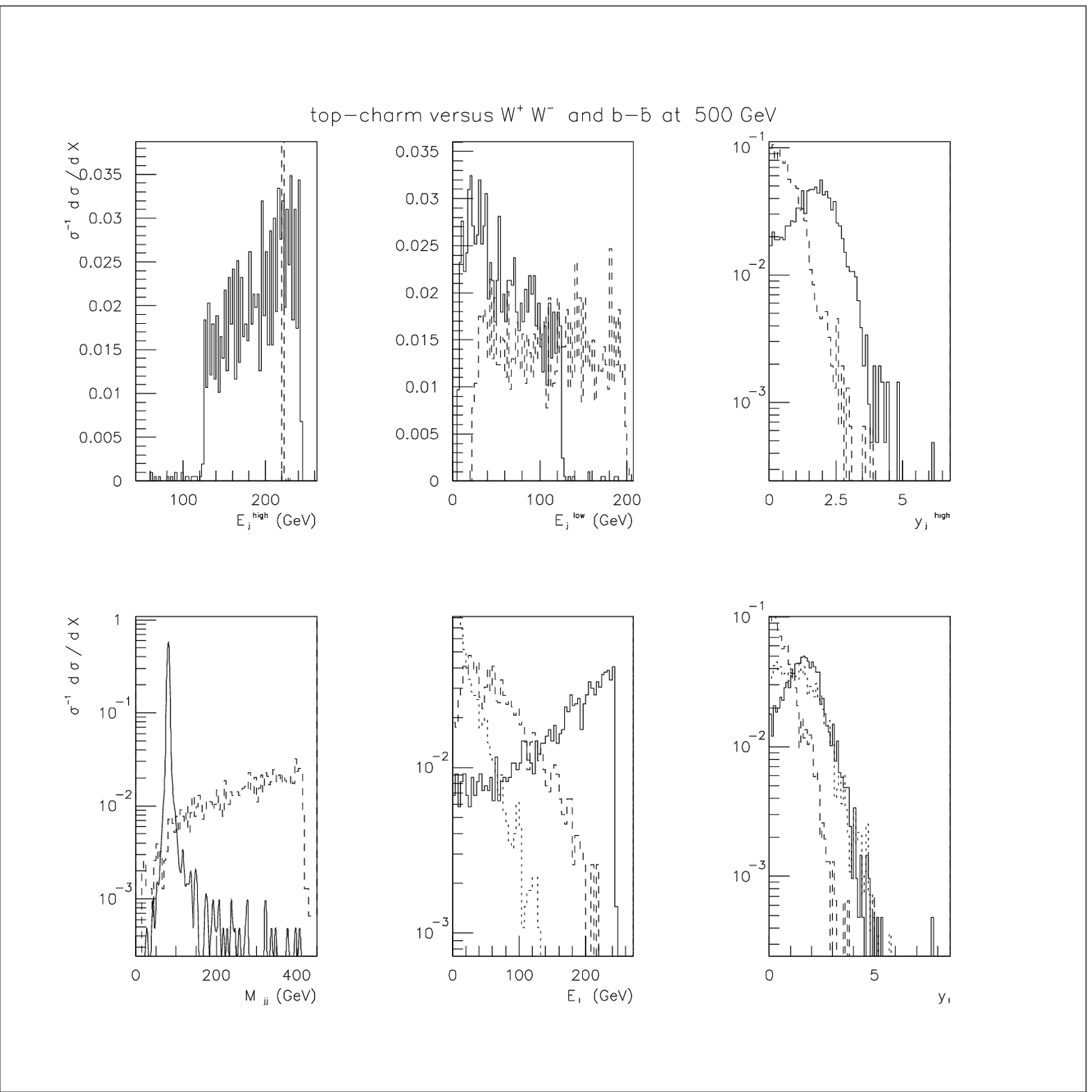} }
\vskip 1 cm
\caption{ Normalized dynamical distributions associated with the
signal events, $l^+l^-\to (t \bar c) +(\bar t c) \to (b l^+ \nu \bar
c) +( \bar b l^-\bar \nu c)$, (dashed line) at $\tilde m =100 \ GeV$,
and the background events, $ l^+l^-\to W^+W^- \to ( l^+ \nu \bar q q'
) + ( l^- \bar \nu q\bar q' ) $, (continuous line) at a center of mass
energy, $ s^\ud = 500 \ GeV$. The kinematical variables in the
histograms, from left to right and up to down, are the jets maximum
energy, the jets minimum energy, the rapidity for the highest energy
jet, the di-jet invariant mass, the charged lepton energy and the
charged lepton rapidity.  The charged lepton energy and rapidity
distributions are also plotted for the $ b-\bar b$ background
production events, $ l^+l^-\to b \bar b \to l^\pm + hadrons $ (dotted
line).  }
\label{figtopp}
\end{figure}

In order to separate the signal from background, we consider the same
set of characteristic final state kinematical variables as proposed in
the study by Han and Hewett. \cite{h:han} These are the maximum and
minimum energy of the two jets, $ E_j^{high}, \ E_j^ {low}$, the dijet
invariant mass, $ M_{jj} $, the charged lepton energy, $E_l,$ and
rapidity, $y_l = \ud \log {E_l +p_{l\parallel } \over E_l-p_{l
\parallel } } $.  The distributions in these $6$ variables for the
signal and background, at a center of mass energy, $ \sqrt s = 0.5
TeV$, are plotted in Fig.\ref{figtopp}.  These numerical results were
obtained by means of the PYTHIA \cite{h:sjostrand} event generator. One
notices marked differences between signal and background.  The maximum
jet energy distribution is uniformly distributed for the background
but sharply peaked for the signal, where the peak position is
determined by the top mass and the incident energy as, $m_t^2 = (m_c^2
-s +2\sqrt s E_p)$.  The minimum jet energy is uniformly distributed
for both signal and background, but happily the corresponding
intervals are very partially overlapping.  The signal events rapidity
distributions for the maximum energy jet are more central for signal
than background. A similar trend holds for the lepton rapidity
distributions.  The dijet invariant mass is a most significant
variable in discriminating against the background due to its
pronounced peak at the $ W $ mass. For the signal, the dijet invariant
mass is uniformly spread out. Although we do not show here the
distributions for the top mass reconstruction, this also features a
strong contrast between a strongly peaked signal and a uniform
background. The lepton energy distributions for the signal and
background are peaked at the opposite low and high energy ends of the
physical interval, respectively. This is a familiar effect associated
with the correlation between the W-boson spin polarization, which is
predominantly longitudinal in the top decay and transverse in the
direct W-boson decay, and the velocity of the emitted charged lepton.
In the signal decay amplitude, $ t \to b\bar l \nu $, the fact that
the left handed b-quark must carry the top polarization, forces the
lepton to travel with opposite velocity to that of top.  In the
background decay amplitude, $ W^-\to l \bar \nu $, the charged lepton
is emitted with a velocity pointing in the same direction as that of
the W-boson. Thus, the Lorentz boost effects on the emitted charged
leptons act in opposite ways for the signal and background events.

\begin{figure} [t]
\centerline{ \psfig{figure=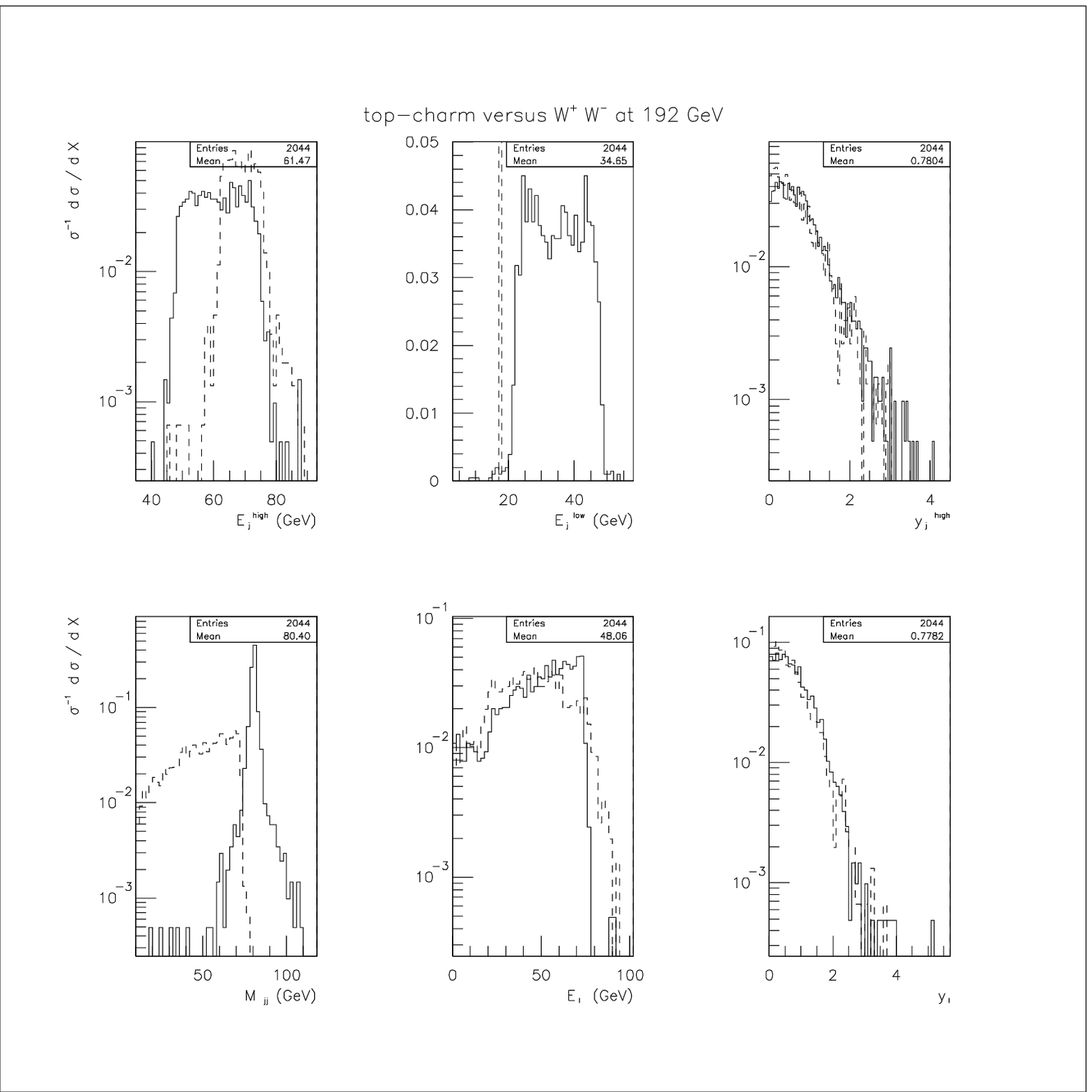} }
\vskip 1 cm
\caption{ Same distributions as in Fig.\ref{figtopp}, at a center of
mass energy $ s^\ud = 192 \ GeV$. }
\label{figtopp192}
\end{figure}

While the above distinctive features between signal and background
events get further pronounced with increasing center of mass energy,
opposite trends occur as the initial energy is lowered.  The
distributions at the LEP-II center of mass energy, $ \sqrt s = 0.192 \
TeV,$ are plotted in Fig.\ref{figtopp192}.  At this energy, the
monovalued distribution for the signal jet, which is now the softer
lower energy jet, is still well separated from the corresponding
background jet distribution.  So, this variable, along with the dijet
invariant mass stand up as useful discrimation tests for the signal.
By contrast, the energy and rapidity distributions for the maximum
signal jet may not be easily distinguished from the
background. Similarly, the lepton energy distributions in the signal
and background are overlapping due to the small Lorentz boost effect.

The distributions obtained with the RPV interactions are rather
similar to those found with the higher dimension operator
mechanism. \cite{h:han} This is due to the formal structure of the RPV
amplitude, involving an effective $u$-channel vector particle
exchange.  In fact, the selection cuts proposed by Han and Hewett
\cite{h:han} appear to be quite appropriate also in the RPV case, and,
for convenience, we recapitulate below the cut conditions used to
characterize the selected events.

$ E_j^{low} < 20,\ E_j^ {high} > 60,\ E_l > 0,\ \d_{jj}> 10,\ \d_t <
5, \quad [ \sqrt s = 192] $

$ E_j^{low} > 20,\ E_j^ {high} > 200,\ E_l < 150, \ \d_{jj}> 10,\ \d_t
< 40, \quad [\sqrt s = 500 ] $

$ E_j^{low} > 20,\ E_j^ {high} > 460,\ E_l< 350,\ \d_{jj} > 10,\ \d_t
< 100.  \quad [\sqrt s = 1000] $

 The above listed variables correspond to the minimum and maximum
energy of the two jets, $ E_j^ {low} , \ E_j^{high}, $ the charged
lepton energy, $E_l, $ the distance between the dijet invariant mass
and W-boson mass, $\d_{jj} = \vert M_{jj} -m_W \vert, $ the distance
of the reconstructed top mass to the true mass, $ \d_t = \vert
m_t^{reconst}-m_t \vert $.  The assigned numerical values are all
expressed in GeV units.  Besides the above cuts, we also impose the
usual detection cuts on energies and rapidities, $ E_{j,l} > 10 \ GeV,
\ \vert \eta_{j,l} \vert < 2$, aimed at removing the particles
travelling too close to the beam pipe.  We allow for the detection
efficiency of the particle energies only in an approximate way,
namely, by accounting for the following approximate uncertainties, $\D
E /E = 40 \%, \ 10 \%$, on the jets and lepton energies, respectively,
at the level of imposing the above selection cuts, rather than by the
usual procedure of performing a Gaussian smearing of the particle
energies.
 
The numerically evaluated efficiencies on the signal and background
events are, $ \e_S \simeq 0.8 , \ \e_B \simeq 3. 10^{-3} $, with a
very weak dependence on the center of mass energy and, for the signal,
a weak dependence on the mass parameter $\tilde m$, which was set at $
\tilde m = 100 \ GeV$ in the numerical simulations.  After applying
the cuts, the background rates are, $ \s_B \e_B = [ 17., \ 5., \ 2. ]
\ fb, $ and the signal rates, $ \s_S\e_S = [ 0.68, \ 0.74, \ 0.21 ] \
fb $, for $\sqrt s = [192., \ 500., \ 1000 ] \ GeV$.  The results for
the cut signal and background rates, as given in Table \ref{h:table1},
show that the background is very significantly reduced by the cuts.
The situation is clearly far more favorable for the future linear
colliders than for LEP-II.  Nevertheless, the number of surviving
signal events is still one order of magnitude below that of the
surviving background, so that the option of cutting down the
background by means of a $b $ quark tagging procedure is to be
preferred since the ensuing reduction would be much more drastic. An
integrated luminosity of $\call = 100 \ fb^{-1}$ would lead to a
number of signal events, $(\l '_{12k} \l'_{13k} /10^{-2} ) \times
O(30) $.

We have also performed an indicative event generator study of the
background, $ l^+l^-\to b \bar b \to l^\pm + hadrons $, restricting
consideration to the emitted charged leptons only. A jet
reconstruction of the partonic level distributions is a task beyond
the scope of the present work. We focus on the first charged lepton
emitted during the semileptonic decays of the produced $B , \ \bar B $
mesons, since this carries the largest velocity. As seen on
Fig.\ref{figtopp}, the distribution for the first emitted charged
lepton energy is peaked at low energies.  One expects that the most
energetic lepton is that produced in the semileptonic decays of the $B
$ mesons.  The rapidity distribution is less central than for the
signal and nearly overlaps with that of the $ W^+W^-$ background.
Therefore, imposing the additional lower bound cut on the lepton
energy, say, at $ E_l > 20 \ GeV$, for $s^\ud = 500 GeV$, should be
sufficient to appreciably suppress the $ b-\bar b$ background without
much affecting the signal.

\begin{figure} [t]
\centerline{ \psfig{figure=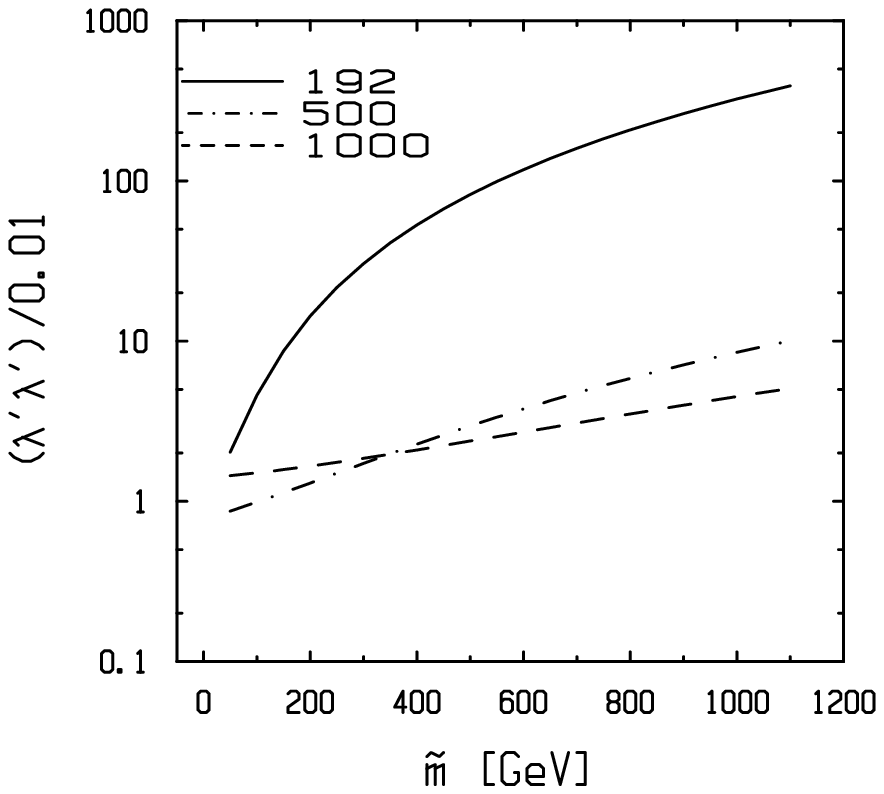}}
\vskip 1.  cm
\caption{ Sensitivity reach plot for the RPV coupling constants
product, $\l '_{12k} \l '_{13k} /0.01$, as a function of the down
squark mass, $\tilde m$, for fixed center of mass energy, $ s^\ud =[
192. \ , 500 , \ 1000 ] \ GeV$, and corresponding fixed integrated
luminosity, $ \call = [2., \ 100., \ 100. ]\ fb^{-1} $, using an
acceptancy for the background, $ \e _B = 3\ 10 ^{-3} $, and an
acceptancy for the signal, $ \ \e_S = 0.8 $, assumed to be independent
of $\tilde m$.}
\label{figsens}
\end{figure}

We may infer the reach with respect to the free parameters by
evaluating the statistical significance ratio for a discovery, as
defined by, $ \hat \sigma = {S\over \sqrt { S+B} } ,\ S = \s_{S} \call
,\ B =\s_{B} \call $, where $ \call $ denotes the integrated
luminosity.  Setting this at the value, $ \hat \sigma = 3 $,
corresponding to a $ 95 \% $ confidence level, one deduces a
dependence of the RPV coupling constant as a function of the
superpartner mass parameters for a fixed initial energy and integrated
luminosity.  The sensitivity reach contour plot for the relevant
parameters, $ \l ' \l '= \l '_{12k} \l _{13k} ^{'\star } $ and $\tilde
m = m_{ \tilde d _{kR} } $, is shown in Fig.\ref{figsens}.  We note
that the sensitivity limit on the product of coupling constants, $ \l
' \l '$, scales with the luminosity approximately as, $ 1/ \sqrt \call
$.  While the reach on the RPV coupling constants products, $\l
'_{12k} \l '_{13k} < O(10^{-1}) $, lies well above the current
indirect bounds, this covers a wide interval of the down squark mass
which extends out to $ 1 TeV$.  To compare with analogous collider
physics processes, we note that while the flavor diagonal fermion pair
production reactions, $ e^-e^+ \to f_J\bar f_J$, may have a higher
sensitivity reach, these are limited to information on the single
coupling constants, $ \l '_{1jk}$ \cite{h:choudhee}.  The special
reaction, $ e^-e^+ \to b\bar b $, proceeding via a sneutrino
$s$-channel resonance, may probe quadratic products such as, $
\l_{131} \l '_{333} $, \cite{h:erler} or $ \l_{131} \l '_{311}$
\cite{h:kalitev} at levels of $ O(10^{-3})$, but this is subject to the
existence of a wide sneutrino resonance.  The $t\bar b$ associated
production at the hadronic Fermilab Tevatron \cite{h:datta,h:oakes1} and
the Cern LHC \cite{h:oakes1} colliders can be initiated via a charged
slepton $ \tilde e_{iL} $ $s$-channel exchange. The sensitivity reach
on the linear combination of quadratic coupling constants products,
$\l'_{i11} \l'_{i33} $, is of order $ 10^{-2} - 10^{-1}$. This
information should prove complementary to that supplied by our study
aimed at the leptonic colliders.  To conclude this brief comparison,
we observe that the information provided by the single top production
reaction appears to be rather unique in view of the very
characteristic signature of the associated events.

\section{Top polarization observables and a test of  CP violation}
\label{sec3}

\setcounter{equation}{0}

Should single top production become experimentally observable in the
future, an important next step to take is in examining top
polarization observables.  In this section, we present an approximate
study for the top semileptonic decay signal in top-charm associated
production aiming at a test of CP violation.  We exploit an idea which
was developed in early studies of $ t-\bar t$
production. \cite{h:kane,h:schmidt} Interesting extensions are currently
pursued \cite{h:toppol,h:bartl,h:rigolin}. The basic observation is that any
CP-odd quantity depending on the top polarization, such as the
difference of rates between the pair of CP conjugate reactions, $ \s (
l^- l^+ \to t_L \bar c ) - \s ( l^- l^+ \to \bar t_R c ) $, can become
observable by analyzing the top polarization through the kinematical
distributions of its emitted decay ($b$ quark or charged lepton)
products.  An especially interesting observable is the charged lepton
energy distribution for a polarized top.  Any finite contribution to
the CP-odd observables must arise through an interference term
involving imaginary parts of loop and tree amplitudes factors, the
loop amplitude factor bringing a CP-even final state interaction
complex phase with the CP-odd relative complex phase arising from the
coupling constants in the product of loop and tree amplitudes.

\subsection{Helicity basis amplitudes} 
\label{sec30}
 Building on our previous work, \cite{h:chemtob} we shall combine the
tree level RPV induced amplitude discussed in Section \ref{sec2} with
the one-loop RPV induced amplitude associated to the photon and
Z-boson exchange diagrams, restricting ourselves to the vertex
corrections in the electroweak neutral current vertices, $ \g \bar f
_J (p) f_{J'} (p') $ and $ \ Z \bar f _J (p) f_{J'} (p') $.  The
Z-boson vertex admits the general Lorentz covariant decomposition,
\begin{eqnarray}
J_\mu ^{ Z }& = & - {g \over 2 \cos \t_ W} \G ^{JJ'} _\mu (Z), \cr \G
^{JJ'} _\mu (Z) & =& \g_\mu ( A ^{JJ'}_L(f) P_L + A ^{JJ'}_ R (f) P_R)
+{1\over m_J +m_{J'}} \s_{\mu \nu } (p+p')^\nu (i a ^{JJ'} +\g_5 d
^{JJ'} ), \cr & &
\label{equ321}
\end{eqnarray}
where the vectorial vertex functions, $ A^{JJ'} _{L , R} = A^{JJ'} _{L
, R}\vert_{tree} + A^{JJ'} _{L , R}\vert_{loop}$, have a tree level
contribution given by, $ A_{L,R} ^{ JJ' }\vert _{ tree} = \d_{JJ'}
a_{L,R} (f), \ a_{L,R} (f) = 2T_3^{L,R} -2Q(f) \sin ^2 \t_W $, and the
tensorial vertex functions, $ a^{JJ'},\ d^{JJ'} $, are associated with
the anomalous transition magnetic moment and the CP-odd, P-odd
electric transition dipole moment, respectively.  An analogous
decomposition applies for the photon, $J_\mu ^{\g } = -{g \sin \t_ W
\over 2 } \G ^{JJ'}_\mu ( \g ),$ with $ a_{L,R} (f) =2Q(f) $,
determined by the electric charge $Q(f)$.  It is convenient to work
with the $ Z f _J\bar f_{J'}$ vertex in the alternate Lorentz
covariant decomposition, $\G ^{JJ'} _\mu (Z) =\g_\mu (\capa -\capb
\g_5 ) + \ud (p-p')_\mu (\capc -\capd \g_5 ),$ where the vertex
functions, $\capa , \ \capb , \ \capc , \ \capd $, (omitting the up
quarks generation indices $ J, \ J'$ for convenience) are related to
the previously defined vectorial and tensorial ones,
eq. (\ref{equ321}), as,
\begin{eqnarray}
\capa & =& {1\over 2} (A ^{JJ'}_L(f) +A ^{JJ'}_R(f) ) +a ^{JJ'} , \
\capb = {1\over 2} (A ^{JJ'}_L (f) -A ^{JJ'}_R (f) ) + { m_J -m_{J'}
\over m_J +m_{J'} }i d ^{JJ'} , \cr \capc & =& -{2 \over m_J +m_{J'} }
a ^{JJ'} , \ \capd = -{2 \over m_J +m_{J'} } i d ^{JJ'}.
\label{equ51}
\end{eqnarray} 
The one-loop Z-boson exchange amplitude may then be written in the
form,
\begin{eqnarray}
M^{JJ'}_l (Z) & = &\bigg ({g\over 2 \cos \t_W}\bigg )^2 \bar v (\vec
k',\mu ')\g_\s \bigg (a(e_L) P_L+a(e_R)P_R \bigg )u(\vec k, \mu )
{1\over s-m_Z^2+im_Z\G_Z} \cr \times &&\bar u (\vec p,\l ) [\g^\s
(\capa -\capb \g_5 ) + \ud (p-p')^\s (\capc -\capd \g_5 ) ] v({\vec p
' }, \l ') .
\label{equ322}
\end{eqnarray}
Combining the above loop amplitude with the RPV tree amplitude,
eq. (\ref{h:eq1pp}), which we rewrite as,
\begin{eqnarray}
M ^{JJ'} _t = \calr \bar v \g_\mu (1-\g_5) u \bar u \g^\mu (1-\g_5) v,
\quad \calr= -{ \l _{1Jk}^{'\star } \l '_{1 J'k} \over 8
(u-m^2_{\tilde d_{kR } } ) } ,
\label{equ323}
\end{eqnarray}
one obtains,
\begin{eqnarray} 
M^{JJ'}&=& M^{JJ'} _t + M^{JJ'} _l(Z) = [ ( G a^+ \capa + \calr )( \g
_\mu )( \g ^\mu ) - ( G a^+ \capb + \calr )( \g _\mu )( \g ^\mu \g_5)
\cr & - & ( G a^- \capa + \calr )( \g _\mu \g_5)( \g ^\mu ) + ( G a^-
\capb + \calr )( \g _\mu \g_5 )( \g ^\mu \g_5) \cr & + & \ud
(p-p')^\mu [ G a^+ \capc ( \g _\mu )( 1) - G a^+ \capd ( \g _\mu )(
\g_5) - G a^- \capc ( \g _\mu \g_5 )( 1) + G a^- \capd ( \g _\mu \g_5
)( \g_5) ], \cr & &
\label{equ52}
\end{eqnarray}
where, $ a^\pm = \ud (a_L(e)\pm a_R(e)),$ and we have omitted writing
the contractions of the Dirac spinors indices for the initial and
final fermions, respectively.  The photon exchange contribution can be
incorporated by treating the parameters $ a^\pm $ as operators acting
on the vertex functions, $\capa ,\ \capb , \ \capc , \ \capd $, by
means of the formal substitutions,
\begin{eqnarray}
G a^\pm \capa & =& G_Z {a_L(e) \pm a_R(e) \over 2} \bigg ( {A^{JJ'} _L
(f) \pm A^{JJ'} _R (f) \over 2} + a^{JJ'} \bigg ) \cr & & + G_\g { 2 Q(f)
\choose 0} \bigg ( { A_L^ { \g JJ'} (f) \pm A_R ^{ \g JJ'} (f) \over
2}+ a^{JJ'} \bigg ) , \cr G_Z & =& ({g\over 2 \cos \t_W })^2 {1\over
s-m_Z^2 +im_Z \G_Z} , \ G_\g = ({g\sin \t_W \over 2 })^2 {1\over s}.
\label{equ53}
\end{eqnarray}
Analogous formulas to the above ones hold for the other products, $ G
a^\pm \capb ,\ G a^\pm \capc ,\ G a^\pm \capd.$ We have labelled the
vertex functions for the photon current by the suffix $\g $. The
formulas expressing the RPV one-loop contributions to the vertex
functions are provided in Appendix \ref{appexa}, quoting the results
derived in our previous work. \cite{h:chemtob} The amplitude $ M^{JJ'}$
in eq.(\ref{equ52}) may be viewed as a $ 4\times 4 $ matrix in the
fermions polarization space, $ (f(p, \l ) \bar f_{J'} (p', \l ') \vert
M \vert l^+ (k',\mu ') l^- (k, \mu ) ) $.  The various products in
eq. (\ref{equ52}) for the matrix elements with respect to the two
pairs of Dirac spinors separate into 8 distinct terms.  The
calculation of the helicity amplitudes is most conveniently performed
with the help of the `Mathematica' package.  Of the $16$
configurations only the $ 8$ helicity off diagonal configurations in
the initial fermions are non vanishing.  The explicit formulas for the
helicity amplitudes are provided in Appendix \ref{appexa}.

\subsection{Charged lepton energy distribution}
\label{sec31}
The differential cross section for top production and decay is
described in the factorization approximation. Ignoring the spin
correlations, which corresponds to dropping the spin non diagonal
contributions between the production and decay stages, yields:
\begin{eqnarray}
d\sigma & =&{\vert p \vert \over 128 \pi s \vert k \vert } {m_t \over
\pi } \int d(\cos \t ) \sum_{ \l } \vert M_{prod} (l^-l^+\to t_\l \bar
c)\vert ^2 \int dp^2 {1\over \vert p^2 -m_t^2 +im_t \G_t\vert ^2 }
d\G_t ,\cr d \G_t& = &{1\over (2\pi )^3 8 m_t }\overline { \sum } _ {
\l ' } \vert M_{dec} (t_{ \l '} \to b l^+ \nu ) \vert ^2 dE ^ \star _l
dE^\star _b .
\label{equ1}
\end{eqnarray}
The production amplitude is denoted, $M_{prod}$, the top decay
amplitude, $ M_{dec}$, and $ \l , \ \l ' = \pm 1 $, are polarization
labels, which will also be written for short as, $\pm $ .  We shall
assume a narrow resonance approximation for the top propagator, $\vert
p^2 -m_t^2 +im_t \G_t\vert ^{-2} \to {\pi \over m_t \G_t } \d (p^2
-m_t^2)$.  For the energies of interest, all the leptons and quarks,
with the exception of the top, may be treated as massless.  Two frames
of interest are the laboratory $(l^-l^+$) rest frame and the top rest
frame.  The letters denoting momentun variables in the $l^-l^+ $
center of mass (laboratory) frame are distinguished from those in the
top rest frame by the addition of a star. Standard kinematical methods
\cite{h:kajan} can be used to transform variables between these frames.
Exploiting the rotational invariance, one may conveniently choose to
work in the spatial frame where the top momentum lies in the $ xOz $
plane ($\theta , \phi =0 $) and the charged lepton points in an
arbitrary direction described by the spherical angles, $ \t_{l},
\phi_{l}$. The relations between angles may be obtained by use of the
spherical triangle identities, for example, the angle between lepton
and top reads, $ \cos \t_{lt} = \cos \t_{l} \cos \t + \sin \t_{l} \sin
\t \cos \phi_l.$ The Lorentz boost from the top rest frame to the
laboratory frame, involves a velocity parameter, $ \vec v= \vec p/E_p,
\ \beta = p/E_p, \ \g = (1-\b ^2)^{-1/2}= E_p/m_t, $ and yields for
the charged lepton momentum four vector and polar angle relative to
the top momentum, $E^\star _l =\g (E_l - \vec v \cdot \vec k_l ),\quad
\vec k^\star _l= \vec k_l +\g \vec v ({\g \vec v \cdot \vec k_l \over
\g +1} -E_l ), \quad \cos \t_{lt} ^\star ={\cos \t_{lt}-\b \over 1- \b
\cos \t_{lt} }$.

The top differential semileptonic decay rate has been thoroughly
studied in the literature.  \cite{h:kuhn} One representation convenient
for our purposes is the double differential rate with respect to the
final charged lepton energy, $ E^\star _l$, and the final lepton and
neutrino invariant mass squared, $ W ^2 = (k_l+k_ \nu )^2$.  The
result for the unpolarized rate carries no dependence on the
scattering angles and reads, quoting from ref.\cite{h:kuhn},
\begin{eqnarray}
d\G _t&= &{N_lG_F^2m_t^5\over 16\pi^3 } dx_l \int dy { x_l
(x_M-x_l)\over (1- y \xi)^2 +\g ^2 }, \cr &=& {N_lG_F^2m_t^5\over
16\pi^3 } {2\over m_t} {x_l (x_M-x_l) \over \g \xi } \tan^{-1} {\g \xi
x_l (x_M-x_l) \over (1+\g ^2) (1-x_l) -\xi x_l (x_M-x_l ) }\
dE_l^\star .
\label{h:equ4}
\end{eqnarray}
The kinematical variables for the emitted charged lepton and neutrino
are defined as, $ x_l =2E_l^\star /m_t, \ y= W^2/m_t^2, \ [ W=k_l+k_
\nu ] $ with the bounds, $ 0< x_l< x_M , \ 0< y < {x_l (x_M-x_l) \over
1-x_l } $ and we employ the following notations, $ N_l $ for the
number of light lepton flavors, $ \g = \G_W/m_W,\ \xi = m_t^2/m_W^2,\
x_M =1-\e^2, \ \e =m_b/m_t , \ \tan ^{-1} A = Artan \vert A \vert +
\pi \t (-A) $.  Recall that the number of light lepton flavors, $ N_l
$, is set to $ N_l =2$ in our analysis.  A useful trick to obtain the
distribution with respect to the laboratory frame lepton energy, $
E_l$, is to choose the top momentum along the $Oz$ axis fixed frame
and introduce the top rest frame electron energy by means of the
change of variable, $ (E _l , \cos \t^\star _l) \to ( E _l, E ^\star
_l)$, associated with the Lorentz boost between the top rest frame and
the laboratory frame, $ E_l= \g E_l^\star (1+\b \cos \t ^\star _l)
$. The result reads,
\begin{eqnarray}
{d \G_t \over dE_l }=\int_{-1}^{+1} d \cos \t_l^\star {d^2 \G_t \over
dE_l d \cos \t _l ^ \star } = {2\over m_t \g \b } \int_{x_l^-}
^{x_l^+} { d x_l\over x_l } {d ^2 \G _t \over d x_l d \cos \t _l ^
\star } ,
\label{equ41}
\end{eqnarray}
where the integration interval over $ x_l $ is bounded at, $x^\pm _l =
{ 2 E_l \over m_t \g (1\pm \b ) }$.

\subsection{Top polarization observables}
\label{sec32} 
 An essential use will be made of the factorization property of the
double differential distribution for the top decay semileptonic rate
with respect to the emitted lepton energy and angle relative to the
top spin polarization vector. This distribution is described at the
tree level as, $ {d ^2\G_t \over dE_l ^\star d \cos \psi_l } = {d\G_t
\over dE_l ^\star } {1+\cos \psi_l \over 2} $, where, $ \cos \psi_l
=-s(p) \cdot k_l $, is the angle between the lepton momentum and the
top spin polarization vector, $s_\mu (p)$, in the top rest frame.
Equivalently, $ {d ^2 \G _t \over d x_l d \cos \t _l ^ \star } = {d \G
_t \over d x_l } { 1+\cos \psi_l \over 2 } { d\cos \psi_l \over d \cos
\t _l ^ \star }$. \cite{h:kuhn} As it turns out, this representation
remains valid to a good approximation when one-loop QCD corrections
are included. \cite{h:kuhn2} We choose to describe the top polarization
in the spin helicity formalism, using techniques familiar from
previous works. \cite{h:kane,h:topfor} The definition for the helicity
basis Dirac spinors is provided in Appendix \ref{appexa}.  Since the
polarization axis coincides then with the top momentum, the dependence
on $\psi_l $ can also be simply rewritten as, $ (1+ \cos \psi_l )/2 =(
1 + \l \cos \t_l ^\star )/2 ,$ such that, $ \l = [- 1, +1] $,
correspond to $[L,R]$ helicity, respectively.

The helicity amplitudes associated to the pair of CP-conjugate
processes are related by the action of $ CP$ as, $ <f_{\l }\bar f'_{\l
' } \vert M \vert l^+_{ \mu '} l^- _{ \mu } > \to <f '_{-\l '} \bar
f_{-\l } \vert M \vert l^+_{- \mu } l^- _{- \mu ' } >.$ Unlike the
process, $ l^+ l^- \to t\bar t$, where both the initial and final
states are self-conjugate under CP, here only the initial state is
self-conjugate, while the action of CP relates the different final
states, $ t\bar c $ and $ c \bar t$.  Let us express the amplitudes
for the pair of CP-conjugate processes as sums of tree and loop terms,
$ M^{JJ'} = a_0 +\sum_\a b_\a f_\a(s+i\e ) , \ \bar M^{JJ'} = a^\star
_0 +\sum_\a b^\star _\a f_\a(s+i\e ) $, where the loop terms, $b_\a
f_\a(s+i\e ) $, are linear combinations with real coefficients of the
vertex functions, $ A^{JJ'}_{L}, \ A^{JJ'}_{R}, \ a^{JJ'} , \ i
d^{JJ'} $, with the energy dependent complex functions, $ f_\a (s+i\e
) $, representing the factors in loop amplitudes which include the
absorptive parts.  In terms of these notations, a CP asymmetry
associated with the difference of rates for the pair of CP-conjugate
processes in some given CP-conjugate configurations of the particles
polarizations, can be written schematically as,
\begin{eqnarray} 
\vert <\l \l ' \vert M \vert \mu ' \mu >\vert^2 & - & \vert <-\l ' -\l
\vert M \vert -\mu - \mu ' >\vert^2 \propto \sum_\a Im (a_0 b_{\a
}^\star ) Im (f_\a (s+i\e ) ) \cr & - & \sum _{\a < \a '} Im (b_\a
b_{\a '} ^\star ) Im (f_\a (s+i\e ) f_{\a '} ^\star (s+i\e ) ) .
\end{eqnarray} 
Thus, the necessary conditions for a non vanishing polarized asymmetry
to arise from the tree-loop interference term are a relative complex
CP-odd phase between the tree and loop coupling constants and an
absorptive part from the loop terms.  The angle integrated production
rates for the CP-conjugate reactions, $ l^+ l^- \to t \bar c$ and $
l^+ l^- \to c \bar t$, for the case of polarized top and antitop,
respectively, are obtained by summing over the polarization of the $c,
\ \bar c $ quarks as,
\begin{eqnarray} 
\sigma (t_{L})& =& \sigma (t_L \bar c _R ) + \sigma (t_L \bar c _L ) ,
\ \sigma (t_{R}) = \sigma (t_R \bar c _L) + \sigma (t_R \bar c _R ),
\cr \sigma (\bar t_{L})& =& \sigma (\bar t_L c _R ) + \sigma (\bar t_L
c _L ) , \ \sigma (\bar t_{R}) = \sigma (\bar t_R c _L) + \sigma (\bar
t_R c _R ).
\label{equ54}
\end{eqnarray}
Forming the half differences and sums of rates, $ \d \sigma = \ud (
\sigma (t_L) -\sigma (t_R) ) , \ \d \bar \sigma =\ud ( \sigma (\bar
t_R) -\sigma (\bar t_L) ), \ \sigma _{av} = \ud ( \sigma (t_L) +\sigma
(t_R) ), \ \bar \sigma _{av}= \ud ( \sigma (\bar t_R) +\sigma (\bar
t_L) )],$ such that, $\sigma (t_{L, R } ) = \sigma _{av} \pm \d \sigma
, \ \bar \sigma (t_{ R, L }) =\bar \sigma _{av} \pm \d \bar \sigma ,$
one can define the following two CP-odd combinations,
\begin{eqnarray}
\cala = {\sigma _{av} -\bar \sigma _{av} \over \sigma _{av} +\bar
\sigma _{av} }= {\sigma _{t\bar c} - \sigma _{\bar t c} \over \sigma
_{t\bar c} + \sigma _{\bar t c} } , \quad \cala ^{pol} & =& { \d
\sigma -\d \bar \sigma \over \sigma _{av} + \bar \sigma _{av} },
\label{equ55}
\end{eqnarray}
which will be designated as unpolarized and polarized integrated rate
asymmetries.  The above definition for the unpolarized asymmetry, $
\cala $, is identical to the one studied in our previous
work. \cite{h:chemtob} The asymmetries depend on the RPV coupling
constants through the ratio of loop to tree amplitudes as, $ Im ( { \l
^{'\star } _{iJk } \l '_{iJ'k} \over \l ^{'\star } _{1Jk'} \l
'_{1J'k'} } ) \propto \sin \psi $, where the dependence on the CP
violation angle parameter, $\psi $, reflects the particular
prescription adopted in this study to include the CP-odd phase.  The
index $ k ' $ refers to the d-squark generation in the tree amplitude
and the indices $ i, k $ to the fermion-sfermion generations for the
internal fermion-sfermion pairs, $ {d_k \choose \tilde e^\star _{iL}}
, \ {e^c_i \choose \tilde d_{kR}} $, in the loop amplitude.

  It is important not to confuse the above analysis with that of the
top-antitop pair production, $ l^- l^+ \to t\bar t$, where a CP-odd
asymmetry observable for a single final state may be defined in terms
of the difference of helicity configurations, $ \s (t_L \bar t_L ) -
\s (t_R \bar t_R ) $.  A non vanishing value for the corresponding
difference of polarized rates can only arise via tree-loop
interference terms involving the absorptive part of the top quark
electric dipole moment, $ Im (d^{JJ}) $. \cite{h:schmidt,h:chang1} One
should note here that the one-loop contribution of the RPV $\l '$
interactions to $ Im (d_t ^{JJ}) $ vanishes.  Two closely related
processes, which are amenable to an analogous treatment, are the
$b\bar b$ quark pair \cite{h:valencia} and $\tau ^+ \ \tau ^-$ lepton
pair production.  Double spin correlation observables for the latter
reaction, $ l^- l^+ \to \tau ^- \tau ^+ $, have been examined in a
recent work.  \cite{h:shalom} We note that the RPV $\l $ interactions
can give a non vanishing contribution to $ Im (d_\tau ^{JJ}) $.

\begin{figure} [t]
\centerline{ \psfig{figure=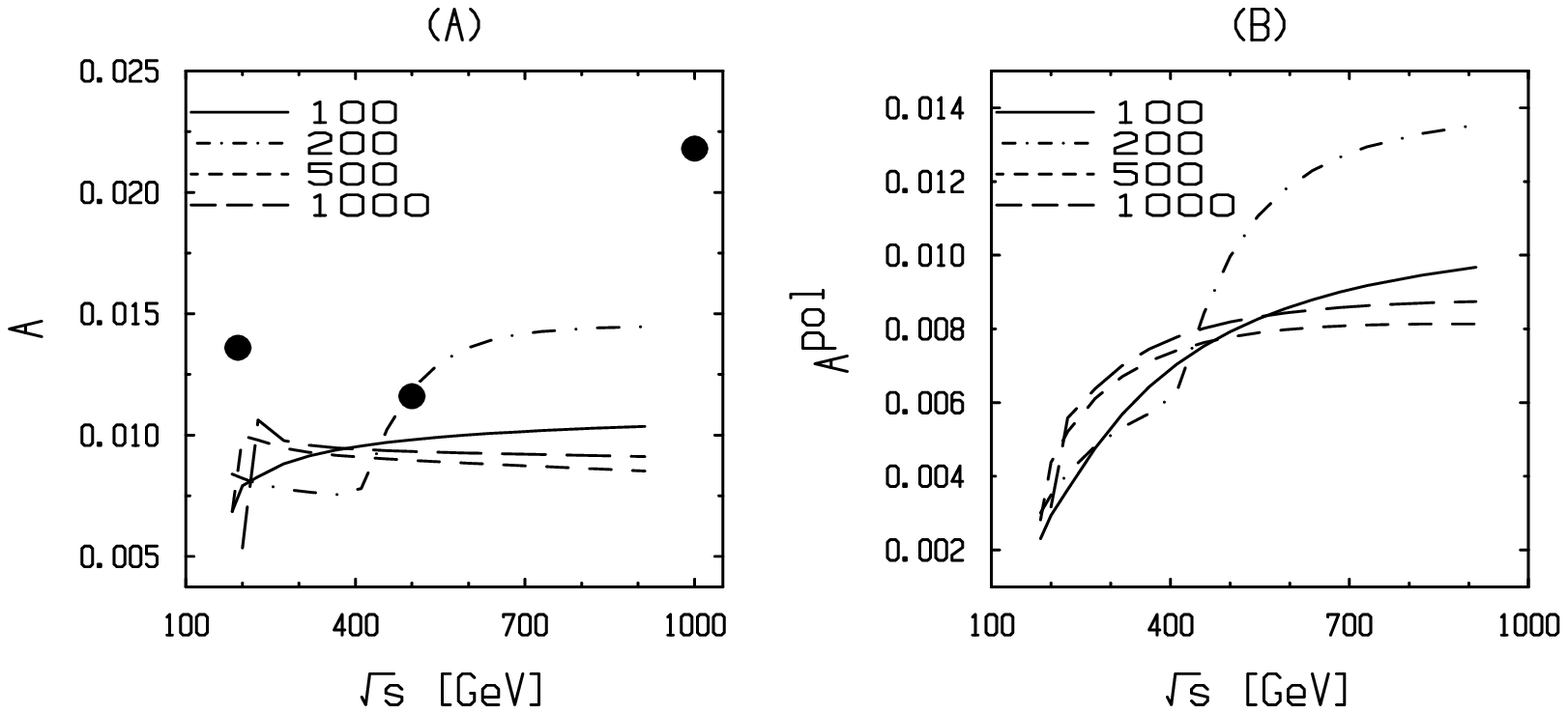,width=17cm} }
\vskip 1 cm
\caption{ The CP-odd production rate asymmetries as a function of the
center of mass energy, $ s^\ud $, for fixed values of the down squark
mass, $\tilde m = [100, \ 200, \ 500 , \ 1000 ] \ GeV$.  The left hand
plot $ (A)$ gives the unpolarized asymmetry, $ \cala = ( \sigma _{av}
- \bar \sigma _{av} ) /(\sigma _{av} + \bar \sigma _{av} )$.  The
upper bounds for the absolute values of the statistical errors on the
asymmetries, as evaluated with $\l '_{12k}\l '_{13k} =0.1, \ \tilde m
= 100 \ GeV$ and integrated luminosities $\call =100.  \ fb^{-1}$ are
shown as full circles.  The right hand plot $ (B) $ gives the spin
polarization dependent asymmetry, $\cala ^{pol} = (\d \sigma -\d \bar
\sigma ) /(\sigma _{av} + \bar \sigma _{av} ) $. }
\label{figbspol}
\end{figure}

The results for the rate asymmetries are displayed in
Fig.\ref{figbspol}.  The numerical results for the unpolarized case
(window $(A)$ in Fig.\ref{figbspol}) update the results presented in
ref. \cite{h:chemtob} since the present calculation includes the
contributions from the Lorentz covariant tensorial ($\s _{\mu \nu }$)
coupling which were ignored in our previous work.  \cite{h:chemtob} The
asymmetry for the polarized case (window $(B)$ in Fig.\ref{figbspol})
involves the difference of the spin helicity asymmetry in the total
production cross sections for the CP mirror conjugate top and antitop
mirror reactions. While this CP-odd polarized asymmetry is not
directly observable, it enters as an important intermediate quantity
in evaluating the measurable kinematic distributions of the top decay
products dependent on the top spin.  We have assumed all the relevant
RPV coupling constants to be equal and set the CP-odd phase at $\sin
\psi =1$. The rapid change in slope for the $\tilde m =200\ GeV$ case
are due to the threshold effect from the imaginary part in the
superpartner one-loop contributions, which set at $ \sqrt s = 400 \
GeV$.  Aside from this large discontinuous contribution, one sees that
both asymmetries comprise another contribution which is nearly
independent of $\tilde m$ and increases smoothly with the initial
energy.  Both asymmetries, $ \cala $ and $ \cala^{pol} $, take values
of order a few $ 10^{-3} $, reaching $ O(10^{-2}) $ at the highest
incident energies.

The statistical uncertainties on the asymmetry may be evaluated in
terms of the signal cross sections and the integrated luminosity by
considering the approximate definition, $ \d \cala = 1/ [\call (\s
_{t\bar c} + \s _{\bar t c})] ^\ud $.  Using the same input value for
the luminosity $ \call = 100.  \ fb ^{-1} $ at the three cm energies,
$\sqrt s =[0.192, \ 0.5 , \ 1. ] \ TeV$, along with the cut signal
rates in Table \ref{h:table1}, we obtain statistical errors on the
asymmetries of order $ O(10^{-1})$. These values lie nearly two order
of magnitudes above the value obtained for the signal. At this point,
it is important to observe that in getting the above estimates for the
rates we have been using somewhat conservative assignments for the RPV
coupling constants.  As already noted, the single top production cross
sections could possibly be two order of magnitudes larger if we were
to use coupling constants values of order, $\l '_{12k} \l '_{13k}
\simeq 10^{-1}$.  Such values are compatible with the indirect bounds
only for the extreme down squark mass $ \tilde m = O(1\ TeV)$ range.
In the hypothetical case where the production rates would be enhanced
by two order of magnitudes, the statistical errors on the asymmetries
would correspondingly get reduced by a factor $ O(10^{-1})$, thereby
reaching the same order of magnitude as the signal asymmetries.
Nevertheless, as plotted in window $(A)$ of Fig.\ref{figbspol}, the
corresponding errors would still be somewhat larger than the signals.
We should note here that the contribution to the one-loop amplitude
from internal sfermion and fermion lines belonging to the third
generation is controlled by the coupling constants quadratic product,
$ \l '_{323} \l '_{333} $, which is subject to weak constraints.
Should the RPV coupling constants exhibit a hierarchical structure
with respect to the quarks and leptons generations, one cannot exclude
the possibility of a factor $10$ enhancement from the ratio, $ Im ( \l
^{'\star }_{323} \l '_{333} / \l ^{'\star } _{123} \l '_{133} )$.
Such an order of magnitude gain on this ratio would raise the
asymmetries up to $ O(10^{-1})$ bringing them well above the
experimental uncertainties.  Lastly, we observe that a more complete
formula for the uncertainties on the asymmetries reads, $ (\d \cala
)^2 = 2 (\d \s_{t\bar c})^2 [1- C +(1+C) \cala ^2 ]/ (\s _{t\bar c} +
\s _{\bar t c})^2 $, where we used equal standard deviations for the
CP conjugate reactions rates, $ \d \s_{t\bar c} = \d \s_{\bar t c} $,
and denoted the correlated error on these two rates as, $ C = <\d
\s_{t\bar c} \d \s_{\bar t c}> / \d \s_{t\bar c}^2 $.  Clearly, an
improvement on the statistical treatment of the $ t\bar c +\bar t c$
events sample, allowing for a positive non vanishing value of the
error correlation associated with the identification of isolated
single negatively and positively charged lepton events, should greatly
help in reducing the experimental uncertainties caused by the small
event rates.

\begin{figure} [t]
\centerline{ \psfig{figure=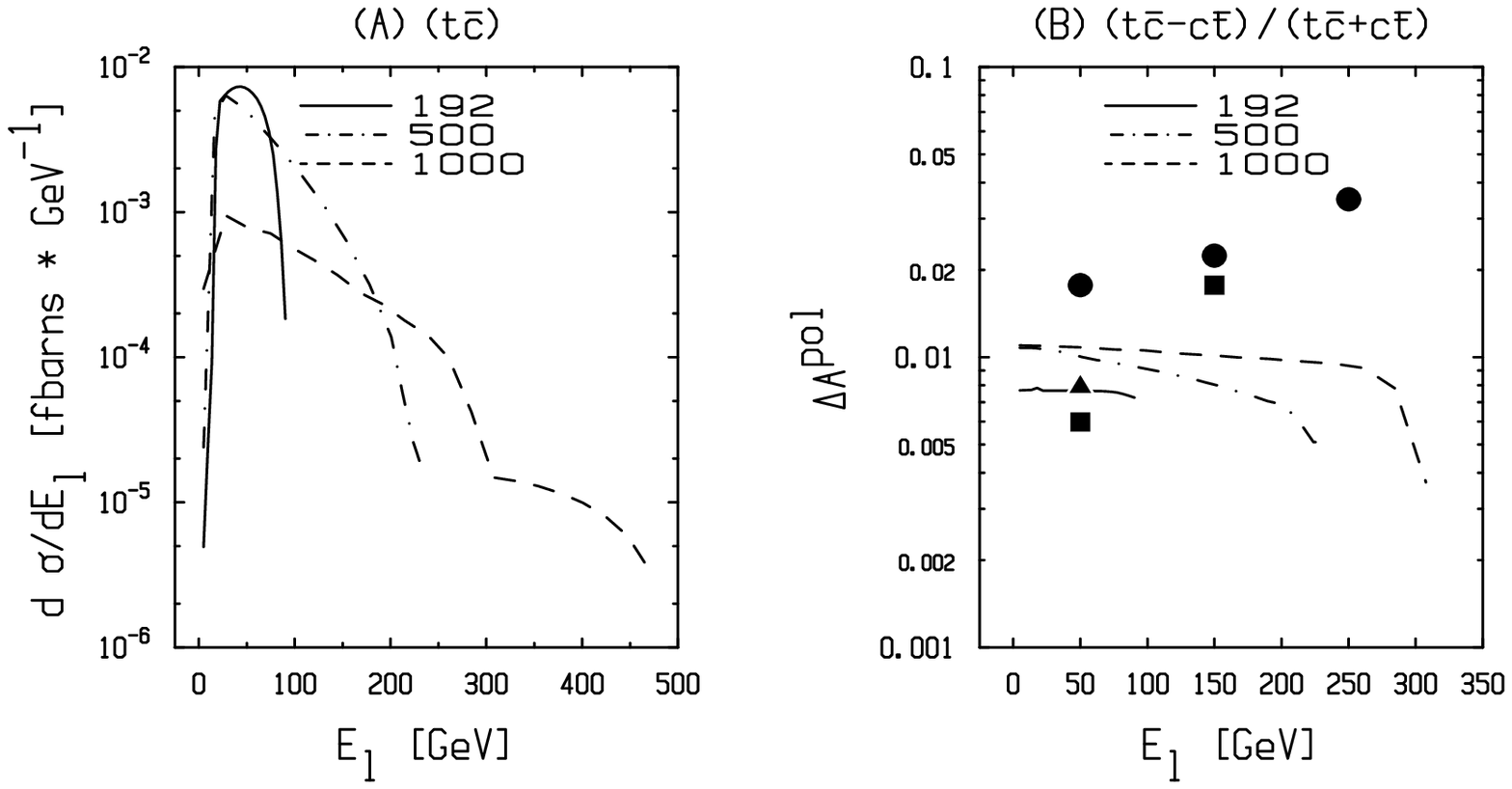,width=16cm} }
\vskip 1 cm
\caption{Energy distribution for the charged lepton as a function of
the laboratory frame lepton energy, for a set of center of mass
energy, $ s^\ud =[ 192, \ 500 , \ 1000] \ GeV$.  The parameters are
set at, $ \l ' =0.1, \ \tilde m =100 \ GeV$.  The left hand plot $ (A)
$ gives the differential lepton energy distribution, $ { d \sigma
\over dE_l }$. The right hand plot $ (B) $ gives the asymmetry in the
energy distribution for leptons of opposite charges in the
CP-conjugate final state channels, $(t\bar c) $ and $ \ (c\bar t)$: $
\D \cala ^{pol}= [ {d \sigma ^+ \over d E_l} - {d \sigma ^- \over d
E_l}]/ [{d \sigma ^+ \over d E_l} + {d \sigma ^- \over d E_l} ].$ The
upper bounds for the absolute values of the statistical errors on the
asymmetries, as evaluated with $\l '_{12k}\l '_{13k} =0.1$ and with
integrated luminosities, $\call =100. \ fb^{-1}$, are shown for three
energy bins of width $ 100 \ GeV$ each, centered at the charged lepton
energies, $ E_l= (50,\ 150,\ 250 ) \ GeV $.  The results for three
values of the center of mass energy, $ s^\ud =[ 192, \ 500 , \ 1000] \
GeV$ are displayed by full triangles, squares and circles. }

\label{figenasy}
\end{figure}

The energy distribution for the negatively and positively charged
leptons in the pair of CP-conjugate reactions may be defined as,
\begin{eqnarray}
<\s ^+> \equiv <{ d \sigma ^+ \over dE_l }> = < \sigma (t_L) f_L +
 \sigma (t_R) f_R > , \cr \quad <\sigma ^- > \equiv < { d \sigma ^- \over
 dE_l }> = < \sigma ( \bar t_R) f_L + \sigma (\bar t_L) f_R >,  
\label{equ56}
\end{eqnarray}
where the correlations between the top spin lepton momentum are
described by the factors, $ f_{L,R}= \ud (1\mp \cos \t^\star _l )$,
and the brackets stand for the angular integration.  The occurrence of
angular correlation factors of opposite signs in the $\bar t$
production case accounts for the kinematical fact that the antitop is
oriented in space with a momentum $ - \vec p$. A CP-odd charge
asymmetry observable with respect to the charged lepton energy
distribution may be defined by considering the following normalized
difference of distributions,
\begin{eqnarray}
\D \cala ^{pol} & =& { <\s ^+ > - <\s ^- > \over <\s ^+ > + <\s ^- > }
={ (\sigma _{av} -\bar \sigma _{av}) +< (\d \sigma - \d \bar \sigma )
(f_L -f_R) > \over (\sigma _{av} +\bar \sigma _{av}) + < (\d \sigma +
\d \bar \sigma ) (f_L -f_R) > }.
\label{equ57}
\end{eqnarray}
The numerical results for the charged lepton energy distributions and
for the above defined charge asymmetry in the lepton energy
distributions are displayed in Fig.\ref{figenasy}.  (Note that the
transverse energy distribution, in the plane orthogonal with respect
to the top momentum, may be simply obtained as, ${ d\G \over dE_{lT}}
= { d\G \over dE_{l}} {1\over \sin \t_{lt} ^\star }$.  The
distribution in the plane orthogonal to the collision axis is less
trivial to evaluate since this requires an additional integration over
the lepton azimuthal angle.)  The energy distributions for the
unpolarized cross section essentially reproduce the results found in
our above quoted event generator predictions, Fig.\ref{figtopp}.  The
energy distributions for the polarized asymmetry lie at values of
order of magnitude, $O( 10^{-3}) $, always retaining the same positive
sign as the lepton energy varies.  For a fixed energy of the emitted
lepton, the asymmetry increases with the initial energy, reaching
values of order $O( 10^{-2}) $.  In window $(B)$ of Fig.\ref{figenasy}
we have plotted the experimental uncertainties using the same inputs
for the luminosities and the rates as in the discussion of the
unpolarized asymmetries given above.  To ease the comparison with
experiment, we divide the charged leptons energy interval into three
bins of width $ 100 GeV$ each, centered at the three lepton energies,
$ E_l= (50,\ 150,\ 250 ) \ GeV $.  The statistical errors on the
asymmetries in the energy distributions lie at the same level as those
associated to the total asymmetries, so that similar conclusions
should apply.  Setting ourselves within the same optimistic scenario
by using $\l '_{12k} \l '_{13k} =10^{-1}$ and $\call= 100 fb^{-1}$, we
obtain expected errors of order $ O(10^{-2})$. These values are
insufficient for a comfortable identification of a signal asymmetry.
However, we reiterate, as in the above discussion, that an enhancement
of the signal asymmetries to an observable level of $ O(10^{-1})$, due
to a hierarchical structure in the generation dependence of the $\l
'_{ijk}$, is a real possibility.

\section{Conclusions}
\label{sec4}

\setcounter{equation}{0}

We have demonstrated that single top production through the RPV
interactions could be observed at the future linear colliders or else
be used to set bounds on the RPV coupling constants, $ \l ' _{12k} \l
'_{13k} < O(10^{-2})$, over a wide interval for the down squark mass,
$ m _{\tilde d_{kR} } < 1. \ TeV$.  The $ b $ quark tagging would help
greatly to overcome the background. Even with an imperfect $ b $ quark
tagging, it is still possible to drastically reduce the background,
from $ WW $ and $ b\bar b$, without much harming the signal. The
analysis of top polarization observables via the semileptonic decay
channel of the top allows to test for the presence of a CP violating
complex phase, embedded in quadratic products of the RPV coupling
constants. We have focused on the asymmetry in the energy
distributions of the charged leptons in the CP-conjugate pair of final
states, $b l^+ \nu \bar c$ and $\bar b l^- \bar \nu c $, obtaining
asymmetries of order $ 10^{-3} \ - \ 10^{-2} $ for the incident
energies expected at the future leptonic colliders.  These values lie
somewhat below the anticipated limits of observability.  However, it
may be possible to obtain enhanced values of order $ 10^{-1}$, should
the RPV coupling constants $\l'_{ijk}$ exhibit large hierarchies with
respect to the quarks or leptons generations.  Future promising
extensions might include analogous reactions accessible with
lepton-photon or photon-photon colliding beams, $ l \g \to t\bar c , \
\g \g \to t\bar c $, where the expected production rates are
substantially larger than those for the $ l^-l^+ $ colliders.

\clearpage

\appendix

\setcounter{subsection}{0} 
\setcounter{equation}{0}

\section{Helicity amplitudes and one-loop \rpv vector boson vertex functions}
\label{appexa}

{\bf Helicity amplitudes.}

 The helicity spin basis Dirac spinors for a fermion or an
antifermion, of mass $ m$ and four momentum, $ k_\mu = (E_k = (k^2
+m^2) ^\ud, \vec k)$, and polar coordinates, $ \vec k = (\t , \phi )$,
can be written in the form of direct products of the Dirac spinor
two-component space with the the two-component space of Pauli helicity
basis spinors, $ \phi _\l (\vec k) , $ satisfying, $ \vec \s \cdot
\hat k \phi _\l (\vec k) = \l \phi _\l (\vec k). $ In the Dirac
representation for the Dirac matrices, $\g_0 = \b ,\ \vec \g = \b \vec
\a ,\ \g_5 = \pmatrix{0 & 1 \cr -1 & 0}$, the spinors read,
\begin{eqnarray}
u(\vec k, \l )&=& \sqrt {\e_k }\pmatrix {1\cr \tilde k \l } \times
\phi _\l (\vec k ), \ v(\vec k, \l )=\sqrt { \e_k } \pmatrix {- \tilde
k \l \cr 1 } \times \phi_{-\l } (\vec k) ,\ \cr \phi _{-1}(\vec k) &
=& \pmatrix{-\sin (\t /2) e^{-i\phi } \cr \cos (\t /2) } , \phi
_{+1}(\vec k) = \pmatrix { \cos (\t /2) \cr \sin (\t /2) e^{+i\phi }
},
\label{equ5}
\end{eqnarray}
where, $\e_k= E_k+m, \ \tilde k= \vert \vec k \vert /(E_k+m), $ and $
\chi_{\l }, \ [\l =\pm 1] $, are the Pauli spinors in the basis with a
fixed quantization axis identified with the spatial three-axis, $Oz$.
The helicity basis spin eigenstates with a space parity reversed three
momentum are defined as, $\phi_{\l } (-\vec k) = e^{ -i (\phi + \pi )
(\l ' -\l )/2 } (e^{-i{(\pi -\t ) \over 2} \s_y })_{\l '\l } \chi_{\l
' } = \phi_{\l } (\vec k)\vert _{[\t \to \pi -\t , \ \phi \to \phi
+\pi ]} $.

The $8$ non vanishing helicity amplitudes for the process, $ l^+(k' ,
\mu ') + l^- (k, \mu) \to u_J (p, \l ) +\bar u_{J'} (p', \l ') $, are
listed in the formulas below:
\begin{eqnarray} M_1= M(+-+-) & = & 4 \calf \, [ - ( ( 1 + {\tilde
p}\, {\tilde p'} ) \, ( {\ya} + {\yb} ) \ + ( {\tilde p} + {\tilde p'}
) \, ( {\yc} + {\yd} ) ] \, {{\sin ^2({{\t } / {2}})}}, \cr M_2= M
(+-++) & = & 2 \calf \, [ ( -1 + {\tilde p}\, {\tilde p'} ) \, (
{\ya}+ {\yb}) + ( {\tilde p} - {\tilde p'} \ ) \, {\yc} + ( {\tilde p}
- {\tilde p'} \ ) \, {\yd} \cr & + & 2\,p\, ( {\tilde p} + {\tilde p'}
) \, ( {\ye} + {\yf}) + 2\,p\, ( 1 + {\tilde p}\, {\tilde p'} ) \,(
{\yg} + {\yh}) ] \,\sin (\t ) ,\cr M_3= M (-++-) & = & -4 \calf \, [ -
( 1 + {\tilde p}\, {\tilde p'} \ ) \, ( -\ya + {\yb} ) - ( {\tilde p}
+ {\tilde p'} \ ) \, ( {\yc} - {\yd} ) ] \, {{\cos^2 ({ {\t } / {2}})}
} ,\cr M_4 = M (-+++) & = & 2 \calf \, [ ( -1 + {\tilde p}\, {\tilde
p'} \ ) \,(-\ya + {\yb} ) + ( - {\tilde p} + {\tilde p'} \ ) \, (
{\yc} - {\yd}) \cr & -& 2\,p\, ( {\tilde p} + {\tilde p'} ) \, ( {\ye}
- {\yf}) - 2\,p\, ( 1 + {\tilde p}\, {\tilde p'} ) \, ( {\yg} - {\yh}
] \,\sin (\t ) ,\cr M_5 = M (+---)& = & 2 \calf \, [ ( -1 + {\tilde
p}\, {\tilde p'} ) \, ( {\ya} + {\yb}) + ( - {\tilde p} + {\tilde p'}
\ ) \, {\yc} + ( - {\tilde p} + {\tilde p'} \ ) \, {\yd} \cr & + &
2\,p\, ( {\tilde p} + {\tilde p'} ) \, ( {\ye} + {\yf}) - 2\,p\, ( 1 +
{\tilde p}\, {\tilde p'} ) \, ( {\yg} - {\yh}) ] \,\sin (\t ) ,\cr
M_6= M (+--+) & = & -4 \calf \, [ (1+ {\tilde p}\, {\tilde p'} ) \, (
{\ya} + {\yb}) + ( {\tilde p} + {\tilde p'} ) \, ( {\yc} + {\yd} ) ]
\, {{\cos ^2({{\t } / {2}})} } ,\cr M_7= M (-+--) & = & 2 \calf \, [ (
-1 + {\tilde p}\, {\tilde p'} \ ) \,( - {\ya} + {\yb} ) + ( {\tilde p}
- {\tilde p'} \ ) \, ( {\yc} - {\yd}) \cr & - & 2\,p\, ( {\tilde p} +
{\tilde p'} ) \,( {\ye} - {\yf}) + 2\,p\, ( 1 + {\tilde p}\, {\tilde
p'} ) \, ( {\yg} - {\yh} ) ] \,\sin (\t ) ,\cr M_8= M ( -+-+) & = & 4
\calf \, [ - ( 1 + {\tilde p}\, {\tilde p'} ) \, ( {\ya} - {\yb} ) - (
{\tilde p} + {\tilde p'} ) \, ( {\yc} - {\yd} ) ] \, {{\sin ^2({{\t }
/ {2}})}}.
\label{equ7}
\end{eqnarray}
The arguments refer to the fermions helicity in the following order, $
M_i((h_{e^+},h_{e^-} ,h_f ,h_{\bar f} ) $.  The remaining helicity
amplitudes, omitted from the above list, are understood to vanish
identically. We denote by $\t $ the top scattering angle, $\cos \t =
\vec k \cdot \vec p $, by $ [E_p, E '_{p }] =(s\pm m_J^2 \mp m_{J'}^2
)/2\sqrt s $, the top and charm quarks energies, and use the following
abbreviated notations, $ \tilde p = {p\over E_p +m_J }, \ \tilde p ' =
{p\over E '_{p } +m_{J'} }, \ \calf = {1\over 2} [s (E_p +m_J) (E
'_{p} + m_{J'} )] ^\ud , $ along with the useful compact notations,
\begin{eqnarray}
X_1& = & G a^ + \capa +\calr, \ X_2= G a^- \capa +\calr, \ X_3 = G a^+
\capb +\calr ,\ X_4 = G a^- \capb + \calr , \ \cr X_5& =& \ud G a^+
\capc , \ X_6= \ud G a^- \capc , \ X_7 = \ud G a^+ \capd , \ X_8=\ud G
a^- \capd ,
\label{equ7p}
\end{eqnarray}
where $ G a^\pm \capa , \cdots $ are defined in eq. (\ref{equ53}),
$\calr $ in eq. (\ref{equ323}), and $\capa , \cdots , \capd $, in
eq. (\ref{equ51}).

\vspace{2.cm}

{\bf One-loop RPV vector boson vertex functions.}

 The one-loop vertex functions, as derived in \cite{h:chemtob}, are
given by the formulas,
\begin{eqnarray}
&& A_L^{JJ' } = \FLU [ {a_L(u) }\, {B_1^{(2)}} + {a(f_L)}\, {{
{m_f}}^2} C_0 + {a(\tilde f')} ( 2 {\tilde C_{24}} + 2 \ m_J^2\, (
{\tilde C_{12}}\, - {\tilde C_{21}}\, + {\tilde C_{23}}\, - {\tilde
C_{11}}\, ) ) \cr &+& {a(f_R)}\, ( {B_0^{(1)}} - 2\, {C_{24}} - {{
{m_{\tilde f'}}}^2} {C_0} + {{ {m_{J}}}^2} ( C_0 +3 C_{11} -2C_{12}
+2C_{21} -2C_{23} ) -m_{J'}^2 C_{12} ) ], \cr && A_R^{JJ'} = \FLU m_J
m_{J'} [ 2 {a(\tilde f')}\, ( - {\tilde C_{23}} + {\tilde C_{22}} ) +
{a(f_R)}\, ( -C_{11} +C_{12} -2C_{23} +2 {C_{22}} ) \, ] \cr && {
a^{JJ'}\choose - id^{JJ'} } = \FLU {m_J +m_{J'} \over 4}\bigg [ \pm
m_J [ a(f_R) (C_{11}-C_{12}+C_{21}-C_{23}) \cr & -& a(\tilde f ')
(\tilde C_{11}+\tilde C_{21}-\tilde C_{12}-\tilde C_{23}) ] + m_{J'}
[a(f_R) (C_{22}-C_{23}) +a(\tilde f ') (\tilde C_{23}-\tilde C_{22}) ]
\bigg ].
\label{h:eqm11}
\end{eqnarray} 
The relevant configurations for the internal fermion and sfermion
propagating in the loop are: $ {f \choose \tilde f'} = {d_k \choose
\tilde e^\star _{iL}} , \ {e^c_i \choose \tilde d_{kR}} $.  The
notations for the Passarino-Veltman two-point and three-point
integrals, as specified in our work, \cite{h:chemtob} are defined
according to the following conventions, $B_A^{(1)}= B_A(-p-p', m_f,
m_f), \ B_A^{(2)}= B_A(-p , m_f, m_{\tilde f'}), \ [A =0,1] $ and $C_A
= C_A (-p,-p',m_f,m_{\tilde f'} , m_f) , \tilde C_A = C_A
(-p,-p',m_{\tilde f'}, m_f,m_{\tilde f'}) .  \ [ A= 0, 11,
12,21,22,23]$ The integral functions with a tilde are associated with
the one-loop diagram for the sfermion current.

\clearpage

\vspace{10 mm}
\begin{center}
{  }
\end{center}
\vspace{10 mm}

\clearpage

\clearpage

\vspace{10 mm}
\begin{center}
{  }
\end{center}
\vspace{10 mm}

\clearpage

\setcounter{chapter}{0}
\setcounter{section}{0}
\setcounter{subsection}{0}
\setcounter{figure}{0}

\chapter*{Publication IX}
\addcontentsline{toc}{chapter}{CP violation 
flavor asymmetries in slepton pair production at leptonic 
colliders from broken R-parity}

\newpage

\vspace{10 mm}
\begin{center}
{  }
\end{center}
\vspace{10 mm}

\clearpage

\begin{center}
{\bf \huge CP violation 
flavor asymmetries in slepton pair production at leptonic 
colliders from broken R-parity}
\end{center} 
\vspace{2cm}
\begin{center}
M. Chemtob, G. Moreau
\end{center} 
\begin{center}
{\em  Service de Physique Th\'eorique \\} 
{ \em  CE-Saclay F-91191 Gif-sur-Yvette, Cedex France \\}
\end{center} 
\vspace{1cm}
\begin{center}
{Phys. Lett. {\bf B448} (1999) 57, hep-ph/9808428}
\end{center}
\vspace{2cm}
\begin{center}
Abstract  
\end{center}
\vspace{1cm}
{\it
We examine the effect of the R parity odd, lepton number violating,
renormalizable interactions on flavor non-diagonal rates  and CP asymmetries  
in the production of slepton pairs,  
$e^-+e^+\to \tilde e_{HJ} +\tilde  e^\star _{H'J'}, \ [J\ne J'] ,\ [H,H'=(L,R)] $
at leptonic colliders. The R parity odd coupling constants are
assumed to incorporate CP odd complex phases. 
The flavor changing rates are controlled  by tree level amplitudes 
and quadratic products of different R parity violating coupling constants 
and  the  CP violating asymmetries by interference terms between tree and loop level  amplitudes and quartic products. The consideration  
of  loop amplitudes  is restricted  to the 
photon and  Z-boson vertex corrections.
We present numerical  results  using a
family and (quarks and leptons)  species
independent mass parameter, $\tilde m$, 
for all the scalar superpartners and making simple assumptions for
the family dependence of the R parity odd coupling constants.
The flavor non-diagonal rates, $\s_{JJ'}$, vary in the range, $({\l \over 0.1})^4 \  2\ - \ 20 $ fbarns, for sleptons masses $\tilde m < 400 $ GeV, 
as one spans the interval of center of mass  energies from 
the Z-boson  pole up to $ 1000$ GeV. For sleptons masses,  
$ \tilde m> 150 $ GeV, these observables  could be of use at NLC energies 
to set useful bounds on the R parity odd coupling constants. 
The  predicted asymmetries are in order of magnitude,
${\cal A}_{JJ'}= { \s_{JJ'} -\s_{J'J}
\over \s_{JJ'} +\s_{J'J}  } \simeq 10 ^{-2}  \ - 10^{-3}$.
}

\newpage

\section{Introduction}
\label{secs1}

\setcounter{equation}{0}

On side of the familiar low  energy tests of CP symmetry non-conservation,
a large number of tests have been developed  over the years 
for high energy colliders \cite{i:cpcoll1,i:cpcoll2,i:cpcoll3}. 
The existing proposals have  dealt with  different types of CP odd
observables (quark and leptons flavor aymmetries \cite{i:cpobs0},
spin polarization asymmetries \cite{i:cpobs1,i:cpobs2,i:cpobs3}, 
heavy quarks or leptons electric dipole moments \cite{i:wermes}, ...)
and covered a wide variety of physical processes, ranging from  decay reactions 
($Z, \  W^\pm $ gauge 
bosons \cite{i:cpobs0,i:bern1}, Higgs bosons  \cite{i:chang1,i:grzad1} 
or  top-quarks \cite{i:grzad2})
to production reactions (leptons-antileptons  and light 
quarks-antiquarks  pairs \cite{i:cpobs0}, single top-quarks \cite{i:atwood1}, top-antitop-quark pairs \cite{i:peskin1,i:bartl,i:baek}, or superpartners pairs, 
$\tilde \chi^+ \tilde \chi^-, $ \cite{i:kizu} $ \tilde q \tilde {\bar q} ,$
\cite{i:pilaf}  and  
$\tilde l^+ \tilde  l^ -$ \cite{i:arkani2,i:bowser}).
For lack of space, we have  referred to those works 
from which one could hopefully trace the extensive  published literature.

One of the primary motivations for these high energy tests 
 is the search for physics beyond the standard model. The supersymmetry 
option is especially attractive in this respect since any slight generalization 
of the minimal model, allowing, say, for  some generational 
non universality  in the soft supersymmetry breaking parameters
or for  an approximate R parity symmetry, would introduce 
several new parameters, with a non trivial structure on quarks and leptons
flavors which could accommodate   extra CP violating phases.
As is known, high energy supercolliders  are expected to provide for
precision determinations  of these supersymmetry parameters.
Regarding the  much studied sleptons pair production reaction 
\cite{i:fujii,i:peskin2,i:thomas}, one can define
a simple spin-independent  CP asymmetry observable
in terms of the difference of  integrated rates, $(\s_{JJ'}-\s_{J'J})$,
with  $\s_{JJ'} = \s (e^-+e^+\to \tilde e^-_J +\tilde e^+_{J'}) $, 
for the  case of sleptons pairs  of different flavors, $J\ne J'$.
Recent works,  based on  the mechanism of sleptons flavor oscillations, have
examined  for correlated slepton pairs production,
the flavor non-diagonal rates \cite{i:krasn,i:arkani1}
and the CP-odd flavor  asymmetries,  defined as, ${\cal A}_{JJ'}=
{\s_{JJ'} -\s_{J'J} \over \s_{JJ'} +\s_{J'J}  } $ \cite{i:arkani2,i:bowser}.  
Encouraging values of  order,   ${\cal A}_{JJ'} \approx 10^{-3} $  were predicted  
at the  next linear colliders (NLC) energies \cite{i:arkani2,i:bowser}.
While the rates, $\s _{JJ'}$, depend on pairwise non-degeneracies  in the
sleptons mass spectra, the asymmetries, ${\cal A}_{JJ'}$, entail the 
much stricter conditions  that both non-degeneracies and mixing angles 
between all slepton flavors, as well as the CP odd phase, must not vanish.
 
Our main  observation in this work  is that the
R parity odd interactions  could  provide an  alternative mechanism
for explaining  flavor non-diagonal CP asymmetries through possible
complex CP odd phases incorporated in 
the   relevant dimensionless coupling constants.
  While  these  interactions can contribute 
to flavor changing changing processes already  at tree  level, 
their  contributions to CP asymmetries  involve
interference  terms between tree and loop amplitudes.
Two important questions then are, first, whether  the contributions from the 
RPV  (R parity violating) interactions, given the known bounds on the 
R parity odd coupling constants,
 could  lead to observable production rates; second,   whether the CP  
 asymmetries could reach observable levels. 
We shall present in this work  a study of 
the contributions to the CP asymmetries, in
the reactions, $e^-+e^+\to \tilde  e_{HJ} +\tilde e^{'\star }_{HJ'} ,
 \ [H=L, R, \ J\ne J']$,
at the high energy  leptonic colliders,  for center of mass energies
from the Z-pole up to $1000 $ GeV.   
The RPV  lepton number violating interactions are defined by the familiar superpotential, 
$W_{R-odd}=\sum_{ijk}[ \ud \l _{ijk} L_iL_j E^c_k+
\l ' _{ijk} Q_i L_j D^c_k ]. $
A comparison with the  oscillations mechanism should enhance 
the impact of  future experimental 
measurements of these  observables at  the future high energy colliders. 

The contents are organized into 3 sections. In Section \ref{secs3}, 
we develop the basic formalism for describing 
the scattering  amplitudes 
at tree and one-loop levels for the production of 
slepton pairs, $\tilde e^-_{L} \tilde e^+_L $ and  
$\tilde e^-_R \tilde e^+_R $.
In Section  \ref{secs4}, we present and discuss  our numerical results
for the integrated cross sections and  the CP asymmetries.
\section{Production of charged sleptons  pairs}
\label{secs3}

\setcounter{equation}{0}

\subsection{General formalism}
\label{subsecs31}
\begin{figure} [t]
\begin{center}
\leavevmode
\psfig{figure=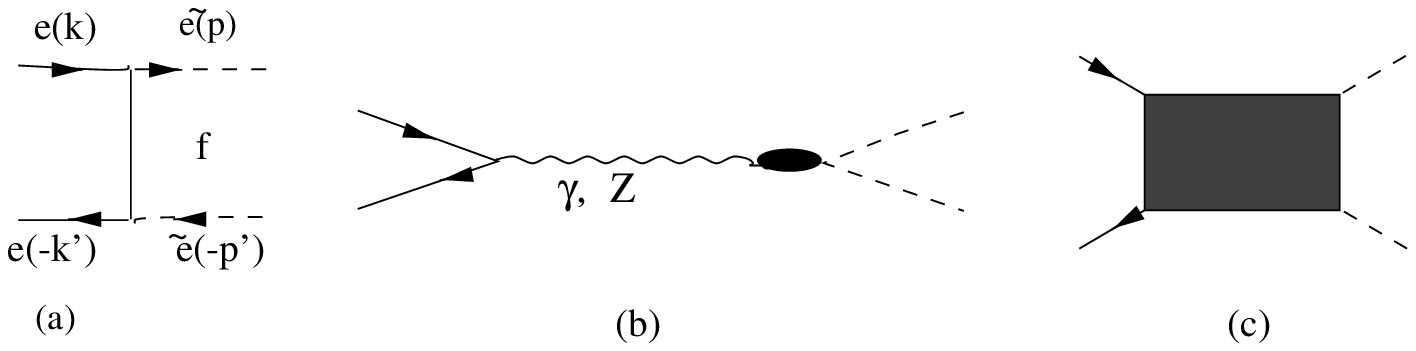}
\end{center}
\caption{Flavor non-diagonal process of $e^-e^+$ production of 
a sfermion-antisfermion pairs, $e^-(k)+e^+(k')\to \tilde e^-_J (p)
+\tilde  e^+_{J'} (p') $. The tree level diagram in $(a)$ represents
a neutrino, $f=\nu , \   t$-channel exchange amplitude. 
The  loop   level diagram in $(b)$ represents 
$\g -$  and $ Z- $ boson exchange amplitudes  with dressed vertices
 and that  in $(c)$ box amplitudes.  }
\label{figs1}
\end{figure}

The evaluation of spin-independent CP asymmetries  in  the production of a pair of sleptons, 
 $e^-(k)+e^+(k')\to \tilde e^-_{HJ}(p)+\tilde e^+_{H' J'} (p')$,  of different 
flavors, $J\ne J'$, with chiralities, $H=(L,R),\  H'=(L,R)   $, 
involves both tree and loop amplitudes.
Let us start  with the  case of  two left-chirality  sleptons, $H=H'=L$. 
At tree level, the R parity odd couplings, $\l _{ijk}$,   give
a non-vanishing contribution which is  
described by a  neutrino,  $\nu_i , \  t $-channel exchange Feynman 
diagram, as displayed  in  $(a)$ of Fig. \ref{figs1}. The associated flavor non-diagonal 
amplitude reads:
\begin{eqnarray}
M^{JJ'}_{tree} (\tilde e_L)= -{\l ^\star _{iJ1} \l _{iJ'1}\over t-m^2_{\nu_{i}}  }
\bar v(k') P_L(\kslash -\pslash )P_R u(k).
\label{eqs1}
\end{eqnarray}
Under our working assumption  that flavor changing effects are absent from  the 
supersymmetry breaking interactions, no other tree level  contributions  arise, 
since the gauge interactions can contribute, through the familiar
neutralinos $t$-channel  and gauge bosons $s$-channel exchanges, to flavor diagonal 
amplitudes, $J=J'$, only.

At one-loop level, there  occurs $\g -$ and  $Z-$boson 
exchange amplitudes with dressed $\g \tilde f\tilde {f}'$  and 
$Z \tilde f\tilde {f}'$  vertices involving three-point  vertex correction 
loop diagrams, as well as  box diagrams, of the type depicted
schematically in  $(b)$ and $(c)$  of  Fig.\ref{figs1}. 
We shall restrict consideration to the one-loop triangle diagrams 
contributions in the gauge bosons exchange  amplitude only.
Defining the dressed vertex functions for the  Z-boson
coupling to sleptons of chirality  $Z_\mu (P)\to 
\tilde f^J_H(p) +\tilde 
f^{J'\star }_H (p'), \ [H=L,R]$,  by the  effective Lagrangian,
\begin{eqnarray}
 L=-{g\over 2\cos \t_W} Z^\mu \G_\mu ^Z (p,p'), \quad 
\G_\mu ^Z(p,p')= (p-p')_\mu    [a_H(\tilde f_H) \d_{JJ'} 
+A^{JJ'}_H (\tilde f, s +i\e ) ] ,
\label{eqs2}
\end{eqnarray}
where, $ a(\tilde f_H)= a(f_H)=a_H(f)= 2T_3^H(f)-2Q(f) x_W, \ [x_W=\sin ^2\t_W]$ 
such that, $a(e_L)= -1+2x_W,\ a(e_R)= 2x_W$, 
we can express the one-loop Z-boson exchange amplitude as:
\begin{eqnarray}
M^{JJ'}_{loop}(\tilde e_H)&=&\bigg ({g\over 2\cos \t_W}\bigg )^2
\bar v(k')\g^\mu \bigg  (a(e_L)P_L+ 
a(e_R)P_R \bigg ) u(k) {1\over s-m_Z^2+im_Z\G_Z } \cr  &\times & (p-p')_\mu 
[a(\tilde e_H)\d_{JJ'}+ A_H^{JJ'} (\tilde e, s+i\e )],
\label{eq14}
\end{eqnarray}
where the shifted complex  argument,  $s+ i\e $, is  incorporated to remind us
 that  the vertex functions 
are complex functions in the complex plane of the 
Z-boson  virtual mass squared,  $s = (k+k')^2=(p+p')^2 $,  to be evaluated at 
the upper lip of the cut along the positive real axis.
In  the dressed  vertex function descibing the coupling,
$Z \tilde f \tilde f^\star $, eq.(\ref{eqs2}), 
we  have omitted the Lorentz covariant proportional to, $P_\mu =(p+p')_\mu 
=(k+k')_\mu $,  since this will give negligibly small lepton mass terms 
upon contraction in the total Z-boson exchange
amplitude, eq.(\ref{eq14}),  with the initial state leptons vertex covariant.
It is most convenient to describe the initial leptons polarizations 
in the helicity eigenvalue basis. 
In the limit of vanishing initial leptons masses, only the two helicity flip
configurations, $e^-_R e^+_L, \ e^-_L e^+_R,$ are non vanishing. While the gauge
bosons $s$-channel  exchange  contributes  to both of these configurations, the
R parity violating neutrino  $s$-channel exchange contributes only to the first. 
The summed tree and loop amplitude, $M^{JJ'}(\tilde e_L) = M_{tree}^{JJ'}(\tilde e_L)
+ M_{loop} ^{JJ'}(\tilde e_L) $,    in the relevant configuration, namely,
$e^-_R +e^+_L= e^-(h=-\ud ) +e^+(\bar h = \ud ) $, reads:
\begin{eqnarray}
M^{JJ'}(\tilde e_L)&=&M(e^-_R+e^+_L\to \tilde e^-_{LJ}+\tilde e^+_{LJ'})= 
-\ud \b s \sin \t \bigg  [
{\l^\star _{iJ1} \l _{iJ'1}\over t-m^2_{\nu_i}} \cr & +&2\bigg (
{g\over 2\cos \t_W }\bigg )^2 {a(e_R) 
A_L^{JJ'} (\tilde e , s+i\e ) \over s-m_Z^2+im_Z\G_Z  } \bigg ].
\label{eq15}
\end{eqnarray}
The Z-boson exchange contribution to the 
other helicity flip configuration, $e^-_L e^+_R$, is simply obtained by the
substitution, $ a(e_R)\to a(e_L)$. 
We also note that the $\g $ exchange contribution
has the same formal structure as that of the Z-boson exchange,
and can be easily  incorporated by adding to the above amplitudes 
the terms obtained by the replacements, 
${g\over 2 \cos \t _W } \to { g \sin \t_W \over 2},
\ a_{L,R}(f)\to 2Q(f), \ (s-m_Z^2  +im_Z \G_Z)^{-1}\to s^{-1},$ along 
with the substitution of Z-boson by photon vertex functions, 
$ A_{L,R}^{JJ'} (\tilde e, s+i\e ) \to A_{L,R}^{\g JJ'} (\tilde e, s+i\e )$.
The kinematical notations here refer to the center of mass system, where $\beta ={p\over k}=
{2p\over \sqrt s}, \ \theta $ is the scattering angle and
the differential cross section for unpolarized initial leptons reads: $d\s /d \cos \t = 
{p\over  128\pi s k } \sum_{pol} \vert M^{JJ'} \vert^2 .$
(For unpolarized beams, one must remove the polarization sums and multiply by a factor
of $4$. Our results agree with those quoted in \cite{i:peskin2}.)
Denoting the amplitude for the charge conjugate process, $e^- +e^+\to 
\tilde e^-_{HJ'} +\tilde e^+_{HJ}, \ [H=L,R]$,  by $\bar M^{JJ'} (\tilde e_H)$ 
and using the simple relationship, $\bar M^{JJ'} (\tilde e_H) 
= M^{J'J} (\tilde e_H) $, one can describe the decomposition into
tree and loop components for the pair of CP conjugate processes as, 
\begin{eqnarray}
M^{JJ'}(\tilde e_H)=a_0^{JJ'}+\sum_\a a_\a ^{JJ'}  F^{JJ'}_\a (s+i\e ), \ \ 
\bar M^{JJ'} (\tilde e_H) = a^{JJ' \star } _0+\sum_\a a^{JJ'\star } _\a  
F^{JJ'} _\a (s+i\e ).
\label{eqstt}
\end{eqnarray}
A spin-independent CP asymmetry 
can be defined in  the familiar way as the normalized difference of rates,
\begin{eqnarray}
{\cal A}_{JJ'}(\tilde e_H)= {\vert M^{JJ'}(\tilde e_H) \vert ^2- \vert \bar  M^{JJ'}
(\tilde e_H) \vert ^2\over 
\vert M^{JJ'}(\tilde e_H) \vert ^2+ \vert \bar  M^{JJ'} (\tilde e_H) \vert ^2 } 
\simeq {2\over \vert a_0\vert ^2}
\sum_\a  Im(a_0a_\a ^\star ) Im (F_\a (s+i\e )),  
\label{eq17}
\end{eqnarray}
where we have assumed  in the second step that the tree level flavor 
non-diagonal amplitude, $a_0$, 
dominates over the loop level amplitude, $a_\a  F_\a $,
and used the index  $\a $ to label the internal states running inside the loop.
\subsection{Loop amplitudes}
\label{subsecs32}
\begin{figure} [t]
\begin{center}
\leavevmode
\psfig{figure=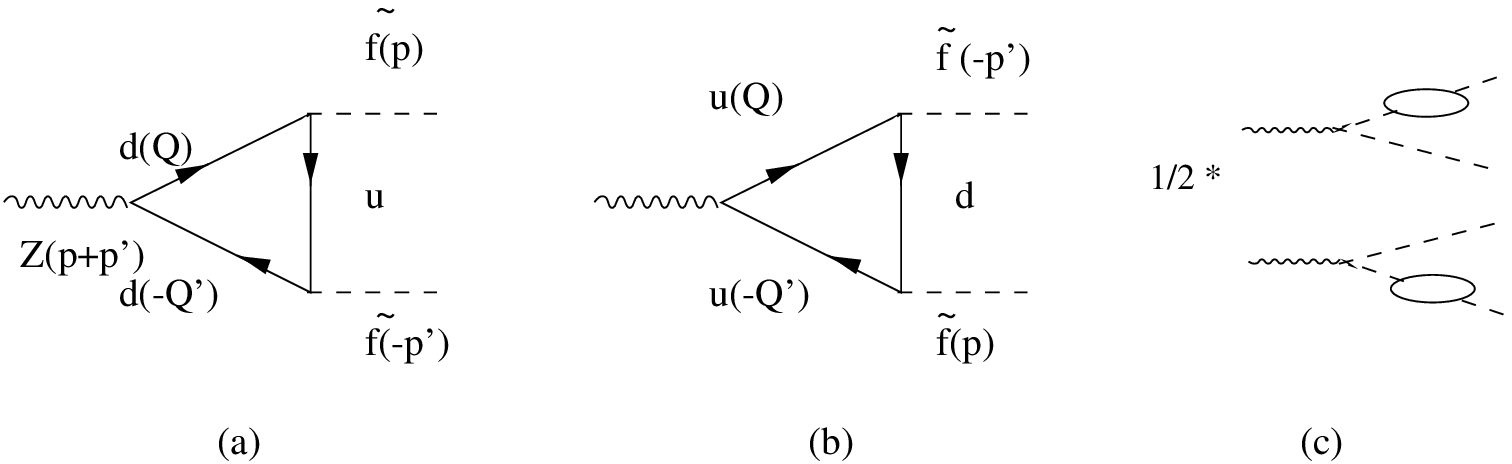}
\end{center}
\vskip 0.5cm
\caption{One-loop diagrams for the dressed $Z\tilde f \tilde f^\star $ vertex.
The flow of four-momenta for the intermediate  fermions  is  denoted  as, 
$ Z (P=k+k') \to f(Q)+\bar f(Q') \to \tilde f_J(p) + \tilde
f^\star _{J'} (p') $.  }
\label{figs2}
\vskip 0.5cm
\end{figure}

The  one-loop  triangle  diagrams,  describing the dressed vertex functions,
$Z\tilde f_L \tilde f_L^{\star } $, arise 
in two distinct charge configurations,   
shown in Fig. \ref{figs2}  by the diagrams $(a) $ and $(b)$, which involve
the d- and u-quark Z-boson  currents, respectively. 
The associated vertex functions read:
\begin{eqnarray}
&&\G_{\mu } ^Z (p,p')\vert _{a}=-i N_c{\l ' }_{Jjk}^\star 
\l ' _{J'jk} \cr  
&\times &\int_Q {  Tr[P_R(\qslash +m_{d_k}) \g_\mu 
(a(d_L)P_L+a(d_R)P_R)(-\Pslash +\qslash +m_{d_k})
P_L(\qslash -\pslash +m_{u_j  }  )]
\over (-Q^2+m_{d_k}^2)(-(Q-p-p')^2+m_{d_k}^2)(-(Q-p)^2+m_{u_j}^2)  }\ , \cr 
&&\G_{\mu } ^Z(p,p')\vert _{b}=-iN_c{\l ' }_{Jjk}^\star \l ' _{J'jk} 
\cr  & \times & 
\int_Q {  Tr[P_R(\qslash  +\pslash +m_{d_k})  P_L (\qslash +\pslash +\pslash '
+m_{u_j} ) \g_\mu (a(u_L)P_L+a(u_R)P_R)(\qslash +m_{u_j})  ]
\over (-Q^2+m_{u_j}^2)(-(Q+p+p')^2+m_{u_j}^2)(-(Q+p)^2+m_{d_k}^2)  }\ . \cr   &&
\label{eq12}
\end{eqnarray}
Applying the formalism  of Passarino-Veltman \cite{i:pasvel}, 
the vertex function from diagram $(a)$   can be expressed in the form:
\begin{eqnarray}
A_L^{JJ'}\vert_a  &=& 
{ {\lambda '}_{Jjk}^\star {\lambda '} _{J'jk}\over 2 (4\pi )^2 } 
N_c \bigg [2 a(d_L)m_d^2(C_0+C_{11}-C_{12})
+a(d_R)\bigg ( B_0^{(2)} +B_0^{(3)}+ 2 P\cdot p (C_{11}-C_{12}) \cr
&+&P^2C_0+2m^2_{\tilde e_J} (-C_{11}+C_{12}) 
-2m_d^2C_0+2 m^2_u(C_{11}-C_{12}) \bigg )\bigg ] \ .
\label{eq13}
\end{eqnarray}
The  conventions of ref.\cite{i:pasvel} are used for the  two-point and three-point 
integral functions, $ B_X \ [X=0,1] $ and $  \ C_X \  [X=0, 11, 12, 21,22,23,24]$. For notational convenience, 
  we have introduced the following abbreviations for the dependence on 
 argument variables:
$B_X^{(1)}= B_X(-p-p', m_d, m_d), \ B_X^{(2)}= B_X(-p , m_d, m_u),
\ B_X ^{(3)}= B_X(-p', m_u , m_d) $ and $C_X(-p,-p',m_d,m_u , m_d). $ 
 The amplitude from diagram $(b)$ can be 
 obtained from that of diagram $(a)$  by performing 
 the following   substitutions:  $ m_{d_k}\to m_{u_j}, \ p\to p', P_L\to P_R, 
 a(d_H) P_H \to a(u_H) P_H, \ [H=L,R]$. 
The self-energy contributions, which are represented 
by the diagrams $(c) $ in Fig. \ref{figs2},
with a single configuration only for the 
 d- and u-quarks  which propagate  inside the loop, 
are most conveniently  calculated through  a consideration of
the scalar  fields renormalization factors $Z_{JJ'}  $.
Starting from the schematic equations for the scalar field
$\phi $  bare Lagrangian density, 
 $L=\phi^\star (p^2-m^2+\Pi(p))\phi , $
where, $\Pi (p)= \Pi_1 p^2 -m^2 \Pi_0+\cdots  $, 
one transfers from bare   to 
renormalized quantities  by applying the substitutions,  
$\phi \to \phi/(1+\Pi_1)^\ud, m^2\to m^2(1+\Pi_1)/(1+\Pi_0)$, such that the renormalization equations for
the fields and mass parameters  read,  $\phi_J= Z_{JJ'} \phi^{ren}_{J'}, \ 
 m^2_{JJ'}=  Z^m_{JK} m^{ren 2}_{KJ'}  $,  with
$Z=(1+\Pi_1)^{-1 },\   Z^m = (1+\Pi_0)(1+\Pi_1)^{-1}$, using a matrix notation 
for the flavor dependence. The 
self-energy contribution in the vertex function becomes then,
\begin{equation} 
A_L^{JJ'}\vert_{SE}= [(Z_{JJ'} Z^\star _{JJ'})^\ud -1] \G_\mu ^Z =
2 N_c 
{ {\lambda '}_{Jjk}^\star {\lambda '} _{J'jk}\over (4\pi )^2 } 
a_L(\tilde e) B_1^{(2)}.
\label{eqself}
\end{equation}
Grouping together  the self-energy and the fermionic triangle   diagram contributions,
such that the total amplitudes read as, 
$A_L^{JJ'} (\tilde e)  =A_L^{JJ'} (\tilde e)_a  + A_L^{JJ'} (\tilde e)_b  $,
yields the final formulas:
\begin{eqnarray}
A_L^{JJ'}(\tilde e)_a&=&{N_c\over 2} 
{ {\lambda '}_{Jjk}^\star {\lambda '} _{J'jk}\over (4\pi )^2 } 
\bigg [2a(d_L)m_d^2 (C_0+C_{11}-C_{12} ) +
a(d_R) \bigg (B_0^{(2)}+B_0^{(3)}+2P\cdot p (C_{11}-C_{12}) \cr 
&+&P^2 C_0
+2m^2_J (-C_{11}+C_{12})-2m^2_dC_0+2m^2_u(C_{11}-C_{12}) \bigg )
+2a(\tilde e_L) B_1^{(2)} \bigg ] ,   \cr
A_L^{JJ'}(\tilde e)_b&=&-{N_c\over 2} 
{ {\lambda '}_{Jjk}^\star {\lambda '} _{J'jk}\over (4\pi )^2 } 
\bigg [2a(u_R)m_u^2 (C_0+C_{11}-C_{12} ) +
a(u_L) \bigg (B_0^{(2)}+B_0^{(3)}+2P\cdot p (C_{11}-C_{12}) \cr 
&+&P^2 C_0
+2m^2_{J} (-C_{11}+C_{12})-2m^2_uC_0+2m^2_d(C_{11}-C_{12})\bigg )
-2a(\tilde e_L) B_1^{(2)}
\bigg ]. 
\label{eq16}
\end{eqnarray}
For notational convenience, we have split the self-energy contribution 
into two equal parts that we absorbed  within the  above two amplitudes, 
distinguished by the suffices $a $ and $  b$. Note that the arguments 
in the  $B$- and $C$-integrals for
the  amplitude $ b$  are deduced from those  of the amplitude $a$   by replacing, 
$d_k \to u_j$. To obtain these results we have used 
the mathematica routine package  ``Tracer" \cite{i:jamin} and, 
for a cross-check, ``FeynCalc" \cite{i:calc}.
 A very useful check concerns the cancellation of the  ultraviolet  divergencies.
We indeed find that  the familiar \cite{i:pasvel} logarithmically  divergent term, $\Delta $, 
enters with the factors, $+a(\tilde e_L) -2a(d_R)$ (amplitude $a$) and 
$a(\tilde e_L) +2a(u_L) $  (amplitude $b$), whose total sum  vanishes 
identically.

The interactions associated with 
the  coupling constants, $\l_{ijk} $, can  also contribute at one-loop order.
Exploiting  the formal similarity between  the $ \l $ and $ \l '$  
interaction terms  in the Lagrangian density,
namely,   $L=-\l' _{ijk} \tilde e_{iL} \bar d_{kR} u_{jL} -  
\l_{ijk} \tilde e_{iL} \bar e_{kR} \nu_{jL} + \cdots $,  dispenses us from 
performing  a new calculation. The results can be derived from those in 
eq.(\ref{eq16}) by substituting   for  the internal lines,
$ d_k \to e_k, \  u_j \to \nu_j $, and for the parameters, 
$a_H(u)\to a_H(\nu ), \ a_H(d)\to a_H(e), \ \ 
 \l ^{'\star } _{Jjk} \l '_{J'jk} \to    \l ^{\star } _{Jjk} \l _{J'jk}  $. 

 Let us now turn to the production of right-chirality sleptons where 
analogous results  can be derived.
The tree level amplitude is related to that in eq.(\ref{eqs1}) 
by a simple  chirality change,
 \begin{eqnarray}
M^{JJ'}_{tree} (\tilde e_R)= -{\l ^\star _{i1J'} \l _{i1J}\over t-m^2_{\nu_{i}}  }
\bar v(k') P_R(\kslash -\pslash )P_L u(k).
\label{eqs17}
\end{eqnarray}
There occurs only one non-vanishing   helicity  flip 
configuration for the initial leptons, namely,  $e^-_L e^+_R$, 
in which the neutrinos  $ t$-channel and  the gauge bosons 
$s$-channel contributions 
interfere. The  amplitude is given by a  formula similar to eq.(\ref{eq15}), except
for  the substitution in the second term,
$ a_R(e) A_L^{JJ'} (\tilde e, s+i\e ) \to 
 a_L(e) A_R^{JJ'} (\tilde e, s+i\e )$. Concerning  the  one-loop contribution 
to  the vertex function  $A_R^{JJ'}(\tilde f)$, 
we find that the RPV interactions with the  coupling constants $\l_{ijk} $ 
can only  contribute, while those with $\l '_{ijk}$ vanish identically.
Diagram $(a)$ in Fig. \ref{figs2} refers to an $e_j $ current 
and diagram $(b)$ 
to a $\nu_i^c$ current.
The results can be derived by  inspection   from  eq.(\ref{eq16}) by substituting,  $ \l '_{J'jk} \l ^{'\star } _{Jjk} \to 
\l_{ijJ} \l^\star _{ijJ'}$,  
 $ d_{jR} \to e_{jL}, u_{jL} \to \nu_{iR}^c, \ \tilde e_L \to 
\tilde e_R $ and, 
accordingly,  $a(d_H)\to a(e_H), \ 
a(u_H)\to a(\nu^c_H), \ [H=L,R], \ a(\tilde e_L)  \to a(\tilde e_R)  $. 
For definiteness, we quote the explicit formulas:
\begin{eqnarray}
A_R^{JJ'}(\tilde e)_a&=&{N_c\over 2} 
{ {\lambda }_{ijJ} {\lambda }_{ijJ'}^\star \over (4\pi )^2 }
\bigg [2a(e_R)m_e^2 (C_0+C_{11}-C_{12} ) +
a(e_L) \bigg (B_0^{(2)}+B_0^{(3)}+2P\cdot p (C_{11}-C_{12}) \cr 
&+&P^2 C_0
+2m^2_J (-C_{11}+C_{12})-2m^2_eC_0+2m^2_{\nu }(C_{11}-C_{12}) \bigg )
+2a(\tilde e_R) B_1^{(2)} \bigg ],  \cr
A_R^{JJ'}(\tilde e)_b&=&-{N_c\over 2} 
{ {\lambda }_{ijJ} {\lambda }_{ijJ'}^\star \over (4\pi )^2 } 
\bigg [2a(\nu_L^c)m_{\nu }^2 (C_0+C_{11}-C_{12} ) +
a(\nu_R^c) \bigg (B_0^{(2)}+B_0^{(3)}+2P\cdot p (C_{11}-C_{12}) \cr 
&+&P^2 C_0
+2m^2_{J} (-C_{11}+C_{12})-2m^2_{\nu }C_0+2m^2_e(C_{11}-C_{12})\bigg )
-2a(\tilde e_R) B_1^{(2)}
\bigg ]. 
\label{eq18}
\end{eqnarray}
The discussion of the mixed chiralities cases, $\tilde e_{LJ}^-\tilde e_{RJ'}^+, \ 
\tilde e_{RJ}^-\tilde e_{LJ'}^+, \ [J\ne J']$ turns out to be quite brief.  
The tree level RPV contributions, which  can only come from the $\l_{ijk}$ interactions,
vanish identically for massless neutrinos. As for the one-loop
contributions to the vertex, $Z\tilde f_L\tilde f_R^\star $, this also vanishes up to mass 
terms in the internal fermions. Since flavor non-diagonal rates arise then from 
loop contributions only and CP asymmetries from  interference of distinct  loop
contributions, one concludes that both observables should be very small. 

Finally, let us add here a general comment  concerning 
the photon vertex functions, $ A_{L,R}^{\g JJ'} $, which  are  given by 
formulas  similar to those in eqs.(\ref{eq16}) or (\ref{eq18}) with the 
appropriate replacements, $a_{L,R}(f) \to  2 Q(f)$. Therefore, to incorporate
the $\g $-exchange contributions in the total
amplitudes (eq.(\ref{eq15}) and related equations)
one needs to substitute, 
$$ a_{R,L}(e) A^{JJ'}_{L,R} \to 
a_{R,L}(e) \sum_f a(f) C_f+ 2Q(e)  \sin^2 \t_W \cos^2 \t_W  
[(s-m_Z^2+im_Z \G_Z) / s] \sum_f 2Q(f) C_f, $$
where we have used the schematic representation, $A^{JJ'}_{L,R}=\sum_f a(f) C_f$.
\begin{figure}
\begin{center}
\leavevmode
\psfig{figure=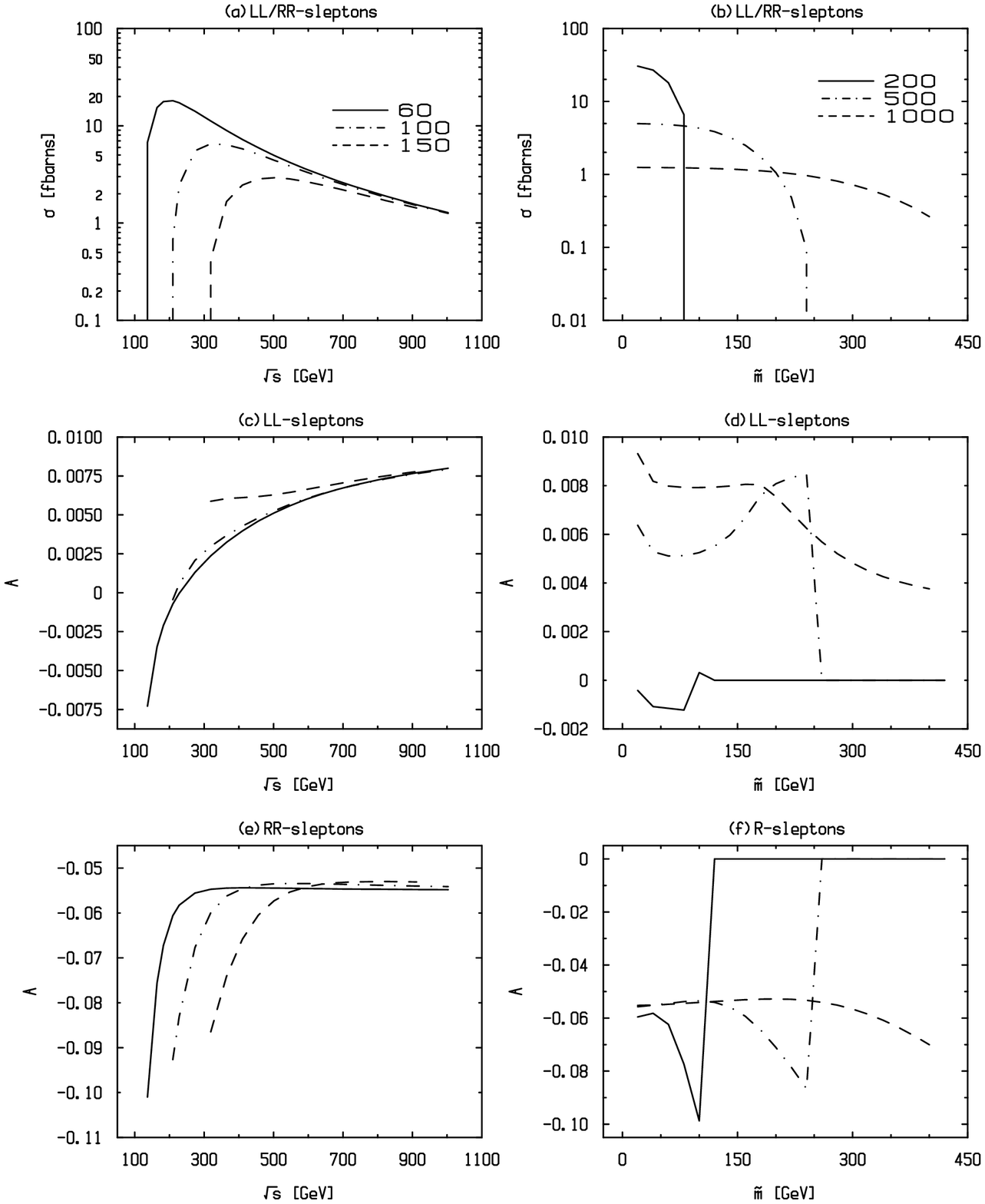,width=14.5cm}
\end{center}
\caption{Integrated flavor non-diagonal cross sections and CP asymmetries  
in the production of  slepton-antislepton pairs  of left-chirality (L) 
(interactions $\l '_{ijk}$ only) and of  right-chirality  (R) 
(interactions $\l _{ijk}$). The  three windows on
 the left-hand side ($(a), \ (c), \ (e) $)  show the variation with center of mass
energy, $s^{1/2}$,  for three choices of the  scalar superpartners mass parameter,  $\tilde m $: 
$ 60 GeV$ (continuous lines), $ 100 GeV$ (dashed-dotted  lines),  
$  150 GeV$ (dashed lines).  The  three windows on the right-hand side  ($(b),\ (d),\ (f) $) 
show the variation with  scalar superpartner mass, $\tilde m$, 
for three  choices  of the center of mass energy
$  s^{1/2} = 200 GeV$ (continuous lines), $ 500 GeV$ (dashed-dotted  lines),
$  1000  GeV$ (dashed lines).  The tree level amplitude includes the  t-channel 
exchange contribution. The one-loop  amplitudes (with both  photon and Z-boson exchanges)  correspond to Case {\bf IV} which 
includes the contributions from all three internal fermions  generations.} 
\label{figg6}
\end{figure}
\section{Results and discussion}
\label{secs4}

\setcounter{equation}{0}

Let us first comment briefly on the experimental  observability 
of flavor non-diagonal sleptons pair production.
One  convenient non degraded signal here  is that which corresponds
to lepton pair final states, $e_J ^-e^+_{J'} $,
which are produced through the  two-body decay channels for sleptons,
$\tilde e_{[J, J']} ^\pm 
\to e^\pm _{[J , J']} +\tilde \chi_1^0$.  Of course,  in the broken 
R parity case, the  produced lightest neutralinos  are unstable  
and could conceivably be  reconstructed 
through  their dominant  decay channels which involve
two leptons, or two jets, together with missing energy. 
We shall not elaborate further on this issue, except to note 
that the efficiency factors at NLC energies  for the flavor diagonal rates, assuming 
a stable $\tilde \chi_1^0$, and including rough detection cuts,
such that the physical rates  for the fermion pairs channels is, $\s _{JJ'} \e $,
are typically set at $\e \approx 30 $ \% \cite{i:arkani1}.

Proceeding to the predictions, we observe that
the  main source of uncertainties concerns the RPV coupling constants.
The sfermion mass eigenvalues are not known, but these
parameters appear explicitly  through the kinematics. 
We shall neglect mass splittings  and mixings  between L- and R-sleptons.
A  unique sleptons mass parameter, $\tilde m$,  will be used  and
varied in  the interval, 
$60 < \tilde m <  400$ GeV.   Regarding the RPV coupling constants,
it is useful  here to catalog the family configurations and intermediate states 
entering the calculations. 
Examining the structure of  the  flavor non-diagonal 
 tree amplitudes, we note  that these involve a onefold 
summation  over leptons  families weighted by the factors,
$ t^i_{JJ'}=  \l _{iJ1}^\star \l_{iJ'1}  $, for L-sleptons and  
$ t^i_{JJ'}=  \l _{i1J}^\star \l_{i1J'}  $, for R-sleptons. 
The loop amplitudes involve a twofold summation  over leptons families of form, 
$\sum_{jk} l^{jk}_{JJ'}  F^{jk} (m_j, m_k, s+i\e ) , \ $ where  $l^{jk}_{JJ'}$ 
depend quadratically on the RPV coupling constants while
the  loop integrals, $F^{jk} $, have a  non-trivial
dependence on the fermions    masses, as exhibited 
on the formulas  derived in Section \ref{secs3} 
[see, e.g., eq. (\ref{eq16})].  The relevant coupling constants, the 
species and family configurations  for the internal fermions are for L-sleptons,
$ l^{jk}_{JJ'}=  \l ^{'\star }_{Jjk}
\l ' _{J'jk} , 
[d_k, \ u_j]; \ \    
 l^{jk}_{JJ'}=  \l ^{\star }_{Jjk}
\l _{J'jk} , 
[e_k, \ \nu _j]$; 
and for R-sleptons,
$ l^{ij}_{JJ'}=  \l _{ijJ}
\l ^{\star } _{ijJ'} ,\  [e_j,\  
\nu ^c_{i}  ] $.
The dependence of rates on the RPV coupling constants 
has the schematic structure,  $ \s _{JJ'}\simeq \sum_i\vert t^i_{JJ'}\vert^2$, 
and that of CP asymmetries, 
$ {\cal A}_{JJ'} \simeq  \sum_{ij k} Im(l^{jk}_{JJ'} 
t^{i\star }_{JJ'})/\sum_l\vert t^l_{JJ'} \vert ^2$ for L-sleptons and
$ {\cal A}_{JJ'} \simeq  \sum_{ij k} Im(l^{ij}_{JJ'} 
t^{k\star }_{JJ'})/\sum_l\vert t^l_{JJ'} \vert ^2$ for R-sleptons. 
Therefore,  rates (asymmetries) are controlled by two (four) RPV
coupling constants in different family configurations. 
Note the  expected invariance of asymmetries under phase redefinitions of the fields.

While the dependence on the mass of the exchanged neutrino family index in 
$t^i_{JJ'}$  can be clearly  ignored, that on the pair  of indices $(i,j)$ in $l^{ij}_{JJ'}$, which involves  the ratios of the masses
of the appropriate internal  fermions, $m_{i,j} $, to the external scale associated with the
center of mass energy, $\sqrt {s}$, can be ignored as long as, 
$\sqrt {s} >> m_{i,j}$. Therefore,  at the energies of interest, the only relevant
fermion mass parameter is  that of the  top-quark. 
Instead of listing the various distinct family configurations 
for the quadratic (tree)  or quartic (loop) 
products of the RPV coupling constants, we shall consider a 
 set of specific assumptions concerning the family dependence.
First, for  the cases involving $[e_j, \nu_i^c]$ or $[e_k, \nu_j]$ internal states, neglecting neutrino masses, 
we  need only account for the masses of charged leptons. 
For the case  with $[d_k, \ u_j]$ internal states, 
we  restrict consideration to  the diagonal family configuration,
namely, $k= j $. Second, we  include a CP odd phase, $\psi $,
between $t^i_{JJ'}$ and  all of the $l^{jk}_{JJ'} $ or $l^{ij}_{JJ'}$, as the case may be. Finally, we consider the  following four 
 discrete  choices for the variation intervals on which 
run the internal fermion indices indices,
$j = k$ or $i, \ j$. 
Case {\bf I}:  $\{ 1 \}$;  Case {\bf II} $\{ 2 \}$;
 Case  {\bf III} $\{ 3 \}$;  Case  {\bf IV} $\{ 1, 2,3\}$. 
In all these four  cases, we set the relevant coupling constants 
at the reference values, $l^{ij}_{JJ'}  = l^{jk}_{JJ'}
=10^{-2},\  t^i_{JJ'}=10^{-2}$ and use a maximal CP odd phase,
 $ arg( l_{JJ'} ^{\star [ij, jk] } t^l_{JJ'} )  \equiv  \psi =\pi/2$. 
Because of the proportionality of asymmetries to  the  imaginary part of the phase factor, the  requisite dependence may be simply
reinstated by inserting a factor, $\sin \psi $. 
\begin{table}[t]
\begin{center} 
\begin{tabular}{cccccc}
\hline 
    & &   $s^{1/2} = 200 $ GeV &    $ s^{1/2} =500 $ GeV &    &   \\
\hline
 & &    $\tilde m =60 $ &      $\tilde m =60$ & 
  $\tilde m =100 $ &  $\tilde m =200$    \\
\hline
${\bf \tilde e_L \ \tilde e_L }$ \\
\\
$\lambda ' \lambda '^\star $  \\
    {\bf  I} &$ Z$ & $ -2.1 d-5     $& $-3.3 d-6    $& $-2.6 d-6     $& $-2.4d-6    $\\
 & $\g +Z$    & $ -7.7 d-5     $& $-1.39 d-5    $& $-1.09 d-5     $& $-1.03d-5    $\\
   \\
 {\bf  III}& $Z$ & $ +2.6 d-4      $& $-1.6 d-3    $& $ -1.8 d-3    $& $-2.3d-3    $\\
&   $\g +Z$ & $ -1.01  d-3      $& $+5.1 d-3    $& $ +5.3 d-3    $& $+8.1d-3    $\\
   \\
   \\
$\lambda  \lambda ^\star $  \\
    {\bf  I} &$Z$ & $ -2.1 d-5      $& $ -3.3 d-6    $& $-2.6d-6     $& $-2.4d-6    $\\
 &   $\g +Z$ & $ -7.69 d-5      $& $ -1.39 d-5    $& $-1.09d-5     $& $-1.03d-5    $\\
   \\
    {\bf  III} &$Z$ & $-2.4 d-5      $& $-5.5 d-6    $& $-3.4 d-6     $& $ -2.7 d-6   $\\
 &   $\g +Z $ & $-6.39 d-5      $& $+2.59 d-6    $& $-5.06 d-6     $& $ -8.32d-6   $\\
   \\
   \\
\hline
${\bf \tilde e_R\   \tilde e_R}$ \\
\\
$\lambda  \lambda ^\star $  \\
    {\bf  I} &$Z$ & $ -7.2 d-3      $& $-5.5 d-3    $& $-5.4 d-3     $& $ -7.2 d-3    $\\
&  $\g +Z$ & $ -2.1 d-2      $& $-1.83 d-2    $& $-1.80 d-2     $& $ -2.40 d-2    $\\
\\
\hline 
\end{tabular}
\label{i:table1}
\end{center}
\caption{CP asymmetries, ${\cal A}_{JJ'}$, 
in sleptons pair production at two values of the 
center of mass energy, $s^{1/2}= 200, \ 500 $ GeV and for 
values of the sleptons mass parameter, $\tilde m = 60, \ 100, \ 200 $ GeV,
appearing in the column fields.  For each case, the first line  ($Z$) 
is associated with the  gauge Z-boson exchange contribution and the second
line ($\g +Z$) with both photon and Z-boson exchanges added in together.
The contributions  to left-chirality ($\tilde e_L\tilde e_L$) 
and right-chirality ($\tilde e_R\tilde e_R$) sleptons,  
induced by the $\l '_{ijk}$ and $\l _{ijk}$ interactions,  are distinguished 
by the labels, $\l ' \l ^{'\star } , \ \l  \l ^\star  $, respectively. 
Cases {\bf I , \ III } correspond to internal fermions belonging to 
the first  and  third families, respectively. 
The notation $ n d-x $ stands for $ n \  10^{-x}$. }
\end{table}
To illustrate the dependence of asymmetries on the internal fermions families
and on the $\l '$ or $\l $  interaction  types, 
we display in Table \ref{i:table1}   a set of representative 
results obtained for selected subsets of Cases {\bf I, \ II , \ III, \ IV}. 
The reason is that the results for Cases {\bf I , \ II} (light families)
are identical in all cases, while those for Case {\bf III} (heavy 
families) differ only  for cases  involving up-quarks. 
As one sees on Table \ref{i:table1}, the  interference between photon and Z-boson  exchange 
contributions  has a significant  effect on the results.  
The  strongly reduced values for the L-sleptons  asymmetries found 
in Cases {\bf I} for the $\l ' \l'^\star $ interactions and in  
all Cases  for the $\l  \l^\star $ interactions, arise from 
the existence of a strong cancellation between  the amplitudes termed $(a)$
and $(b)$ for  nearly  massless  internal  quarks or leptons. 
Case {\bf III} with the $\l ' \l'^\star $ interactions 
is relatively enhanced thanks to the top-quark contribution (configuration 
$\bar t \ b$).  That the above  cancellation is  not 
generic  to the RPV contributions 
is verified on the results  for R-sleptons production, 
 where all three  families of leptons give nearly 
equal, unsuppressed  contributions to loop amplitudes.

In the currently favored  situation  where the RPV
 coupling constants  are assumed to exhibit a strong 
hierarchical structure, the peculiar rational 
dependence of CP asymmetries on ratios of
quartic products of the coupling constants, might  lead to strong enhancement
factors.  We recall the schematical structure of this dependence,
${\cal A}_{JJ'}\propto  [ \sum_{ijk}  Im (\l ^{'\star } _{Jjk} \l '_{J'jk} \l '_{iJ1} \l ^{'\star } _{iJ'1} ) 
/ \sum _l \vert  \l _{lJ1} \l ^\star _{lJ'1}\vert ^2   ] $, 
and note that the coupling constants involving  third family  indices are 
amongst those that are the least  strongly constrained. Therefore,  assuming 
that the coupling constants take the values given by the 
current bounds from low energy constraints \cite{i:reviews}, 
one  would  obtain,  
$${\cal A}_{J=3,J'=2} 
\simeq [ Im (\l ^{' \star }_{333} \l '_{323} \l '_{331} \l ^{'\star }_{321})
/ \vert \l _{131} \l _{121} ^{'\star }  \vert ^2 ]\approx 90 \sin \psi  .$$

The  dependence  of rates and asymmetries on center of mass energy and
sleptons masses are displayed  for Case {\bf IV} in 
Figure \ref{figg6}. Regarding the variation with energy (figure (a)), 
after a rapid rise at  threshold (with the expected $\b ^3  \ p$-wave like
behavior)  the rates settle, roughly as $\tilde m^2/s$, 
to constant values with growing energy,   and
vary  inside the range, $ ({\l ^\star \l \over 0.01} )^2 \ 20 \ - \
2$ fbarns,  as one sweeps through the interval,   $\tilde m \in [60, 400]$ GeV.
The variation with $\tilde m$ (figure (b))  is rather smooth.
For the envisaged integrated luminosities,
 $ {\cal L} \simeq 50 \ - \ 100  
fbarns ^{-1}  /yr $,  these results indicate that reasonably sized samples of 
order $100 $ events could be collected at NLC.  
Noting that the dependence of rates on energy
 rapidly saturates for $\sqrt s > \tilde m$, we conclude that the
relevant  bounds that could be inferred on quadratic  products
of different  the RPV coupling constants, should, for increasing sleptons 
masses, become competitive   with those deduced from 
low energy constraints, which scale typically as,
$[\l   \l  , \ \l '  \l ' ] <  0.1 (100 GeV /\tilde m)^2$. 
The results in Fig.\ref{figg6} (c,d,e,f) for the  CP  asymmetries,
${\cal A}_{JJ'}$,   indicate the existence of a wide, nearly 
one order of magnitude, gap between L-sleptons with 
$\l ^{' \star } \l '$ interactions 
 and R-sleptons with  $\l ^{ \star } \l $ interactions,  with values that lie 
at a few times $10^{-3}$  and $10^{-2}$, respectively.

In our  prescription of using  equal numerical  values for the RPV coupling constants ($t^i _{JJ'} $ and $l^{ij}_{JJ'}$)  which
control tree and loop contributions,
the asymmetries are independent of the specific reference values chosen.
In the event that the rates would be dominated by some alternative mechanism,
say, lepton flavor  ocillations, whereas RPV effects would 
remain significant in
asymmetries, these would then  scale as, $Im(t^{i\star}_{JJ'} l^{jk}_{JJ'})$.  
It is instructive in view of such a possibility  to compare with 
predictions  found in  the flavors  oscillation approach.
Scanning over wide intervals of values for  the  relevant  parameters,
$[\cos 2\t_R , x=\D \tilde m/\G ] $, 
associated with the common values for all three  
mixing angles  and ratios  of  families mass  differences  to the total
sleptons decay widths, respectively,  the authors of \cite{i:arkani1} found 
flavor non-diagonal rates  which ranged between 
$ 250  $ and $0.1 $ fbarns for $ \sqrt s =190 $ GeV and 
$ 100 $ and $ 0.01 $ fbarns for $\sqrt s =500 $ GeV.
Our predictions,  $ \s _{JJ'} \simeq ({\l \over 0.1 })^4 \  2 \ - 20 $ fbarns, 
which hold  approximately for energies, $\sqrt  s > \tilde m$, 
lie roughly in between these extreme values. 
On the other hand, the authors of \cite{i:arkani2}
found CP  asymmetry rates, $S_{JJ'} =
\s_{JJ'}-\s _{J'J} \approx  3-16 $ fbarns.  
For comparison,  our predicted asymmetry rates for the same quantity, 
namely,  $S_{JJ'}= 2\s_{JJ'} {\cal A}_{JJ'}
\approx 10^{0} - 10^{-1}$ fbarns, lie around one order of magnitude 
below these values.
It should be said, however,  that  the flavor oscillation contributions
could have a stronger model dependence  than  the 
variation range exhibited by the above predictions, and that 
these  predictions were 
obtained subject to assumptions that tend to maximize CP violation effects.
The existing constraints, \cite{i:fcnc} which are mostly derived from low energy
phenomenology, constrain only a small subset of the parameters
describing the scalar superpartners  mass spectra and generational mixings.

To summarize, we have shown that moderately small contributions to 
flavor non-diagonal  rates  and CP violating  spin-independent asymmetries 
in sleptons pair production  could arise from the RPV interactions. 
These contributions  seem to be of smaller 
size than those  currently associated with flavor oscillations, 
although  the model dependence of predictions  
in the flavor oscillation approach is  far from being under control.
An experimental observation of the  non-diagonal  slepton production rates 
would give information on  quadratic products of
different coupling constants, $\l \l^\star $. Owing to the smooth dependence 
of rates on the slepton masses, already for  masses, $\tilde m > 100 $ GeV,  
it should be possible  here to deduce  stronger bounds than  the current ones inferred from 
low energy constraints. The  observation of CP  violating 
asymmetries  requires  the presence of non vanishing CP odd phases in 
quartic products of the coupling constants,
$ Im (\l ^{\star } _{Jjk} \l _{J'jk} \l _{iJ1} \l ^{\star } _{iJ'1} ) $, 
(and similarly with $\l \to \l '$)
which remain largely unconstrained so far.  The peculiar rational dependence, 
$Im (\l \l ^\star  \l \l  ^\star )/\l ^4$, leaves room for possible
strong enhancement factors.

\clearpage

\vspace{10 mm}
\begin{center}
{  }
\end{center}
\vspace{10 mm}

\clearpage

\clearpage

\vspace{10 mm}
\begin{center}
{  }
\end{center}
\vspace{10 mm}

\clearpage 


\newpage

\addtocontents{toc}{\protect \vspace{1cm} \protect \begin{center} 
\protect {\bf Annexes} \protect \end{center} }

\appendix

\chapter{Dimensions}
\label{dims}

\renewcommand{\thesubsection}{A.\arabic{subsection}}
\renewcommand{\theequation}{A.\arabic{equation}}
\setcounter{subsection}{0}
\setcounter{equation}{0}

Notons $M$, $L$ et $T$ les dimensions de masse, de longueur et de temps.
L'action ${\cal S}$ a la dimension de la constante de Planck r\'eduite $\bar
h=h/2 \pi$,
\`a savoir une dimension d'\'energie ($M L^2 T^{-2}$) fois une dimension de
temps:
\begin{eqnarray}
[{\cal S}]=[\bar h]=M L^2 T^{-1}.
\label{dimeq1}
\end{eqnarray}
Le lagrangien ${\cal L}$ \'etant reli\'e \`a l'action par,
\begin{eqnarray}
{\cal S}=\int {\cal L} d^4 x,
\label{dimeq2}
\end{eqnarray}
il a la dimension suivante,
\begin{eqnarray}
[{\cal L}]=M L^{-1} T^{-2},
\label{dimeq3}
\end{eqnarray}
puisque,
\begin{eqnarray}
[d^4 x]=L^3 T.
\label{dimeq4}
\end{eqnarray}

Tout au long de cette th\`ese, nous adoptons les 2 hypoth\`eses
simplificatrices $c=1$,
$c$ \'etant la c\'el\'erit\'e de la lumi\`ere,
et $\bar h=1$ qui ont respectivement pour cons\'equence que $L=T$ et
$M=L^{-2} T$, d'apr\`es
Eq.(\ref{dimeq1}). Dans le cadre de ces 2 hypoth\`eses, nous avons donc,
\begin{eqnarray}
M = L^{-1} =T^{-1}.
\label{dimeq5}
\end{eqnarray}
Prenant l'unit\'e de masse comme unit\'e de r\'ef\'erence,
les dimensions de l'action, de $d^4 x$ et du lagrangien sont, selon
Eq.(\ref{dimeq5}),
\begin{eqnarray}
[{\cal S}]=0, \ [d^4 x]=-4 \ et \ [{\cal L}]=4.
\label{dimeq6}
\end{eqnarray}
Les termes de masse d'un champ de spin 1/2 $\psi$ (\`a 4 composantes) et
d'un champ scalaire $z$
s'\'ecrivant ${\cal L}=m \psi \bar \psi$ et ${\cal L}=m^2 z^{\star} z$,
nous d\'eduisons de Eq.(\ref{dimeq6}) les dimensions suivantes,
\begin{eqnarray}
[\psi] = {3 \over 2}, \ [z] = 1 \  et \  [\partial_{\mu}]=[{\partial  \over
\partial x^{\mu}}] = 1.
\label{dimeq7}
\end{eqnarray}

\clearpage

\vspace{10 mm}
\begin{center}
{  }
\end{center}
\vspace{10 mm}

\clearpage

\clearpage 

\begin{titlepage}

\vspace{10 mm}
\begin{center}
{  }
\end{center}
\vspace{30 mm}

\end{titlepage}

\clearpage

\begin{titlepage}

\vspace{10 mm}
\begin{center}
{  }
\end{center}
\vspace{30 mm}

\begin{center}
\underline{R\'esum\'e:}
\end{center}

\vspace{10 mm}

L'extension supersym\'etrique du Mod\`ele Standard peut contenir des 
interactions violant la sym\'etrie dite de R-parit\'e. La pr\'esence de tels
couplages engendrerait une violation des nombres leptonique et/ou baryonique
et modifierait en profondeur la ph\'enom\'enologie de la supersym\'etrie aupr\`es
des futurs collisionneurs de particules. Nous avons d\'evelopp\'e des tests des
interactions violant la R-parit\'e. D'une part, nous avons \'etudi\'e des signaux
clairs de la production d'un seul partenaire supersym\'etrique de particule du
Mod\`ele Standard, qui implique les couplages violant la R-parit\'e, dans le
cadre de la physique aux prochains collisionneurs leptoniques et
hadroniques. Les r\'esultats montrent que de fortes sensibilit\'es pourront \^etre
obtenues sur les param\`etres de brisure douce de la supersym\'etrie et
indiquent la possibilit\'e d'une am\'elioration d'un \`a deux ordres de grandeur
des limites indirectes actuelles sur les valeurs de plusieurs constantes de
couplage violant la R-parit\'e. De plus, il a \'et\'e v\'erifi\'e que l'analyse de la
production d'un seul superpartenaire offre l'opportunit\'e de reconstruire de
fa\c{c}on ind\'ependante du mod\`ele th\'eorique diverse masses de superpartenaires
avec une grande pr\'ecision. D'autre part, nous avons \'etudi\'e des effets de
violation de la sym\'etrie CP aux futurs collisionneurs leptoniques dans la
production de paire de fermions de saveurs diff\'erentes, ou de leur
superpartenaire, permettant de mettre en \'evidence d'\'eventuelles phases
complexes des constantes de couplage violant la R-parit\'e. Nous avons vu
notamment que la production d'un quark top accompagn\'e d'un quark charm\'e
permet de tester la violation  de CP dans le secteur hadronique. La
conclusion de ces travaux est que les phases complexes de certaines
constantes de couplage violant la R-parit\'e pourraient \^etre observ\'ees, et
plus particuli\`erement dans un sc\'enario o\`u ces couplages exhiberaient une
grande hi\'erarchie dans l'espace des saveurs.


\end{titlepage}

\end{document}